\documentclass[]{aa}  
\usepackage{graphicx}
\usepackage{txfonts}
\usepackage{fixltx2e}
\usepackage{lscape}
\begin{document} 
   \title{Star formation in $z>1$ 3CR host galaxies as seen 
	      by \textit{Herschel}\thanks{{\it Herschel} is an ESA space observatory 
	      with science instruments provided by European-led Principal 
	      Investigator consortia and with important participation from NASA.}
          }
   \author{P.~Podigachoski\inst{\ref{inst1}}     
      \and P.~D.~Barthel\inst{\ref{inst1}}
      \and M.~Haas\inst{\ref{inst2}}
      \and C.~Leipski\inst{\ref{inst3}}
      \and B.~Wilkes\inst{\ref{inst4}}
      \and J.~Kuraszkiewicz\inst{\ref{inst4}}
      \and C.~Westhues\inst{\ref{inst2}}
      \and S.~P.~Willner\inst{\ref{inst4}}
      \and M.~L.~N.~Ashby\inst{\ref{inst4}}
      \and R.~Chini\inst{\ref{inst2}} 
      \and D.~L.~Clements\inst{\ref{inst5}}
      \and G.~G.~Fazio\inst{\ref{inst4}}
      \and A.~Labiano\inst{\ref{inst6}}
      \and C.~Lawrence\inst{\ref{inst7}}
      \and K.~Meisenheimer\inst{\ref{inst3}}
      \and R.~F.~Peletier\inst{\ref{inst1}}
      \and R.~Siebenmorgen\inst{\ref{inst8}}
      \and G.~Verdoes~Kleijn\inst{\ref{inst1}}
          }

   \institute{Kapteyn Astronomical Institute, University of Groningen, 9747 AD Groningen, The Netherlands\\ \email{podigachoski@astro.rug.nl}\label{inst1}
         \and Astronomisches Institut, Ruhr Universit\"{a}t, D-44801 Bochum, Germany\label{inst2}
         \and Max-Planck Institut f\"{u}r Astronomie (MPIA), D-69117 Heidelberg, Germany\label{inst3}
         \and Harvard-Smithsonian Center for Astrophysics, Cambridge, MA 02138, USA\label{inst4}
         \and Astrophysics Group, Imperial College London, Blackett Laboratory, Prince Consort Road, London SW7 2AZ, UK\label{inst5}
         \and Institute for Astronomy, Department of Physics, ETH Zurich, Wolfgang-Pauli-Strasse 27, CH-8093 Zurich, Switzerland\label{inst6}
         \and Jet Propulsion Laboratory, California Institute of Technology, 4800 Oak Grove Drive, Pasadena CA 91109, USA\label{inst7}
         \and European Southern Observatory, Karl-Schwarzschild-Str. 2, 85748 Garching b. München, Germany\label{inst8}
         }
   \date{Received ; accepted}
   
   \abstract{We present \textit{Herschel} (PACS and SPIRE) far-infrared 
             (FIR) photometry of a complete sample of $z>1$ 3CR sources, 
             from the \textit{Herschel} GT project \textit{The Herschel 
             Legacy of distant radio-loud AGN} (PI: Barthel). Combining 
             these with existing \textit{Spitzer} photometric data, we 
             perform an infrared (IR) spectral energy distribution (SED) 
             analysis of these landmark objects in extragalactic research 
             to study the star formation in the hosts of some of the brightest 
             active galactic nuclei (AGN) known at any epoch. Accounting 
             for the contribution from an AGN-powered warm dust component 
             to the IR SED, about 40\% of our objects undergo episodes of 
             prodigious, ULIRG-strength star formation, with rates of hundreds 
             of solar masses per year, coeval with the growth of the central 
             supermassive black hole. Median SEDs imply that the quasar 
             and radio galaxy hosts have similar FIR properties, in 
             agreement with the orientation-based unification for 
             radio-loud AGN. The star-forming properties of the AGN hosts 
             are similar to those of the general population of equally 
             massive non-AGN galaxies at comparable redshifts, thus there 
             is no strong evidence of universal quenching of star formation 
             (negative feedback) within this sample. Massive galaxies at 
             high redshift may be forming stars prodigiously, regardless of
             whether their supermassive black holes are accreting or not. 
             }

   \maketitle
\section{Introduction}
   The understanding that most (if not all) galaxies in the Universe 
   host a supermassive black hole (SMBH) is among the most important 
   findings of modern astronomy. The growth of a SMBH through mass 
   accretion generates large amounts of energy during a phase in the 
   evolution of the galaxy known as an active galactic nucleus (AGN) 
   phase. Although there is a difference of a factor of $\sim$~$10^9$ 
   in their physical size scales, the SMBHs and their host galaxies 
   exhibit strong scaling relations \citep[e.g.][]{Magorrian98,Tremaine02,Gueltekin09}, 
   suggesting a link between the growth of the SMBHs and that of their 
   host galaxies. Moreover, both these processes are thought to peak at 
   redshifts $z\sim2$ \citep[e.g.][]{Hopkins&Beacom06,Alexander08}.  
   The symbiosis of black hole and global galaxy growth is intriguing 
   because of the possible feedback effects: positive (AGN inducing 
   star formation) and/or negative (AGN quenching of star formation). 
   These feedback processes are of paramount importance for our 
   understanding of galaxy formation \citep[e.g.][]{Croton06,Hopkins08}. 
   However, neither the feedback mechanisms nor the overall impact of 
   feedback on the host galaxies is known. Other big unknowns are the 
   duration and frequency of AGN accretion and host galaxy star formation 
   phases. 

   High redshift radio-loud AGN 
   (P$_{\mathrm{1.4~GHz}}$ > $10^{25}$ W~Hz$^{-1}$ and $z > 1$)
   provide a unique opportunity to probe the interplay between the growth 
   of the black hole and the hosting stellar bulge. They are invariably 
   associated with massive galaxies having 
   M$_{\mathrm{stellar}}$ $\gtrsim$ 10$^{11}$ M$_{\odot}$ 
   \citep{Best98,Seymour07,DeBreuck10}, and have edge-brightened, 
   double-lobed, FRII morphologies \citep{Fanaroff&Riley74} that permit 
   estimates of the duration of the episode of strong AGN activity. In 
   addition to being used in studies of massive galaxy evolution, 
   radio-loud AGN are being used extensively in unification studies, 
   where, distant radio-loud galaxies and quasars are believed to make up 
   one and the same population \citep{Barthel89,Antonucci93,Urry&Padovani95}, 
   hence have equally massive hosts. Ultraviolet or visible opaque circumnuclear 
   dust is an essential element of this scenario; distant 3CR quasars and 
   radio galaxies are indeed luminous mid-infrared (MIR) emitters 
   \citep{Siebenmorgen04,Haas08,Leipski10}.   
  
   It has long been suspected that hosts of powerful high-redshift radio-loud 
   AGN undergo episodes of vigorous (dust obscured) star formation 
   \citep[e.g.][]{Archibald01}. Huge reservoirs of molecular gas, have 
   been deduced in several objects from submillimetre (submm) studies 
   \citep[e.g.][]{Reuland04}.
   Such studies were mainly based on one submm flux measurement and were limited to the 
   highest redshift objects for which the obscured newborn star radiation re-emitted 
   by the ubiquitous cold (30-50 K) dust is redshifted to submm wavelengths. 
   However, quantification of the cold dust emission (e.g. constraining the cold 
   dust temperature) requires sampling the full rest-frame infrared-submm spectral 
   energy distribution (SED) of the studied objects.
   Earlier far-infrared (FIR) studies failed to provide strong constraints 
   on the FIR properties for relatively large samples of powerful radio-loud AGN 
   because of their small detection fractions and only limited rest-frame 
   FIR wavelength coverage \citep{Heckman92,Hes95,Meisenheimer01,Siebenmorgen04,Haas04,Cleary07}.

   The \textit{Herschel Space Observatory} \citep{Pilbratt10}, with its 
   unprecedented FIR sensitivity and wavelength coverage, explored terra incognita 
   (caelum incognitum...) allowing studies which have revolutionized the 
   understanding of the connection between AGN and star formation activity. 
   Several studies utilizing deep X-ray and \textit{Herschel} data revealed 
   that the hosts of moderately luminous radio-quiet AGN out to $z \sim 3$ form stars at 
   rates comparable to the general non-AGN population 
   \citep{Shao10,Mullaney12,Rosario12}. 
   For high (radio-quiet) AGN luminosities (L$_{\mathrm{X}} > 10^{44}$ erg s$^{-1}$), 
   \citet{Page12} reported suppression of star formation, consistent 
   with the expectations from theoretical models, while \citet{Harrison12} 
   found no clear evidence of suppression of star formation by extending 
   the analyses to samples larger by an order of magnitude. Moreover, at the 
   highest AGN luminosities (in excess of 10$^{46}$ erg s$^{-1}$), recent 
   studies based on decomposition of the IR emission to AGN 
   and star formation contributions, have shown star formation rates (SFRs) 
   of the order of several hundred solar masses per year in the hosts of 
   some of the most powerful radio galaxies 
   \citetext{\citealp{Barthel12}, - B12 hereafter; \citealp{Seymour12}; \citealp{Drouart14}} 
   and (radio-quiet) quasars \citep{Leipski13,Leipski14}.
      
   In order to quantify the energetics of AGN at the peak of their 
   activity as well as their star formation characteristics, we obtained 
   five-band \textit{Herschel} photometry of the 3CR sample using the 
   \textit{Photodetector Array Camera} (PACS) at 70 and 160~$\mu$m and 
   the \textit{Spectral and Photometric Imaging Receiver} (SPIRE) at 250, 
   350, and 500~$\mu$m on-board the \textit{Herschel Space Observatory}. 
   The first results, dealing with 3 archetypal objects of that sample 
   were presented in B12. Here we analyse the FIR properties of the complete 
   (flux-limited) sample of objects spanning the redshift range $1 < z < 2.5$. 
   This paper is organized as follows. Section~\ref{section:Data} describes 
   the sample selection, the data obtained, and the steps used for measuring 
   flux densities in the five \textit{Herschel} bands. Section~\ref{section:SEDs} 
   addresses the procedure for fitting the observed IR SEDs of the objects. 
   Results are then presented and discussed in Sect.~\ref{section:Results} 
   and Sect.~\ref{section:Discussion}, respectively, and the paper is briefly 
   summarized in Sect.~\ref{section:Conclusions}. Throughout this paper we use 
   a flat cosmology with H$_{0}$ = 70 km s$^{-1}$ Mpc$^{-1}$ and 
   $\Omega_{\Lambda}$ = 0.7, and we follow the conversion in \citet{Kennicutt98}
   \citetext{which assumes a \citealp{Salpeter55} initial mass function} when 
   deriving SFRs.  
\section{Data}
\label{section:Data}
   \subsection{Sample selection}
   \begin{figure}
      \centering
      \includegraphics[width=\hsize]{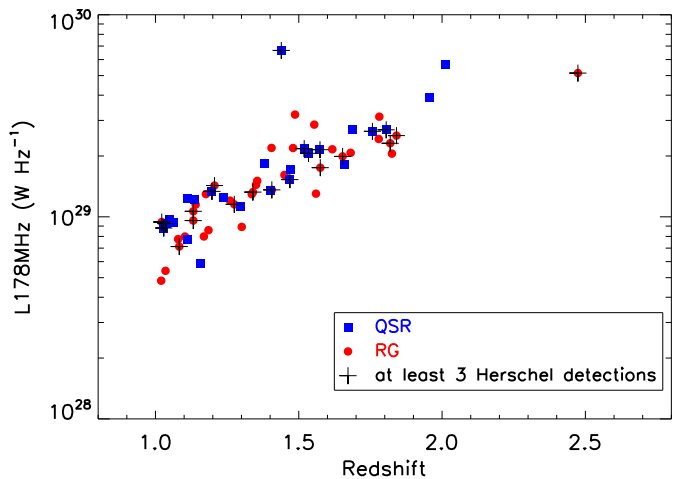}
      \caption{Observed radio (178~MHz) luminosity as a function of 
               redshift for the 3CR sample considered in this work. 
               Circles (red) indicate radio galaxies, and squares 
               (blue) indicate quasars. The plus symbols mark the 
               objects detected in at least three \textit{Herschel} 
               bands (typically the two PACS and the SPIRE 250~$\mu$m 
               bands.}
      \label{figure:sampleDescr}
   \end{figure} 
   With this study, we target the well known, complete flux-limited sample of the 
   brightest (F$_{\mathrm{178~MHz}}~>$~10~Jy), high-redshift 
   ($z>1$) radio-loud AGN sample in the northern hemisphere: the 
   Revised Third Cambridge Catalogue of radio sources 
   \citetext{hereafter 3CR; \citealp{Spinrad85}}. The extremely high 
   luminosities (Fig.~\ref{figure:sampleDescr}) of these double-lobed 
   radio galaxies (RGs) and quasars (QSRs) are produced by some of the 
   most powerful accreting SMBHs. The low-frequency (178 MHz) radio 
   selection ensures no bias with respect to orientation: the steep-spectrum 
   lobes of radio-loud AGN emit optically thin and isotropic 
   synchrotron radiation, making the 3CR sample ideal for testing the 
   orientation-based unification scenario of radio-loud AGN. As shown in 
   Fig.~\ref{figure:sampleDescr}, both the RGs and QSRs are homogeneously 
   distributed in redshift. The $z>1$ 3CR objects show mostly FRII 
   morphologies; this is well-established from, for instance, 
   high-resolution VLA maps. Compact, presumably young, morphologies 
   \citep{Odea98} are also found within the sample. 

   The $z>1$ 3CR sample is (spectroscopically) completely identified 
   using 3 to 5~m-class telescopes in the 1960s-1980s \citep{Spinrad85}. 
   The objects in the sample almost universally accrete at high Eddington 
   rate, i.e. in quasar-mode \citep[e.g.][]{Best&Heckman12}. 
   The total number of objects in the $z>1$ 3CR sample is 64\footnote{The 
   sample includes two 4C objects, 4C~13.66 and 4C~16.49, which formally 
   match the selection criteria of, and are included in, the 3CR sample.}.
   The highest redshift 3CR source is 3C~257 ($z = 2.47$). Two $z>1$ 3CR 
   sources, 3C~287 and 3C~300.1, have been observed in other \textit{Herschel} 
   observing modes, and thus have been dropped from this work. The remaining 
   62 sources, 37 RGs and 25 QSRs, comprise the \textit{Herschel} sample 
   studied in this work. An overview of selected properties is provided 
   in Table\ref{table:VarProp}. 
 
   The high-$z$ 3CR sample has been observed with many space telescopes 
   \citep[including \textit{Hubble}, \textit{Spitzer}, and 
   \textit{Chandra};][]{Best98,Haas08,Leipski10,Wilkes13}: objects from this 
   sample represent landmarks in our study of active galaxies through cosmic time.  
   \begin{longtab}
   \renewcommand{\arraystretch}{1.2}
   \begin{longtable}{lccccccc}
      \caption{\label{table:VarProp} Summary of selected properties of the high-$z$ 3CR sample studied in this work.}\\
      \hline\hline
       Name & Type & $z$ & RA (J2000) & Dec (J2000) & log (L$_{\mathrm{178MHz}}$ (W Hz$^{-1}$)) & PACS OBSIDs & SPIRE OBSIDs \\  
  	   (1) & (2) & (3) & (4) & (5) & (6) & (7) & (8) \\   
      \hline
      \endfirsthead
      \caption{continued.}\\
      \hline\hline
       Name & Type & $z$ & RA (J2000) & Dec (J2000) & log (L$_{\mathrm{178MHz}}$ (W Hz$^{-1}$)) & PACS OBSIDs & SPIRE OBSIDs \\  
	   (1) & (2) & (3) & (4) & (5) & (6) & (7) & (8) \\   
      \hline
      \endhead
      \hline
      \endfoot
\object{3C~002} & QSR & 1.04 & 00:06:22.58 & -00:04:24.69 & 29.0 & 1342221796/1342221797 & 1342212374   \\
\object{3C~009} & QSR & 2.01 & 00:20:25.21 & +15:40:54.59 & 29.8 & 1342222444/1342222445 & 1342213198   \\
\object{3C~013} & RG & 1.35 & 00:34:14.55 & +39:24:16.65 & 29.2 & 1342223179/1342223180 & 1342213491   \\
\object{3C~014} & QSR & 1.47 & 00:36:06.44 & +18:37:59.23 & 29.2 & 1342222429/1342222428 & 1342213196   \\
\object{3C~036} & RG & 1.30 & 01:17:59.48 & +45:36:21.75 & 29.0 & 1342223508/1342223509 & 1342203613   \\
\object{3C~043} & QSR & 1.47 & 01:29:59.80 & +23:38:20.28 & 29.2 & 1342223506/1342223507 & 1342213488   \\
\object{3C~065}\tablefootmark{a} & RG & 1.18 & 02:23:43.19 & +40:00:52.45 & 29.1 & 1342238005/1342238006 & 1342239821   \\
\object{3C~068.1} & QSR & 1.24 & 02:32:28.87 & +34:23:46.79 & 29.1 & 1342223870/1342223871 & 1342226628   \\
\object{3C~068.2} & RG & 1.58 & 02:34:23.85 & +31:34:17.46 & 29.2 & 1342223866/1342223867 & 1342224971   \\
\object{3C~119} & RG & 1.02 & 04:32:36.50 & +41:38:28.44 & 29.0 & 1342227975/1342227976 & 1342216924   \\
\object{3C~124} & RG & 1.08 & 04:41:59.10 & +01:21:01.91 & 28.9 & 1342226718/1342226719 & 1342216939   \\
\object{3C~173} & RG & 1.03 & 07:02:20.58 & +37:57:23.50 & 28.7 & 1342219418/1342219419 & 1342206177   \\
\object{3C~181} & QSR & 1.38 & 07:28:10.30 & +14:37:36.24 & 29.3 & 1342220573/1342220574 & 1342204852   \\
\object{3C~186} & QSR & 1.06 & 07:44:17.45 & +37:53:17.15 & 29.0 & 1342220127/1342220128 & 1342206178   \\
\object{3C~190} & QSR & 1.20 & 08:01:33.55 & +14:14:42.94 & 29.1 & 1342205262/1342205263 & 1342205052   \\
\object{3C~191} & QSR & 1.96 & 08:04:47.97 & +10:15:23.69 & 29.6 & 1342220655/1342220656 & 1342205072   \\
\object{3C~194} & RG & 1.18 & 08:10:03.61 & +42:28:04.31 & 28.9 & 1342220123/1342220124 & 1342206180   \\
\object{3C~204} & QSR & 1.11 & 08:37:44.95 & +65:13:34.92 & 28.9 & 1342220115/1342220116 & 1342206190   \\
\object{3C~205} & QSR & 1.53 & 08:39:06.45 & +57:54:17.12 & 29.3 & 1342220117/1342220118 & 1342206188   \\
\object{3C~208.0} & QSR & 1.11 & 08:53:08.60 & +13:52:54.98 & 29.1 & 1342220790/1342220791 & 1342206221   \\
\object{3C~208.1} & RG & 1.02 & 08:54:39.28 & +14:05:52.56 & 28.7 & 1342220788/1342220789 & 1342206220   \\
\object{3C~210} & RG & 1.17 & 08:58:09.96 & +27:50:51.57 & 28.9 & 1342220797/1342220796 & 1342230768   \\
\object{3C~212} & QSR & 1.05 & 08:58:41.49 & +14:09:43.97 & 29.0 & 1342220786/1342220787 & 1342206219   \\
\object{3C~220.2} & QSR & 1.16 & 09:30:33.47 & +36:01:24.17 & 28.8 & 1342220798/1342220799 & 1342222125   \\
\object{3C~222} & RG & 1.34 & 09:36:32.01 & +04:22:10.30 & 29.1 & 1342221142/1342221143 & 1342210521   \\
\object{3C~225A} & RG & 1.56 & 09:42:08.48 & +13:51:54.23 & 29.1 & 1342221258/1342221259 & 1342210518   \\
\object{3C~230} & RG & 1.49 & 09:51:58.82 & -00:01:27.23 & 29.5 & 1342221136/1342221137 & 1342210520   \\
\object{3C~238} & RG & 1.40 & 10:11:00.37 & +06:24:39.72 & 29.3 & 1342221144/1342221145 & 1342210519   \\
\object{3C~239}\tablefootmark{a} & RG & 1.78 & 10:11:45.41 & +46:28:19.75 & 29.5 & 1342231241/1342231242 & 1342230739   \\
\object{3C~241} & RG & 1.62 & 10:21:54.52 & +21:59:30.71 & 29.3 & 1342221152/1342221153 & 1342198253   \\
\object{3C~245} & QSR & 1.03 & 10:42:44.60 & +12:03:31.26 & 28.9 & 1342221264/1342221265 & 1342210516   \\
\object{3C~249} & RG & 1.55 & 11:02:03.84 & -01:16:17.39 & 29.5 & 1342221853/1342221854 & 1342198569   \\
\object{3C~250} & RG & 1.26 & 11:08:52.12 & +25:00:54.61 & 29.1 & 1342221154/1342221155 & 1342210509   \\
\object{3C~252} & RG & 1.10 & 11:11:32.99 & +35:40:41.64 & 28.9 & 1342221160/1342221161 & 1342210508   \\
\object{3C~255} & RG & 1.36 & 11:19:25.23 & -03:02:51.50 & 29.2 & 1342221851/1342221852 & 1342210515   \\
\object{3C~256} & RG & 1.82 & 11:20:43.02 & +23:27:55.22 & 29.4 & 1342221262/1342221263 & 1342210510   \\
\object{3C~257} & RG & 2.47 & 11:23:09.17 & +05:30:19.47 & 29.7 & 1342221966/1342221967 & 1342210514   \\
\object{3C~266}\tablefootmark{a} & RG & 1.27 & 11:45:43.36 & +49:46:08.24 & 29.1 & 1342222695/1342222696 & 1342222663   \\
\object{3C~267} & RG & 1.14 & 11:49:56.56 & +12:47:19.07 & 29.1 & 1342222448/1342222449 & 1342200236   \\
\object{3C~268.4} & QSR & 1.40 & 12:09:13.61 & +43:39:20.96 & 29.1 & 1342221162/1342221163 & 1342210501   \\
\object{3C~270.1} & QSR & 1.52 & 12:20:33.87 & +33:43:12.05 & 29.3 & 1342221952/1342221953 & 1342200238   \\
\object{3C~280.1} & QSR & 1.66 & 13:00:33.30 & +40:09:07.72 & 29.3 & 1342212393/1342212394 & 1342210498   \\
\object{3C~294}\tablefootmark{a} & RG & 1.78 & 14:06:53.20 & +34:11:21.10 & 29.4 & 1342211098/1342211099 & 1342206200   \\
\object{3C~297} & RG & 1.41 & 14:17:23.99 & -04:00:47.54 & 29.1 & 1342223834/1342223835 & 1342203577   \\
\object{3C~298} & QSR & 1.44 & 14:19:08.18 & +06:28:34.80 & 29.8 & 1342223664/1342223665 & 1342213464   \\
\object{3C~305.1} & RG & 1.13 & 14:47:09.56 & +76:56:21.80 & 29.0 & 1342220952/1342220953 & 1342206193   \\
\object{3C~318} & QSR & 1.57 & 15:20:05.44 & +20:16:05.76 & 29.3 & 1342223844/1342223845 & 1342204107   \\
\object{3C~322} & RG & 1.68 & 15:35:01.23 & +55:36:52.87 & 29.3 & 1342199131/1342199132 & 1342206196   \\
\object{3C~324} & RG & 1.21 & 15:49:48.89 & +21:25:38.06 & 29.2 & 1342202562/1342202563 & 1342213461   \\
\object{3C~325} & QSR & 1.13 & 15:49:58.42 & +62:41:21.66 & 29.1 & 1342219034/1342219035 & 1342206195   \\
\object{3C~326.1} & RG & 1.83 & 15:56:10.06 & +20:04:20.44 & 29.3 & 1342224482/1342224483 & 1342213462   \\
\object{3C~356} & RG & 1.08 & 17:24:19.04 & +50:57:40.14 & 28.9 & 1342219036/1342219037 & 1342206197   \\
\object{3C~368} & RG & 1.13 & 18:05:06.45 & +11:01:35.06 & 29.0 & 1342216599/1342216600 & 1342216954   \\
\object{3C~418} & QSR & 1.69 & 20:38:37.03 & +51:19:12.66 & 29.4 & 1342219032/1342219033 & 1342210542   \\
\object{3C~432} & QSR & 1.80 & 21:22:46.32 & +17:04:37.95 & 29.4 & 1342211499/1342211500 & 1342210541   \\
\object{3C~437} & RG & 1.48 & 21:47:25.10 & +15:20:37.49 & 29.3 & 1342211497/1342211498 & 1342210540   \\
\object{3C~454.0} & QSR & 1.76 & 22:51:34.73 & +18:48:40.12 & 29.4 & 1342210949/1342210950 & 1342210539   \\
\object{3C~454.1} & RG & 1.84 & 22:50:32.93 & +71:29:19.18 & 29.4 & 1342211436/1342211437 & 1342212365   \\
\object{3C~469.1} & RG & 1.34 & 23:55:23.32 & +79:55:19.60 & 29.1 & 1342221170/1342221171 & 1342220543   \\
\object{3C~470}\tablefootmark{a} & RG & 1.65 & 23:58:35.89 & +44:04:45.55 & 29.3 & 1342237858/1342237859 & 1342236248   \\
\object{4C~13.66} & RG & 1.45 & 18:01:38.95 & +13:51:23.85 & 29.2 & 1342216597/1342216598 & 1342216956   \\
\object{4C~16.49} & QSR & 1.30 & 17:34:42.61 & +16:00:31.21 & 29.1 & 1342216595/1342216596 & 1342216955   \\
      \hline                                   
   \end{longtable}
   \tablefoot{(1) Name of object; (2) AGN type; (3) Redshift; (4) Right Ascension; 
              (5) Declination; (6) Log of 178~MHz luminosity (in the observer frame); 
              (7) PACS ObsID; (8) SPIRE ObsID \\
   \tablefoottext{a}{\textit{Herschel} observations taken from OT1$\_$nseymour$\_$1 (PI: Seymour)}
             }                       
   \end{longtab}
   \subsection{\textit{Herschel} photometry}
   
   The data for this work were obtained as part of our \textit{Herschel} 
   Guaranteed Time project \textit{The Herschel Legacy of distant 
   radio-loud AGN} (PI: Barthel, 38 hours of observations). Five objects 
   (see Table~\ref{table:VarProp}) were observed as part of another 
   \textit{Herschel} programme (PI: Seymour). The raw data for these objects 
   were retrieved from the \textit{Herschel Science Archive} (HSA), and 
   the data reduction was performed as detailed below.
\subsubsection{PACS}
\label{subsection:PACS}
   Photometric observations were carried out with PACS \citep{Poglitsch10} 
   in the scan-map observational mode, both in the blue (70~$\mu$m, 5$\arcsec$ 
   angular resolution) and in the red (160~$\mu$m, 11$\arcsec$ angular 
   resolution) bands. A concatenated pair of coextensive scan maps at two different 
   orientations was obtained for each source. Data reduction was performed 
   within the \textit{Herschel Interactive Processing Environment} 
   \citep[HIPE,][version 11.0.0]{Ott10}, following the standard procedures for 
   deep field observations. Maps were created by employing the high-pass 
   filtering method, using an appropriate source masking step to avoid 
   significant flux losses due to the high-pass filter. A first data 
   reduction resulted in a preliminary map, created after combining the 
   individually (for each scan orientation) processed scan maps. Source 
   masking was performed by hand, using the preliminary created map as 
   an input. This method in particular allowed us to minimize the flux 
   losses of the observed sources \citep{Popesso12}. The final data 
   reduction and mosaicking were then performed using the mask generated 
   in the previous step.
   
   Photometry (using appropriate aperture corrections) was performed 
   within HIPE, using the \textit{annularSkyAperturePhotometry} task. 
   Apertures of 6$\arcsec$ and 10$\arcsec$ radius for PACS blue and PACS 
   red, respectively, were centred on the known radio core position of 
   the object in the map. PACS maps suffer from correlated noise, thus 
   pixel-to-pixel variations cannot yield robust photometric uncertainties. 
   Instead, we opted for the well-established procedure of placing 
   apertures at random positions on the sky \citep{Lutz11,Popesso12}. 
   Following \citet{Leipski13}, we placed 500 apertures of 6$\arcsec$ 
   (for blue) and 10$\arcsec$ (for red) radii at locations avoiding the 
   noisy edges of the map, requiring that the central pixel of the 
   random aperture has at least 75\% of the integration time of that 
   of the source of interest. The resulting distribution of the flux 
   densities measured in these 500 apertures was then fitted with a 
   Gaussian, and the sigma value of the Gaussian was taken to be the 
   1$\sigma$ photometric uncertainty of the map. Measured flux densities and 
   associated 1$\sigma$ uncertainties, together with 3$\sigma$ upper limits 
   for the non-detections are provided in Table~\ref{table:HerscPhot}.
   The PACS photometric uncertainties provided in Table~\ref{table:HerscPhot} 
   do not include the 5\% uncertainty on the absolute flux calibration 
   \citep{Balog13}. Postage stamps of the resulting PACS maps, centred 
   on the radio position of the objects, are included in Appendix~\ref{appendix:stamps}.
  \begin{longtab}
  \renewcommand{\arraystretch}{1.2}
  \begin{landscape}
   \begin{longtable}{c c c c c c c c c c c c}
      \caption{\label{table:HerscPhot} \textit{Herschel} and \textit{Spitzer} photometry of the 3CR objects studied in this work. Photometric uncertainties are 1$\sigma$ values, and upper limits are 3$\sigma$ values.}\\
      \hline\hline
       Object & F$_{3.6 \mu \mathrm{m}}$ & F$_{4.5 \mu \mathrm{m}}$ & F$_{5.8 \mu \mathrm{m}}$ & F$_{8.0 \mu \mathrm{m}}$ & F$_{16 \mu \mathrm{m}}$ & F$_{24 \mu \mathrm{m}}$ & F$_{70 \mu \mathrm{m}}$ & F$_{160 \mu \mathrm{m}}$ & F$_{250 \mu \mathrm{m}}$ & F$_{350 \mu \mathrm{m}}$ & F$_{500 \mu \mathrm{m}}$ \\
		      &	($\mu$Jy)			     & ($\mu$Jy)		        & ($\mu$Jy)			       & ($\mu$Jy)		          & ($\mu$Jy)               & ($\mu$Jy)               & (mJy)                   & (mJy)                    & (mJy)                    & (mJy)                    & (mJy) \\
	   (1) & (2) & (3) & (4) & (5) & (6) & (7) & (8) & (9) & (10) & (11) & (12) \\      
      \hline
      \endfirsthead
      \caption{continued.}\\
      \hline\hline
       Object & F$_{3.6 \mu \mathrm{m}}$ & F$_{4.5 \mu \mathrm{m}}$ & F$_{5.8 \mu \mathrm{m}}$ & F$_{8.0 \mu \mathrm{m}}$ & F$_{16 \mu \mathrm{m}}$ & F$_{24 \mu \mathrm{m}}$ & F$_{70 \mu \mathrm{m}}$ & F$_{160 \mu \mathrm{m}}$ & F$_{250 \mu \mathrm{m}}$ & F$_{350 \mu \mathrm{m}}$ & F$_{500 \mu \mathrm{m}}$ \\
		      &	($\mu$Jy)			     & ($\mu$Jy)		        & ($\mu$Jy)			       & ($\mu$Jy)		          & ($\mu$Jy)               & ($\mu$Jy)               & (mJy)                   & (mJy)                    & (mJy)                    & (mJy)                    & (mJy) \\
	   (1) & (2) & (3) & (4) & (5) & (6) & (7) & (8) & (9) & (10) & (11) & (12) \\
      \hline
      \endhead
      \hline
      \endfoot
3C~002 & 283$\pm$42 & 330$\pm$50 & 530$\pm$80 & 809$\pm$121 & 1550$\pm$233 & 2970$\pm$446 & 10.0$\pm$2.6 & 21.9$\pm$4.4 & 16.1$\pm$5.1 & 19.6$\pm$4.8 & $\textless$ 17.0   \\
3C~009 & 884$\pm$133 & 1080$\pm$162 & 1590$\pm$239 & 2220$\pm$333 & 3330$\pm$500 & 3470$\pm$520 & 13.7$\pm$2.6 & 18.1$\pm$4.9 & $\textless$ 15.4 & $\textless$ 13.8 & $\textless$ 17.2   \\
3C~013 & 133$\pm$20 & 133$\pm$20 & 147$\pm$22 & 283$\pm$42 & 375$\pm$56 & 2060$\pm$309 & 24.6$\pm$2.1 & 30.7$\pm$5.8 & $\textless$ 18.8 & $\textless$ 15.3 & $\textless$ 20.7   \\
3C~014 & 1040$\pm$156 & 1710$\pm$257 & 2740$\pm$411 & 4150$\pm$623 & 7070$\pm$1061 & 10300$\pm$1545 & 20.9$\pm$2.6 & 21.9$\pm$6.8 & 20.0$\pm$6.3 & $\textless$ 15.4 & $\textless$ 19.8   \\
3C~036 & 163$\pm$24 & 205$\pm$31 & 256$\pm$38 & 360$\pm$54 & 560$\pm$84 & 874$\pm$131 & $\textless$  4.2 & $\textless$ 10.6 & $\textless$ 11.2 & $\textless$ 11.8 & $\textless$ 14.2   \\
3C~043 & 193$\pm$29 & 270$\pm$41 & 356$\pm$53 & 445$\pm$67 & 1010$\pm$152 & 1610$\pm$242 & ...\tablefootmark{b} & ...\tablefootmark{b} & $\textless$ 14.1 & $\textless$ 17.9 & $\textless$ 15.8   \\
3C~065 & 202$\pm$30 & 233$\pm$35 & 299$\pm$45 & 418$\pm$63 & 798$\pm$120 & 1700$\pm$255 & $\textless$  6.7 & $\textless$ 12.1 & $\textless$ 20.8 & $\textless$ 18.1 & $\textless$ 20.2   \\
3C~068.1 & 967$\pm$145 & 1430$\pm$215 & 2040$\pm$306 & 2780$\pm$417 & 3800$\pm$570 & 7760$\pm$1164 & 22.7$\pm$2.3 & $\textless$ 17.8 & $\textless$ 18.2 & $\textless$ 15.9 & $\textless$ 20.1   \\
3C~068.2 & 105$\pm$16 & 129$\pm$19 & 137$\pm$21 & 112$\pm$17 & 1340$\pm$201 & 1170$\pm$176 & 27.4$\pm$2.6 & 39.6$\pm$5.9 & 42.0$\pm$7.2 & 38.7$\pm$7.0 & 29.0$\pm$7.2   \\
3C~119 & 802$\pm$120 & 878$\pm$132 & 1280$\pm$192 & 1850$\pm$278 & 4820$\pm$723 & 8260$\pm$1239 & 24.9$\pm$2.2 & 32.6$\pm$8.9 & 28.9$\pm$15.4\tablefootmark{a} & $\textless$ 56.1 & $\textless$ 44.1   \\
3C~124 & 144$\pm$22 & 120$\pm$18 & 188$\pm$28 & 310$\pm$47 & 1840$\pm$276 & 3560$\pm$534 & 34.2$\pm$2.1 & 55.7$\pm$6.7 & 52.1$\pm$7.3 & 31.5$\pm$7.9 & $\textless$ 24.3   \\
3C~173 & 163$\pm$24 & 172$\pm$26 & 197$\pm$30 & 227$\pm$34 & 374$\pm$56 & 710$\pm$107 & 6.9$\pm$1.6 & $\textless$ 10.2 & $\textless$ 11.6 & $\textless$ 13.0 & $\textless$ 16.5   \\
3C~181 & 348$\pm$52 & 485$\pm$73 & 722$\pm$108 & 1110$\pm$167 & 2180$\pm$327 & 4260$\pm$639 & 12.2$\pm$2.3 & $\textless$ 13.6 & $\textless$ 20.4 & $\textless$ 16.5 & $\textless$ 21.8   \\
3C~186 & 791$\pm$119 & 1020$\pm$153 & 1410$\pm$212 & 1960$\pm$294 & 3660$\pm$549 & 6660$\pm$999 & 18.9$\pm$2.6 & $\textless$ 18.0 & $\textless$ 13.7 & $\textless$ 17.1 & $\textless$ 21.0   \\
3C~190 & 739$\pm$111 & 908$\pm$136 & 1290$\pm$194 & 1740$\pm$261 & 3310$\pm$497 & 6690$\pm$1004 & 46.1$\pm$2.5 & 72.5$\pm$4.8 & 74.1$\pm$6.5 & 54.1$\pm$4.7 & $\textless$ 20.9   \\
3C~191 & 333$\pm$50 & 399$\pm$60 & 655$\pm$98 & 1010$\pm$152 & 2270$\pm$341 & 3810$\pm$572 & 26.4$\pm$2.7 & 21.9$\pm$4.9 & $\textless$ 18.9 & $\textless$ 17.1 & $\textless$ 18.1   \\
3C~194 & 201$\pm$30 & 176$\pm$26 & 164$\pm$25 & 208$\pm$31 & 509$\pm$76 & 885$\pm$133 & $\textless$  4.6 & $\textless$  9.3 & $\textless$ 20.2 & $\textless$ 18.9 & $\textless$ 20.7   \\
3C~204 & 917$\pm$138 & 1250$\pm$188 & 1920$\pm$288 & 2540$\pm$381 & 4730$\pm$710 & 7360$\pm$1104 & $\textless$  6.1 & $\textless$ 13.5 & $\textless$ 17.9 & $\textless$ 14.5 & $\textless$ 18.2   \\
3C~205 & 1460$\pm$219 & 2080$\pm$312 & 2920$\pm$438 & 4090$\pm$614 & 7320$\pm$1098 & 12800$\pm$1920 & 62.7$\pm$2.6 & 66.2$\pm$6.8 & 56.2$\pm$6.8 & 31.6$\pm$4.8 & 19.9$\pm$6.7\tablefootmark{a}   \\
3C~208.0 & 660$\pm$99 & 803$\pm$120 & 1160$\pm$174 & 1620$\pm$243 & 2980$\pm$447 & 5870$\pm$881 & $\textless$  7.4 & $\textless$ 16.7 & $\textless$ 20.5 & $\textless$ 21.7 & $\textless$ 20.3   \\
3C~208.1 & 331$\pm$50 & 430$\pm$65 & 656$\pm$98 & 954$\pm$143 & 1360$\pm$204 & 2110$\pm$317 & $\textless$  8.1 & $\textless$ 15.7 & $\textless$ 17.1 & $\textless$ 16.0 & $\textless$ 20.2   \\
3C~210 & 256$\pm$38 & 336$\pm$50 & 489$\pm$73 & 1090$\pm$164 & 3410$\pm$512 & 4430$\pm$665 & 31.6$\pm$2.4 & 56.0$\pm$4.0 & ...\tablefootmark{b} & ...\tablefootmark{b} & ...\tablefootmark{b}   \\
3C~212 & 925$\pm$139 & 1430$\pm$215 & 2340$\pm$351 & 3400$\pm$510 & 6710$\pm$1007 & 10800$\pm$1620 & 16.6$\pm$2.6 & $\textless$ 16.7 & $\textless$ 38.5 & $\textless$ 56.0 & $\textless$ 39.9   \\
3C~220.2 & 592$\pm$89 & 870$\pm$131 & 1330$\pm$200 & 2000$\pm$300 & 4150$\pm$623 & 6720$\pm$1008 & 26.6$\pm$2.2 & 22.5$\pm$5.3 & $\textless$ 14.1 & $\textless$ 14.6 & $\textless$ 15.7   \\
3C~222 & 83$\pm$12 & 91$\pm$14 & 73$\pm$11 & 65$\pm$10 & 331$\pm$50 & 229$\pm$34 & 14.3$\pm$1.9 & 50.8$\pm$4.7 & 48.4$\pm$4.8 & 50.9$\pm$3.5 & 28.7$\pm$5.4   \\
3C~225A & 47$\pm$7 & 49$\pm$7 & 71$\pm$11 & 108$\pm$16 & 321$\pm$48 & $\textless$ 1070 & $\textless$  7.8 & $\textless$ 21.4 & $\textless$ 21.3 & $\textless$ 16.3 & $\textless$ 22.7   \\
3C~230 & 1040$\pm$156 & 672$\pm$101 & 438$\pm$66 & 317$\pm$48 & 1150$\pm$173 & 1560$\pm$234 & 11.9$\pm$2.3 & $\textless$ 19.8 & $\textless$ 40.7 & $\textless$ 36.1 & $\textless$ 34.7   \\
3C~238 & 65$\pm$10 & 77$\pm$12 & 84$\pm$12 & $\textless$   92 & $\textless$  283 & 266$\pm$40 & $\textless$  4.0 & $\textless$  8.2 & $\textless$ 13.4 & $\textless$ 14.2 & $\textless$ 17.3   \\
3C~239 & 96$\pm$14 & 111$\pm$17 & 130$\pm$20 & 142$\pm$21 & 651$\pm$98 & 1450$\pm$218 & $\textless$  7.3 & $\textless$ 15.3 & $\textless$ 19.1 & $\textless$ 14.9 & $\textless$ 21.0   \\
3C~241 & 92$\pm$14 & 101$\pm$15 & 116$\pm$17 & 161$\pm$24 & 389$\pm$58 & 591$\pm$89 & 7.9$\pm$1.1 & $\textless$  8.2 & $\textless$ 15.5 & $\textless$ 12.5 & $\textless$ 17.6   \\
3C~245 & 1420$\pm$213 & 1900$\pm$285 & 3350$\pm$503 & 5270$\pm$790 & 10400$\pm$1560 & 20400$\pm$3060 & 47.5$\pm$2.4 & 35.3$\pm$6.4 & 35.2$\pm$5.9 & $\textless$ 16.1 & $\textless$ 22.3   \\
3C~249 & 54$\pm$8 & 52$\pm$8 & 42$\pm$6 & 47$\pm$7 & 194$\pm$29 & $\textless$  516 & $\textless$  3.2 & $\textless$ 10.7 & $\textless$ 11.5 & $\textless$ 10.4 & $\textless$ 14.4   \\
3C~250 & 61$\pm$9 & 59$\pm$9 & 46$\pm$7 & 29$\pm$4 & 162$\pm$24 & $\textless$  147 & $\textless$  3.7 & $\textless$ 10.7 & $\textless$ 11.8 & $\textless$ 15.0 & $\textless$ 13.9   \\
3C~252 & 225$\pm$34 & 382$\pm$57 & 787$\pm$118 & 1390$\pm$209 & 3900$\pm$585 & 7000$\pm$1050 & 21.4$\pm$2.4 & $\textless$ 21.3 & $\textless$ 17.2 & $\textless$ 17.4 & $\textless$ 23.1   \\
3C~255 & 85$\pm$13 & 86$\pm$13 & 57$\pm$9 & 22$\pm$3 & $\textless$  116 & $\textless$  241 & $\textless$  3.2 & $\textless$  8.4 & $\textless$ 14.5 & $\textless$ 17.0 & $\textless$ 19.1   \\
3C~256 & 34$\pm$5 & 37$\pm$6 & 43$\pm$7 & 75$\pm$11 & 743$\pm$111 & 1900$\pm$285 & 17.8$\pm$2.3 & 31.9$\pm$5.3 & 28.2$\pm$6.9 & $\textless$ 19.1 & $\textless$ 21.6   \\
3C~257 & 85$\pm$13 & 111$\pm$17 & 194$\pm$29 & 322$\pm$48 & ... & 1360$\pm$204 & 8.1$\pm$1.0 & 15.6$\pm$2.5 & 33.1$\pm$4.7 & 31.8$\pm$5.5 & 32.3$\pm$8.6   \\
3C~266 & 68$\pm$10 & 73$\pm$11 & 45$\pm$7 & 102$\pm$15 & 370$\pm$56 & 980$\pm$147 & 7.6$\pm$2.4 & 29.4$\pm$4.1 & 19.5$\pm$5.6 & $\textless$ 15.9 & $\textless$ 18.9   \\
3C~267 & 153$\pm$23 & 218$\pm$33 & 414$\pm$62 & 739$\pm$111 & 2370$\pm$356 & 3730$\pm$560 & 15.3$\pm$2.3 & $\textless$ 12.7 & $\textless$ 16.8 & $\textless$ 14.1 & $\textless$ 19.1   \\
3C~268.4 & 1060$\pm$159 & 1560$\pm$234 & 2220$\pm$333 & 3330$\pm$500 & 7580$\pm$1137 & 11600$\pm$1740 & 30.3$\pm$2.1 & $\textless$ 17.0 & $\textless$ 18.7 & $\textless$ 18.2 & $\textless$ 24.3   \\
3C~270.1 & 606$\pm$91 & 944$\pm$142 & 1430$\pm$214 & 2260$\pm$339 & 3910$\pm$587 & 5470$\pm$821 & 30.0$\pm$2.2 & 38.9$\pm$4.7 & 27.9$\pm$5.5 & $\textless$ 17.5 & $\textless$ 19.6   \\
3C~280.1 & 378$\pm$57 & 512$\pm$77 & 777$\pm$116 & 1170$\pm$176 & 1680$\pm$252 & 2160$\pm$324 & $\textless$  5.0 & $\textless$  9.4 & $\textless$ 17.0 & $\textless$ 13.4 & $\textless$ 17.2   \\
3C~294 & $\textless$   93 & $\textless$  103 & 68$\pm$10 & 67$\pm$10 & ... & 348$\pm$52 & $\textless$  6.6 & $\textless$ 22.1 & $\textless$ 16.6 & $\textless$ 14.5 & $\textless$ 21.9   \\
3C~297 & 119$\pm$18 & 126$\pm$19 & 122$\pm$18 & 121$\pm$18 & $\textless$  288 & 432$\pm$65 & 12.6$\pm$1.2 & 15.4$\pm$2.4 & 24.5$\pm$4.3 & $\textless$ 13.8 & $\textless$ 17.2   \\
3C~298 & 1600$\pm$240 & 2390$\pm$359 & 3710$\pm$556 & 5510$\pm$827 & 9160$\pm$1374 & 12600$\pm$1890 & 78.7$\pm$2.4 & 96.8$\pm$4.3 & 96.0$\pm$6.9 & 51.5$\pm$5.9 & 24.1$\pm$6.7   \\
3C~305.1 & 181$\pm$27 & 282$\pm$42 & 495$\pm$74 & 972$\pm$146 & 2410$\pm$362 & 2490$\pm$374 & 24.0$\pm$2.3 & 40.4$\pm$4.3 & 34.9$\pm$6.0 & $\textless$ 18.1 & $\textless$ 18.9   \\
3C~318 & 343$\pm$51 & 427$\pm$64 & 571$\pm$86 & 806$\pm$121 & 1960$\pm$294 & 3400$\pm$510 & 18.3$\pm$2.3 & 43.8$\pm$5.9 & 42.9$\pm$6.1 & ...\tablefootmark{b} & ...\tablefootmark{b}   \\
3C~322 & 128$\pm$19 & 135$\pm$20 & 94$\pm$14 & 120$\pm$18 & 411$\pm$62 & 804$\pm$121 & $\textless$  5.0 & $\textless$ 11.0 & $\textless$ 19.0 & $\textless$ 17.5 & $\textless$ 21.9   \\
3C~324 & 165$\pm$25 & 160$\pm$24 & 178$\pm$27 & 450$\pm$68 & 2580$\pm$387 & 2820$\pm$423 & 23.5$\pm$2.3 & 31.7$\pm$5.6 & 21.0$\pm$6.2 & $\textless$ 18.2 & $\textless$ 21.4   \\
3C~325 & 472$\pm$71 & 565$\pm$85 & 708$\pm$106 & 1200$\pm$180 & 1990$\pm$299 & 3030$\pm$455 & $\textless$  7.6 & $\textless$ 18.4 & $\textless$ 15.4 & $\textless$ 16.1 & $\textless$ 18.3   \\
3C~326.1 & 29$\pm$4 & 34$\pm$5 & 26$\pm$4 & 72$\pm$11 & 829$\pm$124 & 1430$\pm$215 & $\textless$  6.7 & $\textless$ 14.5 & 29.8$\pm$5.9 & 21.0$\pm$6.9 & $\textless$ 20.8   \\
3C~356 & 108$\pm$16 & 110$\pm$16 & 122$\pm$18 & 434$\pm$65 & 2270$\pm$341 & 4060$\pm$609 & 11.6$\pm$2.5 & 19.7$\pm$4.9 & $\textless$ 18.4 & $\textless$ 15.3 & $\textless$ 20.4   \\
3C~368 & 126$\pm$19 & 112$\pm$17 & 112$\pm$17 & 210$\pm$32 & 1370$\pm$206 & 3250$\pm$488 & 29.9$\pm$2.0 & 61.5$\pm$4.8 & 44.4$\pm$7.4 & 23.8$\pm$6.2 & $\textless$ 21.3   \\
3C~418 & 1130$\pm$170 & 1630$\pm$245 & 2470$\pm$371 & 3900$\pm$585 & 6680$\pm$1002 & 13600$\pm$2040 & 95.1$\pm$2.6 & 200.1$\pm$16.1 & 173.8$\pm$40.0 & 259.2$\pm$43.7 & 387.4$\pm$28.6   \\
3C~432 & 420$\pm$63 & 526$\pm$79 & 857$\pm$129 & 1490$\pm$224 & 2710$\pm$407 & 3940$\pm$591 & $\textless$  7.3 & $\textless$ 13.8 & 34.7$\pm$5.1 & 25.2$\pm$5.1 & 30.7$\pm$5.8   \\
3C~437 & 82$\pm$12 & 85$\pm$13 & 97$\pm$15 & 80$\pm$12 & 384$\pm$58 & 941$\pm$141 & $\textless$  6.4 & $\textless$ 18.4 & $\textless$ 17.0 & $\textless$ 14.3 & $\textless$ 18.9   \\
3C~454.0 & 339$\pm$51 & 481$\pm$72 & 811$\pm$122 & 1220$\pm$183 & 2490$\pm$374 & 4150$\pm$623 & 15.7$\pm$2.2 & 39.5$\pm$5.2 & 31.0$\pm$5.6 & 35.3$\pm$5.2 & 28.0$\pm$7.2   \\
3C~454.1 & 77$\pm$12 & 76$\pm$11 & 112$\pm$17 & 135$\pm$20 & 612$\pm$92 & 1500$\pm$225 & 13.7$\pm$2.5 & 37.0$\pm$4.7 & 50.2$\pm$8.7 & 26.7$\pm$10.4\tablefootmark{a} & $\textless$ 50.0   \\
3C~469.1 & 160$\pm$24 & 244$\pm$37 & 509$\pm$76 & 1090$\pm$164 & 3270$\pm$491 & 1970$\pm$296 & 10.9$\pm$2.3 & 24.3$\pm$4.6 & $\textless$ 20.9 & $\textless$ 19.1 & $\textless$ 21.8   \\
3C~470 & 50$\pm$7 & 75$\pm$11 & 72$\pm$11 & 266$\pm$40 & 1510$\pm$227 & 2650$\pm$398 & 16.0$\pm$2.7 & 29.3$\pm$5.1 & 48.0$\pm$6.5 & 36.3$\pm$5.2 & $\textless$ 21.5   \\
4C~13.66 & 24$\pm$4 & 24$\pm$4 & 21$\pm$3 & 18$\pm$3 & $\textless$  260 & 276$\pm$41 & $\textless$  5.5 & $\textless$ 13.4 & $\textless$ 15.6 & $\textless$ 13.9 & $\textless$ 17.9   \\
4C~16.49 & 329$\pm$49 & 420$\pm$63 & 573$\pm$86 & 743$\pm$111 & 1070$\pm$161 & 1830$\pm$275 & $\textless$  5.0 & $\textless$ 17.3 & $\textless$ 18.5 & $\textless$ 18.7 & $\textless$ 23.2   \\
  \end{longtable}
     \tablefoot{\\
   \tablefoottext{a}{Less than $3\sigma$ detection entering our SED fitting routine.}\\
   \tablefoottext{b}{Photometric measurement hindered by the presence of a nearby source.}
             }
   \end{landscape}
   \end{longtab}
\subsubsection{SPIRE}
\label{subsection:SPIRE}
   SPIRE \citep{Griffin10} photometric observations were carried out in 
   small scan-map observational mode, at 250 (18.2$\arcsec$ angular 
   resolution), 350 (24.9$\arcsec$ angular resolution) and 500~$\mu$m 
   (36.3$\arcsec$ angular resolution). Data reduction was performed in 
   HIPE following standard procedures for SPIRE data. Source extraction 
   on the fully reduced map was performed using the 
   \textit{sourceExtractorSussextractor} task \citep{Savage&Oliver07}. 
   Extracted sources located within half the Full Width at Half Maximum 
   (FWHM) of the given SPIRE array (measured from the radio core position 
   of the sources) were selected as tentative detections. 

   While SPIRE does not suffer from correlated noise, SPIRE observations 
   are dominated by confusion noise, of the order of 6-7 mJy beam$^{-1}$, 
   as estimated from deep extragalactic observations \citep{Nguyen10}. 
   Our adopted procedure for the determination of the photometric uncertainties 
   in the SPIRE maps is fully described by \citet{Leipski13}, 
   which in turn follows the procedures presented by \citet{Elbaz11} and 
   \citet{Pascale11}. Initially, an artificial source free 
   map was created by removing the extracted sources in a SPIRE map 
   from the map itself. Then, the pixel-to-pixel rms in this source 
   free map was calculated using a box centred on the nominal position 
   of the target object. The size of the box was chosen as a compromise 
   between avoiding the noisy edges of the SPIRE map and obtaining 
   proper statistics of the immediate environment of the target object.

   As indicated in Table~\ref{table:HerscPhot}, three SPIRE 
   detections are formally below the estimated 3$\sigma$ values. These 
   particular measurements were included in the subsequent analyses 
   because, upon visual inspection, they showed obvious emission at the 
   known position of the target object. The availability of ancillary 
   multi-wavelength data at shorter wavelengths (to check for 
   source confusion), and the understanding of the overall shape of the 
   object's SED, further support the inclusion of these flux densities 
   in the subsequent analyses. While the formal signal-to-noise ratio 
   of one of these three detections is very close to three, the other 
   two detections do not reach this ratio only because the associated 
   SPIRE maps are less clean than other maps in the sample, 
   leading to significantly larger photometric uncertainties.
   The SPIRE 500~$\mu$m photometry should be considered tentative 
   because the beam at this particular wavelength is large, and 
   undetected sources in the region surrounding the AGN may contribute 
   to the measured flux density. The SPIRE photometric uncertainties 
   provided in Table~\ref{table:HerscPhot} do not include the 4\% 
   uncertainty on the absolute flux calibration \citep{Bendo13}. Postage 
   stamps of the resulting SPIRE maps, centred on the radio position of 
   the objects, are included in Appendix~\ref{appendix:stamps}.
   \subsection{Supplementary Data}
   The FIR photometry of all 3CR sources in our work was supplemented 
   with MIR photometry obtained with the \textit{Spitzer Space Telescope} 
   \citep{Werner04} during three \textit{Spitzer} GT observing programmes 
   (PI: G. Fazio) in six bands, using the instruments IRAC \citep{Fazio04}, 
   IRS-16 peak-up array \citep{Houck04}, and MIPS \citep{Rieke04}. 
   Details on the \textit{Spitzer} data reduction and photometry have 
   previously been published by \citet{Haas08}. Table~\ref{table:HerscPhot} 
   lists the \textit{Spitzer} photometry. When available, additional 
   850~$\mu$m data were collected from the literature. The 850~$\mu$m 
   emission in quasars can be heavily contaminated by synchrotron 
   contribution (see Sect.~\ref{subsection:synchrotron}). The quasar 
   850~$\mu$m thermal flux densities utilized in this work were taken 
   from \citet{Haas06}. To obtain the 850~$\mu$m thermal flux densities of 
   quasars, \citet{Haas06} extrapolated the synchrotron contribution at 
   850~$\mu$m using the measured radio core flux densities, and subtracted 
   it from the total flux density at 850~$\mu$m. Table~\ref{table:submmtable} 
   lists the radio galaxy and quasar thermal submm flux densities used 
   in this work.  
\section{Spectral energy distributions}
\label{section:SEDs}
   \subsection{Fitting components}
   \begin{figure*}
      \centering
      \includegraphics[width=9.66cm, bb = 54 360 558 719, clip]{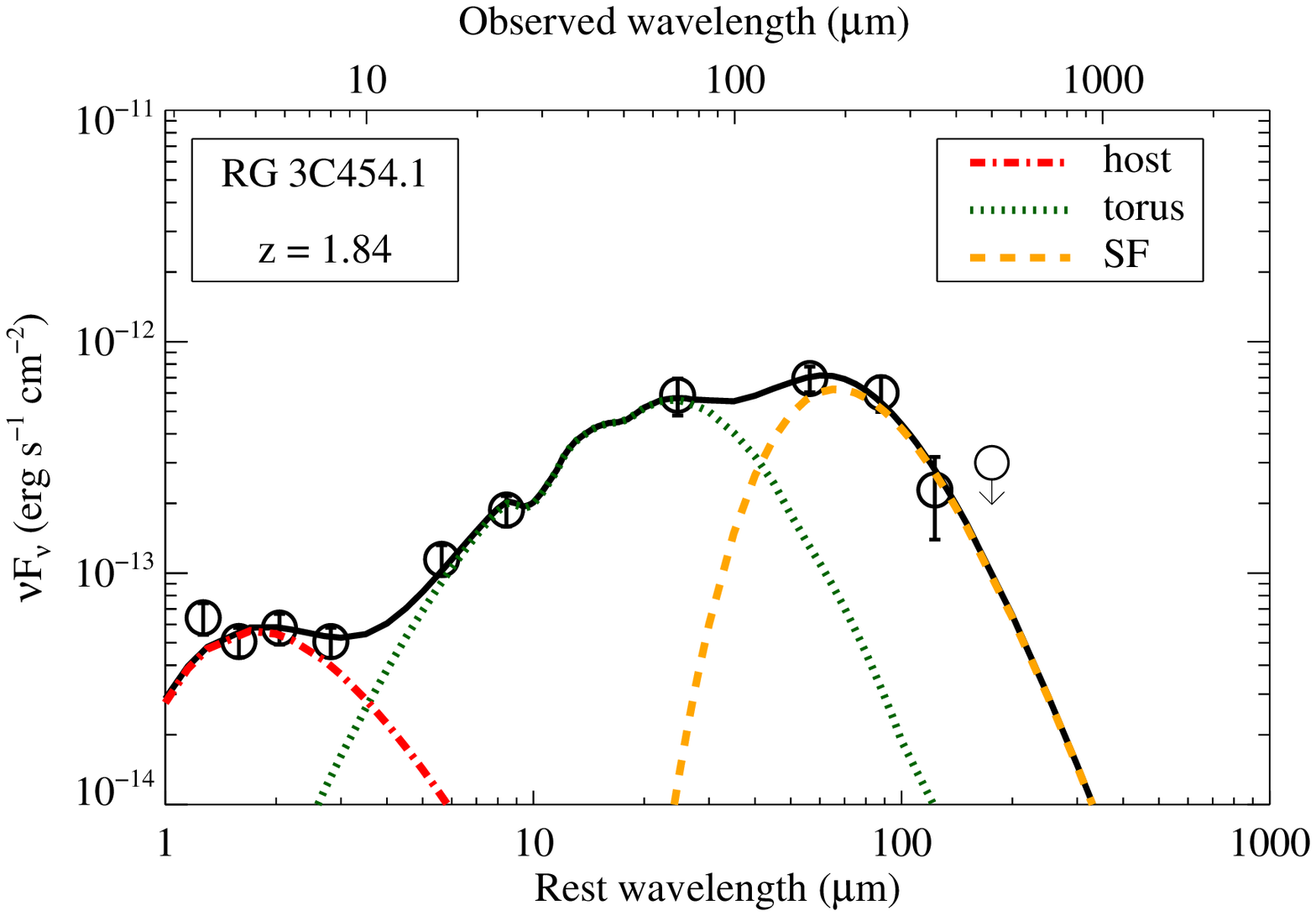}
      \includegraphics[width=8.34cm, bb = 124 360 558 719, clip]{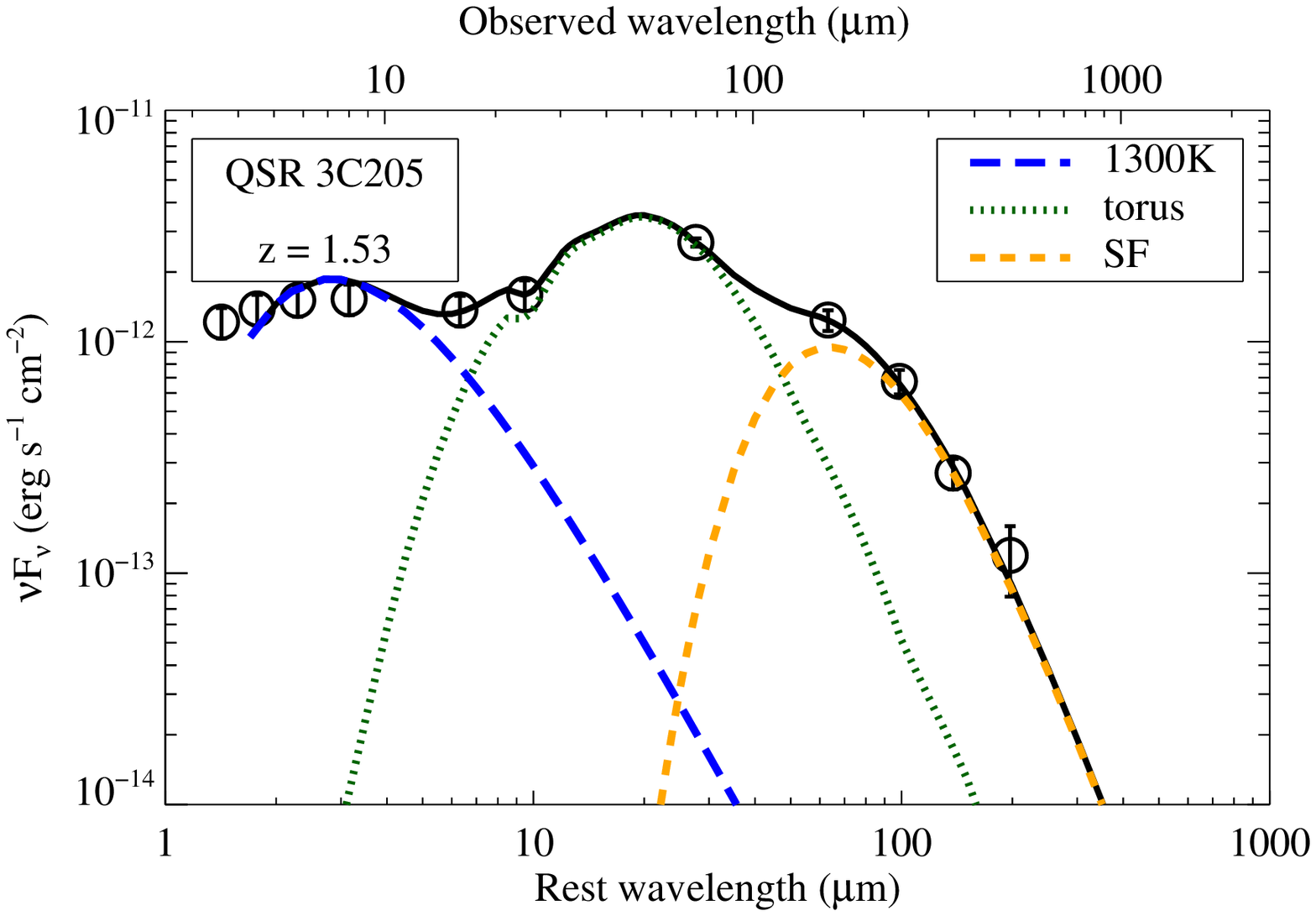}
      \caption{IR spectral energy distributions (SEDs, solid black) for two 
               representative objects from this work. Open circles show the 
               photometric data. Error bars correspond to 1~$\sigma$ 
               photometric uncertainties. Arrows indicate 3~$\sigma$ upper limits. 
               Left panel: 3C~454.1, a radio galaxy at $z$=1.84. The three 
               components used to fit the SEDs of radio galaxies account for 
               emission from host galaxy (old) stars (dash-dotted red), from 
               an AGN-heated torus (dotted green), and from dust heated by 
               star formation (dashed yellow). Right panel: 3C~205, a quasar 
               at $z$=1.53. The multi-component approach used to fit quasars 
               accounts for emission from hot (1300 K) dust (long-dashed blue), 
               from the AGN-heated torus (dotted green) and from the star formation 
               heated dust (dashed yellow). 
               }
      \label{figure:exampleSEDs}
   \end{figure*}
   The estimation of physical properties for the active galaxies was 
   performed using an SED fitting technique. Our fitting routine 
   is based on a combination of several distinct components, responsible 
   for the emission from active galaxies in different wavelength regimes. 
   Below, we describe this multi-component approach of fitting the 
   observed IR SEDs. 
   
   The presence of circumnuclear dust surrounding the broad line regions in 
   AGN and blocking their UV/visible emission is central to orientation-based 
   unification of powerful FRII radio galaxies and quasars \citep{Barthel89,Antonucci93}.
   Given its proximity to the AGN, the emission from this AGN-illuminated 
   warm dust peaks at rest-frame MIR (10-40~$\mu$m) wavelengths 
   \citep[e.g.][]{Rowan-Robinson95}. \textit{Spitzer} photometric and 
   spectroscopic data have shown that the majority of high-$z$ 3CR objects 
   are luminous MIR emitters \citep{Haas08,Leipski10}, with observed MIR 
   luminosities L$_{15 \mu \mathrm{m}}$ much higher than 
   $8~\times~10^{43}$~ergs~s$^{-1}$, the value separating hidden quasars 
   from mid-IR weak radio galaxies at intermediate redshifts \citep{Ogle06}. 
   There exists broad agreement that the AGN-heated 
   nuclear dust is mainly located in clumps which are distributed in a 
   toroidal pattern altogether referred to as the AGN torus 
   \citep[e.g.][]{Nenkova02,Kuraszkiewicz03,Hoenig06,Schartmann08}. 
   To account for the emission from the AGN heated dust, we chose the 
   library of torus models from \citet{Hoenig&Kishimoto10}. The parameters 
   considered when creating this library include the (1) radial dust 
   distribution of dust clumps; (2) geometric thickness of the torus; 
   (3) number of clumps along an equatorial line of sight; (4) optical 
   depth of the individual clumps; and (5) outer radius of the torus. 
   There are 240 sets of parameters in the library, each computed for 
   seven viewing-angles ranging from face-on ($i = 0^{\degr}$) to edge-on 
   ($i = 90^{\degr}$) in steps of 15 degrees, resulting in a total of 
   1680 torus models. Detailed information on the model parameters and 
   the adopted strategy in generating the tori SEDs is provided by 
   \citet{Hoenig&Kishimoto10}. In addition to the parameters listed above, 
   the overall flux normalization of the model is another free parameter 
   throughout the fitting procedure (outlined below).  

   The rest-frame FIR emission (40-500~$\mu$m) is largely generated by cold 
   dust, heated by star formation on kpc-scales in the AGN host 
   \citep[e.g.][]{Rowan-Robinson95,Schweitzer06,Netzer07}. Following these 
   authors, we interpret any FIR emission (in excess of the emission of the 
   AGN-heated dust) as being powered by star formation, and we represent it 
   with an optically thin modified blackbody component, i.e., a blackbody 
   modified by frequency-dependent emissivity, given by 
   \begin{equation}
   S_{\nu} \propto B_{\nu}(T)\nu^{\beta}.
   \end{equation} 
   We reduce the number of free parameters 
   in this component by fixing the dust emissivity index $\beta$ to a typical 
   value of 1.6 as found in studies of high-$z$ AGN \citep[e.g.][]{Beelen06}. 
   The remaining two free parameters here are the cold dust temperature and 
   the flux normalization of the modified blackbody component. The use of 
   a modified blackbody, as opposed to starburst templates \citep[e.g.][]{Drouart14}, 
   might slightly underestimate the star formation luminosities because one 
   misses the starburst MIR emission, but offers the unique possibility of 
   constraining the cold dust temperatures.

   The two components described above feature in the SED fitting of both 
   radio galaxies and quasars. We include additional SED components to 
   the fitting depending on the type of the studied object. For radio 
   galaxies, we added a blackbody component peaking in the near-IR to 
   account for the emission from the old stellar population in the AGN host 
   \citep[e.g.][]{Seymour07}. The temperature of the blackbody and its 
   flux normalization are the two free parameters for this SED component.  
   For quasars, we added a blackbody component to account for the hot 
   (graphite) dust close to the sublimation temperature. This component 
   is often empirically required to fit the observed SEDs of quasars 
   \citep[e.g.][]{Mor12,Leipski13}. Following these authors, we fixed 
   the temperature of the blackbody to 1300 K, leaving its flux 
   normalization as the only free parameter during the fitting. Such a 
   component is also needed in the fitting of the SEDs of a few radio 
   galaxies (see also B12) whose observed photometry in the 
   NIR/MIR could not be well represented with the components described 
   above. These radio galaxies, indicated in Table~\ref{table:PhysProp}, 
   might be viewed along lines of sight at which the nuclear region is 
   only partly obscured, thus resulting in somewhat elevated MIR 
   luminosities. The inclusion of the hot dust component to the SEDs of 
   some of the radio galaxies might lower the estimates of the mass of 
   the evolved stellar populations, but this is outside the scope of 
   this work. For the quasars we also considered an additional 
   power-law component representing the emission from an accretion disk 
   in the UV/visible. However, as demonstrated in 
   Appendix~\ref{appendix:UV/optical}, the inclusion of this power-law 
   component had little influence on the results obtained from the FIR 
   part of the SED, therefore it was excluded from the fitting procedure. 
   \subsection{Fitting procedure}
   While our physically motivated fitting approach results in a close 
   approximation of the observed SEDs of the sample objects, it is not 
   primarily designed to yield precise models of their SEDs. In particular, 
   we are not interested in constraining the properties of the dusty torus 
   with our multi-wavelength broad-band photometry. This kind of analysis 
   remains challenging even at lower redshifts \citep[e.g.][]{Ramos-Almeida09}. 
   Torus models are used to separate AGN-heated dust emission (peaking 
   in MIR) from star-formation-heated dust emission (peaking in FIR), and 
   to determine for the first time the star-formation-dominated FIR energetics 
   of the high-$z$ 3CR sources. When fitting the observed SEDs of our objects, 
   we used a chi-square minimization technique based on the MPFIT routine 
   \citep{Markwardt09}. In practice, we started with a torus model from the 
   library of \citet{Hoenig&Kishimoto10} and added a linear combination of 
   the remaining SED components (depending on the object type) to minimize 
   the overall chi-square. We repeated the procedure for each torus model 
   in the library.  
   
   Example best-fit SEDs, along with their individual SED components, 
   are shown in Fig.~\ref{figure:exampleSEDs}. At the redshifts of our 
   sample, the PACS 70~$\mu$m band is crucial in our adopted fitting 
   approach, as it strongly constrains the longer wavelengths of the torus 
   emission. On the other hand, the SPIRE 250~$\mu$m band is the most 
   important measurement for constraining the component representing the 
   cold dust emission. Therefore, our fitting approach was applied to all 
   objects that are detected in at least three \textit{Herschel} photometric 
   bands (typically the three shortest \textit{Herschel} bands). These 
   objects are homogeneously distributed in the redshift range ($1<z<2.5$) 
   studied in this work (Fig.~\ref{figure:sampleDescr}). Best-fit SEDs 
   for these objects, together with images centred on the radio positions 
   of the AGN, are presented in Appendix~\ref{appendix:FIR-detectedSEDs} 
   and in Appendix~\ref{appendix:stamps}, respectively. Occasional SED 
   mismatches at observed-frame 16~$\mu$m and/or 24~$\mu$m are most 
   probably due to luminous polycyclic aromatic hydrocarbon emission 
   and/or the 10~$\mu$m silicate absorption \citep{Haas08,Leipski10}. 
   At FIR wavelengths, the fixed $\beta$ approach may be the main reason 
   behind the failure of the fitting routine to exactly reproduce some 
   SPIRE data points. 
         
   The physical parameters constrained by our fitting method include 
   IR star formation and IR AGN luminosity, and the temperature and 
   mass of the cold dust component (Fig.~\ref{figure:MixedHist}, 
   Table~\ref{table:PhysProp}). Uncertainties in the derived parameters 
   were calculated by resampling the observed SEDs, allowing the individual 
   photometric measurements to vary within their 1$\sigma$ ranges of uncertainty. 
   More precisely, we generated 100 mock observed SEDs and studied the 
   distributions of the parameters derived from the corresponding best 
   fits. We inspected the best fits to the mock SEDs to confirm their 
   overall quality in the different wavelength regimes. From the 
   distributions, we retained the median values as the best estimates of 
   the parameters, and the 16th-84th percentile ranges as their associated 
   uncertainties (which in case of a Gaussian distribution would correspond 
   to $\pm 1\sigma $ values). 
   
   For objects with fewer than three \textit{Herschel} detections, we 
   estimated upper limits for the IR star formation and IR AGN luminosities 
   using two different approaches. In the first approach, we fitted the 
   (longer wavelength) \textit{Herschel} upper limits using a modified 
   blackbody with fixed $\beta = 1.6$ value and fixed cold dust temperature 
   ($T_{\mathrm{dust}} = 37$~K), typically found from the fits of the objects 
   detected in at least three \textit{Herschel} bands (see below). We then 
   integrated under the blackbody component to estimate upper limits for 
   the star formation luminosities. In the second approach, we took the 
   70/160/250~$\mu$m upper limits as tentative detections, and estimated upper 
   limits for the star formation luminosities using the procedure adopted 
   for the objects detected in at least three \textit{Herschel} bands. Both 
   approaches yielded similar results (within 10\%) for the star formation 
   luminosities, but we retained the second approach because it allowed us to 
   estimate the IR AGN luminosities for objects with only PACS 70~$\mu$m 
   detections. Best-fit SEDs for objects with fewer than three \textit{Herschel} 
   detections, together with images centred on the radio positions of the AGN, 
   are presented in Appendix~\ref{appendix:nondetectedSEDs} and in 
   Appendix~\ref{appendix:stamps}, respectively. 
   Systematically demanding three \textit{Herschel} detections 
   when fitting the observed photometry regardless of redshift means that 
   the FIR results for the objects detected in only the two PACS bands are 
   treated as upper limits. However, depending on the object's redshift, 
   the PACS 160~$\mu$m band alone may probe the peak of the cold dust 
   emission, allowing robust constraints for the physical parameters 
   estimated in this work.
   \subsection{Synchrotron contribution}
   \label{subsection:synchrotron}
   Earlier submillimetre and millimetre studies of high-redshift 3CR 
   sources presented clear evidence for a synchrotron contribution to the 
   observed flux densities \citep{VanBemmel97,Archibald01,Willott02,Haas06}. 
   While negligible for radio galaxies, extrapolation from core radio data 
   shows that synchrotron emission can account for up to 80\% of the 
   observed submm flux densities from quasars \citep{Haas06}. However, 
   the longest rest-frame wavelengths probed by the SPIRE 500~$\mu$m 
   band in our study are $\sim 200~\mu$m: emission at these wavelengths 
   is completely dominated by dust and therefore free from any synchrotron 
   contribution. The only exception is 3C~418. This source appears to be 
   flat-spectrum-core-dominated (in an otherwise steep-spectrum selected 
   sample), and as such its FIR SED is clearly dominated by non-thermal 
   (synchrotron) radiation from its core. The power-law like IR SED of 
   this source is shown in Appendix~\ref{appendix:FIR-detectedSEDs}. 
   3C~418 was removed from the subsequent analyses. 
   \begin{table}
      \renewcommand{\arraystretch}{1.2}
      \centering                       
      \caption[]{\label{table:submmtable}Objects with significant thermal submillimetre flux densities.}
      \begin{tabular}{lcc}
      \hline \hline
      Object & Thermal F$_{850 \mu \mathrm{m}}$ & Reference \\
             & (mJy)                            & \\ 
      \hline
      3C~191   & 2.95 & 1, 2 \\
      3C~257   & 5.40 & 3 \\
      3C~280.1 & 2.48 & 1, 2 \\
      3C~298   & 7.25 & 1, 2 \\
      3C~368   & 4.08 & 3 \\
      3C~432   & 6.33 & 1, 2 \\      
      3C~470   & 5.64 & 3 \\
      4C~13.66 & 3.53 & 3 \\
      \hline
      \end{tabular}
      \tablebib{(1)~\citet{Willott02}; (2) \citet{Haas06}; \\
      (3) \citet{Archibald01}.}
   \end{table}  
\section{Results}
\label{section:Results}
   \subsection{Detection statistics}
   The \textit{Herschel} detection rate throughout our sample ranges 
   from 67\% in the PACS 70~$\mu$m band to 13\% in the SPIRE 500~$\mu$m 
   band. In particular, 7 objects have robust detections in all five 
   \textit{Herschel} bands. Furthermore, a total of 24 objects are 
   detected in at least three \textit{Herschel} bands, most importantly 
   in the SPIRE 250~$\mu$m band, which for the highest redshift of our 
   sample (3C~257: $z=2.47$) corresponds to $\sim 70~\mu$m rest-frame 
   emission\footnote{For the median redshift of our sample, ($z_{med}$=1.38), 
   the SPIRE 250~$\mu$m band samples the peak ($\sim 100~\mu$m) of the 
   typical cold dust SED, allowing strong constraints on the modified 
   blackbody component used in the fitting approach.}. Excluding 3C~418 
   from these 24 objects, results in 13 radio galaxies and 10 quasars. 
   The \textit{Spitzer} detection rate throughout our sample ranges from 
   94\% in the IRS 16~$\mu$m and MIPS 24~$\mu$m bands to 100\% in the 
   IRAC 5.8~$\mu$m band. Comments on selected individual objects are 
   included in Appendix~\ref{appendix:Comments}.  
   \subsection{Physical properties obtained from the SED fitting}
\label{subsection:PhysProp}
   \begin{figure*}
      \centering
      \includegraphics[width=17cm]{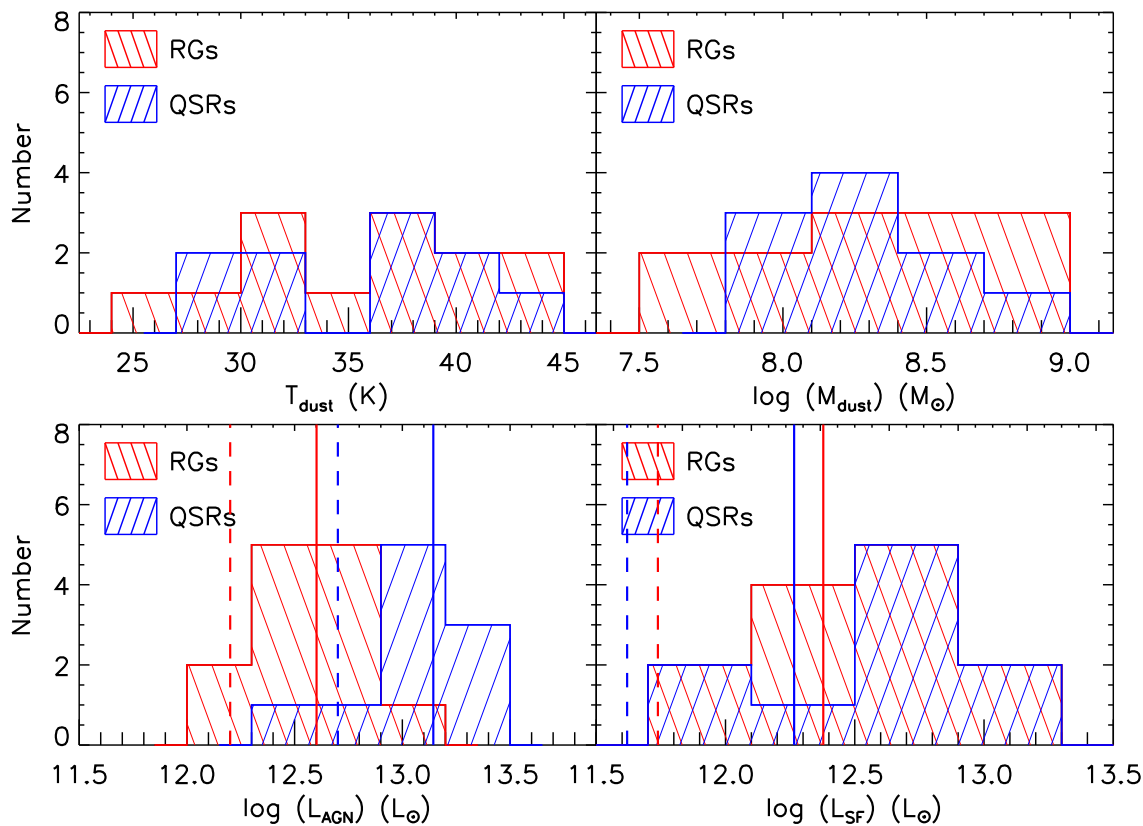}
      \caption{Distributions of individual physical parameters obtained 
               from the SED fits, for radio galaxies (red) and quasars 
               (blue). The values plotted are those for the objects 
               detected in at least three \textit{Herschel} bands.
               Upper left panel: temperature of the cold dust component 
               emitting in the FIR, T\textsubscript{dust}. Upper right 
               panel: mass of the FIR emitting cold dust component, 
               M\textsubscript{dust}. Lower left panel: AGN-powered IR 
               luminosity, L$_{\mathrm{AGN}}$. Lower right panel: 
               star-formation-powered IR luminosity, L\textsubscript{SF}.
               The vertical lines in the lower panels correspond to the 
               average values of the FIR-detected (solid lines) and 
               non-detected (dashed lines) stacked subsamples discussed 
               in Sect.~\ref{subsection:Stacking}. 
               }
      \label{figure:MixedHist}
   \end{figure*}
\begin{longtab}
\renewcommand{\arraystretch}{1.2}
\begin{longtable}{lccccc}
\caption{\label{table:PhysProp} Physical parameters estimated from the SED fitting.} \\
\hline\hline
 Object & L$_{\mathrm{AGN}}$      & L$_{\mathrm{SF}}$       & SFR                              & M$_{\mathrm{d}}$       & T$_{\mathrm{d}}$ \\ 
       	& (10$^{12}$ L$_{\odot}$) & (10$^{12}$ L$_{\odot}$) & (10$^{2}$ M$_{\odot}$ yr$^{-1}$) & (10$^{8}$ M$_{\odot}$) & (K) \\  
    (1) & (2) & (3) & (4) & (5) & (6) \\
\hline
\endfirsthead
\caption{continued.}\\
\hline\hline
Object & L$_{\mathrm{AGN}}$      & L$_{\mathrm{SF}}$       & SFR                              & M$_{\mathrm{d}}$       & T$_{\mathrm{d}}$ \\ 
       & (10$^{12}$ L$_{\odot}$) & (10$^{12}$ L$_{\odot}$) & (10$^{2}$ M$_{\odot}$ yr$^{-1}$) & (10$^{8}$ M$_{\odot}$) & (K) \\  
   (1) & (2) & (3) & (4) & (5) & (6) \\
\hline
\endhead
\hline
\endfoot
\multicolumn{6}{c}{Objects detected in at least three \textit{Herschel} bands} \\
\hline
3C~002 & $ 2.1\substack{+ 0.1 \\ - 0.2}$ & $ 0.6\substack{+ 0.2 \\ - 0.1}$ & $ 1.1\substack{+ 0.4 \\ - 0.2}$ & $ 1.1\substack{+ 0.7 \\ - 0.7}$ & $31.6\substack{+ 9.2 \\ - 3.1}$\\
3C~014 & $16.1\substack{+ 0.5 \\ - 0.5}$ & $ 1.3\substack{+ 0.3 \\ - 0.2}$ & $ 2.3\substack{+ 0.5 \\ - 0.3}$ & $ 0.9\substack{+ 1.1 \\ - 0.4}$ & $37.5\substack{+ 5.1 \\ - 5.7}$\\
3C~068.2 & $ 6.6\substack{+ 0.4 \\ - 0.3}$ & $ 2.1\substack{+ 0.2 \\ - 0.3}$ & $ 3.7\substack{+ 0.4 \\ - 0.4}$ & $ 7.0\substack{+ 2.2 \\ - 2.0}$ & $28.5\substack{+ 2.3 \\ - 1.7}$\\
3C~119\tablefootmark{a} & $ 4.4\substack{+ 0.3 \\ - 0.4}$ & $ 0.9\substack{+ 0.2 \\ - 0.1}$ & $ 1.5\substack{+ 0.4 \\ - 0.3}$ & $ 0.7\substack{+ 2.8 \\ - 0.3}$ & $37.6\substack{+ 5.4 \\ - 9.7}$\\
3C~124 & $ 3.7\substack{+ 0.1 \\ - 0.1}$ & $ 1.5\substack{+ 0.1 \\ - 0.1}$ & $ 2.7\substack{+ 0.2 \\ - 0.3}$ & $ 2.6\substack{+ 0.6 \\ - 0.6}$ & $31.9\substack{+ 2.2 \\ - 1.3}$\\
3C~190 & $ 9.0\substack{+ 0.4 \\ - 0.4}$ & $ 2.7\substack{+ 0.1 \\ - 0.1}$ & $ 4.7\substack{+ 0.2 \\ - 0.1}$ & $ 4.9\substack{+ 0.2 \\ - 0.5}$ & $31.7\substack{+ 0.6 \\ - 0.3}$\\
3C~205 & $27.9\substack{+ 0.8 \\ - 0.8}$ & $ 4.1\substack{+ 0.3 \\ - 0.3}$ & $ 7.0\substack{+ 0.5 \\ - 0.6}$ & $ 1.8\substack{+ 0.3 \\ - 0.2}$ & $40.7\substack{+ 1.1 \\ - 1.8}$\\
3C~222 & $ 1.5\substack{+ 0.1 \\ - 0.1}$ & $ 2.5\substack{+ 0.1 \\ - 0.1}$ & $ 4.3\substack{+ 0.2 \\ - 0.2}$ & $ 5.4\substack{+ 0.7 \\ - 0.7}$ & $30.6\substack{+ 1.0 \\ - 0.8}$\\
3C~245 & $11.2\substack{+ 0.5 \\ - 0.4}$ & $ 0.8\substack{+ 0.1 \\ - 0.1}$ & $ 1.4\substack{+ 0.2 \\ - 0.2}$ & $ 2.4\substack{+ 1.0 \\ - 0.8}$ & $28.8\substack{+ 3.4 \\ - 2.0}$\\
3C~256 & $ 6.9\substack{+ 0.5 \\ - 0.5}$ & $ 2.6\substack{+ 0.4 \\ - 0.2}$ & $ 4.5\substack{+ 0.7 \\ - 0.4}$ & $ 0.9\substack{+ 0.7 \\ - 0.2}$ & $43.0\substack{+ 1.4 \\ - 4.3}$\\
3C~257\tablefootmark{a}\tablefootmark{d} & $ 8.0\substack{+ 0.4 \\ - 0.6}$ & $ 5.4\substack{+ 0.3 \\ - 0.3}$ & $ 9.2\substack{+ 0.6 \\ - 0.5}$ & $ 3.4\substack{+ 0.4 \\ - 0.4}$ & $38.2\substack{+ 1.1 \\ - 0.9}$\\
3C~266 & $ 1.0\substack{+ 0.2 \\ - 0.3}$ & $ 1.5\substack{+ 0.2 \\ - 0.2}$ & $ 2.7\substack{+ 0.3 \\ - 0.3}$ & $ 0.4\substack{+ 0.1 \\ - 0.1}$ & $44.2\substack{+ 0.7 \\ - 1.9}$\\
3C~270.1 & $12.7\substack{+ 0.5 \\ - 0.4}$ & $ 2.3\substack{+ 0.4 \\ - 0.2}$ & $ 3.9\substack{+ 0.6 \\ - 0.4}$ & $ 0.8\substack{+ 0.2 \\ - 0.2}$ & $43.5\substack{+ 1.2 \\ - 2.8}$\\
3C~297 & $ 2.2\substack{+ 0.1 \\ - 0.2}$ & $ 0.9\substack{+ 0.1 \\ - 0.1}$ & $ 1.6\substack{+ 0.2 \\ - 0.2}$ & $ 6.5\substack{+ 3.7 \\ - 2.2}$ & $24.9\substack{+ 1.7 \\ - 1.8}$\\
3C~298\tablefootmark{d} & $29.2\substack{+ 0.8 \\ - 0.8}$ & $ 5.4\substack{+ 0.2 \\ - 0.2}$ & $ 9.3\substack{+ 0.4 \\ - 0.3}$ & $ 3.8\substack{+ 0.3 \\ - 0.4}$ & $37.5\substack{+ 0.8 \\ - 0.8}$\\
3C~305.1\tablefootmark{a} & $ 3.4\substack{+ 0.2 \\ - 0.3}$ & $ 1.3\substack{+ 0.3 \\ - 0.1}$ & $ 2.2\substack{+ 0.4 \\ - 0.2}$ & $ 1.4\substack{+ 0.6 \\ - 0.7}$ & $34.7\substack{+ 6.7 \\ - 2.6}$\\
3C~318 & $ 7.6\substack{+ 0.4 \\ - 0.4}$ & $ 3.4\substack{+ 0.4 \\ - 0.3}$ & $ 5.8\substack{+ 0.6 \\ - 0.5}$ & $ 1.7\substack{+ 0.6 \\ - 0.4}$ & $39.6\substack{+ 2.7 \\ - 1.9}$\\
3C~324 & $ 3.6\substack{+ 0.2 \\ - 0.2}$ & $ 1.0\substack{+ 0.2 \\ - 0.2}$ & $ 1.8\substack{+ 0.4 \\ - 0.3}$ & $ 0.5\substack{+ 0.5 \\ - 0.1}$ & $39.6\substack{+ 3.3 \\ - 4.4}$\\
3C~368\tablefootmark{d} & $ 3.3\substack{+ 0.2 \\ - 0.2}$ & $ 2.0\substack{+ 0.2 \\ - 0.2}$ & $ 3.5\substack{+ 0.4 \\ - 0.3}$ & $ 1.4\substack{+ 0.4 \\ - 0.3}$ & $37.3\substack{+ 2.6 \\ - 1.6}$\\
3C~432\tablefootmark{d} & $ 9.3\substack{+ 0.7 \\ - 0.4}$ & $ 2.4\substack{+ 0.2 \\ - 0.2}$ & $ 4.2\substack{+ 0.4 \\ - 0.4}$ & $ 6.4\substack{+ 1.3 \\ - 1.4}$ & $29.8\substack{+ 1.1 \\ - 1.6}$\\
3C~454.0 & $10.5\substack{+ 0.4 \\ - 0.5}$ & $ 3.6\substack{+ 0.3 \\ - 0.2}$ & $ 6.2\substack{+ 0.6 \\ - 0.4}$ & $ 2.3\substack{+ 0.6 \\ - 0.5}$ & $38.3\substack{+ 2.0 \\ - 2.0}$\\
3C~454.1 & $ 5.2\substack{+ 0.5 \\ - 0.7}$ & $ 4.4\substack{+ 0.2 \\ - 0.4}$ & $ 7.5\substack{+ 0.4 \\ - 0.7}$ & $ 2.4\substack{+ 0.7 \\ - 0.7}$ & $39.3\substack{+ 2.2 \\ - 2.0}$\\
3C~470\tablefootmark{d} & $ 5.8\substack{+ 0.3 \\ - 0.3}$ & $ 3.0\substack{+ 0.2 \\ - 0.2}$ & $ 5.1\substack{+ 0.4 \\ - 0.3}$ & $ 4.5\substack{+ 0.5 \\ - 0.7}$ & $32.8\substack{+ 1.1 \\ - 1.0}$\\
\hline
\multicolumn{6}{c}{Objects detected in fewer than three \textit{Herschel} bands} \\
\hline
3C~009 & 17.1 & $<$ 1.8 & $<$ 3.1 &  & \\
3C~013 &  4.0 & $<$ 1.2 & $<$ 2.0 &  & \\
3C~036\tablefootmark{a}\tablefootmark{b} & $<$ 1.2 & $<$ 0.5 & $<$ 0.9 &  & \\
3C~043 &  2.8 & $<$ 1.2 & $<$ 2.0 &  & \\
3C~065\tablefootmark{a}\tablefootmark{b} & $<$ 1.4 & $<$ 0.6 & $<$ 1.0 &  & \\
3C~068.1\tablefootmark{c} &  8.6 & $<$ 0.7 & $<$ 1.1 &  & \\
3C~173\tablefootmark{b} &  0.4 & $<$ 0.3 & $<$ 0.5 &  & \\
3C~181\tablefootmark{c} &  5.2 & $<$ 0.8 & $<$ 1.5 &  & \\
3C~186\tablefootmark{c} &  4.7 & $<$ 0.5 & $<$ 0.8 &  & \\
3C~191 & 17.0 & $<$ 1.7 & $<$ 3.0 &  & \\
3C~194\tablefootmark{b} & $<$ 0.7 & $<$ 0.6 & $<$ 0.9 &  & \\
3C~204\tablefootmark{c} & $<$ 4.8 & $<$ 0.5 & $<$ 0.9 &  & \\
3C~208.0 & $<$ 3.6 & $<$ 0.6 & $<$ 1.0 &  & \\
3C~208.1\tablefootmark{a} & $<$ 1.6 & $<$ 0.4 & $<$ 0.7 &  & \\
3C~210\tablefootmark{a} &  5.2 & $<$ 1.9 & $<$ 3.3 &  & \\
3C~212 &  6.2 & $<$ 0.8 & $<$ 1.3 &  & \\
3C~220.2 &  6.3 & $<$ 0.7 & $<$ 1.2 &  & \\
3C~225A\tablefootmark{b} & $<$ 1.5 & $<$ 1.8 & $<$ 3.1 &  & \\
3C~230 &  3.1 & $<$ 1.9 & $<$ 3.3 &  & \\
3C~238\tablefootmark{b} & $<$ 0.7 & $<$ 0.5 & $<$ 0.9 &  & \\
3C~239\tablefootmark{b} & $<$ 3.0 & $<$ 1.7 & $<$ 2.9 &  & \\
3C~241\tablefootmark{b} &  2.1 & $<$ 0.8 & $<$ 1.4 &  & \\
3C~249\tablefootmark{b} & $<$ 0.9 & $<$ 0.8 & $<$ 1.5 &  & \\
3C~250 & $<$ 0.6 & $<$ 0.5 & $<$ 0.9 &  & \\
3C~252\tablefootmark{a} &  4.5 & $<$ 0.6 & $<$ 1.0 &  & \\
3C~255\tablefootmark{b} & $<$ 0.5 & $<$ 0.5 & $<$ 0.9 &  & \\
3C~267\tablefootmark{a} &  3.1 & $<$ 0.4 & $<$ 0.7 &  & \\
3C~268.4 & 16.2 & $<$ 0.8 & $<$ 1.4 &  & \\
3C~280.1\tablefootmark{c} & $<$ 5.4 & $<$ 1.1 & $<$ 1.9 &  & \\
3C~294\tablefootmark{b} & $<$ 2.3 & $<$ 1.7 & $<$ 2.9 &  & \\
3C~322 & $<$ 1.5 & $<$ 1.3 & $<$ 2.2 &  & \\
3C~325 & $<$ 2.5 & $<$ 0.7 & $<$ 1.2 &  & \\
3C~326.1 & $<$ 3.7 & $<$ 3.0 & $<$ 5.2 &  & \\
3C~356 &  1.9 & $<$ 0.8 & $<$ 1.3 &  & \\
3C~437\tablefootmark{b} & $<$ 1.6 & $<$ 1.2 & $<$ 2.1 &  & \\
3C~469.1\tablefootmark{a} &  3.3 & $<$ 1.3 & $<$ 2.2 &  & \\
4C~13.66 & $<$ 1.0 & $<$ 0.8 & $<$ 1.4 &  & \\
4C~16.49\tablefootmark{c} & $<$ 2.4 & $<$ 0.9 & $<$ 1.5 &  & \\
\end{longtable}
\tablefoot{(1) Name of object; (2) IR luminosity (integrated between 1~$\mu$m and 1000~$\mu$m) of the AGN powered dust emission, 
           i.e. torus component and sum of torus and hot dust components for RGs and QSRs, respectively;
           (3) IR luminosity (integrated between 8~$\mu$m and 1000~$\mu$m) of the modified blackbody component ($\beta = 1.6$) presumably 
           powered by star formation activity in the AGN host galaxy; 
           (4) star formation rate determined from the IR luminosity in (3), using the calibration derived in \citet{Kennicutt98}; 
           (5) mass of the modified blackbody component;
           (6) temperature of the modified blackbody component. \\
\tablefoottext{a}{For this radio galaxy, a blackbody component (1300 K) was included in the SED fitting.}\\
\tablefoottext{b}{This object was included in the stacking of non-detected radio galaxies.}\\
\tablefoottext{c}{This object was included in the stacking of non-detected quasars.}\\
\tablefoottext{d}{The emissivity index of the cold dust component, $\beta$, was estimated for this object.}
           } 
\end{longtab}
   Studies investigating the cold dust temperatures in high-$z$ objects 
   prior to \textit{Herschel} were often uncertain because they relied 
   heavily on observations in a single photometric broadband 
   \citep[e.g.][]{Benford99,Beelen06}. In our current study, cold dust 
   temperatures estimated for the FIR-detected objects range from $\sim$~25 
   to $\sim$~45~K (Fig.~\ref{figure:MixedHist}, Table~\ref{table:PhysProp}). 
   Radio galaxies and quasars span the same range in cold dust temperatures, 
   This range is similar to that obtained for $z > 5$ quasars \citep{Leipski14} 
   and to that estimated for distant submm galaxies \citep[e.g.][]{Magnelli12}. 
   Similarly to \citet{Leipski13}, the inclusion of the 1300 K hot dust 
   component in the SED fitting lowers the estimates of the cold dust 
   temperatures by $\sim$5~K. By including the hot dust component, we 
   preferentially select torus models that emit more of their energy at 
   longer wavelengths. As a consequence, the cold dust components are 
   also shifted to colder temperatures.
   
   Central to the subsequent discussion are the star formation  
   luminosities, L\textsubscript{SF}, which we computed (or estimated 
   upper limits) by integrating the best-fit modified blackbody components 
   from 8~$\mu$m through 1000~$\mu$m. Figure~\ref{figure:MixedHist} 
   shows the distribution of L\textsubscript{SF} for the radio galaxies 
   and quasars detected in at least three \textit{Herschel} bands. Both 
   types of objects show similarly broad distributions, with many objects 
   having L\textsubscript{SF} > 10$^{12}$ L$_{\odot}$, characterizing them 
   as ultra-luminous infrared galaxies (ULIRGs) 
   (see also Table~\ref{table:PhysProp}). The median star formation 
   luminosities of the RGs and QSRs plotted in Fig.~\ref{figure:MixedHist} 
   are $2.0 \times 10^{12}$ L$_{\odot}$ and $2.6 \times 10^{12}$ L$_{\odot}$, 
   respectively. Converting the star formation luminosities into SFRs 
   using the calibration derived by \citet{Kennicutt98} gives 
   100 M$_{\odot}$ yr$^{-1}$ < SFR < 1000 M$_{\odot}$ yr$^{-1}$, 
   consistent with SFRs obtained for typical submm galaxies (SMGs) at 
   comparable redshifts \citep[e.g.][]{Magnelli12}.   

   To compute the IR luminosities of the components powered by the AGN, 
   L$_{\mathrm{AGN}}$, we integrated the best-fit torus 
   component for the RGs and the sum of the best-fit hot dust and torus 
   component for the QSRs\footnote{A few RGs also require an additional 
   hot dust component to better fit their observed photometry. The 
   computation of L$_{\mathrm{AGN}}$ for these objects takes into 
   account this component as well.} between 1~$\mu$m and 1000~$\mu$m. 
   Figure~\ref{figure:MixedHist} shows that the RGs and QSRs occupy 
   different ranges in the distributions of L$_{\mathrm{AGN}}$, with the 
   distribution for QSRs shifted to higher values compared to that for 
   RGs. \citet{Haas08} already established this result by investigating 
   the (energetically important) rest 1.6-10~$\mu$m wavelength range for 
   the high-$z$ 3CR objects, finding the QSRs to be, on average, 3-10 
   times more luminous than RGs. We confirm their finding by including 
   rest-frame wavelengths longer than 10~$\mu$m. The median IR AGN 
   luminosities of the RGs and QSRs plotted in Fig.~\ref{figure:MixedHist} 
   are $3.7 \times 10^{12}$ L$_{\odot}$ and $1.1 \times 10^{13}$ L$_{\odot}$, respectively.  
   \begin{figure}
      \centering
      \includegraphics[width=\hsize]{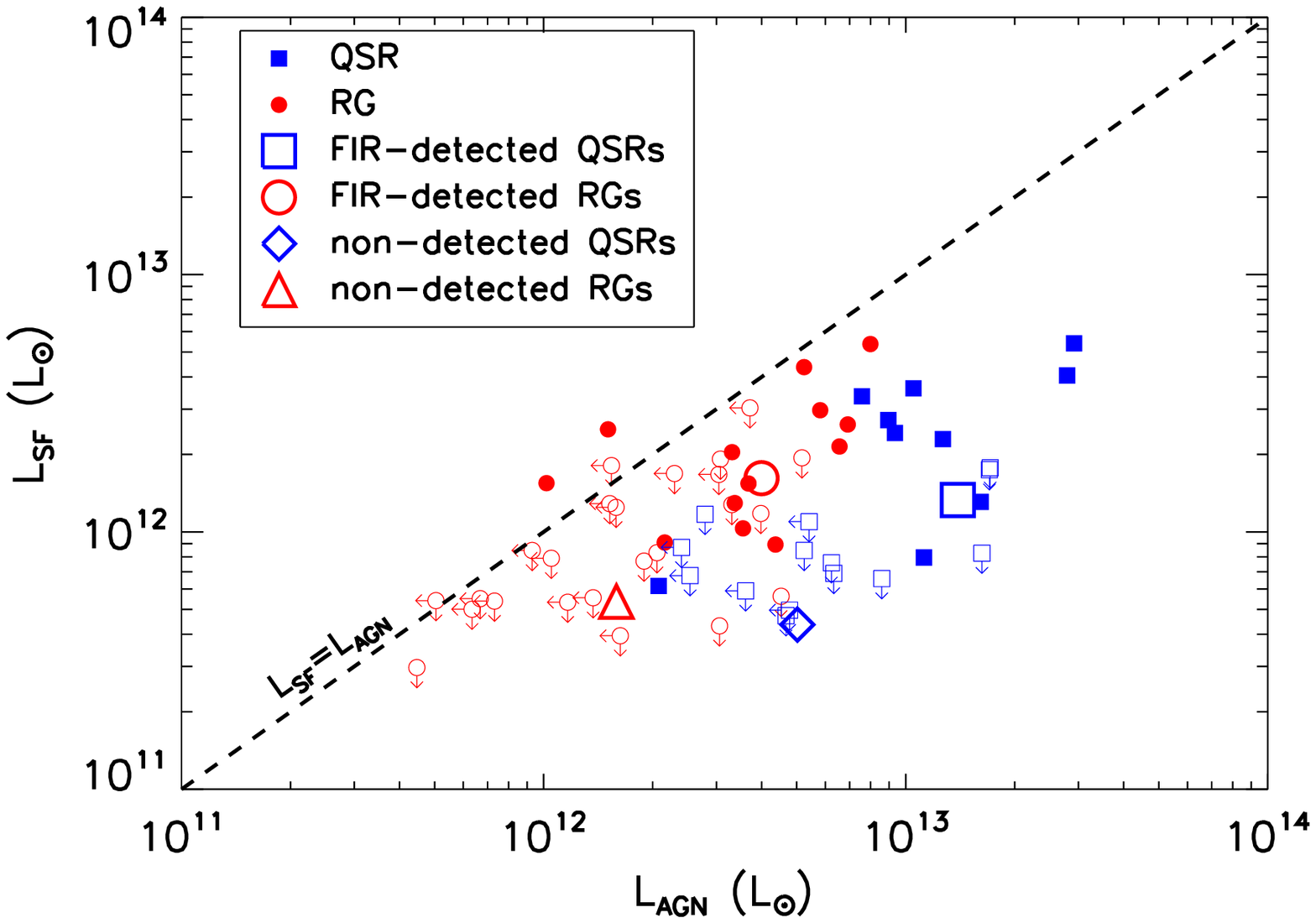}
      \caption{IR emission from star-formation-heated cold dust, 
               L\textsubscript{SF}, versus IR emission from AGN-powered 
               dust, L$_{\mathrm{AGN}}$, for radio galaxies (filled red 
               circles) and quasars (filled blue squares). Upper limits 
               have been estimated as explained in the text. The large empty 
               symbols correspond to the subsamples of FIR-detected RGs 
               (circle) and QSRs (square), and non-detected RGs (triangle) 
               and QSRs (diamond) discussed in Sect.~\ref{subsection:Stacking}. 
               The dashed line marks L\textsubscript{SF} $=$ L$_{\mathrm{AGN}}$.  
               }
      \label{figure:LIRAGN}
   \end{figure}
 
   As outlined in previous sections, we attribute the FIR emission in 
   excess of the AGN-powered dust emission to emission from 
   star-formation-heated dust. Figure~\ref{figure:LIRAGN} shows L\textsubscript{SF} 
   as a function of L$_{\mathrm{AGN}}$ for all objects in our sample. 
   The presence/absence of correlation between these two parameters 
   depends on both redshift and AGN luminosity, and is still 
   debated in the literature \citep{Lutz14}. Given the data, and taking into 
   account only the FIR-detected objects, the two plotted parameters show 
   at most a weak correlation, in part introduced by the dependence 
   of both L\textsubscript{SF} and L$_{\mathrm{AGN}}$ on redshift. 
   Moreover, the numerous upper limits, together with the fact that 
   both parameters span only a limited range ($\sim$ 2 orders of 
   magnitude), make it difficult to establish any such correlation (or 
   lack of) in our sample. Nevertheless, we observe a range of L\textsubscript{SF} 
   from weak (if not absent) to very strong, coeval with the growth of 
   the black hole. Figure~\ref{figure:LIRAGN} also shows that the hosts 
   of even the strongest AGN can have significant star formation activity, 
   unlike the trends found by \citet{Page12} for radio-quiet AGN. In general, 
   the total IR emission from the 3CR AGN is predominantly AGN powered, 
   despite the frequently accompanying strong star formation activity. 
   
   We estimated the mass of the FIR emitting dust component, M$_{\mathrm{dust}}$, using
   \begin{equation}
   M_{\mathrm{dust}} = \frac{S_{250\mu \mathrm{m}}D_{L}^2}{\kappa_{250\mu m}B_{\nu}(250 \mu \mathrm{m}, T_{\mathrm{dust}})}, 
   \end{equation} 
   where $S_{250\mu \mathrm{m}}$ is the flux at 250~$\mu$m rest-frame 
   found from the best-fit, D\textsubscript{L} is the luminosity distance, 
   $\kappa_{250\mu \mathrm{m}}$ is the dust absorption coefficient at 250~$\mu$m 
   \citep[$\kappa_{250\mu \mathrm{m}}$ = 4 $\mathrm{cm^2 g^{-1}}$ from the models of][]{Draine03}, 
   and $B_{\nu}(250 \mu \mathrm{m}, T_{\mathrm{dust}})$ is the value of the Planck function 
   at the corresponding rest-frame wavelength and temperature. 
   Results are shown in Table~\ref{table:PhysProp}. Given that the RGs 
   and QSRs in our sample cover roughly the same redshift range and show 
   similar star formation properties, it is no surprise that their cold 
   dust masses are comparable as well. More interestingly, the masses of 
   the cold dust component in the hosts of radio-loud AGN detected in 
   at least three \textit{Herschel} bands are comparable to those obtained 
   for SMGs at redshifts similar to those of our sample \citep[e.g.][]{Santini10}.
   The dust masses of radio-loud AGN provide clues to the triggering of the 
   starburst event \citep[and also that of the black hole activity,][]{Tadhunter14}. 
   Given that the high-$z$ SMGs are likely undergoing strong merger-induced 
   starburst events \citep[e.g.][]{Kartaltepe12}, their similar dust content suggests that that FIR-luminous 
   radio-loud AGN also build up their stellar mass in major gas-rich mergers. 
   \subsection{Median SEDs} 
   \label{subsection:Average SEDs}
   \begin{figure}
      \centering
      \includegraphics[width=\hsize]{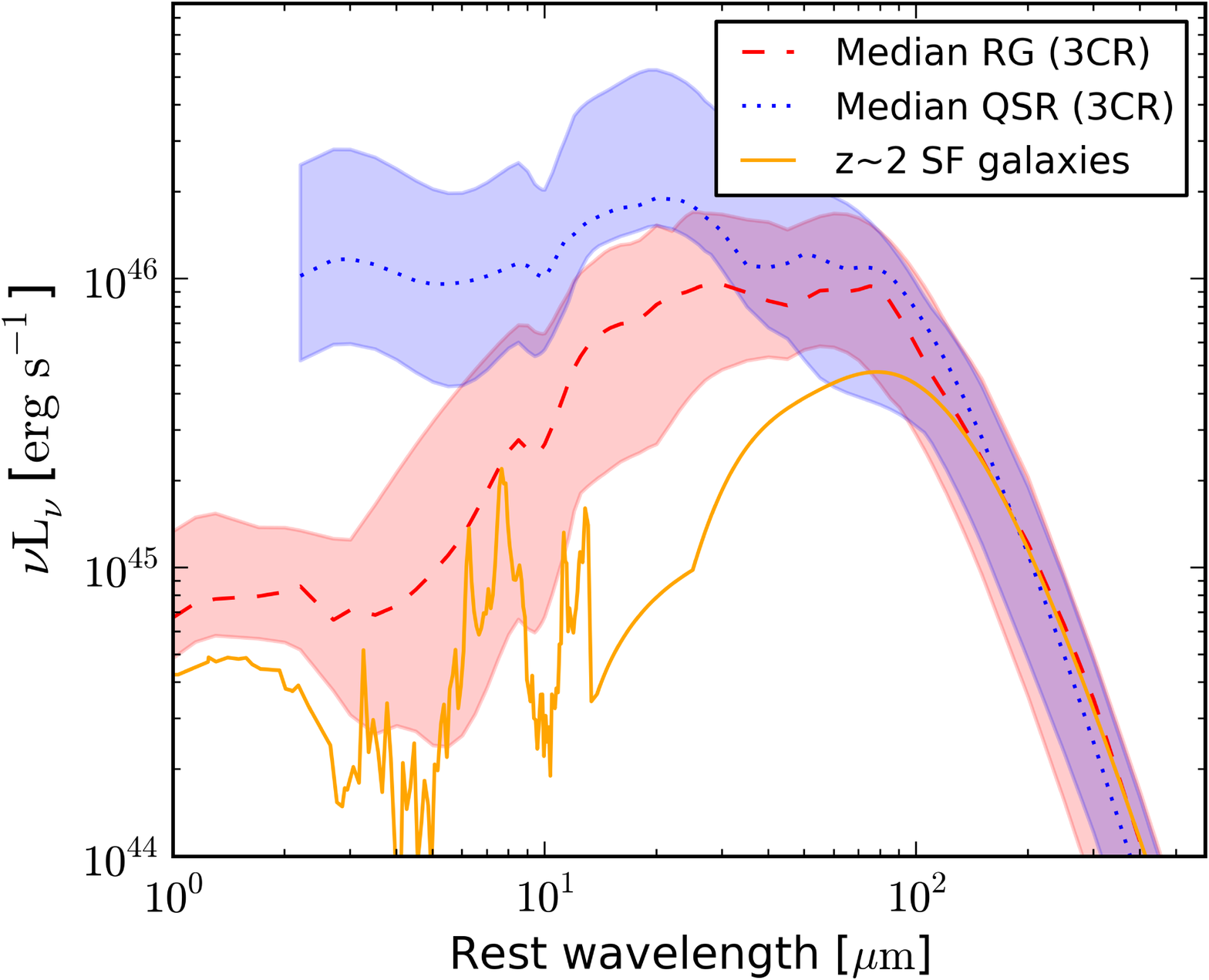}
      \caption{Median spectral energy distributions (SEDs) for the radio 
               galaxies (dashed red) and quasars (dotted blue) detected in at 
               least three \textit{Herschel} bands, and for $z\sim2$ 
               star-forming galaxies (solid yellow) from \citet{Kirkpatrick12}. 
               Shades areas, red for radio galaxies and blue for quasars, 
               correspond to the associated 16th-84th percentile ranges.  
               }
      \label{figure:goodSEDs}
   \end{figure}
   
   The best-fit SEDs for the objects detected in at least three 
   \textit{Herschel} bands are shown in Appendix~\ref{appendix:FIR-detectedSEDs}.
   The SEDs show a considerable range of shapes and absolute scaling, with 
   all QSR (and a few RG) SEDs peaking at wavelengths around 20~$\mu$m, and 
   most RG SEDs peaking at longer wavelengths. 
   Furthermore, the SEDs show that the AGN-powered and 
   star-formation-powered dust emission switch dominance typically at 50~$\mu$m, 
   (but this can happen at all wavelengths between 35 and 65~$\mu$m). 
   Given that all SEDs were computed on the same rest-frame wavelength grid, 
   we created median SEDs for the two types of objects to prevent the most 
   extreme objects dominating the average SEDs. When creating the median SEDs, 
   we refrained from applying any normalization, in order to preserve the 
   absolute luminosities of the individual SEDs. The median\footnote{The QSRs 
   median SED is given only for rest-frame wavelengths longer than 2~$\mu$m 
   because the emission at wavelengths shorter than 2~$\mu$m is dominated by 
   the hot accretion disk, which was not included in our SED fitting (as 
   explained in Appendix~\ref{appendix:UV/optical}).} SEDs are 
   shown in Fig.~\ref{figure:goodSEDs}.  

   The median SEDs of RGs and QSRs differ strongly at rest-frame 3-10~$\mu$m, 
   with the QSRs being a few times more luminous than the RGs. 
   Such an observational difference in this wavelength regime 
   has already been explained by \citet{Haas08} in the context of unification 
   by orientation. In this scenario, the observed luminosity differences 
   result from viewing the QSRs and RGs along different angles, such that the hot 
   inner regions of the dusty torus are observed directly in the case of QSRs but 
   are obscured in the case of RGs. As found by \citet{Leipski10}, similar 
   observational differences correlate with orientation indicators, such as the 
   radio core dominance. The median SED of RGs can therefore be viewed as the 
   median SED of reddened QSRs \citep{Haas08,Leipski10}. 
   At rest-frame wavelengths between 10 and 40~$\mu$m, the median RG and 
   QSR SEDs still show a considerable anisotropy, with the QSRs being 
   a factor of two more luminous compared to the RGs at 20~$\mu$m. At 
   rest-frame FIR wavelengths ($\gtrsim$ 40~$\mu$m), however, the median 
   RG and QSR SEDs appear to be remarkably similar both in shape and 
   absolute scale, arguing for, on average, similar star formation 
   properties for the hosts of both types of AGN.  
      
   Figure~\ref{figure:goodSEDs} also shows the average SED for the 
   subsample of the $z\sim2$ star-forming galaxies from \citet{Kirkpatrick12}. 
   This average SED is composed of 30 (U)LIRG galaxies, initially selected based 
   on their 24~$\mu$m flux density (F$_{24 \mu \mathrm{m}} \gtrsim$ 100~$\mu$Jy), 
   for which good multi-wavelength coverage is available throughout much 
   of the IR regime. Representing the star-formation-heated cold dust emission with 
   a modified blackbody, \citet{Kirkpatrick12} estimate the average SFR 
   and cold dust temperature of these galaxies to be 
   344$\pm$122 M$_{\odot}$ yr$^{-1}$ and 28$\pm$2 K, respectively. 
   Comparing these (U)LIRG numbers to those obtained for the 
   FIR-detected 3CRs in our work, we find comparable SFRs but on average 
   higher cold dust temperatures. Furthermore, Fig.~\ref{figure:goodSEDs} 
   confirms the marked difference between the (U)LIRGs and 3CRs in the 
   NIR/MIR luminosity: while the $z\sim2$ star-forming galaxies are characterized by 
   pronounced polycyclic aromatic hydrocarbon (PAH) features, the powerful 
   emission from the warm dusty torus completely outshines these features 
   in the MIR SED of the 3CR host galaxies. Finally, the average stellar 
   mass of the $z\sim2$ star-forming galaxies is a factor of $\sim3$ smaller 
   than that of the RG hosts as seen from the offset in the respective SEDs 
   at around 1.6~$\mu$m. While AGN contamination to the 1.6~$\mu$m is not 
   accounted for in this discussion, the very steep median SED of radio 
   galaxies between 4 and 6 $\mu$m suggests that any contribution from hot 
   dust to the 1.6~$\mu$m flux is negligible. AGN contamination at 1.6~$\mu$m
   likely is present in individual cases, but does not appear to be crucial 
   when discussing average properties. 
   Overall, the comparison between the 3CR and the \citet{Kirkpatrick12} sample 
   shows that massive galaxies in the high-$z$ Universe may be actively forming 
   stars regardless of whether their supermassive black holes are accreting or not.       
   \subsection{Stacking of non-detected objects}
   \label{subsection:Stacking}
   \begin{figure*}
      \centering
      \includegraphics[width=3cm]{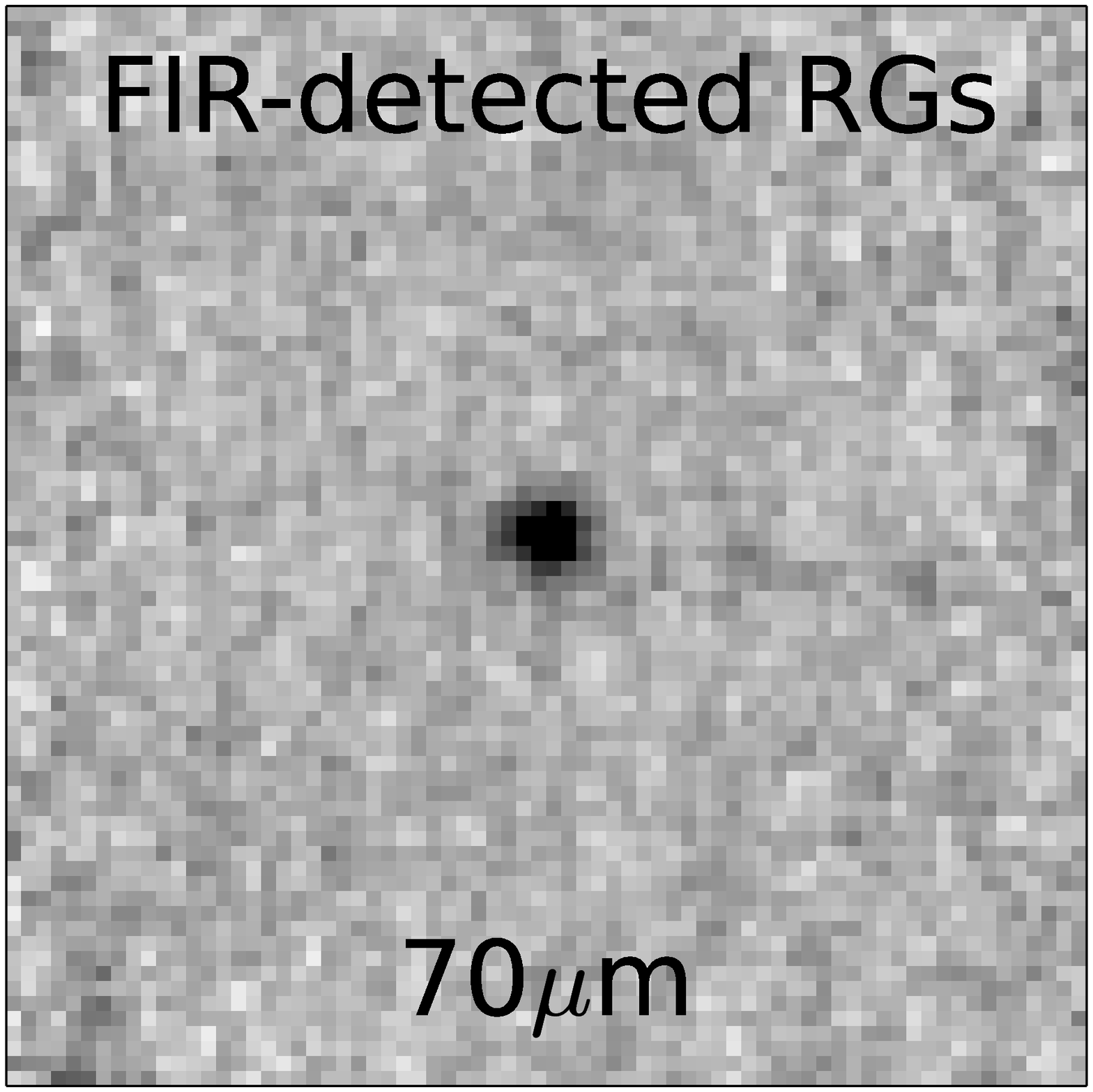}
      \includegraphics[width=3cm]{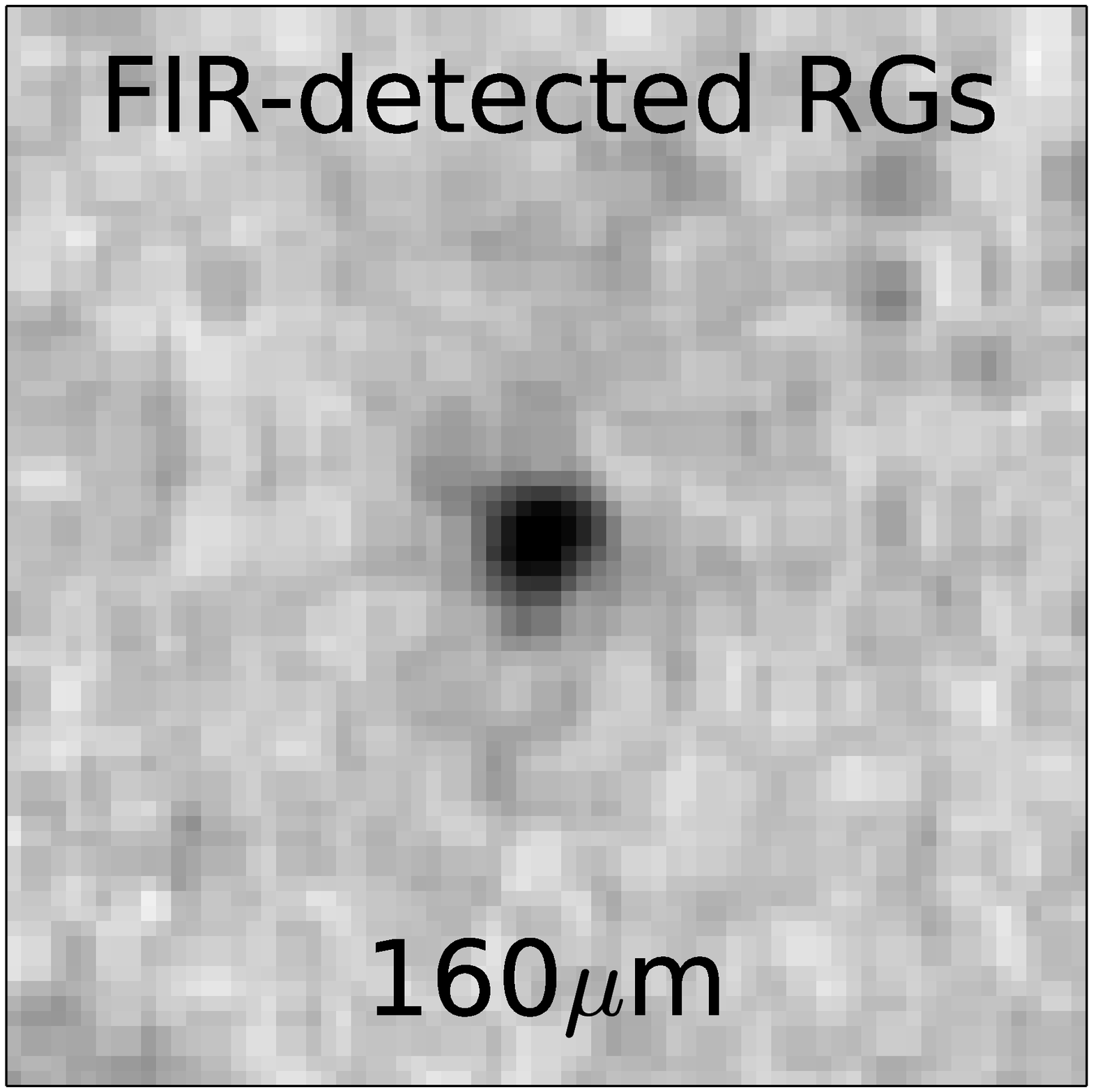}
      \includegraphics[width=3cm]{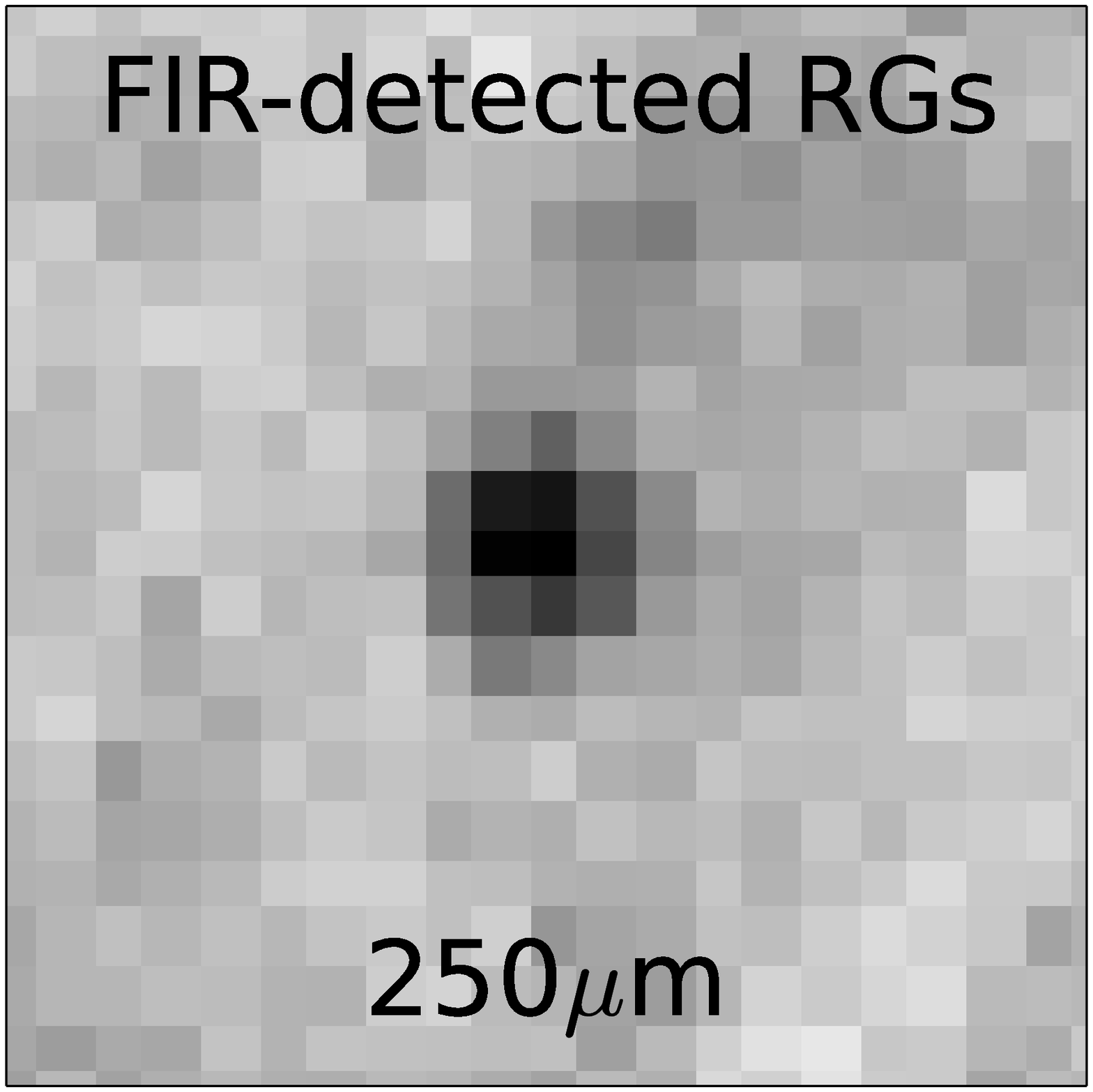}
      \includegraphics[width=3cm]{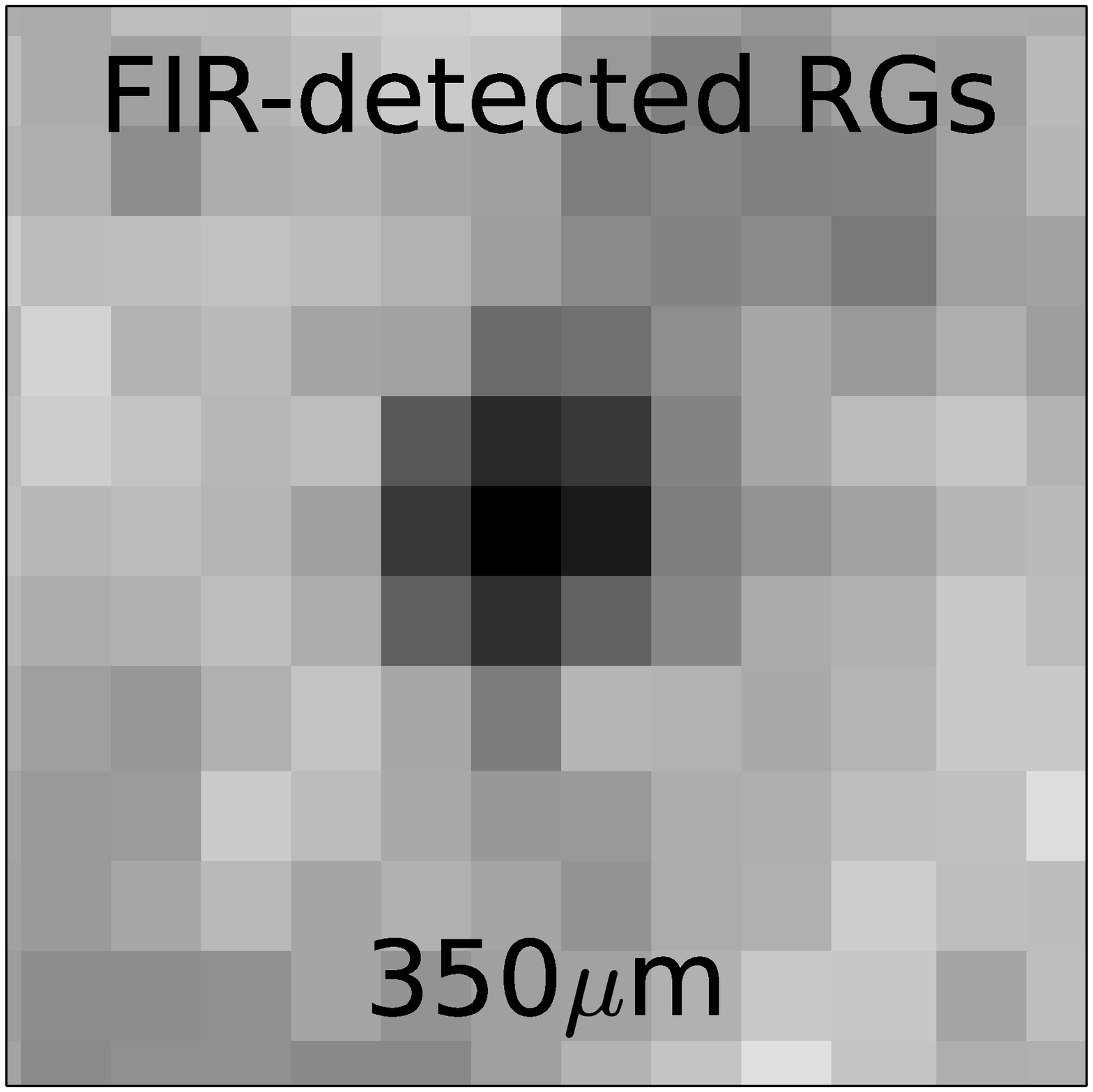}
      \includegraphics[width=3cm]{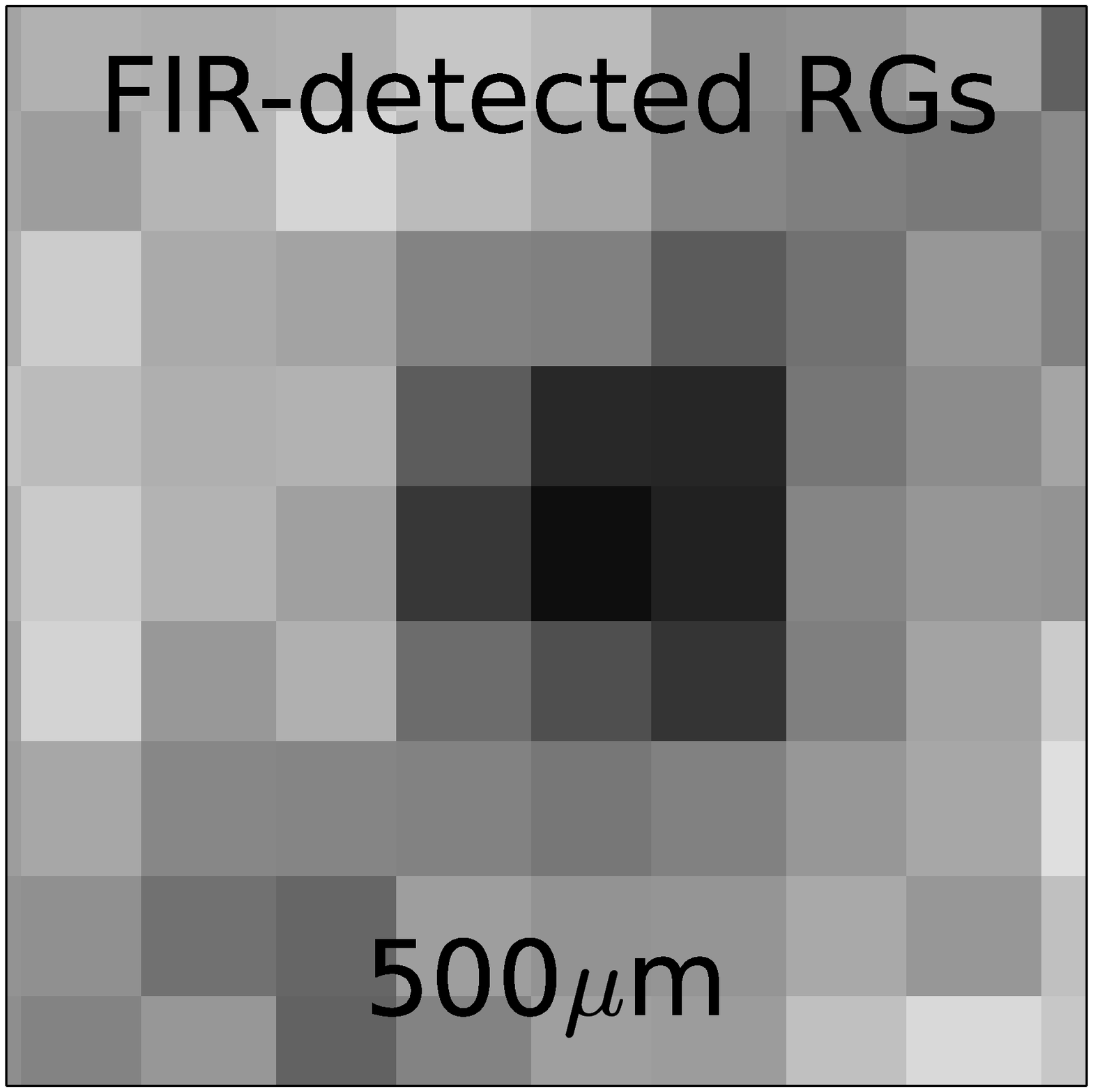}
      \includegraphics[width=3cm]{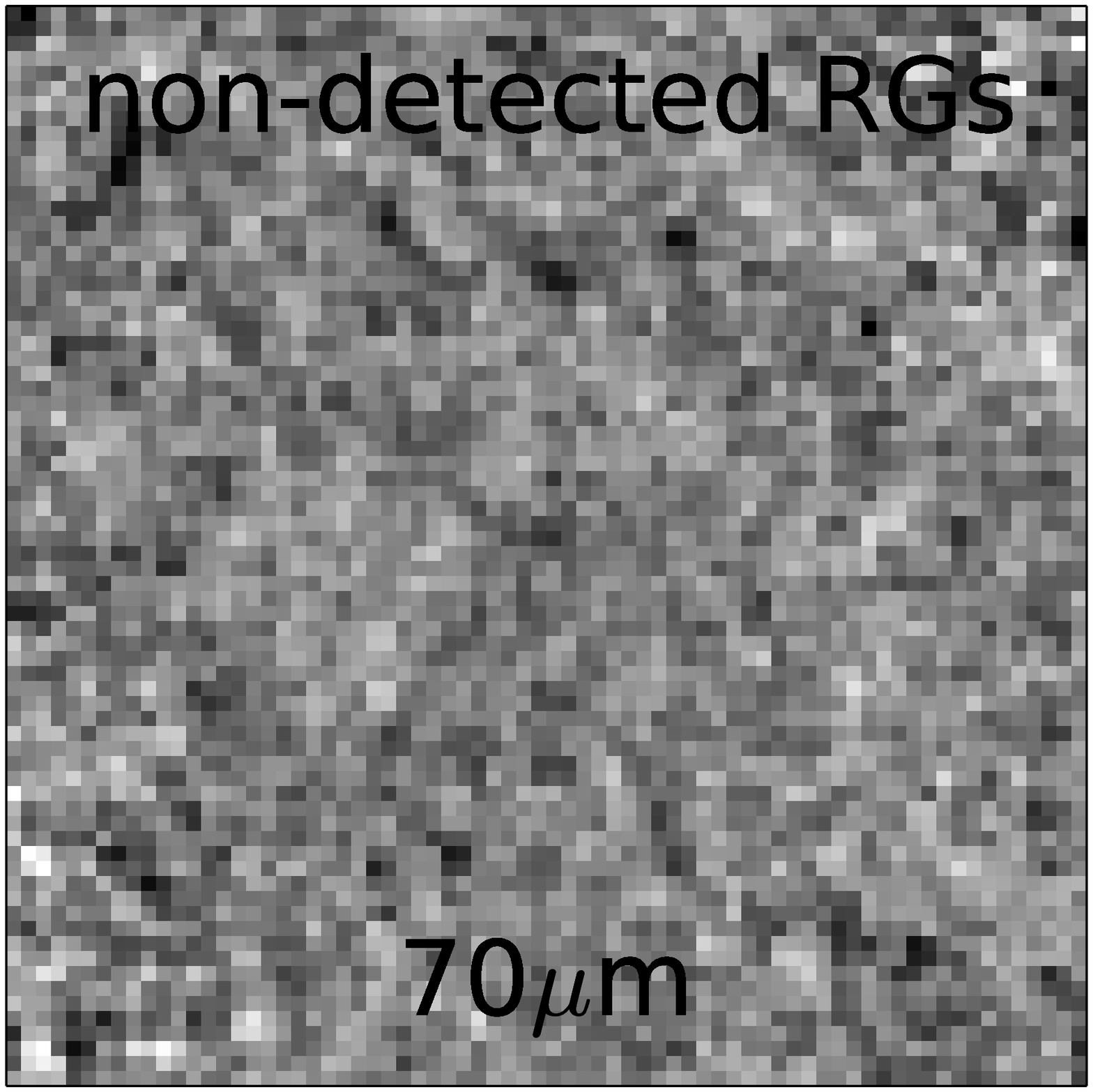}
      \includegraphics[width=3cm]{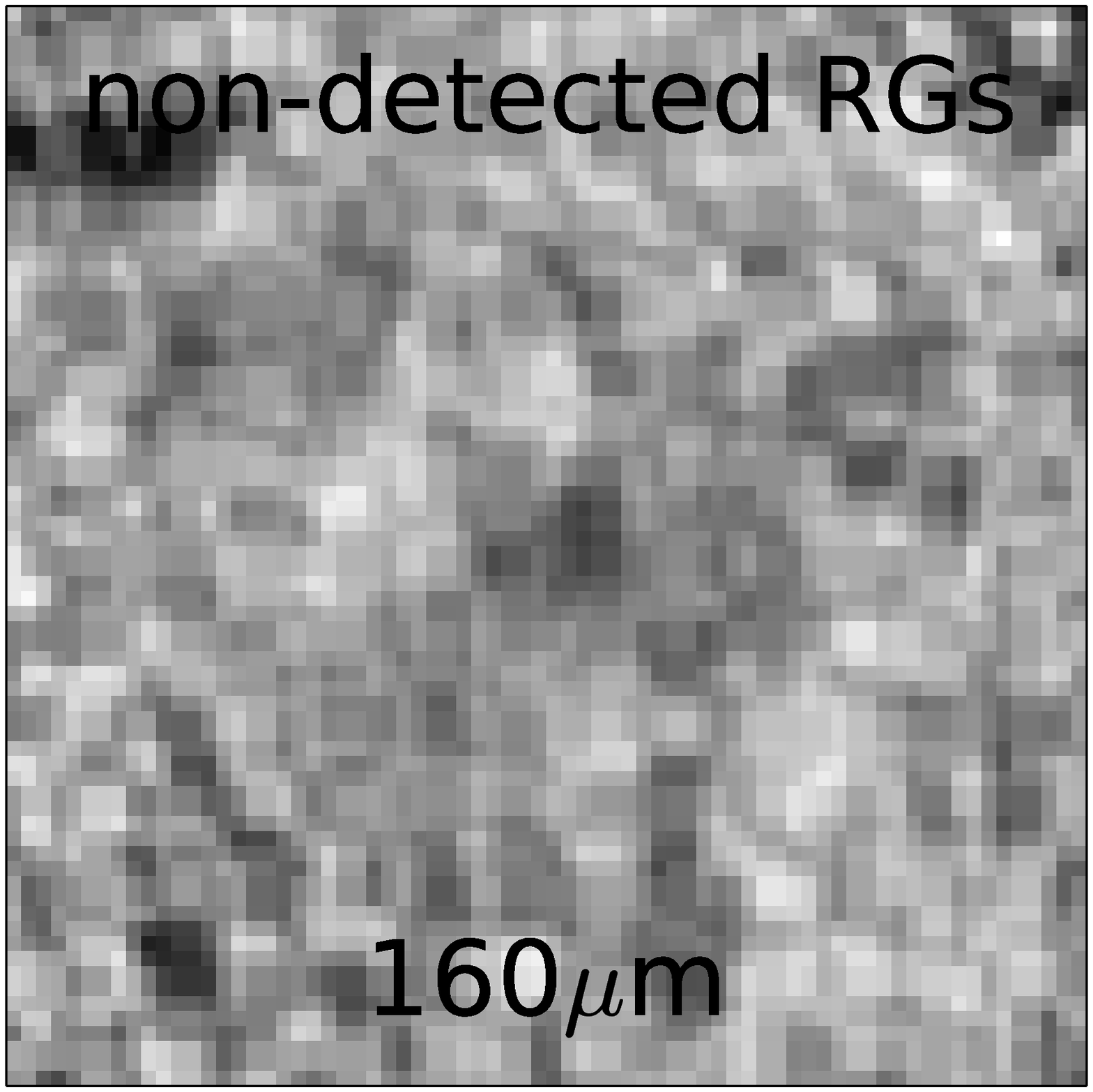}
      \includegraphics[width=3cm]{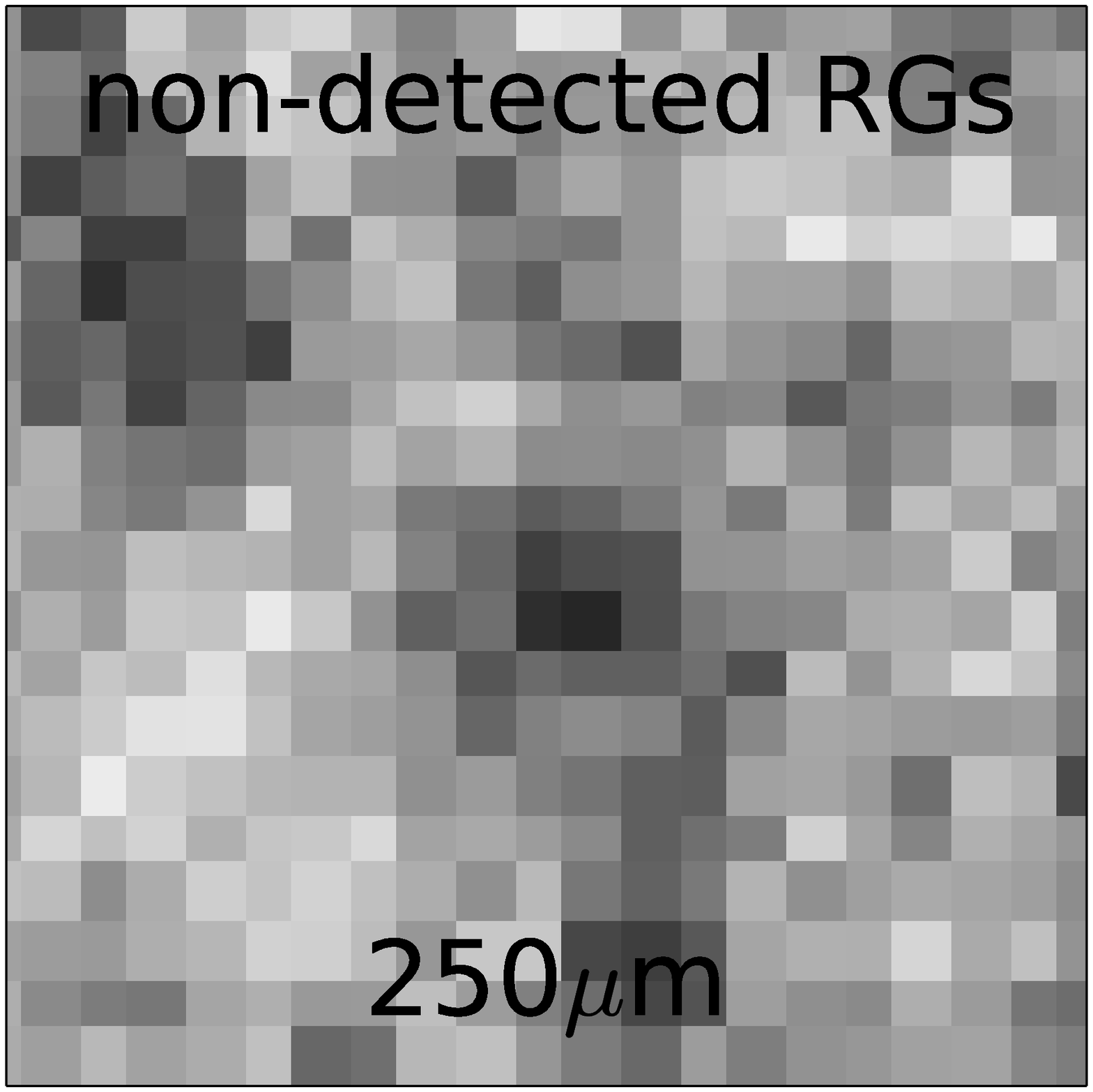}
      \includegraphics[width=3cm]{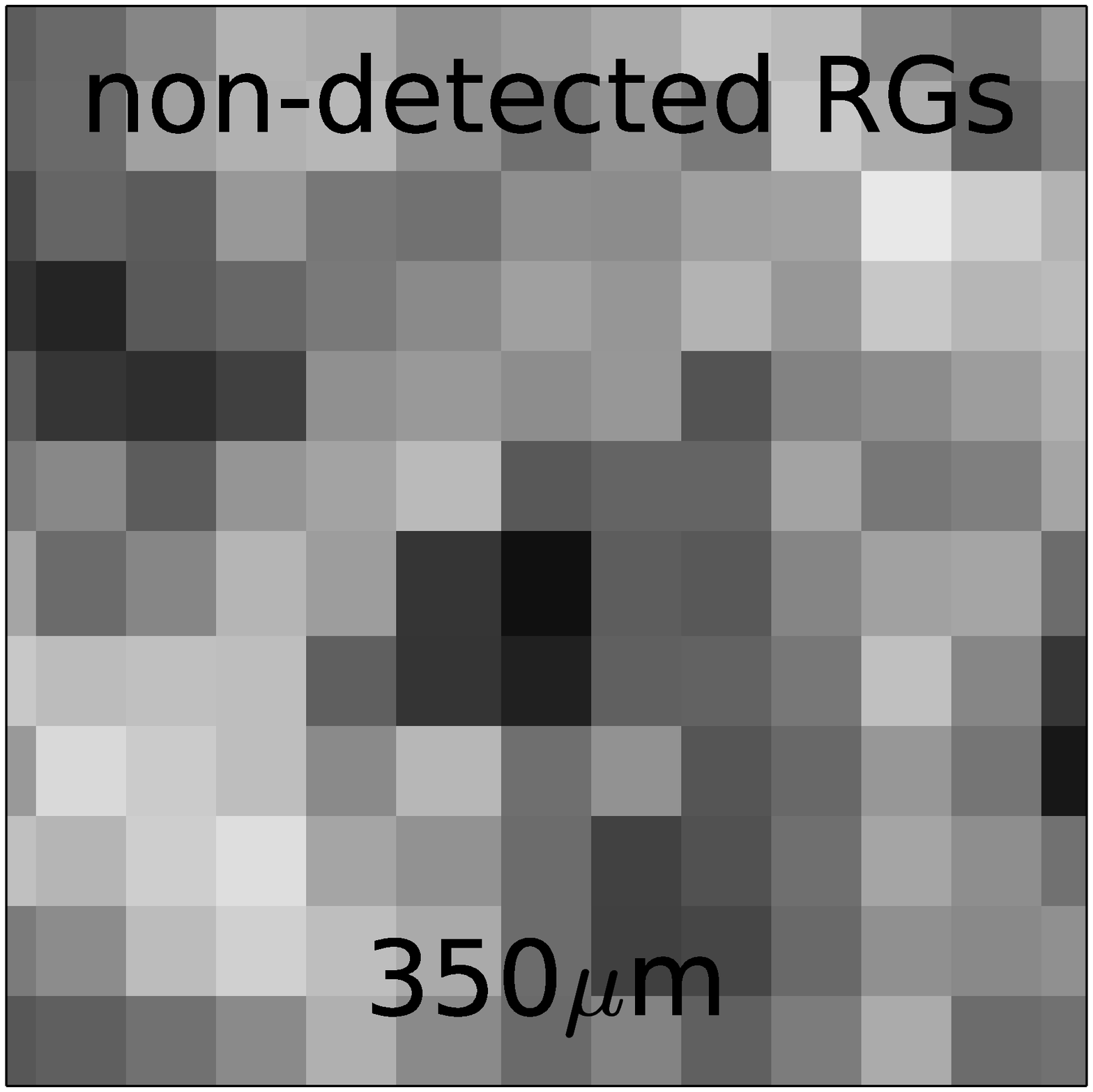}
      \includegraphics[width=3cm]{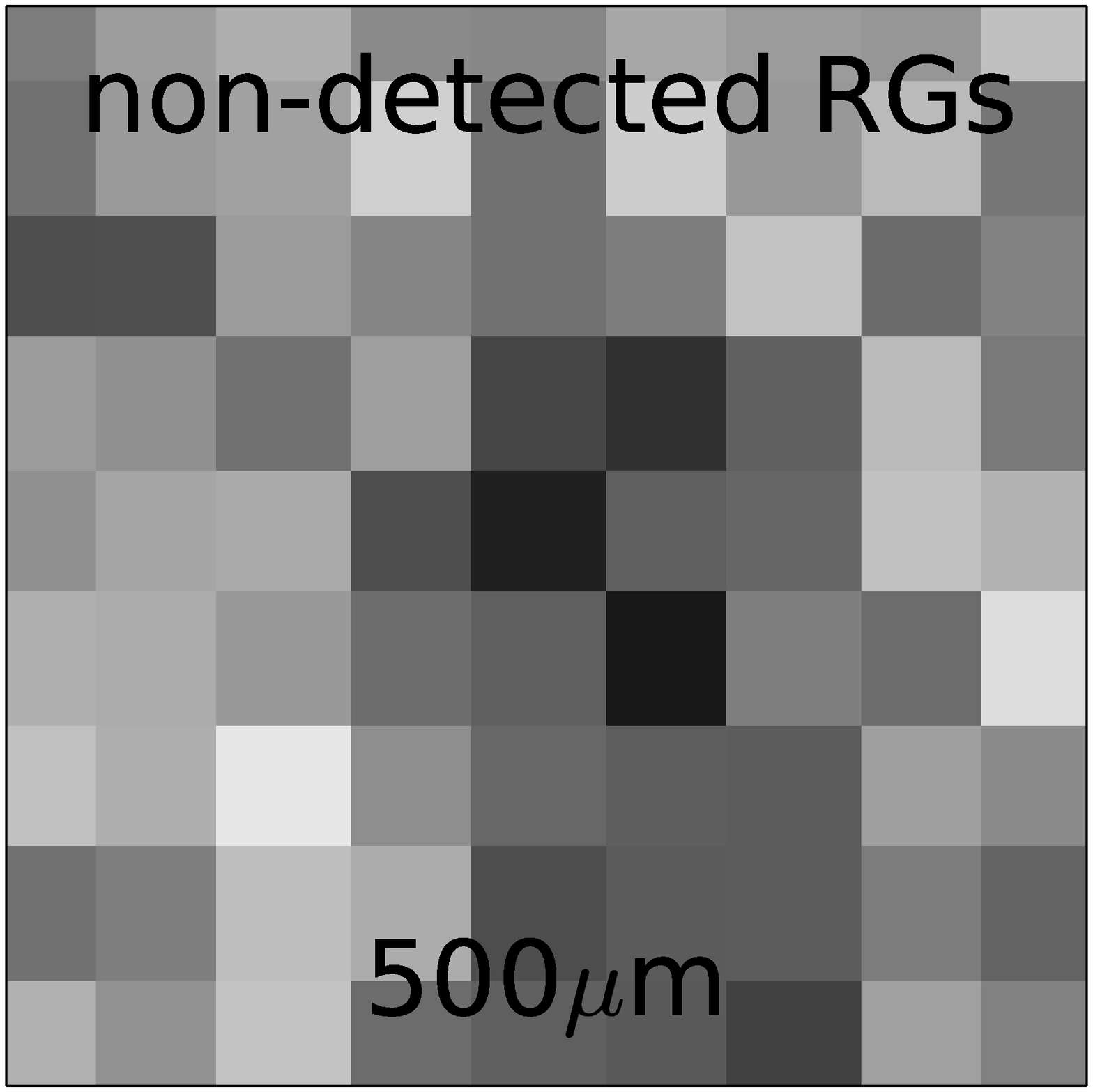}
      \includegraphics[width=3cm]{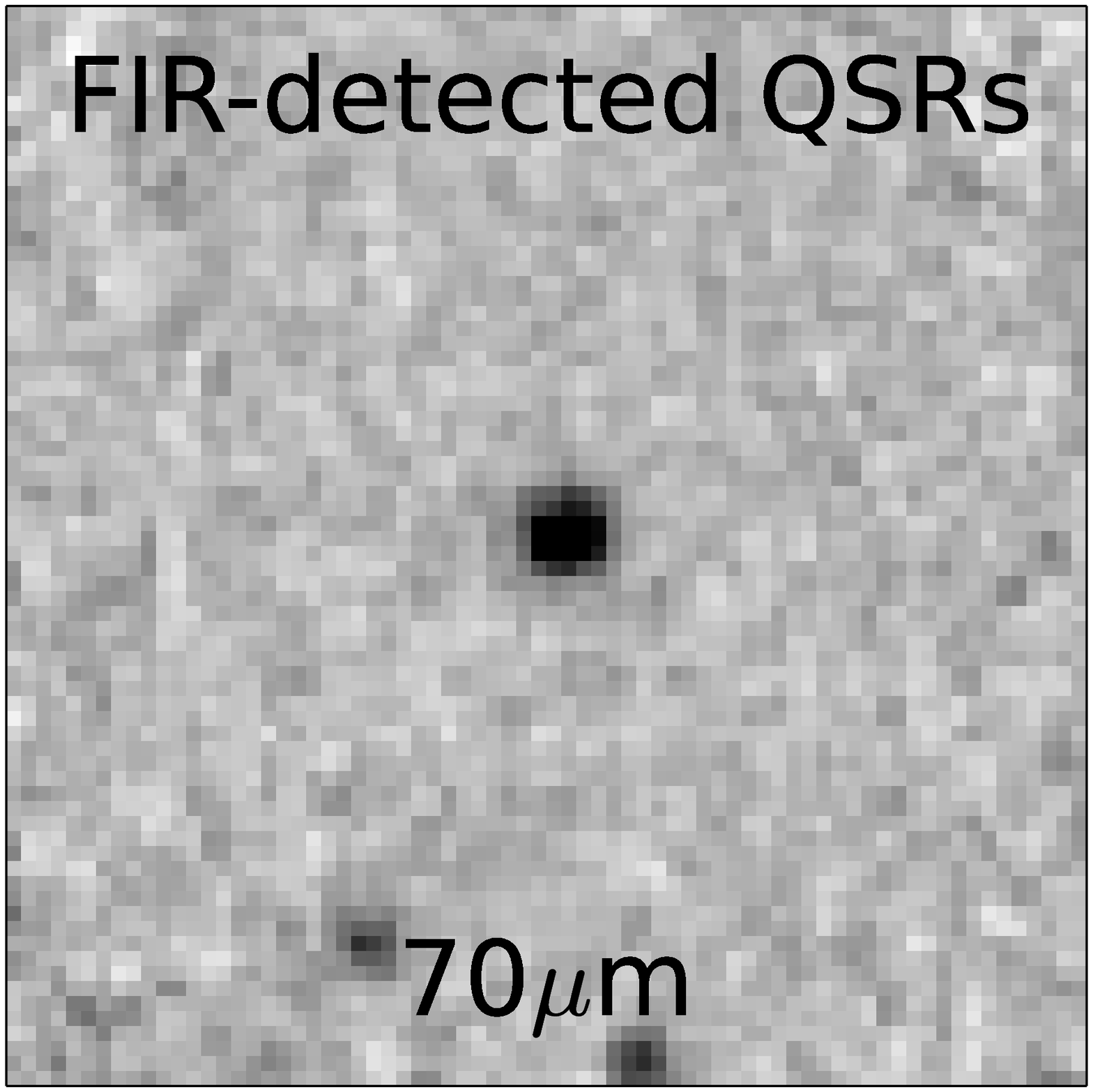}
      \includegraphics[width=3cm]{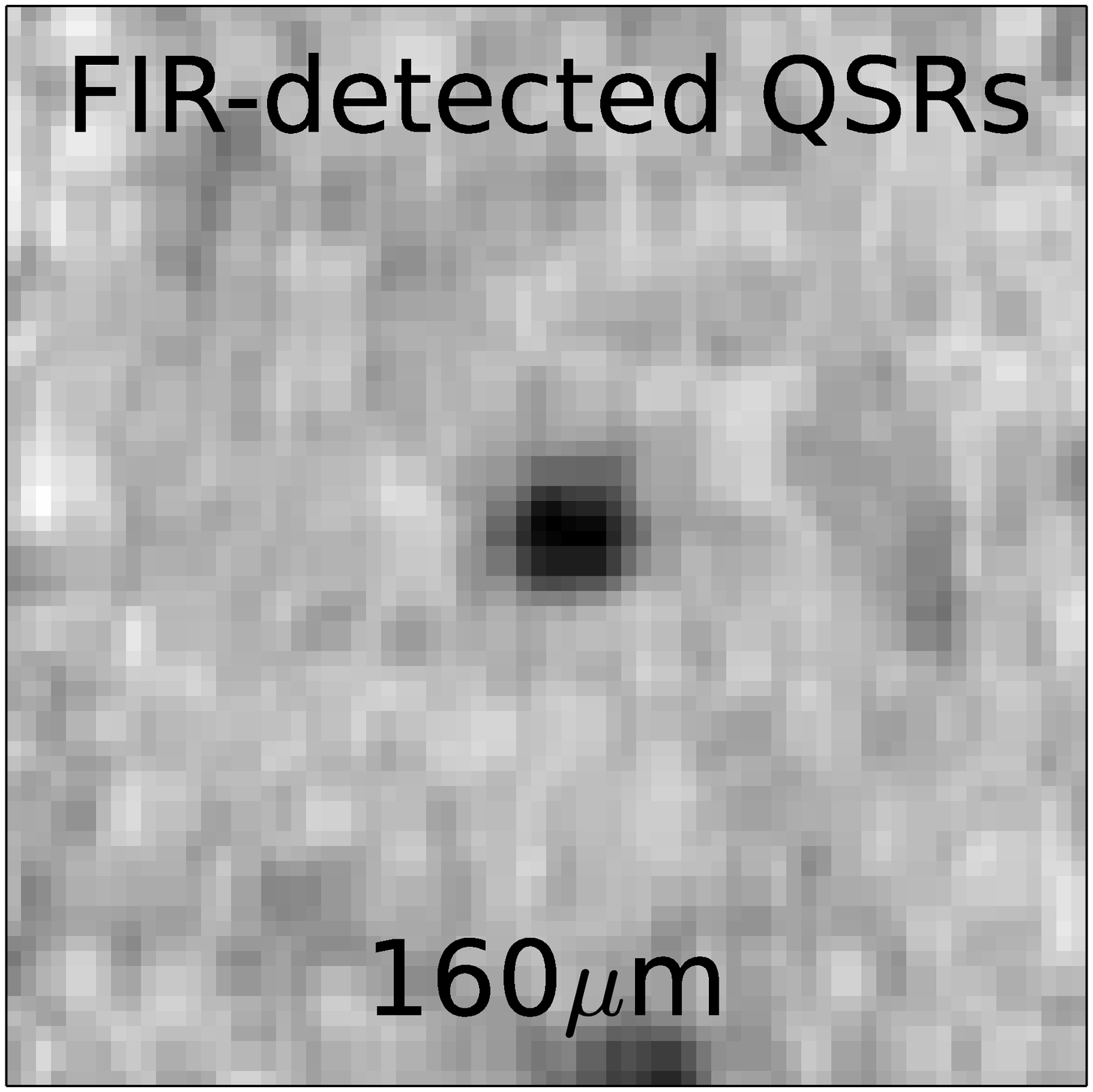}
      \includegraphics[width=3cm]{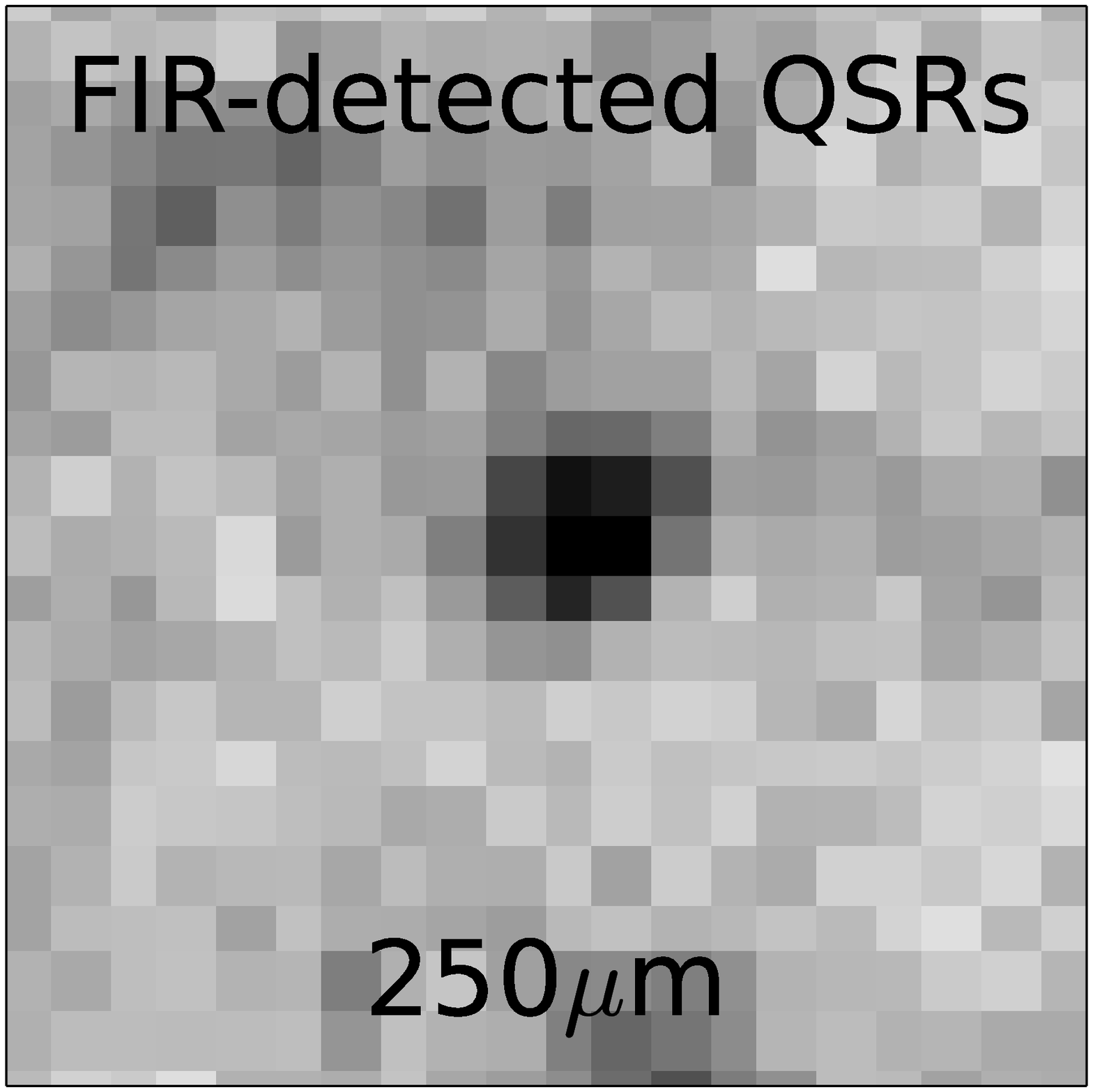}
      \includegraphics[width=3cm]{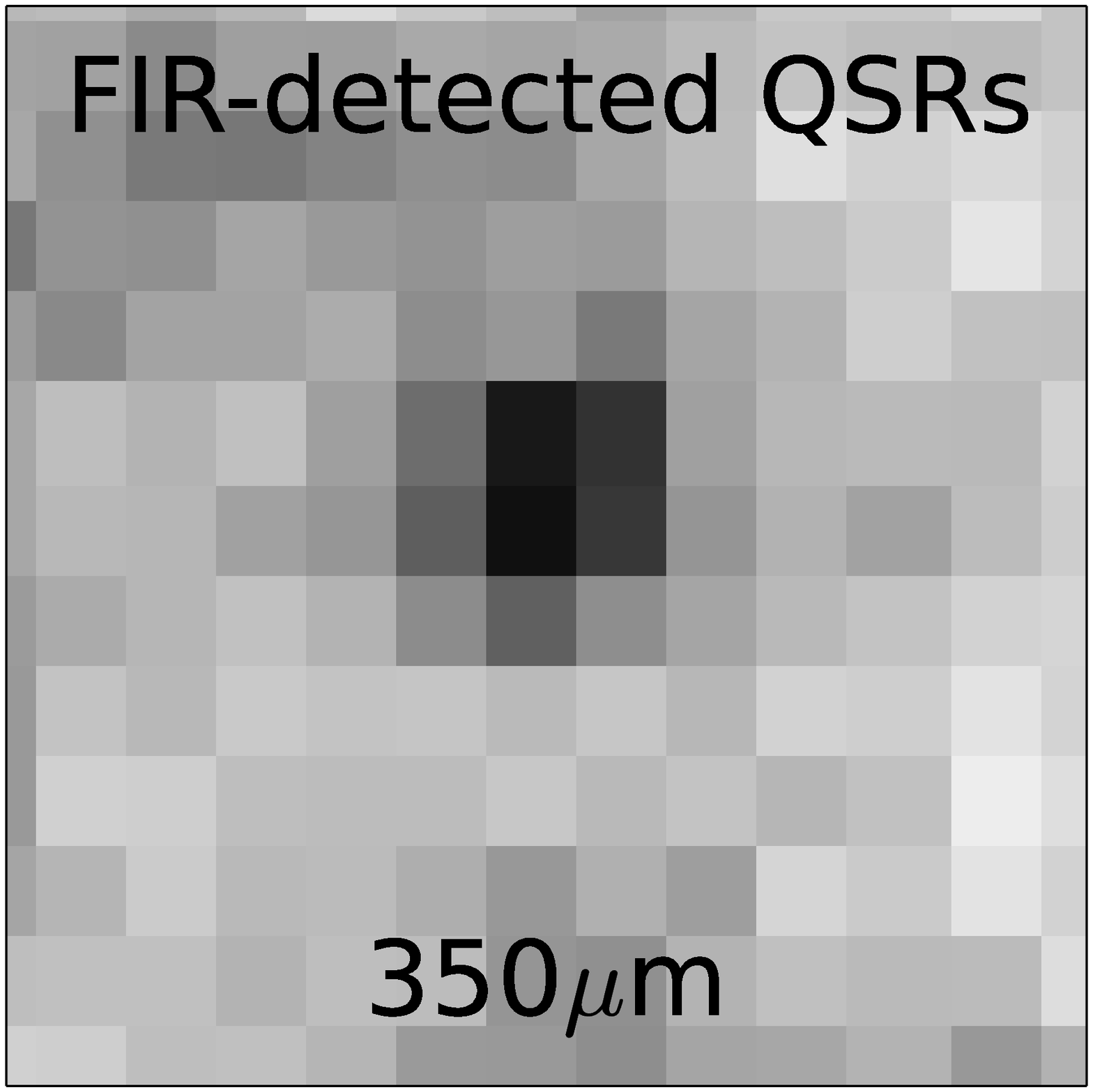}
      \includegraphics[width=3cm]{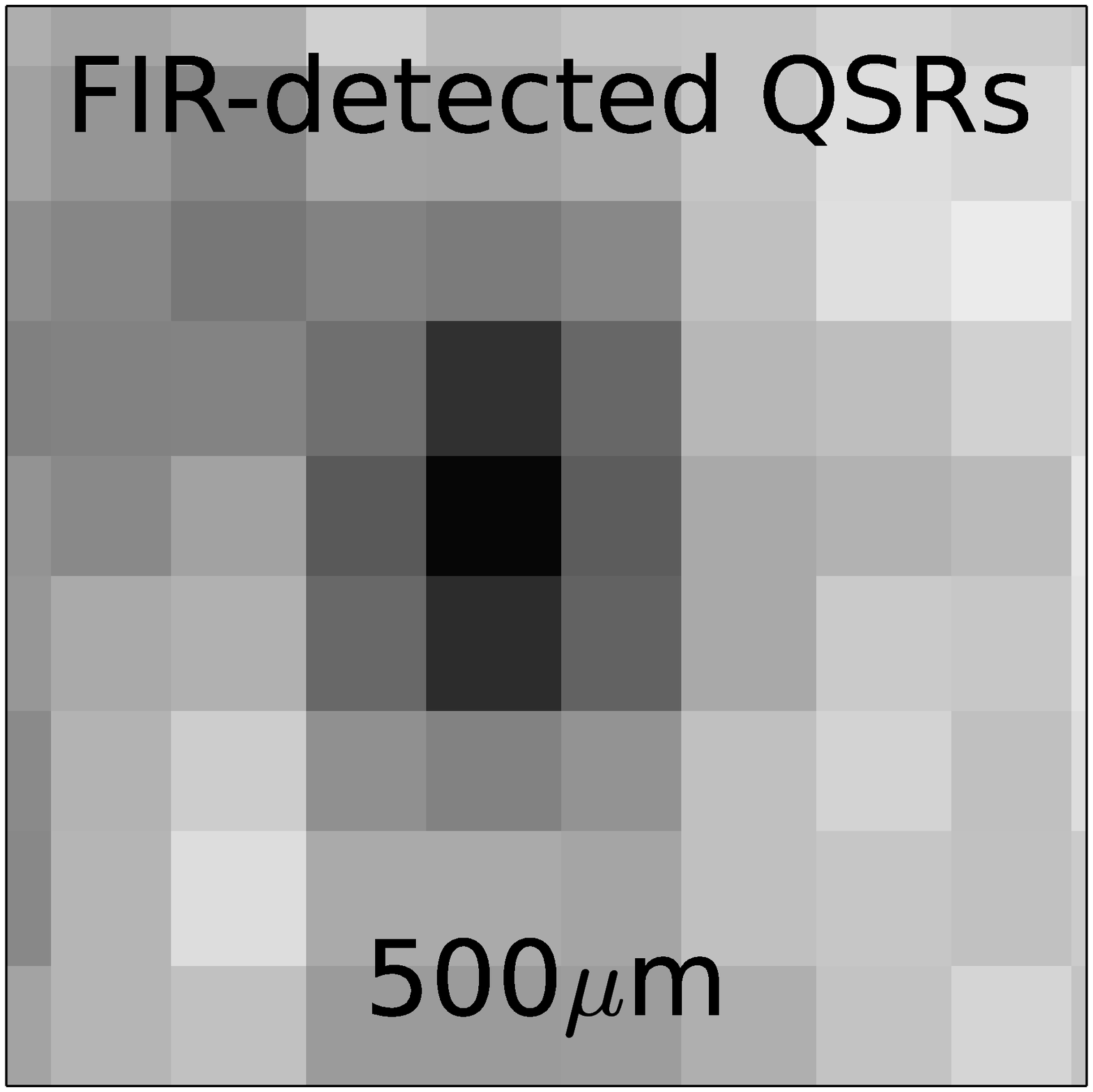}
      \includegraphics[width=3cm]{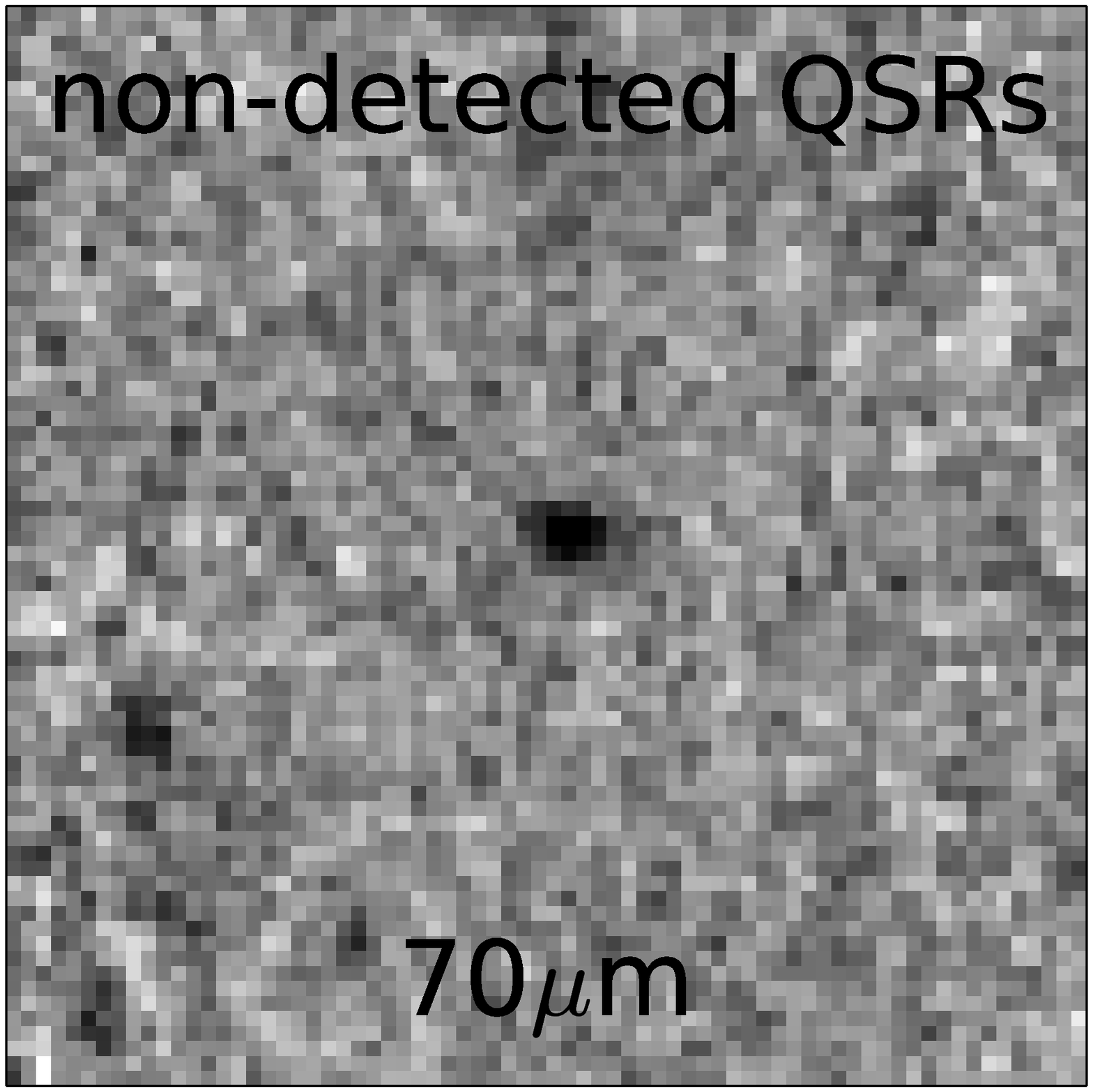}
      \includegraphics[width=3cm]{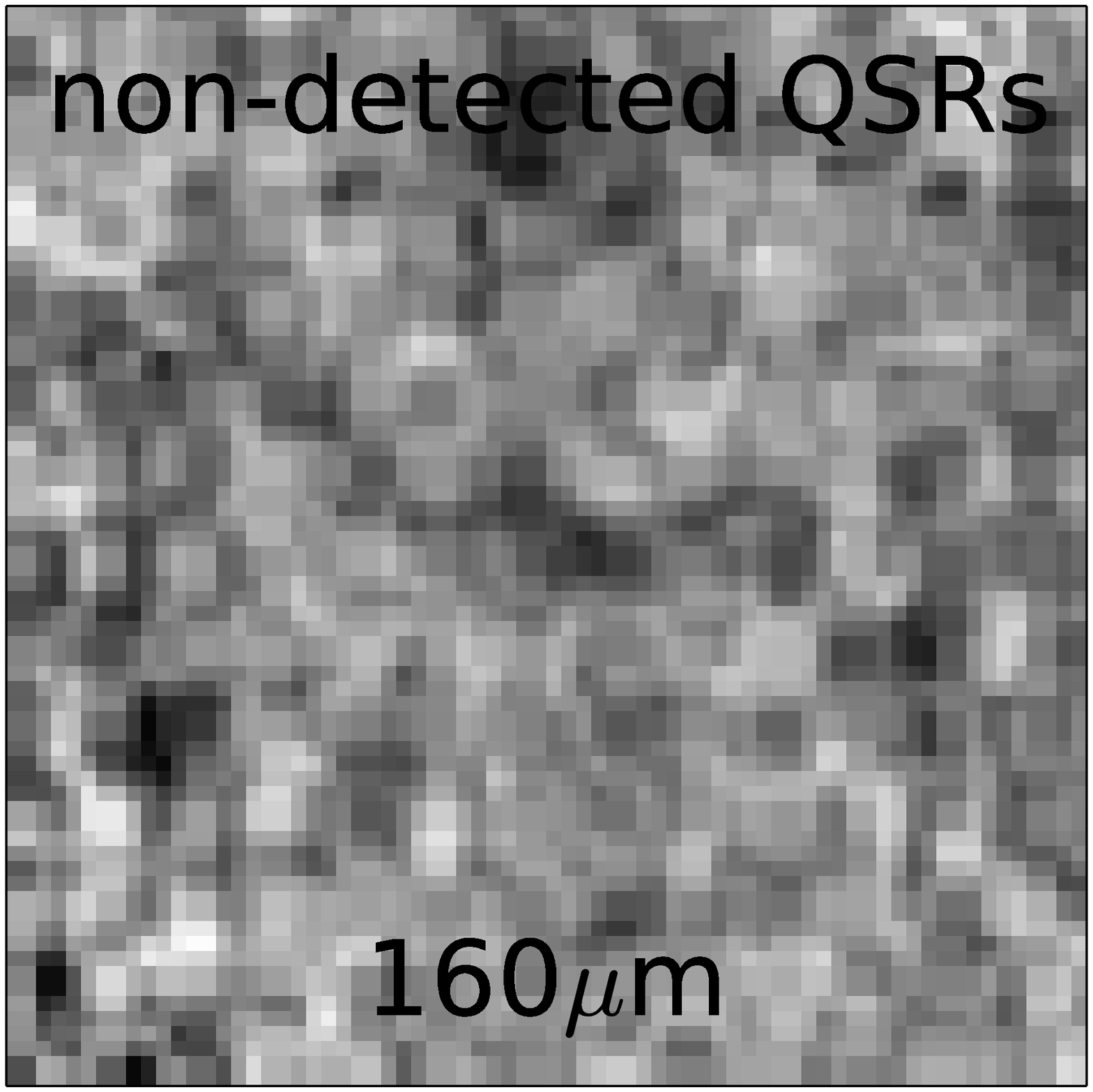}
      \includegraphics[width=3cm]{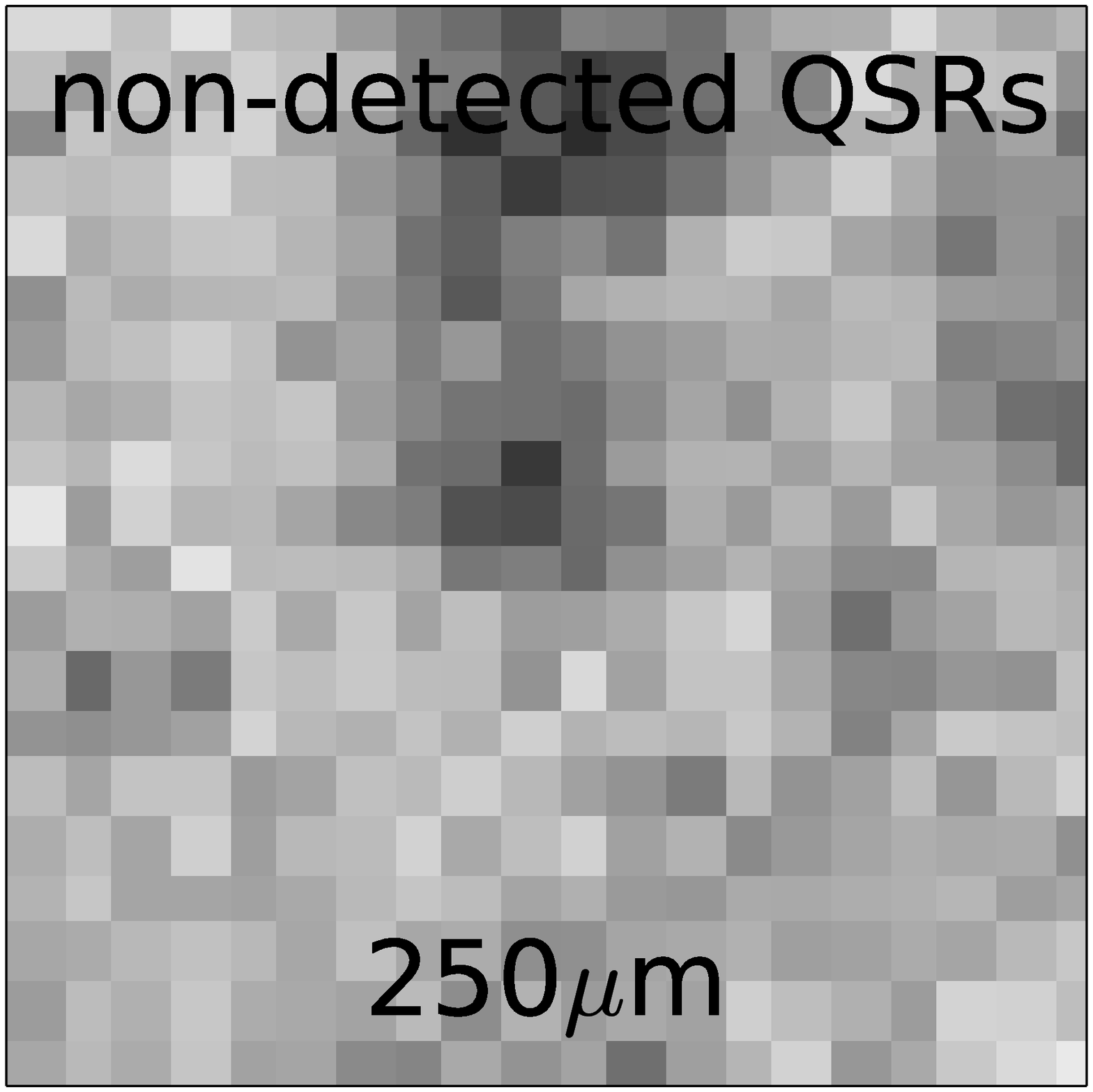}
      \includegraphics[width=3cm]{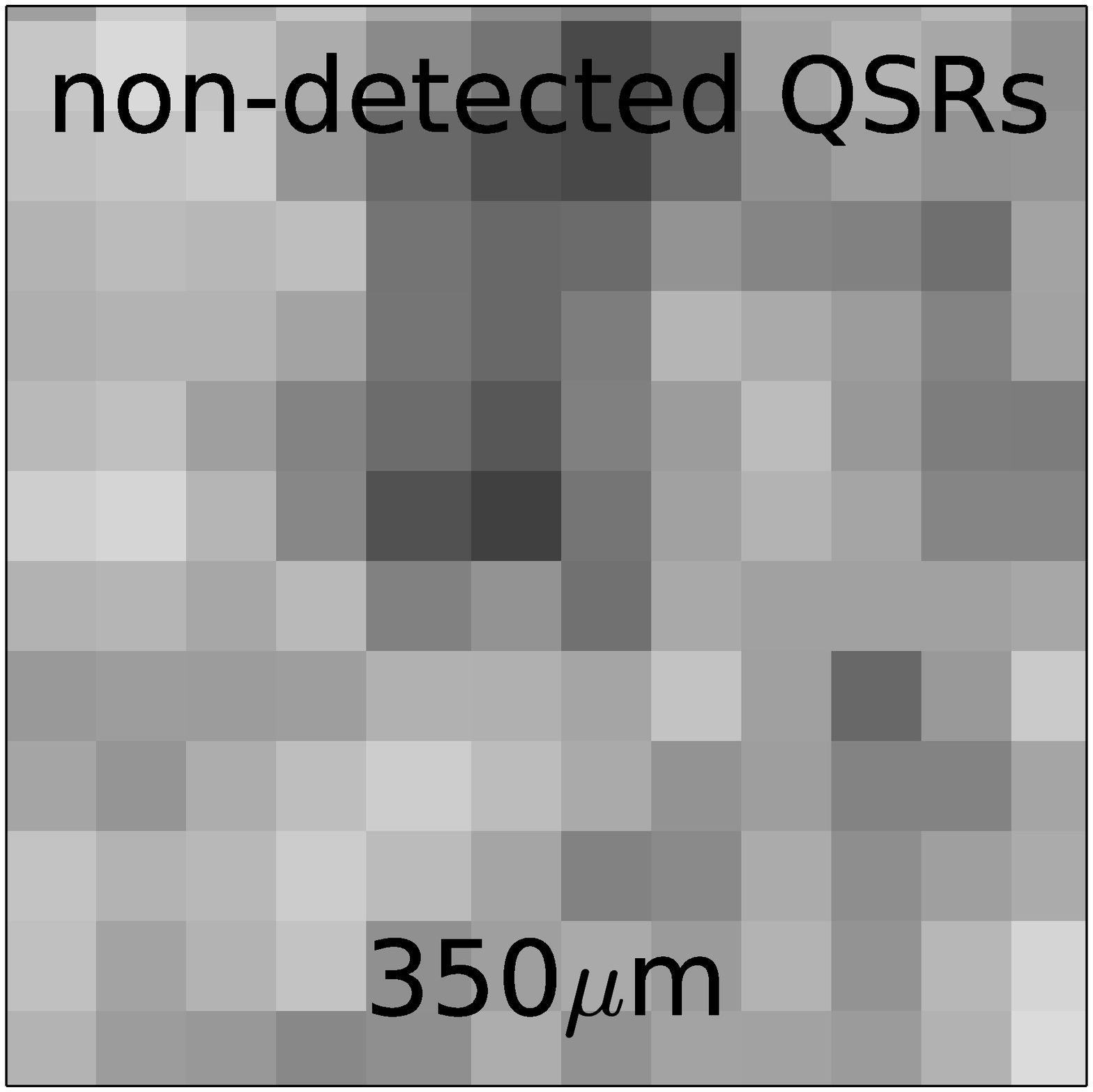}
      \includegraphics[width=3cm]{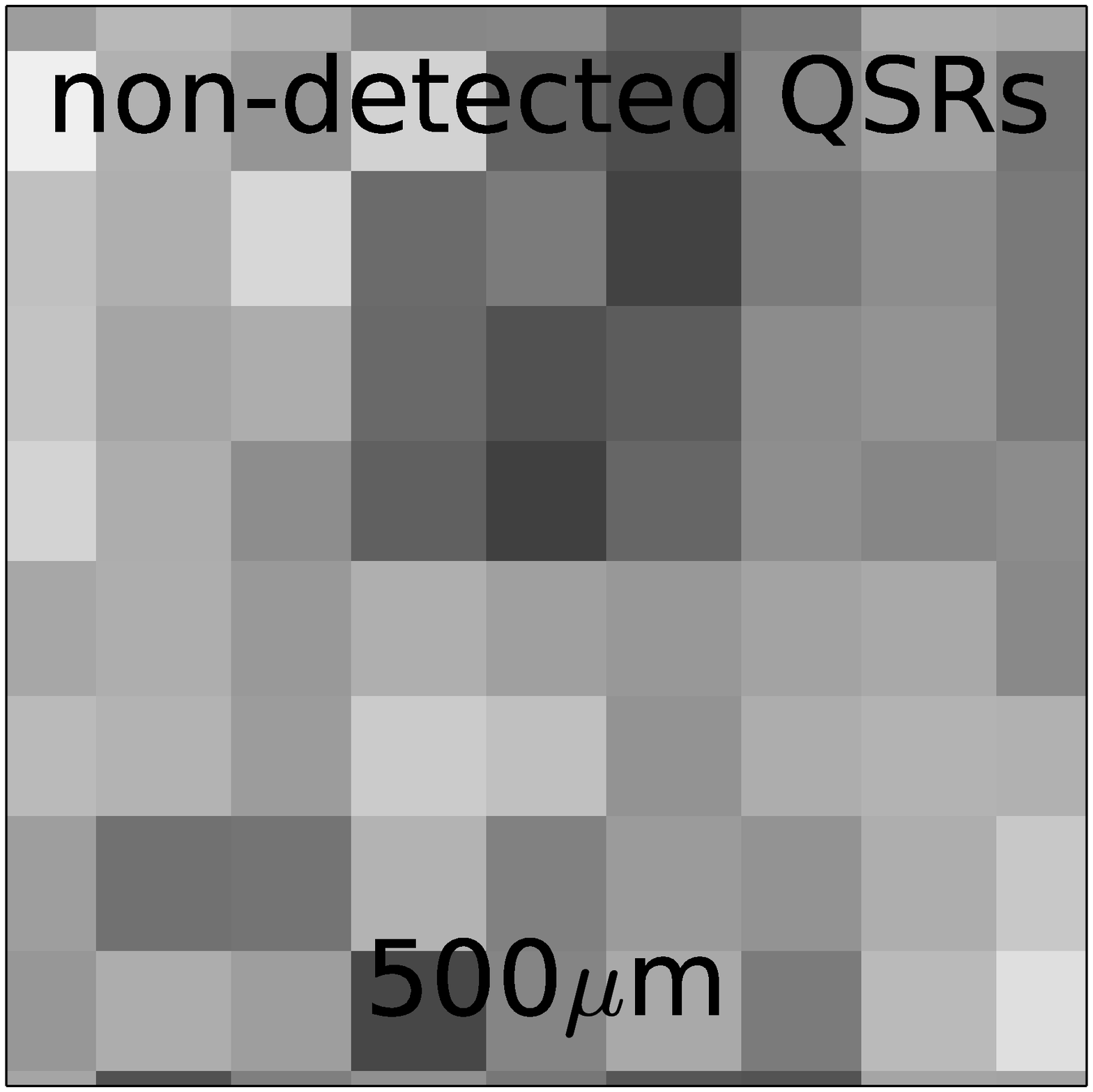}
      \caption{Stacked images in all \textit{Herschel} bands, for the 
               four subsamples discussed here. From left to right: PACS 
               70~$\mu$m, PACS 160~$\mu$m, SPIRE 250~$\mu$m, SPIRE 350~$\mu$m, 
               and SPIRE 500~$\mu$m bands, respectively. From top to bottom: 
               FIR-detected radio galaxies, non-detected radio galaxies, 
               FIR-detected quasars, and non-detected quasars. Each 
               stacked image shown here has dimensions of 2x2 arcmin 
               and is centred on the known radio positions of the objects 
               entering the stack. 
               }
      \label{figure:stackIMAGE}
   \end{figure*}
   \begin{figure} 
      \centering
      \includegraphics[width=\hsize]{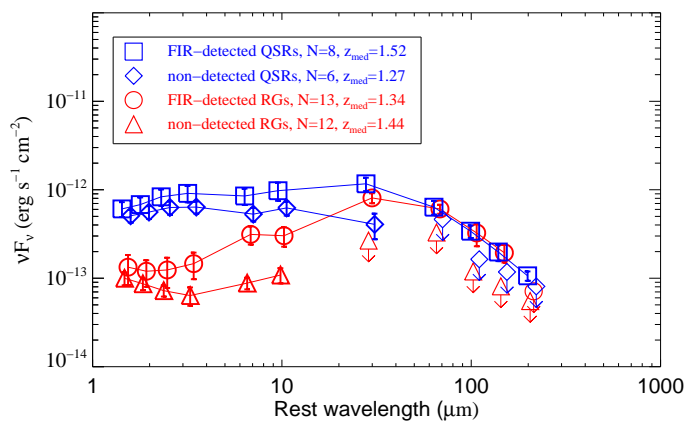}
      \caption{Spectral energy distributions of stacked subsamples 
	       selected from the hosts of the 3CR sources studies here. 
	       The four subsamples presented are: FIR-detected radio 
	       galaxies (red circles), non-detected radio galaxies 
	       (red triangles), FIR-detected quasars (blue squares), 
	       and non-detected quasars (blue diamonds). Arrows 
	       mark 3$\sigma$ upper limits as explained in the text.    
	       }
      \label{figure:stackPLOT}
   \end{figure}
   \begin{table}
   \renewcommand{\arraystretch}{1.2}
   \caption{Results from the SED fitting of the stacked subsamples.}
   \label{table:fitStack}    
   \centering                       
   \begin{tabular}{l c c}        
   \hline\hline                 
   Subsample & log (L$_{\mathrm{AGN}}$ (L$_{\odot}$)) & log (L$_{\mathrm{SF}}$ (L$_{\odot}$)) \\
   \hline                        
      FIR-detected RGs & 12.6 & 12.2 \\
      non-detected RGs & $<$12.2 & $<$11.7 \\
      FIR-detected QSRs & 13.2 & 12.1 \\
      non-detected QSRs & 12.7 & $<$11.6 \\
   \hline                             
   \end{tabular}
   \end{table}
   In addition to probing the general properties of the objects that are 
   individually detected in at least three \textit{Herschel} bands, we 
   attempt a stacking analysis to extract an average signal for the objects 
   which lack significant emission in the \textit{Herschel} bands. For 
   convenience, we refer to the former objects as FIR-detected, and to the 
   letter as non-detected. Our aim is to discuss average properties for 
   each radio-loud AGN type, therefore we split the non-detected objects into 
   RGs and QSRs, respectively. In order to retain decent number statistics 
   in the two subsamples, we decided not to further divide the RGs and QSRs 
   in redshift bins, despite the considerable redshift range of our sample. 
   In practice, we take the non-detected objects to be those that are 
   detected in at most one, namely the PACS 70~$\mu$m band, which given the 
   redshifts of our sources often probes the peak emission from the dusty 
   torus. Several RGs and QSRs were not included in the stacking analysis 
   owing to the presence of potentially confusing sources in the corresponding 
   maps, close to the known radio position of the object in question.  
   Furthermore, two RGs (3C~252 and 3C~267) with a strong detection in 
   only the PACS 70~$\mu$m band were not included in the stacking because 
   they are inconsistent with forming a single population with the objects 
   which have no \textit{Herschel} detections.
   As a result, stacking was performed on subsamples of 12 RGs 
   and 6 QSRs with median redshifts of $z_{med}$=1.44 and $z_{med}$=1.27, 
   respectively. The objects entering the subsamples are flagged in 
   Table~\ref{table:PhysProp}. For comparison, we also selected two subsamples 
   from the FIR-detected objects: a subsample with 13 RGs ($z_{med}$=1.34) 
   and another one with 8 QSRs ($z_{med}$=1.52; 3C~298 and 3C~318, sources 
   with strongly emitting nearby objects, were not taken into consideration).
      
   We stacked equal areas extracted from the individual \textit{Herschel} 
   maps, centred on the known radio position. Photometry on the stacked 
   map was performed following the same procedures as adopted in the case 
   of the individually detected objects (Sect.~\ref{section:Data}). We 
   examined the diversity within the given subsample by bootstrapping with 
   1000 realizations. In practice, from the original subsamples identified 
   above, for each bootstrapping realization we selected a random subsample (with 
   the same number of objects, allowing for repetitions), stacked the 
   \textit{Herschel} maps, and performed the photometry. The centroid and 
   the dispersion of the resulting distribution were taken to be the mean 
   flux density of the on-source stack and its associated uncertainty. 
   Additionally, we stacked random positions in order to inspect the overall 
   significance of the on-source stacked signal. If the mean value of 
   the on-source stack distribution was at least three times larger than 
   the mean value of the corresponding background stack distribution, we 
   concluded that the on-source signal is significant. In cases of a 
   non-significant signal, we took three times the mean value of the background 
   stack distribution to be our on-source stack upper limit value. 
   
   The mean stacks for the four different subsamples selected in our 
   study are presented in Fig.~\ref{figure:stackIMAGE}. 
   The non-detected quasars have a significant stacked signal in only 
   the PACS 70~$\mu$m band, whereas the non-detected radio galaxies have 
   no significant signal in any of the \textit{Herschel} bands.
   On the other hand, the FIR-detected quasars and radio galaxies have 
   significant stacked signals in all, and in all but the SPIRE 500~$\mu$m 
   band, respectively. In order to study the full IR SED of a stacked 
   subsample, we calculated mean flux densities in the \textit{Spitzer} 
   bands, where virtually all objects have been individually detected. 
   All obtained average fluxes are in the observed frame: these were 
   transferred to the rest-frame, using the median redshift of the 
   corresponding subsample. The SEDs of the stacked subsamples are shown 
   in Fig.~\ref{figure:stackPLOT}.  

   The 1-10~$\mu$m SEDs of radio galaxies are composed of emission from 
   an evolved stellar population and from a heavily reddened AGN \citep{Haas08}. 
   At the rest-frame wavelengths (given the comparable median redshifts of the 
   various subsamples) probed by the two shortest wavelength \textit{Spitzer} 
   bands ($<$~2~$\mu$m), the FIR-detected and non-detected subsamples of 
   RGs are similar. Thus, it is likely that the stellar masses of the host
   galaxies for the two RG subsamples are, on average, similar as well. 
   Identifying differences in the host galaxy stellar masses requires additional NIR 
   photometry, and is outside the scope of this work. At wavelengths between  
   2 and 10~$\mu$m, the SEDs of the two RG subsamples differ from each other. 
   At these wavelengths, AGN-heated hot dust emission appears to be present/absent in 
   the SED of the FIR-detected/non-detected subsample, respectively, reflecting 
   different levels of dust obscuration, as inferred from the diversity in the 
   individual MIR SEDs of RGs \citep{Haas08}. Particularly interesting is the 
   non-significant stacked signal for non-detected RGs at rest-frame 30~$\mu$m, 
   which could indicate that the two subsamples have different intrinsic AGN 
   luminosities. Indeed, fitting the SEDs following the routine taken in Sect.~\ref{section:SEDs} 
   gives (at least) a factor of two difference between the estimated L$_{\mathrm{AGN}}$ 
   in the two RGs subsamples (Table~\ref{table:fitStack}, Fig.~\ref{figure:LIRAGN}).
   A clear difference between the SEDs of the two subsamples of RGs is observed 
   at rest-frame wavelengths longer than 60~$\mu$m, suggesting marked differences 
   in the star formation properties of the two subsamples. This results in (at least) 
   a factor of three difference between the estimated SFRs 
   in the two RGs subsamples (Table~\ref{table:fitStack}, Fig.~\ref{figure:LIRAGN}). 
   The SEDs of the two QSR subsamples generally follow the same trends as 
   those observed for the RG subsamples. While the non-detected QSRs could 
   overall be at the fainter end of the QSR population, the most striking 
   difference in the QSR SEDs is at FIR wavelengths, with the FIR-detected 
   QSRs having SFRs (at least) a factor of three higher than the 
   non-detected QSRs. While more analysis is required to pin down the potential differences 
   among the subsamples in wavelengths other than FIR, it is beyond doubt that 
   the formation of new stars can be prodigious in some, but modest or weak in 
   other radio-loud AGN hosts. 
\section{Discussion}
   \label{section:Discussion}
   \subsection{Star formation in hosts of powerful AGN}
   \label{subsection:discSF}
   Starbursts powering the FIR emission in some hosts of radio-loud AGN
   have been found at low-to-intermediate redshifts \citep[e.g.][]{Dicken10}.  
   In the high-$z$ Universe, high levels of star formation in some 
   radio-loud AGN have been estimated using several observational 
   indicators. These include the usage of rest-frame UV spectroscopy to 
   detect huge Ly$\alpha$ halos surrounding the AGN 
   \citep[e.g.][]{VillarMartin99}, submm photometry to probe the cool 
   dust and molecular gas content of AGN hosts \citep[e.g.][]{Archibald01,Reuland04}, 
   and \textit{Spitzer} MIR spectroscopy to detect strong PAH 
   features \citep[e.g.][]{Rawlings13}. The latter, however, are seen only 
   in rare cases because the AGN-powered hot/warm dust emission in powerful 
   radio-loud AGN hosts usually outshines the PAH emission. 

   UV/visible data have been used to infer SFRs for several 3CR radio 
   galaxies studied in this work \citep{Chambers&Charlot90}. These include 
   3C~065, 3C~068.2, 3C~266, 3C~267, 3C~324, 3C~356 and 3C~368. 
   The ratio between the SFR obtained using the approach taken in this work 
   and the UV/visible SFR ranges between 4 and 40, therefore we conclude that the 
   star formation is often strongly obscured in the UV/visible. This comparison 
   demonstrates that rest-frame FIR data are crucial in quantifying the 
   dust-enshrouded star formation in the hosts of these sources, especially 
   the ones hosting quasars because the emission from the accretion disk 
   complicates the SFRs estimates from UV/visible data. 
   Using \textit{Herschel} data, \citet{Seymour11} found mean SFRs for 
   $1.2 < z < 3.0$ radio-selected AGN to range between 80 and 581 
   $\mathrm{M_{\odot} yr^{-1}}$. Similarly, \citet{Drouart14} studied a 
   sample of $\sim$~70 $1 < z < 5$ radio galaxies, estimating SFRs of a 
   few hundred to a few thousand solar masses per year. While the authors 
   suggest that the highest values reported in their study are likely 
   overestimated and should be treated as upper limits, the 
   overall idea that the hosts of high-$z$ radio-loud AGN can be prodigious 
   star-formers is consistent with our findings for the hosts of the 
   powerful 3CR objects. A crucial point in these analyses is that the high 
   star formation luminosities obtained are coeval with the 
   growth of the SMBHs residing in the nuclei of the host galaxies. As such, 
   these results argue for a scenario whereby on average the SMBH has not 
   quenched the star formation in the host galaxy, which is at odds with 
   results presented by \citet{Page12}. A series of coeval episodes of 
   strong star formation and black hole activity may have formed the 
   massive host galaxies and their massive black holes (see B12). 
   
   The SFRs estimated for the FIR-detected radio-loud AGN are comparable 
   to those obtained for SMGs at similar redshifts \citep[e.g.][]{Magnelli12}. 
   Other parameters, such as the temperatures and masses of the cold dust 
   component and total stellar masses agree with each other, at least on average, 
   as well \citep{Santini10,Michalowski12,Swinbank14}. It is widely thought 
   that high-$z$ SMGs form stars in starburst events, induced as a result 
   of a variety of different processes including mergers and tidal 
   interactions. Radio-loud AGN are known to be at the centres of over-densities 
   in the high-$z$ Universe \citep[e.g.][]{Venemans07,Wylezalek13}.   
   Thus, the merger scenario appears to be an attractive way of 
   producing at least some of the extremely high SFRs estimated in our study.
   A similar conclusion has been drawn in a study of a few high-$z$ radio 
   galaxies using both \textit{Herschel} and CO observations \citep{Ivison12}. 
   
   Jet-induced star formation, also known as positive feedback, is yet 
   another possible way to trigger high SFRs \citep[e.g.][]{Dey97,VanBreugel98}
   in hosts of radio-loud AGN. In this scenario, the outgoing radio jet 
   shocks the surrounding interstellar material, which subsequently 
   cools down to form new stars. Recently, based on UV-to-submm templates 
   built with the evolutionary code PEGASE.3, \citet{RoccaVolmerange13} 
   showed that the star formation timescales from stellar population 
   synthesis agree well with the ages of the radio episodes for two 
   high-$z$ radio galaxies. Similar studies of larger samples are 
   required to better understand the details of jet-induced star formation.   
     
   Despite evidence for strong star formation in the 3CR hosts detected 
   in at least three \textit{Herschel} bands, $\sim60\%$ of 3CR sources 
   are detected in fewer than three \textit{Herschel} bands, arguing for 
   significantly lower star formation activity in these hosts. While the 
   upper limits in the SPIRE bands still allow energetically significant 
   SFRs of up to 300 $\mathrm{M_{\odot} yr^{-1}}$ in some of these hosts, 
   the overwhelming majority have SFRs of at most 100 $\mathrm{M_{\odot} yr^{-1}}$. 
   This raises an important question: what leads to the significantly lower 
   star formation activity in these hosts? If the star formation activity 
   is indeed merger-induced, one possible answer to the above 
   question is that occurrences of minor- and/or gas-poor mergers are 
   likely to lead to relatively low levels of star formation activity. 
   The relevance of the merger scenario in the context of the 
   triggering of starbursts, but also of the AGN activity, has been 
   discussed by \citet{Tadhunter11}. They find that the strongly 
   starbursting radio galaxies in their intermediate-redshift sample have 
   optical morphological features consistent with the idea that they are 
   triggered in major mergers. Another possibility is that we are 
   observing some hosts after the star formation has been quenched 
   \citep[e.g.][]{Farrah12}. If 
   the AGN activity is responsible for the quenching, then we argue that 
   this negative feedback is not universal, even if it acts over a very 
   short timescale. First, given the strong star formation in the 3CR 
   hosts detected in at least three \textit{Herschel} bands, it is unlikely 
   that the quenching of star formation occurs before the radio-loud 
   AGN phase. Second, quenching taking place during the radio-loud phase is 
   not supported with the finding in Sect.~\ref{subsection:Stacking} 
   that the $<$~2~$\mu$m rest-frame stacked SEDs of the two RG subsamples 
   are similar. In particular, if star formation in the non-detected RGs 
   is quenched, then they are expected to be brighter than the FIR-detected 
   RGs because the star formation in the latter is heavily dust obscured.
   Third, because the radio-loud phase probably marks the end of the AGN 
   phase, it is unlikely that the quenching occurs after the radio-loud phase. 
   To conclude, while the hosts of the quasar-mode radio-loud AGN studied 
   here may have a wide range of star formation rates, it is unlikely that 
   the triggering or quenching of their star formation activity is associated 
   with a uniform scenario met in all objects.   
   \subsection{RG and QSR unification in the FIR}
   \label{subsection:Unification}
   Unification theories of radio-loud AGN \citep{Barthel89,Antonucci93} 
   ascribe observed differences in the properties of radio galaxies and 
   quasars solely to orientation effects. Using \textit{Spitzer} photometric 
   observations of the sample of 3CR objects studied here, \citep{Haas08} 
   (see also the median SEDs from our current work) found the mean 1-10~$\mu$m 
   rest-frame radio galaxy SED to be consistent with a sum of an underlying 
   host and a heavily obscured quasar. Unification among the high-z 3CR 
   objects was further corroborated with \textit{Spitzer} spectroscopic 
   \citep{Leipski10} and \textit{Chandra} X-ray observations \citep{Wilkes13}.
   Our results reveal that the optically thin (i.e. isotropic) FIR emission 
   is similar for radio galaxies and quasars, thus in line with the 
   predictions of radio-loud AGN unification by orientation.  
   \subsection{AGN versus non-AGN host galaxies}
   \label{subsection:ActvsNon}
   \begin{figure}
      \centering
      \includegraphics[width=\hsize]{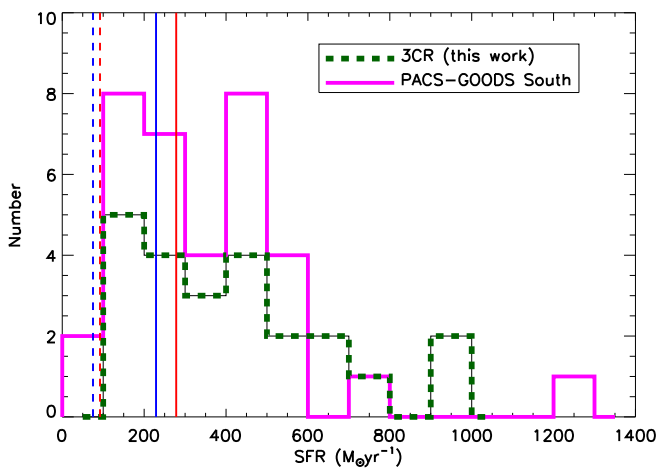}
      \caption{Comparison of star formation rates of AGN and non-AGN host 
               galaxies. Plotted are the objects detected in at least 
               three \textit{Herschel} bands (dashed green). A comparison 
               sample of objects (solid magenta) is selected from some of 
               the deepest \textit{Herschel} data, from PACS-GOODS South 
               field \citep{Rodighiero11}. The vertical lines correspond 
               to the average values of the FIR-detected (solid lines) and 
               non-detected (dashed lines) stacked subsamples, discussed 
               in Sect.~\ref{subsection:Stacking}.
			   }
      \label{figure:Rodighiero}
   \end{figure} 
   It is important to check how the SFRs obtained for the FIR-detected 
   objects in this work compare to those for the non-AGN galaxy population. 
   SFRs in star-forming galaxies are strongly correlated with both 
   stellar mass and redshift \citep[e.g.][]{Elbaz07}, therefore, the comparison
   must be made with a sample having a comparable range of stellar mass and 
   redshift. About a dozen 3CR radio galaxies studied in this 
   work have stellar mass estimates primarily based on SED-fitting of visible/NIR 
   \citep{Best98} or MIR broad-band photometry \citep{Seymour07,DeBreuck10}. 
   The estimated stellar masses range between $1.5 \times 10^{11} M_{\odot}$ and 
   $6 \times 10^{11} M_{\odot}$, and are not expected to be a function of 
   the redshift/luminosity of the sources \citep{DeBreuck10}. 
   Measurements of stellar masses in high-$z$ quasar hosts are problematic because 
   the strong continuum emission from the accretion disk often outshines their host 
   galaxies. However, assuming that unification of powerful radio galaxies and 
   quasars holds, the masses of quasars and radio galaxies hosts are often taken to 
   be (at least on average) similar \citep[e.g.][]{McLure06}. Thus, for the discussion 
   below, we assume that the stellar masses of our 3CR hosts range between 
   $1.5 \times 10^{11} M_{\odot}$ and $6 \times 10^{11} M_{\odot}$
     
   As a control sample of non-AGN galaxies, we selected $1.5 < z < 2.5$ galaxies 
   within the stellar mass range indicated above, whose 
   star-forming properties were estimated from deep PACS data of the GOODS 
   South field \citep[cyan points in Fig.1 from][]{Rodighiero11}. The stellar 
   masses of these galaxies were estimated from SED-fitting as explained by 
   \citet{Rodighiero10}. While the majority of the selected galaxies lie 
   on the main sequence of star-forming galaxies, a fraction are located above 
   it, most likely characterized by the starbursting nature of the ongoing 
   star formation \citep{Rodighiero11}. 
   Figure~\ref{figure:Rodighiero} shows the SFR histograms of the 3CRs 
   and the selected control sample. It is clear that the 3CR FIR-detected 
   objects have, on average, SFRs comparable to those of their equally massive 
   non-AGN counterparts. The majority of FIR-detected objects are thus also 
   located near the main sequence of star-forming galaxies, similarly to what 
   has been found in deep \textit{Herschel}/PACS studies of less powerful, X-ray 
   selected, high-$z$ AGN \citep{Mullaney12,Santini12,Rosario13}. 
   In comparison to the FIR-detected objects, the non-detected objects have 
   similar stellar mass but significantly lower SFRs, placing them below the 
   main sequence of star-forming galaxies. We cannot exclude the possibility 
   that the star formation activity in some of these objects has been quenched.
   Better estimates of the stellar masses of our high-$z$ 3CR sources will 
   allow a more robust statistical study of their exact location with 
   respect to the main sequence of star-forming galaxies.
   \subsection{Subgalactic versus supergalactic radio sources}
   \label{subsection:CSSvsEXT}  
\begin{figure} 
      \centering
      \includegraphics[width=\hsize]{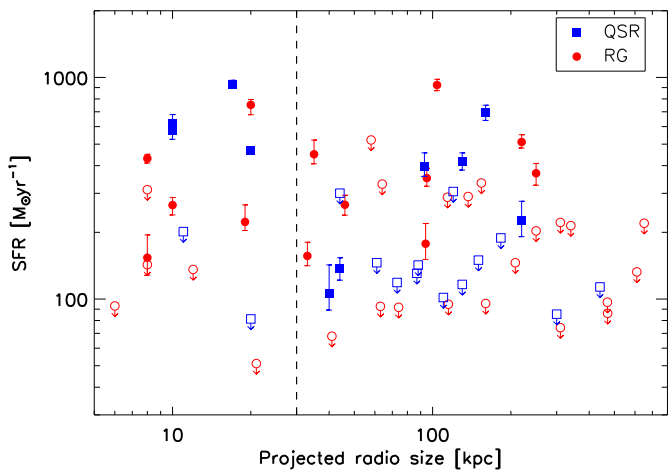}
      \caption{Estimated star formation rate as a function of projected 
               radio size (measured lobe to lobe) for the 3CR galaxies 
               (red circles) and quasars (blue squares). The dashed line 
               indicates the value taken to differentiate between subgalactic 
               (<~30 kpc) and supergalactic (>~30 kpc) radio sources, 
               respectively. Arrows indicate star formation rate upper limits 
               for objects with fewer than three \textit{Herschel} detections. 
	       }
      \label{figure:SFRvssize}
   \end{figure}

   The radio morphologies of radio-loud AGN present a unique opportunity 
   to estimate the duration of the AGN episode by assuming a typical 
   speed of radio jet expansion of 10\%-20\% of the speed of light. 
   The projected radio sizes of our high-$z$ 3CR sources, measured 
   lobe-to-lobe, have been measured from high-resolution radio images. 
   Based on their projected radio sizes, we divided the 3CR sources 
   into two groups, subgalactic (<~30~kpc) and supergalactic 
   radio (>~30~kpc) sources. The subgalactic sources, typically 
   contained within their host galaxies, account for 25\% of our 
   high-$z$ 3CR sample. Figure~\ref{figure:SFRvssize} shows the objects' 
   estimated SFRs as a function of their projected radio size. Both the 
   subgalactic and supergalactic FIR-detected 3CR hosts have comparable 
   SFRs. In contrast, the majority of the non-detected objects have larger, 
   i.e. older radio sources. As such, the ratio of FIR-detected versus 
   non-detected objects appears to be a function of projected radio size, 
   changing from 1.3 for subgalactic to 0.5 for supergalactic sources. 
   
   There are at least two different effects contributing to the findings 
   presented in Fig.~\ref{figure:SFRvssize}. First, the finding 
   that many quiescent galaxies turn up when the radio sources are large 
   is consistent with the fact that star formation depends heavily on the 
   availability of cold gas. Indeed, the process of exhausting the available 
   fuel for star formation has timescales similar to the age of large 
   radio sources. Second, our result that many small radio sources are hosted 
   by strongly star-forming galaxies is consistent with the 
   observational/theoretical finding that radio jets may induce bursts of 
   star formation \citep[positive feedback, e.g.][]{Silk13} in the hosts of 
   high-$z$ \citep{Dey97,VanBreugel98} and low-$z$ radio-loud AGN 
   \citep{Tadhunter11,Dicken12}. \citet{Best96} found that smaller radio 
   sources show stronger alignment effect\footnote{The alignment effect is 
   the co-spatial extent of radio and UV/visible/NIR emission in radio-loud 
   AGN, partly due to the interaction between the jet and the interstellar 
   matter of the host galaxy \citep{McCarthy93,Miley&deBreuck08}.} providing 
   further support for the incidence of positive feedback in hosts of small 
   radio sources. A step forward in probing the incidence of positive feedback 
   within the hosts of 3CR sources may be achieved by correlating the ages 
   of the young stellar components and those of the current radio episode, 
   similarly to the study by \citet{RoccaVolmerange13}. An issue which 
   complicates the overall picture, however, is whether the link between 
   the smaller radio sources and star formation activity is a consequence
   of an observational bias. Namely, as pointed out by \citet{Tadhunter11}, 
   the interaction between the jet and the host ISM in subgalactic sources 
   may boost the radio emission, leading to a preferential selection of such 
   sources in flux-limited samples, like the 3CR studied here. 
   \subsection{Model fit limitations}
   \label{subsection:Limitations}
   Our estimates of the cold dust temperature may suffer from a bias due 
   to the known degeneracy between the cold dust temperature and the dust 
   emissivity index, $\beta$: a lower fixed $\beta$ value will lead to a 
   higher dust temperature. Constraining the emissivity index within our 
   sample is a difficult task because (1) the peak of the dust emission 
   is not well isolated in the SED, and (2) only a few data points probe 
   the Rayleigh-Jeans tail of the dust emission. Because of this, and to 
   limit the number of free parameters throughout the fitting, we fixed 
   the emissivity index to a value of 1.6. However, the detections in all 
   \textit{Herschel} bands, in addition to the availability of submm data, 
   allow $\beta$ to be constrained for a few selected objects. The $\beta$ 
   values for these five objects (flagged in Table~\ref{table:PhysProp}) 
   range from 1 to 2.3, with increased/decreased estimated cold dust 
   temperatures for lower/higher $\beta$ values, respectively. Nevertheless, 
   the star formation luminosities, and consequently SFRs, remain the same 
   (within 10\%). 
   
   While we attribute the FIR emission to star formation on the scale of 
   the host galaxy, we stress that the FIR emission considered is the excess 
   emission after the AGN-powered torus emission \citep[using the models of][]{Hoenig&Kishimoto10} 
   has been accounted for. From our SEDs, we concluded that star formation 
   is the dominant process yielding emission at rest-frame wavelengths 
   longer than $\lambda > 50~\mu\mathrm{m}$ \citep[see also][]{Leipski13}. 
   Nevertheless, should future AGN models demonstrate that the AGN-powered 
   emission continues to dominate at rest-frame wavelengths longer than 
   $\lambda \sim 50~\mu\mathrm{m}$, then our SFR estimates are likely to 
   be upper limits.           
\section{Conclusions}
   \label{section:Conclusions}  
   We present \textit{Herschel} photometry of the complete sample of $z>1$ 
   3CR radio galaxies and quasars. The 3CR sample is a flux-limited sample, 
   consisting of some of the most powerful radio-loud AGN accreting in 
   quasar-mode. Combining the \textit{Herschel} photometry with available 
   \textit{Spitzer} data, we performed a full IR SED analysis, separating 
   the contribution from the AGN and from the star formation activity in 
   the host galaxy. We summarize our findings below: 
   \begin{enumerate}
      \setlength{\itemindent}{1em}
      \item About 40\% of the studied objects have robust PACS and SPIRE 
      		detections, translating into ULIRG-like star formation 
			luminosities, i.e. of the order of SFRs of hundreds of solar masses 
			per year. Such prodigious levels of star formation have 
			recently also been inferred for other high-$z$ radio galaxies 
			\citep[e.g.][]{Drouart14}. Merger induced and/or jet 
			triggered star formation activity are both possible 
			mechanisms leading to the SFRs obtained for these objects.
      \\
	  \item The SFRs of the FIR-detected objects are comparable to those 
	        of mass-matched, non-AGN galaxies, selected from deep 
	        \textit{Herschel} surveys. There is no clear evidence that 
	        the star formation has been quenched in the hosts of the 
	        FIR-detected objects. 
      \\
      \item The total IR (1-1000~$\mu$m) emission from the high-$z$ 3CR radio sources 
			is predominantly powered by the AGN, despite the 
			frequently strong starburst activity coeval with the AGN 
			episode. Furthermore, no strong correlation between 
			the AGN- and star formation powered IR luminosities is found. 
	  \\
	  \item The median SEDs of the FIR-detected objects show that RGs 
	        and QSRs are quite different in the MIR, but remarkably 
	        similar in the FIR. Thus, while the MIR emission is anisotropic, 
	        the FIR emission is isotropic and optically thin. These findings 
			are consistent with the orientation-based unification of radio-loud AGN.
	  \\
	  \item Splitting the sample into subgalactic (<~30 kpc) and 
	        supergalactic (>~30 kpc) radio sources, the fraction of 
	        \textit{Herschel} detected objects is a function of the 
	        projected radio size of the sources. In particular, the hosts 
	        of subgalactic radio sources are more likely to be detected 
	        by \textit{Herschel}, arguing for a possible link between 
	        radio size and star formation activity, i.e. jet-induced 
	        star formation (positive feedback), or for fading 
                of star formation in mature AGN.  
	  \\
      \item Stacking of the \textit{Herschel} non-detected objects reveals a 
			class of MIR/FIR faint objects. While ongoing star formation 
			episodes (at significantly lower levels than those discussed 
			above) cannot be ruled out, star formation has largely ceased 
			in the hosts of these objects. As such, the radio-selected, 
			high-$z$ 3CR hosts appear to be a heterogeneous mixture of 
			MIR/FIR bright and faint objects.
   \end{enumerate} 
   
   Upcoming instruments with better sensitivity/resolution, such as ALMA, 
   will likely help us pinpoint the exact location of the ongoing star formation 
   in high-$z$ galaxies. This will lead to a further understanding of the interplay 
   between the AGN and star formation activity within high-$z$ AGN hosts.      
\begin{acknowledgements}
   The authors acknowledge the expert referee for useful comments which 
   improved the paper, and thank Giulia Rodighiero for kindly providing 
   the PEP data from the deep GOODS South field. PP acknowledges the 
   Nederlandse Organisatie voor Wetenschappelijk Onderzoek (NWO) 
   for a PhD fellowship. MH and CW are supported by the Akademie der 
   Wissenschaften und der K\"unste Nordrhein-Westfalen and by Deutsches 
   Zentrum f\"ur Luft-und Raumfahrt (DLR). 
   The \textit{Herschel} spacecraft was designed, built, tested, and launched under a 
   contract to ESA managed by the \textit{Herschel}/\textit{Planck} Project team by an industrial 
   consortium under the overall responsibility of the prime contractor Thales 
   Alenia Space (Cannes), and including Astrium (Friedrichshafen) responsible 
   for the payload module and for system testing at spacecraft level, Thales 
   Alenia Space (Turin) responsible for the service module, and Astrium (Toulouse) 
   responsible for the telescope, with in excess of a hundred subcontractors.
   PACS has been developed by a consortium of institutes led by MPE 
   (Germany) and including UVIE (Austria); KU Leuven, CSL, IMEC 
   (Belgium); CEA, LAM (France); MPIA (Germany); INAF-IFSI/OAA/OAP/OAT, 
   LENS, SISSA (Italy); IAC (Spain). This development has been supported 
   by the funding agencies BMVIT (Austria), ESA-PRODEX (Belgium), 
   CEA/CNES (France), DLR (Germany), ASI/INAF (Italy), and CICYT/MCYT (Spain).
   SPIRE has been developed by a consortium of institutes led 
   by Cardiff University (UK) and including Univ. Lethbridge (Canada); 
   NAOC (China); CEA, LAM (France); IFSI, Univ. Padua (Italy); IAC 
   (Spain); Stockholm Observatory (Sweden); Imperial College London, RAL, 
   UCL-MSSL, UKATC, Univ. Sussex (UK); and Caltech, JPL, NHSC, Univ. 
   Colorado (USA). This development has been supported by national 
   funding agencies: CSA (Canada); NAOC (China); CEA, CNES, CNRS 
   (France); ASI (Italy); MCINN (Spain); SNSB (Sweden); STFC, UKSA (UK); 
   and NASA (USA).
   HIPE is a joint development by the \textit{Herschel} Science Ground Segment 
   Consortium, consisting of ESA, the NASA \textit{Herschel} Science Center, and 
   the HIFI, PACS and SPIRE consortia.
   This work is partly based on observations made with the \textit{Spitzer} Space 
   Telescope, which is operated by the Jet Propulsion Laboratory, 
   California Institute of Technology under a contract with NASA. 
   This research has made use of the NASA/IPAC Extragalactic Database (NED) which 
   is operated by the Jet Propulsion Laboratory, California Institute of Technology, 
   under contract with the National Aeronautics and Space Administration.
   This research made use of APLpy, an open-source plotting package for 
   Python hosted at http://aplpy.github.com.
\end{acknowledgements}

\appendix
\section{UV/visible SEDs of quasars}
\label{appendix:UV/optical}

   \begin{figure*}
      \centering
      \includegraphics[width=9.66cm, bb = 54 360 558 719, clip]{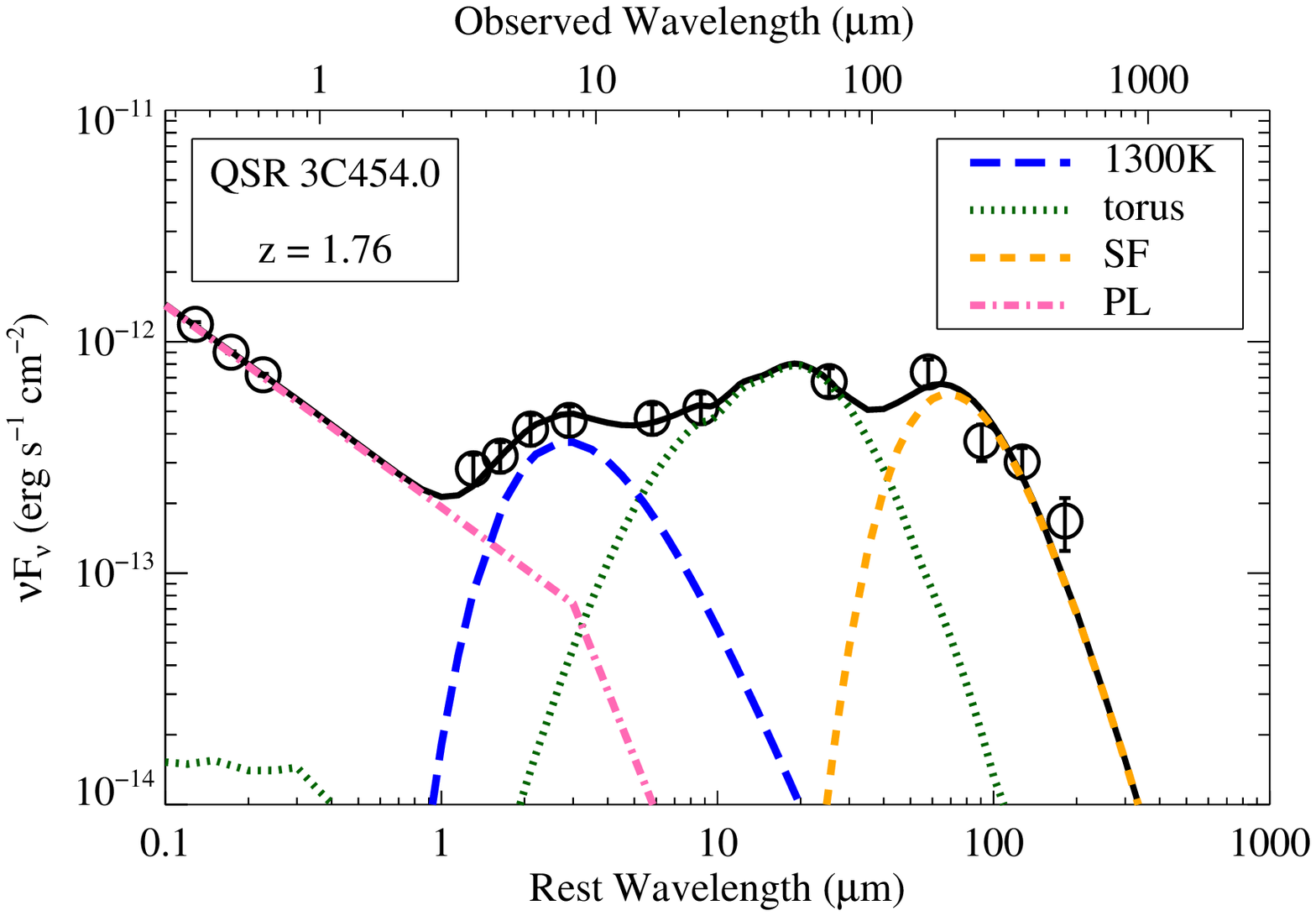}
      \includegraphics[width=8.34cm, bb = 124 360 558 719, clip]{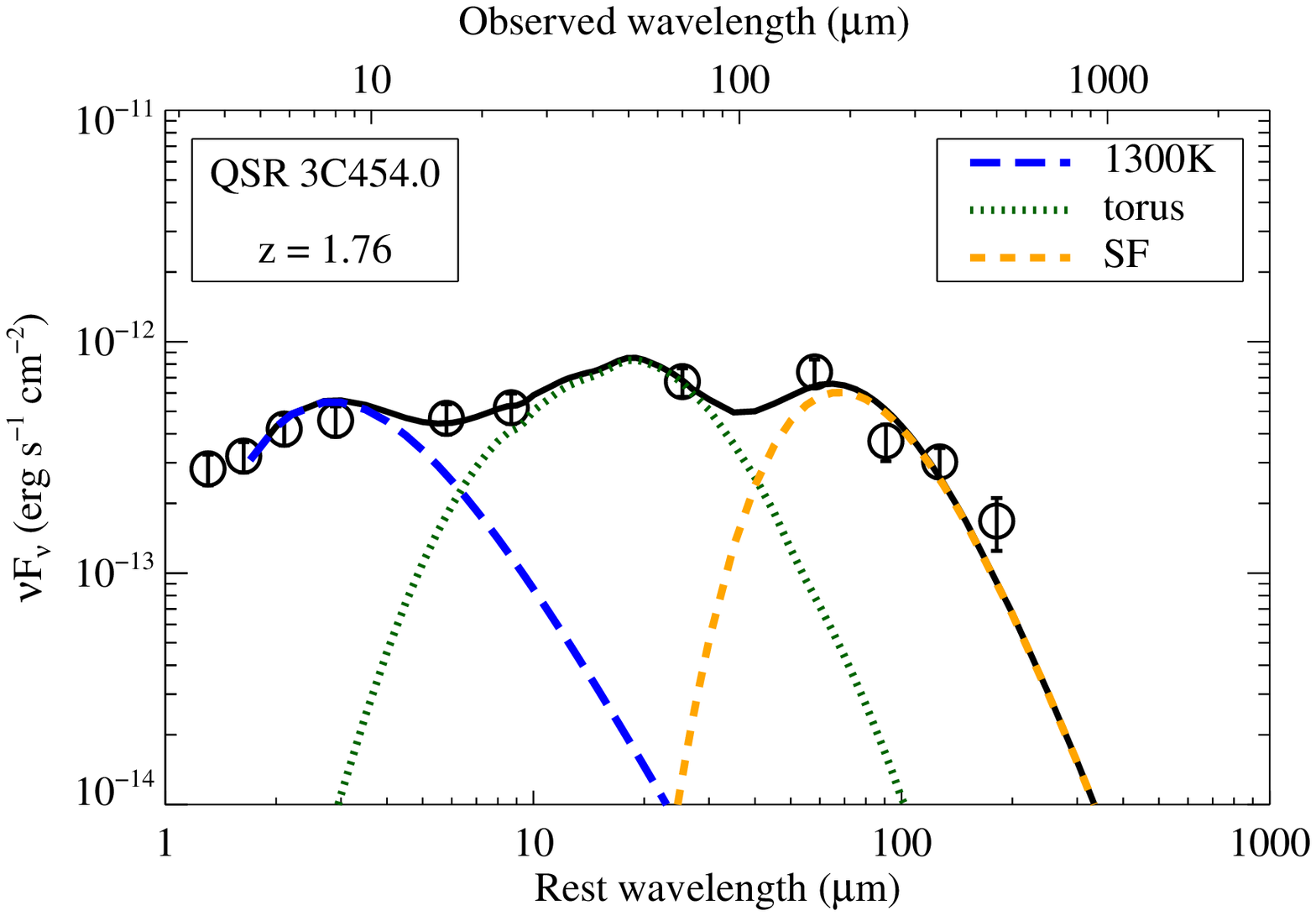} 
      \includegraphics[width=9.66cm, bb = 54 360 558 719, clip]{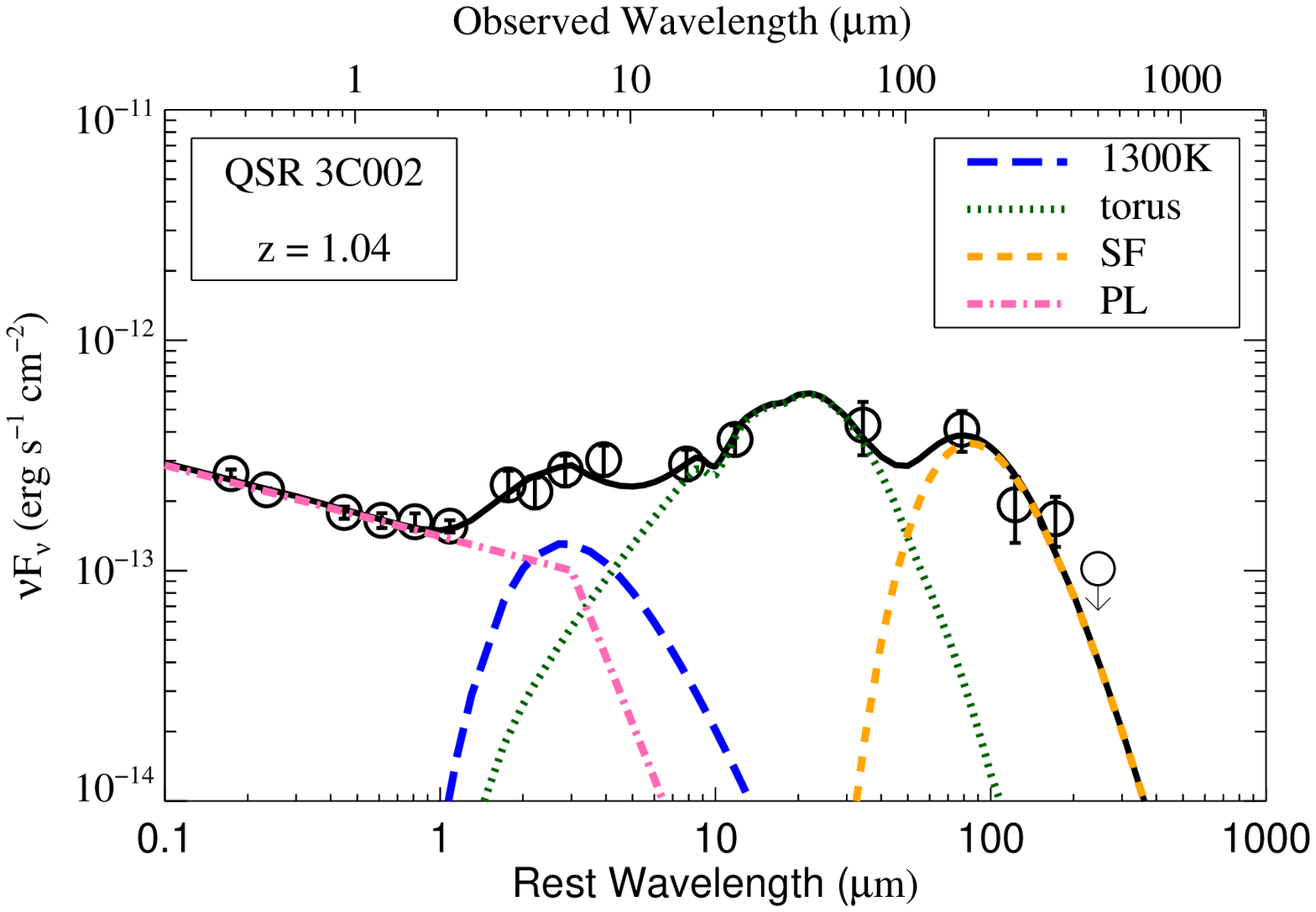}
      \includegraphics[width=8.34cm, bb = 124 360 558 719, clip]{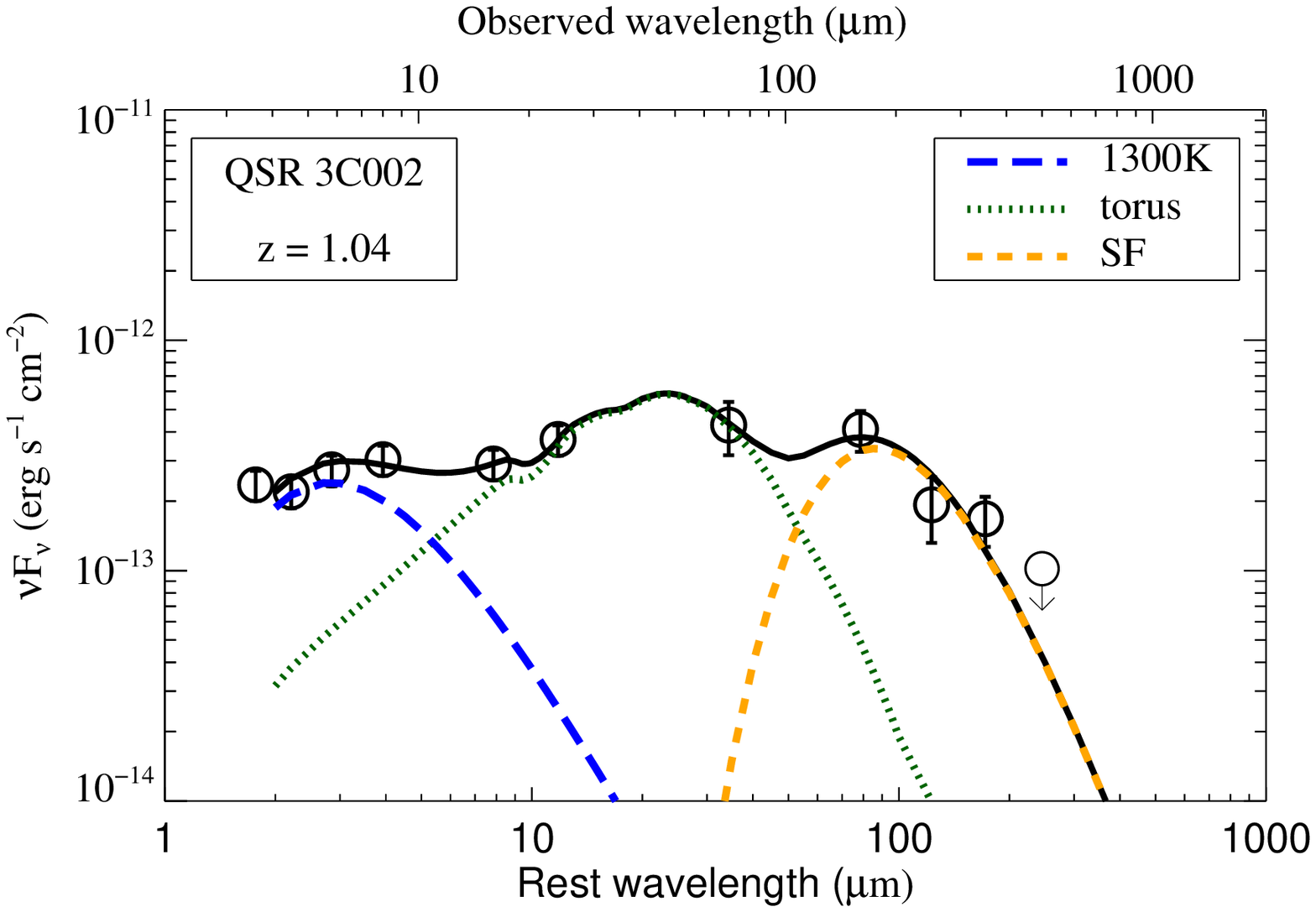}
      \caption{Spectral energy distributions of two quasars with good 
               visible/UV detections. Left panels include a power-law 
               component (dash-dotted pink) representing the emission 
               from the hot accretion disk. Right panels exclude that 
               component, and show the components considered when fitting 
               the quasars in this work. Other individual components as 
               described in Fig.~\ref{figure:exampleSEDs}.
               }
      \label{figure:UVoptical}
   \end{figure*}
   The emission from powerful quasars at UV/visible wavelengths comes 
   predominantly from the hot accretion disk. In quasar SED fitting, this 
   component is typically represented by a power-law (PL), extended to the 
   NIR by introducing a PL index of $-2$ ($F_{\nu} \propto \nu^2$) 
   \citep[e.g.][]{Leipski13} In the current work, we do not fit the 
   rest-frame UV/visible SEDs of quasars because it has a negligible 
   influence on the FIR part of the SED. We demonstrate this using two 
   objects that have good coverage in the UV/visible: 3C~454.0 and 3C~002.

   Firstly, we consider 3C~454.0 (upper panels of Fig.~\ref{figure:UVoptical}), 
   for which we have SDSS photometry (\textit{ugriz}) in addition to the 
   \textit{Spitzer} and \textit{Herschel} photometry, and compare the 
   results when we include/exclude the PL component. In a first attempt, 
   we fit the data using a PL component in addition to the 1300~K 
   blackbody, torus, and star-formation-heated cold dust components. Given the redshift 
   of 3C~454.0, strong MgII emission at rest-frame $\sim$ 0.3 $\mu$m, 
   typical for optical spectra of quasars, significantly contributes to 
   its broad-band $i$ and $z$ photometry. The SED fitting including the PL 
   component was therefore performed using only the $u$, $g$, and $r$ SDSS 
   photometry. In a second attempt, we fit only data at 
   rest-frame wavelengths longer than 2~$\mu$m, excluding the PL component 
   from the fit. As seen in Fig.~\ref{figure:UVoptical}, the best-fit 
   SEDs longward of 2~$\mu$m rest-frame wavelength are very similar for 
   the two fitting approaches. The only difference is the flux normalization 
   of the hot dust component, which turns out to be slightly lower when 
   including the UV/visible part of the SED. Understandably, this is a 
   result of the PL component contributing to the emission in the shorter 
   \textit{Spitzer} wavelengths. While the fitting procedures prefer 
   different torus models, the luminosity of this component changes by 
   less than 10\%. Most importantly, the most relevant physical parameters 
   for the current work: the star formation rate, and the temperature and 
   mass of the cold dust component remain within 10\%. 
      
   Secondly, we consider 3C~002 (lower panels of Fig.~\ref{figure:UVoptical}), 
   for which in addition to SDSS, \textit{Spitzer} and \textit{Herschel} 
   photometry, we have photometry from 2MASS. We fit the full SED of this 
   object, after discarding data points potentially contaminated by strong 
   emission lines. Once again we reach the same results as in the case for 
   3C~454.0, thus we conclude that a detailed treatment of the emission in 
   the UV/visible part of the quasars' SEDs has a negligible effect on the 
   results inferred from the IR part of the SED. Furthermore, many 3CR 
   quasars lack good UV/visible/NIR photometric data, making the inclusion 
   of the PL component to the fitting procedure impossible. Consequently, 
   the results presented in our work were obtained using the best-fit SEDs 
   without fitting a PL to the UV/visible part of the SEDs of quasars. 
\section{Comments on individual objects}
\label{appendix:Comments}
   3C~036 -   This source requires an additional hot dust component to 
              better fit the data. \\
   3C~043 -   There is a bright nearby object dominating the emission in the 
              PACS 160~$\mu$m band. We report no measurement in this band. \\
   3C~065 -   This source requires an additional hot dust component to 
              better fit the data. \\
   3C~068.2 - IRS 16~$\mu$m data point is removed from the fitting procedure. \\
   3C~119 -   Diffuse emission present in the PACS 160~$\mu$m and SPIRE maps 
              owing to the object's low galactic latitude. This source requires 
              an additional hot dust component to better fit the data. \\
   3C~208.1 - This source requires an additional hot dust component to 
              better fit the data. \\
   3C~210 -   Bright nearby object present in SPIRE bands. Deblending is not 
              possible. We report no measurements in the SPIRE bands. Photometry 
              in PACS 160~$\mu$m was performed with an aperture of 6$\arcsec$ 
              radius. This source requires an additional hot dust component to 
              better fit the data. IRAC 8~$\mu$m and IRS 16~$\mu$m data are 
              not well fitted. \\
   3C~222 -   IRS 16~$\mu$m data point is removed from the fitting procedure. \\
   3C~230 -   The \textit{Spitzer} photometry probably includes a star located very 
              close to the radio galaxy. IRAC 3.6~$\mu$m, IRAC 4.5~$\mu$m, and IRAC 5.8~$\mu$m 
              data points are removed from the fitting procedure. \\
   3C~252 -   This source requires an additional hot dust component to 
              better fit the data. \\
   3C~255 -   The IRAC points are not well-fitted with the blackbody component
              representing emission from old stars in the host galaxy. \\ 
   3C~257 -   This source requires an additional hot dust component to 
              better fit the data. \\
   3C~267 -   This source requires an additional hot dust component to 
              better fit the data. \\
   3C~305.1 - This source requires an additional hot dust component to 
              better fit the data. IRAC 8~$\mu$m and IRS 16~$\mu$m data are 
              not well fitted. \\
   3C~318 -   Bright nearby object present in PACS and SPIRE bands. Deblending is 
              not possible for SPIRE 350~$\mu$m and SPIRE 500~$\mu$m. We report 
              no measurement in these two bands. \\
   3C~324 -   IRAC 8~$\mu$m and IRS 16~$\mu$m data are not well fitted. \\
   3C~418 -   The only flat-spectrum-core-dominated object within the high-$z$ 
              3CR sample; completely synchrotron dominated. We do not 
              include this source in the analysis. \\
   3C~454.1 - Diffuse emission present in SPIRE maps. \\
   3C~469.1 - The PACS 160~$\mu$m flux density might be slightly contaminated by 
              a nearby object. This source requires an additional hot dust 
              component to better fit the data. IRAC 8~$\mu$m and IRS 16~$\mu$m 
              data are not well fitted. \\

\Online


\section{Best-fit SEDs of objects detected in at least three \textit{Herschel} bands}
\label{appendix:FIR-detectedSEDs}
   \begin{figure*}
      \includegraphics[width=4.5cm]{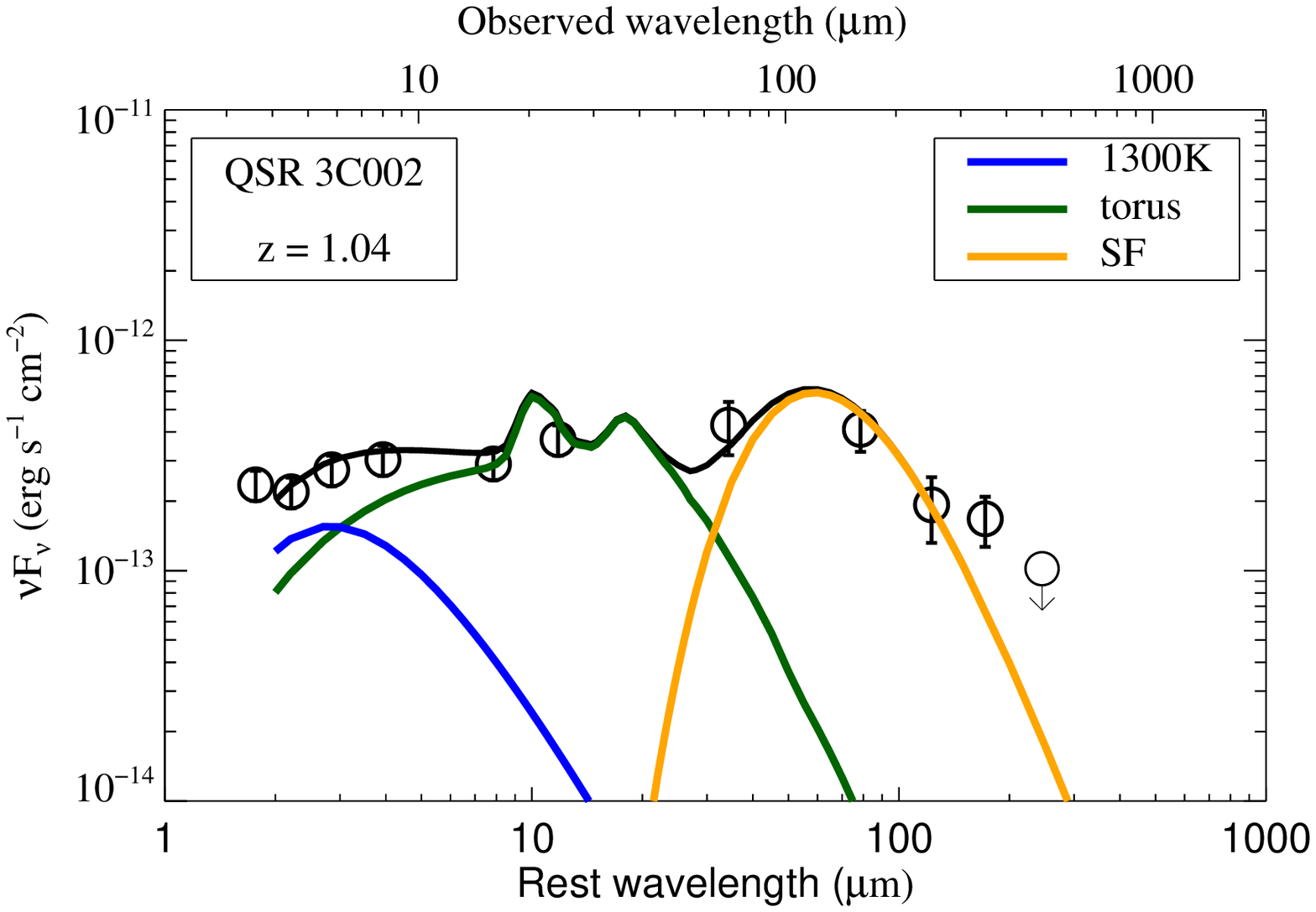}
      \includegraphics[width=4.5cm]{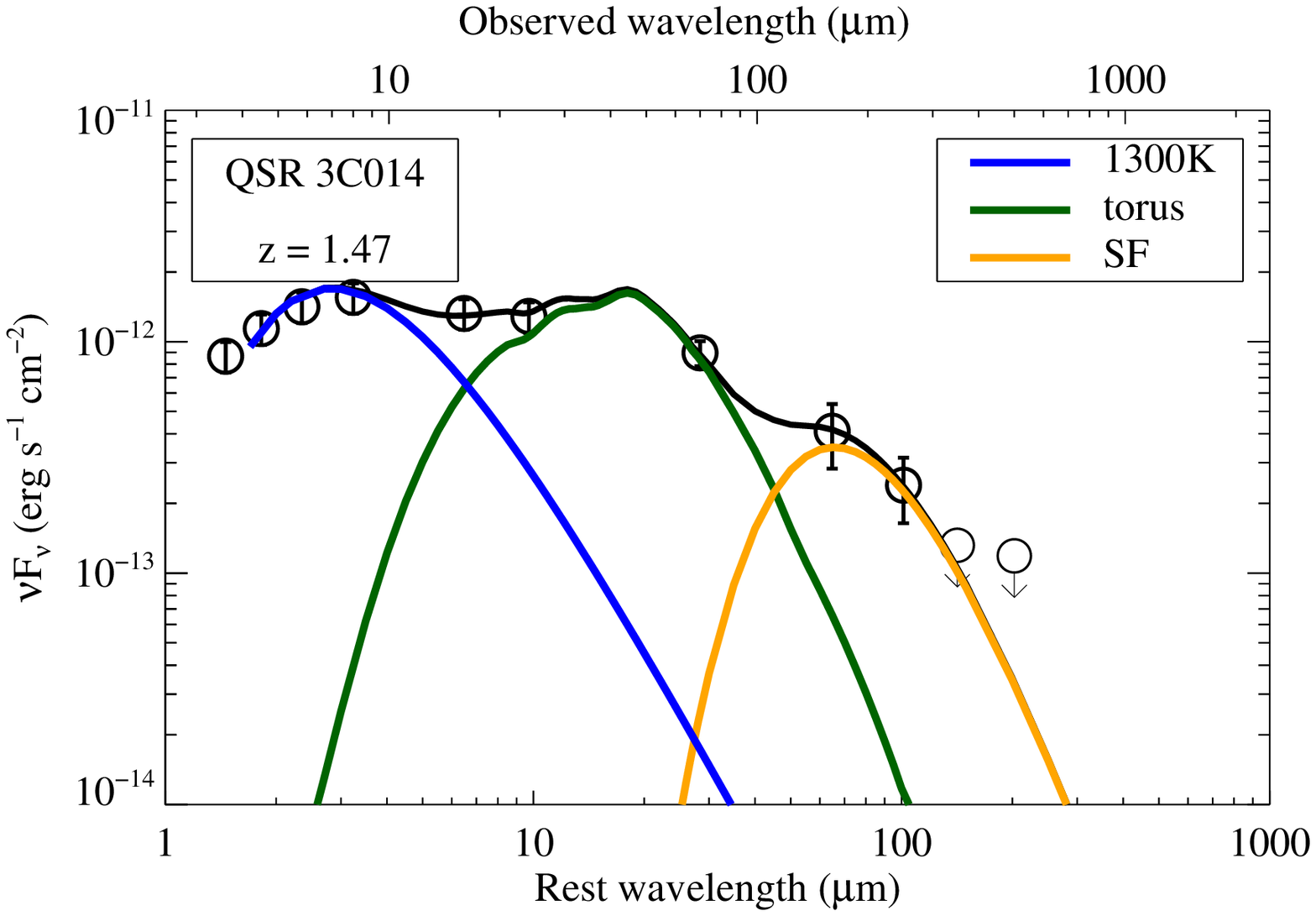}
      \includegraphics[width=4.5cm]{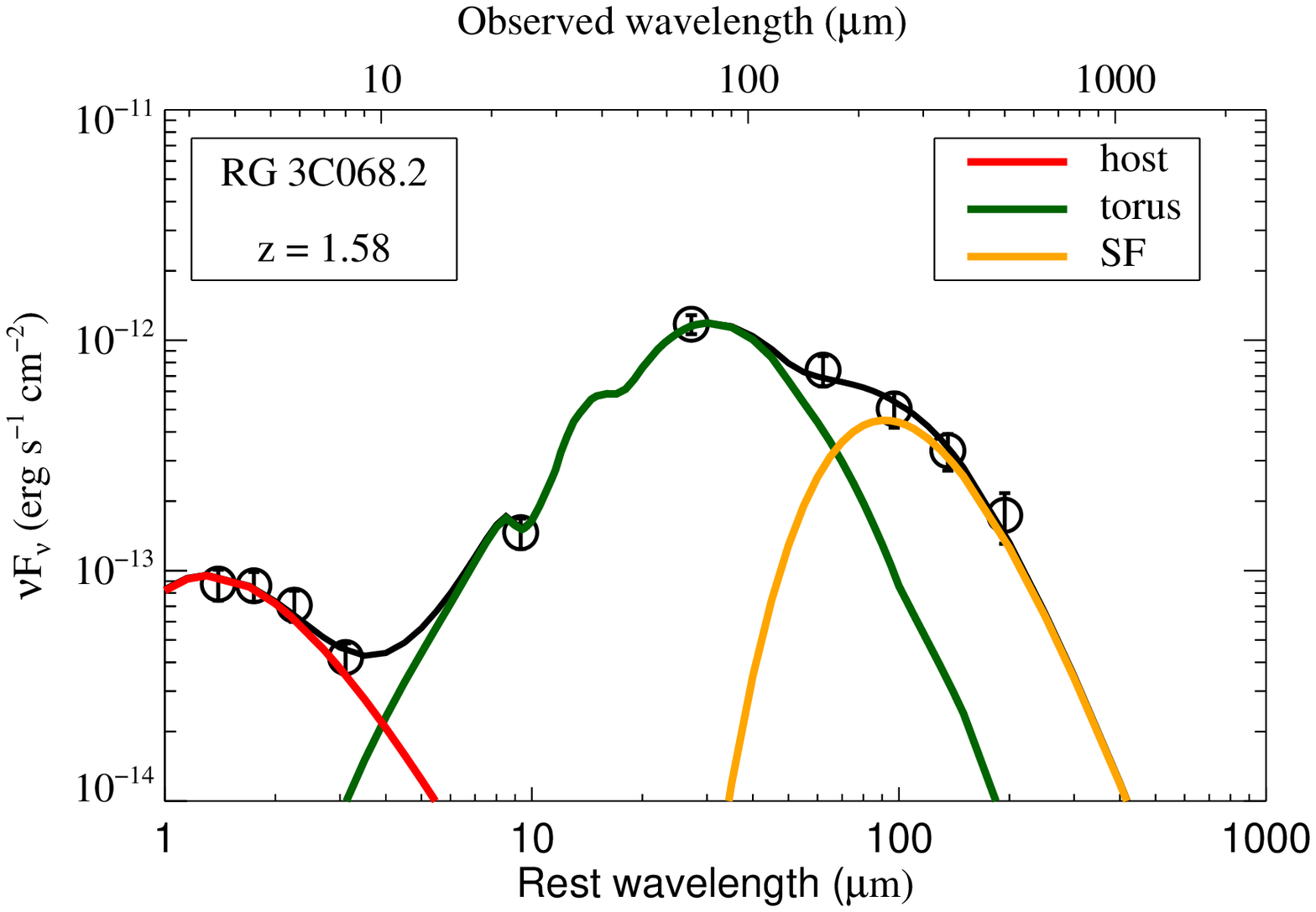}
      \includegraphics[width=4.5cm]{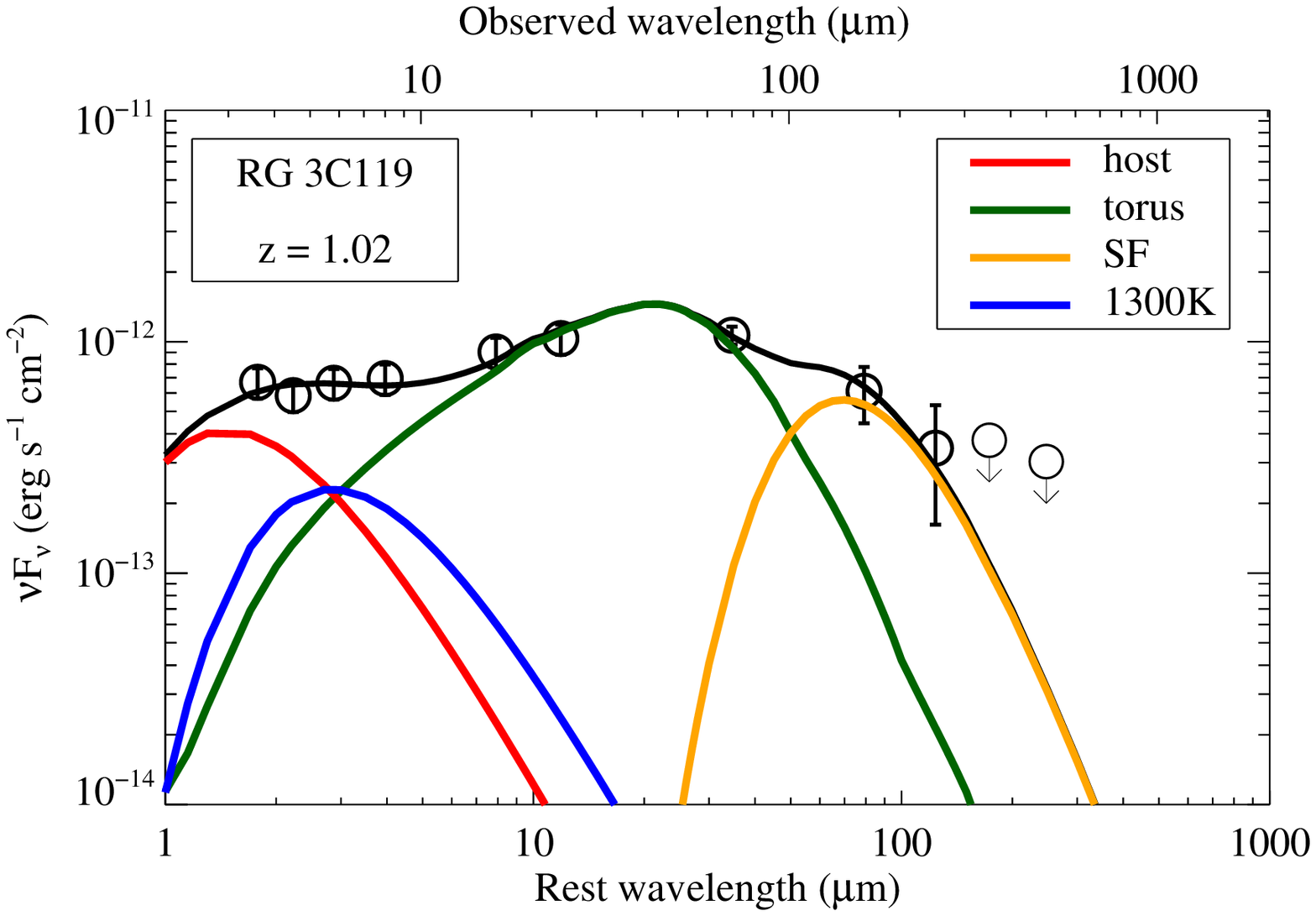}
      \includegraphics[width=4.5cm]{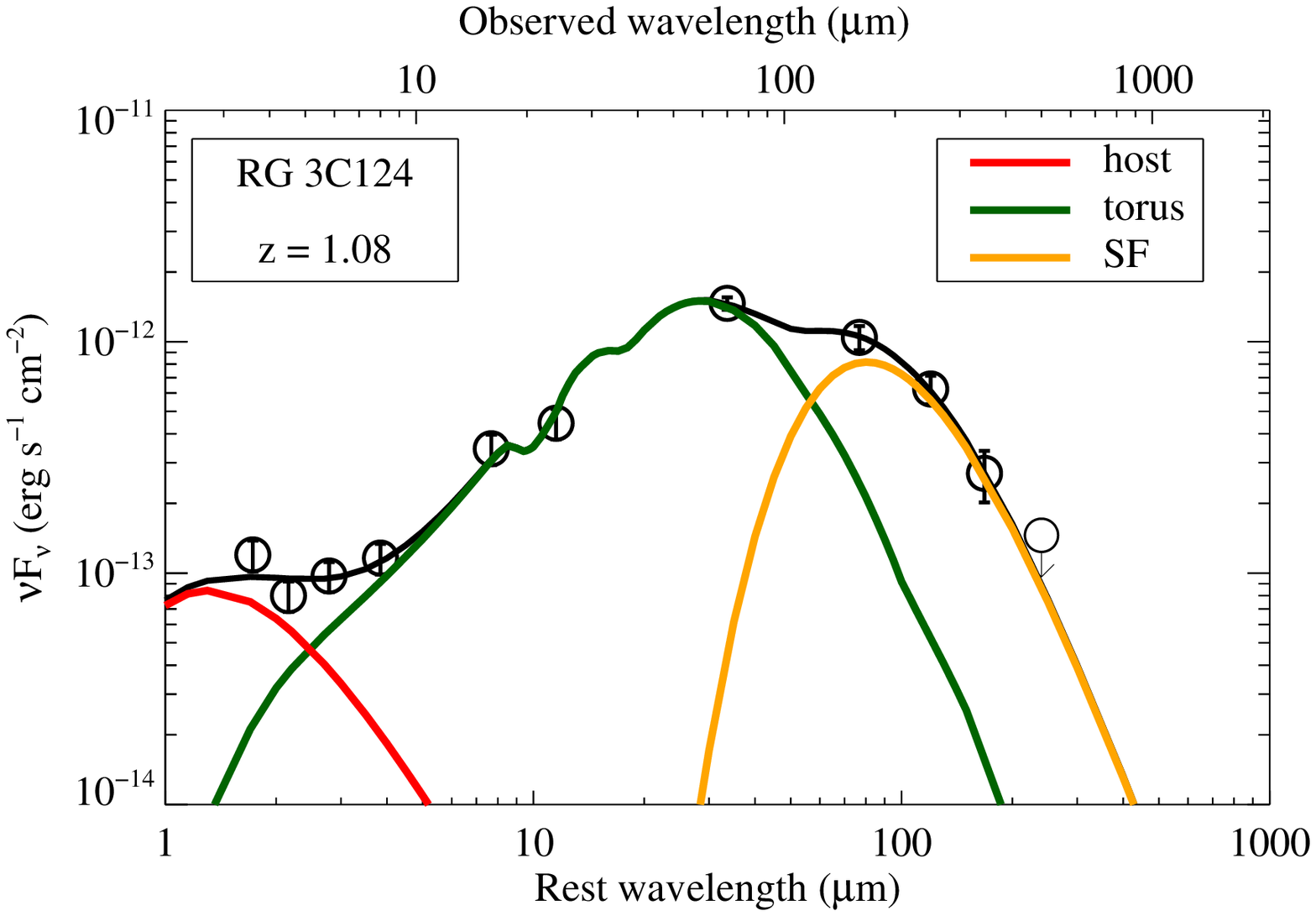}
      \includegraphics[width=4.5cm]{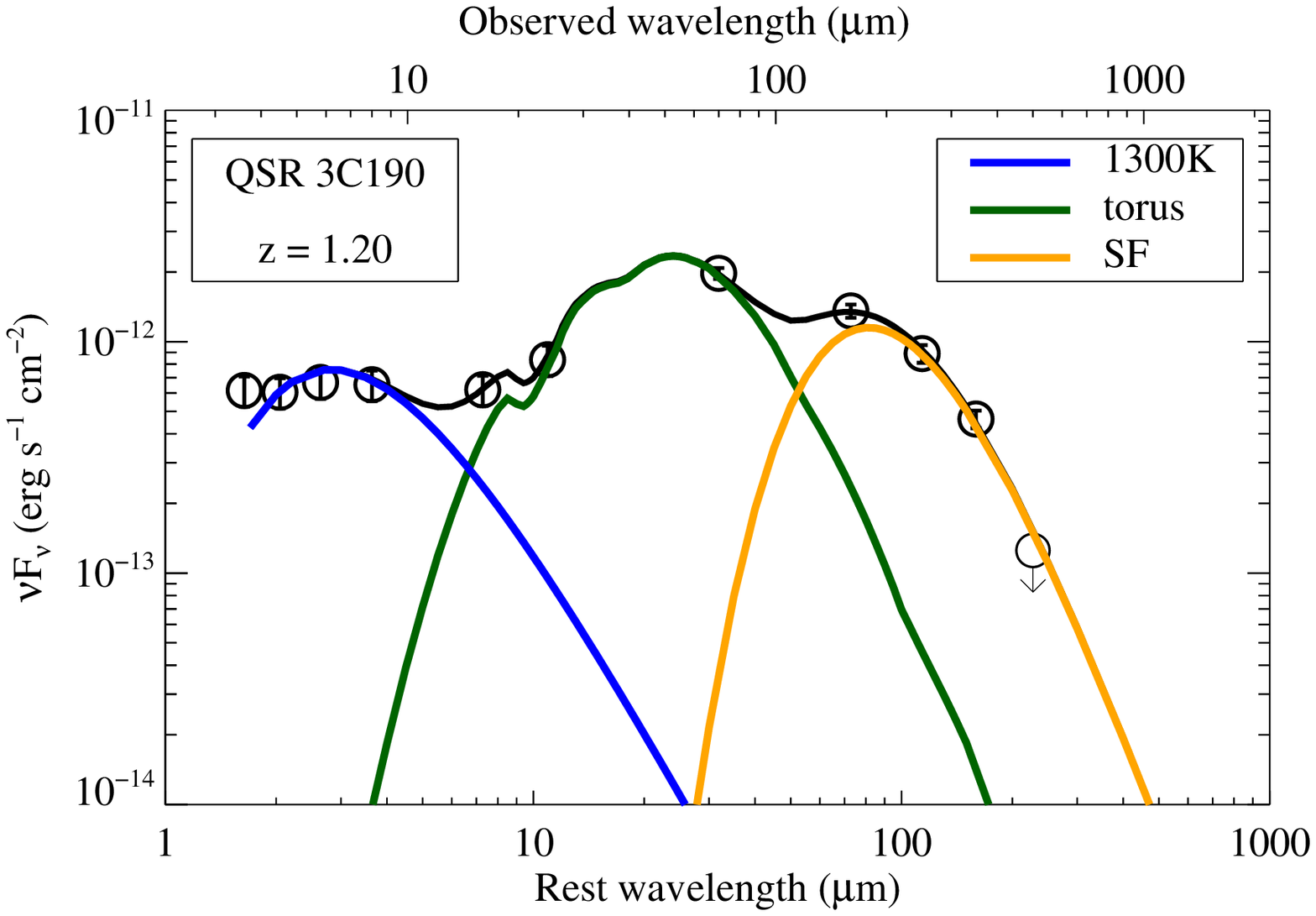}
      \includegraphics[width=4.5cm]{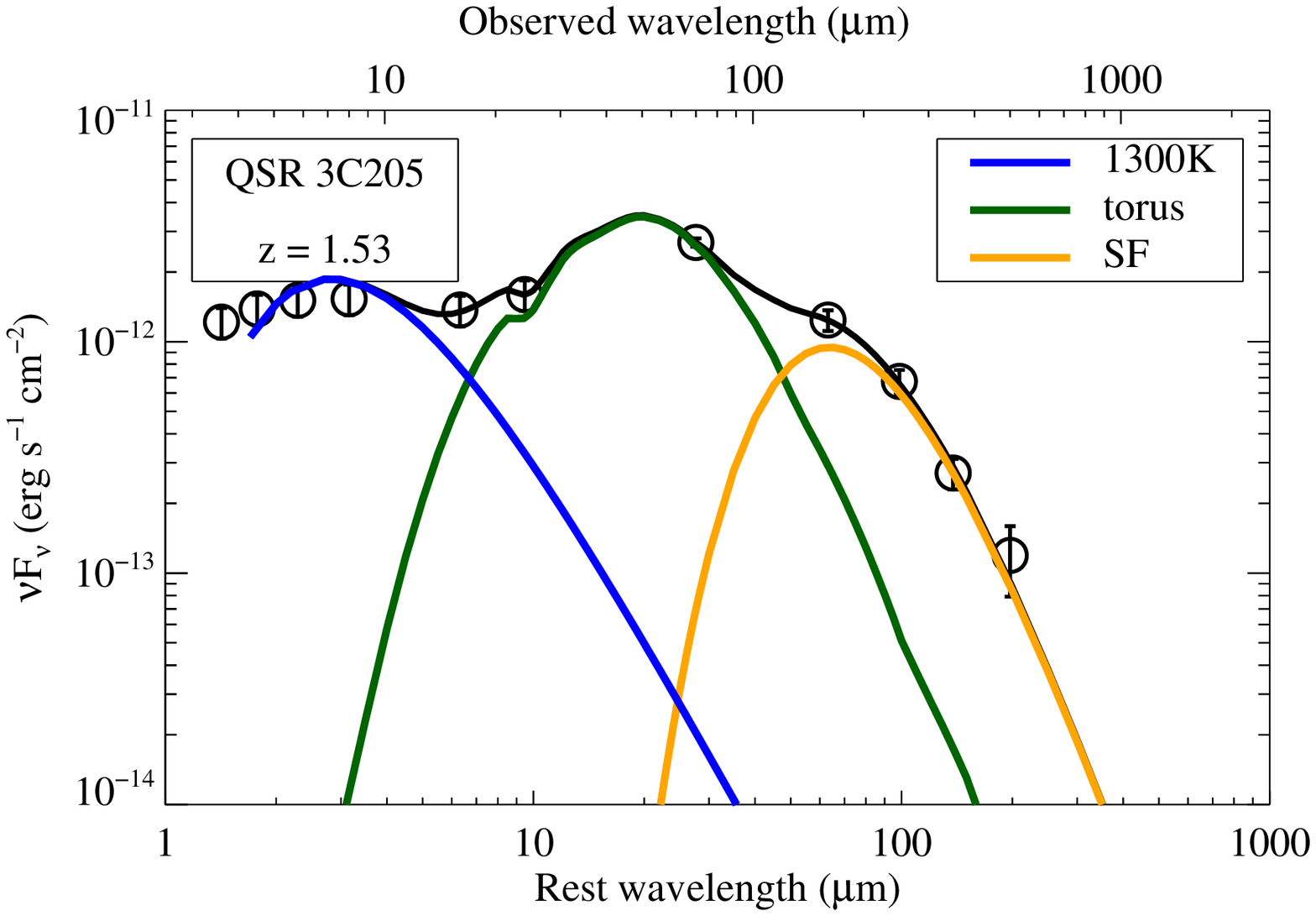}
      \includegraphics[width=4.5cm]{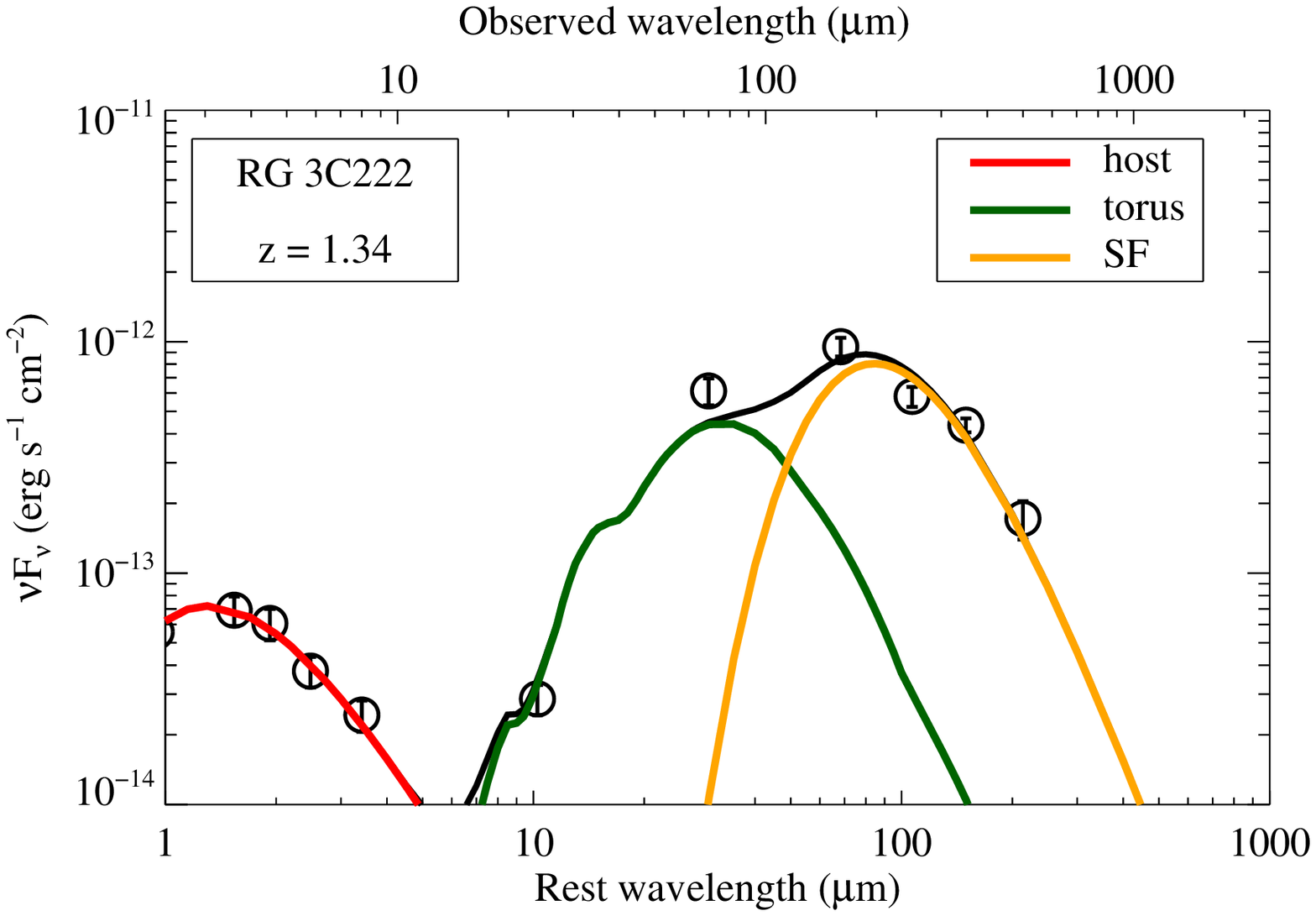}
      \includegraphics[width=4.5cm]{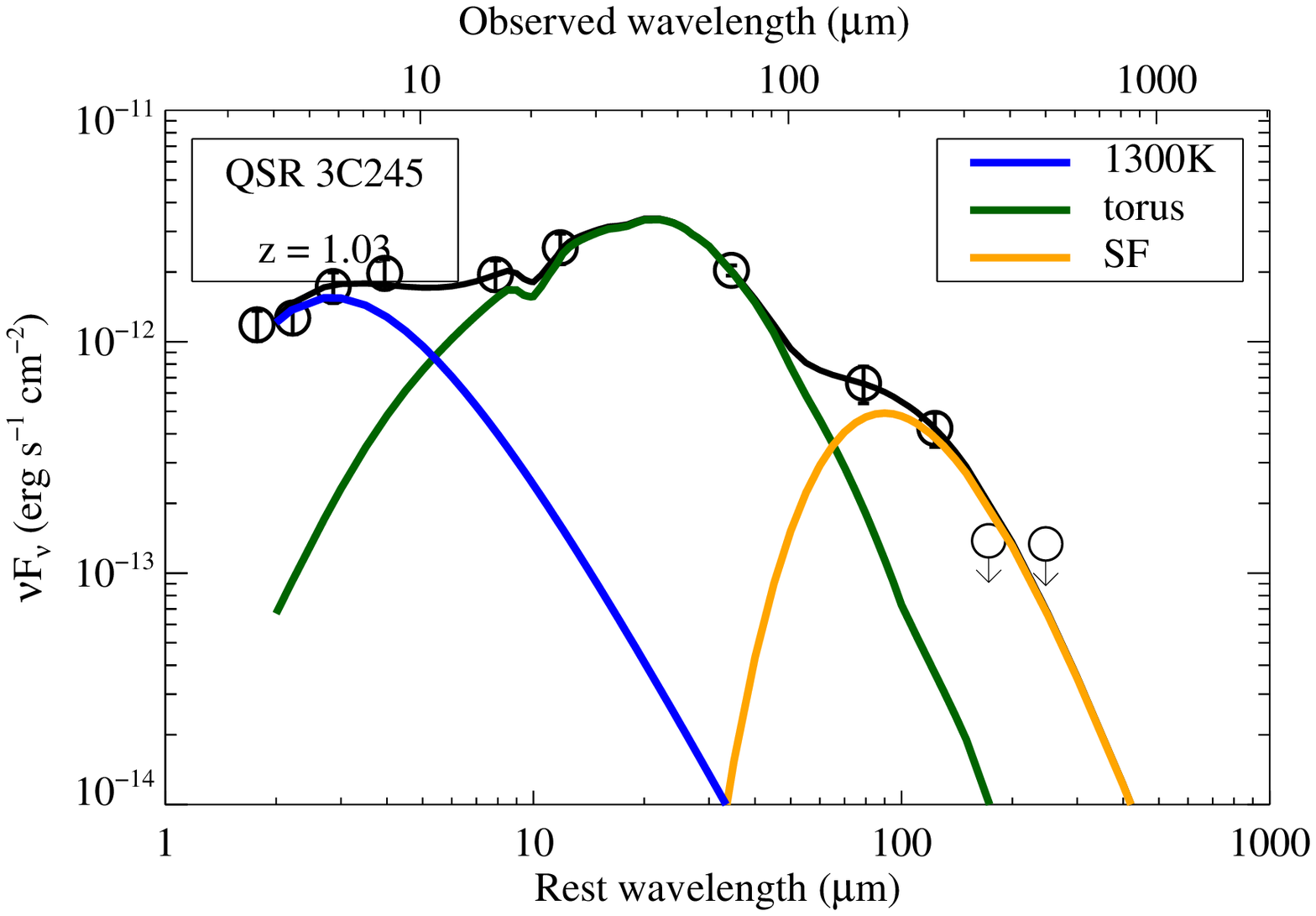}
      \includegraphics[width=4.5cm]{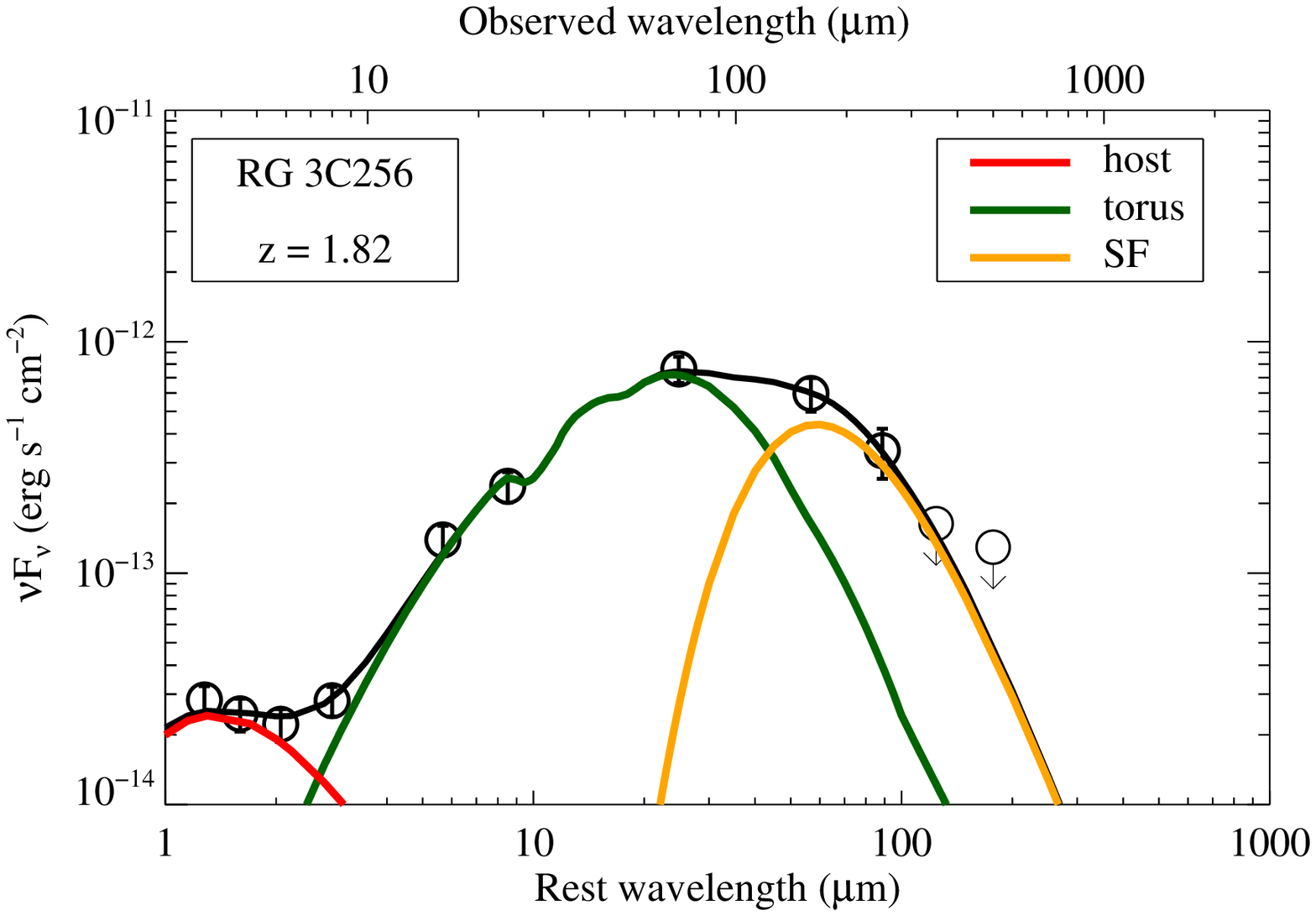}
      \includegraphics[width=4.5cm]{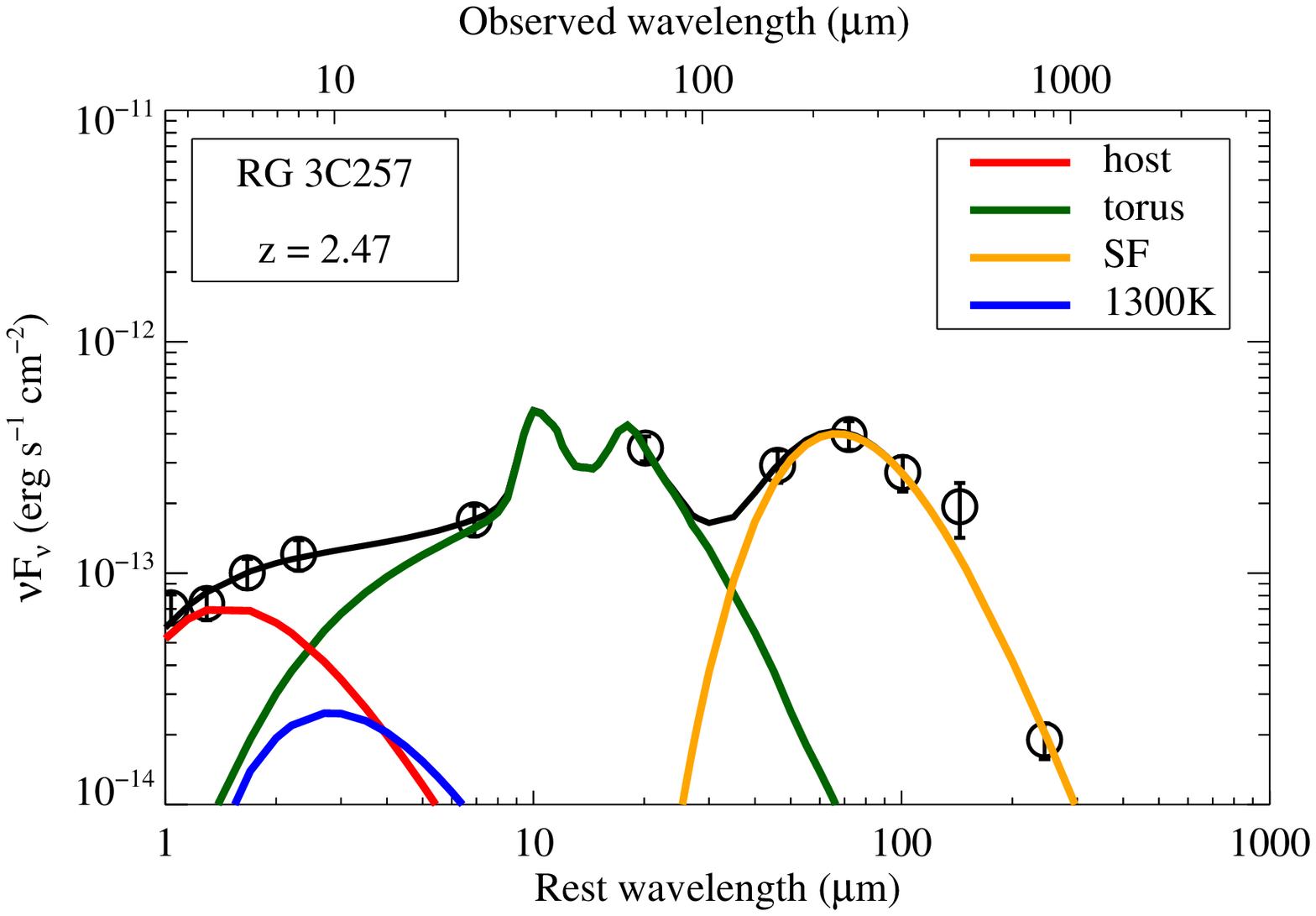}
      \includegraphics[width=4.5cm]{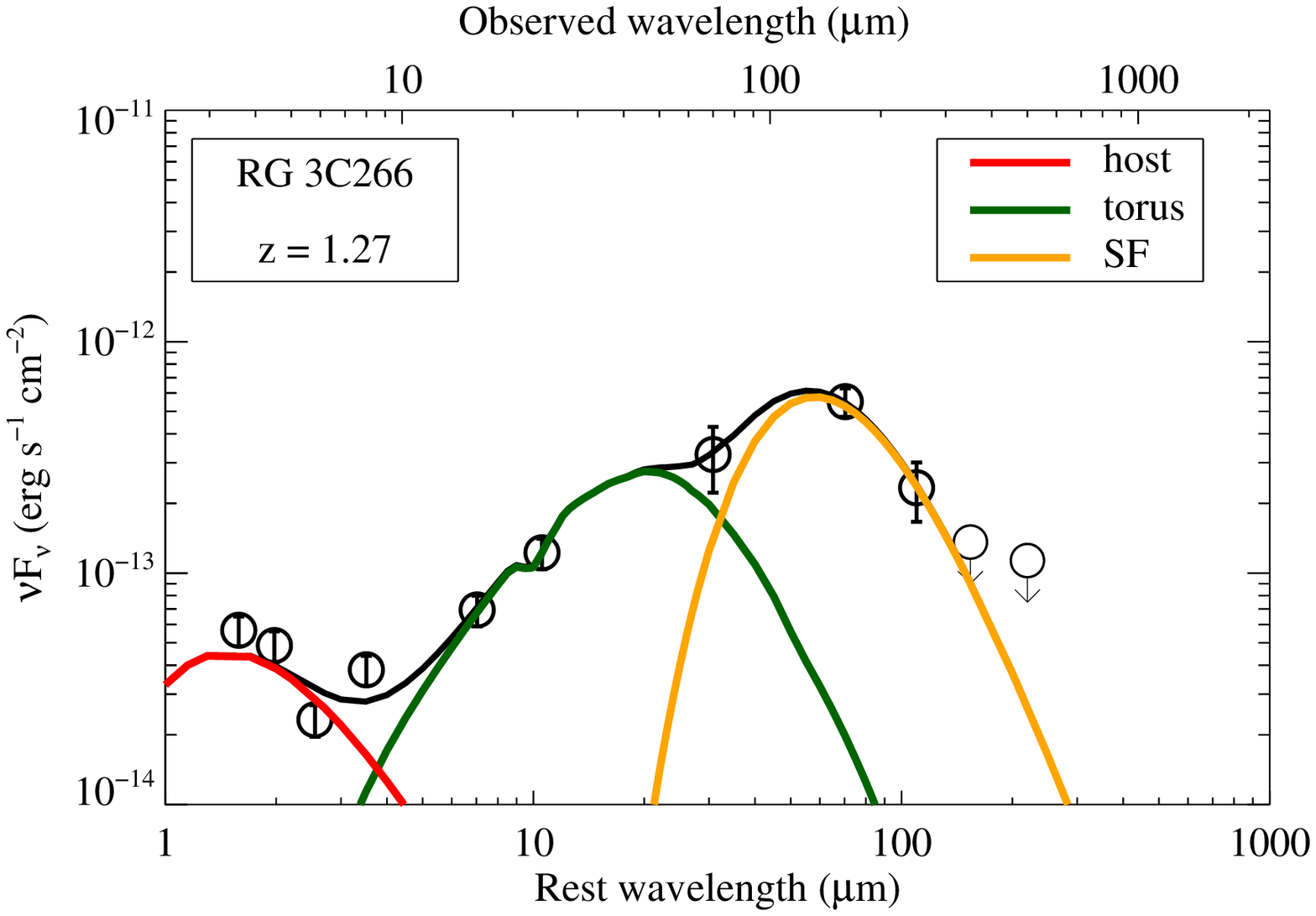}
      \includegraphics[width=4.5cm]{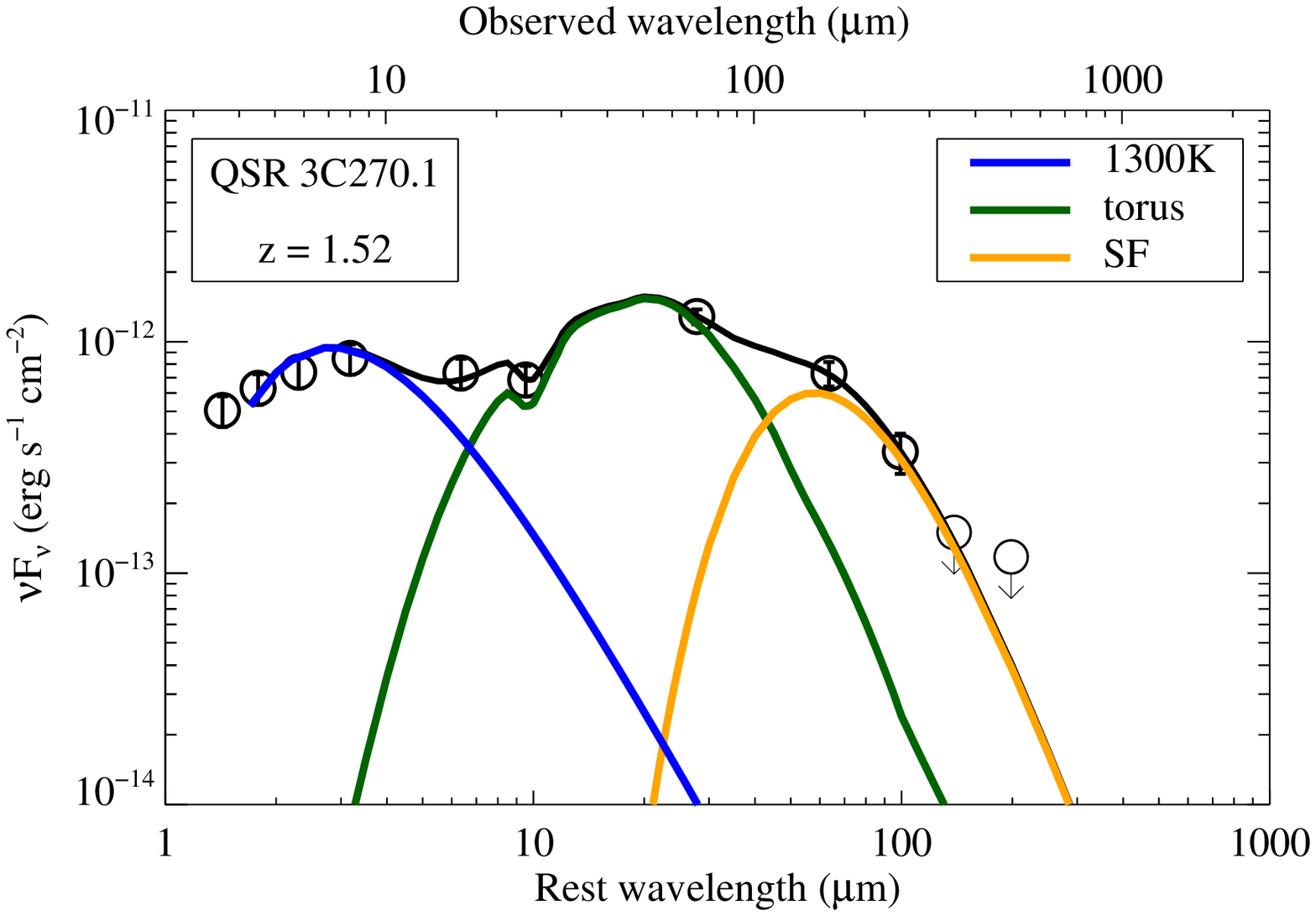}
      \includegraphics[width=4.5cm]{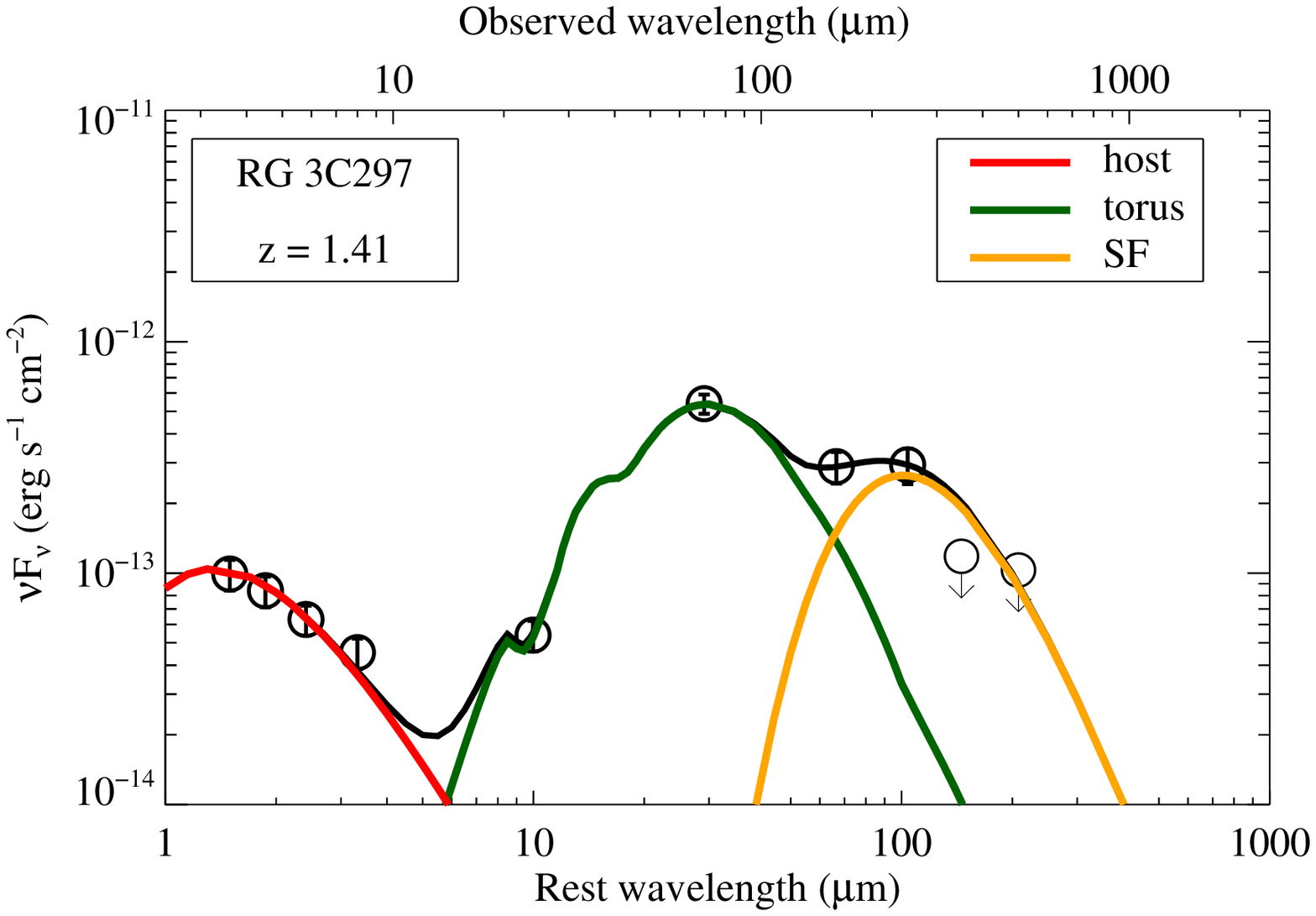}
      \includegraphics[width=4.5cm]{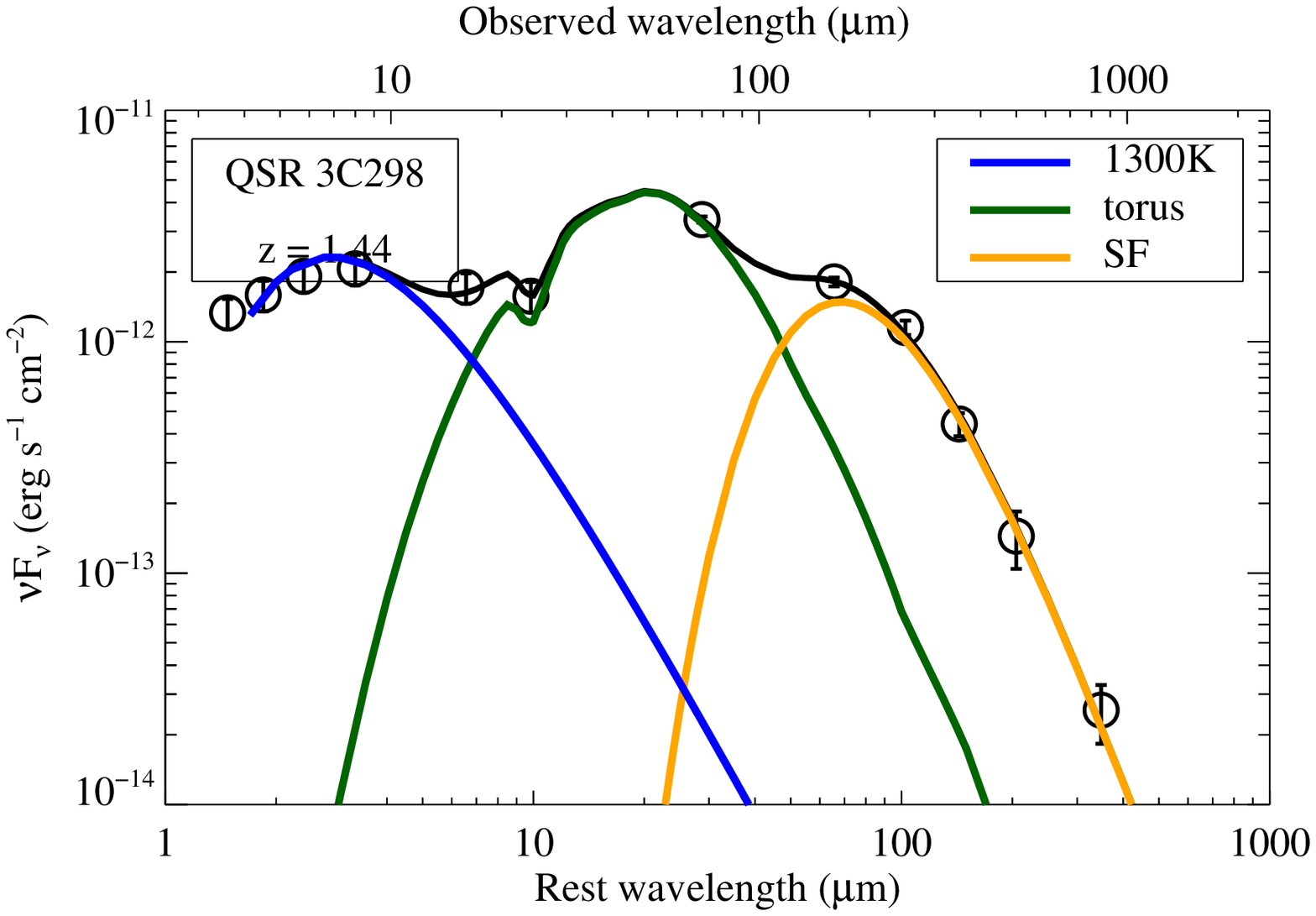}
      \includegraphics[width=4.5cm]{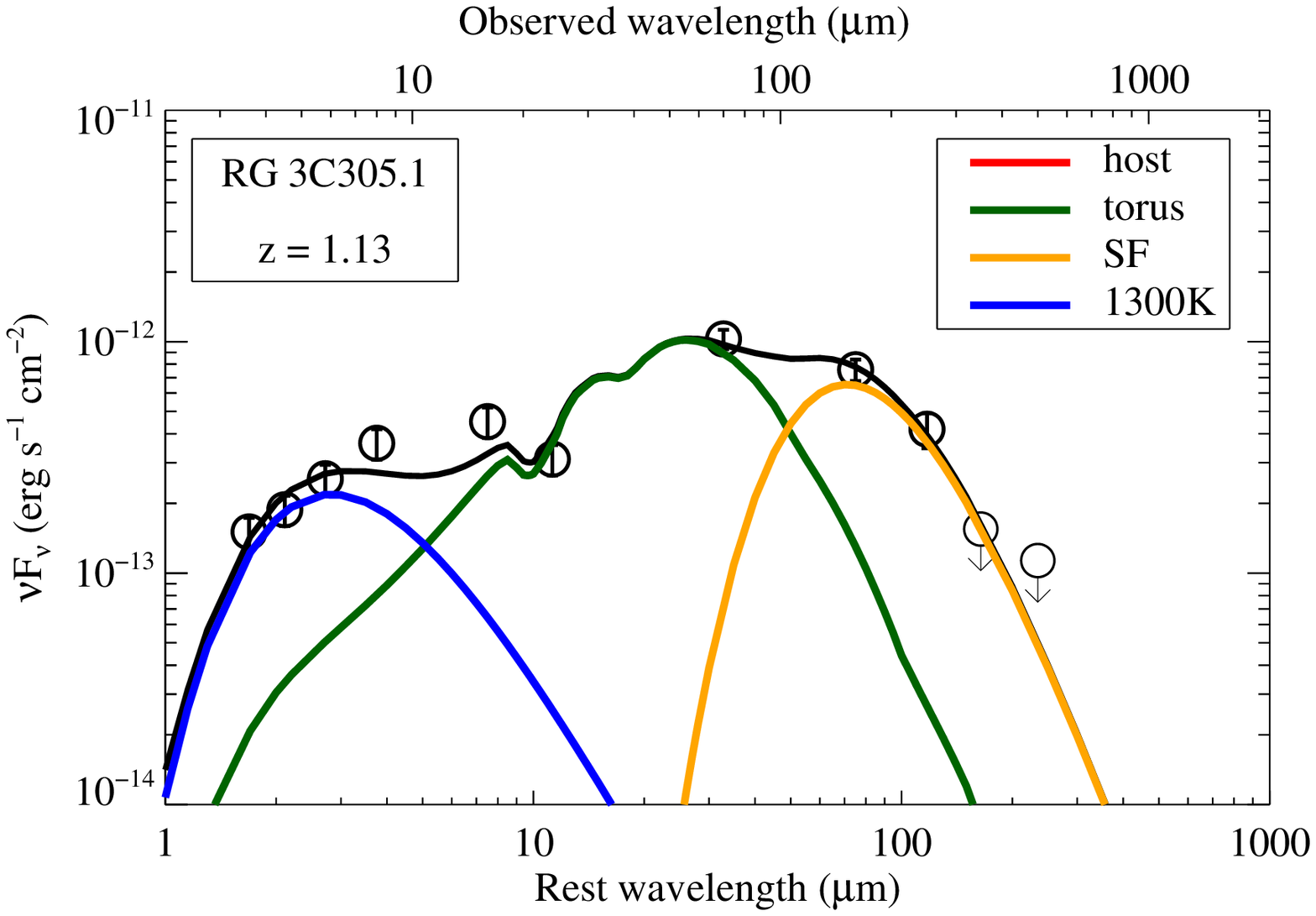}
      \includegraphics[width=4.5cm]{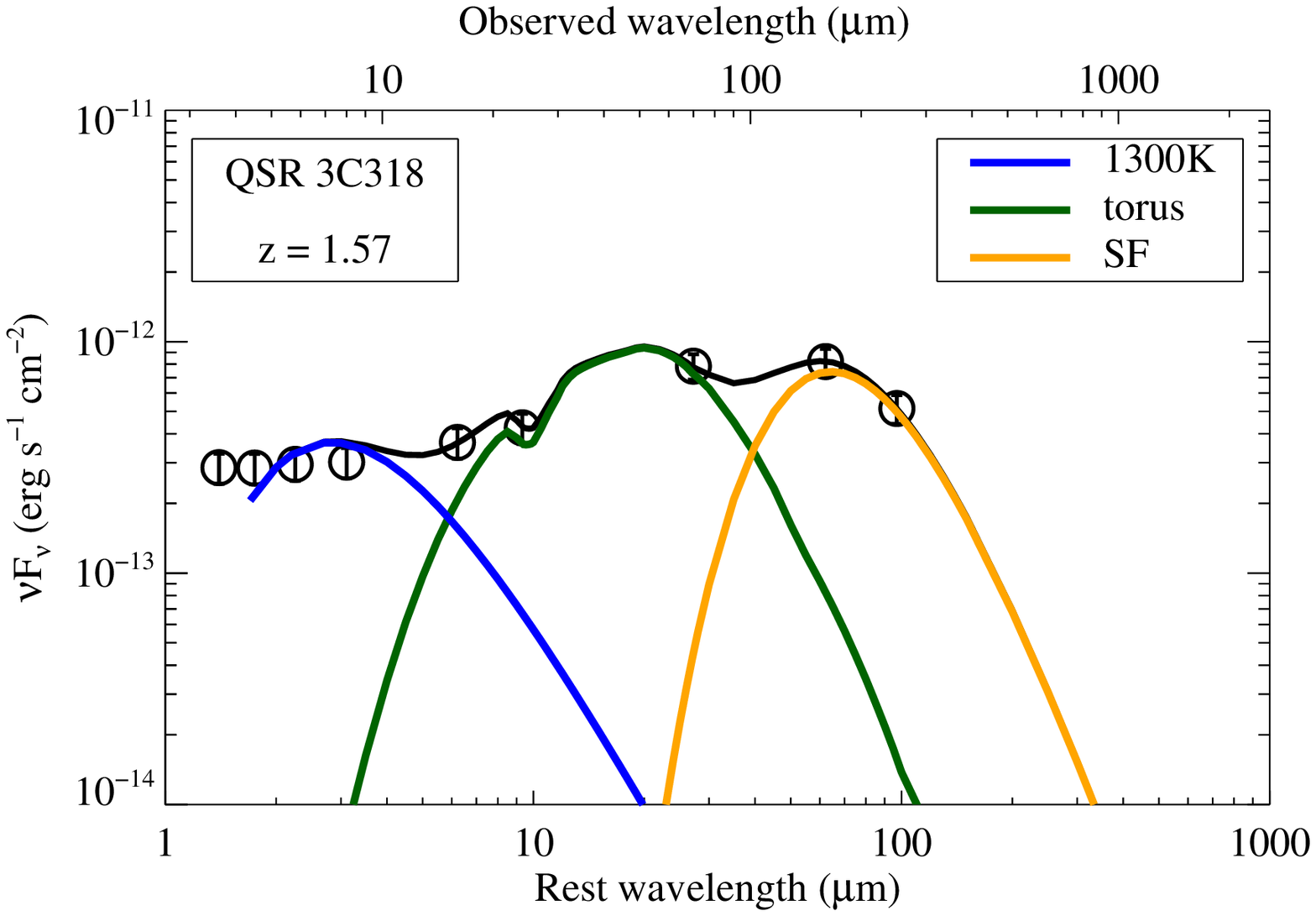}
      \includegraphics[width=4.5cm]{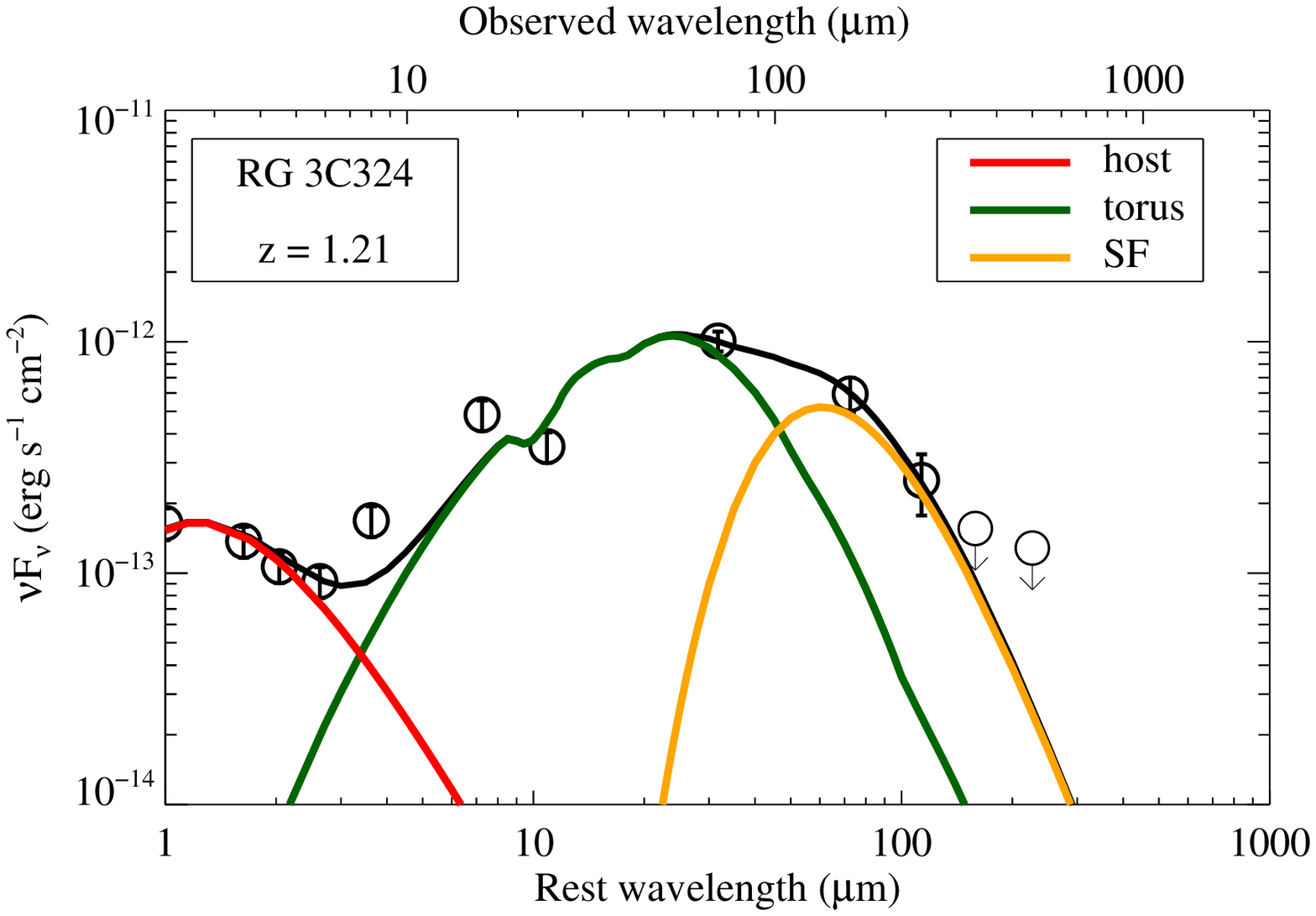}
      \includegraphics[width=4.5cm]{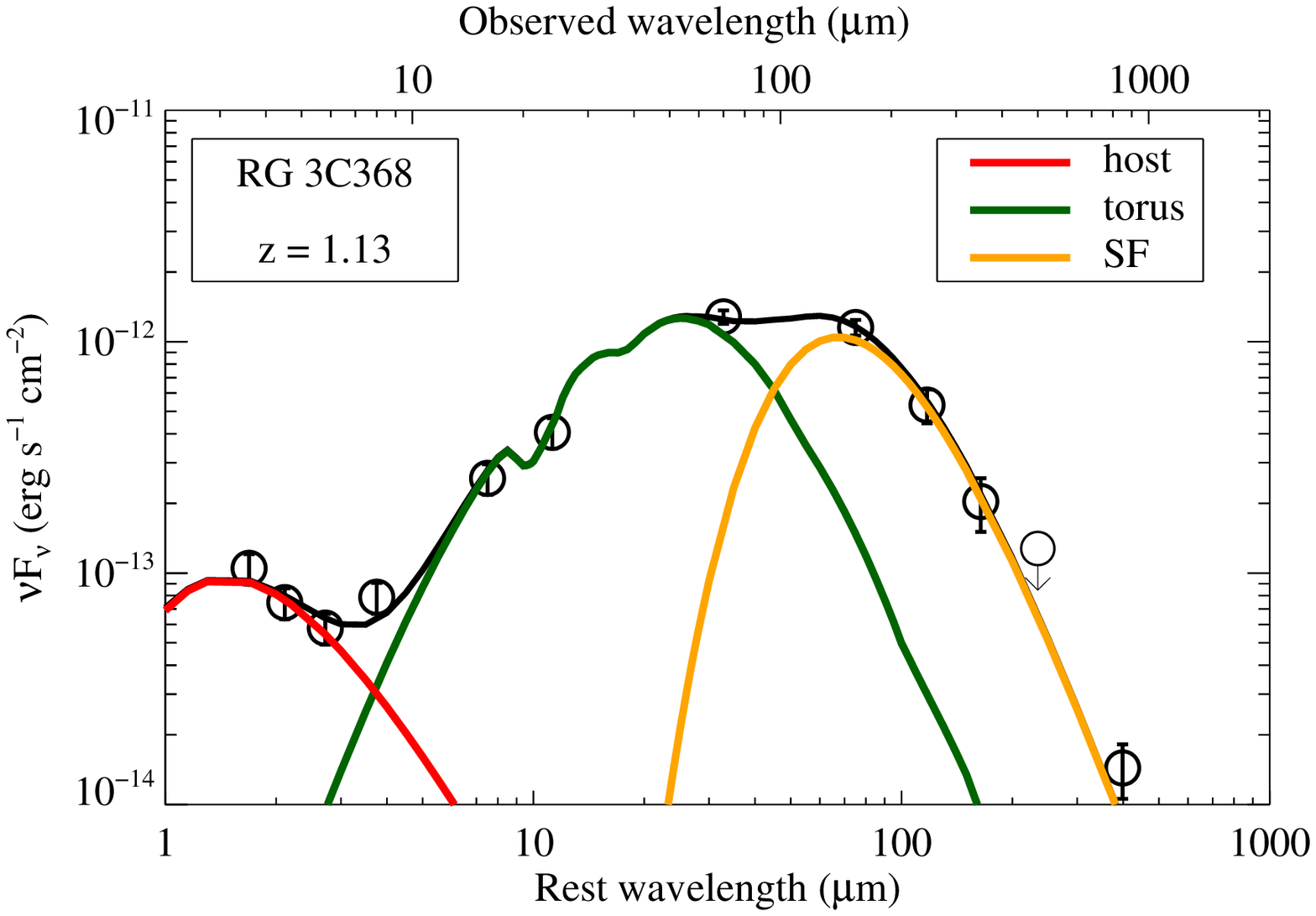}
      \includegraphics[width=4.5cm]{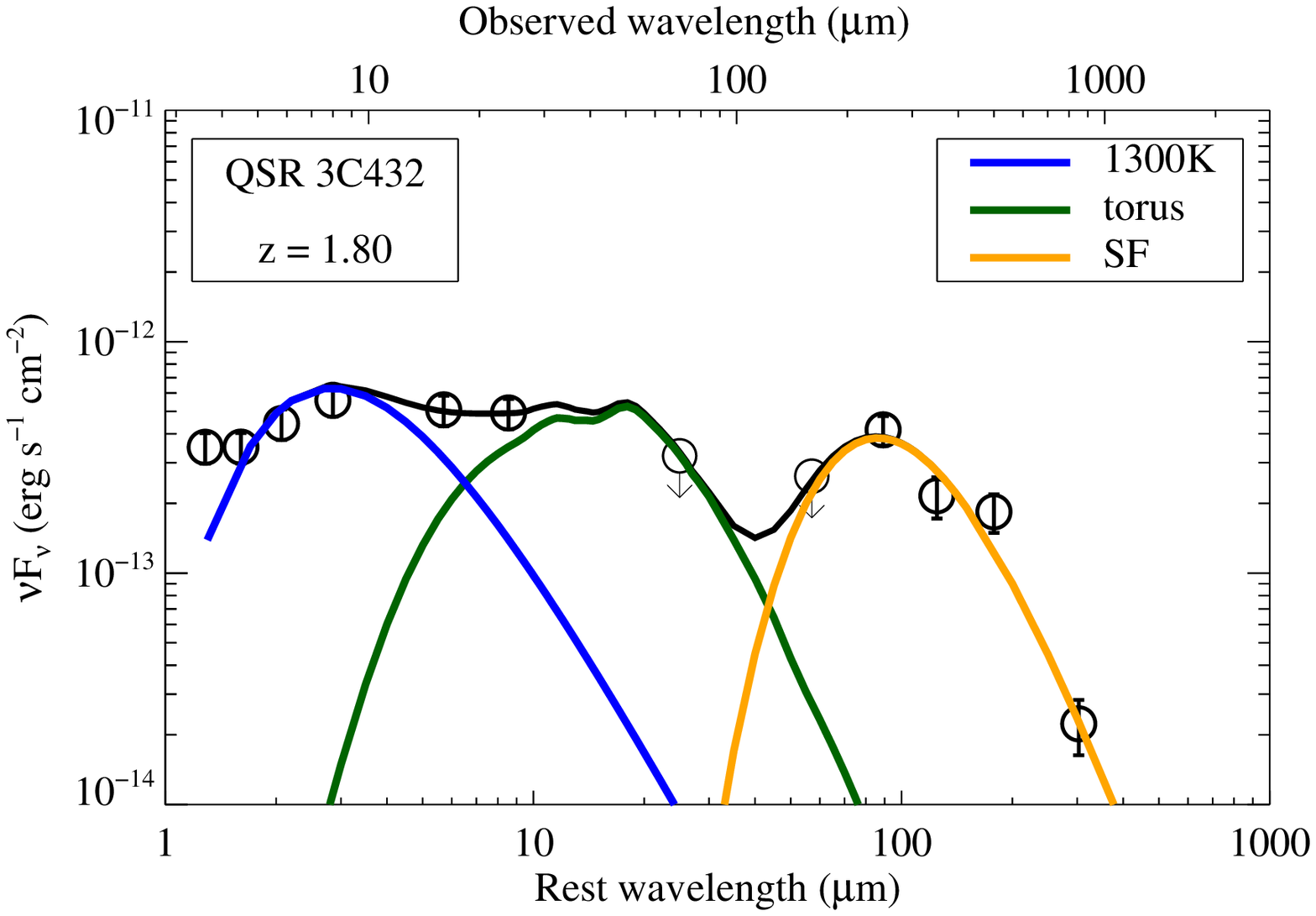}
      \includegraphics[width=4.5cm]{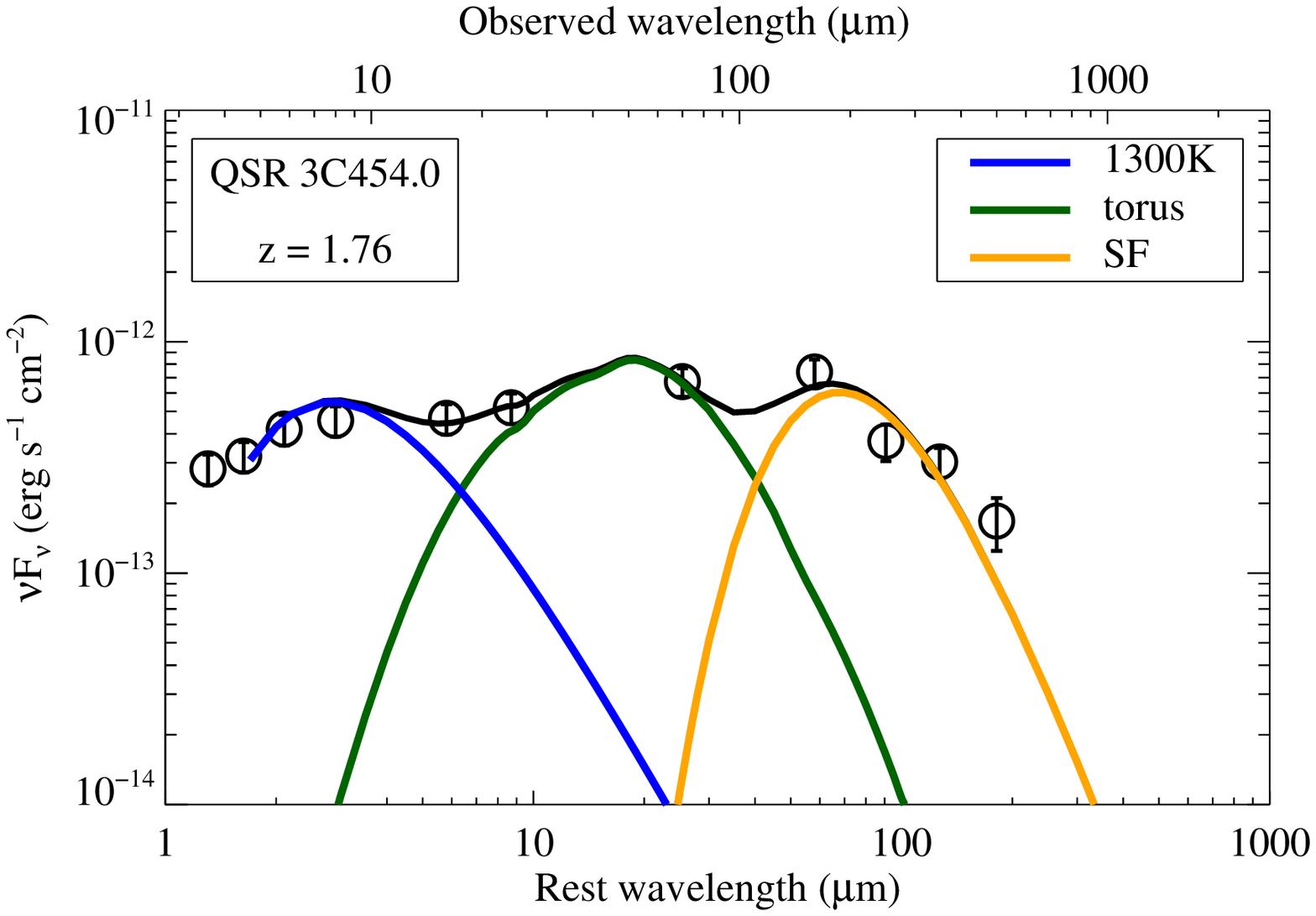}
      \includegraphics[width=4.5cm]{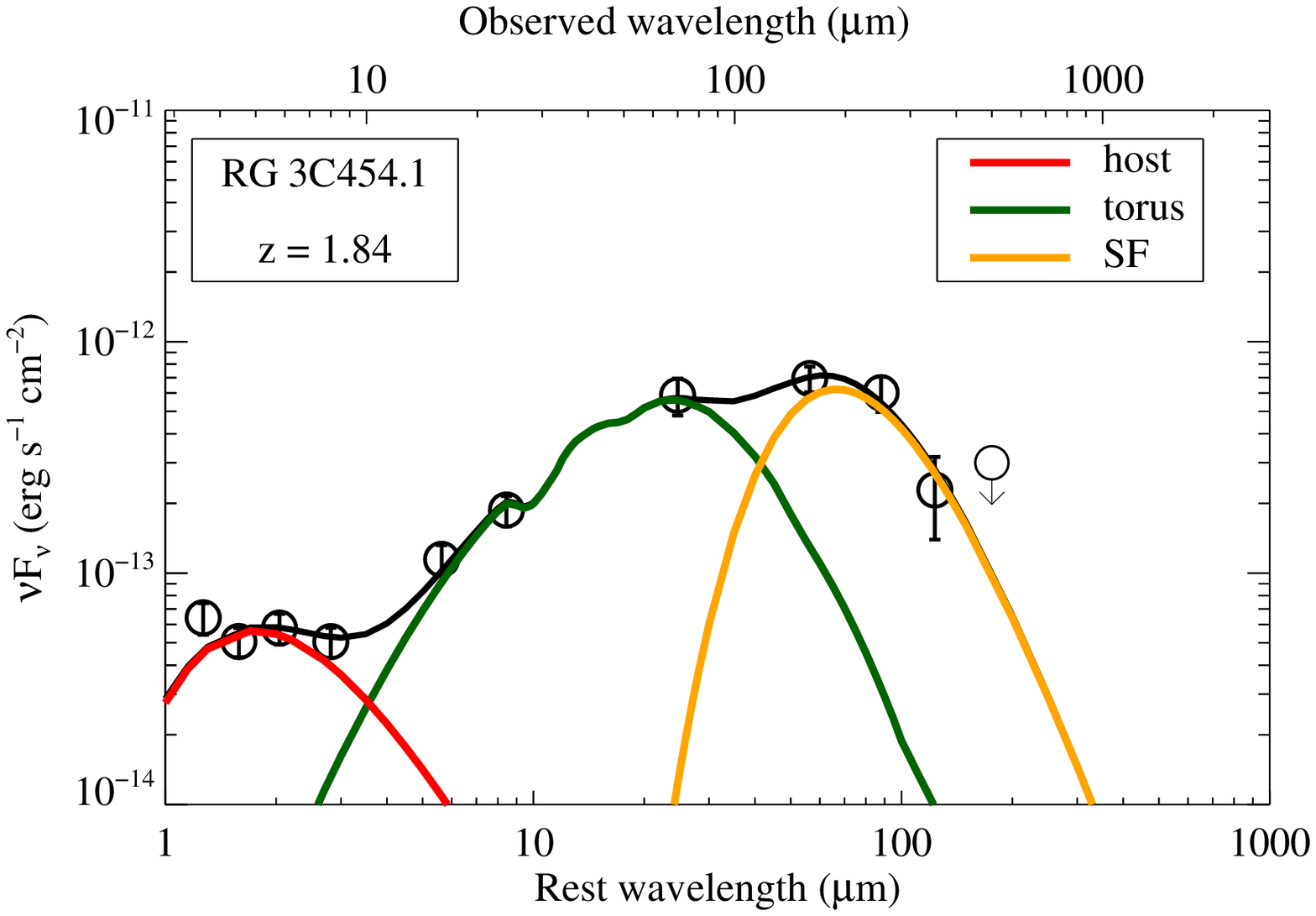}
      \includegraphics[width=4.5cm]{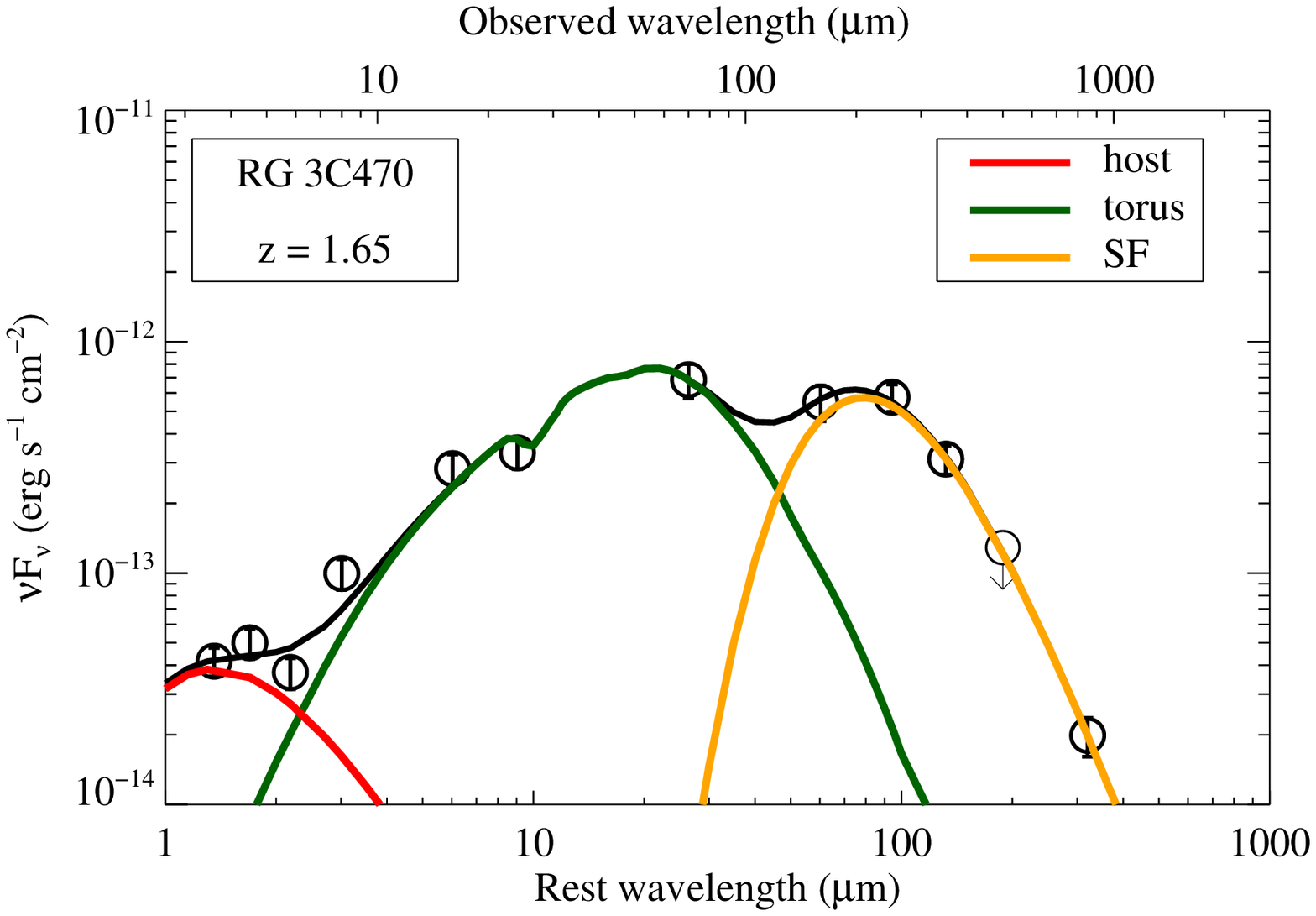}
      \includegraphics[width=4.5cm]{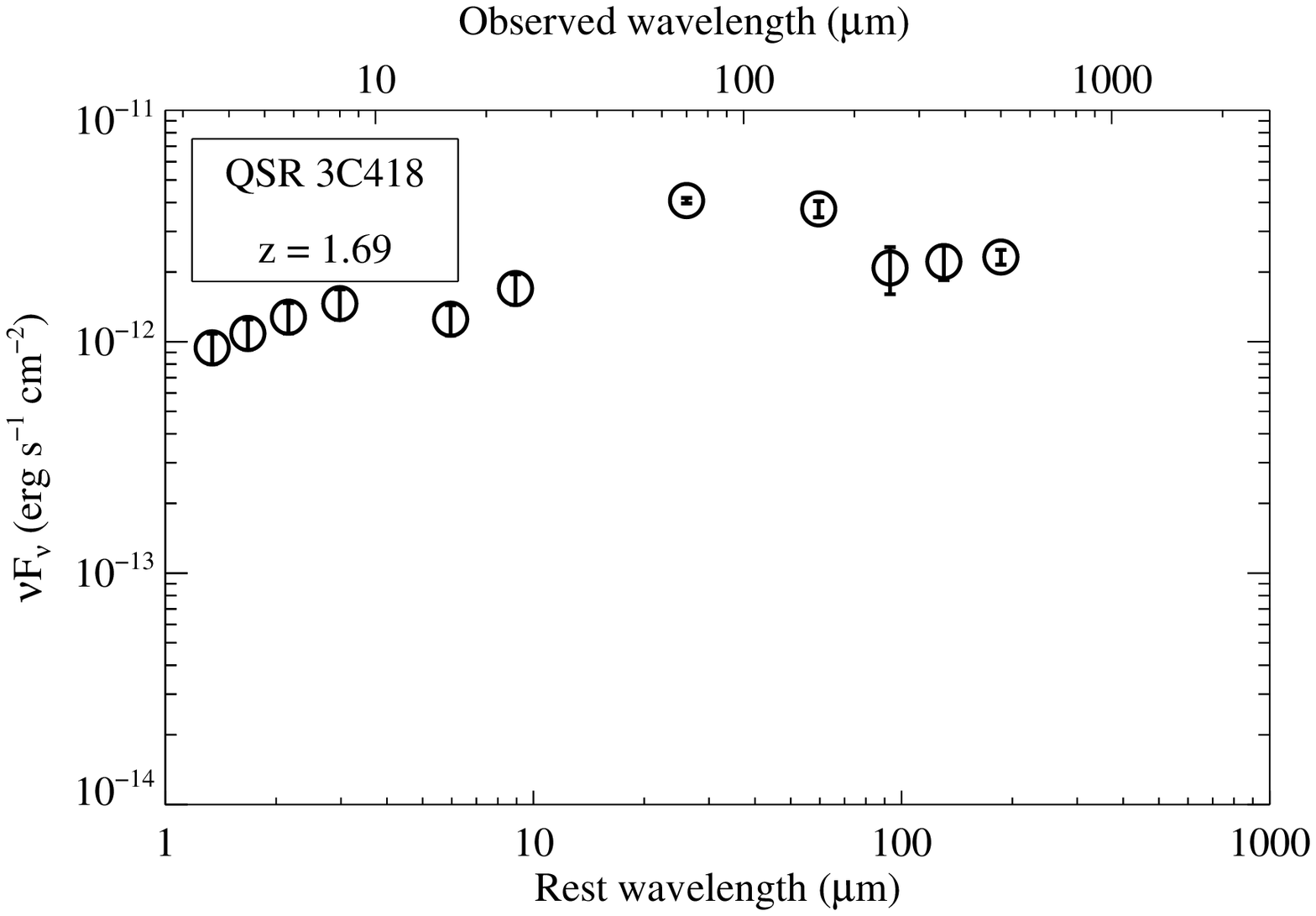}
      \caption{Spectral energy distributions of the 24 objects detected 
               in at least three \textit{Herschel} bands. Individual 
               components as described in Fig.~\ref{figure:exampleSEDs}. 
               3C~418 was not included in the analysis, as discussed 
               in Sect.~\ref{subsection:synchrotron}
               }
      \label{figure:FIRDetectedBestFitSEDs}
   \end{figure*}
\section{Best-fit SEDs of objects detected in fewer than three \textit{Herschel} bands}
\label{appendix:nondetectedSEDs}
   \begin{figure*}
      \includegraphics[width=4.5cm]{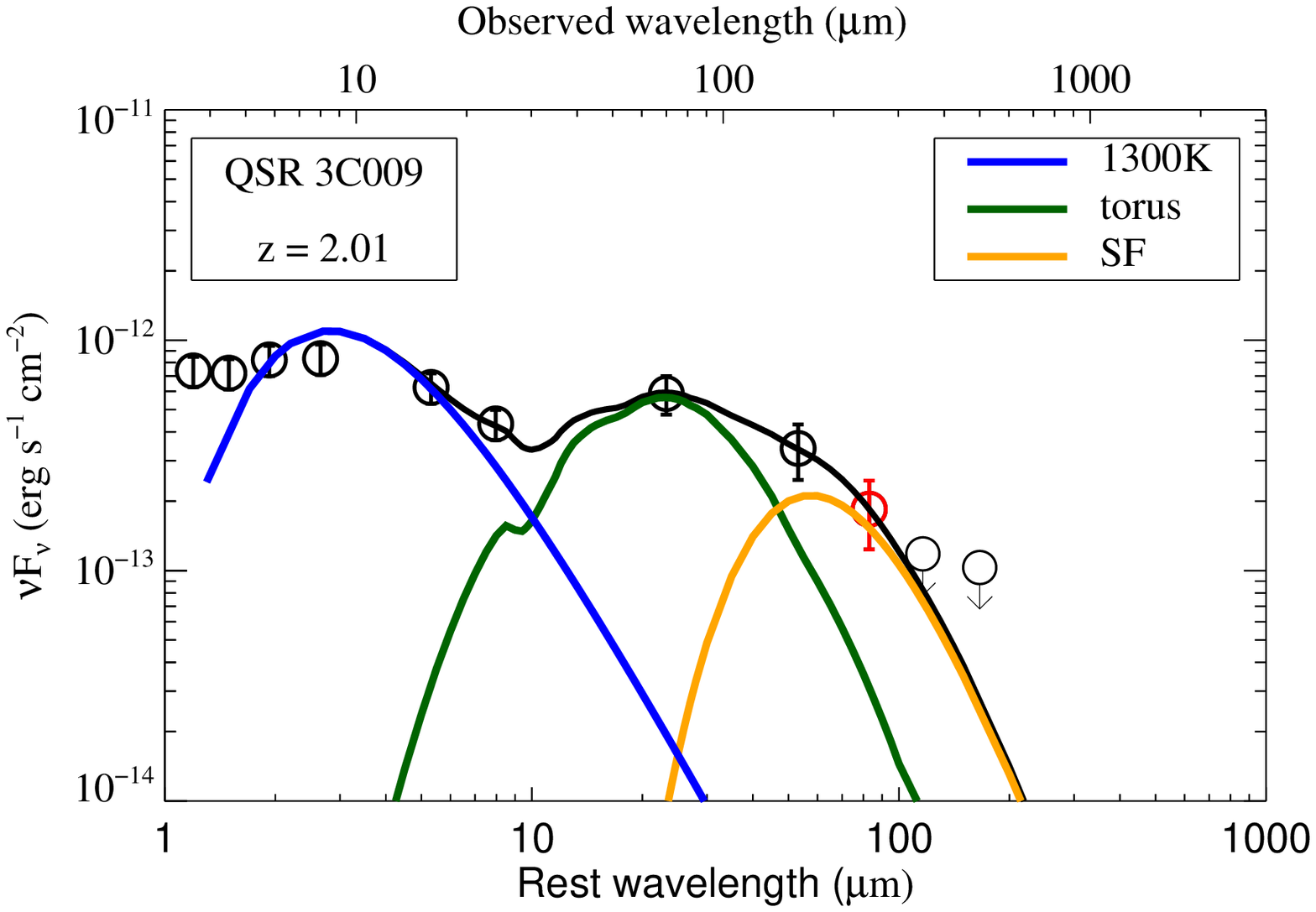}
      \includegraphics[width=4.5cm]{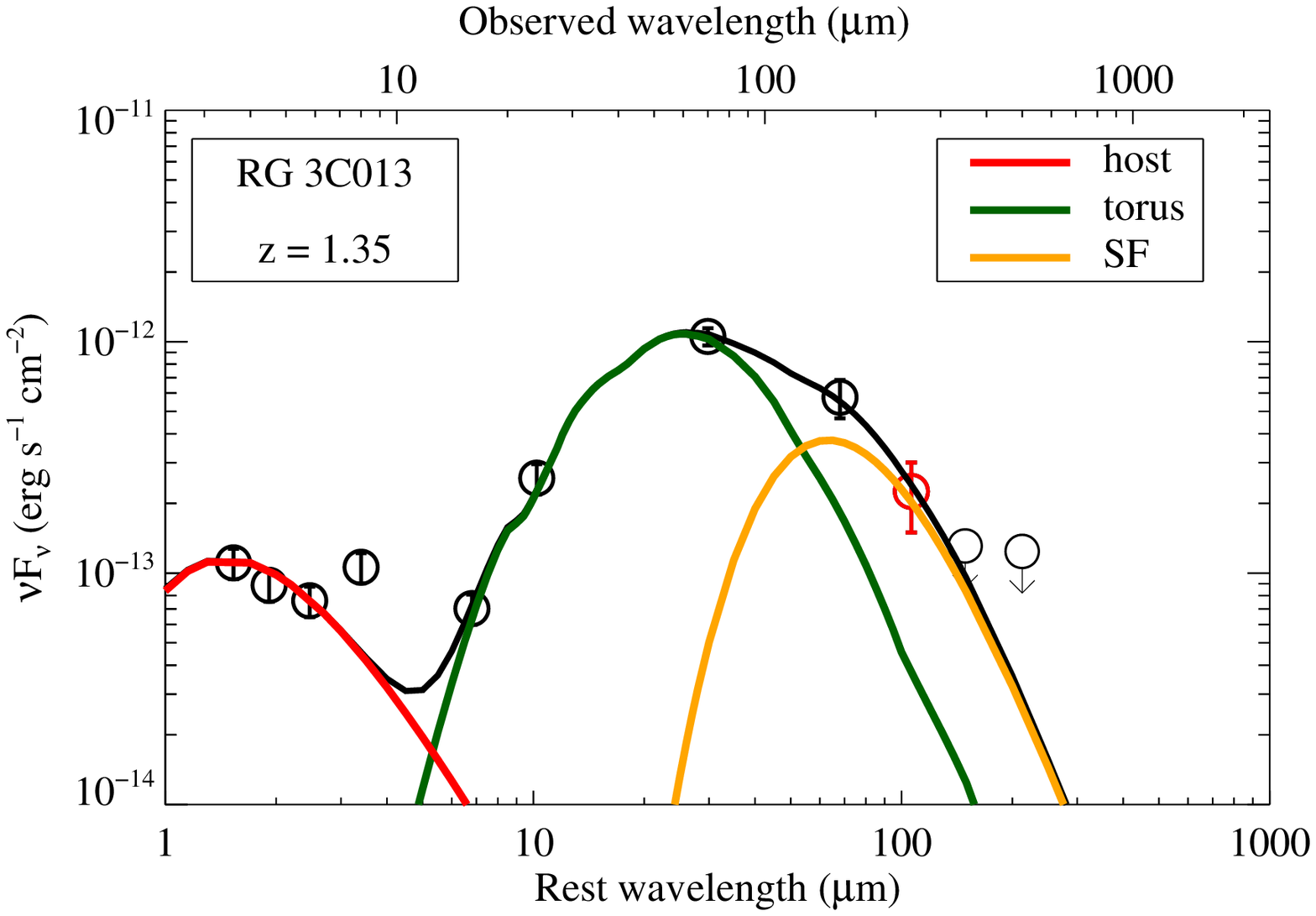}
      \includegraphics[width=4.5cm]{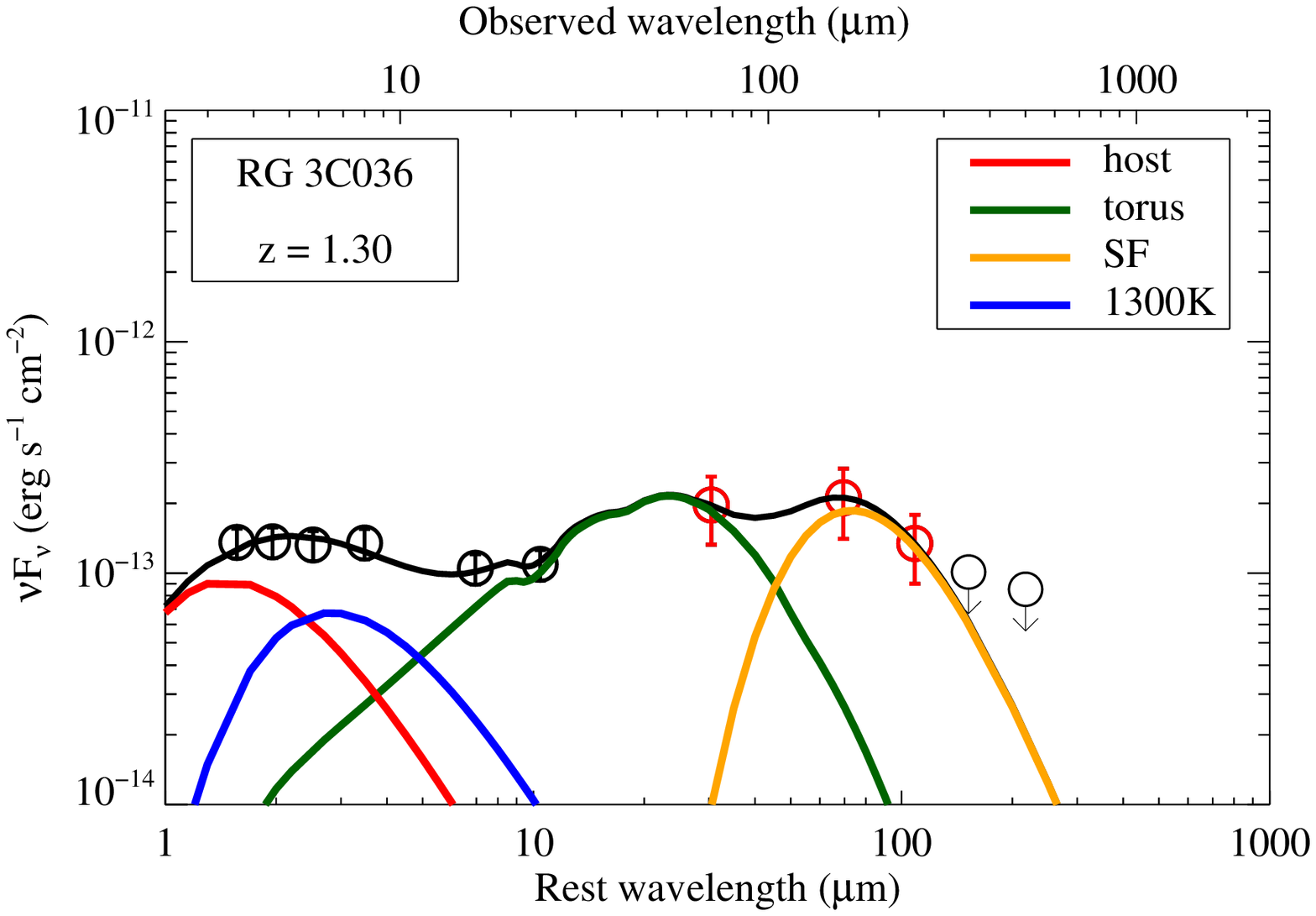}
      \includegraphics[width=4.5cm]{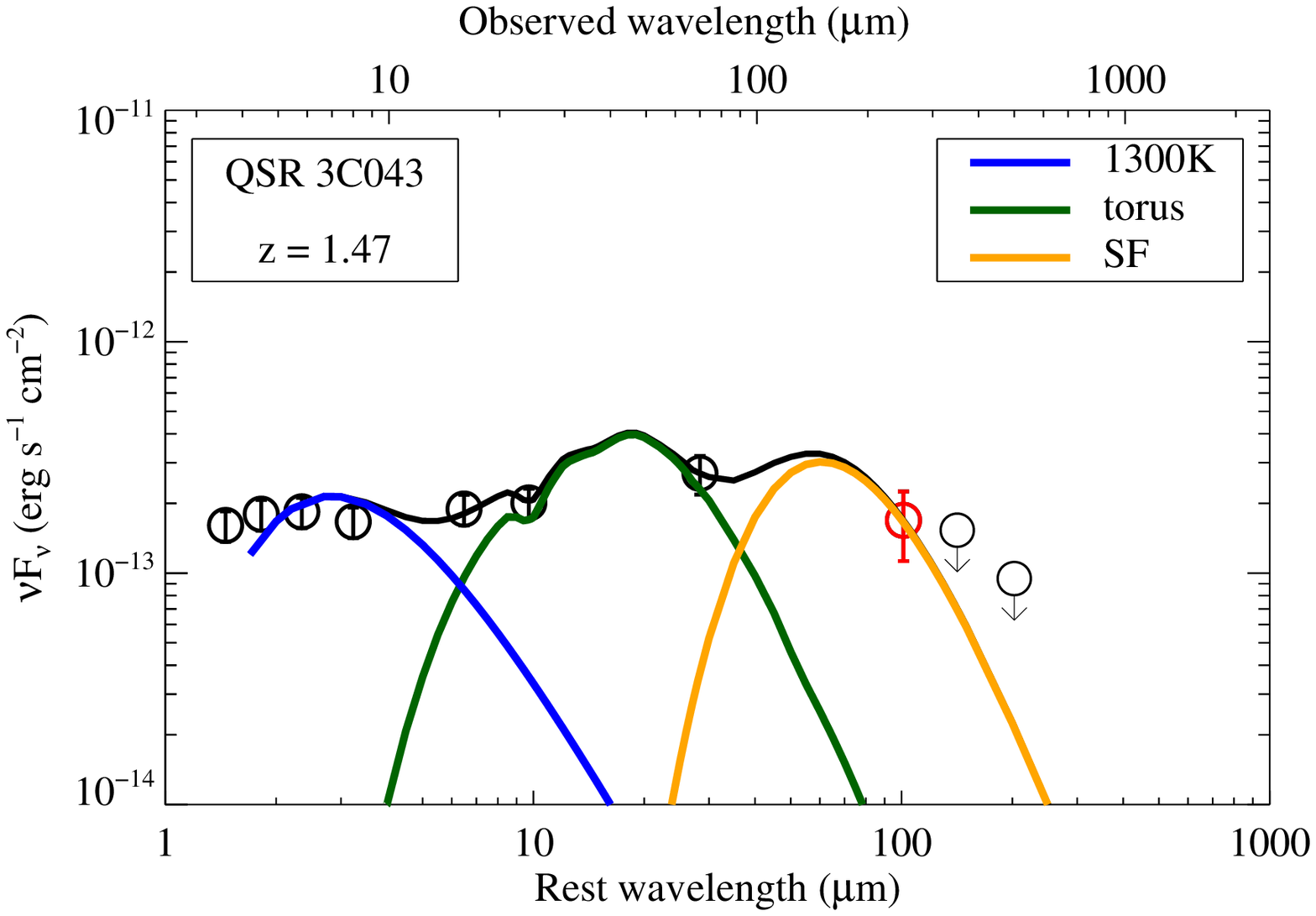}
      \includegraphics[width=4.5cm]{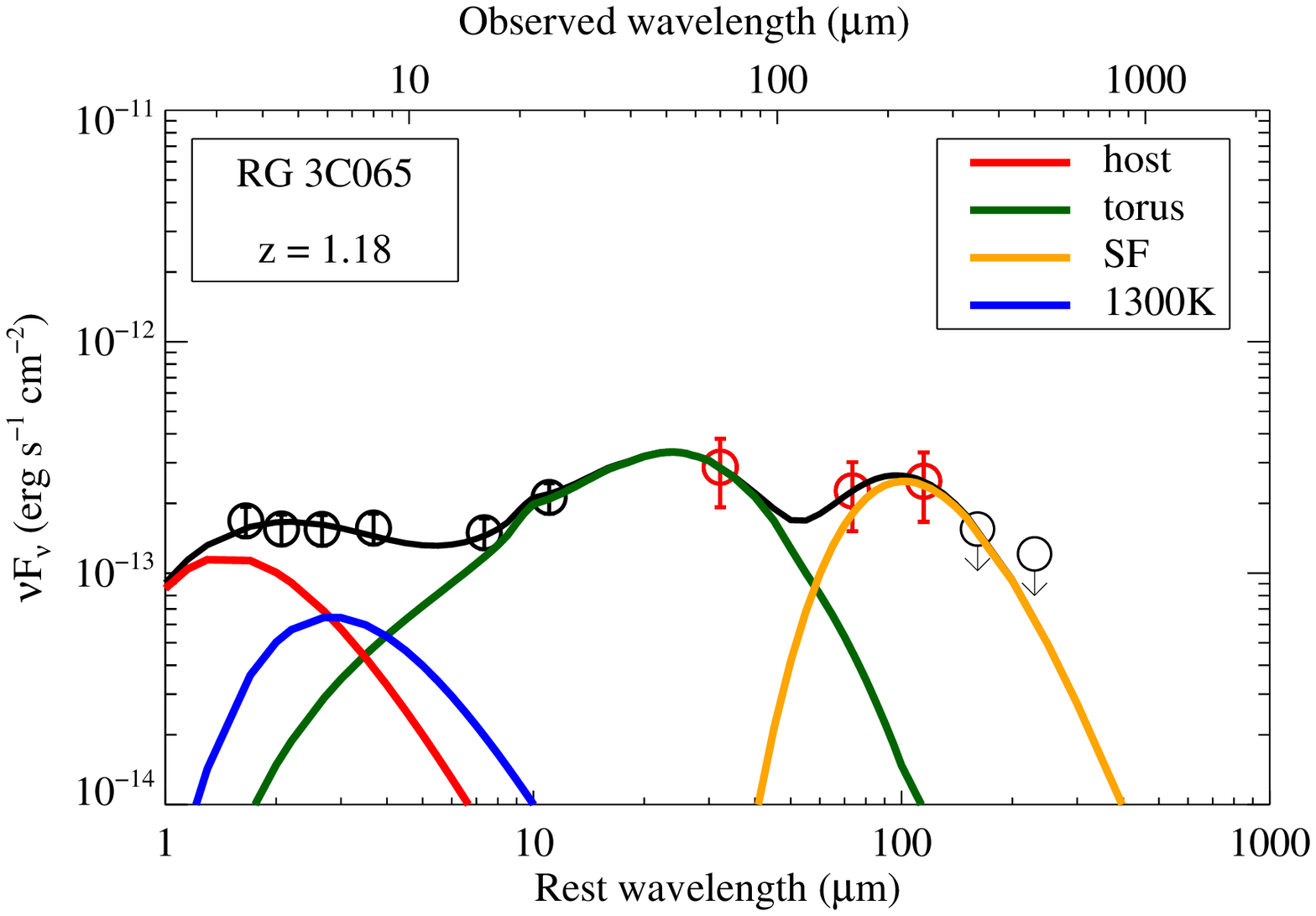}
      \includegraphics[width=4.5cm]{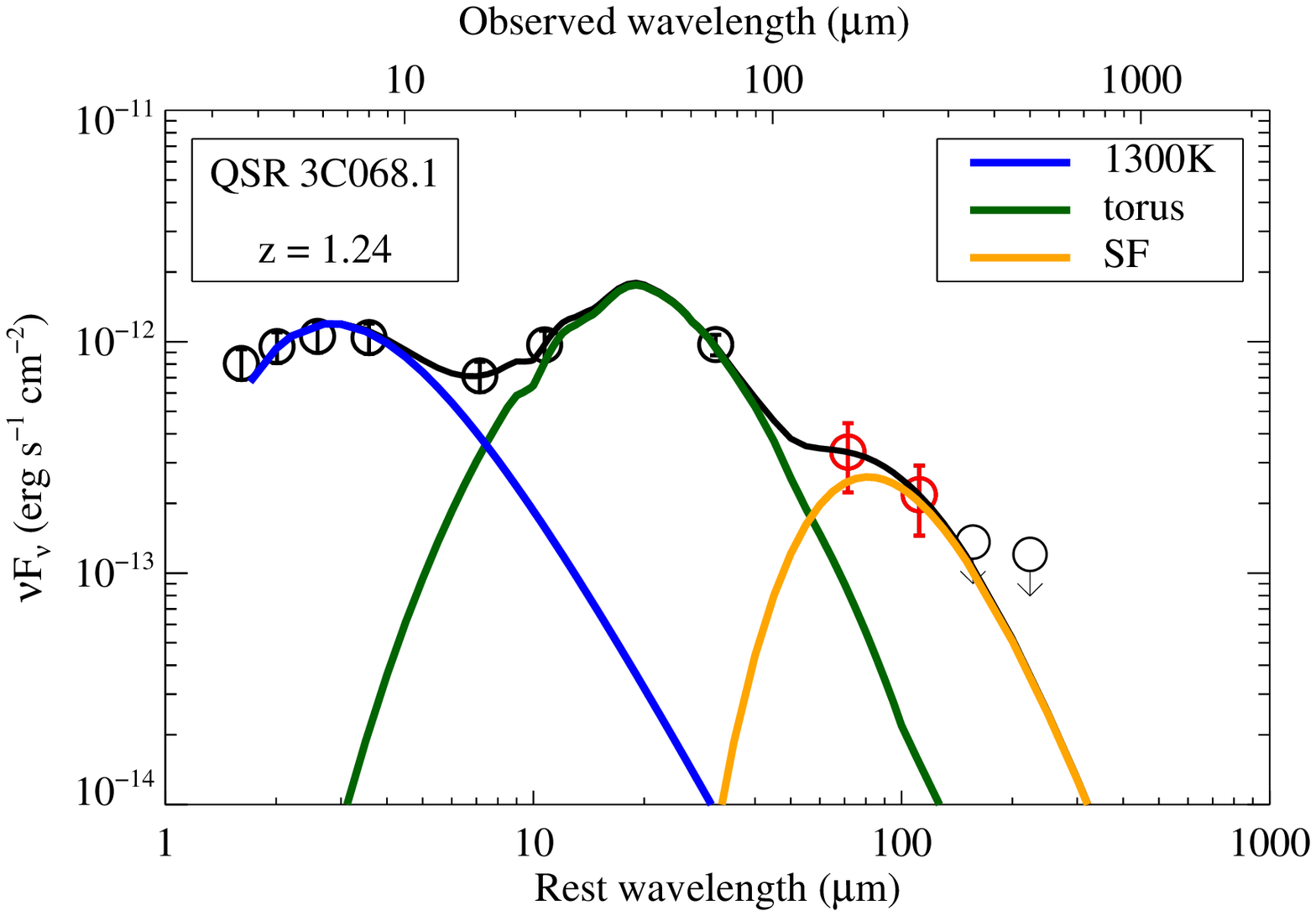}
      \includegraphics[width=4.5cm]{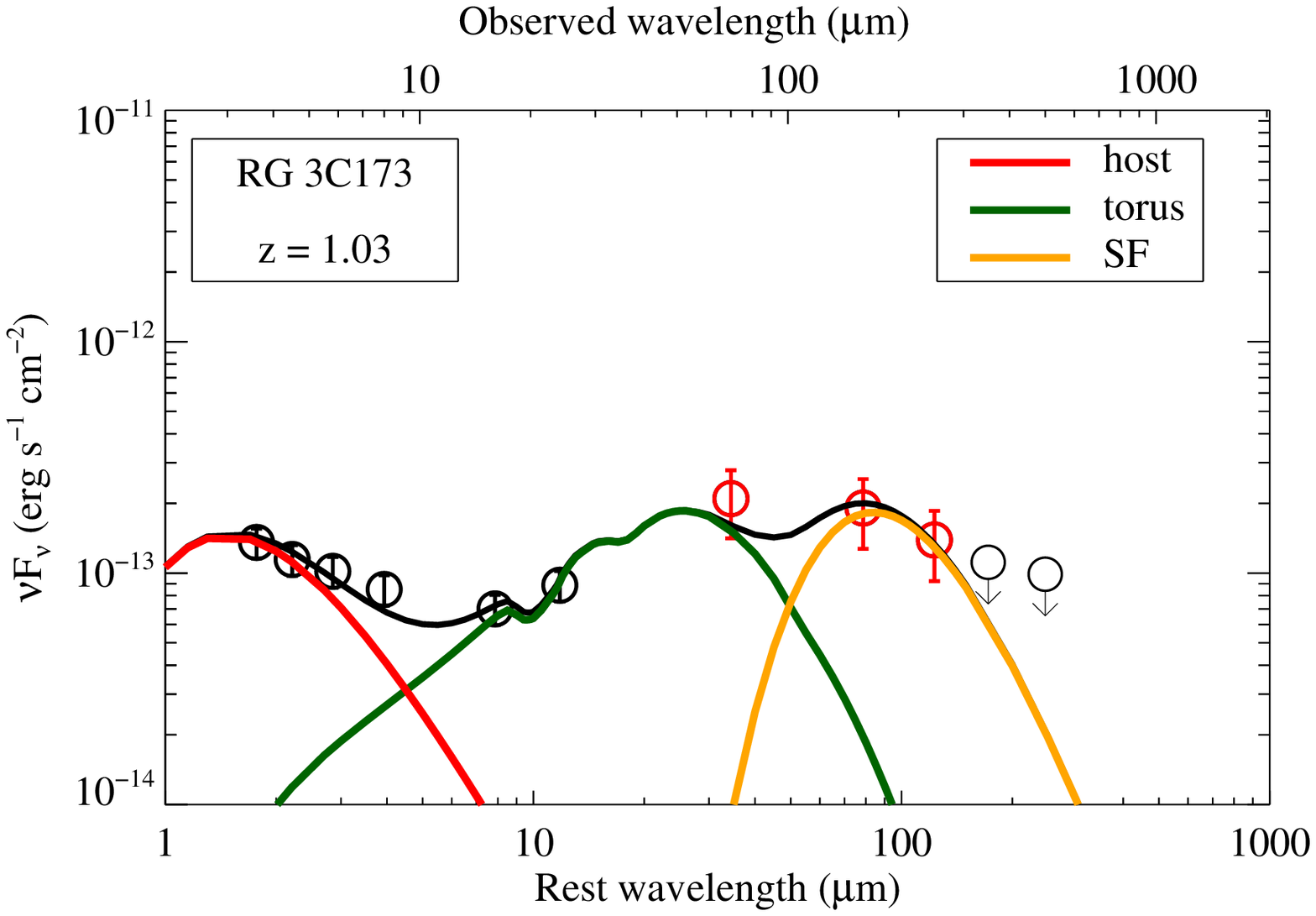}
      \includegraphics[width=4.5cm]{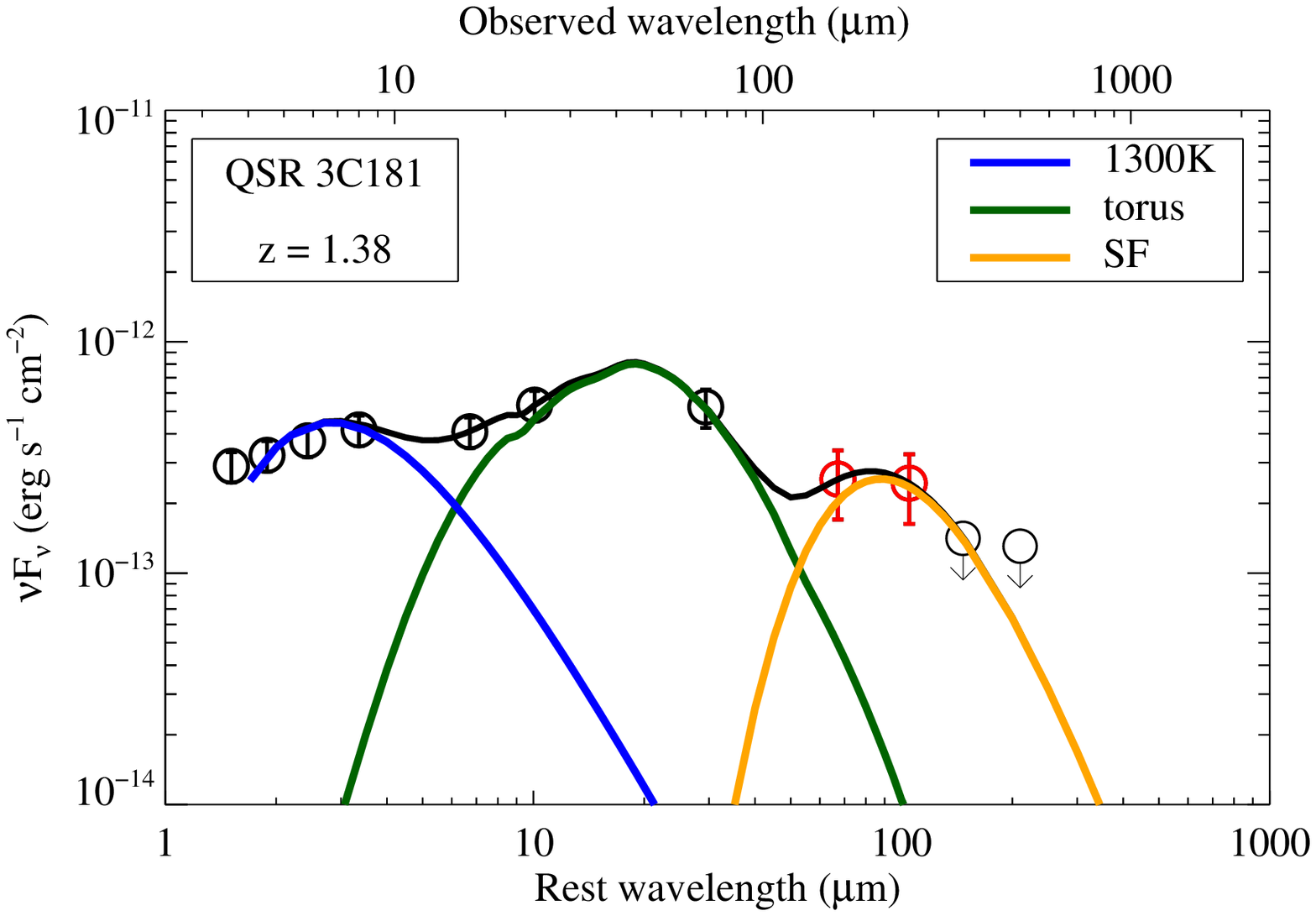}
      \includegraphics[width=4.5cm]{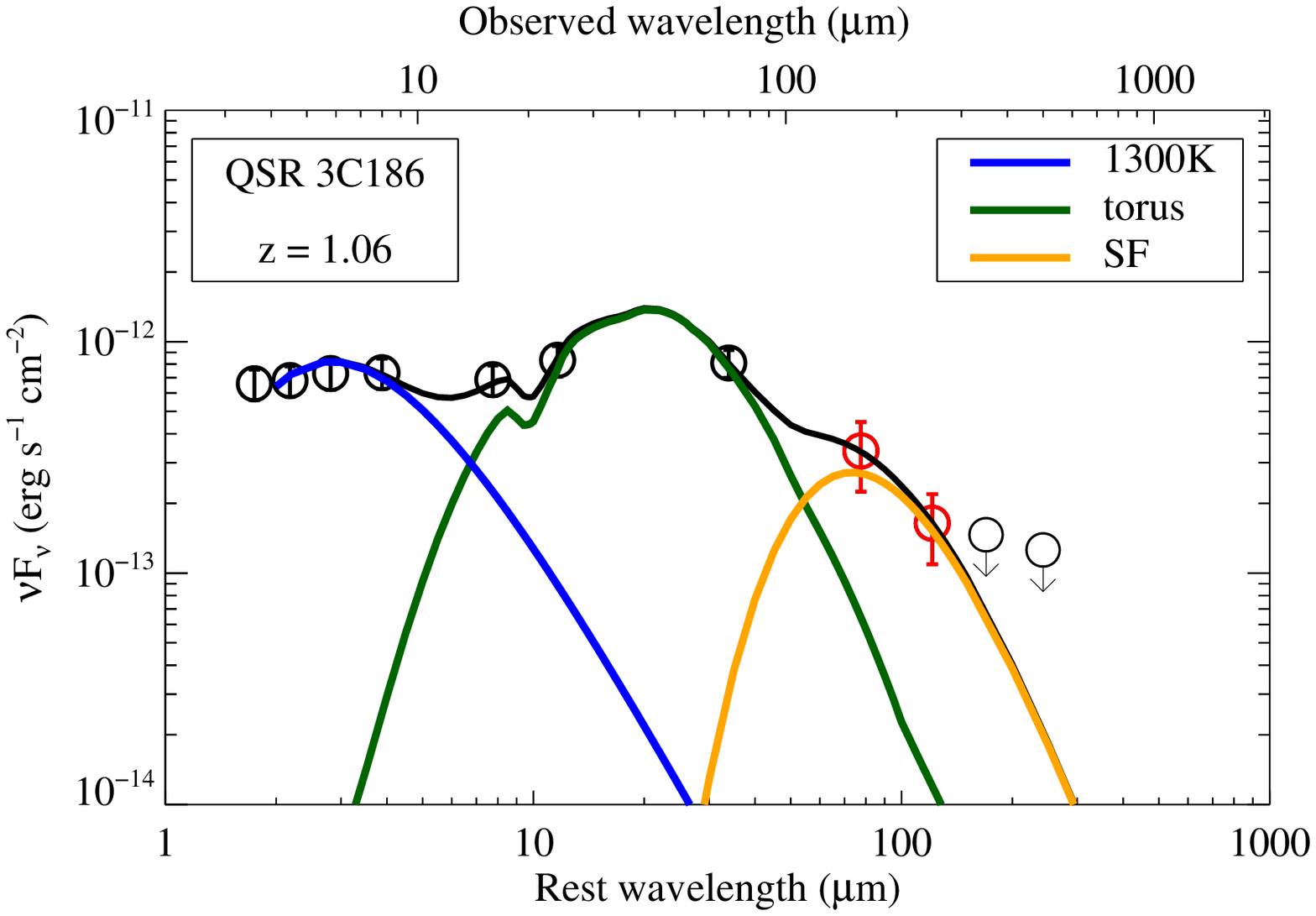}
      \includegraphics[width=4.5cm]{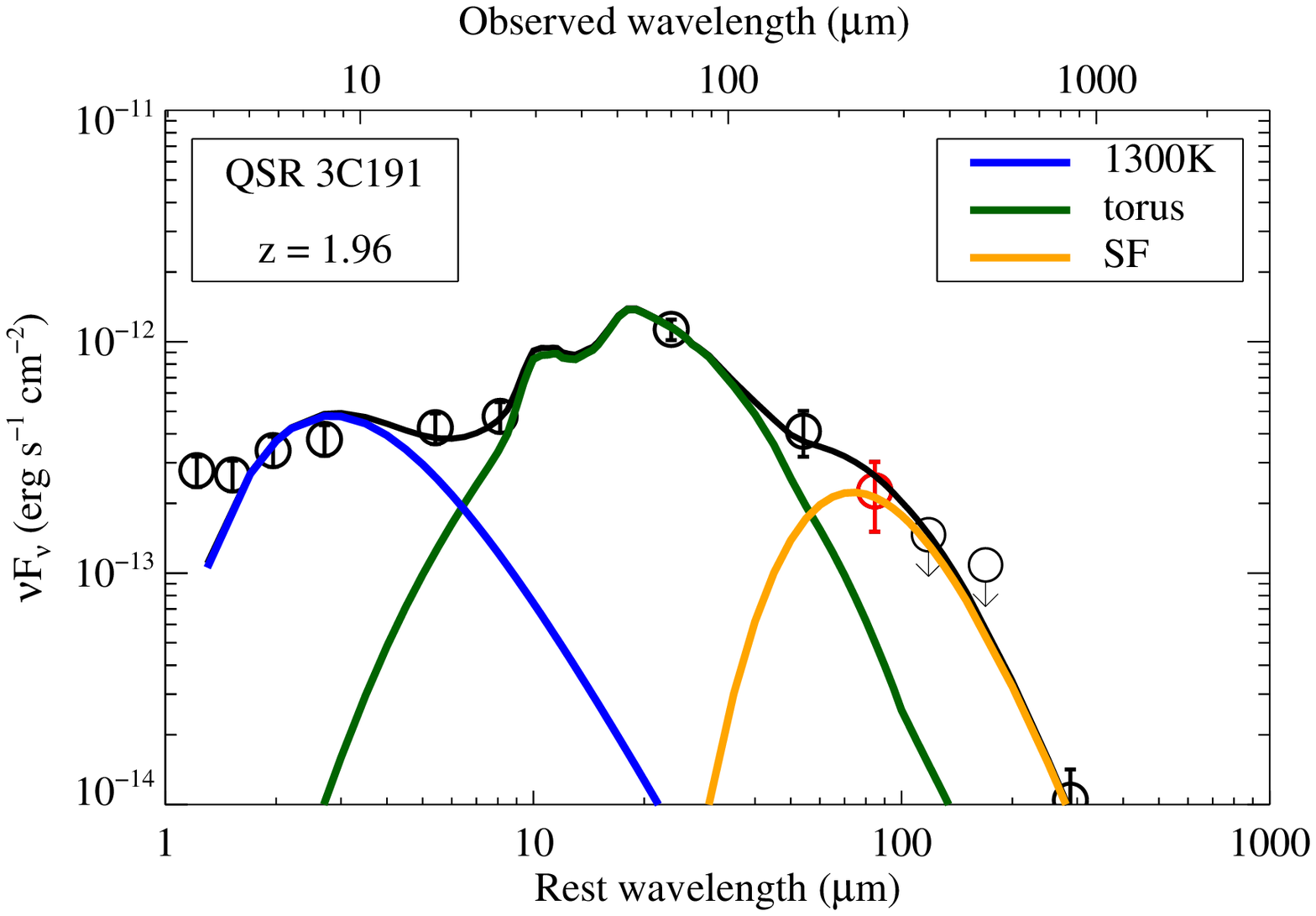}
      \includegraphics[width=4.5cm]{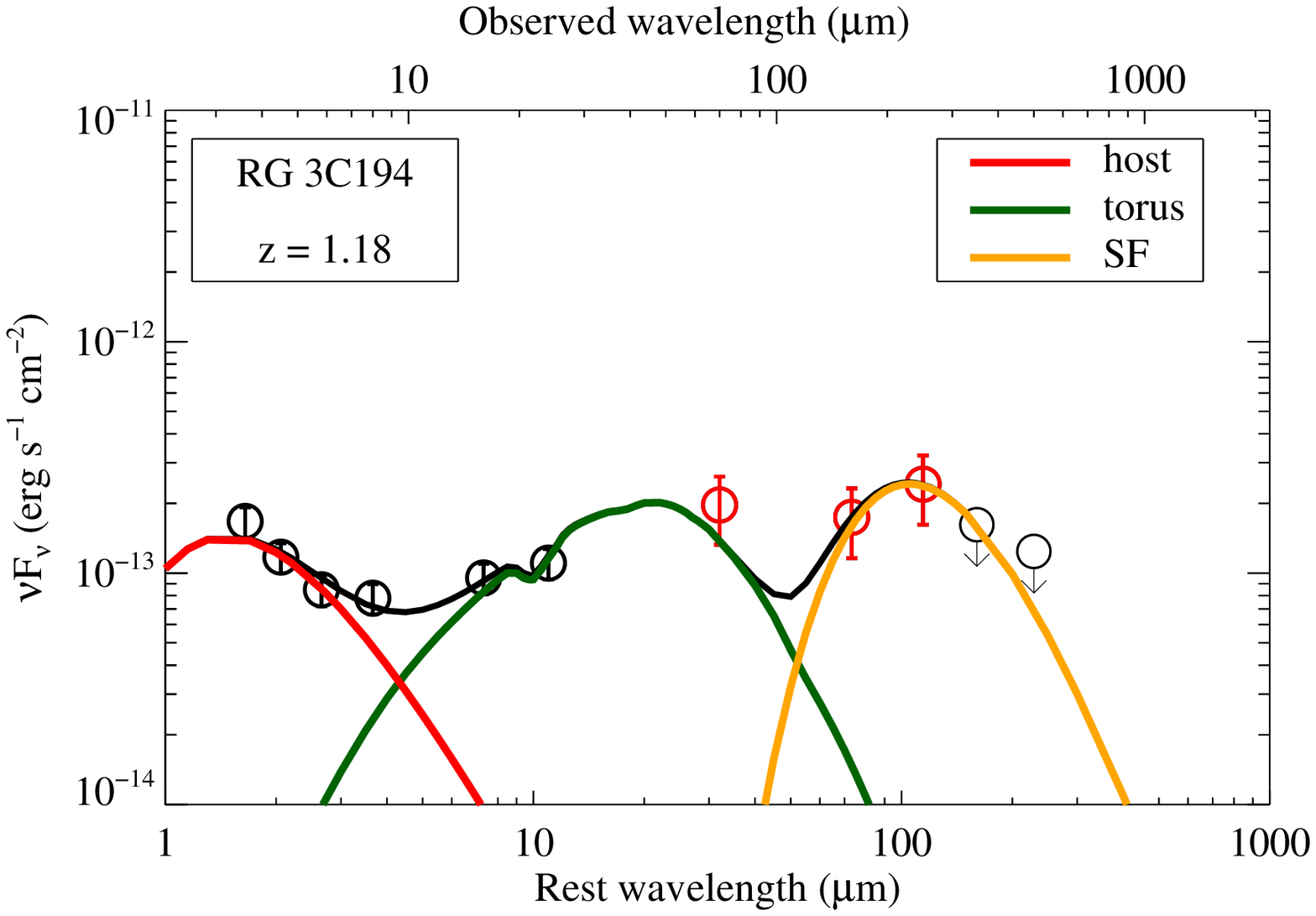}
      \includegraphics[width=4.5cm]{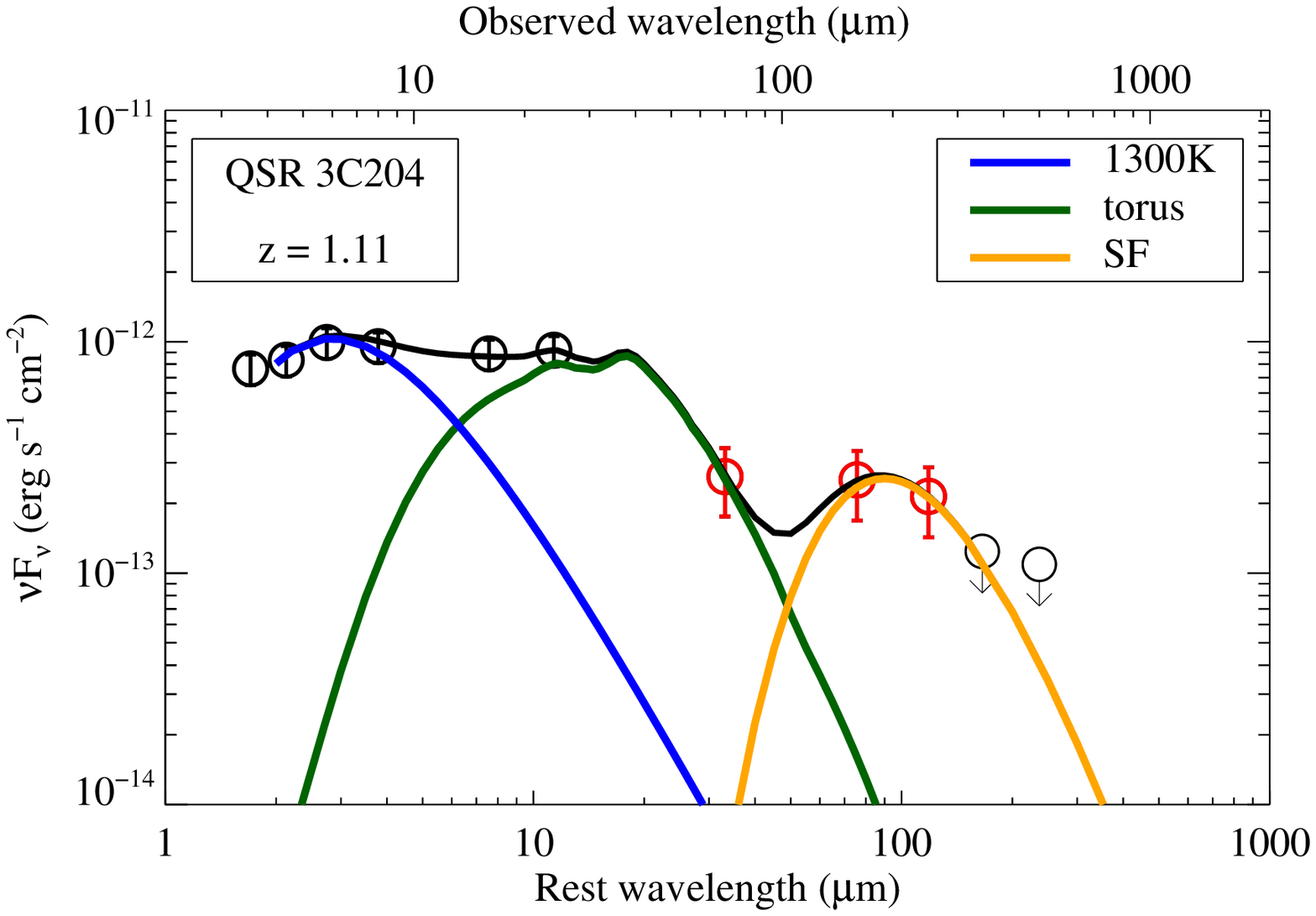}
      \includegraphics[width=4.5cm]{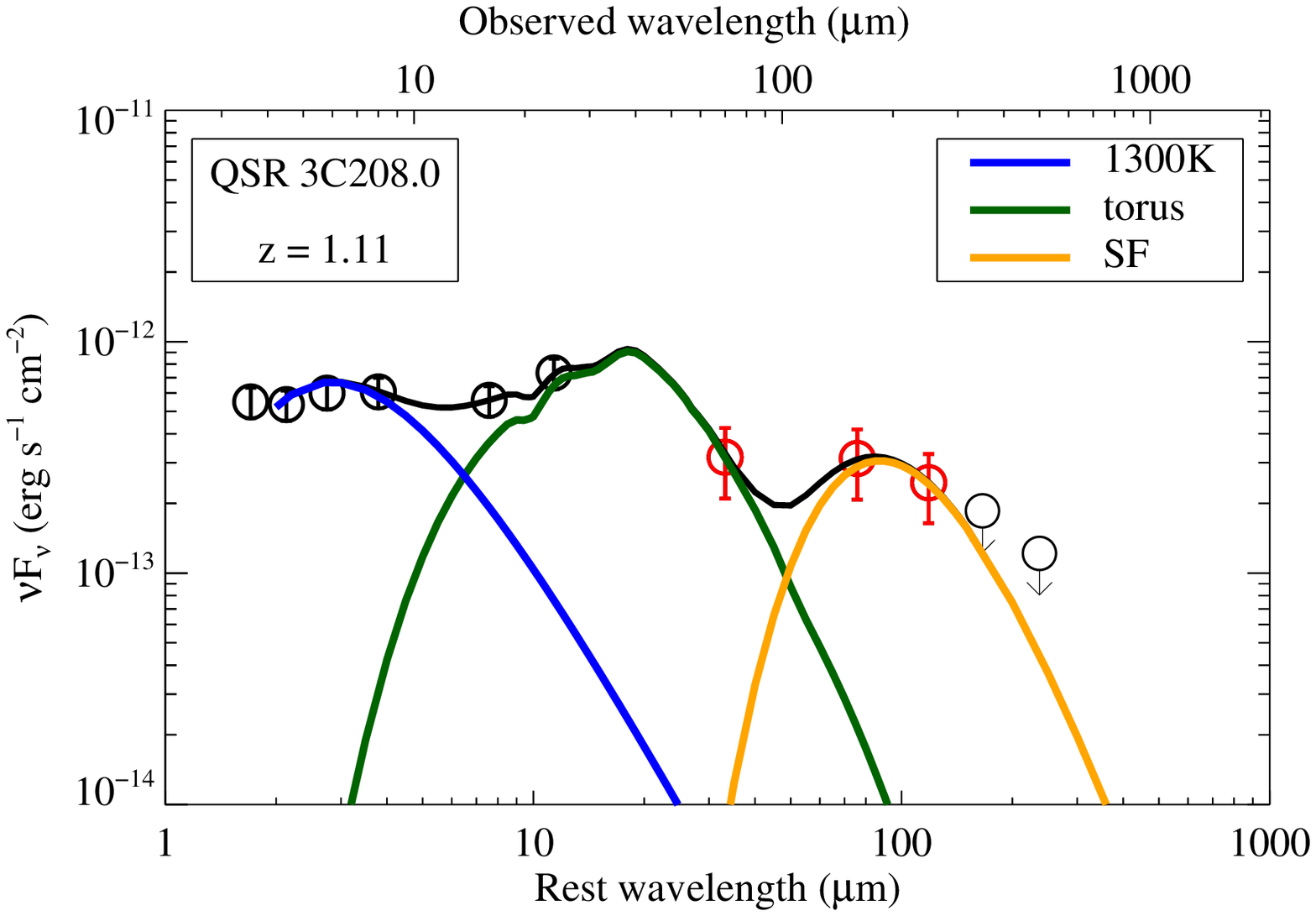}
      \includegraphics[width=4.5cm]{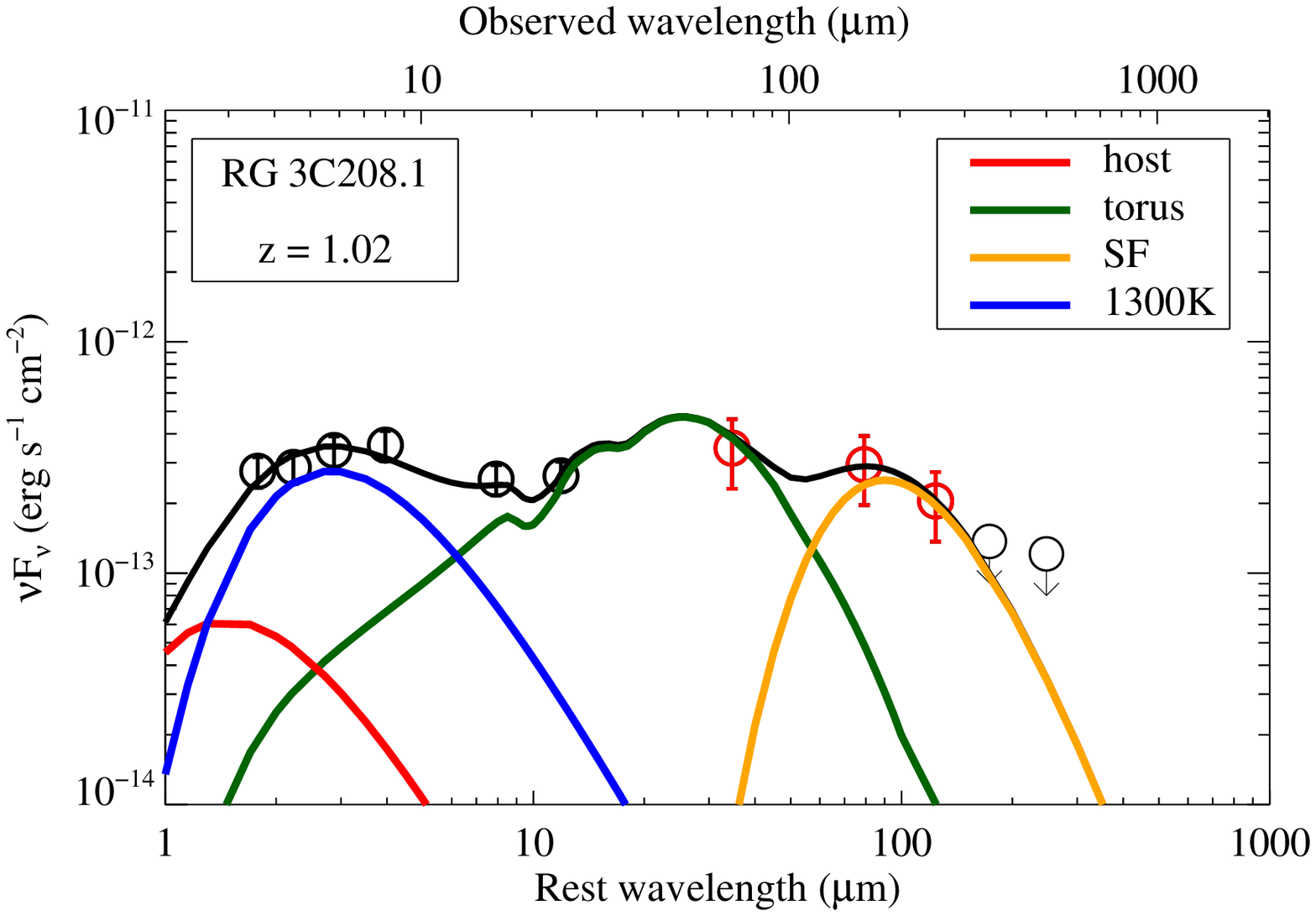}
      \includegraphics[width=4.5cm]{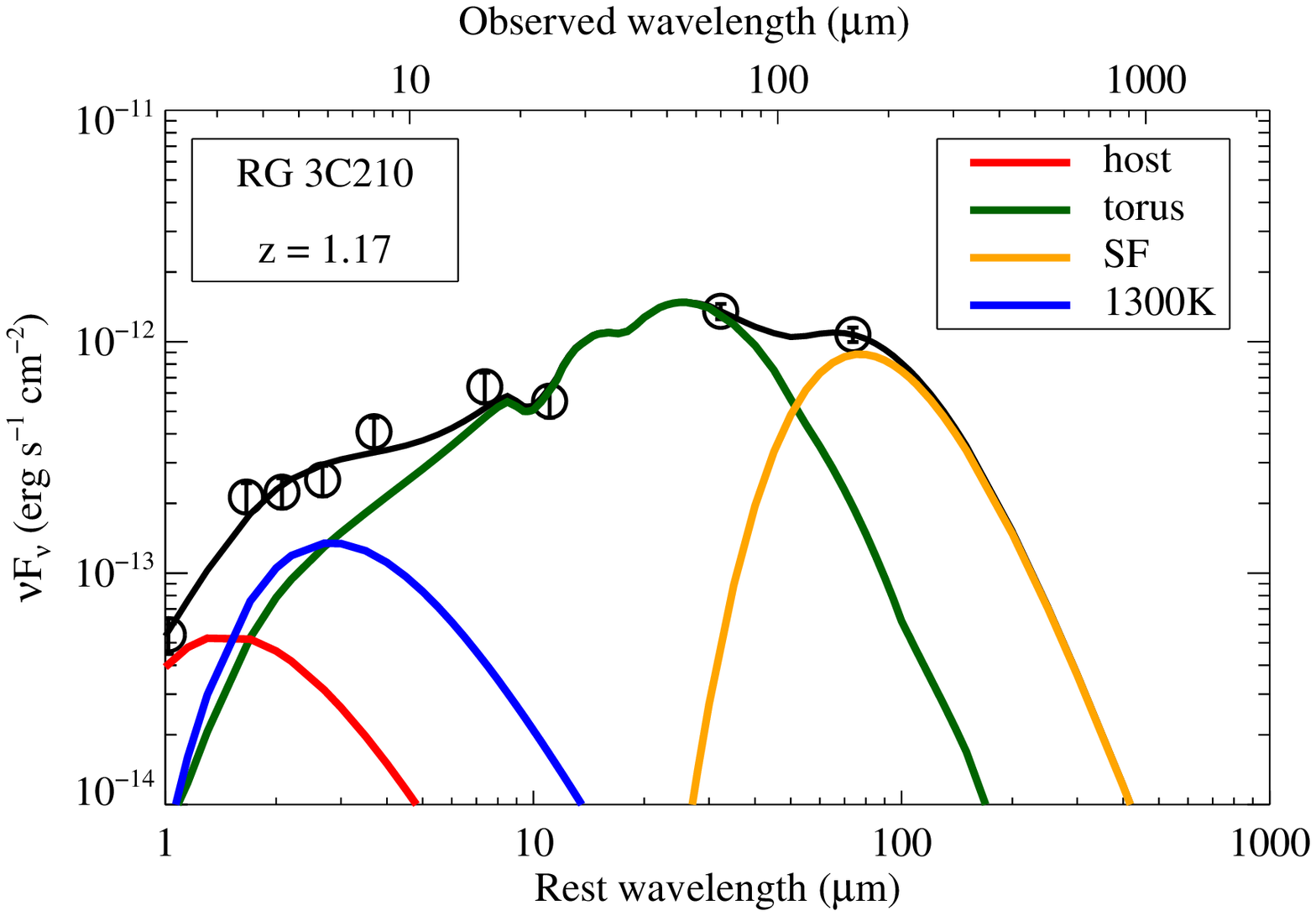}
      \includegraphics[width=4.5cm]{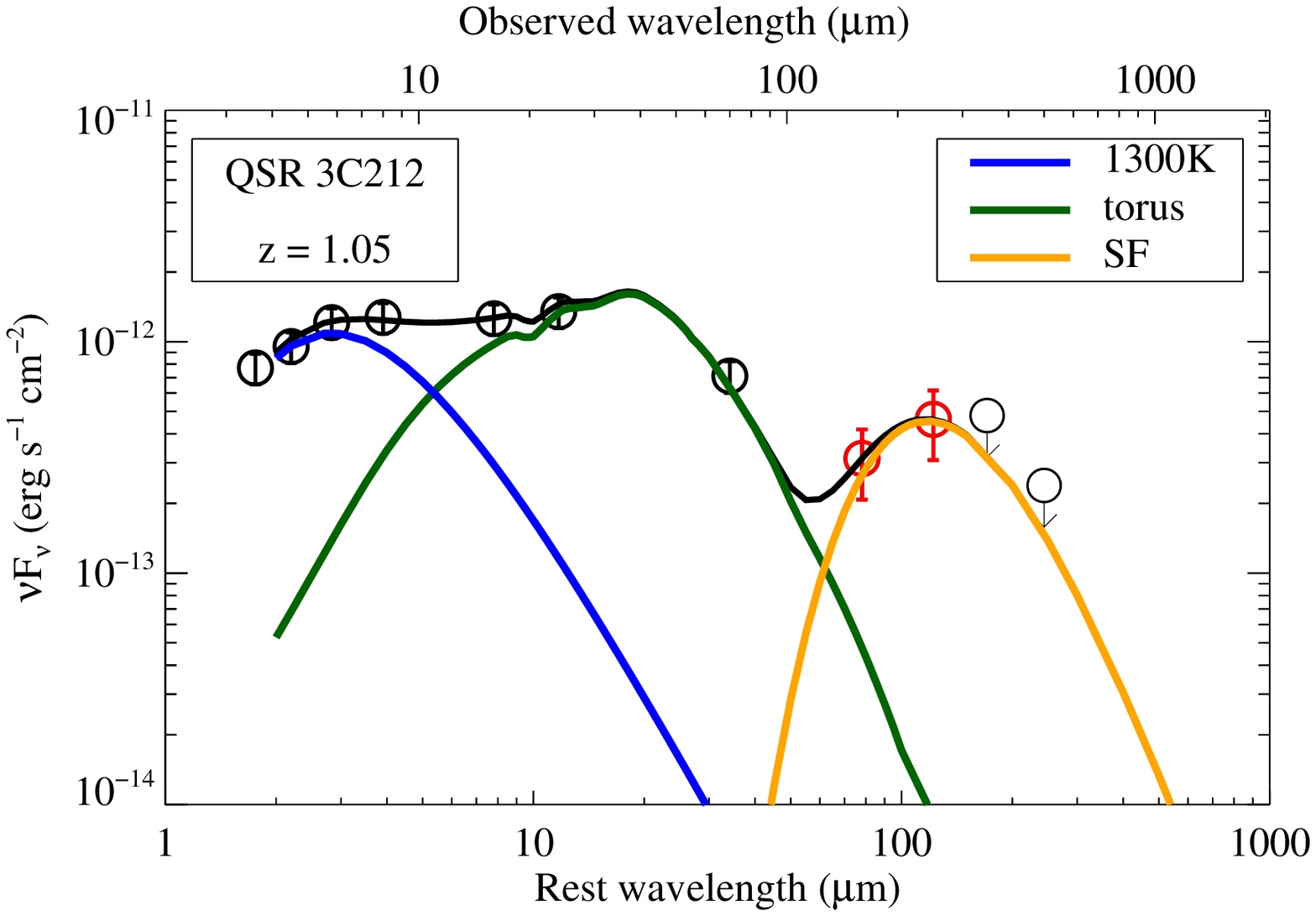}
      \includegraphics[width=4.5cm]{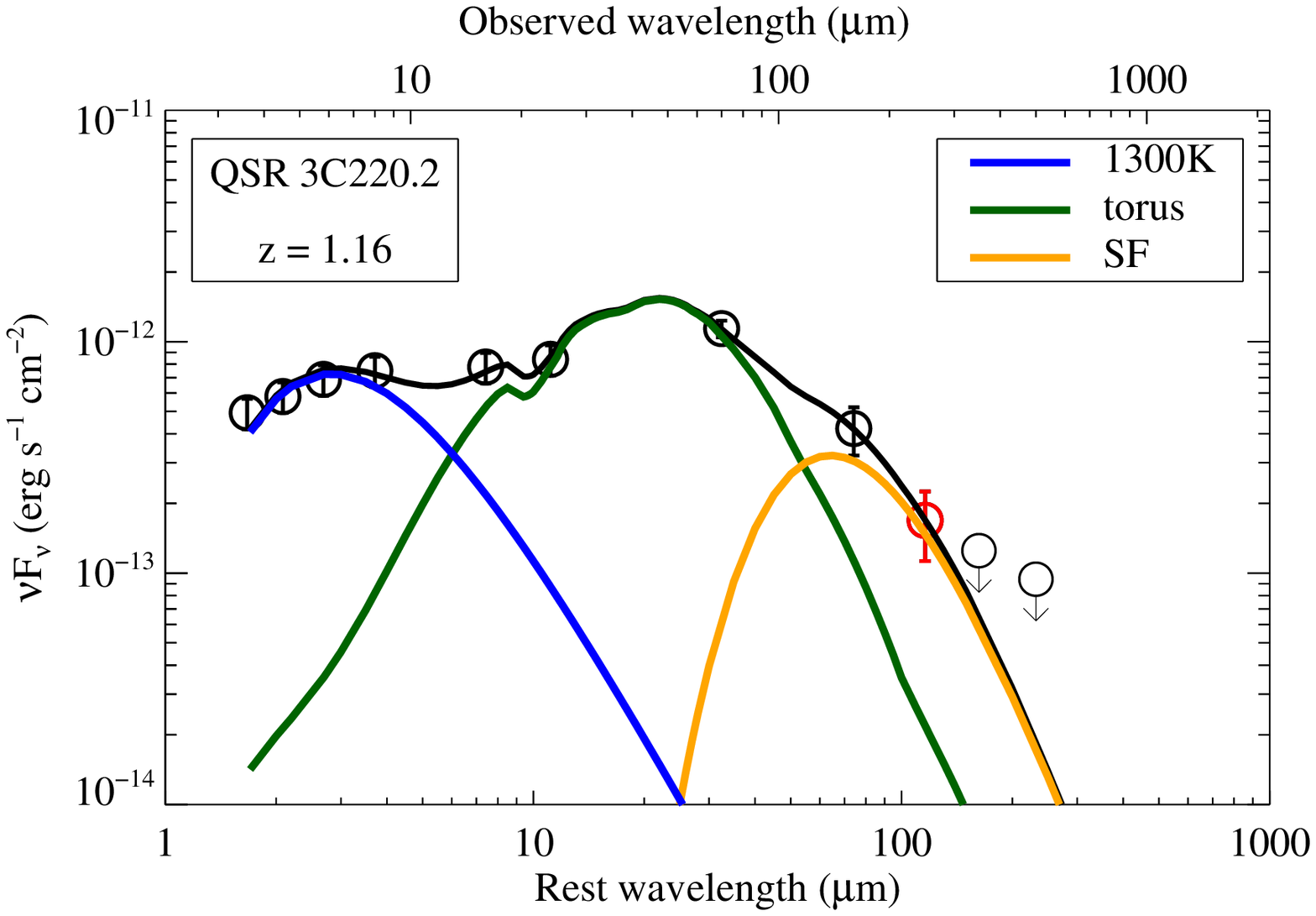}
      \includegraphics[width=4.5cm]{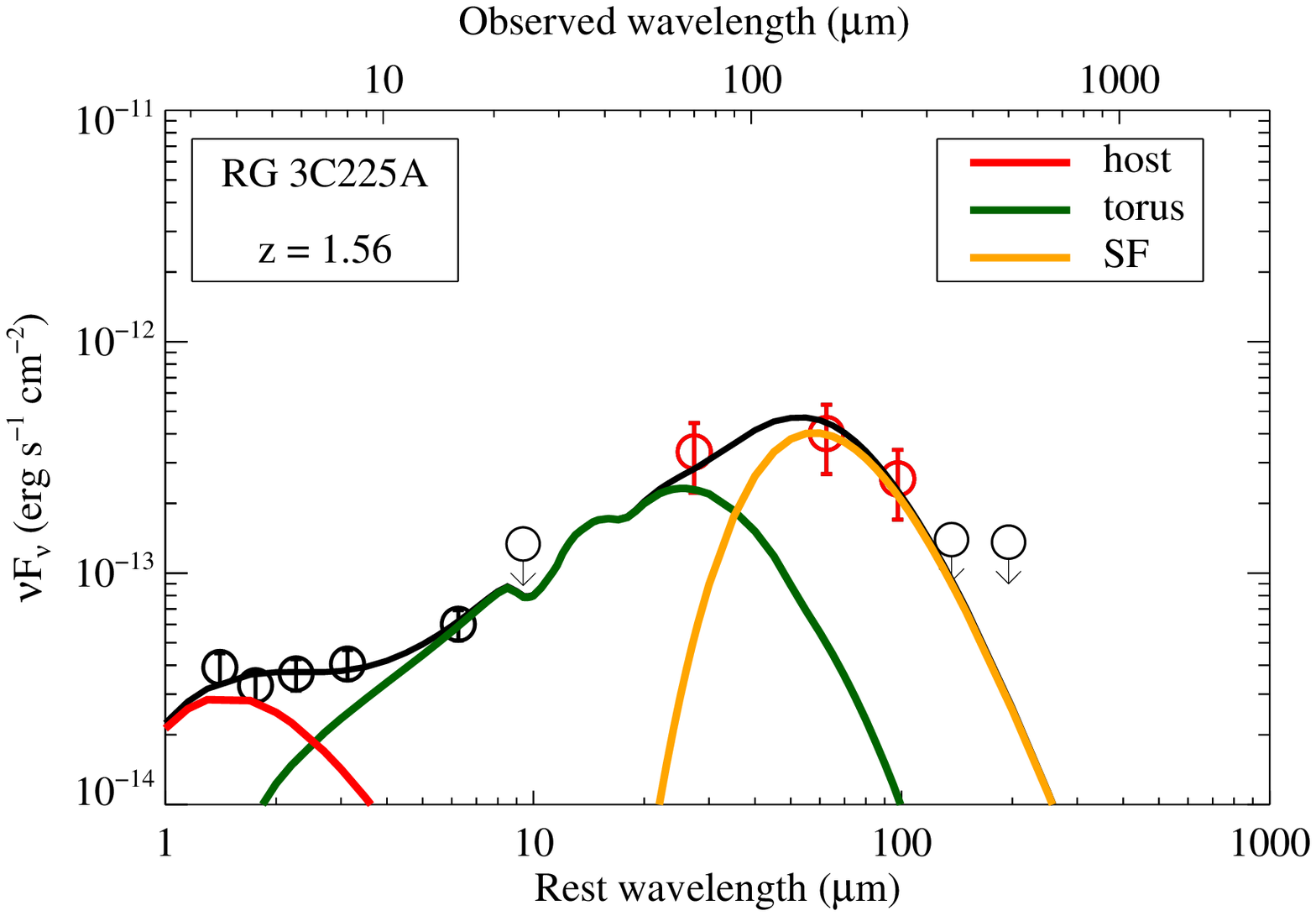}
      \includegraphics[width=4.5cm]{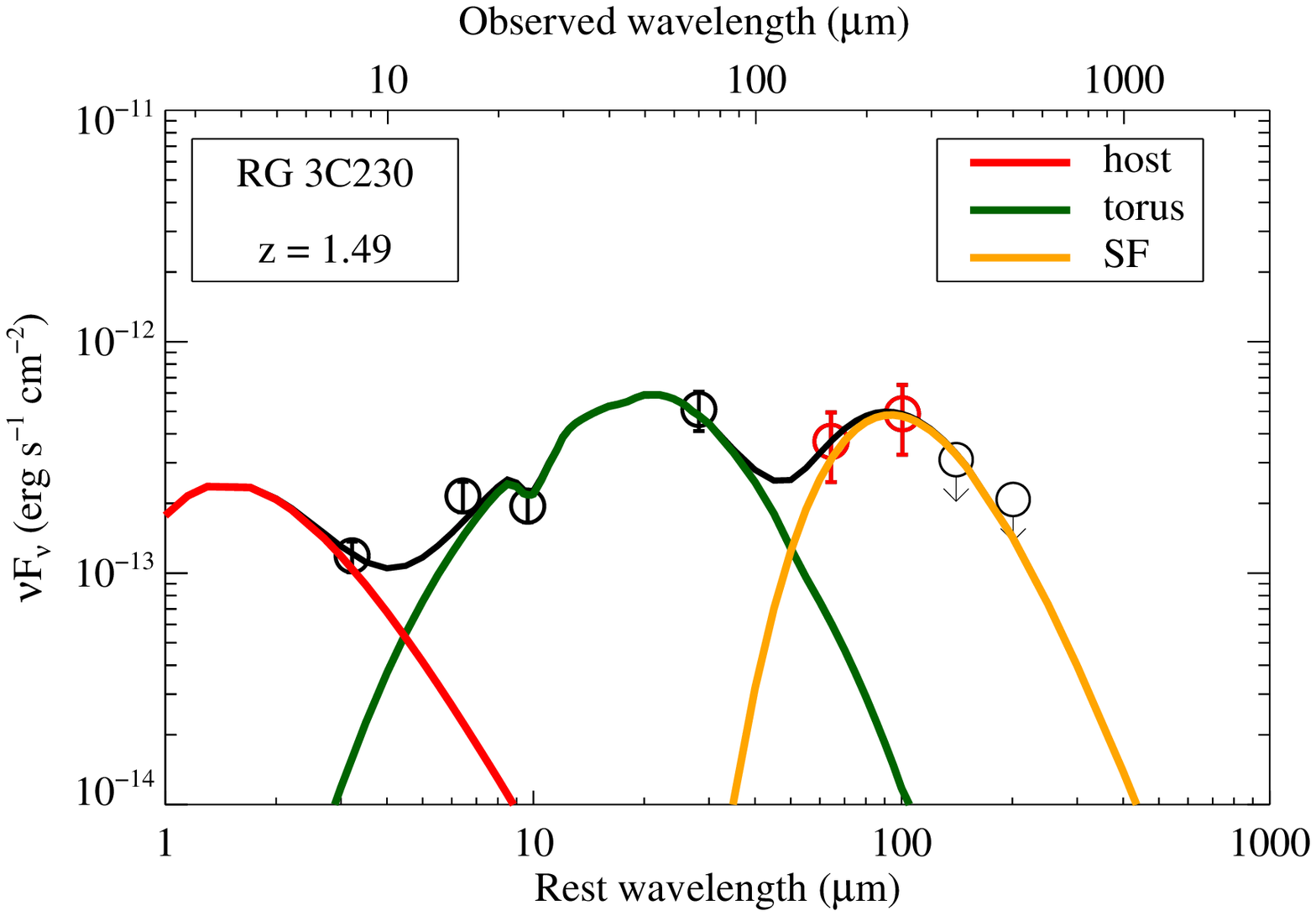}
      \includegraphics[width=4.5cm]{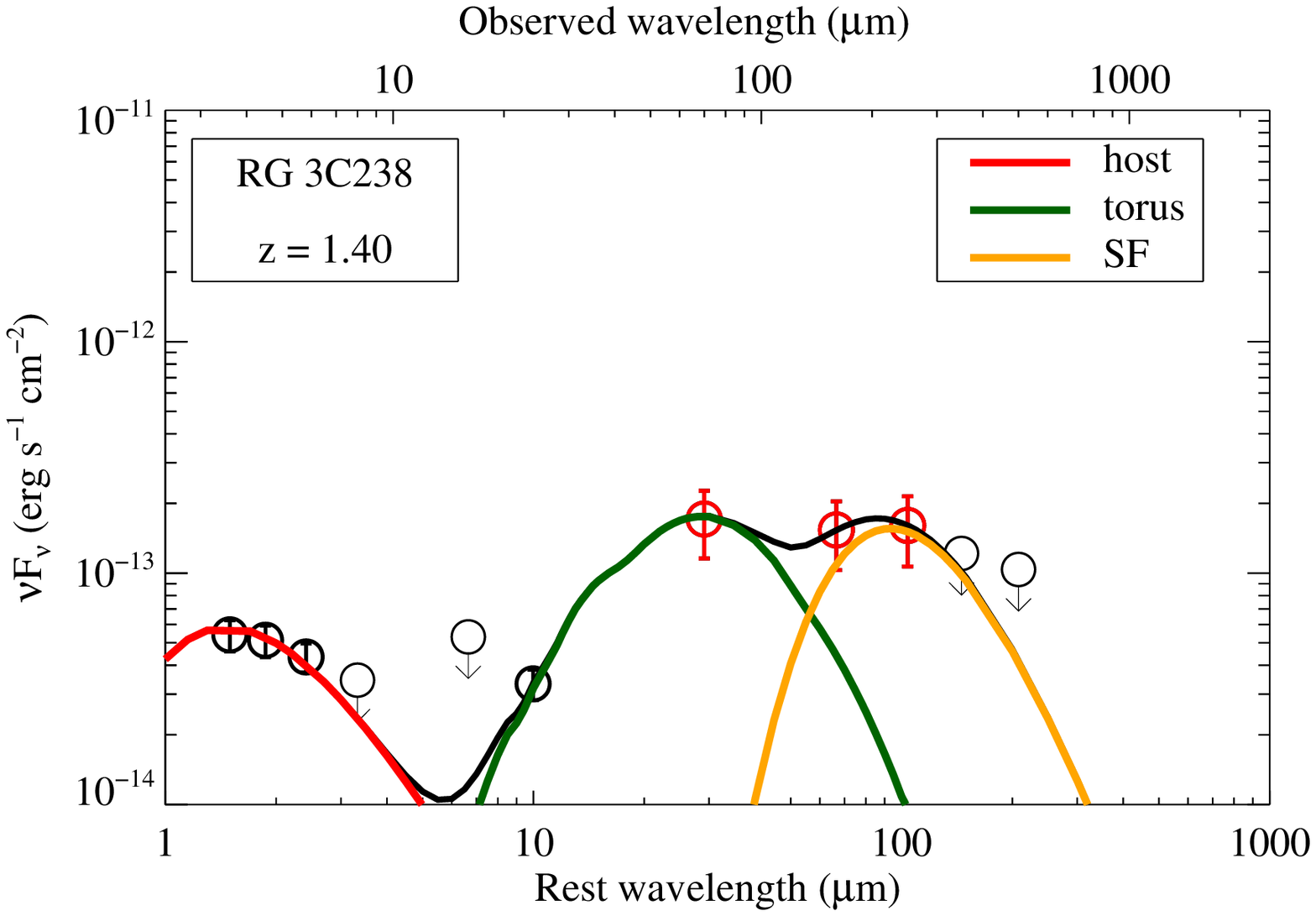}
      \includegraphics[width=4.5cm]{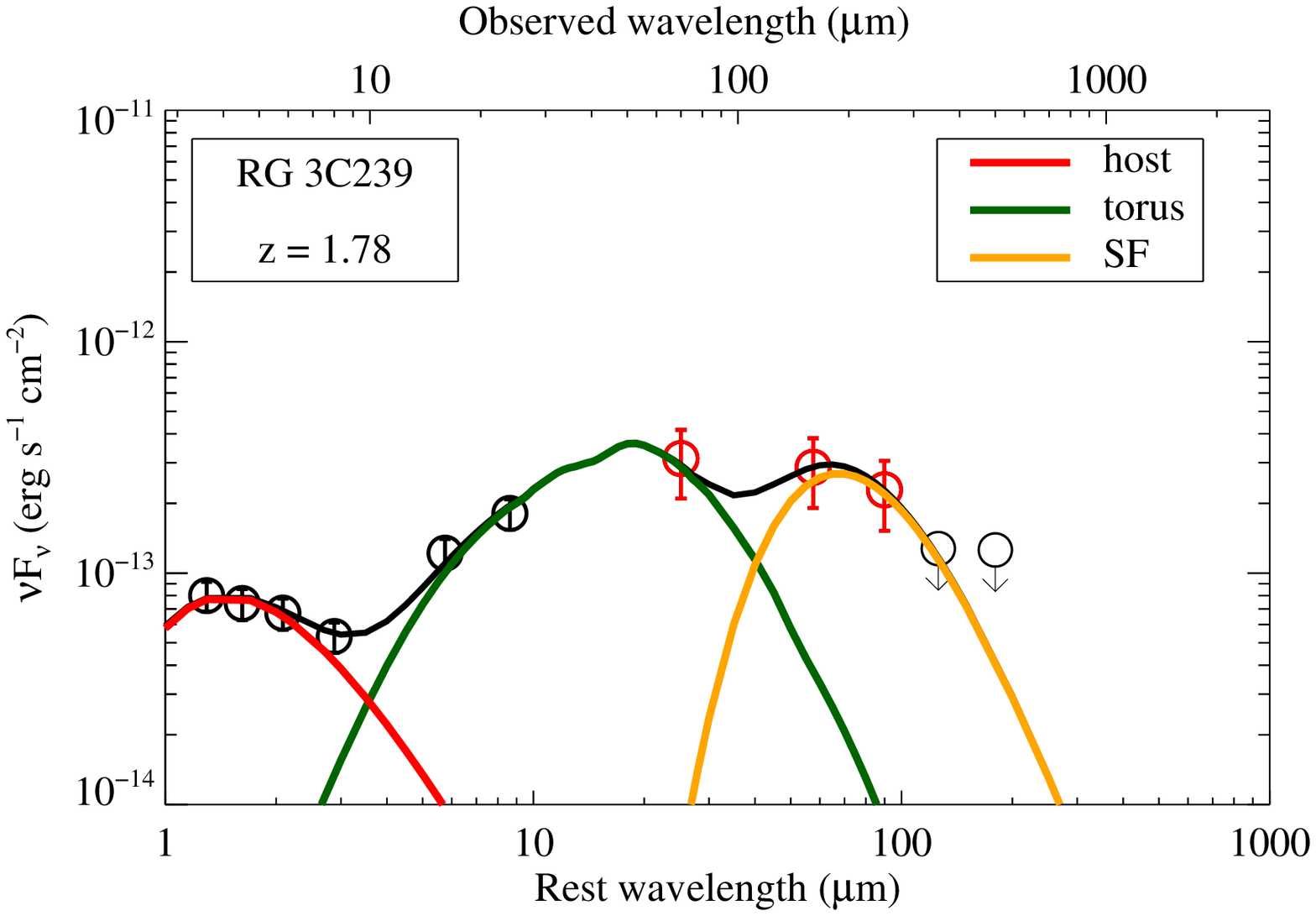}
      \includegraphics[width=4.5cm]{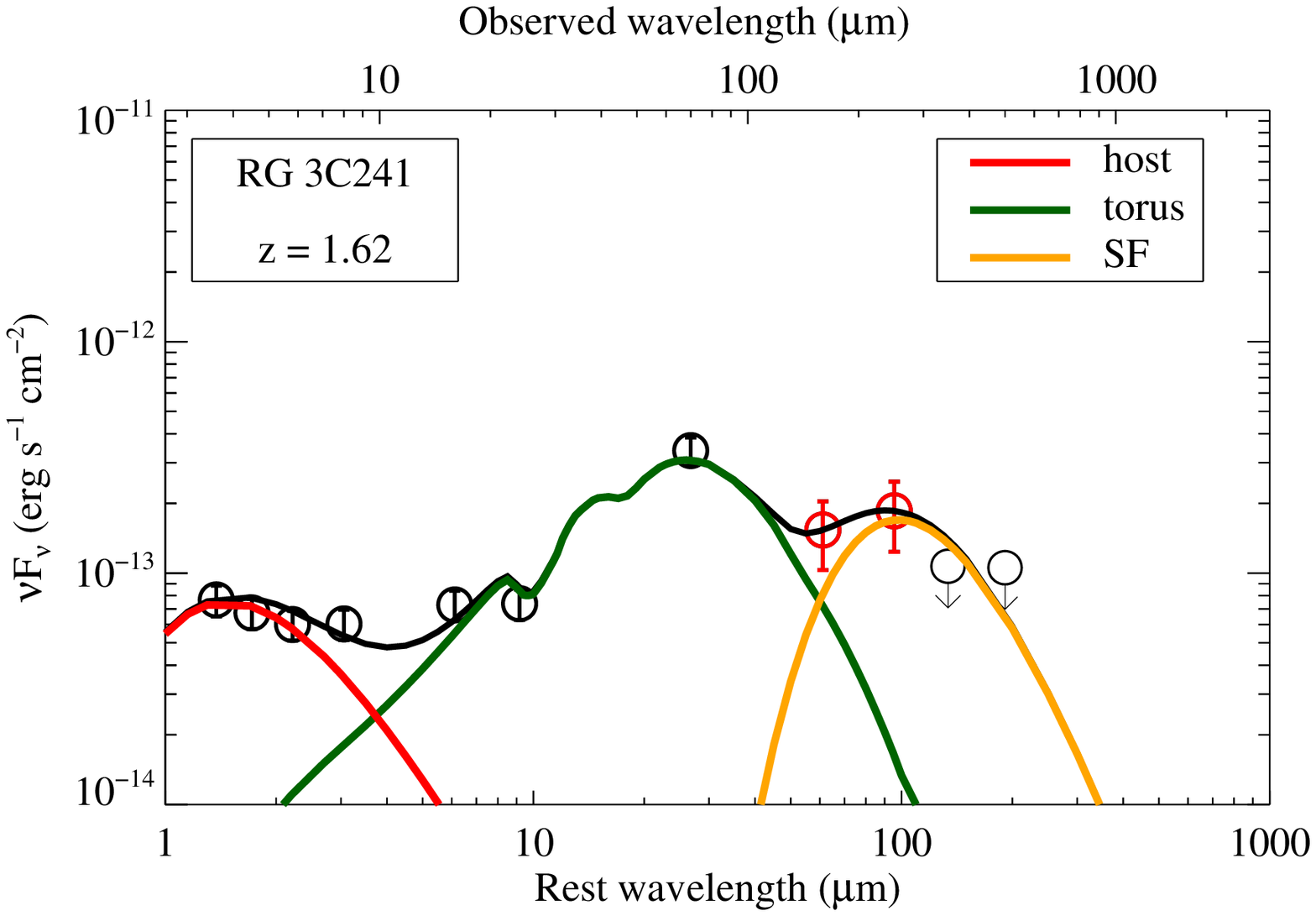}
      \includegraphics[width=4.5cm]{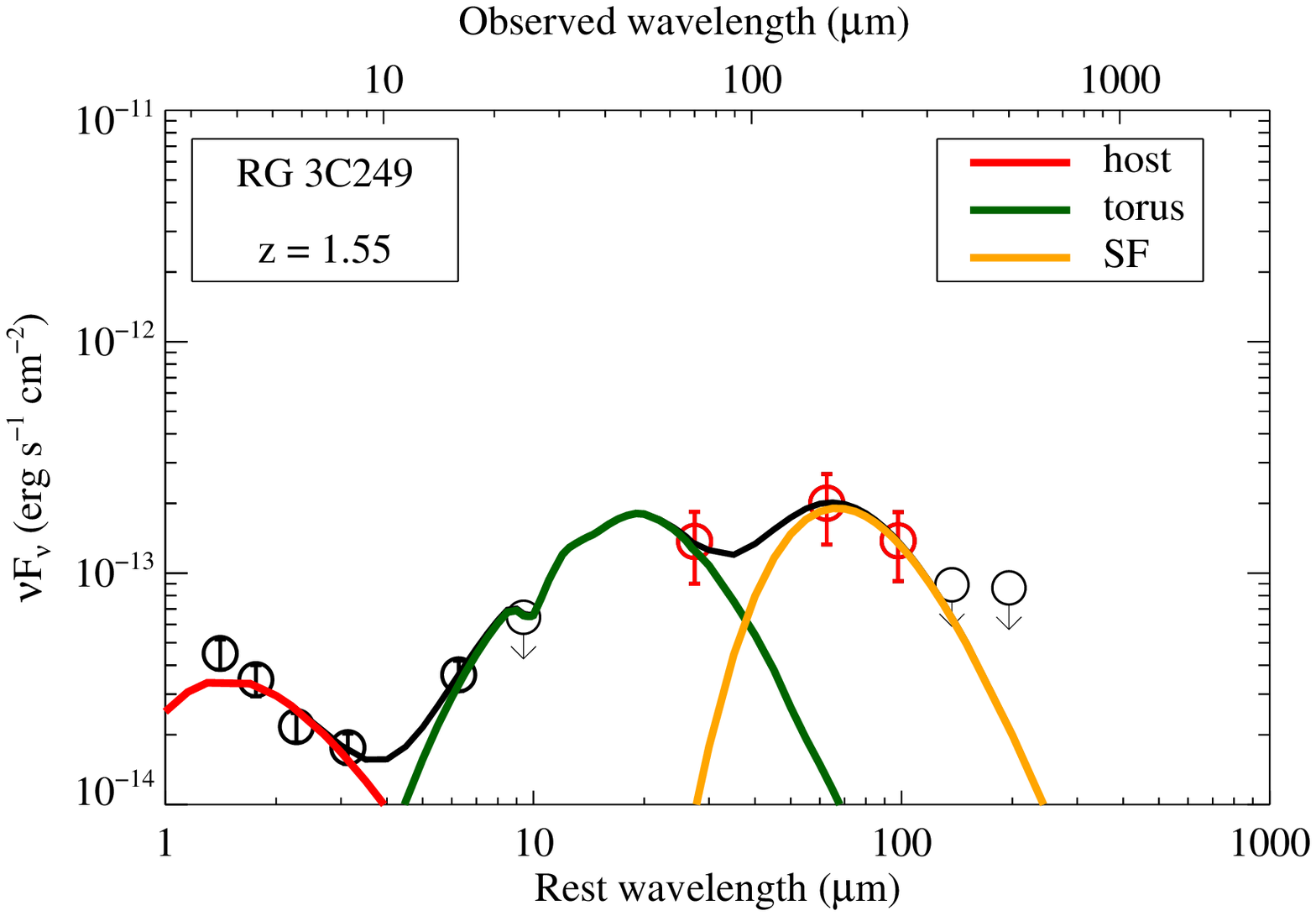}
      \includegraphics[width=4.5cm]{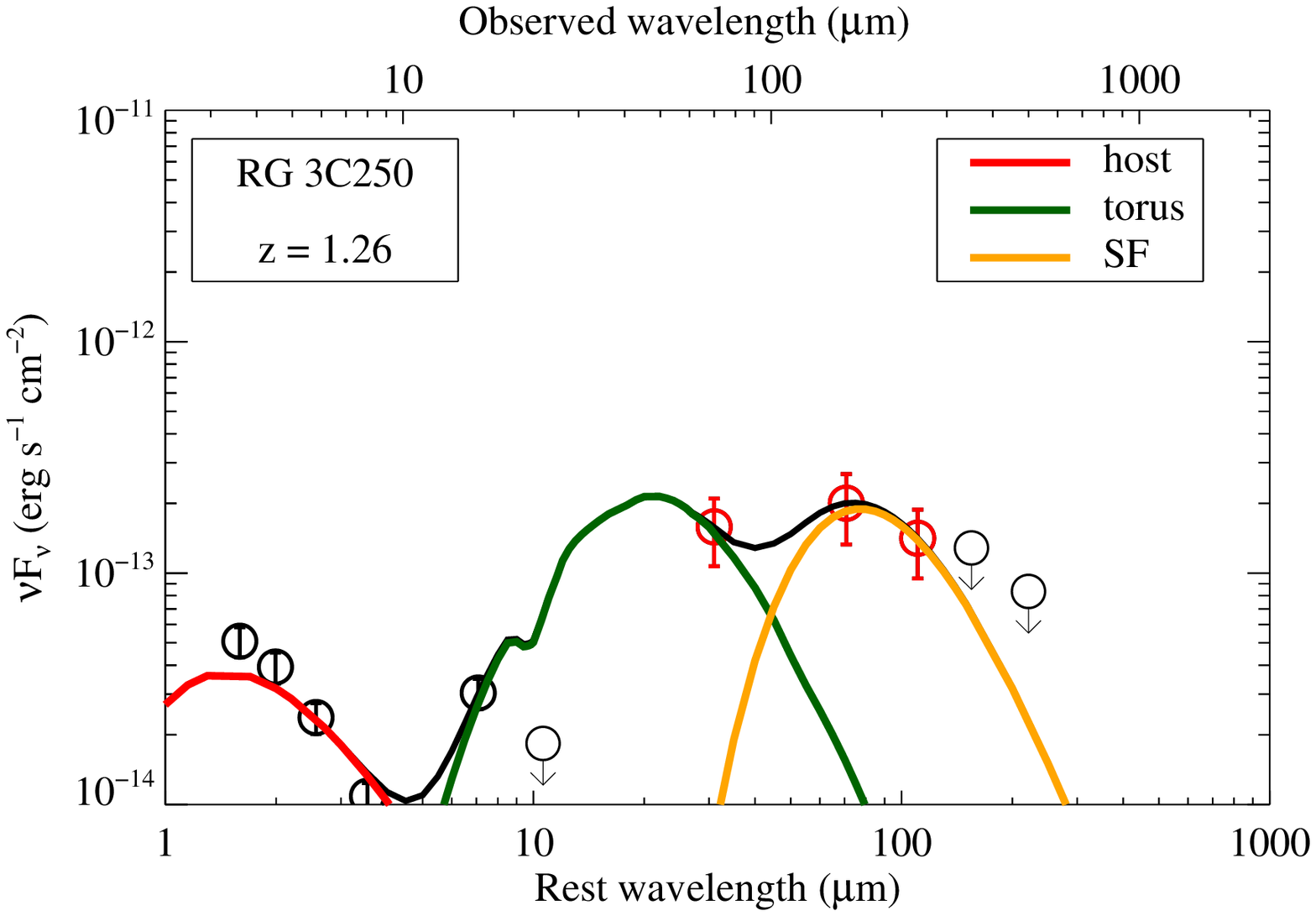}
      \includegraphics[width=4.5cm]{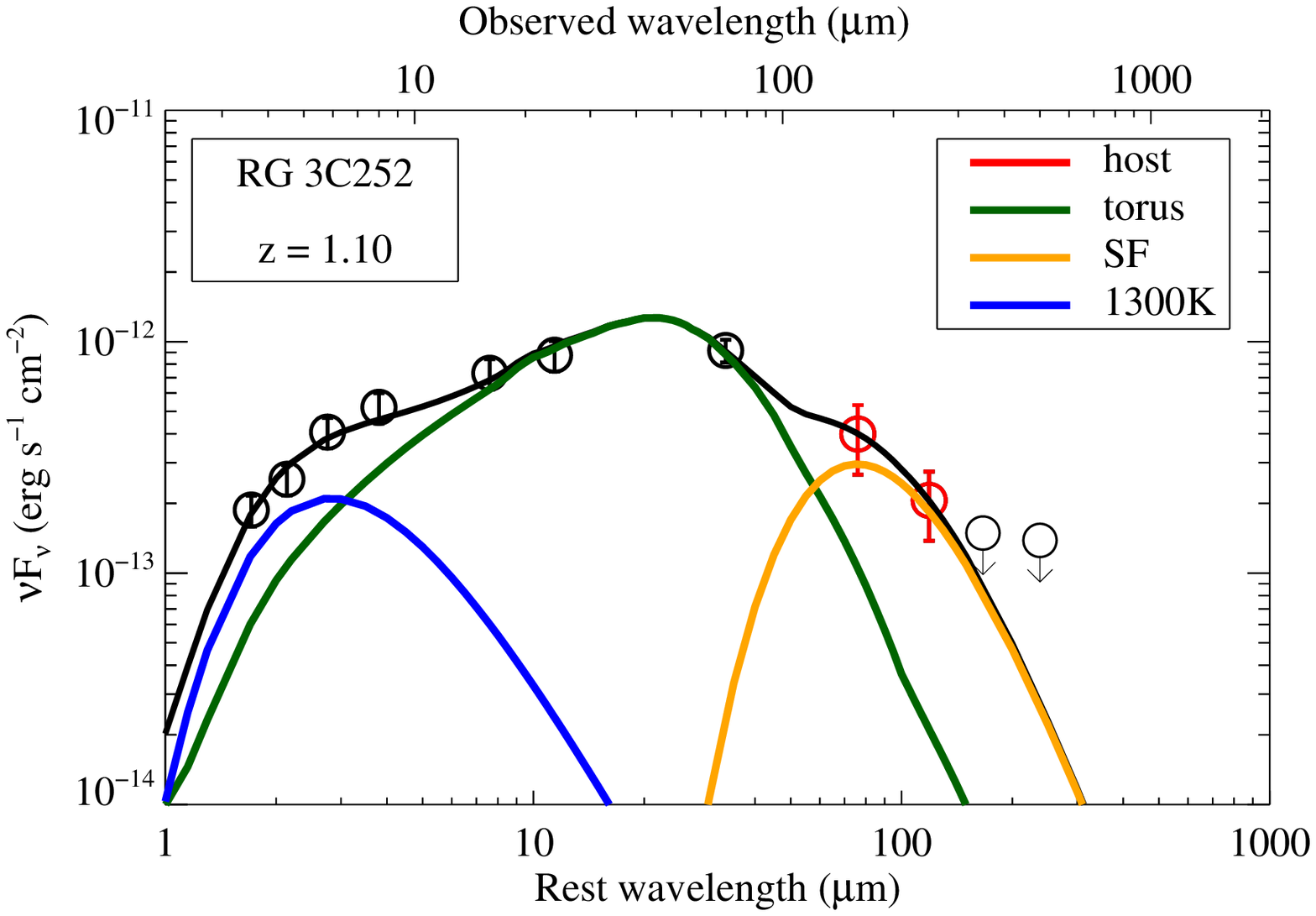}
      \includegraphics[width=4.5cm]{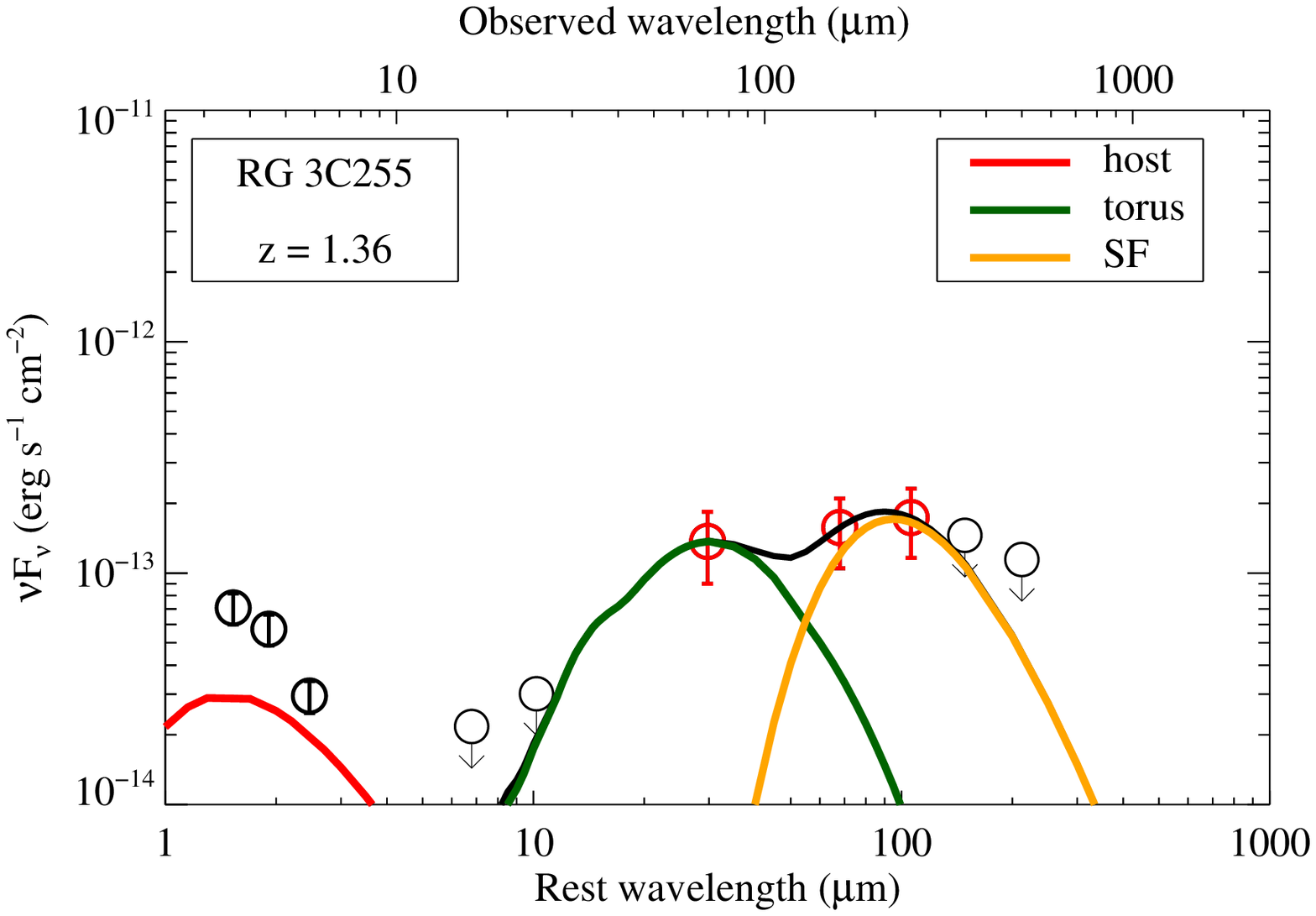}
      \includegraphics[width=4.5cm]{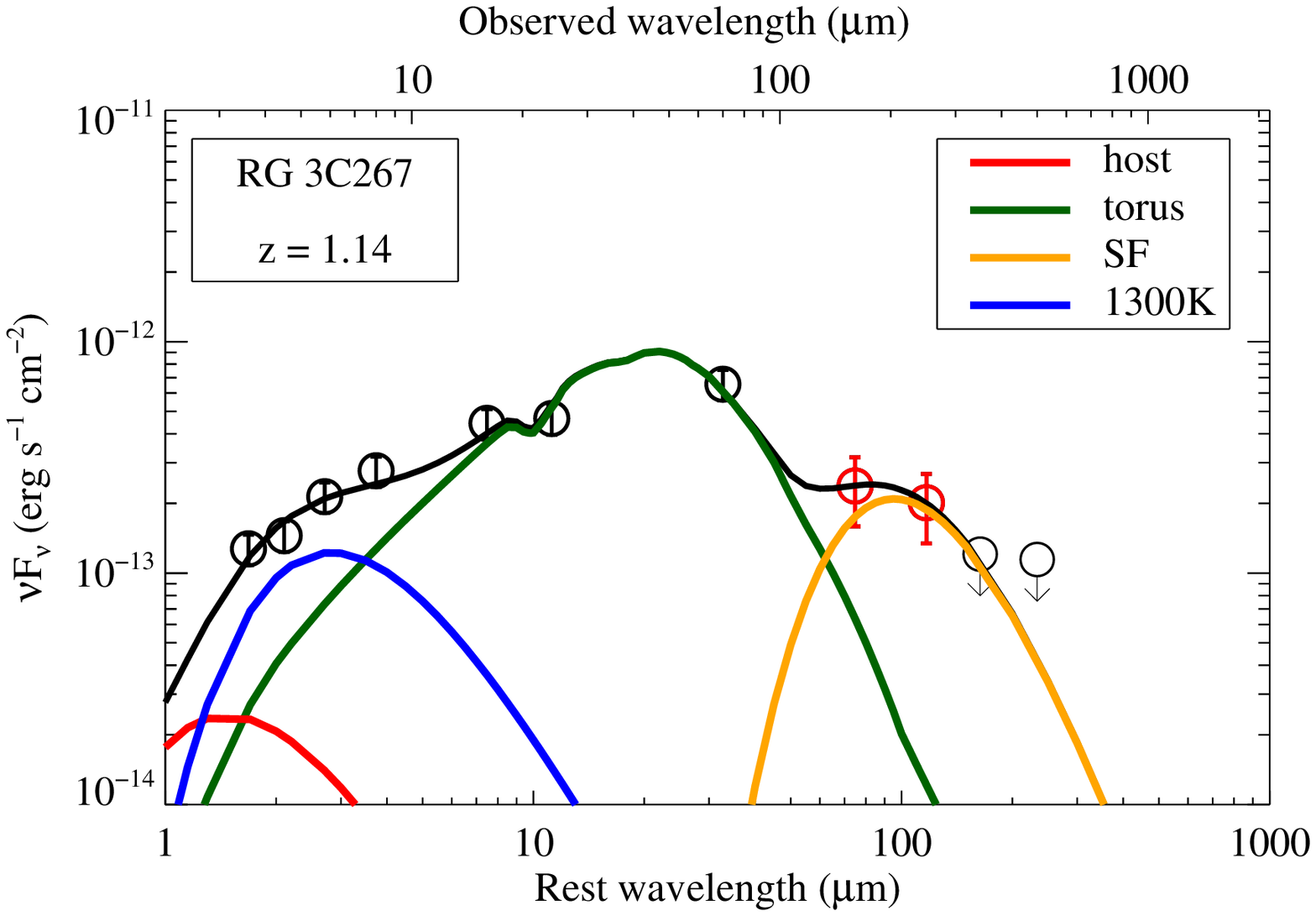}
      \includegraphics[width=4.5cm]{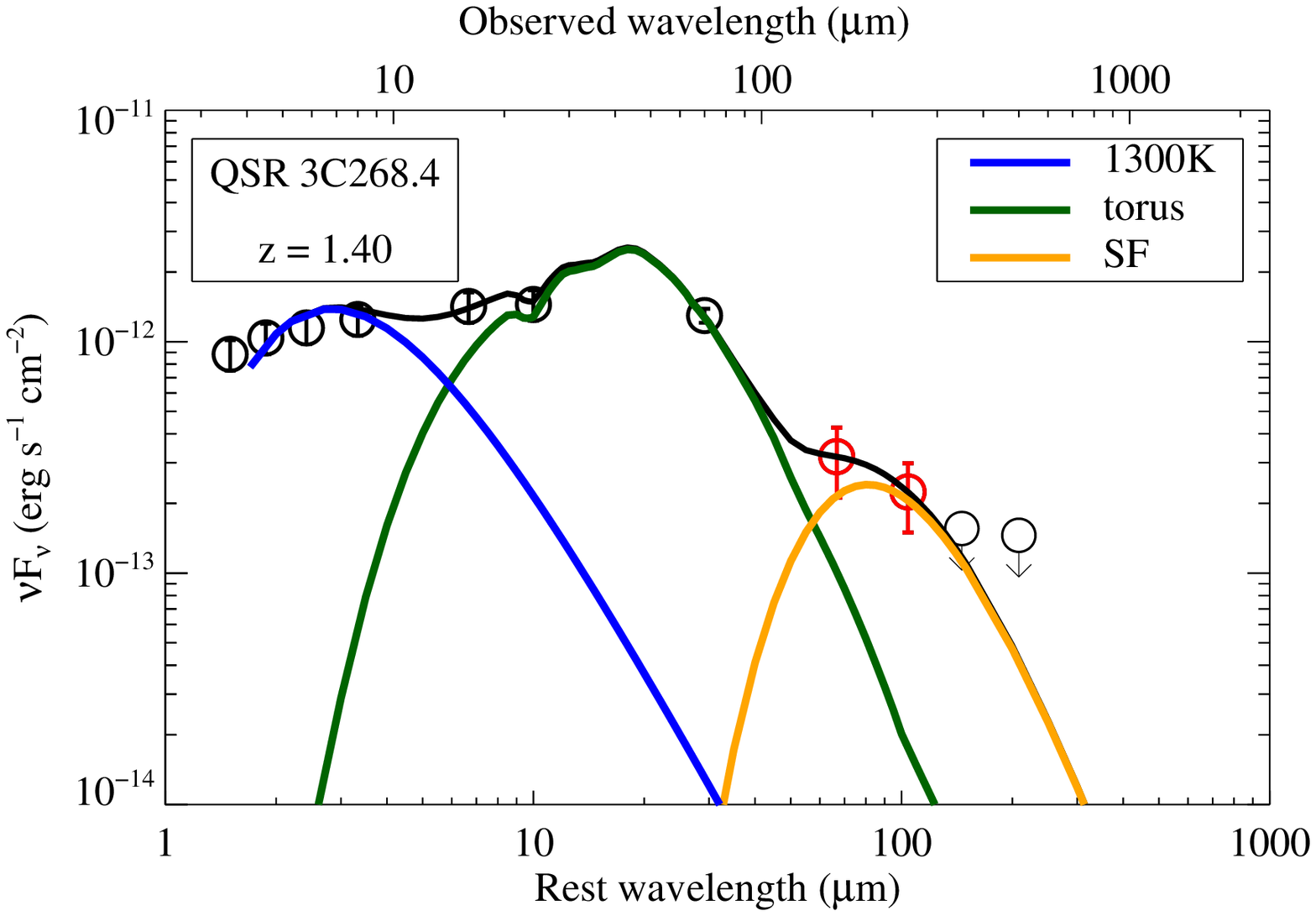}
      \caption{Spectral energy distributions of the objects detected in 
               fewer than three \textit{Herschel} bands. Individual components 
               as described in Fig.~\ref{figure:exampleSEDs}. 
               Red circles denote $3\sigma$ upper limits taken to be 
               tentative detections when calculating upper limits of 
               physical parameters, as explained in Sect.~\ref{section:SEDs}. 
               }
      \label{figure:nonDetectedBestFitSEDs}      
   \end{figure*}
   \addtocounter{figure}{-1}
   \begin{figure*}
      \includegraphics[width=4.5cm]{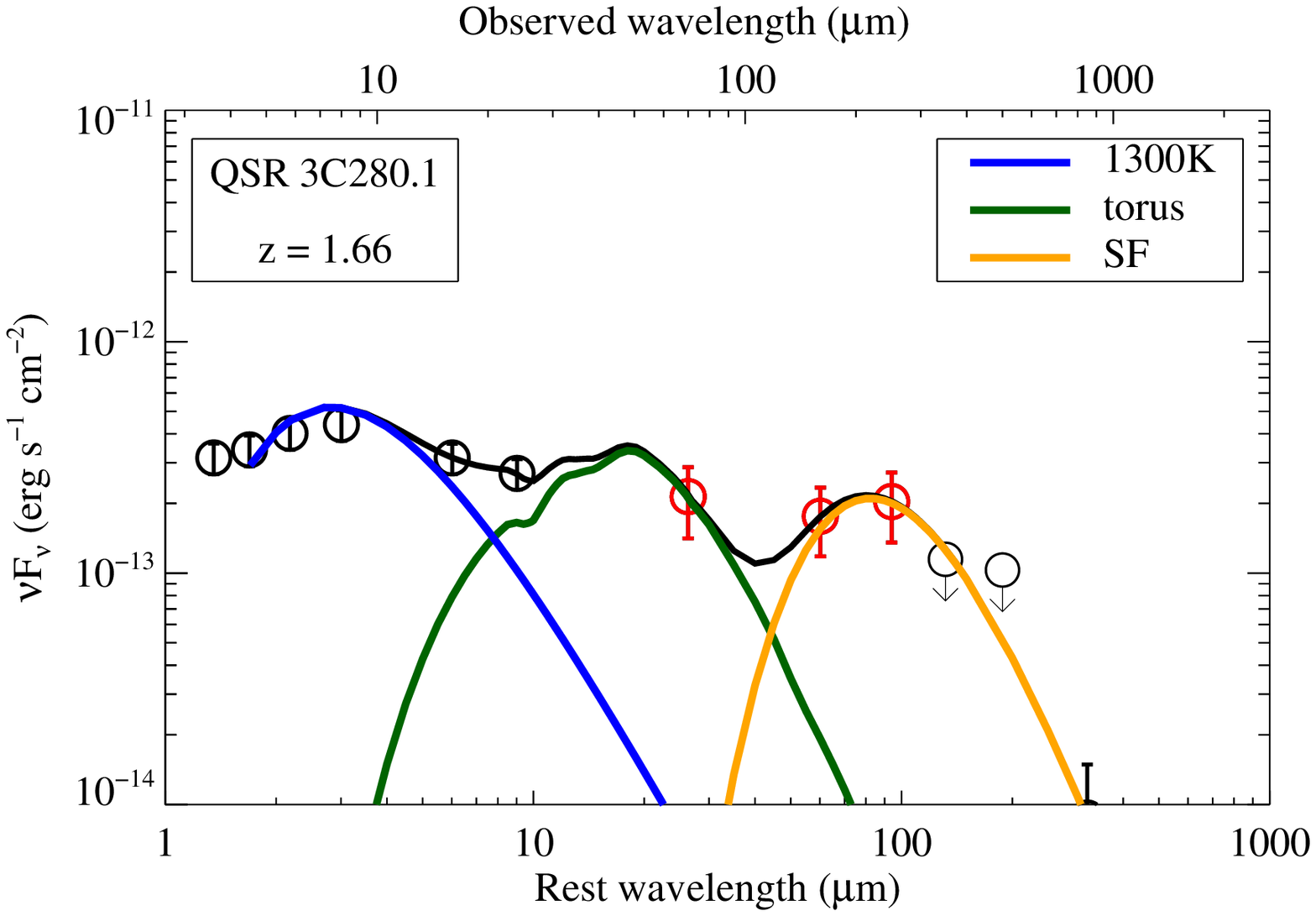}
      \includegraphics[width=4.5cm]{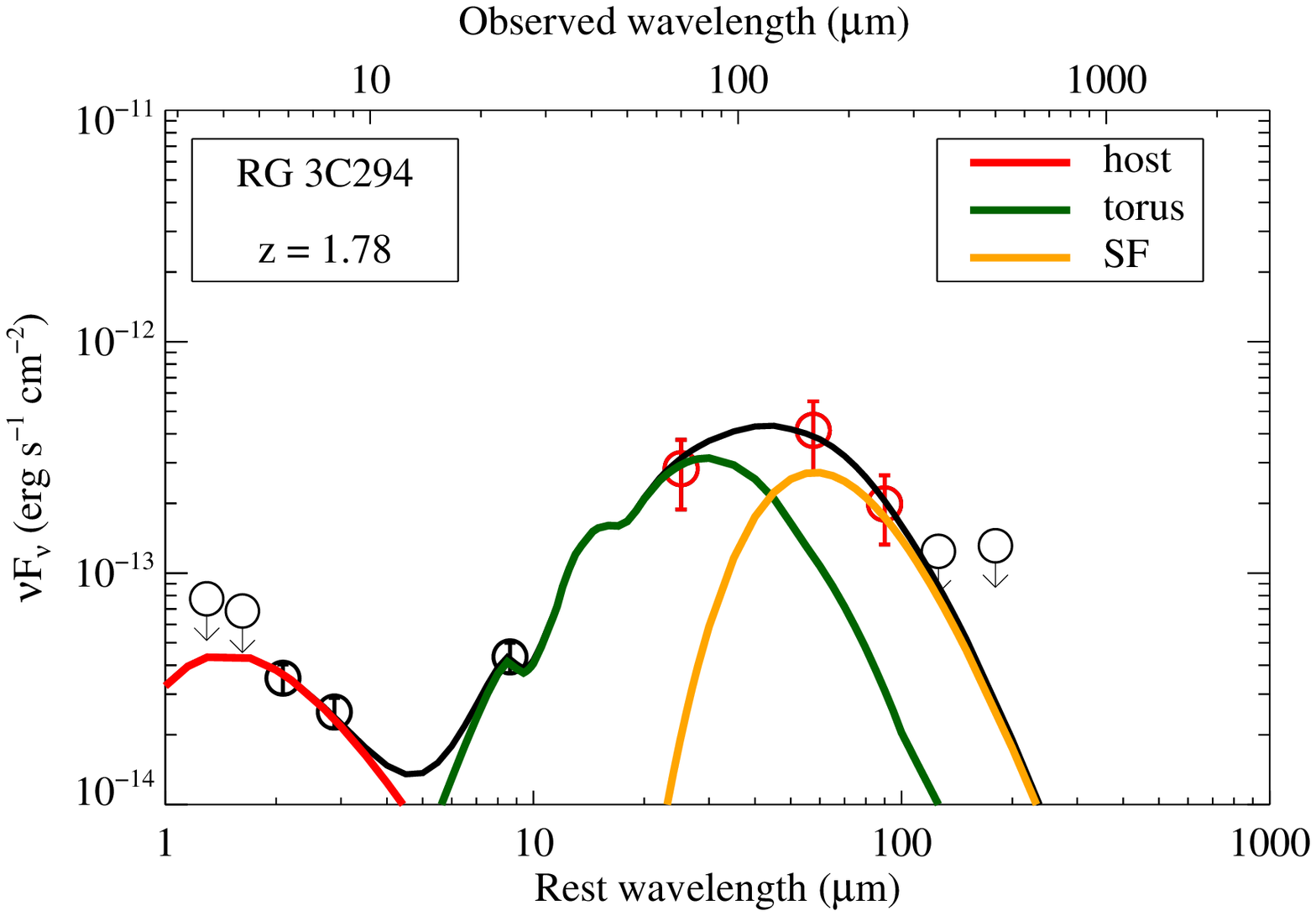}
      \includegraphics[width=4.5cm]{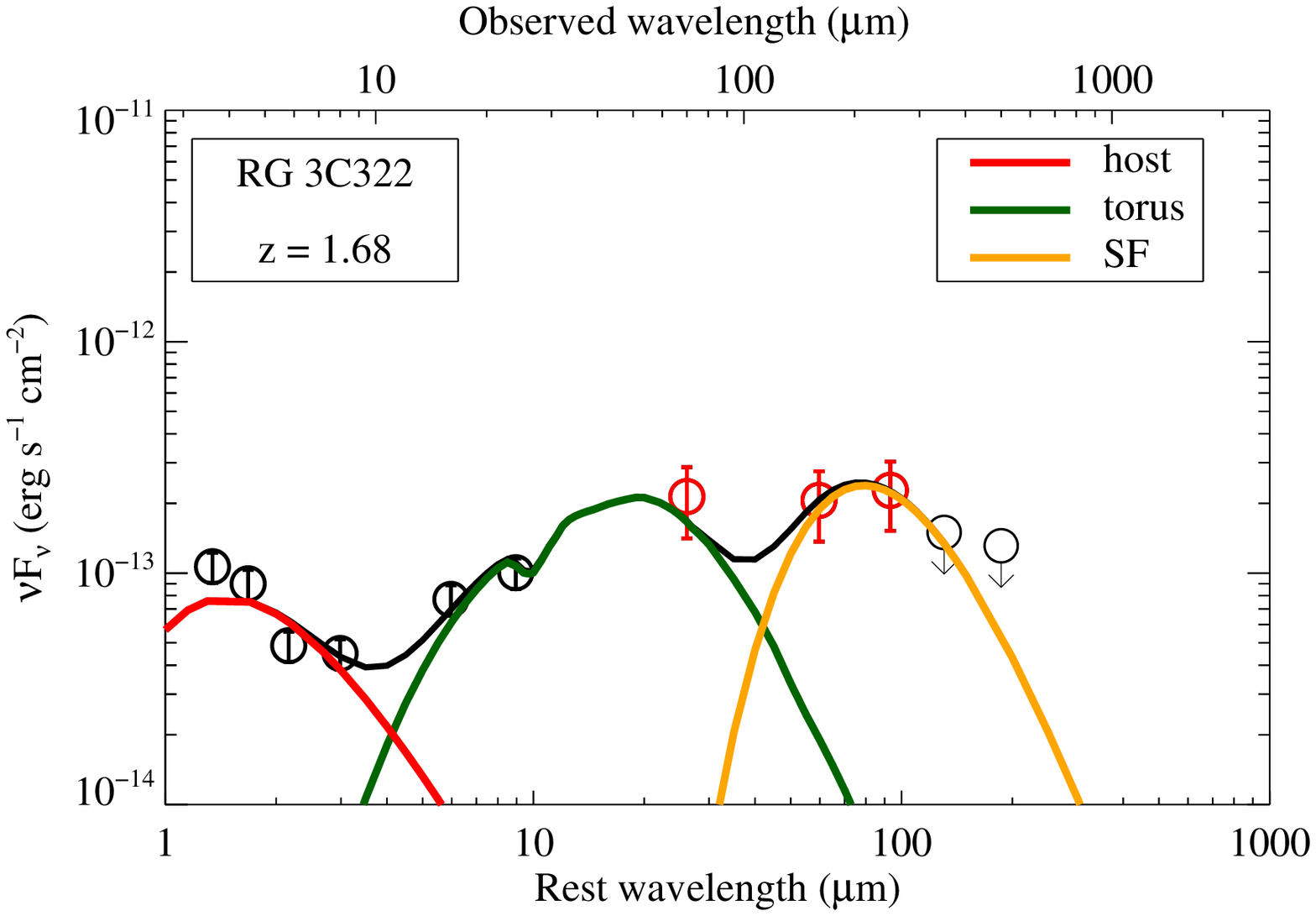}
      \includegraphics[width=4.5cm]{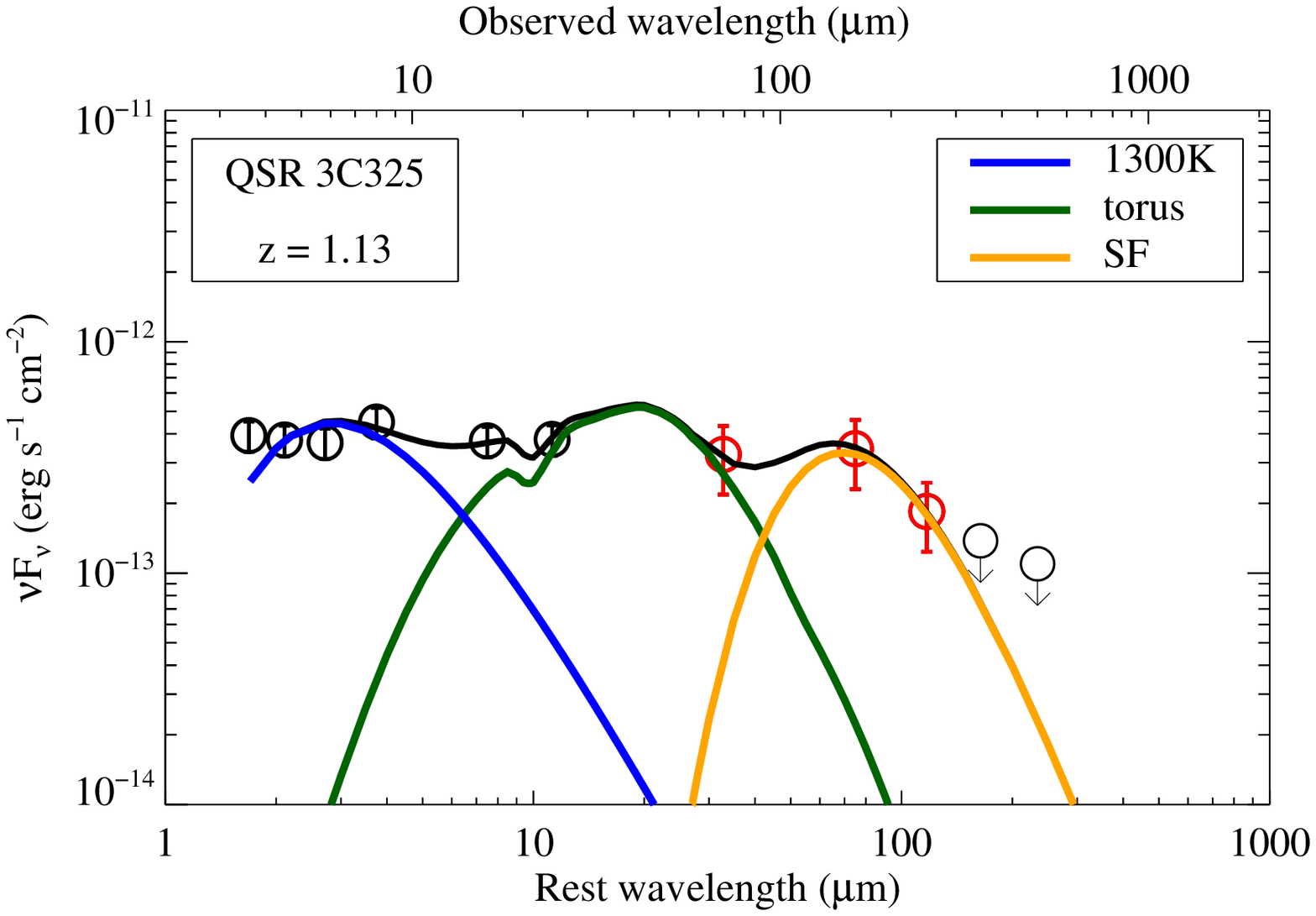}
      \includegraphics[width=4.5cm]{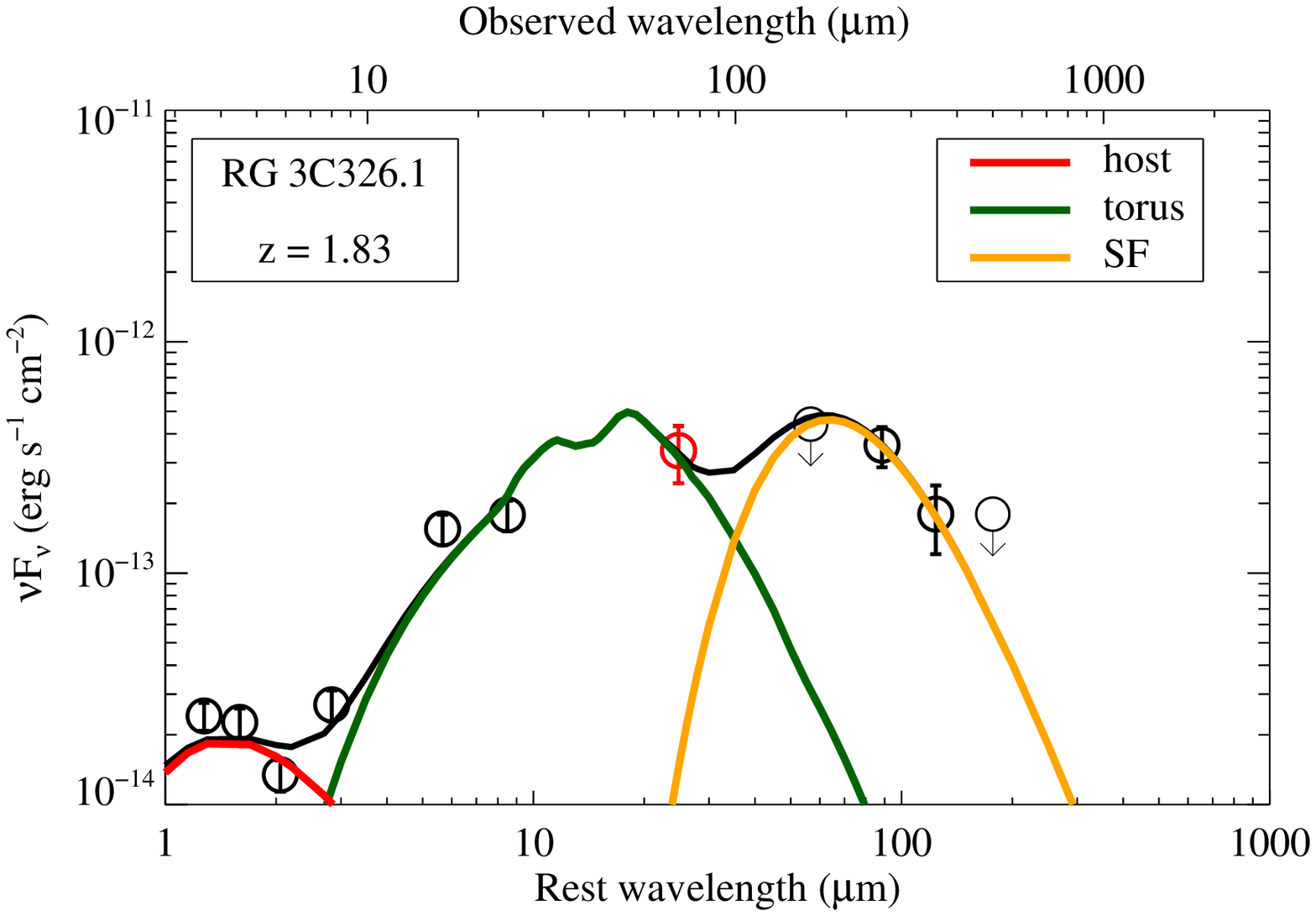}
      \includegraphics[width=4.5cm]{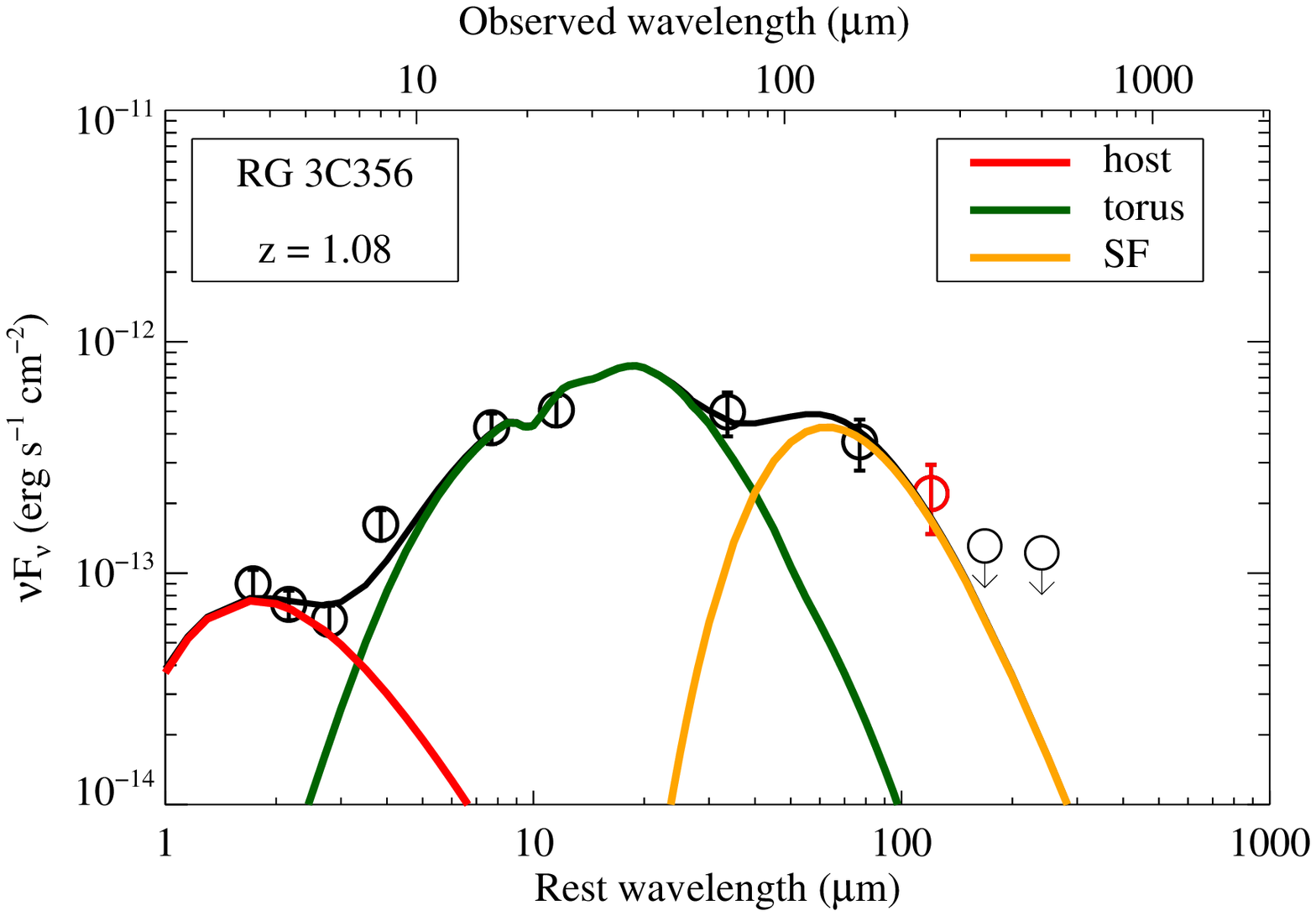}
      \includegraphics[width=4.5cm]{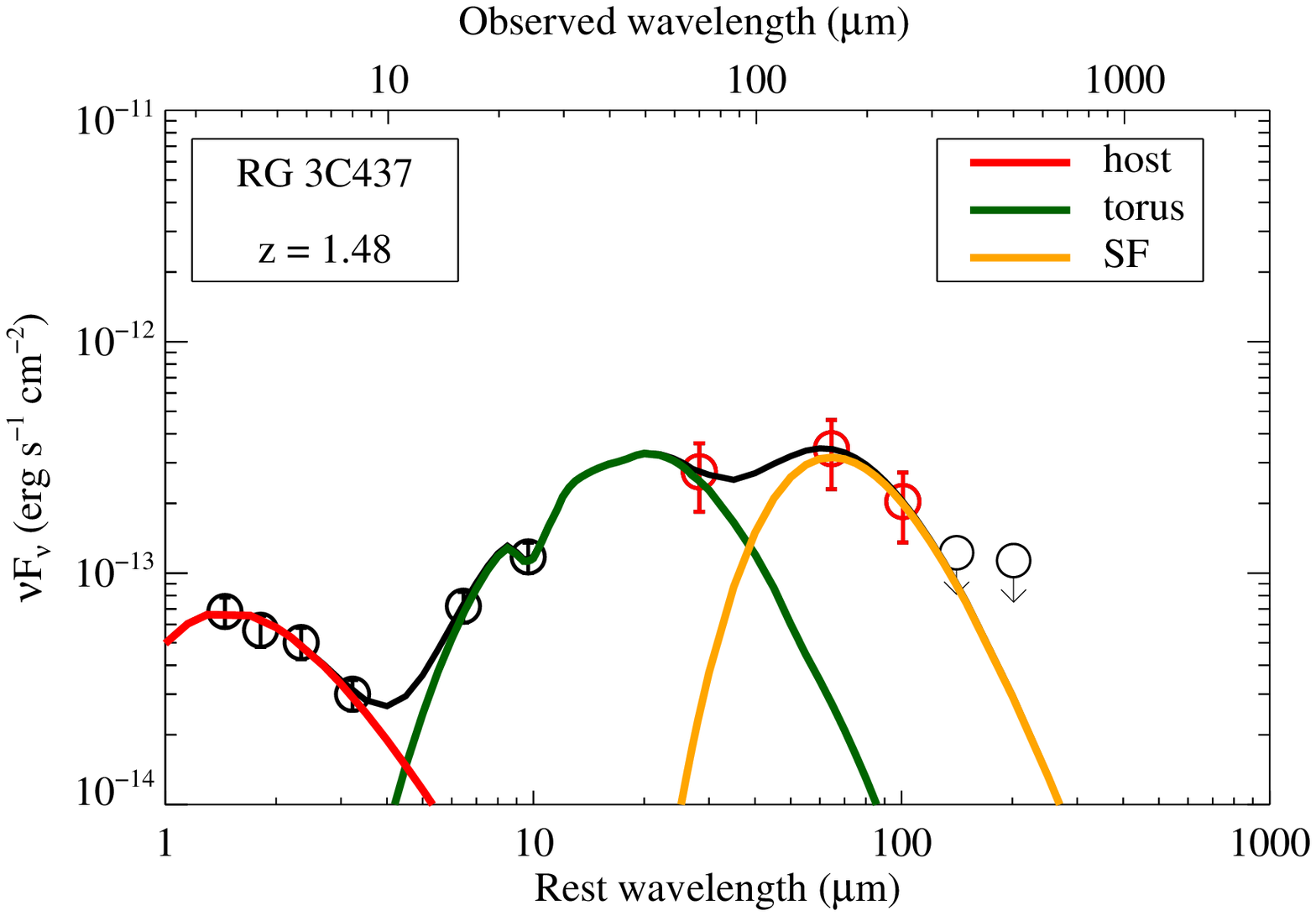}
      \includegraphics[width=4.5cm]{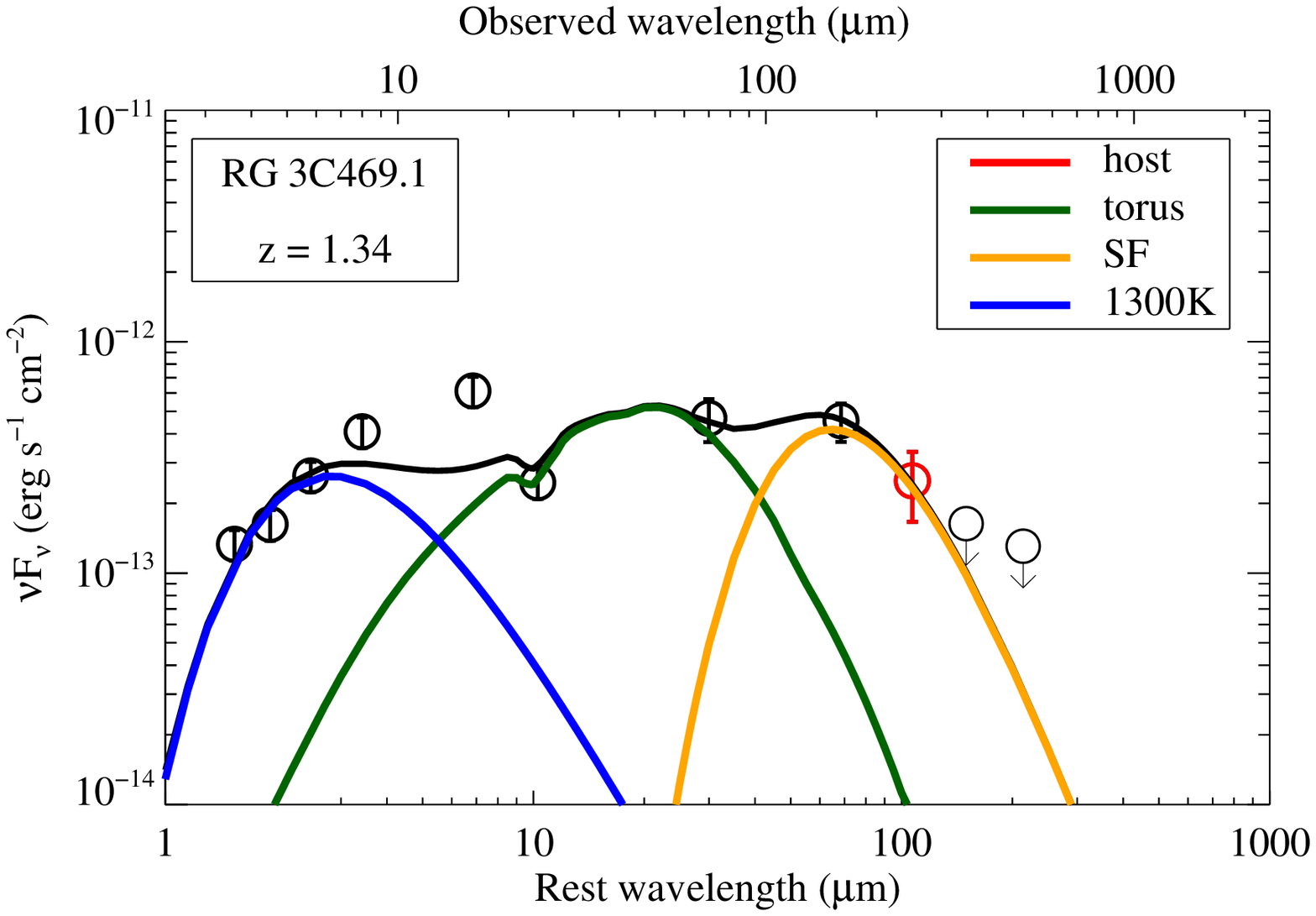}
      \includegraphics[width=4.5cm]{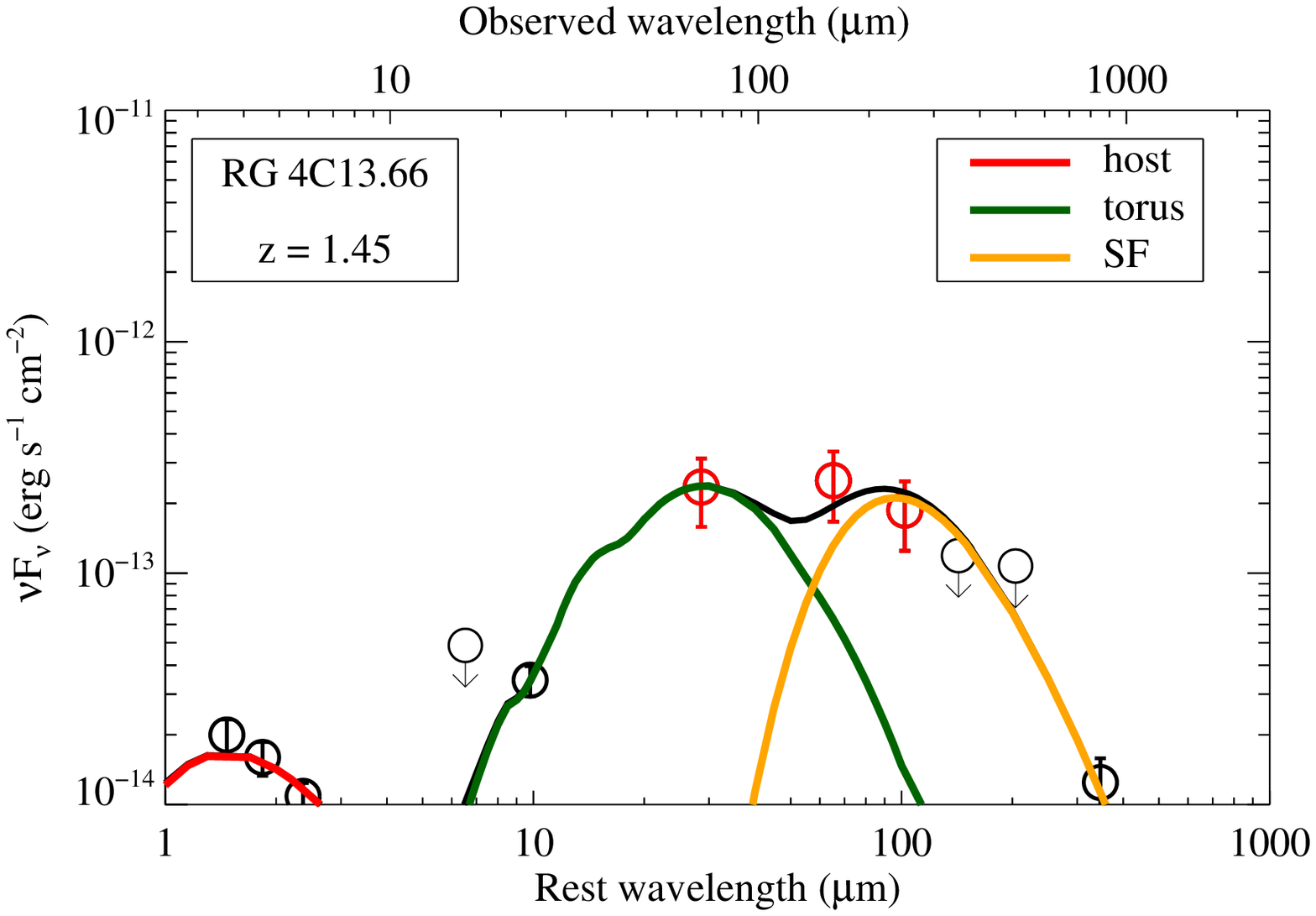}
      \includegraphics[width=4.5cm]{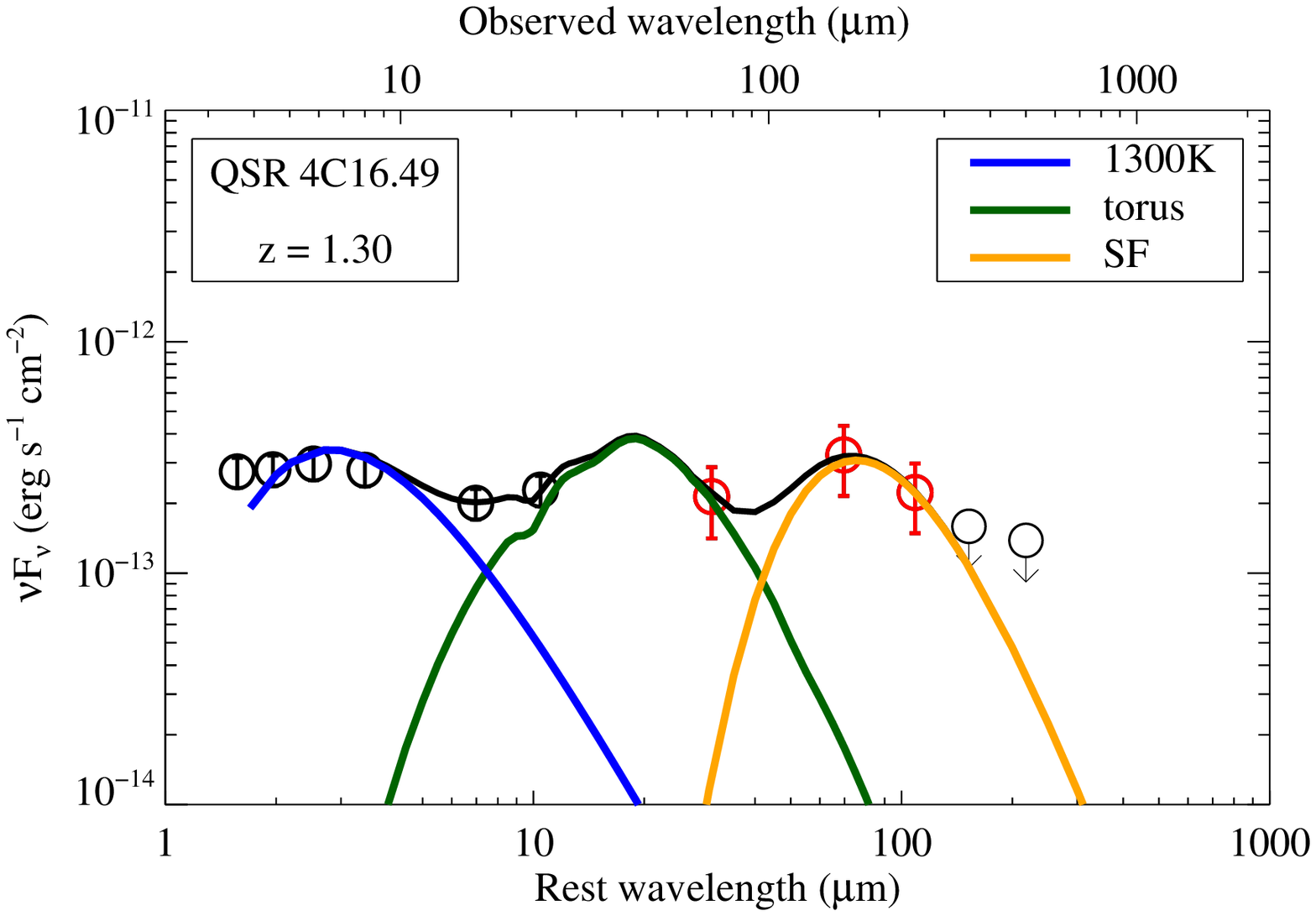}
      \caption{Continued.}
   \end{figure*}
\section{Postage stamps}
\label{appendix:stamps}
   \begin{figure*}
      \includegraphics[width=1.5cm]{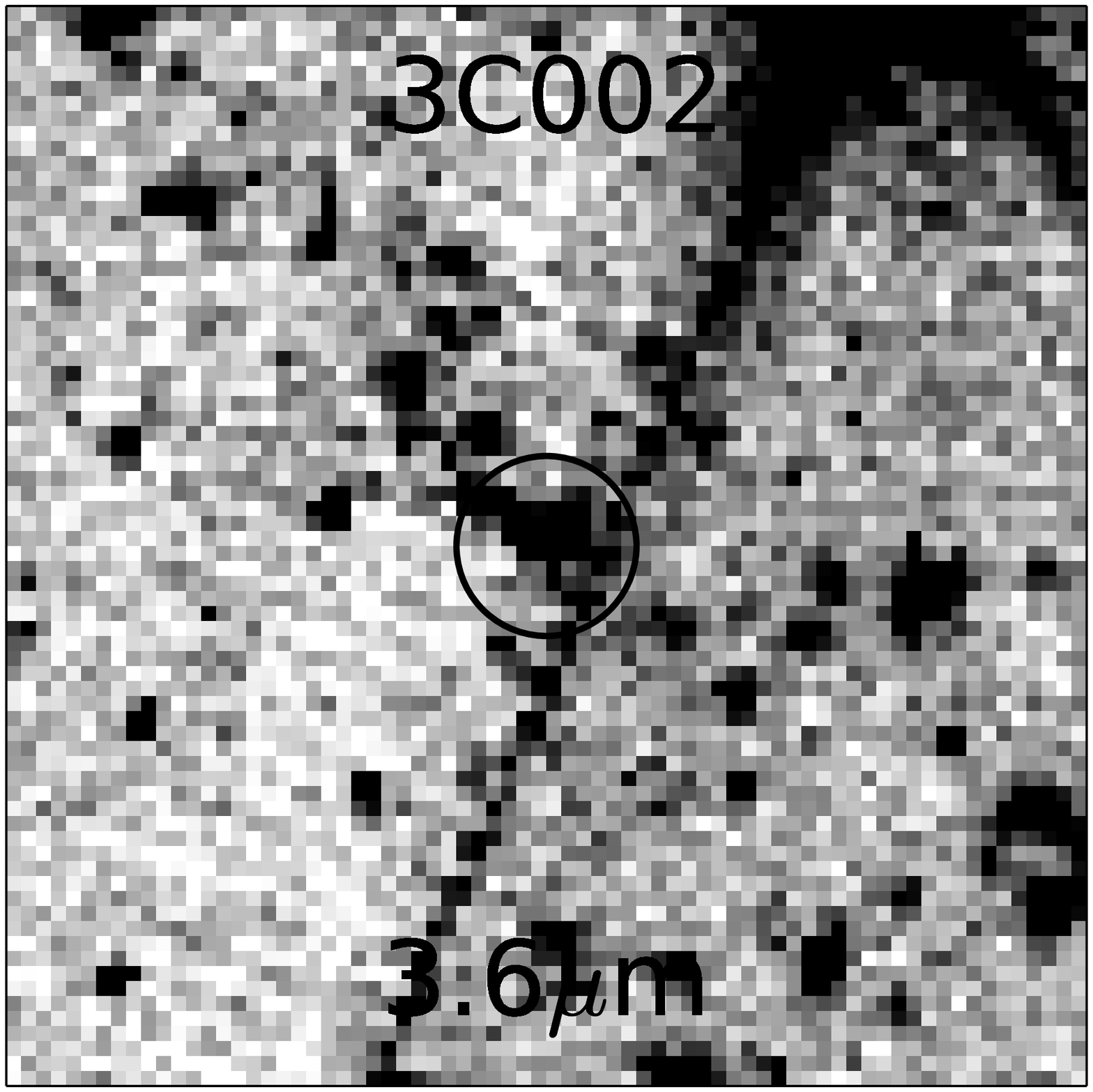}
      \includegraphics[width=1.5cm]{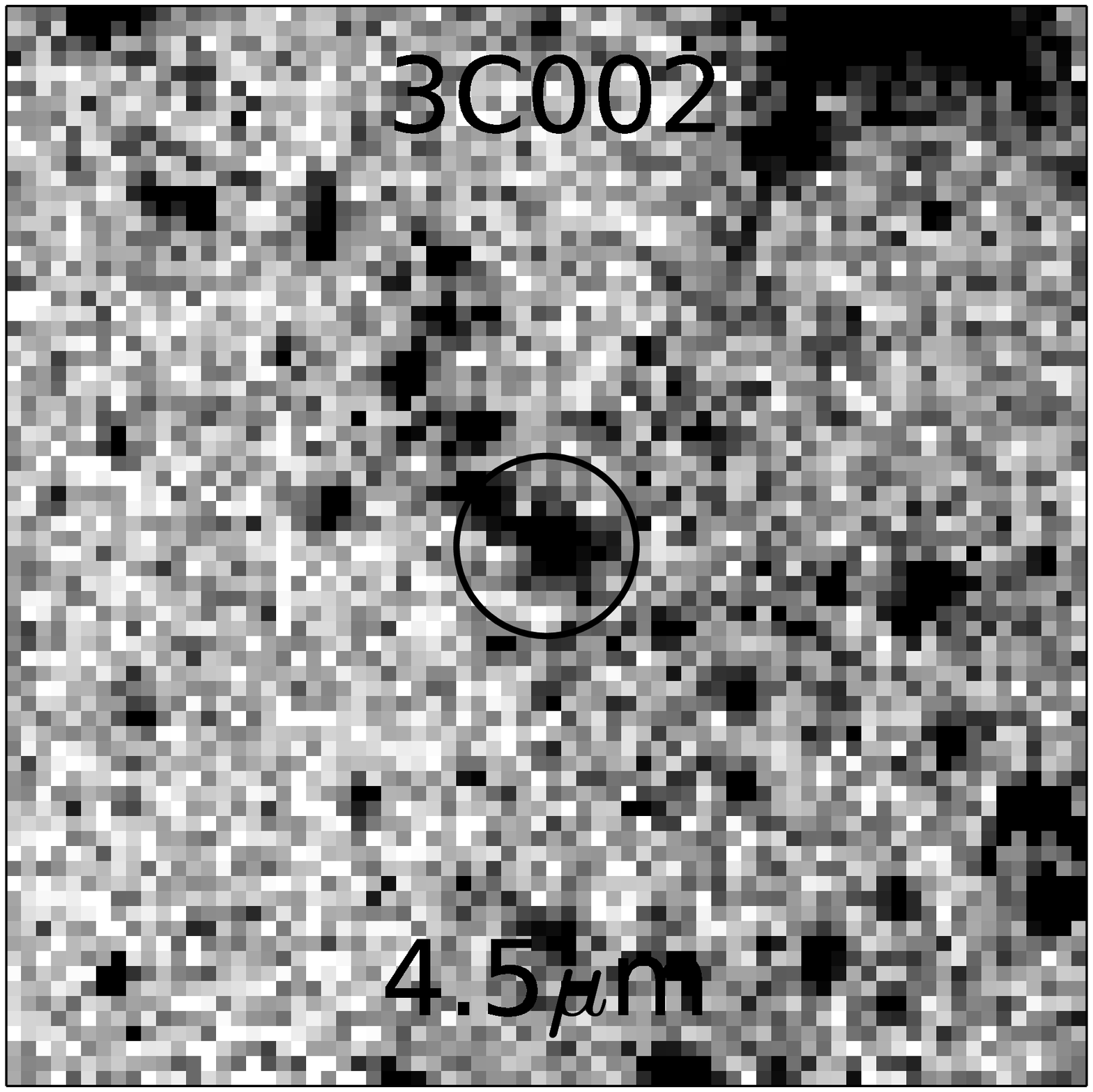}
      \includegraphics[width=1.5cm]{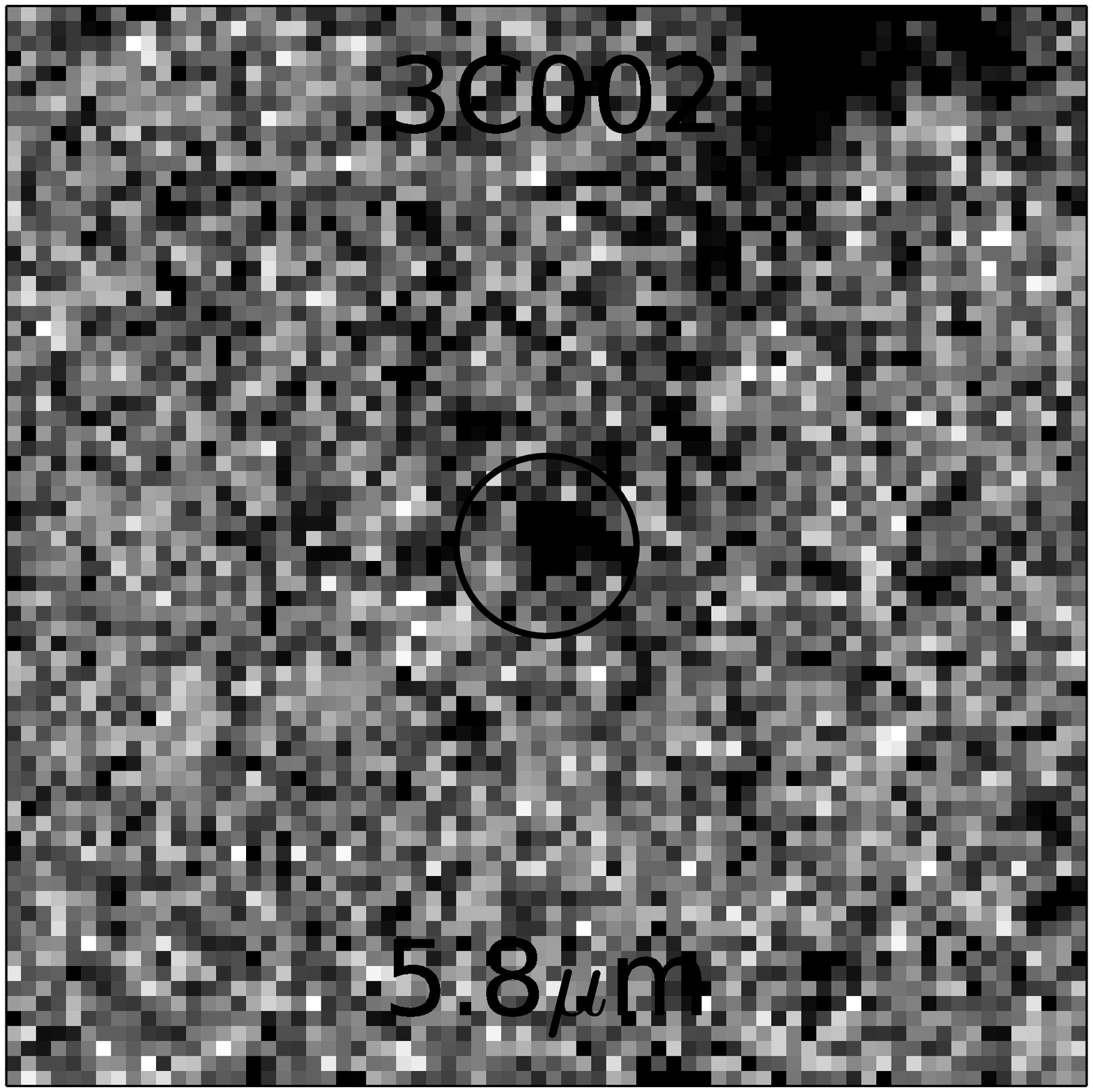}
      \includegraphics[width=1.5cm]{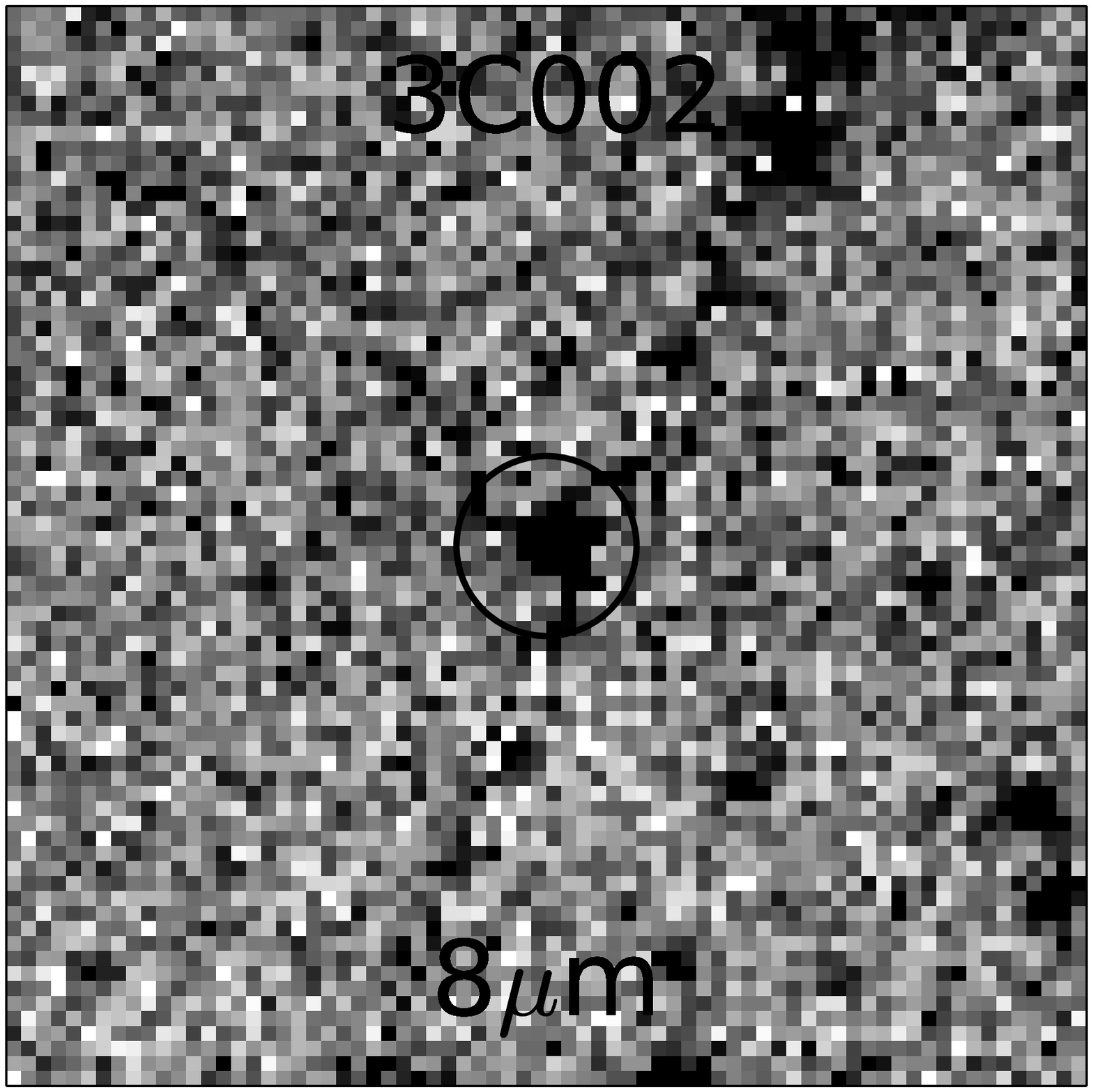}
      \includegraphics[width=1.5cm]{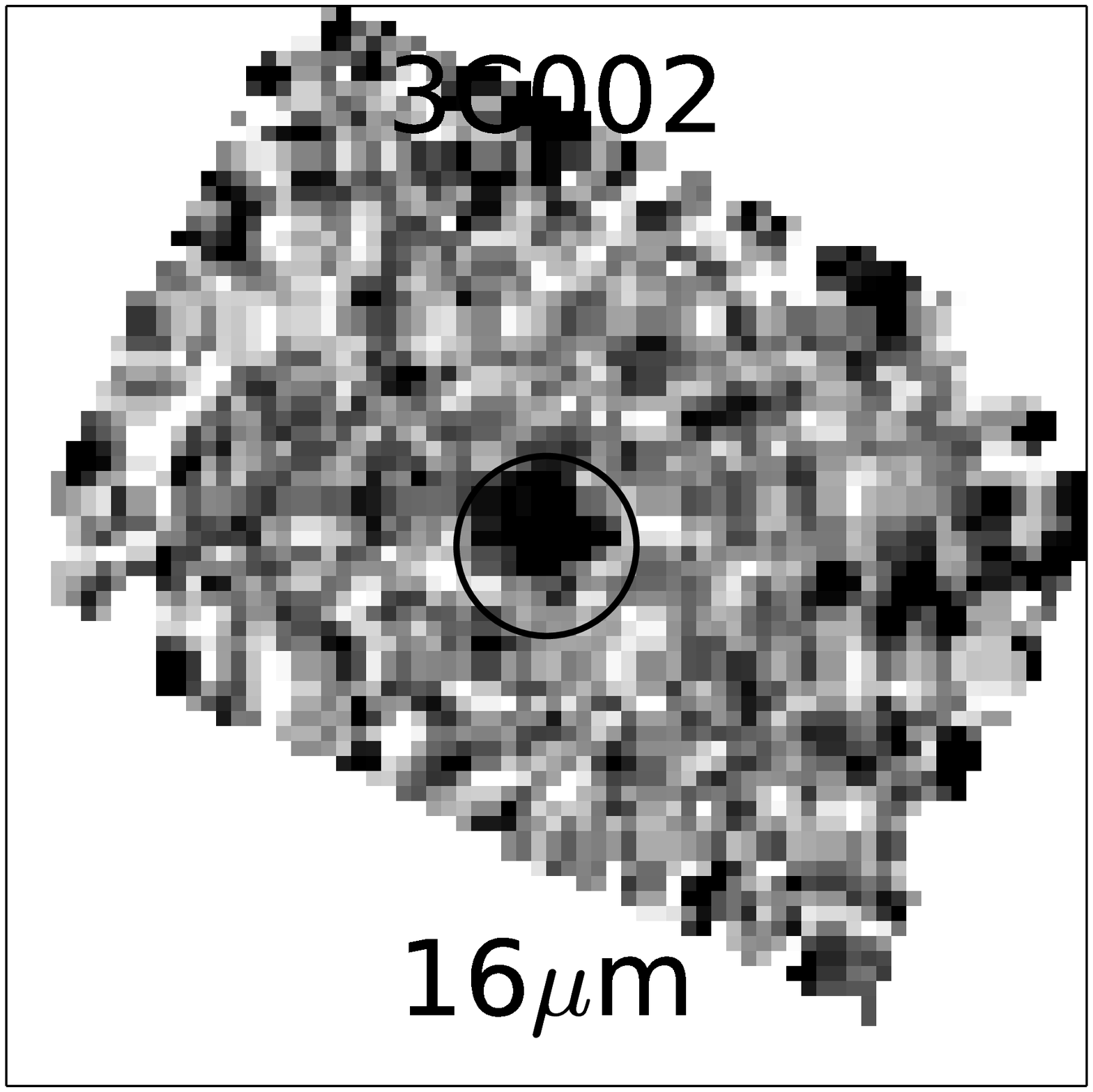}
      \includegraphics[width=1.5cm]{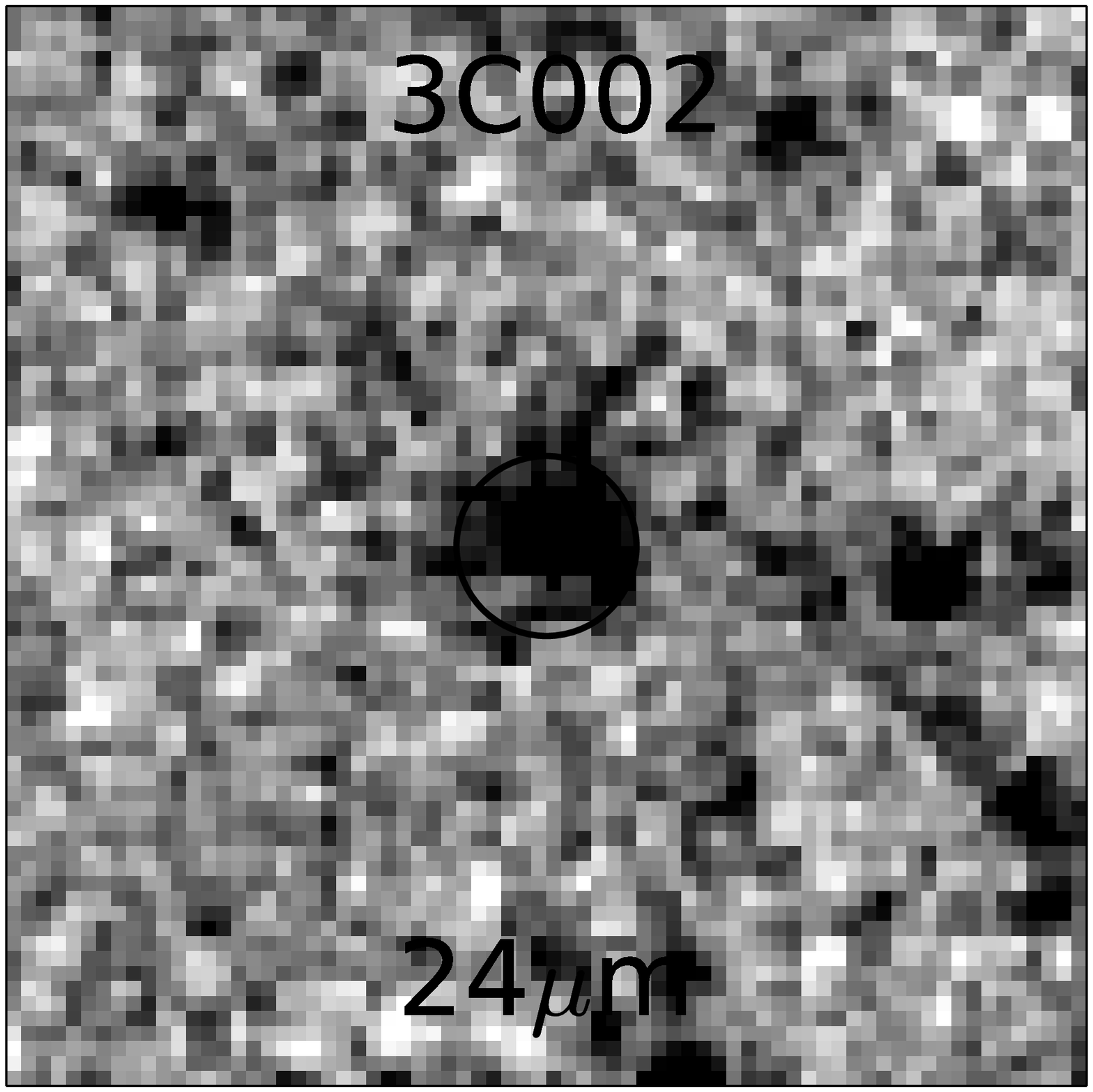}
      \includegraphics[width=1.5cm]{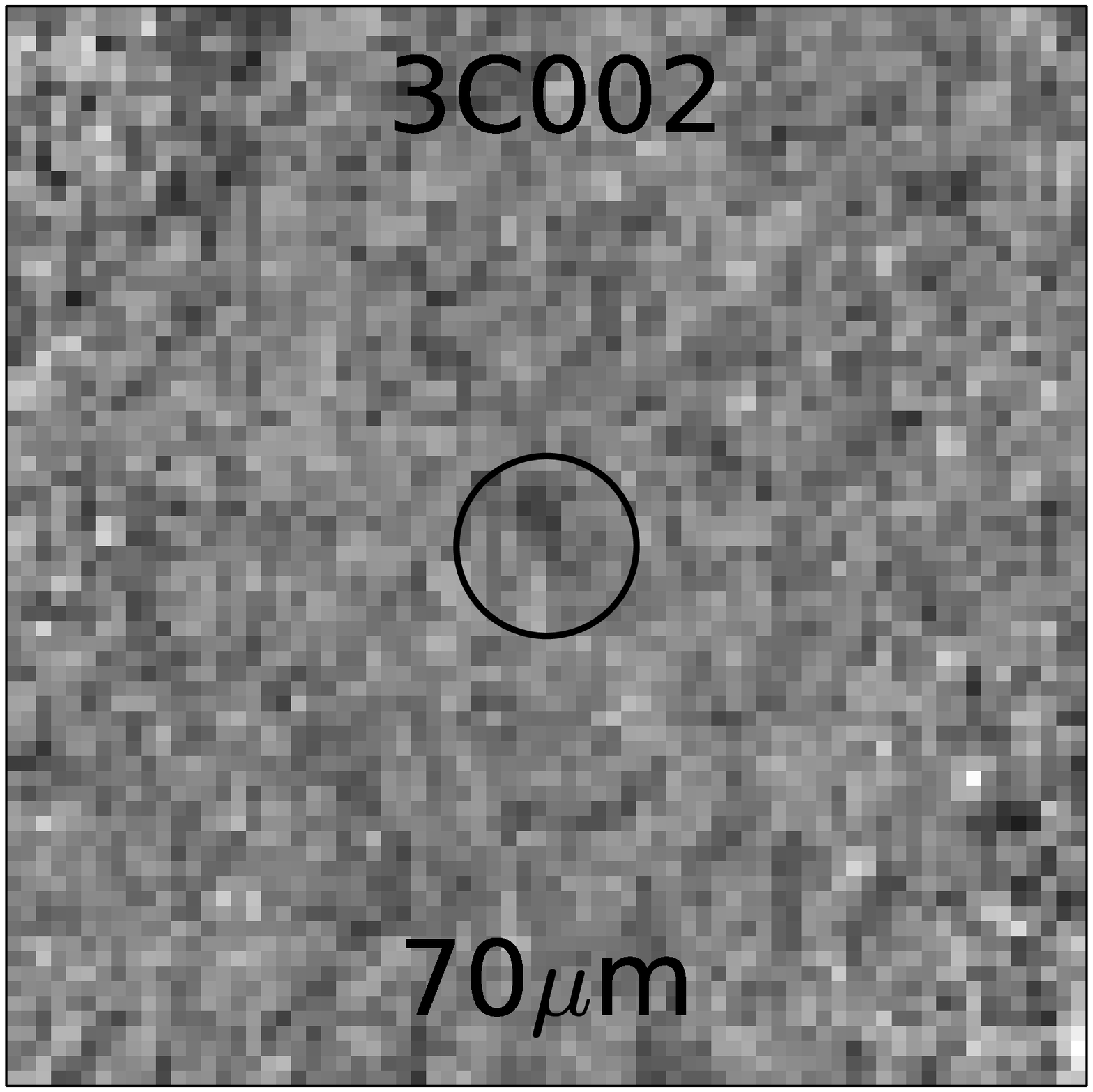}
      \includegraphics[width=1.5cm]{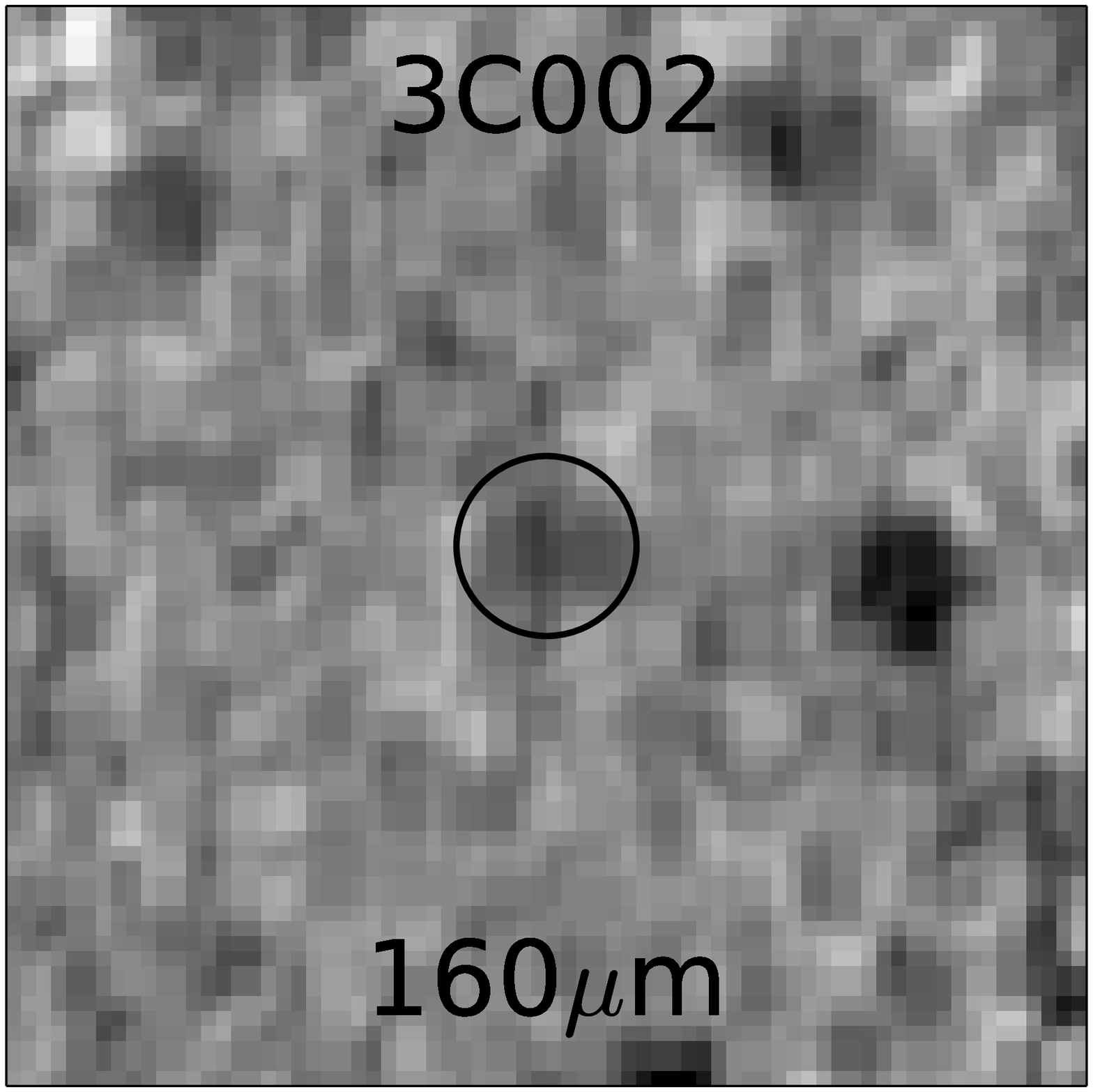}
      \includegraphics[width=1.5cm]{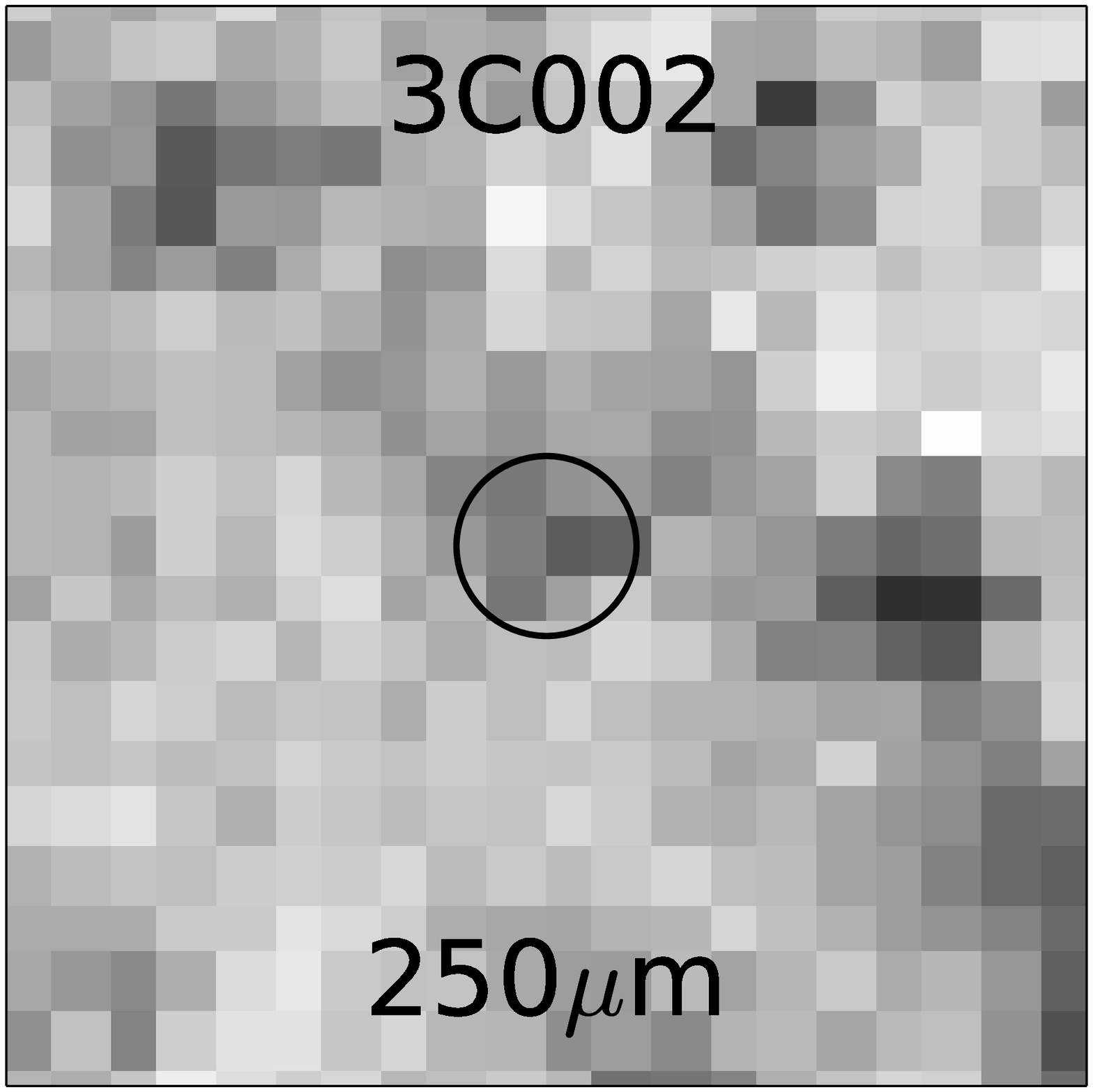}
      \includegraphics[width=1.5cm]{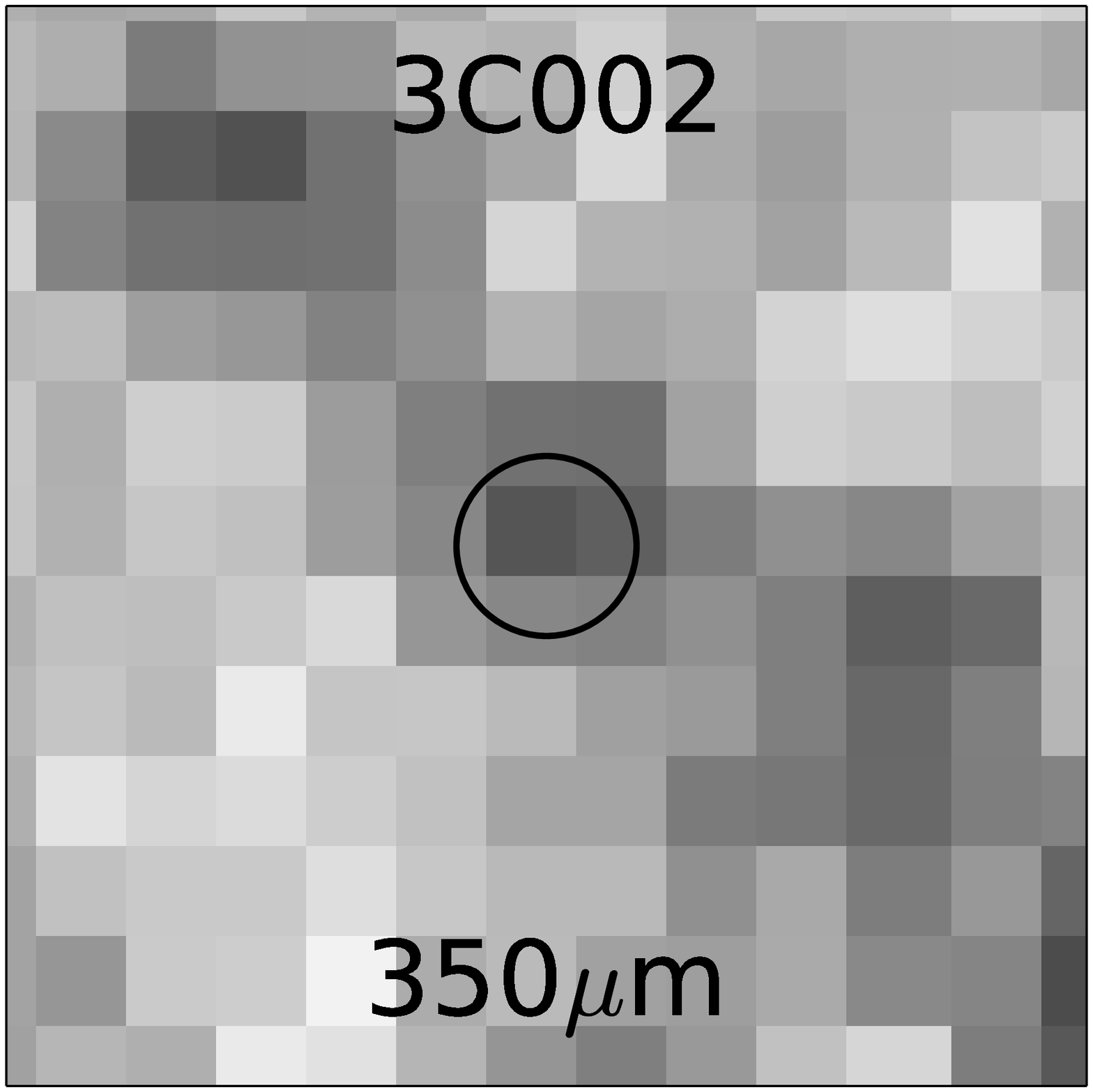}
      \includegraphics[width=1.5cm]{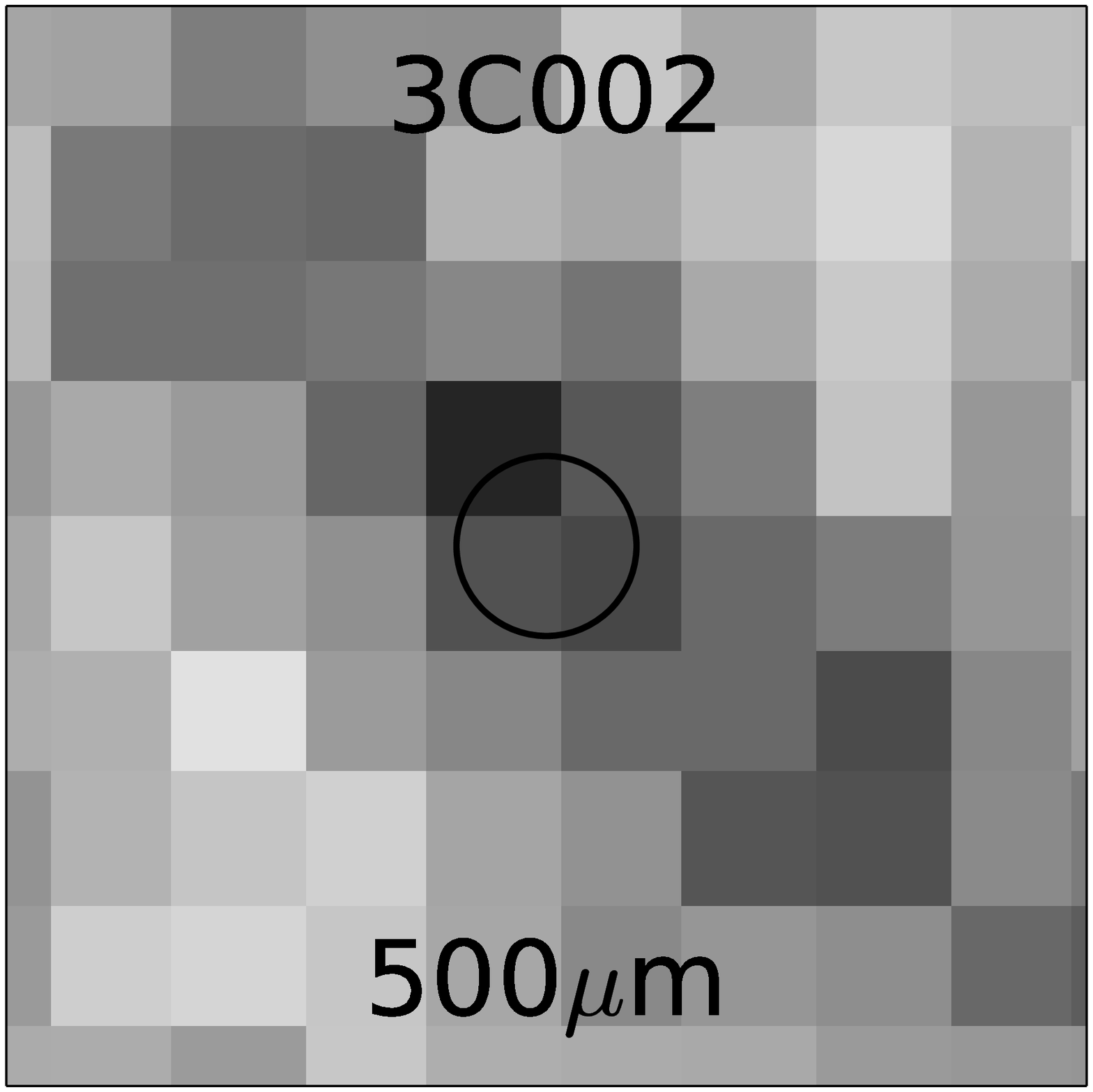}
      \\
      \includegraphics[width=1.5cm]{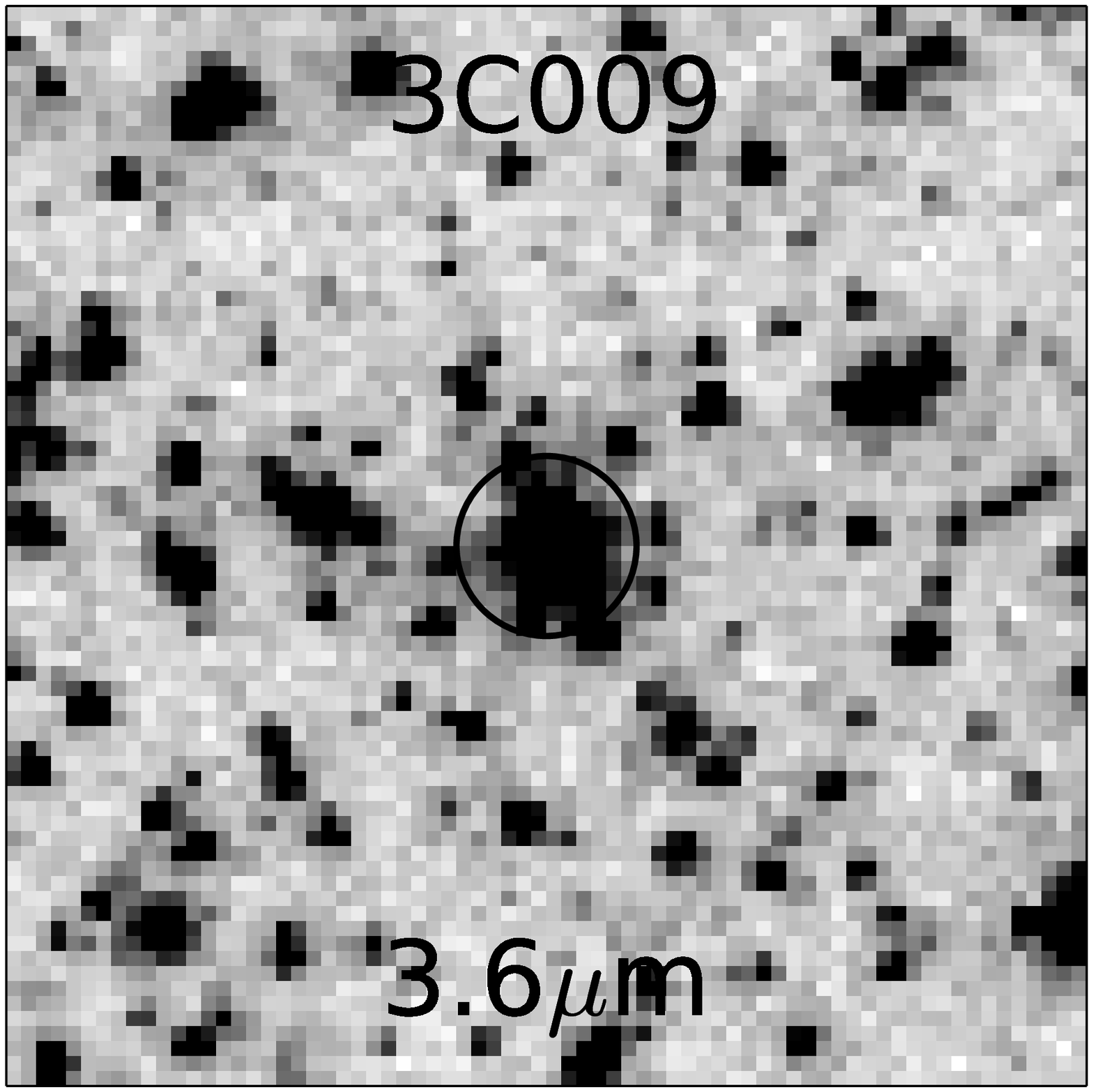}
      \includegraphics[width=1.5cm]{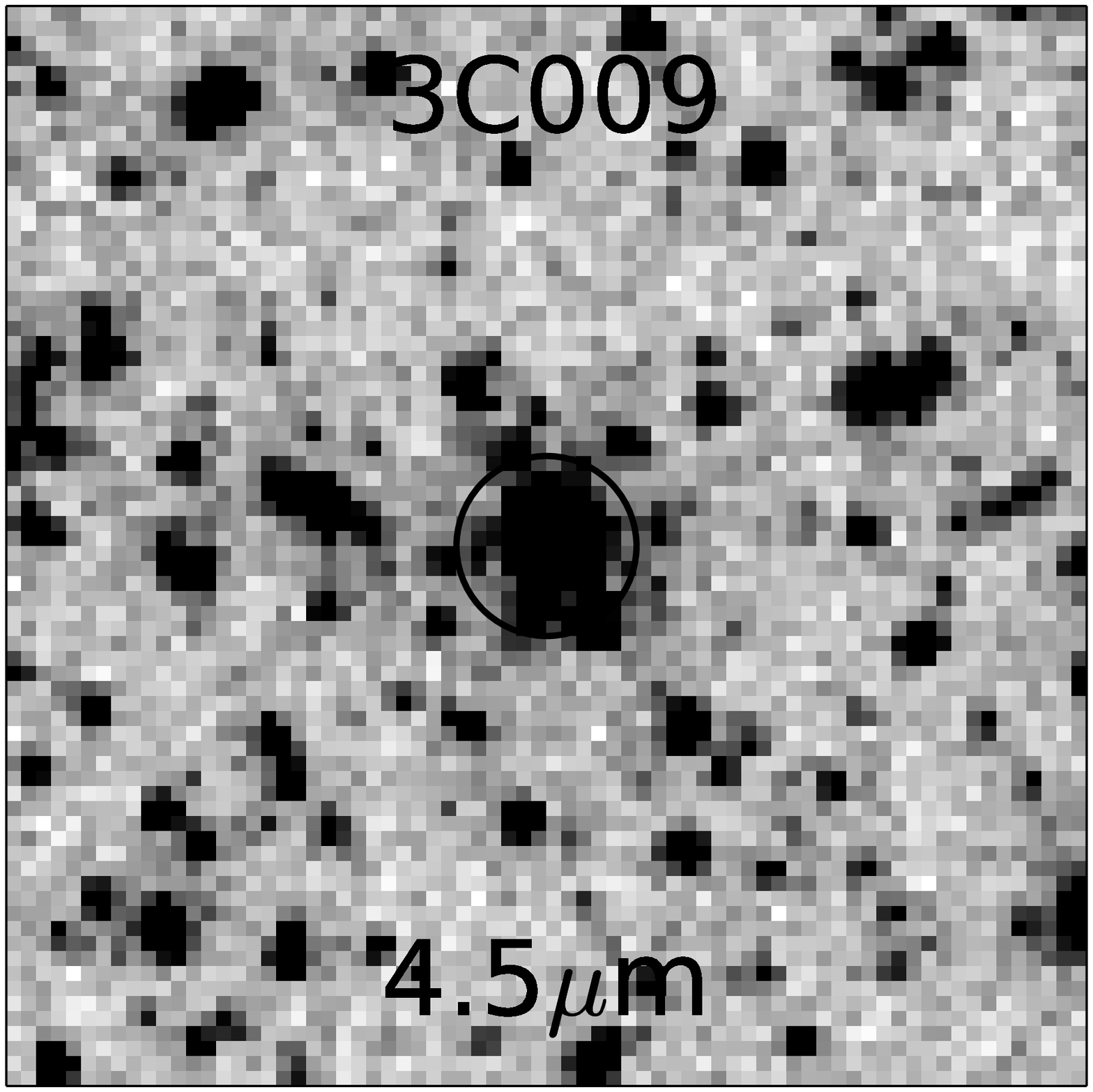}
      \includegraphics[width=1.5cm]{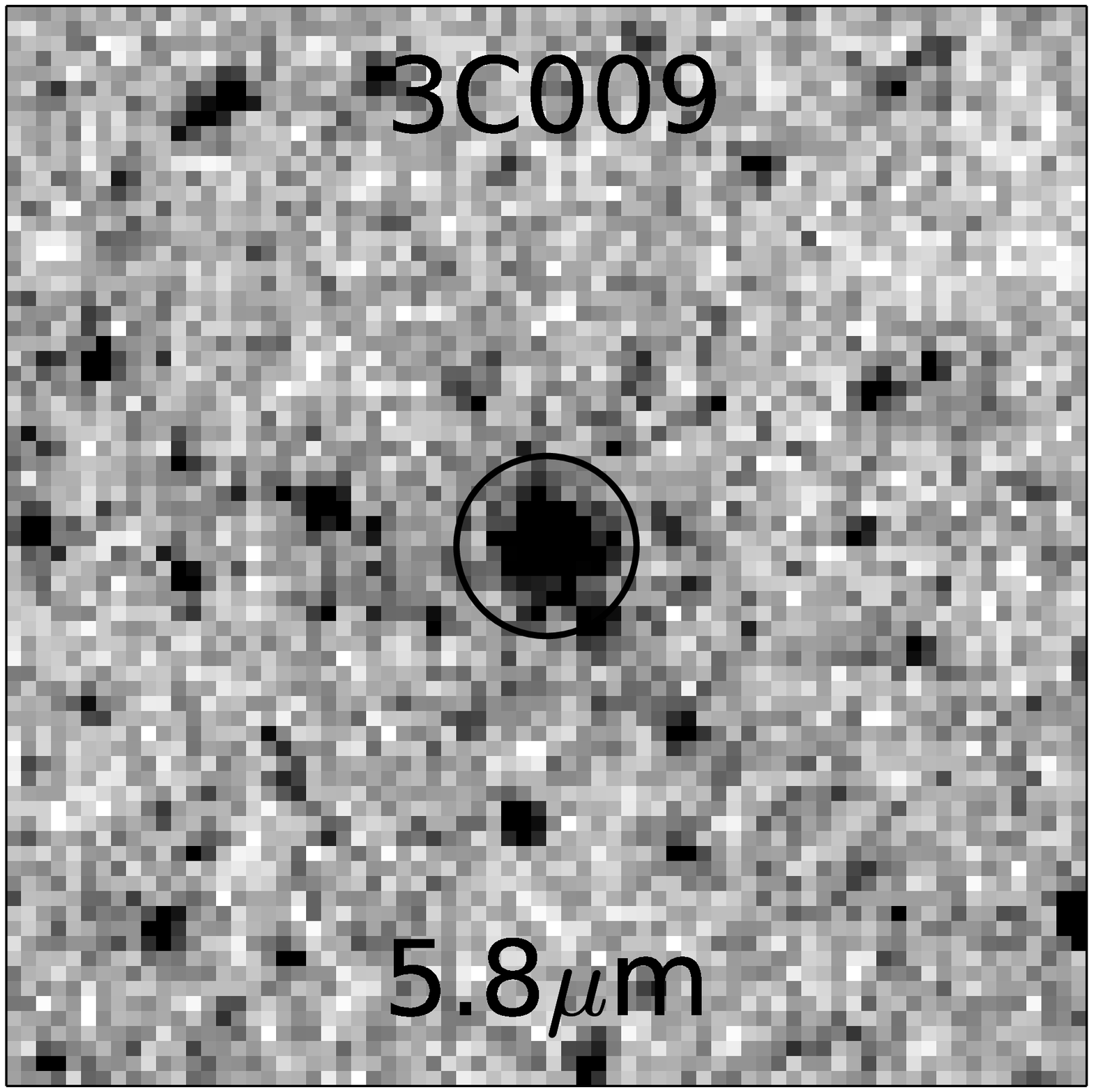}
      \includegraphics[width=1.5cm]{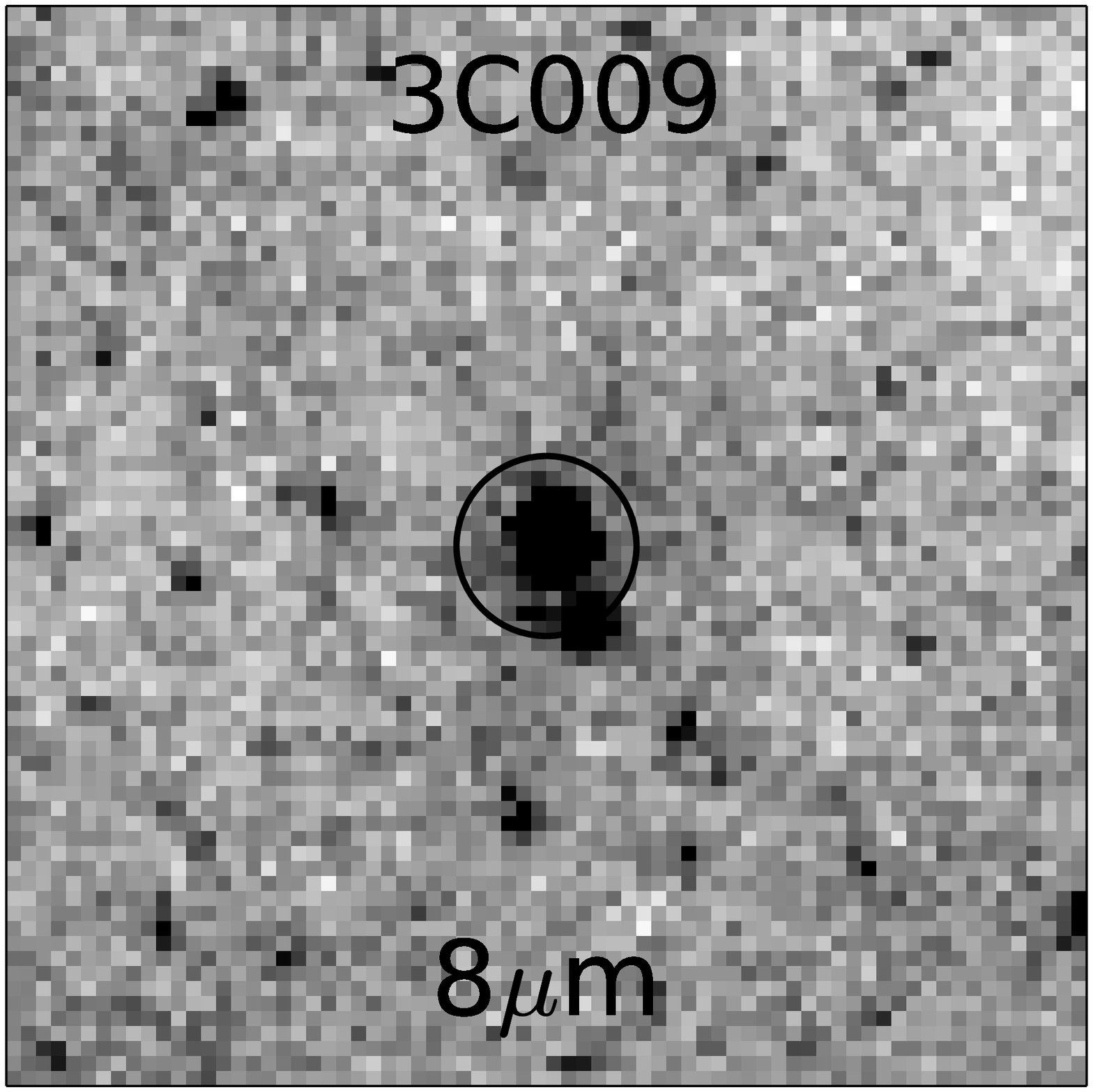}
      \includegraphics[width=1.5cm]{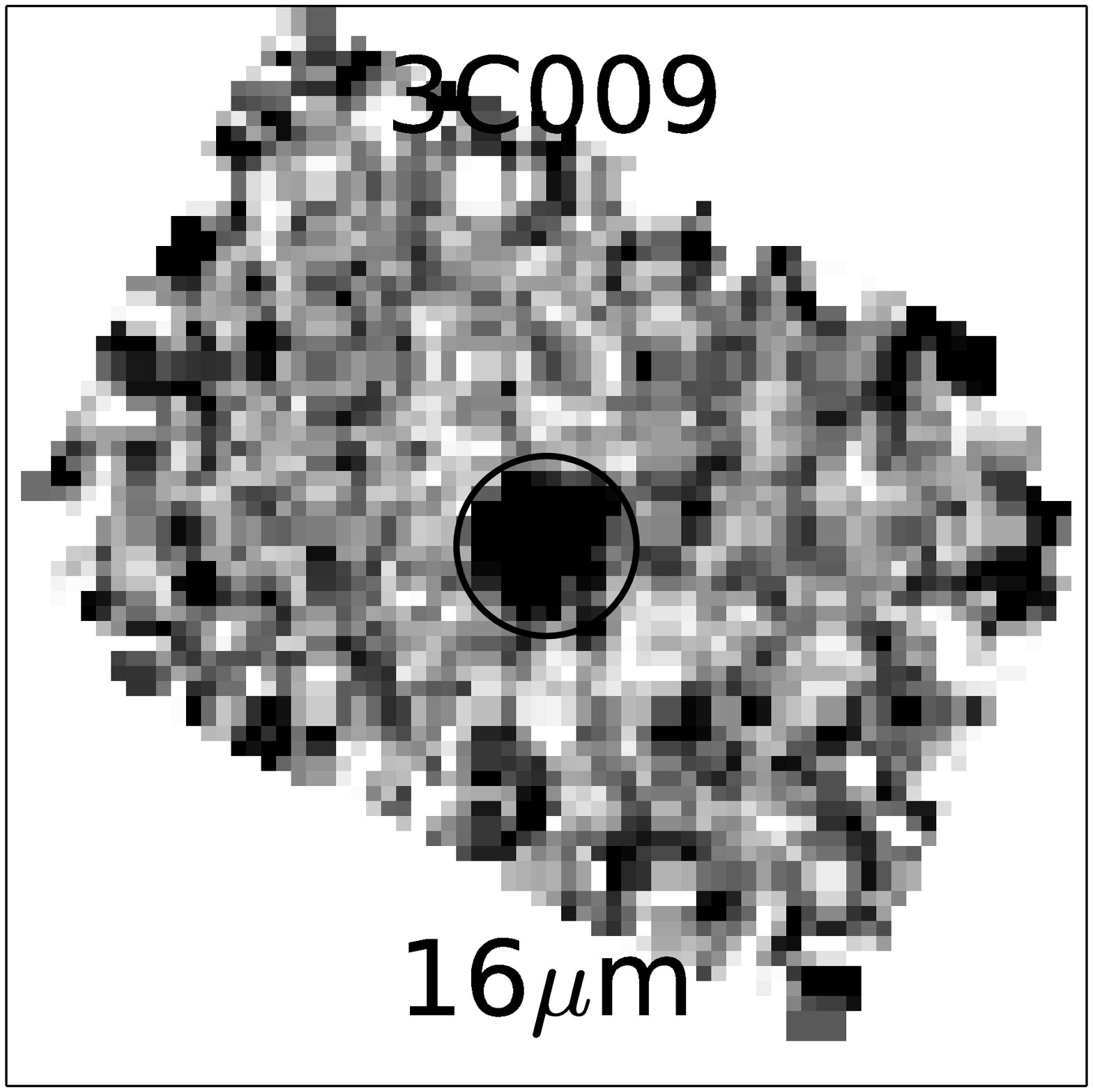}
      \includegraphics[width=1.5cm]{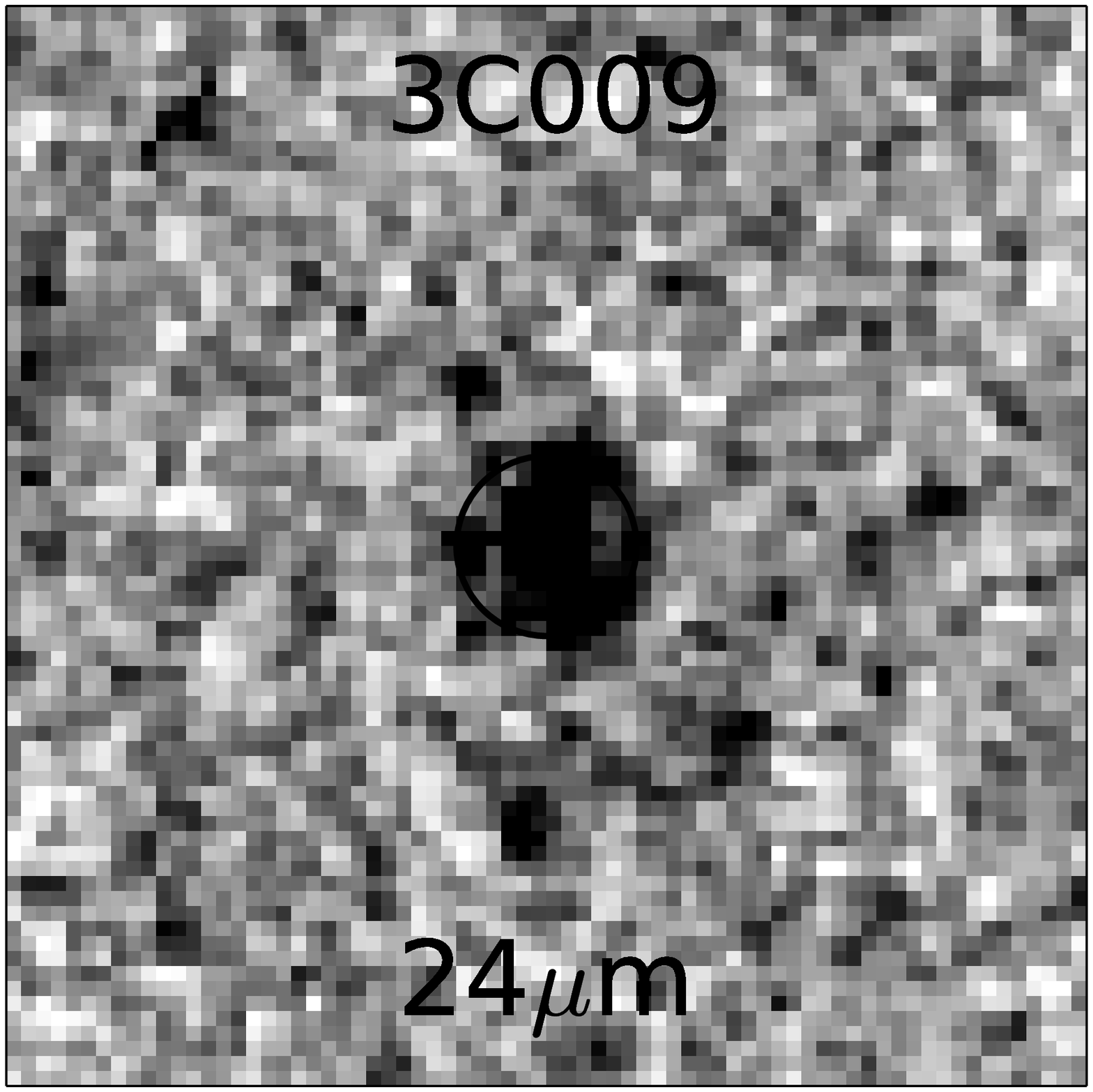}
      \includegraphics[width=1.5cm]{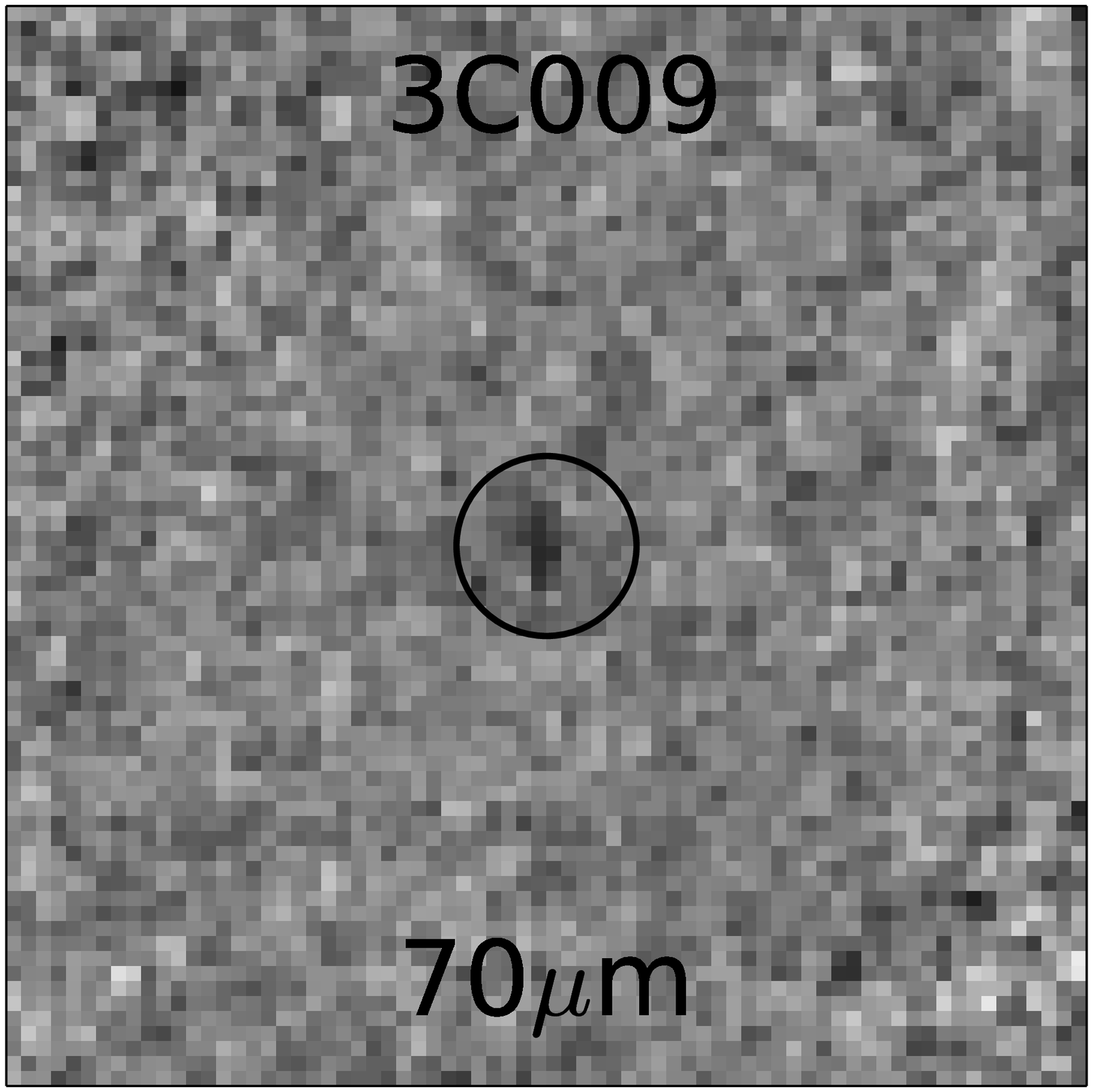}
      \includegraphics[width=1.5cm]{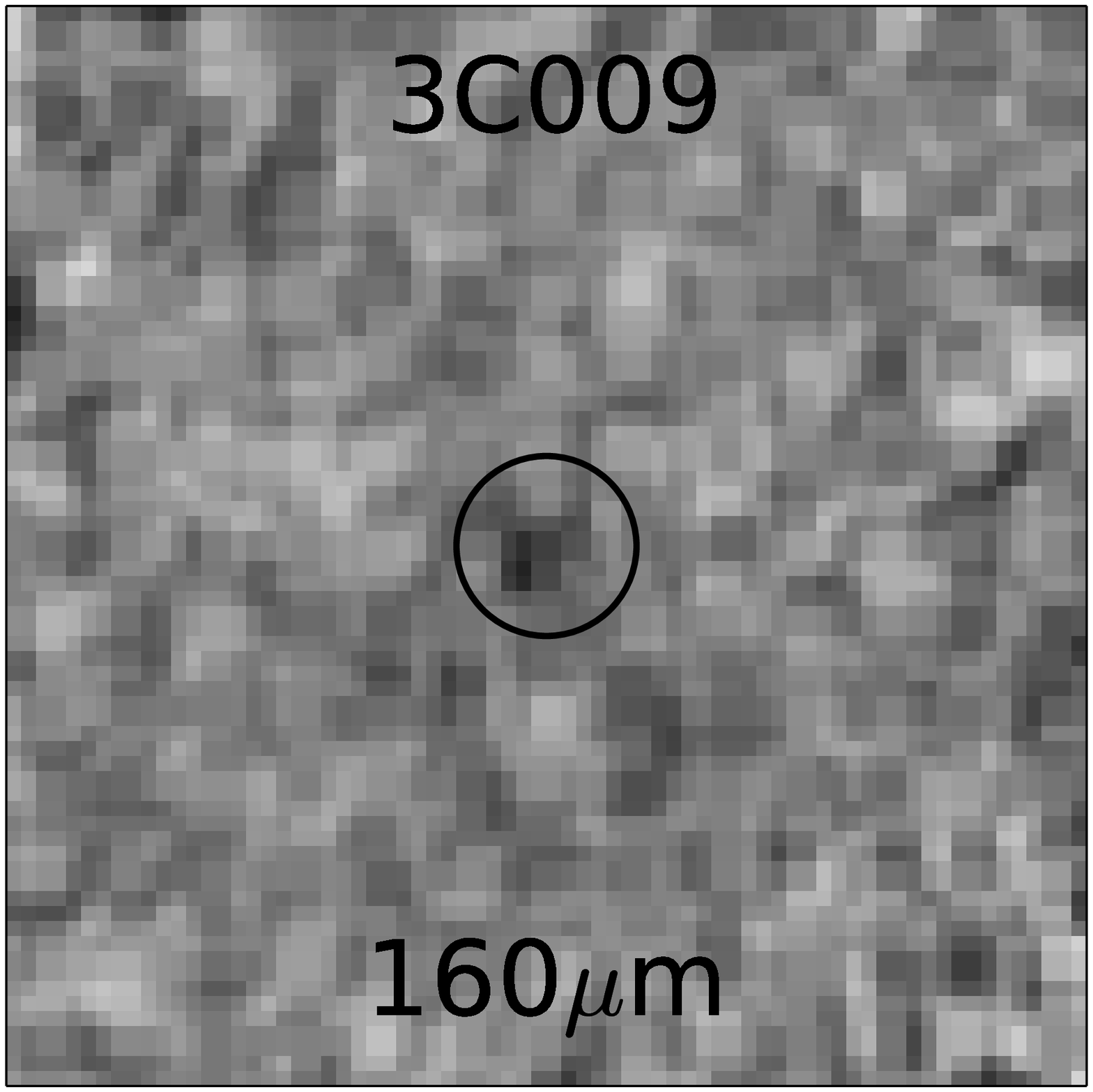}
      \includegraphics[width=1.5cm]{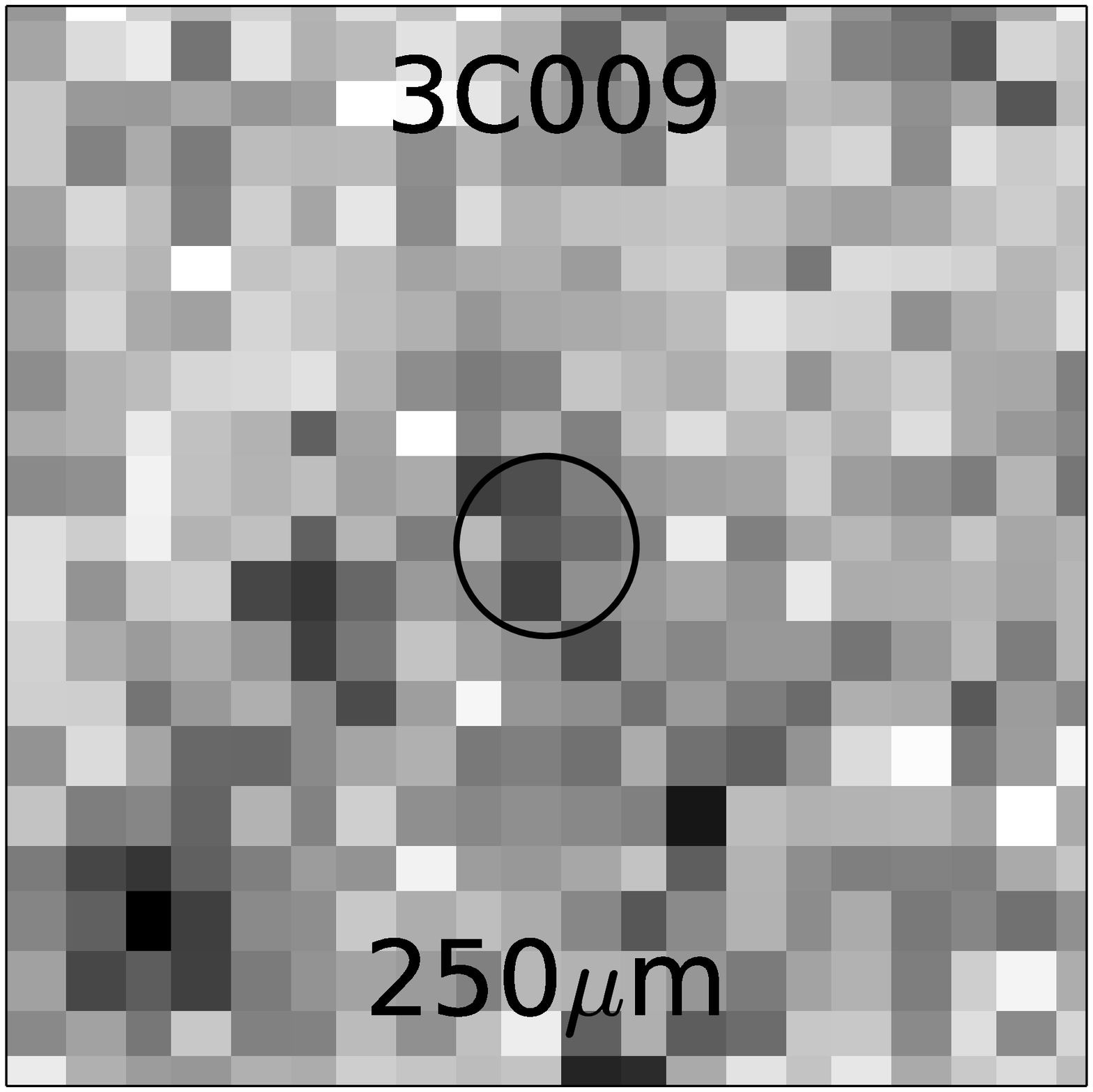}
      \includegraphics[width=1.5cm]{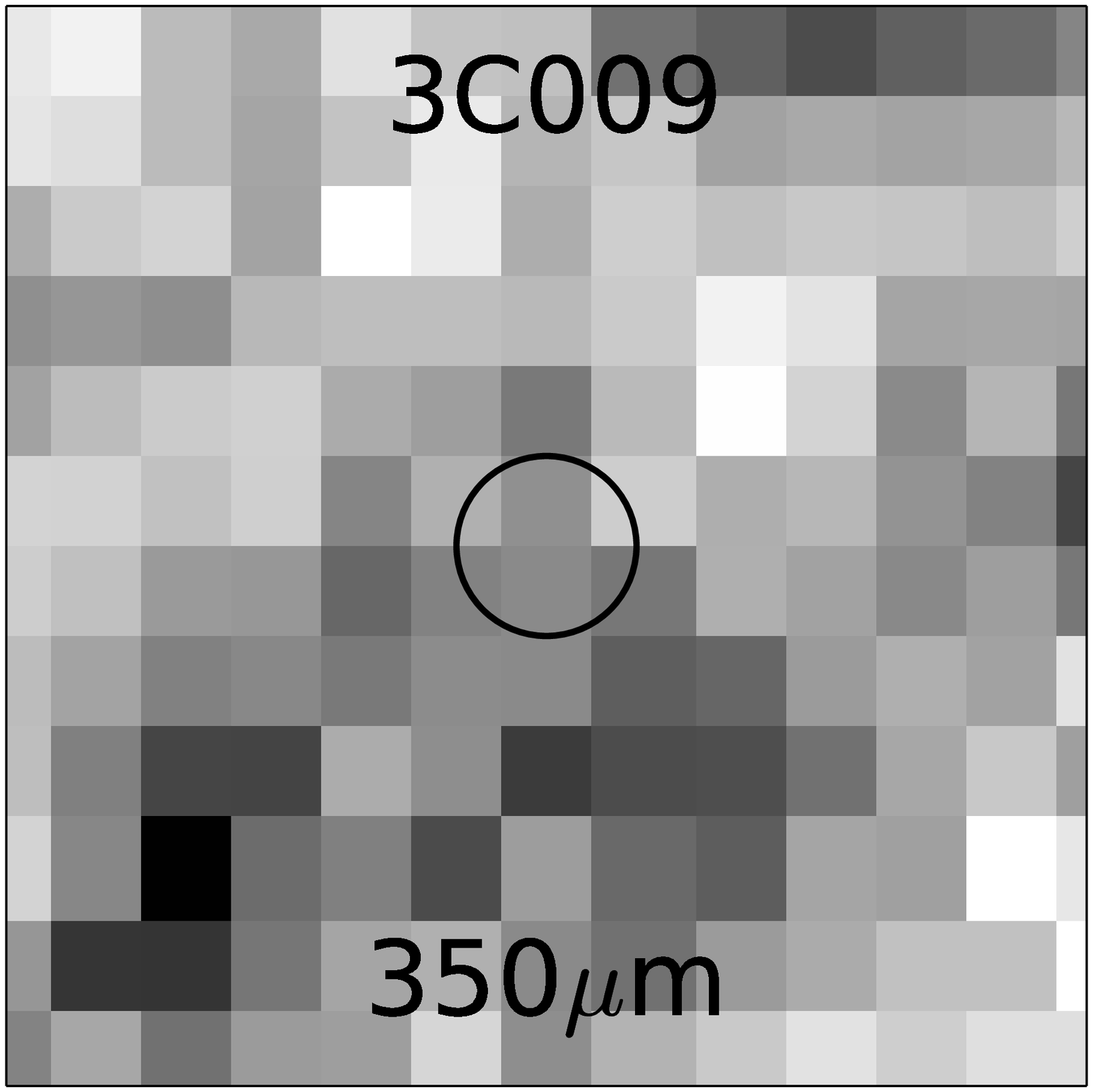}
      \includegraphics[width=1.5cm]{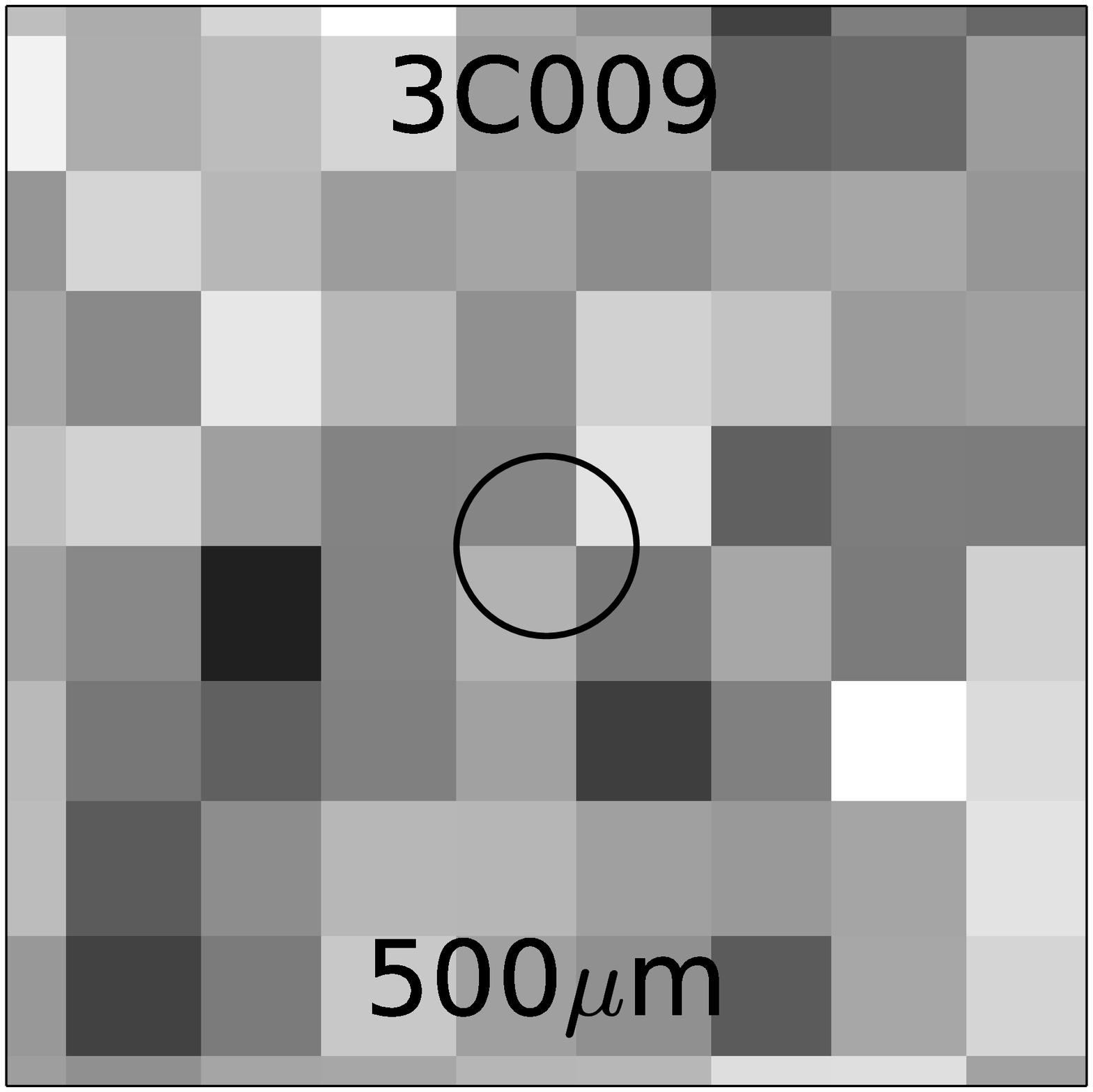}
      \\
      \includegraphics[width=1.5cm]{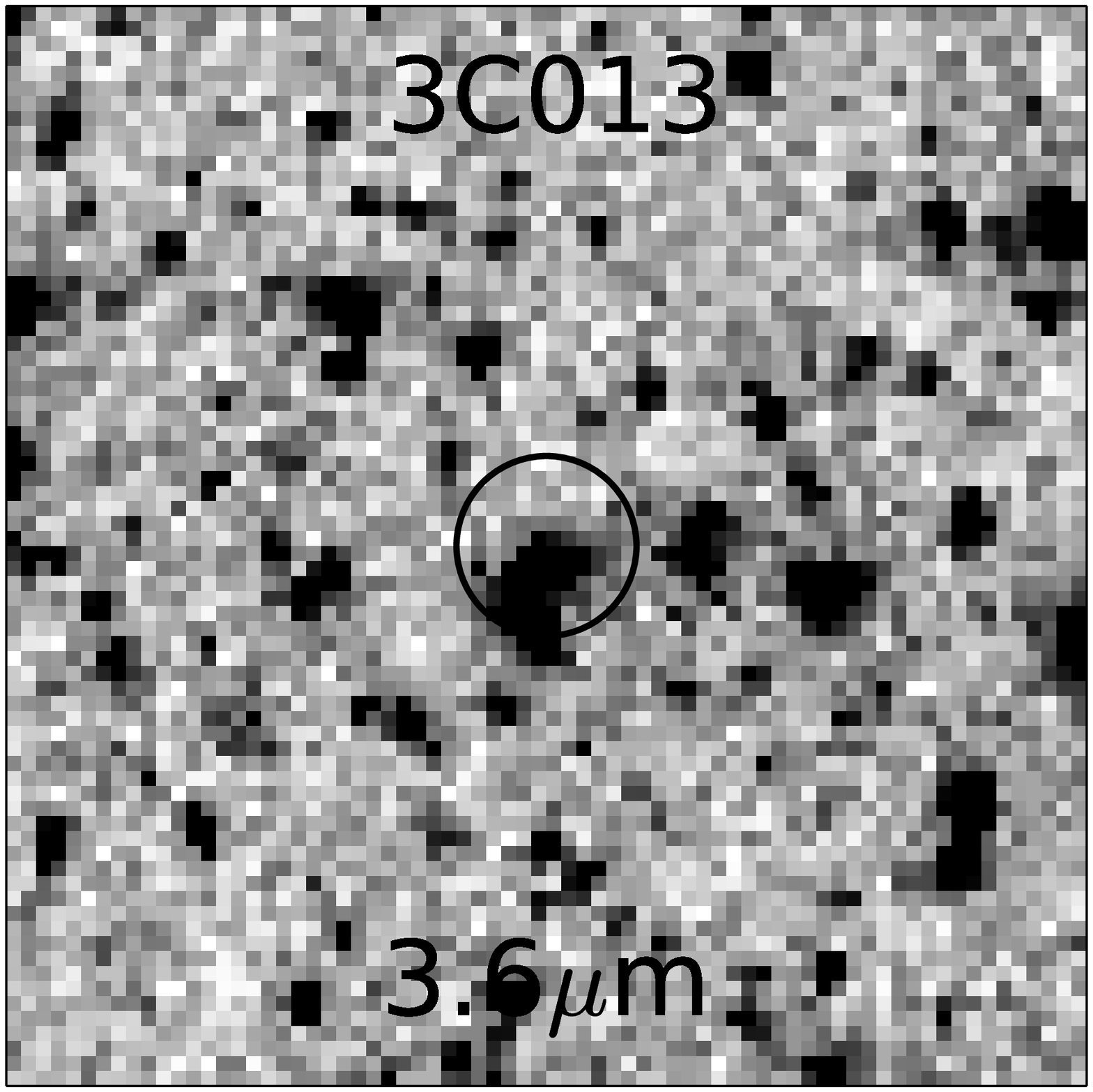}
      \includegraphics[width=1.5cm]{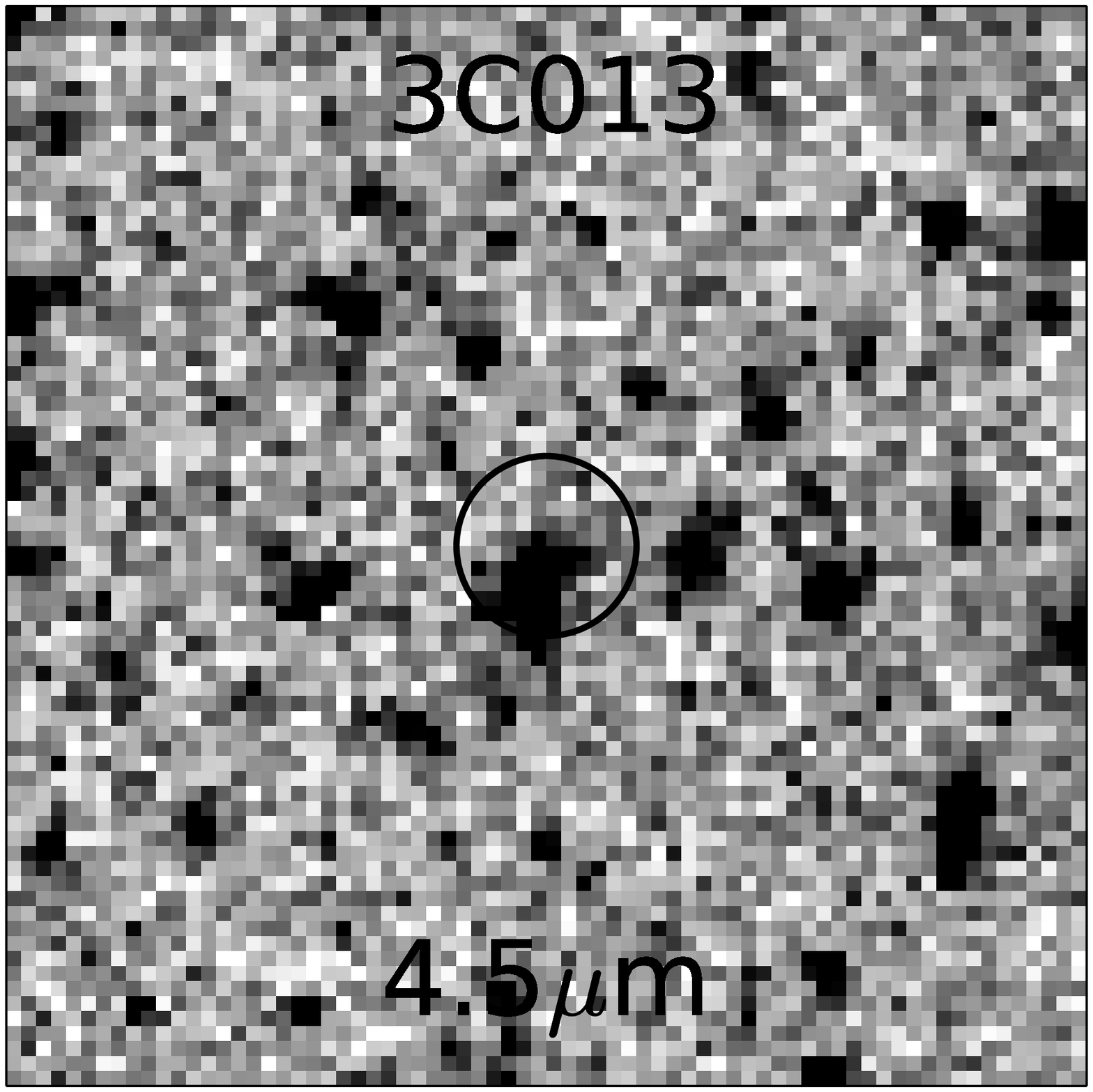}
      \includegraphics[width=1.5cm]{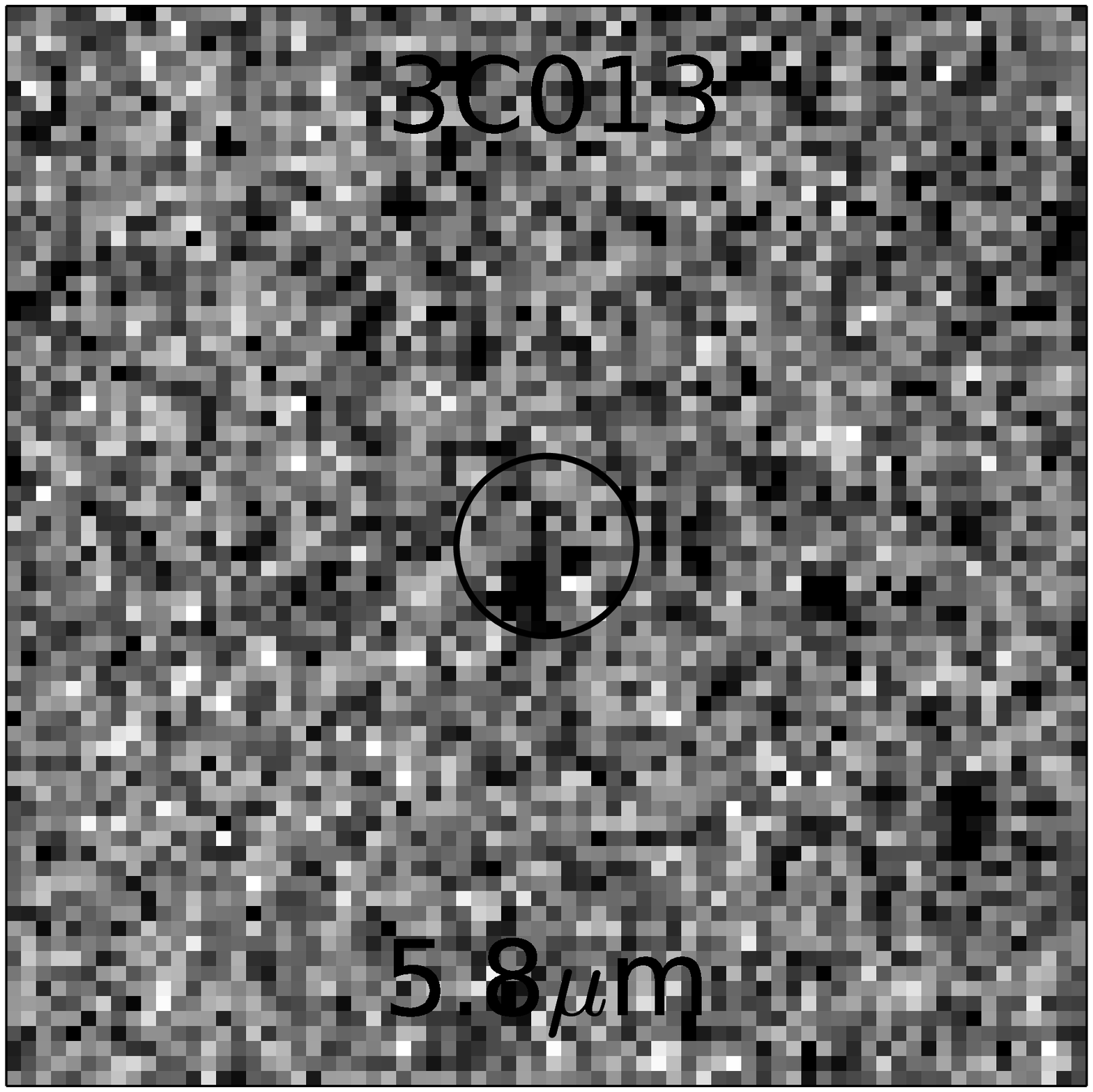}
      \includegraphics[width=1.5cm]{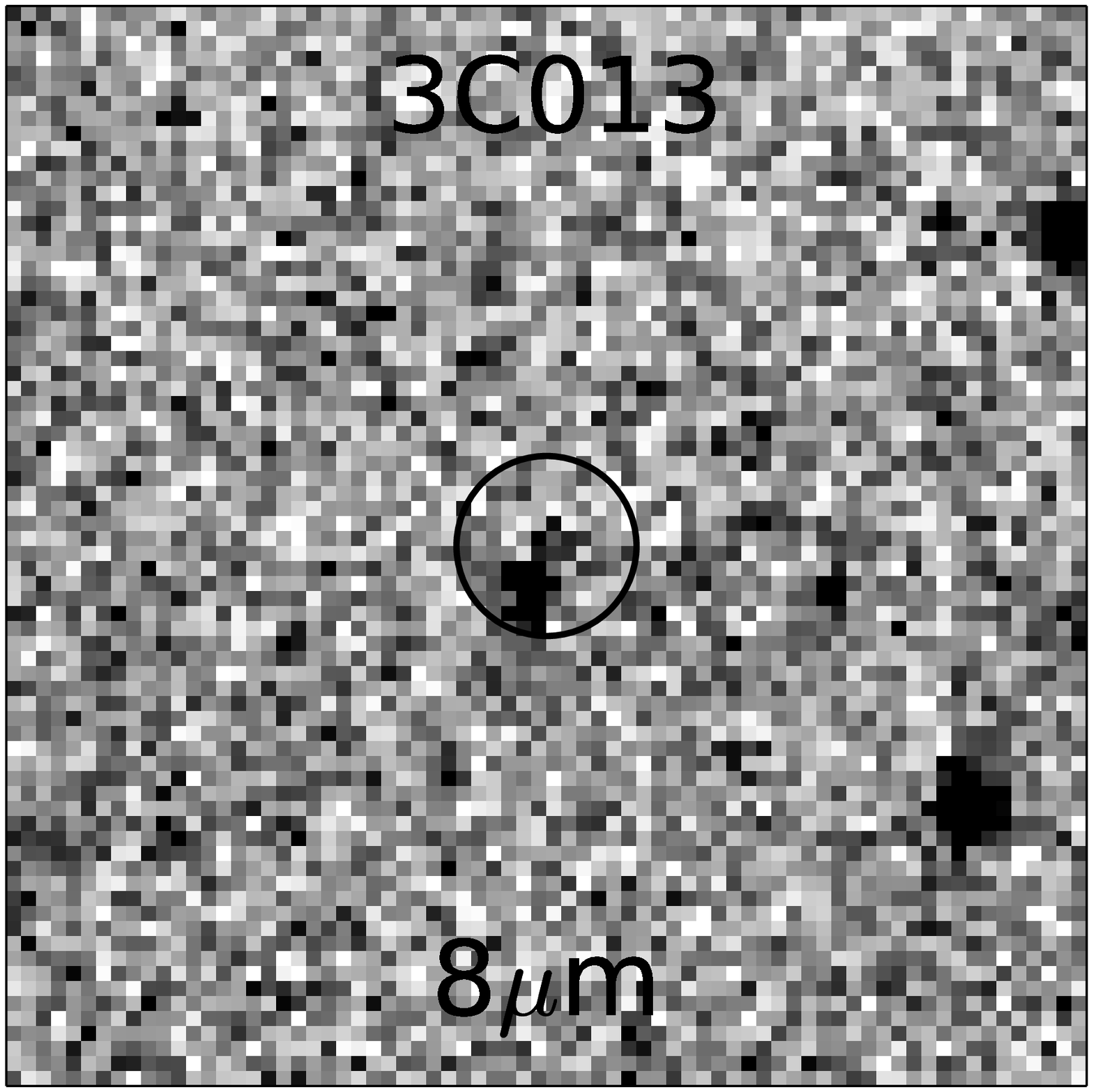}
      \includegraphics[width=1.5cm]{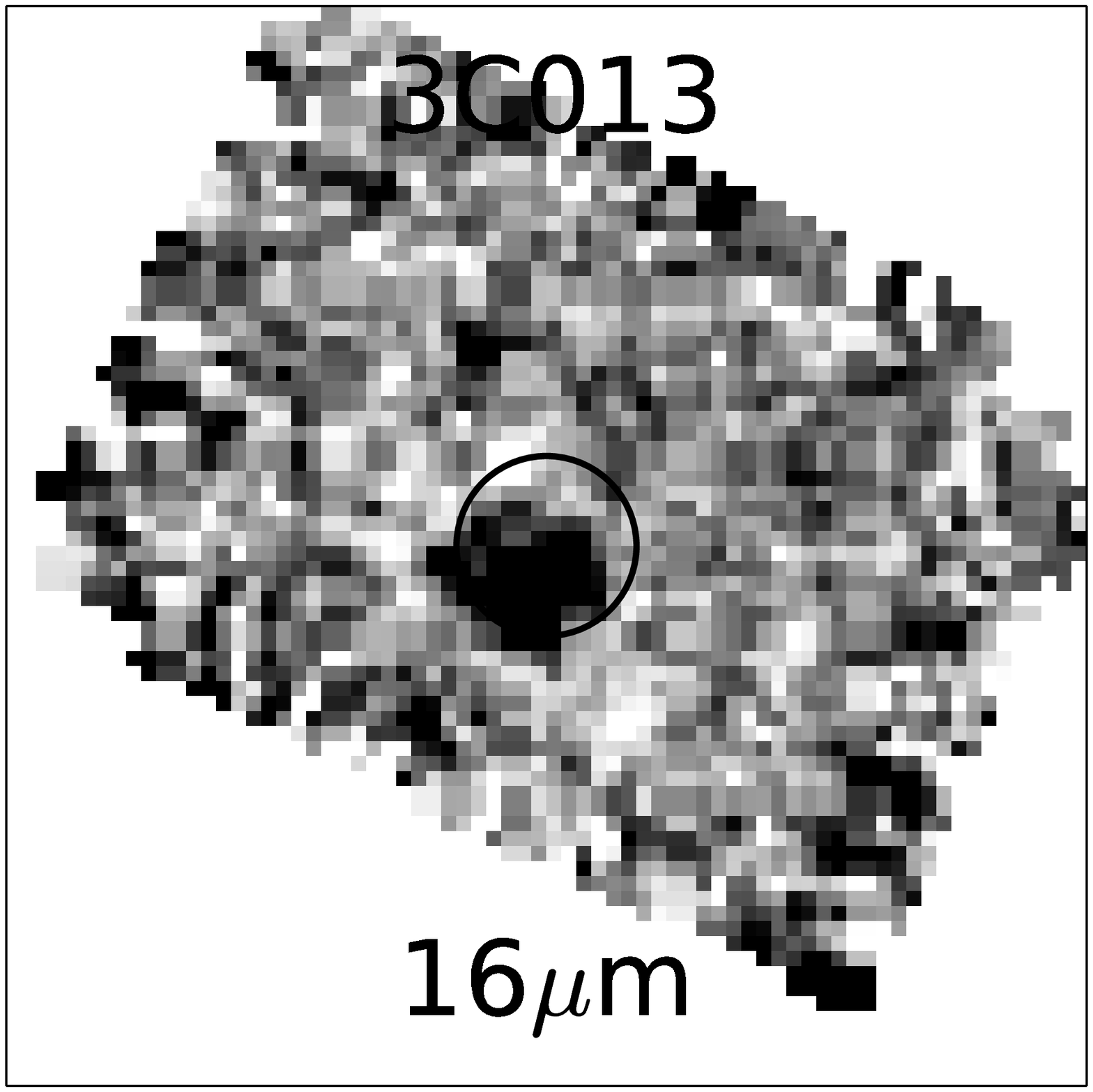}
      \includegraphics[width=1.5cm]{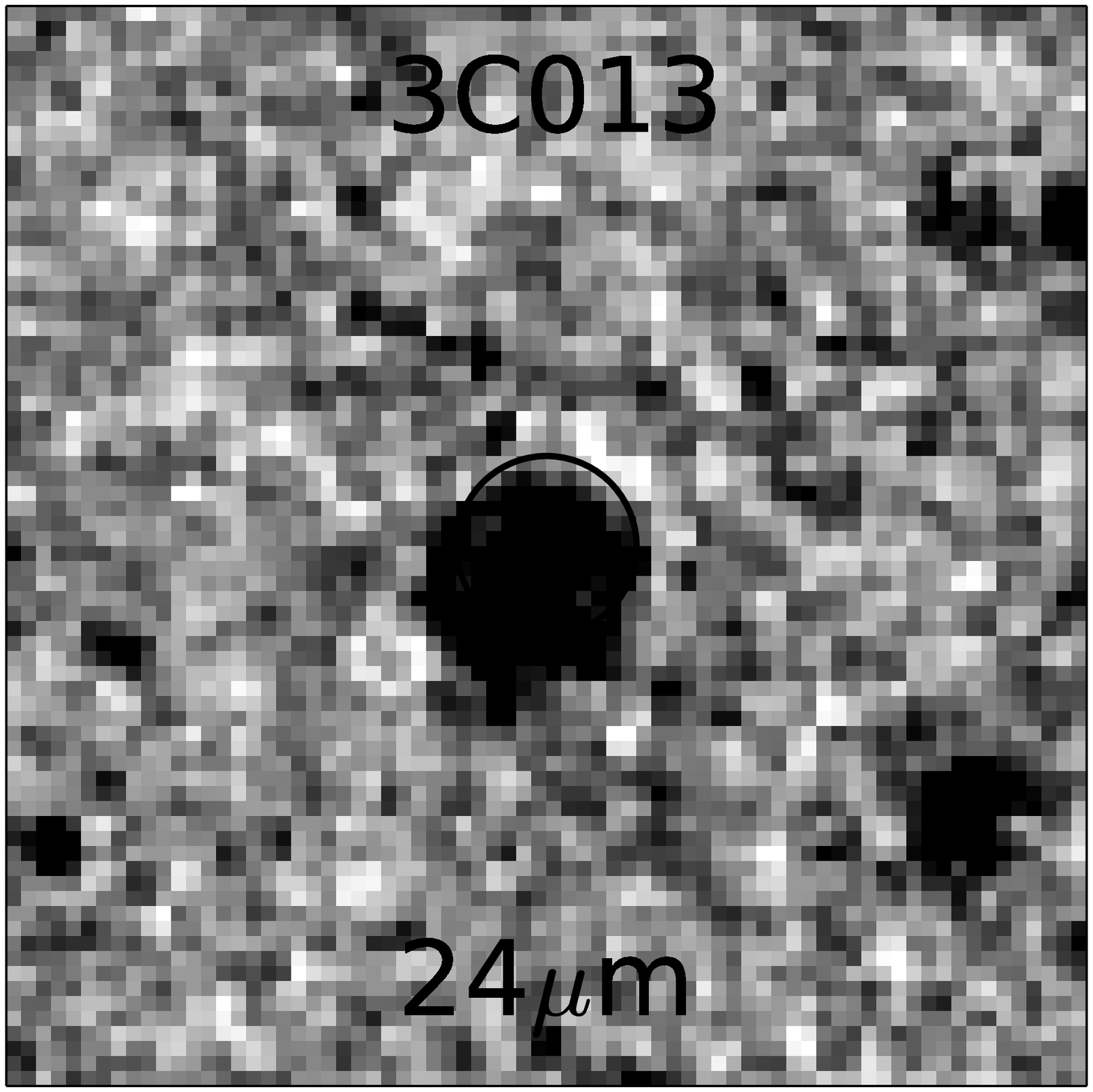}
      \includegraphics[width=1.5cm]{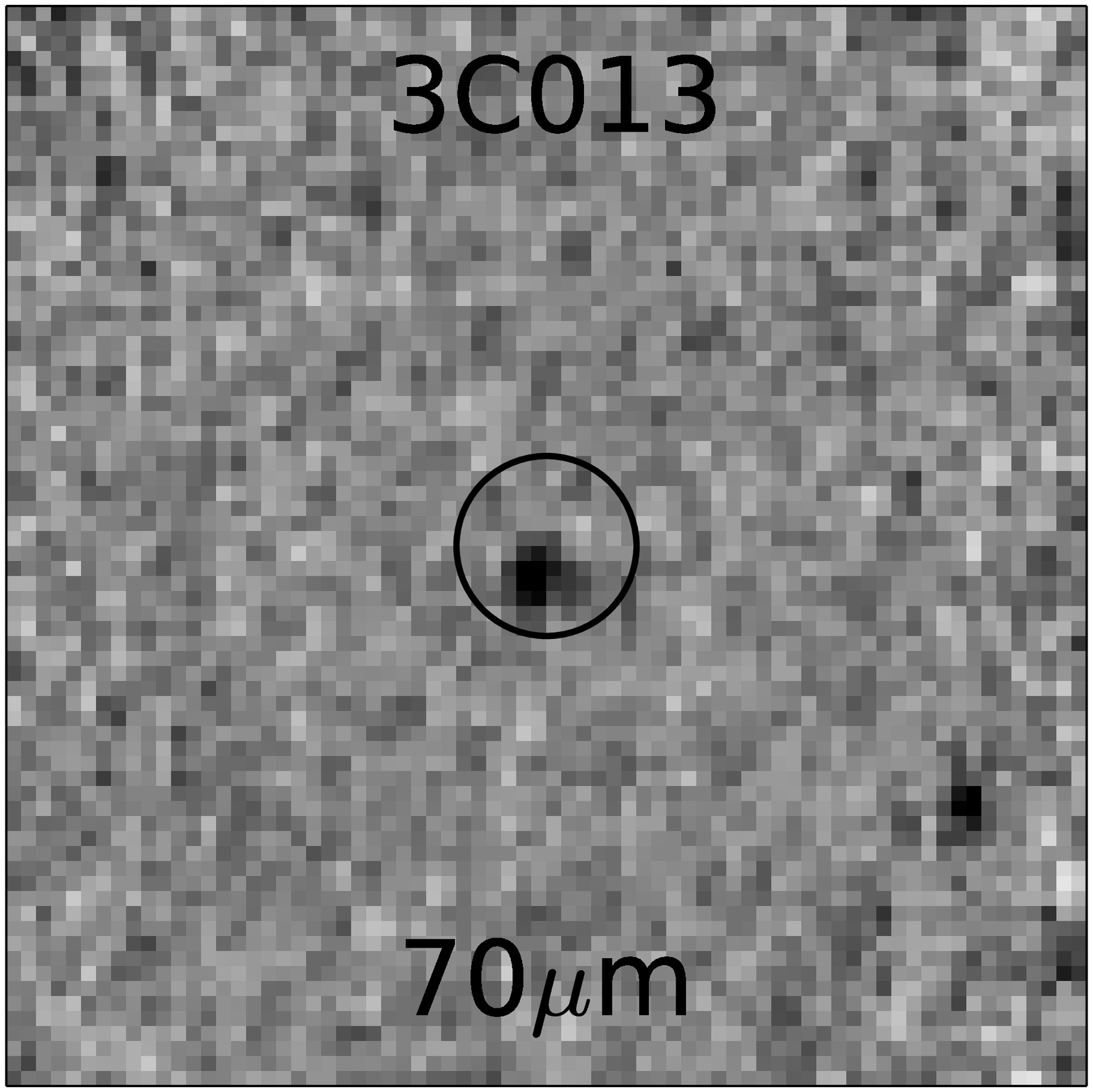}
      \includegraphics[width=1.5cm]{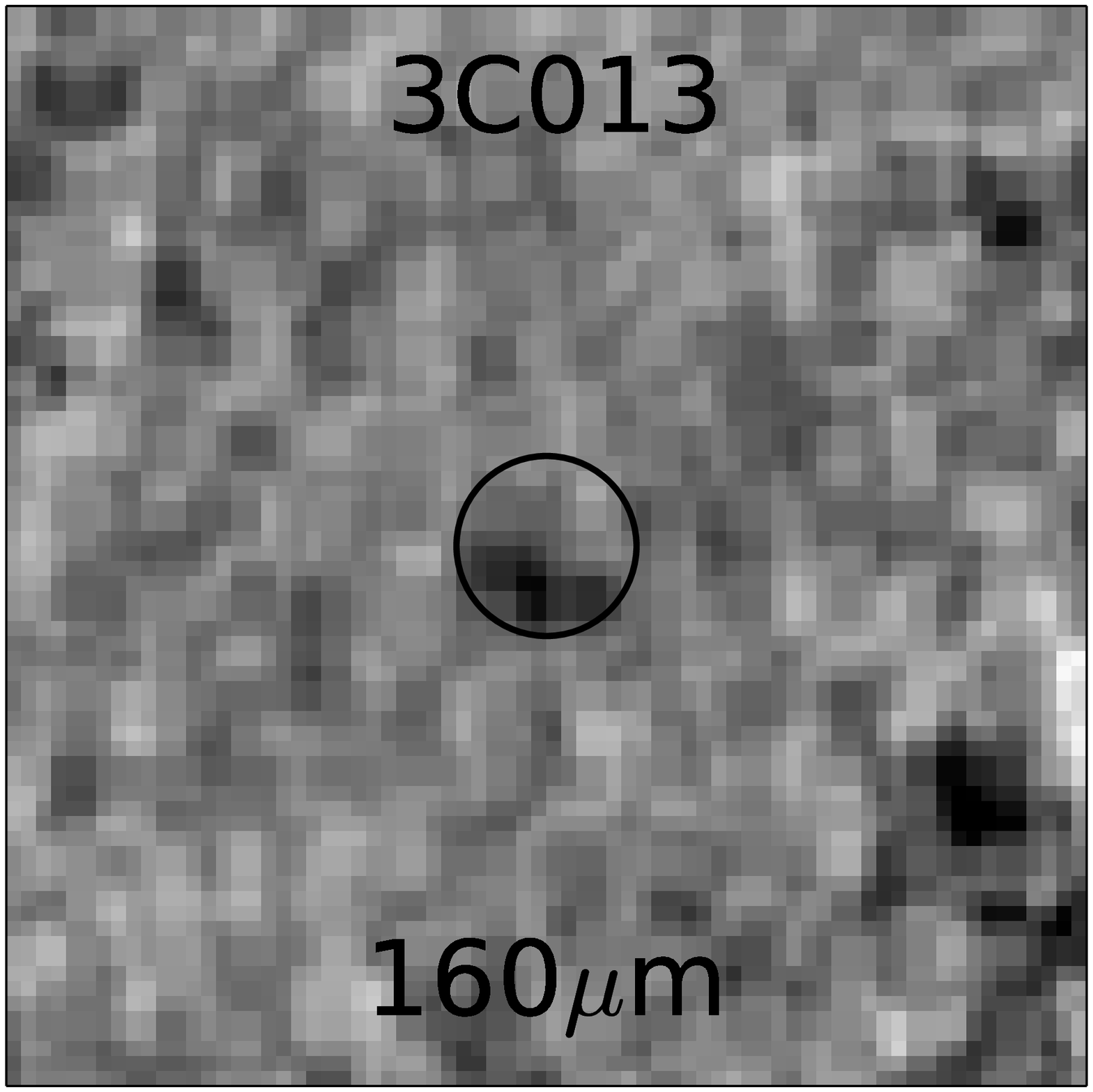}
      \includegraphics[width=1.5cm]{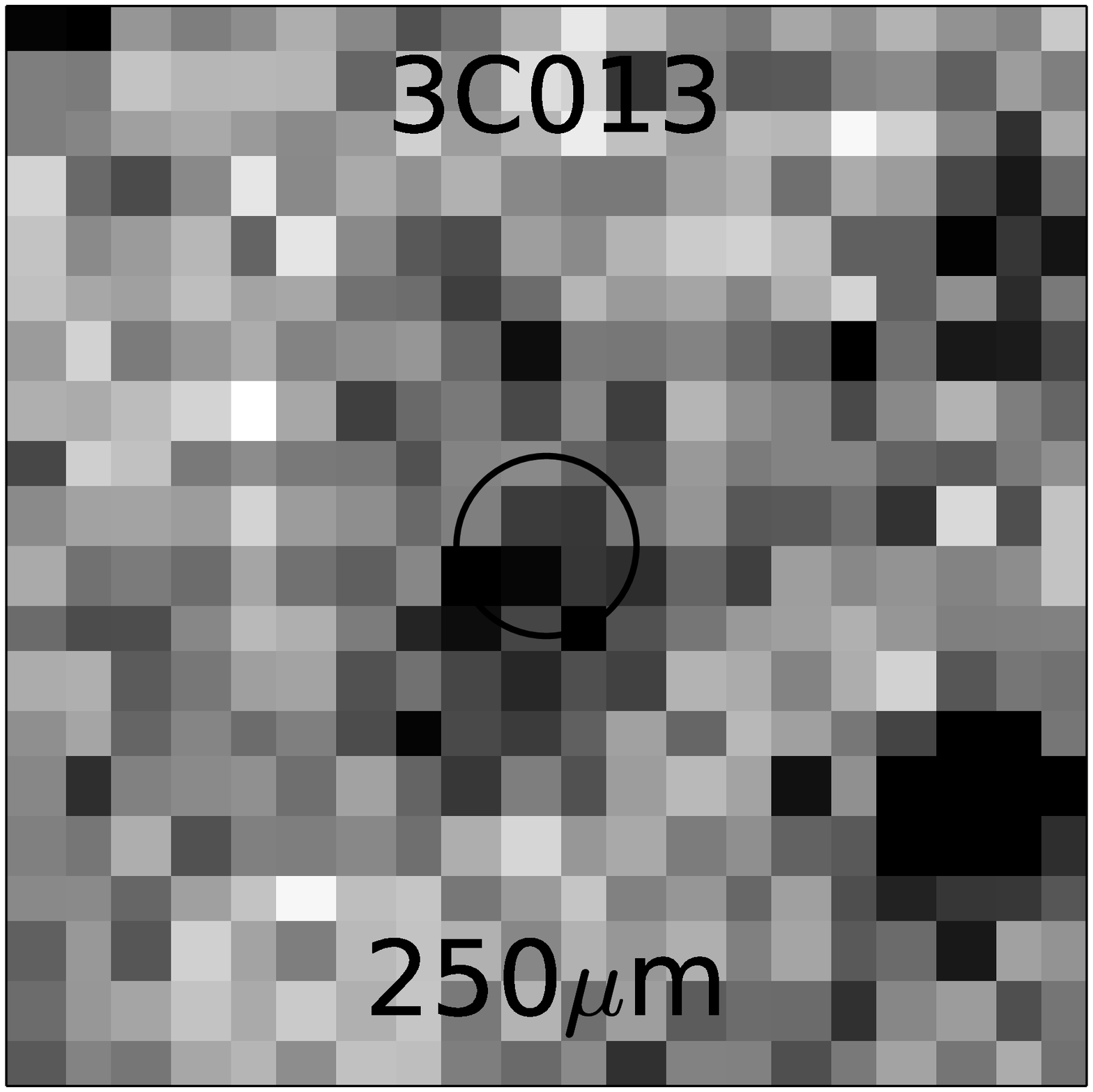}
      \includegraphics[width=1.5cm]{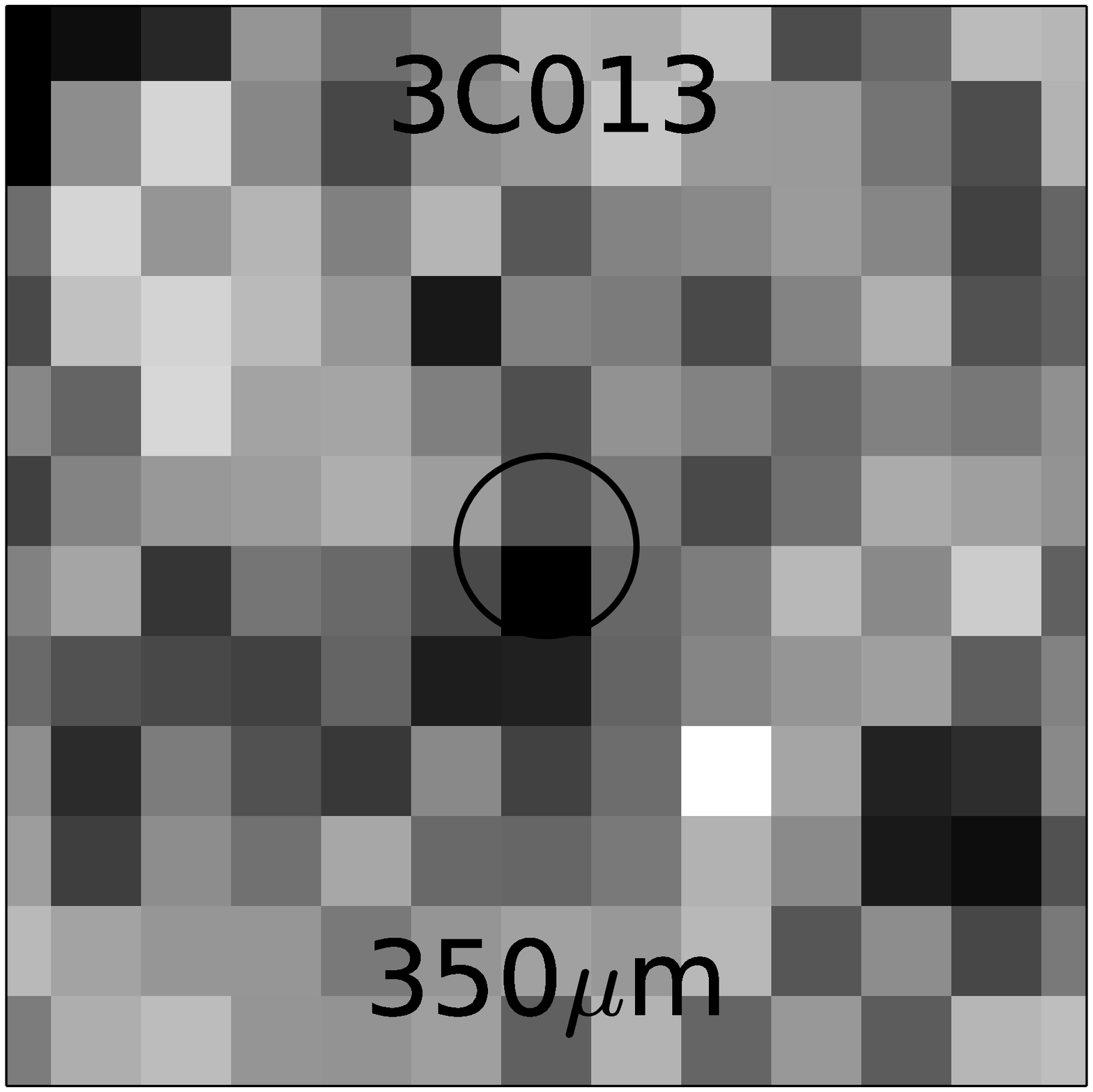}
      \includegraphics[width=1.5cm]{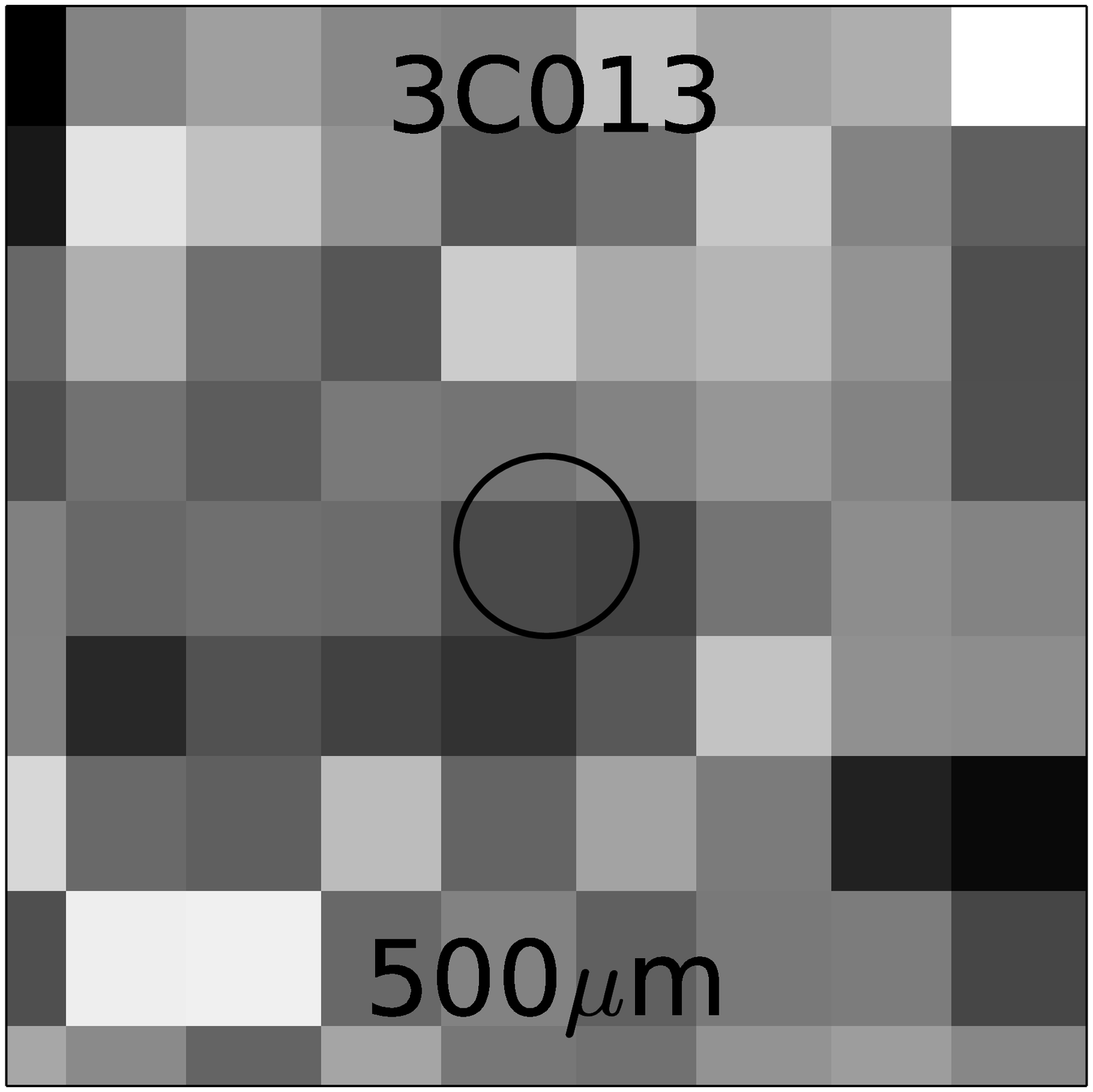}
      \\
      \includegraphics[width=1.5cm]{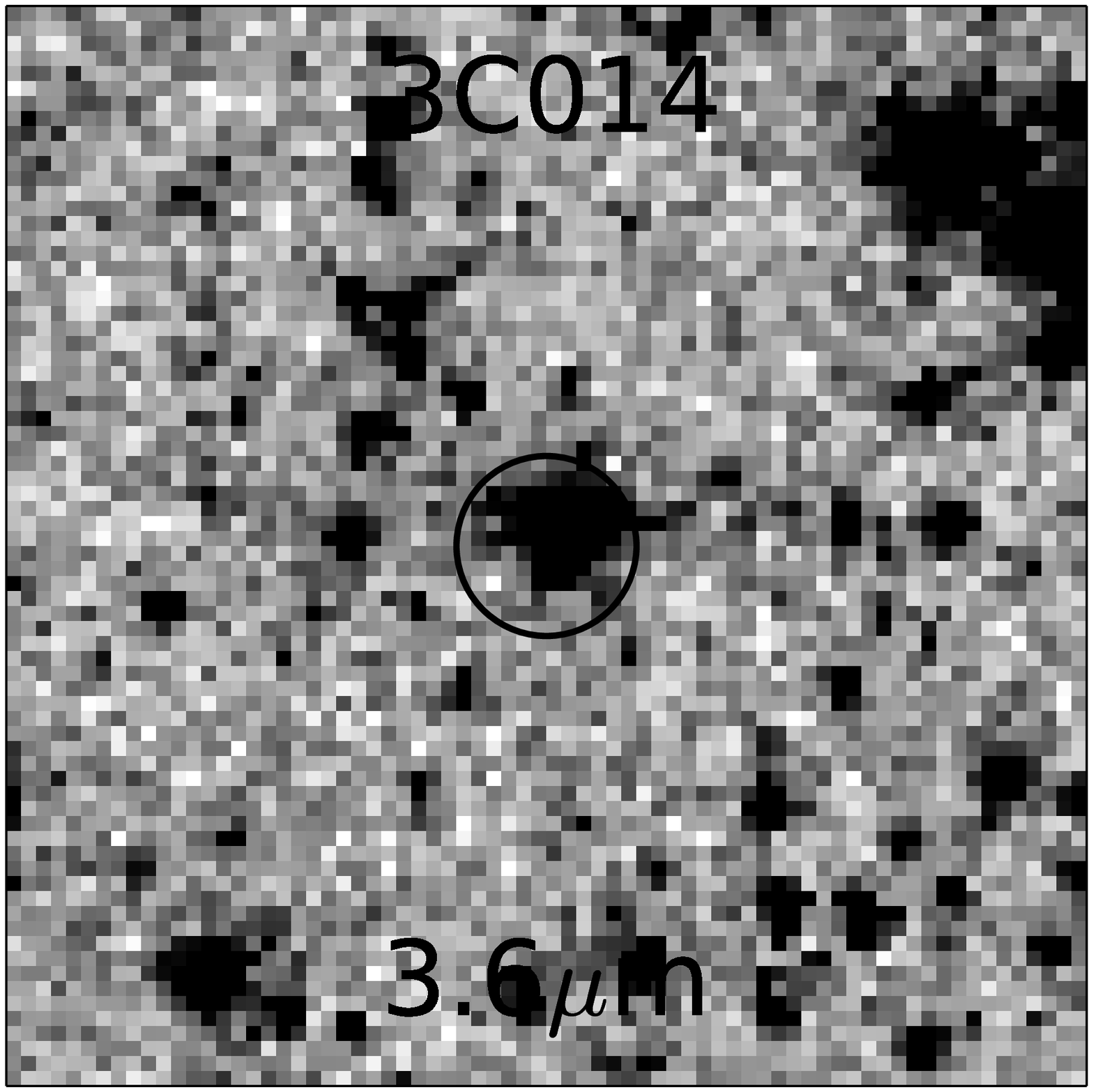}
      \includegraphics[width=1.5cm]{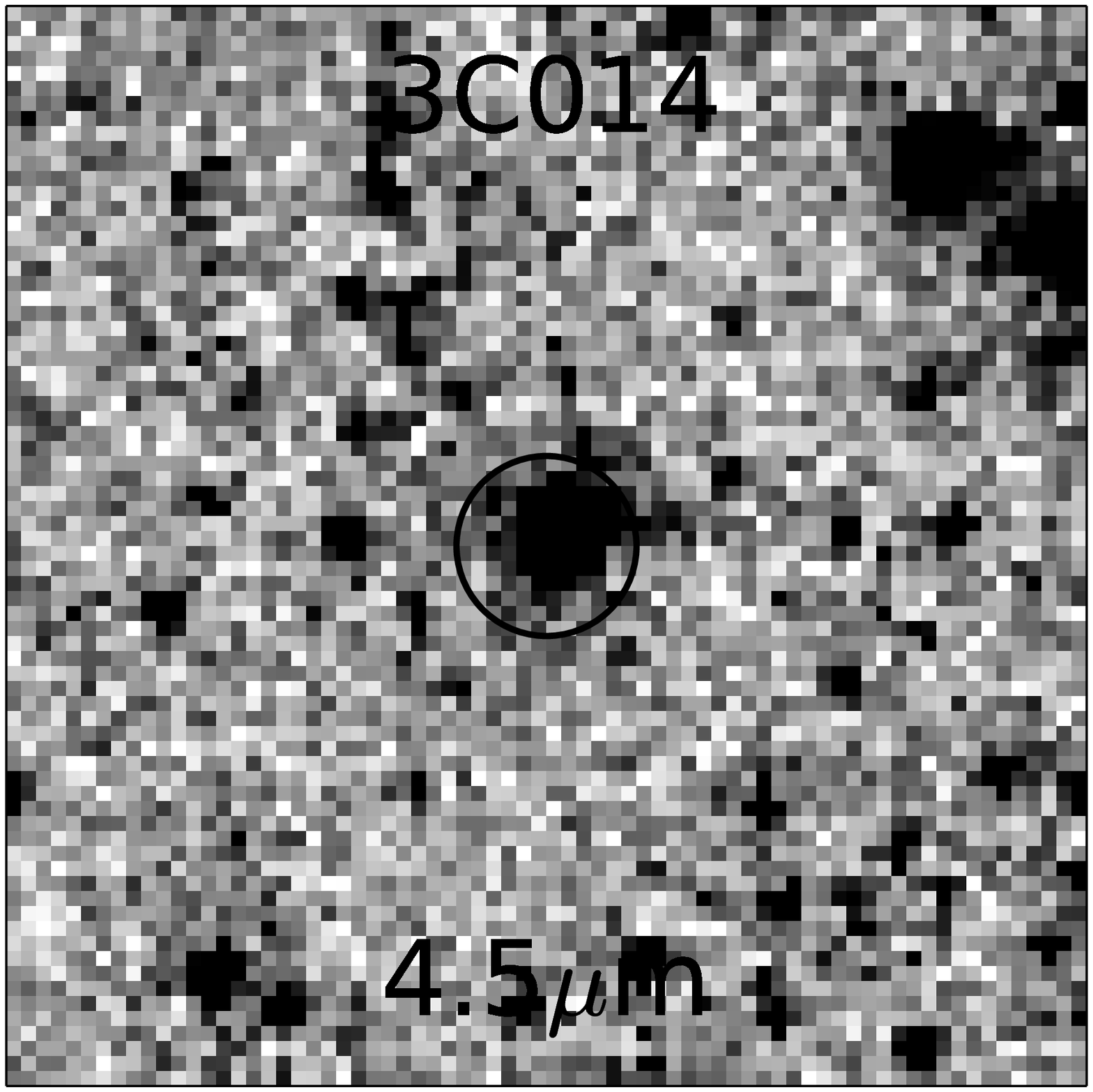}
      \includegraphics[width=1.5cm]{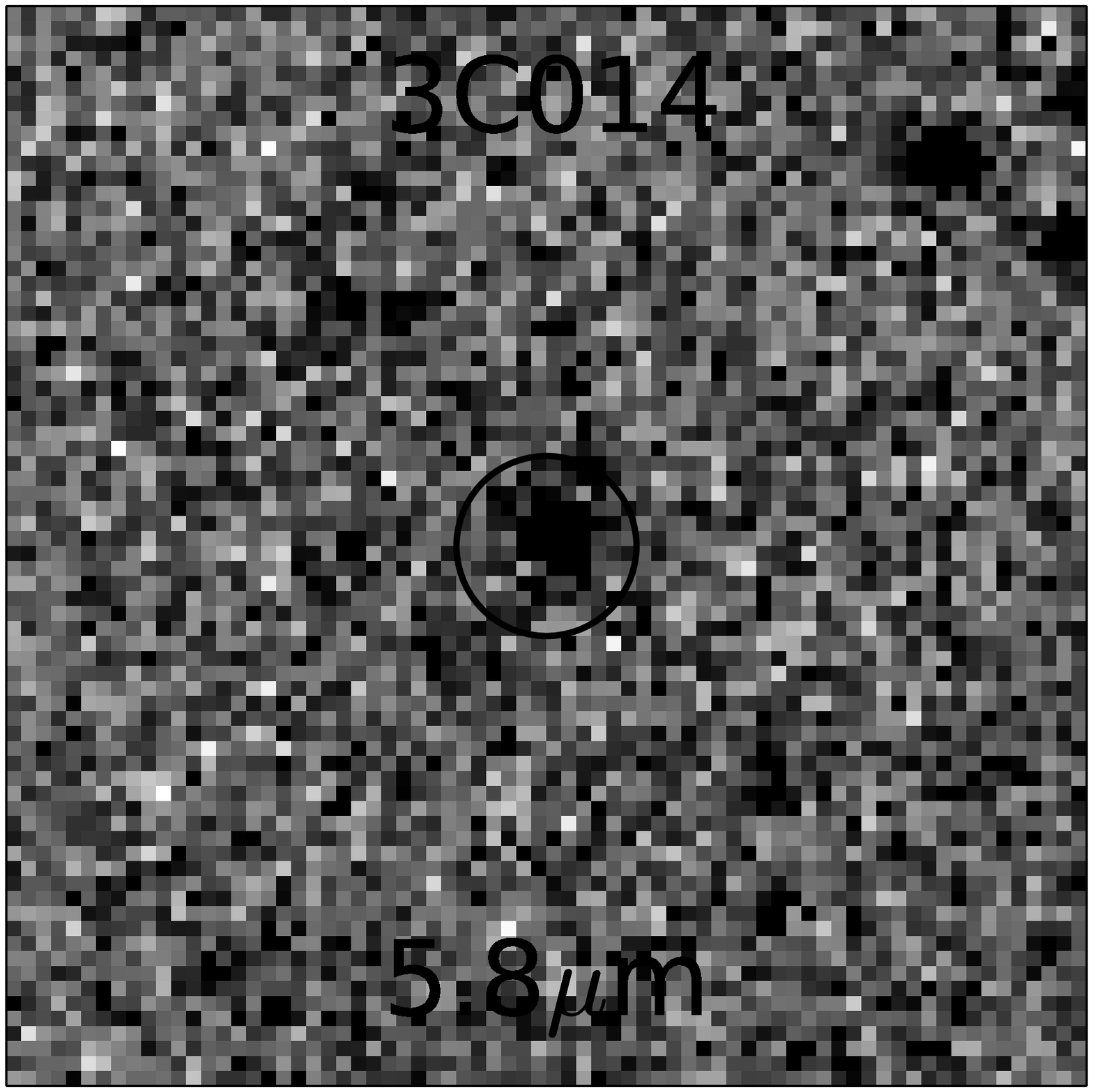}
      \includegraphics[width=1.5cm]{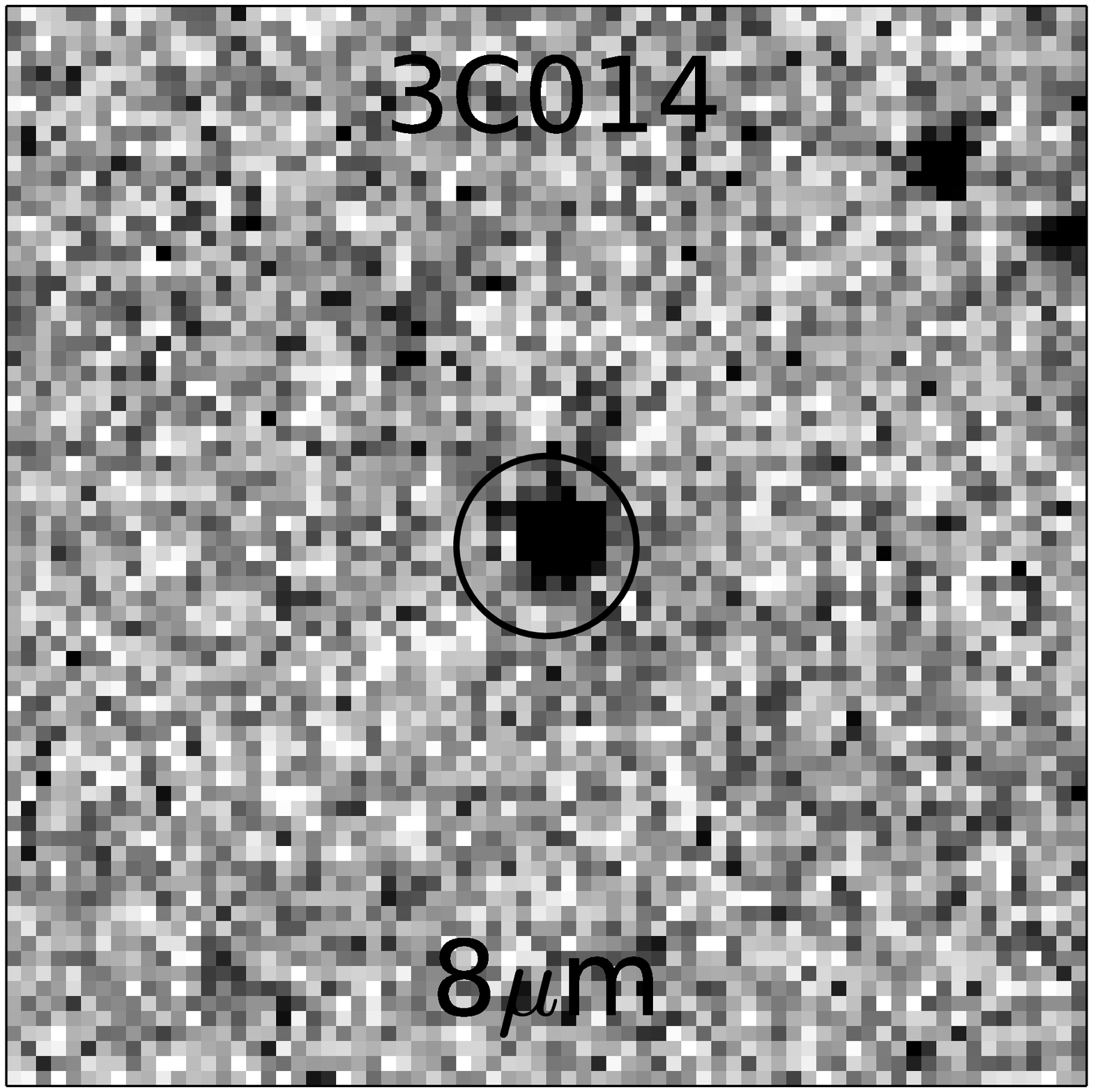}
      \includegraphics[width=1.5cm]{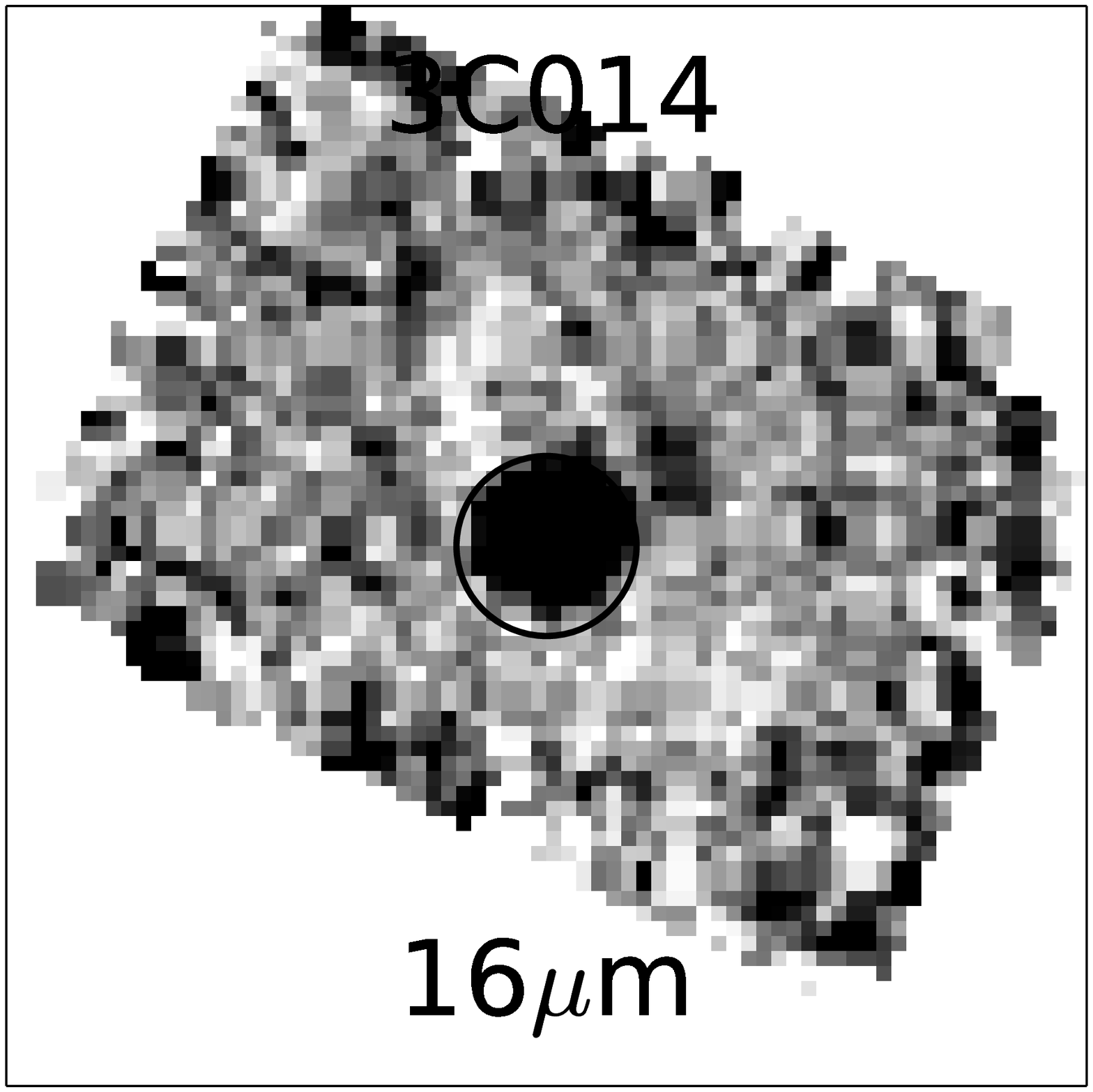}
      \includegraphics[width=1.5cm]{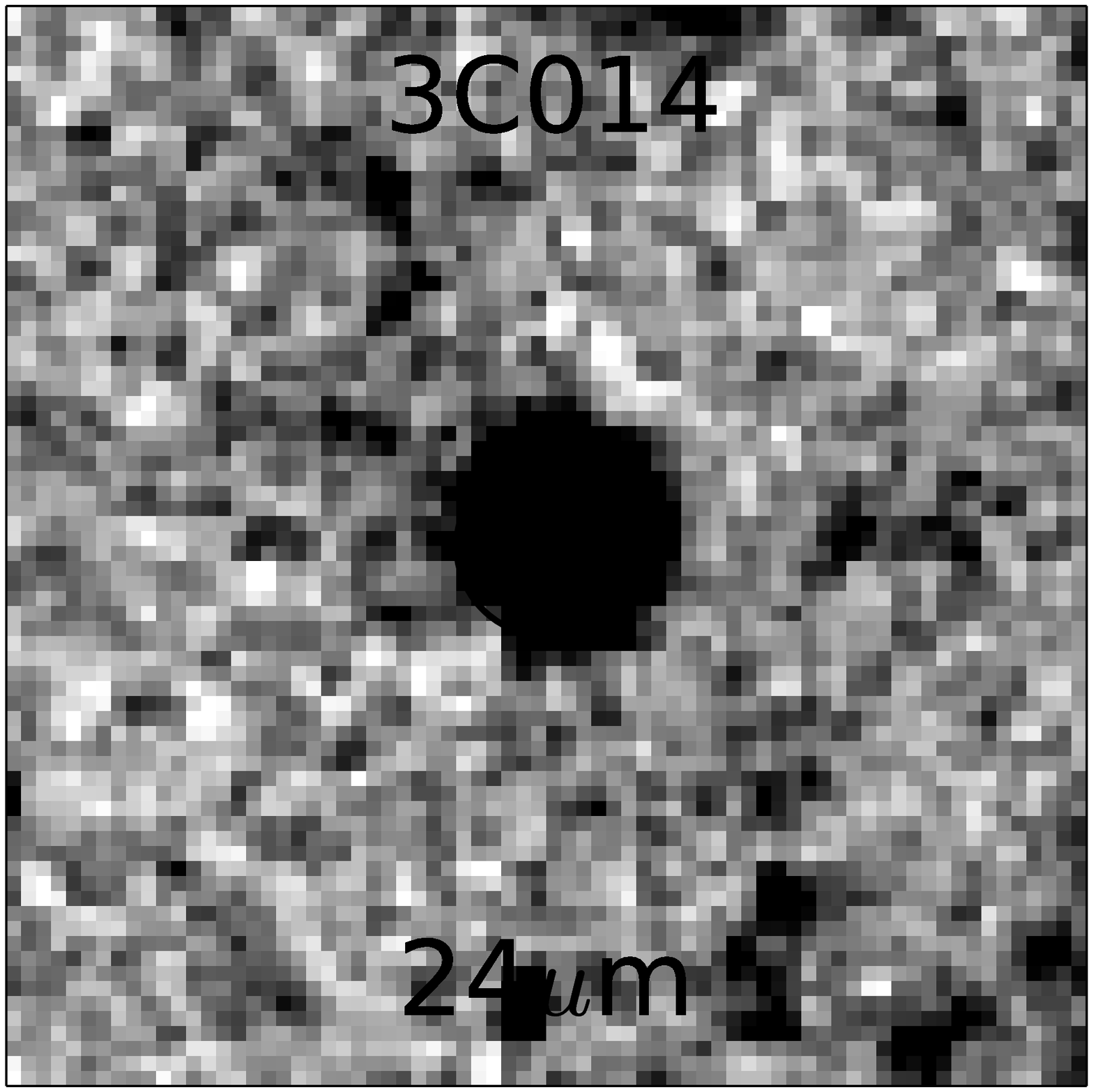}
      \includegraphics[width=1.5cm]{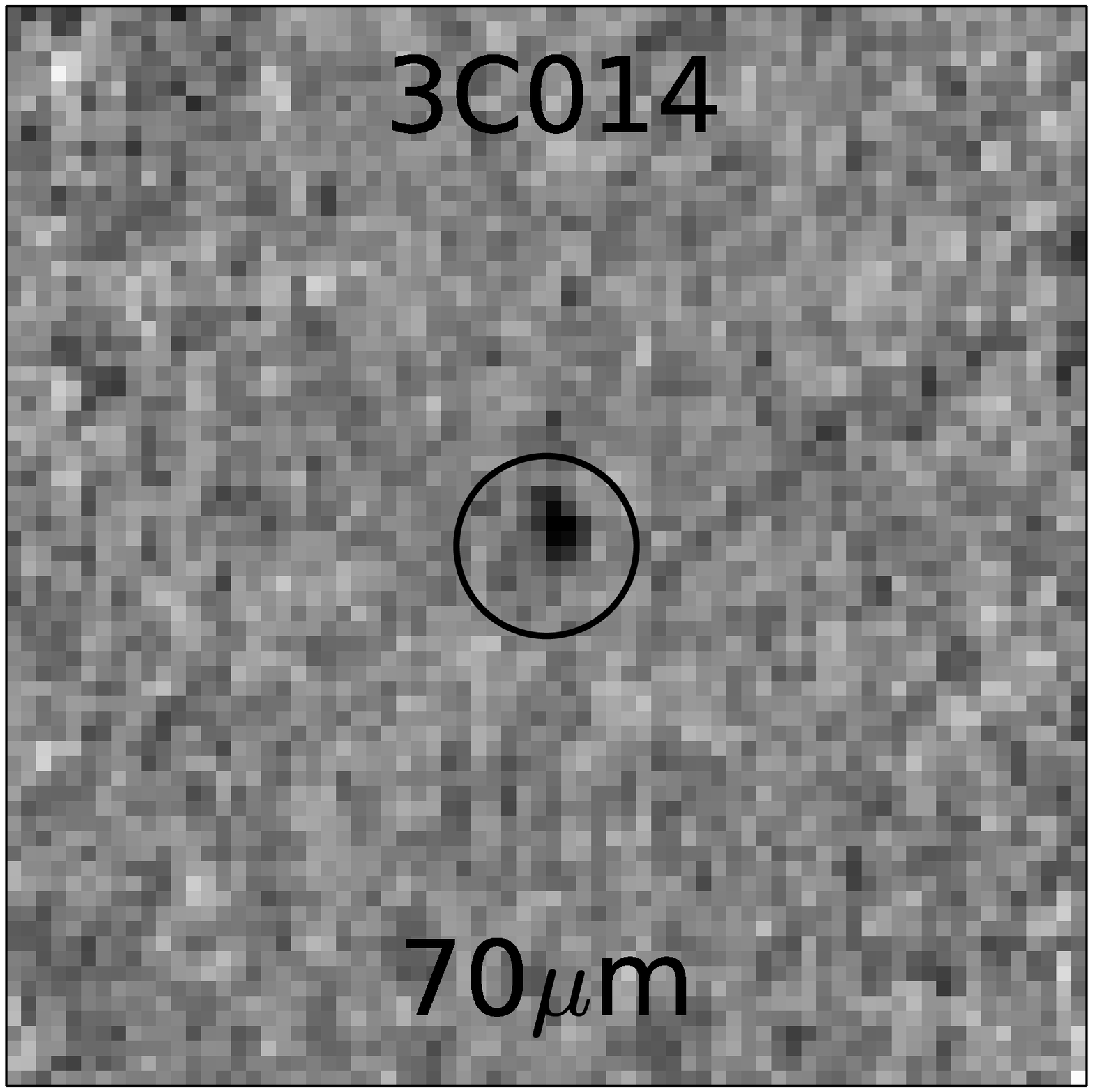}
      \includegraphics[width=1.5cm]{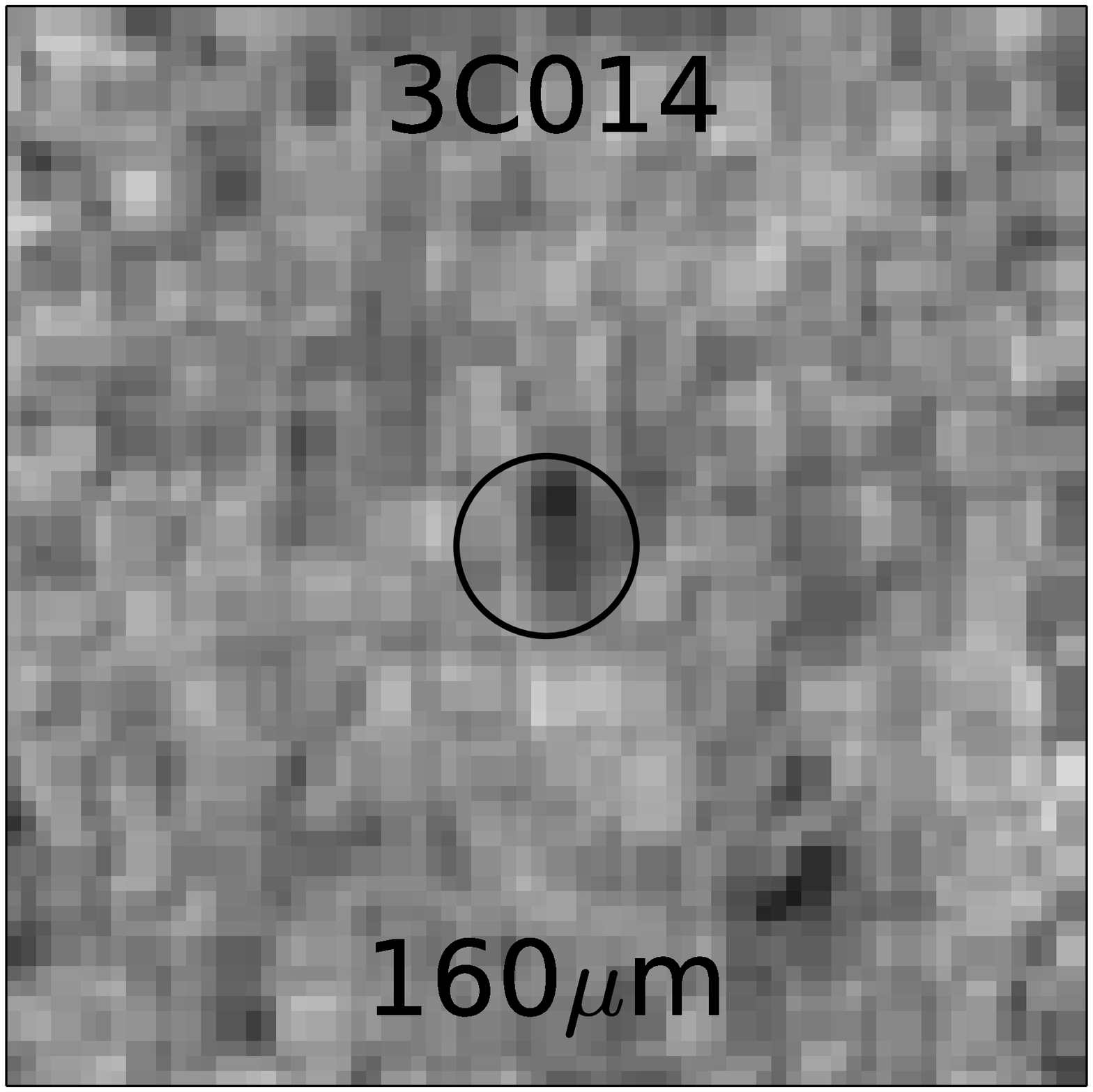}
      \includegraphics[width=1.5cm]{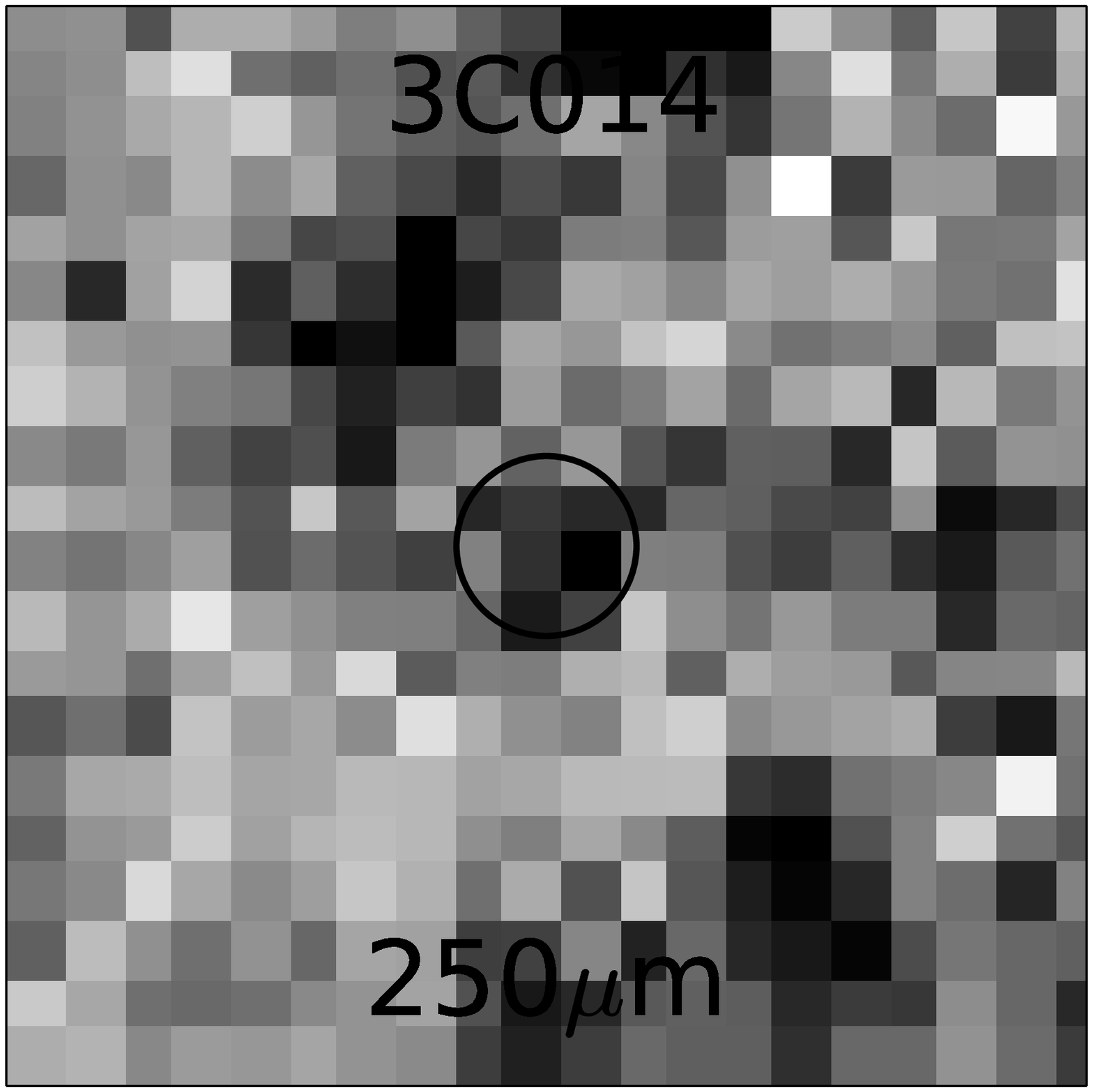}
      \includegraphics[width=1.5cm]{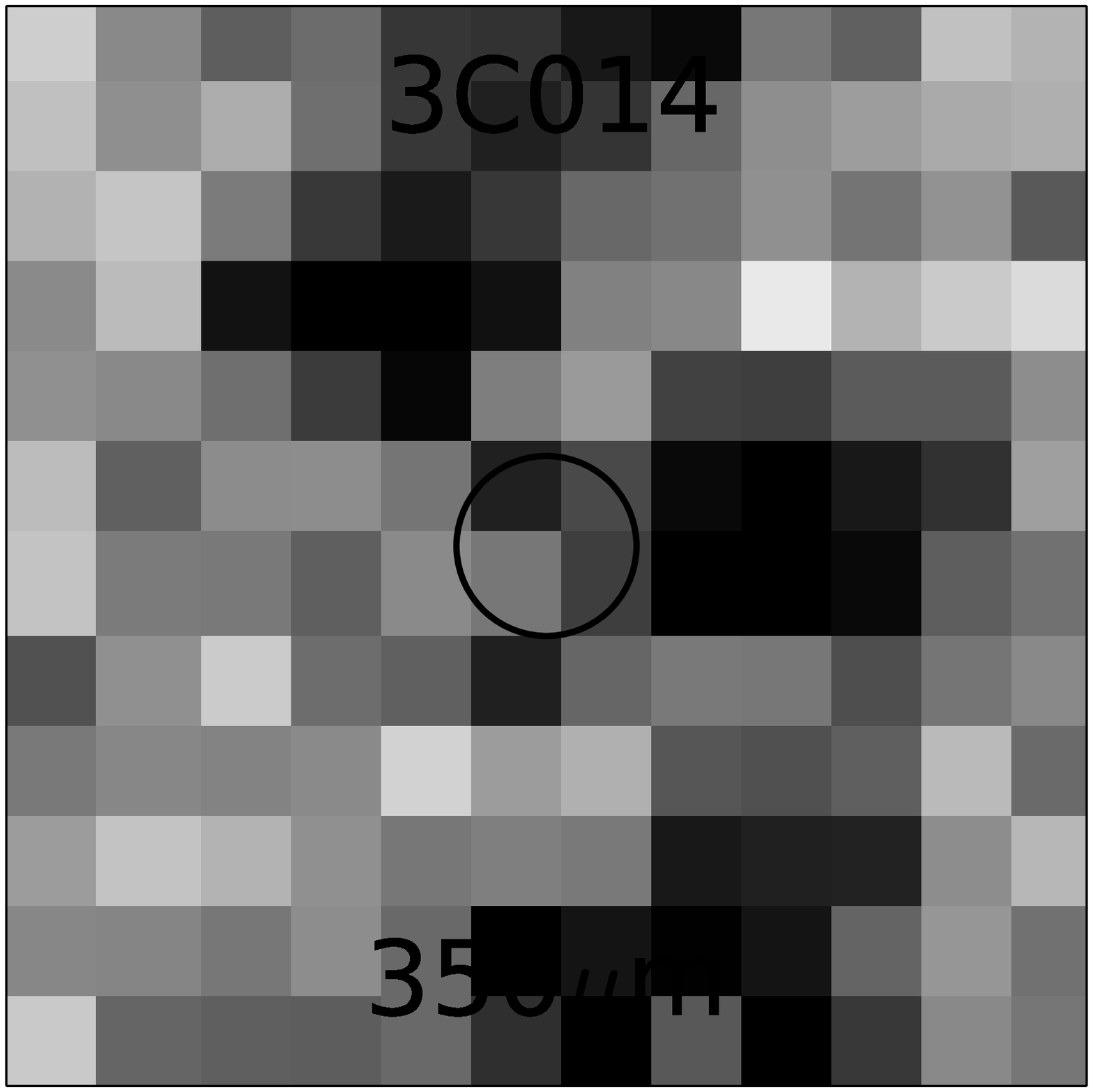}
      \includegraphics[width=1.5cm]{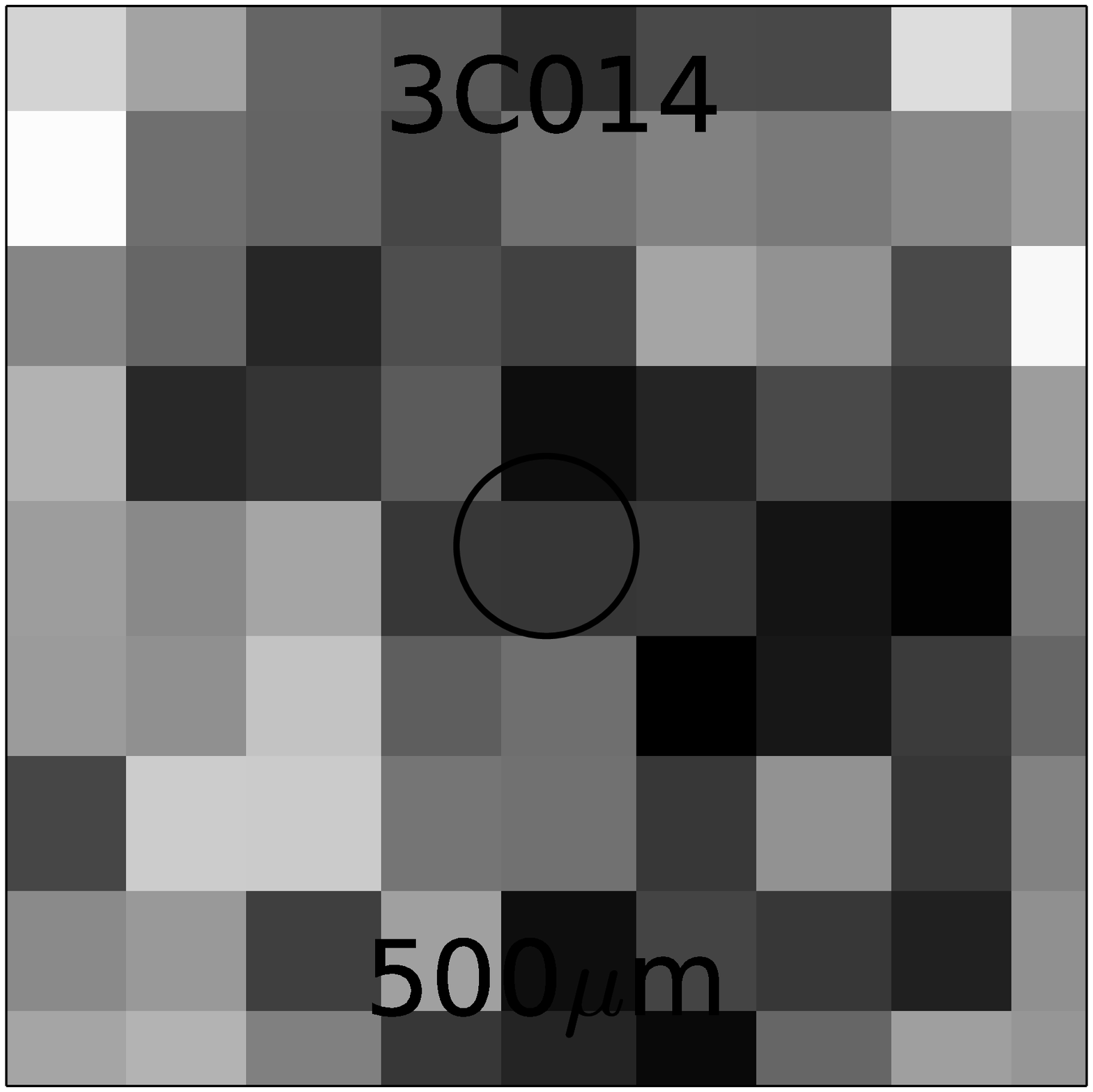}
      \\
      \includegraphics[width=1.5cm]{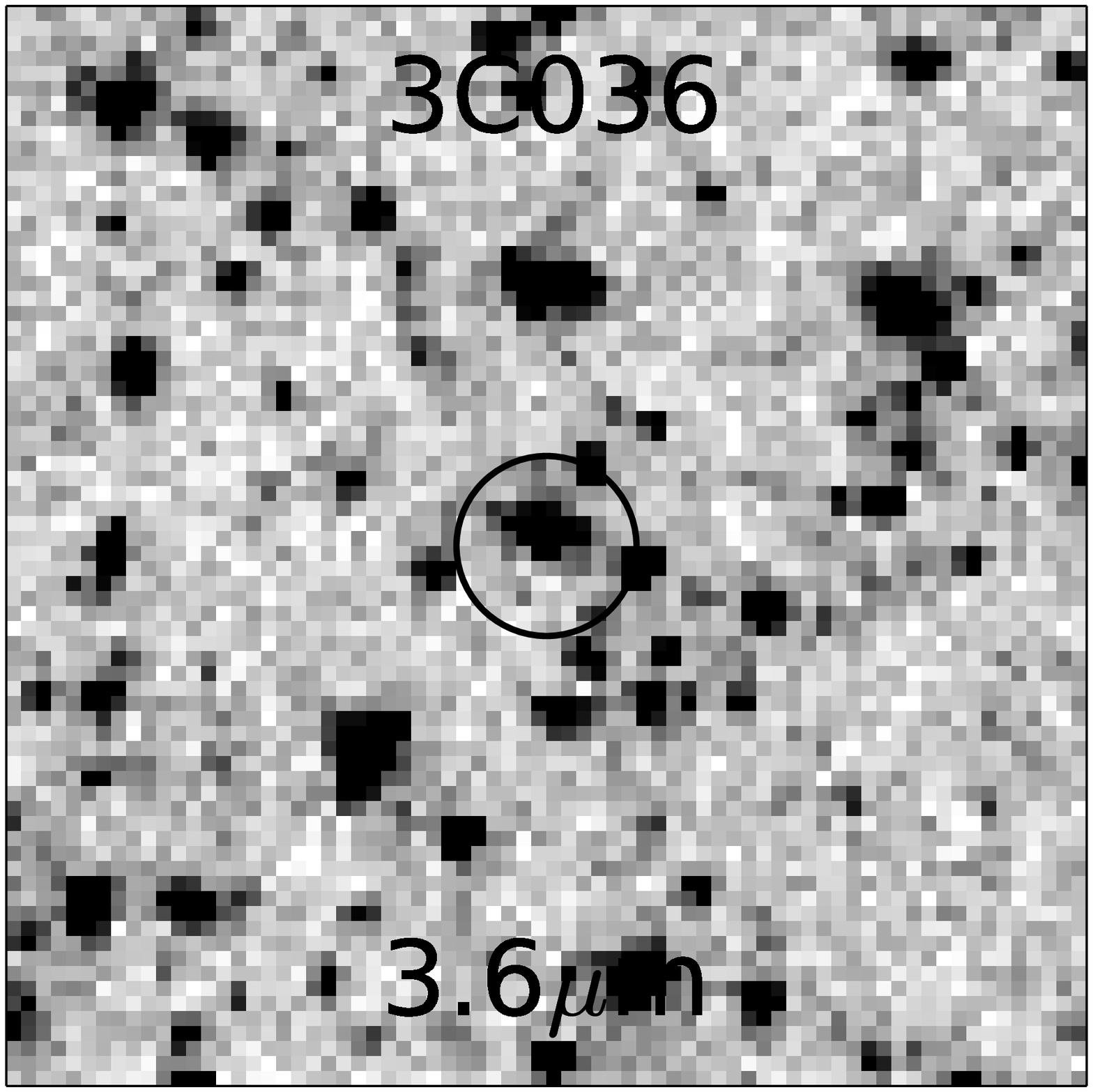}
      \includegraphics[width=1.5cm]{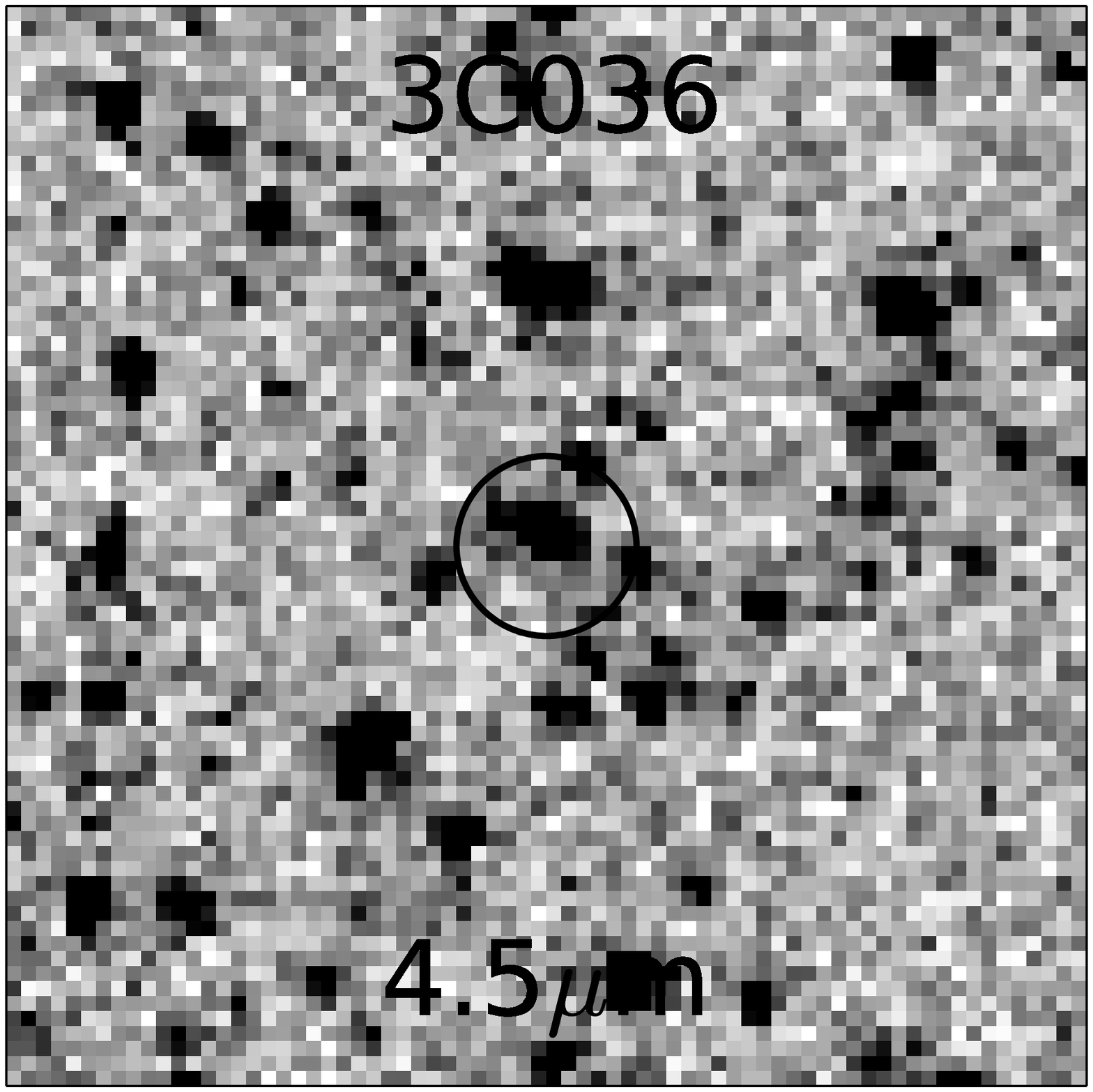}
      \includegraphics[width=1.5cm]{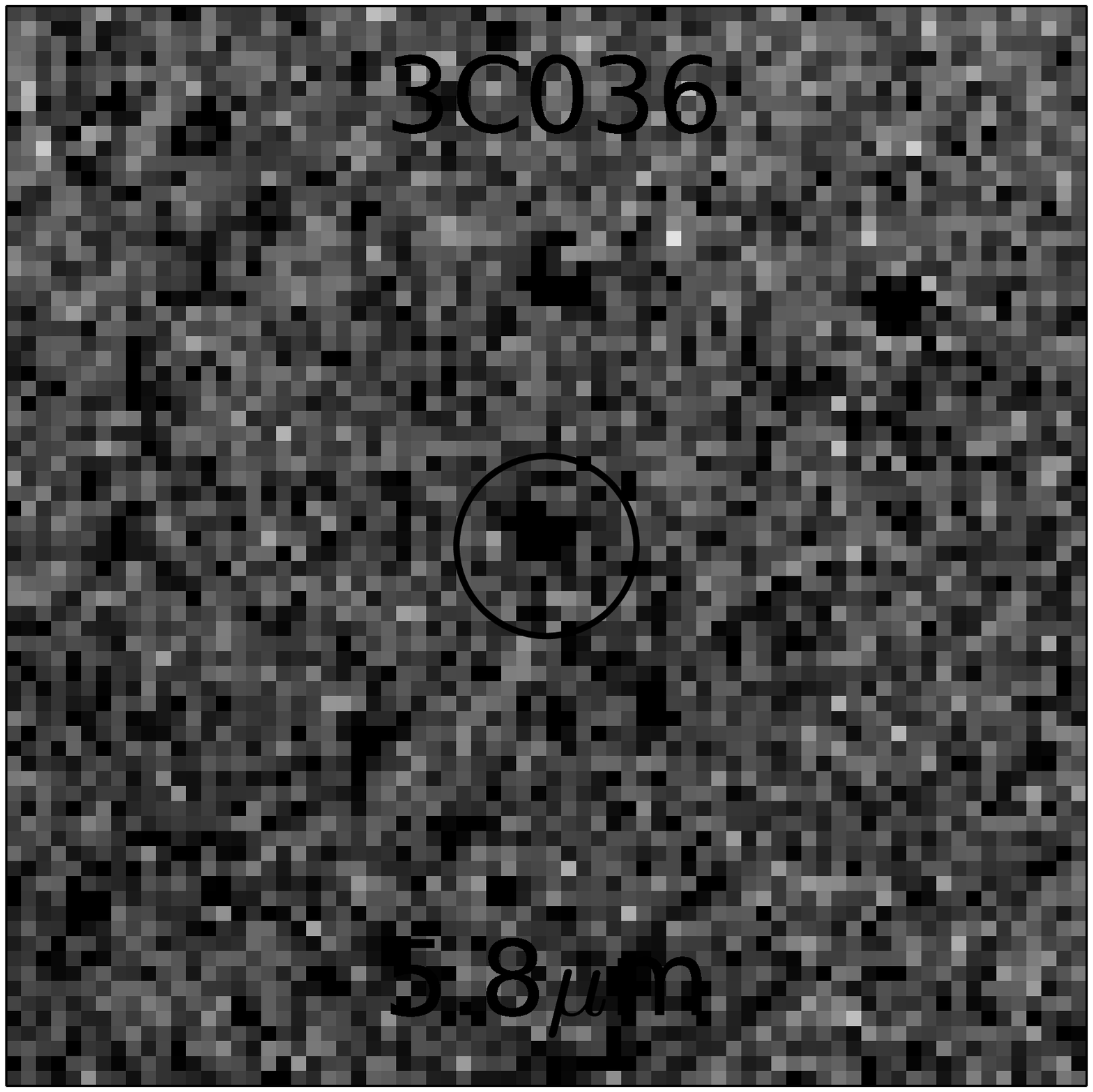}
      \includegraphics[width=1.5cm]{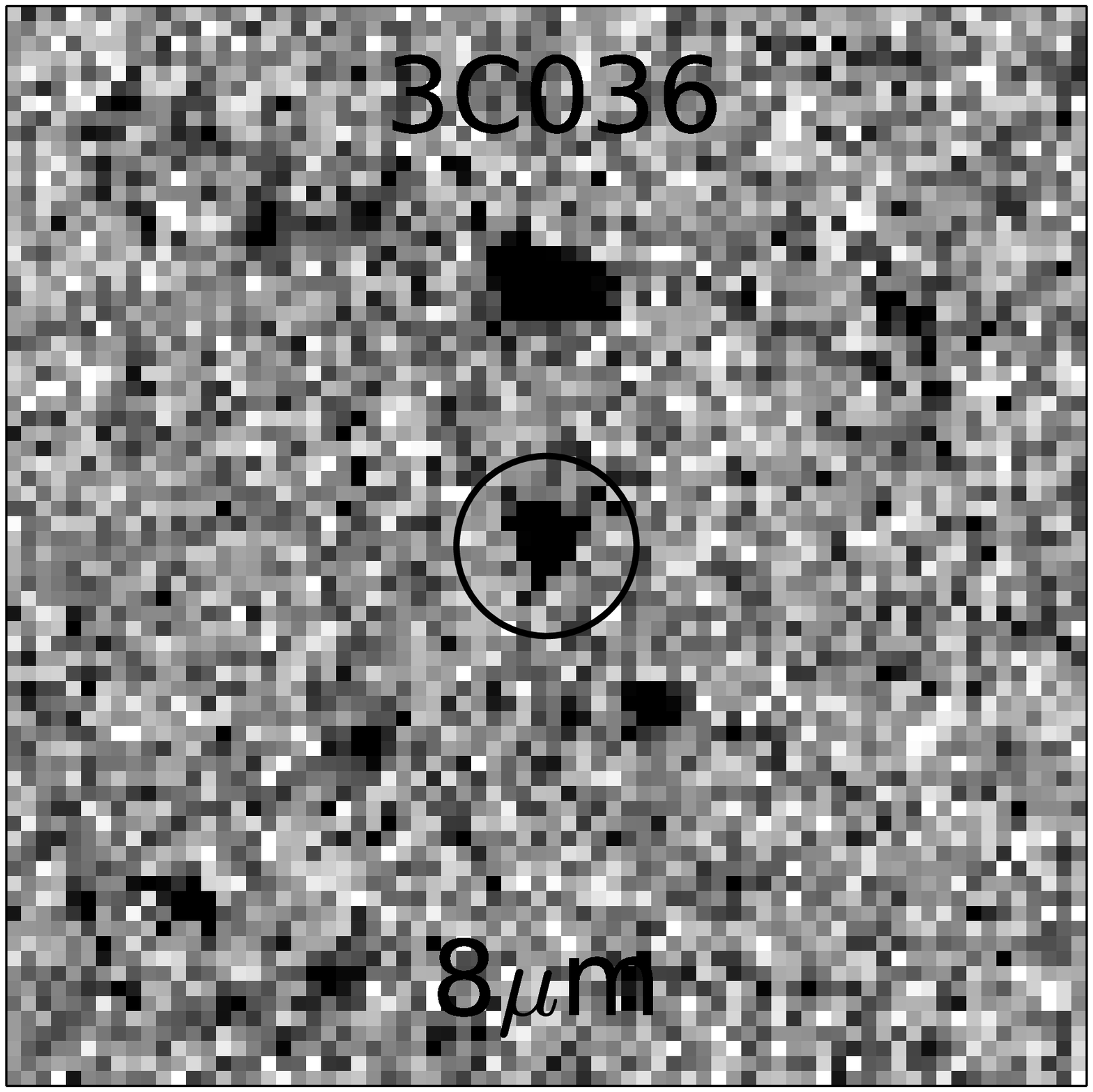}
      \includegraphics[width=1.5cm]{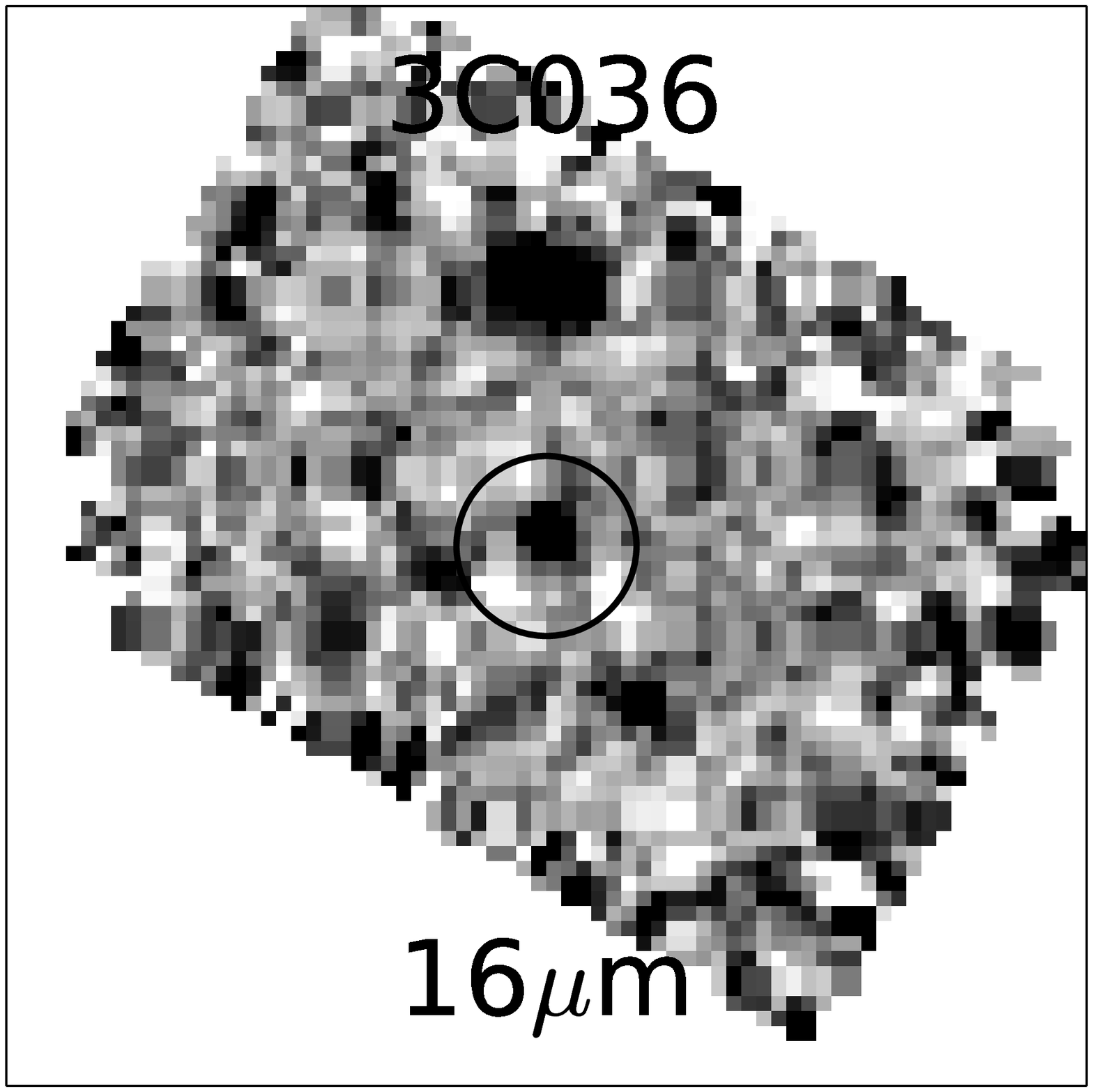}
      \includegraphics[width=1.5cm]{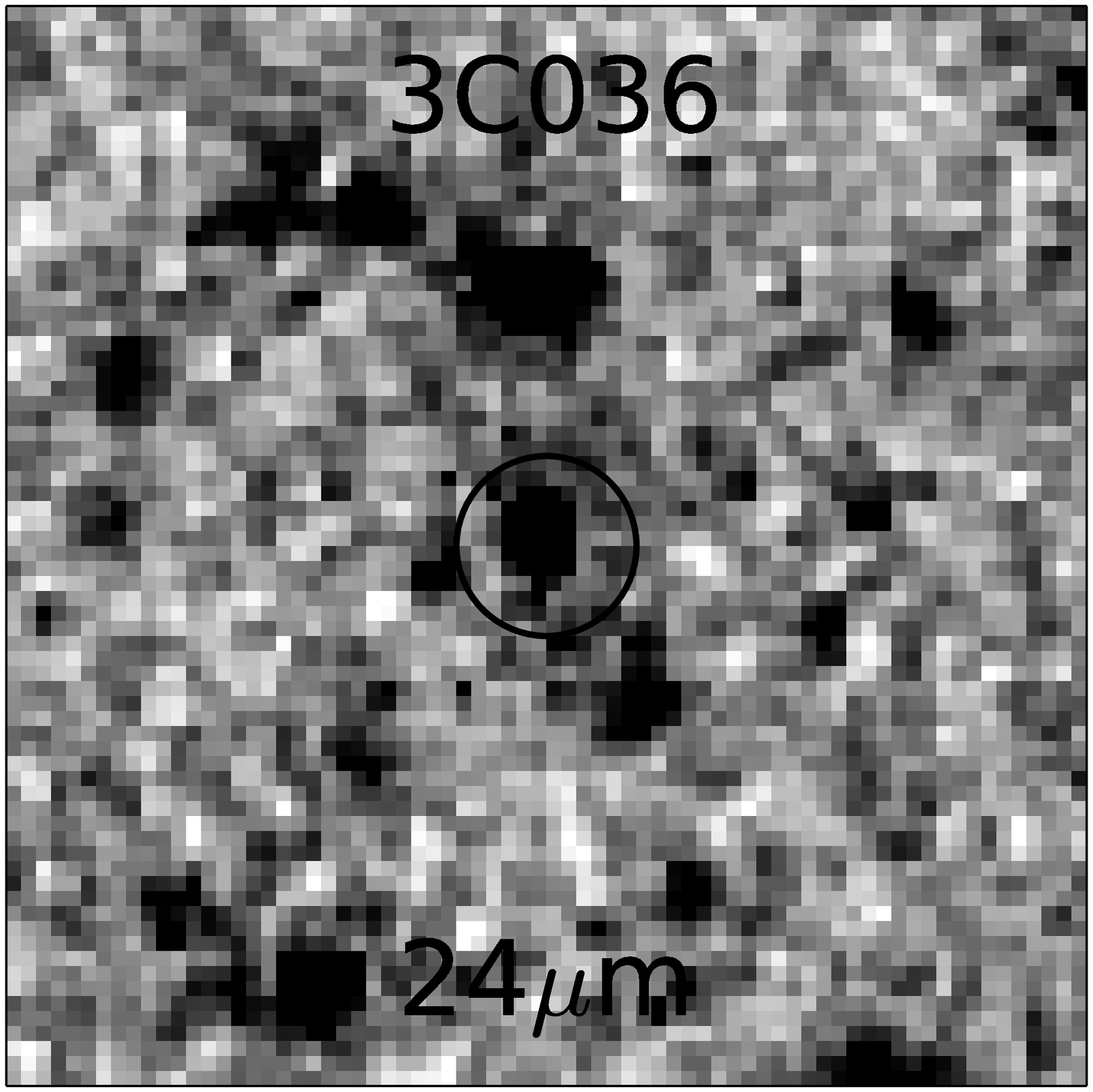}
      \includegraphics[width=1.5cm]{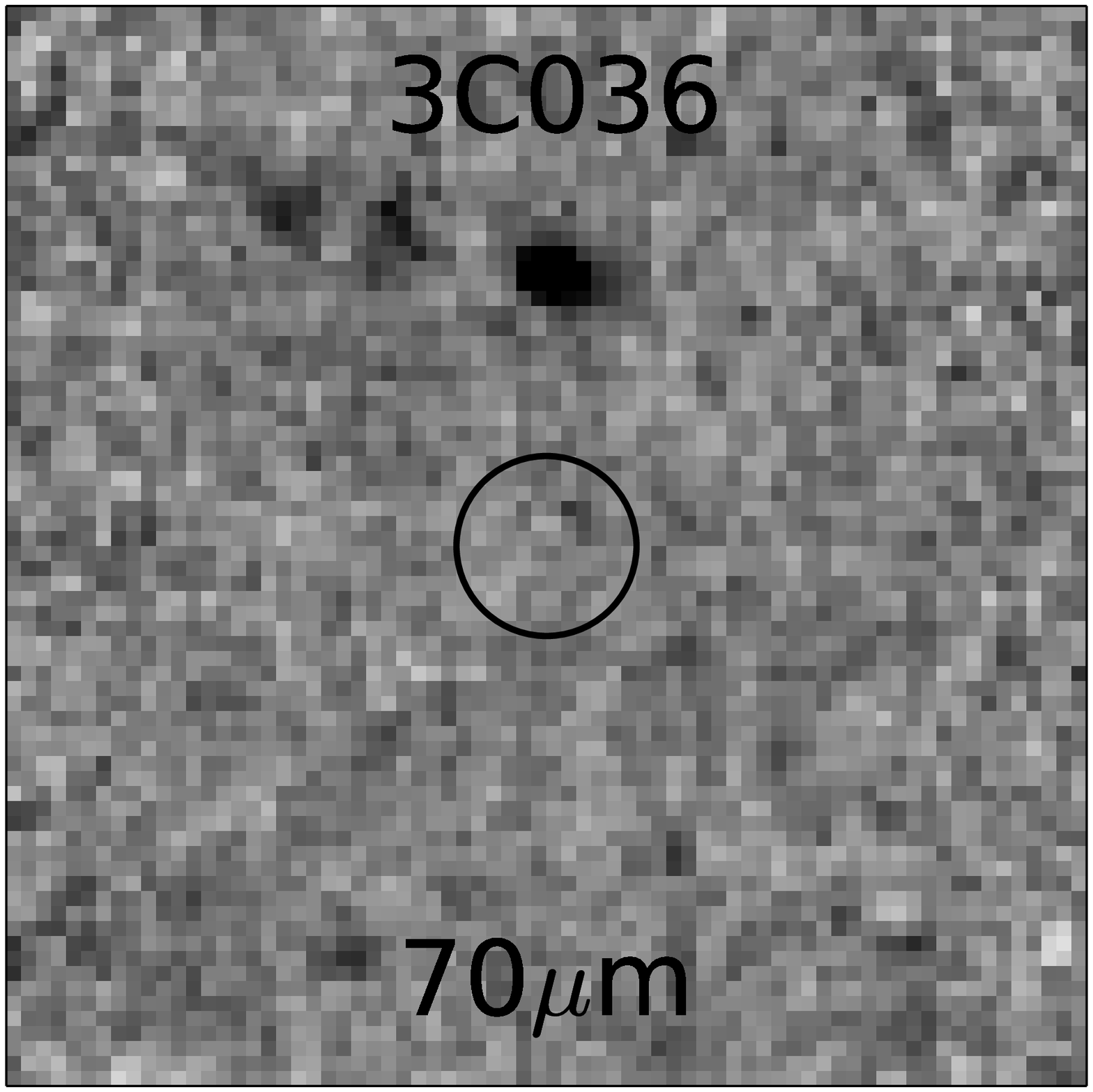}
      \includegraphics[width=1.5cm]{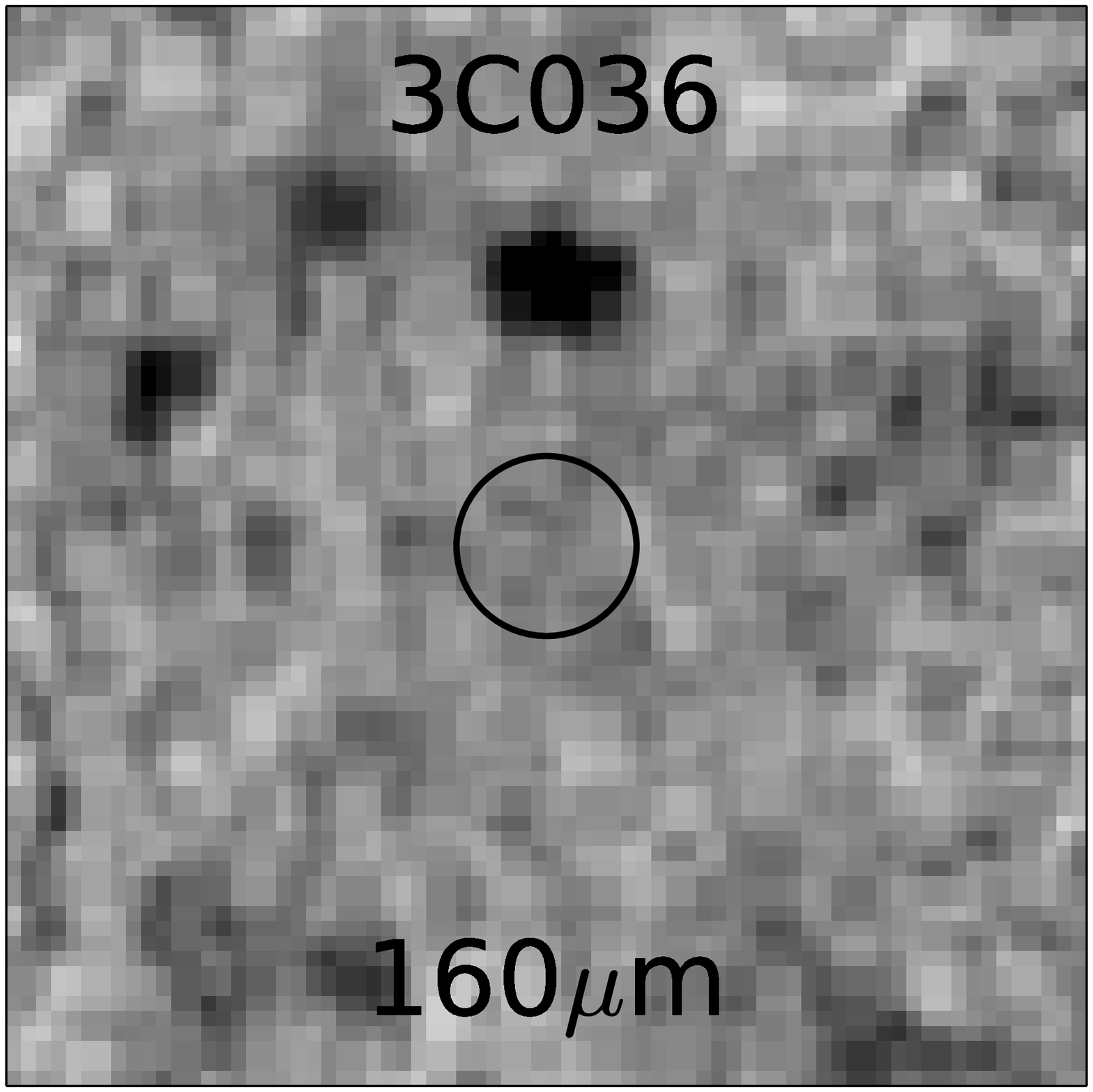}
      \includegraphics[width=1.5cm]{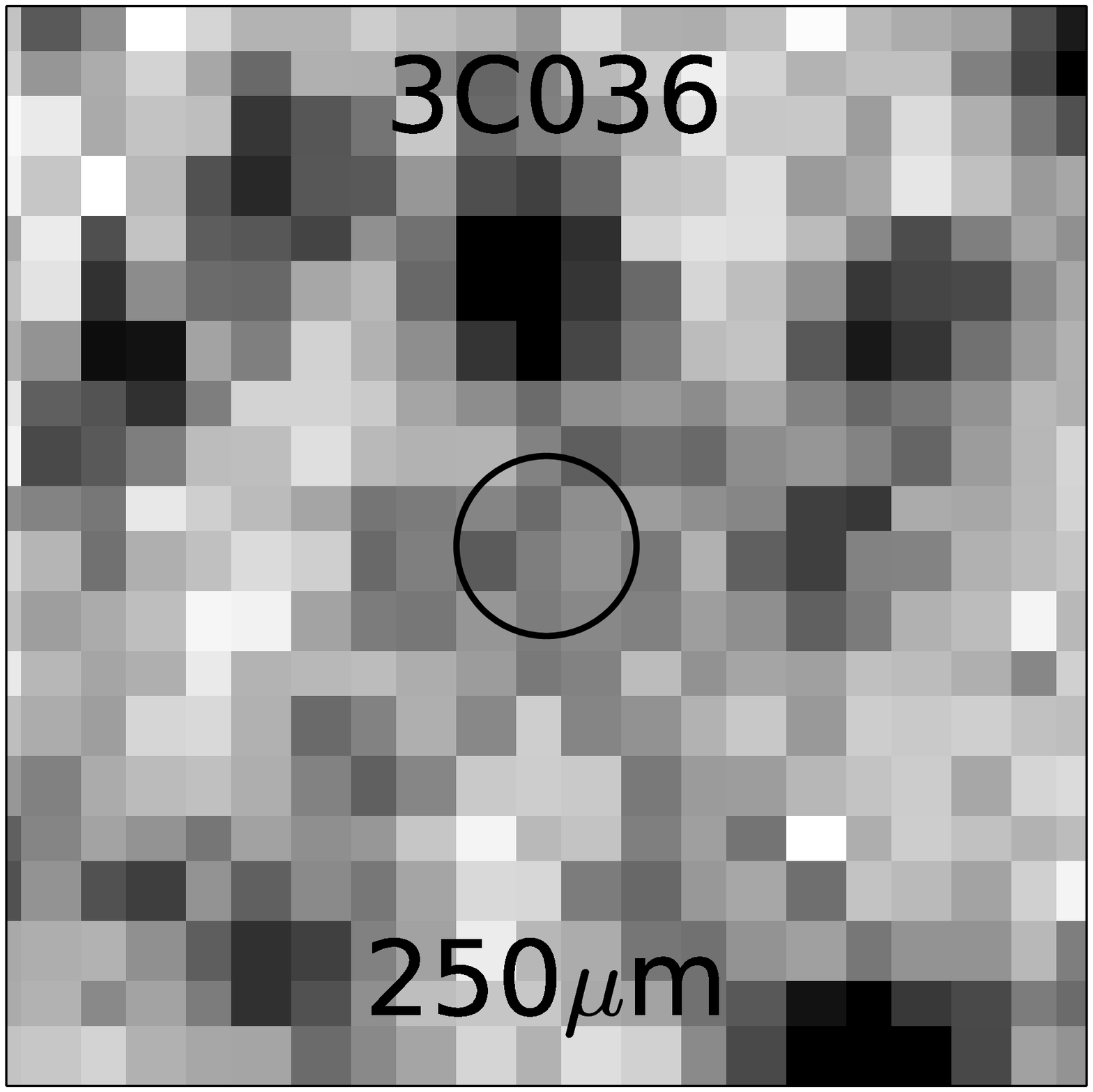}
      \includegraphics[width=1.5cm]{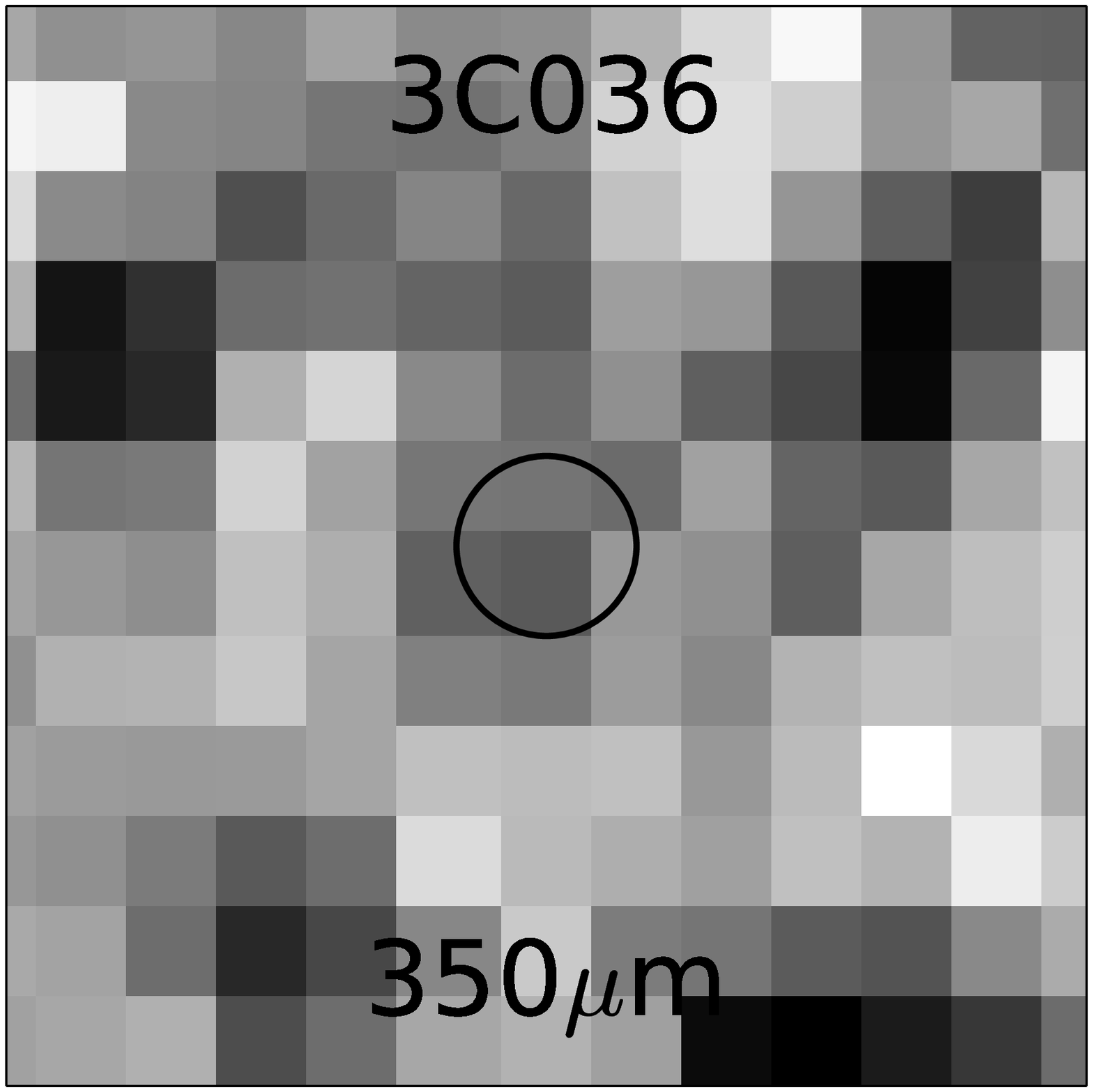}
      \includegraphics[width=1.5cm]{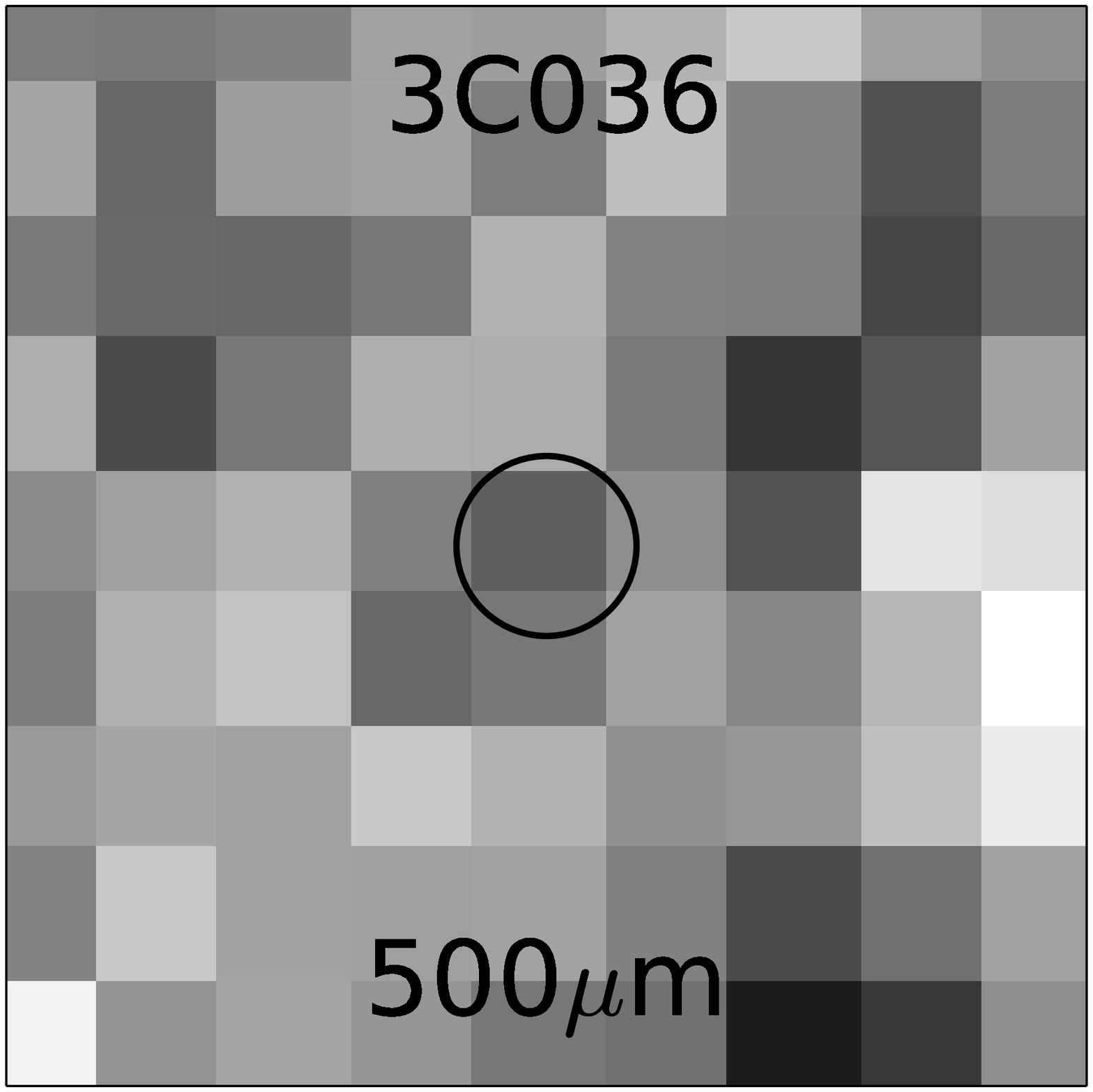}
      \\
      \includegraphics[width=1.5cm]{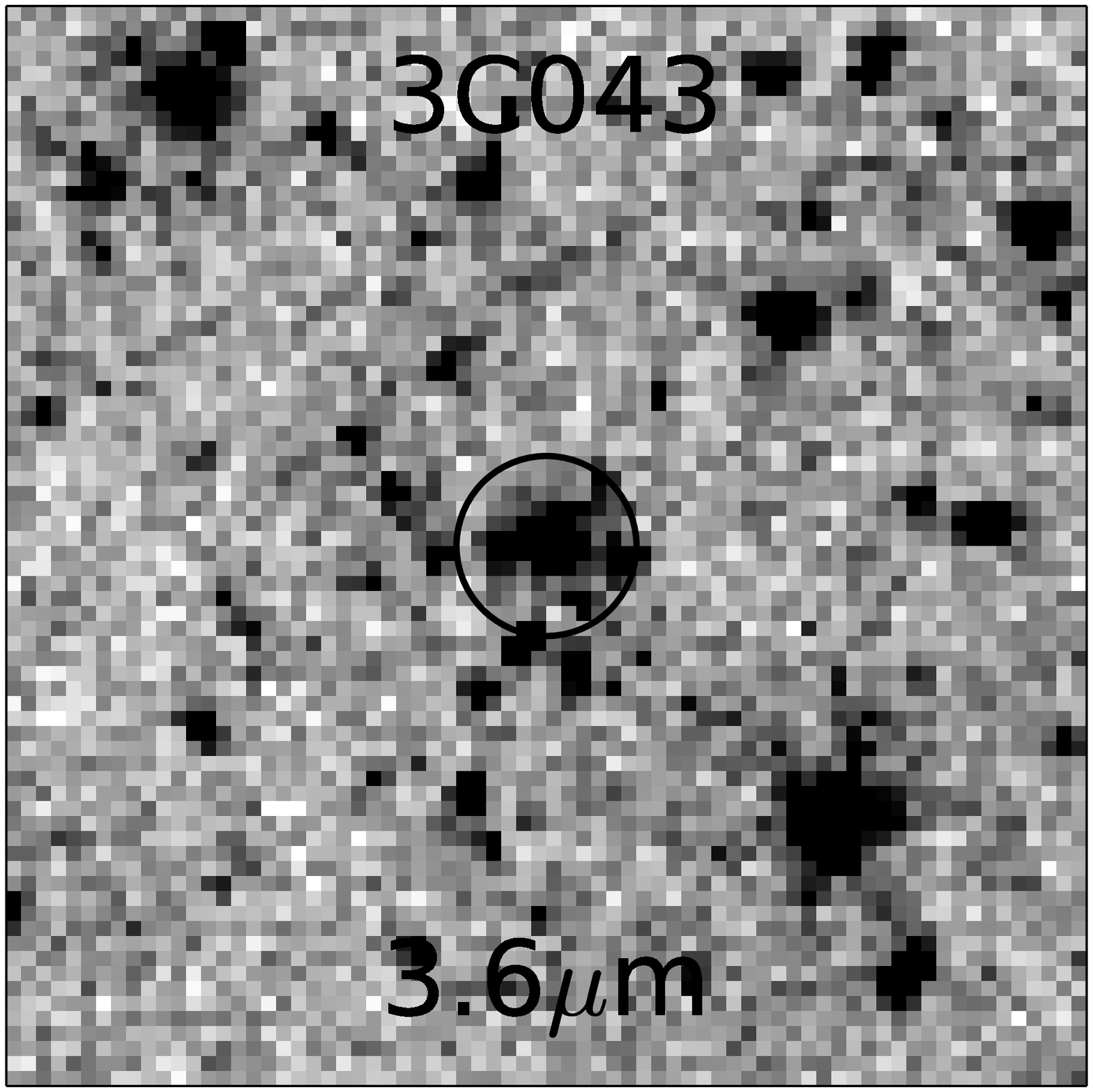}
      \includegraphics[width=1.5cm]{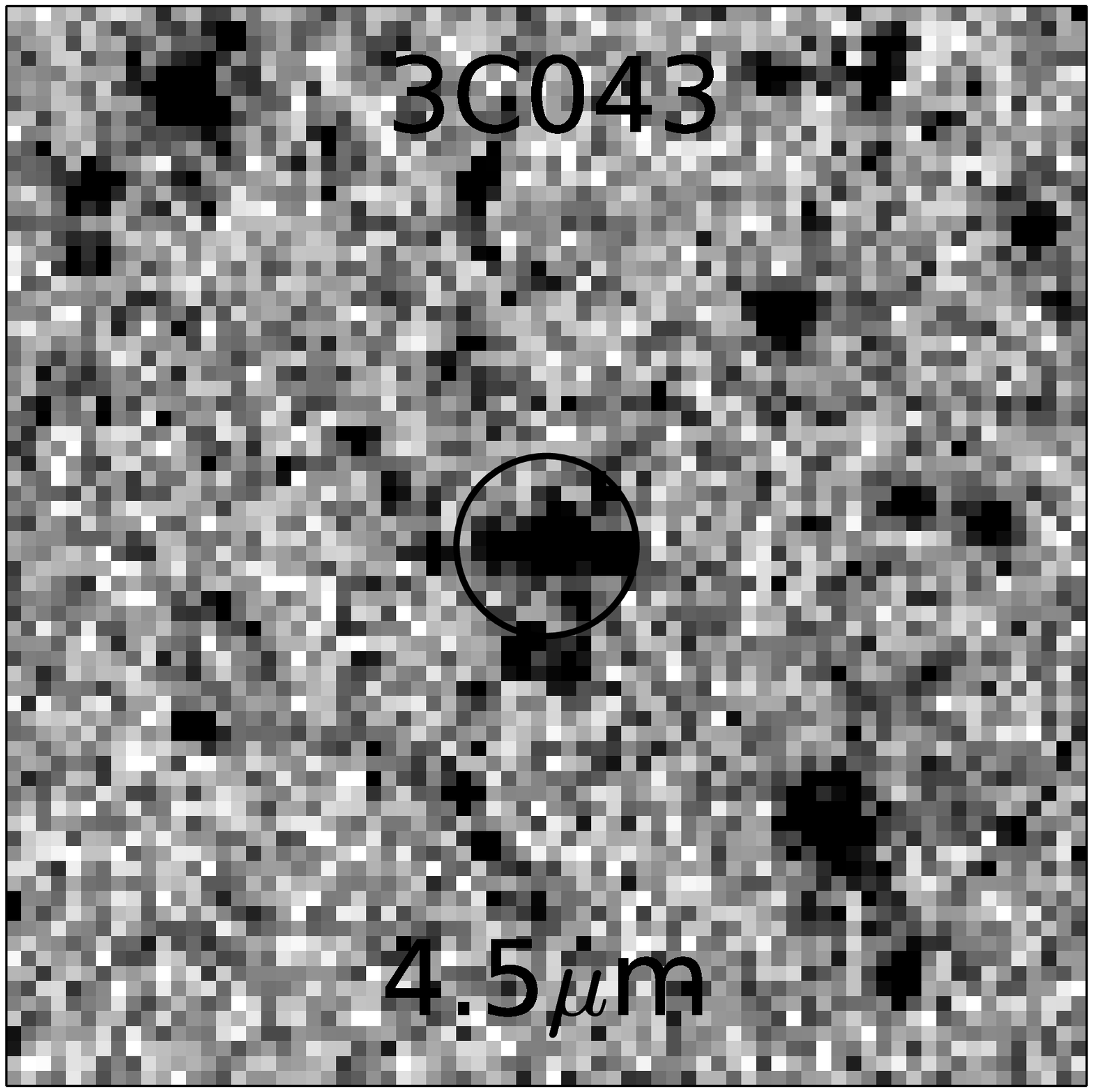}
      \includegraphics[width=1.5cm]{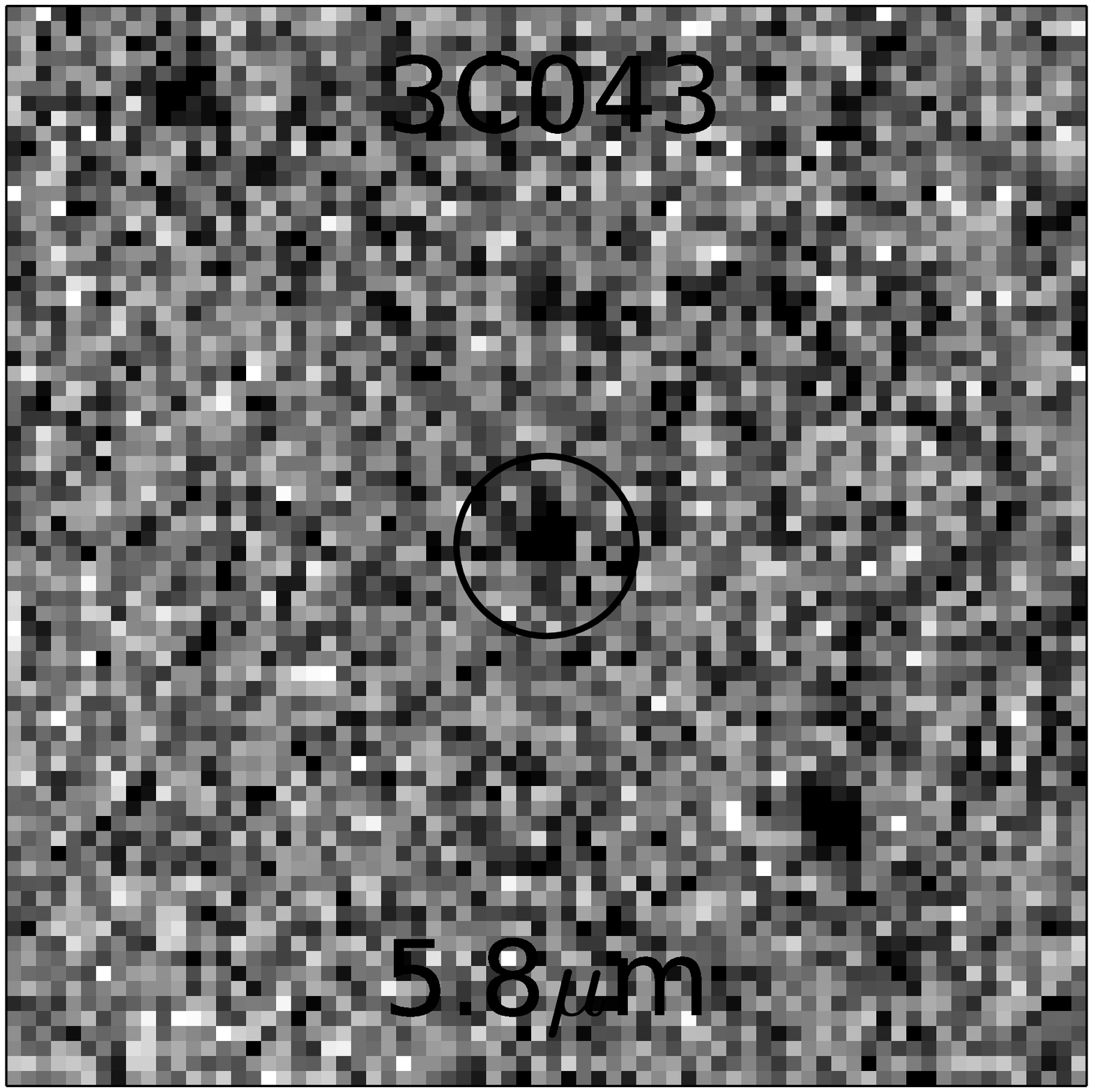}
      \includegraphics[width=1.5cm]{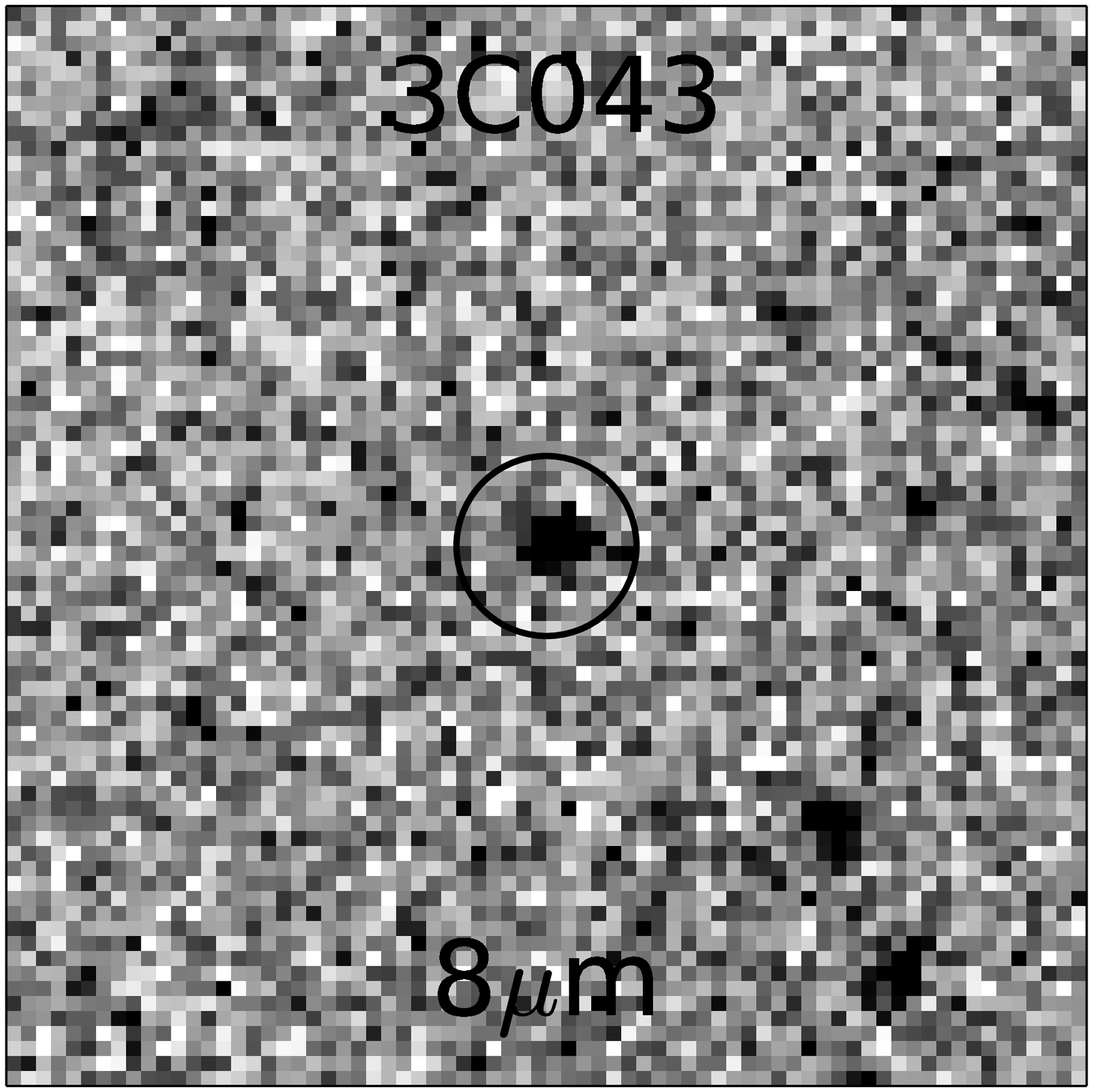}
      \includegraphics[width=1.5cm]{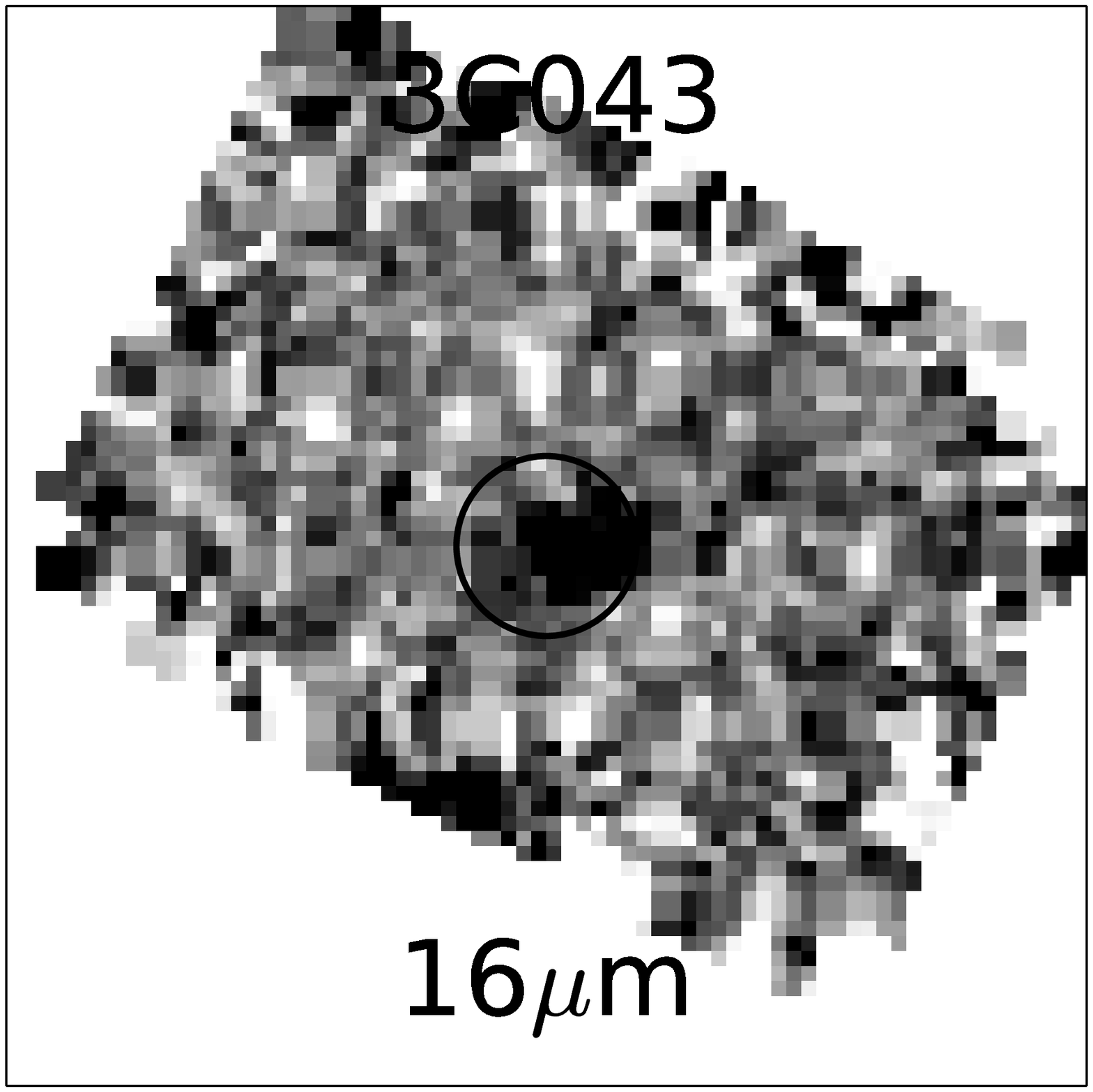}
      \includegraphics[width=1.5cm]{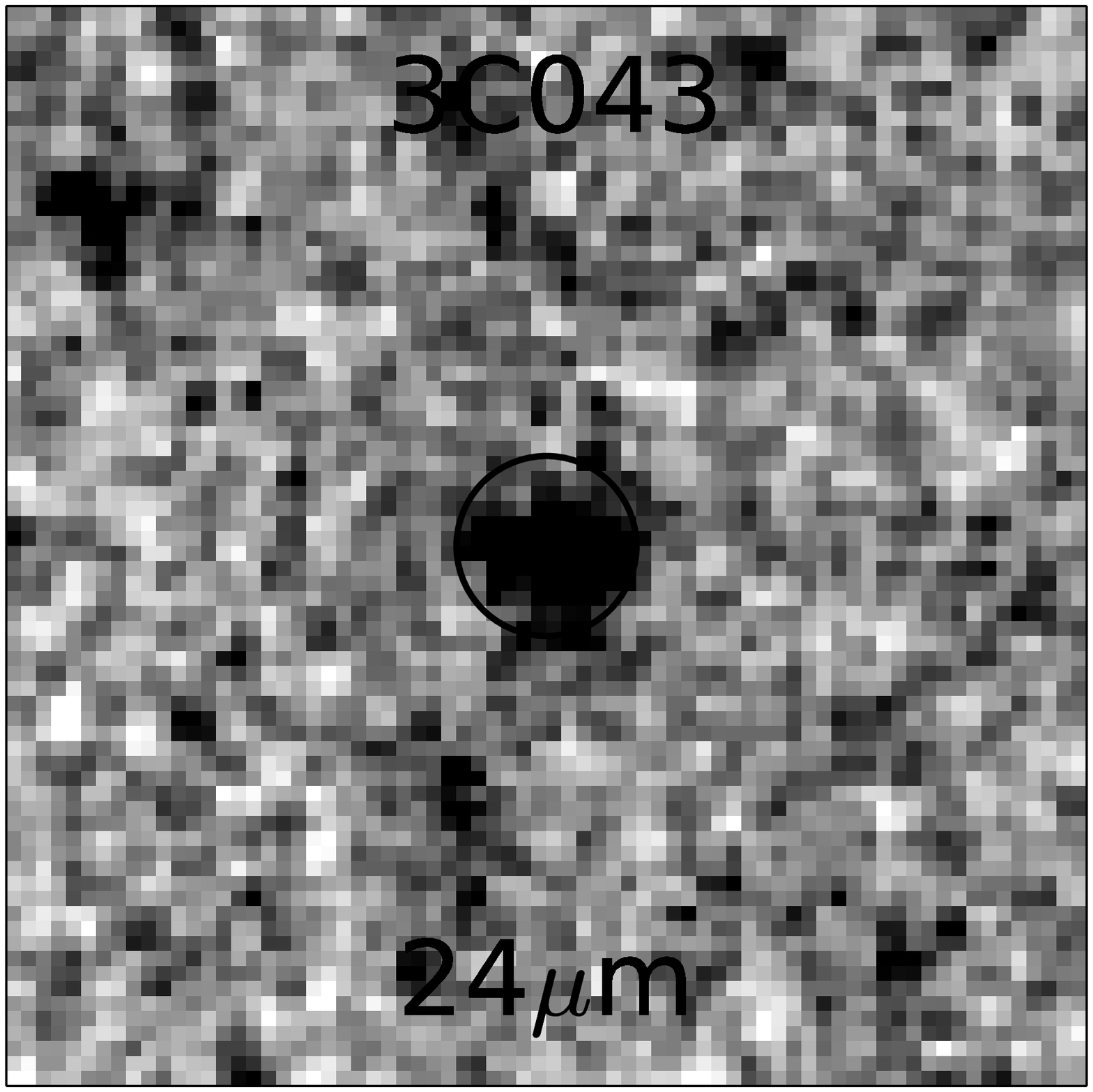}
      \includegraphics[width=1.5cm]{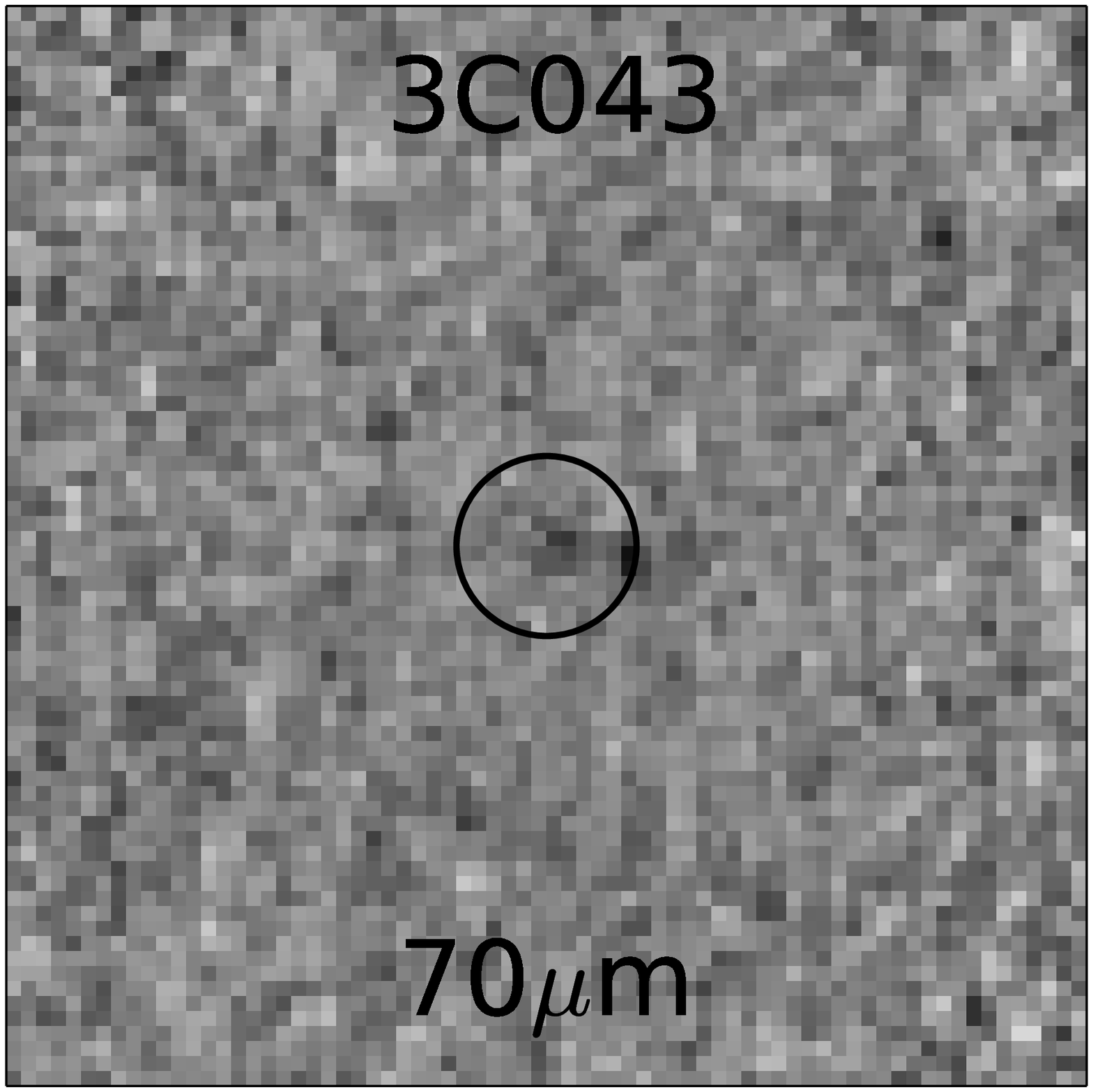}
      \includegraphics[width=1.5cm]{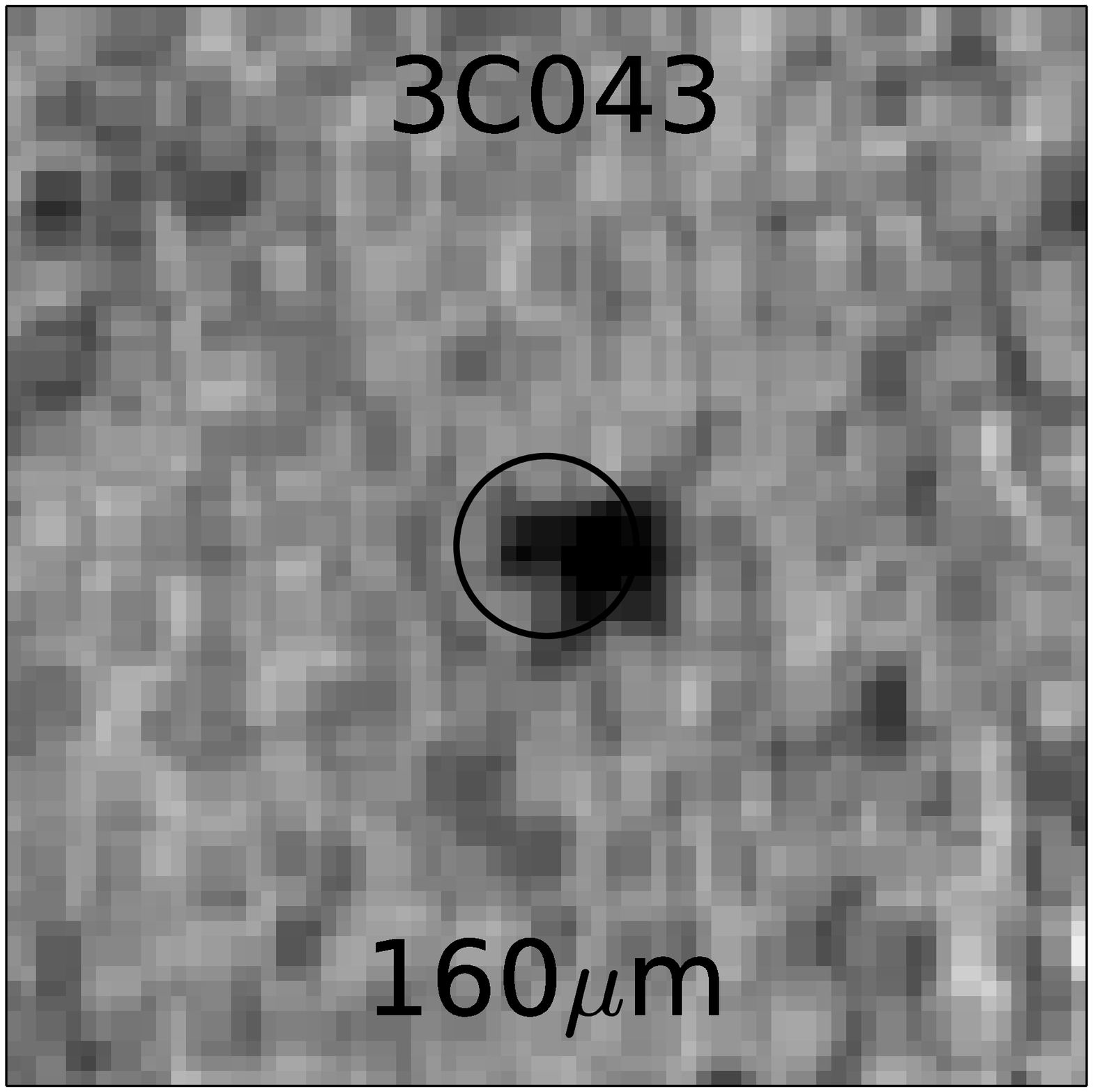}
      \includegraphics[width=1.5cm]{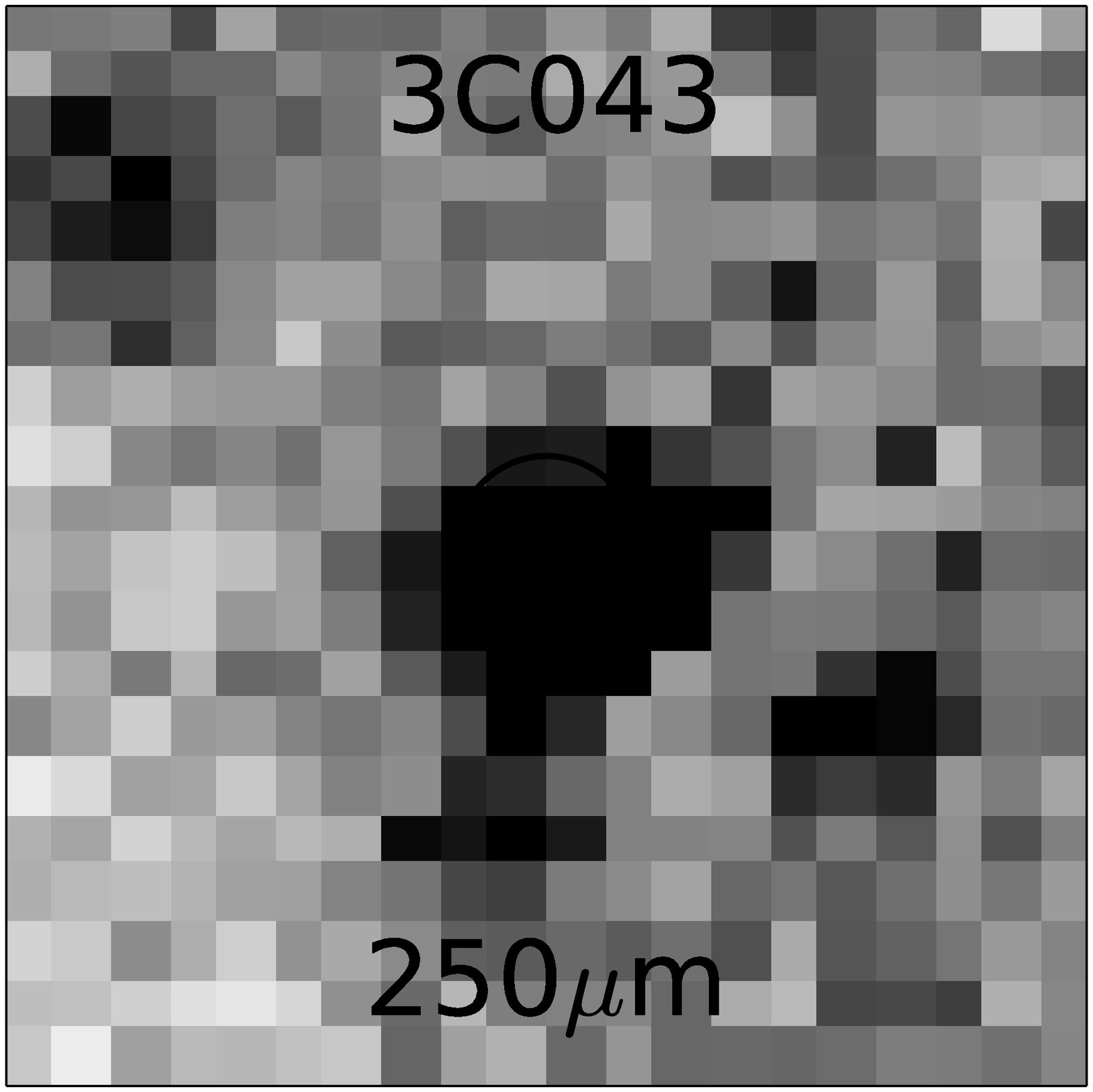}
      \includegraphics[width=1.5cm]{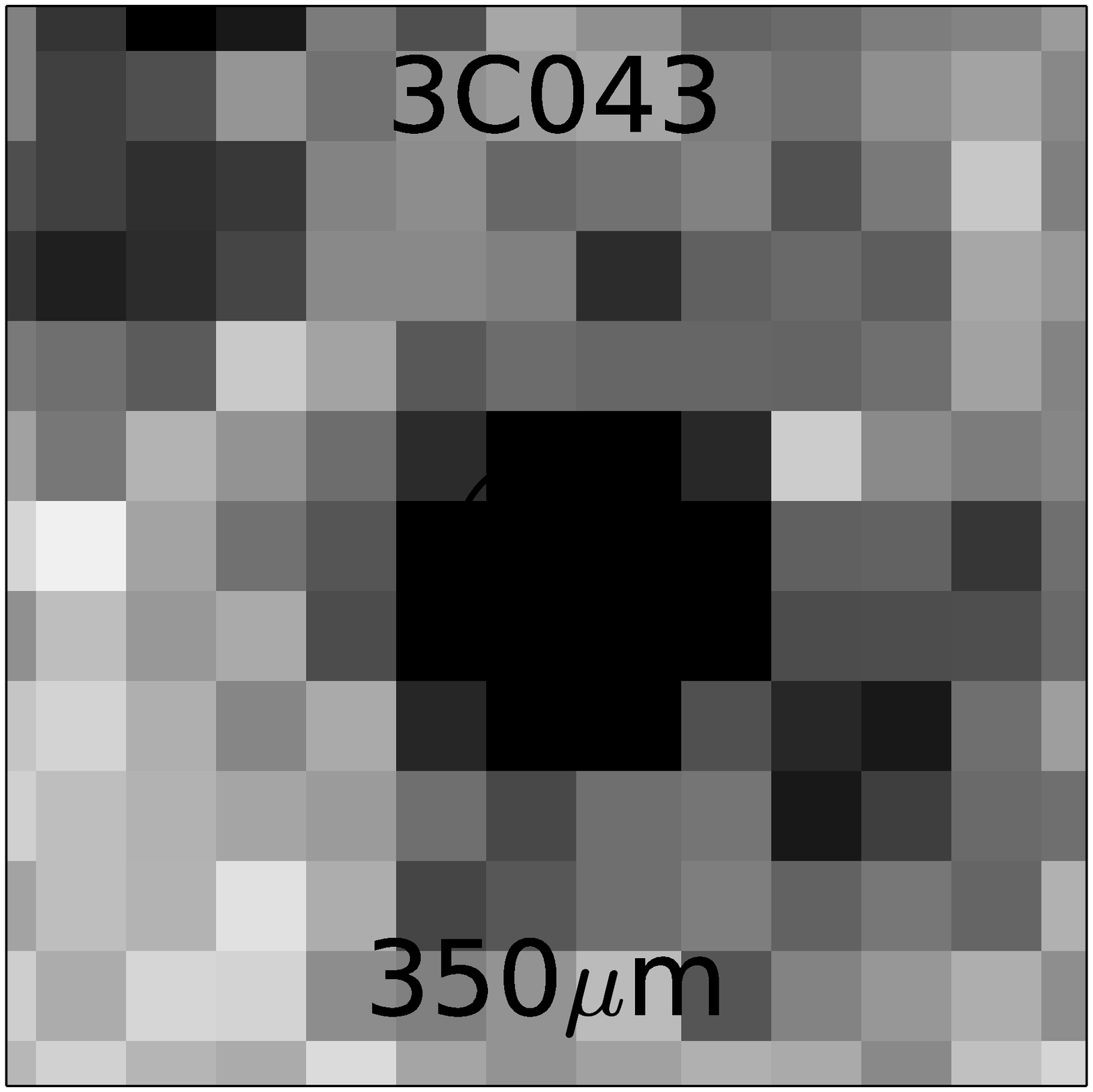}
      \includegraphics[width=1.5cm]{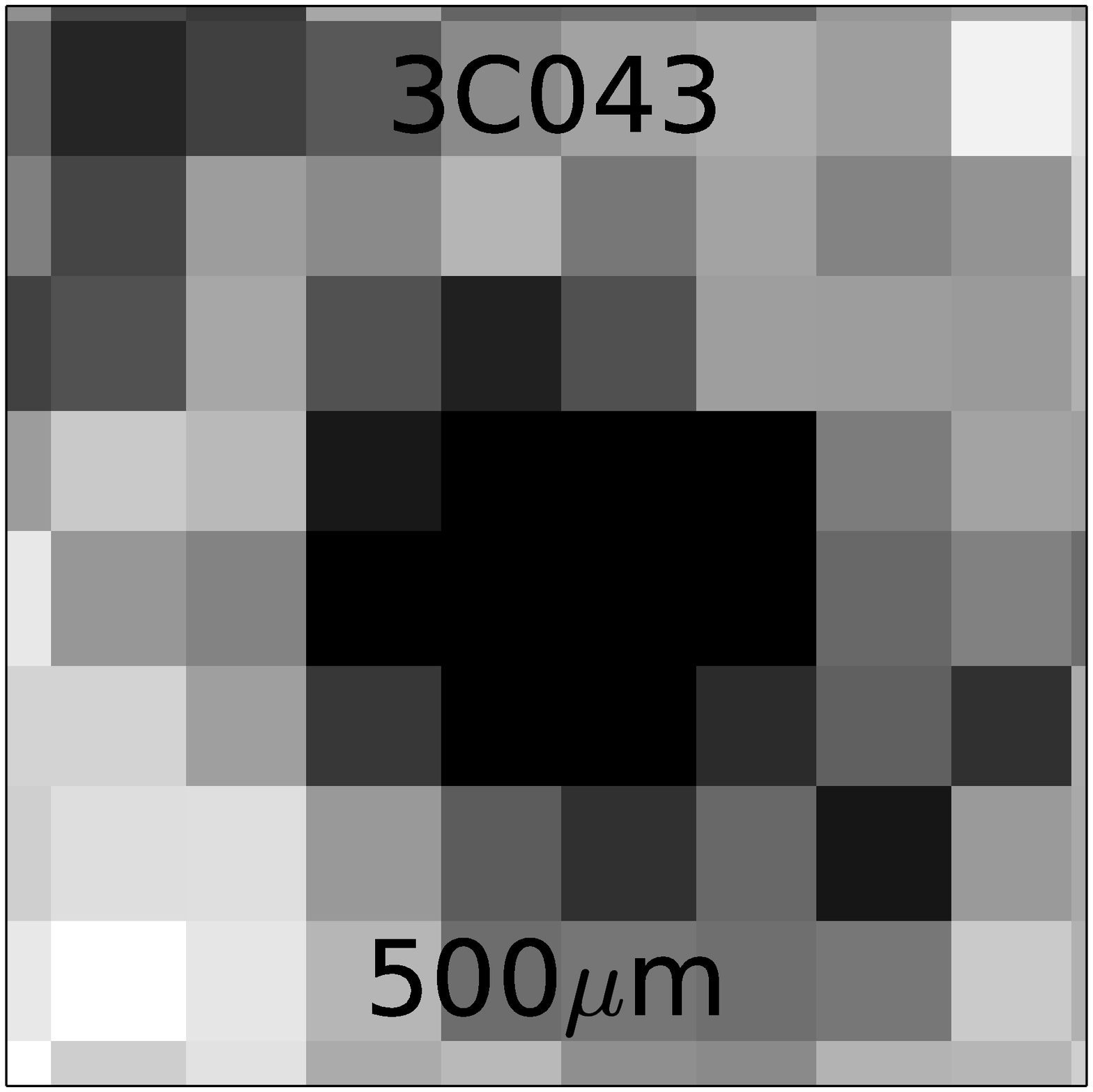}
      \\
      \includegraphics[width=1.5cm]{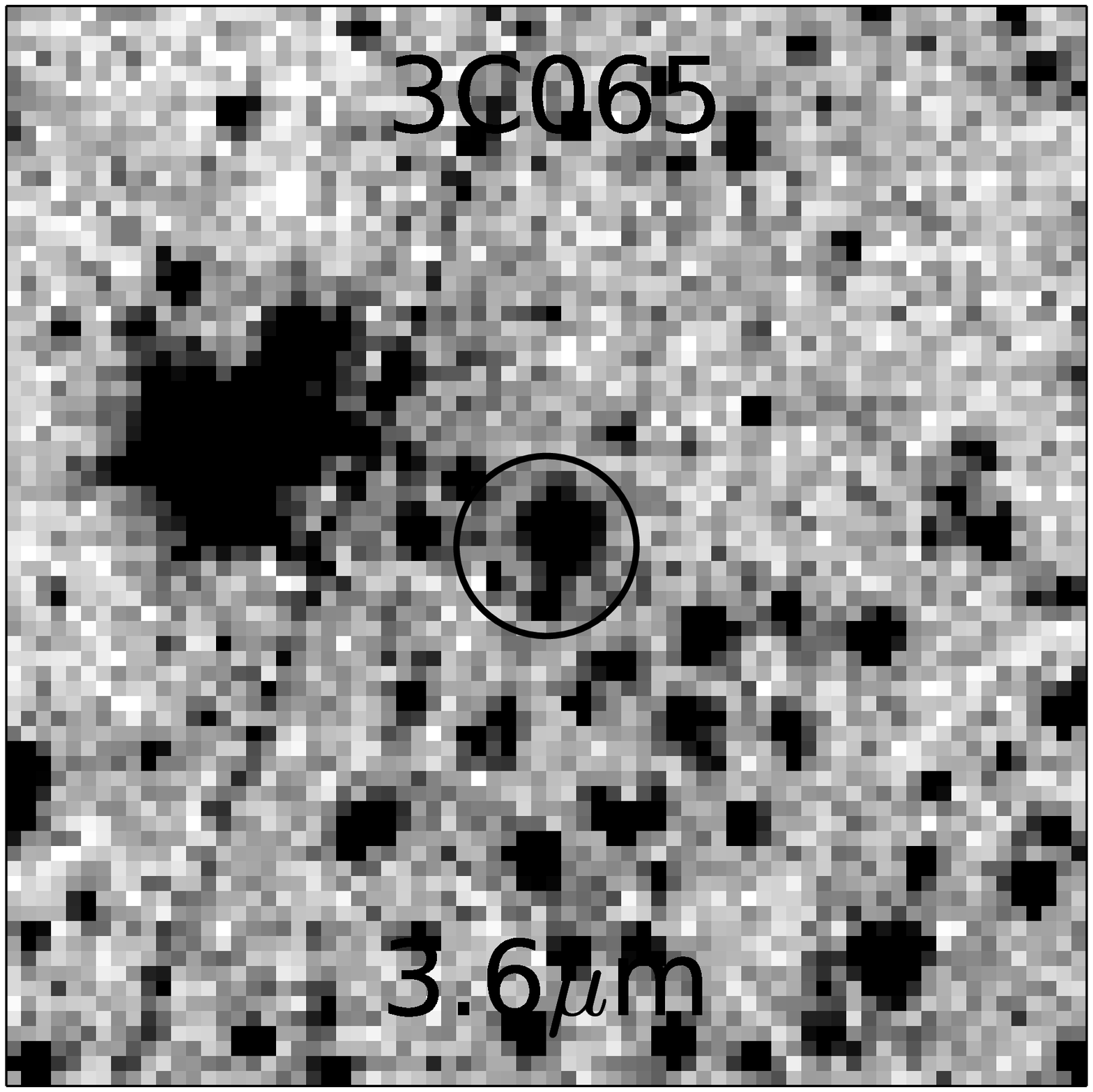}
      \includegraphics[width=1.5cm]{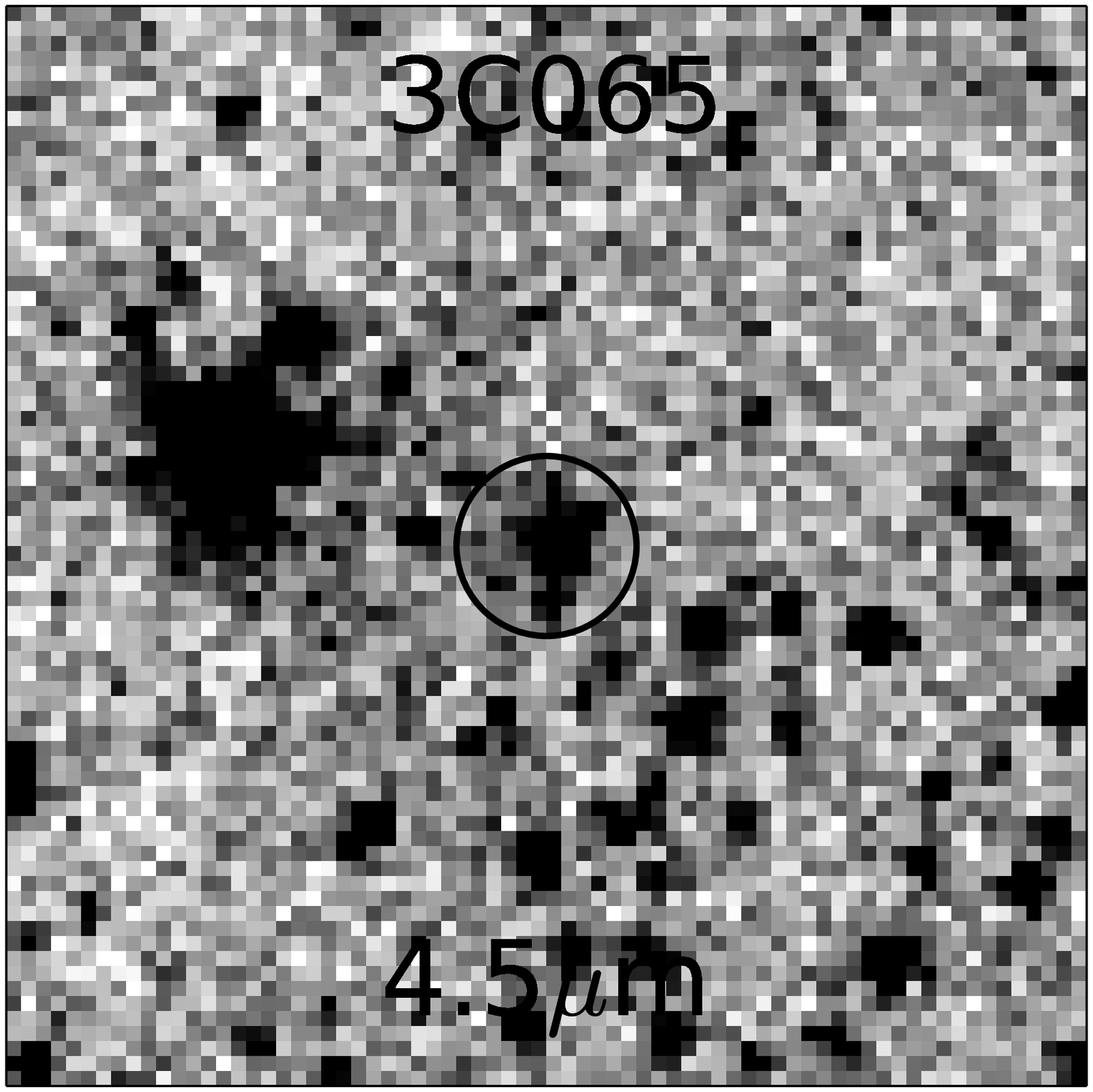}
      \includegraphics[width=1.5cm]{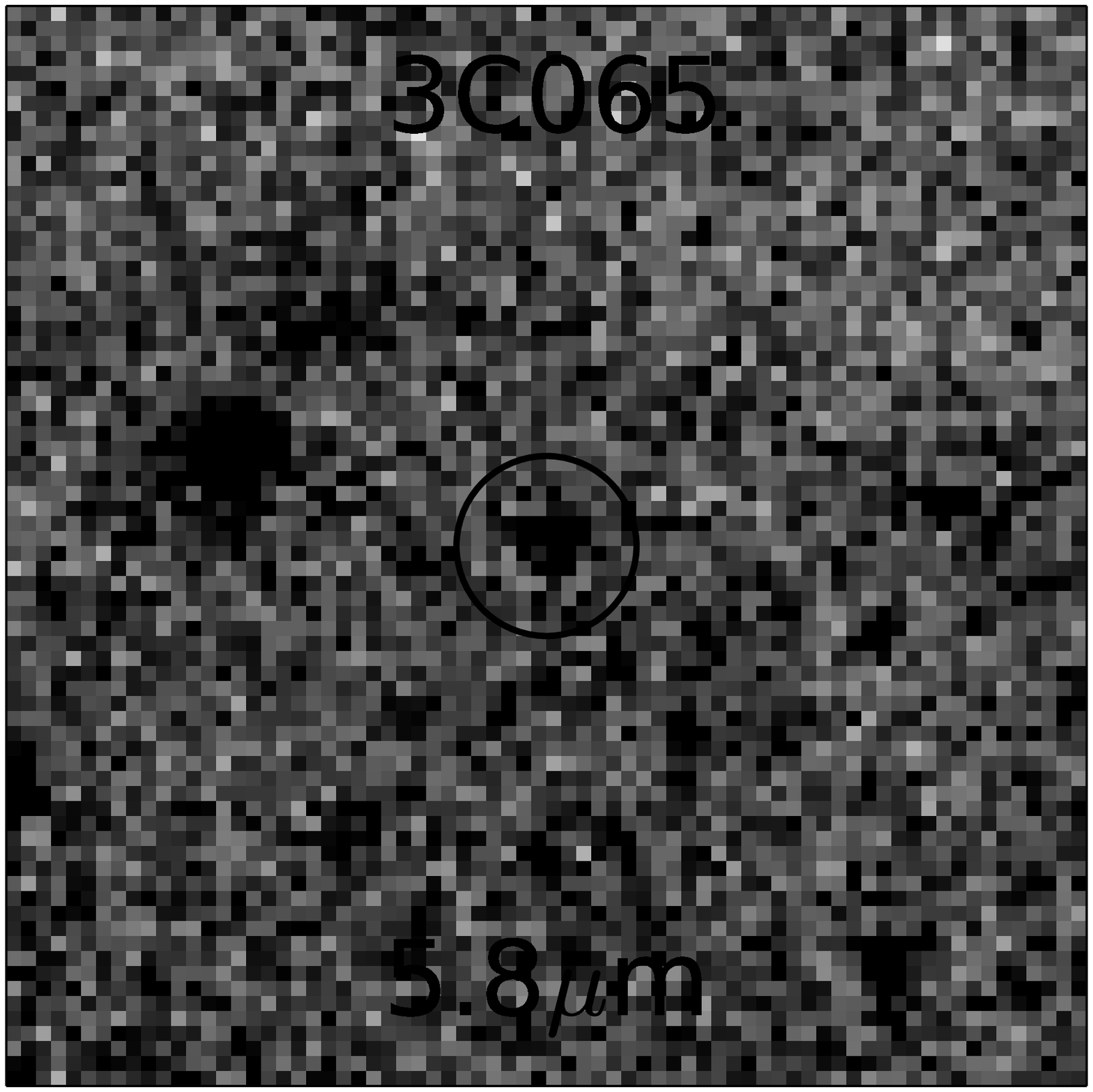}
      \includegraphics[width=1.5cm]{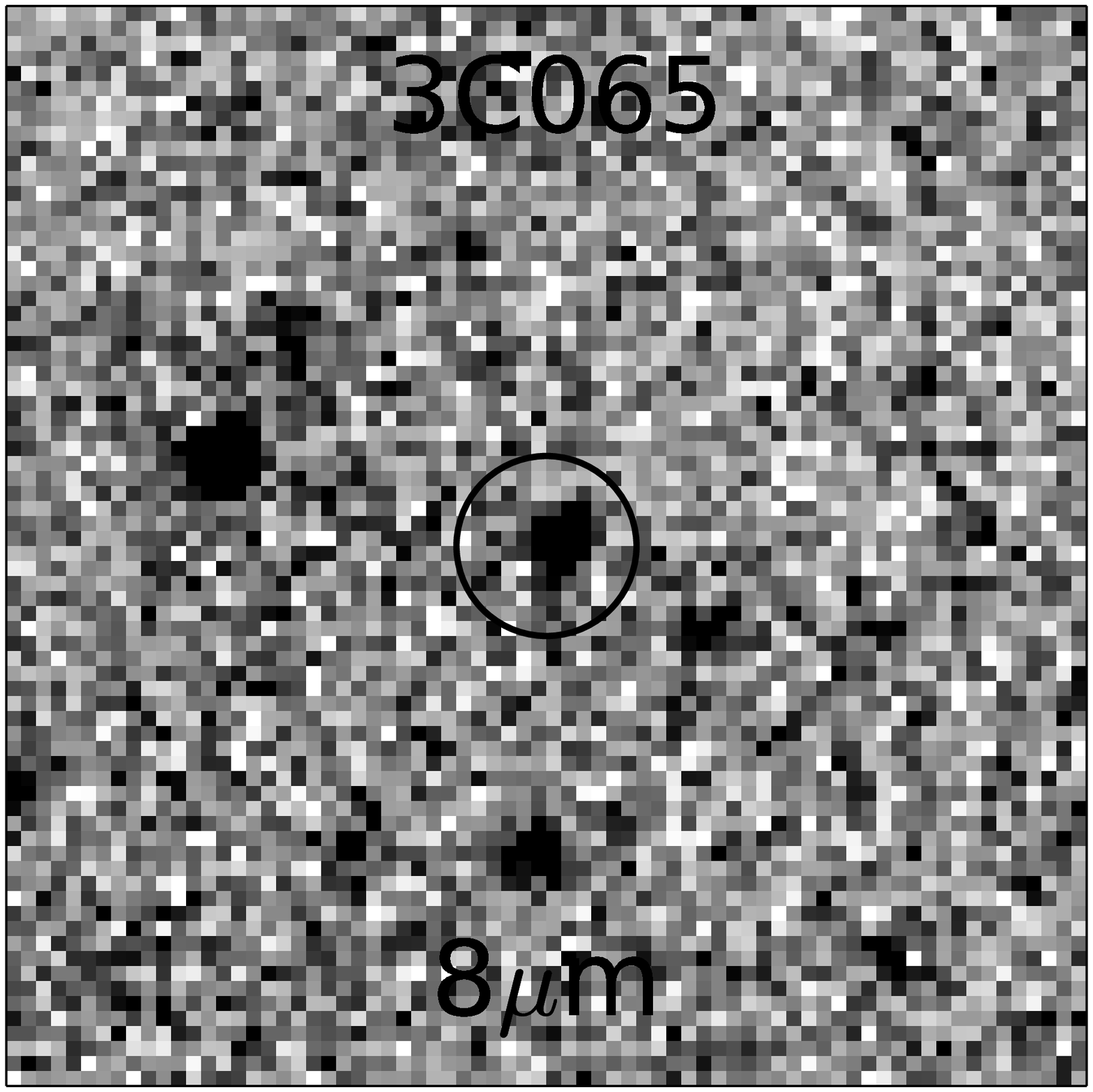}
      \includegraphics[width=1.5cm]{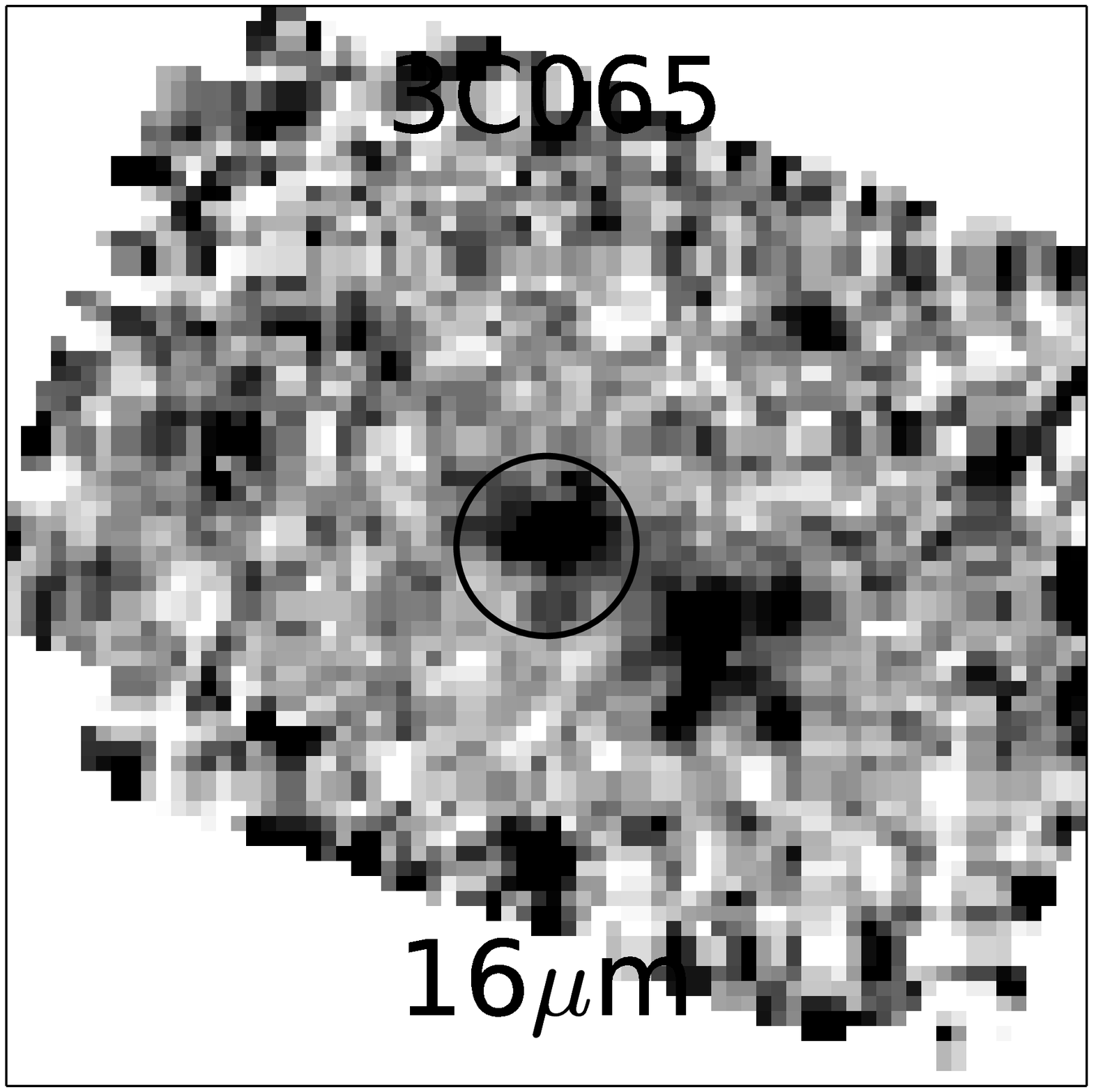}
      \includegraphics[width=1.5cm]{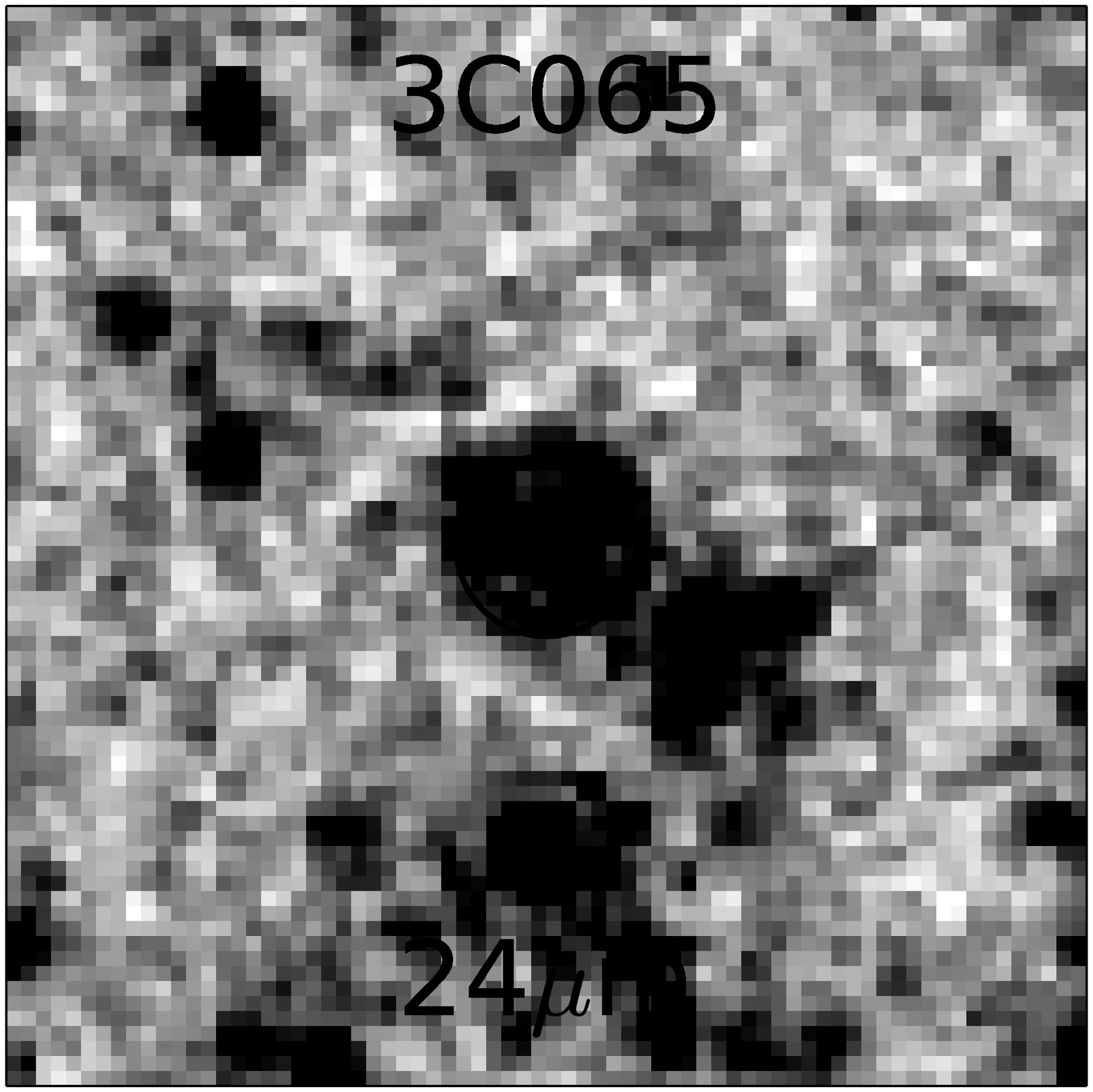}
      \includegraphics[width=1.5cm]{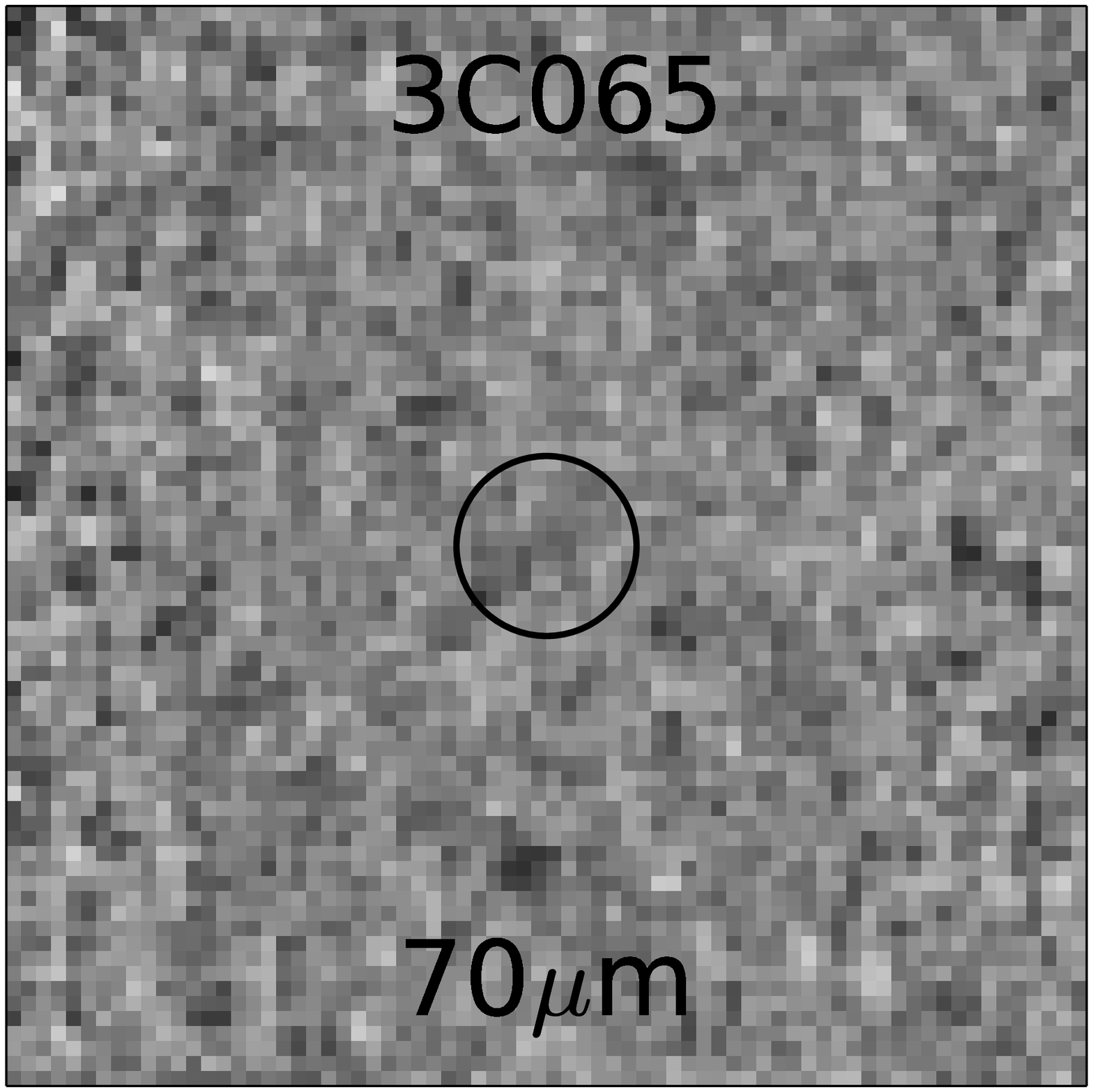}
      \includegraphics[width=1.5cm]{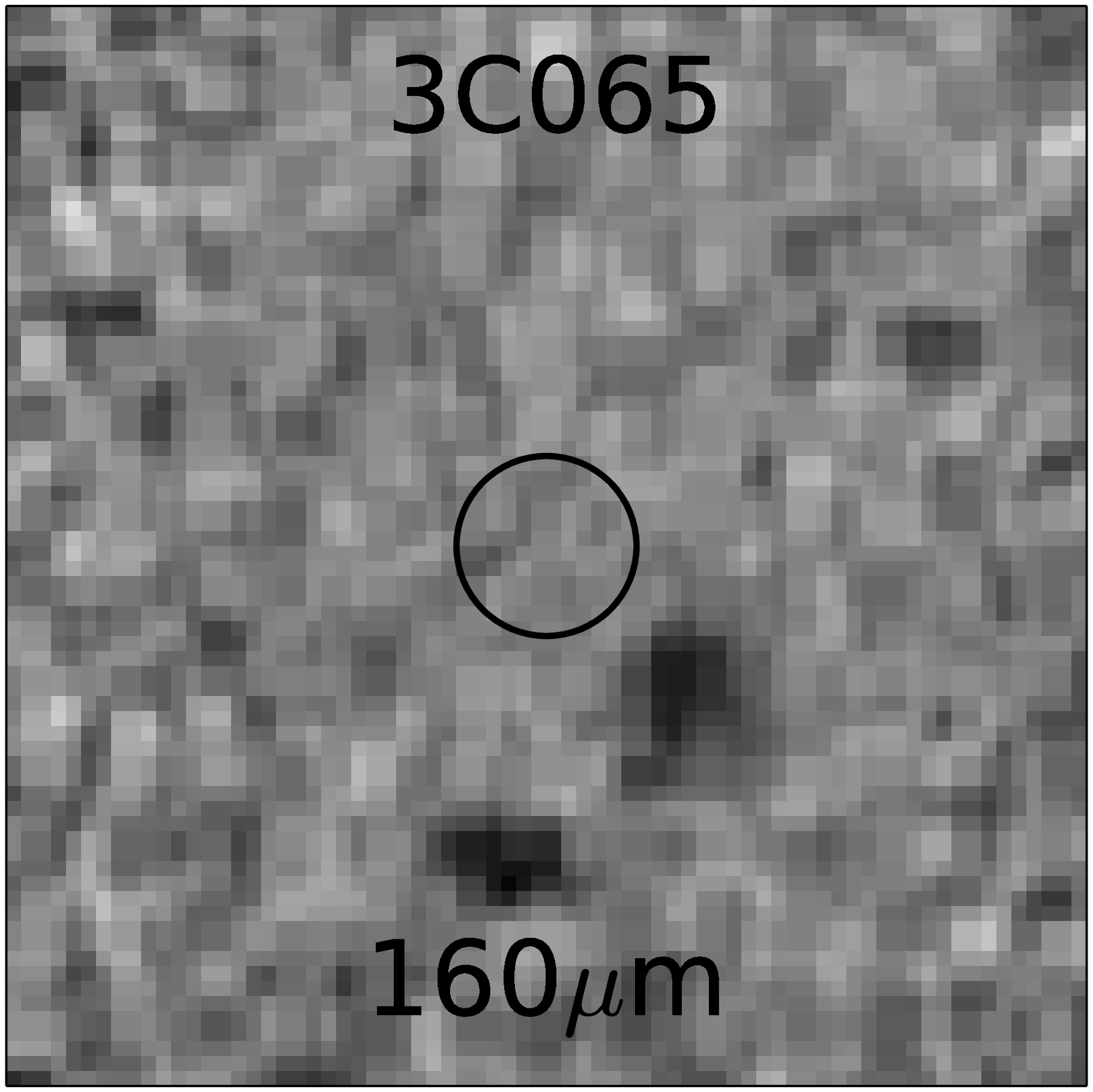}
      \includegraphics[width=1.5cm]{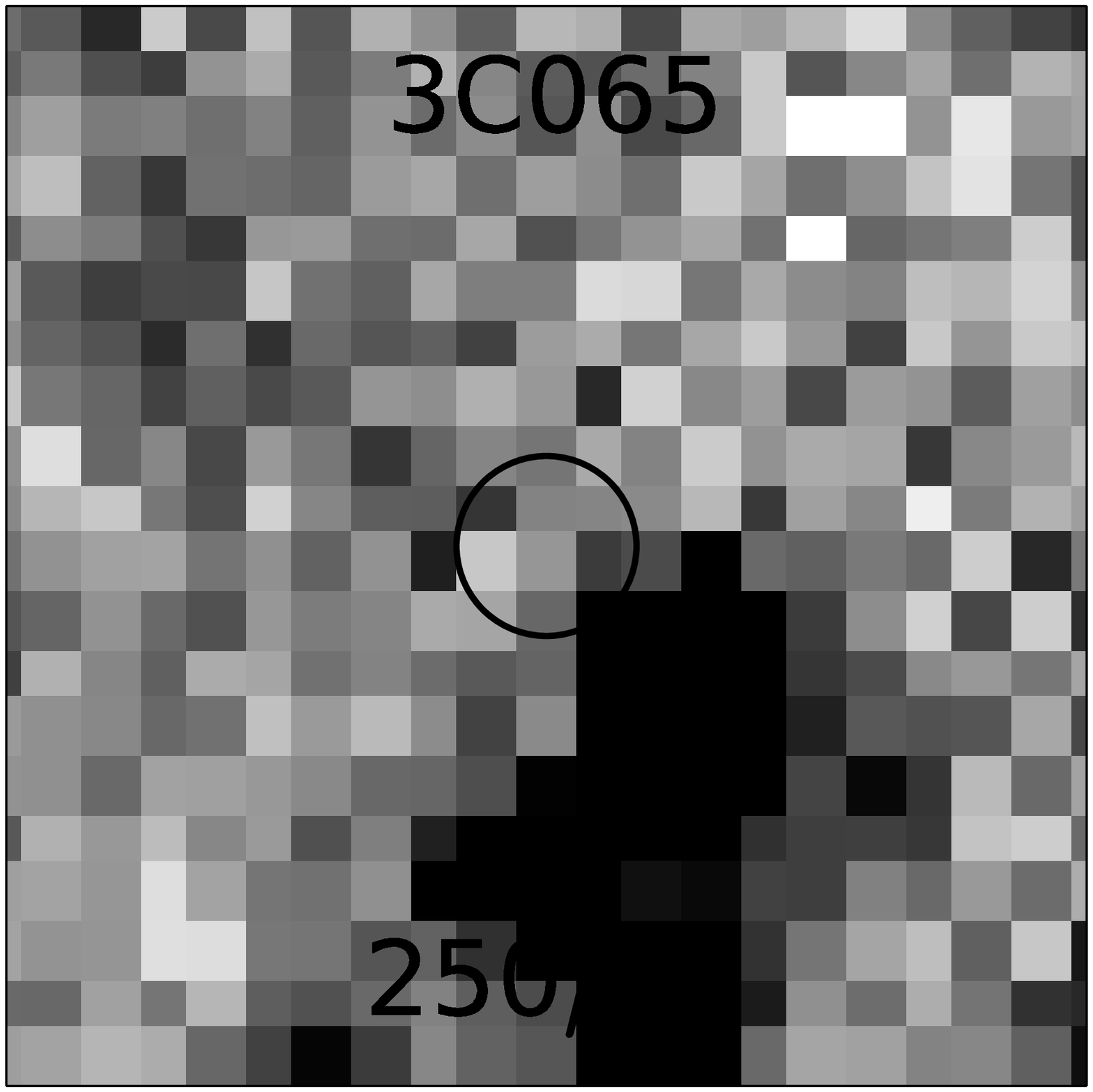}
      \includegraphics[width=1.5cm]{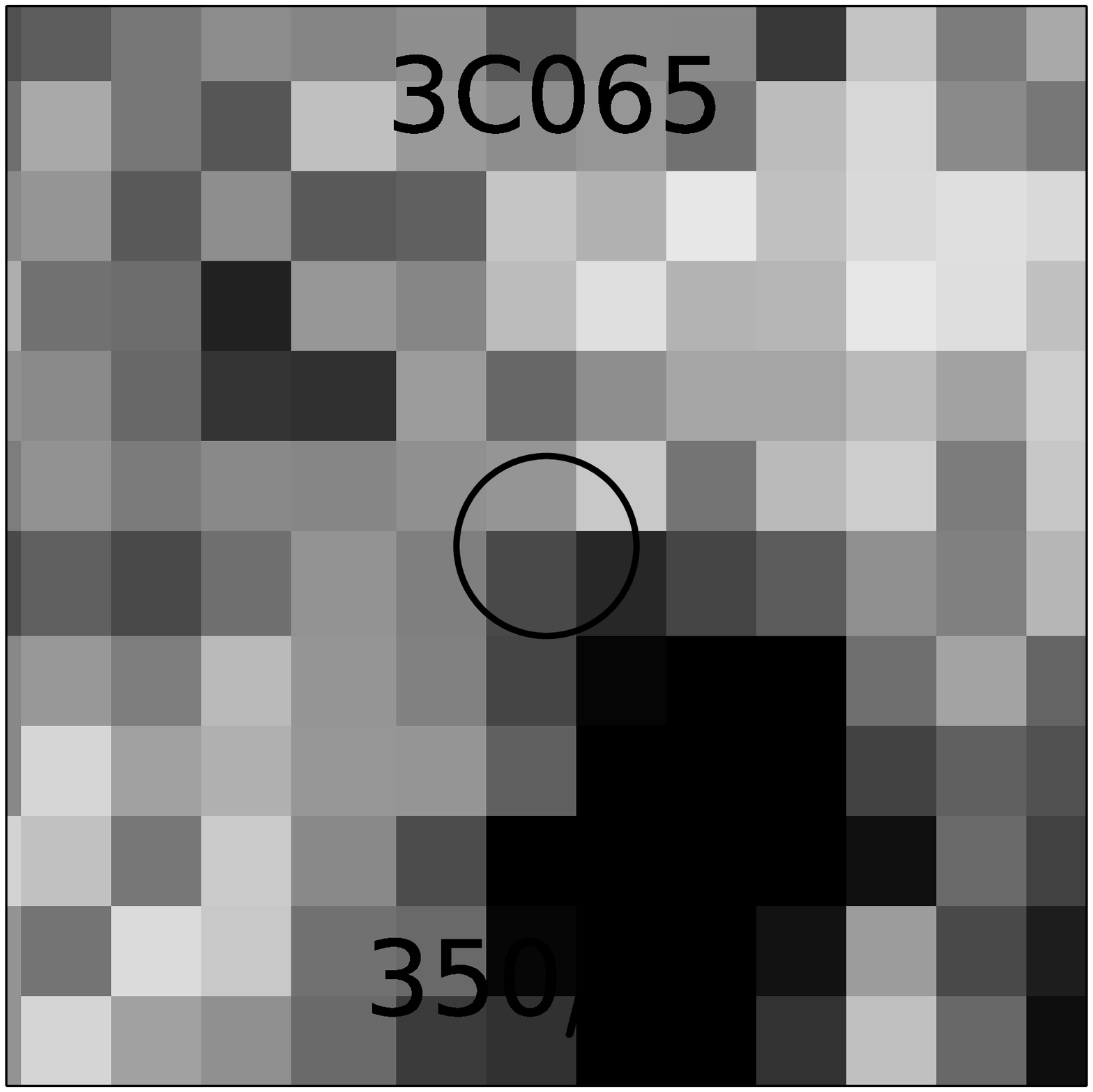}
      \includegraphics[width=1.5cm]{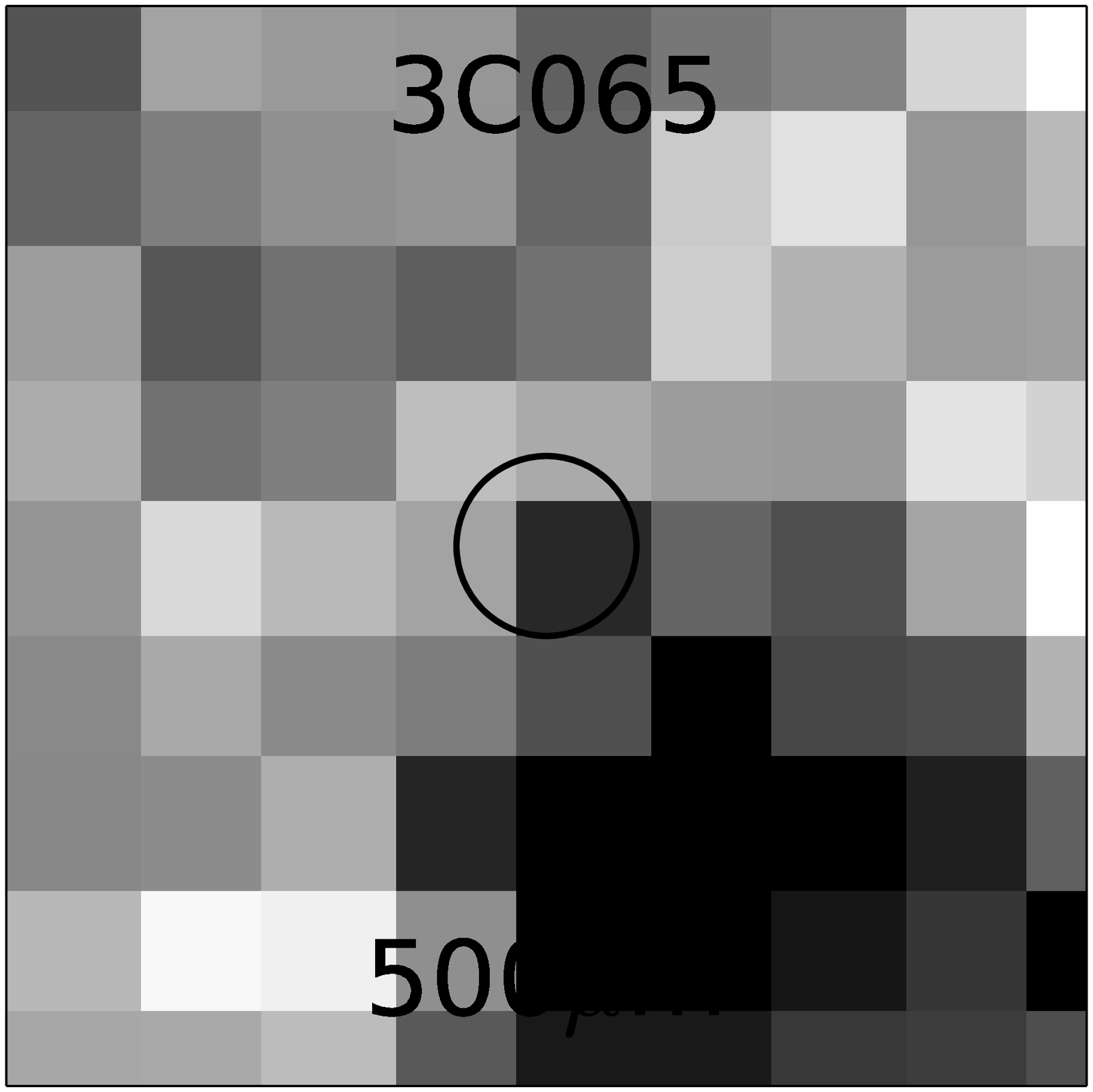}
      \\
      \includegraphics[width=1.5cm]{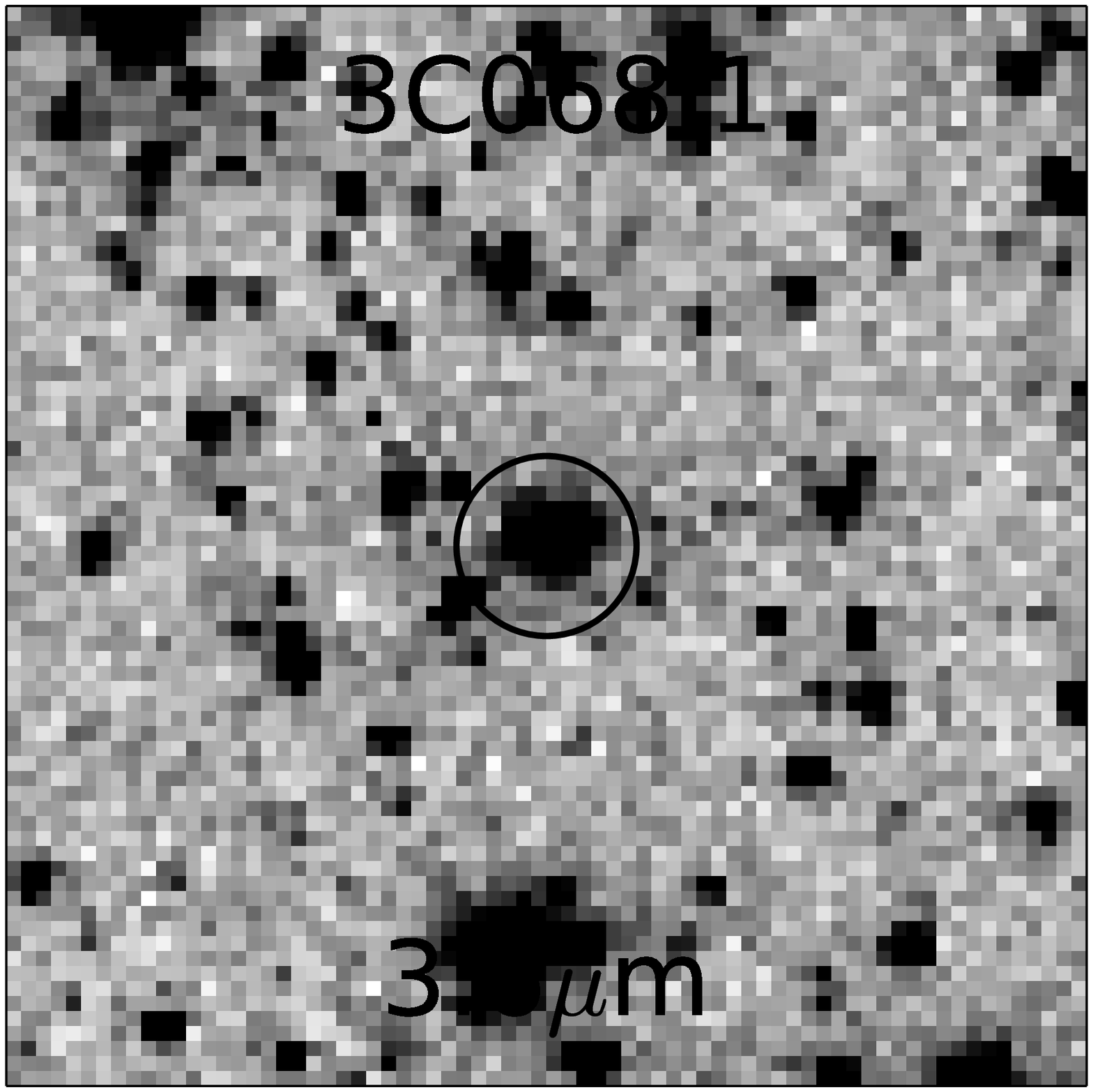}
      \includegraphics[width=1.5cm]{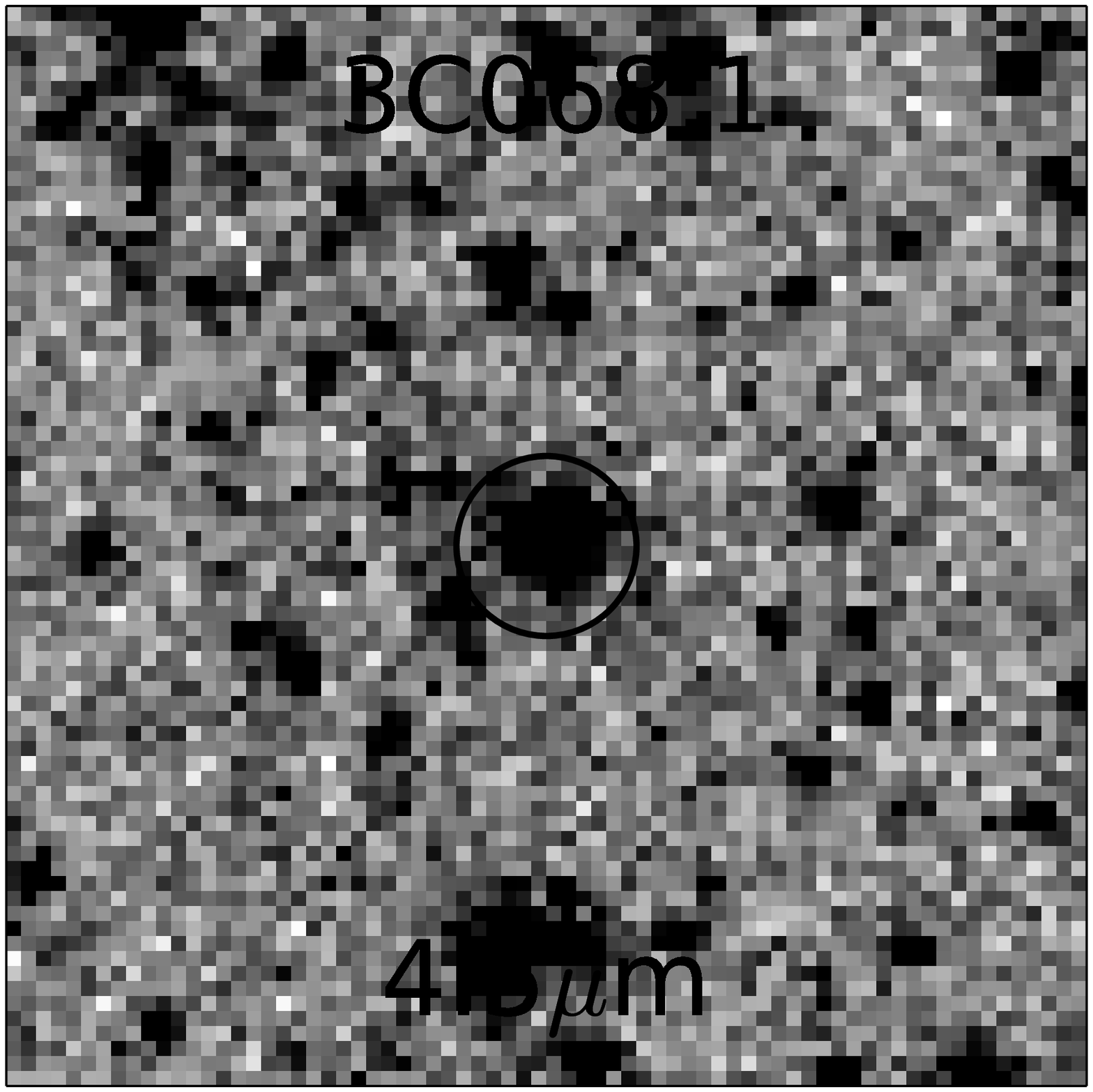}
      \includegraphics[width=1.5cm]{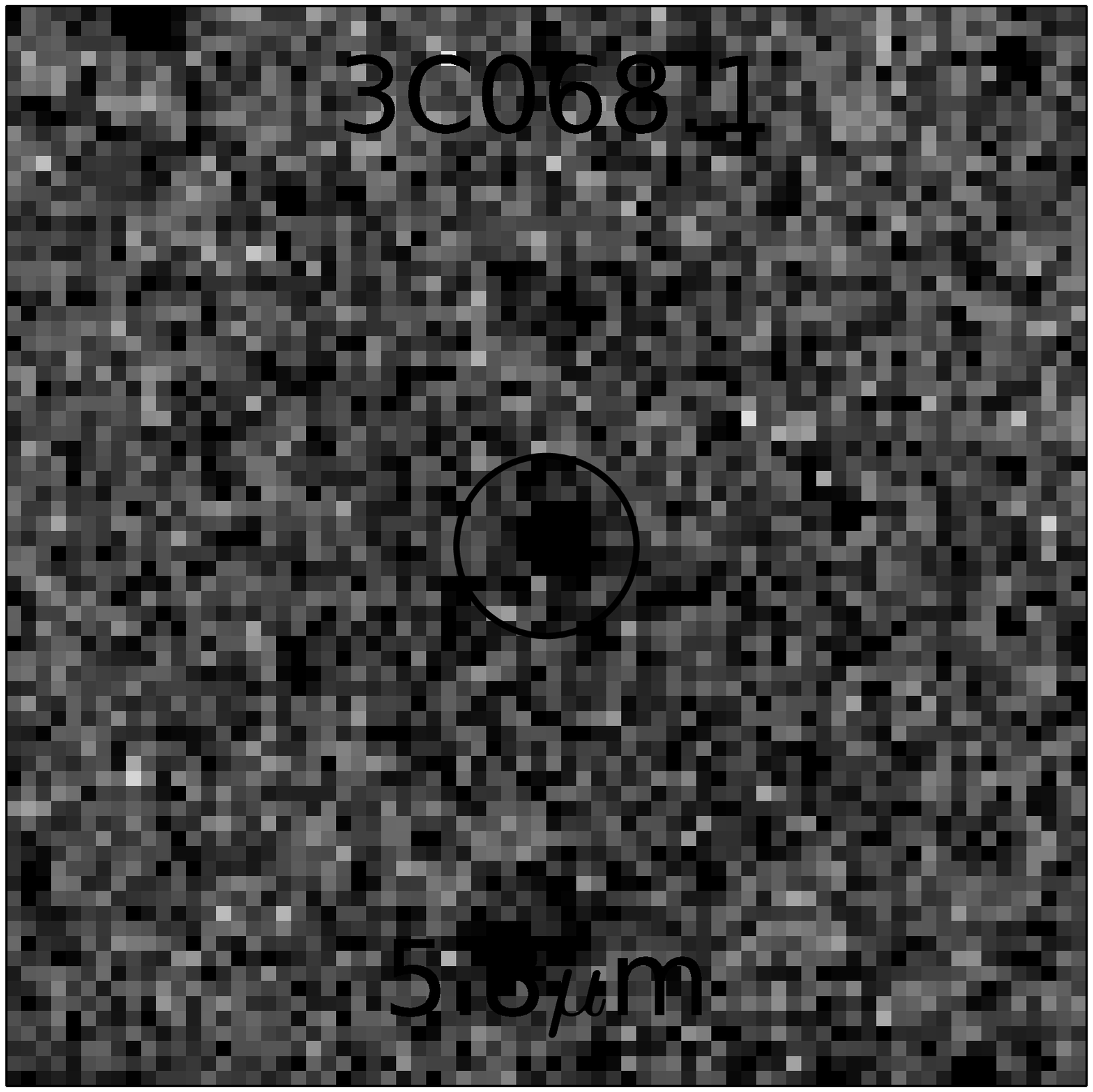}
      \includegraphics[width=1.5cm]{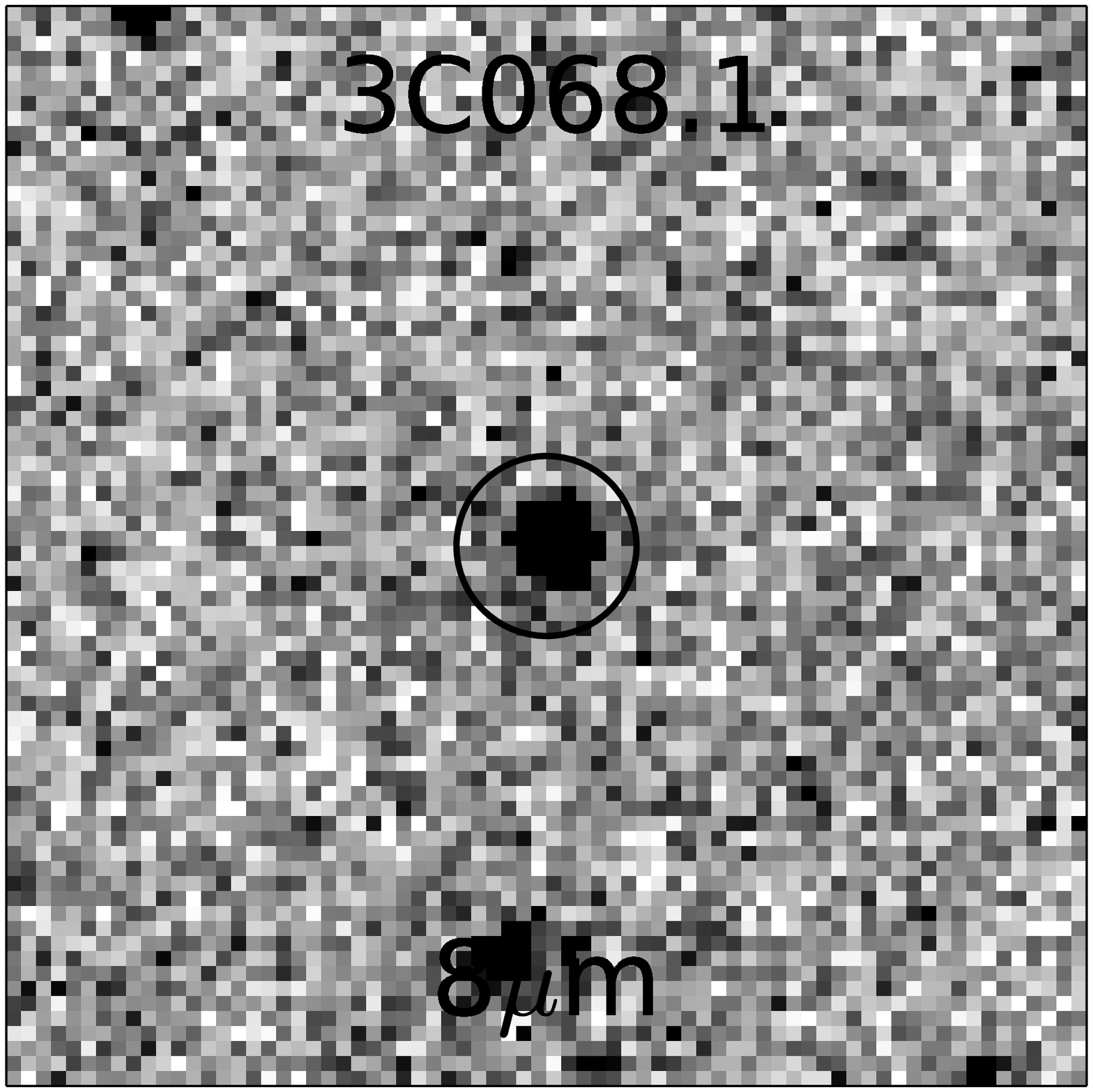}
      \includegraphics[width=1.5cm]{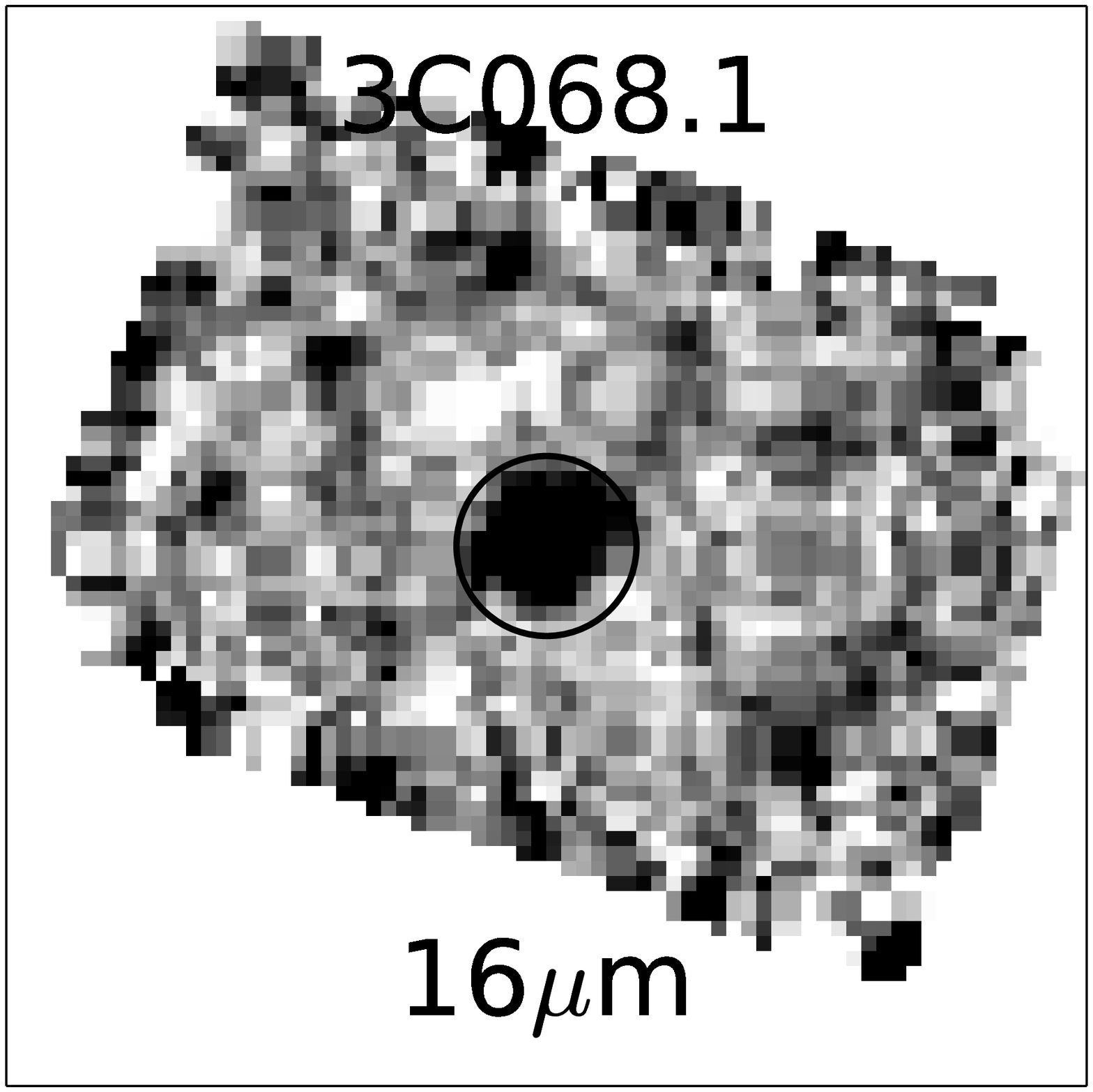}
      \includegraphics[width=1.5cm]{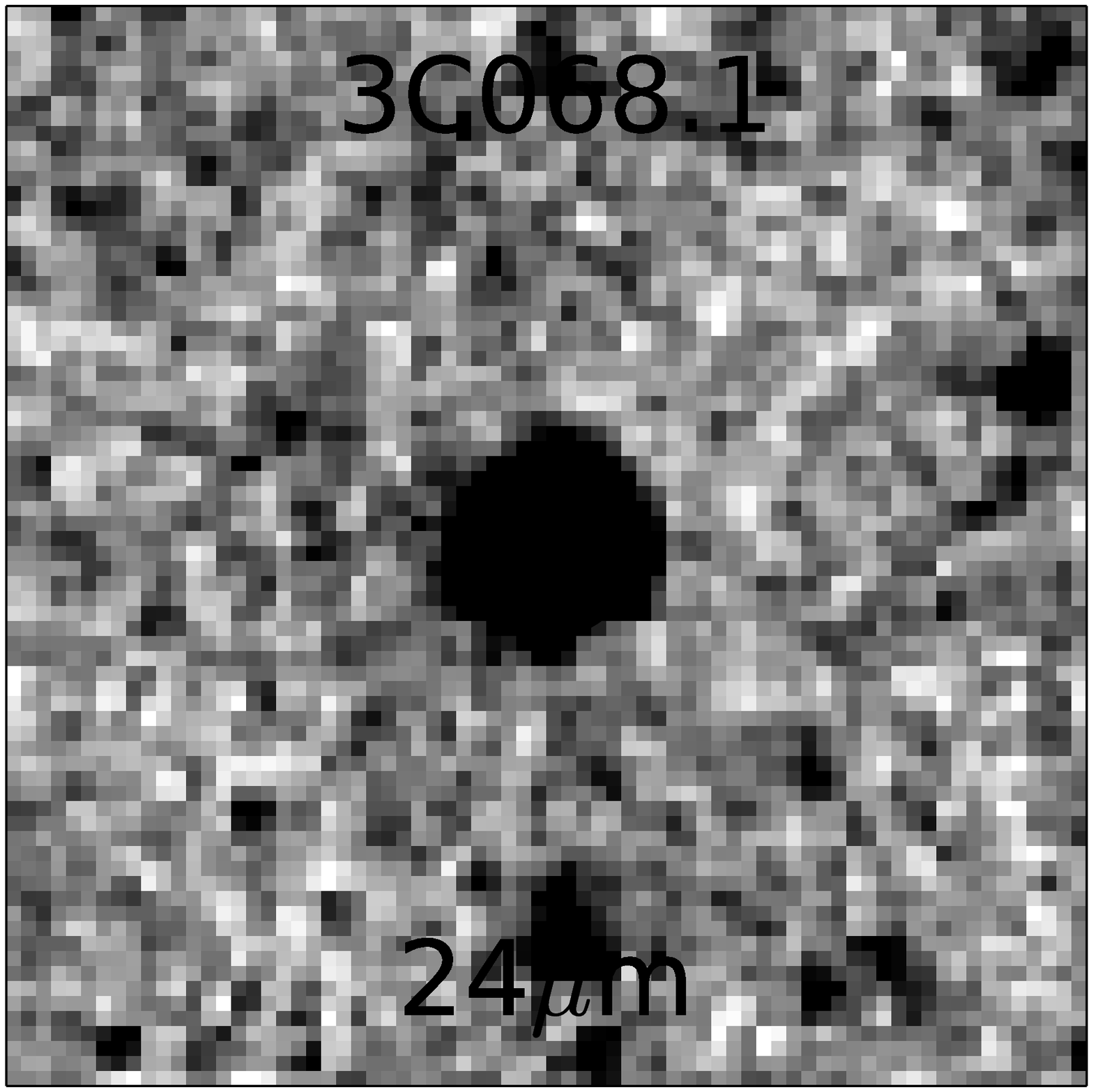}
      \includegraphics[width=1.5cm]{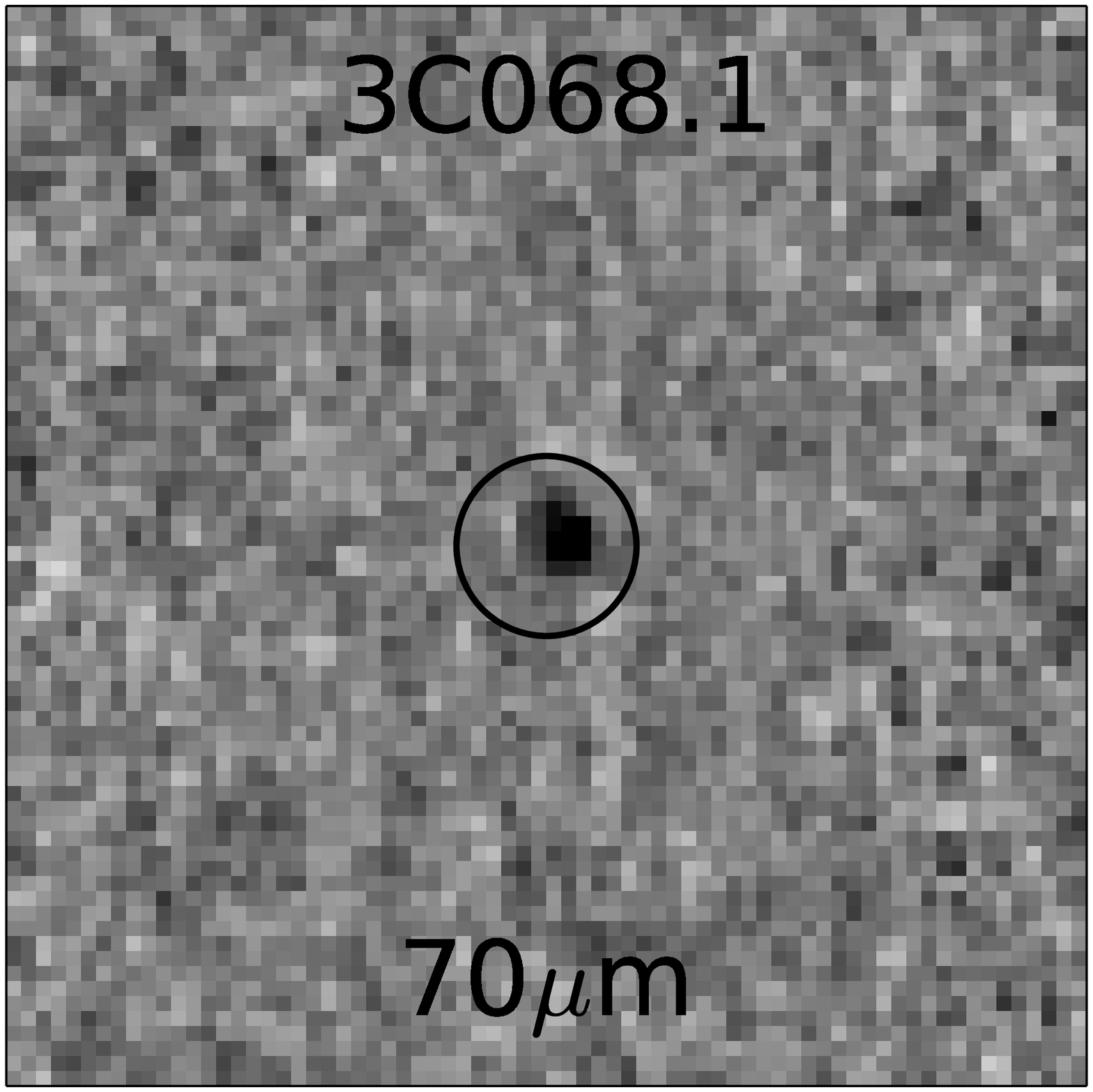}
      \includegraphics[width=1.5cm]{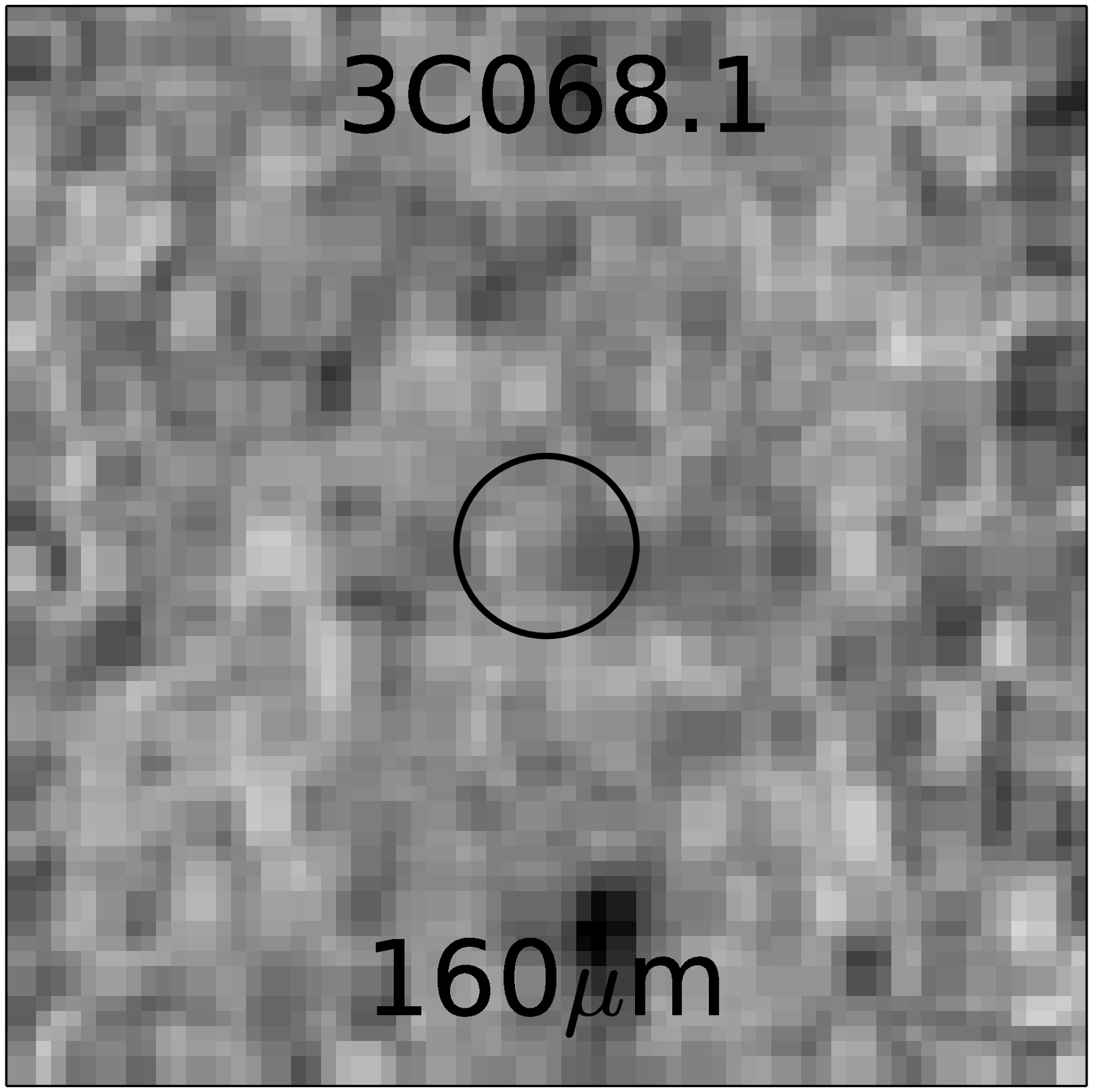}
      \includegraphics[width=1.5cm]{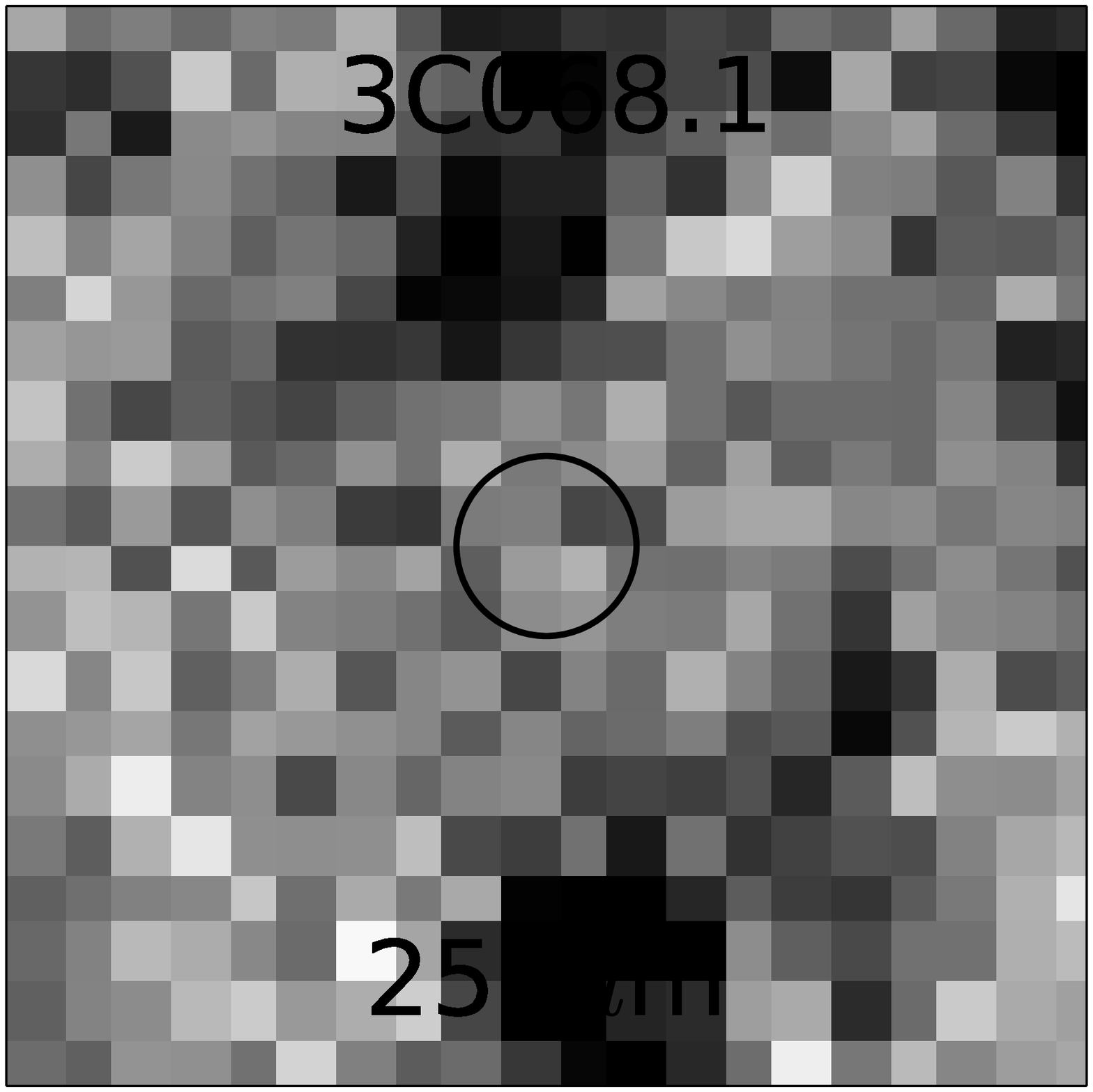}
      \includegraphics[width=1.5cm]{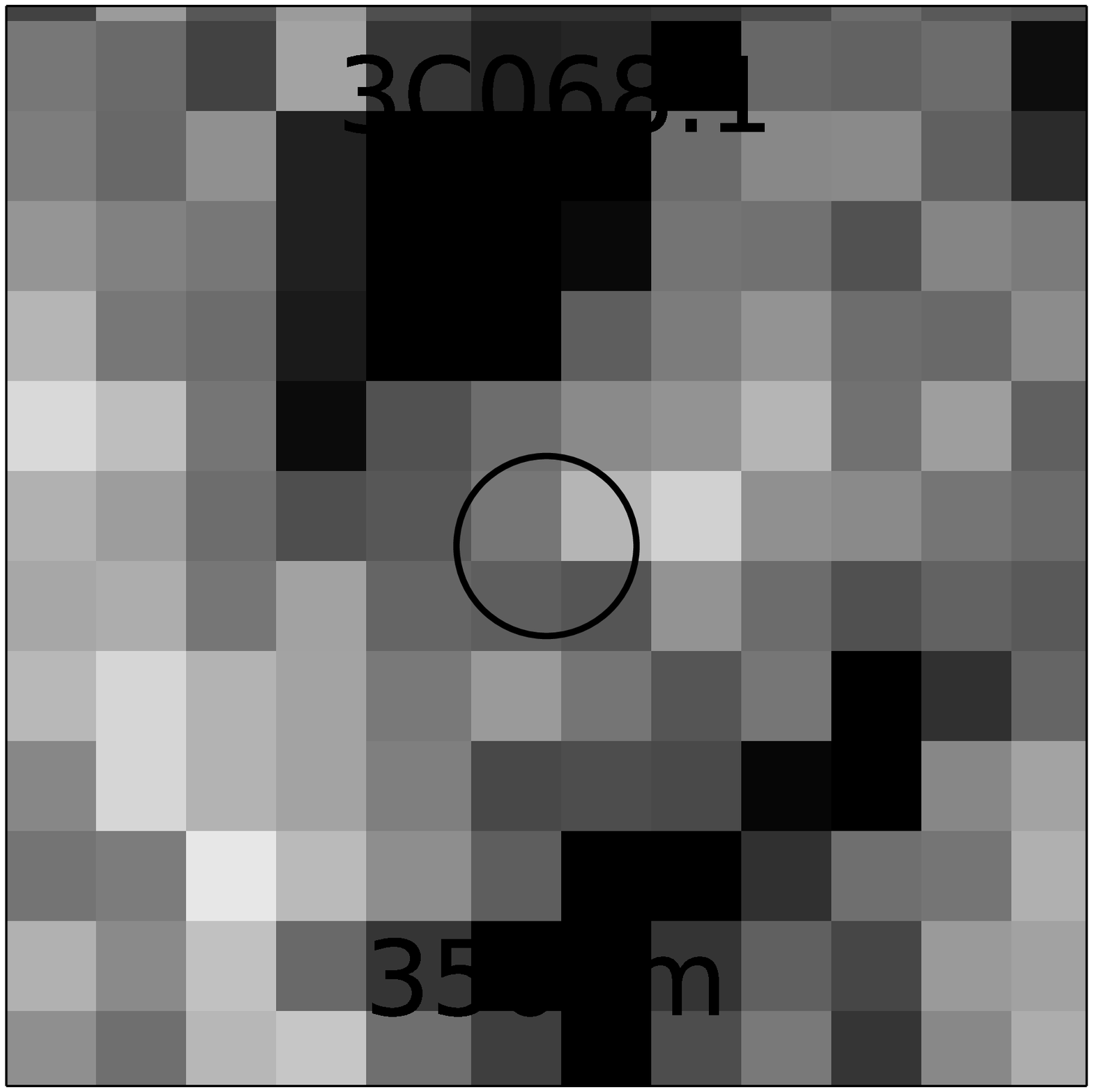}
      \includegraphics[width=1.5cm]{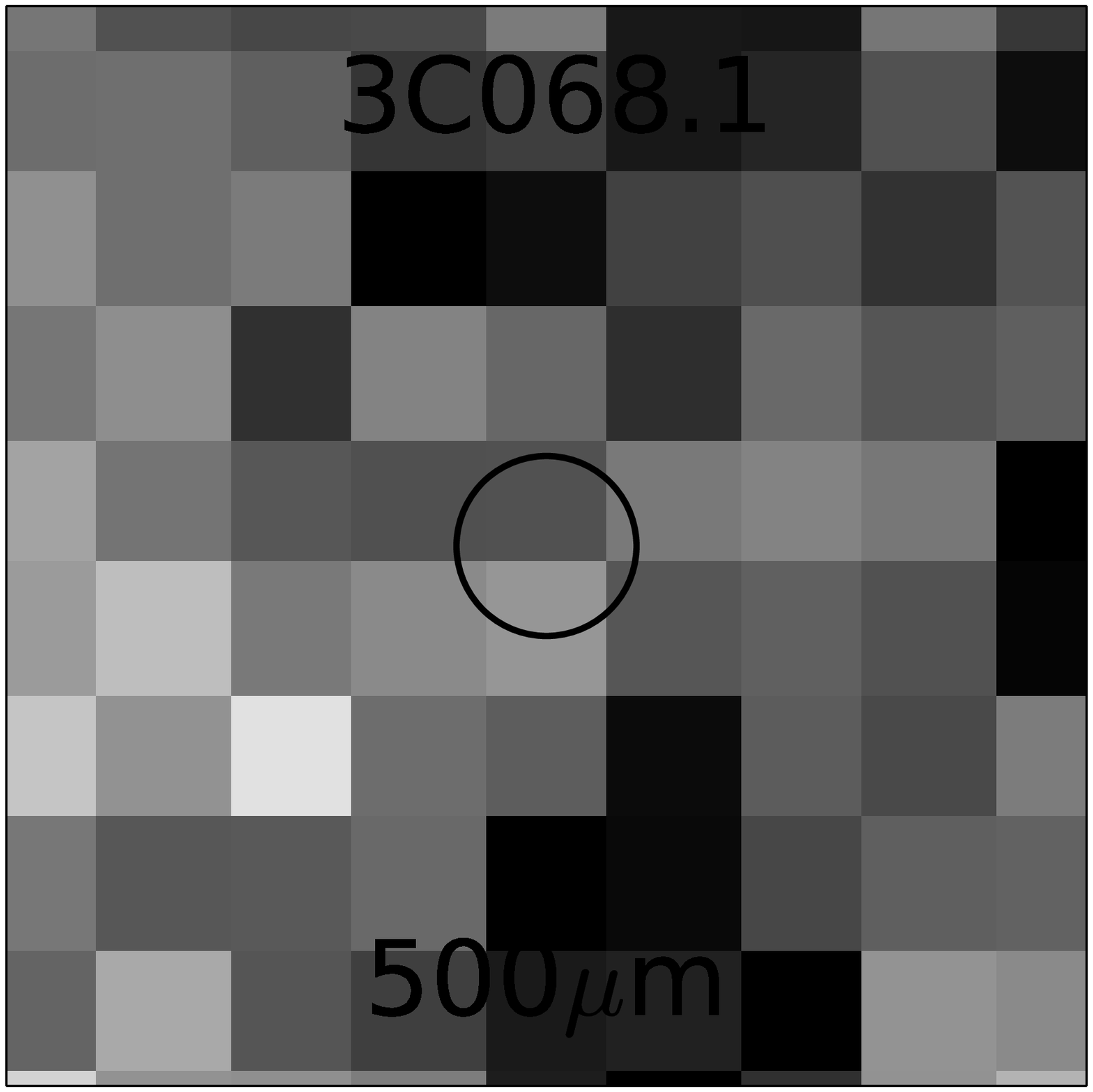}
      \\
      \includegraphics[width=1.5cm]{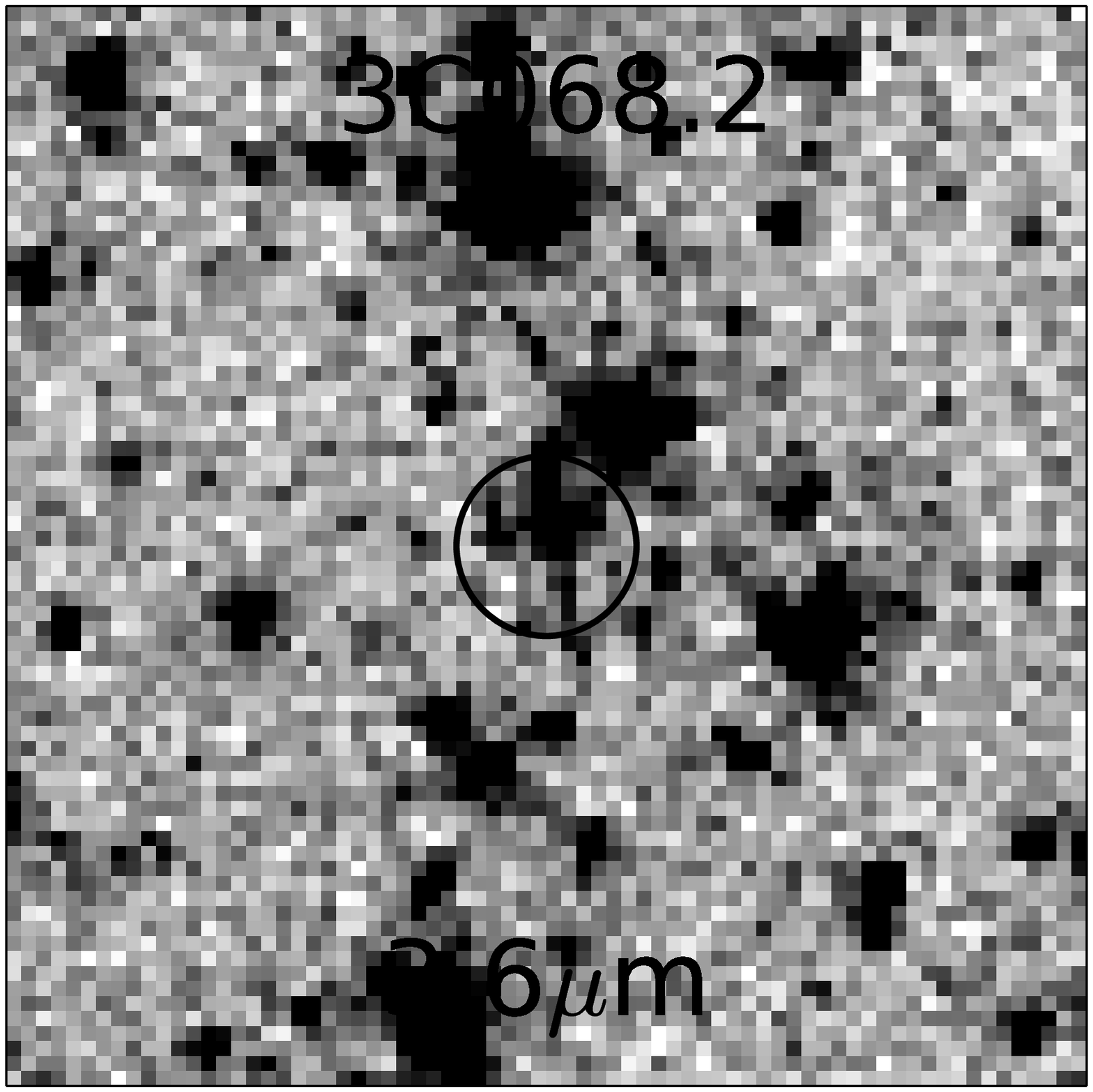}
      \includegraphics[width=1.5cm]{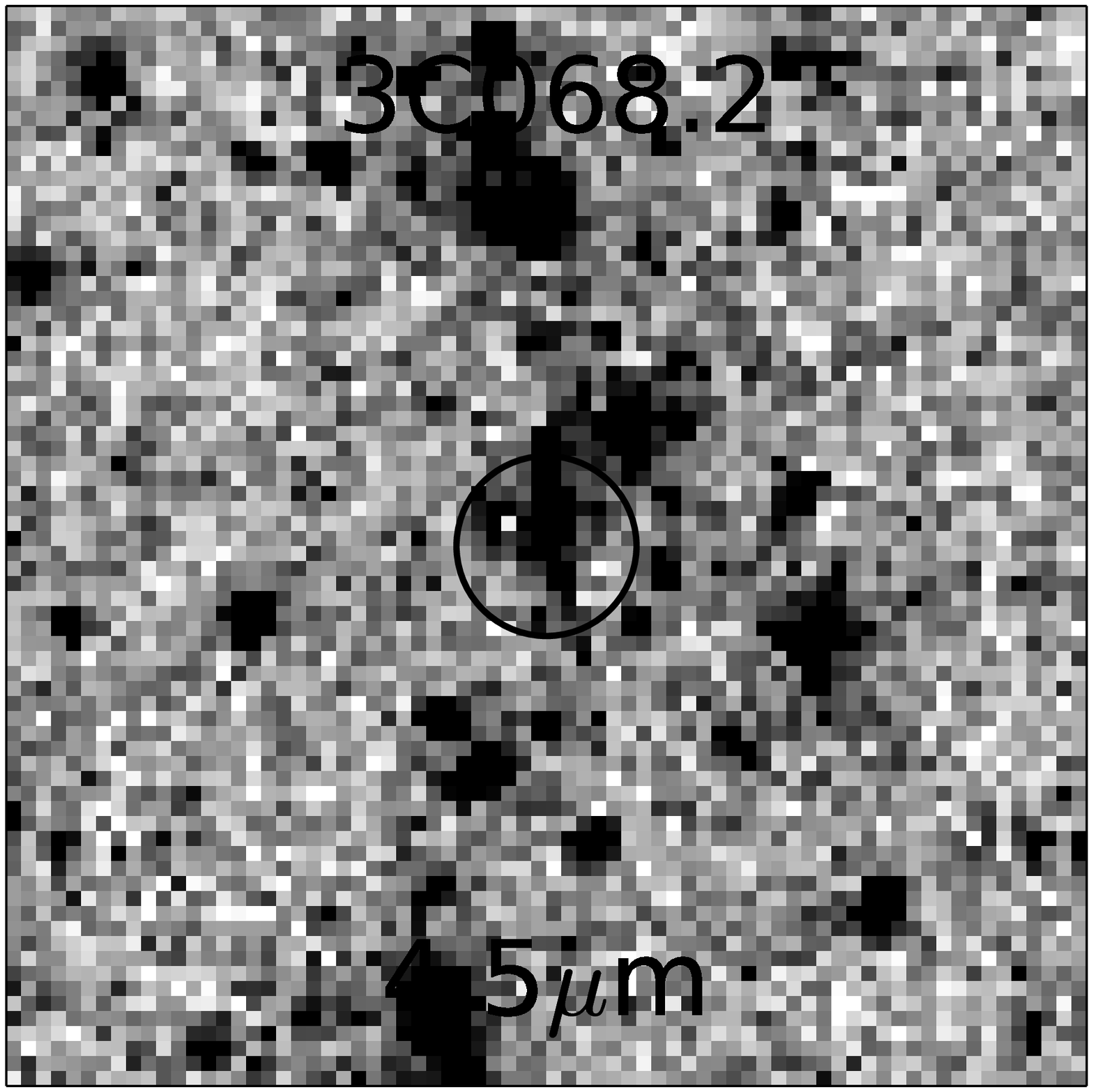}
      \includegraphics[width=1.5cm]{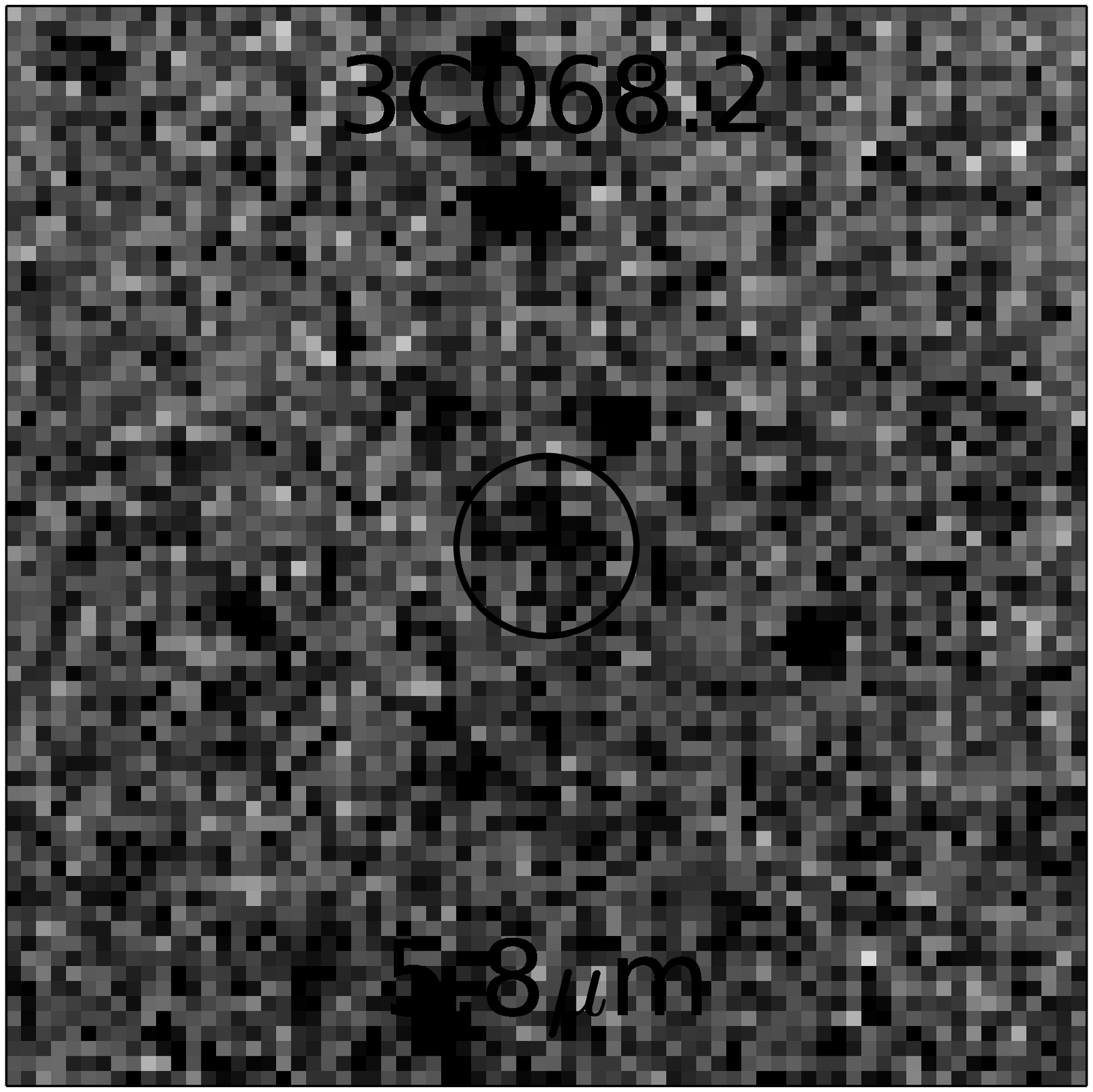}
      \includegraphics[width=1.5cm]{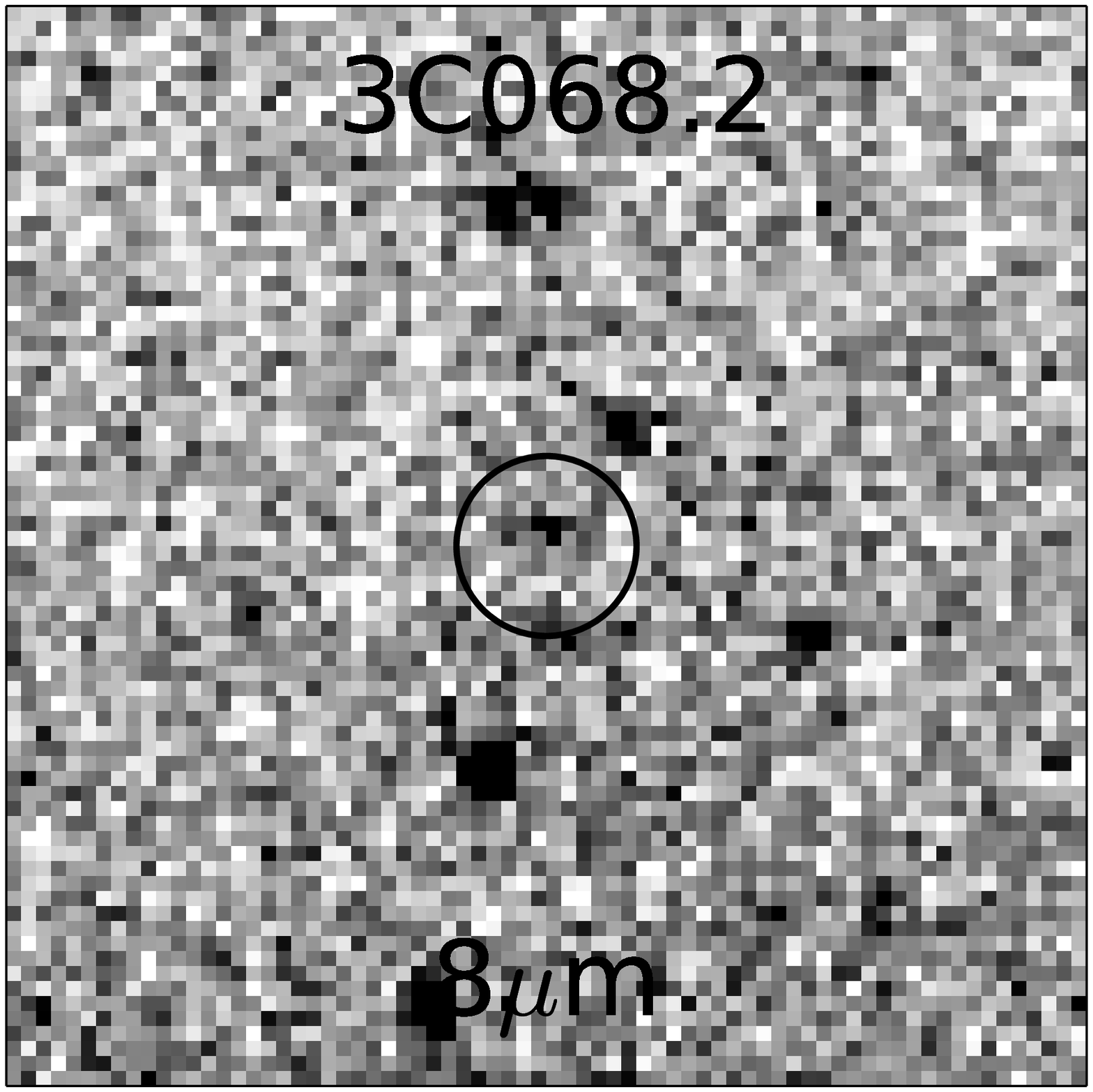}
      \includegraphics[width=1.5cm]{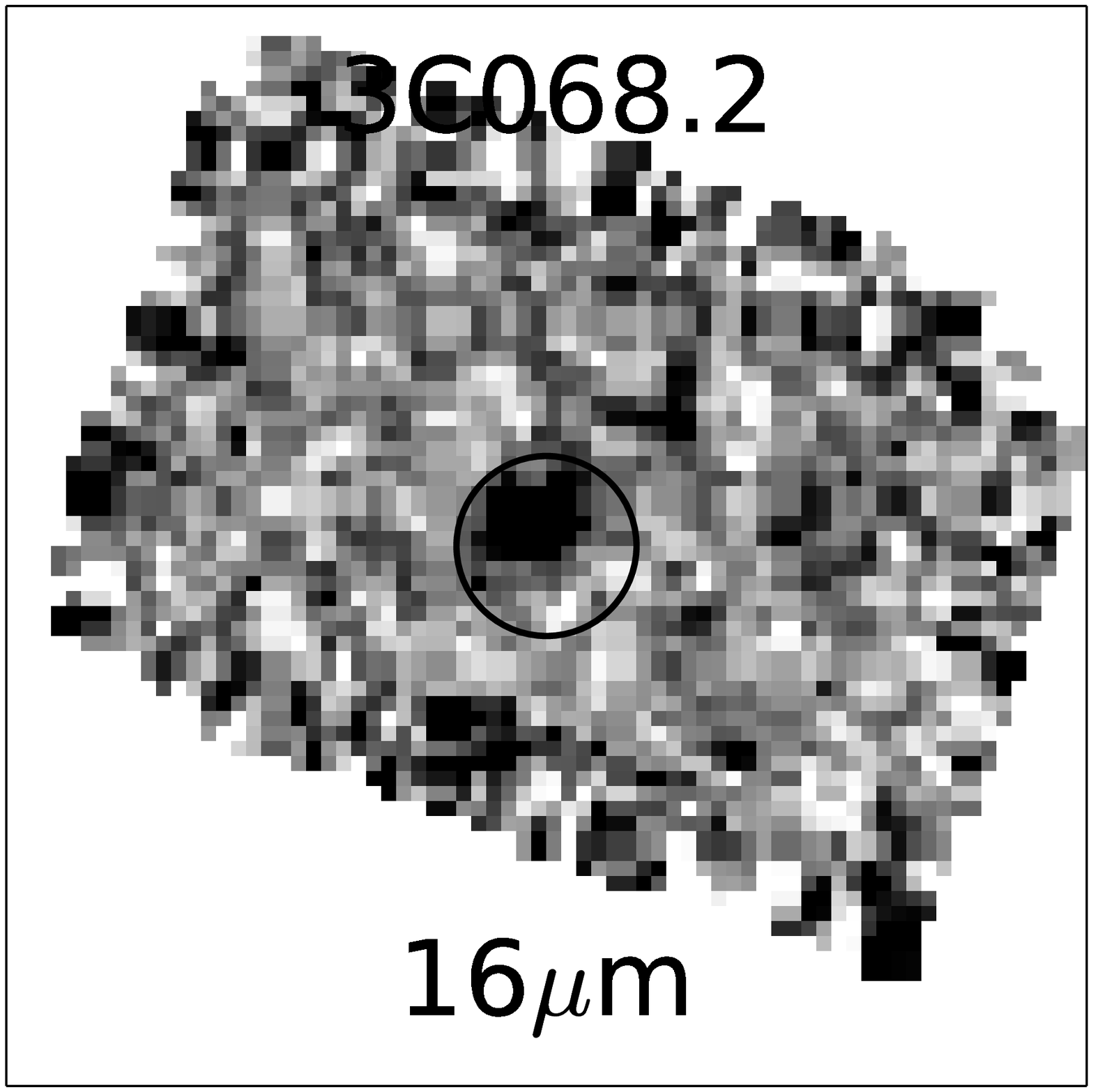}
      \includegraphics[width=1.5cm]{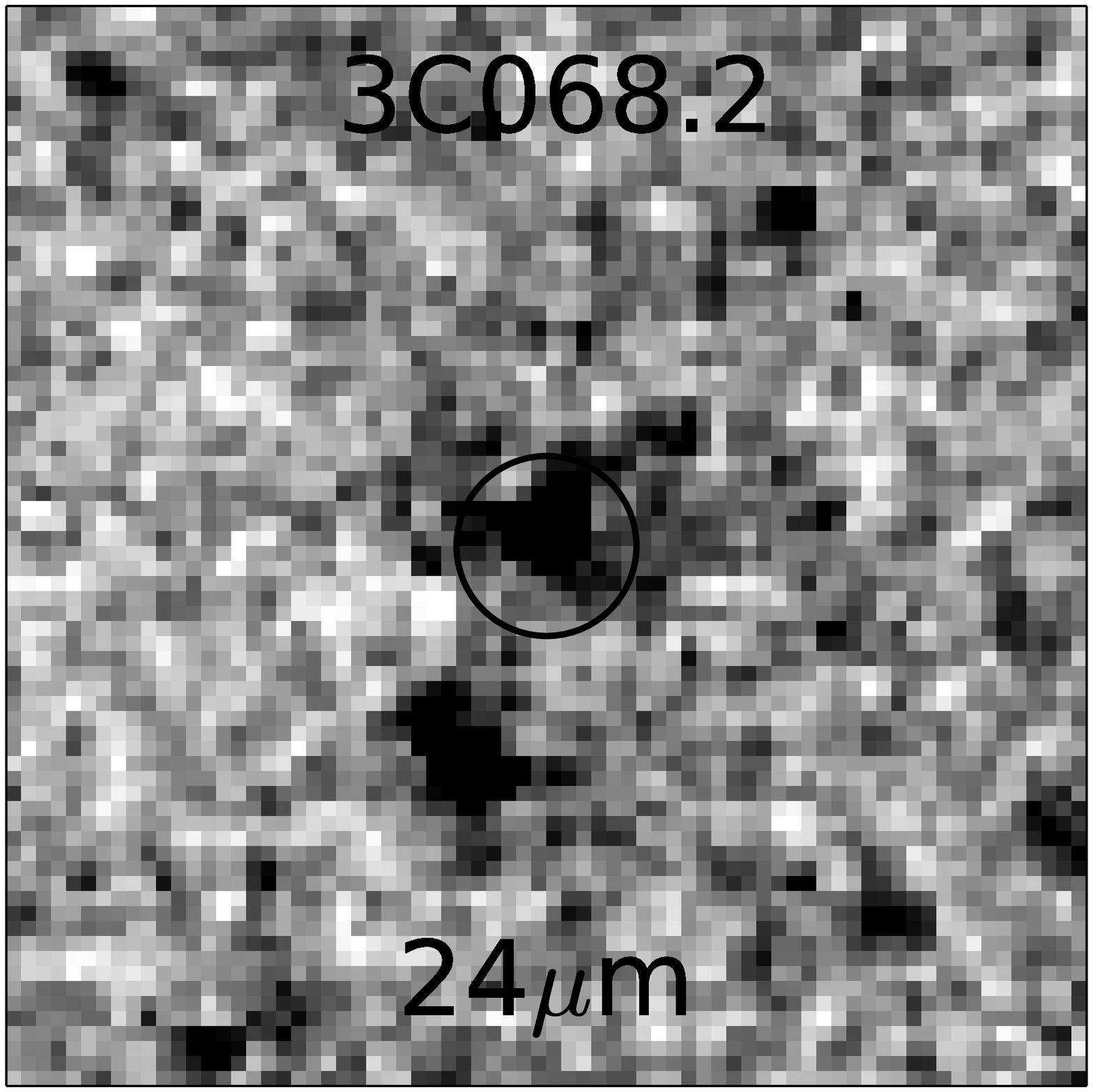}
      \includegraphics[width=1.5cm]{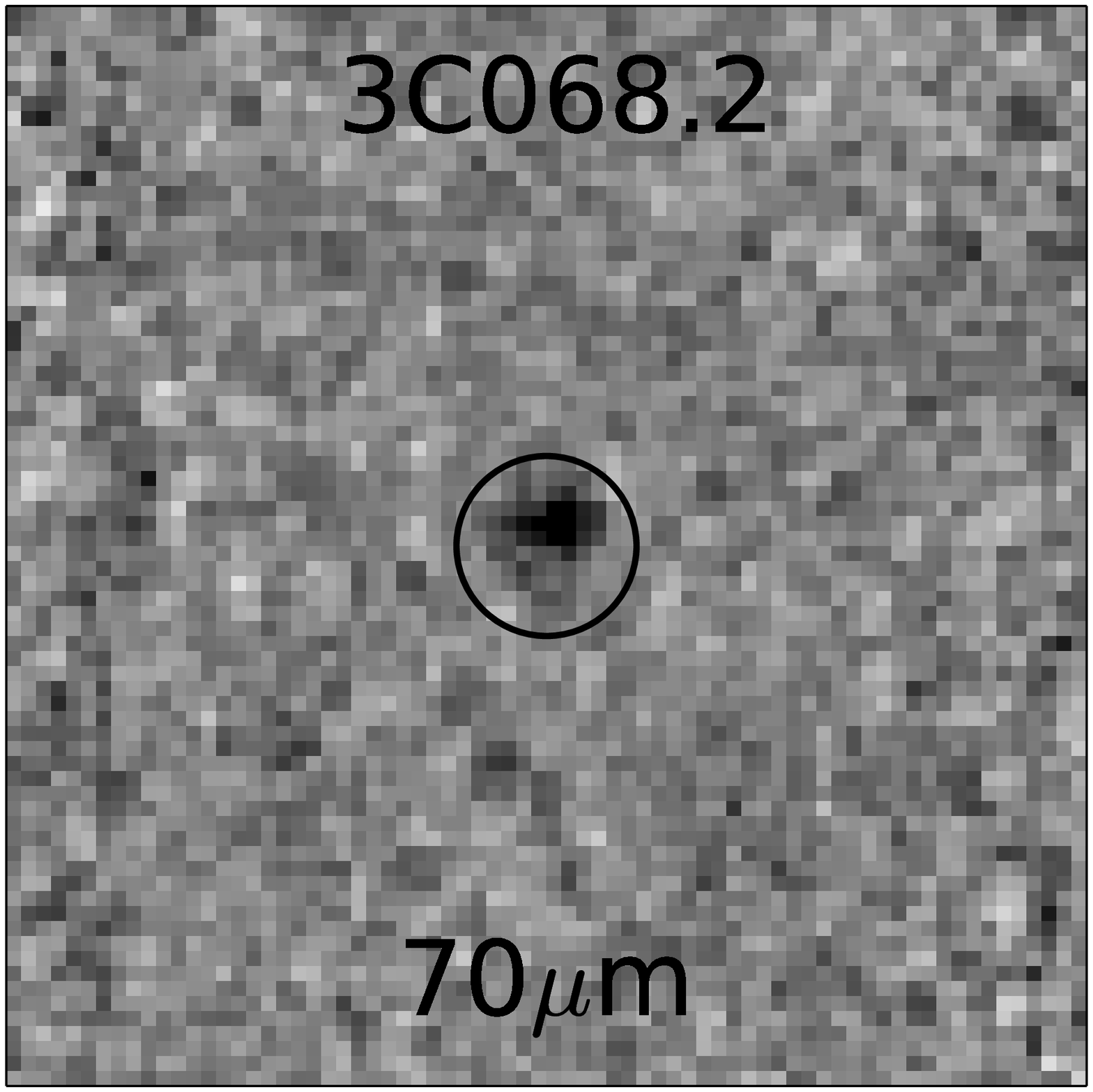}
      \includegraphics[width=1.5cm]{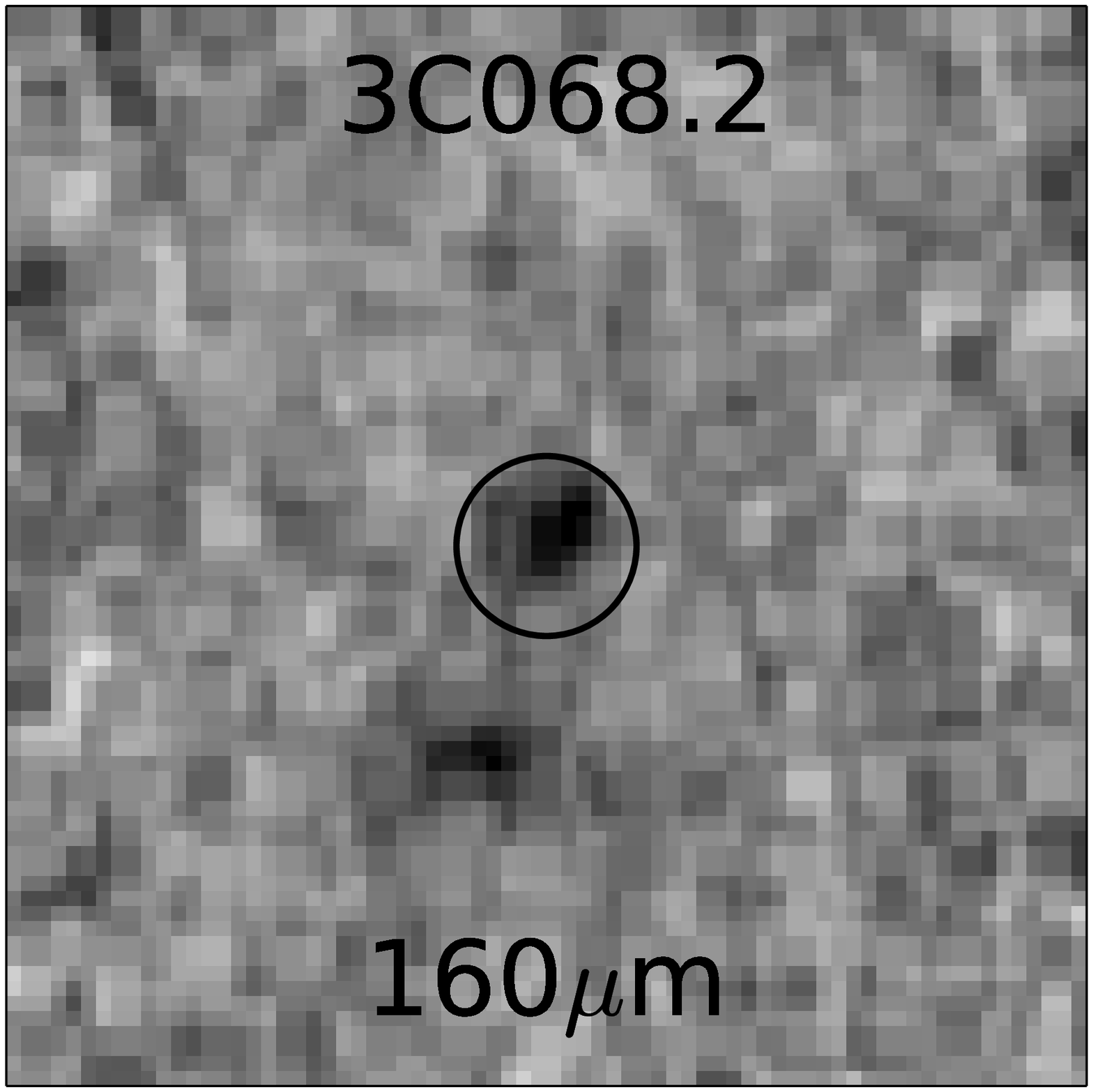}
      \includegraphics[width=1.5cm]{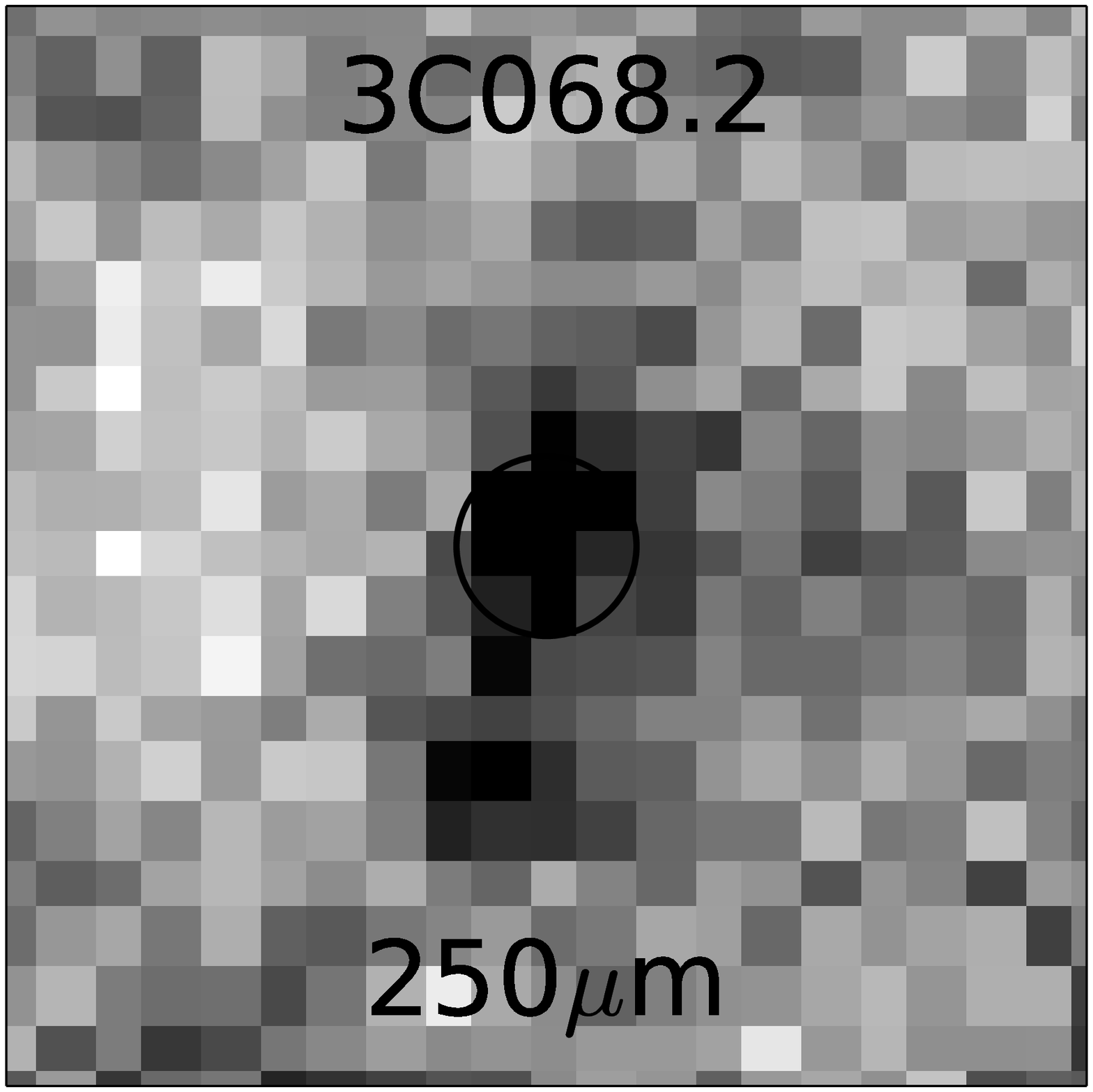}
      \includegraphics[width=1.5cm]{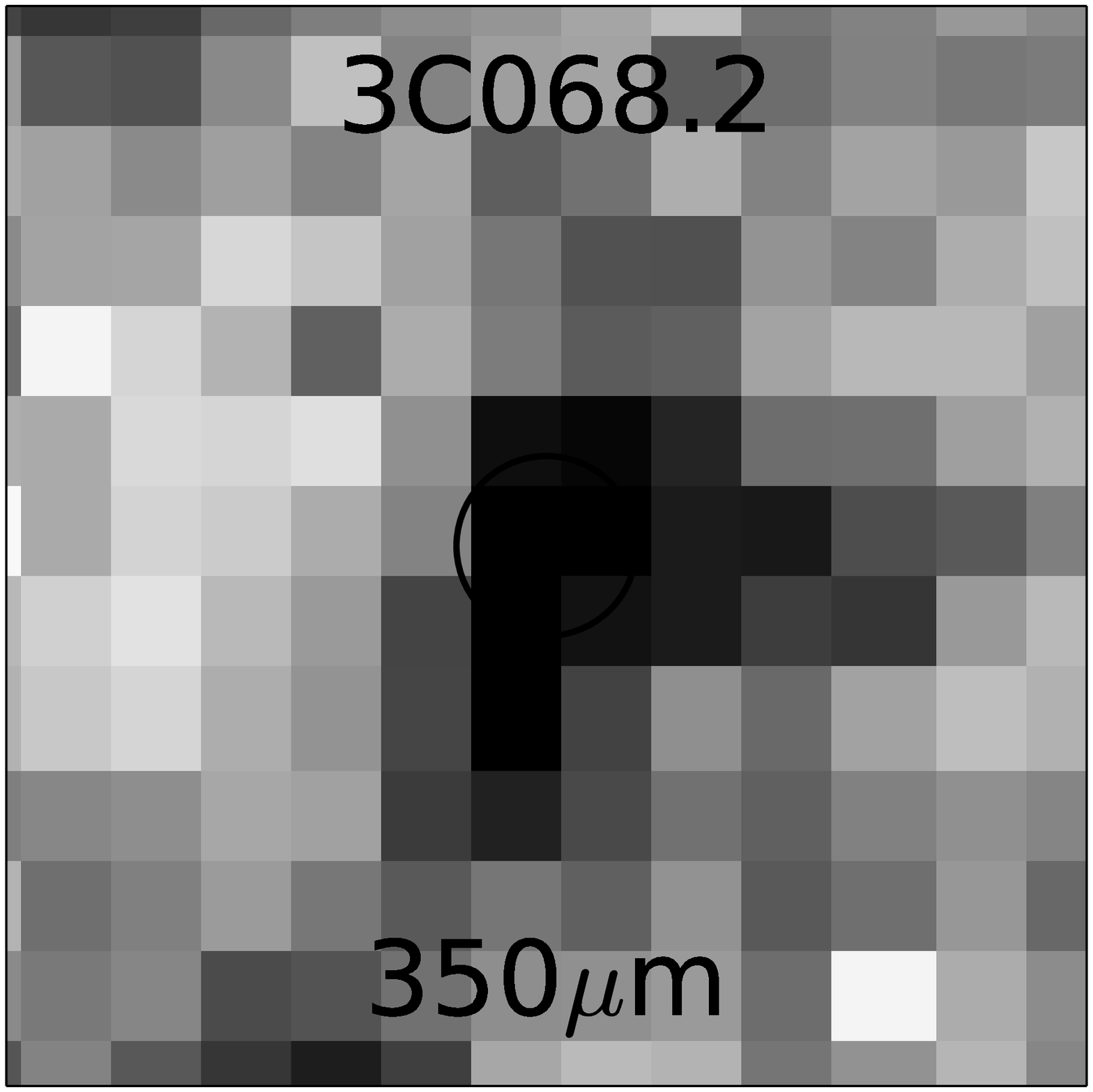}
      \includegraphics[width=1.5cm]{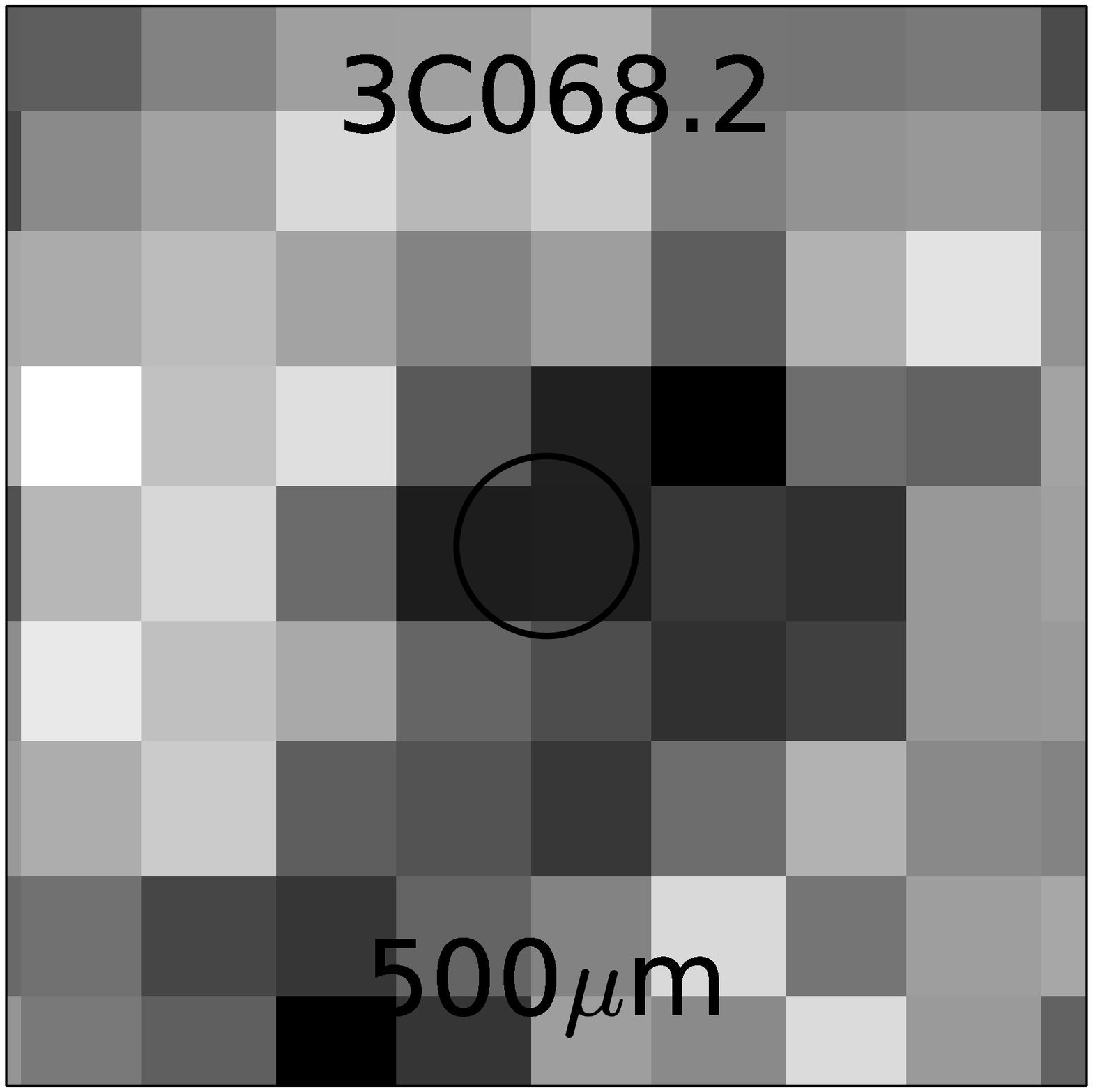}
      \\
      \includegraphics[width=1.5cm]{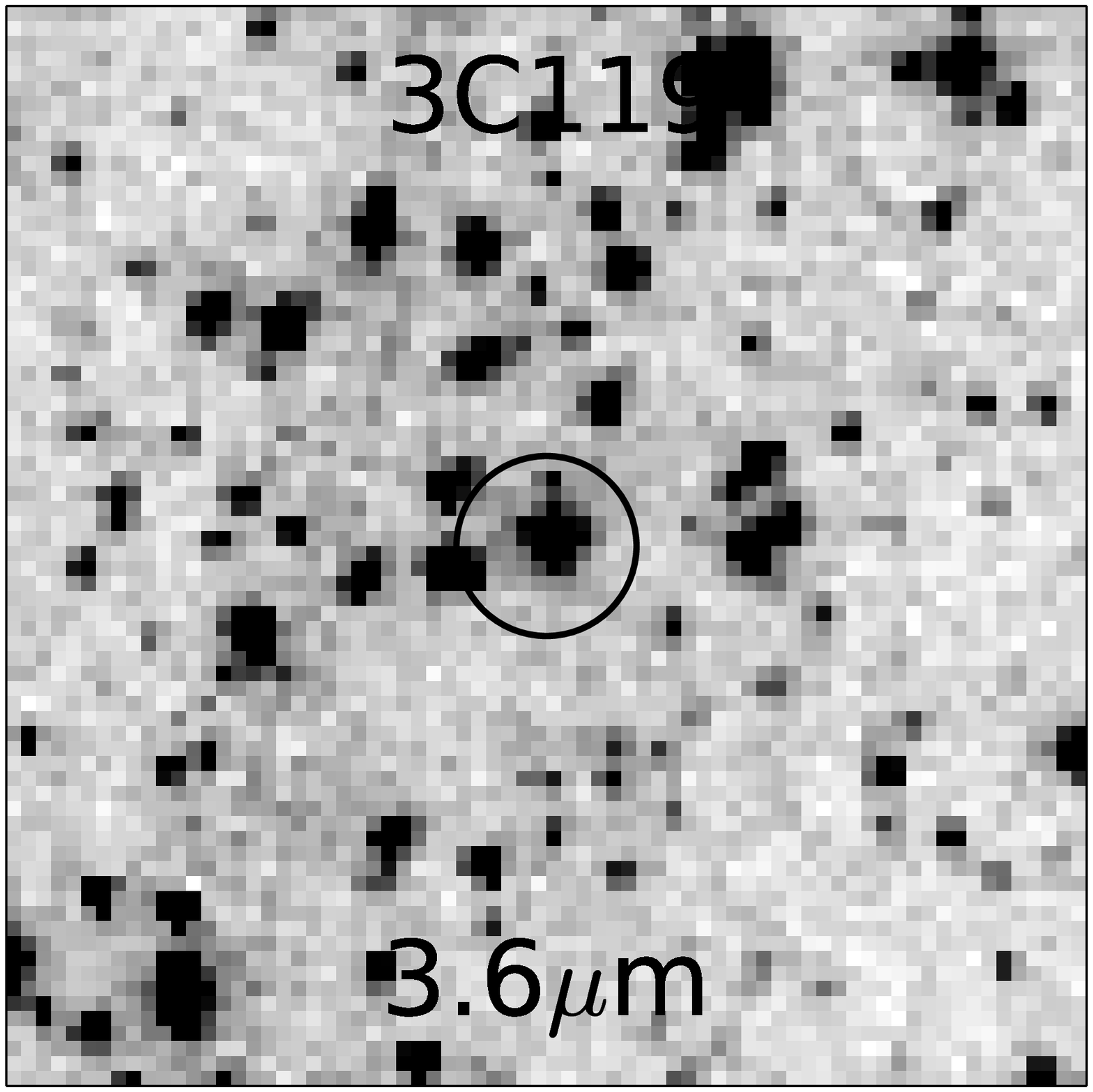}
      \includegraphics[width=1.5cm]{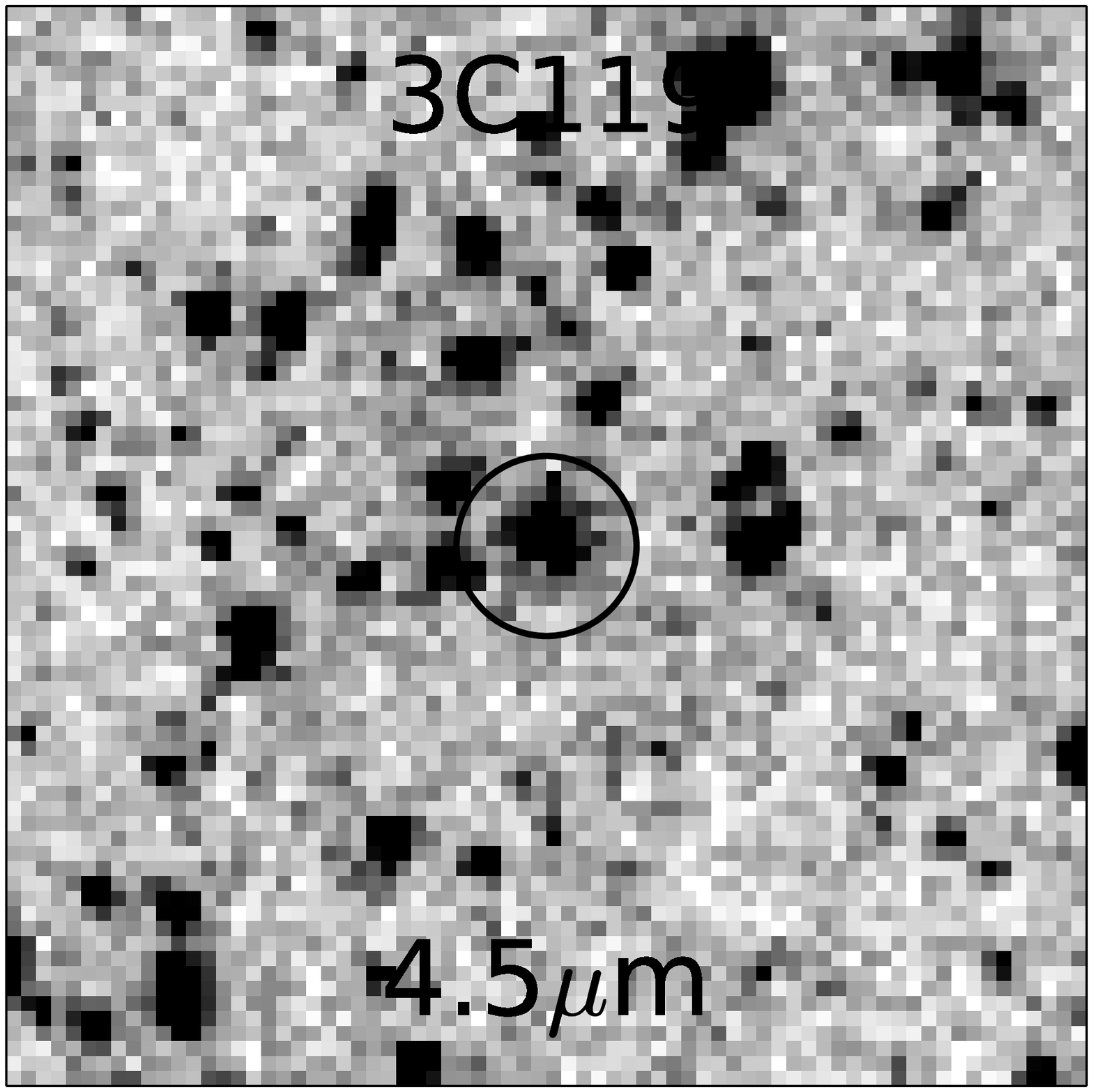}
      \includegraphics[width=1.5cm]{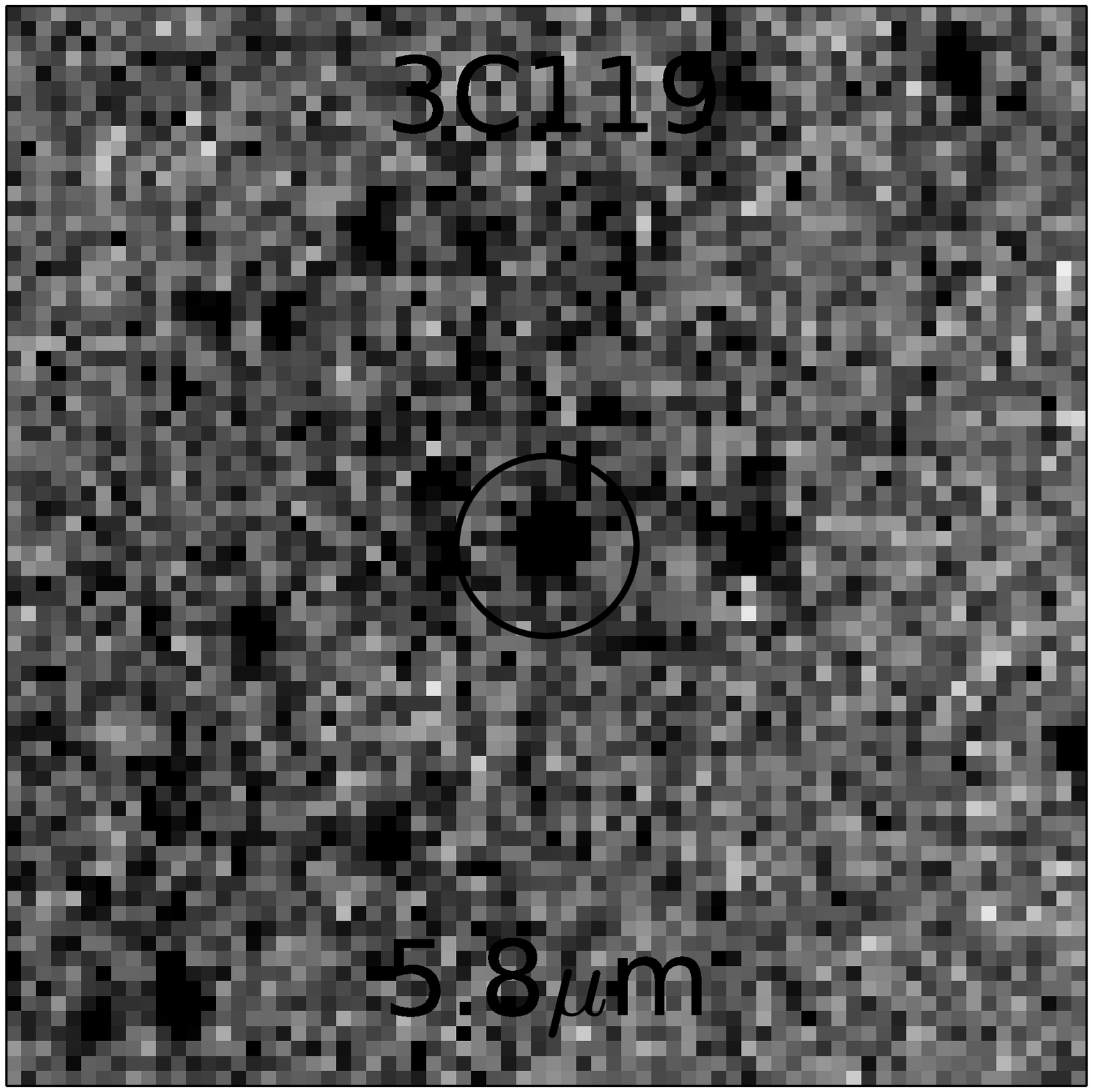}
      \includegraphics[width=1.5cm]{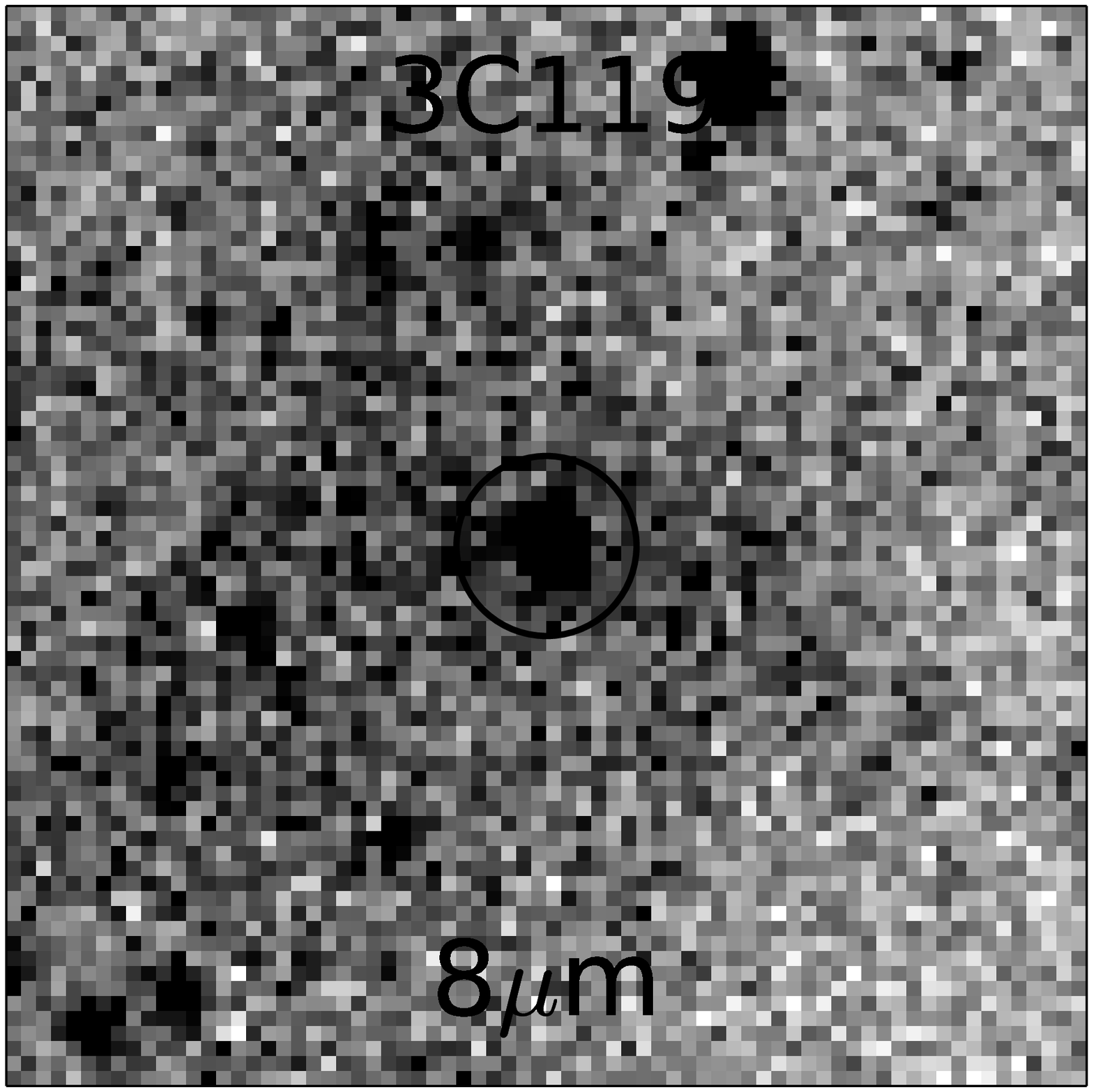}
      \includegraphics[width=1.5cm]{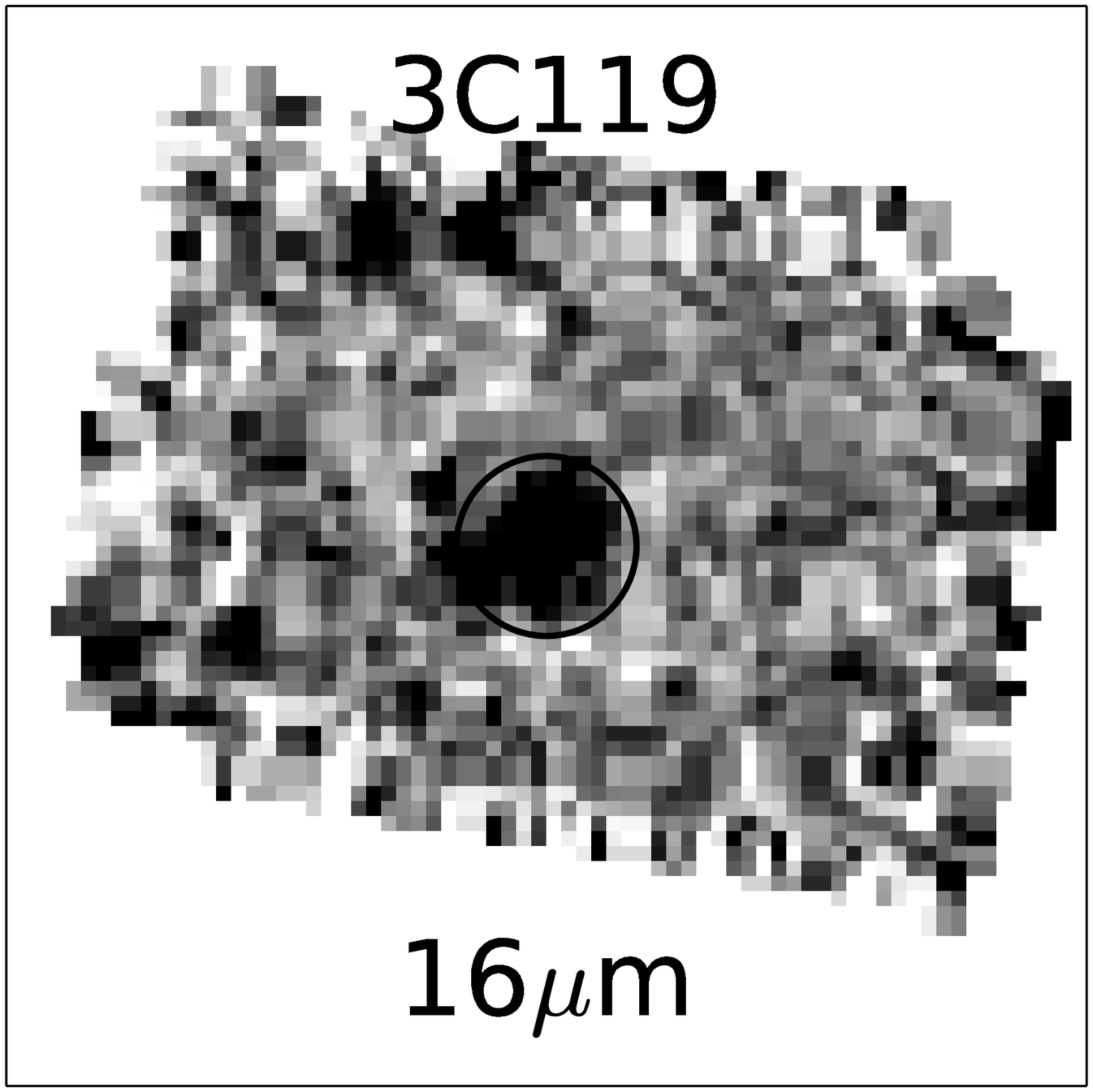}
      \includegraphics[width=1.5cm]{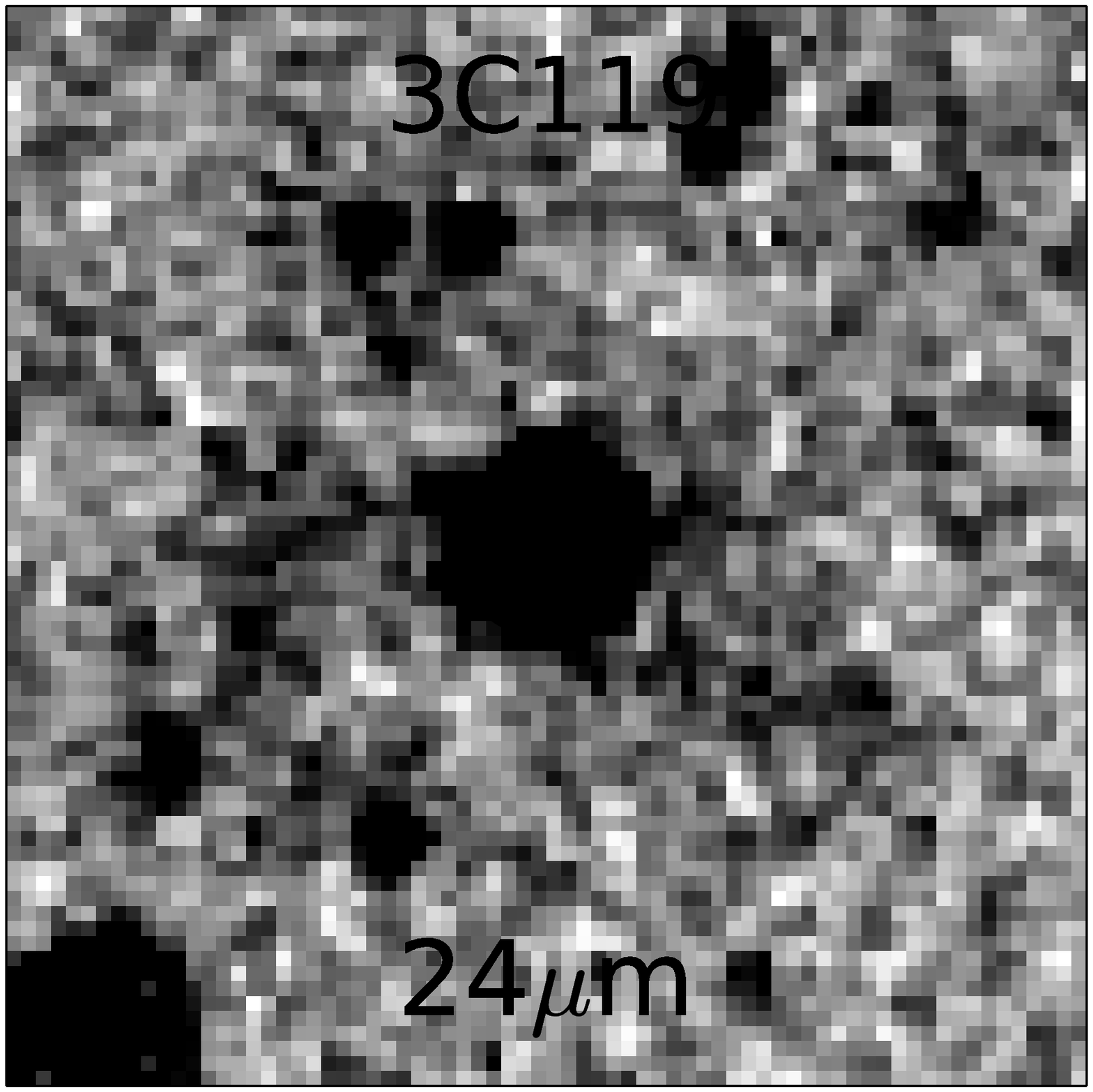}
      \includegraphics[width=1.5cm]{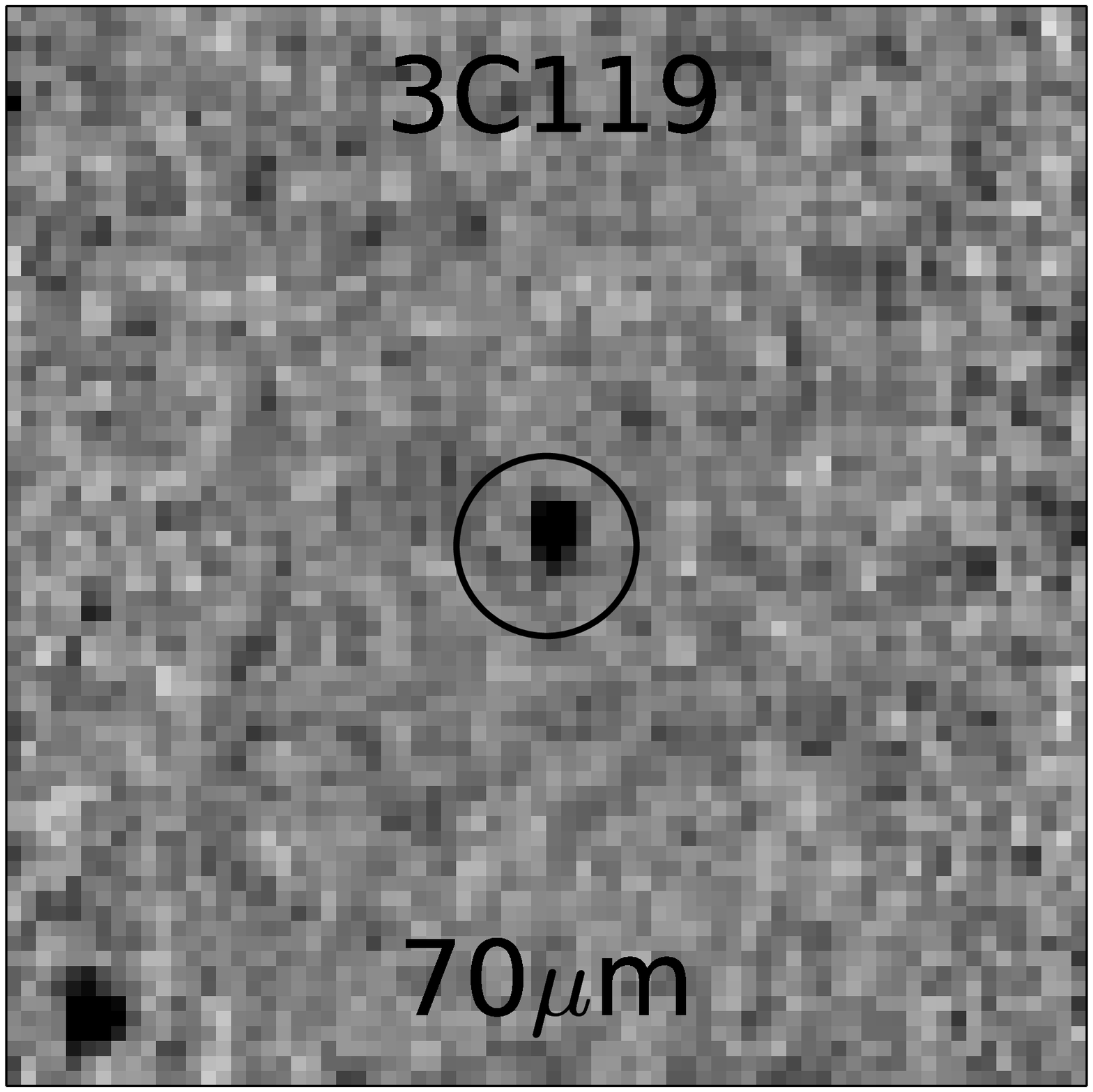}
      \includegraphics[width=1.5cm]{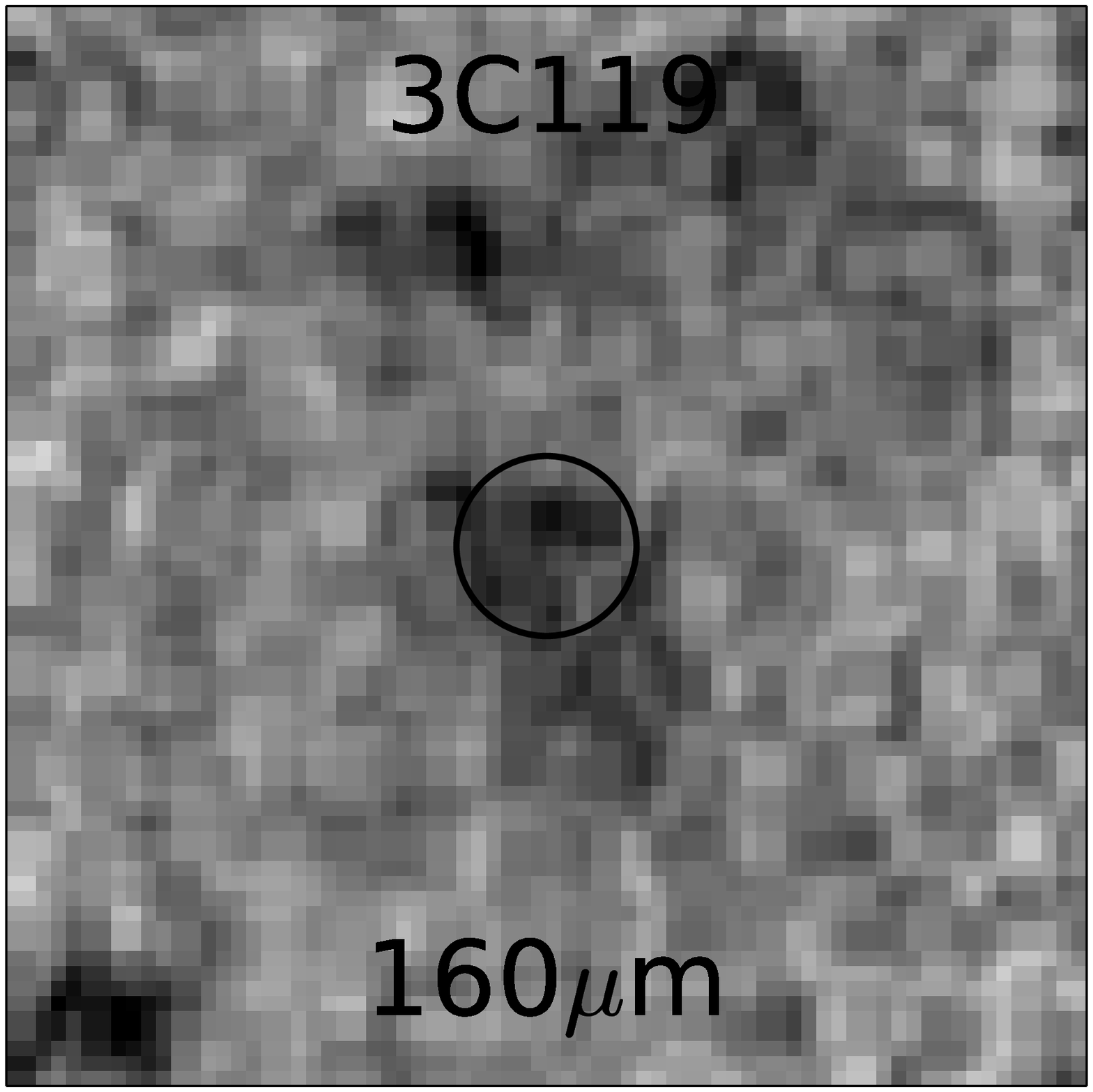}
      \includegraphics[width=1.5cm]{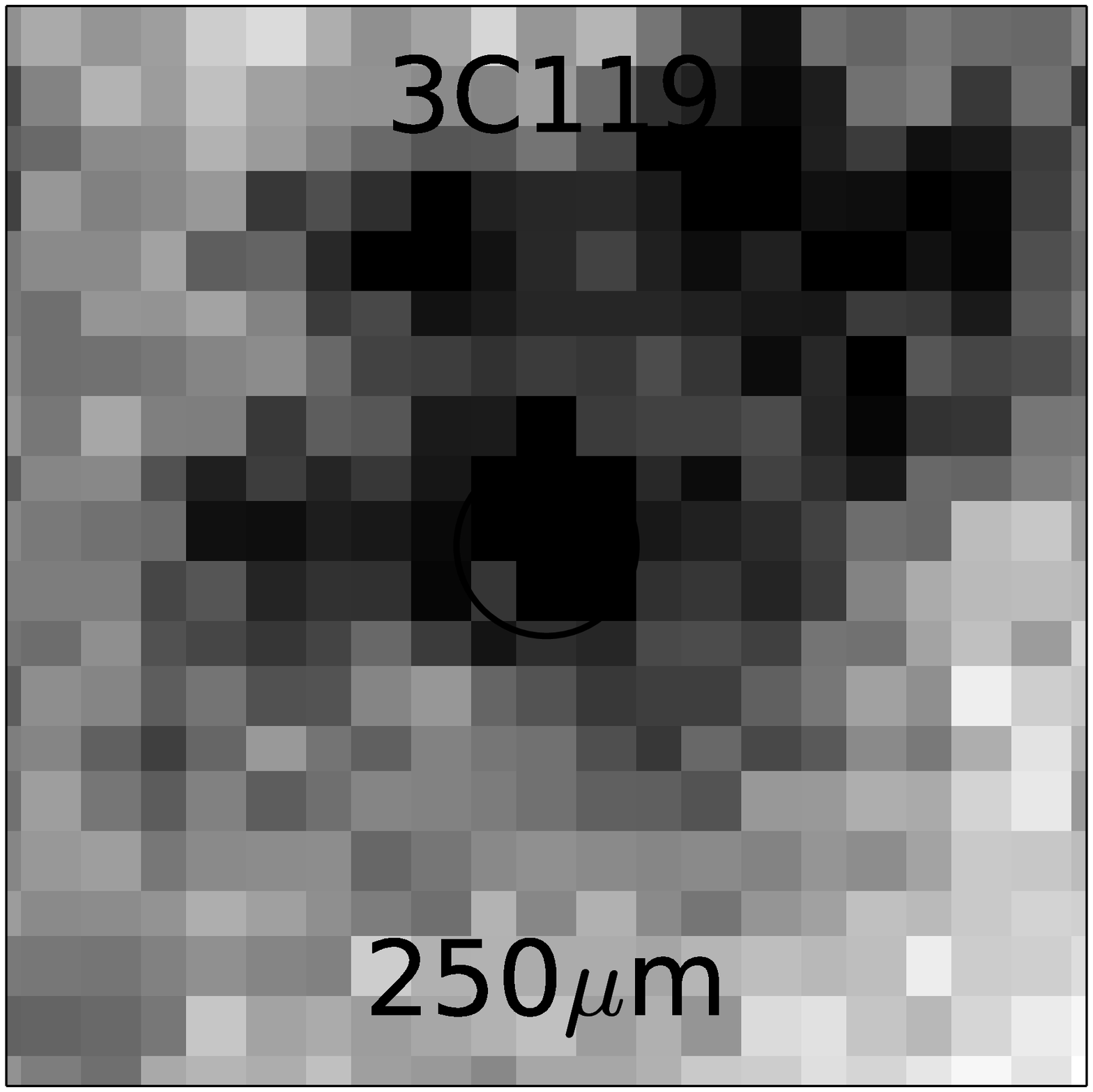}
      \includegraphics[width=1.5cm]{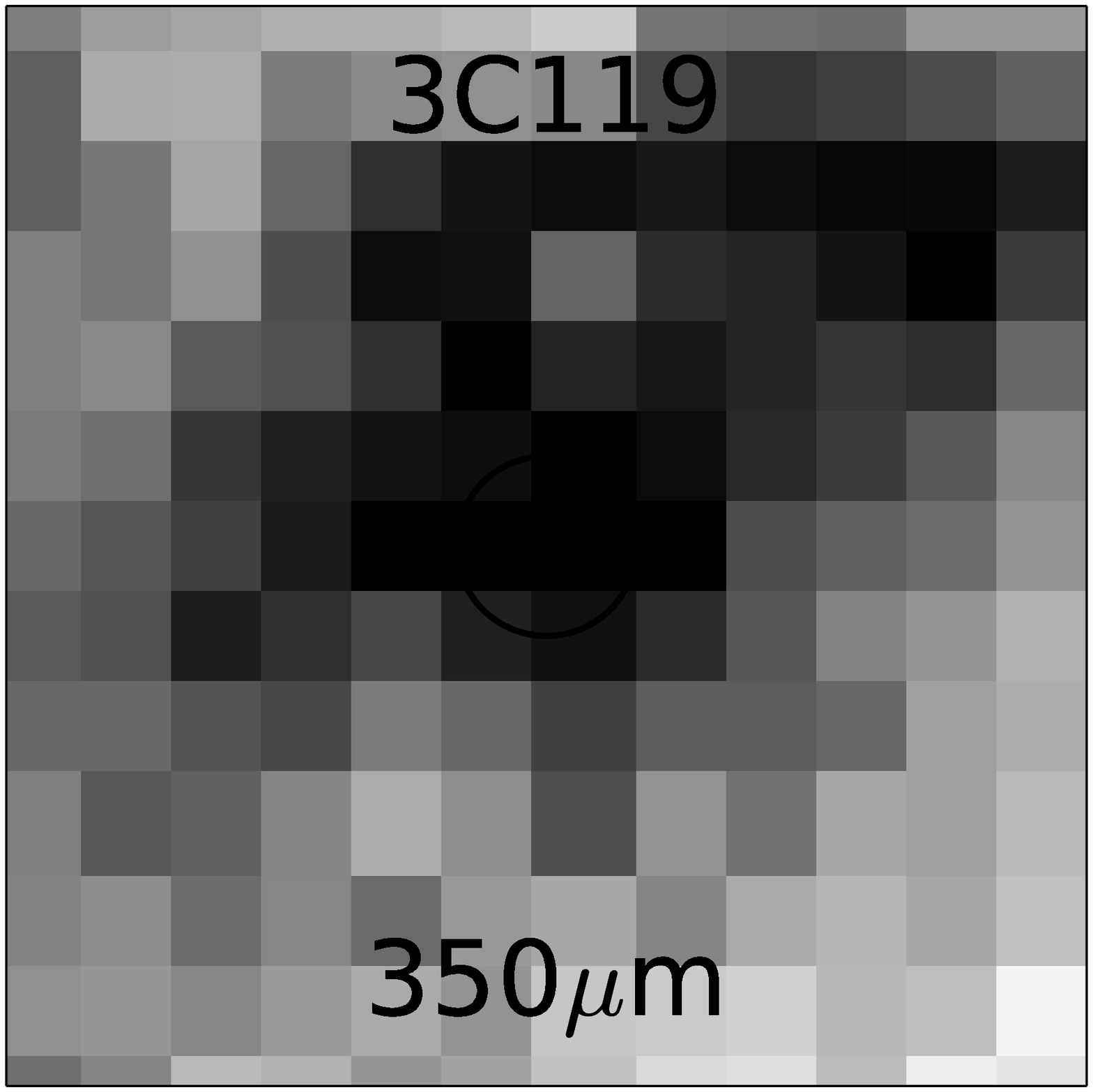}
      \includegraphics[width=1.5cm]{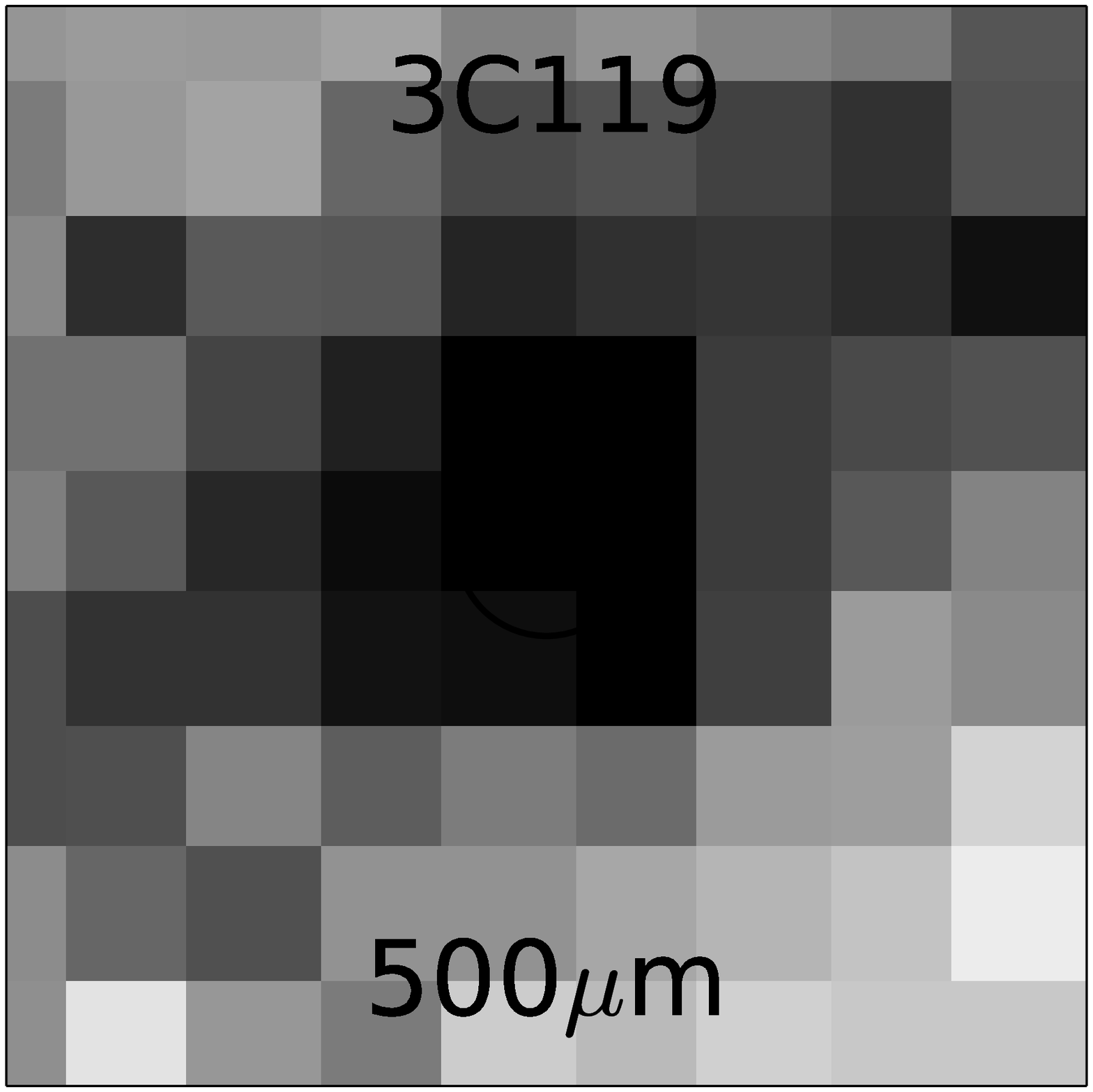}
      \\
      \includegraphics[width=1.5cm]{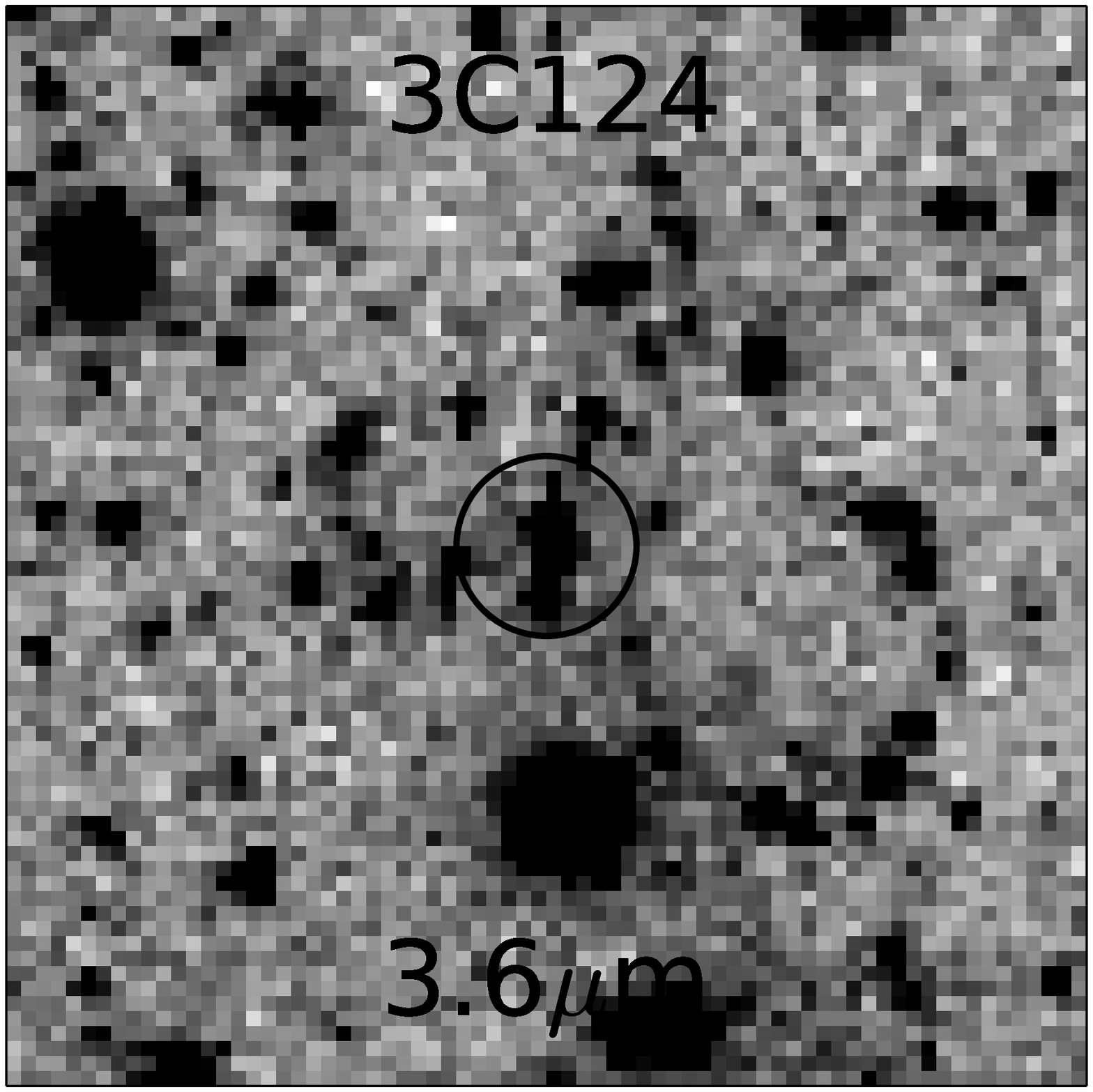}
      \includegraphics[width=1.5cm]{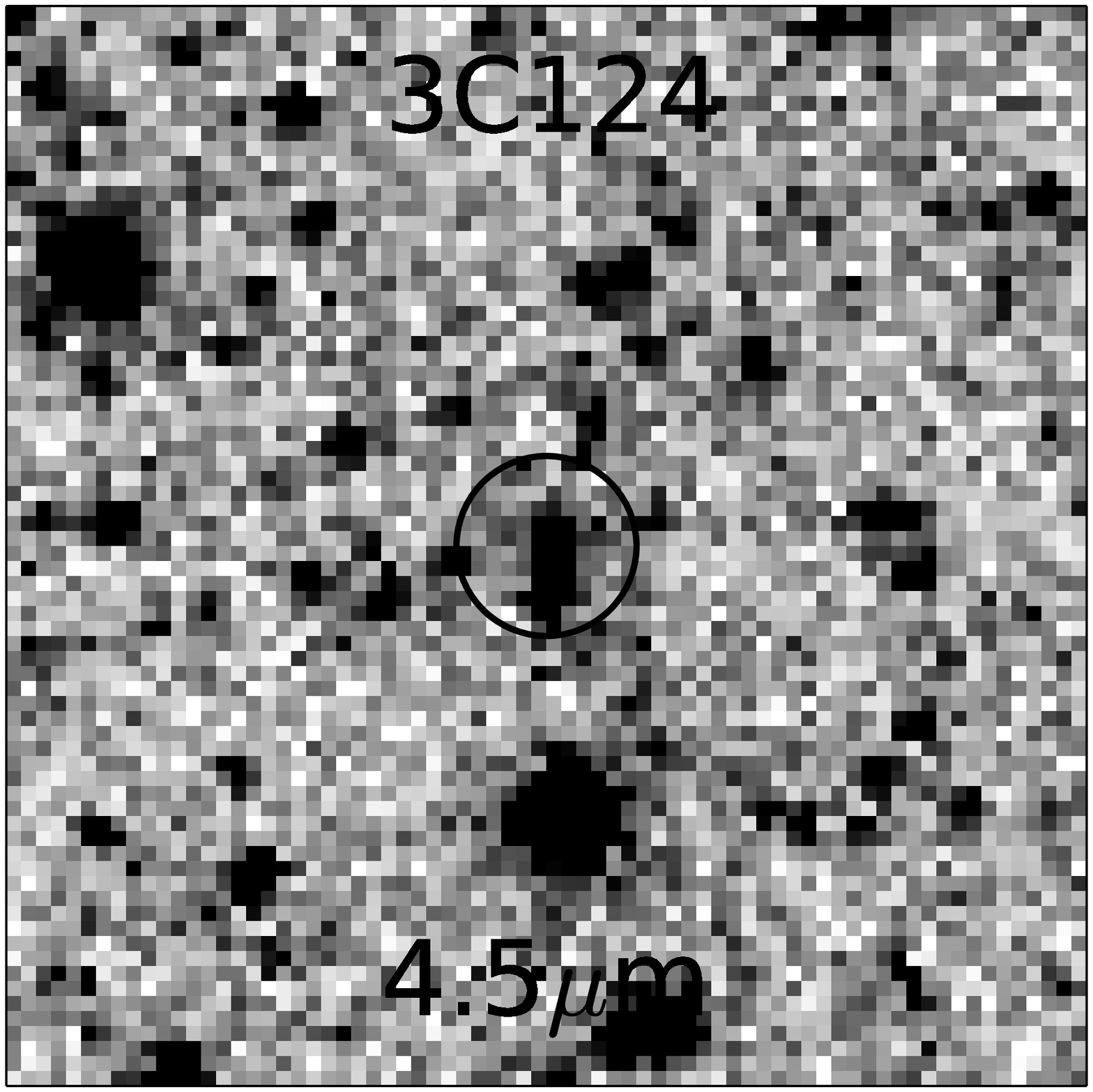}
      \includegraphics[width=1.5cm]{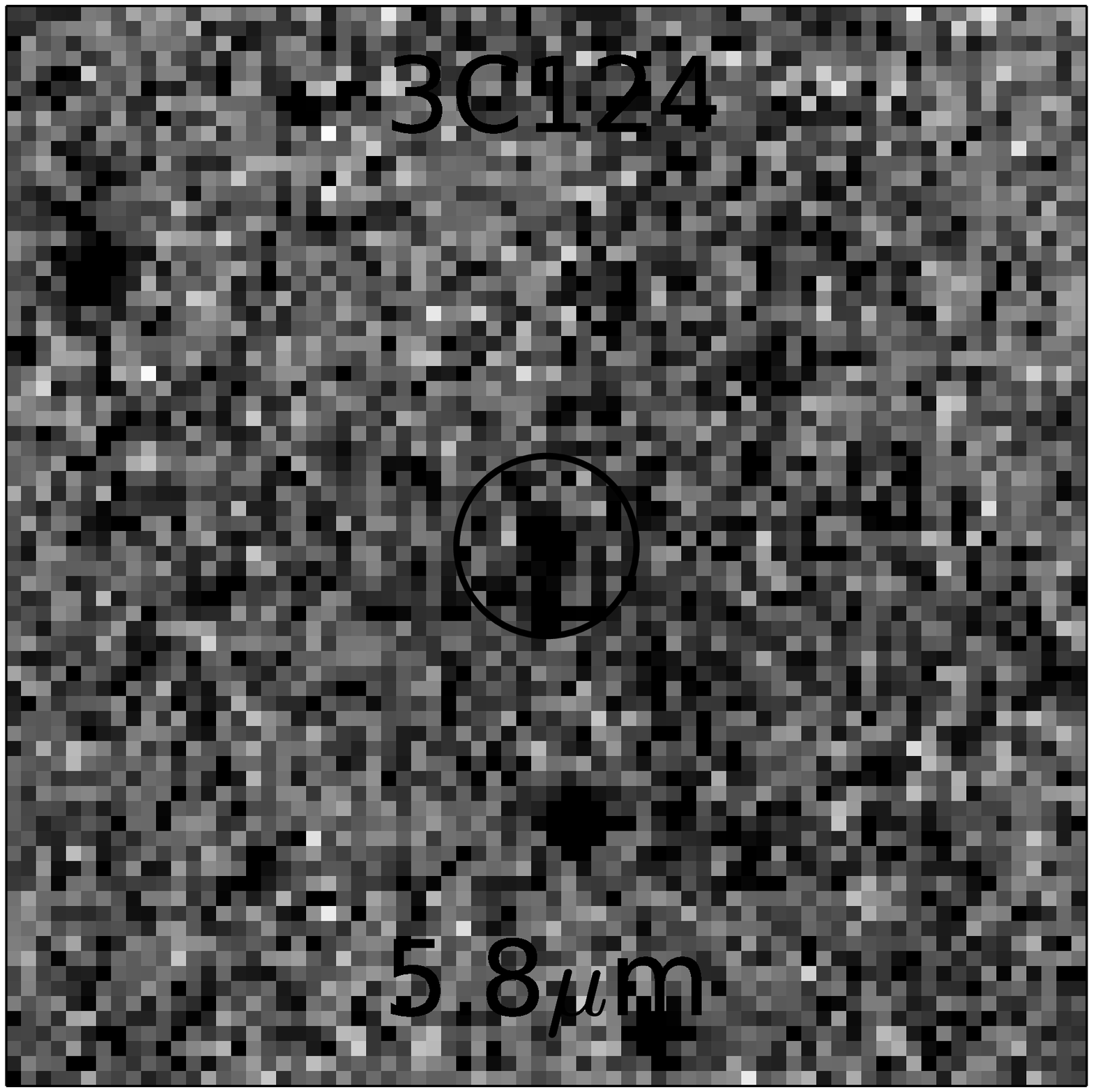}
      \includegraphics[width=1.5cm]{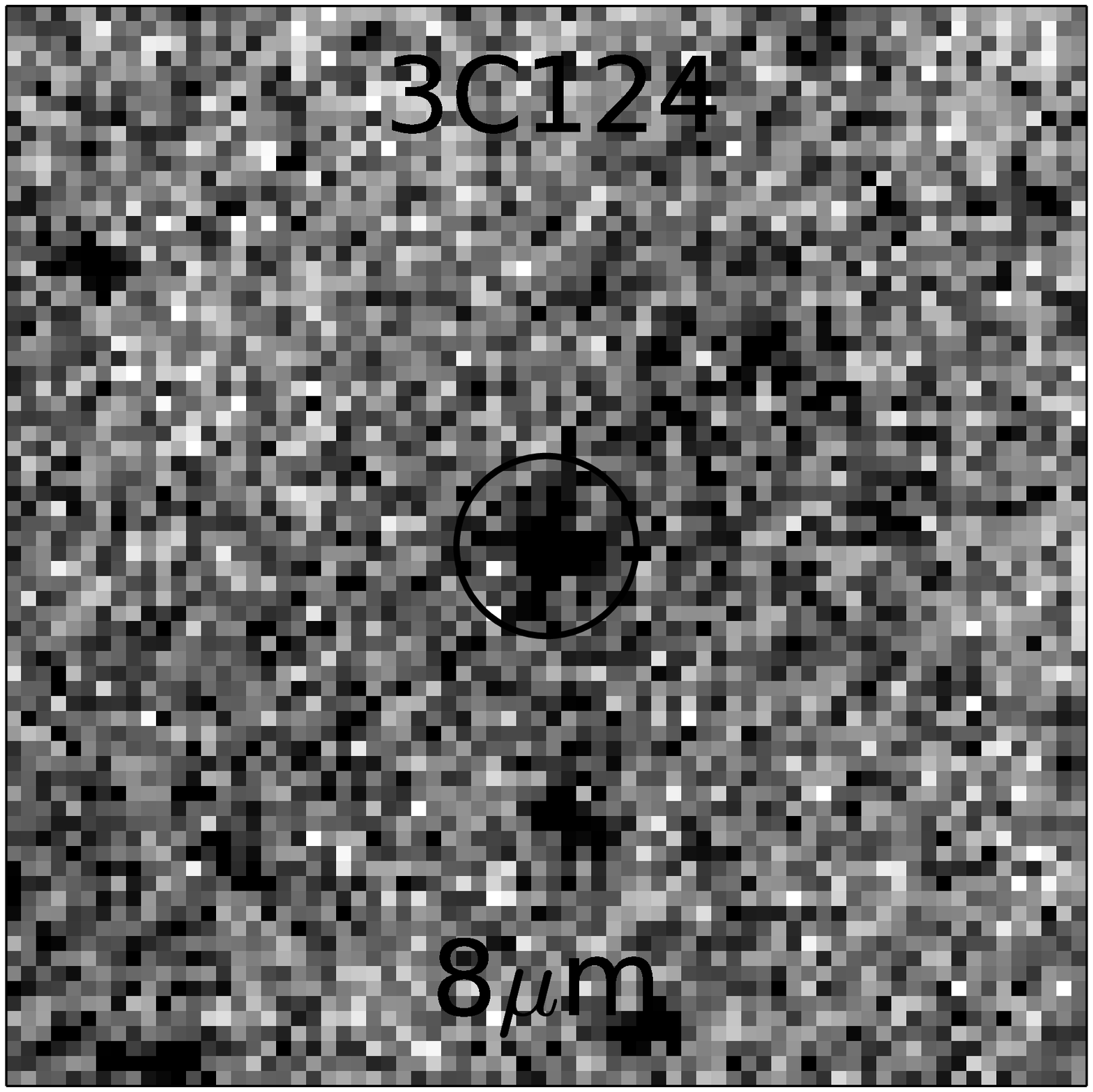}
      \includegraphics[width=1.5cm]{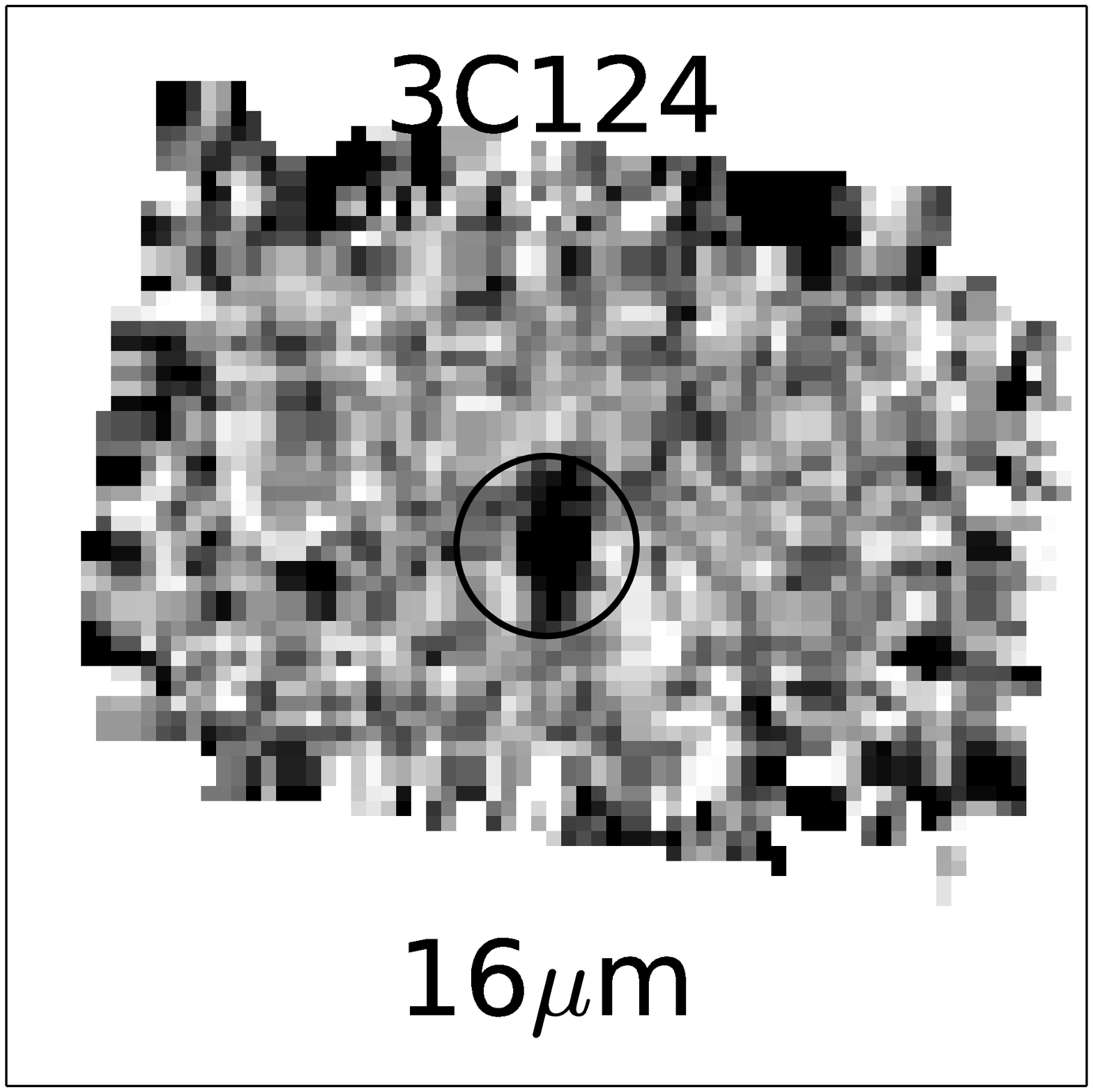}
      \includegraphics[width=1.5cm]{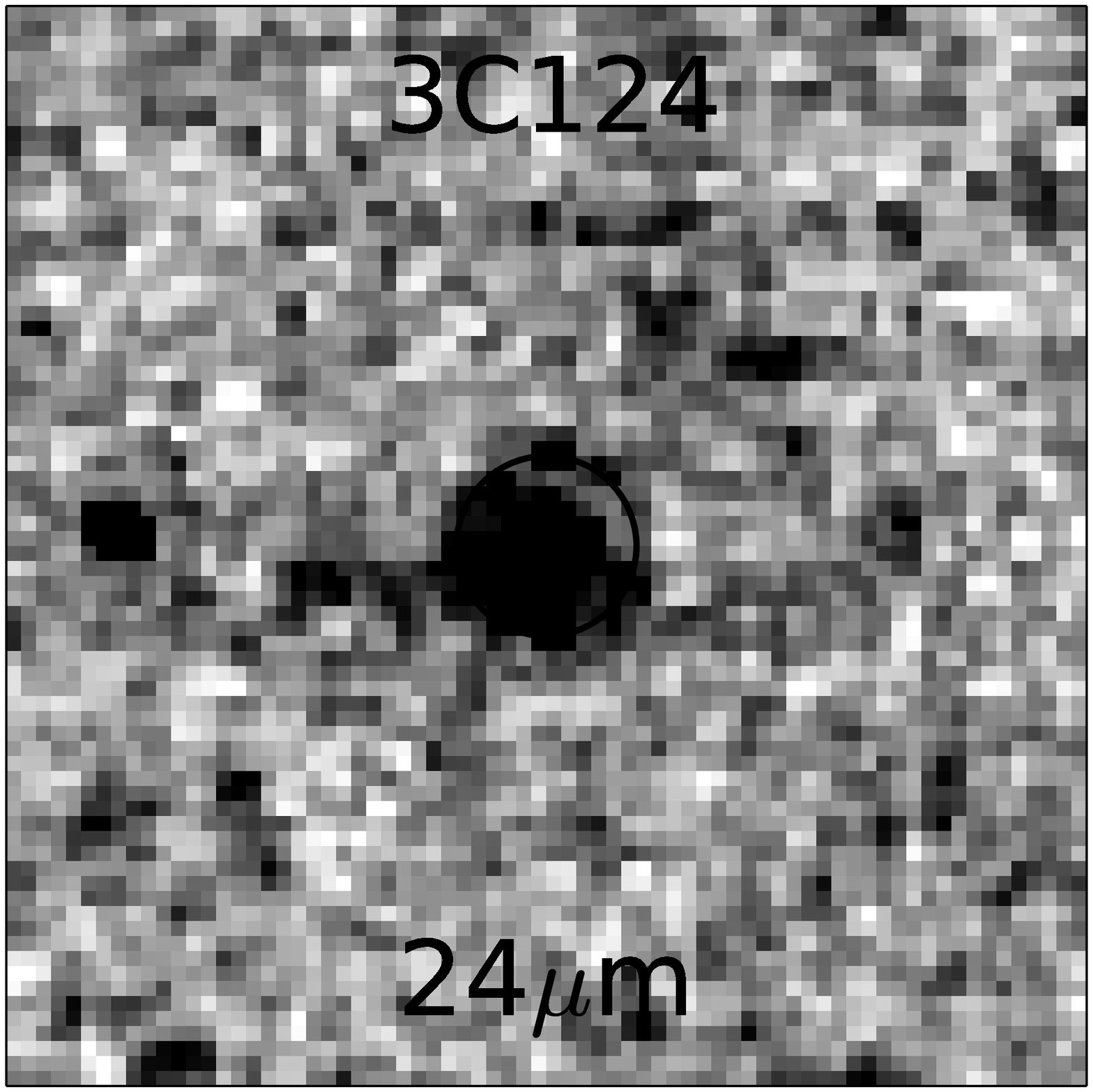}
      \includegraphics[width=1.5cm]{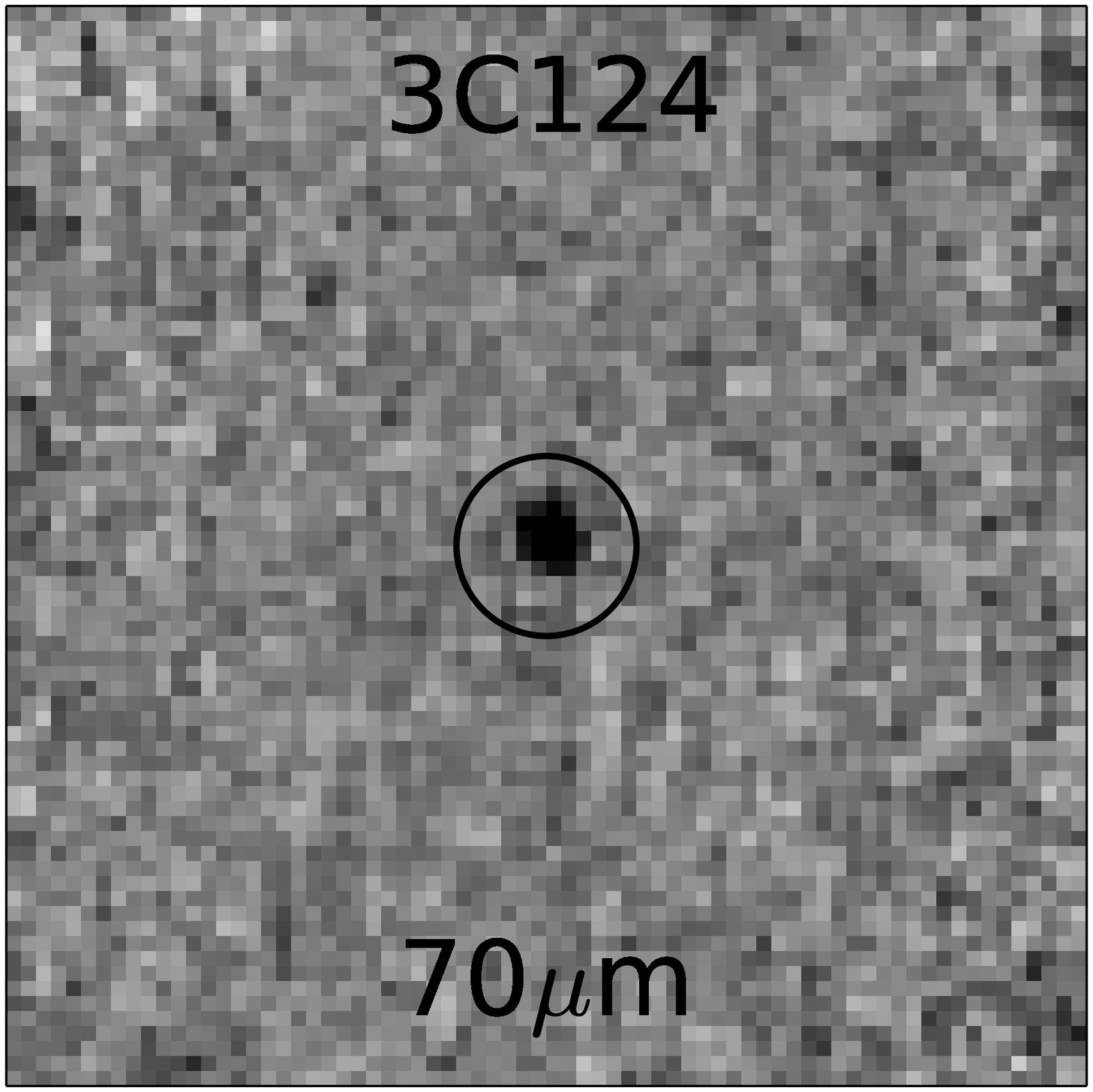}
      \includegraphics[width=1.5cm]{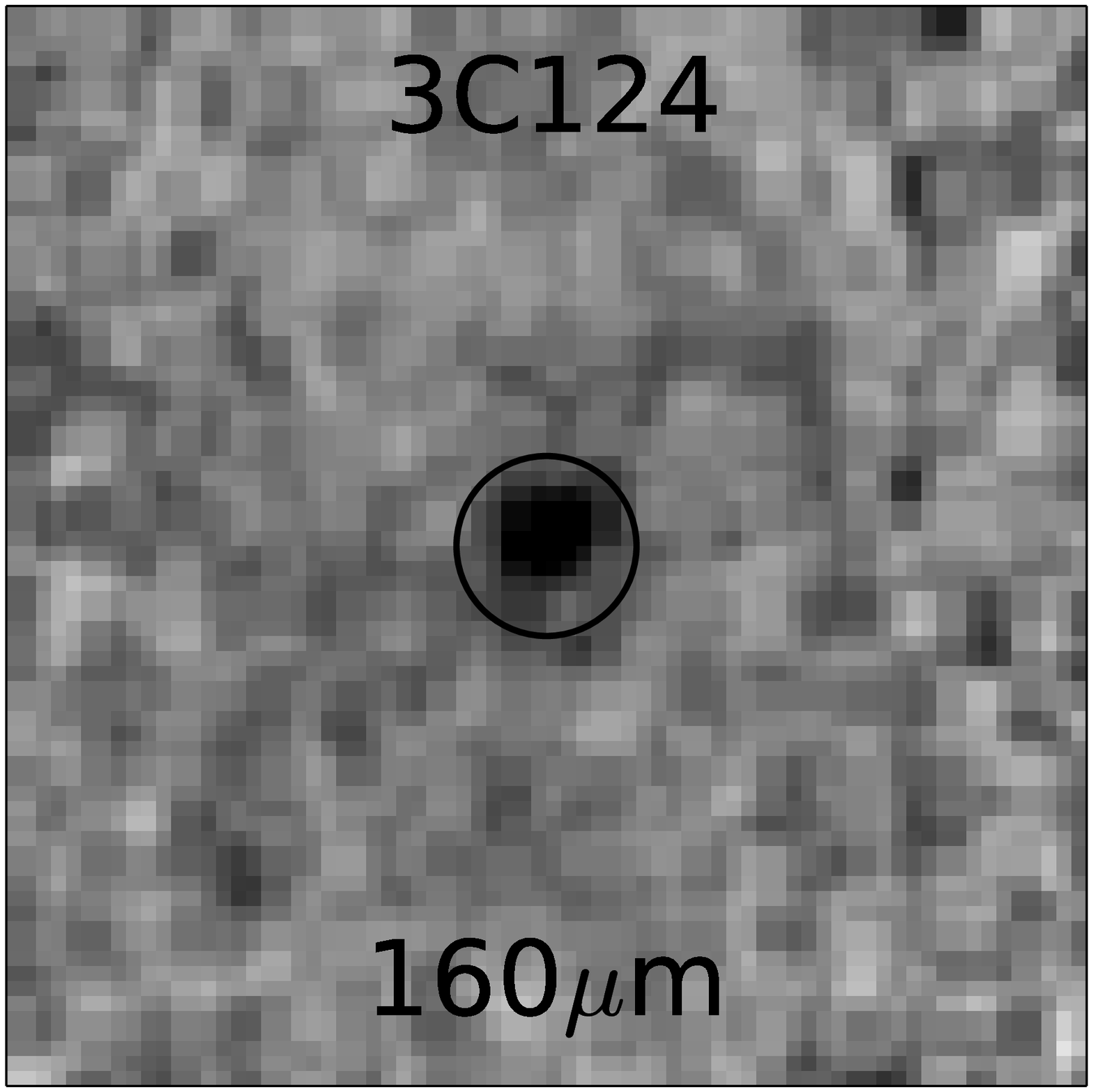}
      \includegraphics[width=1.5cm]{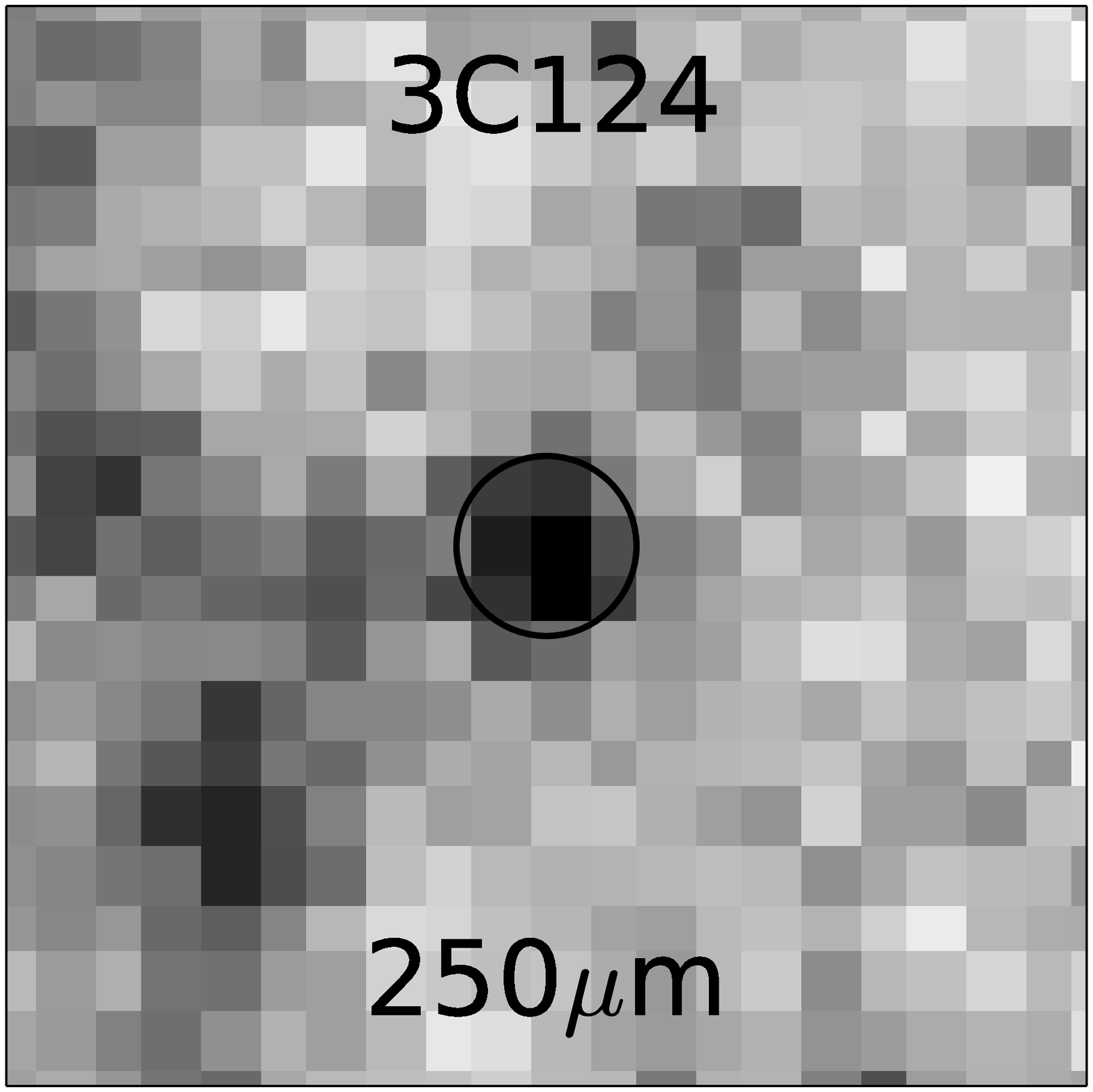}
      \includegraphics[width=1.5cm]{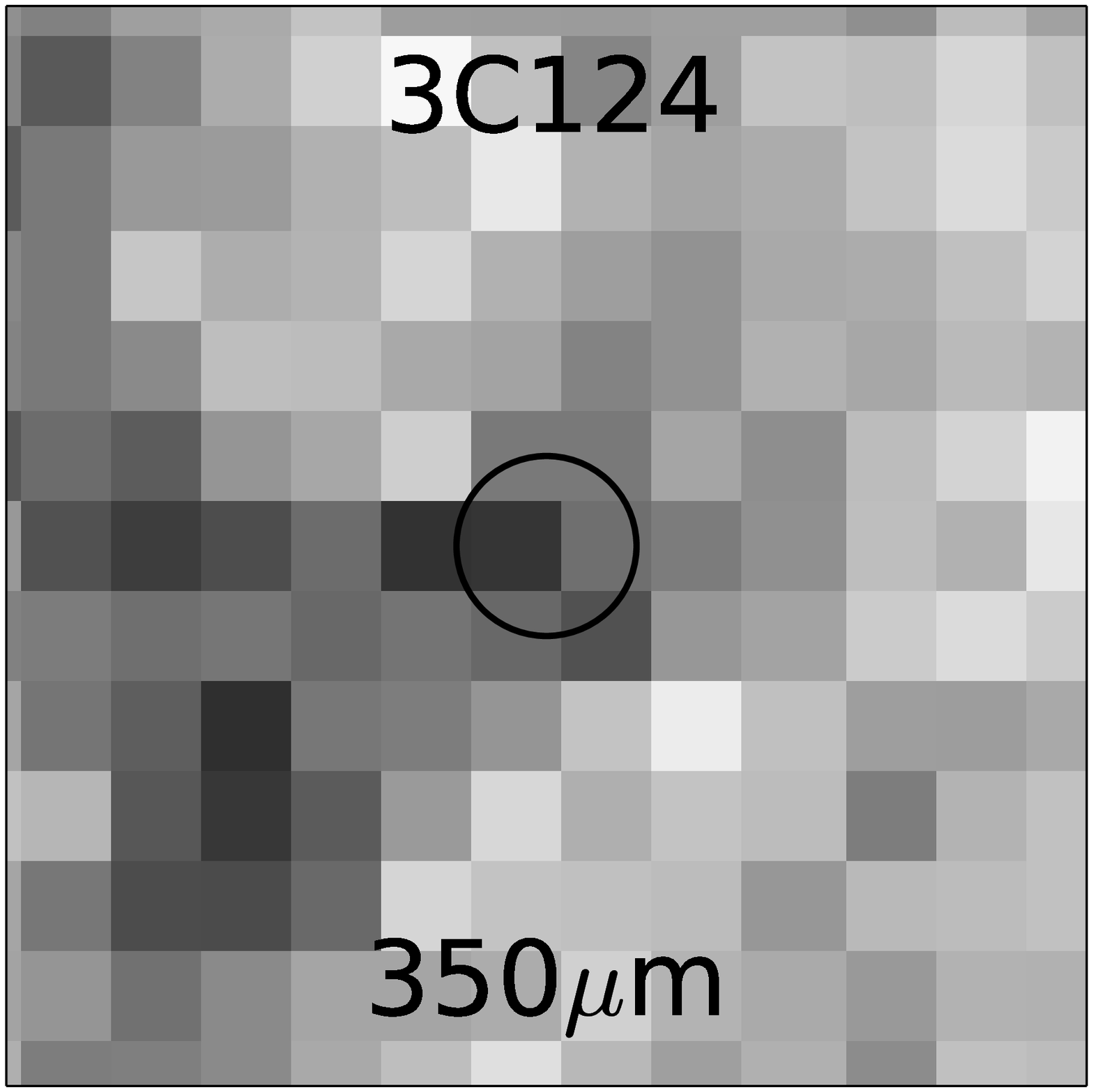}
      \includegraphics[width=1.5cm]{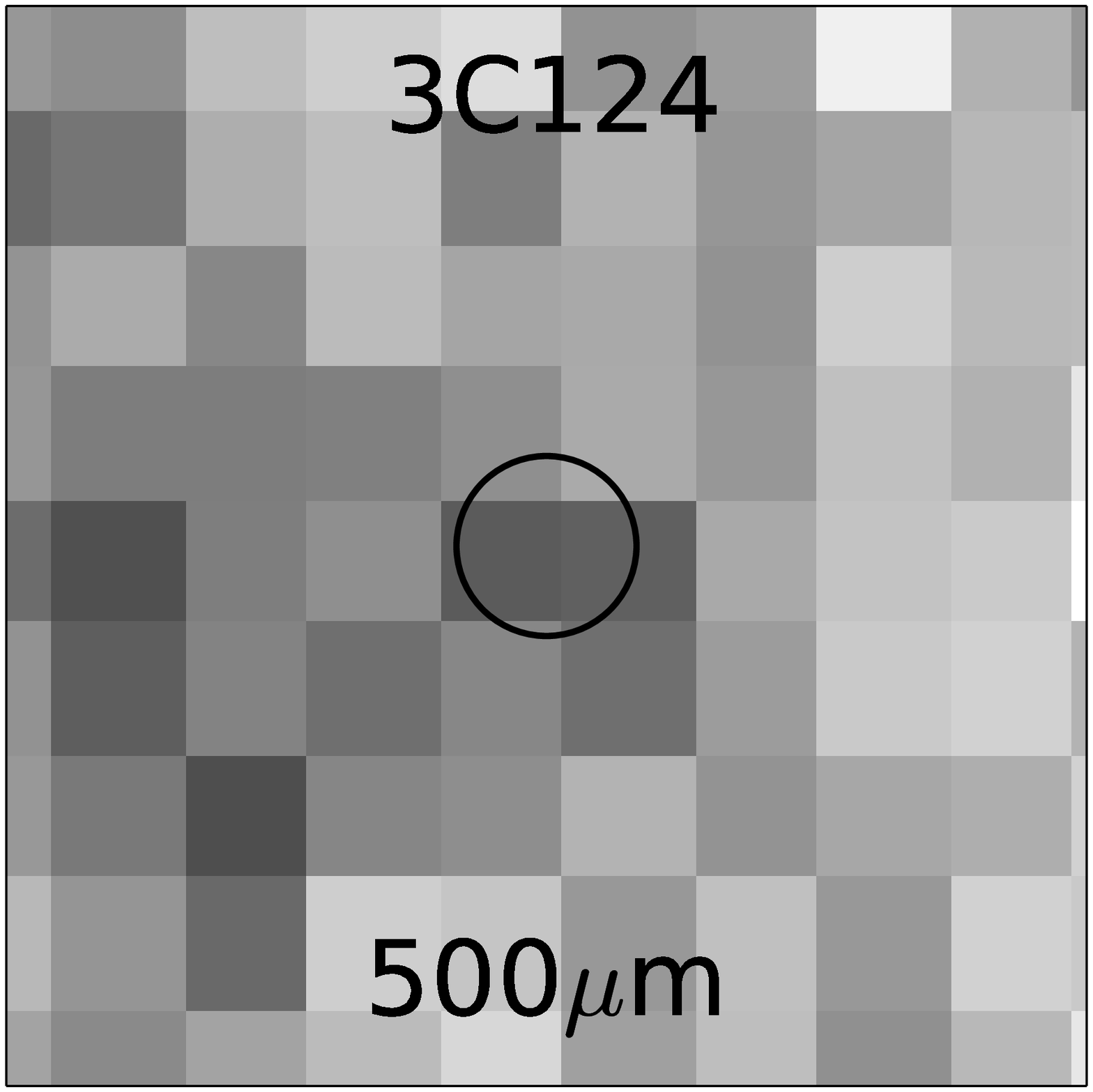}
      \\
      \includegraphics[width=1.5cm]{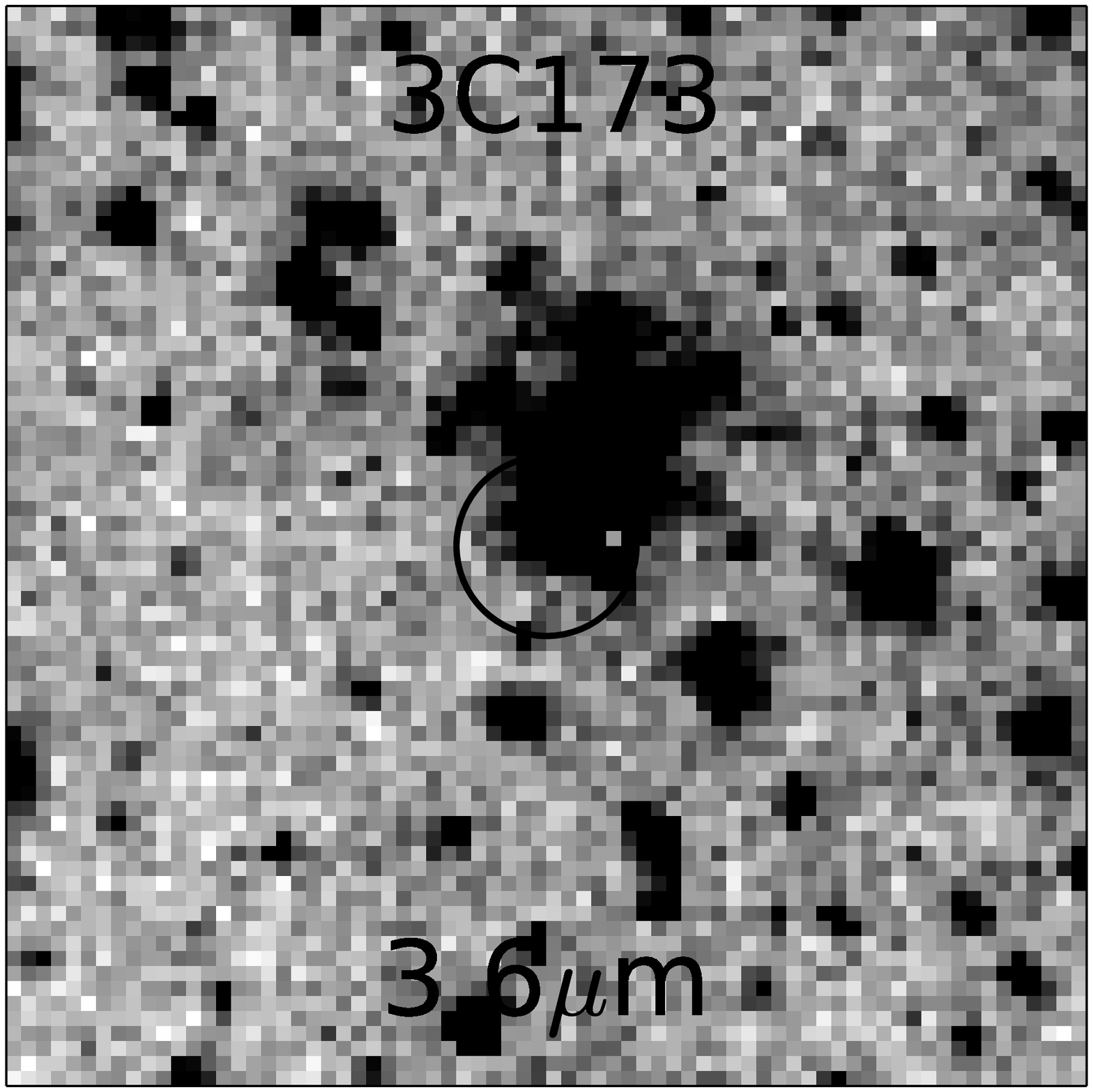}
      \includegraphics[width=1.5cm]{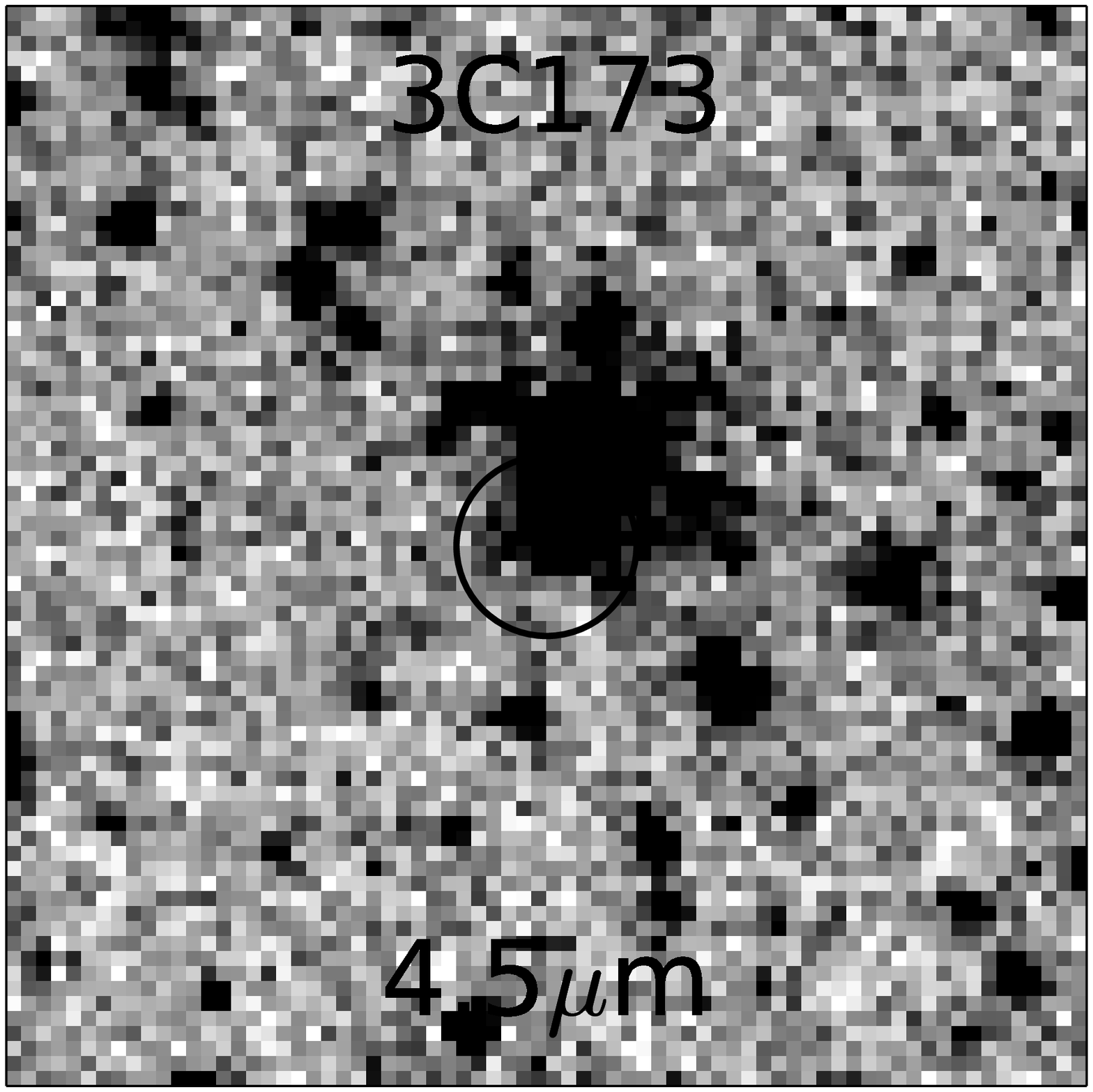}
      \includegraphics[width=1.5cm]{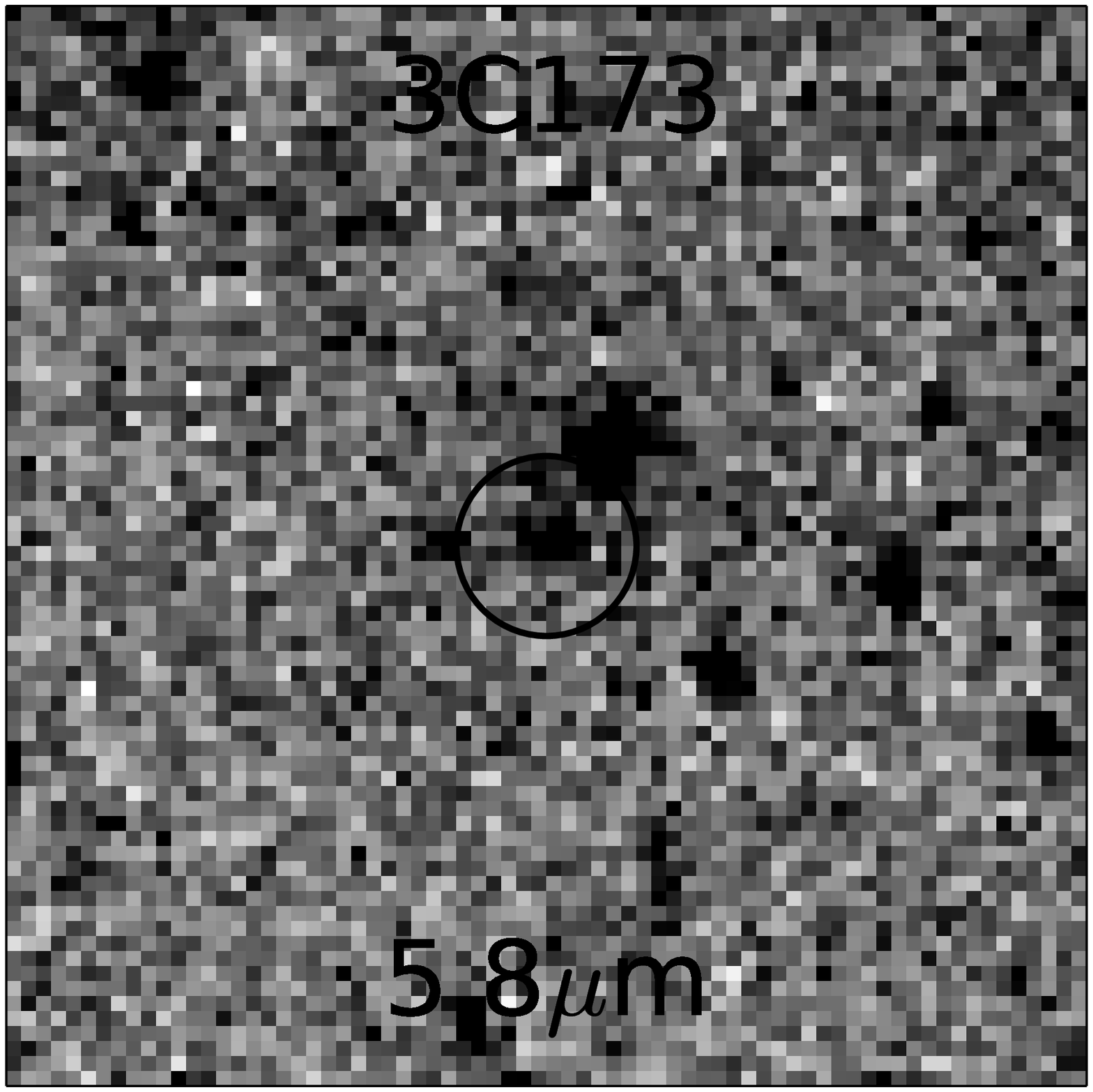}
      \includegraphics[width=1.5cm]{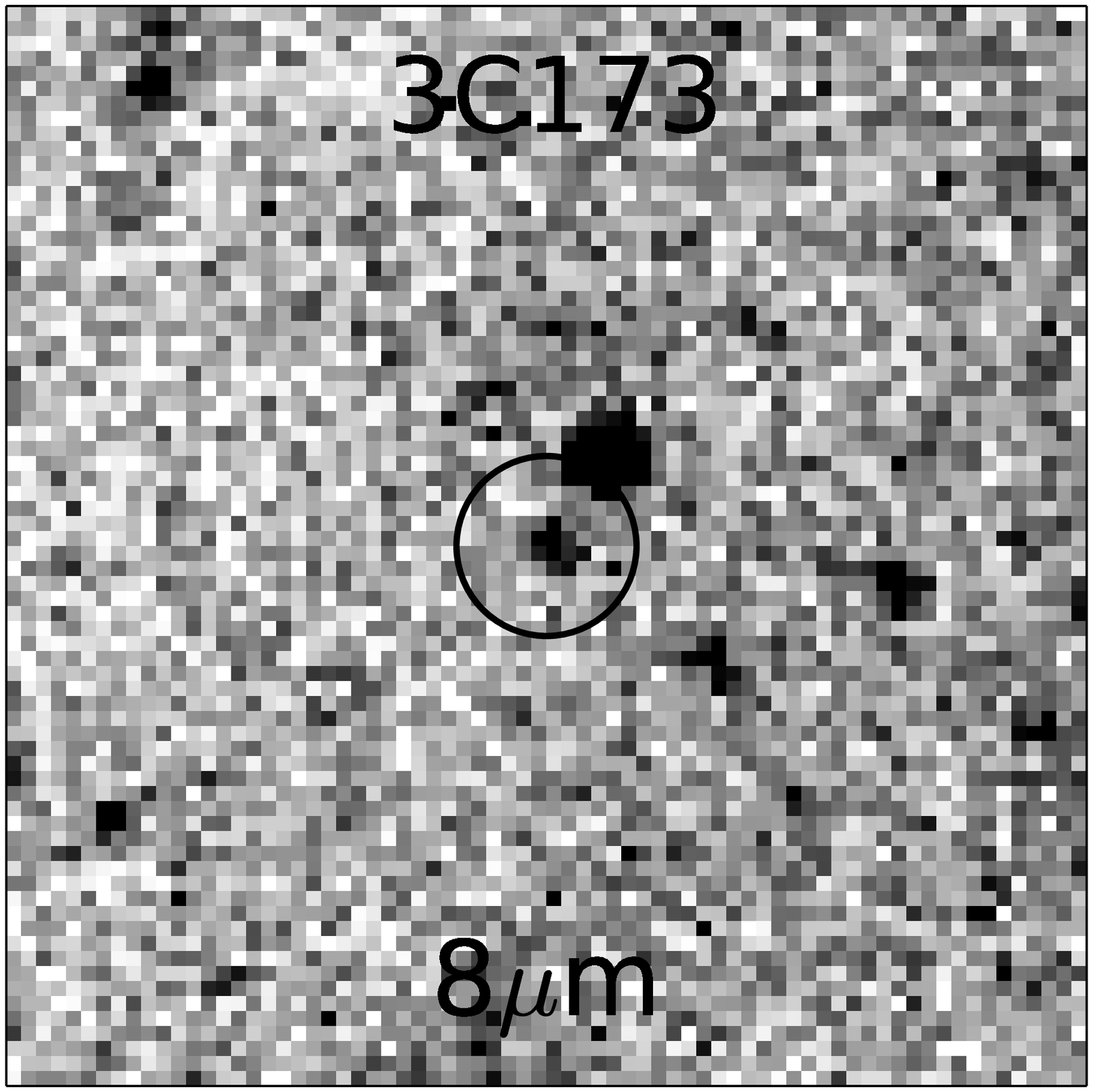}
      \includegraphics[width=1.5cm]{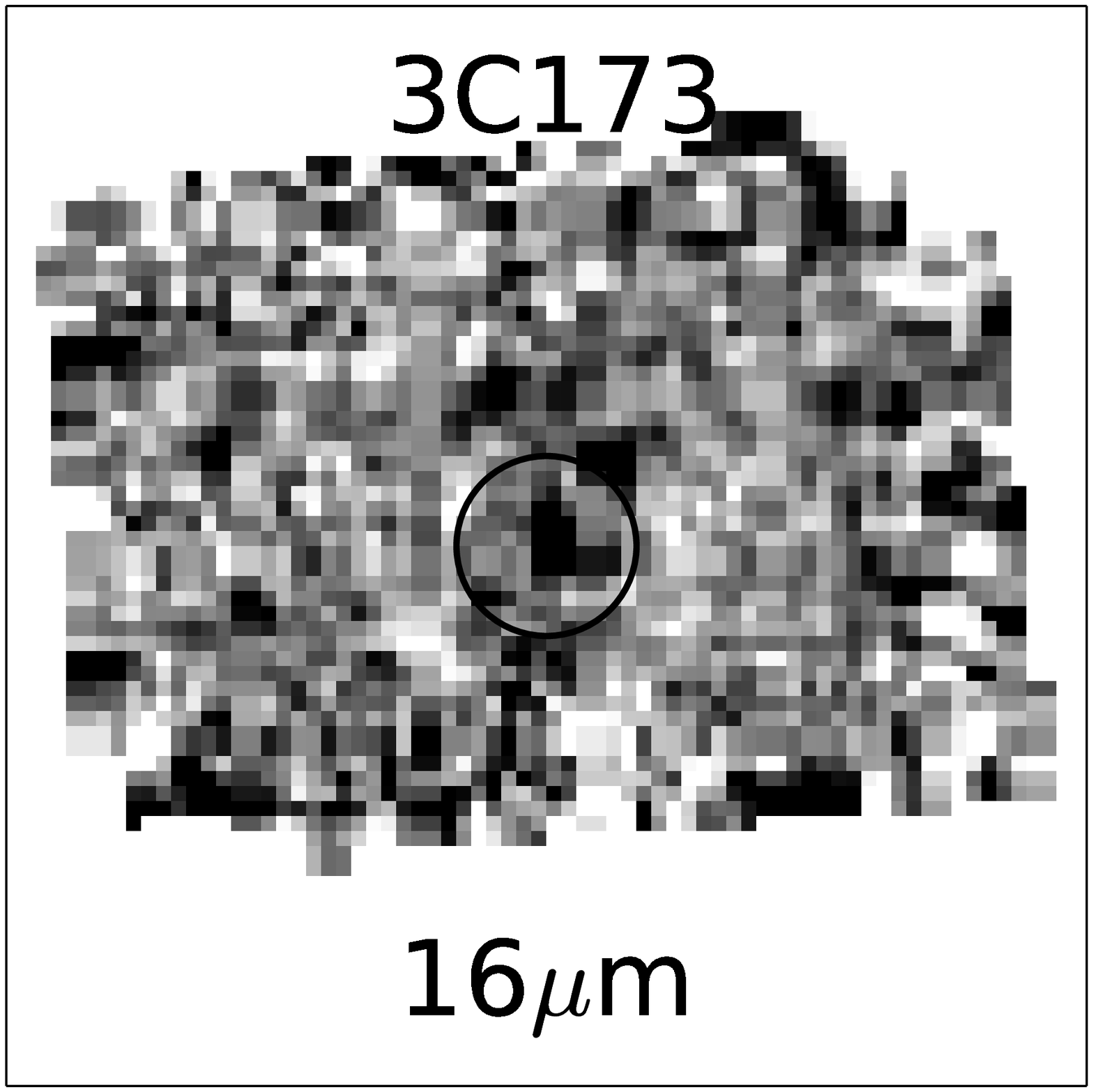}
      \includegraphics[width=1.5cm]{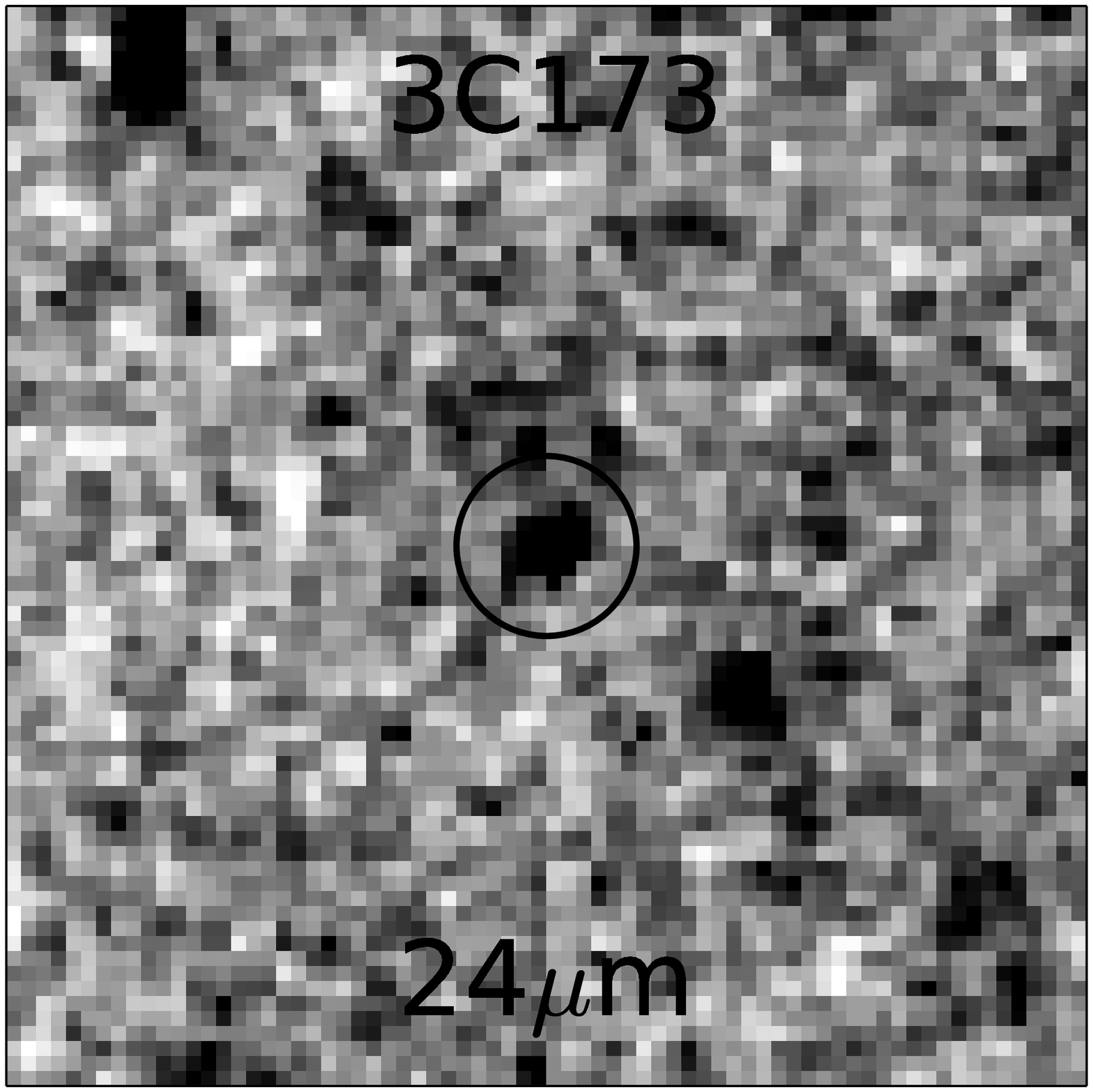}
      \includegraphics[width=1.5cm]{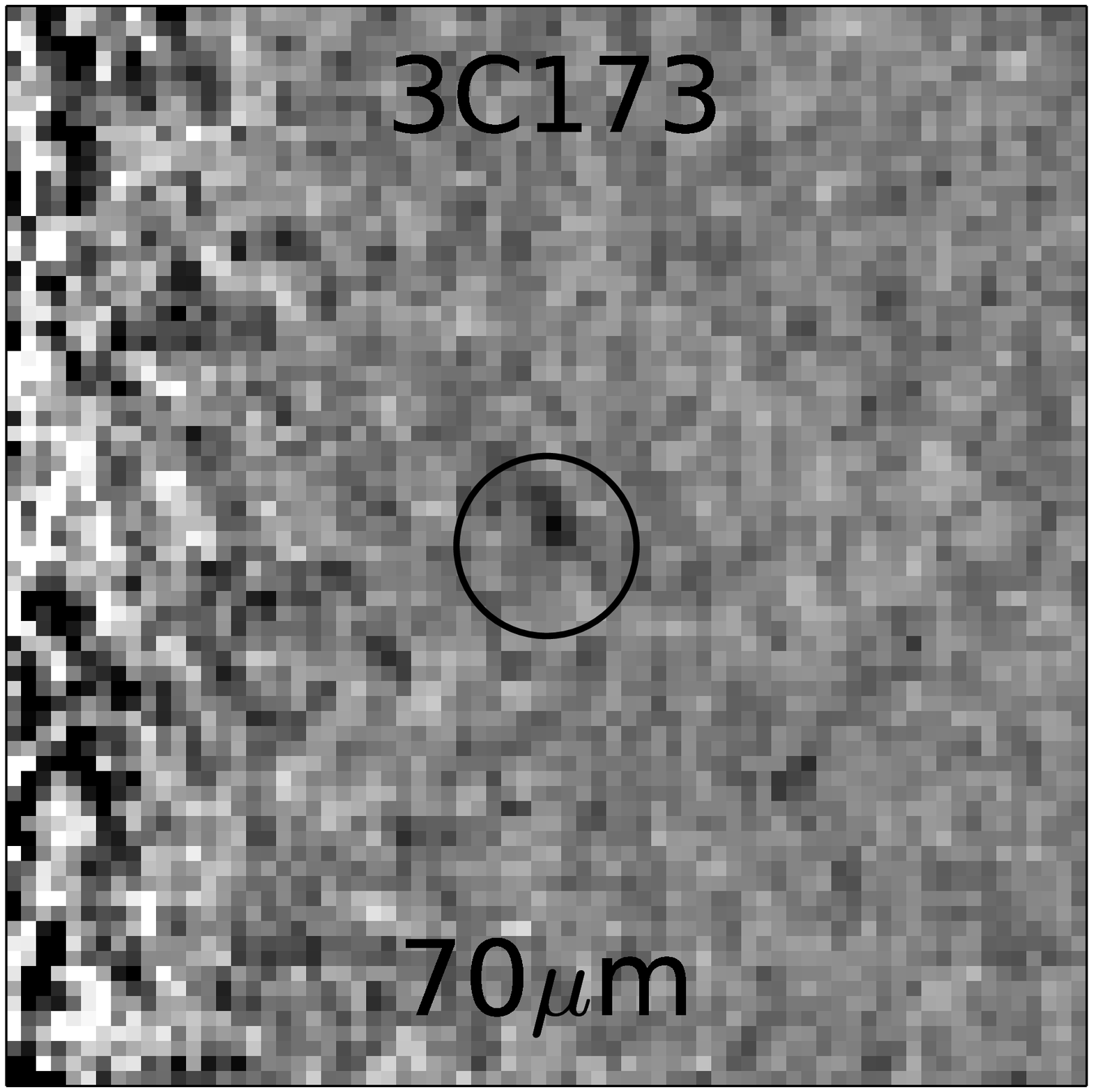}
      \includegraphics[width=1.5cm]{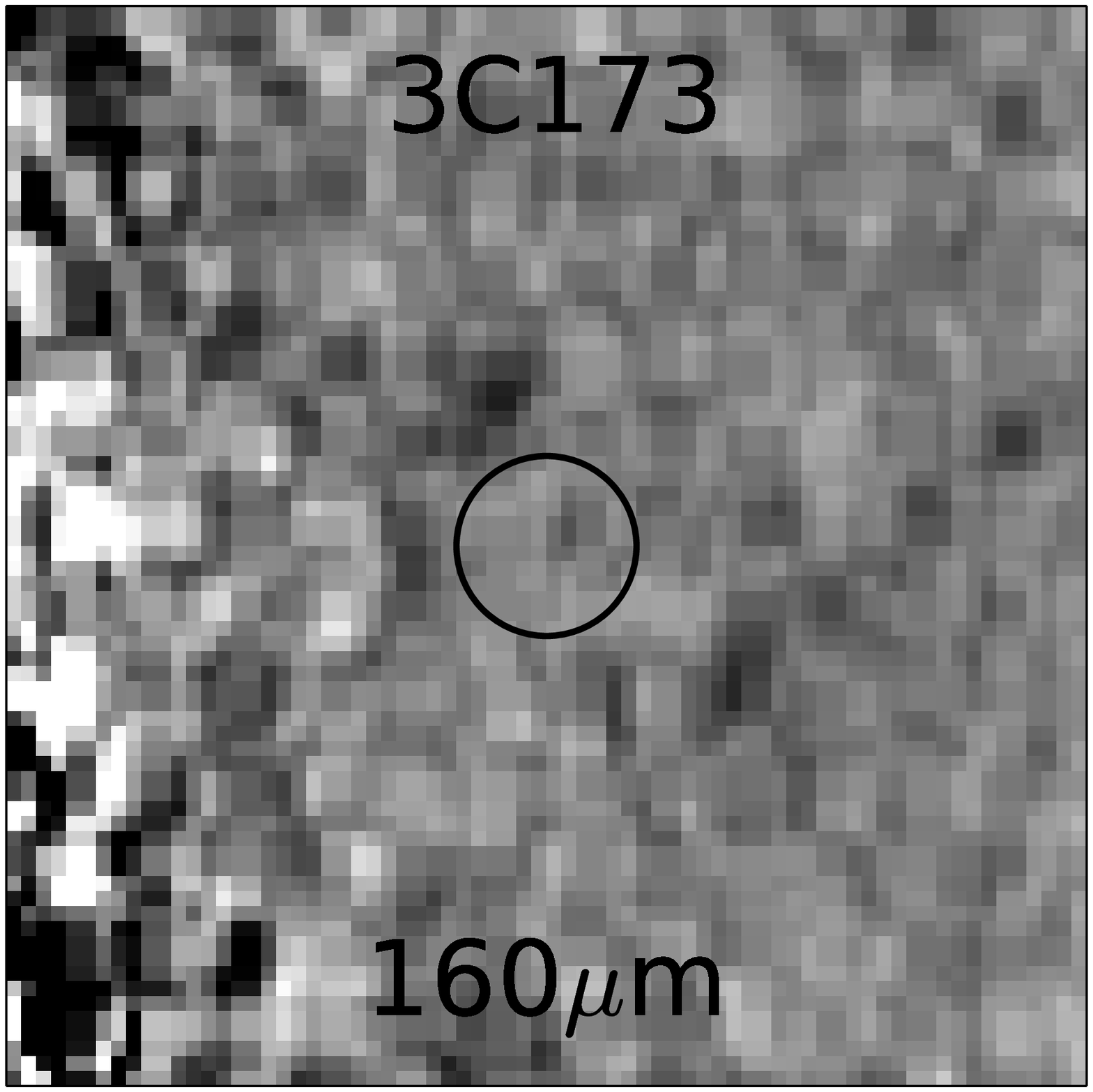}
      \includegraphics[width=1.5cm]{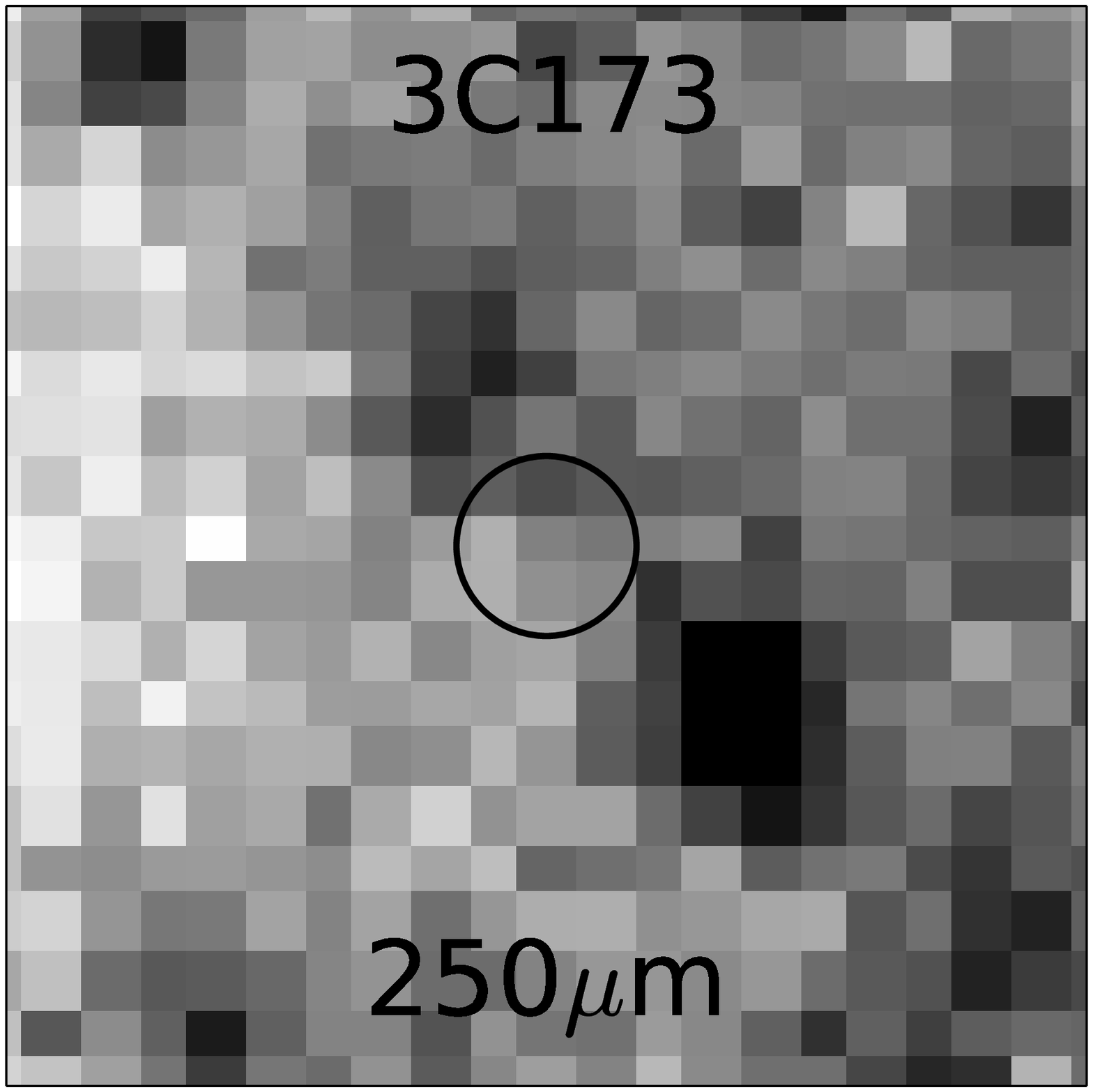}
      \includegraphics[width=1.5cm]{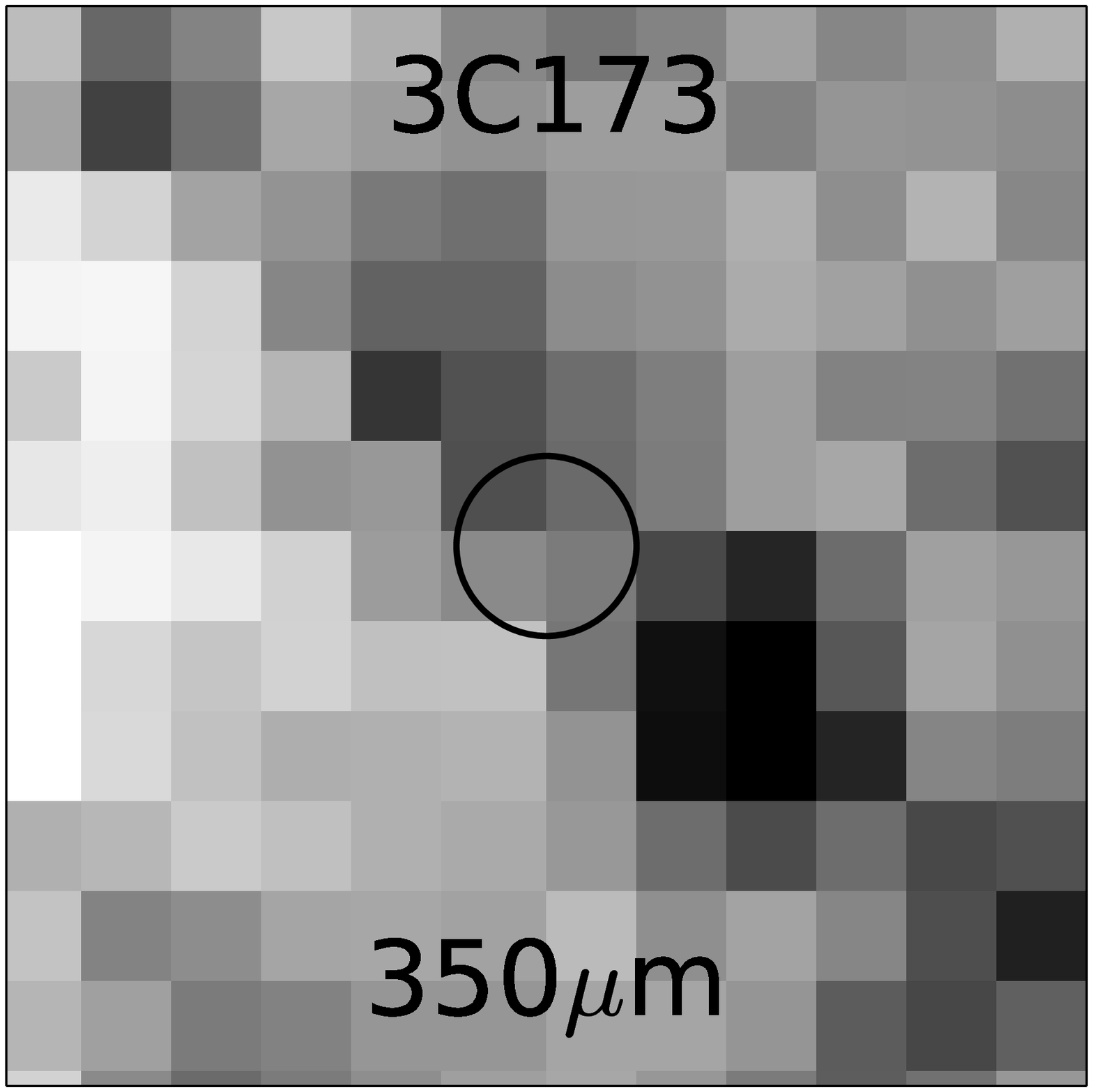}
      \includegraphics[width=1.5cm]{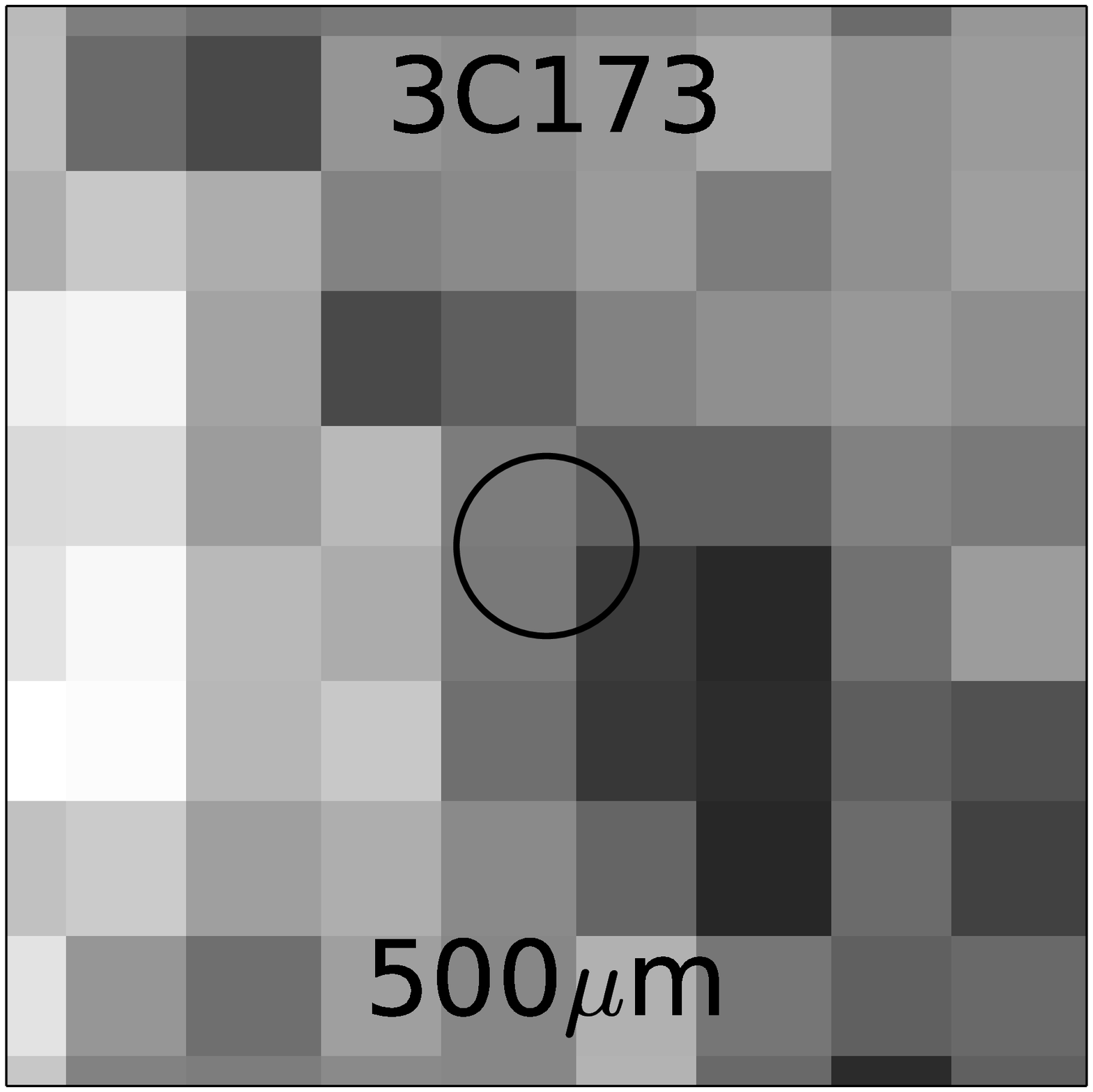}
      \\
      \includegraphics[width=1.5cm]{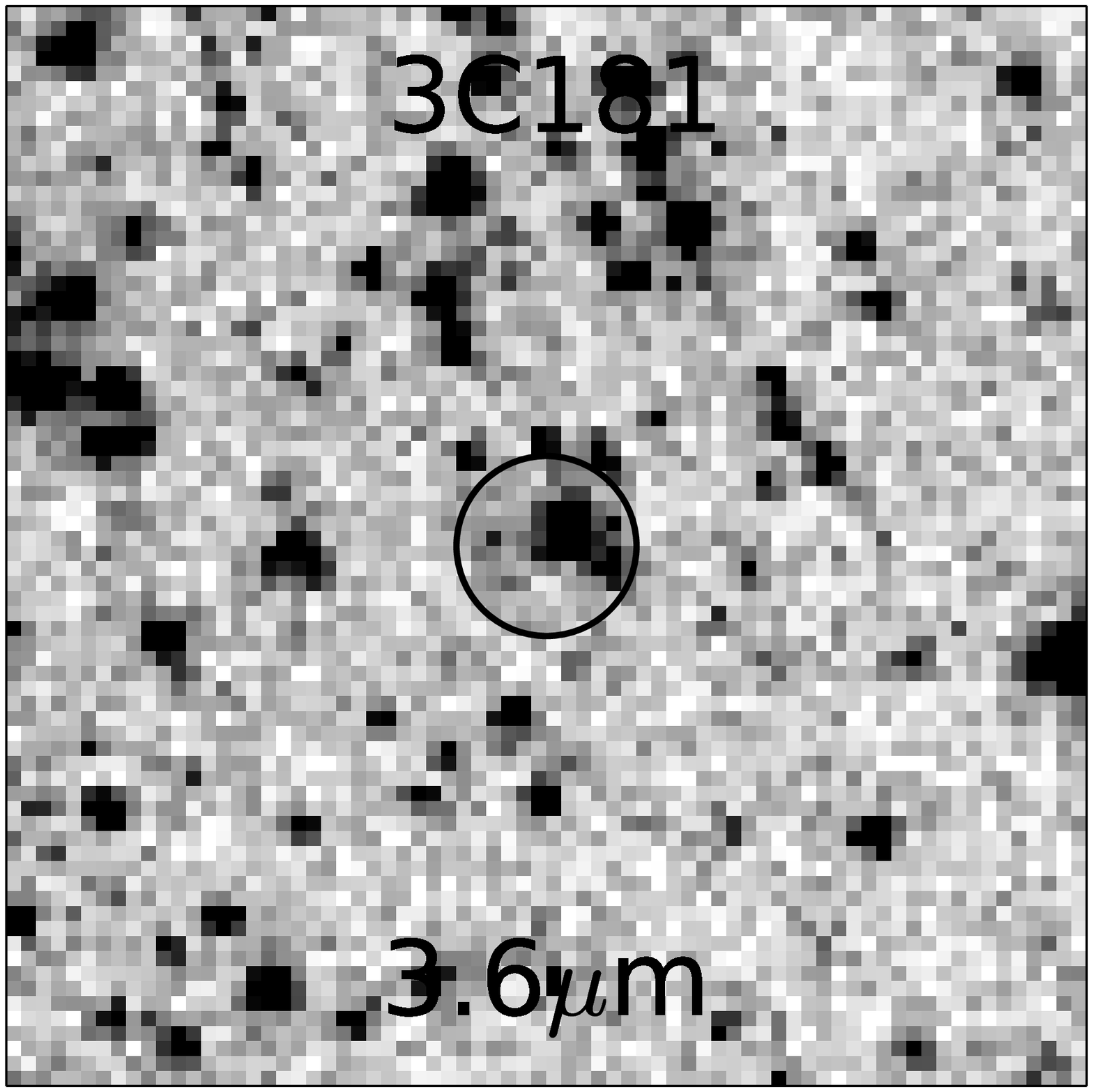}
      \includegraphics[width=1.5cm]{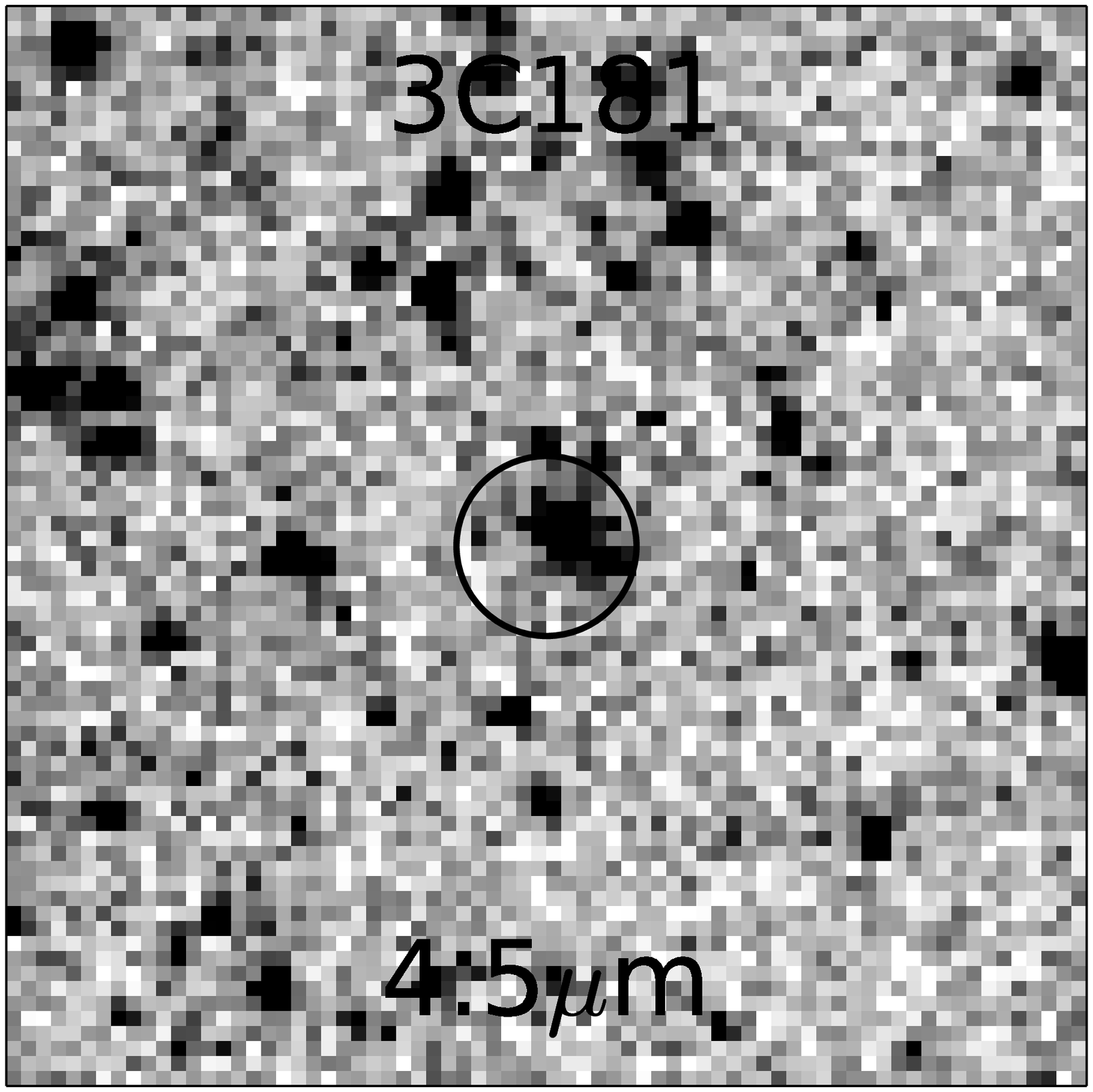}
      \includegraphics[width=1.5cm]{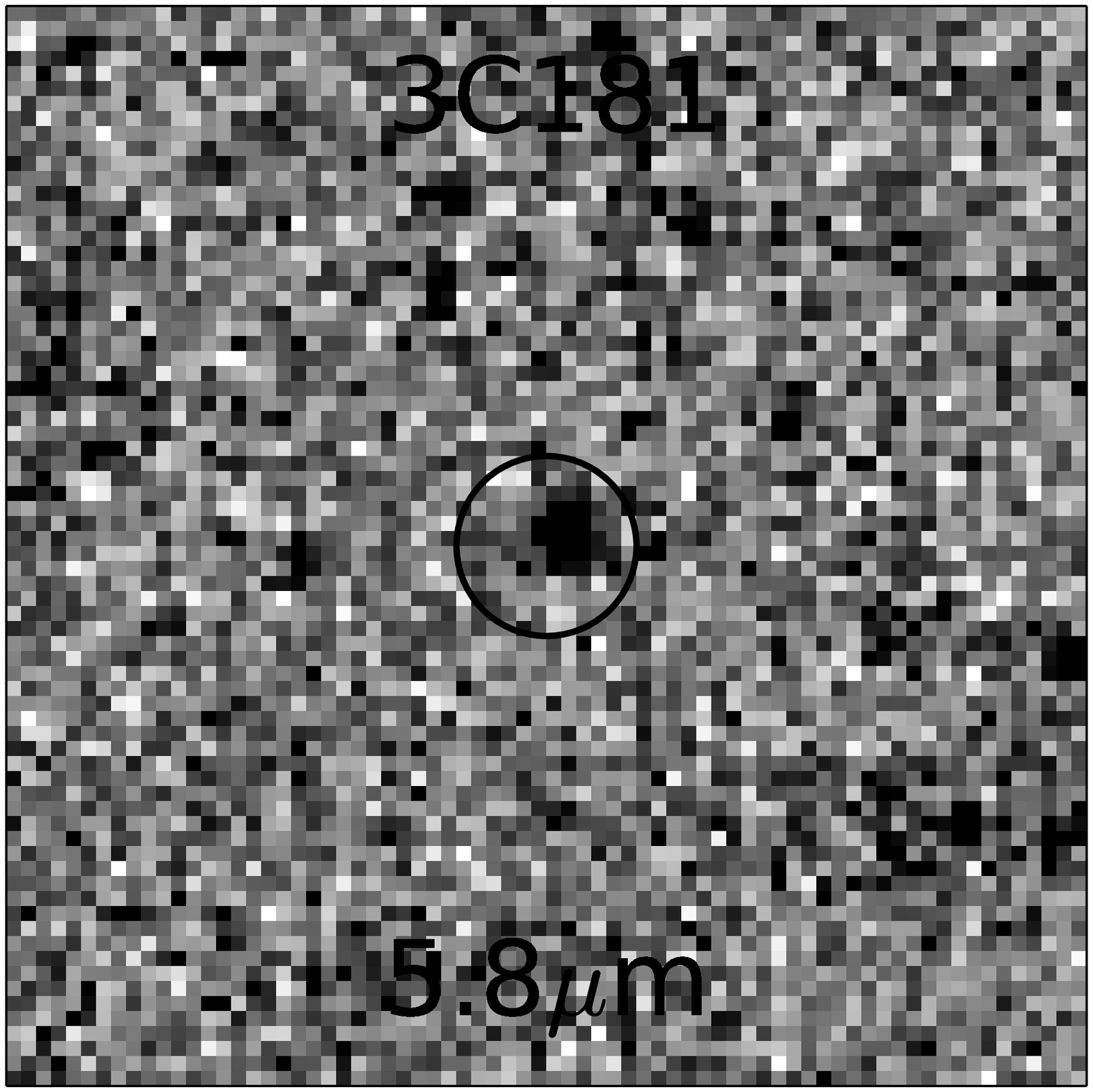}
      \includegraphics[width=1.5cm]{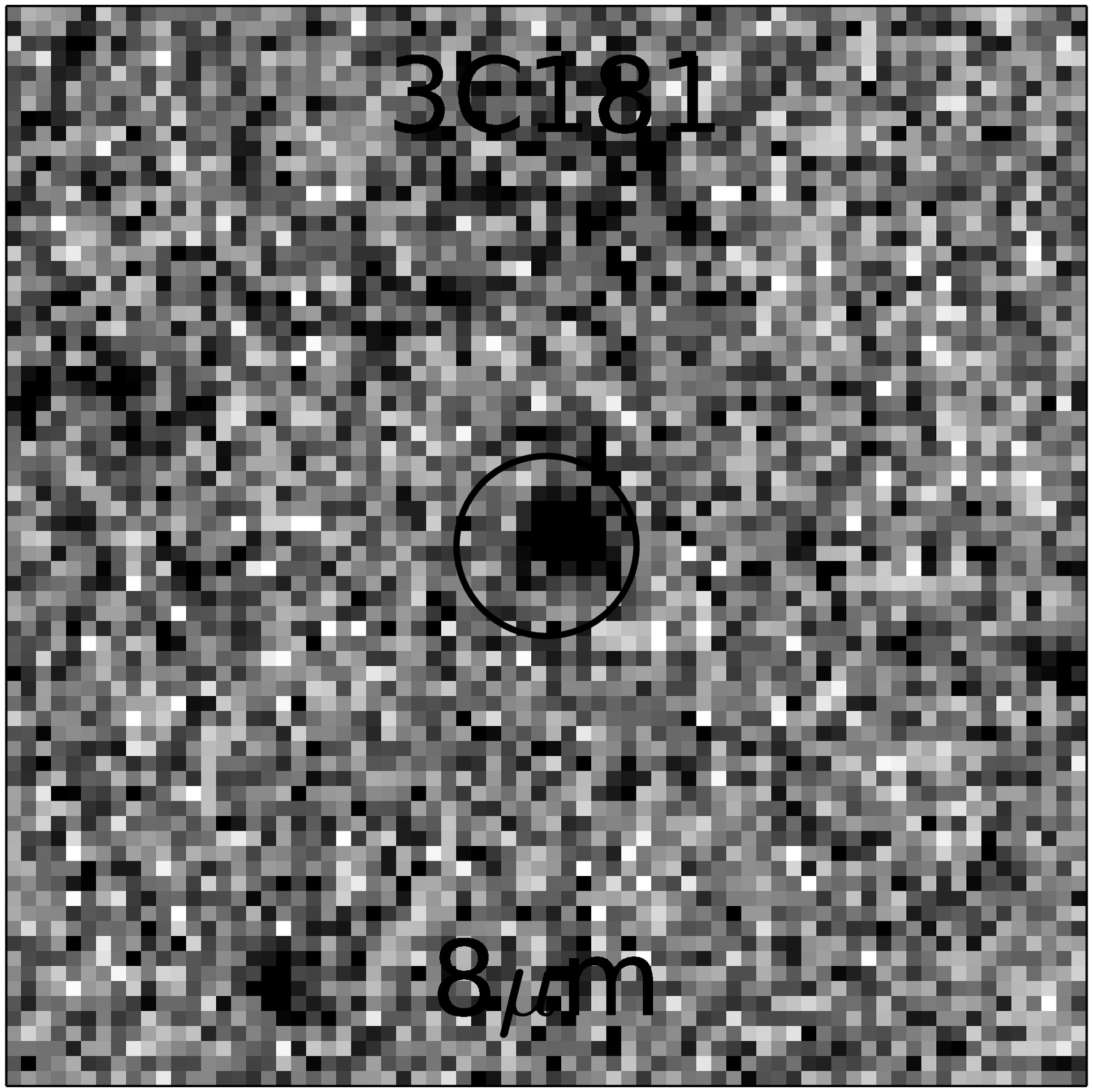}
      \includegraphics[width=1.5cm]{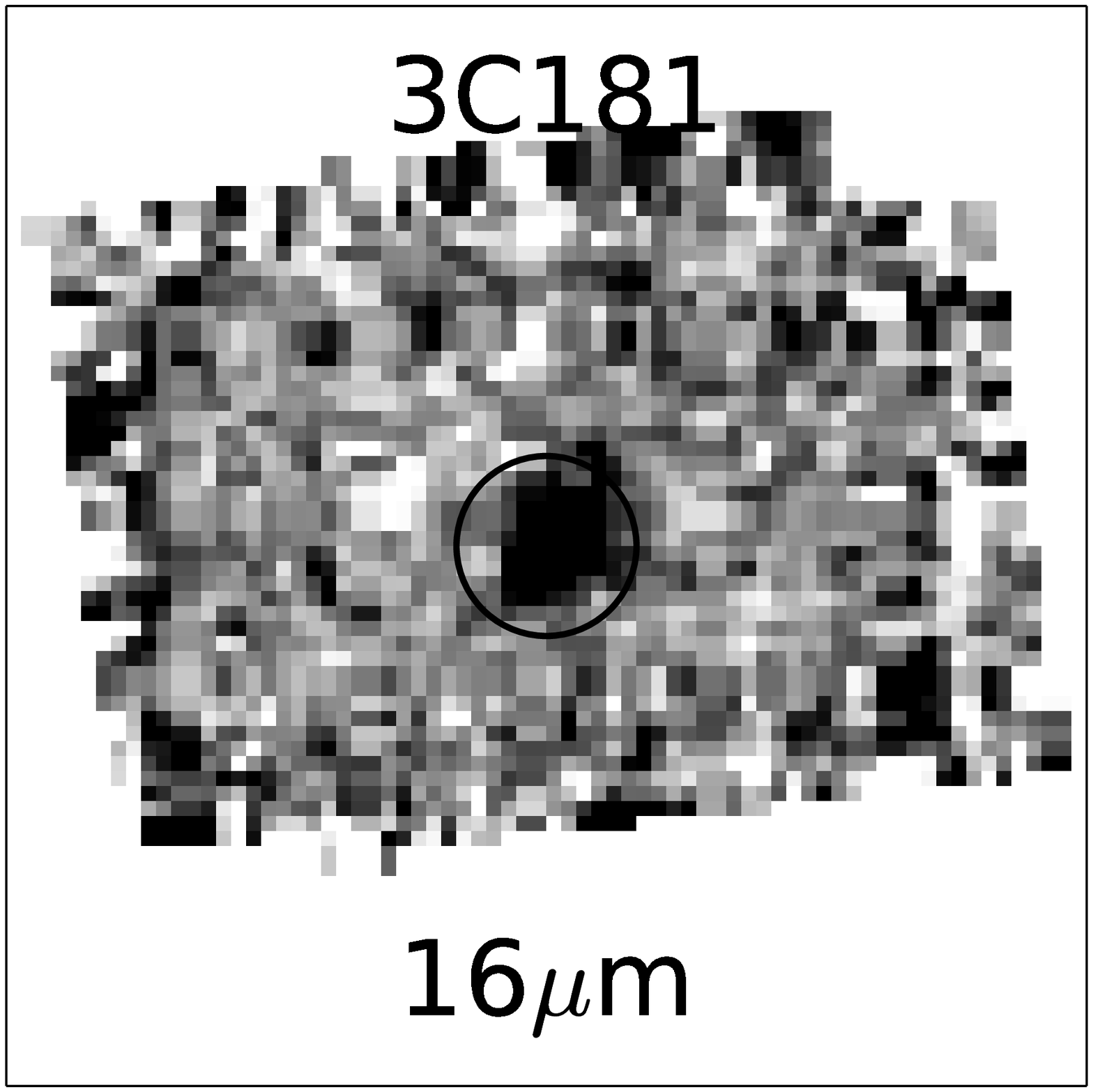}
      \includegraphics[width=1.5cm]{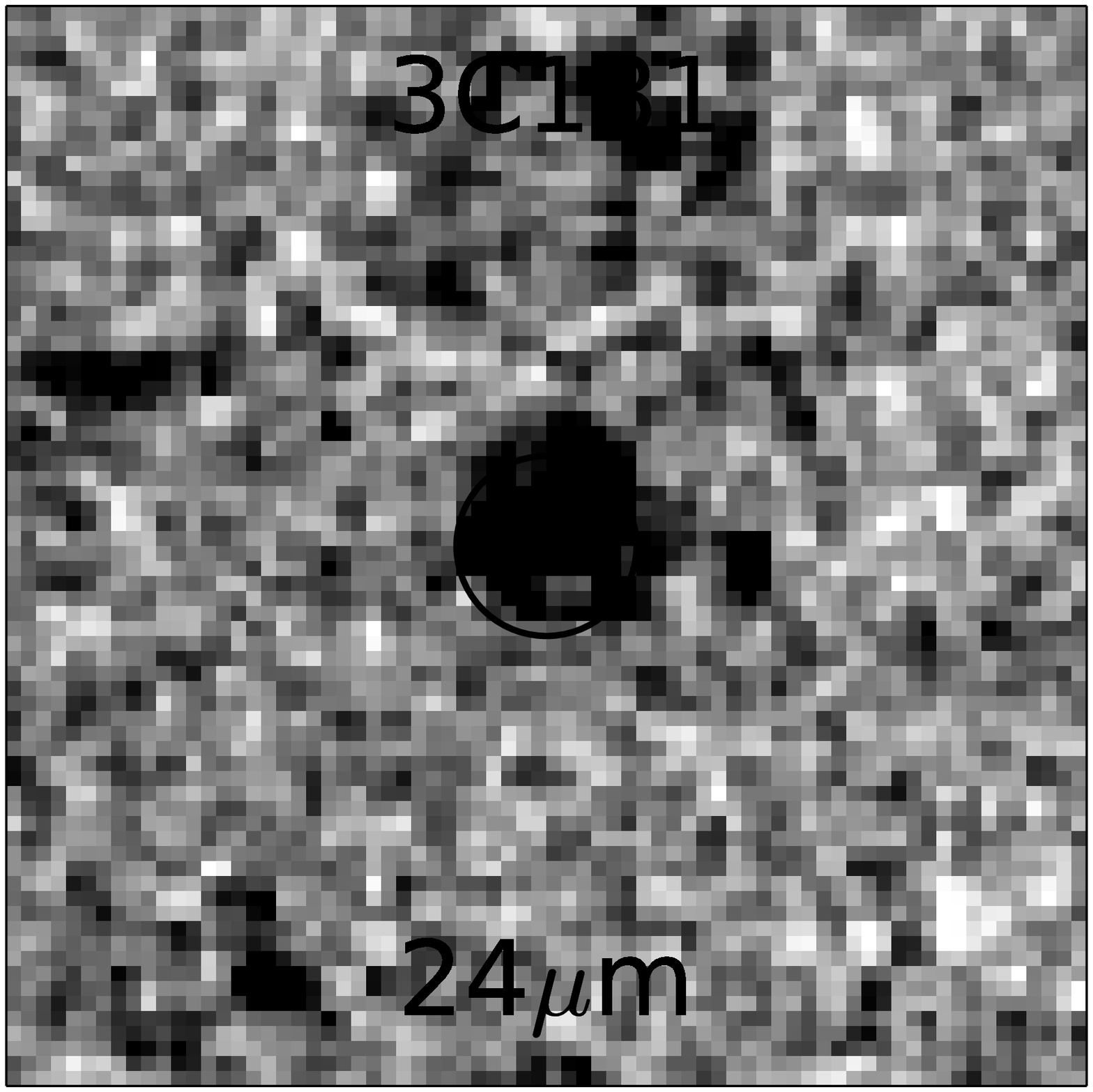}
      \includegraphics[width=1.5cm]{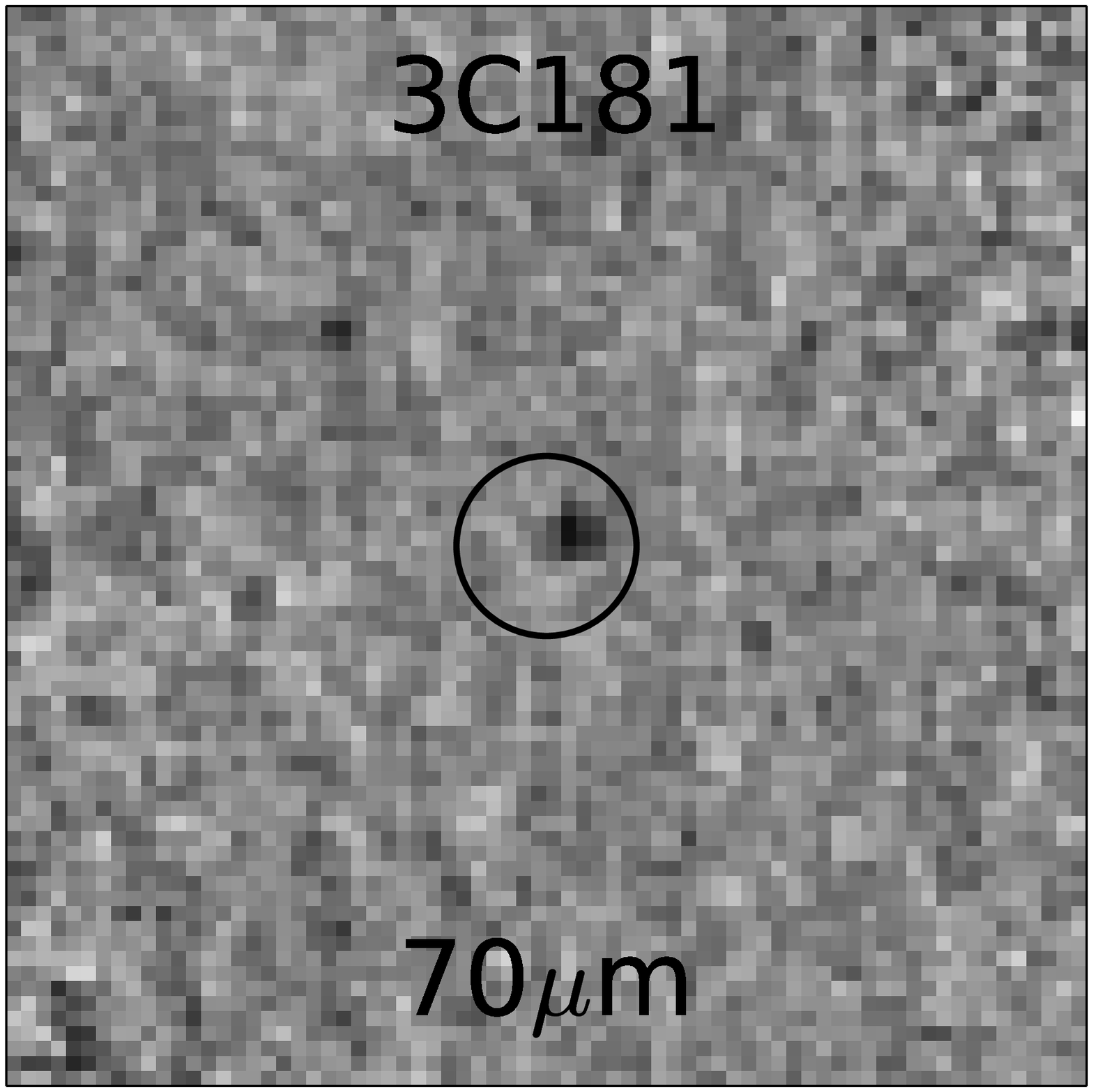}
      \includegraphics[width=1.5cm]{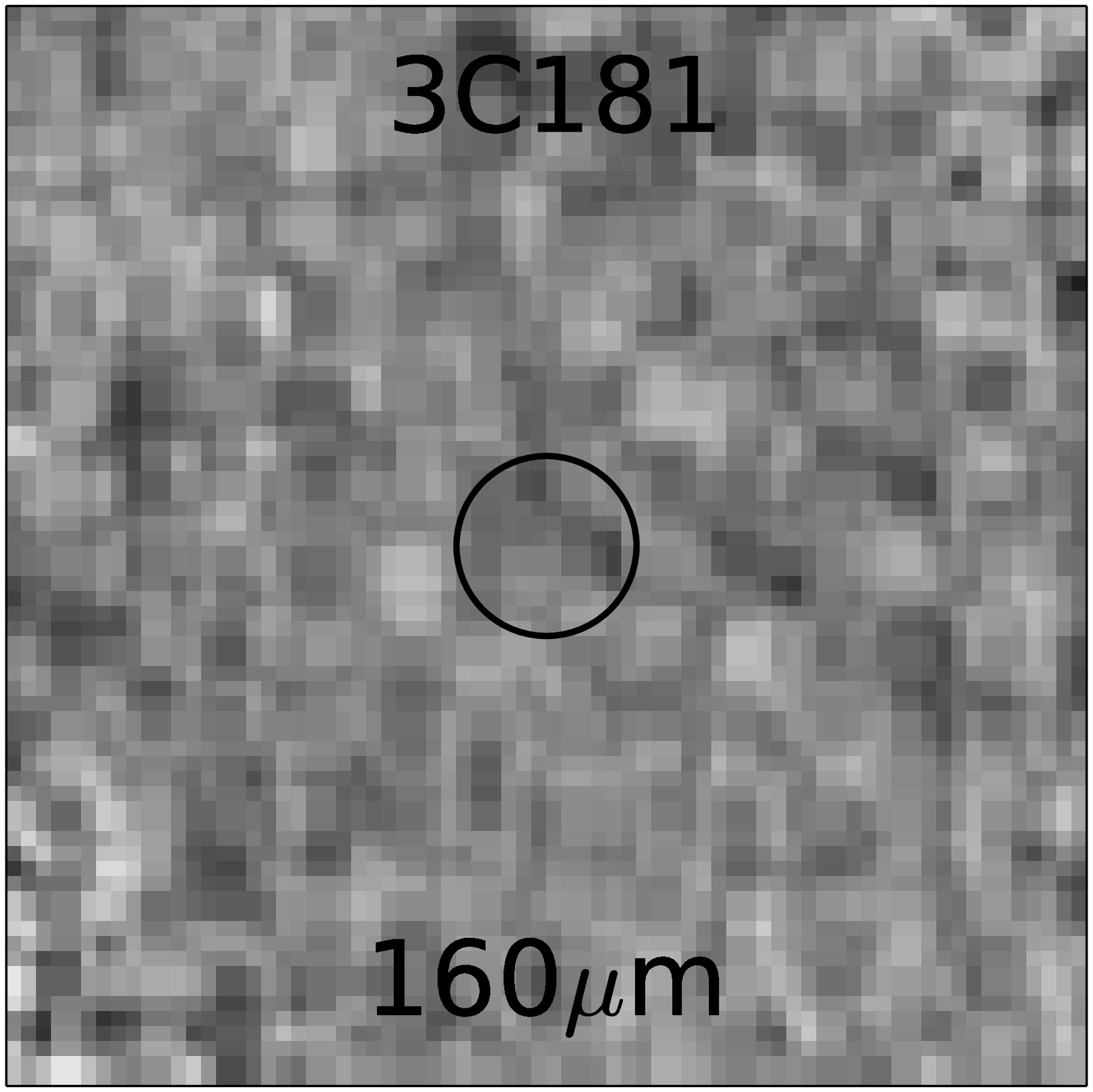}
      \includegraphics[width=1.5cm]{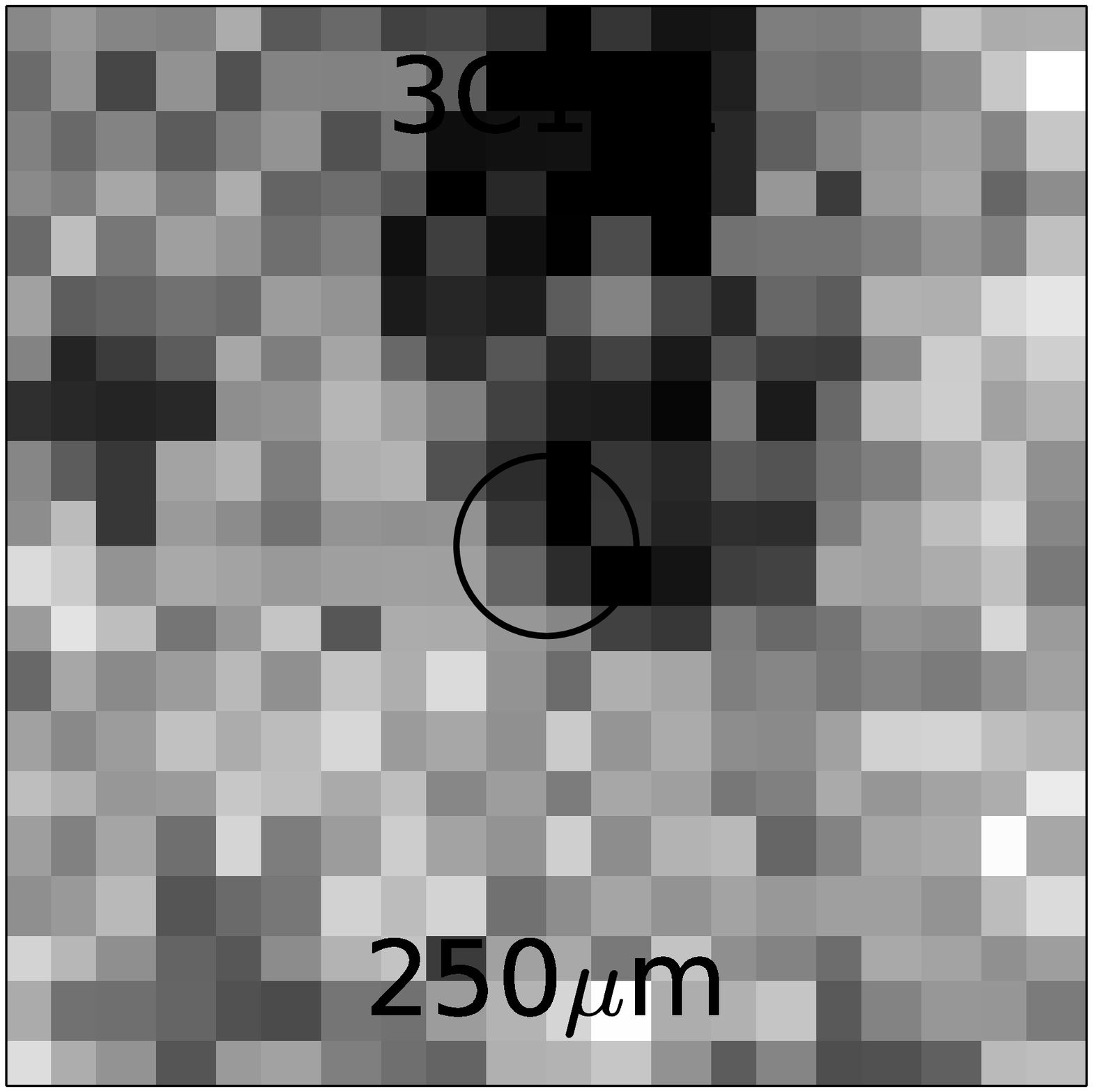}
      \includegraphics[width=1.5cm]{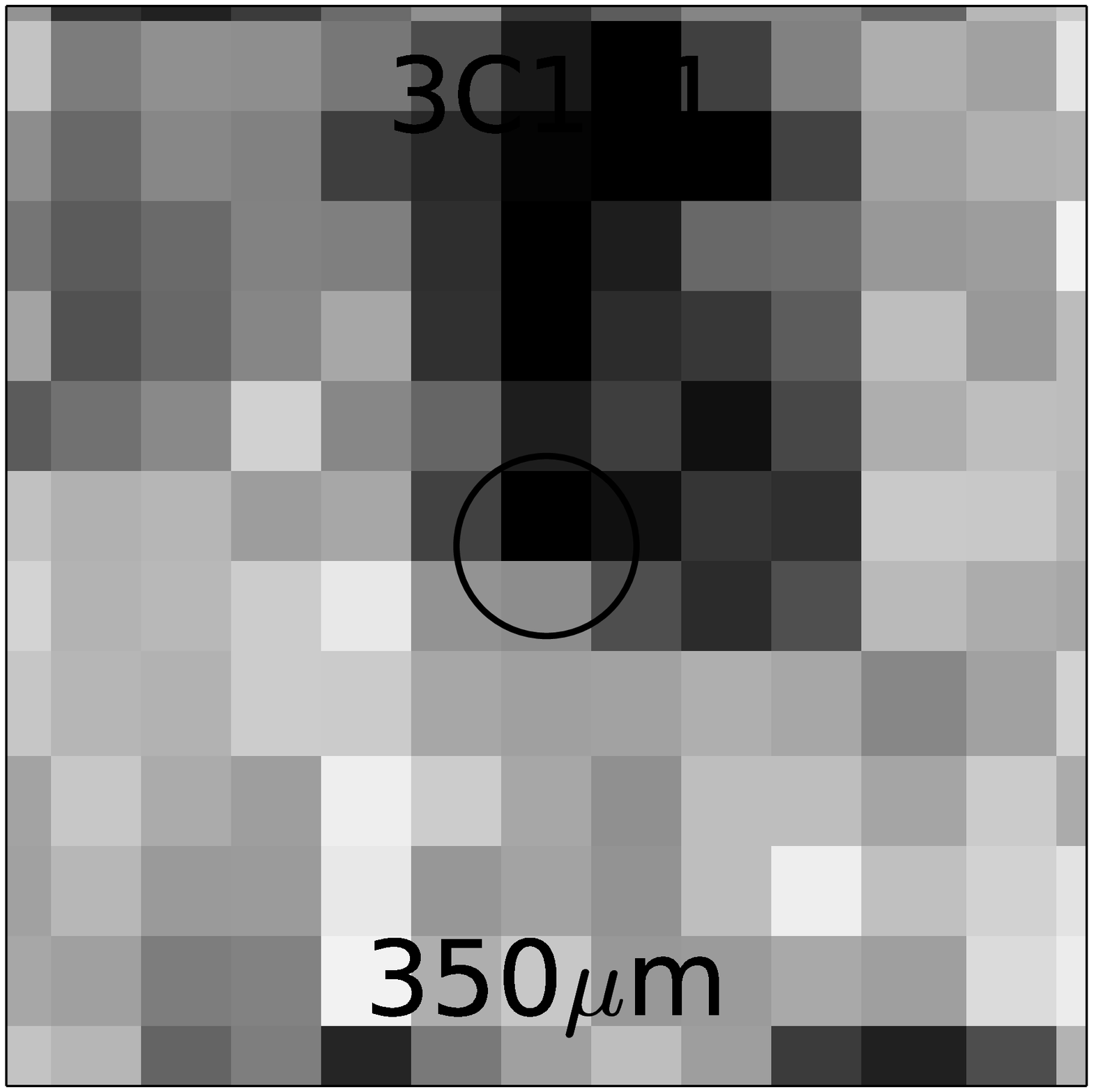}
      \includegraphics[width=1.5cm]{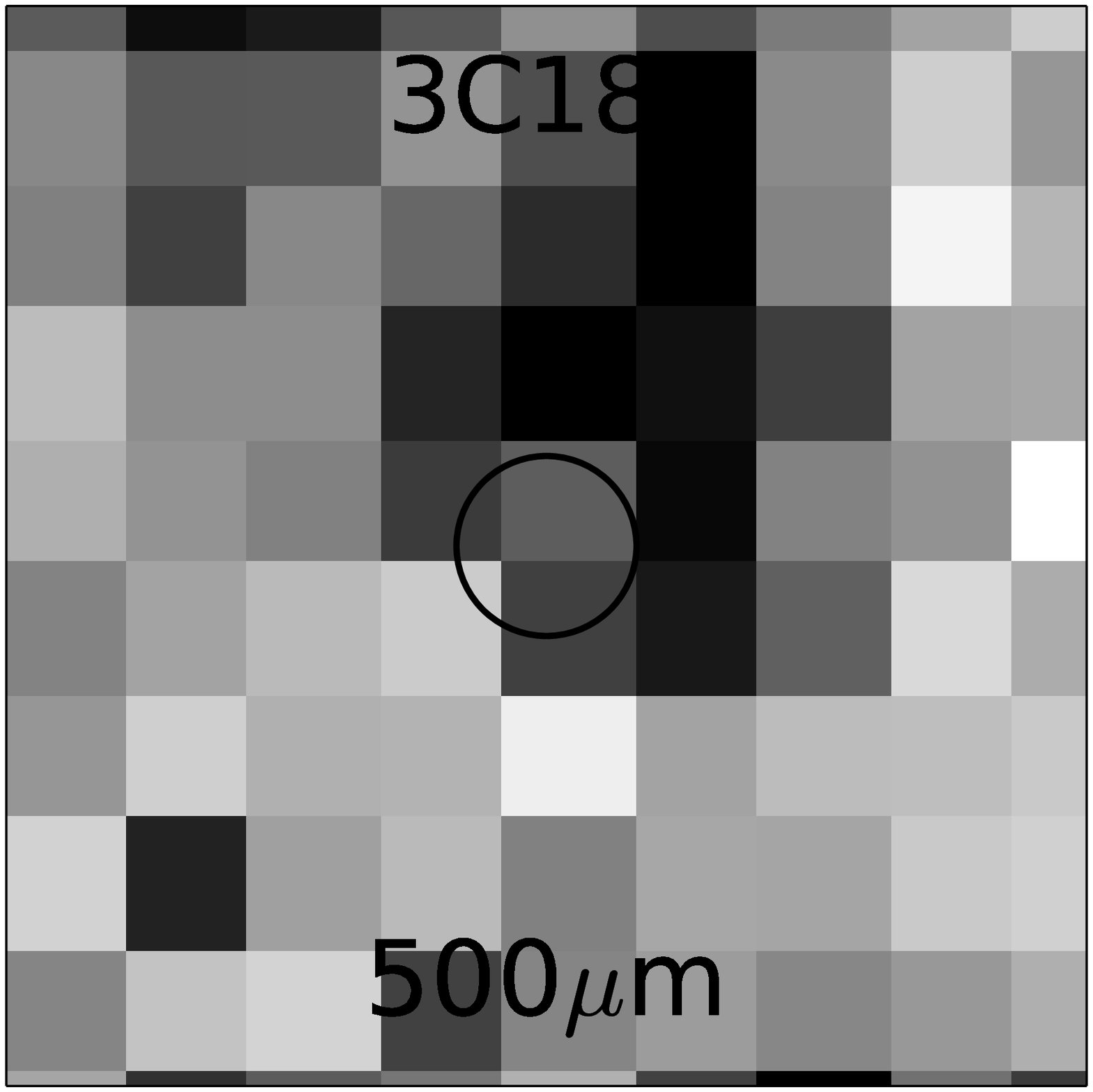}
      \caption{Postage stamps of the high-$z$ 3CR objects studied in this work. 
               From left to right: IRAC 3.6~$\mu$m, IRAC 4.5~$\mu$m, 
               IRAC 5.8~$\mu$m, IRAC 8~$\mu$m, IRS 16~$\mu$m, MIPS 24~$\mu$m, 
               PACS 70~$\mu$m, PACS 160~$\mu$m, SPIRE 250~$\mu$m, 
               SPIRE 350~$\mu$m, and SPIRE 500~$\mu$m bands, respectively. 
               Each image shown here has dimensions of 2x2 arcmin. The 
               circle (10$\arcsec$ in radius) indicates the central 
               position of the source.   
               }
      \label{figure:stamps}
   \end{figure*}
   \addtocounter{figure}{-1}
   \begin{figure*}
      \includegraphics[width=1.5cm]{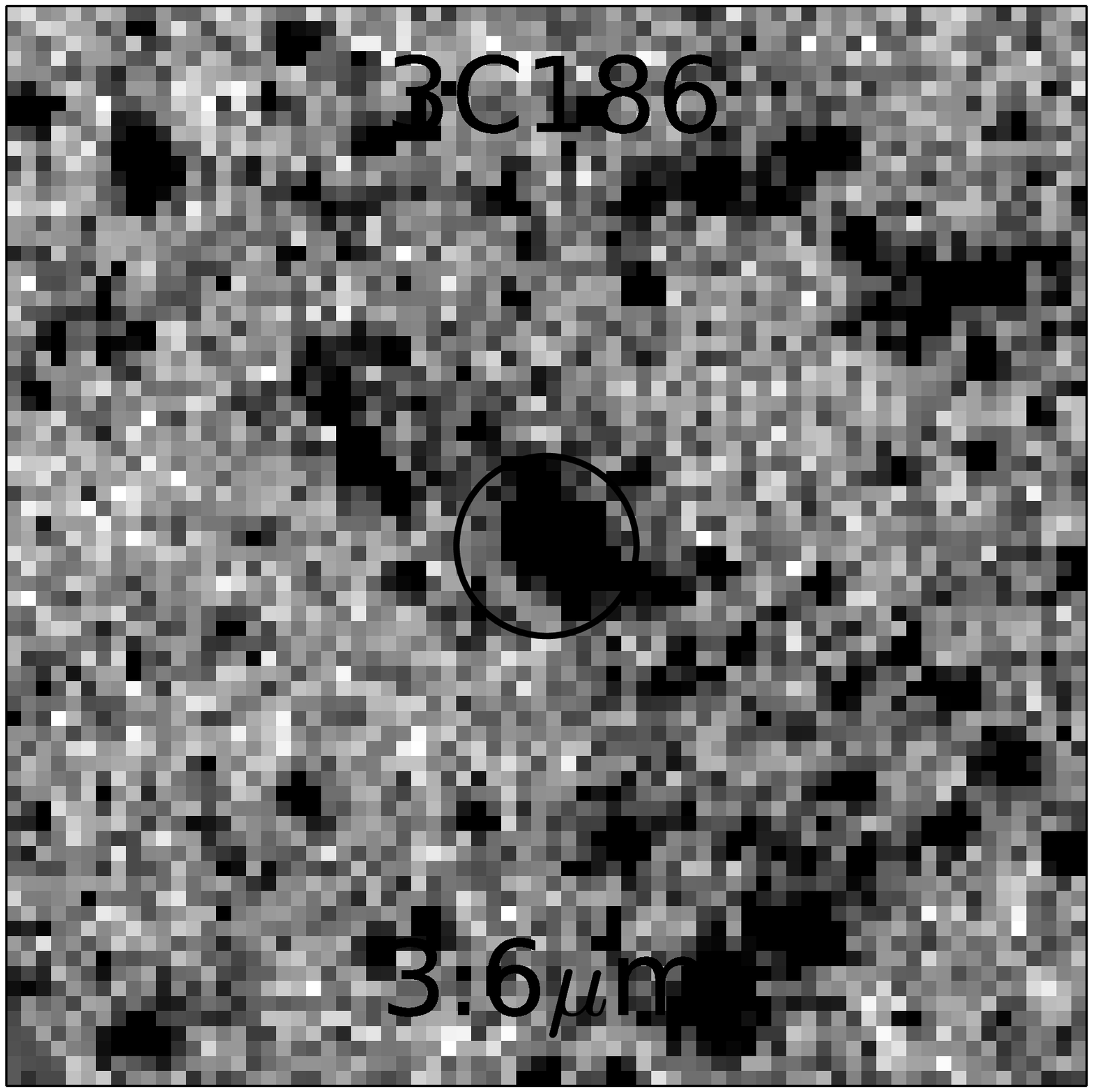}
      \includegraphics[width=1.5cm]{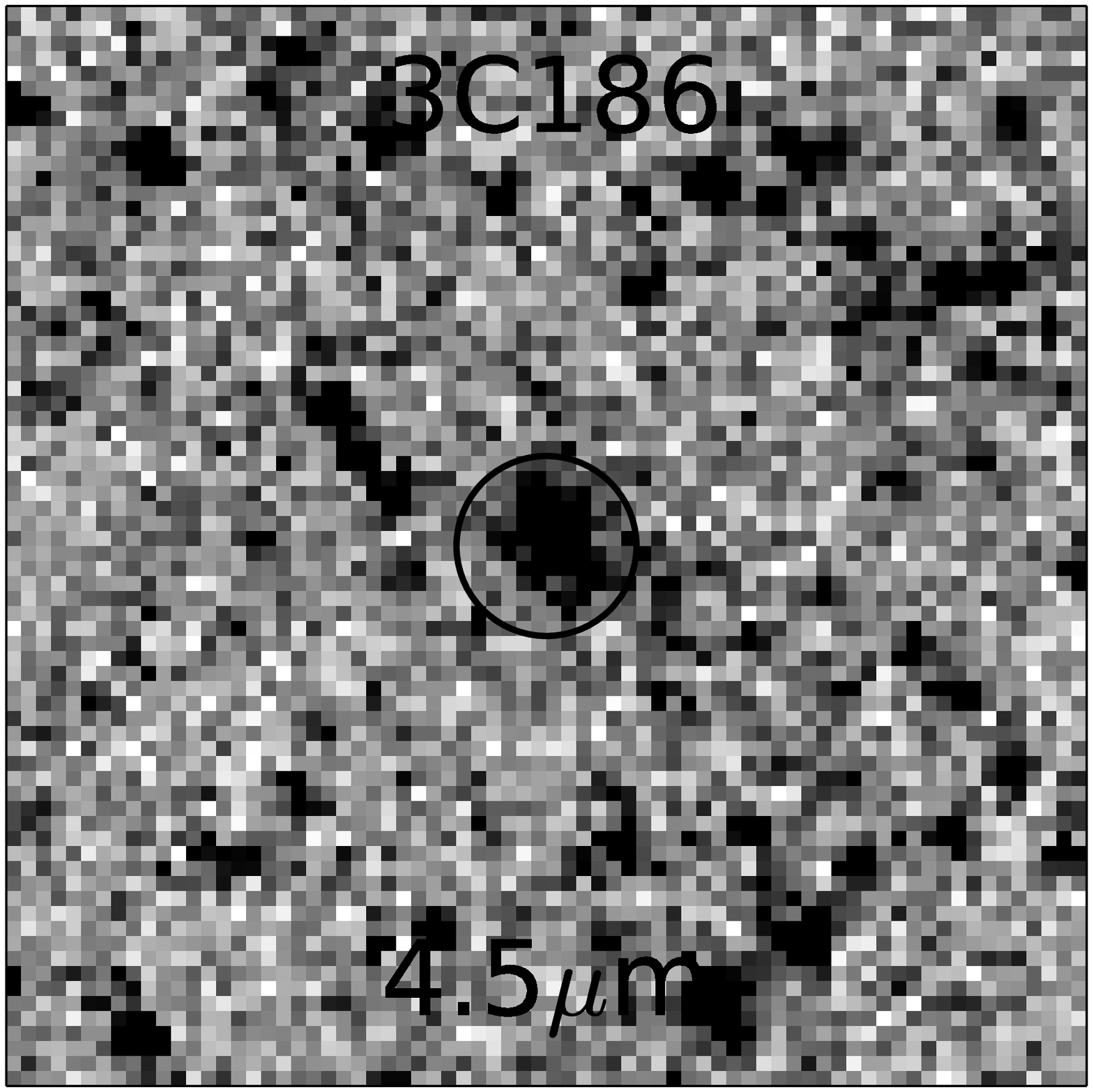}
      \includegraphics[width=1.5cm]{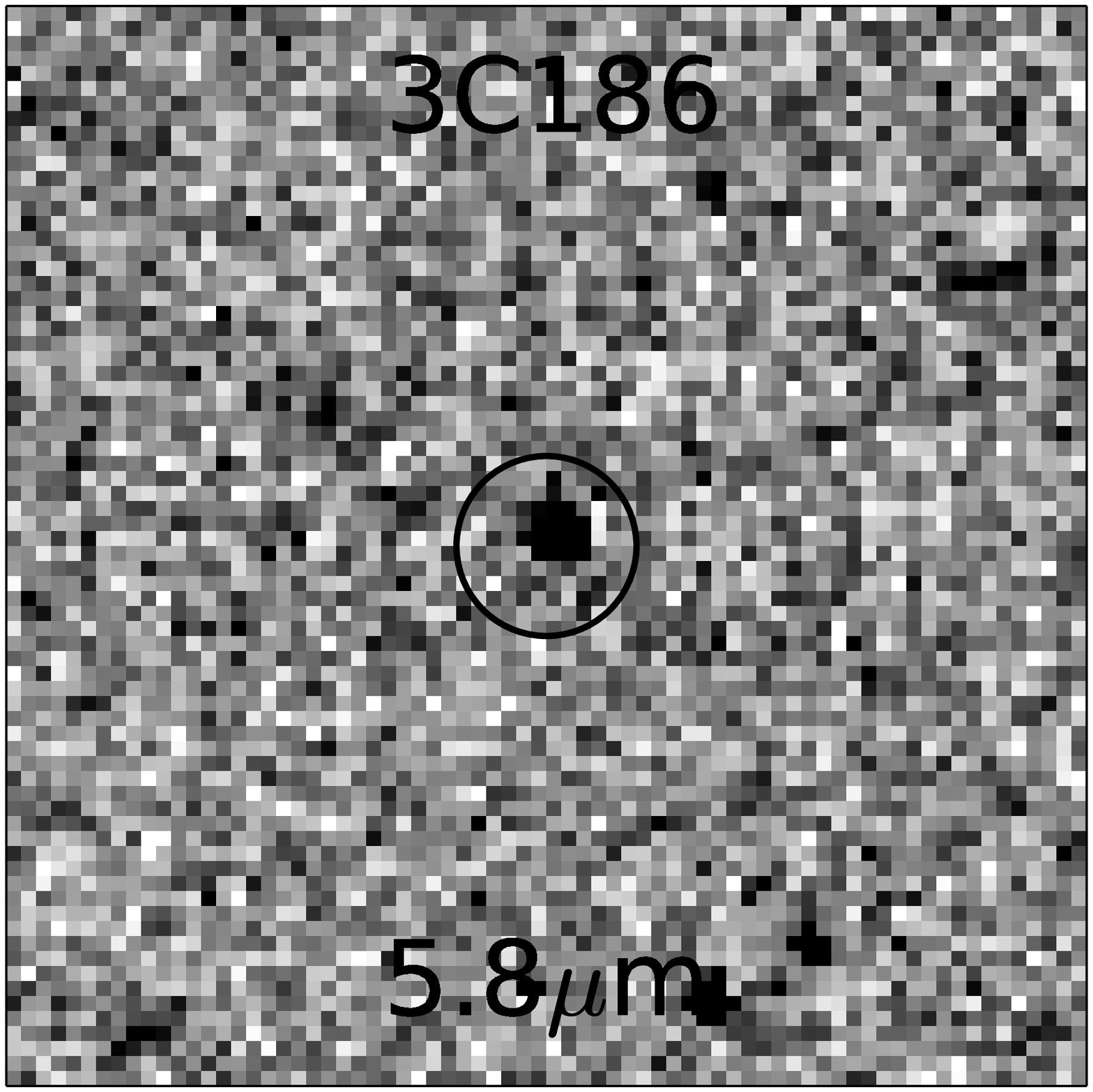}
      \includegraphics[width=1.5cm]{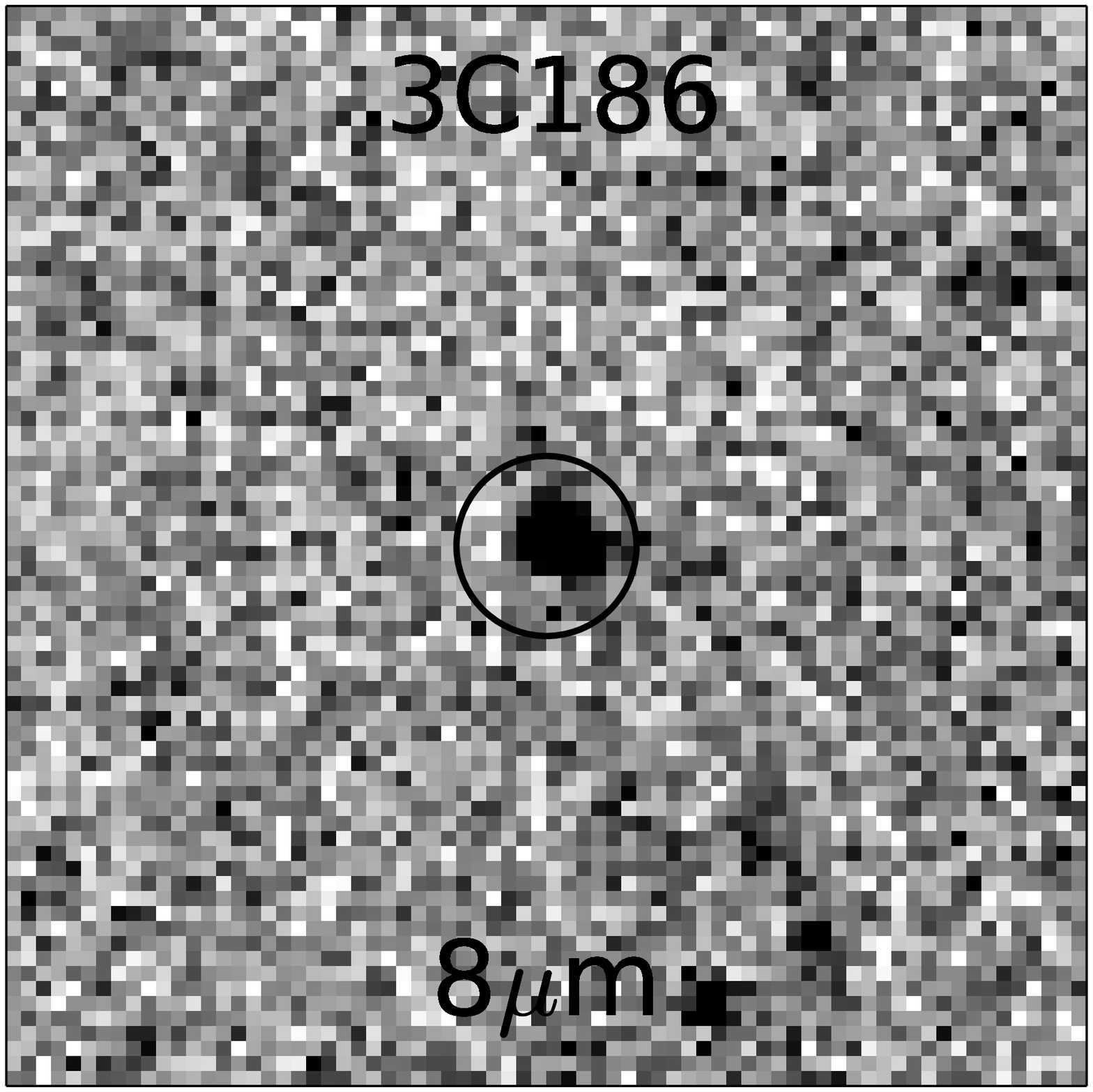}
      \includegraphics[width=1.5cm]{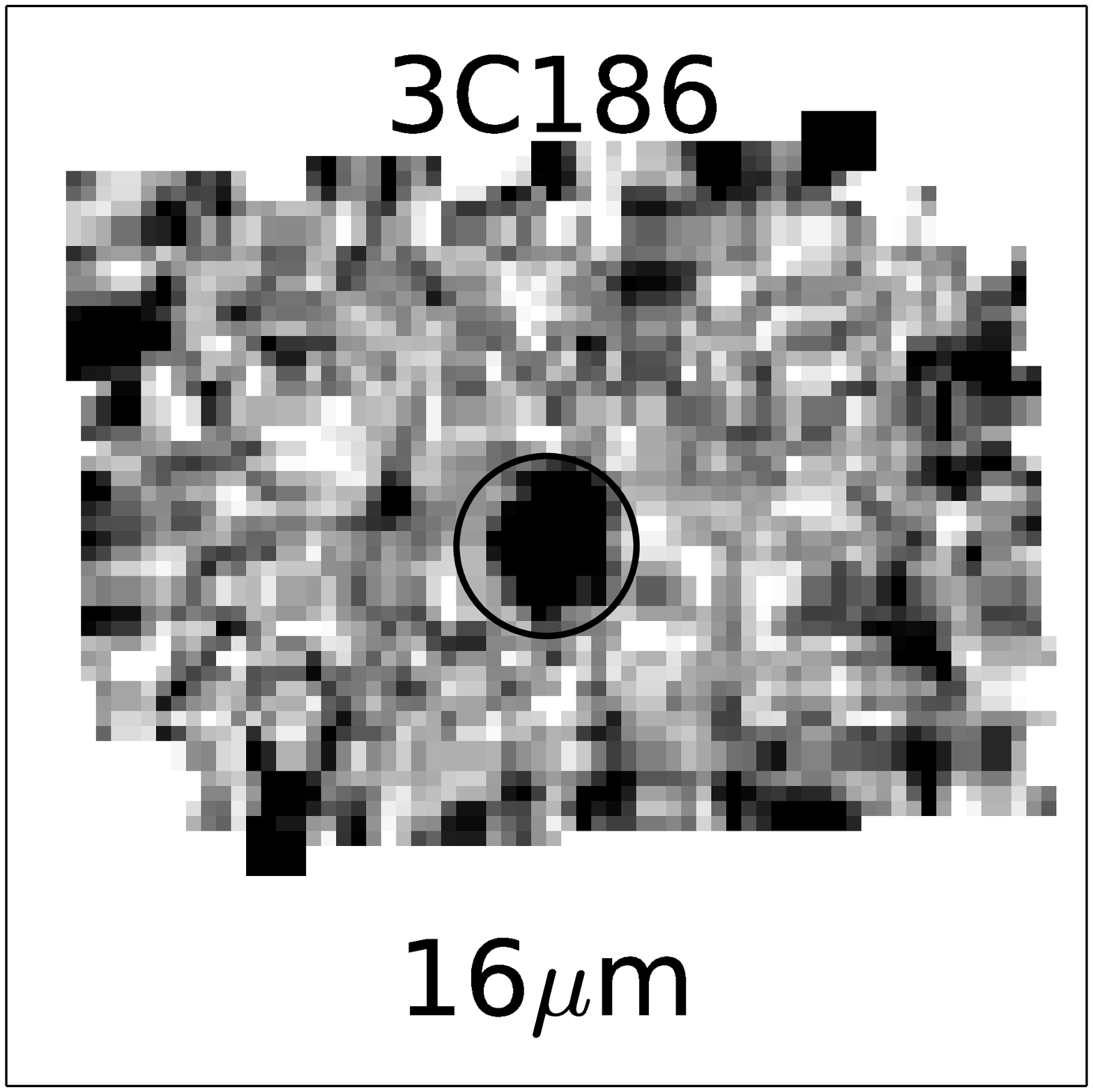}
      \includegraphics[width=1.5cm]{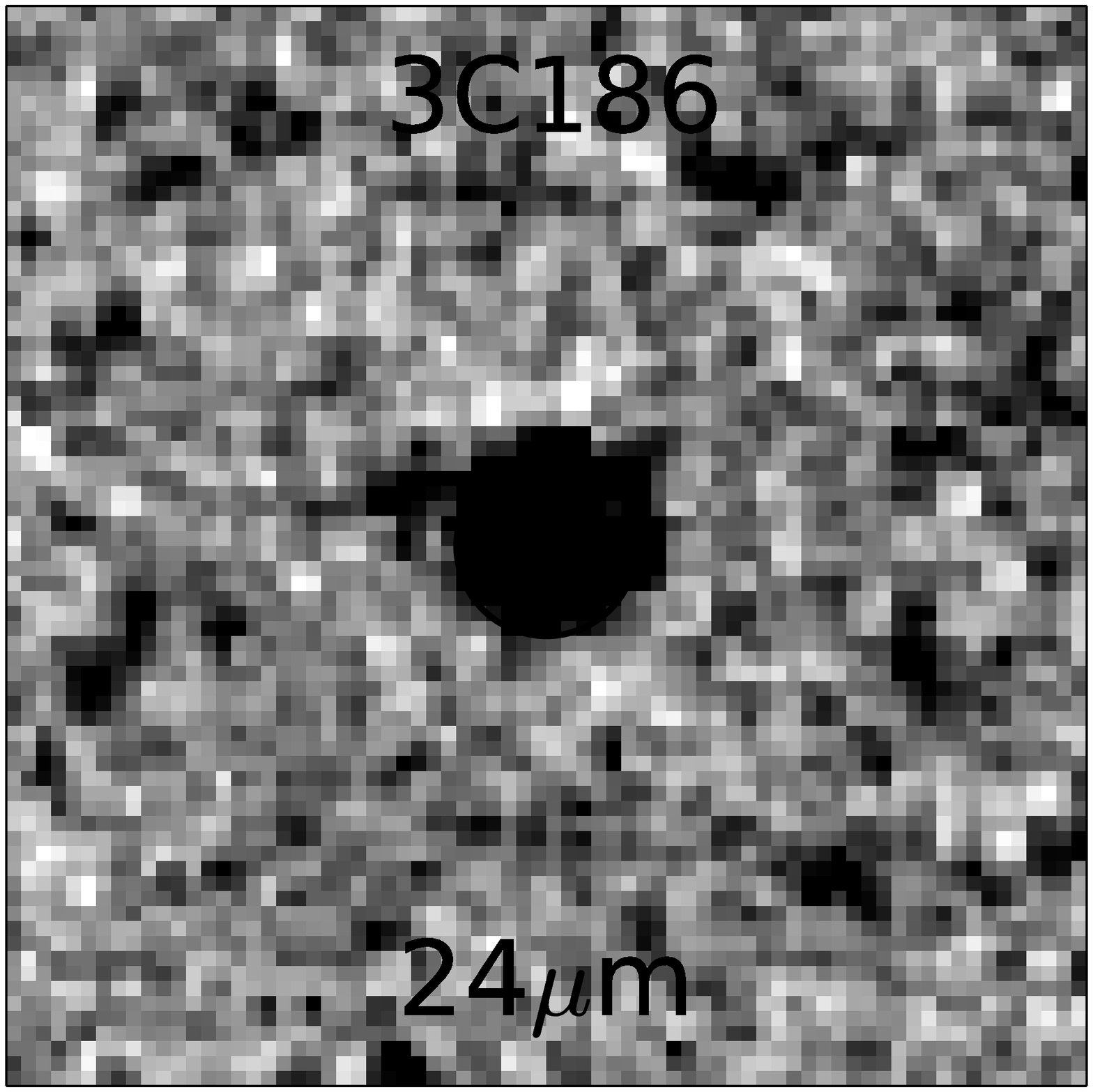}
      \includegraphics[width=1.5cm]{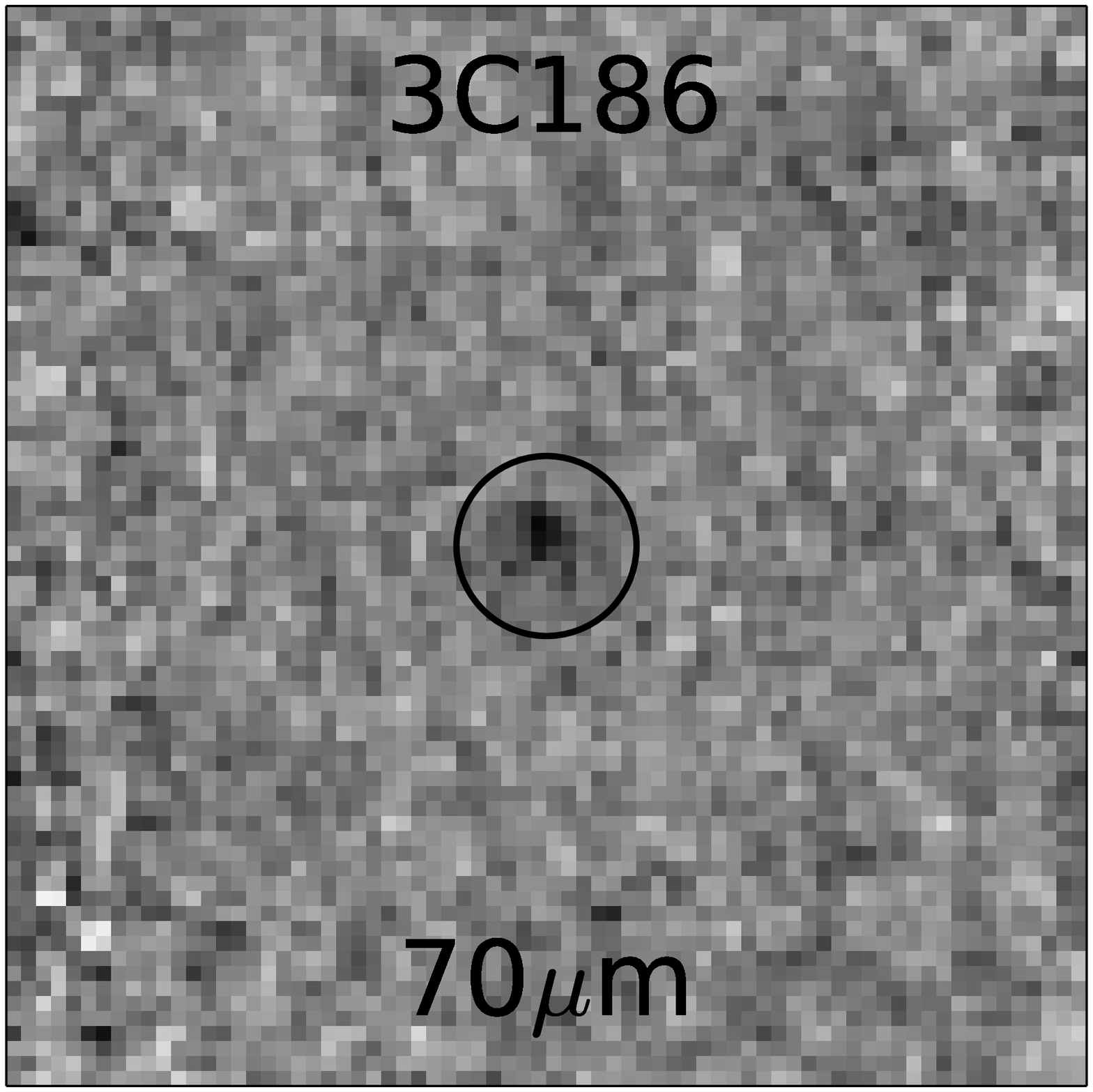}
      \includegraphics[width=1.5cm]{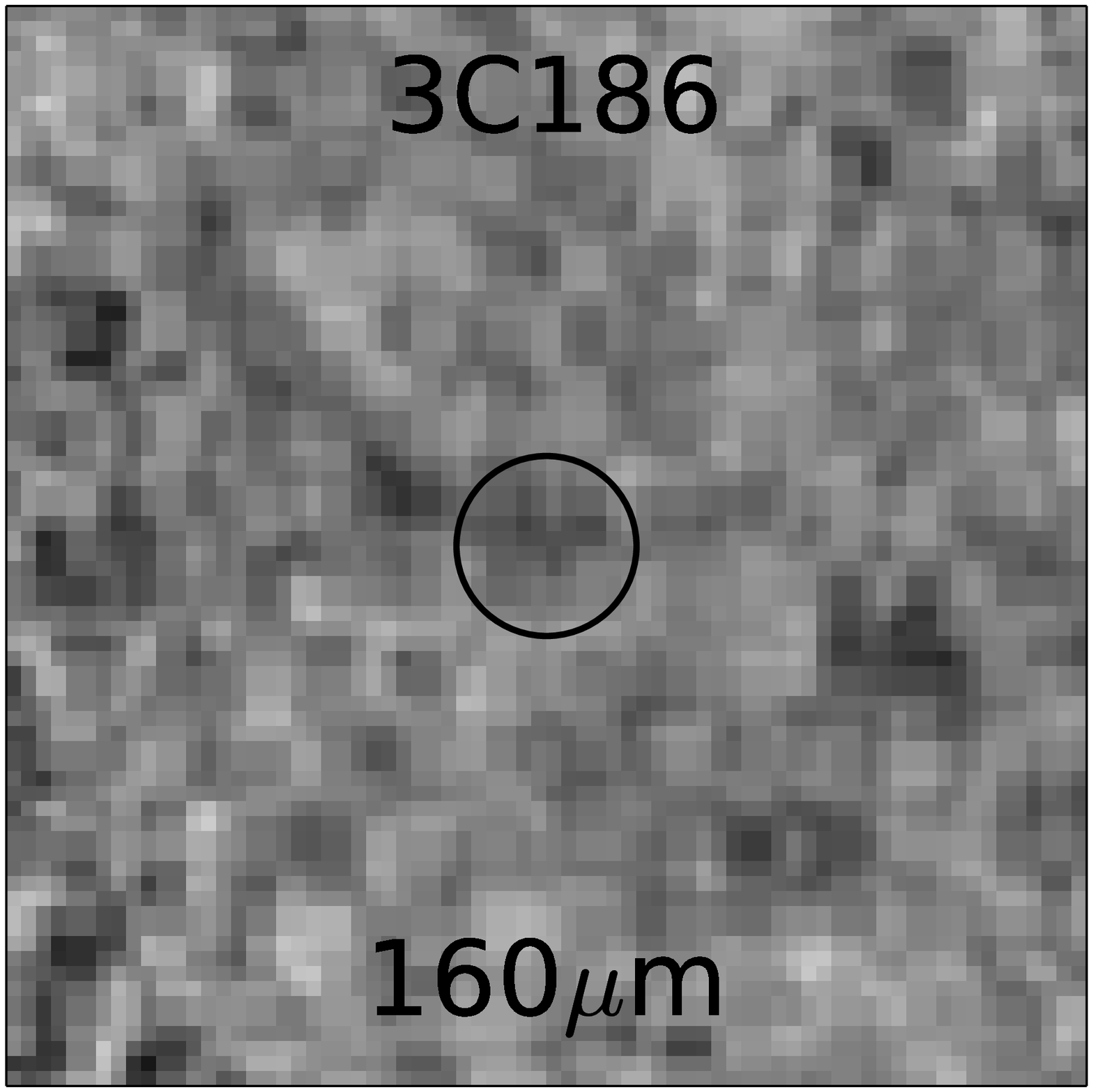}
      \includegraphics[width=1.5cm]{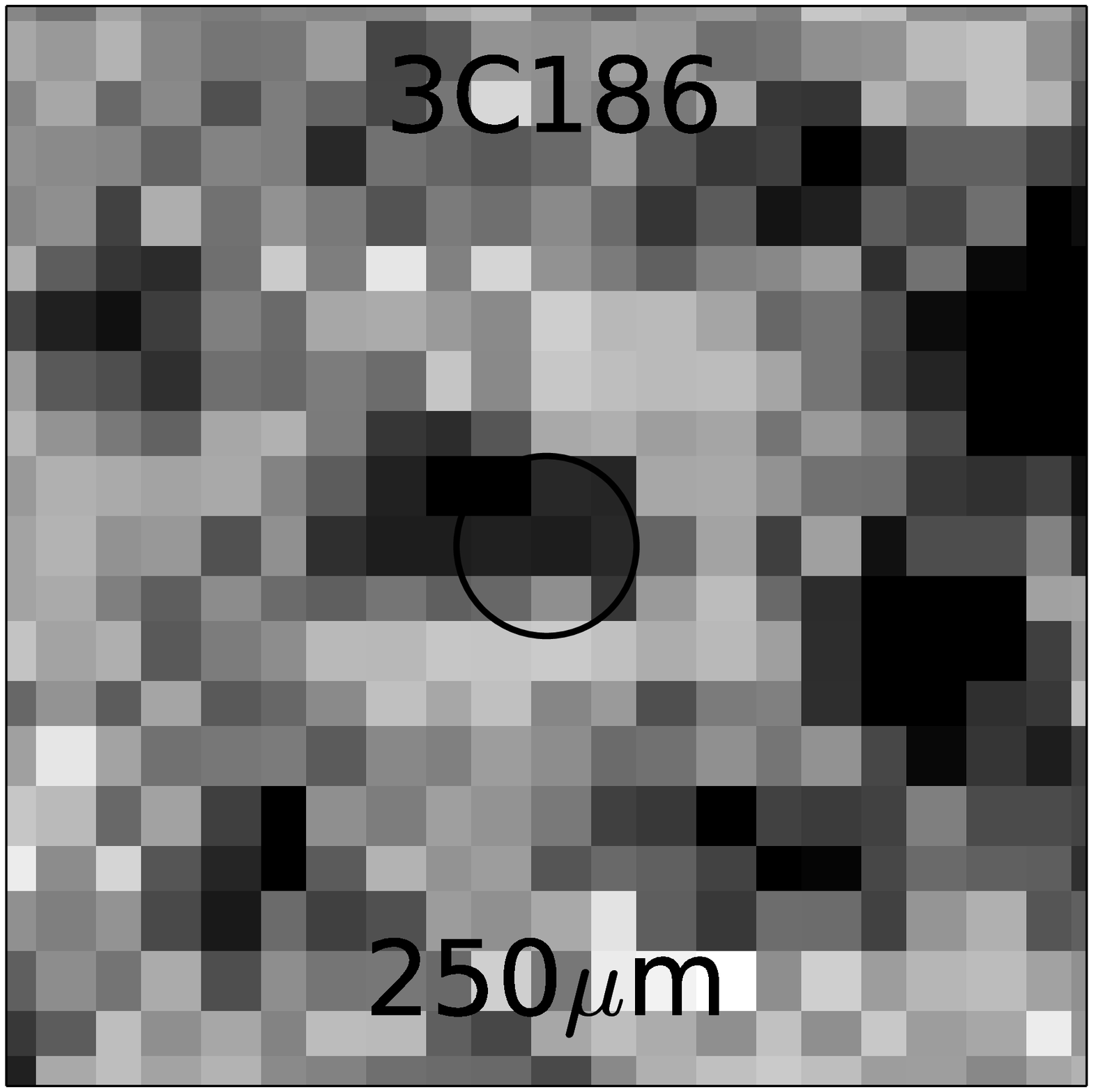}
      \includegraphics[width=1.5cm]{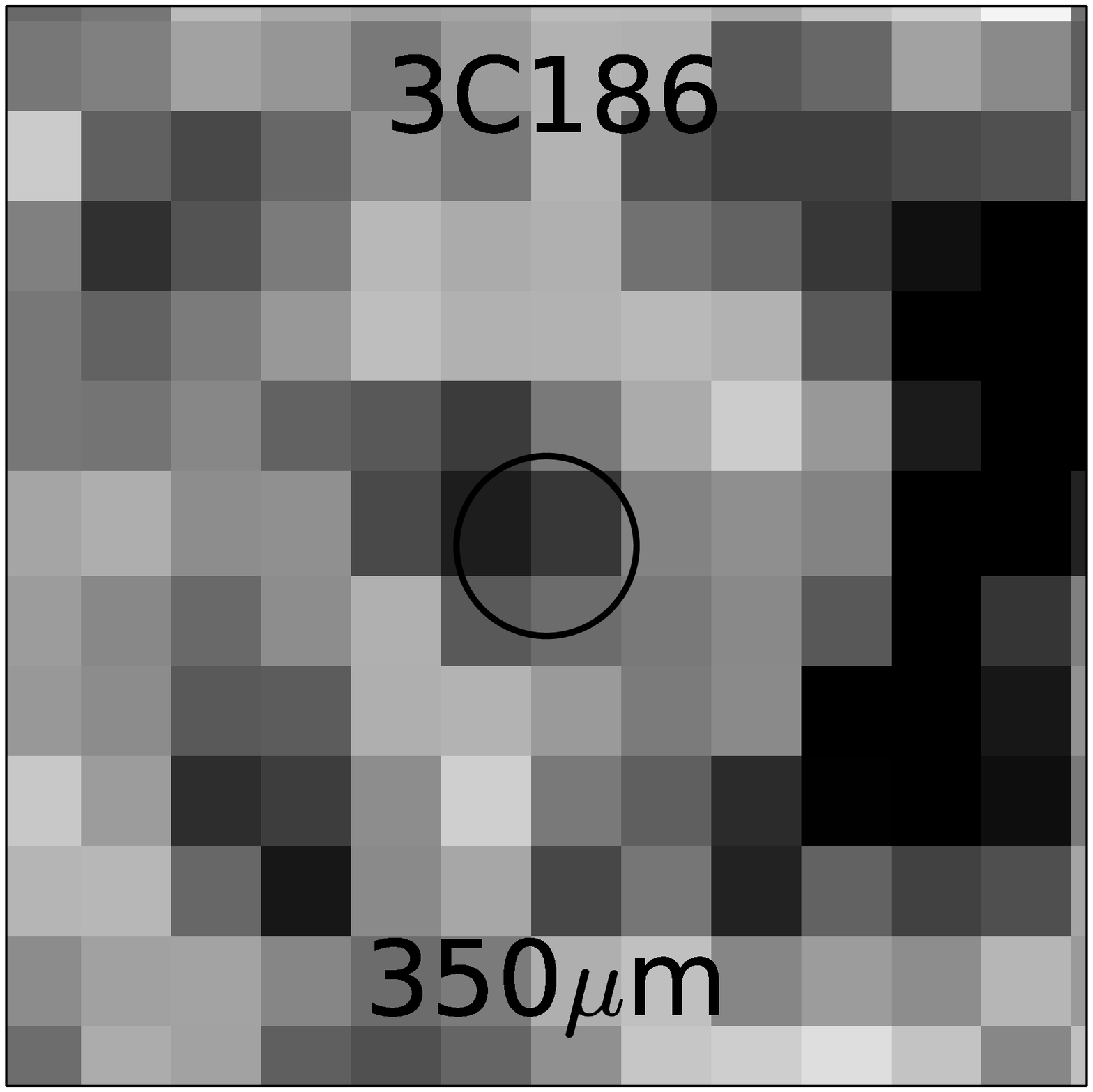}
      \includegraphics[width=1.5cm]{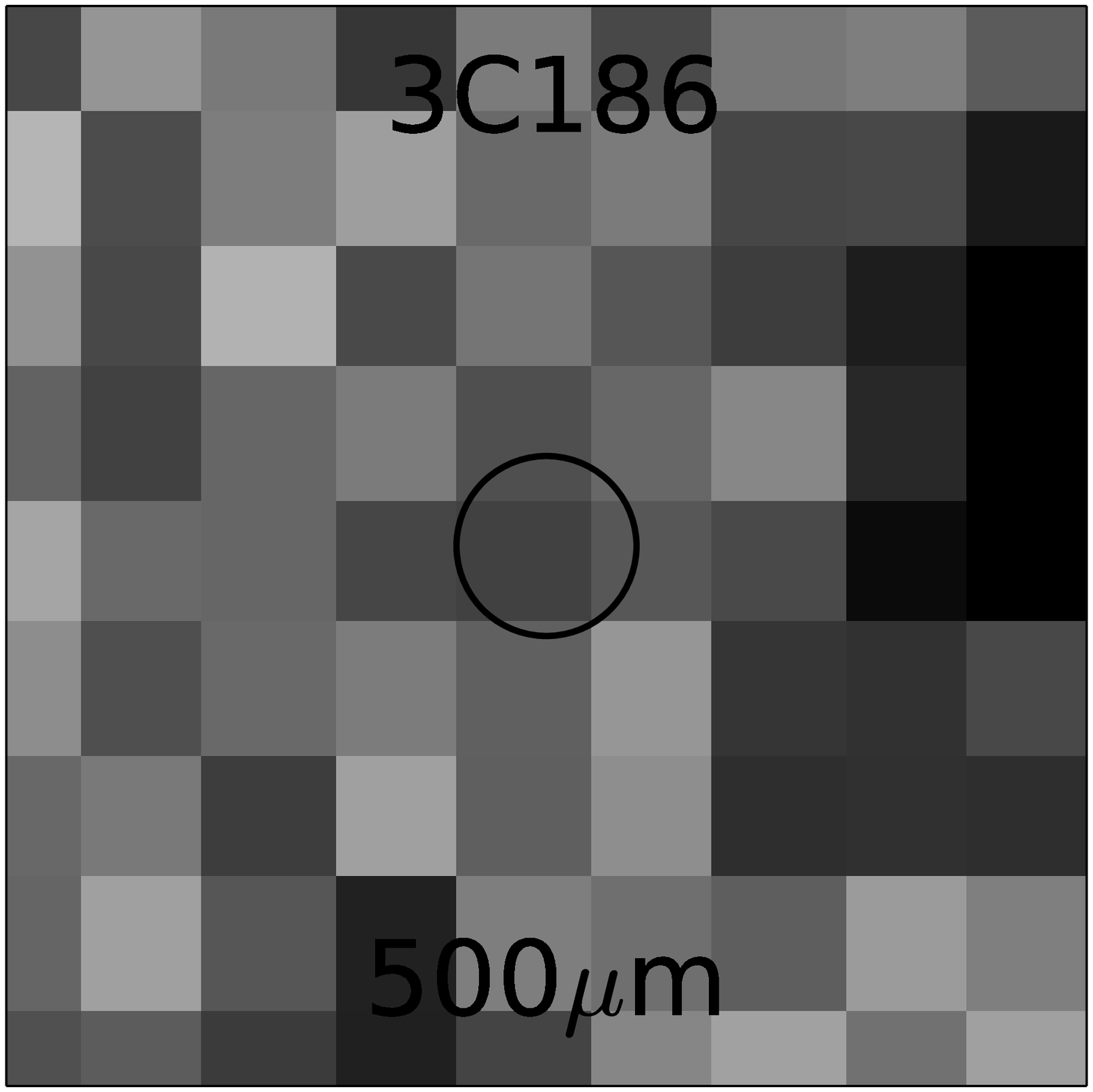}
      \\
      \includegraphics[width=1.5cm]{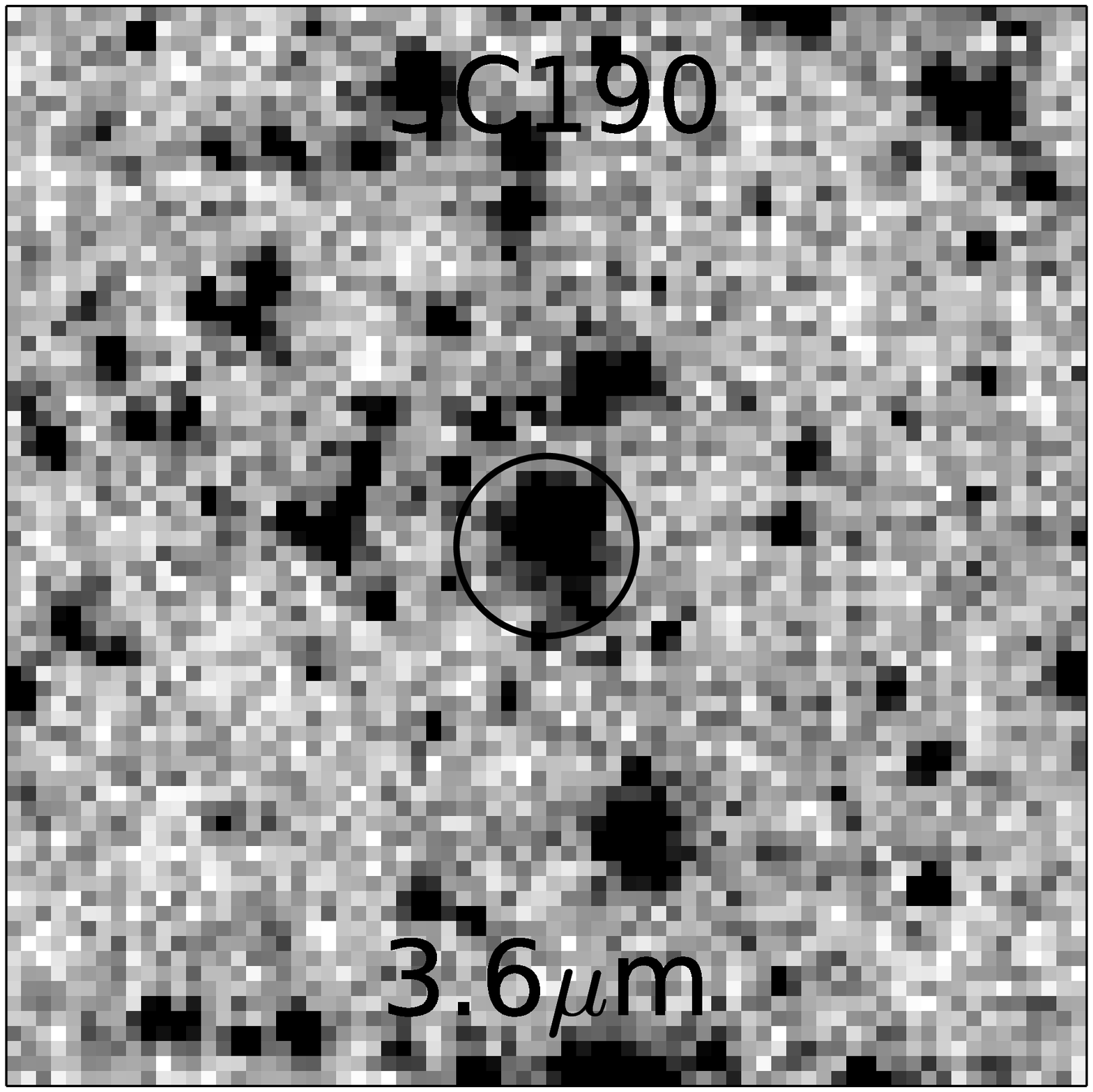}
      \includegraphics[width=1.5cm]{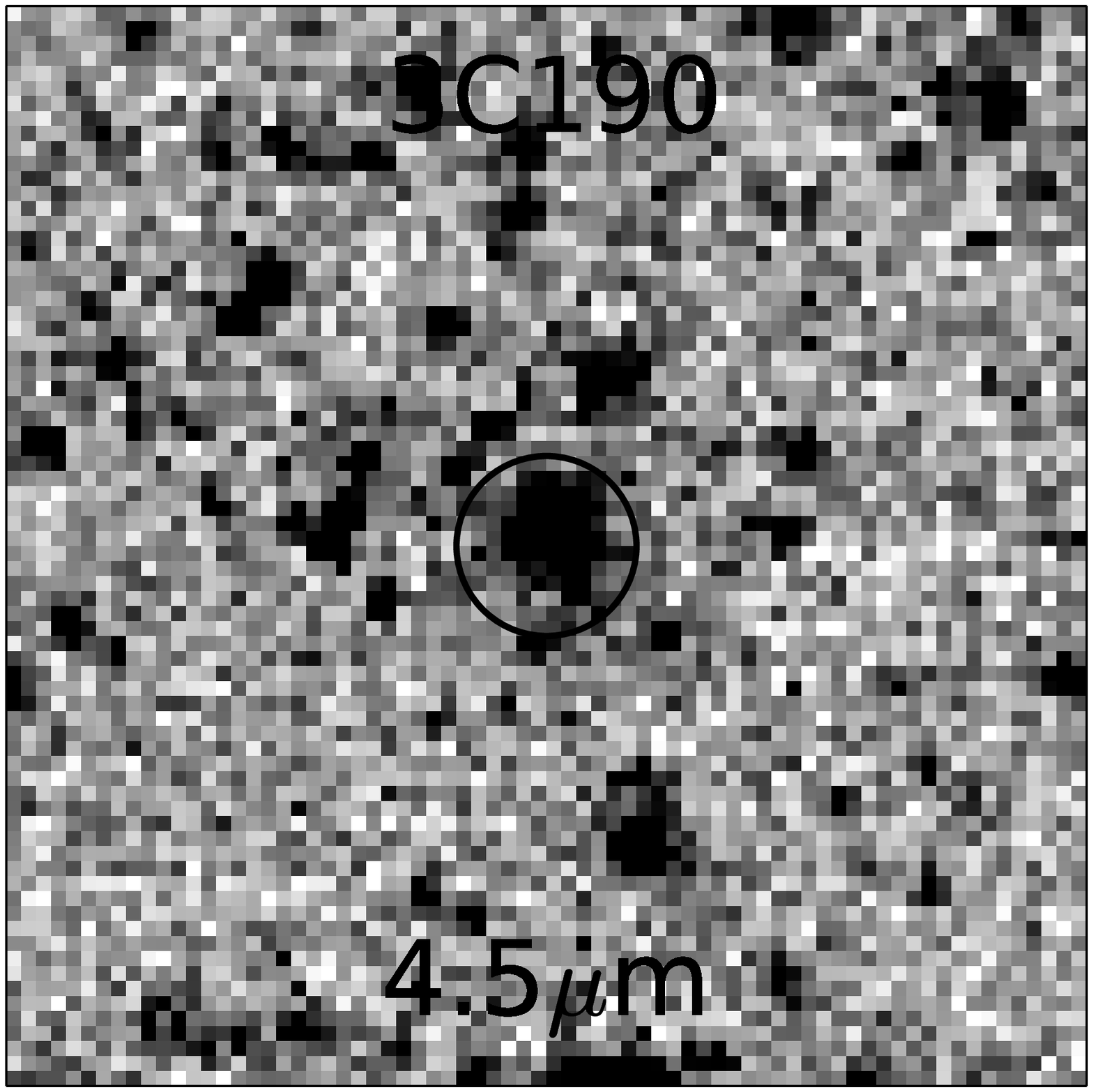}
      \includegraphics[width=1.5cm]{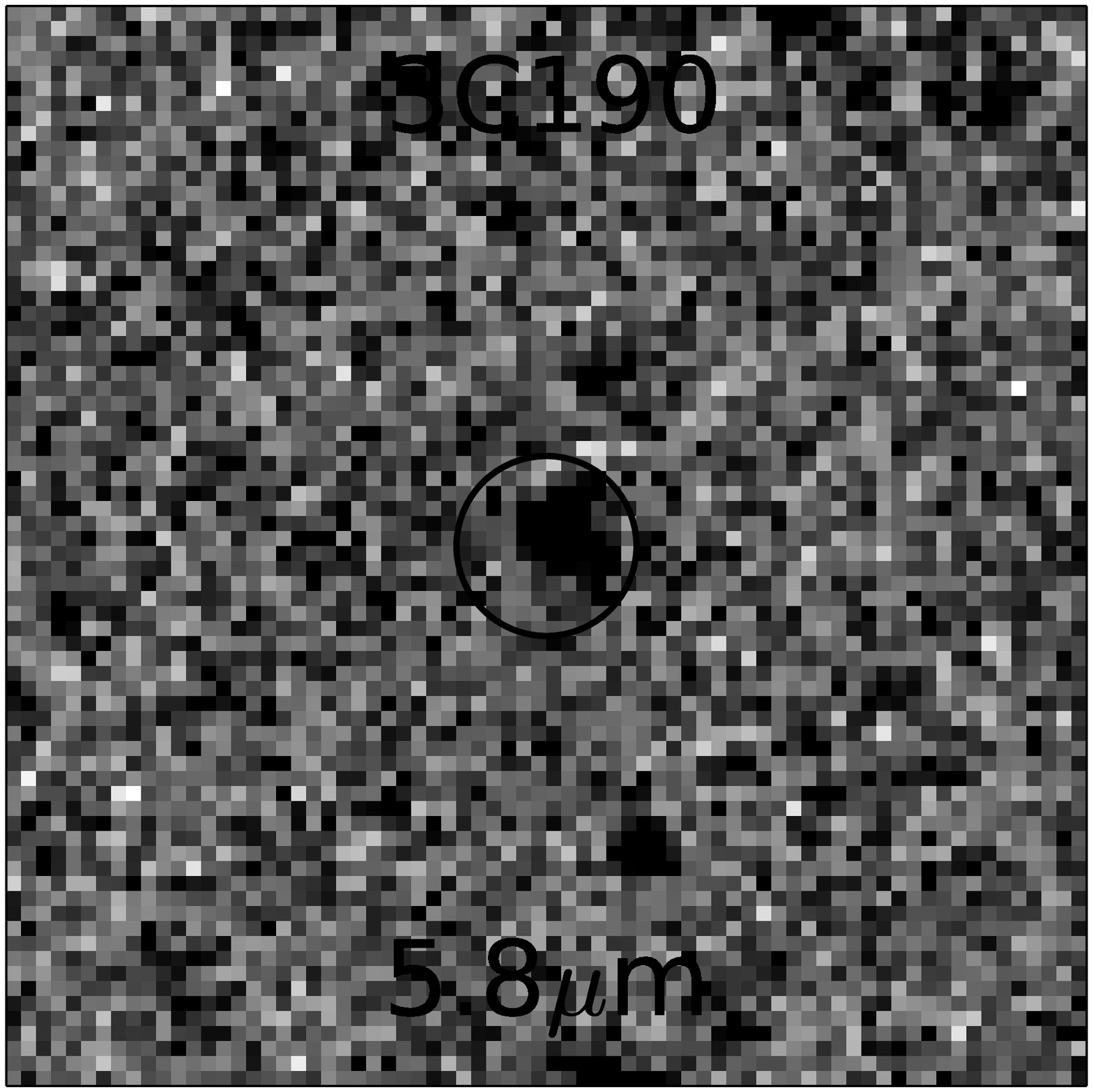}
      \includegraphics[width=1.5cm]{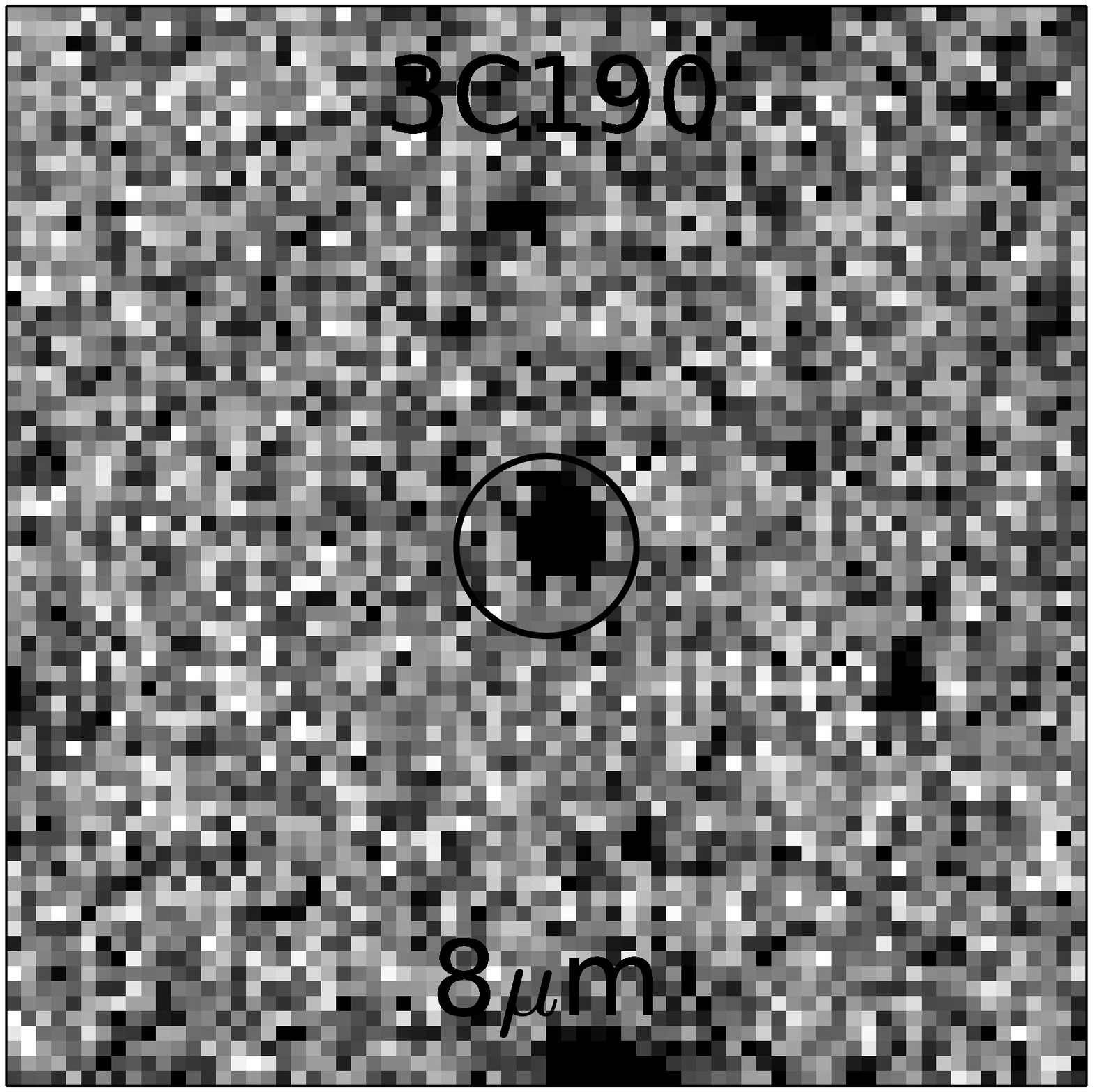}
      \includegraphics[width=1.5cm]{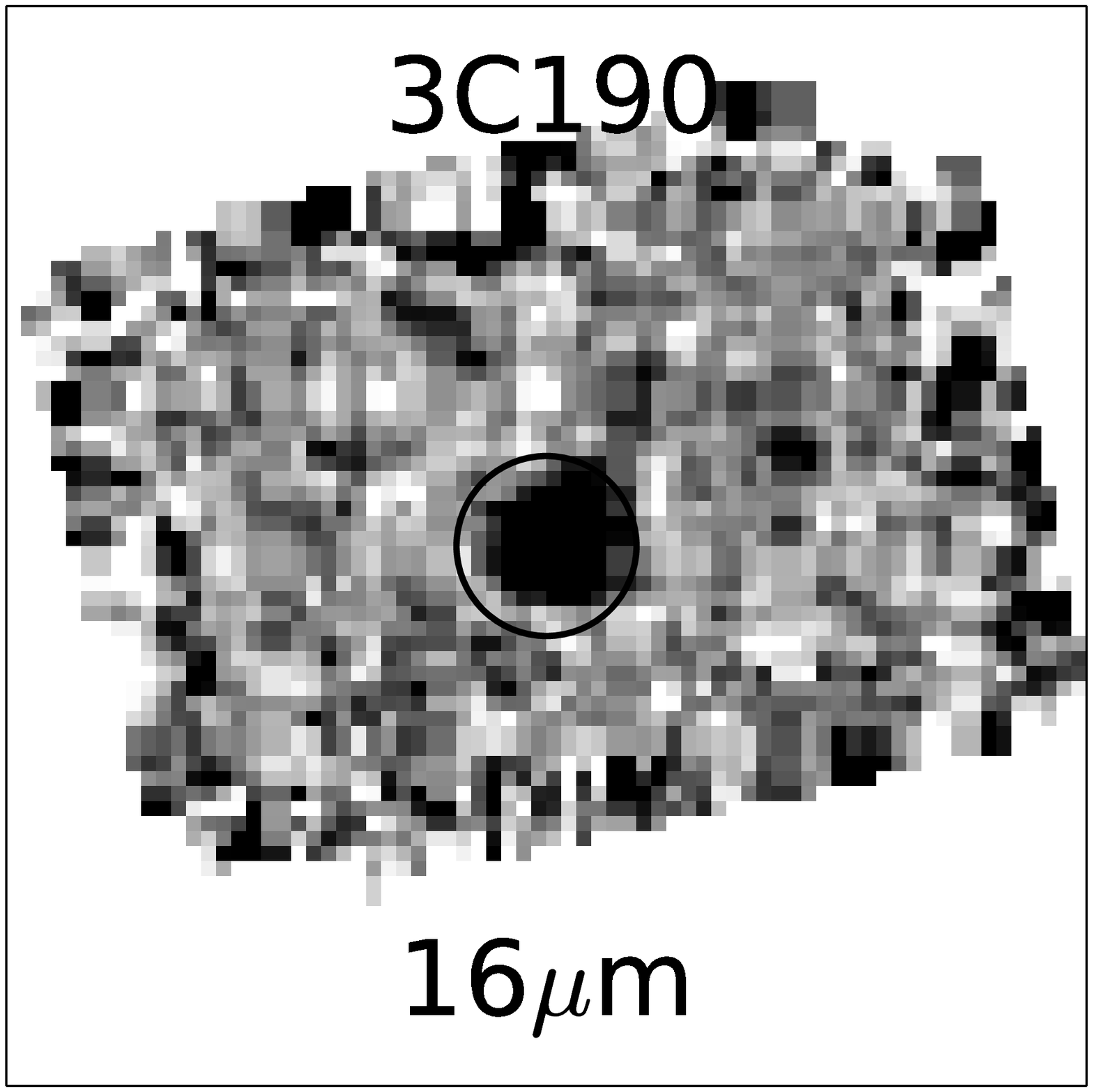}
      \includegraphics[width=1.5cm]{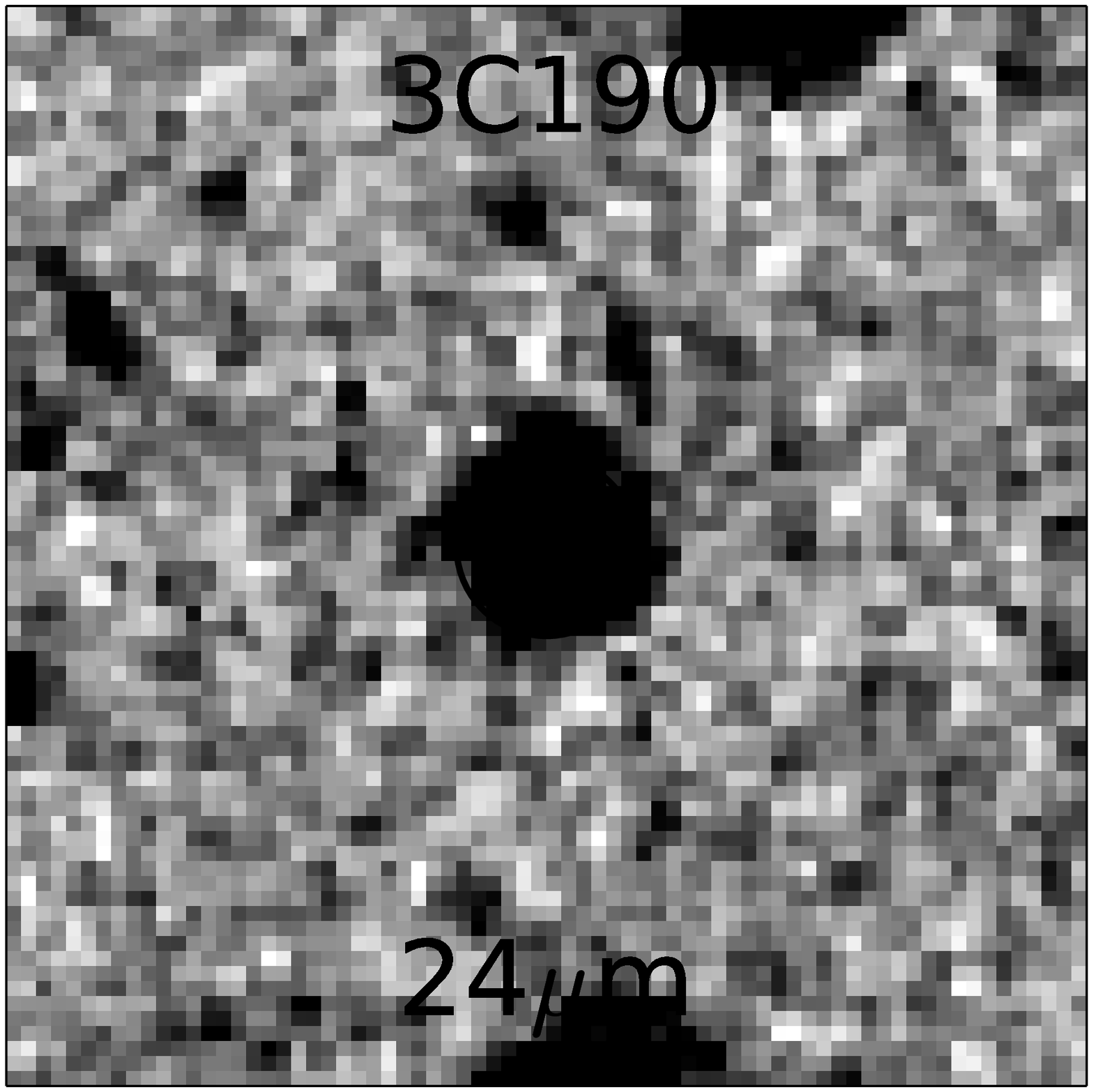}
      \includegraphics[width=1.5cm]{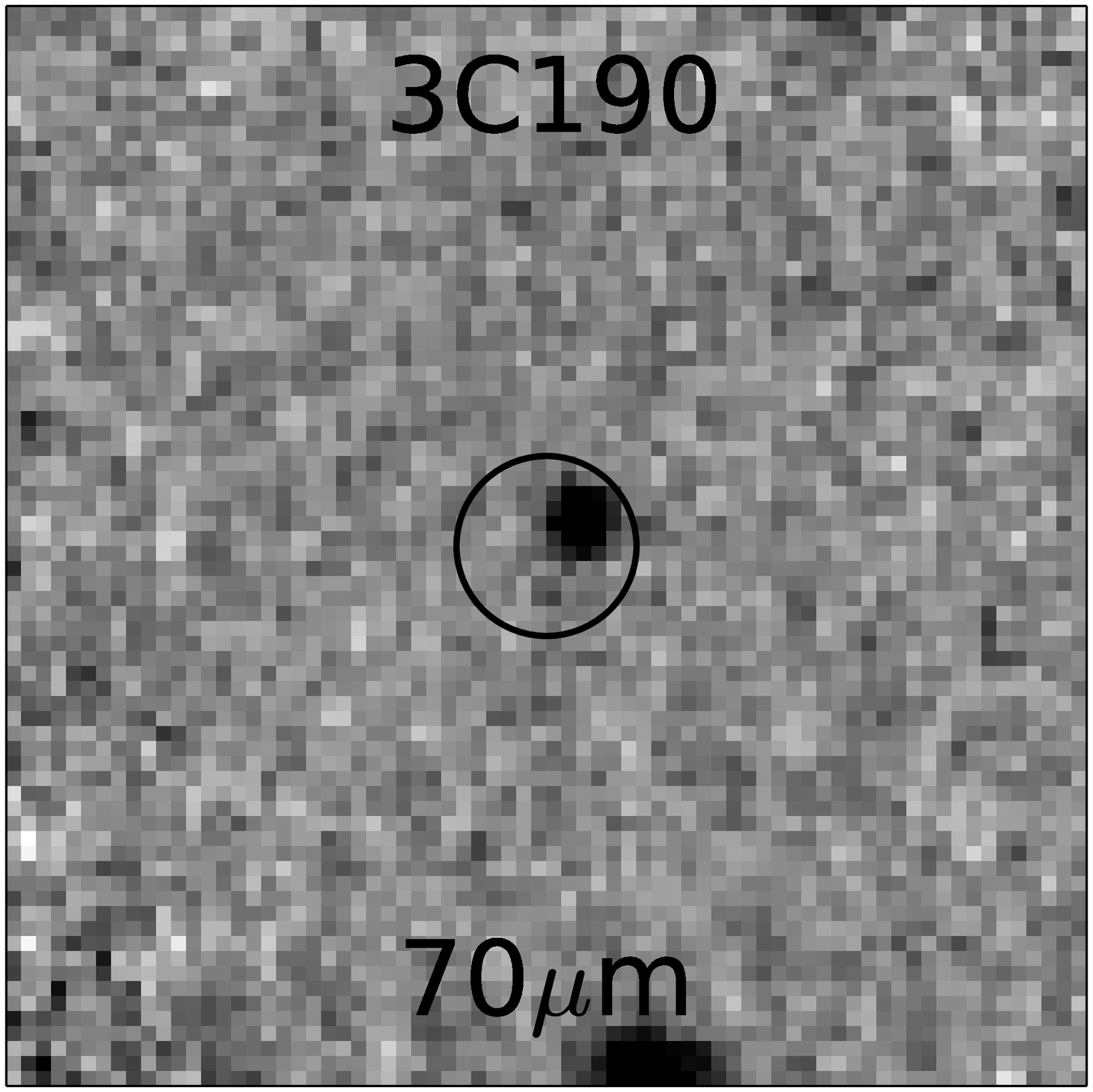}
      \includegraphics[width=1.5cm]{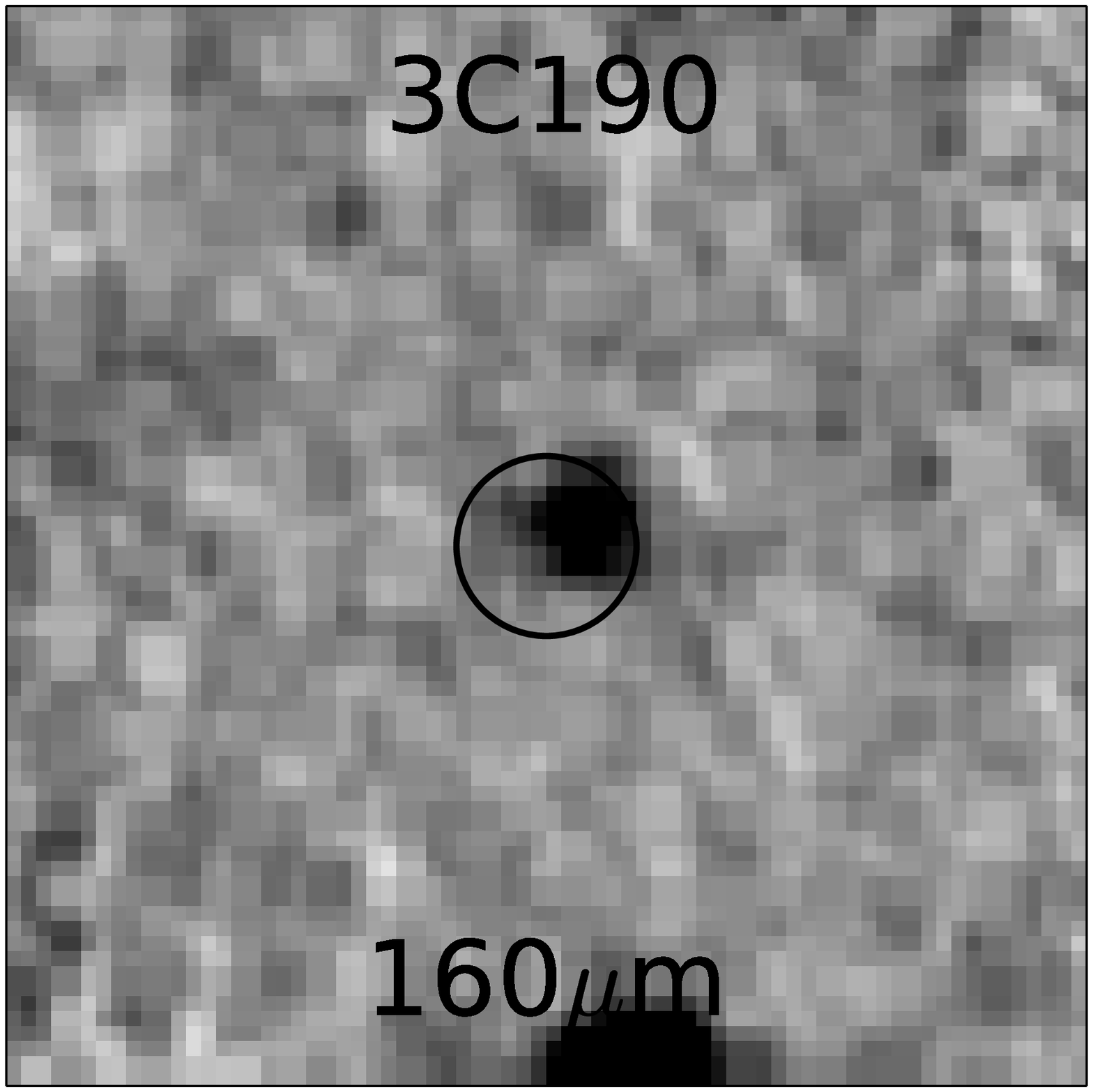}
      \includegraphics[width=1.5cm]{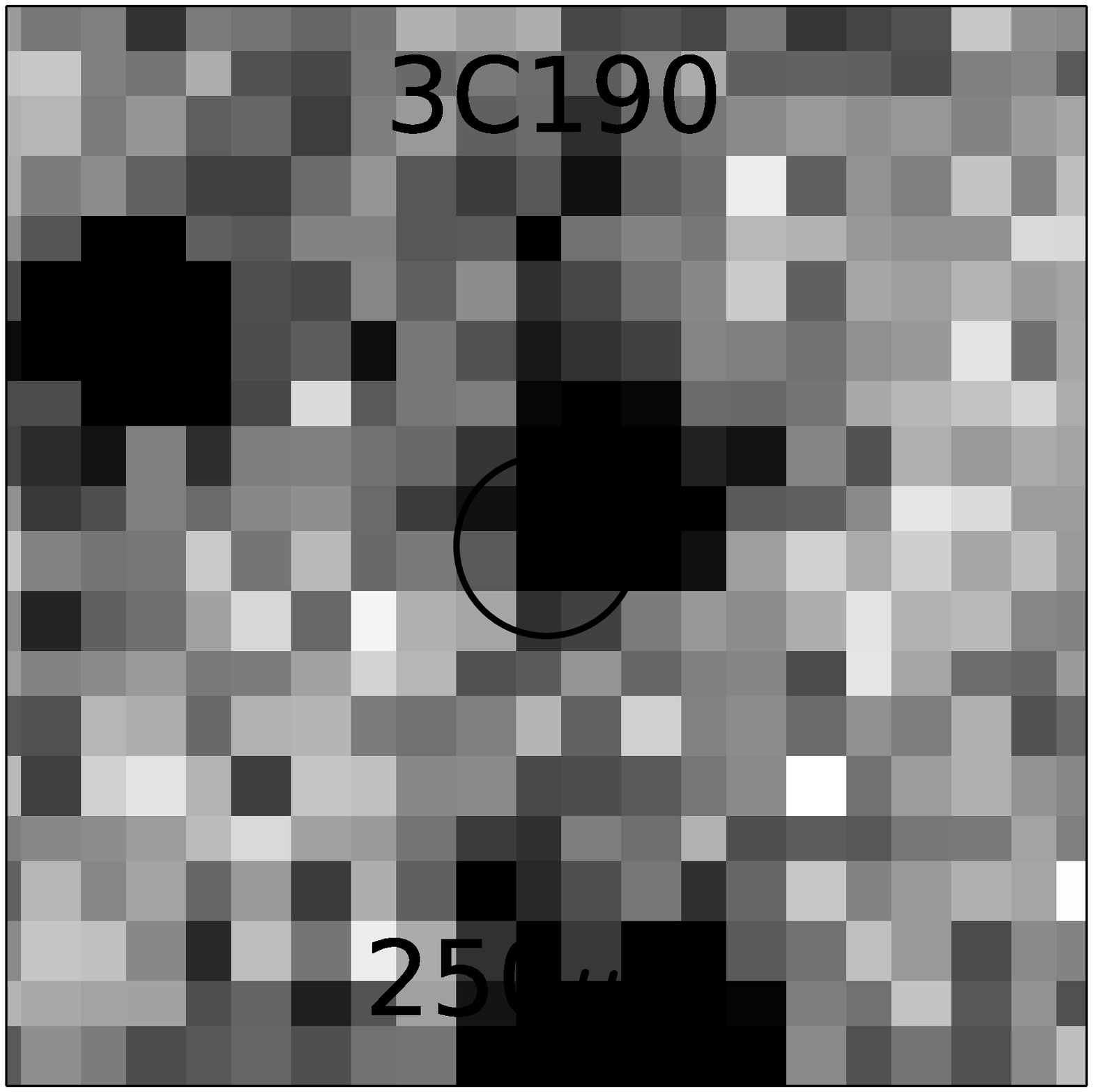}
      \includegraphics[width=1.5cm]{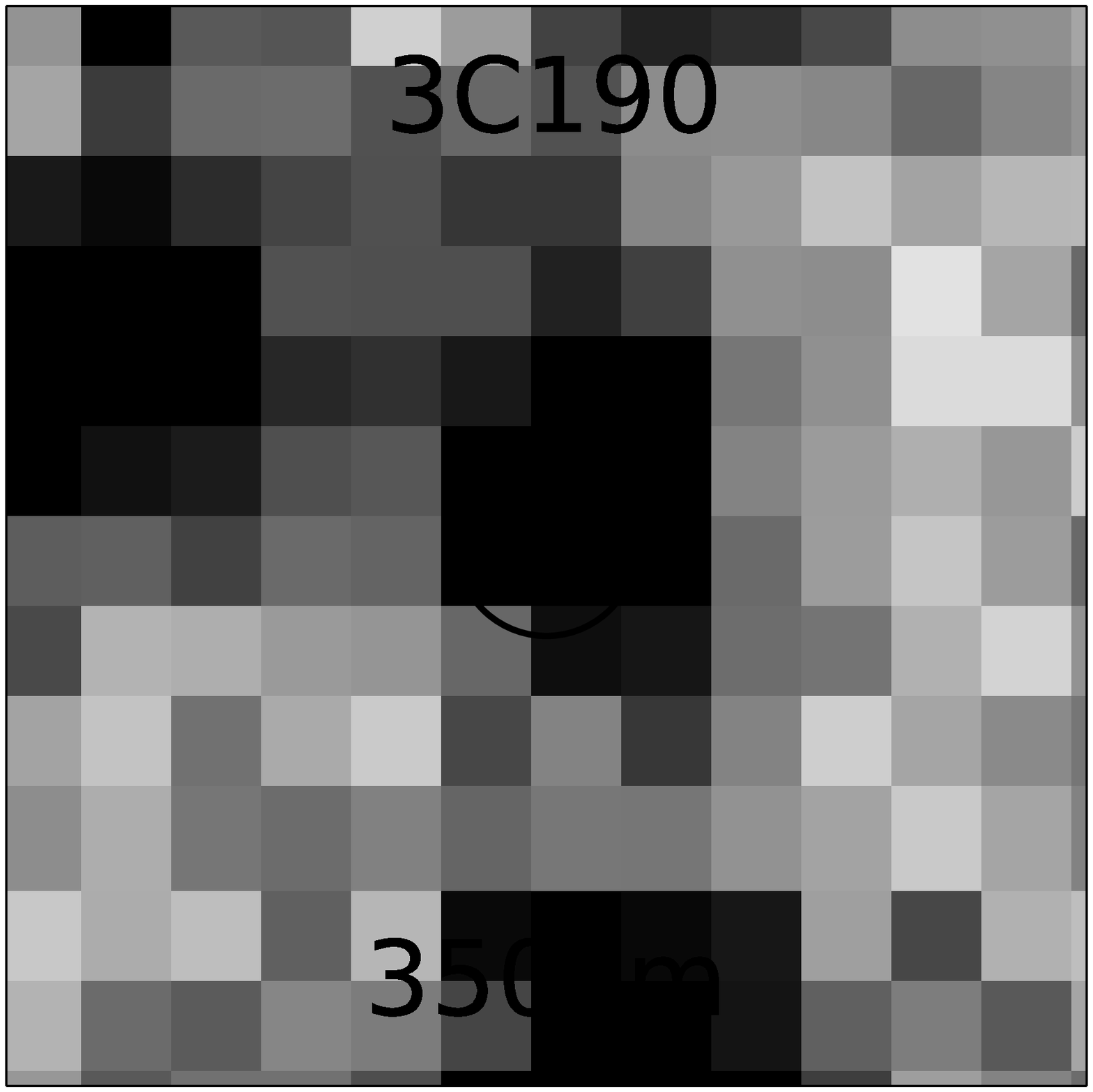}
      \includegraphics[width=1.5cm]{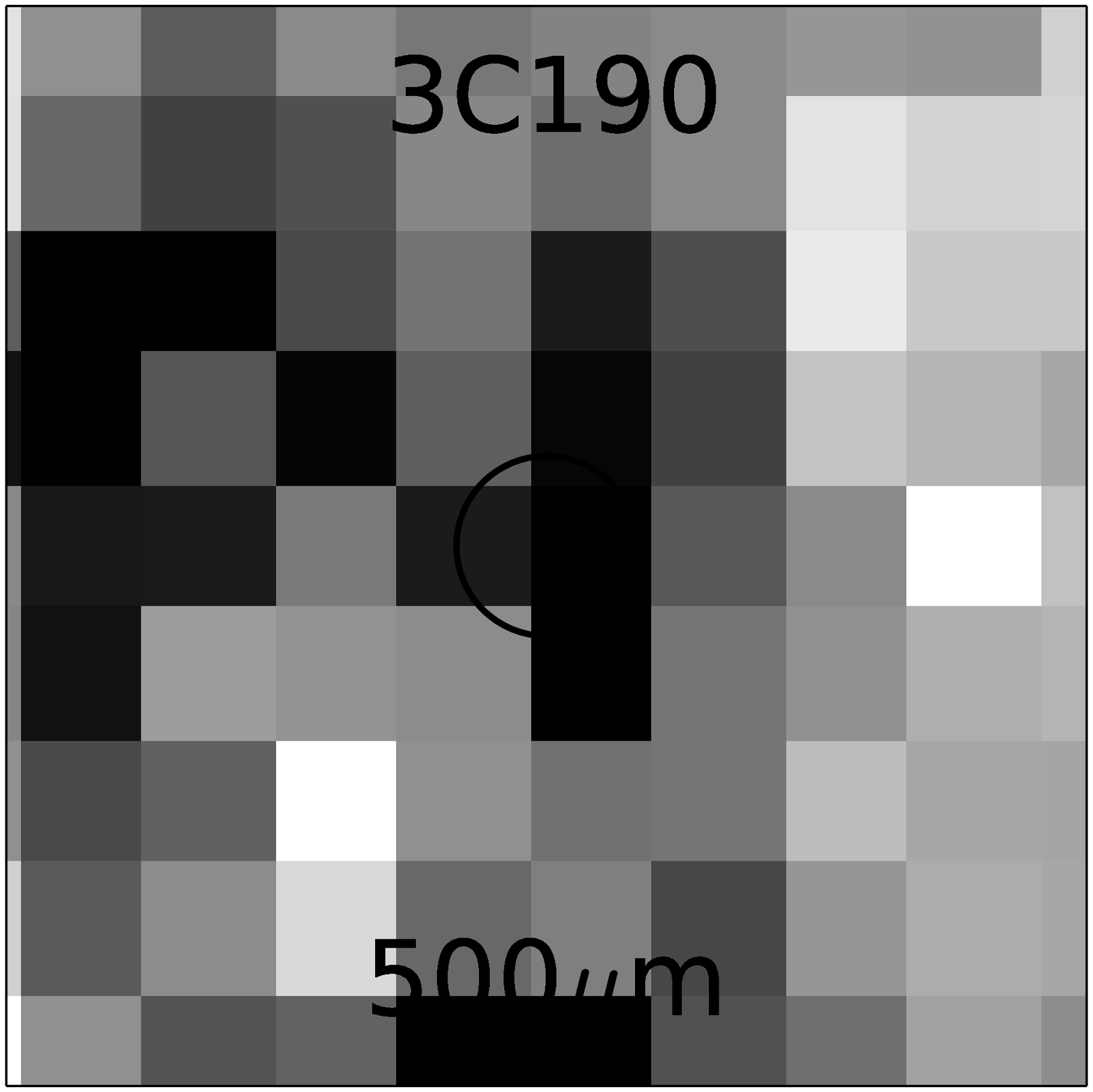}
      \\
      \includegraphics[width=1.5cm]{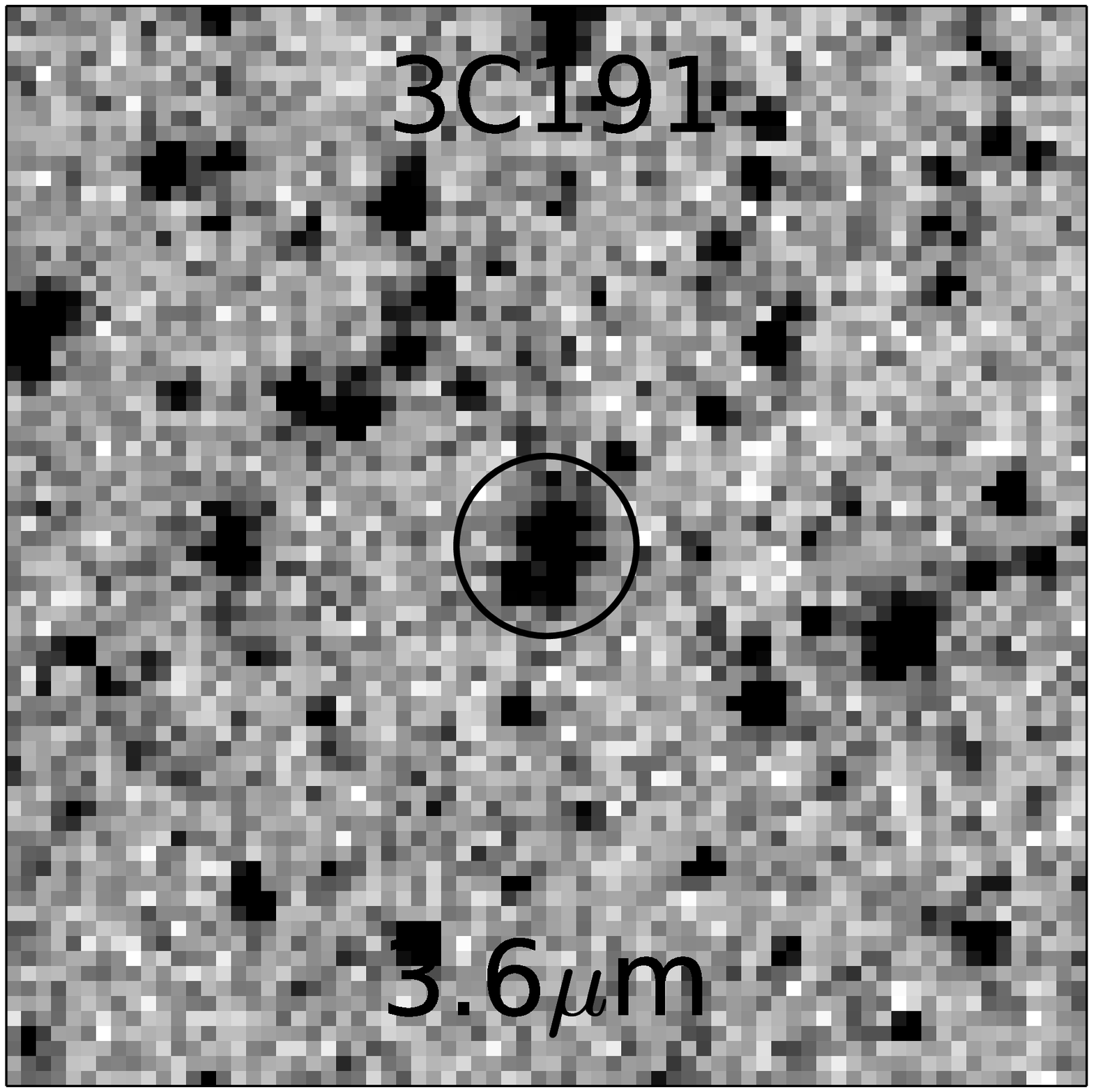}
      \includegraphics[width=1.5cm]{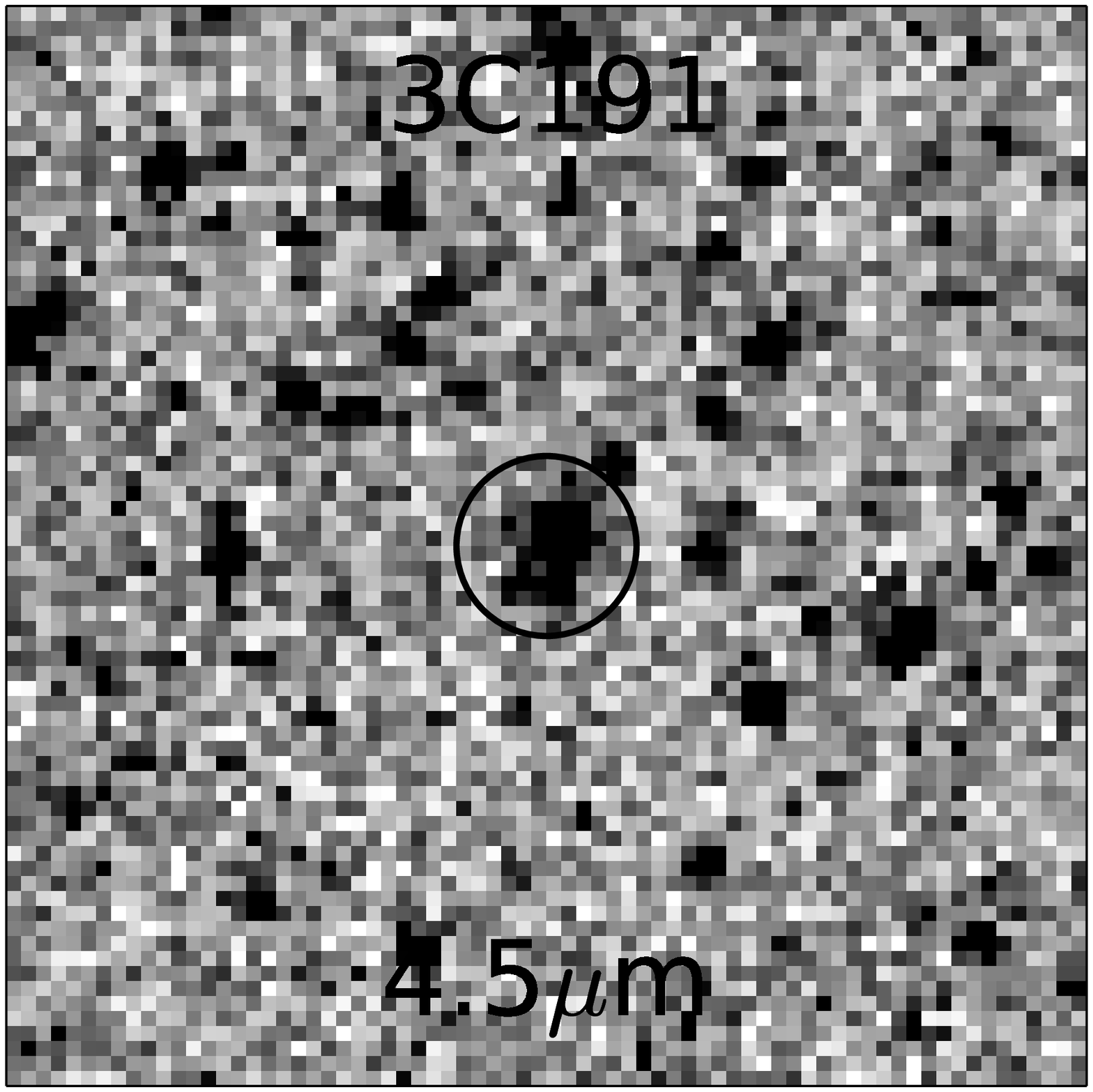}
      \includegraphics[width=1.5cm]{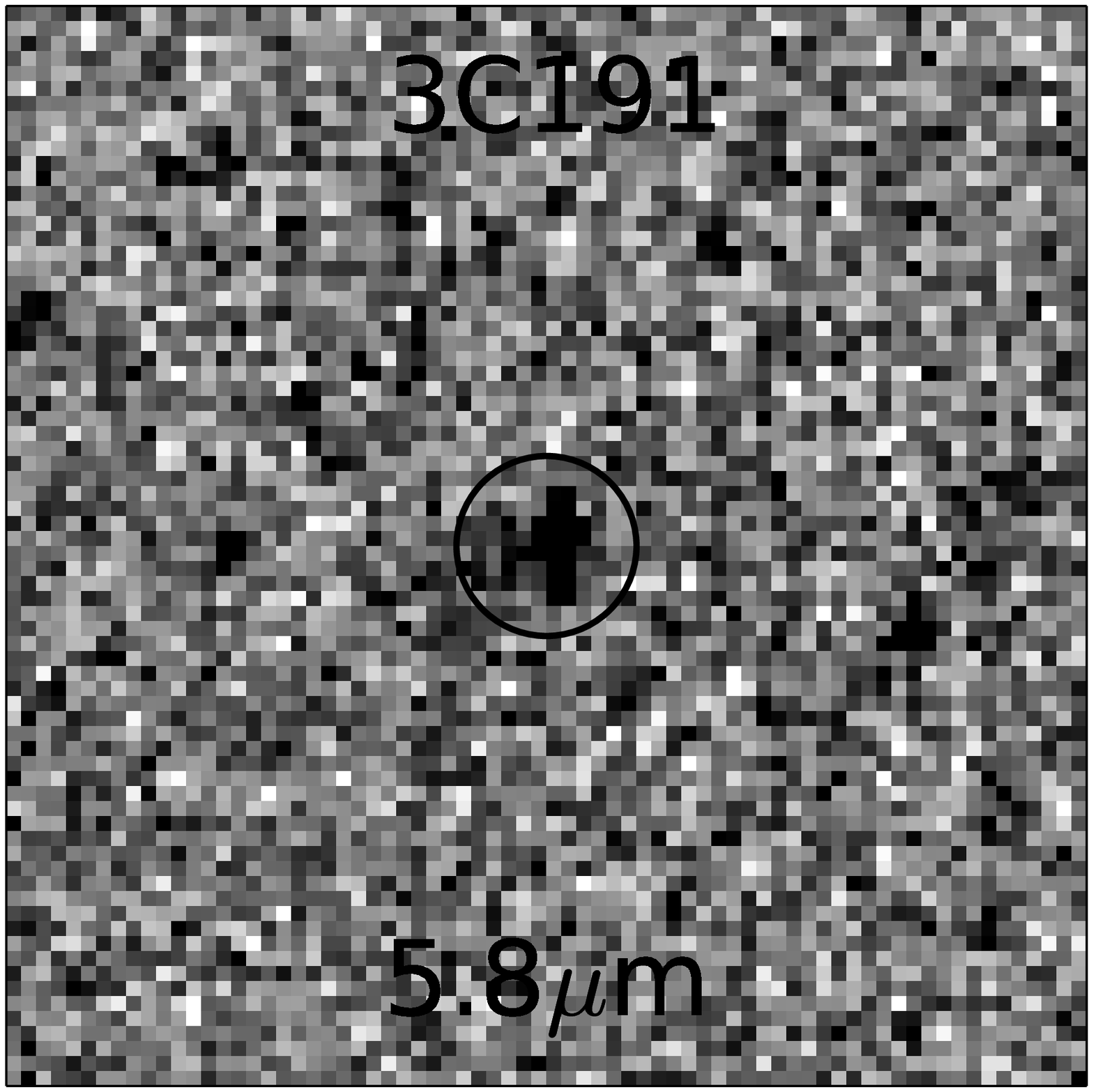}
      \includegraphics[width=1.5cm]{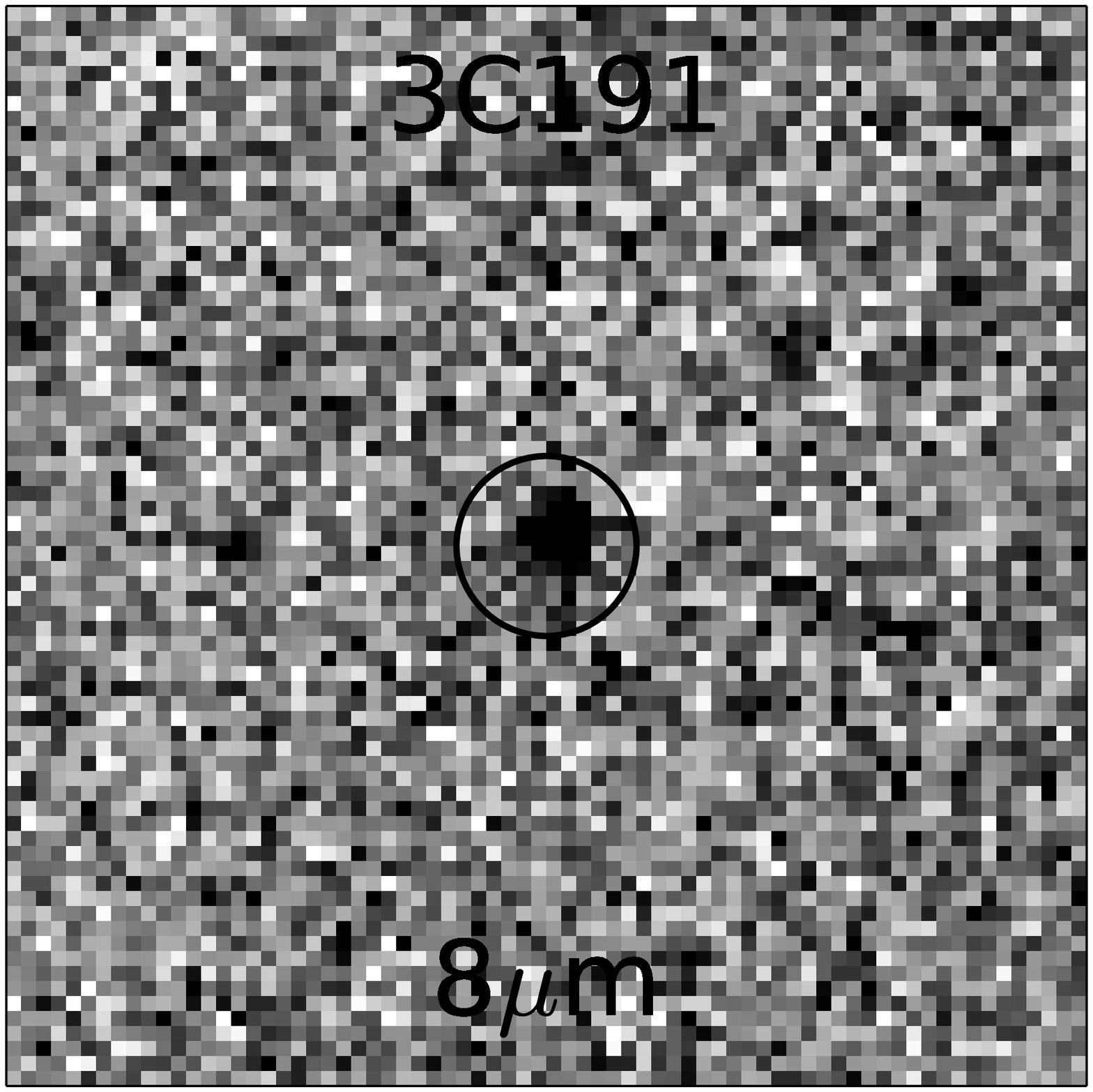}
      \includegraphics[width=1.5cm]{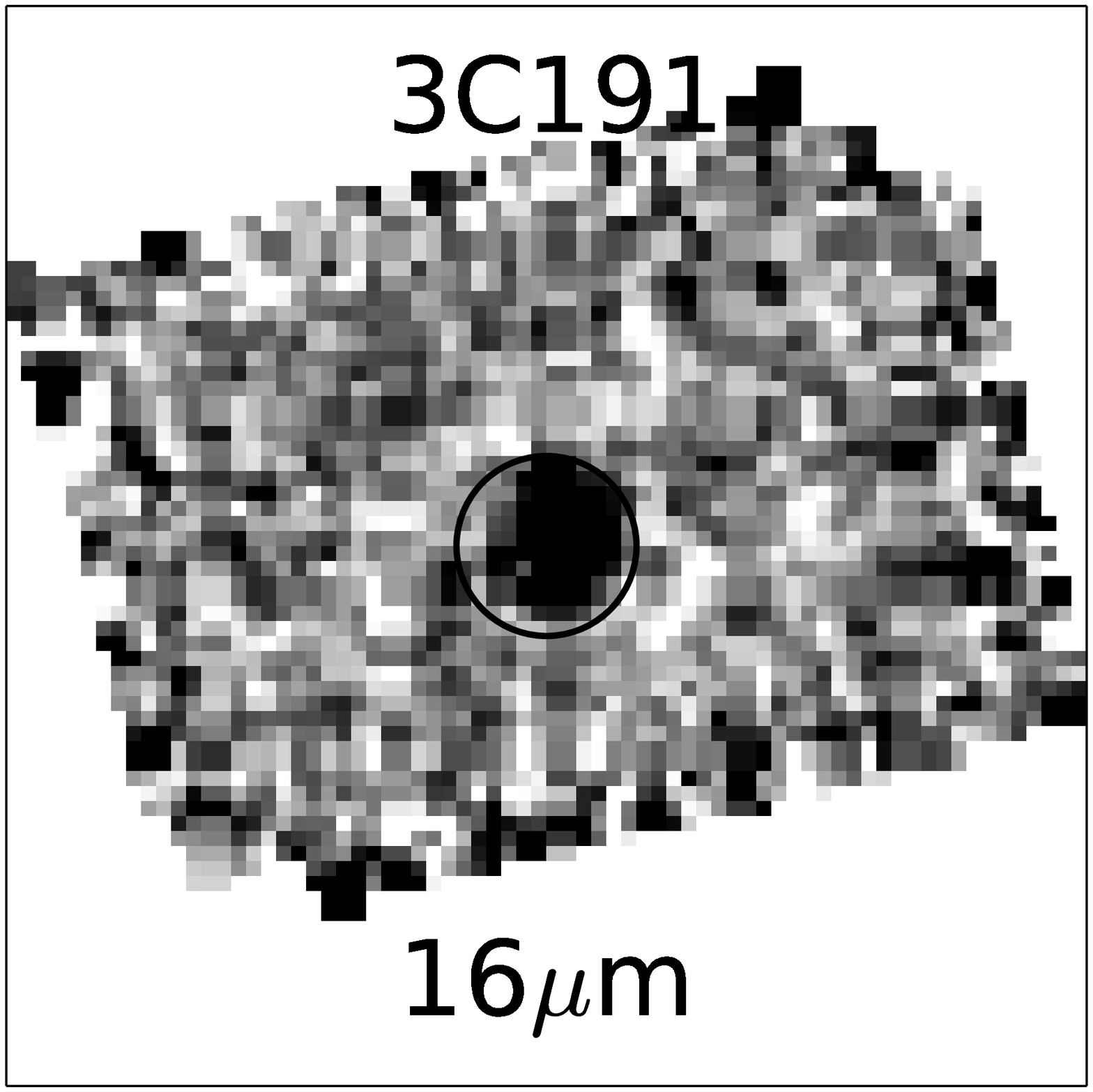}
      \includegraphics[width=1.5cm]{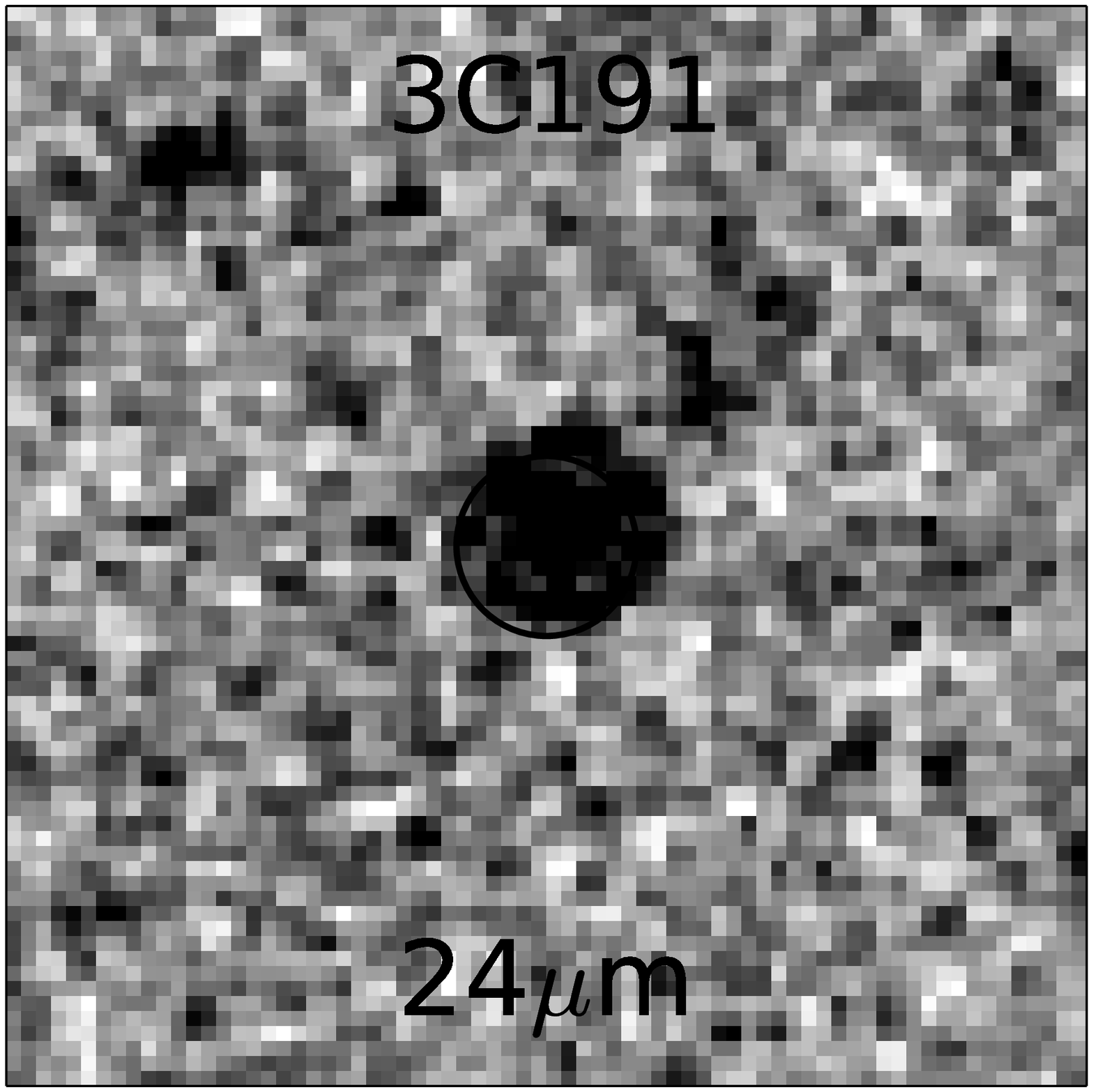}
      \includegraphics[width=1.5cm]{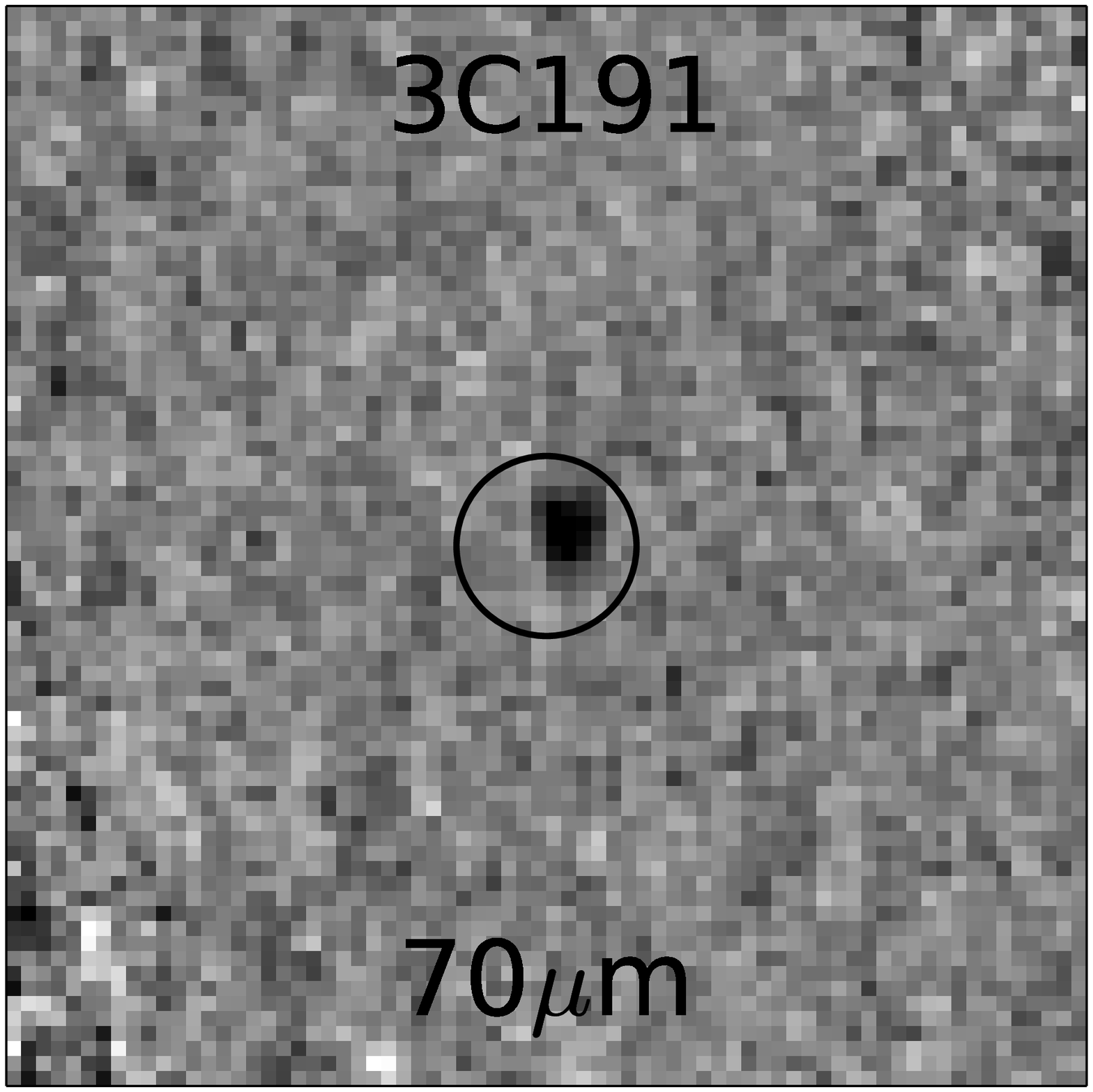}
      \includegraphics[width=1.5cm]{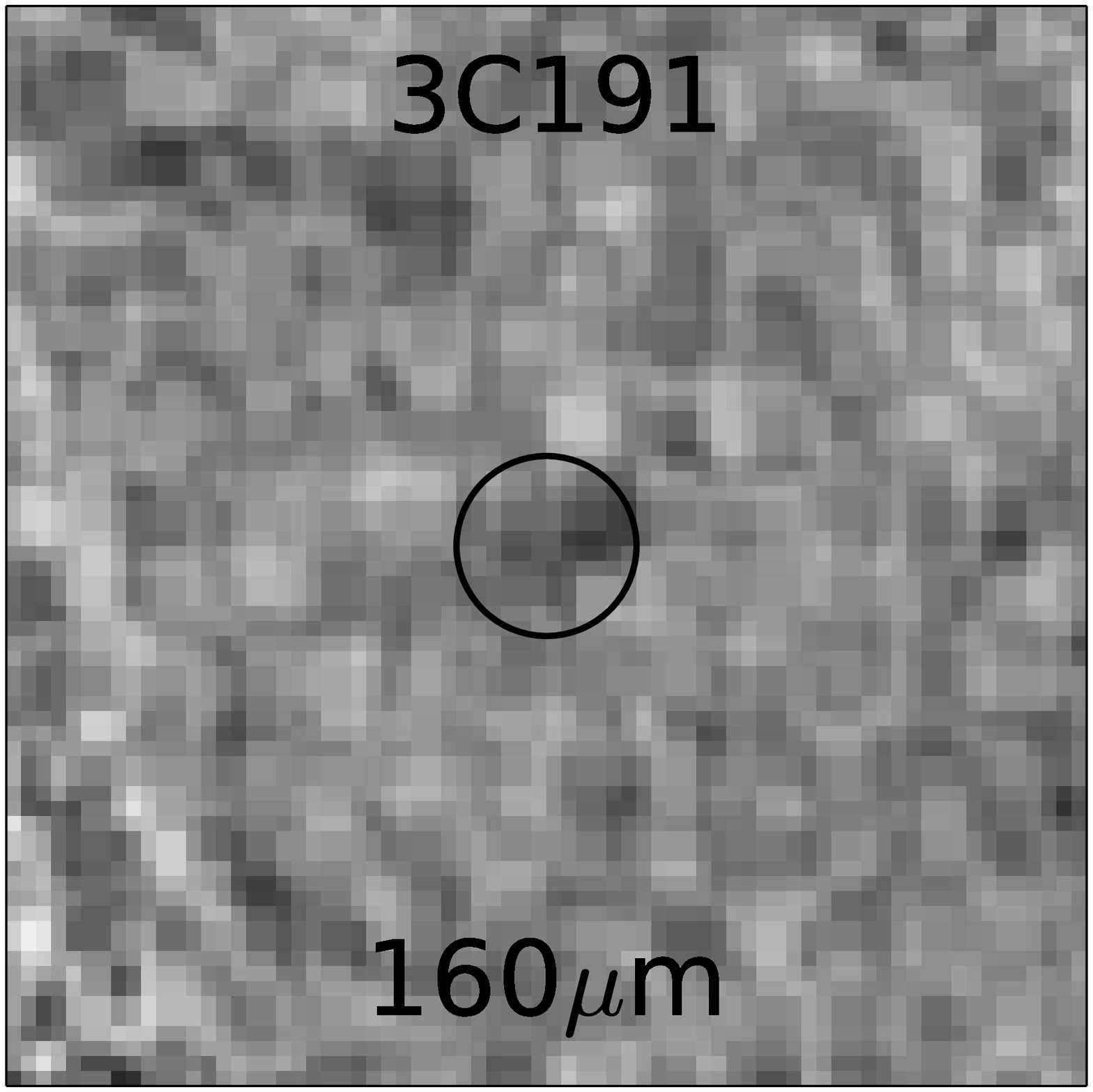}
      \includegraphics[width=1.5cm]{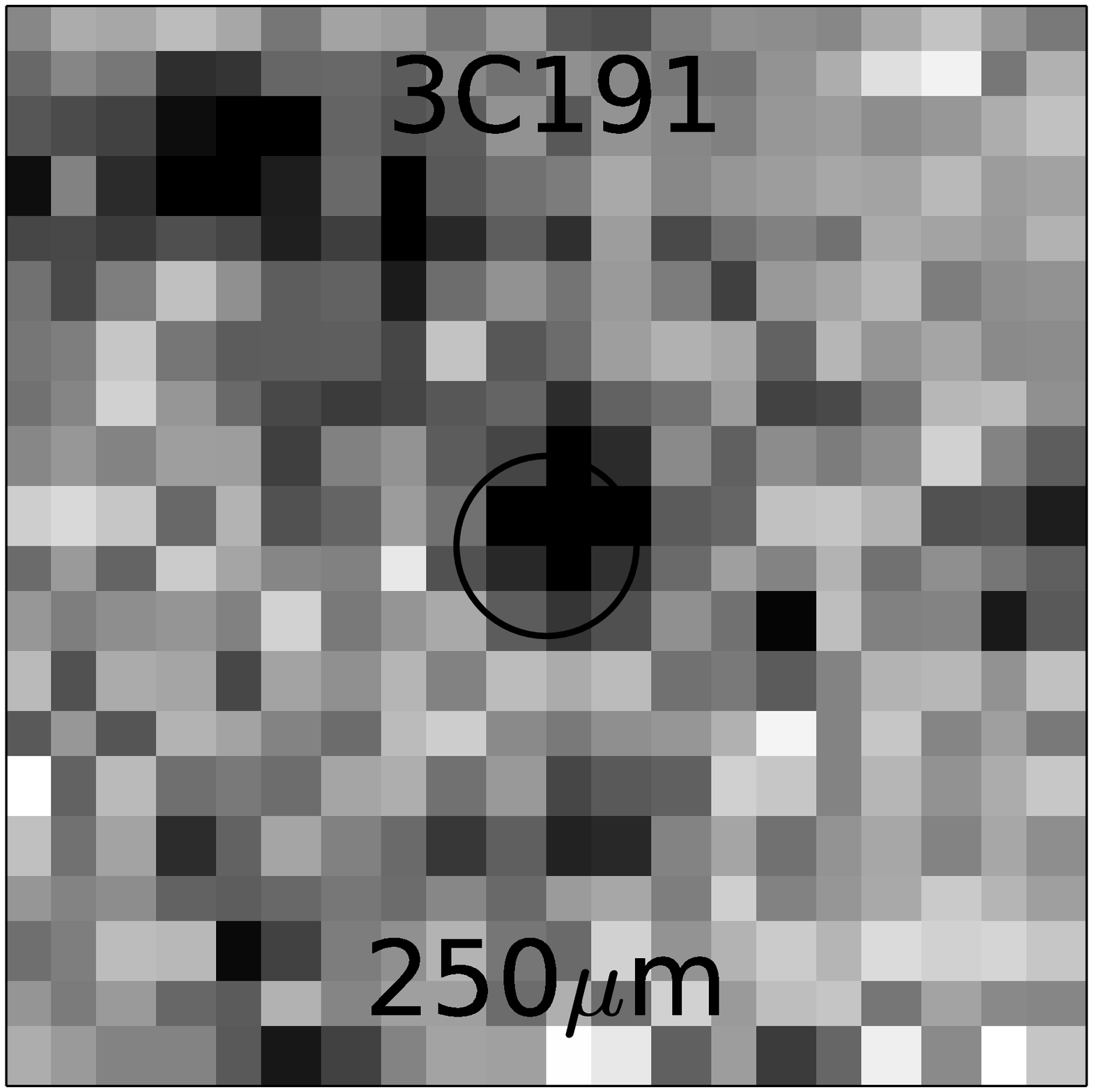}
      \includegraphics[width=1.5cm]{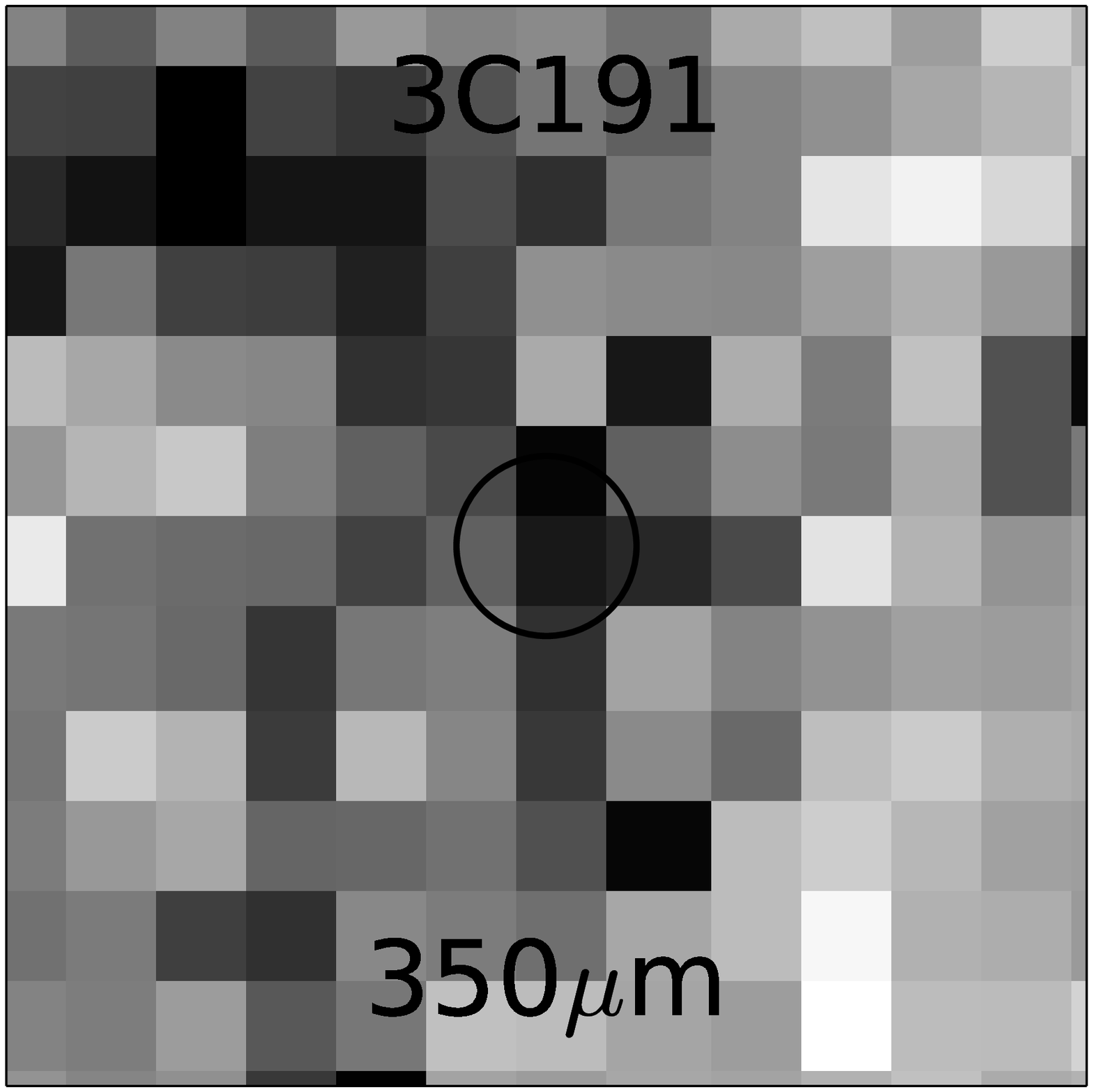}
      \includegraphics[width=1.5cm]{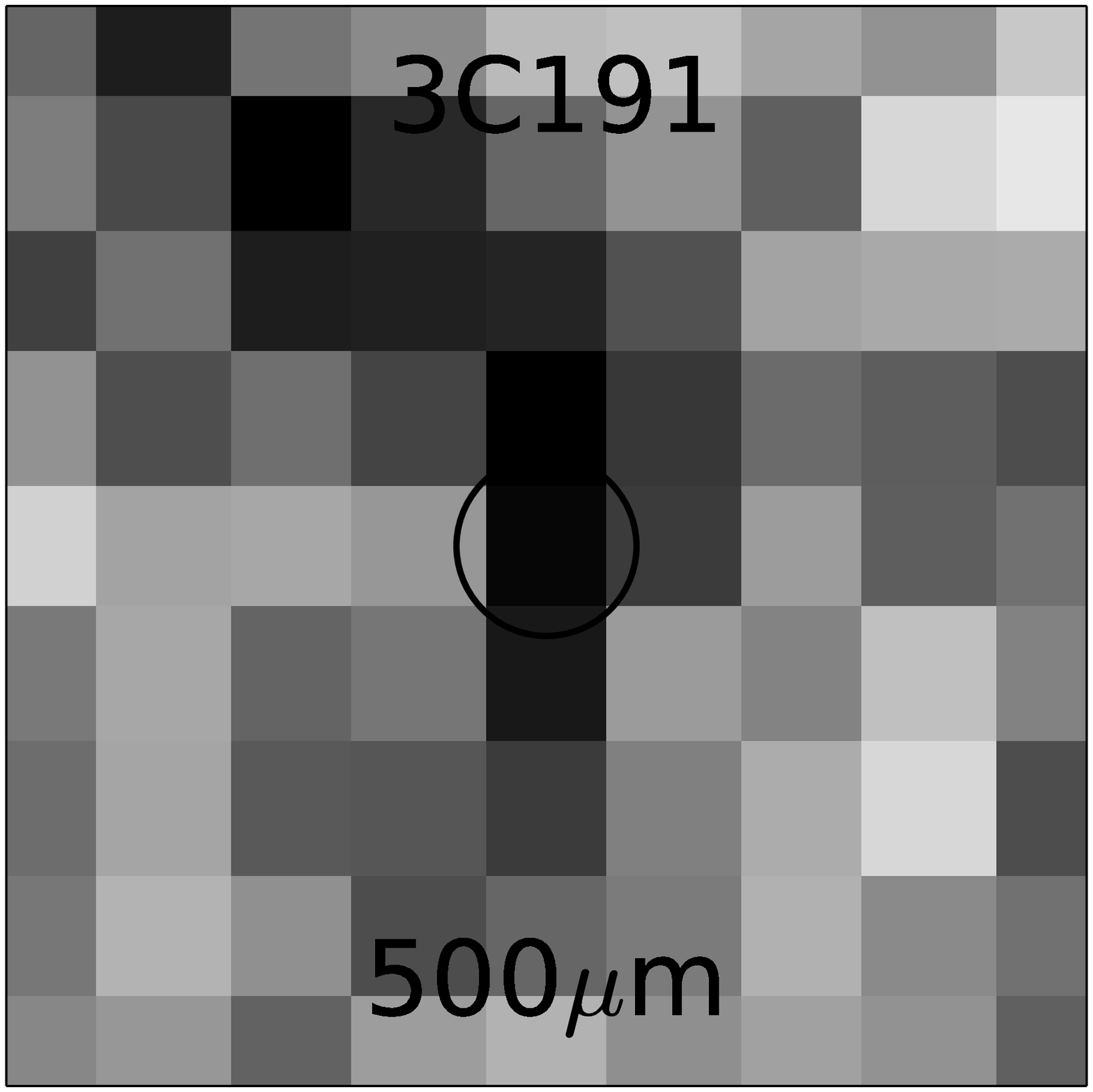}
      \\
      \includegraphics[width=1.5cm]{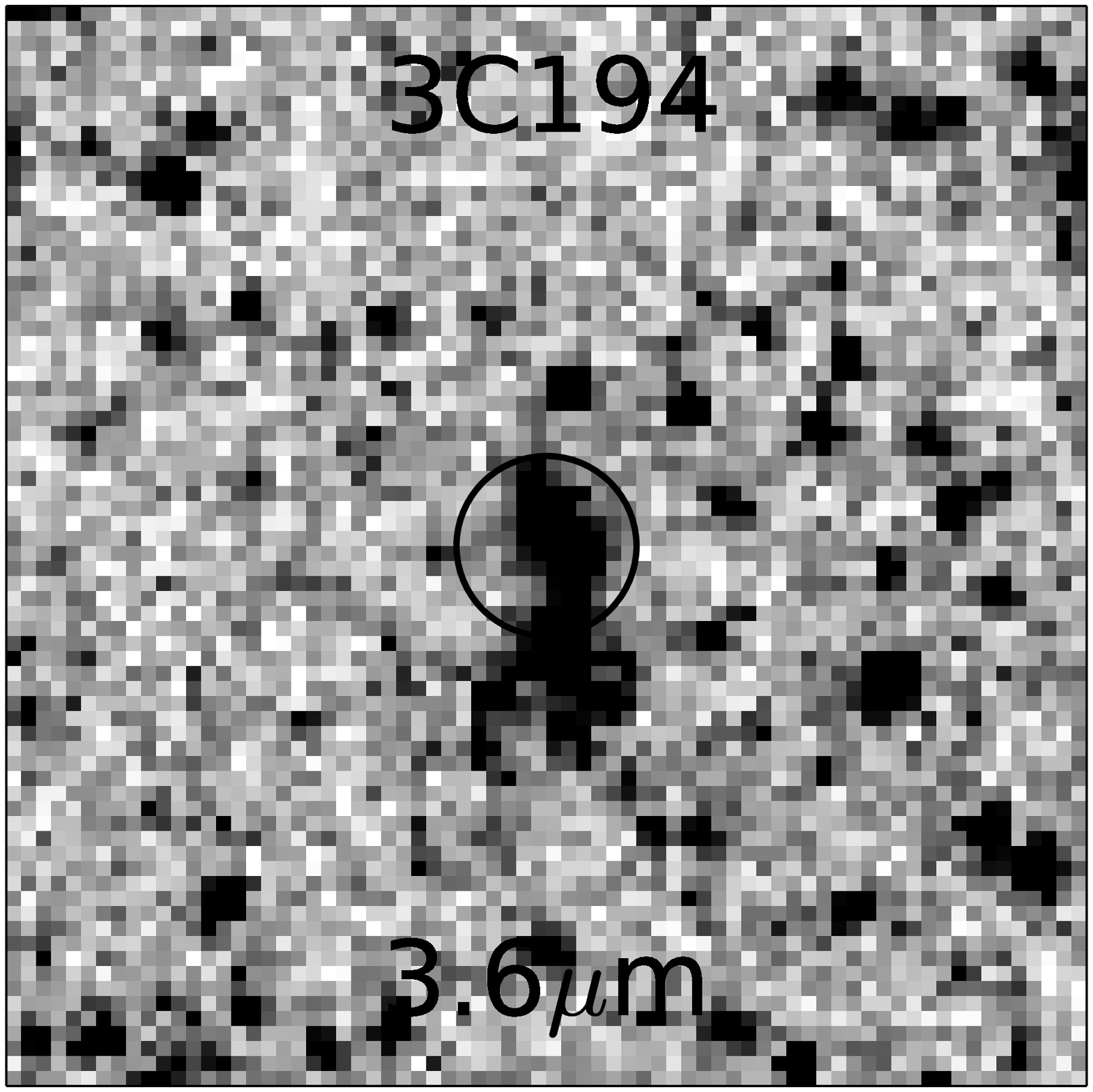}
      \includegraphics[width=1.5cm]{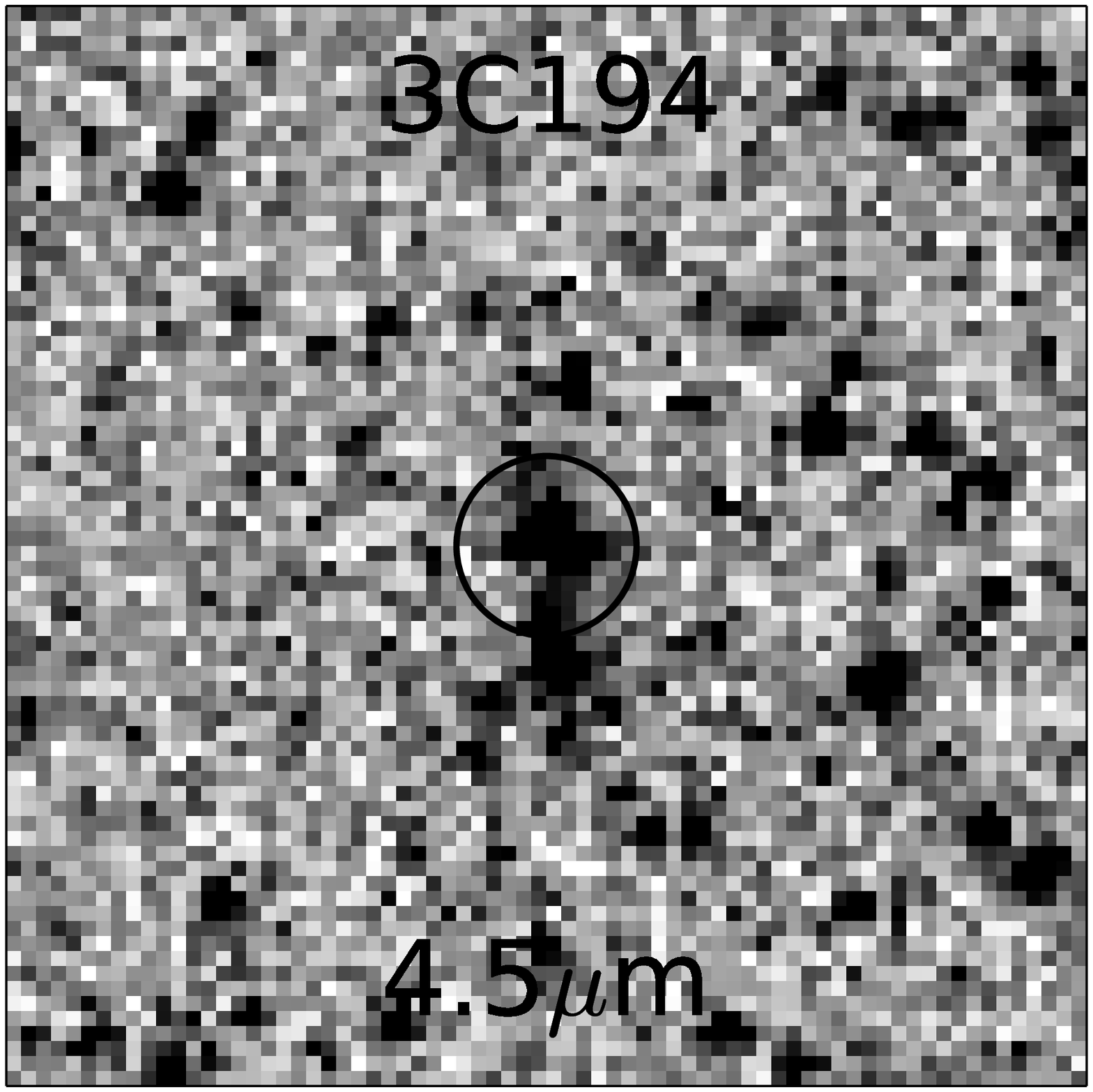}
      \includegraphics[width=1.5cm]{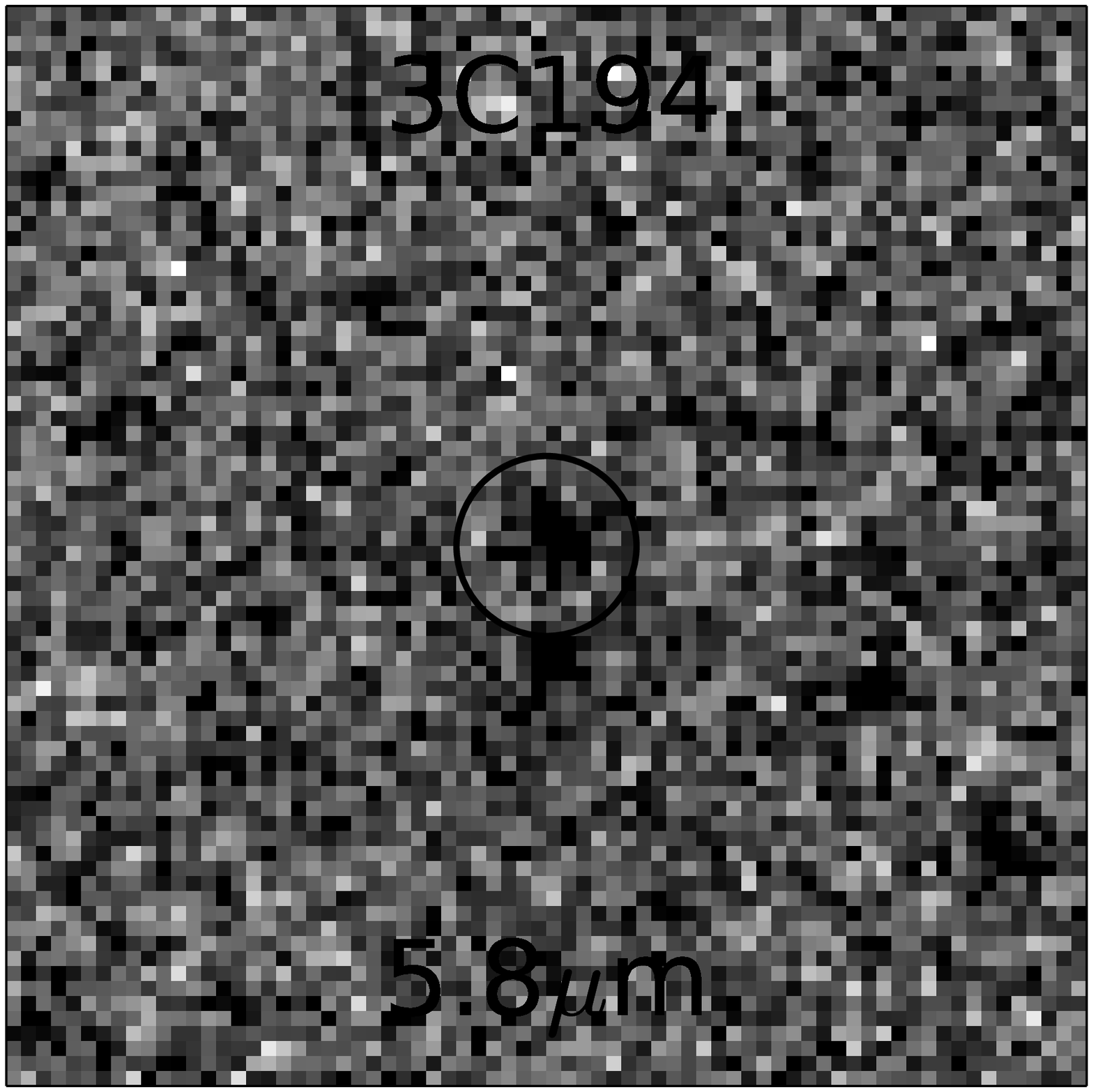}
      \includegraphics[width=1.5cm]{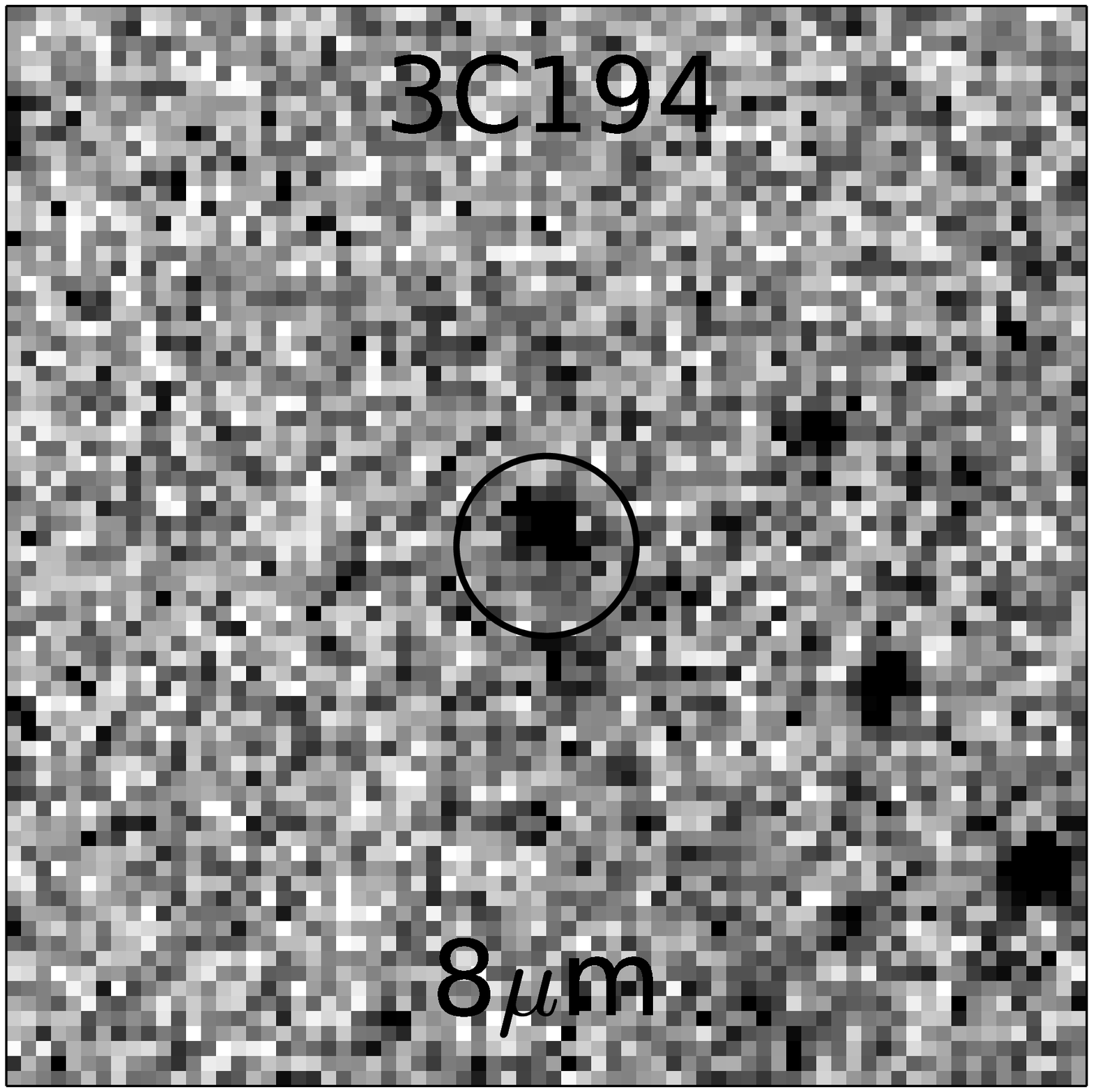}
      \includegraphics[width=1.5cm]{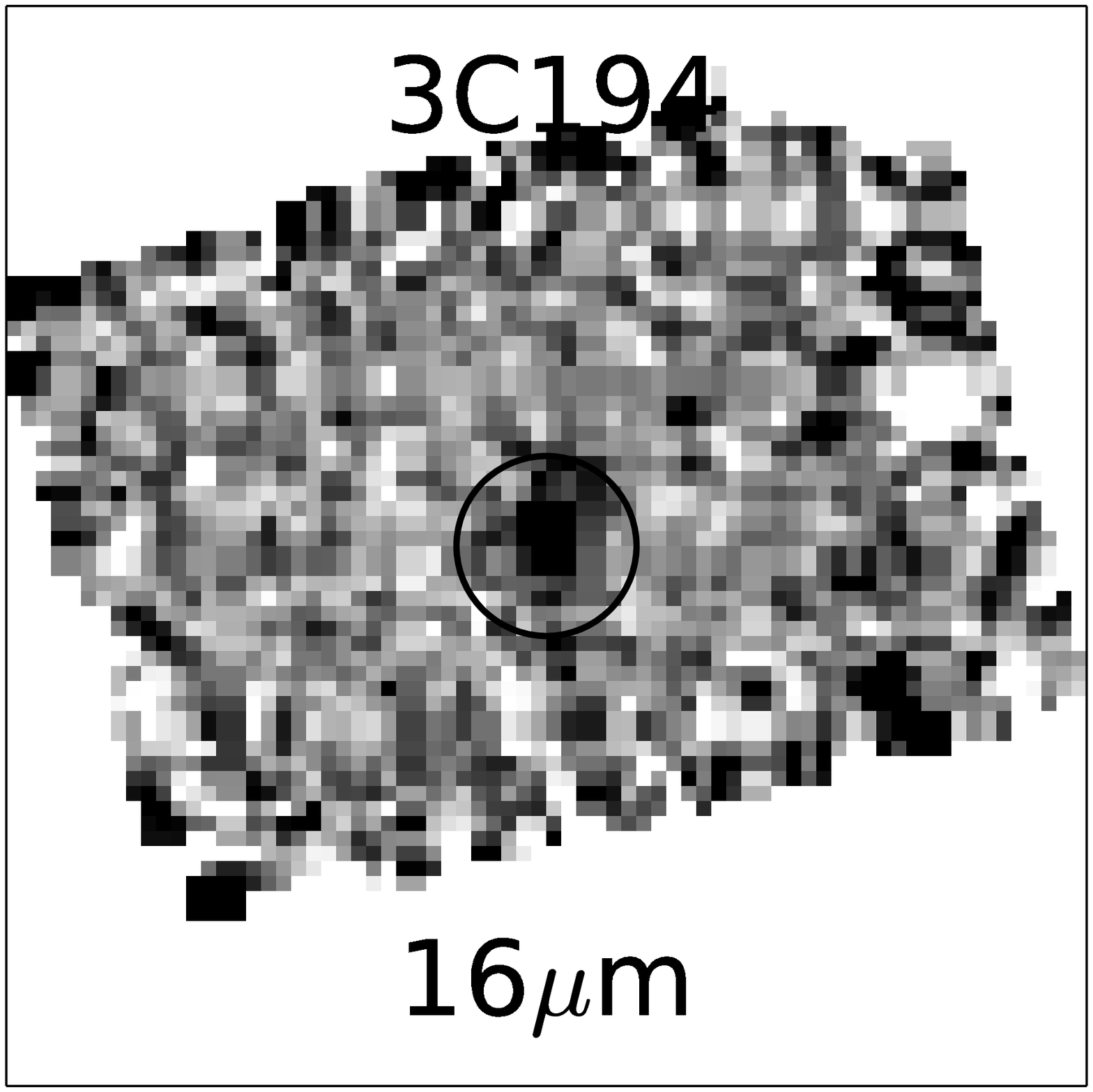}
      \includegraphics[width=1.5cm]{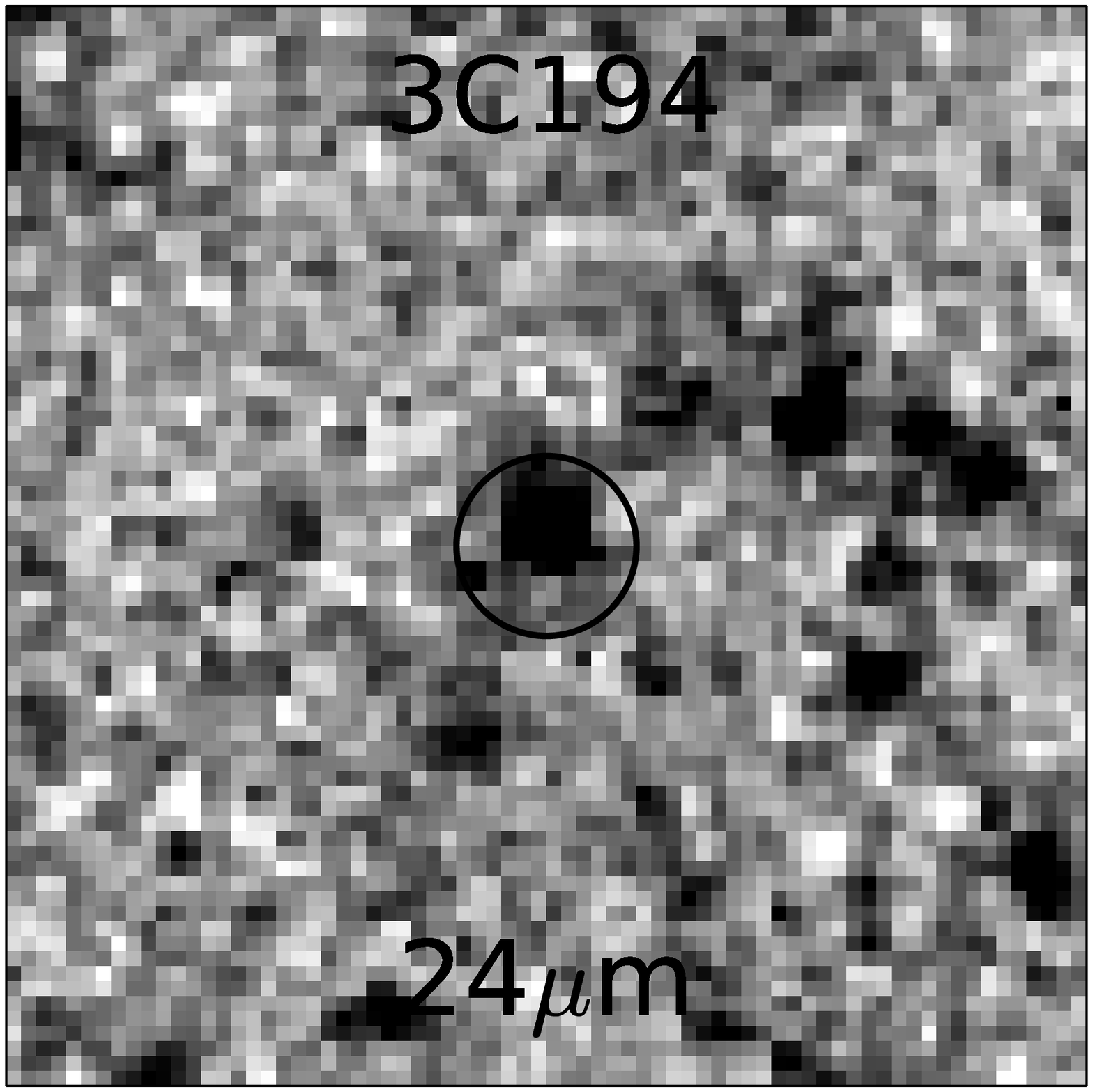}
      \includegraphics[width=1.5cm]{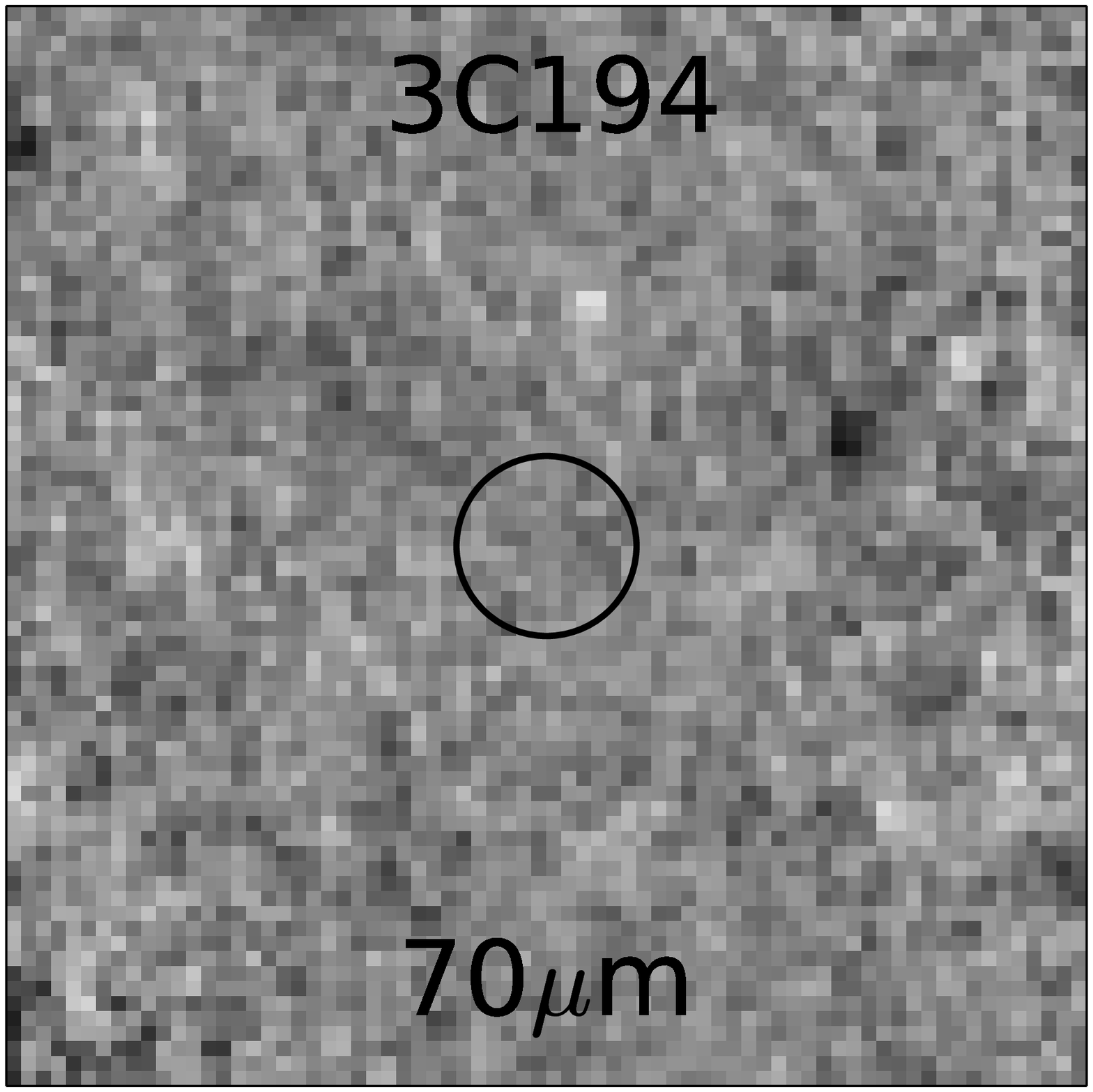}
      \includegraphics[width=1.5cm]{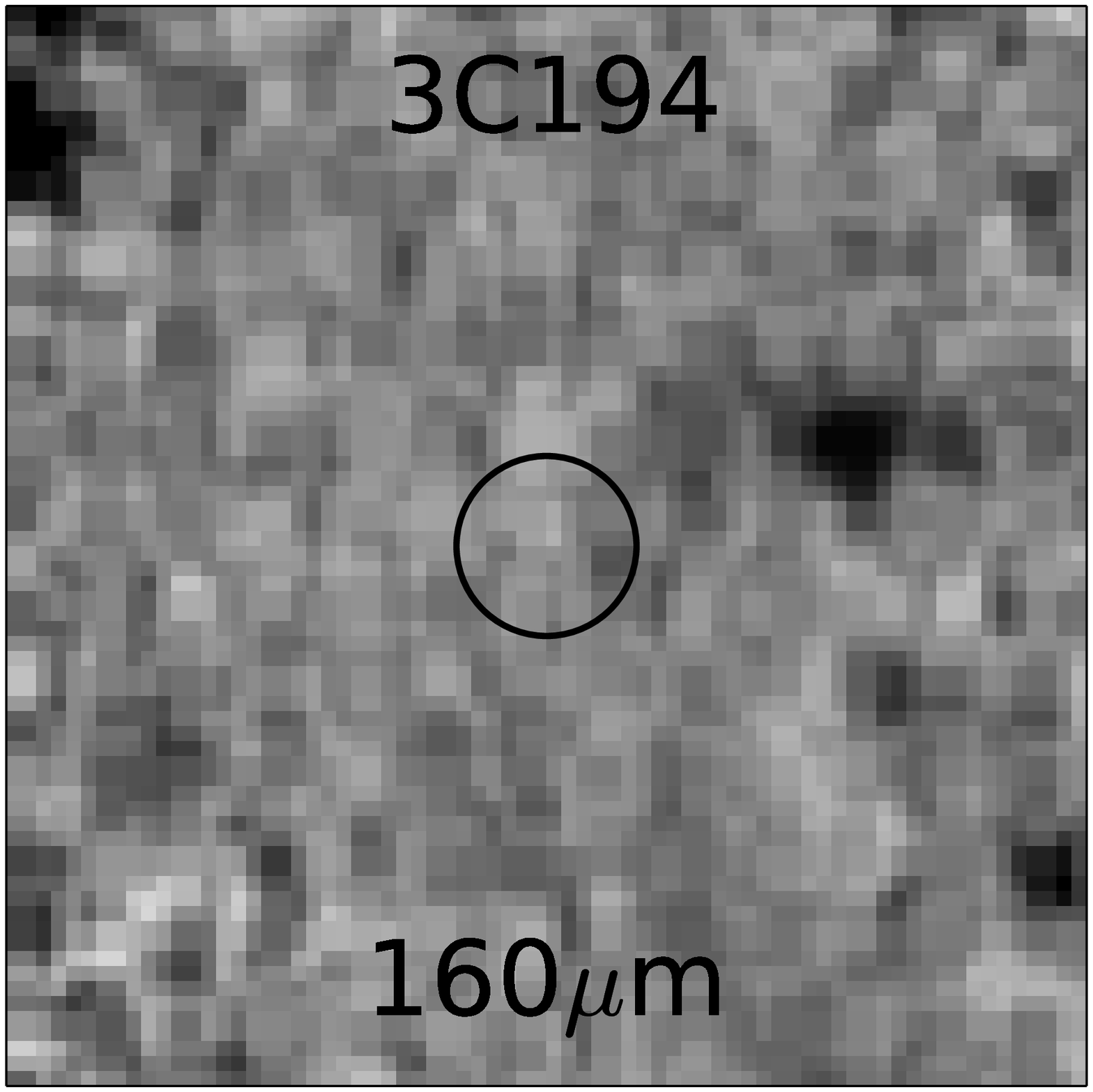}
      \includegraphics[width=1.5cm]{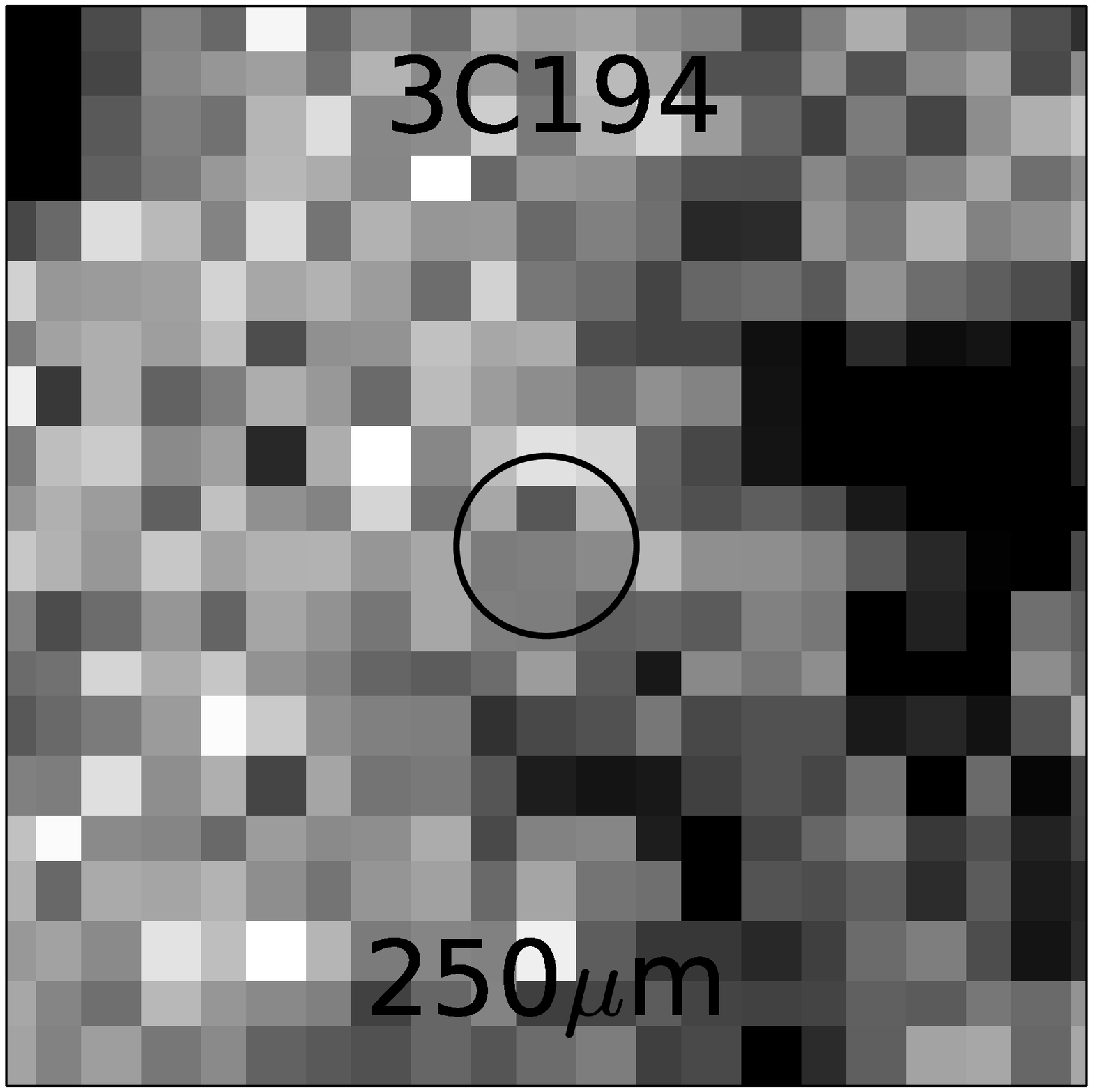}
      \includegraphics[width=1.5cm]{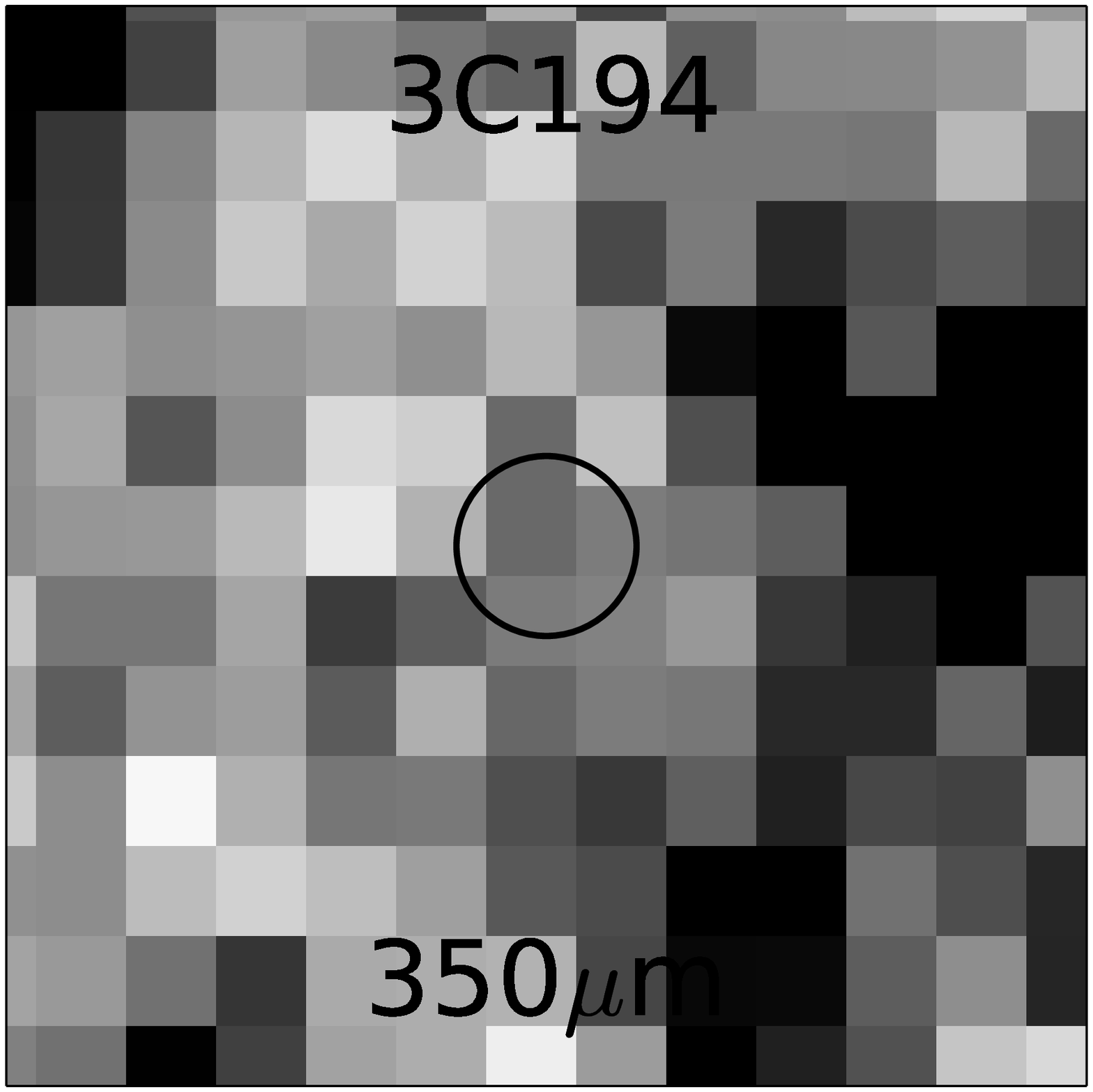}
      \includegraphics[width=1.5cm]{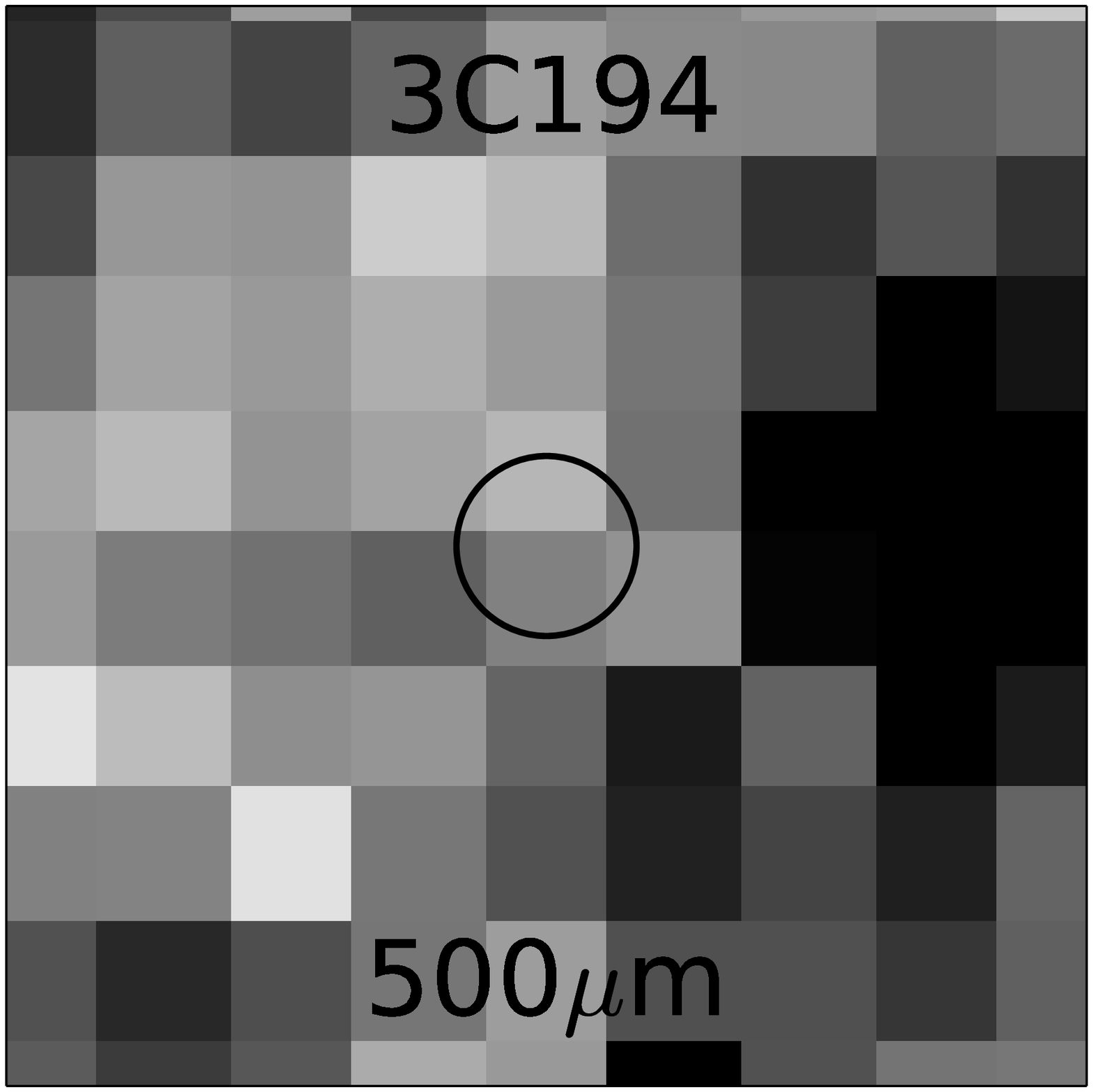}
      \\
      \includegraphics[width=1.5cm]{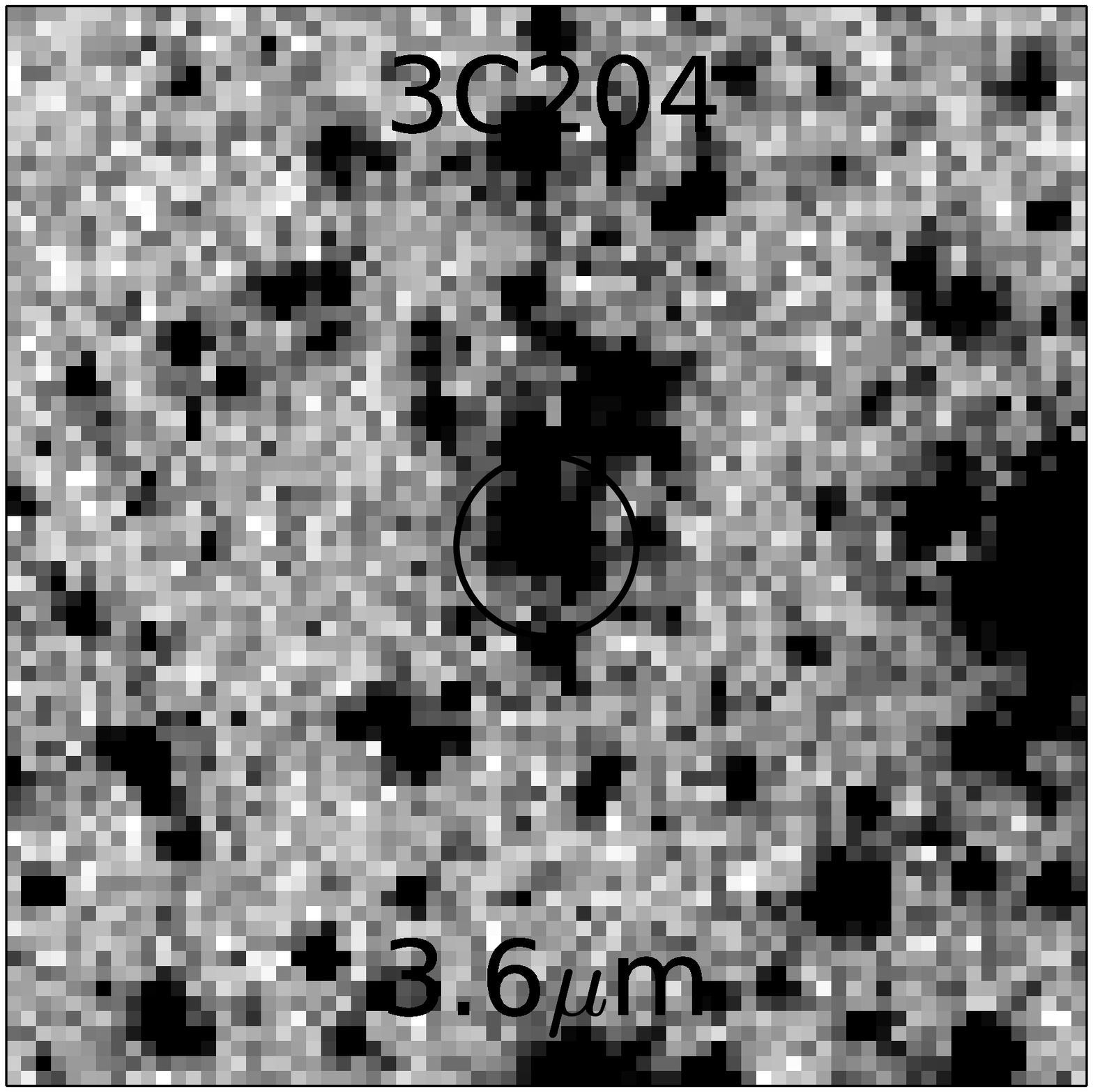}
      \includegraphics[width=1.5cm]{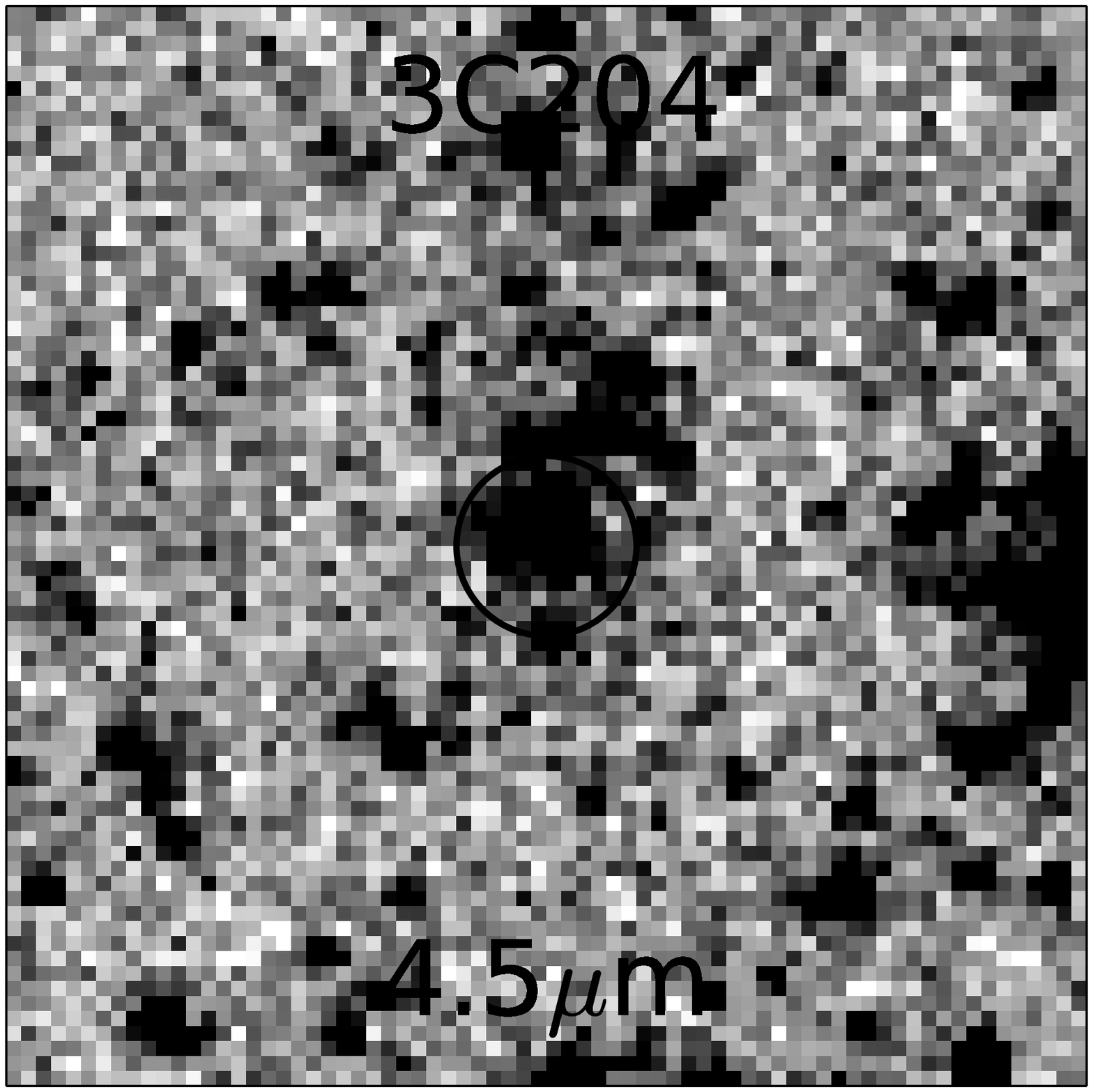}
      \includegraphics[width=1.5cm]{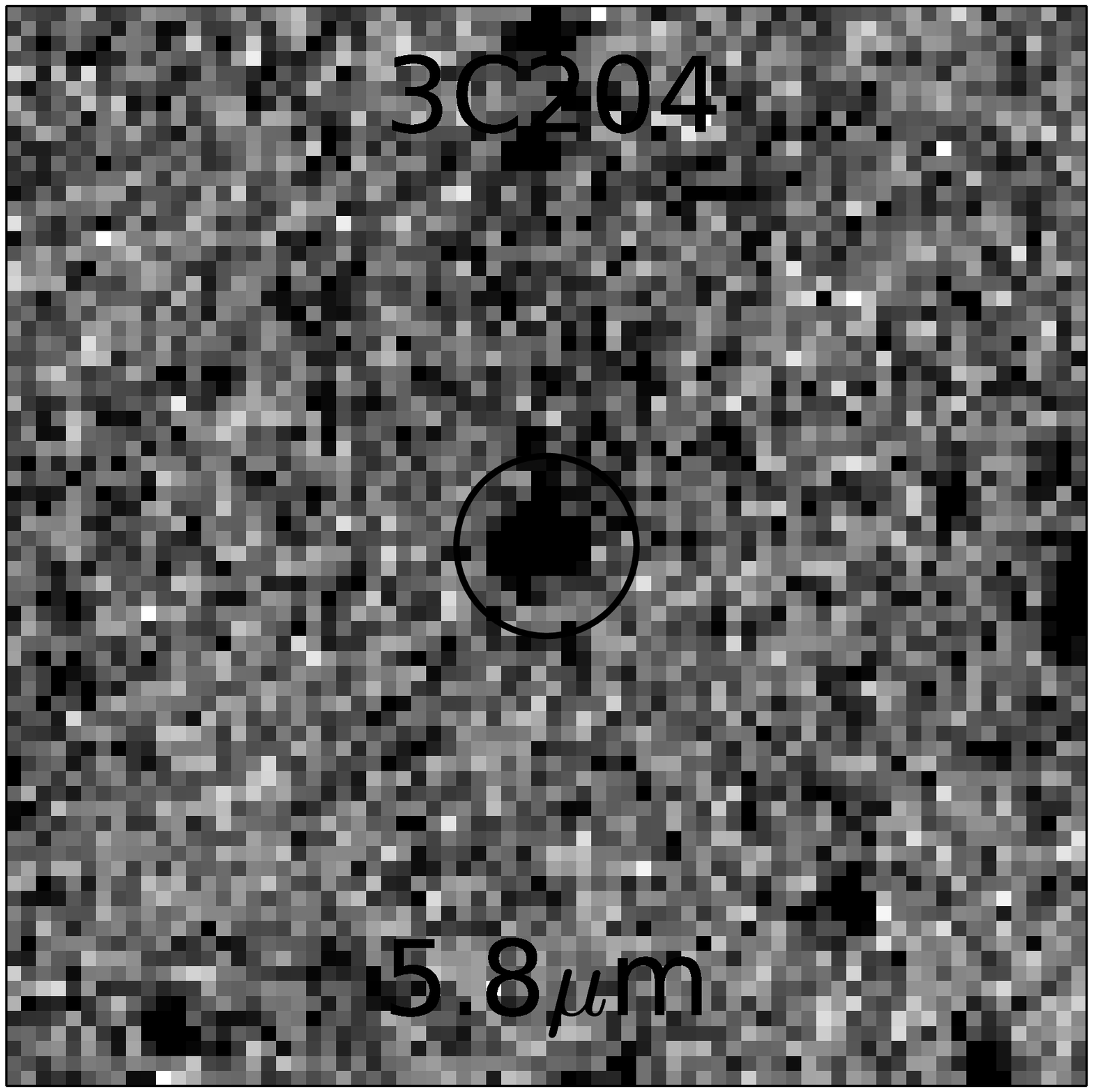}
      \includegraphics[width=1.5cm]{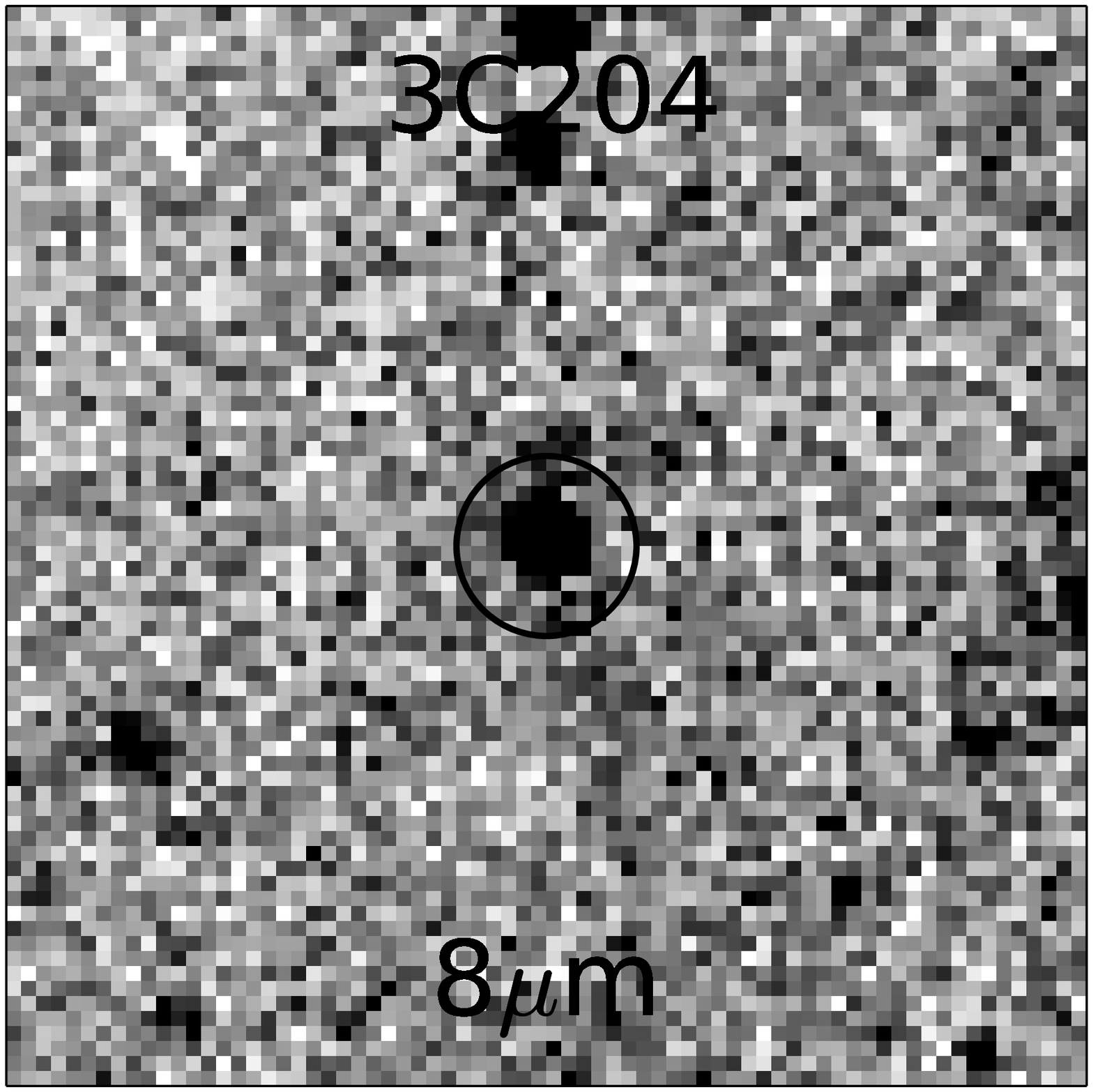}
      \includegraphics[width=1.5cm]{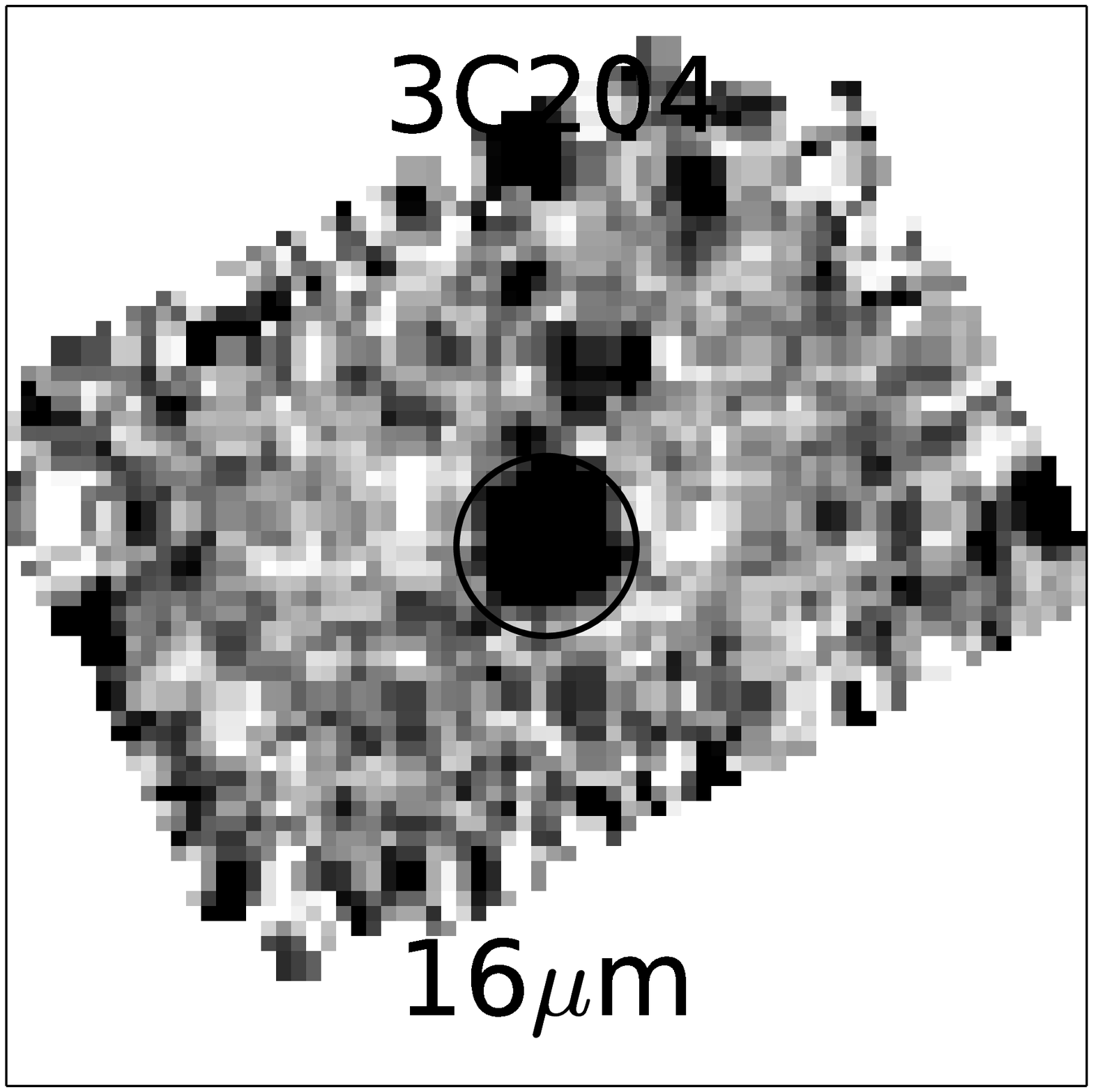}
      \includegraphics[width=1.5cm]{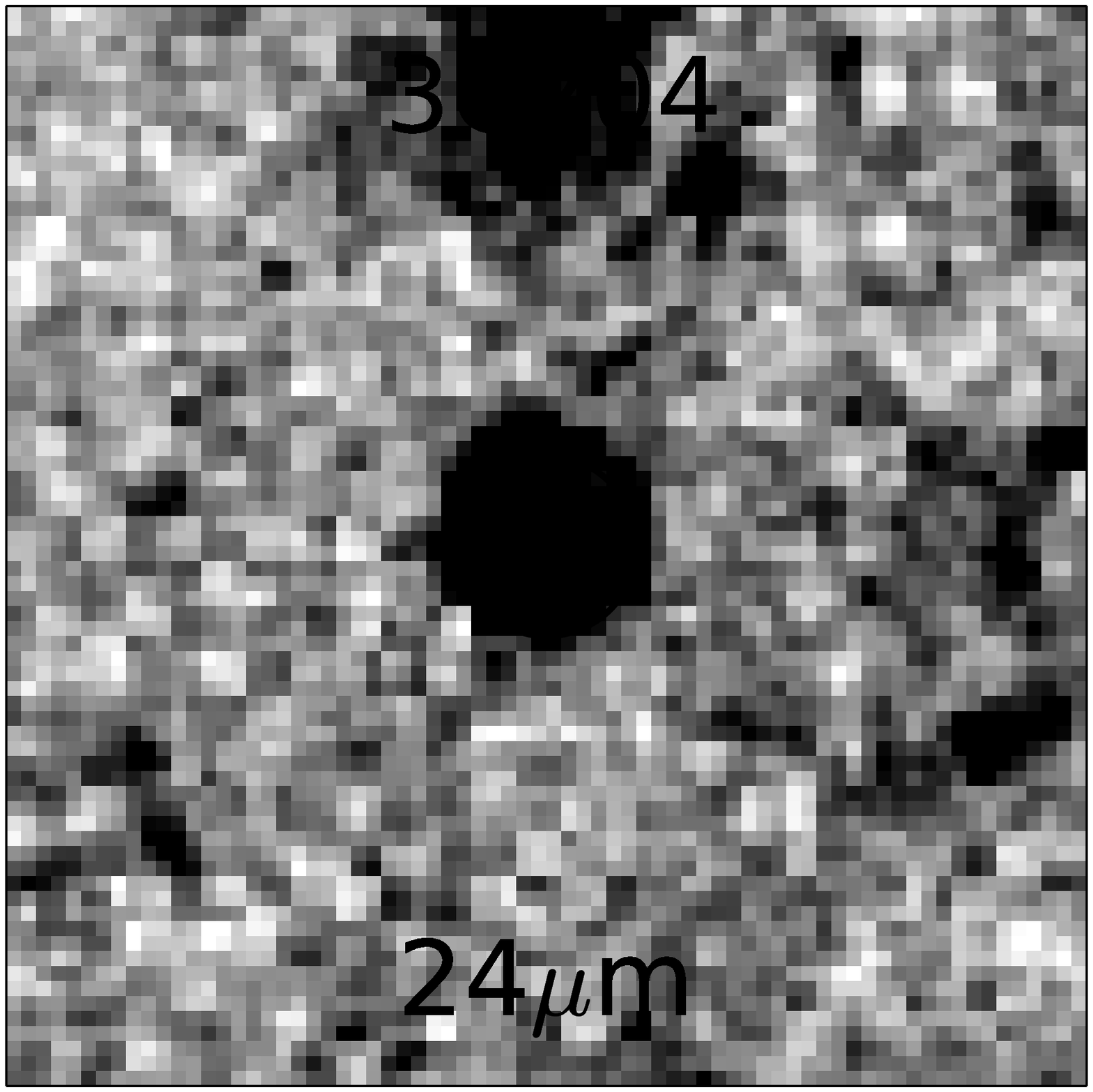}
      \includegraphics[width=1.5cm]{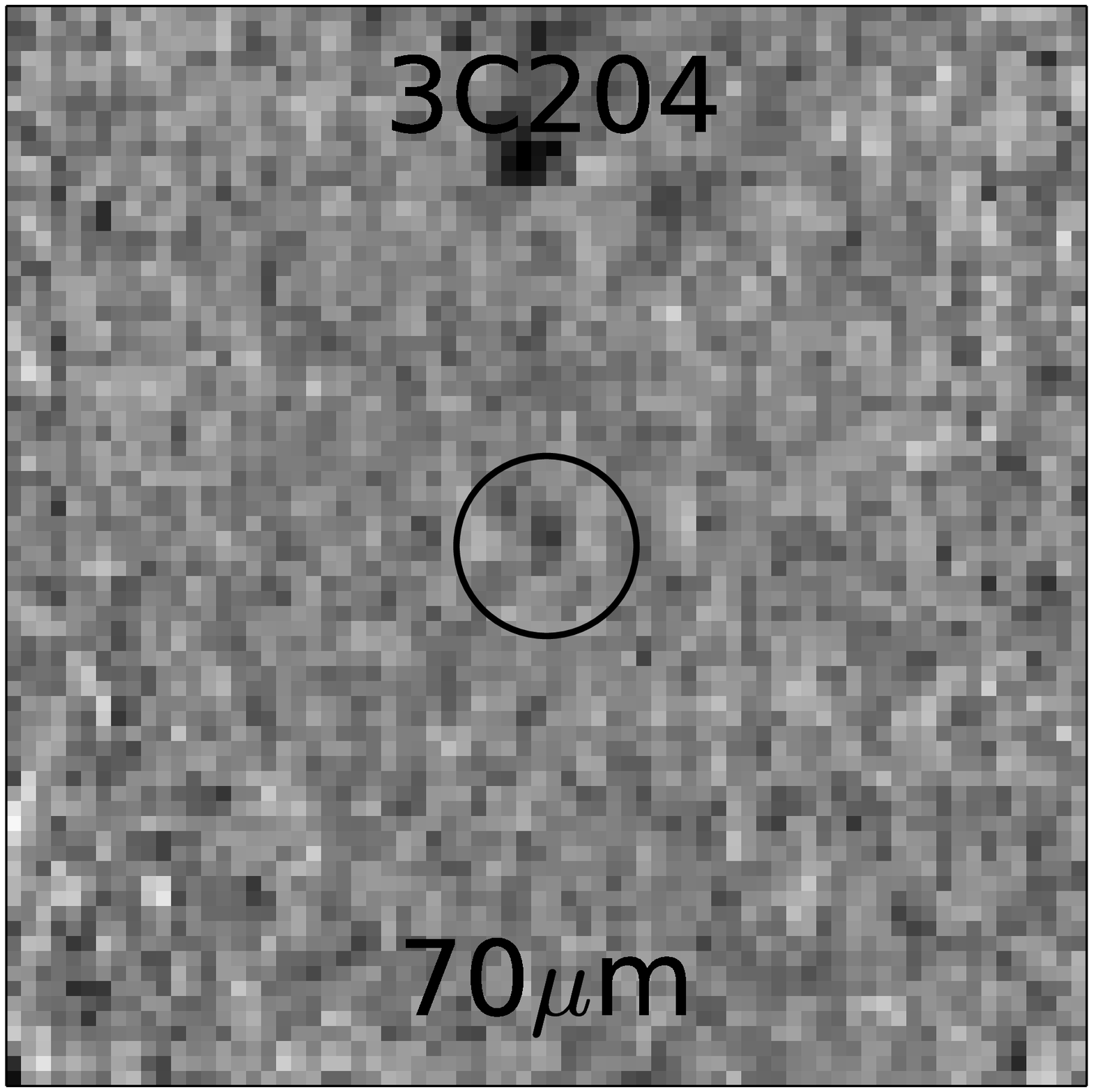}
      \includegraphics[width=1.5cm]{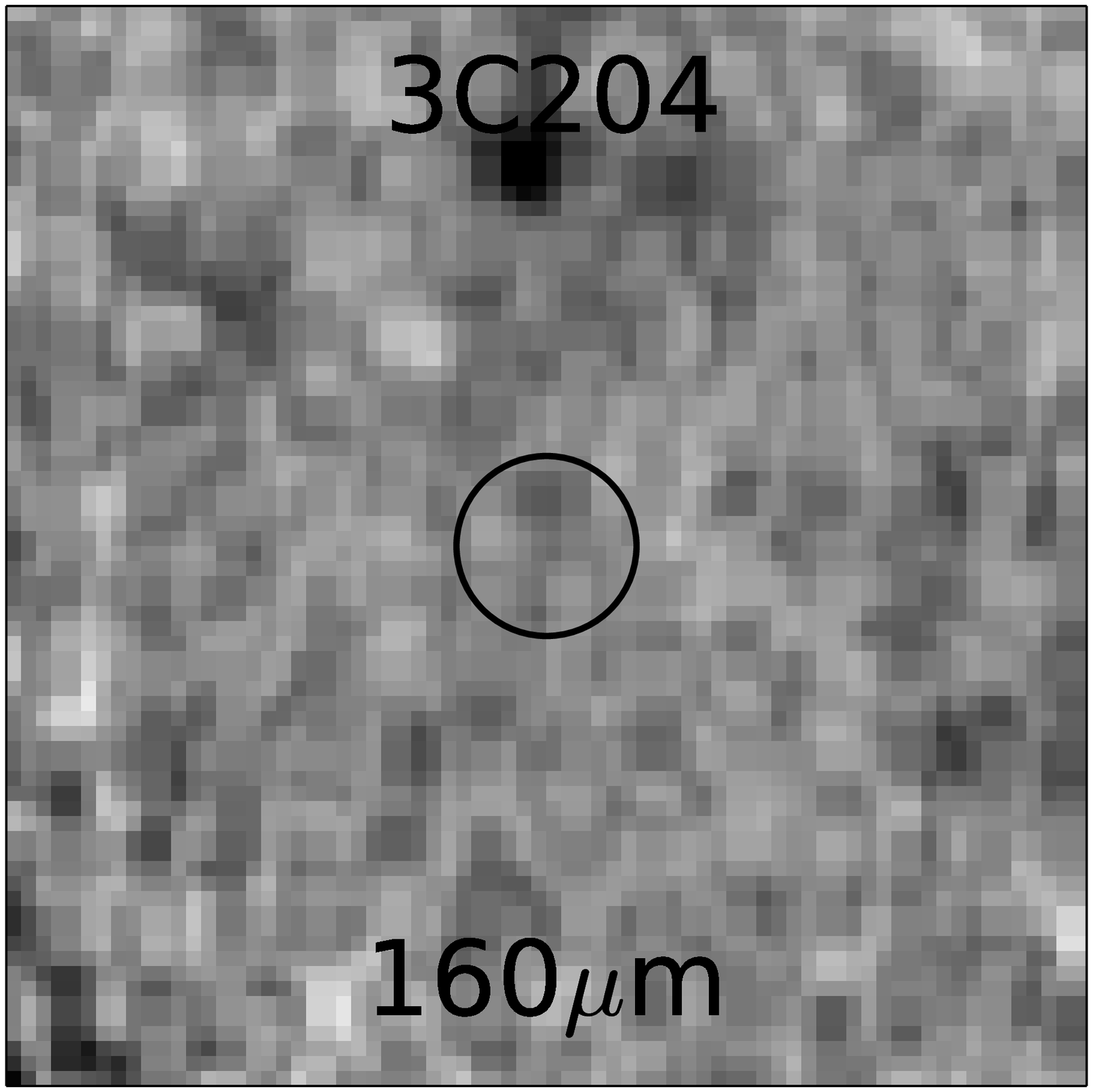}
      \includegraphics[width=1.5cm]{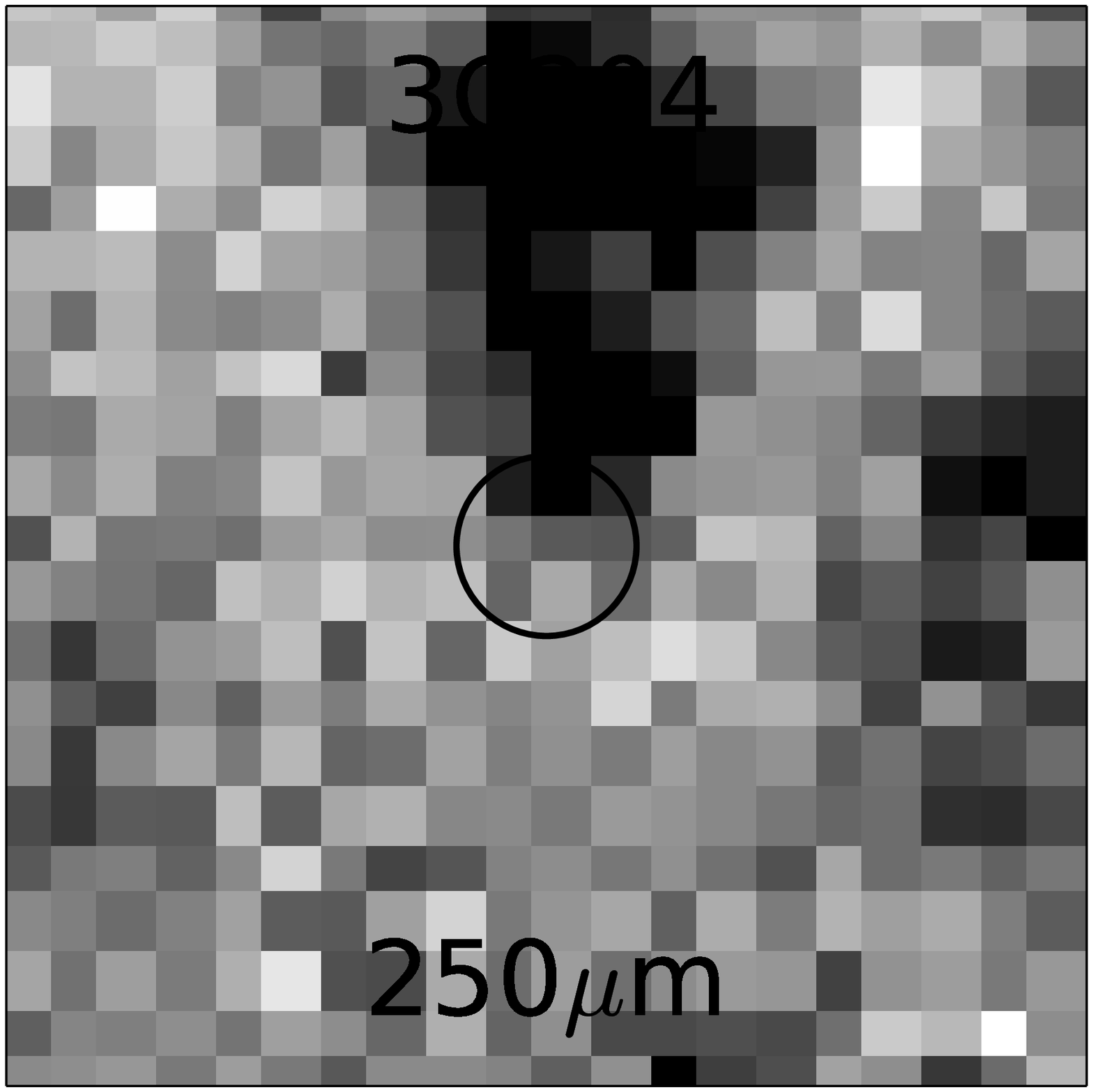}
      \includegraphics[width=1.5cm]{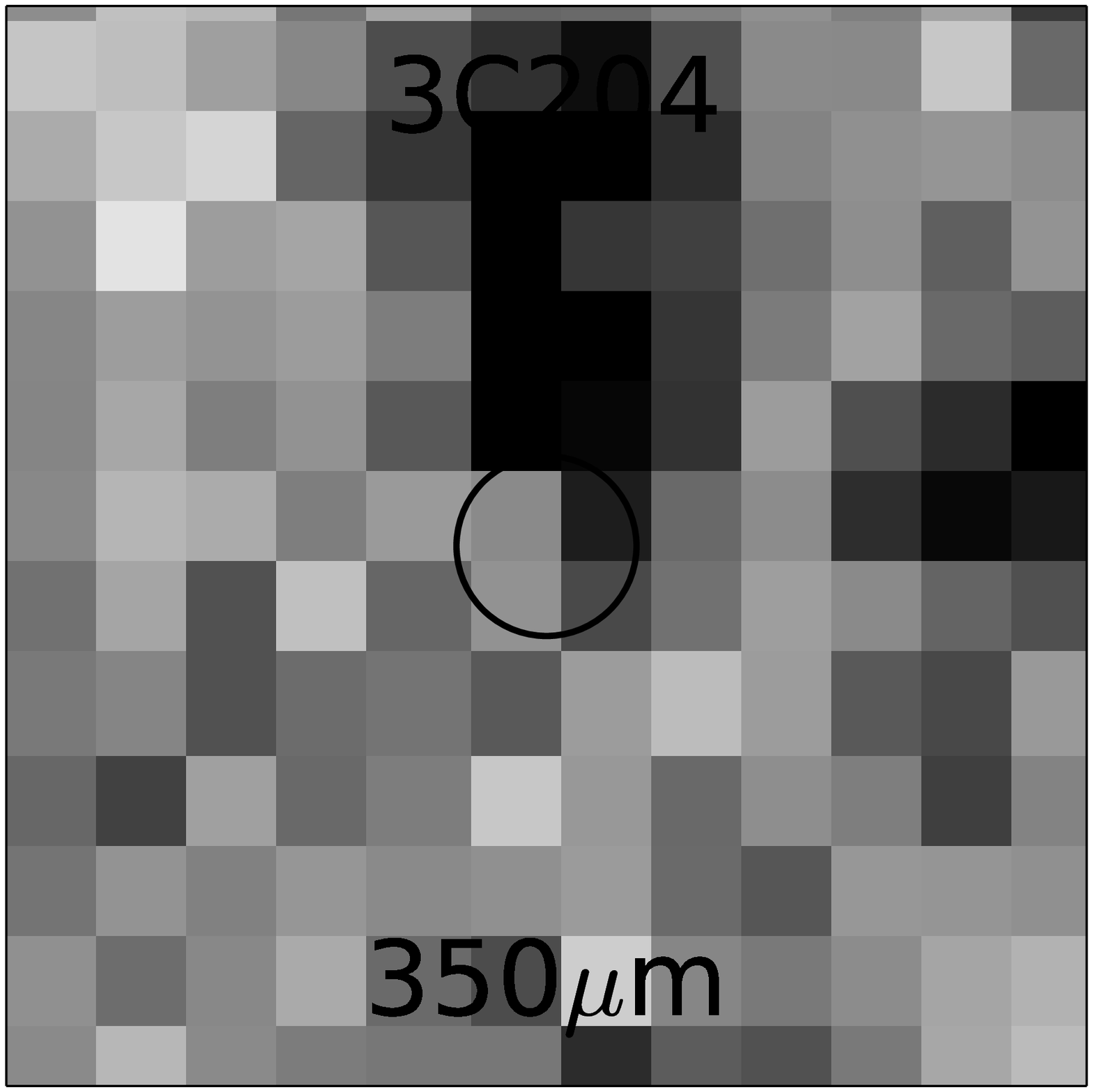}
      \includegraphics[width=1.5cm]{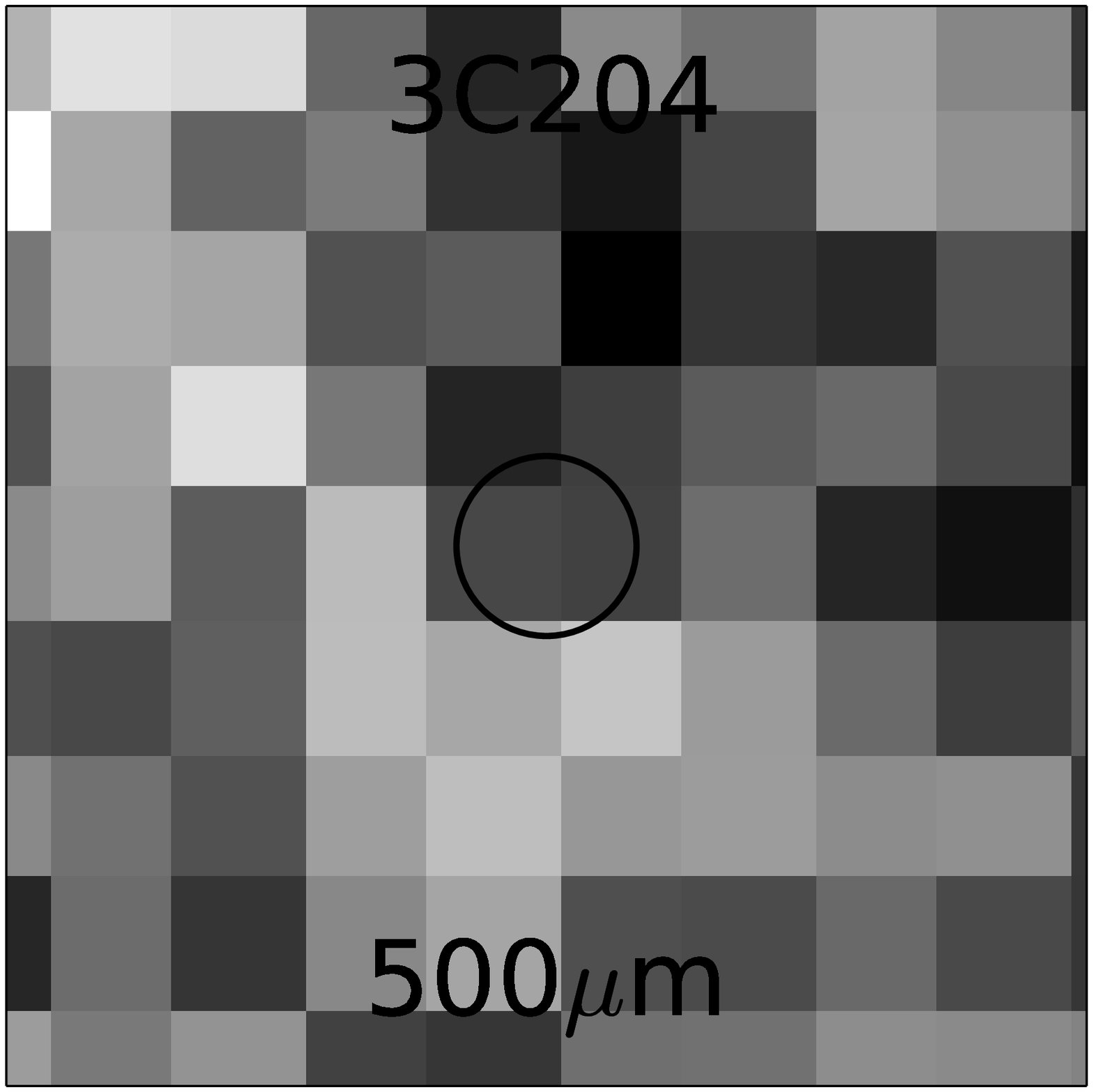}
      \\
      \includegraphics[width=1.5cm]{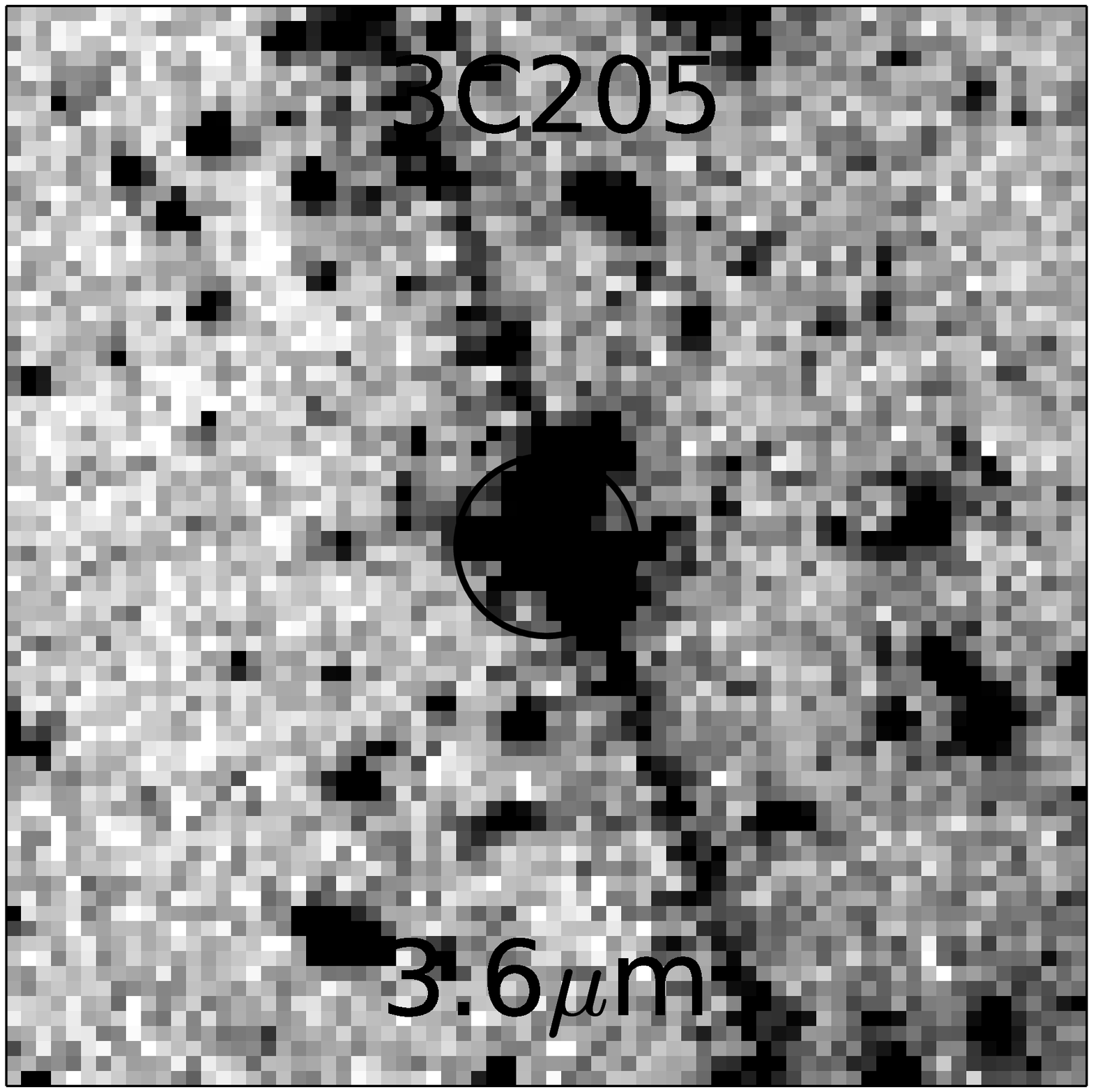}
      \includegraphics[width=1.5cm]{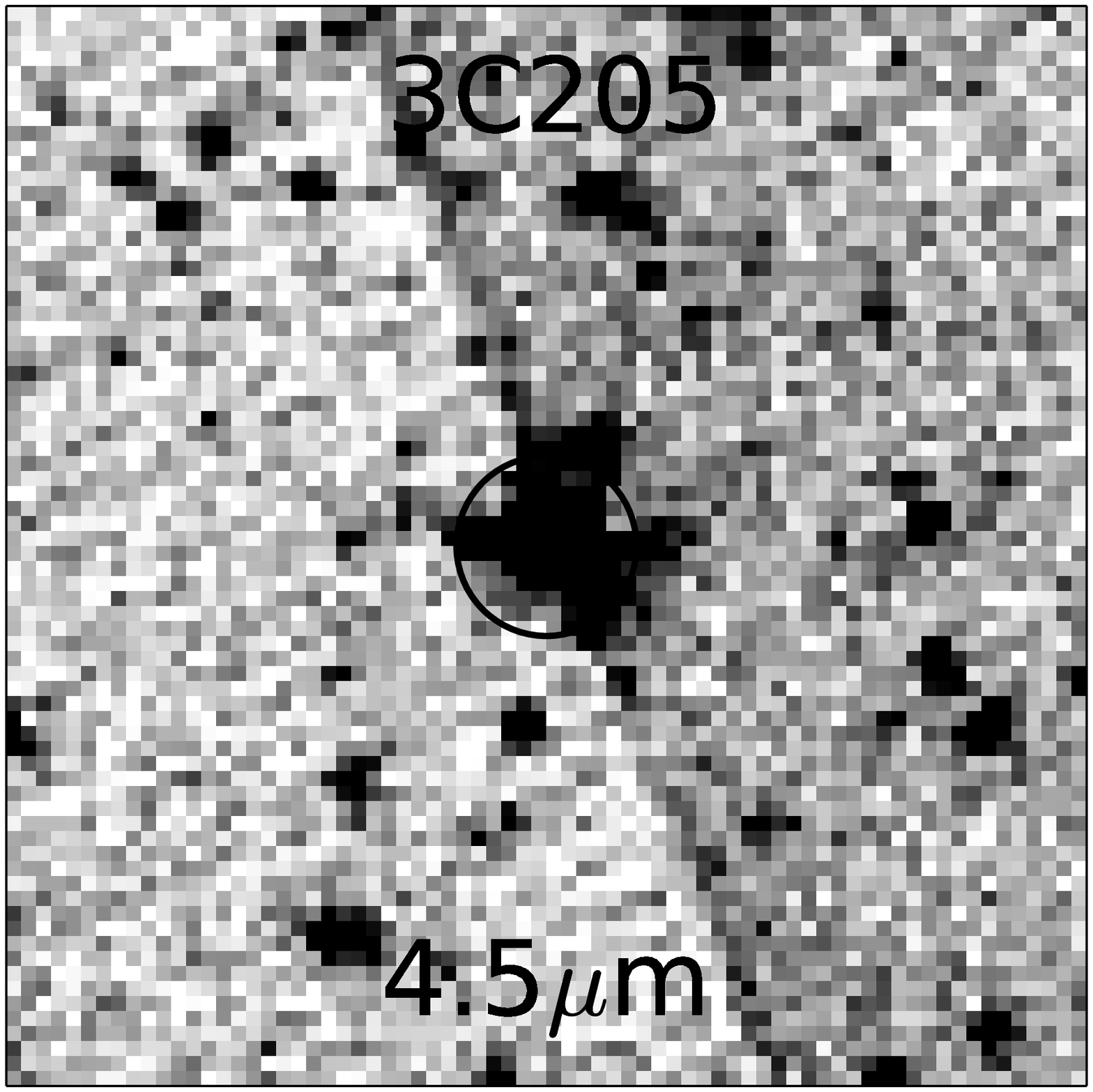}
      \includegraphics[width=1.5cm]{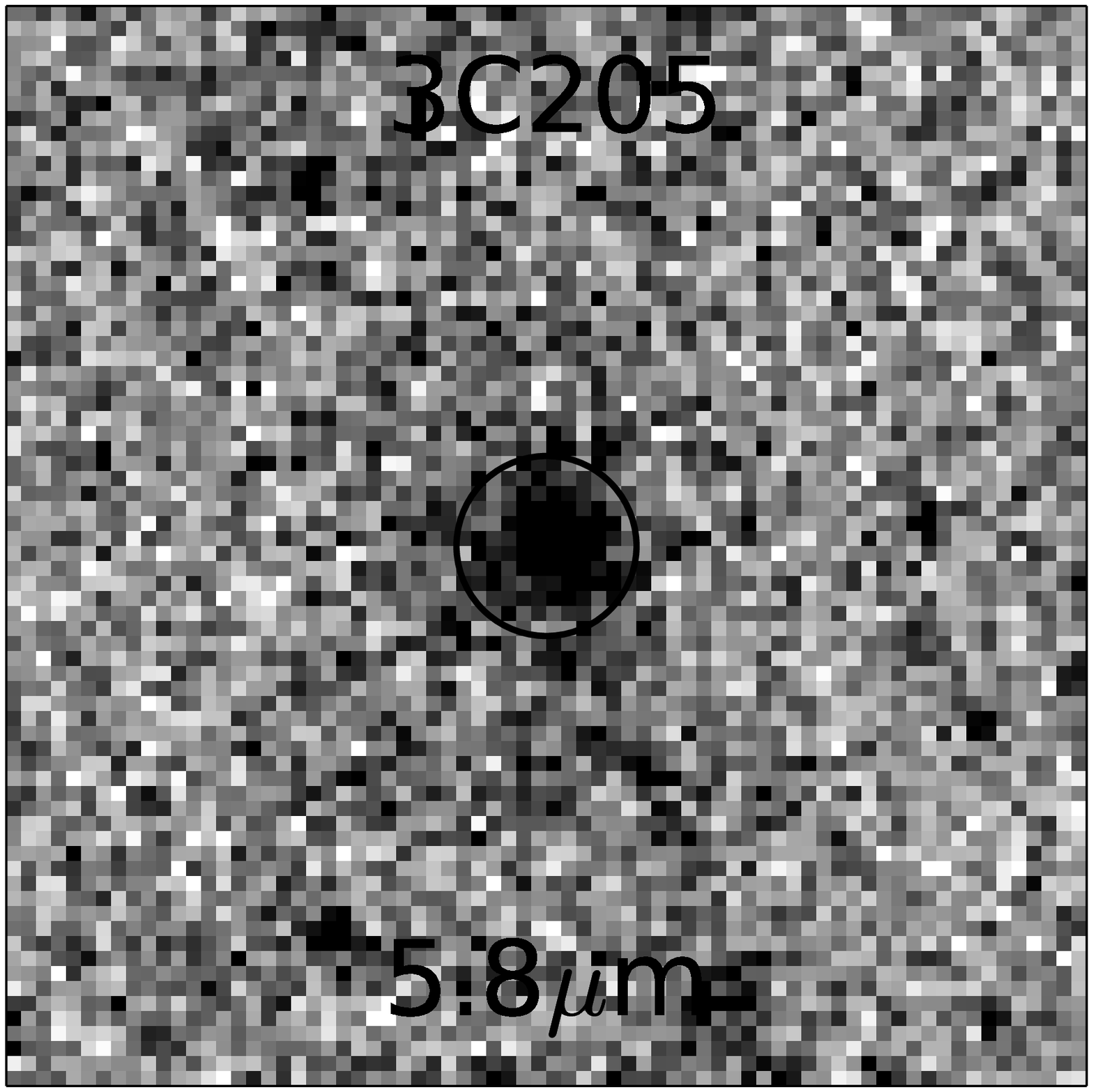}
      \includegraphics[width=1.5cm]{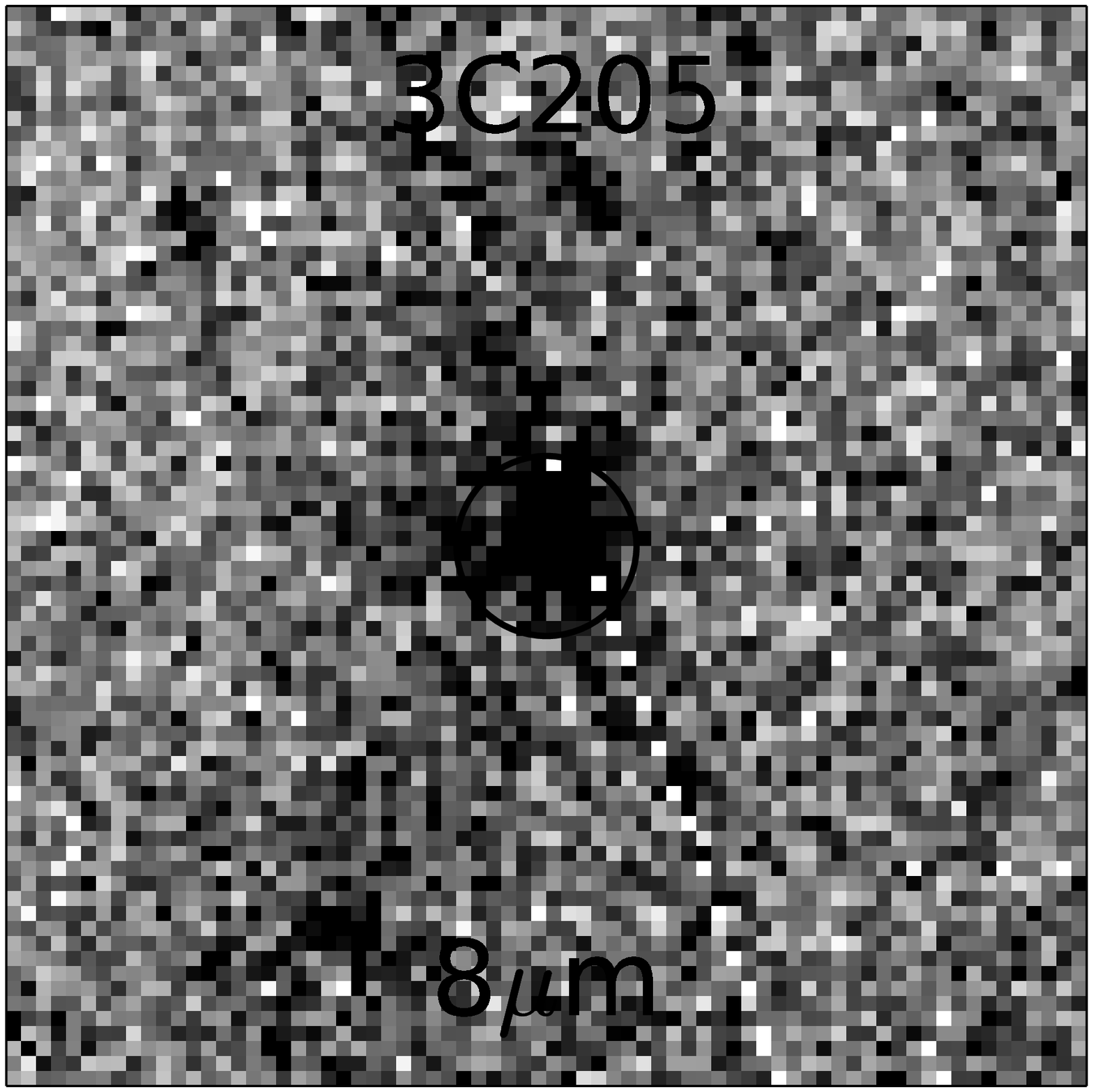}
      \includegraphics[width=1.5cm]{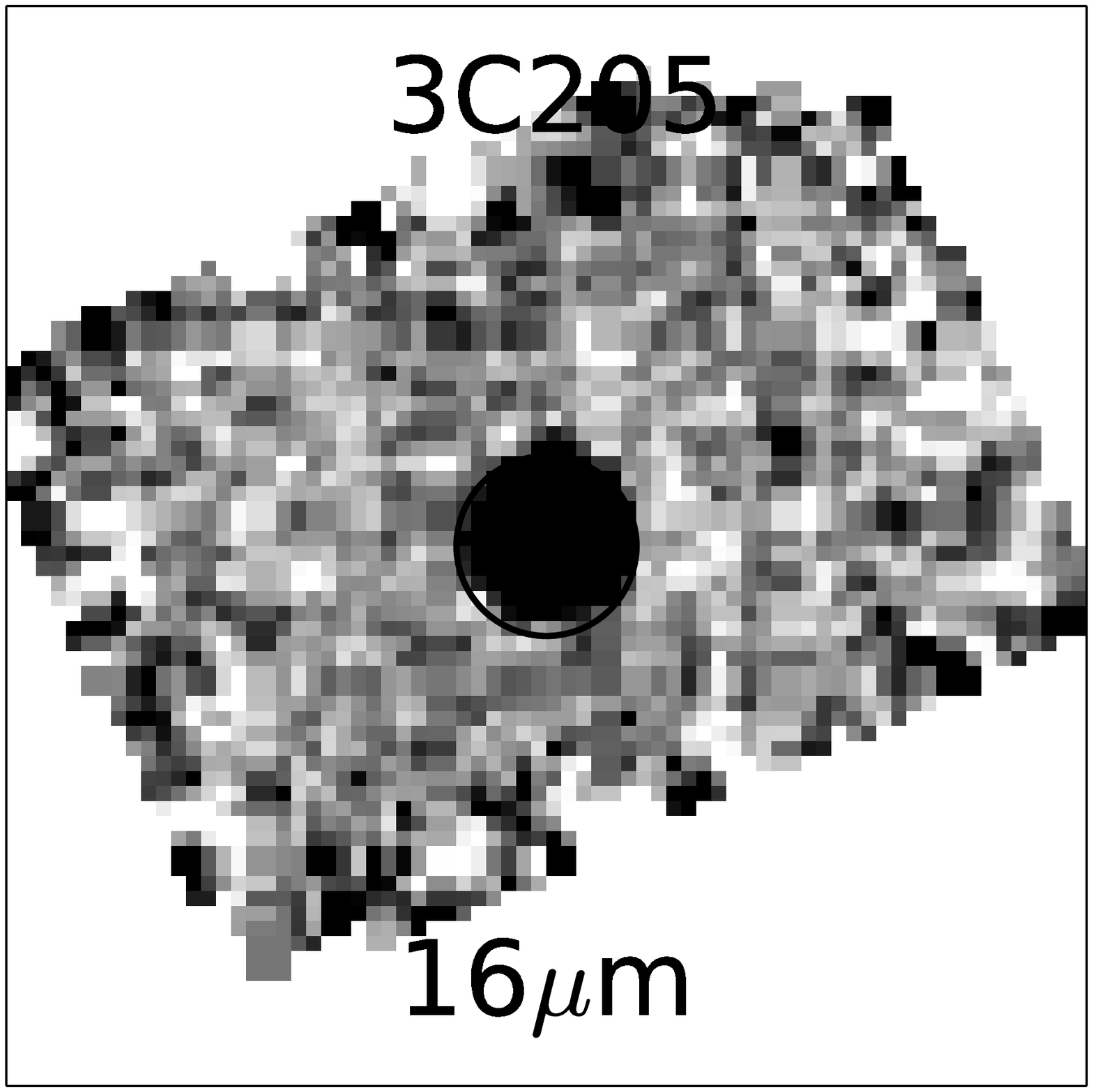}
      \includegraphics[width=1.5cm]{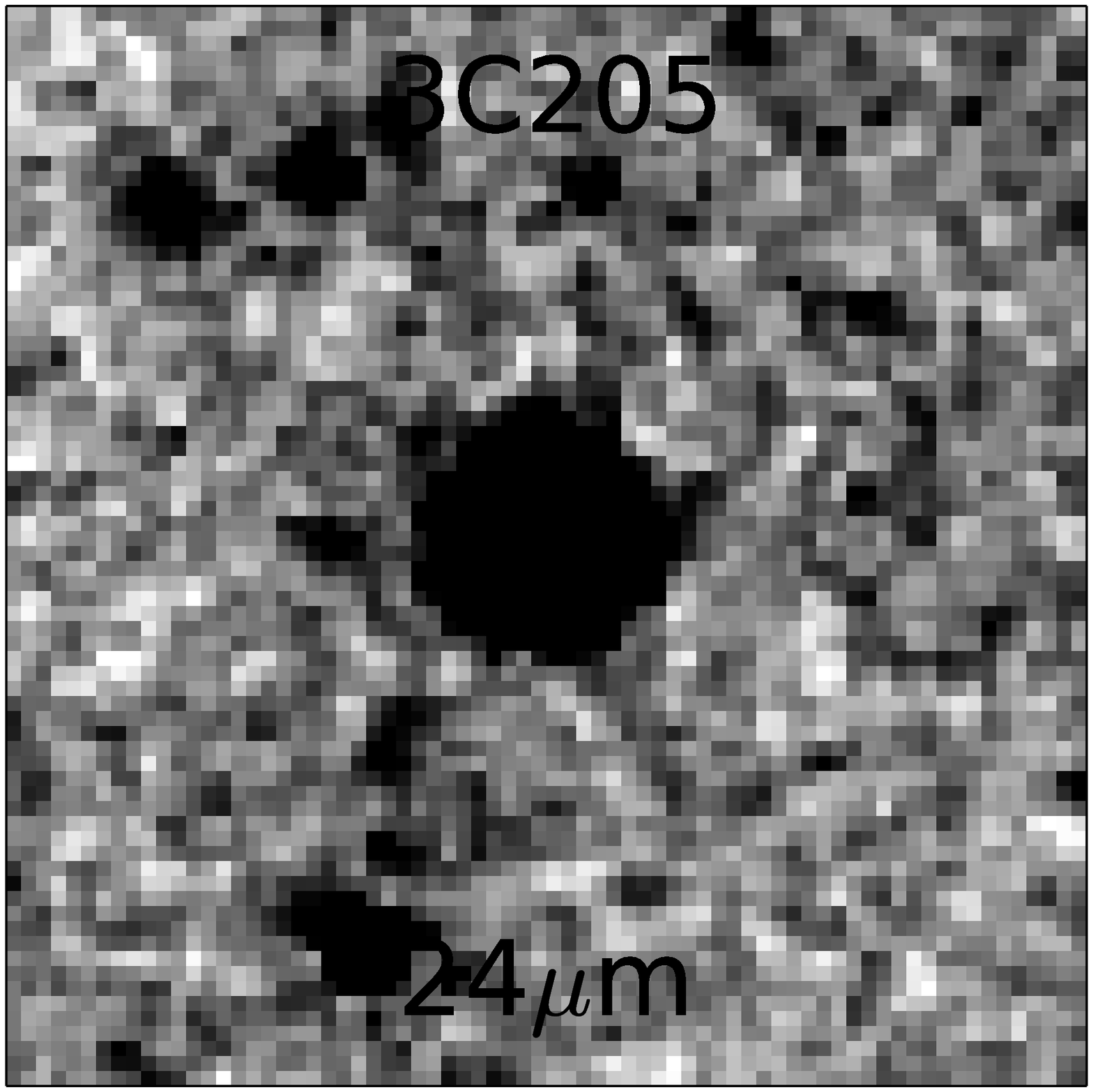}
      \includegraphics[width=1.5cm]{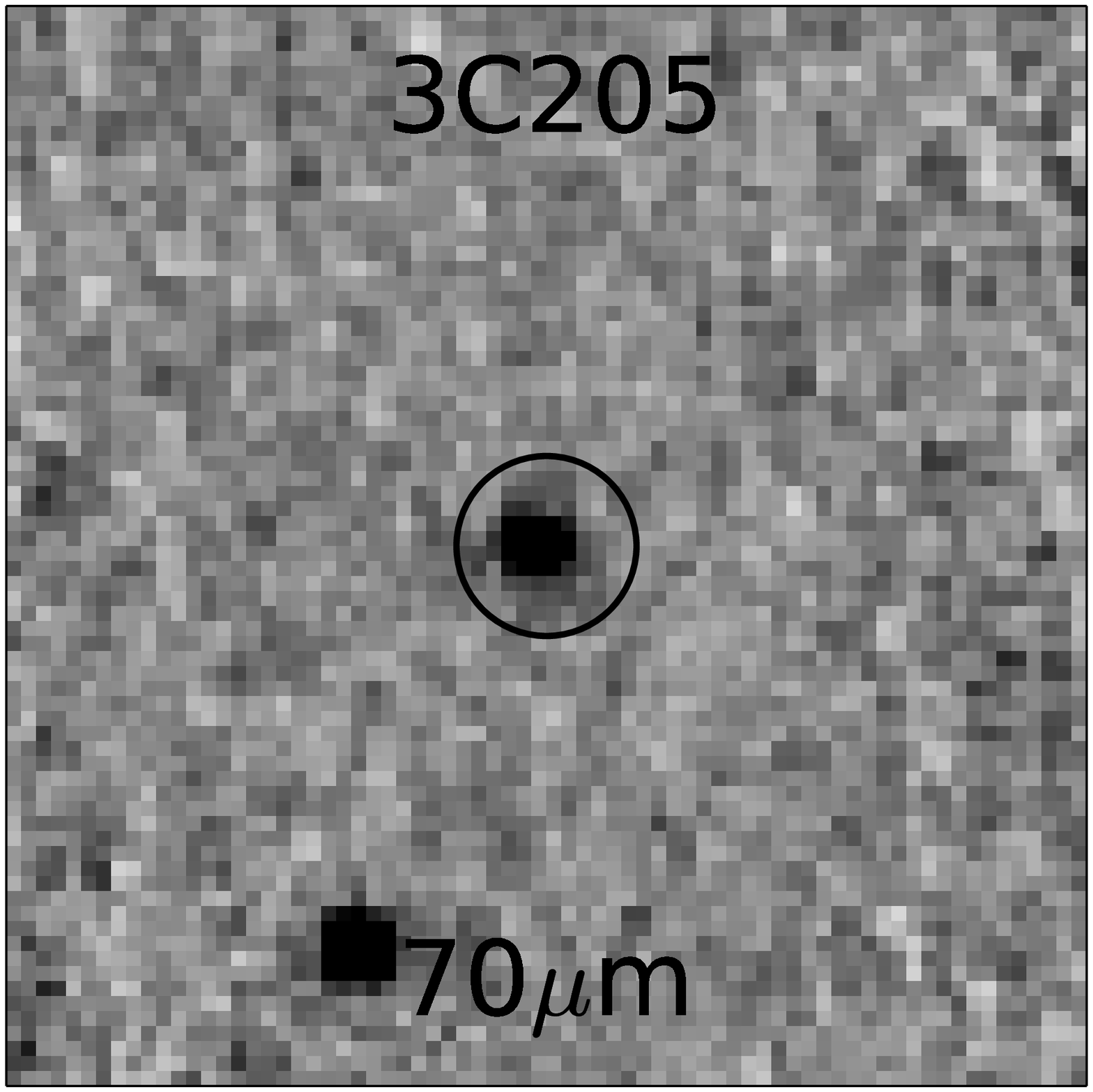}
      \includegraphics[width=1.5cm]{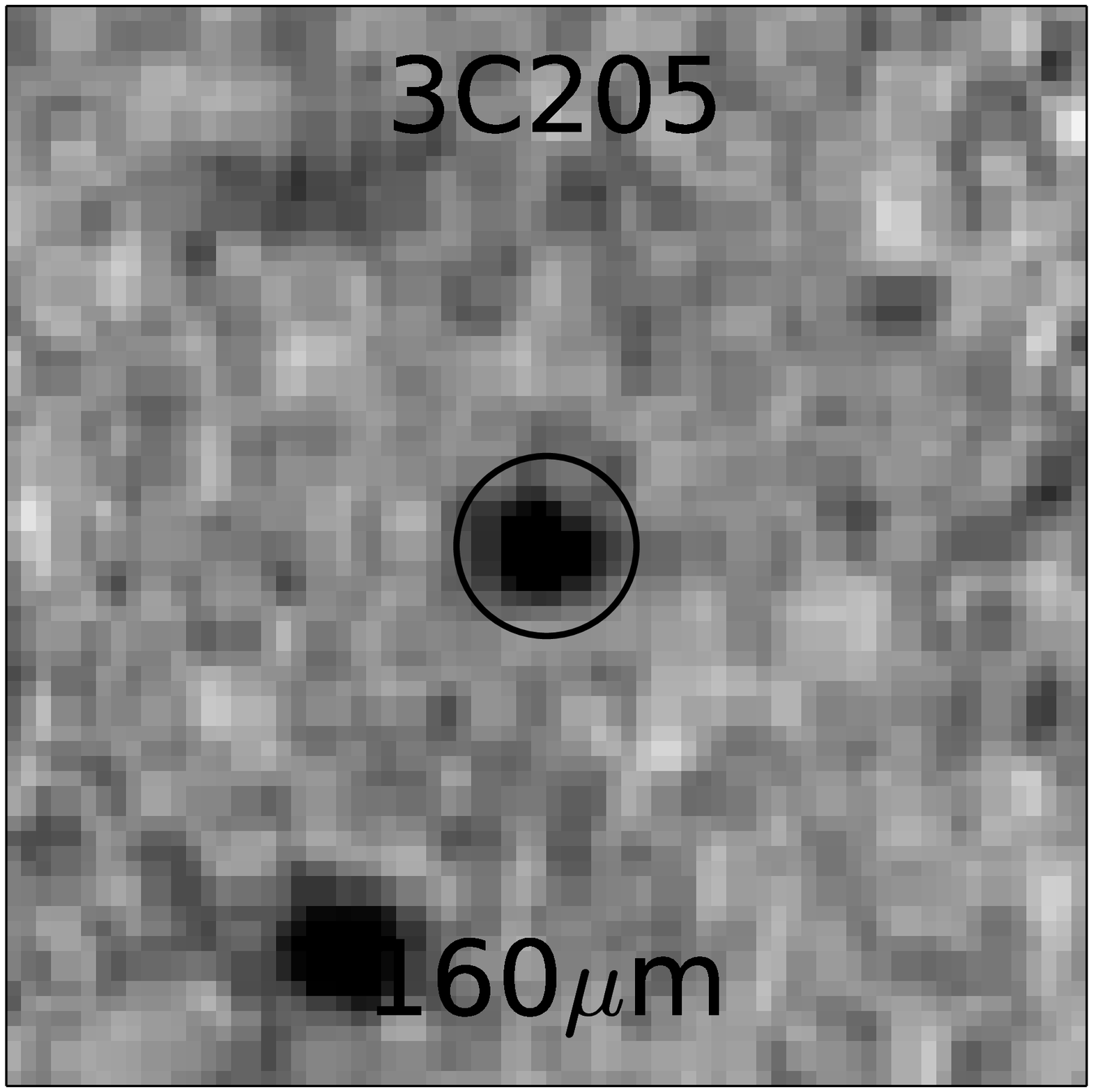}
      \includegraphics[width=1.5cm]{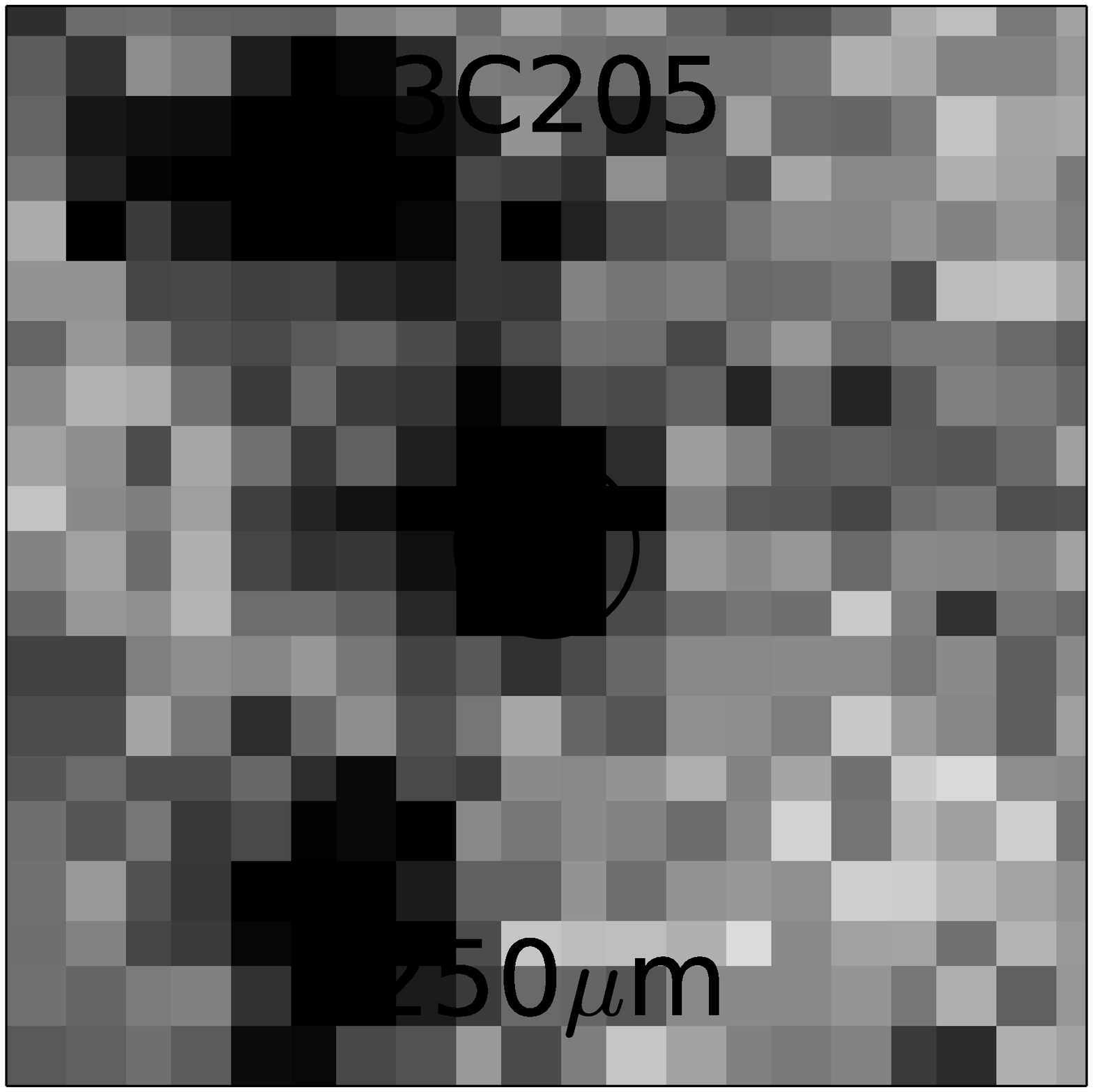}
      \includegraphics[width=1.5cm]{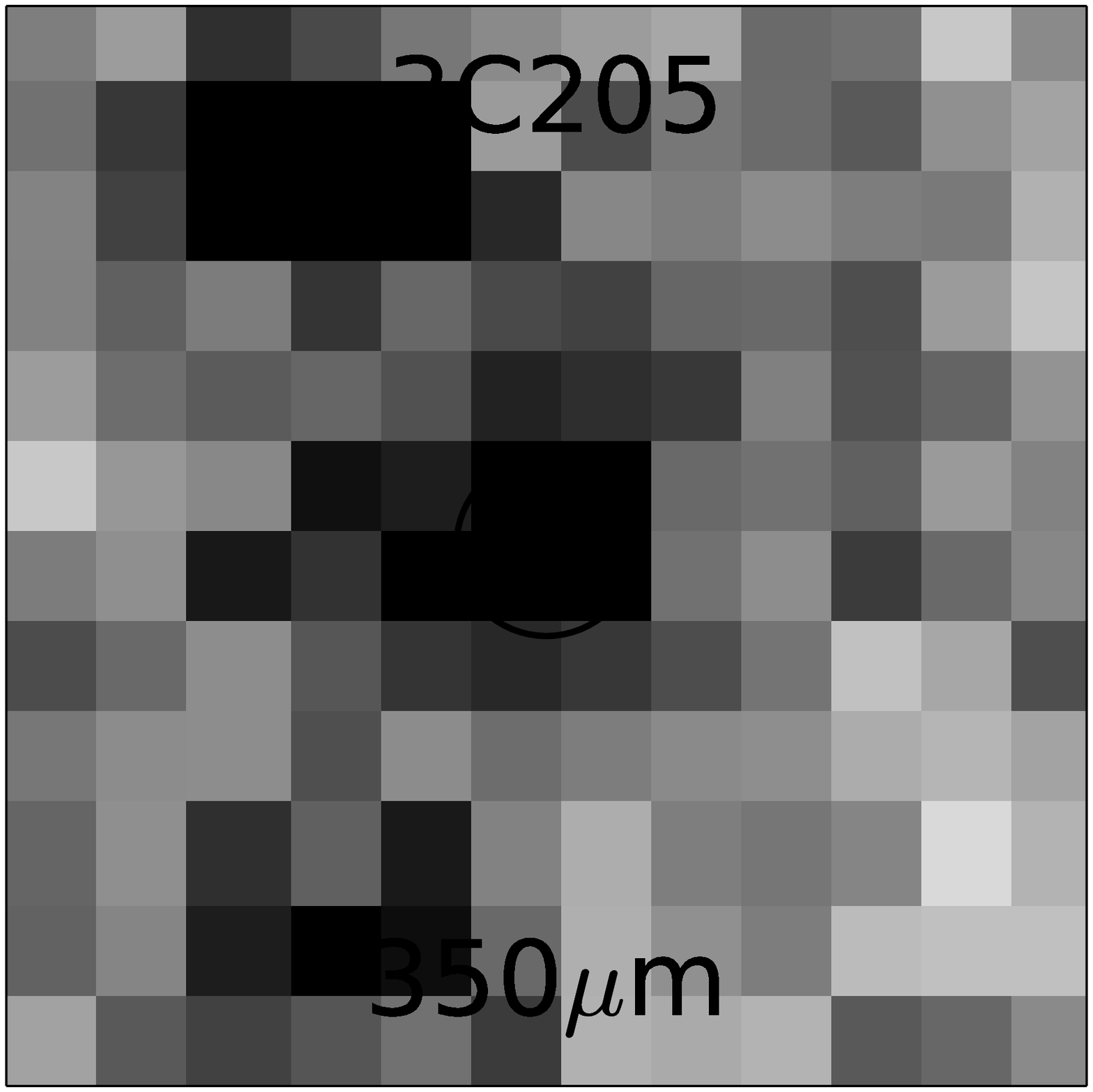}
      \includegraphics[width=1.5cm]{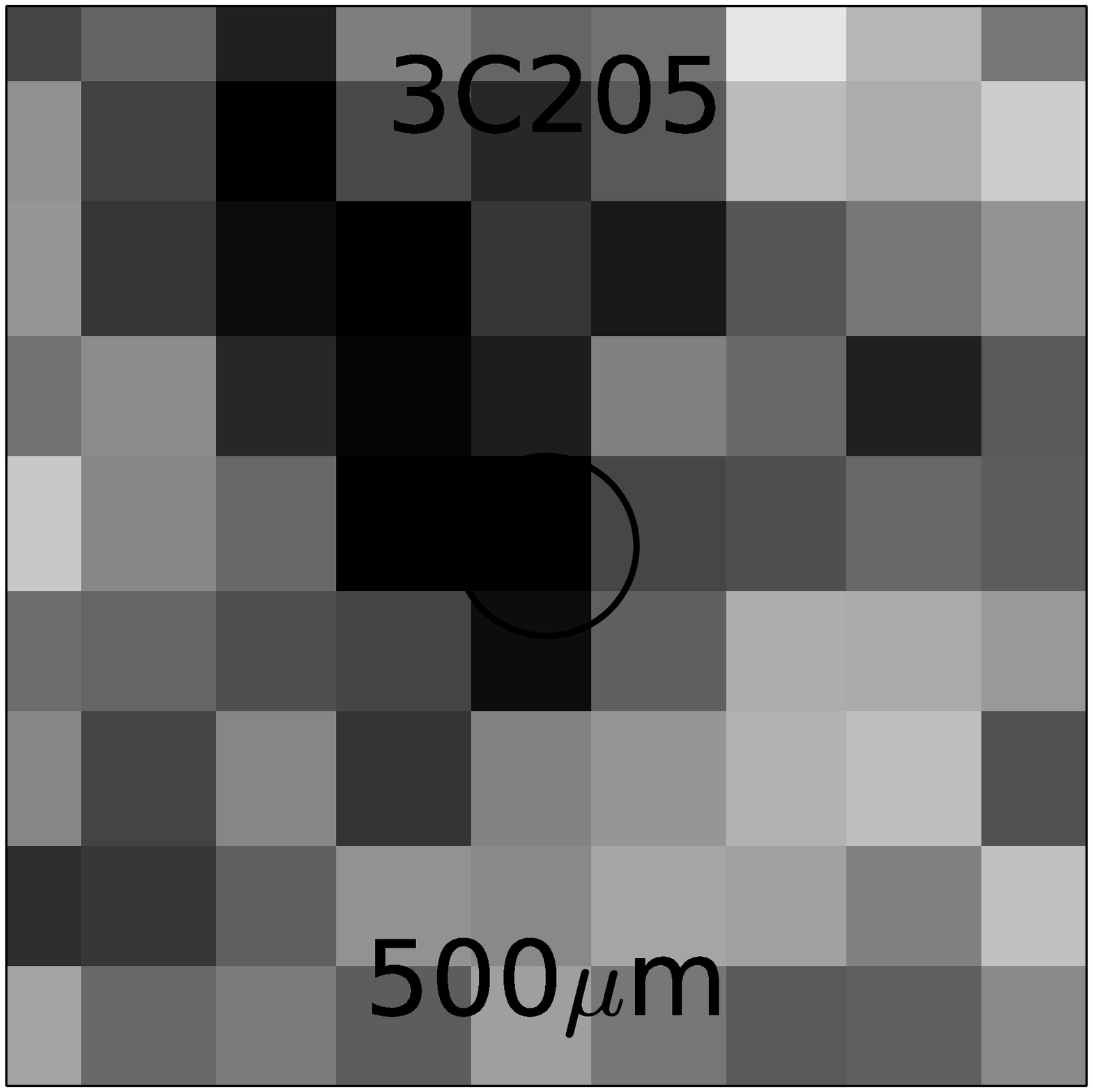}
      \\
      \includegraphics[width=1.5cm]{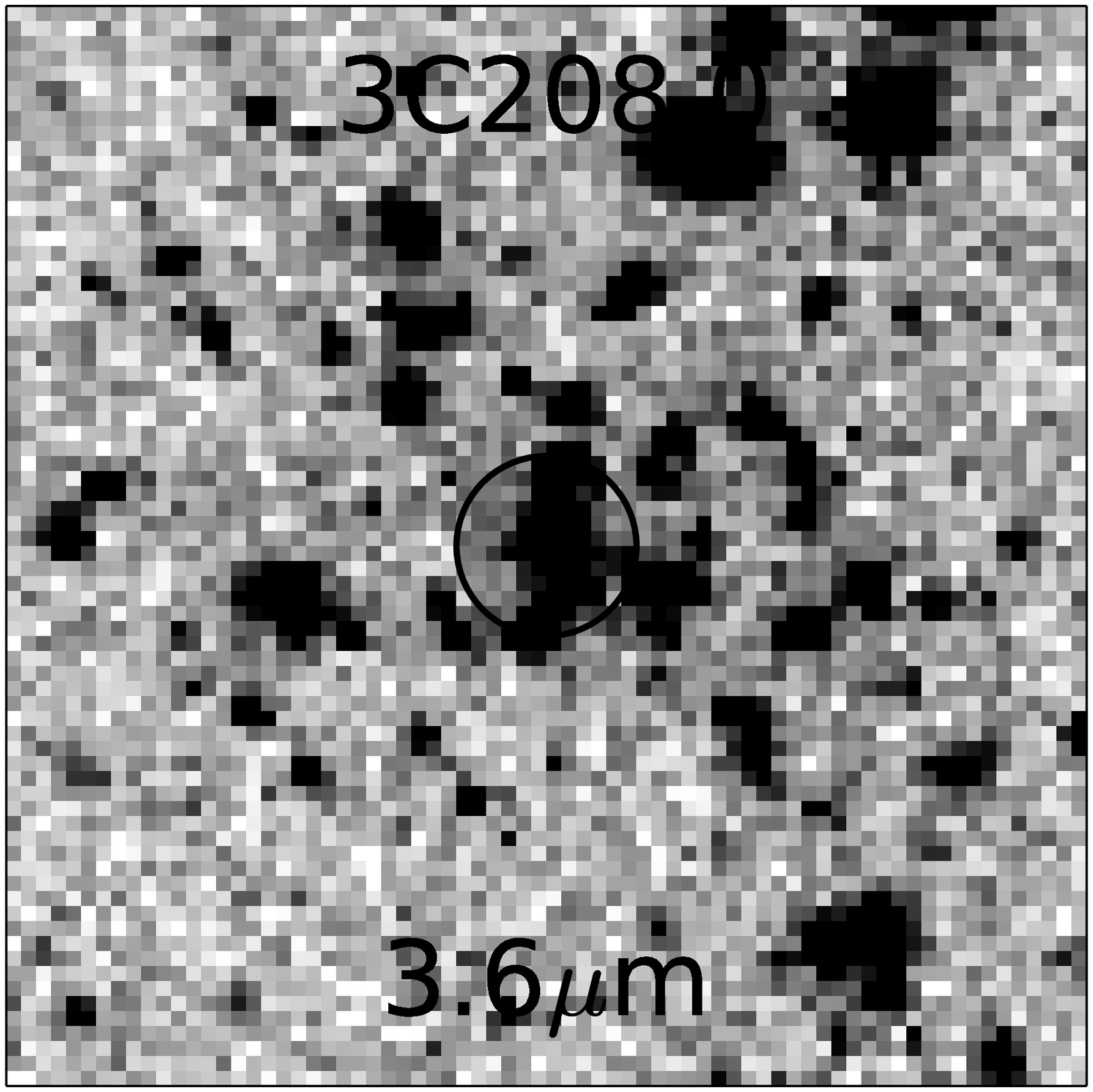}
      \includegraphics[width=1.5cm]{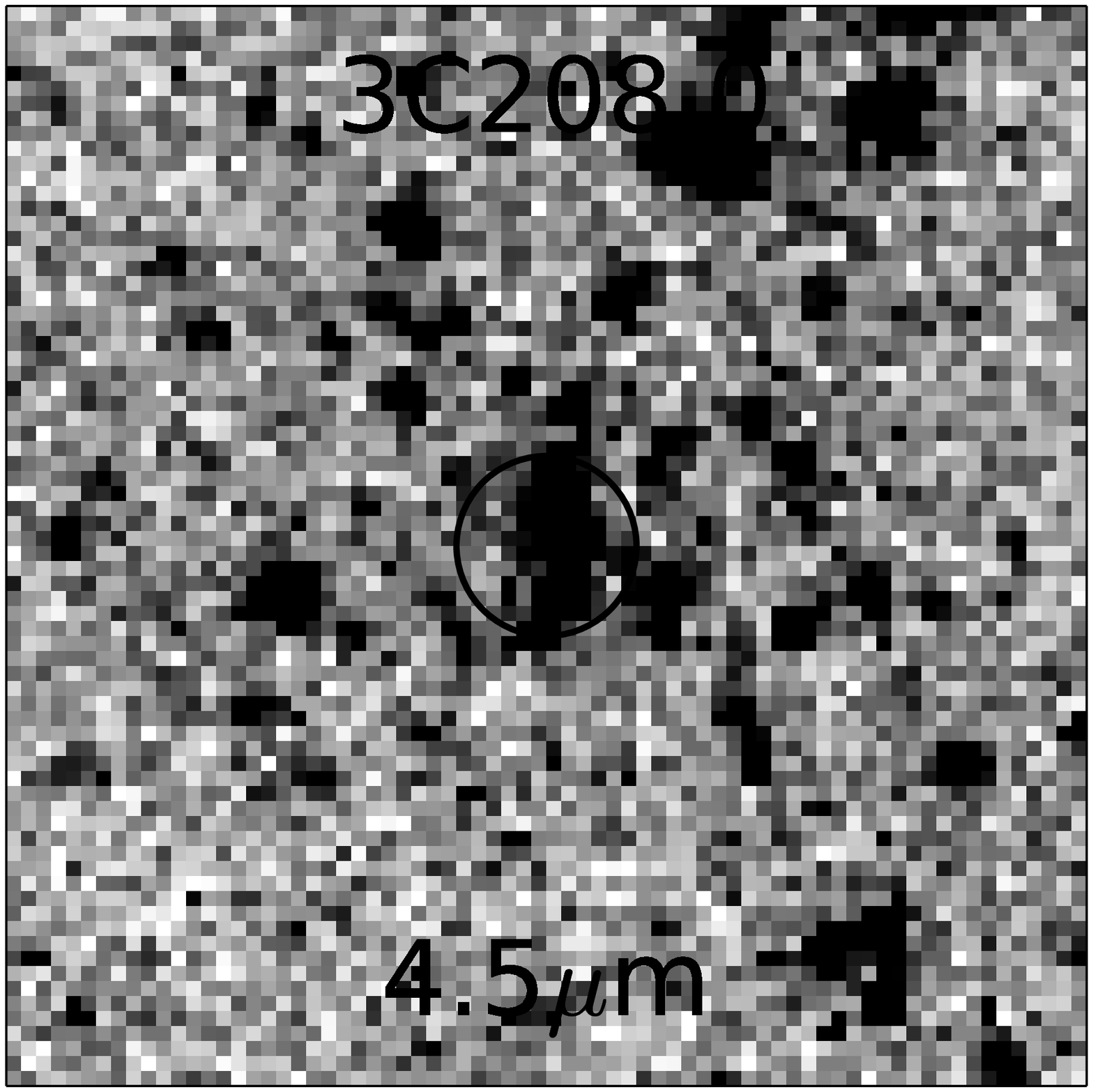}
      \includegraphics[width=1.5cm]{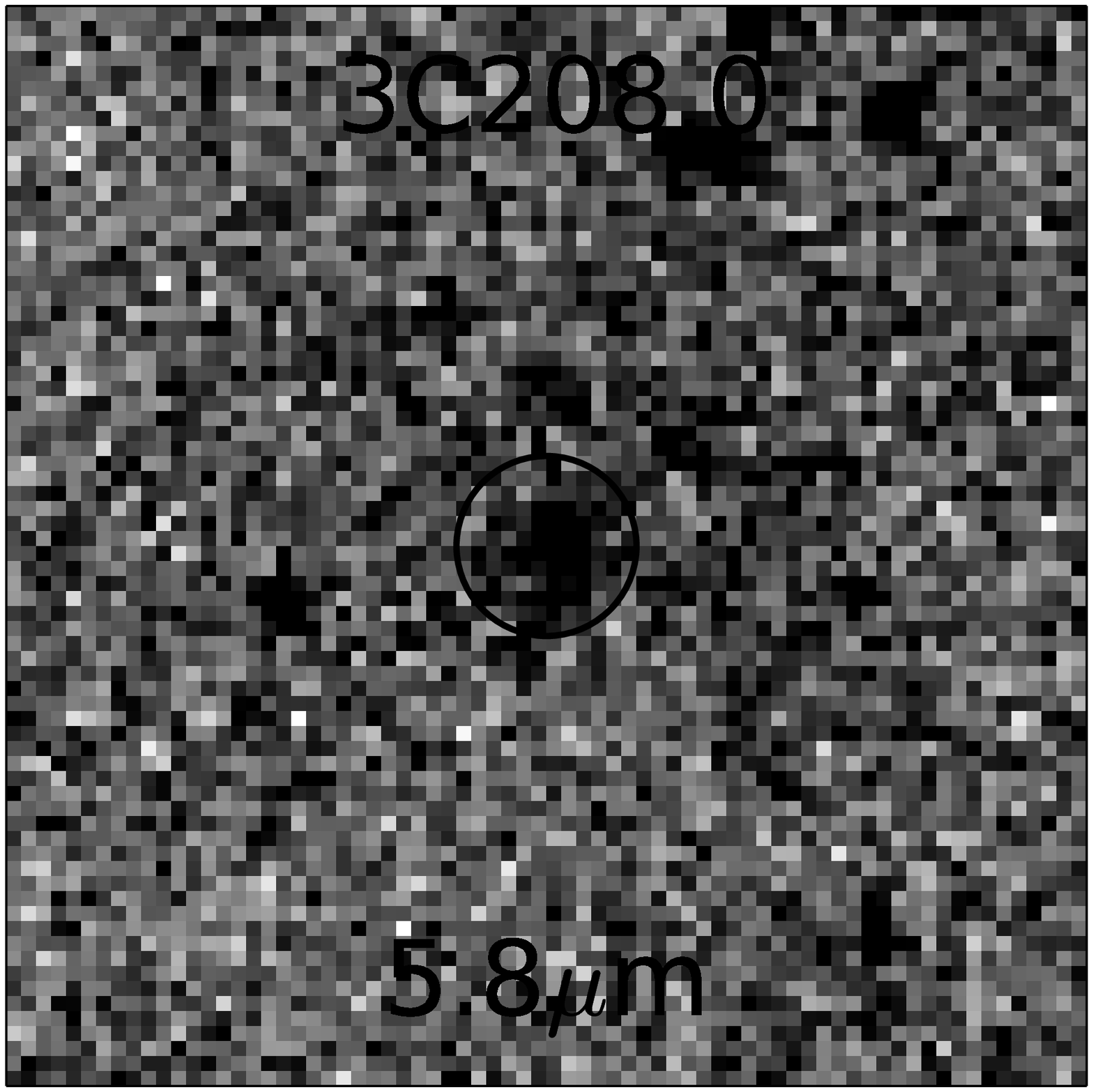}
      \includegraphics[width=1.5cm]{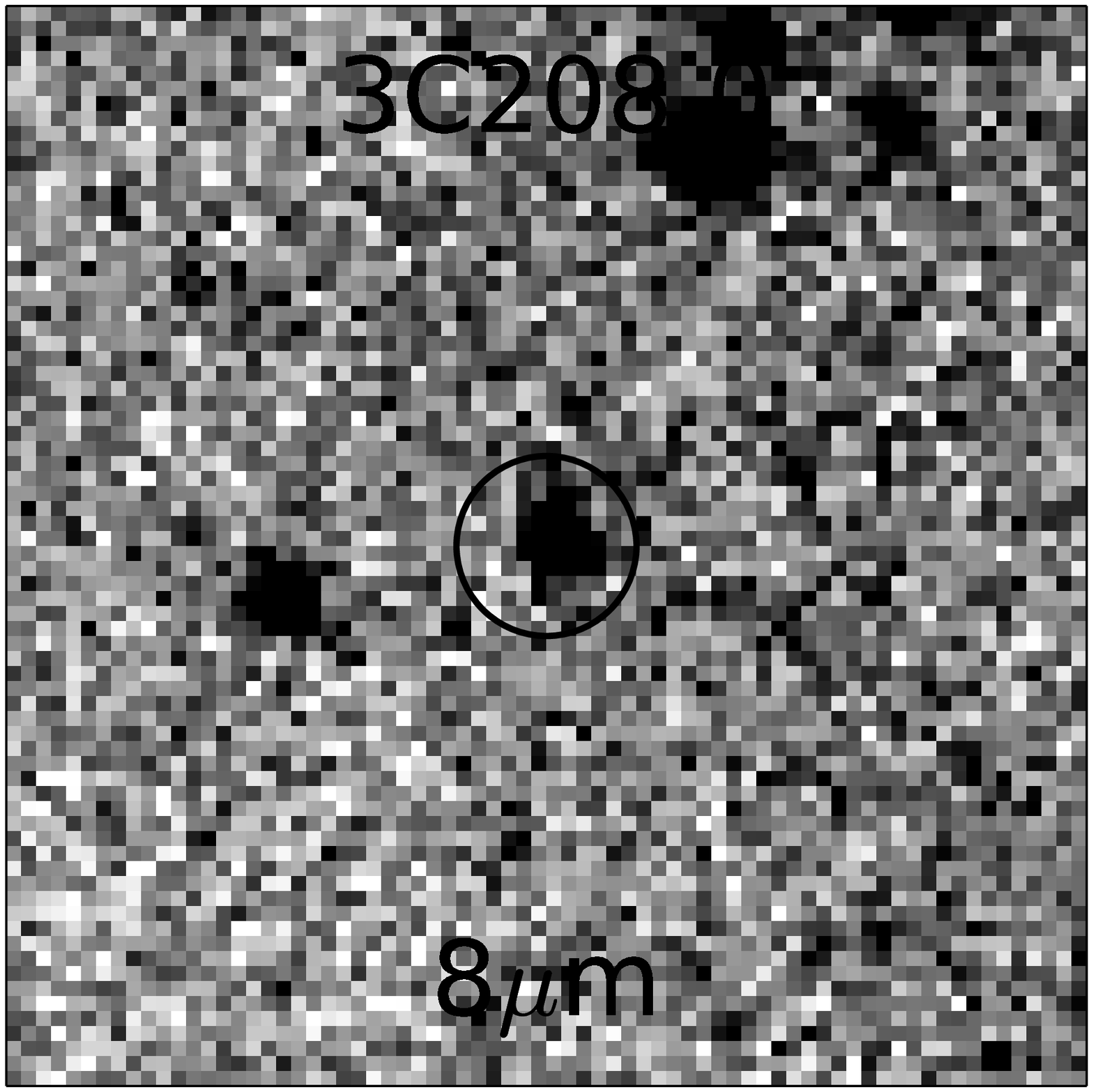}
      \includegraphics[width=1.5cm]{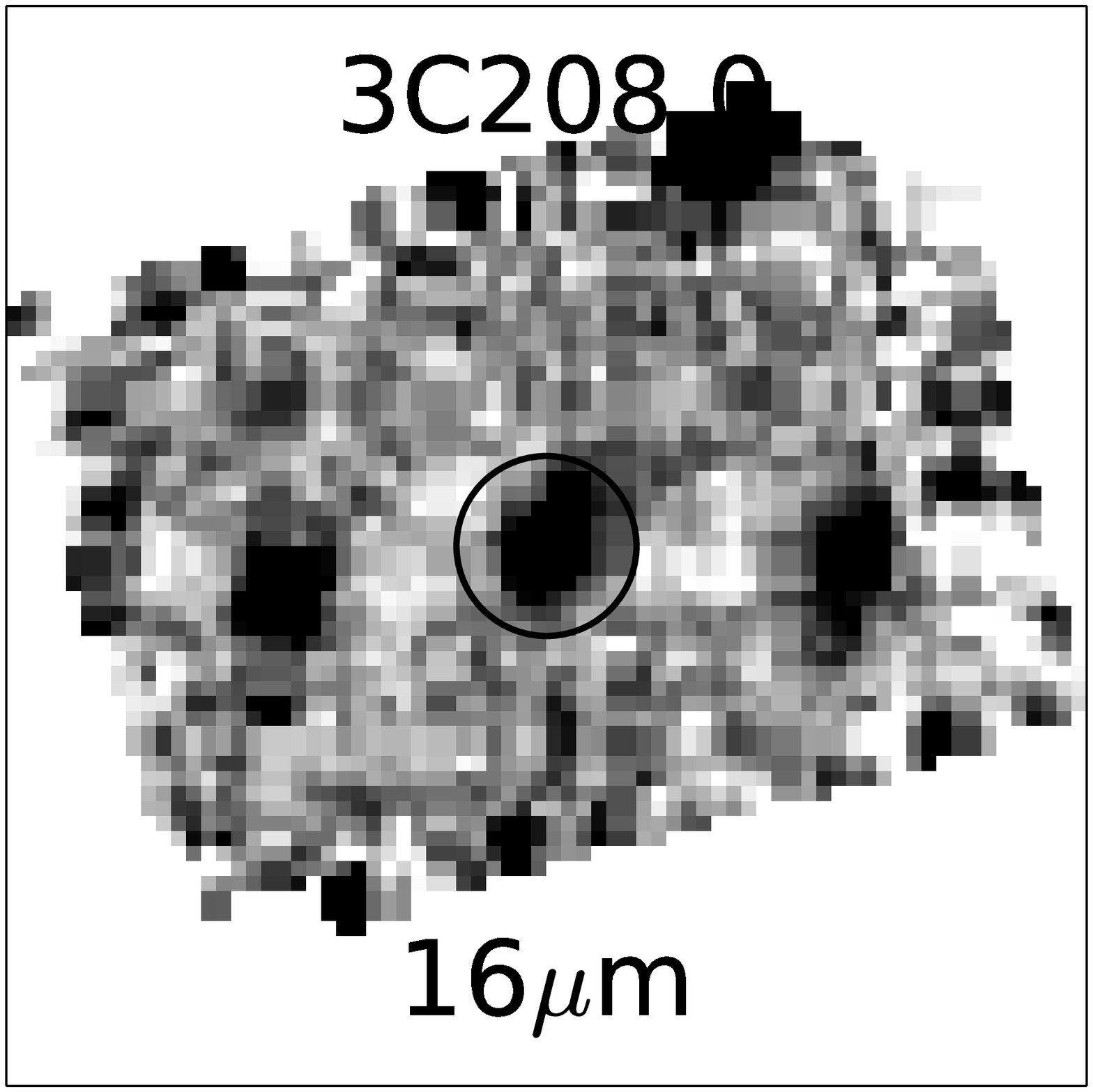}
      \includegraphics[width=1.5cm]{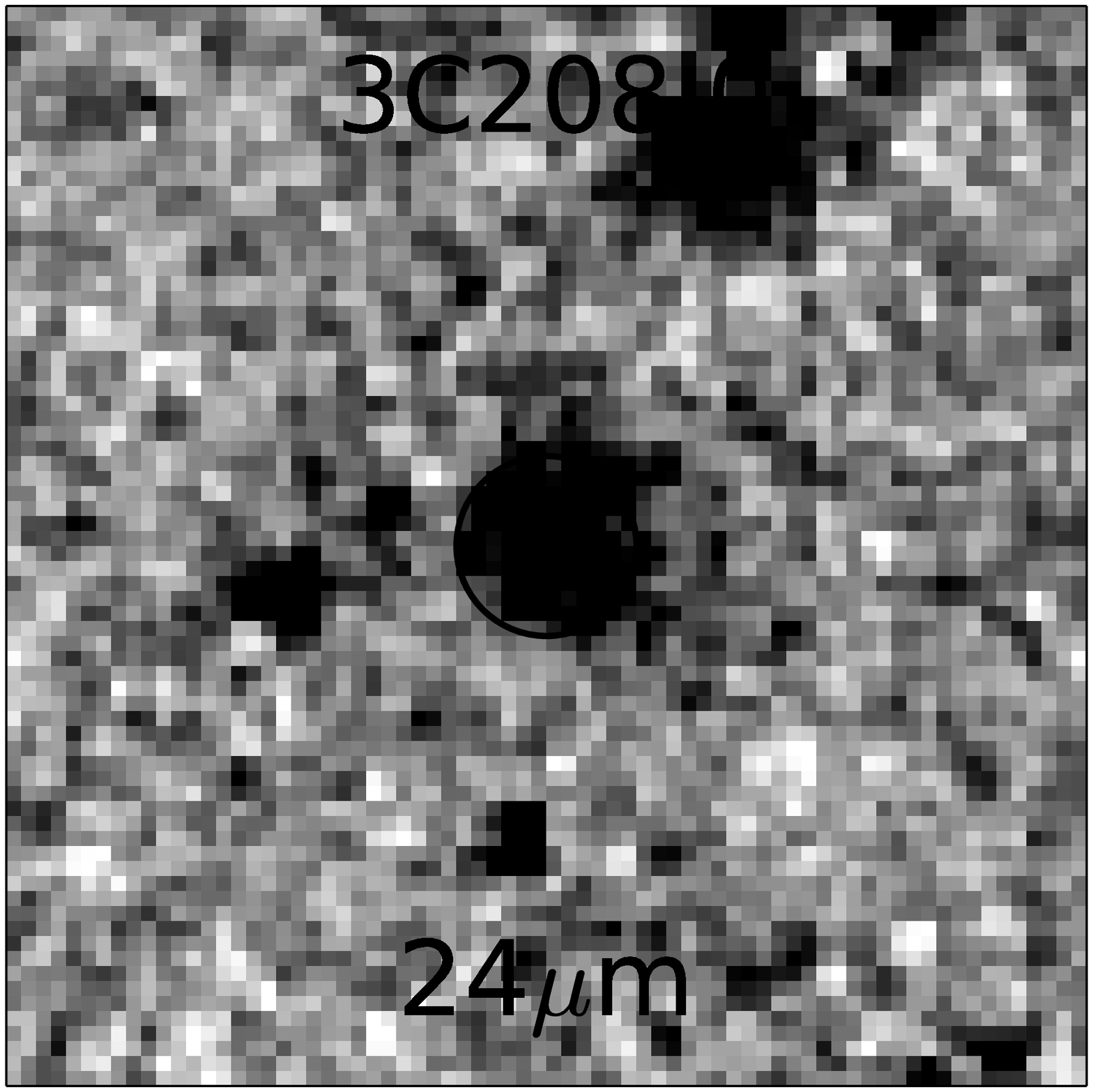}
      \includegraphics[width=1.5cm]{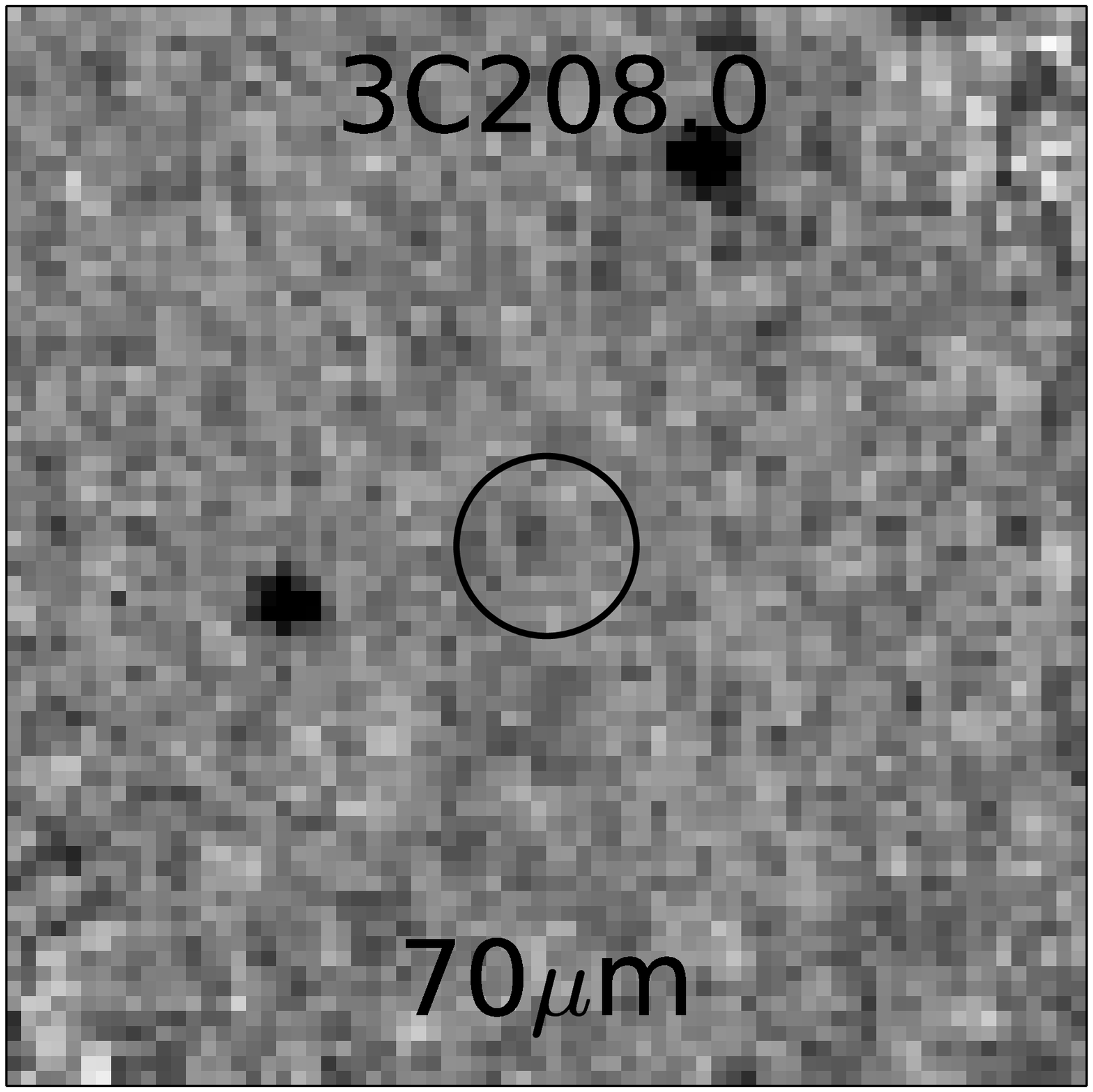}
      \includegraphics[width=1.5cm]{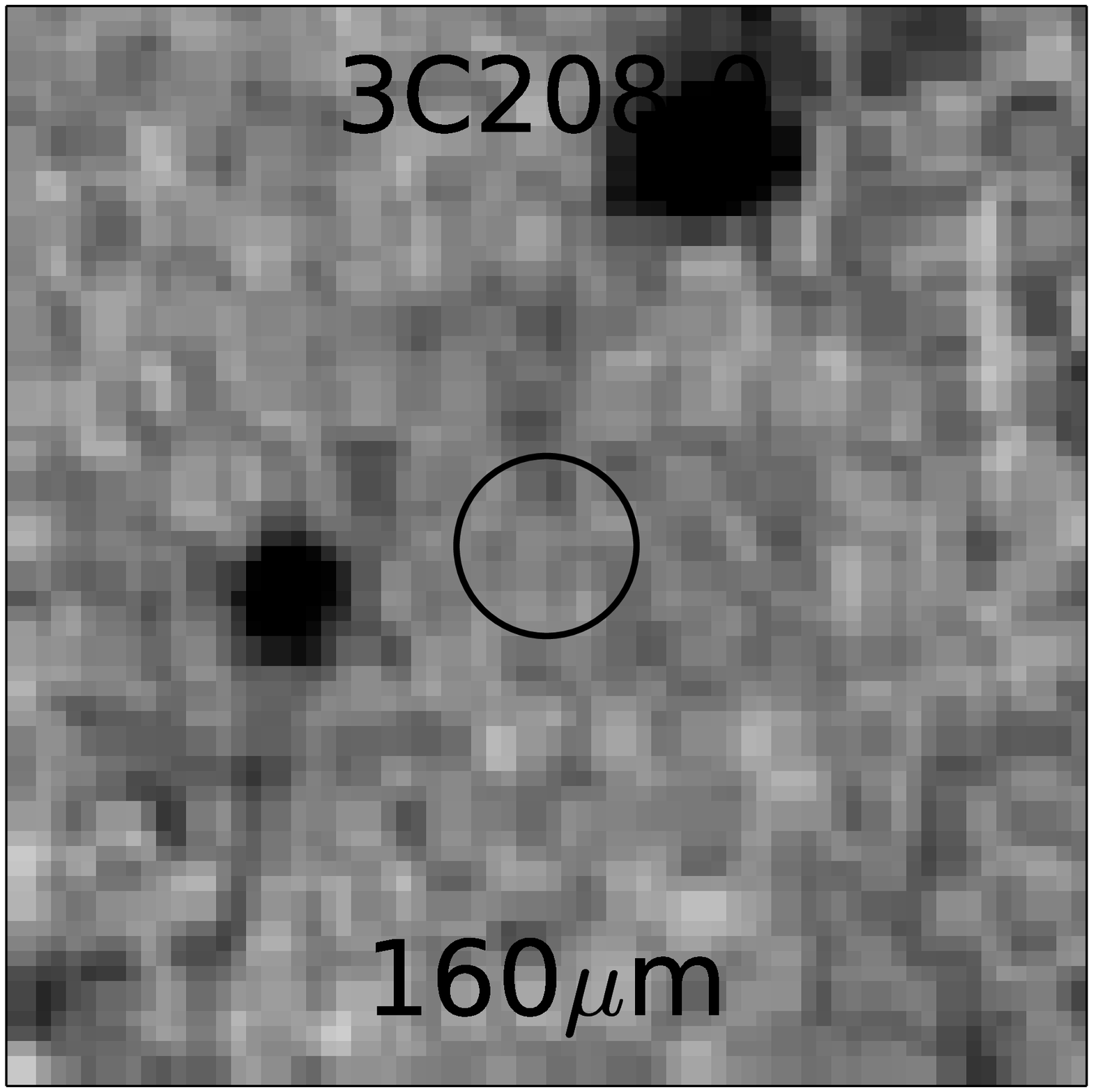}
      \includegraphics[width=1.5cm]{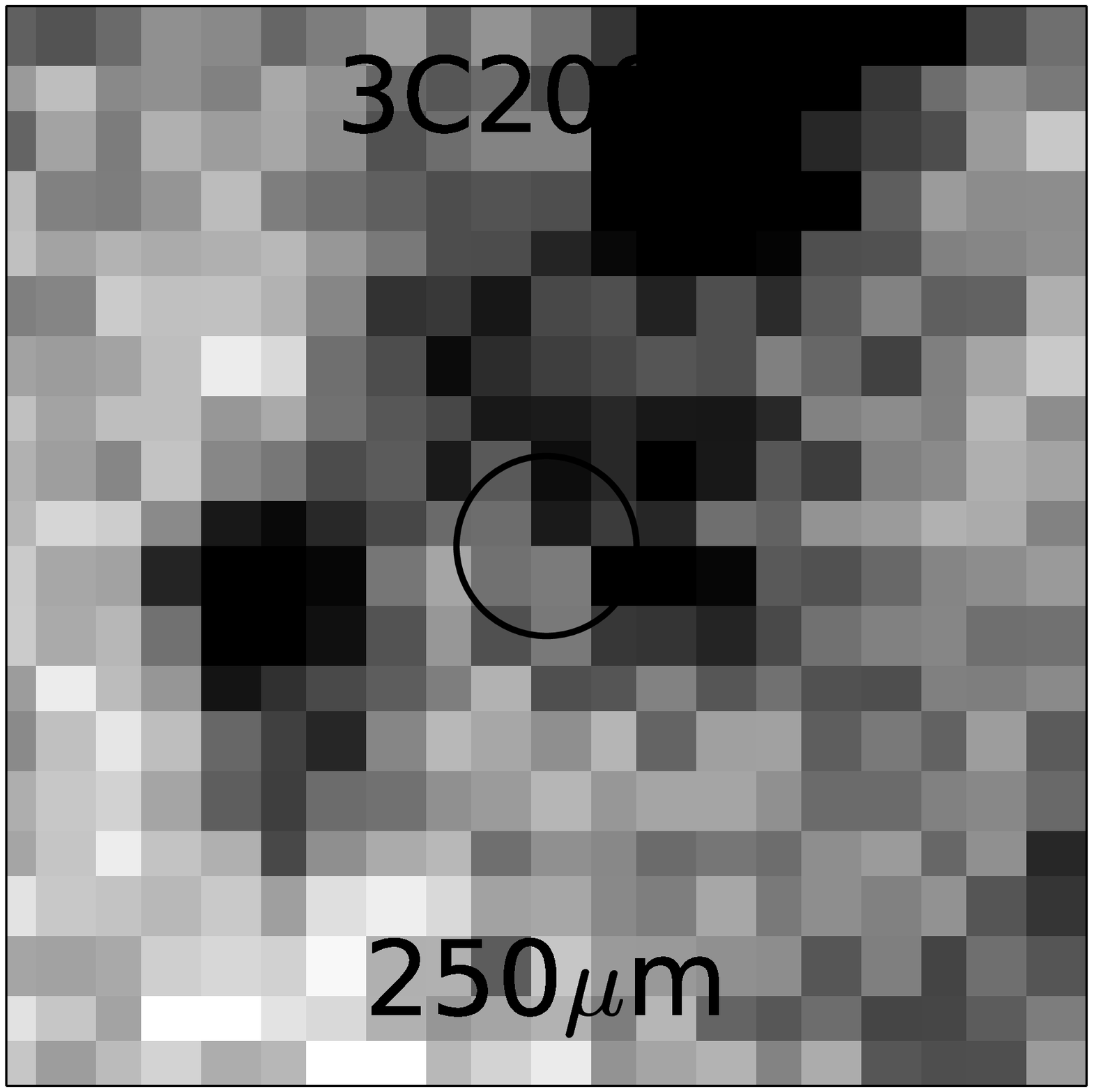}
      \includegraphics[width=1.5cm]{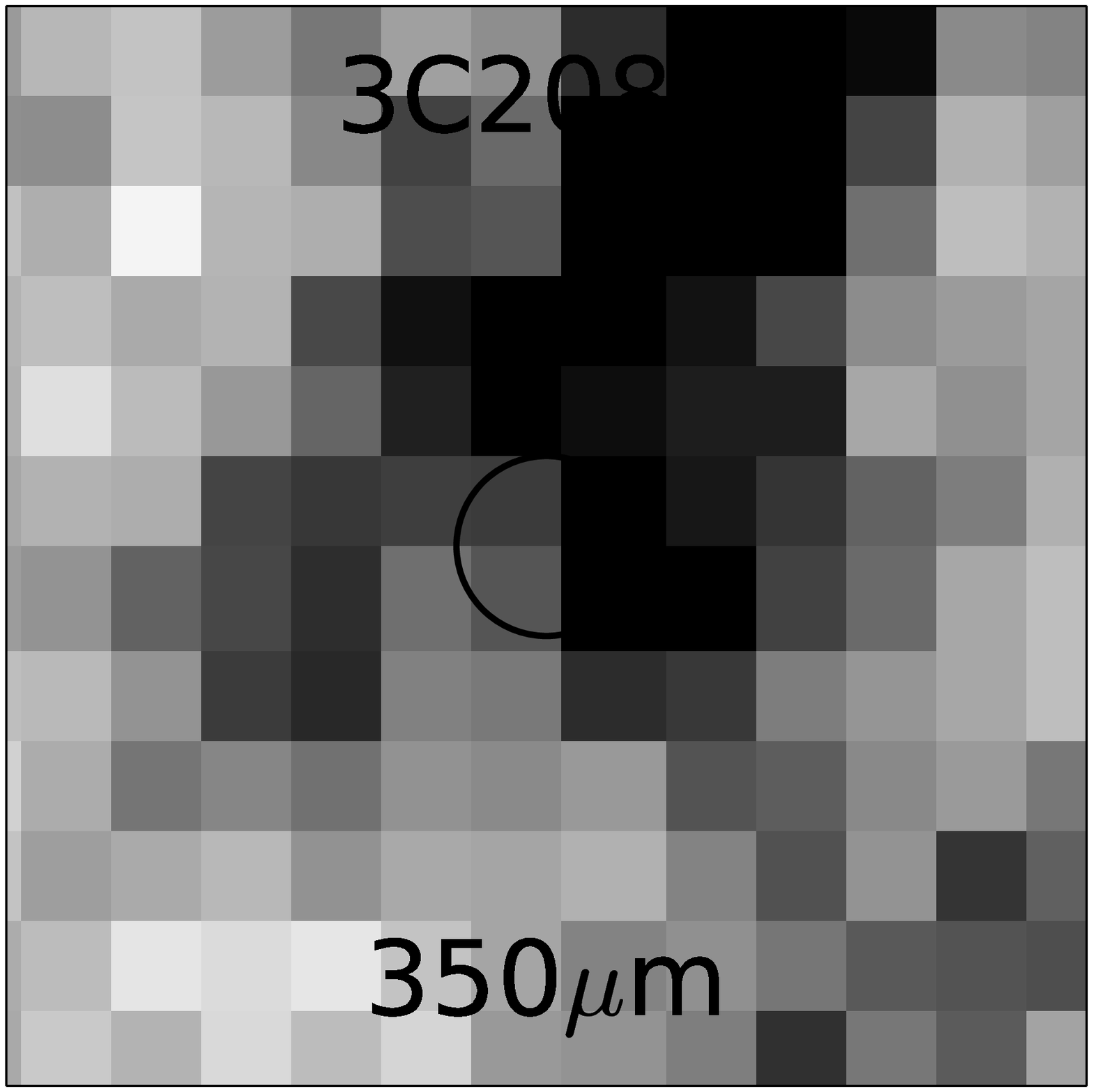}
      \includegraphics[width=1.5cm]{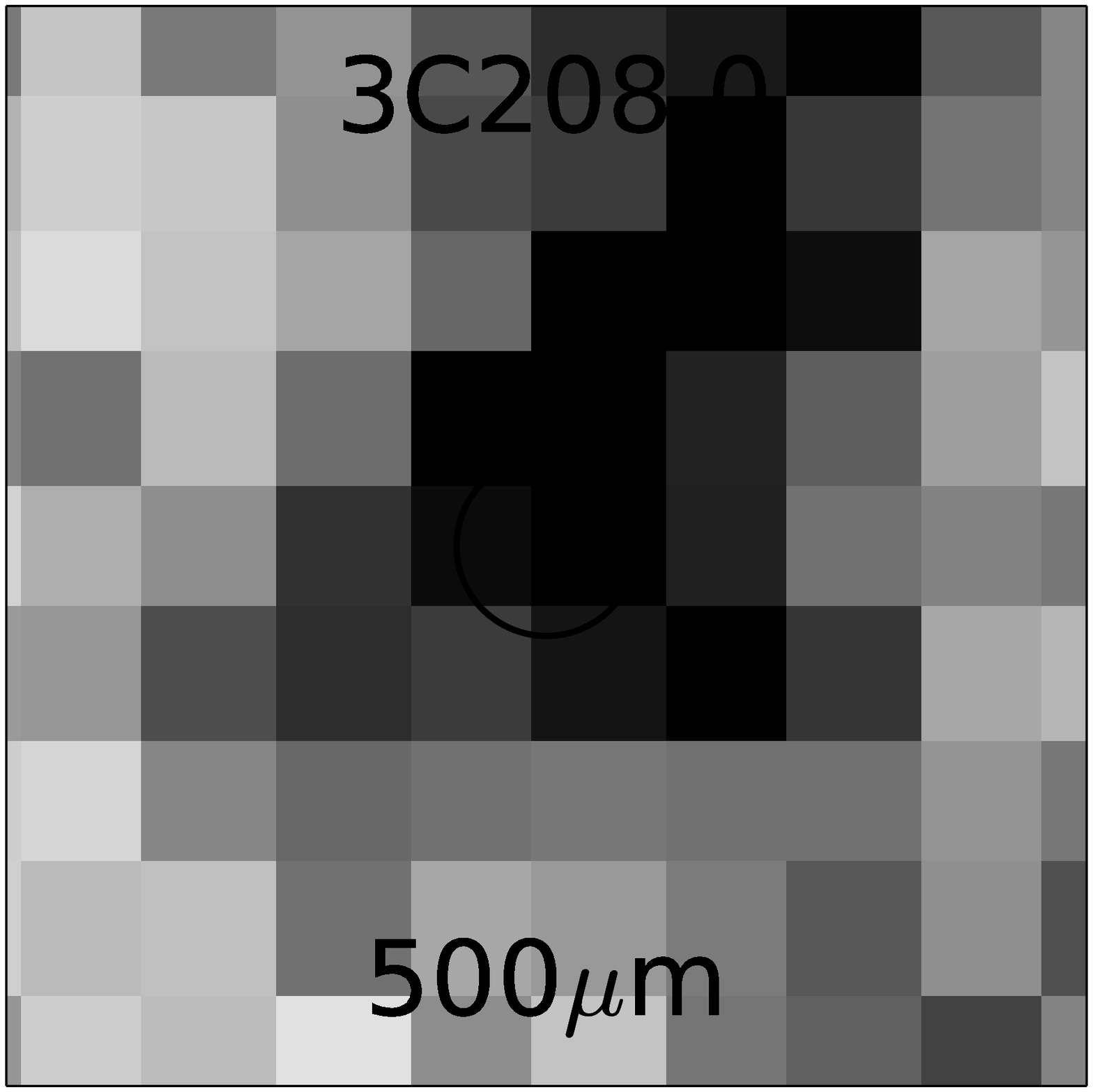}
      \\
      \includegraphics[width=1.5cm]{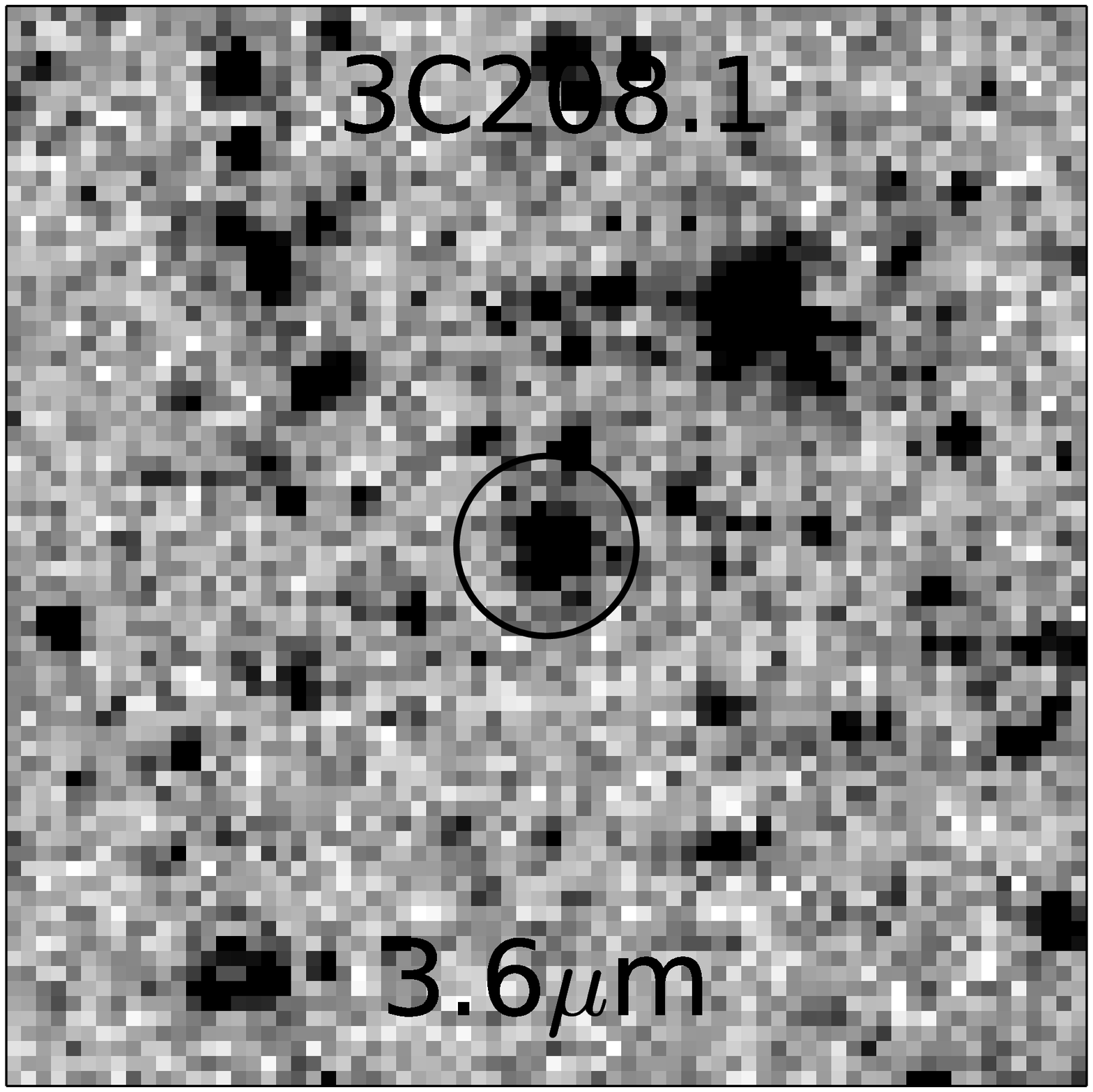}
      \includegraphics[width=1.5cm]{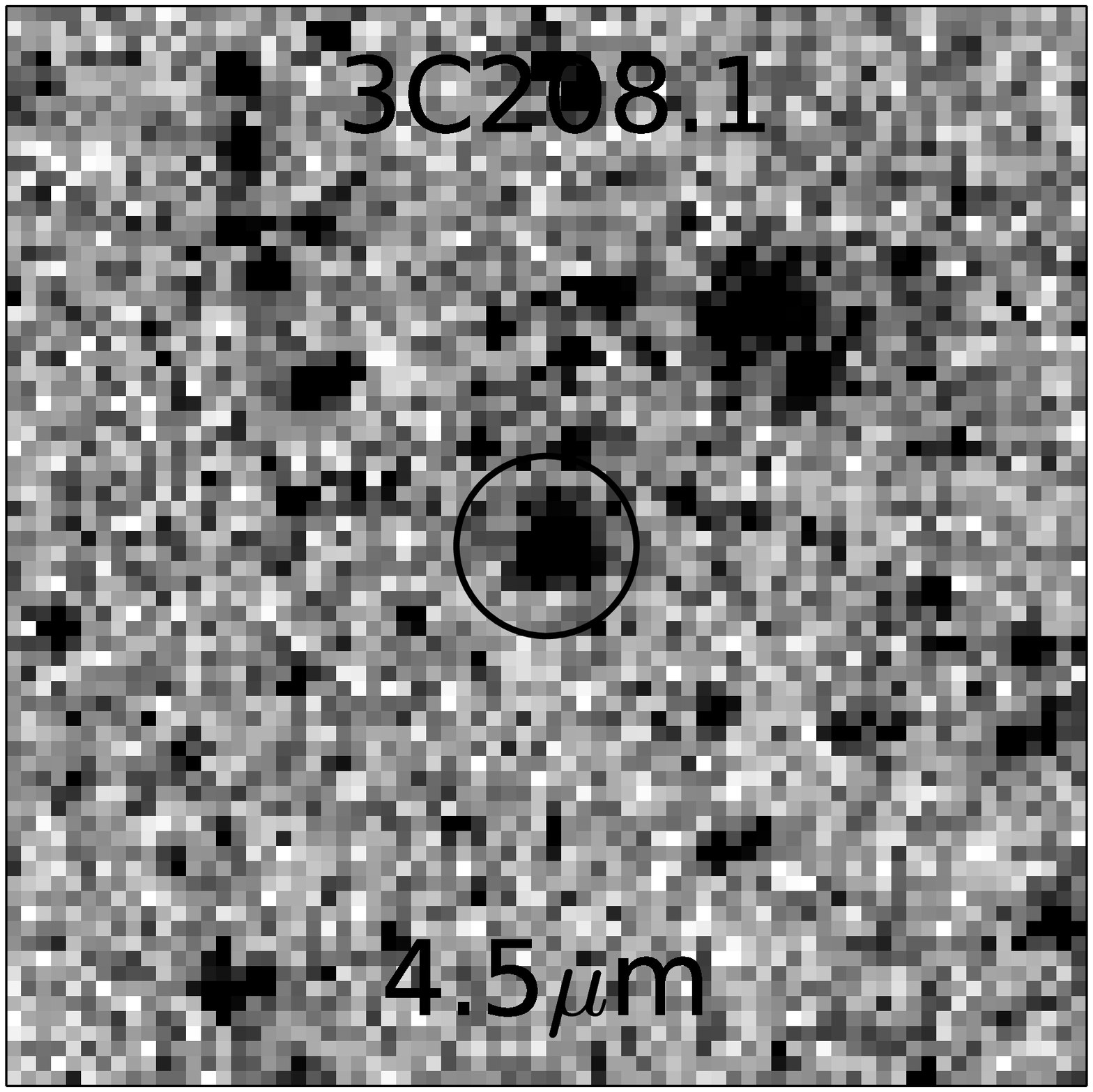}
      \includegraphics[width=1.5cm]{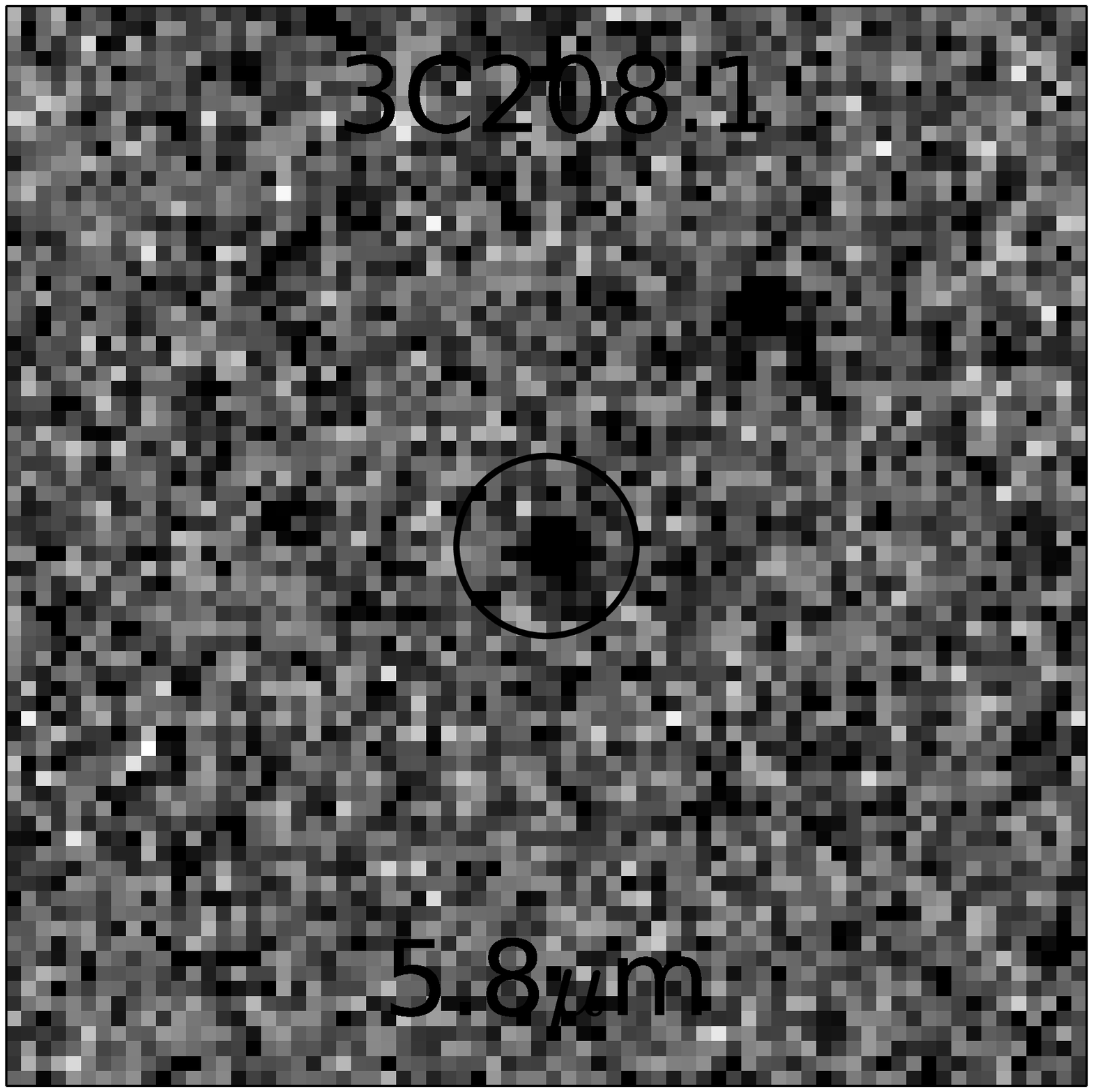}
      \includegraphics[width=1.5cm]{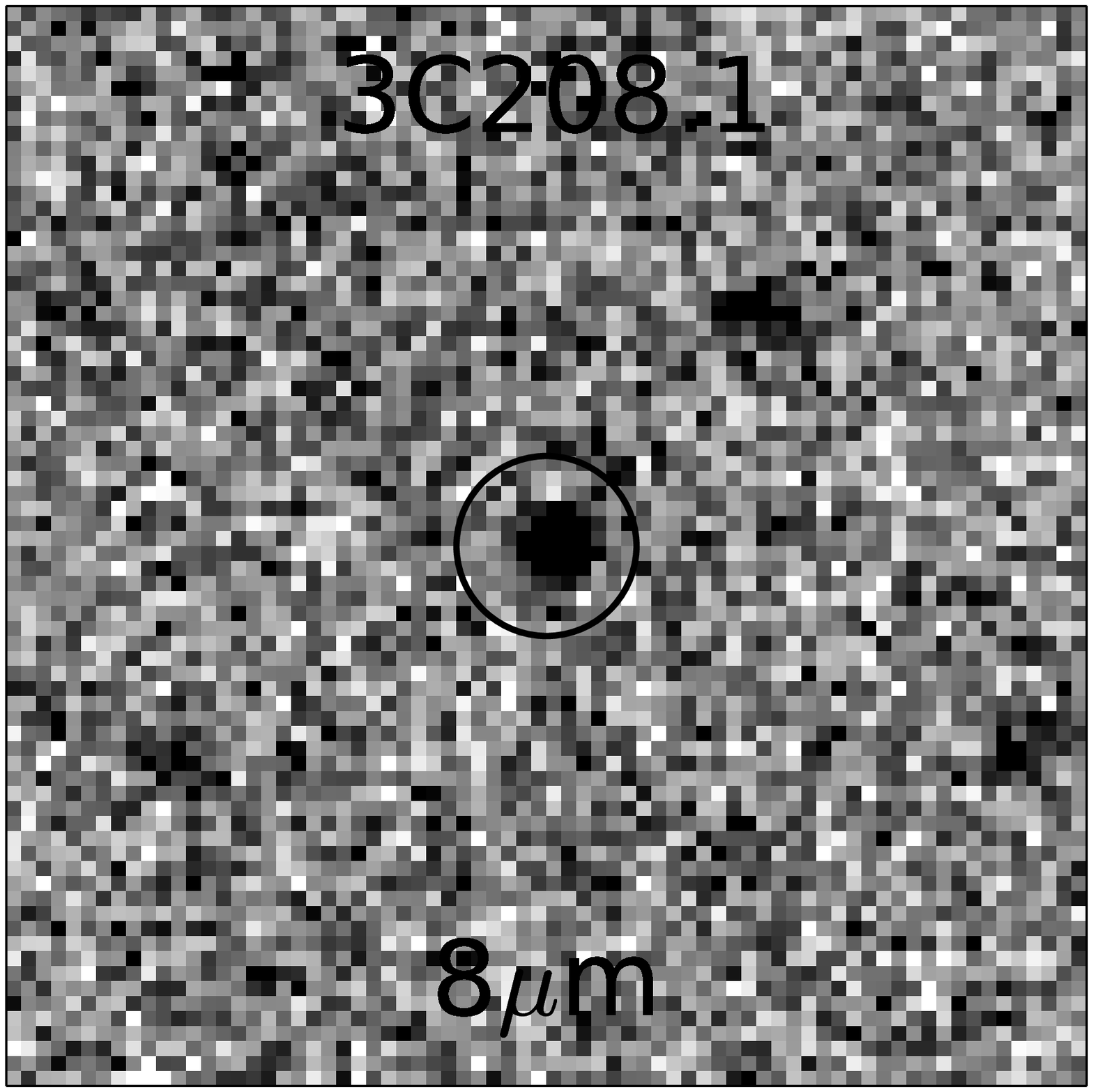}
      \includegraphics[width=1.5cm]{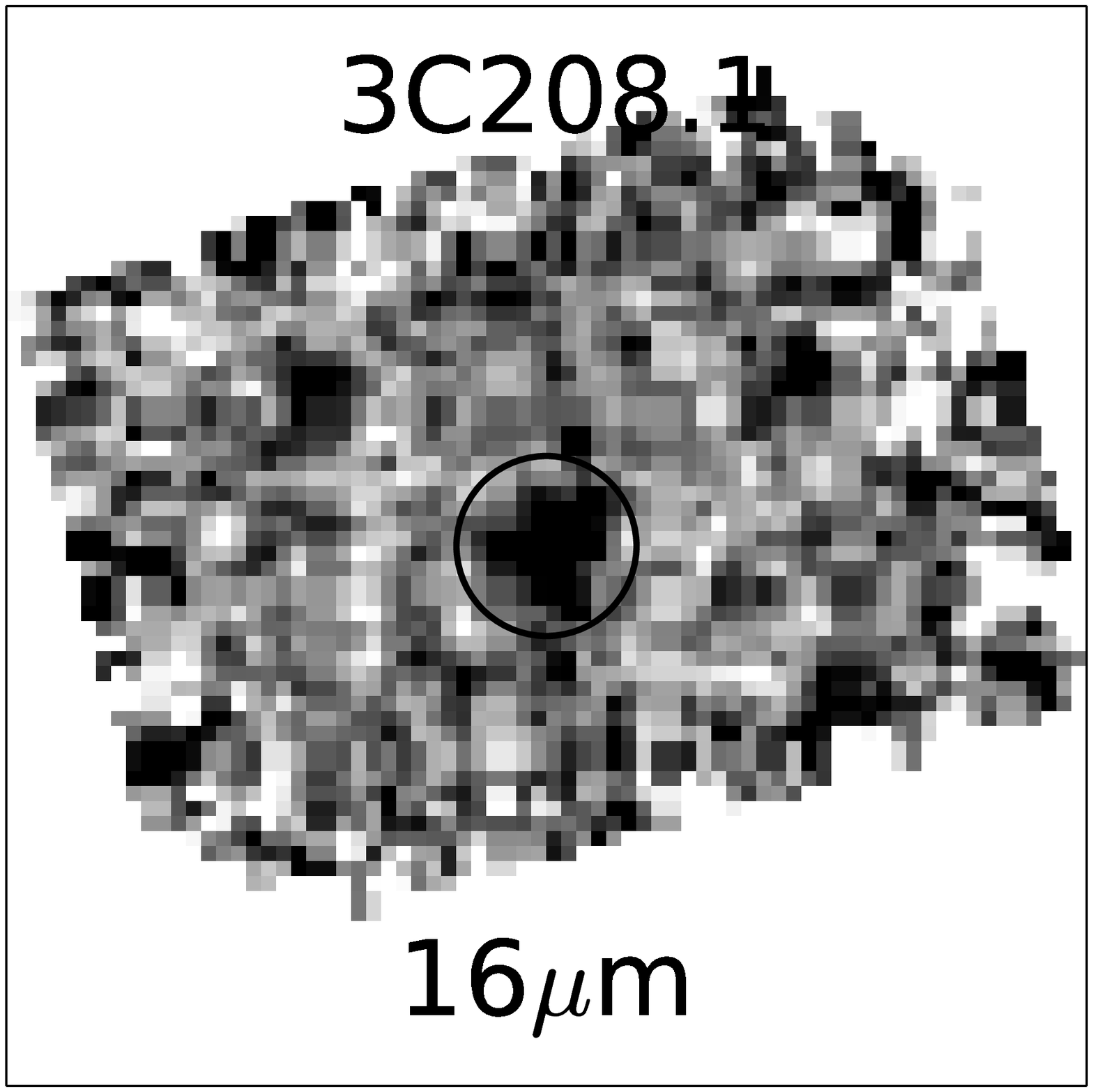}
      \includegraphics[width=1.5cm]{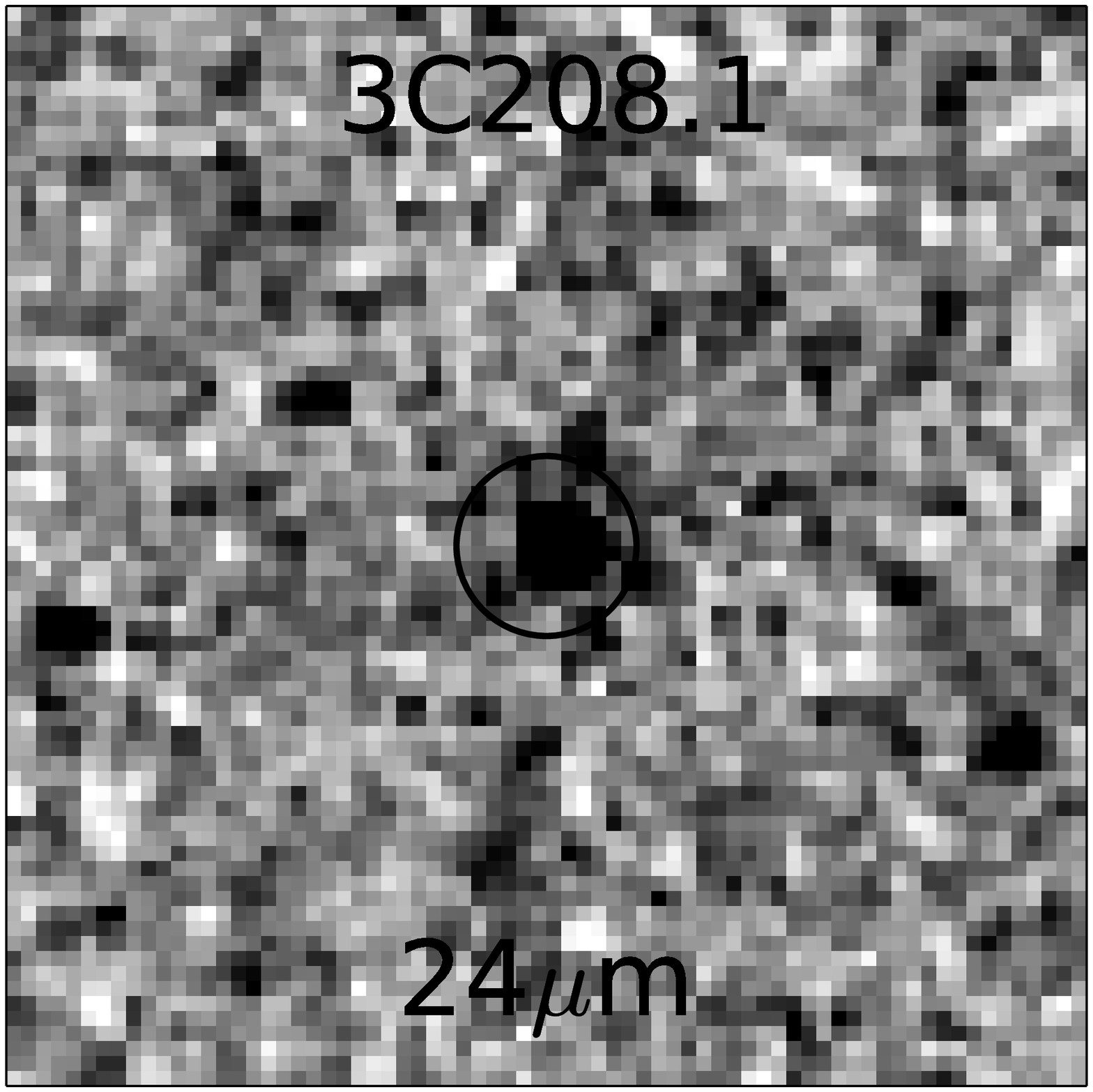}
      \includegraphics[width=1.5cm]{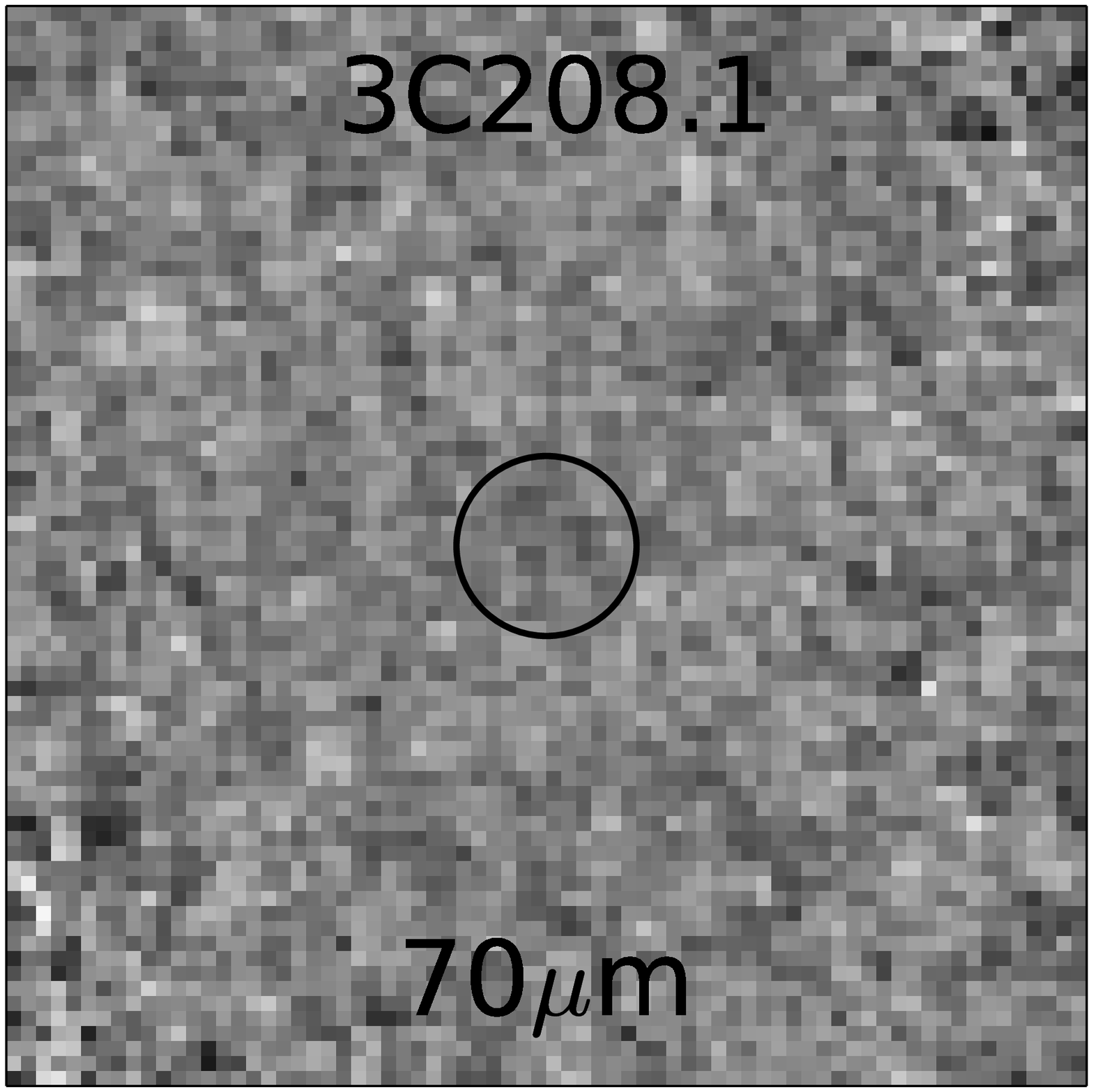}
      \includegraphics[width=1.5cm]{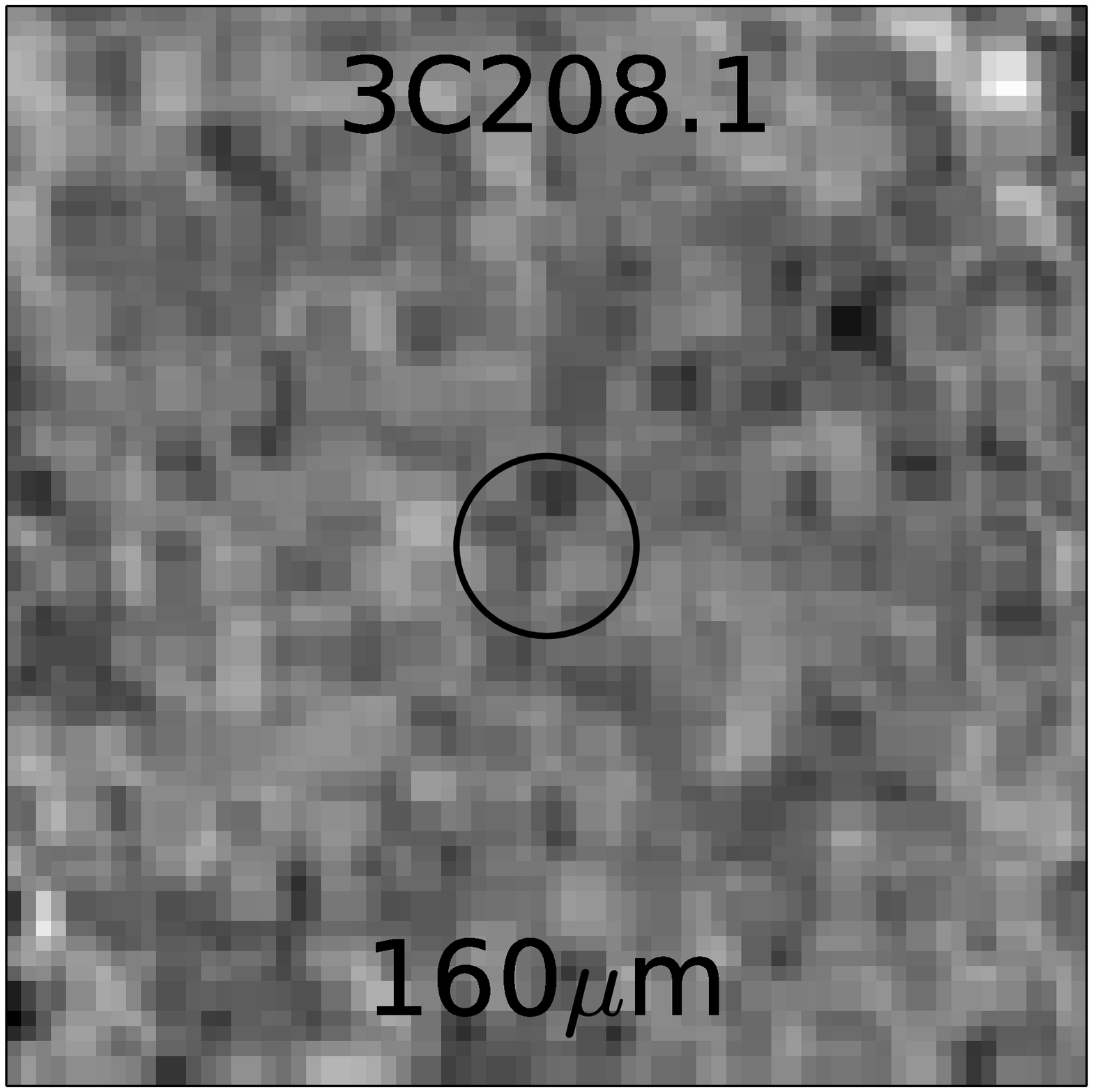}
      \includegraphics[width=1.5cm]{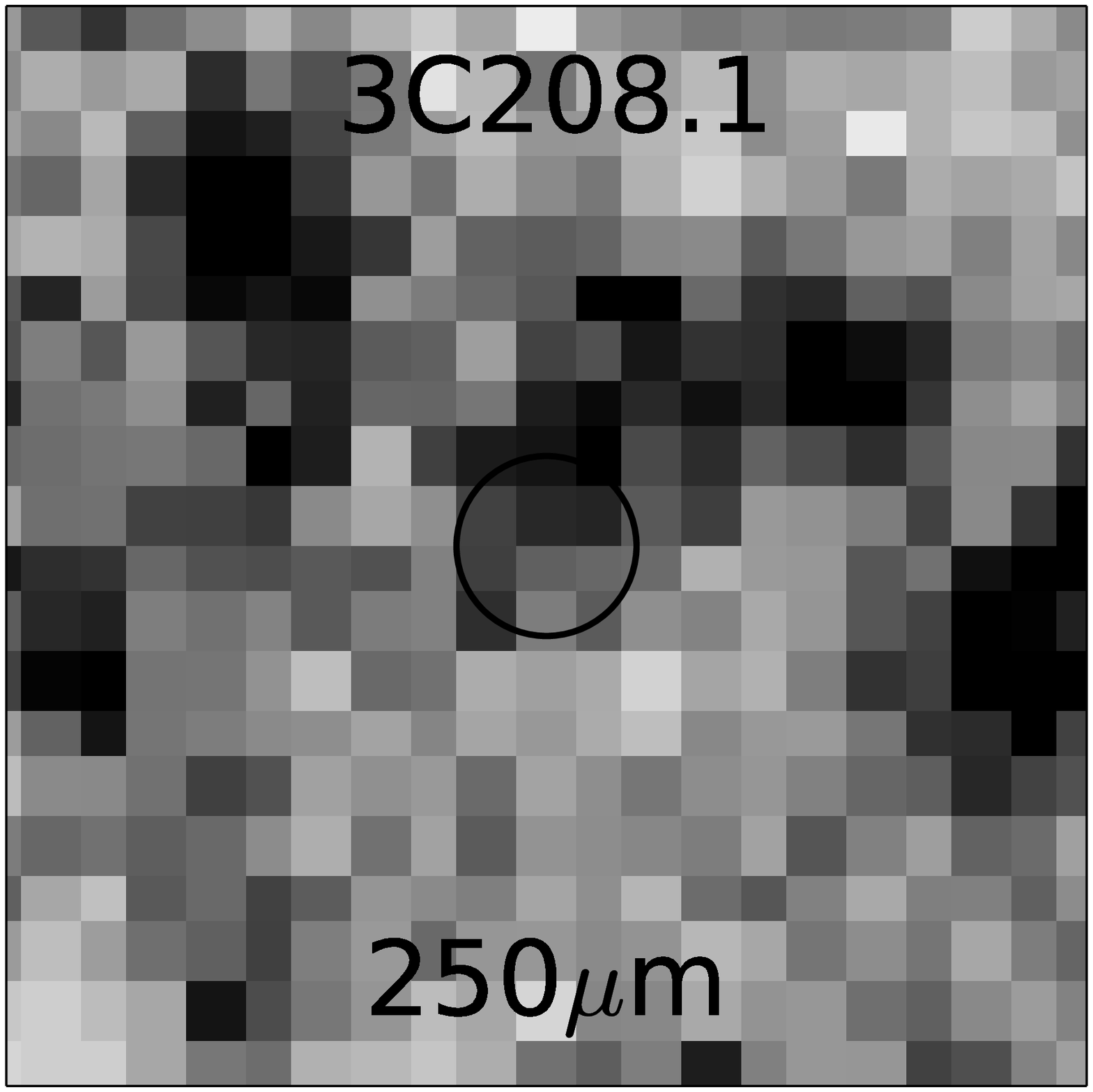}
      \includegraphics[width=1.5cm]{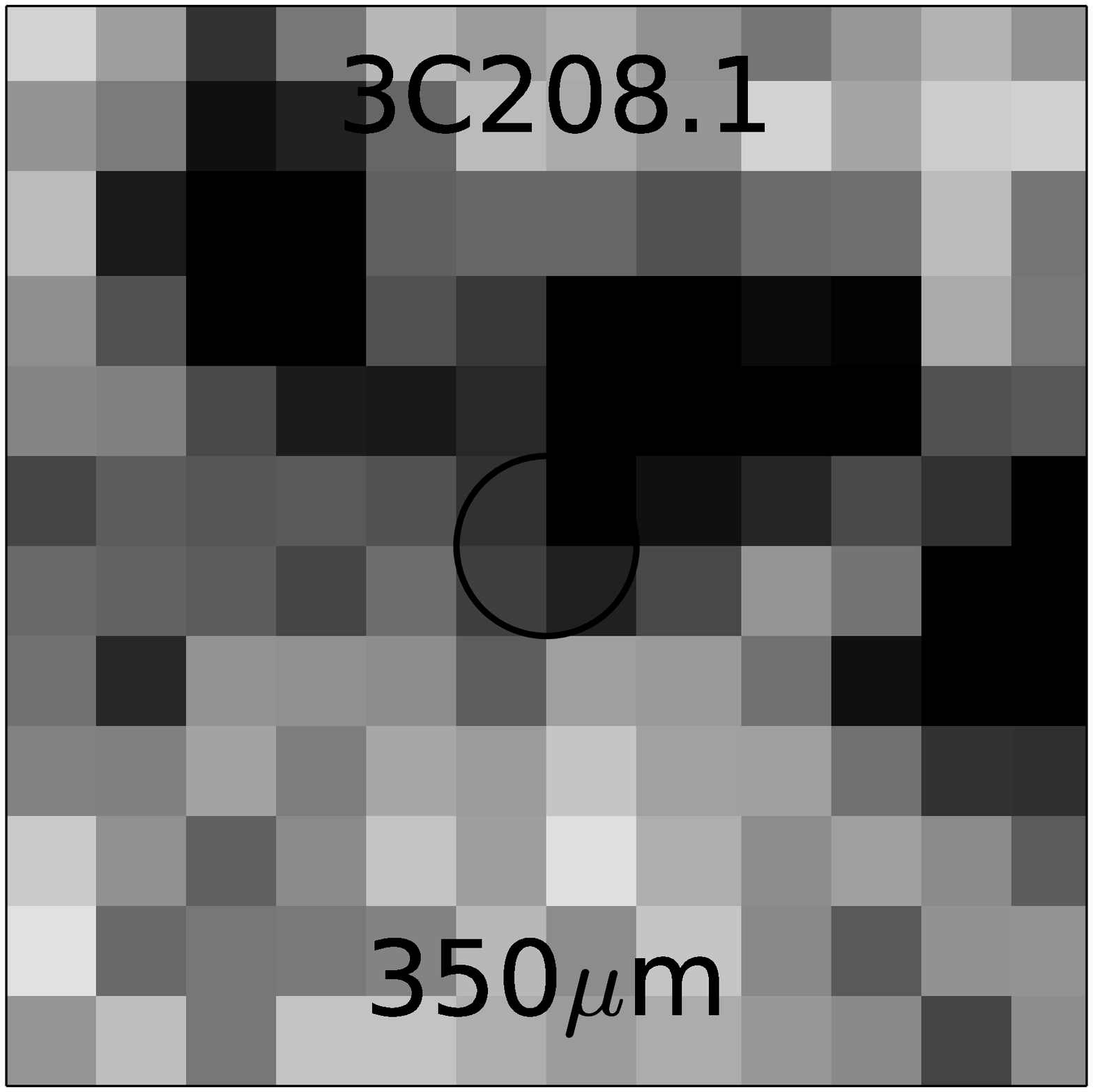}
      \includegraphics[width=1.5cm]{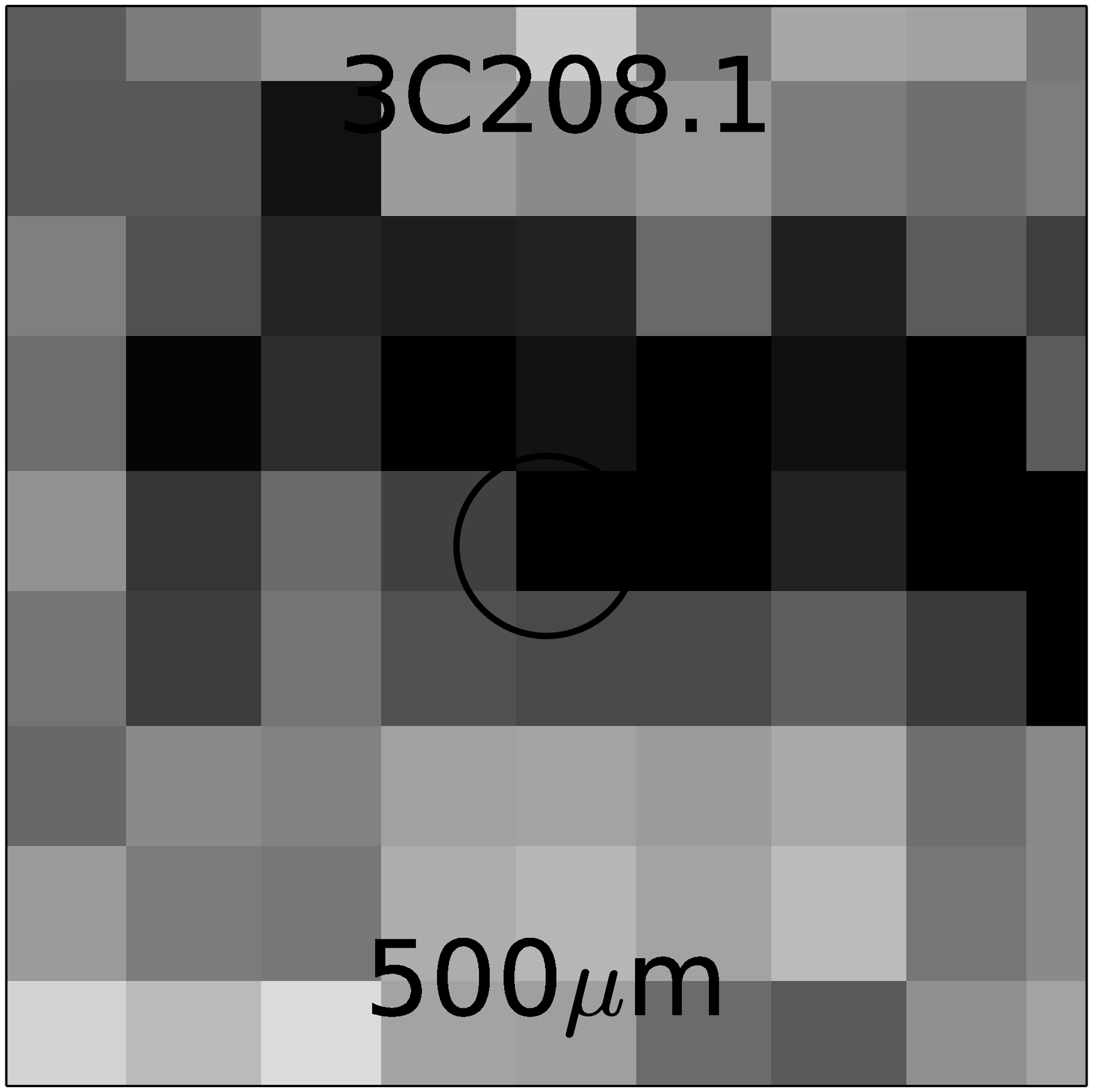}
      \\
      \includegraphics[width=1.5cm]{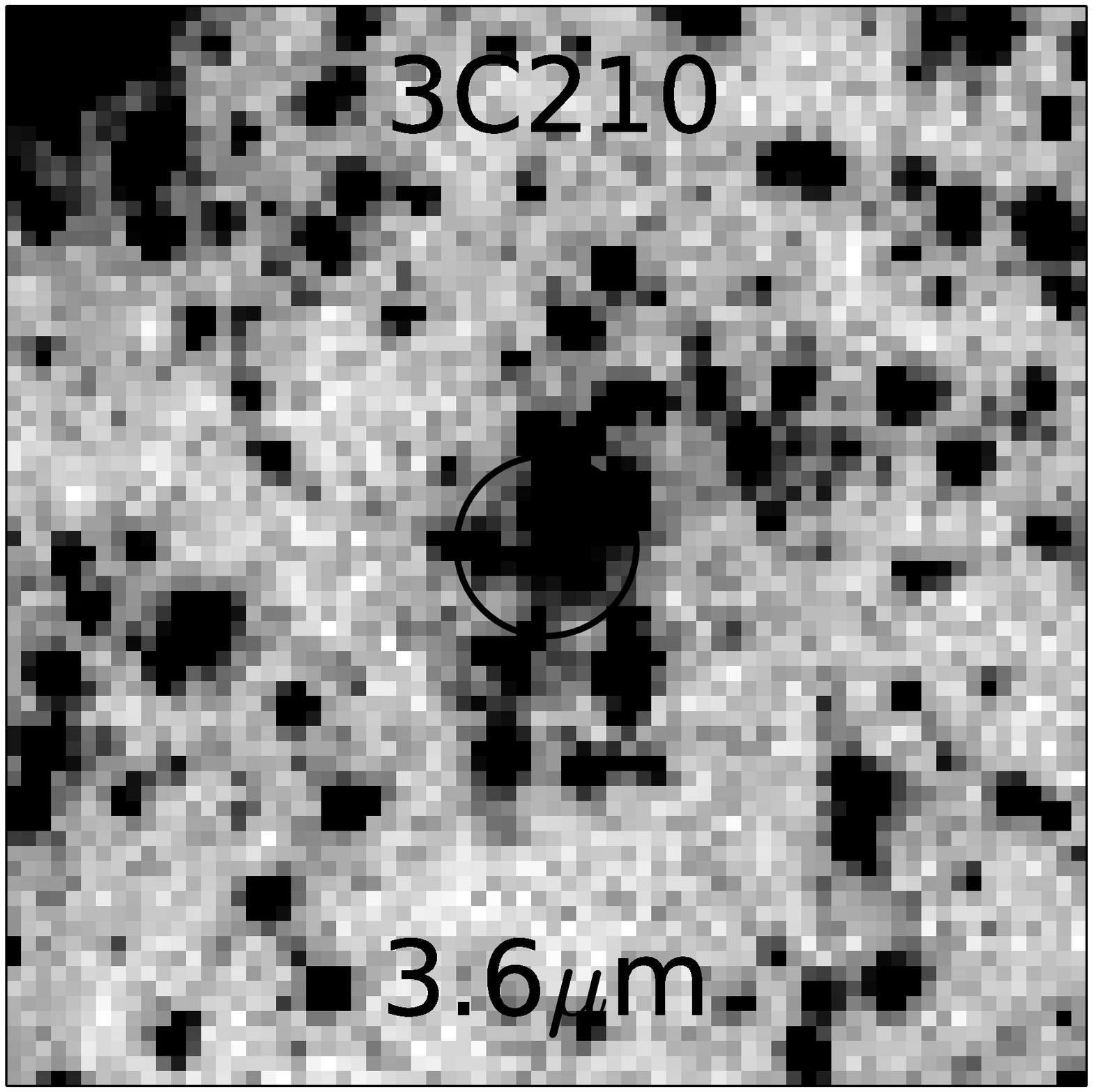}
      \includegraphics[width=1.5cm]{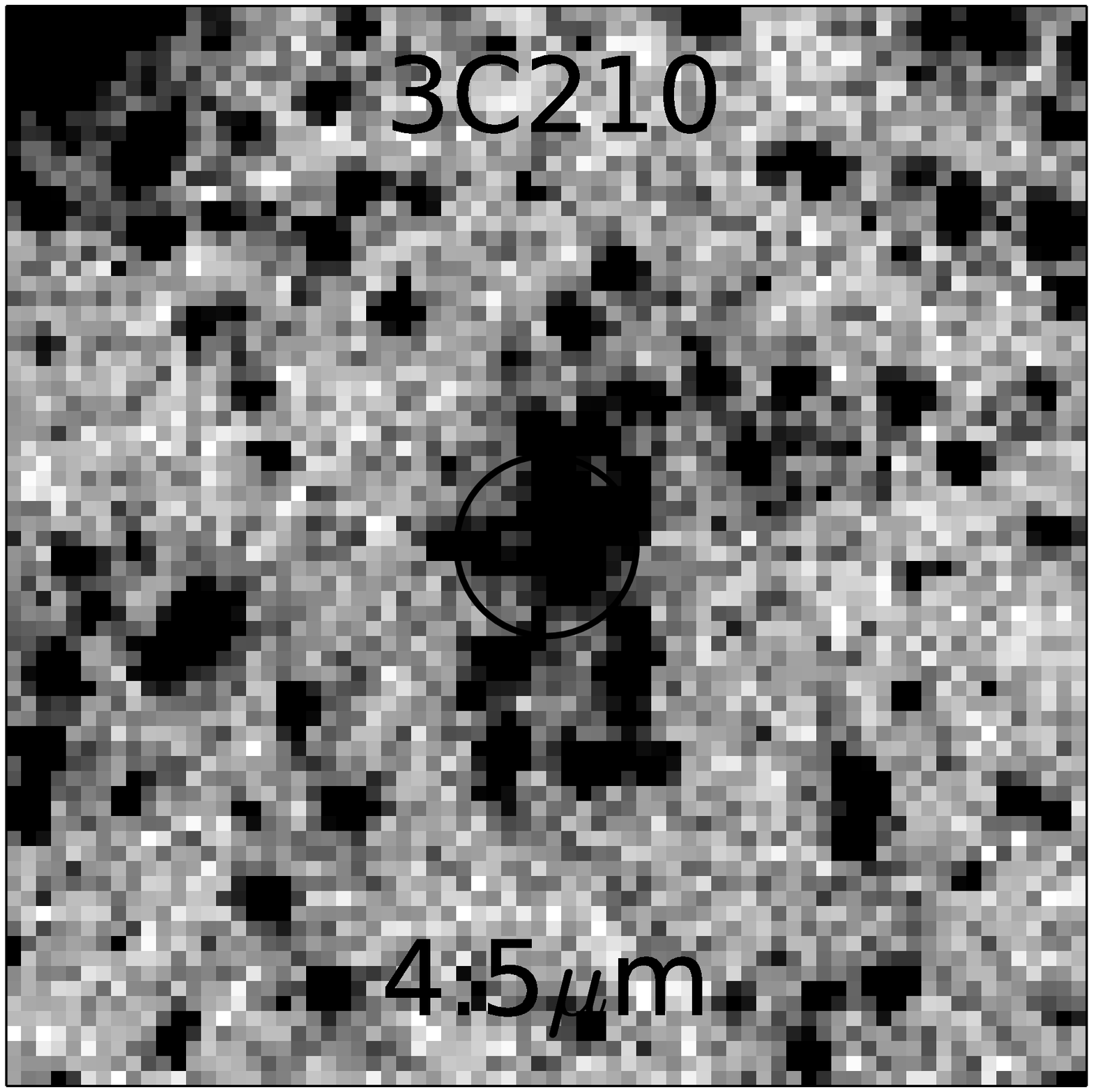}
      \includegraphics[width=1.5cm]{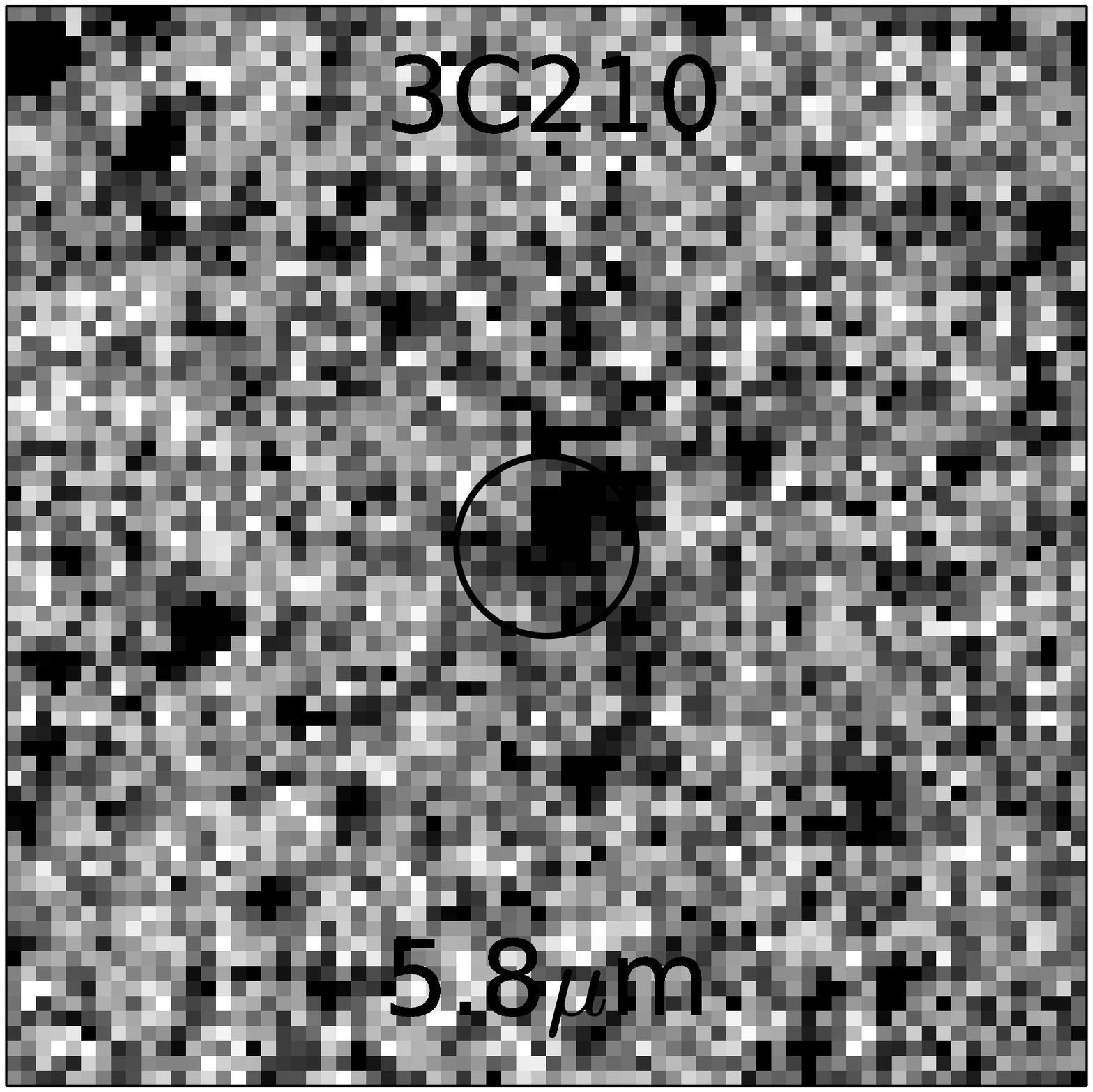}
      \includegraphics[width=1.5cm]{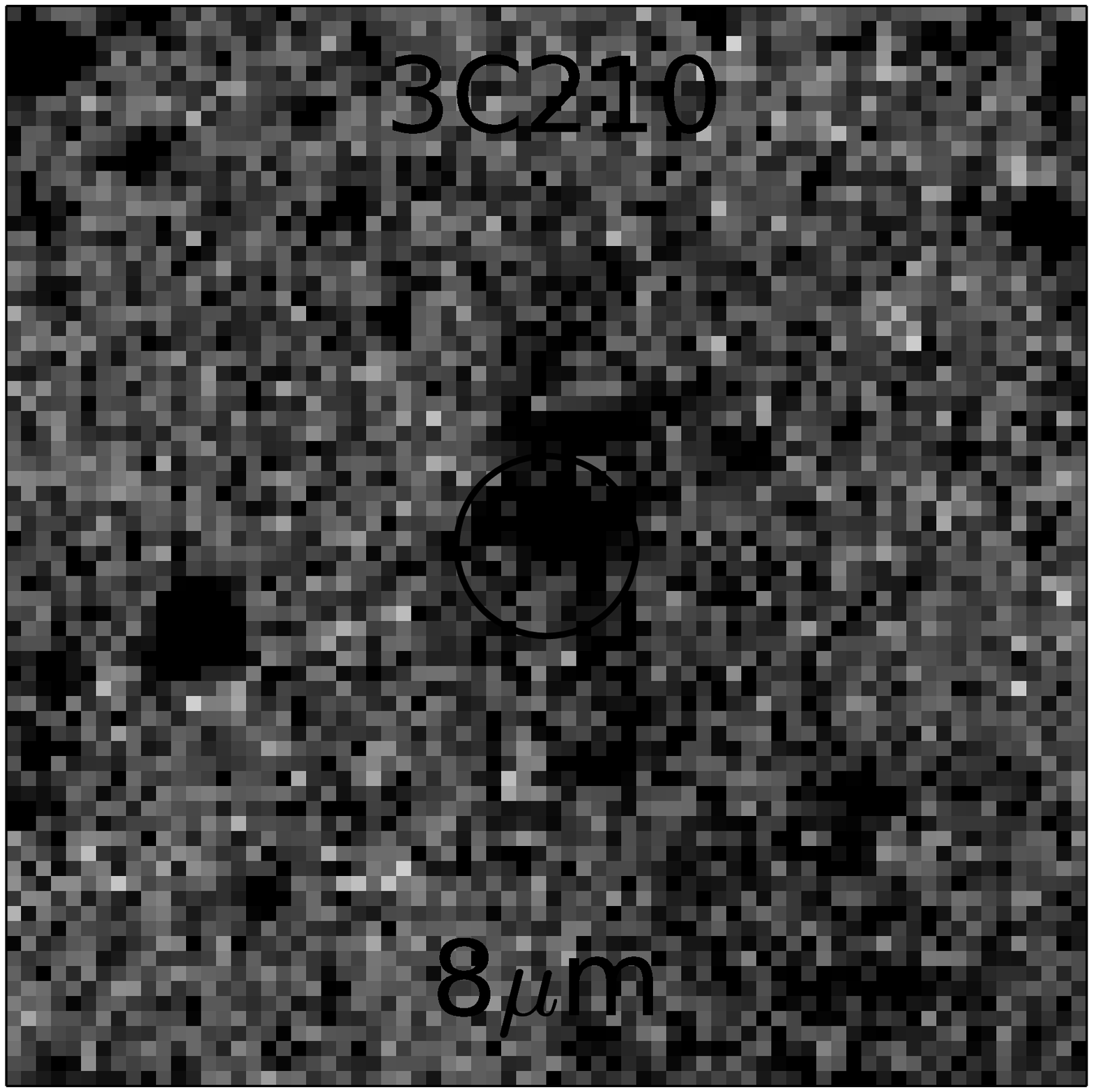}
      \includegraphics[width=1.5cm]{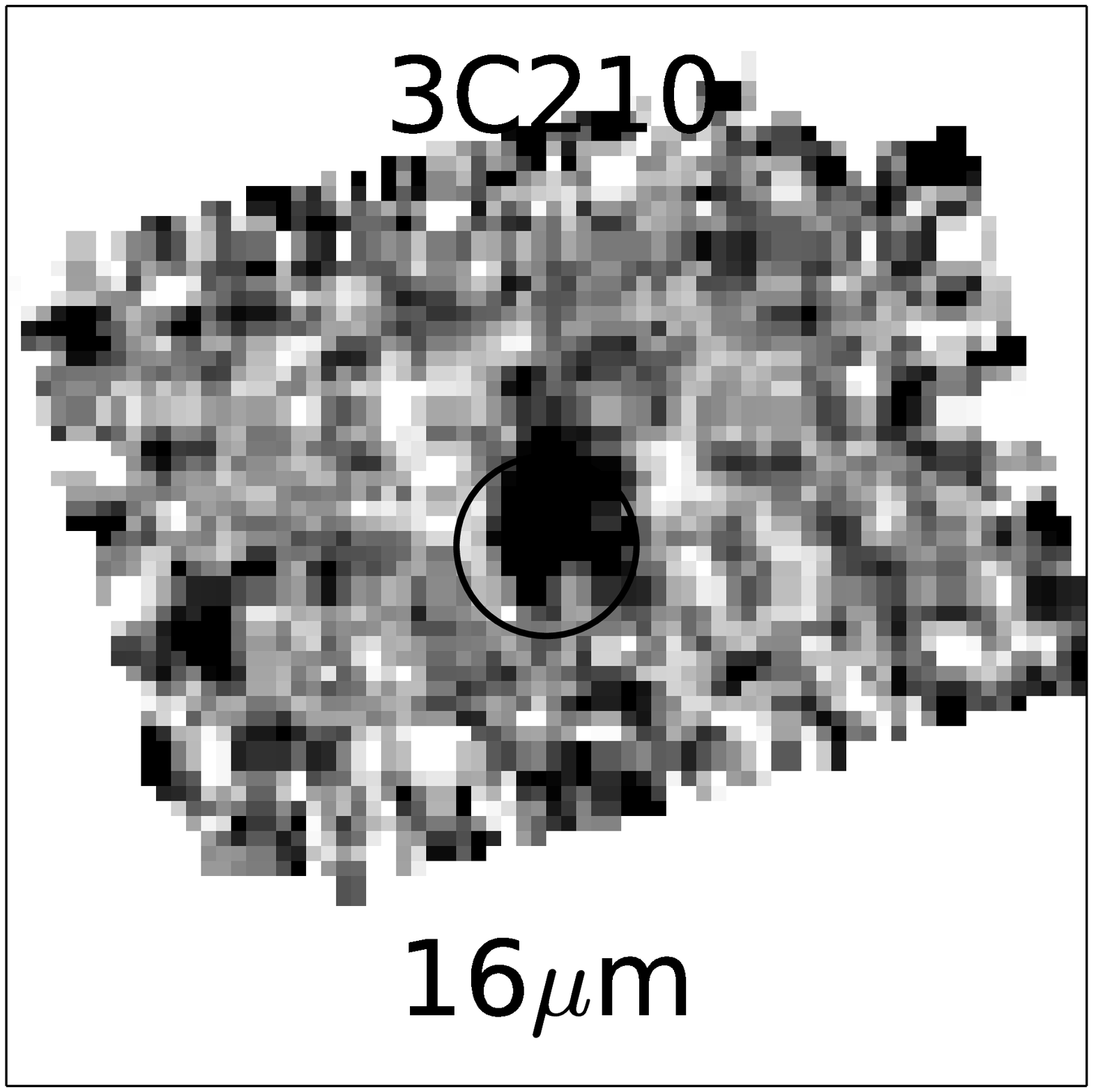}
      \includegraphics[width=1.5cm]{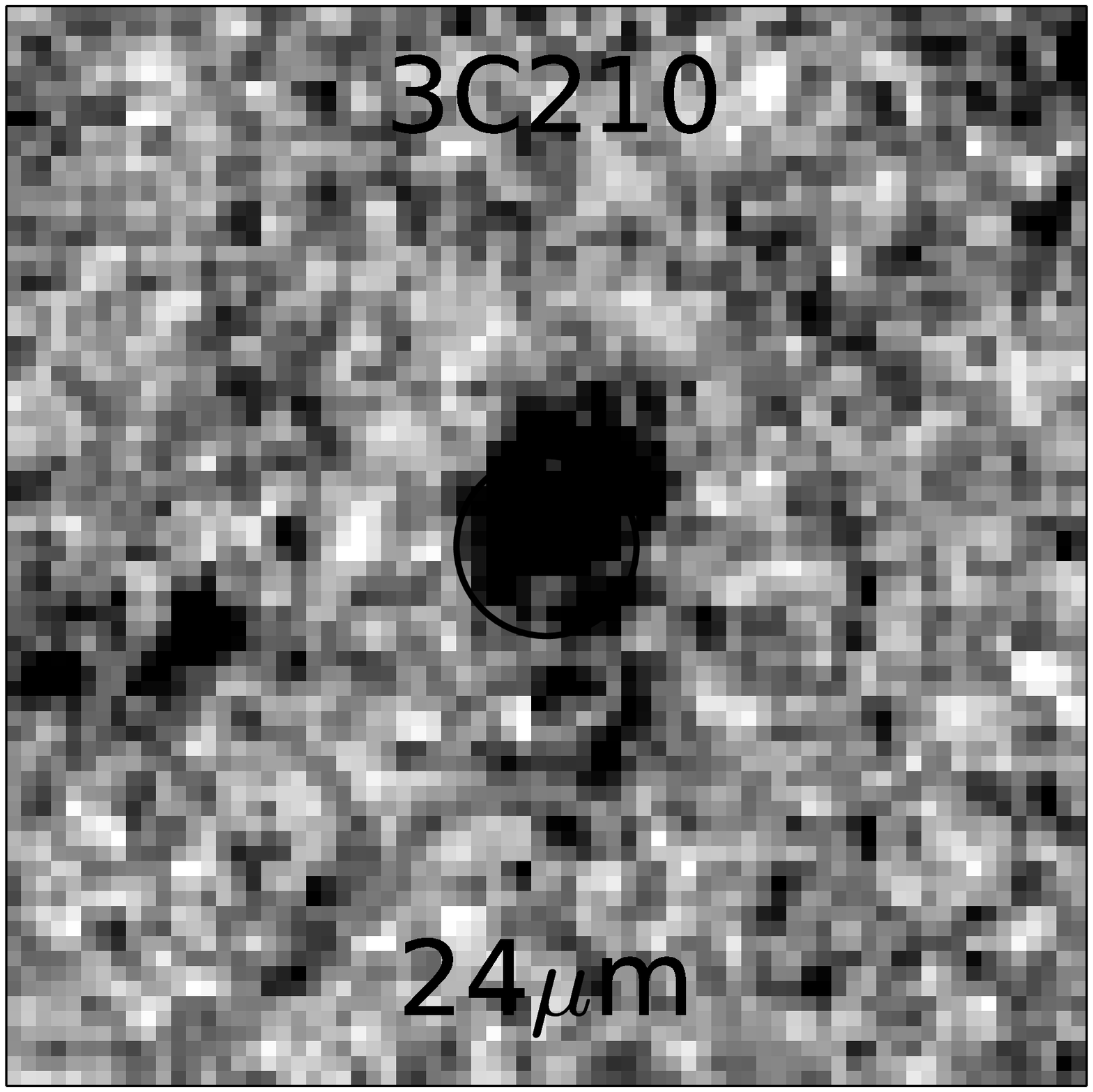}
      \includegraphics[width=1.5cm]{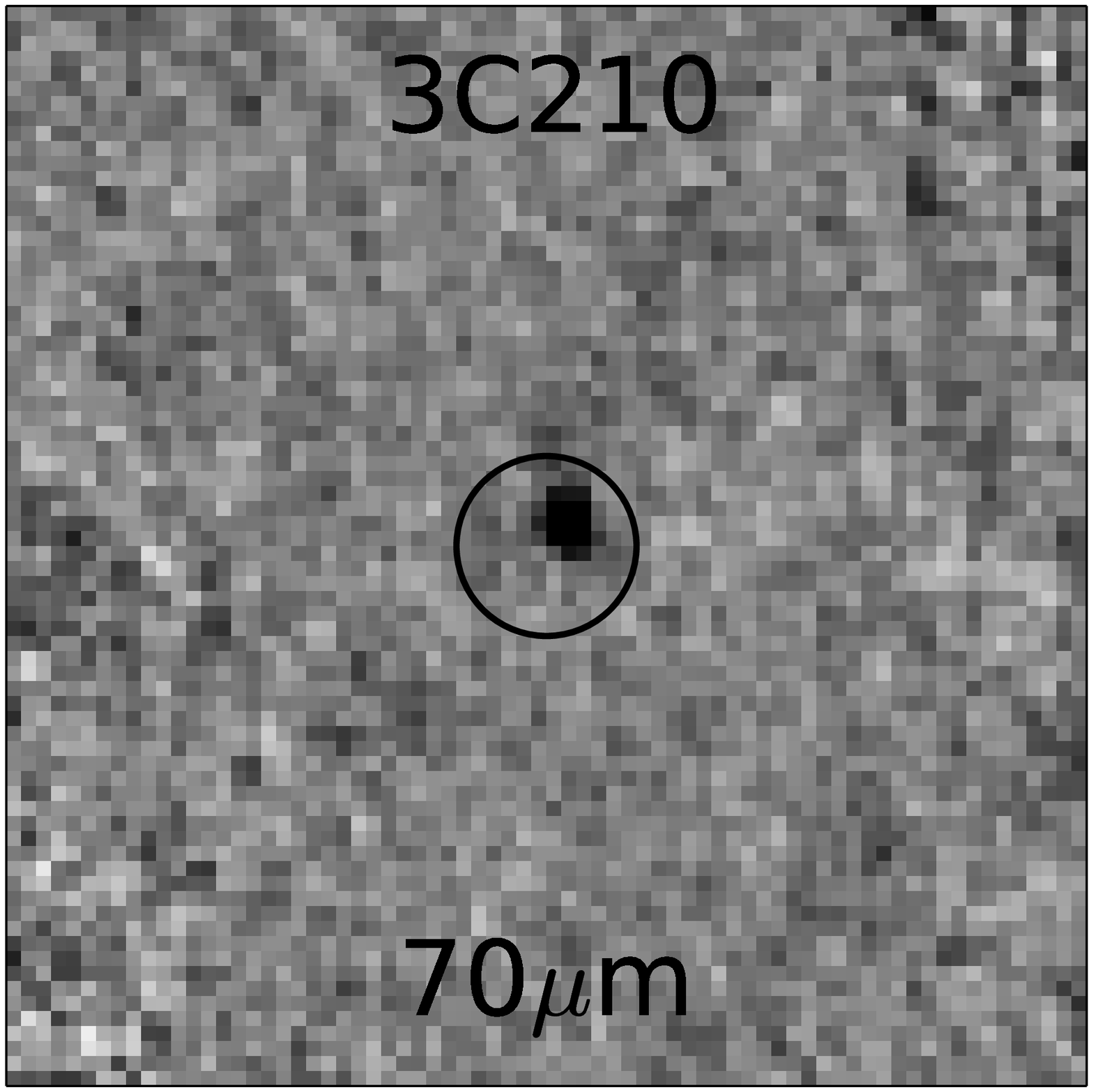}
      \includegraphics[width=1.5cm]{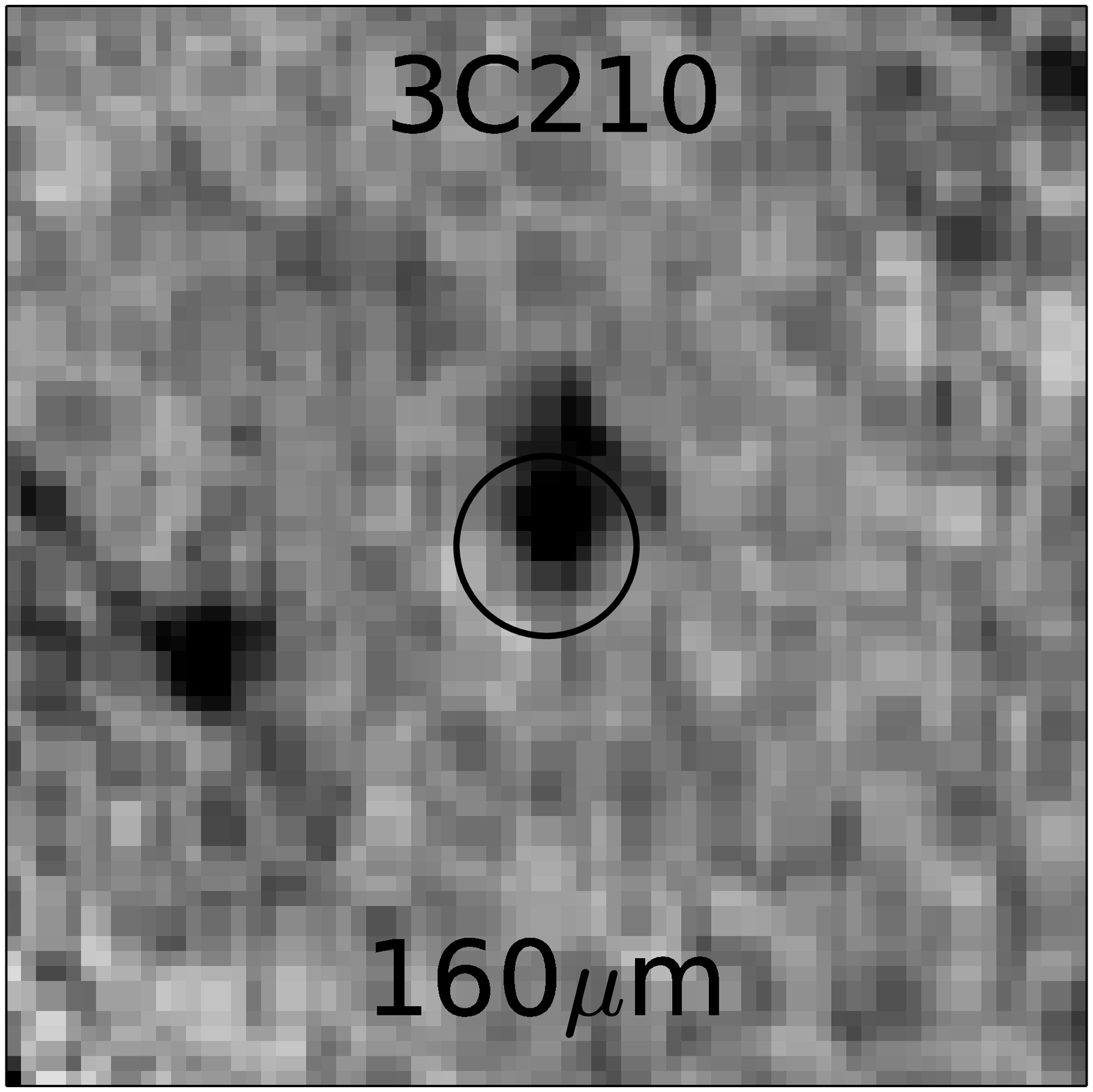}
      \includegraphics[width=1.5cm]{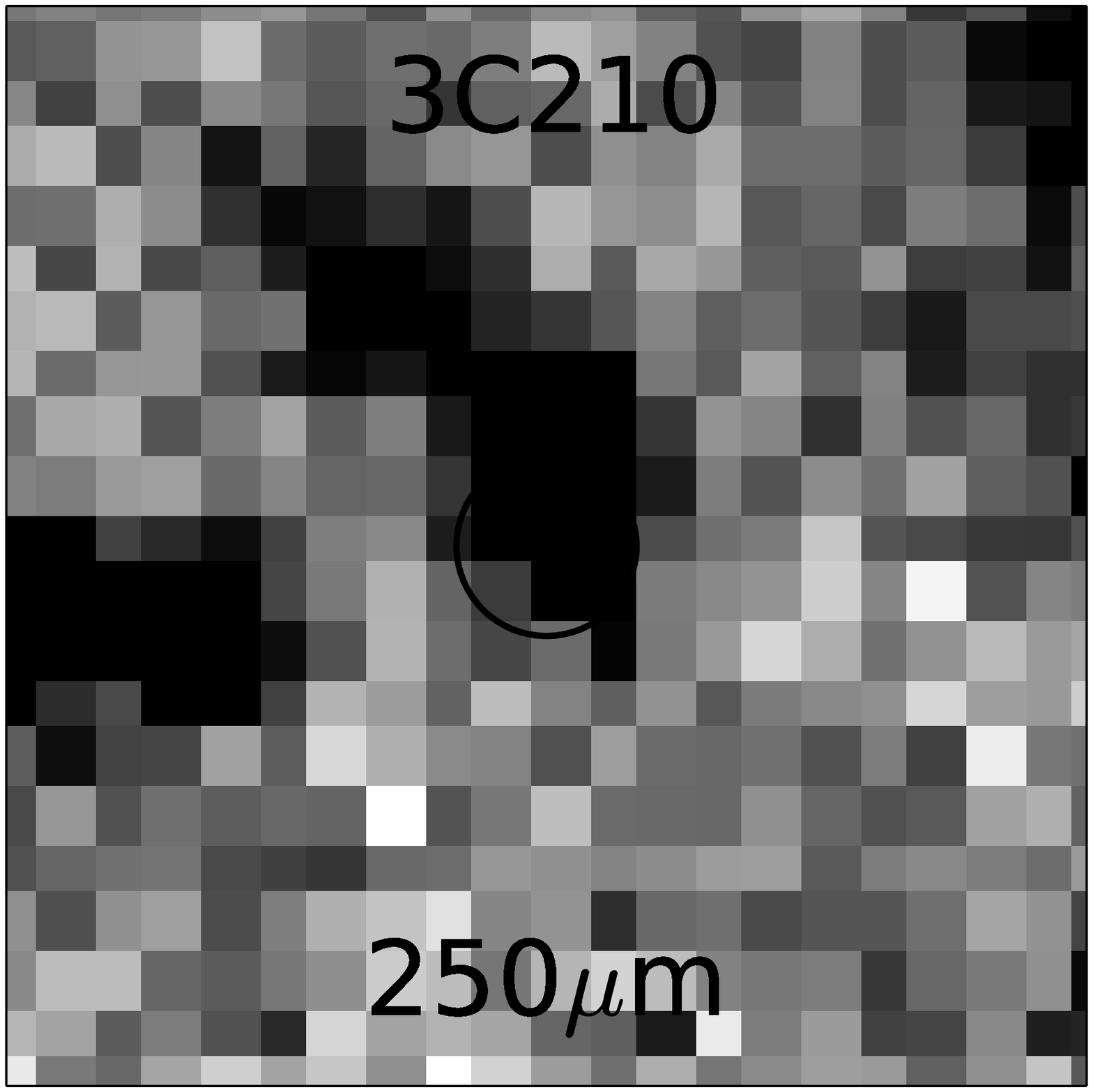}
      \includegraphics[width=1.5cm]{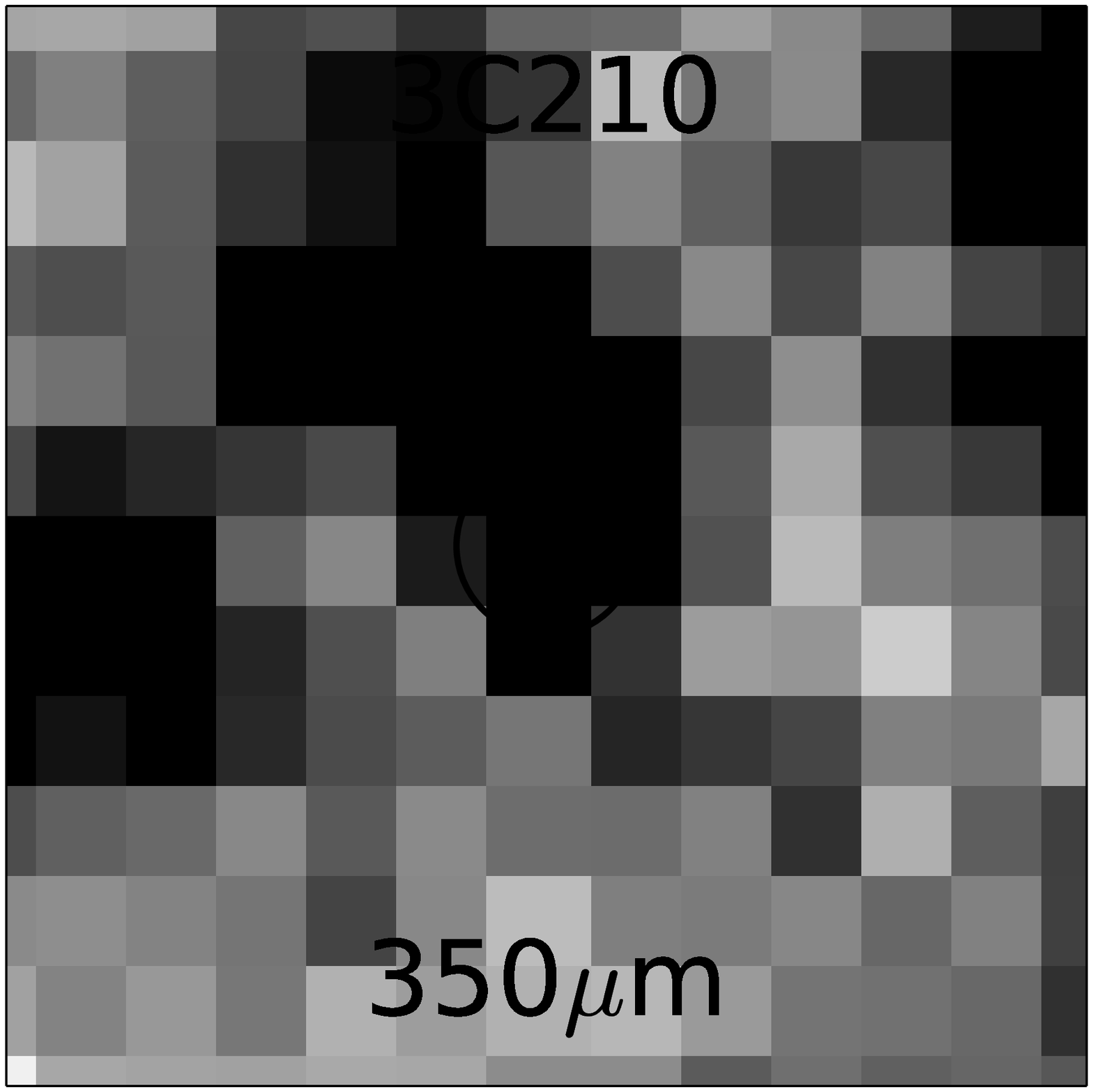}
      \includegraphics[width=1.5cm]{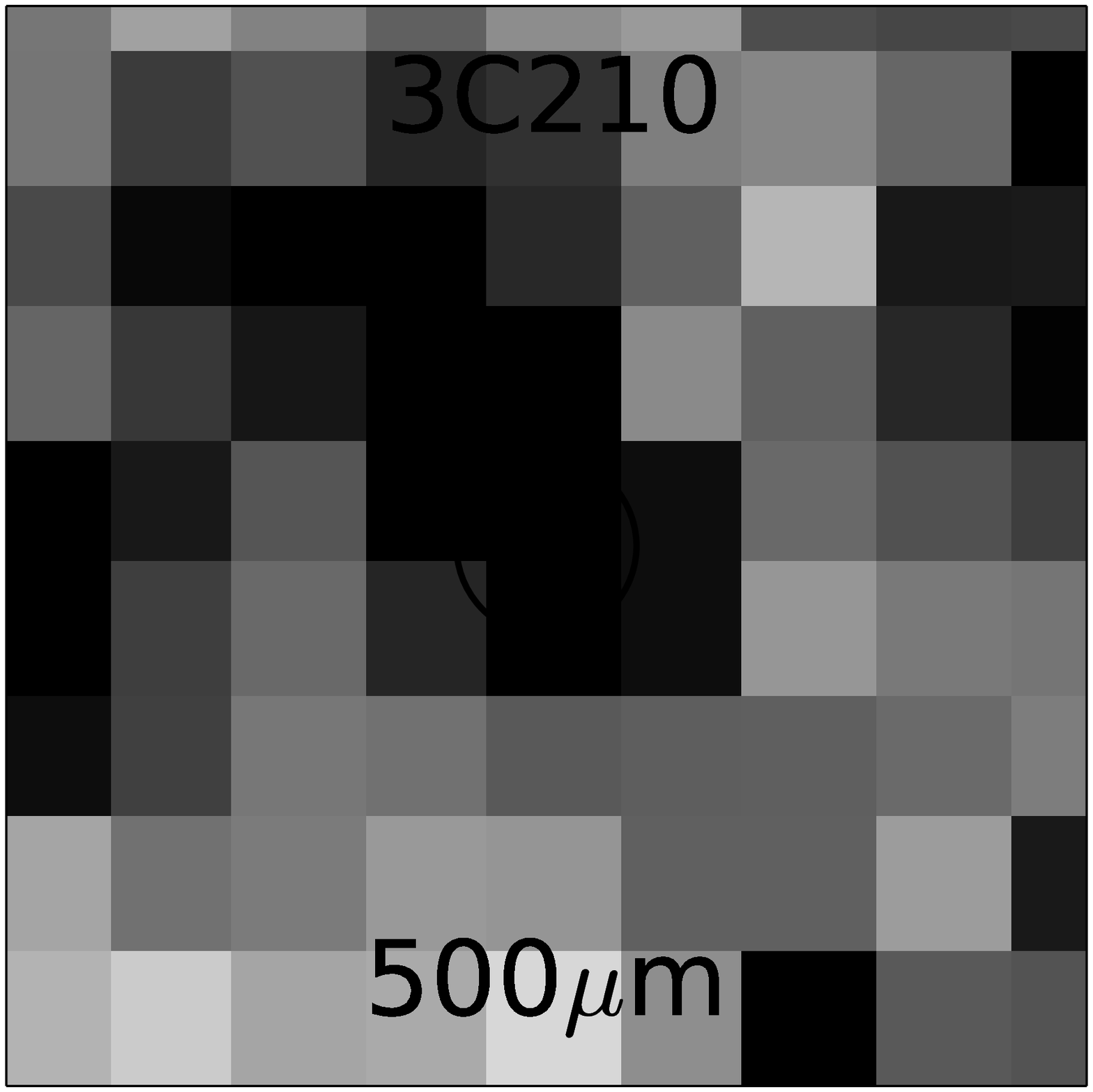}
      \\
      \includegraphics[width=1.5cm]{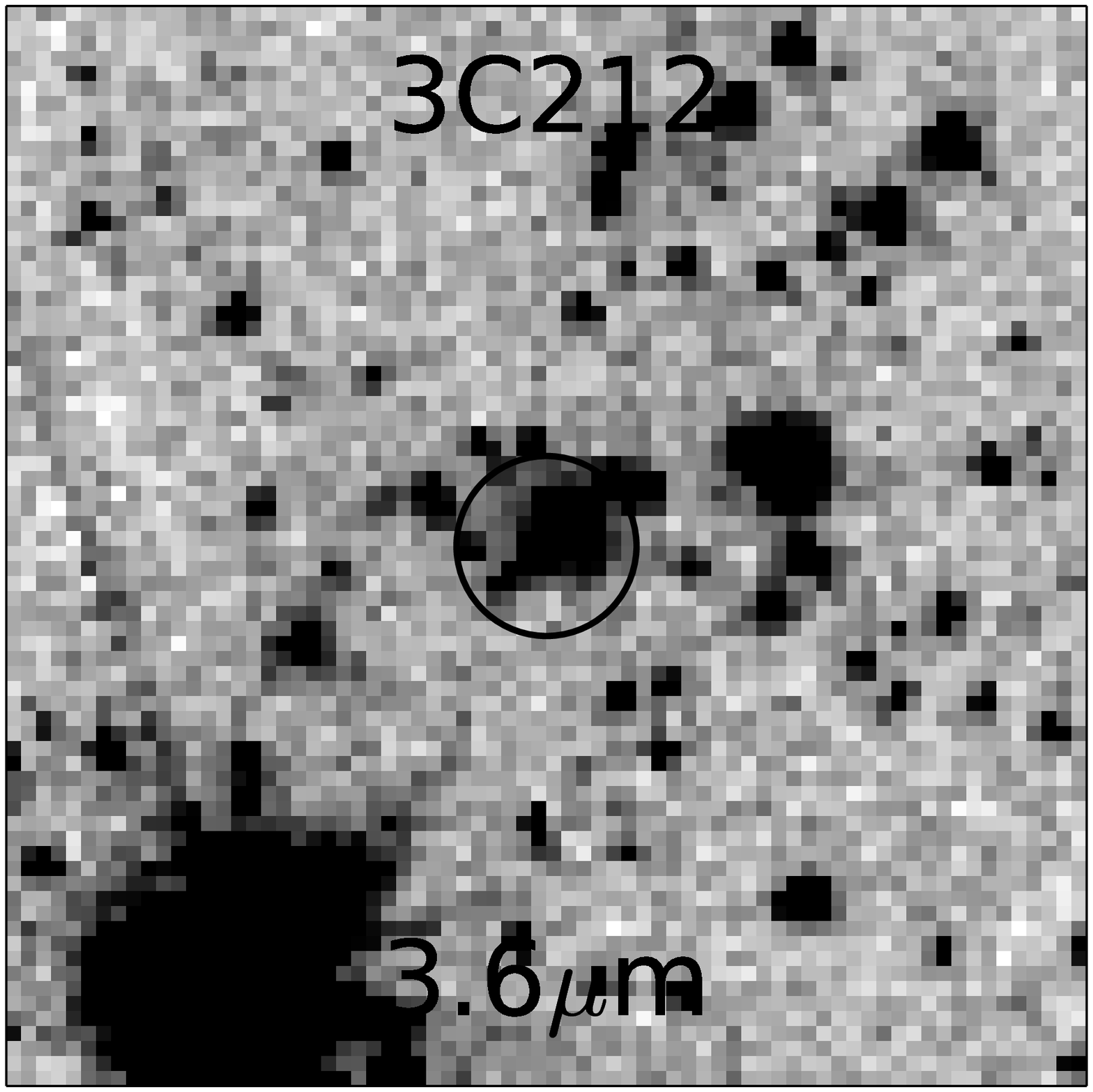}
      \includegraphics[width=1.5cm]{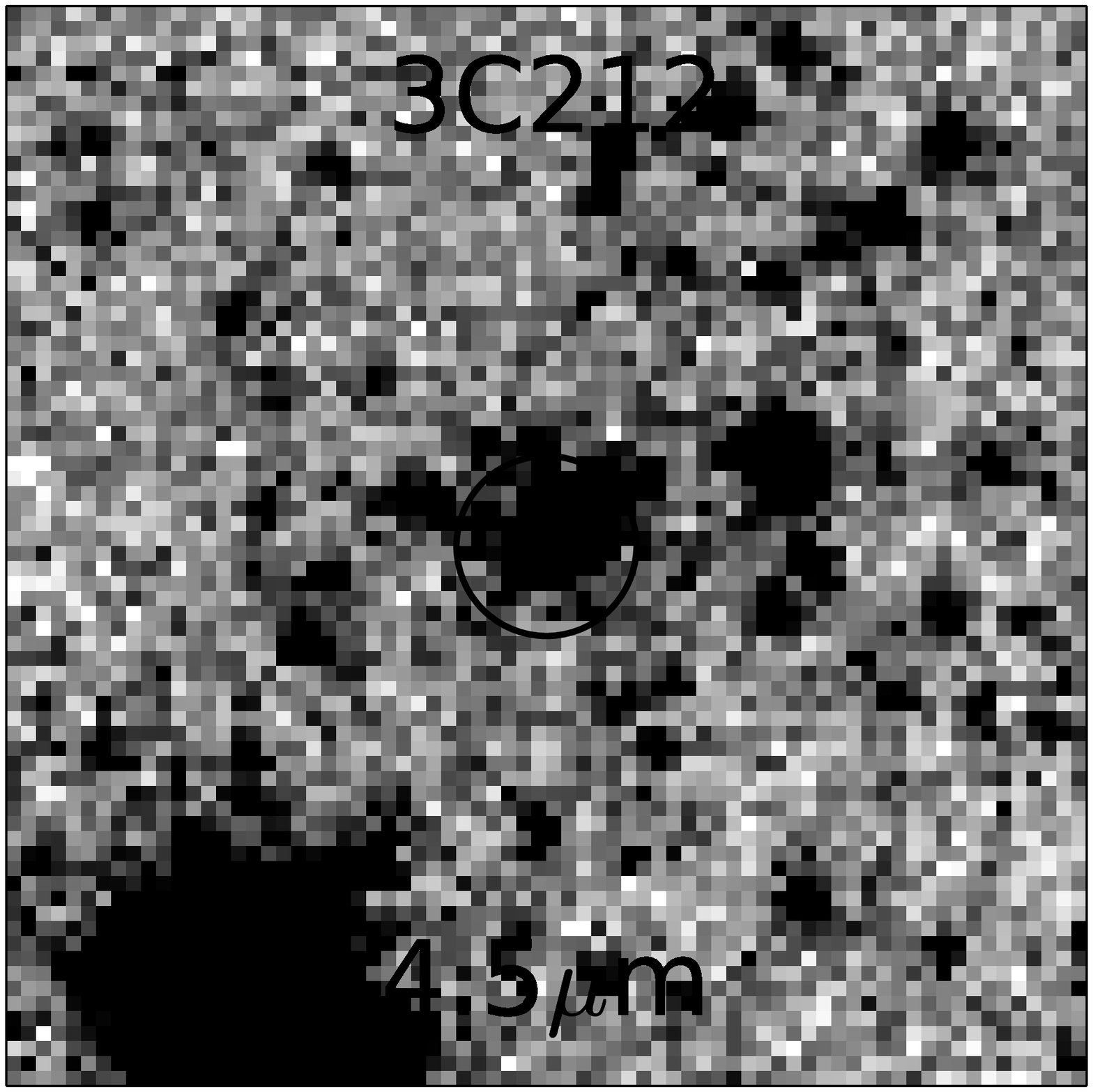}
      \includegraphics[width=1.5cm]{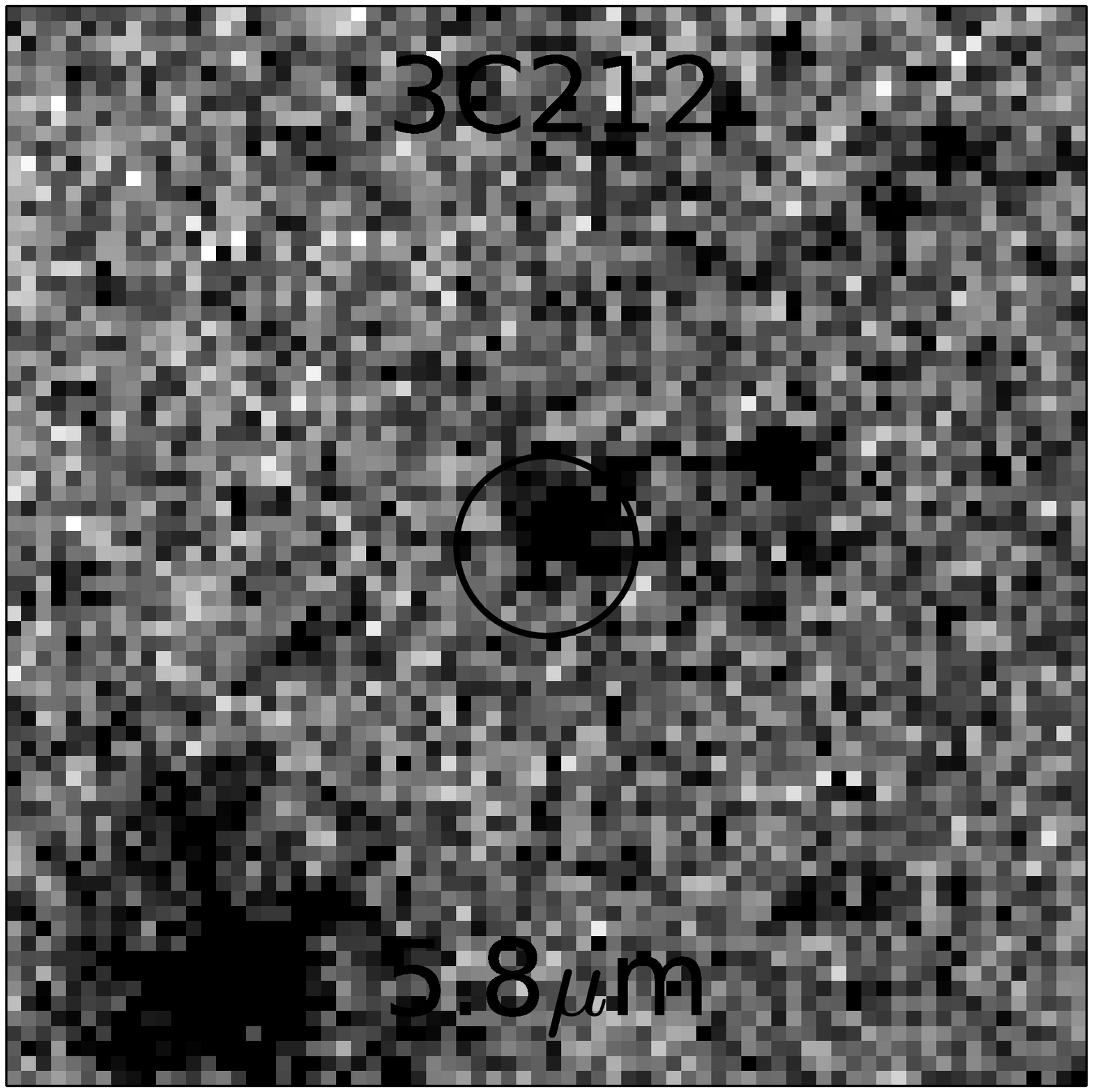}
      \includegraphics[width=1.5cm]{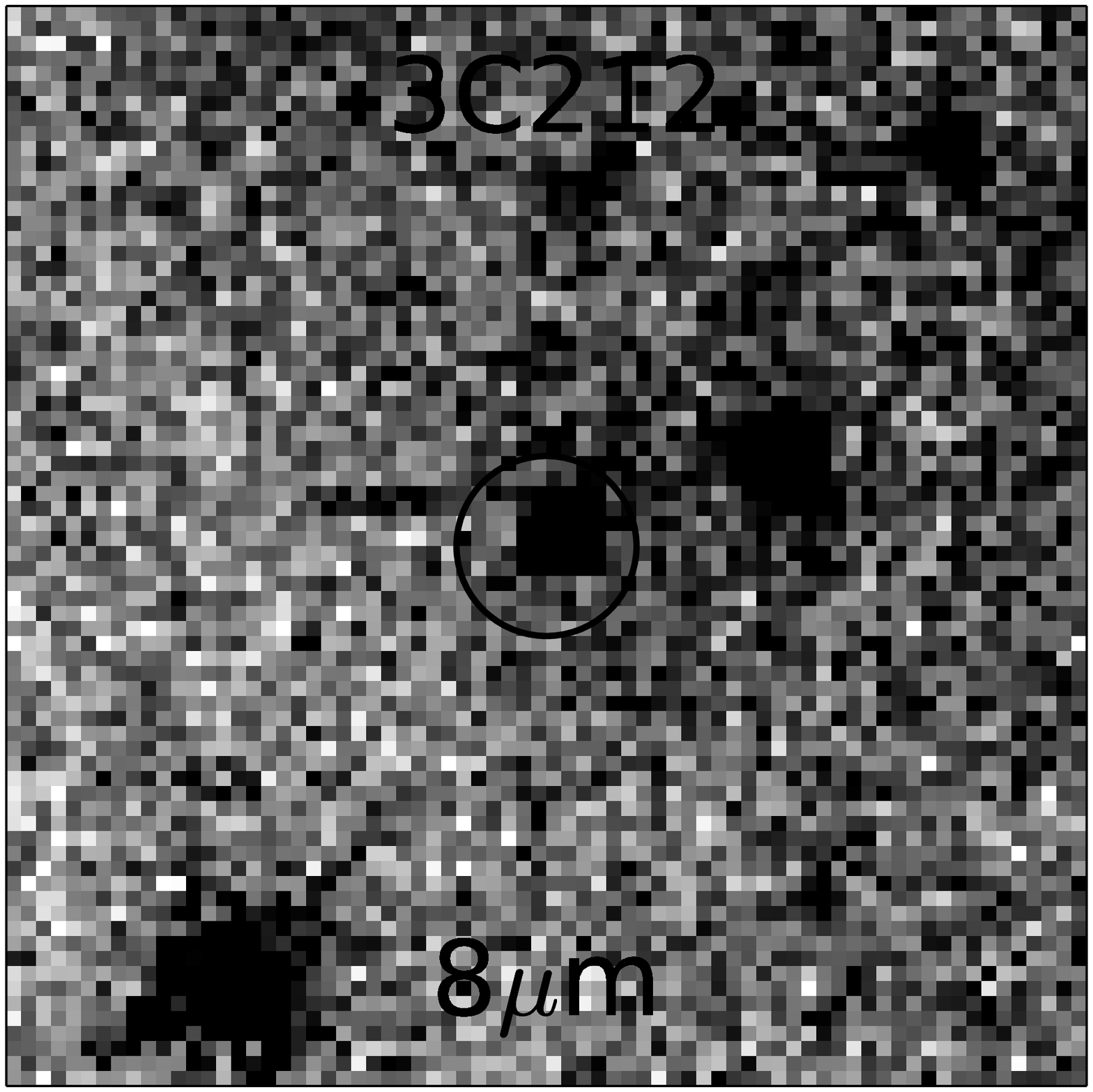}
      \includegraphics[width=1.5cm]{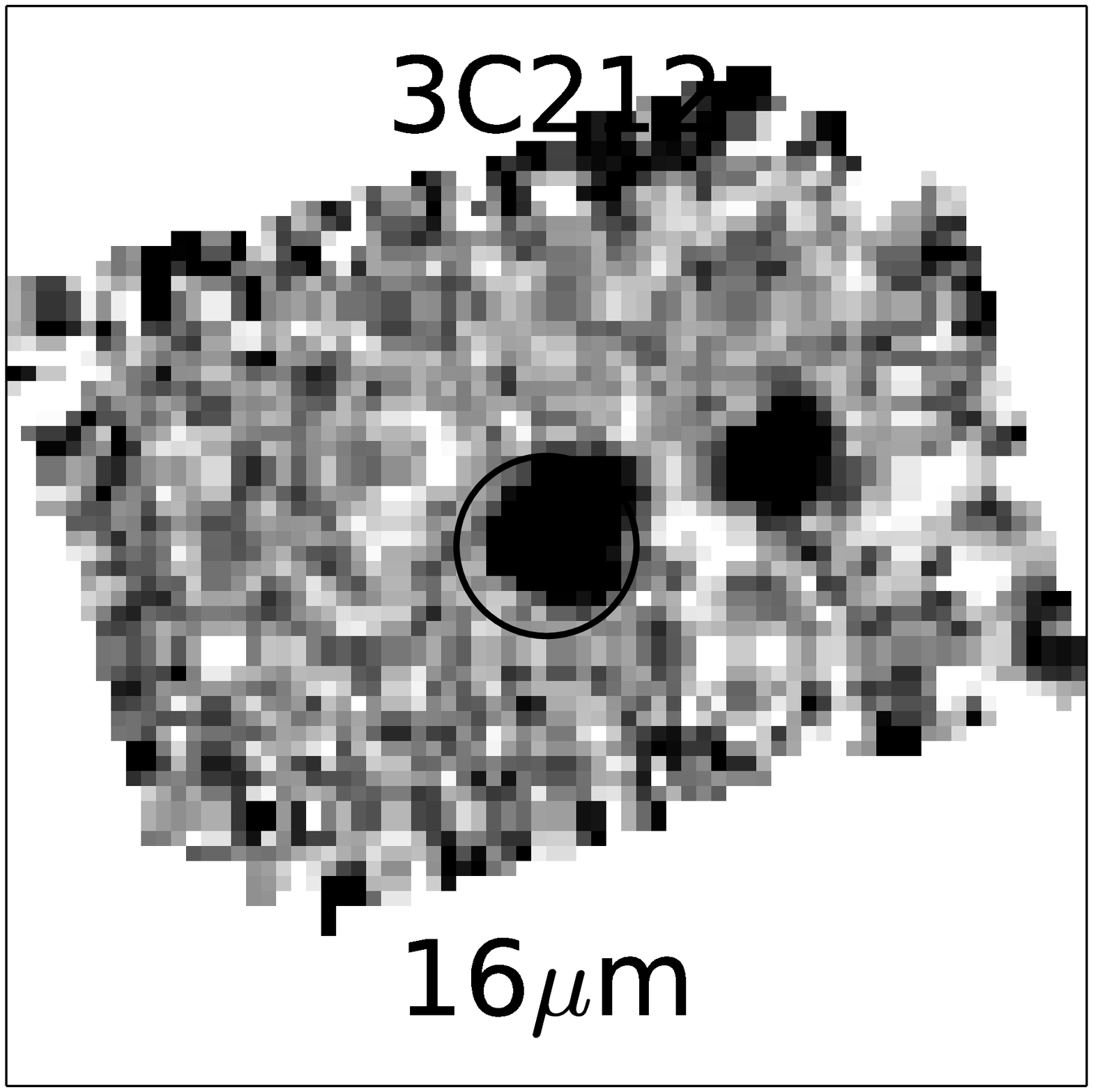}
      \includegraphics[width=1.5cm]{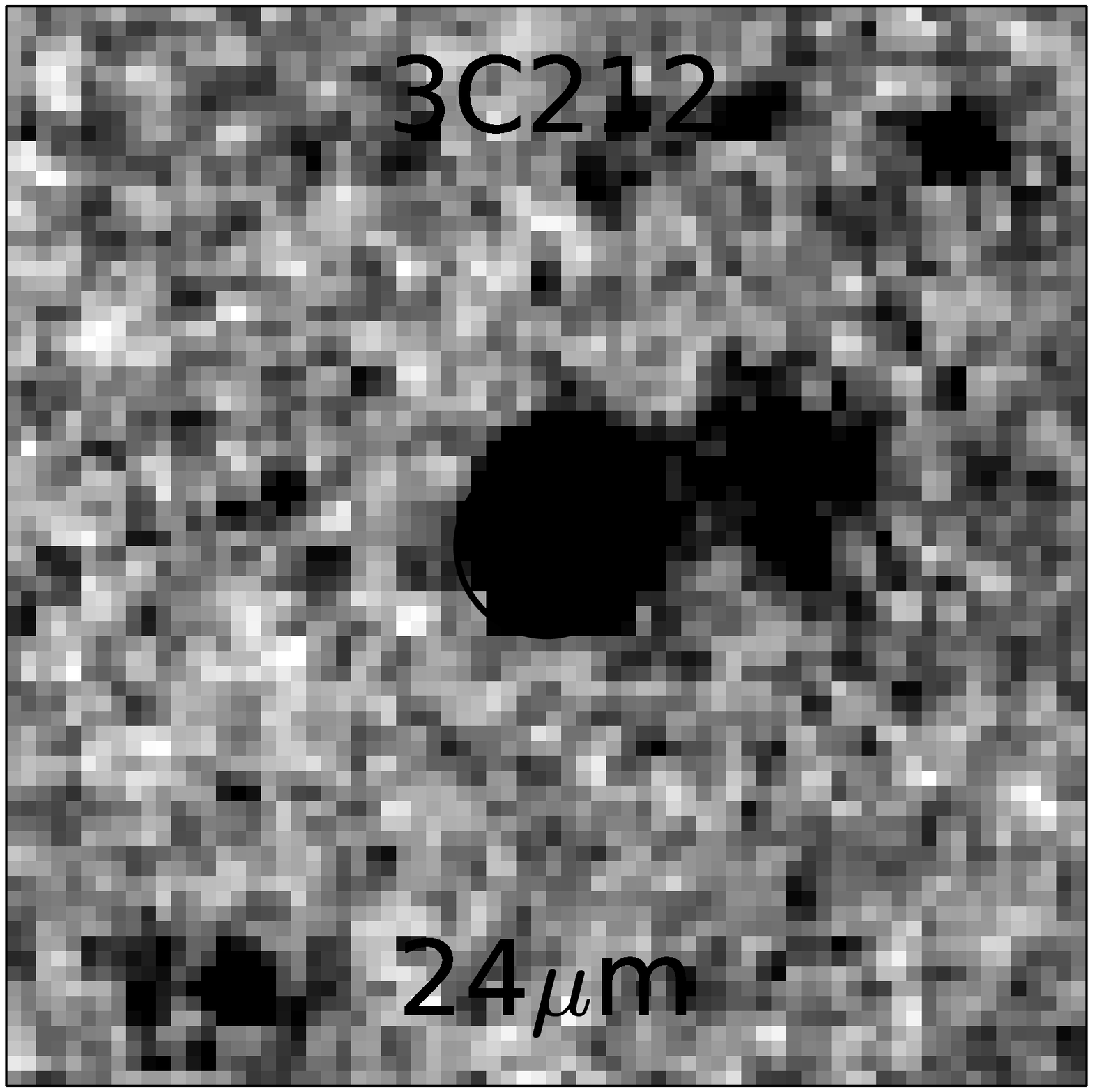}
      \includegraphics[width=1.5cm]{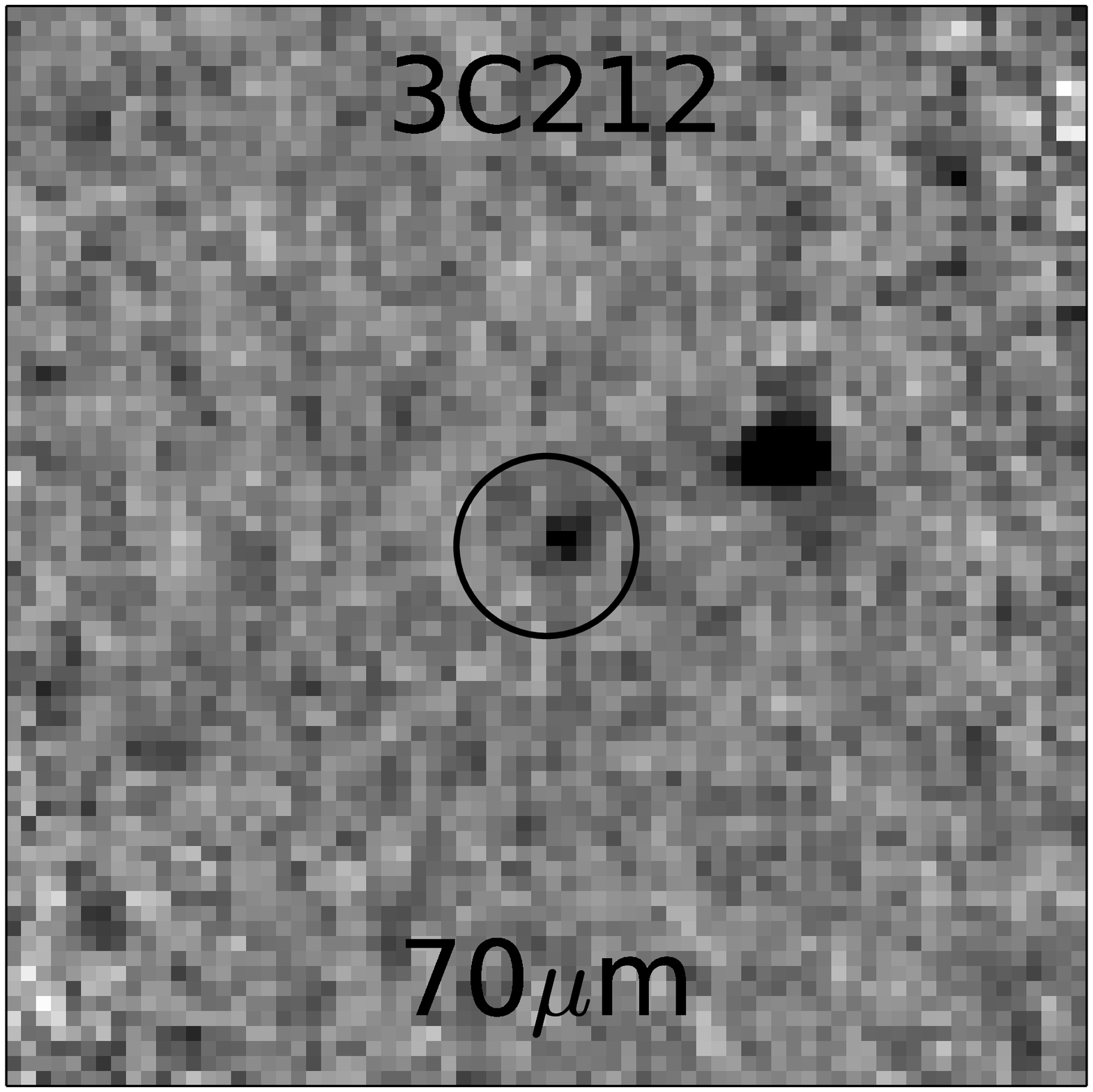}
      \includegraphics[width=1.5cm]{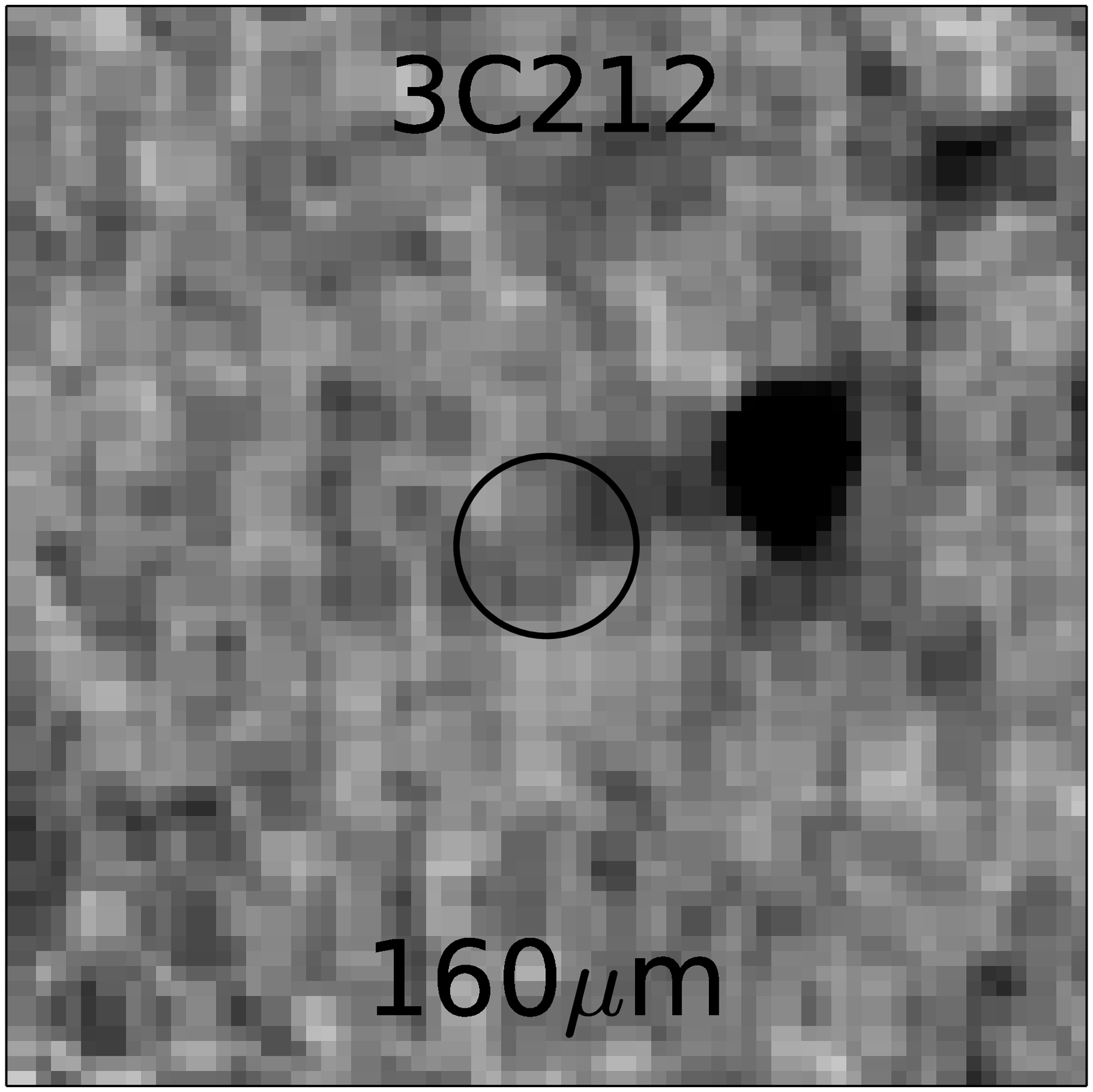}
      \includegraphics[width=1.5cm]{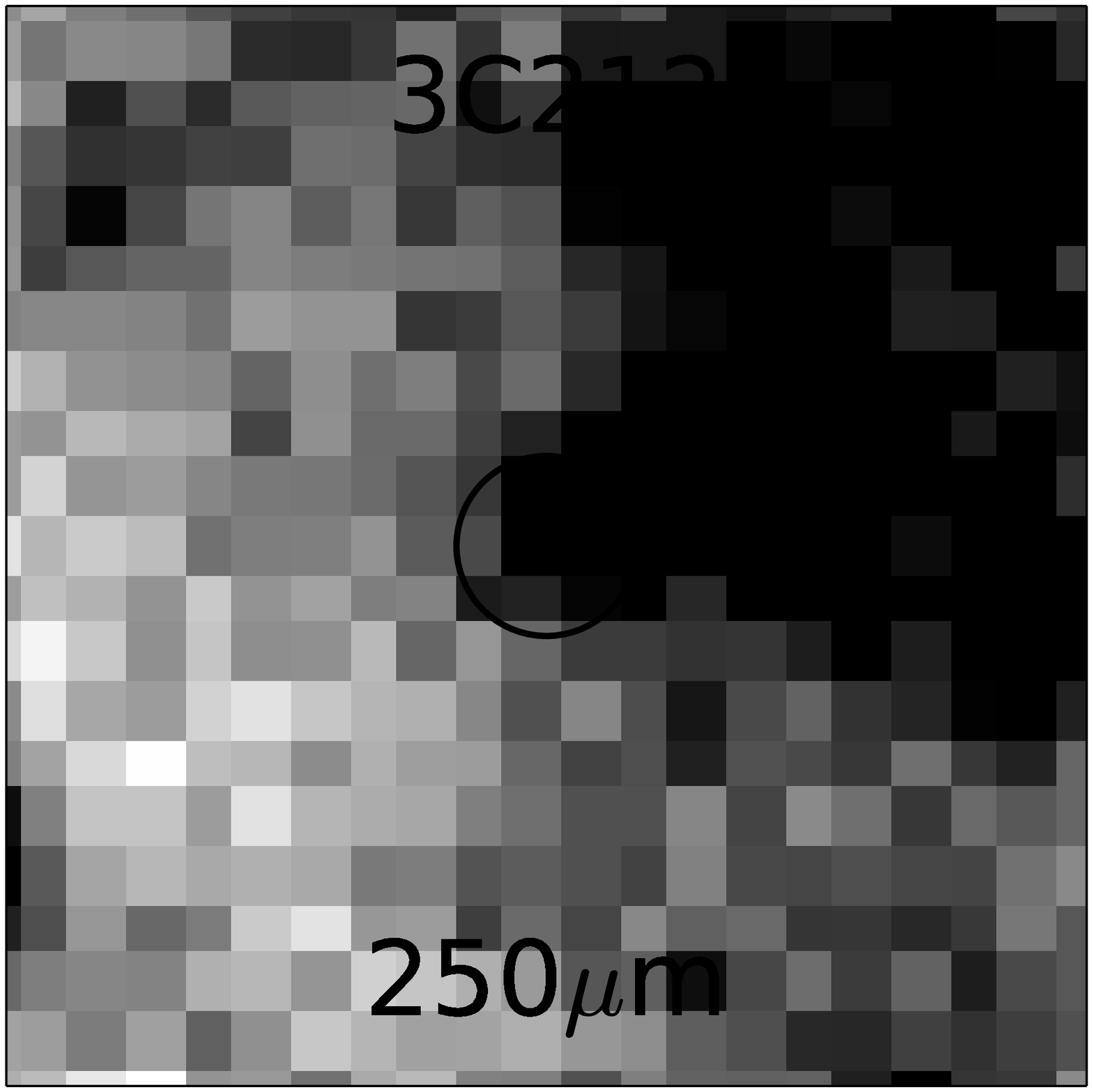}
      \includegraphics[width=1.5cm]{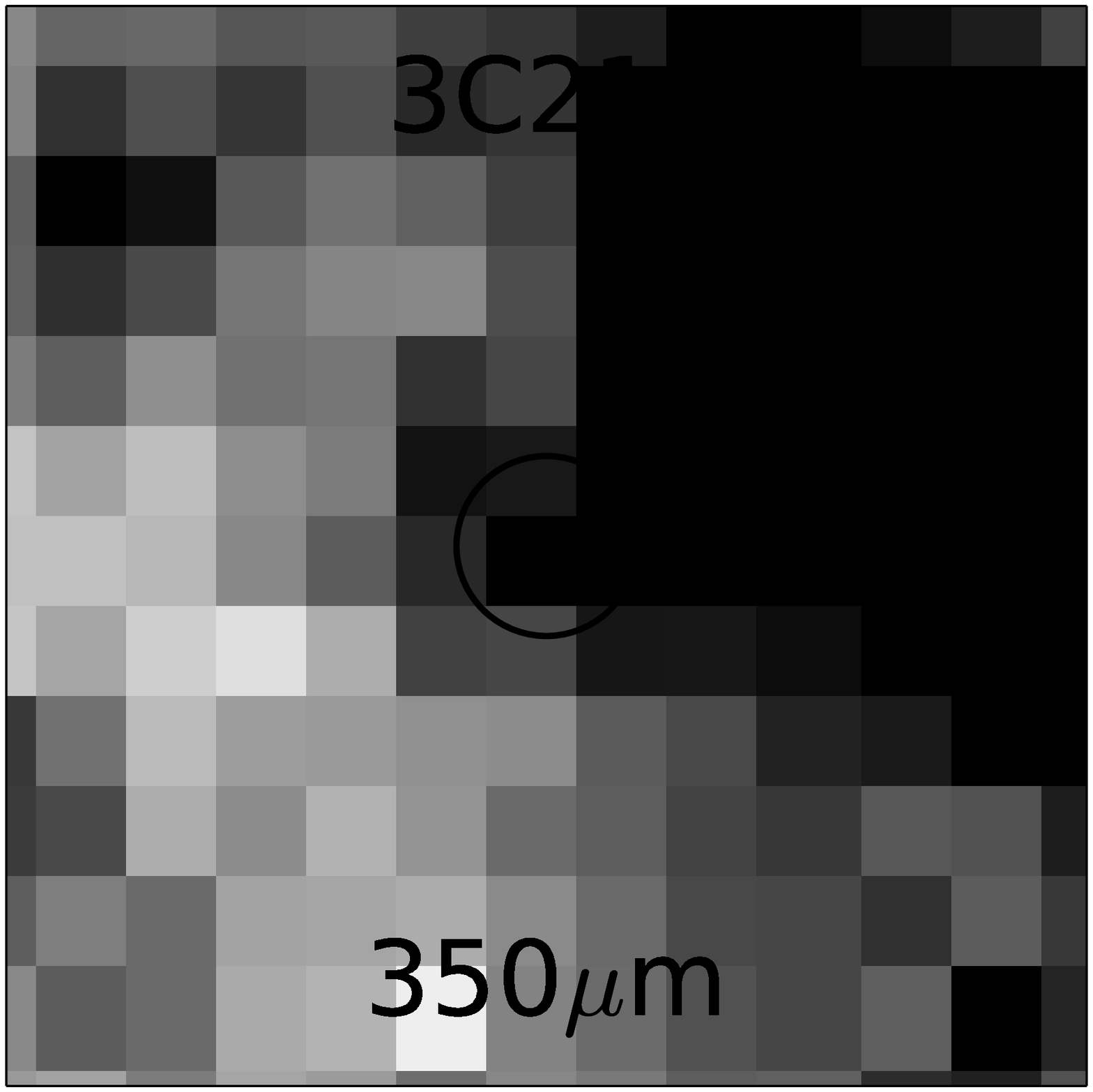}
      \includegraphics[width=1.5cm]{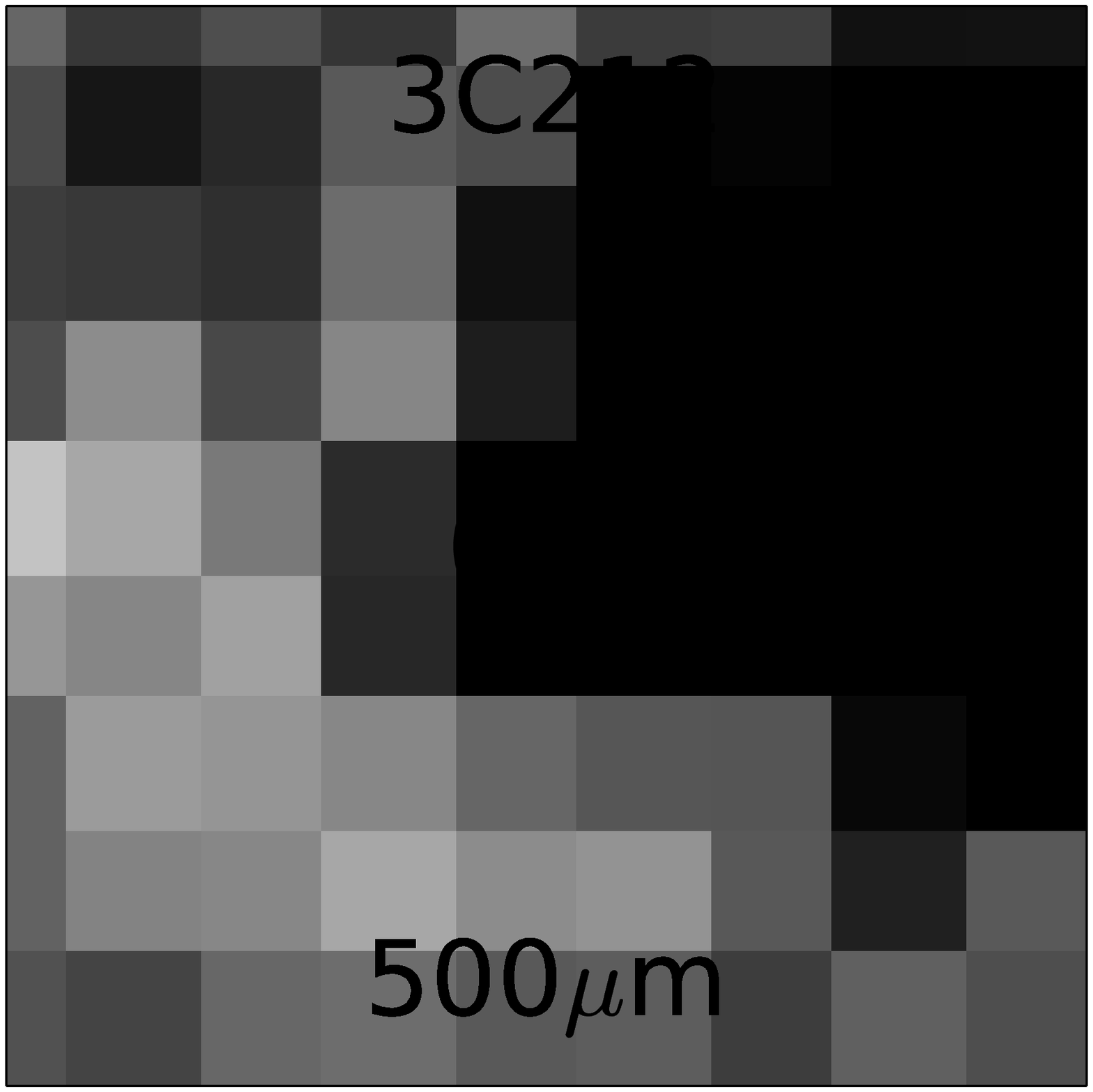}
      \\
      \includegraphics[width=1.5cm]{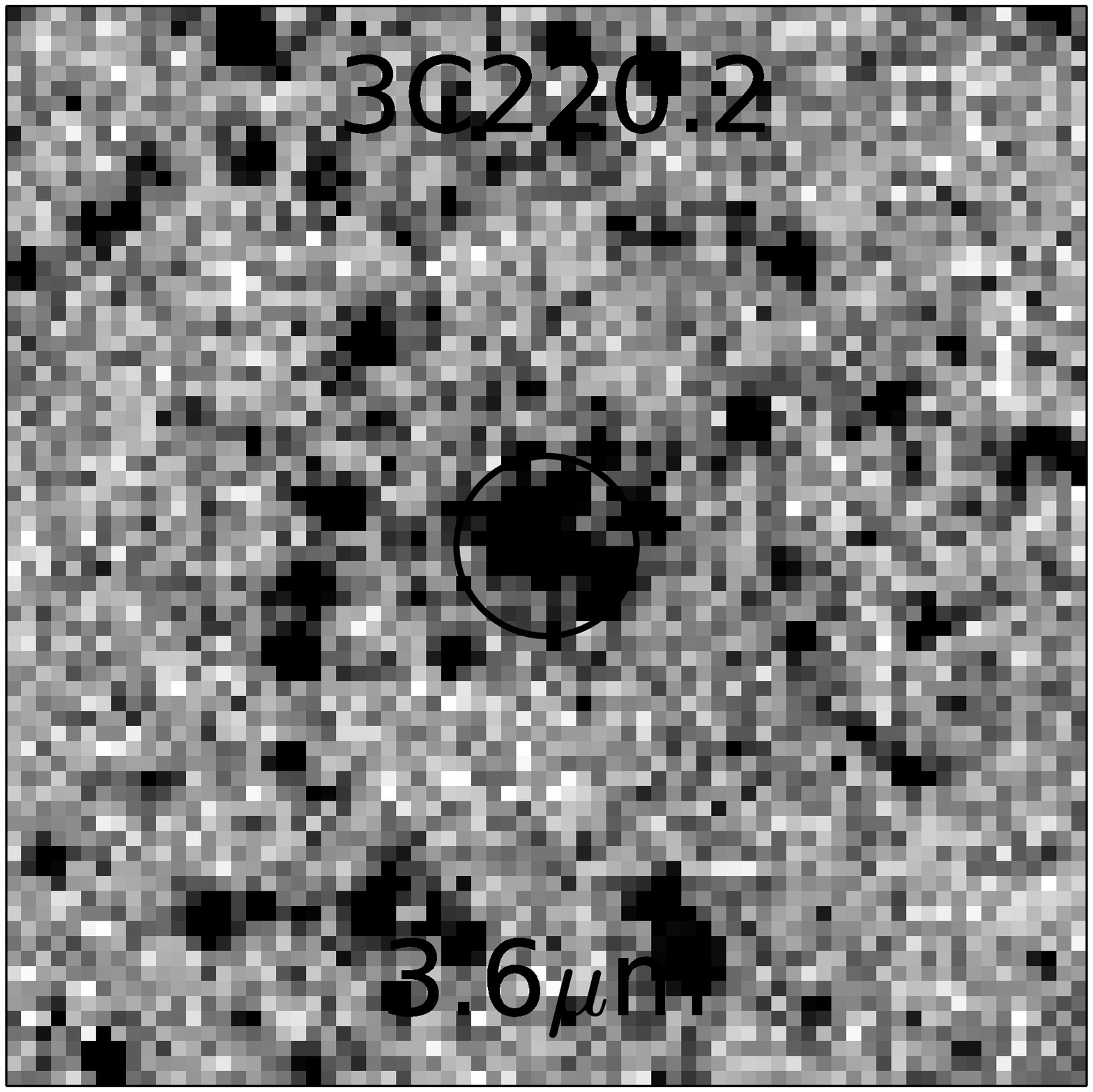}
      \includegraphics[width=1.5cm]{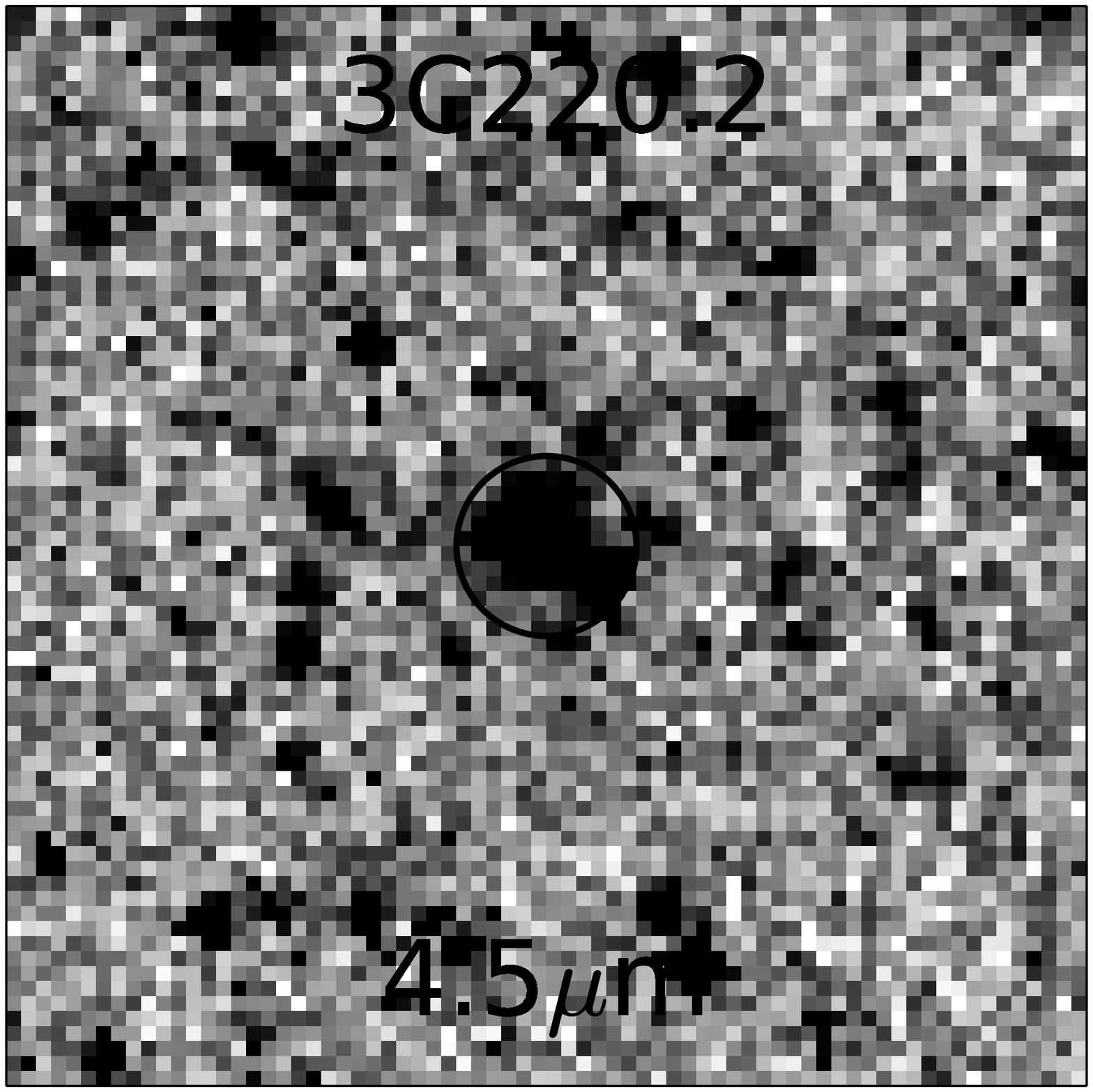}
      \includegraphics[width=1.5cm]{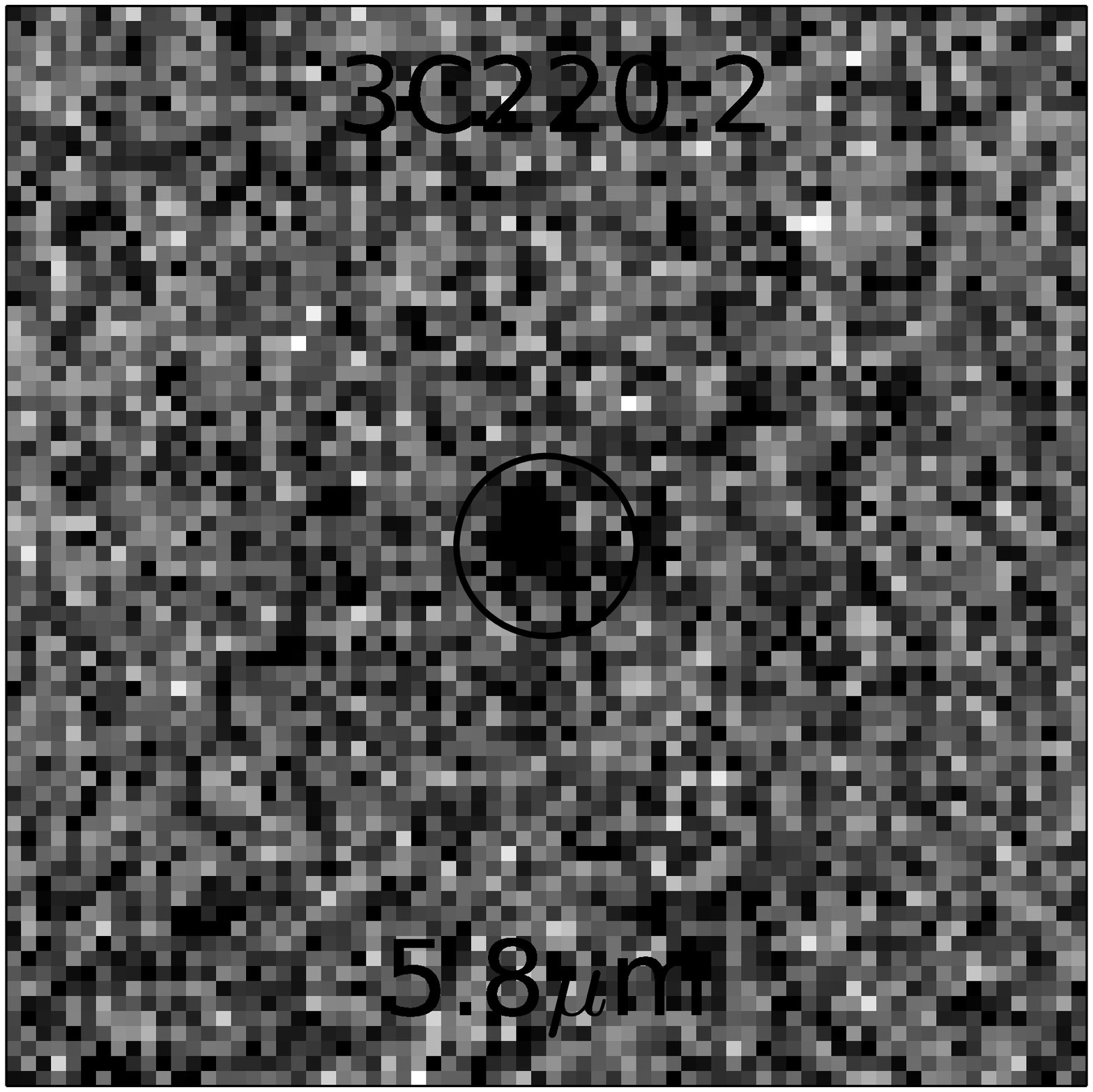}
      \includegraphics[width=1.5cm]{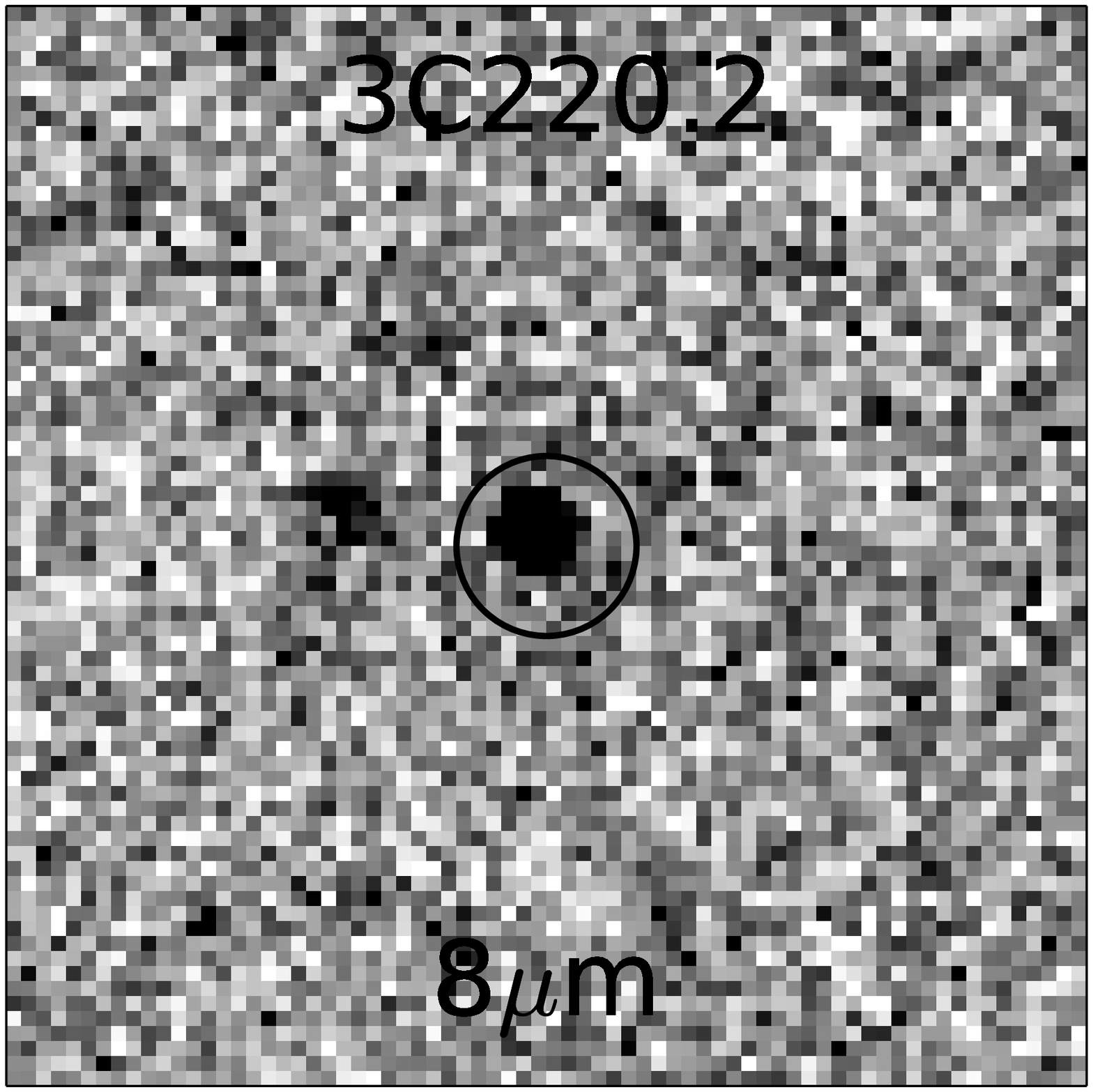}
      \includegraphics[width=1.5cm]{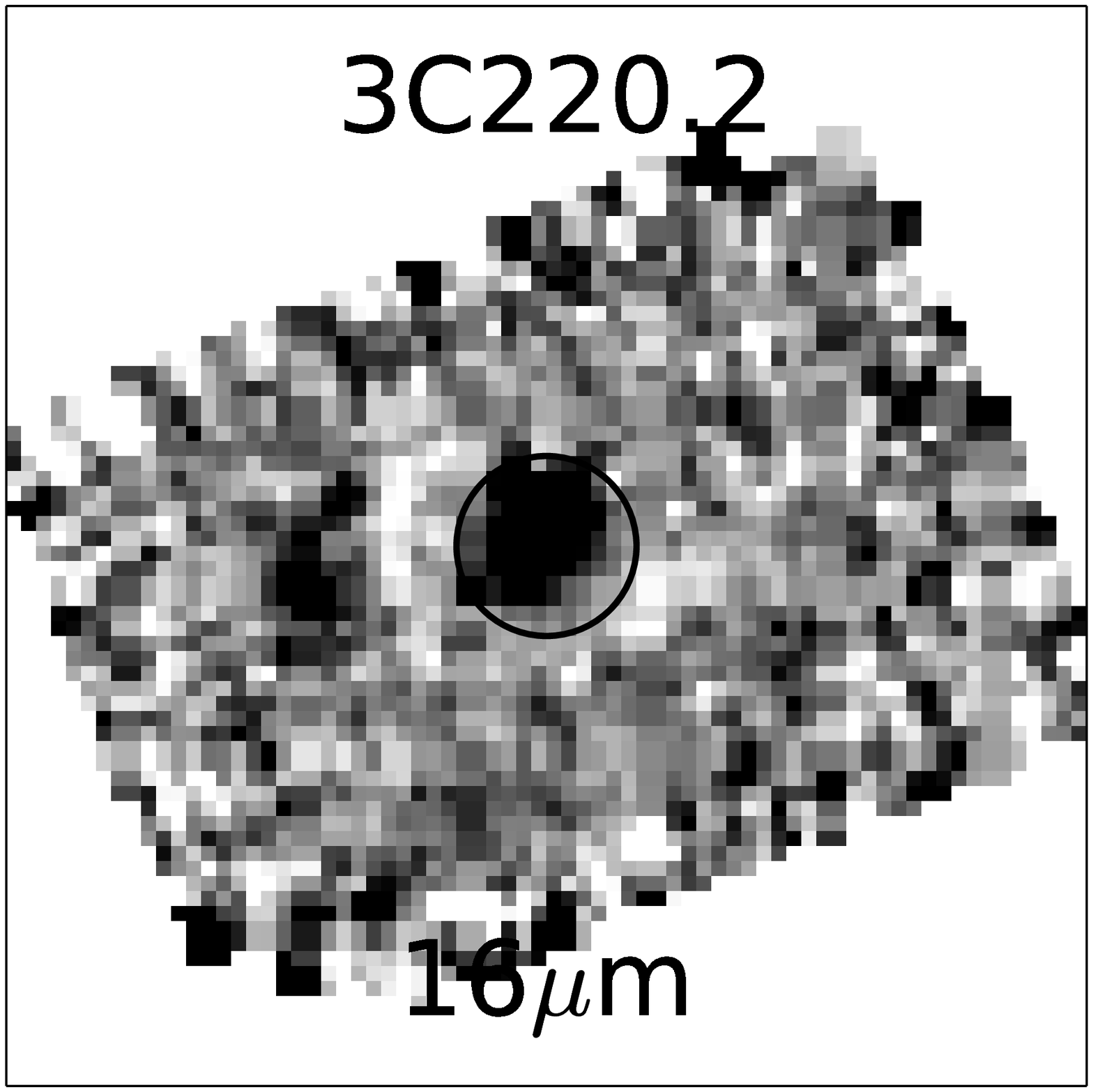}
      \includegraphics[width=1.5cm]{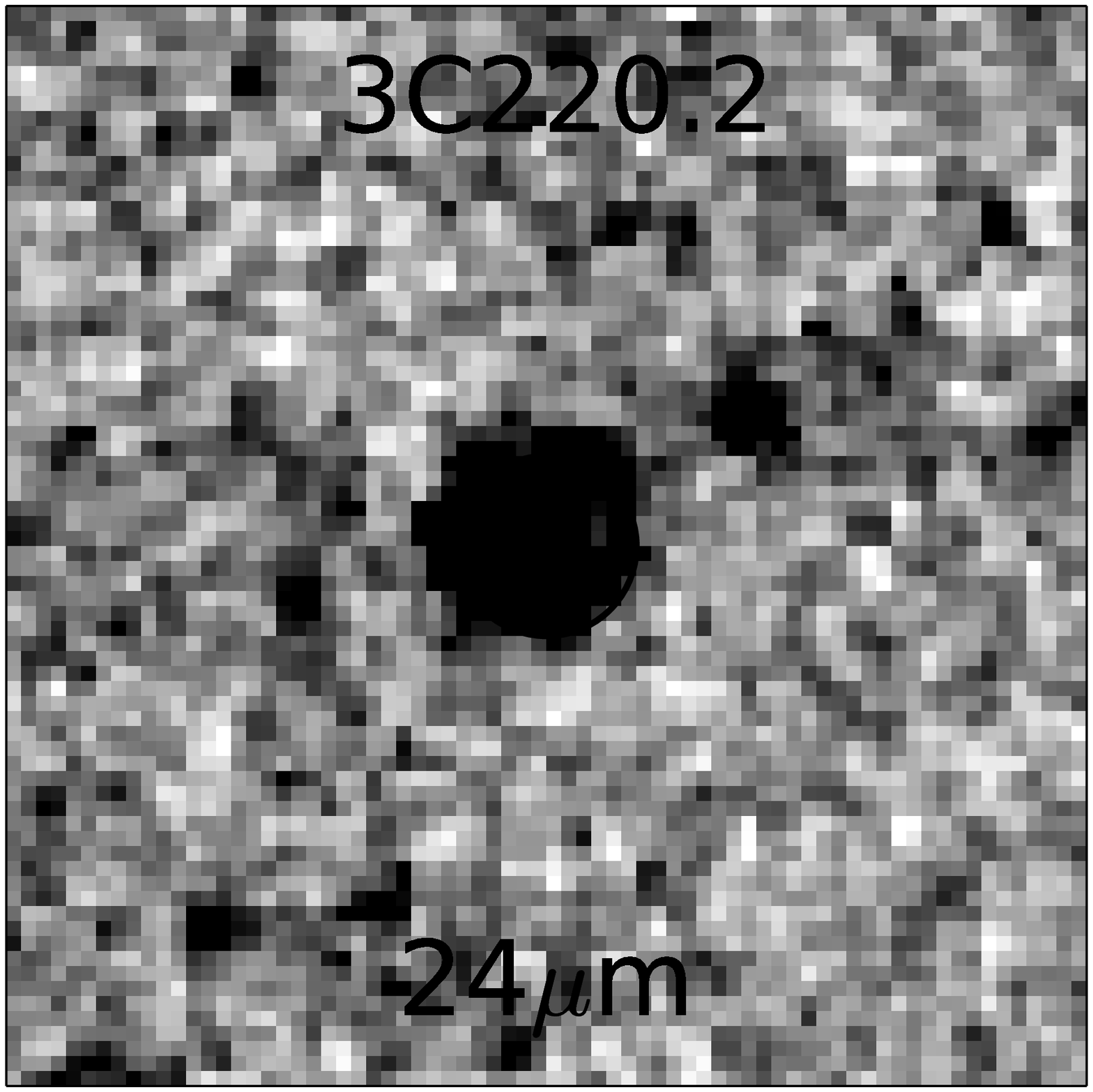}
      \includegraphics[width=1.5cm]{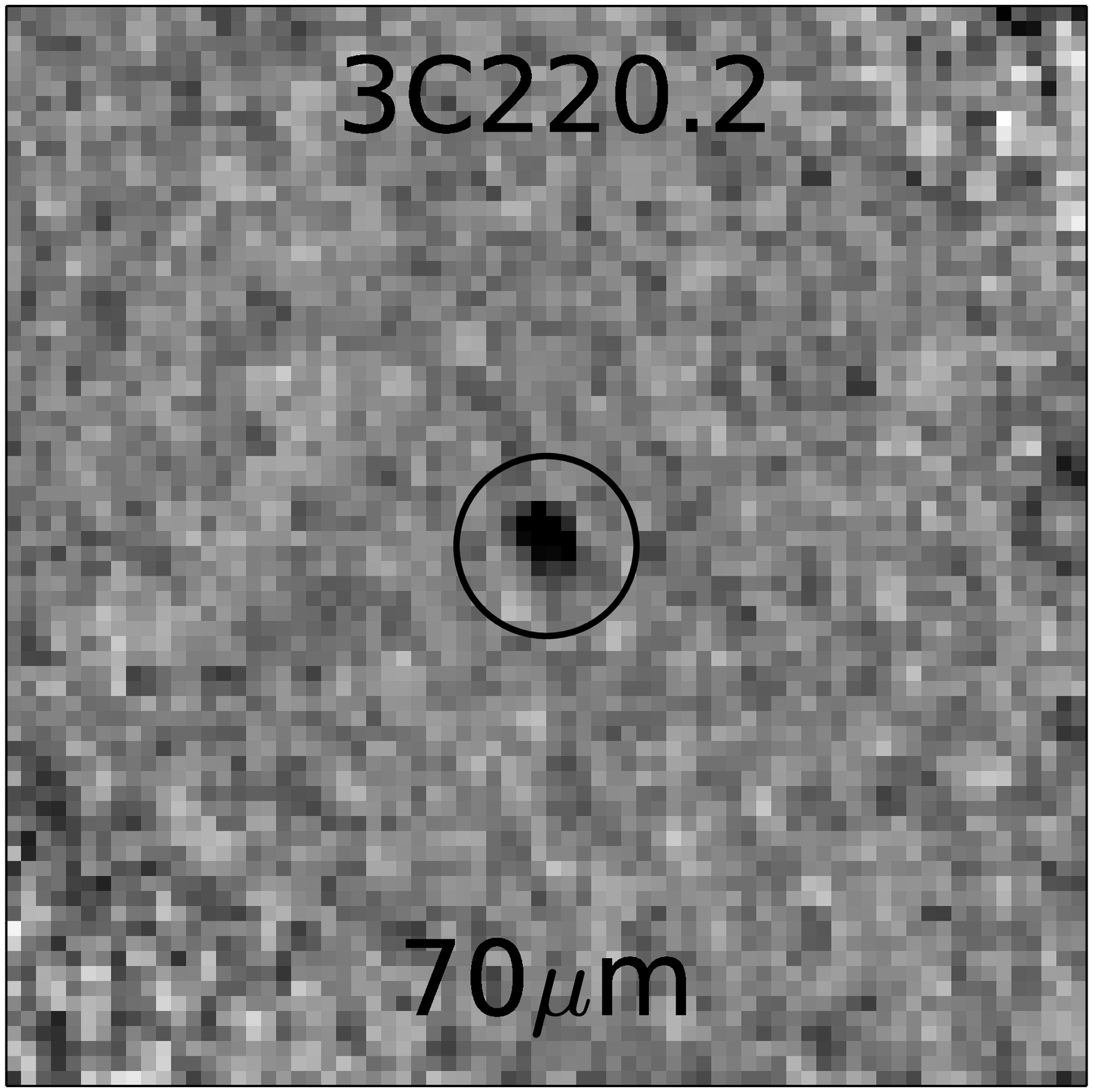}
      \includegraphics[width=1.5cm]{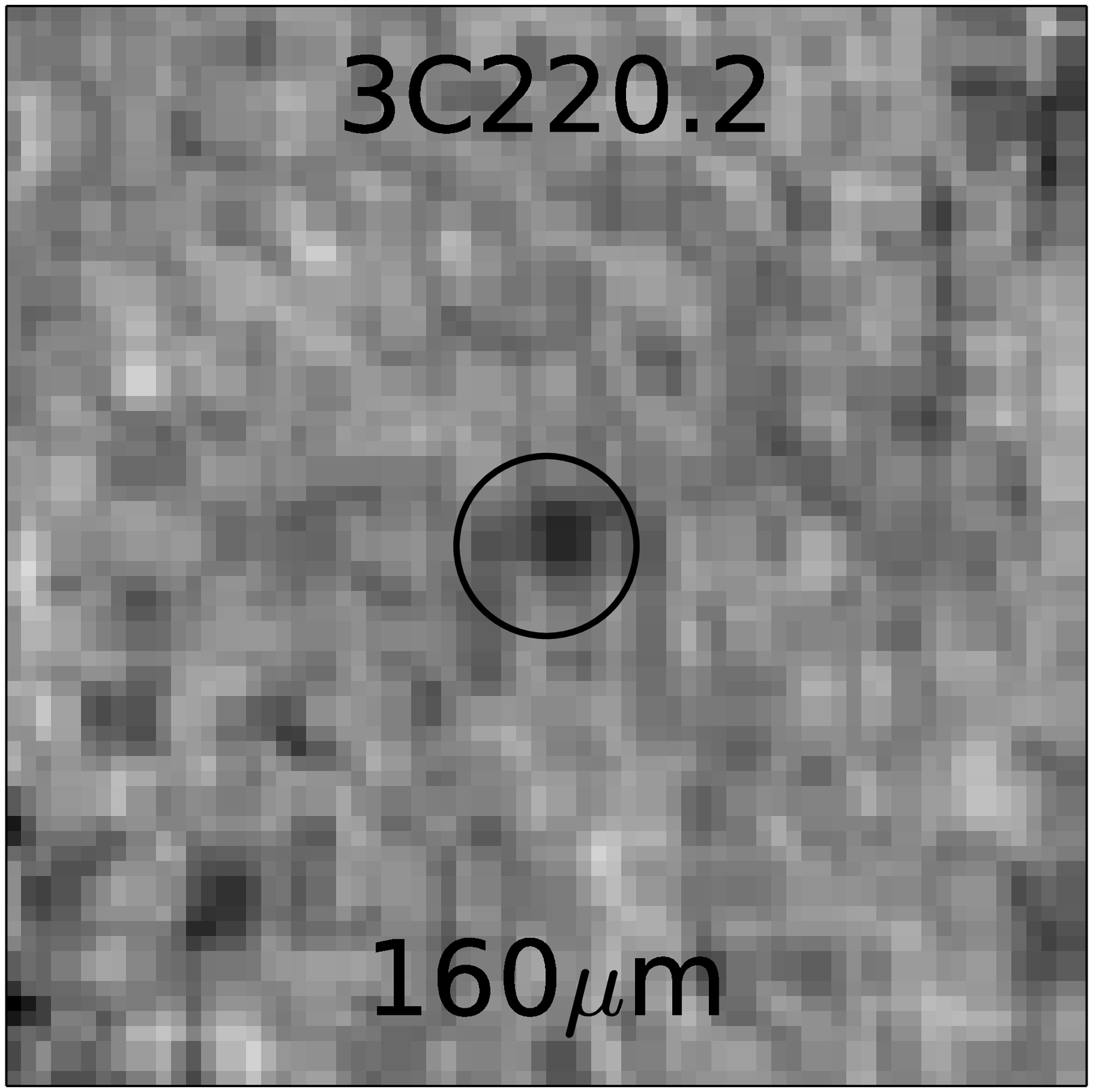}
      \includegraphics[width=1.5cm]{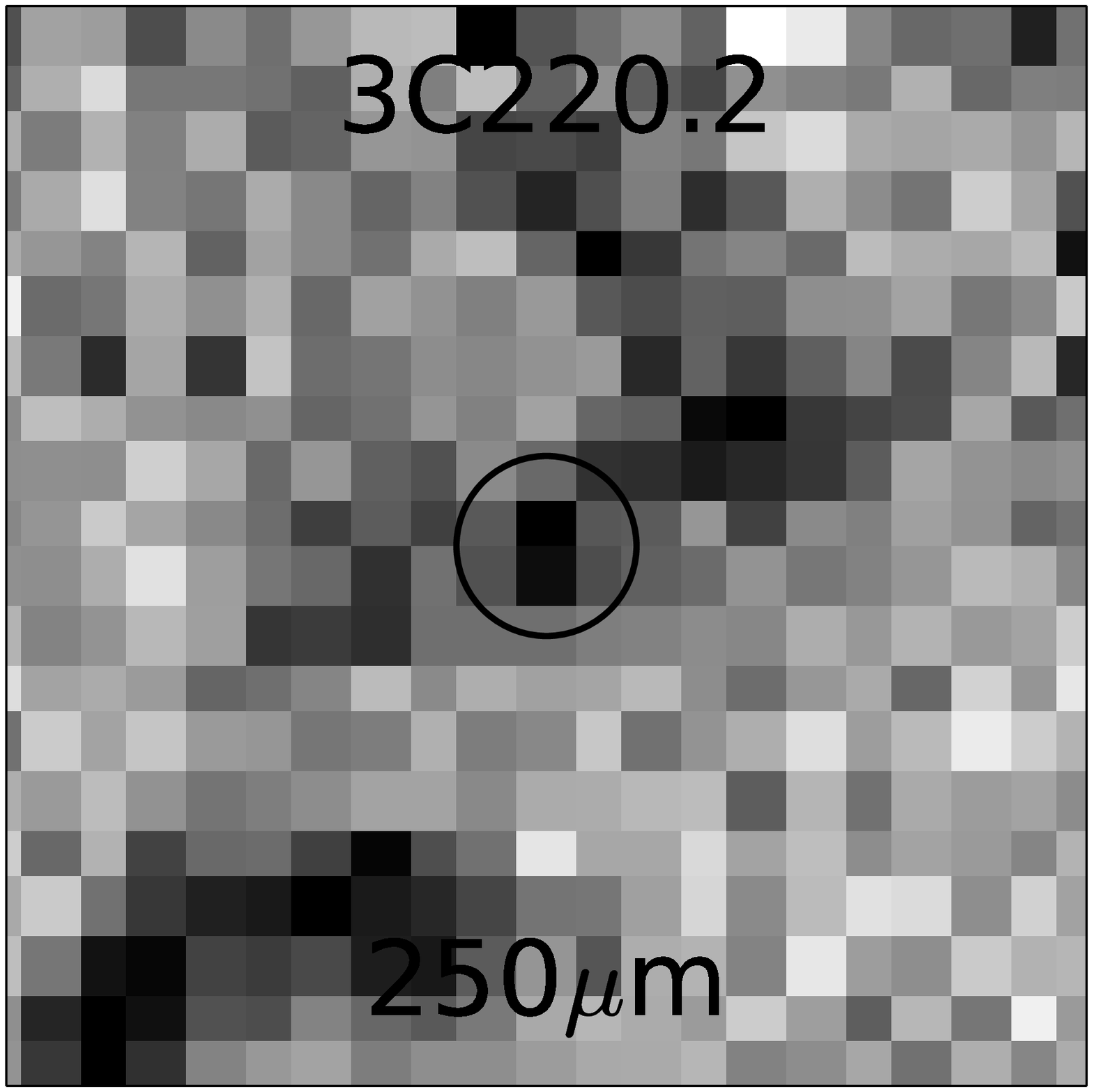}
      \includegraphics[width=1.5cm]{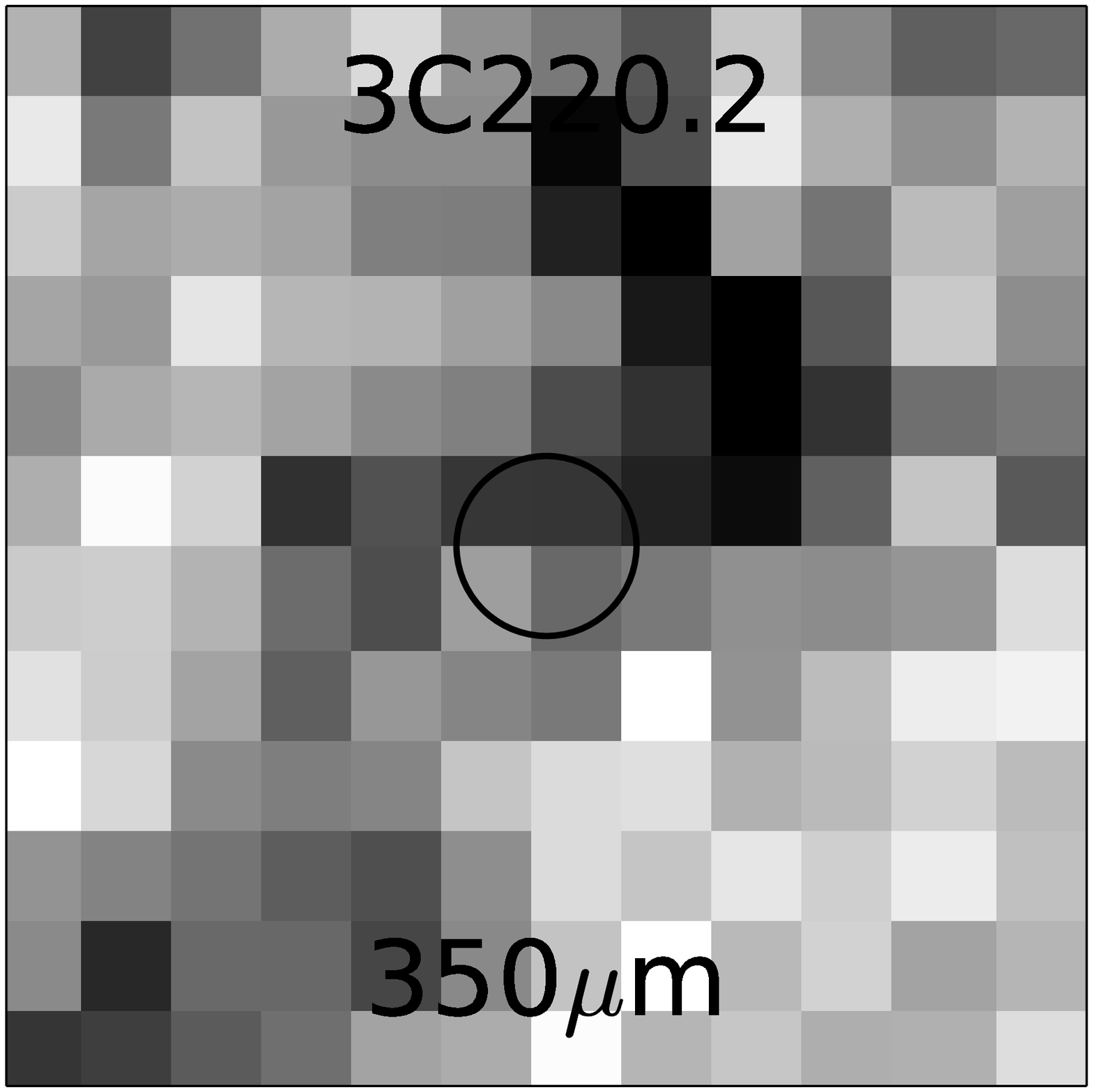}
      \includegraphics[width=1.5cm]{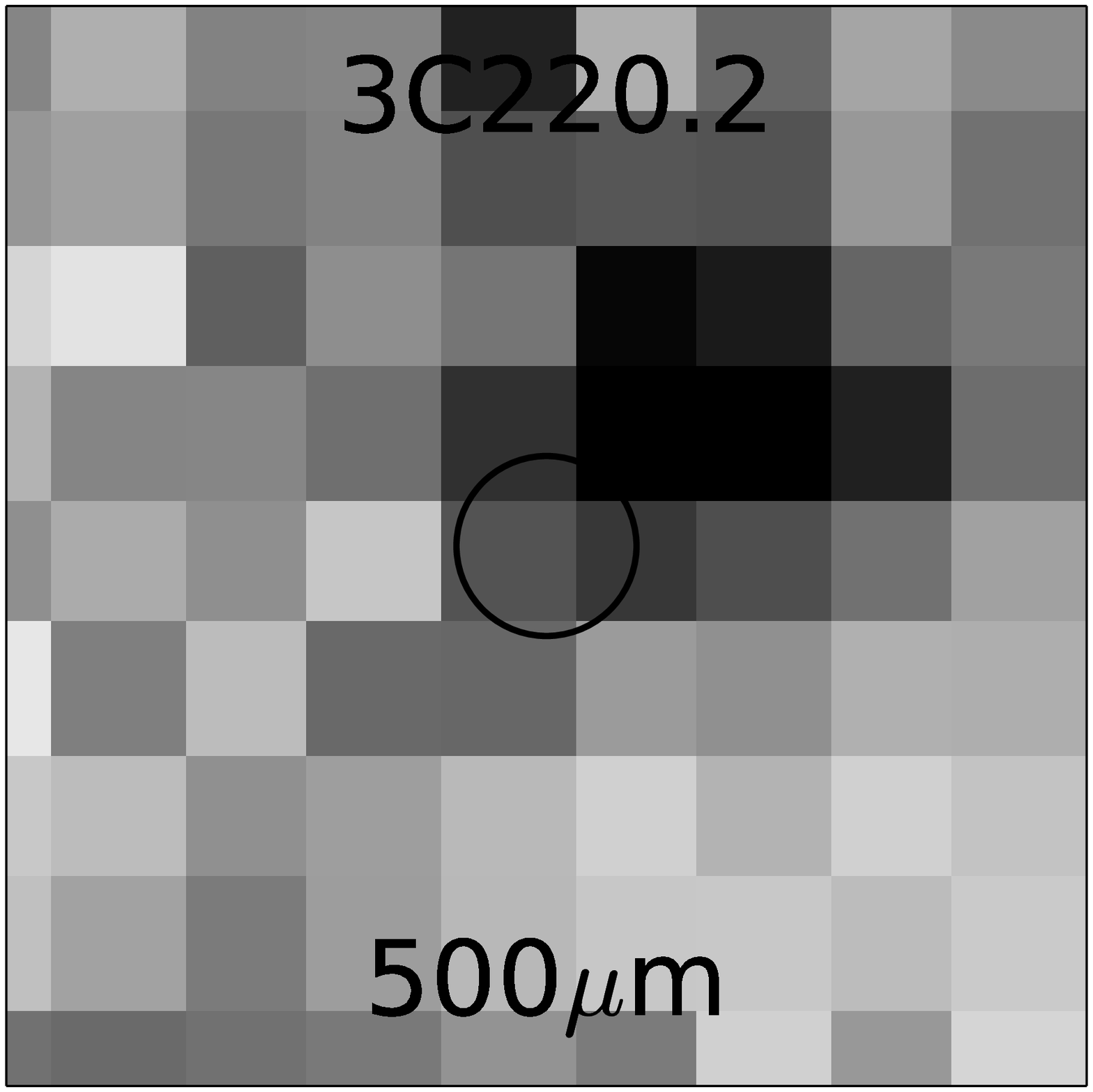}
      \\
      \includegraphics[width=1.5cm]{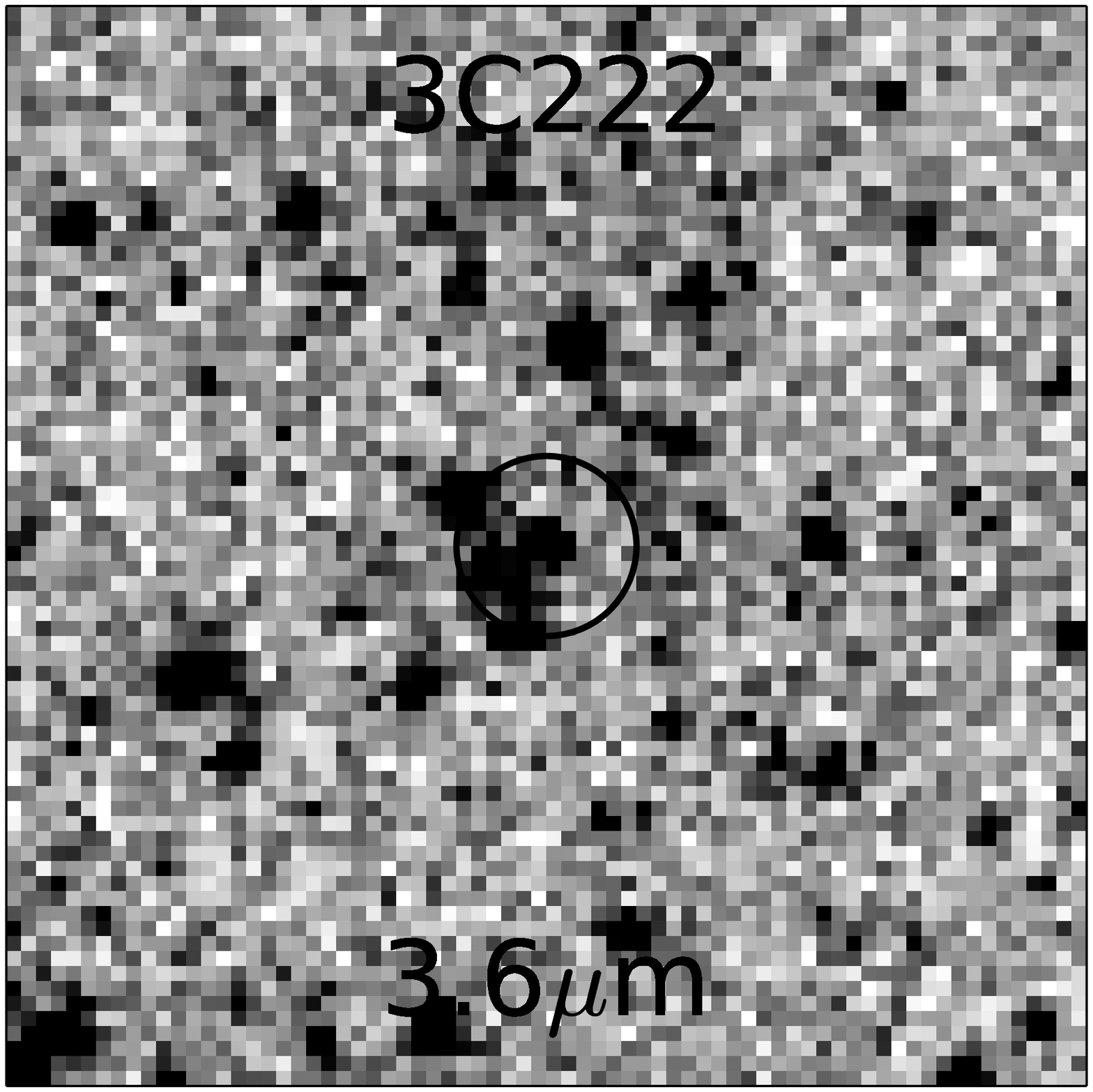}
      \includegraphics[width=1.5cm]{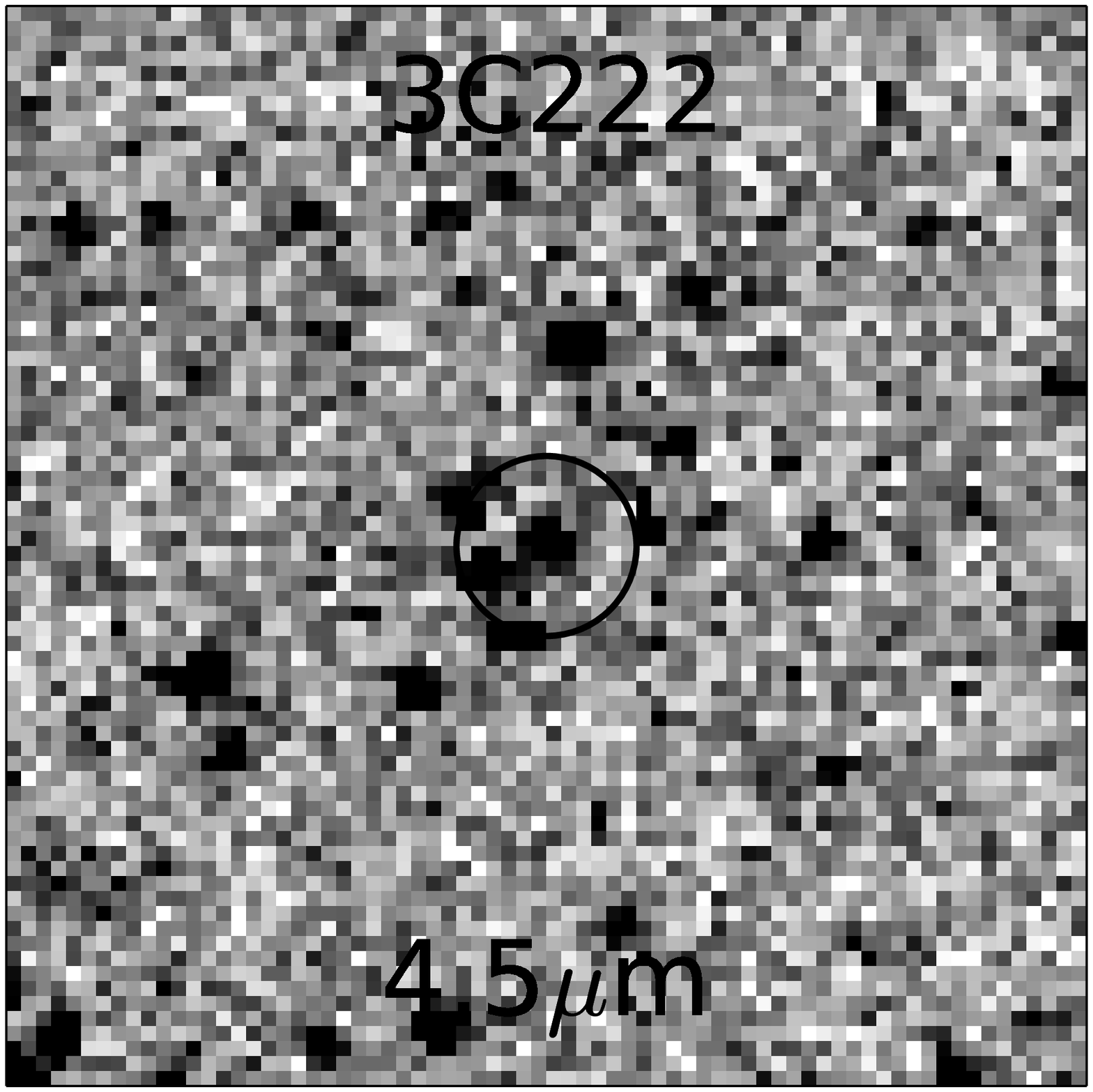}
      \includegraphics[width=1.5cm]{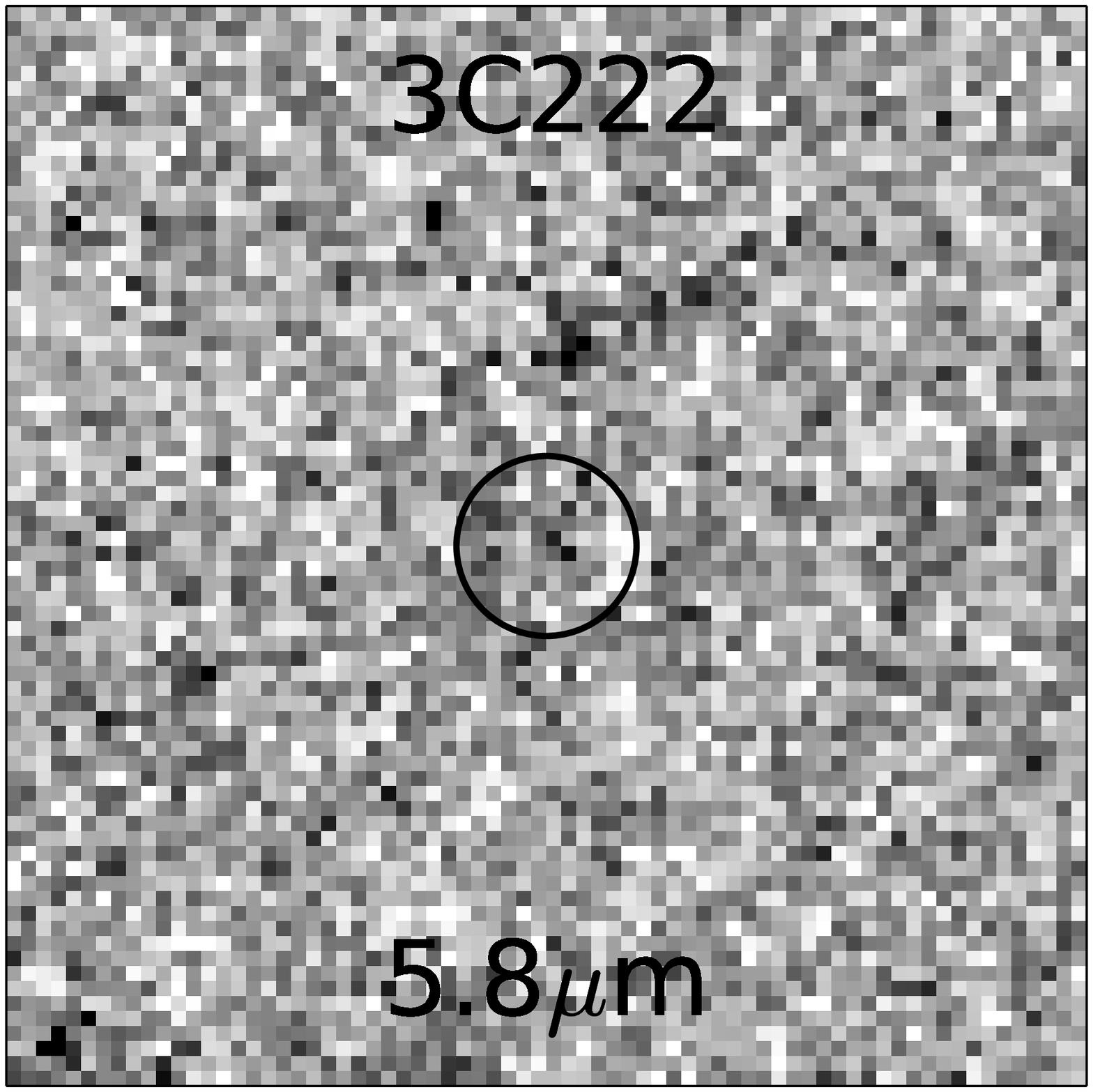}
      \includegraphics[width=1.5cm]{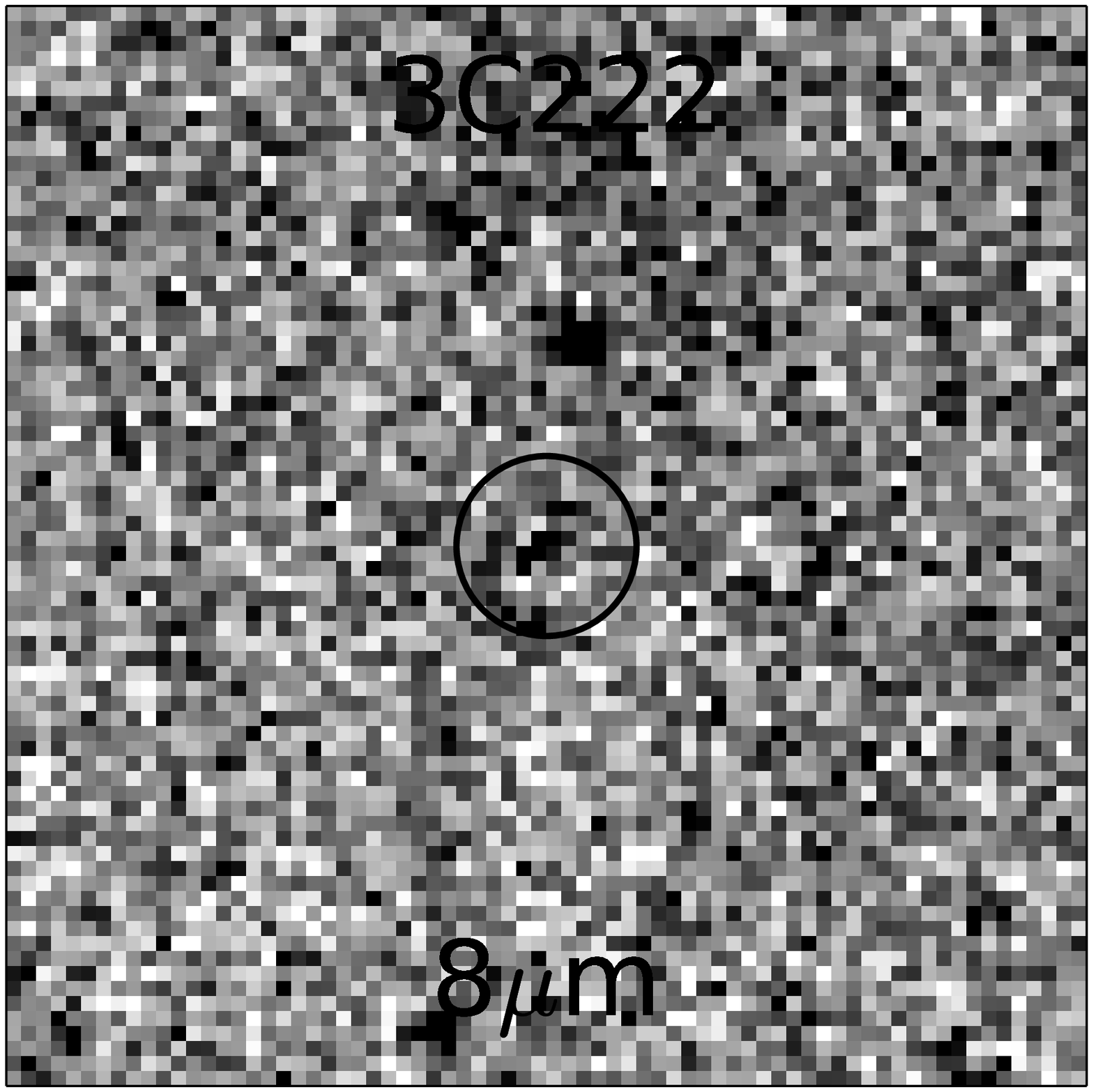}
      \includegraphics[width=1.5cm]{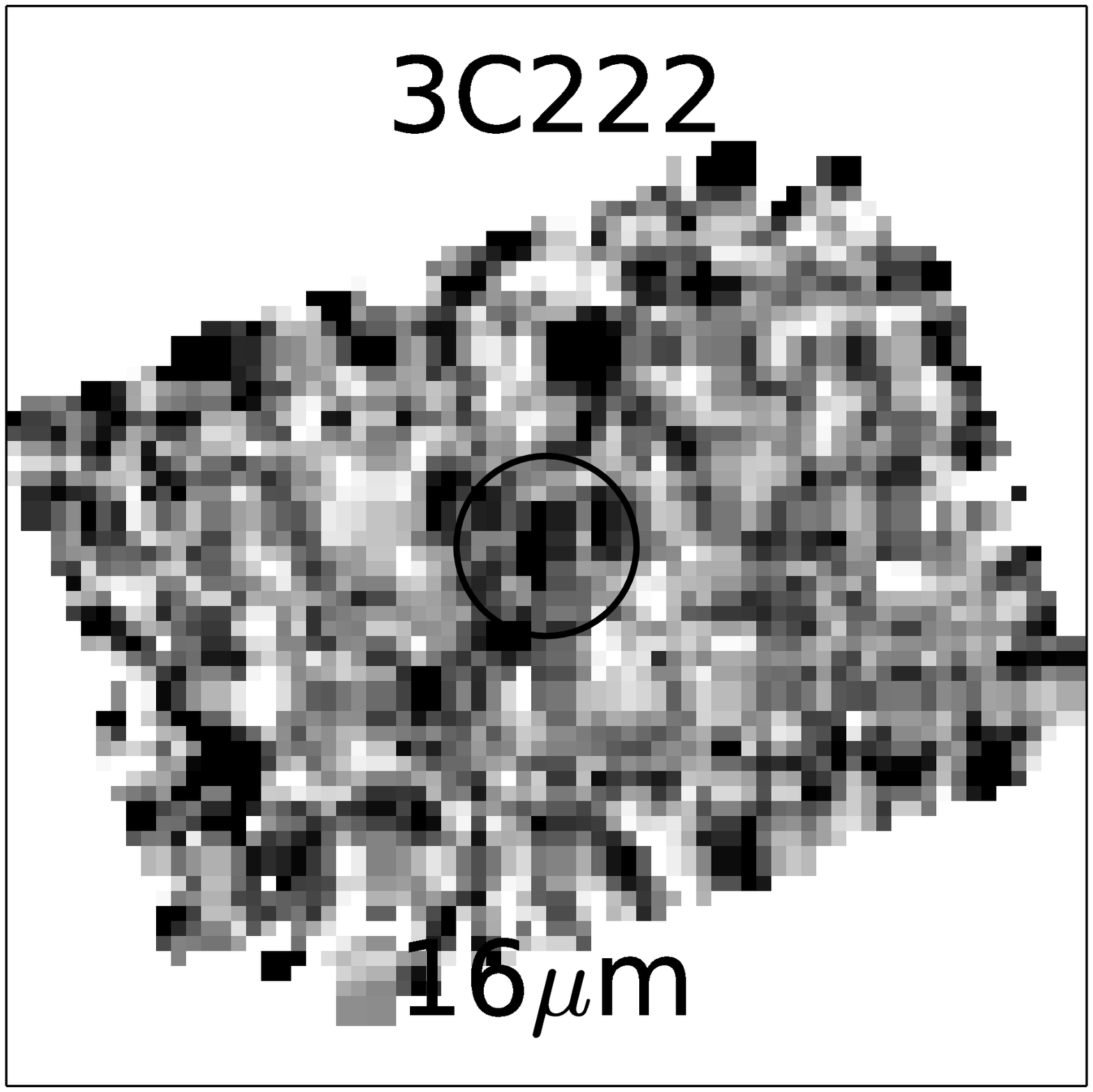}
      \includegraphics[width=1.5cm]{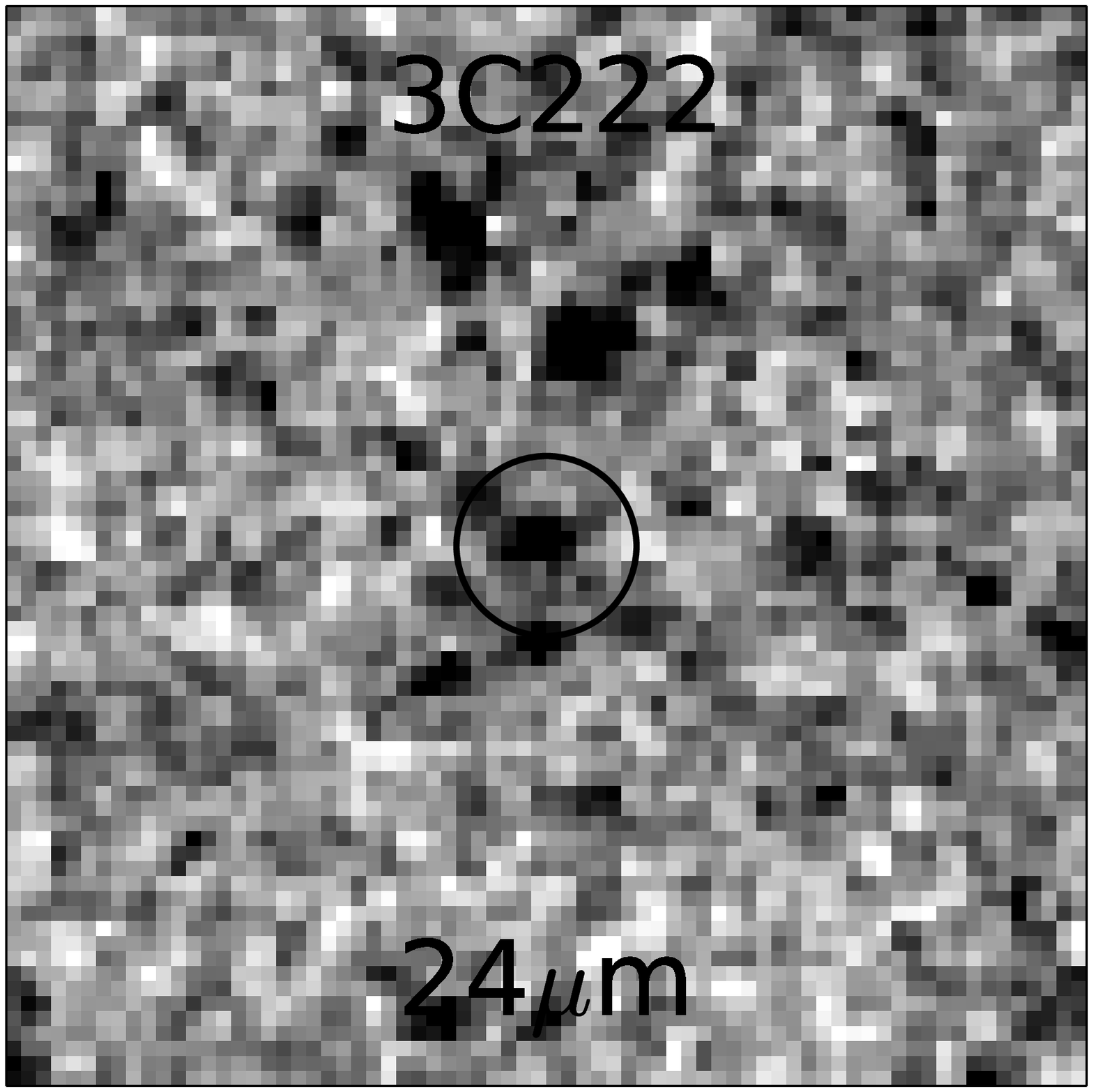}
      \includegraphics[width=1.5cm]{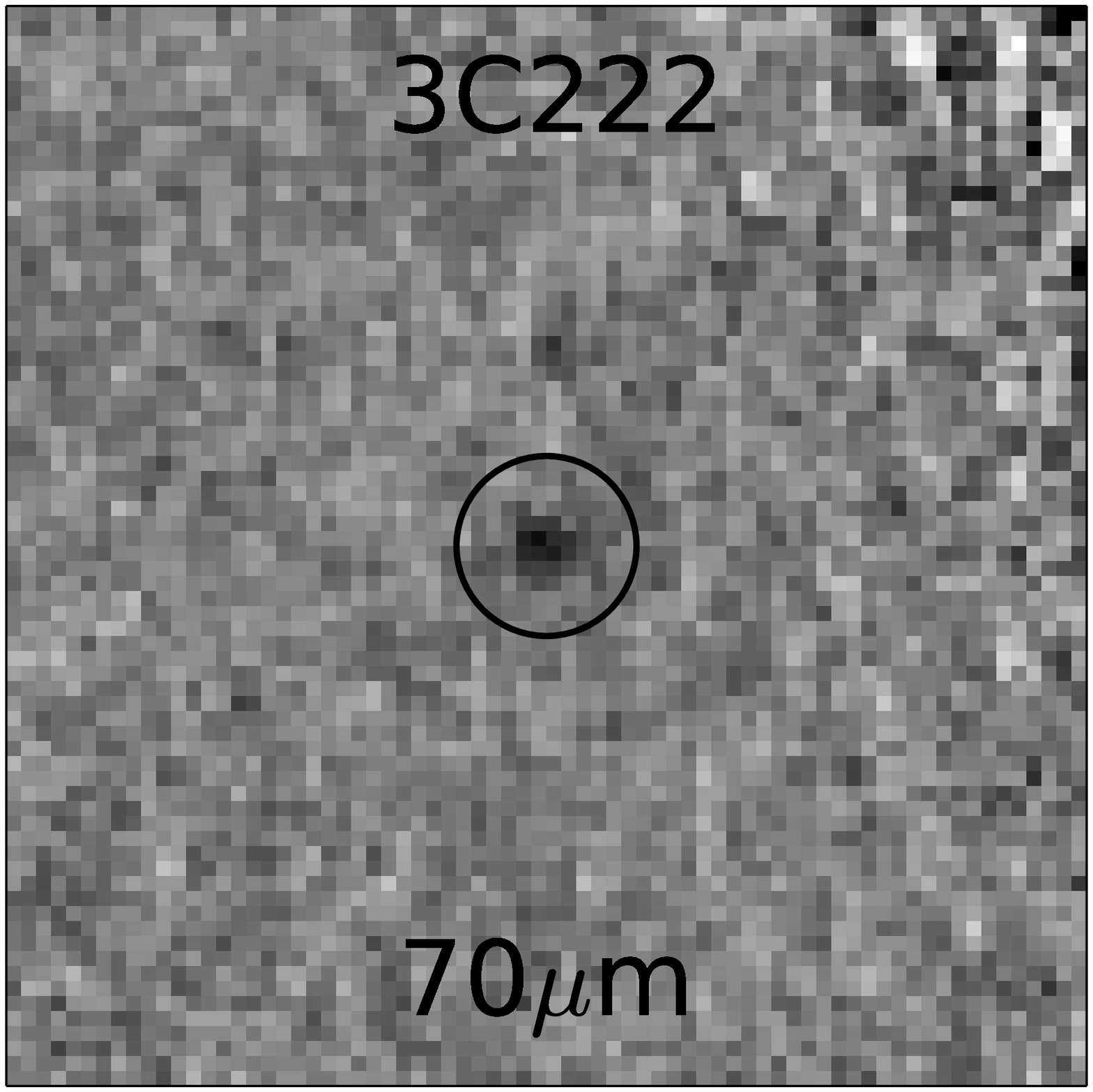}
      \includegraphics[width=1.5cm]{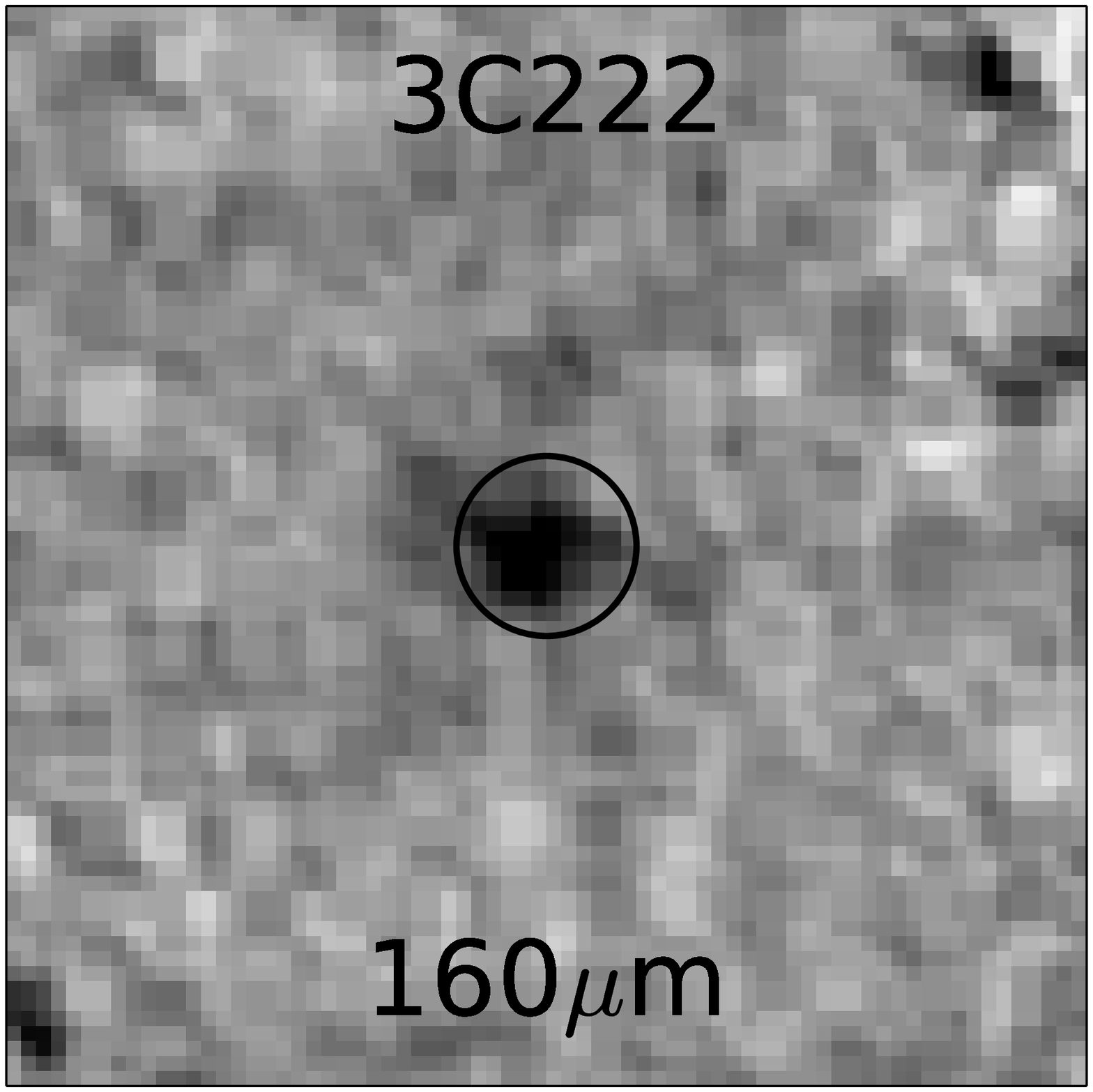}
      \includegraphics[width=1.5cm]{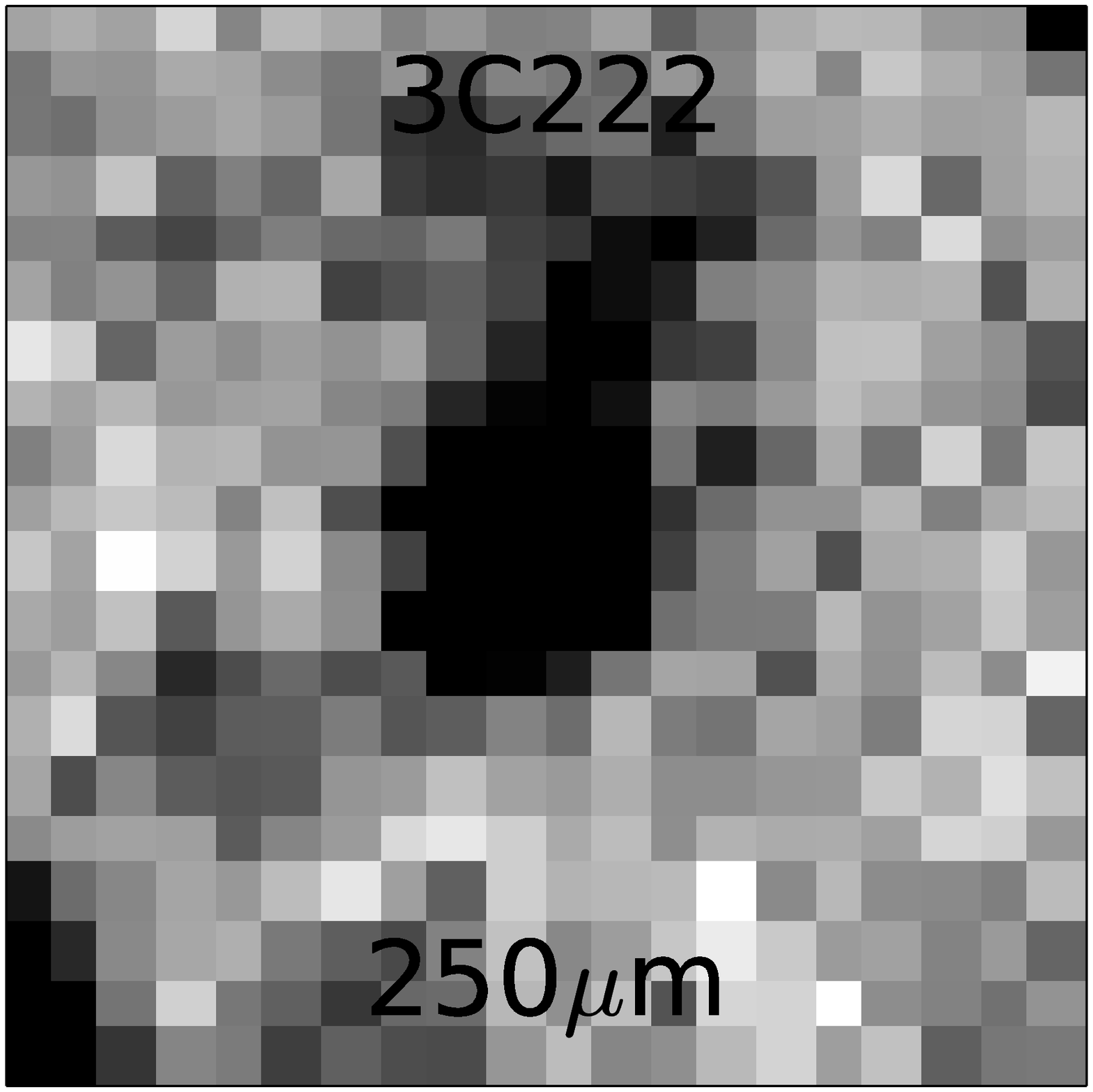}
      \includegraphics[width=1.5cm]{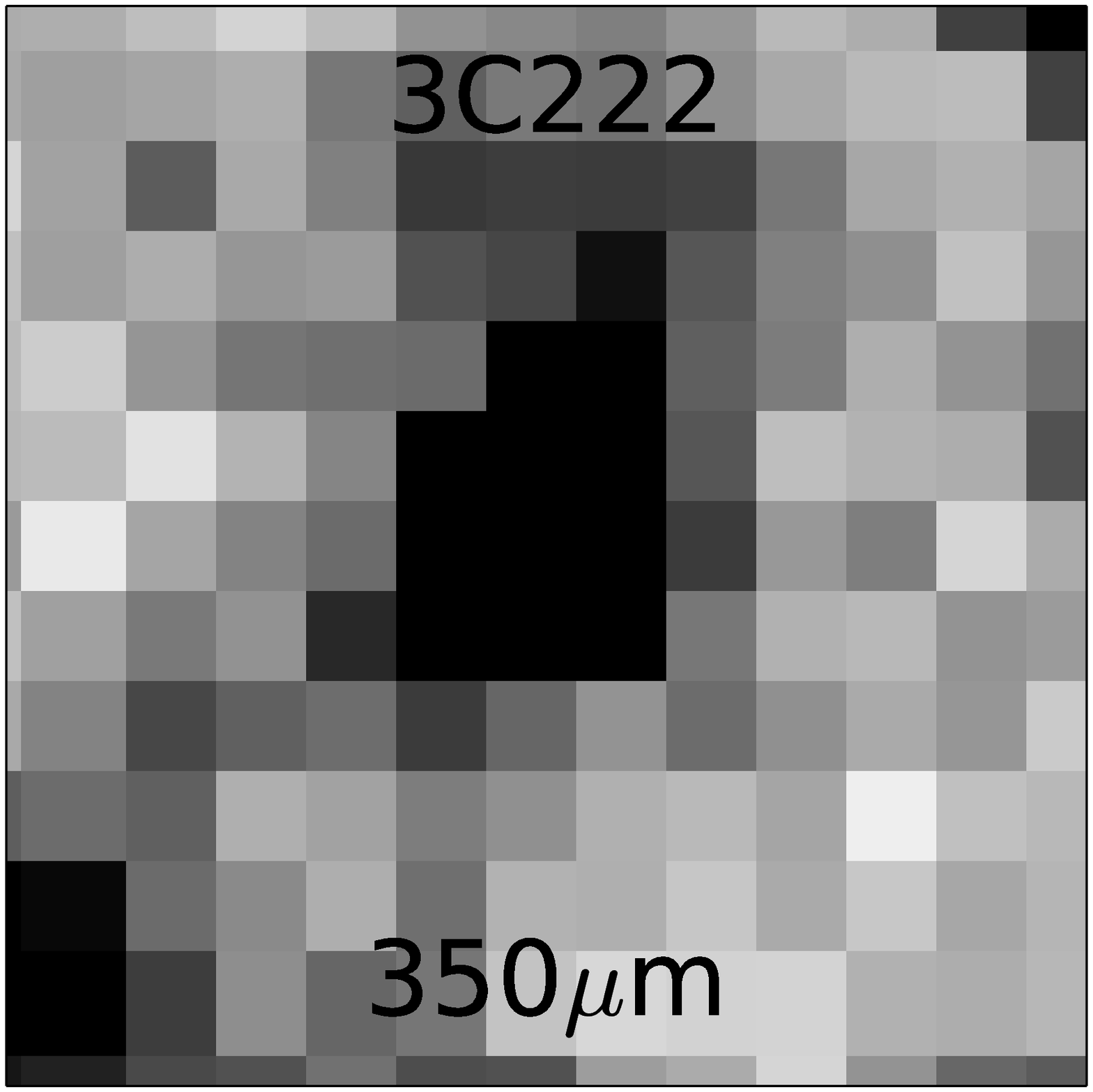}
      \includegraphics[width=1.5cm]{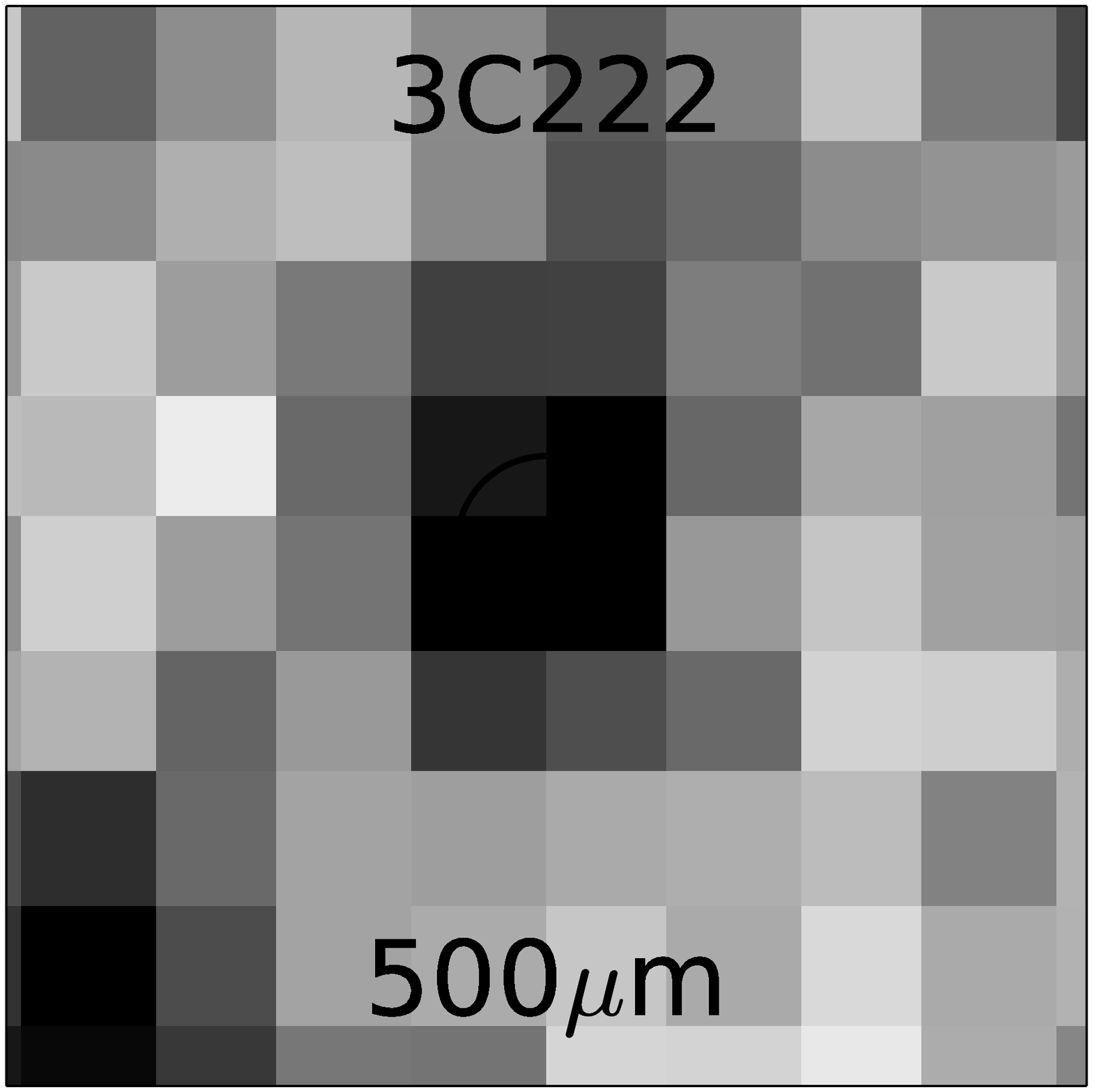}
      \\
      \includegraphics[width=1.5cm]{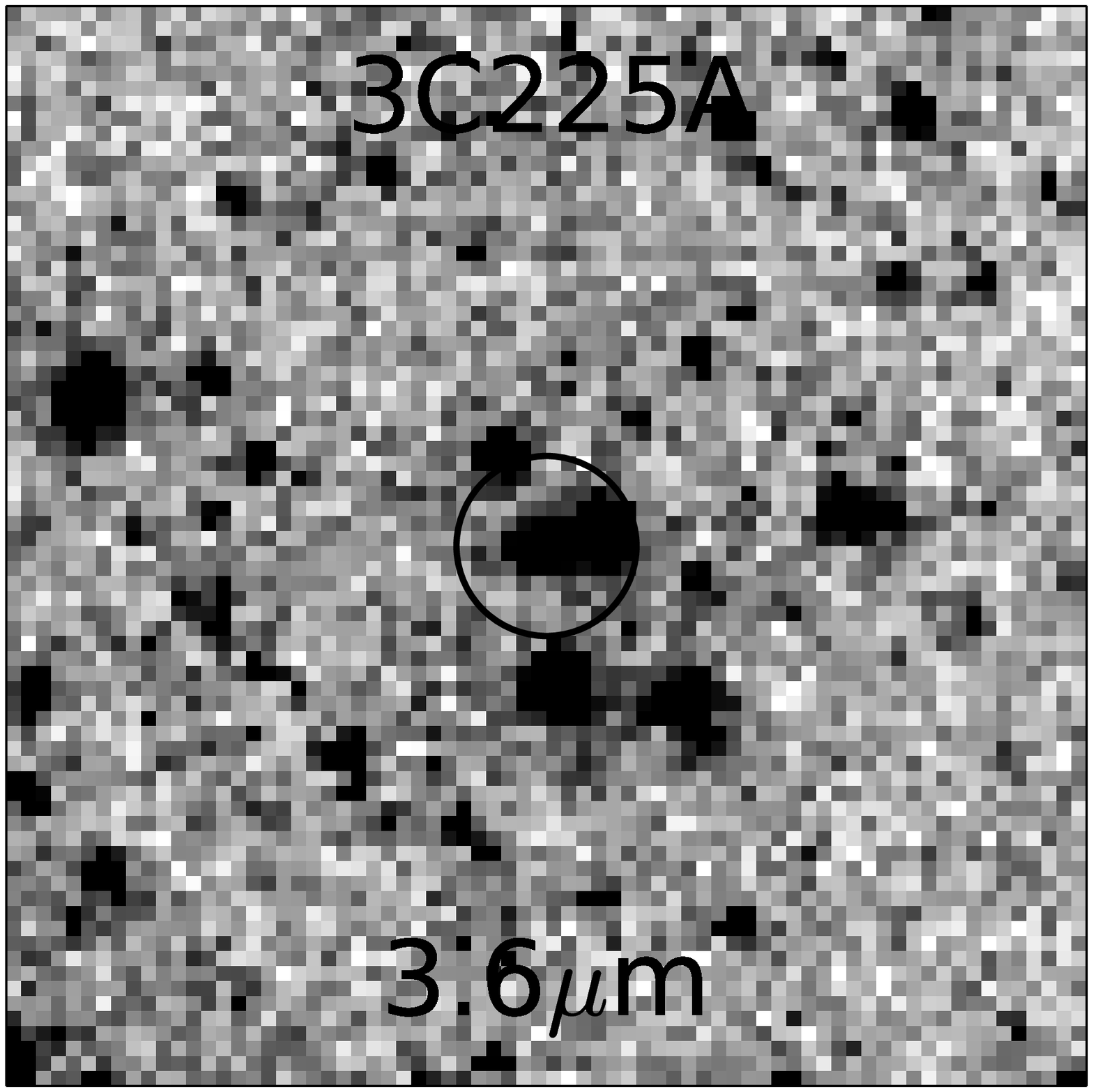}
      \includegraphics[width=1.5cm]{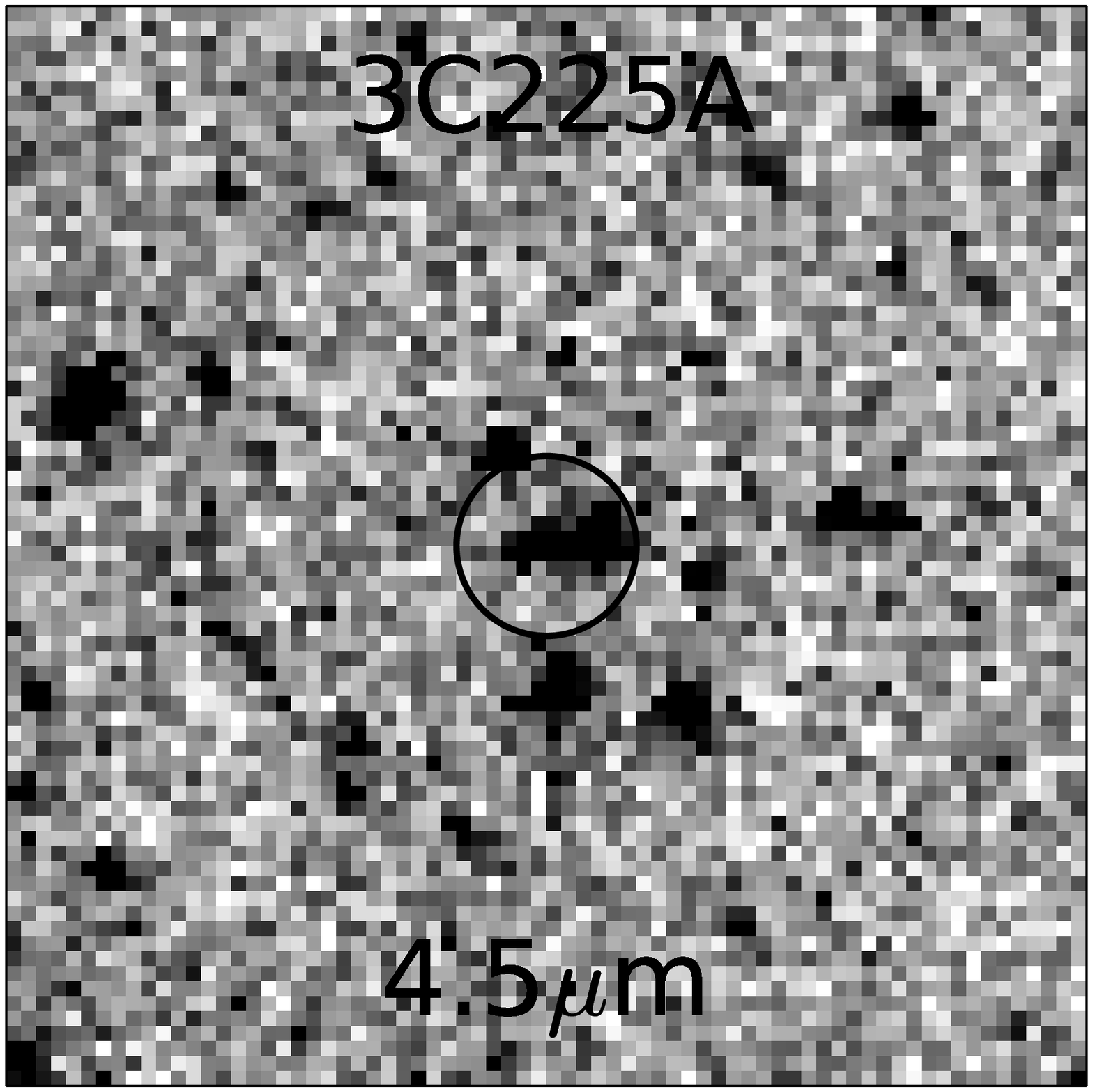}
      \includegraphics[width=1.5cm]{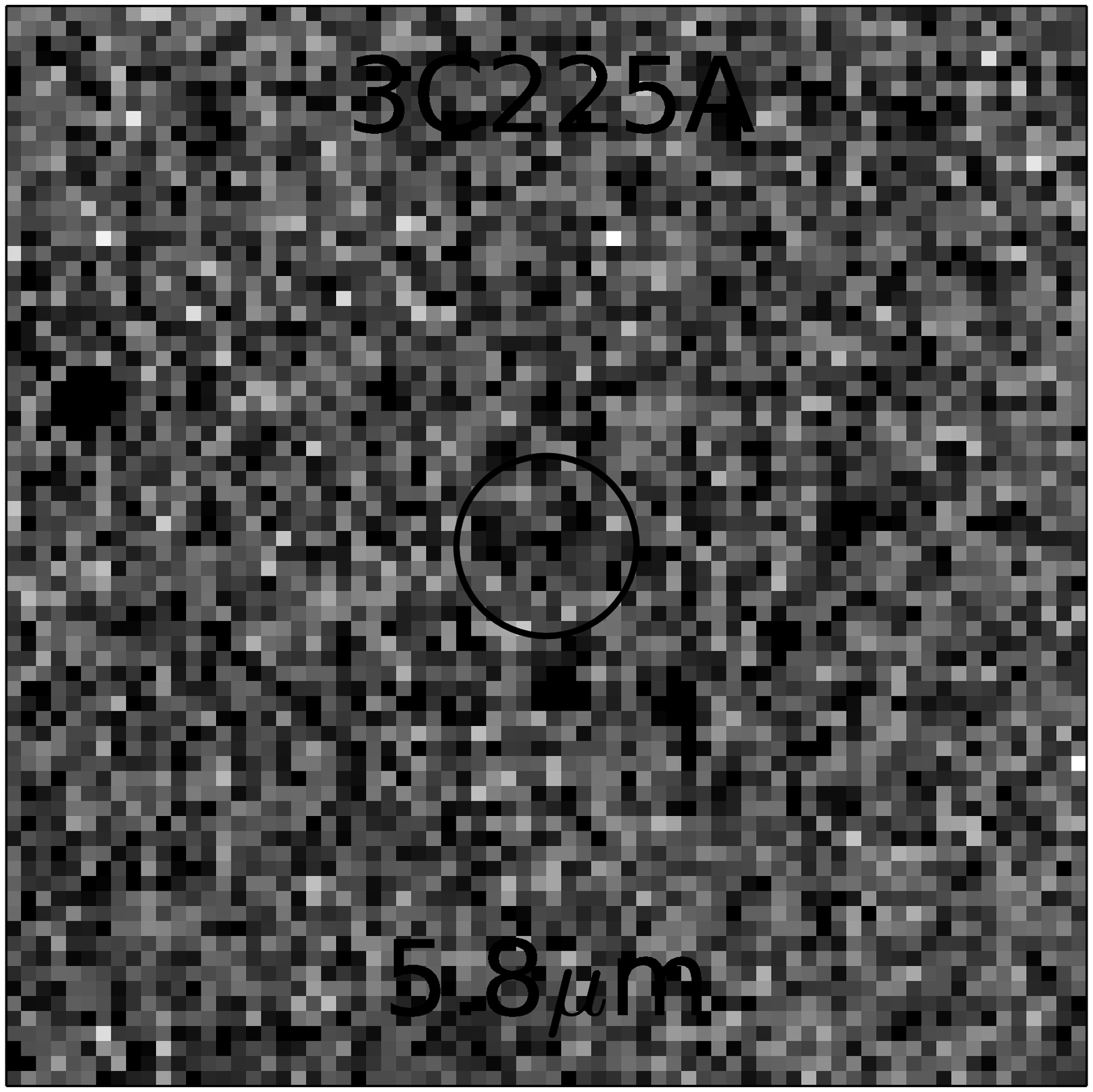}
      \includegraphics[width=1.5cm]{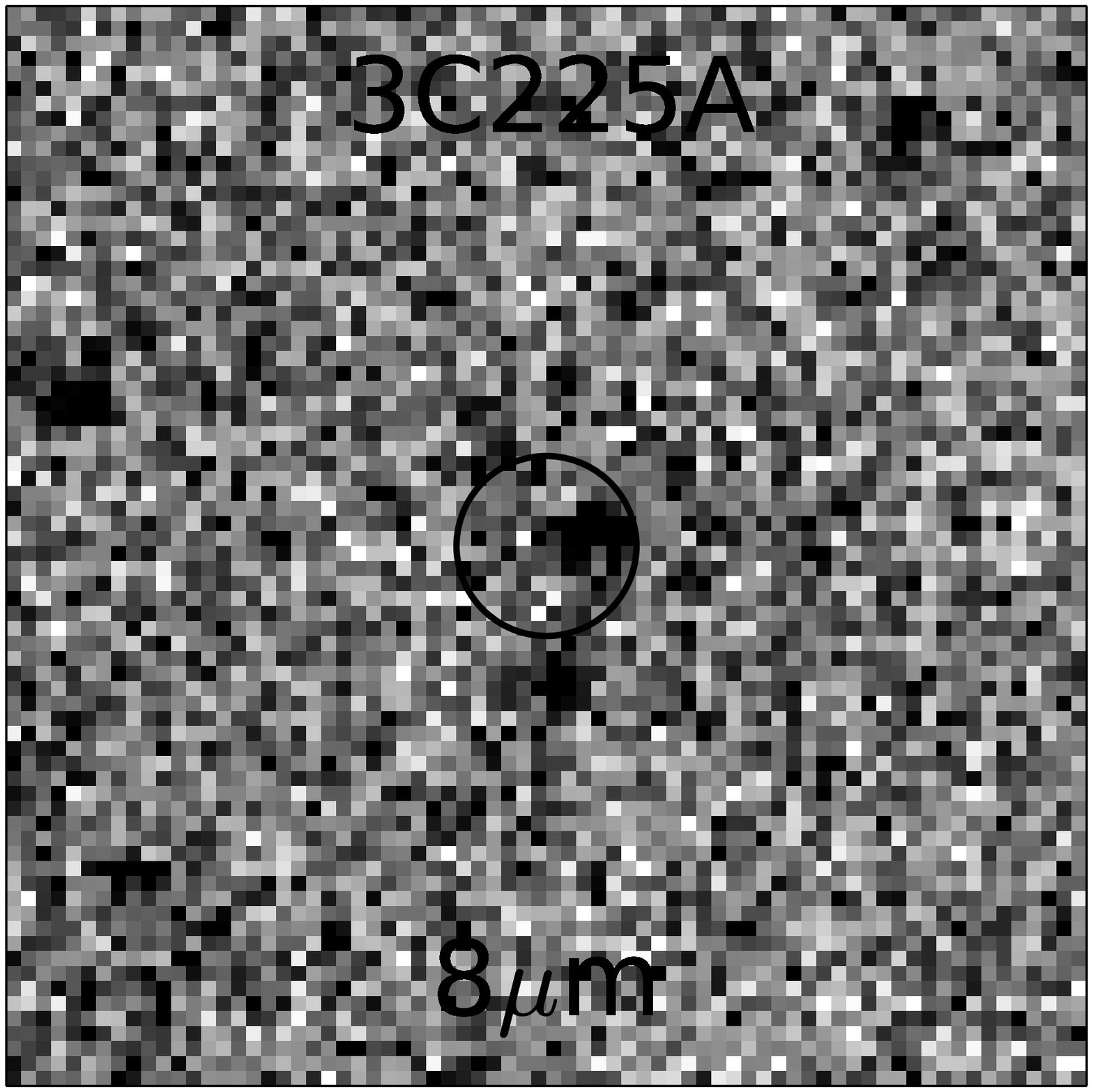}
      \includegraphics[width=1.5cm]{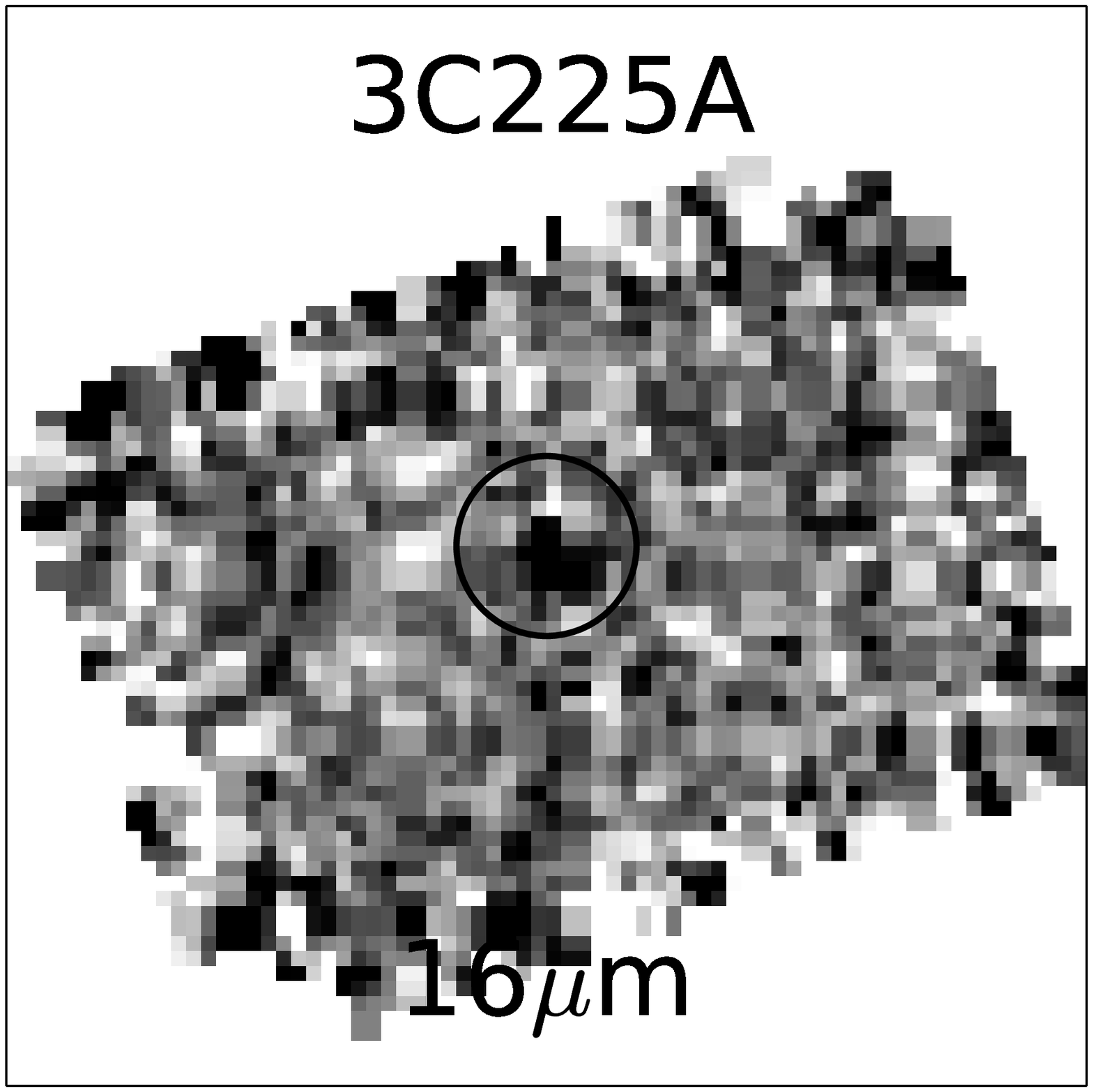}
      \includegraphics[width=1.5cm]{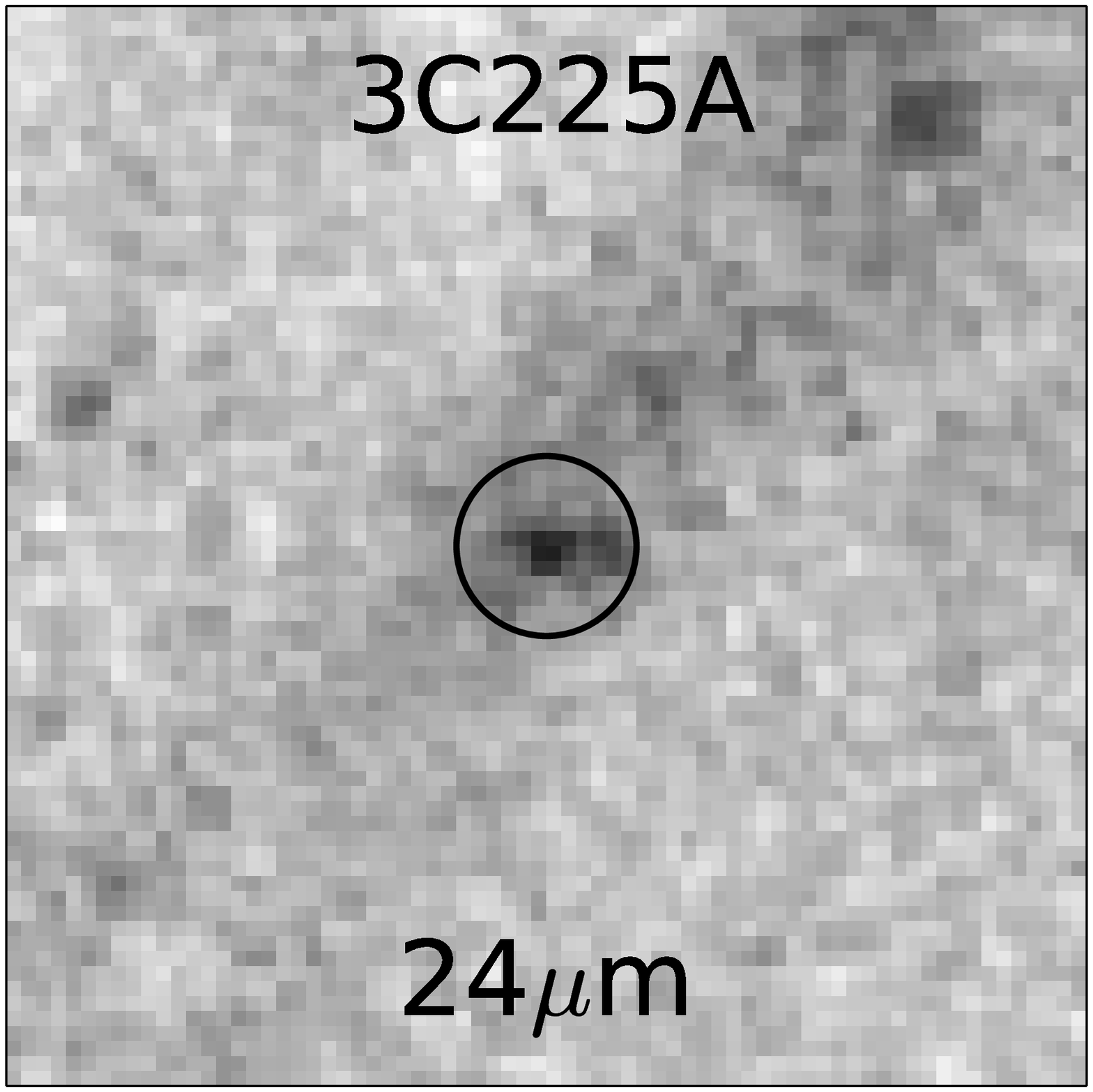}
      \includegraphics[width=1.5cm]{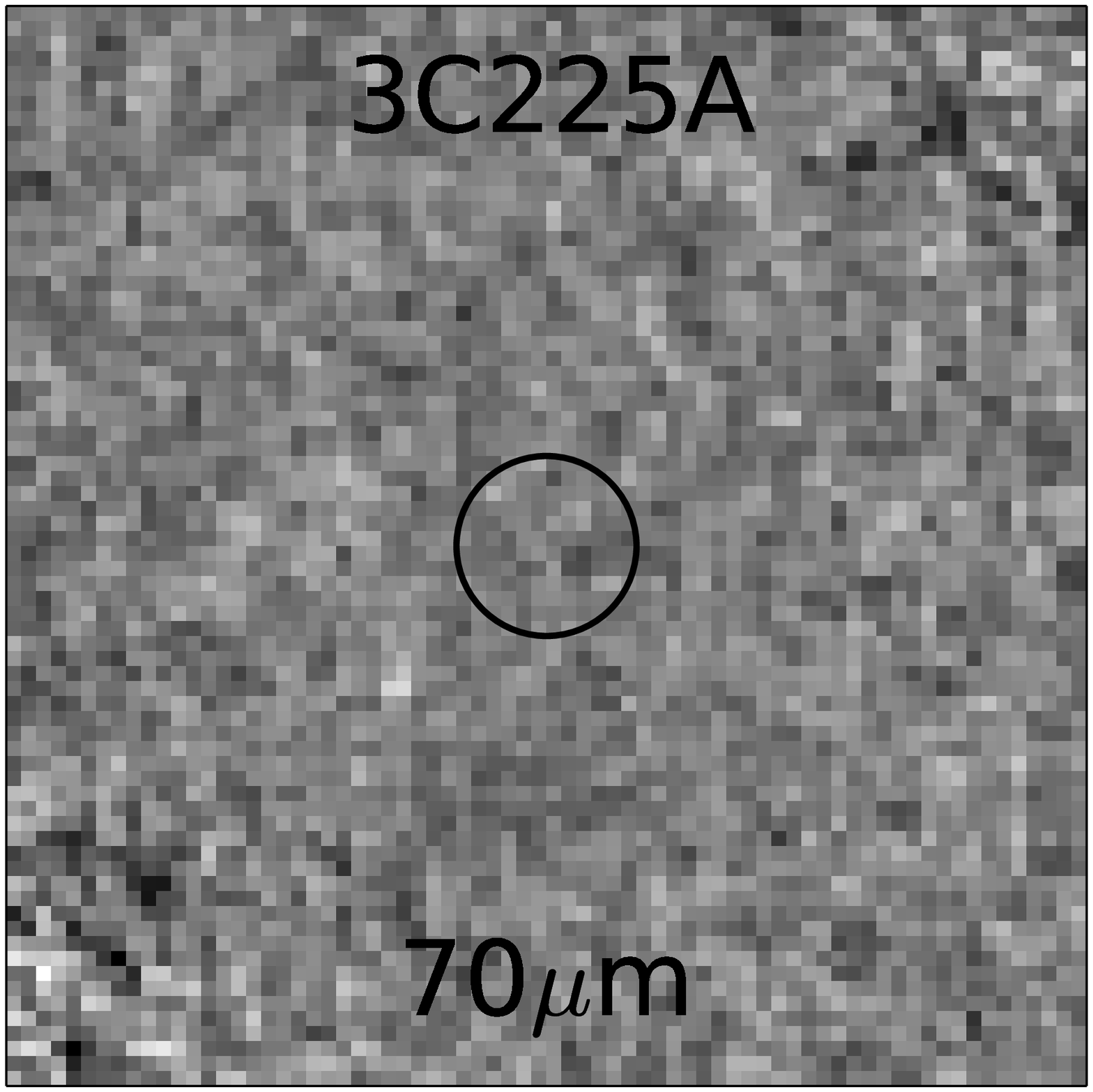}
      \includegraphics[width=1.5cm]{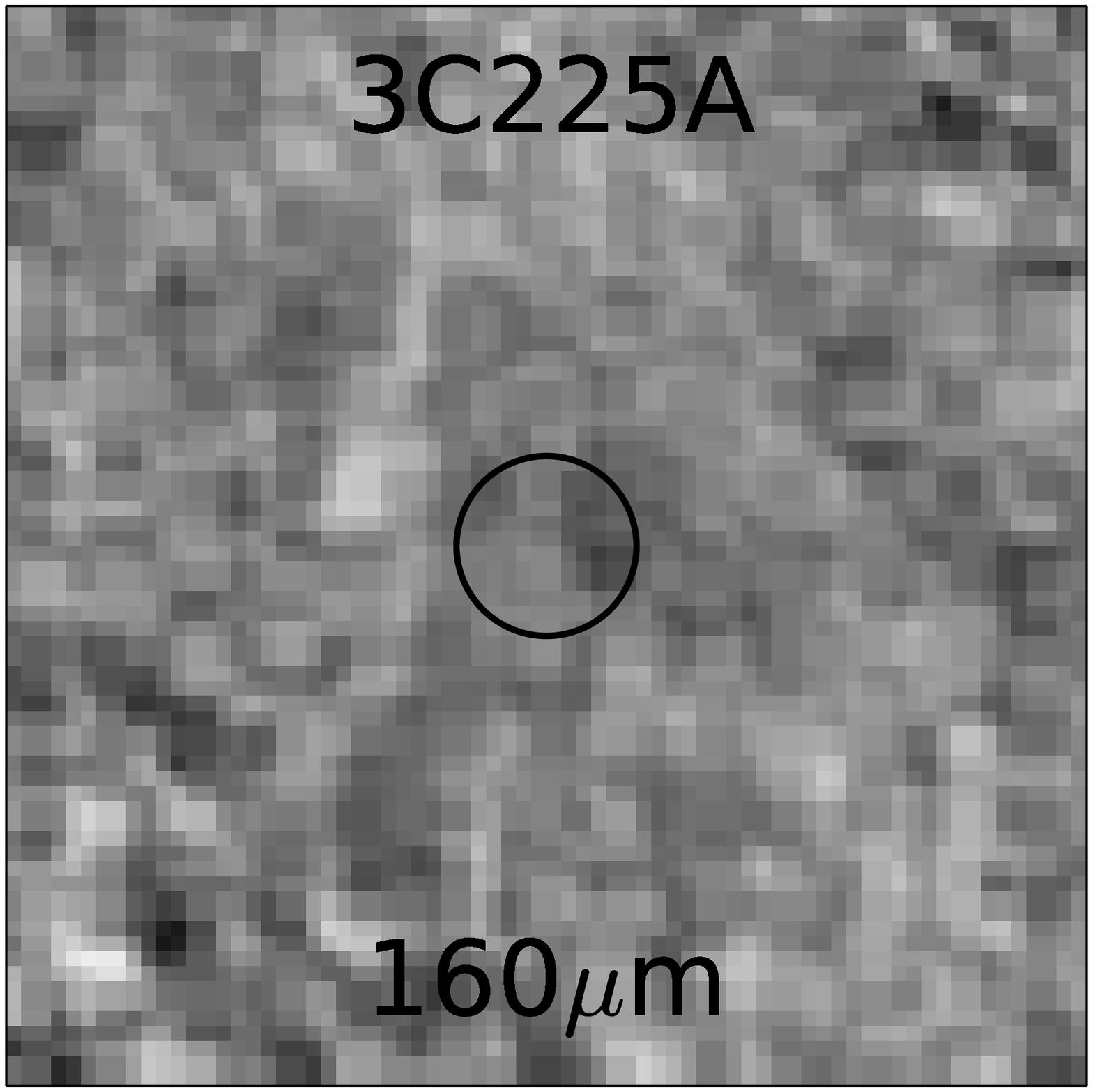}
      \includegraphics[width=1.5cm]{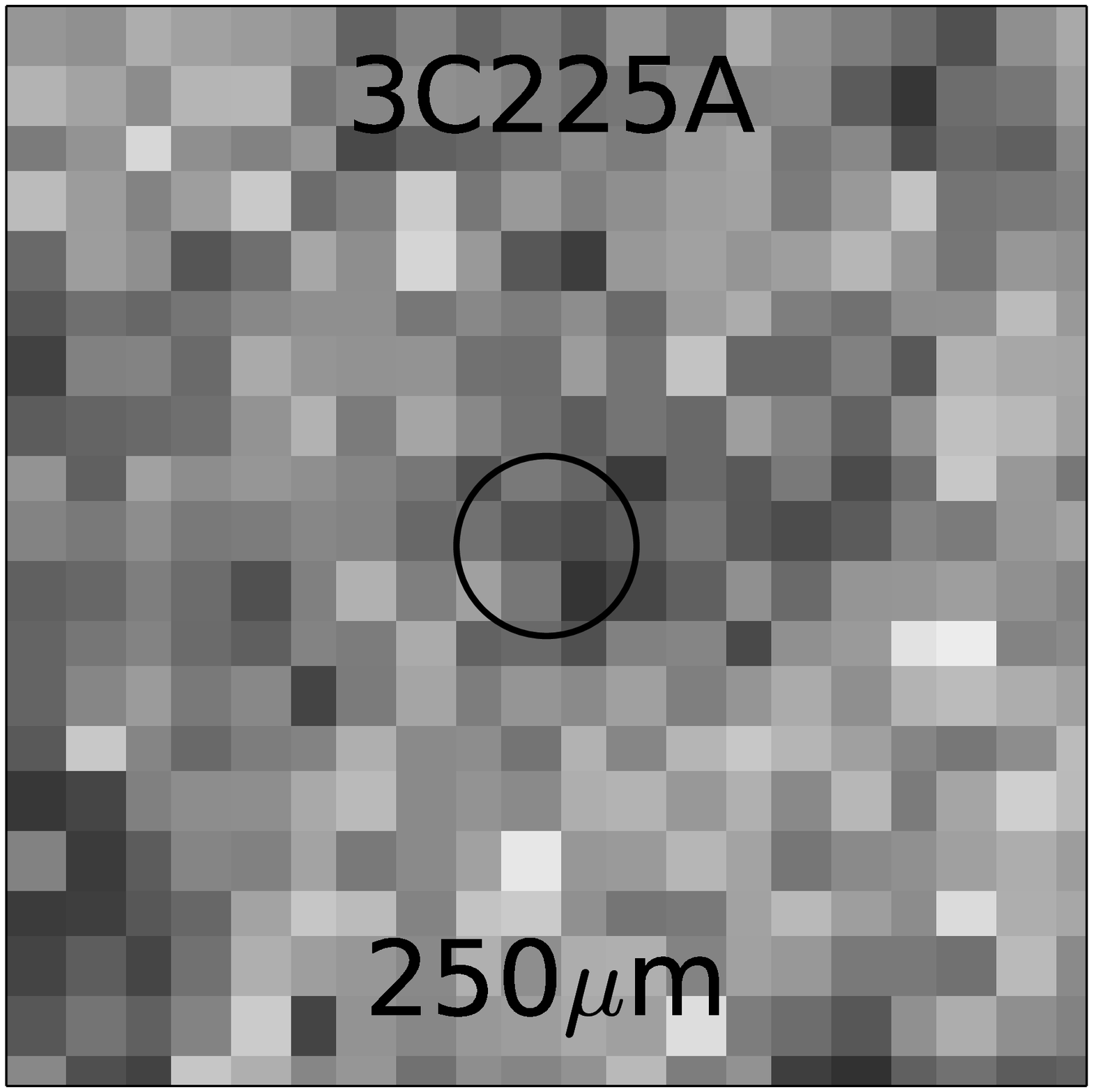}
      \includegraphics[width=1.5cm]{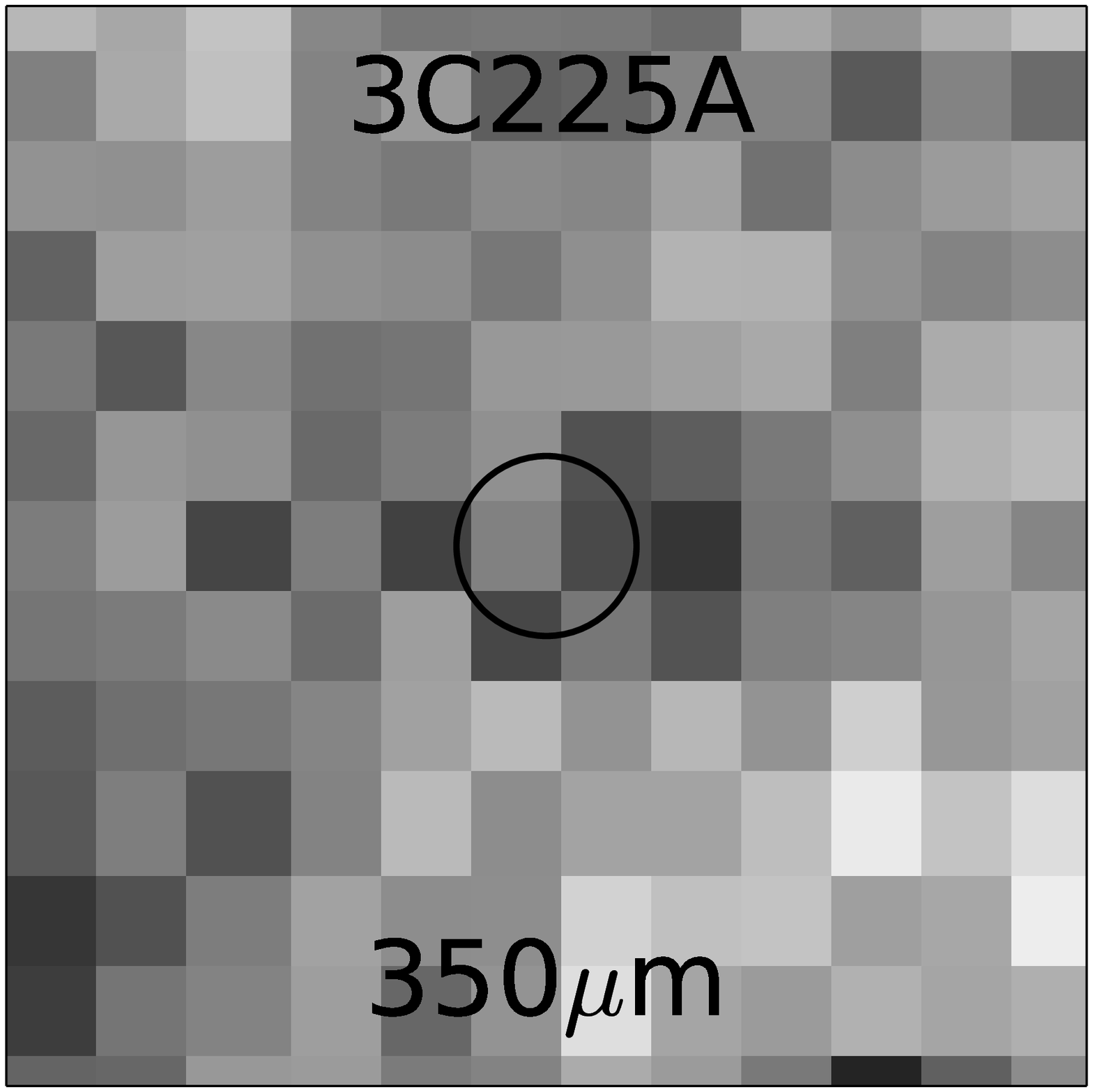}
      \includegraphics[width=1.5cm]{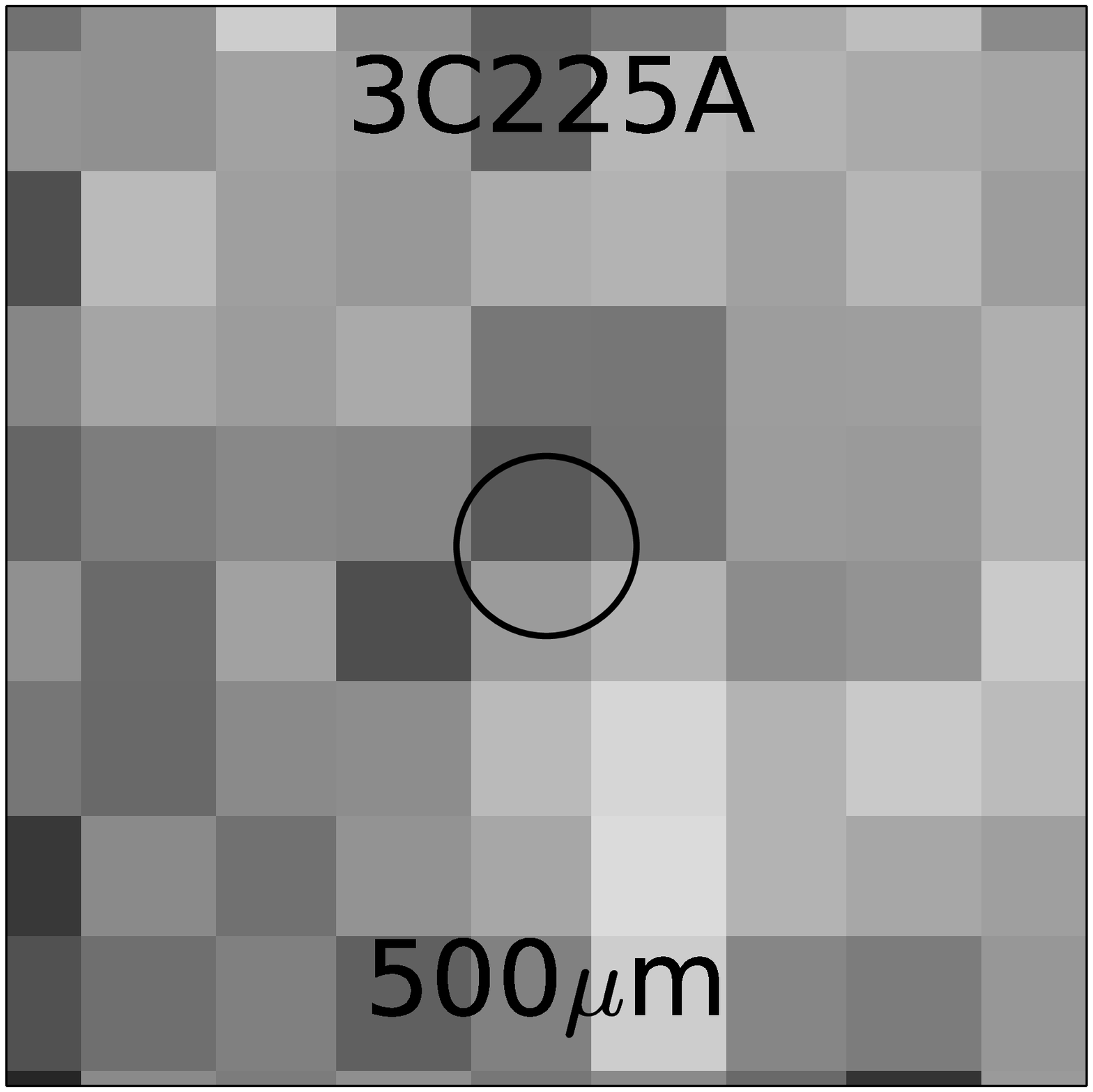}
      \caption{Continued.}
   \end{figure*}    
   \addtocounter{figure}{-1}
   \begin{figure*}
      \includegraphics[width=1.5cm]{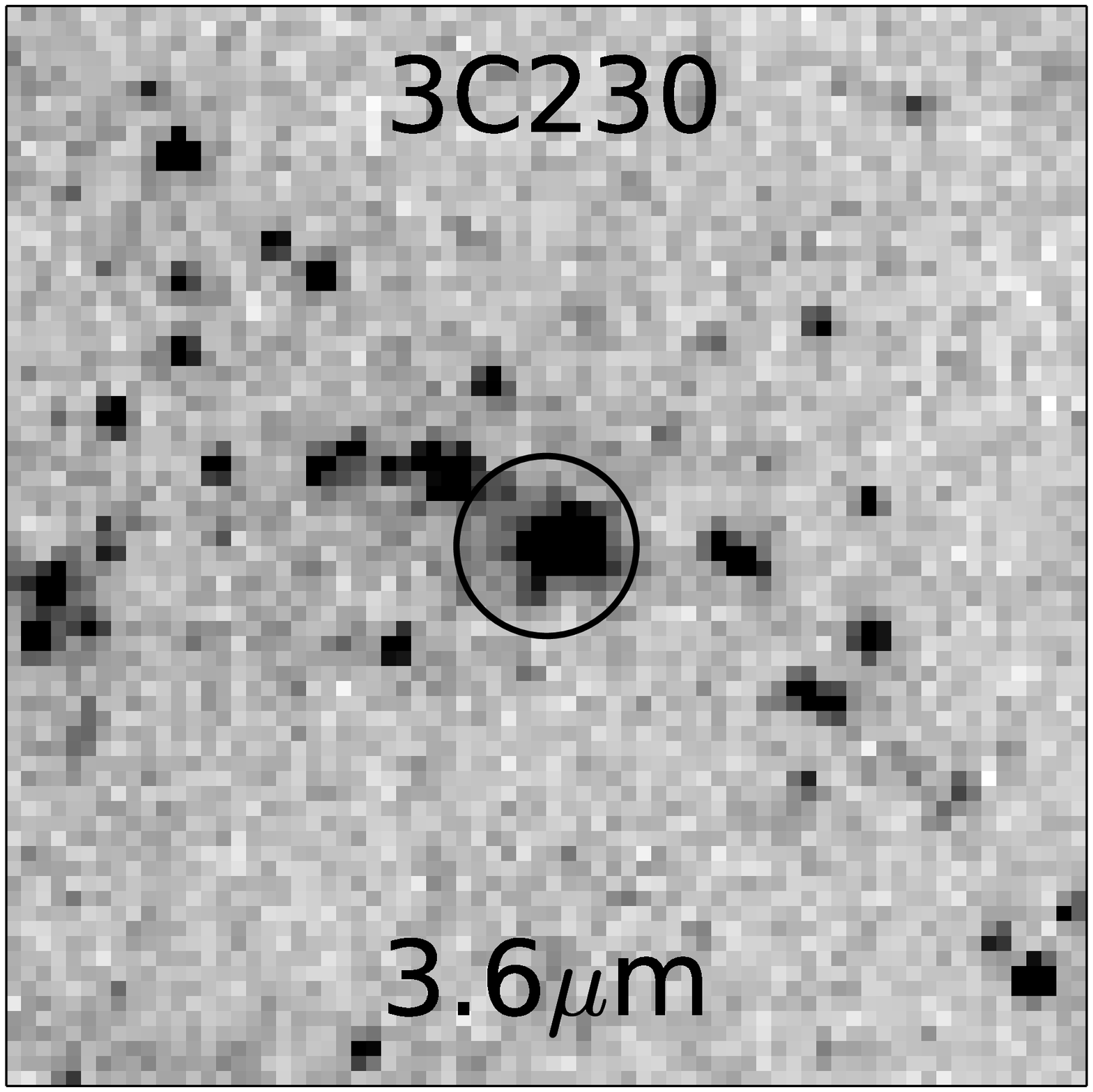}
      \includegraphics[width=1.5cm]{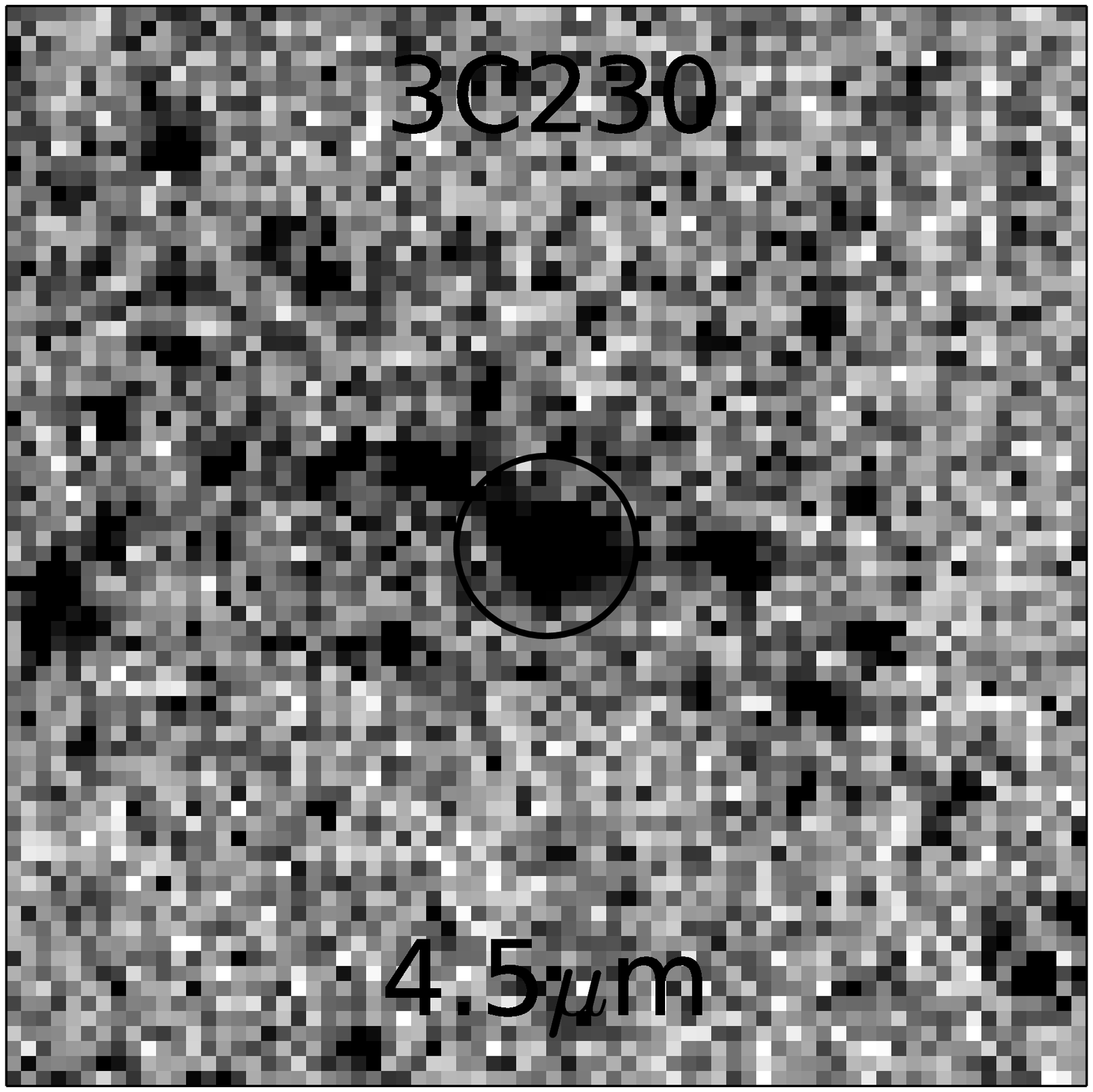}
      \includegraphics[width=1.5cm]{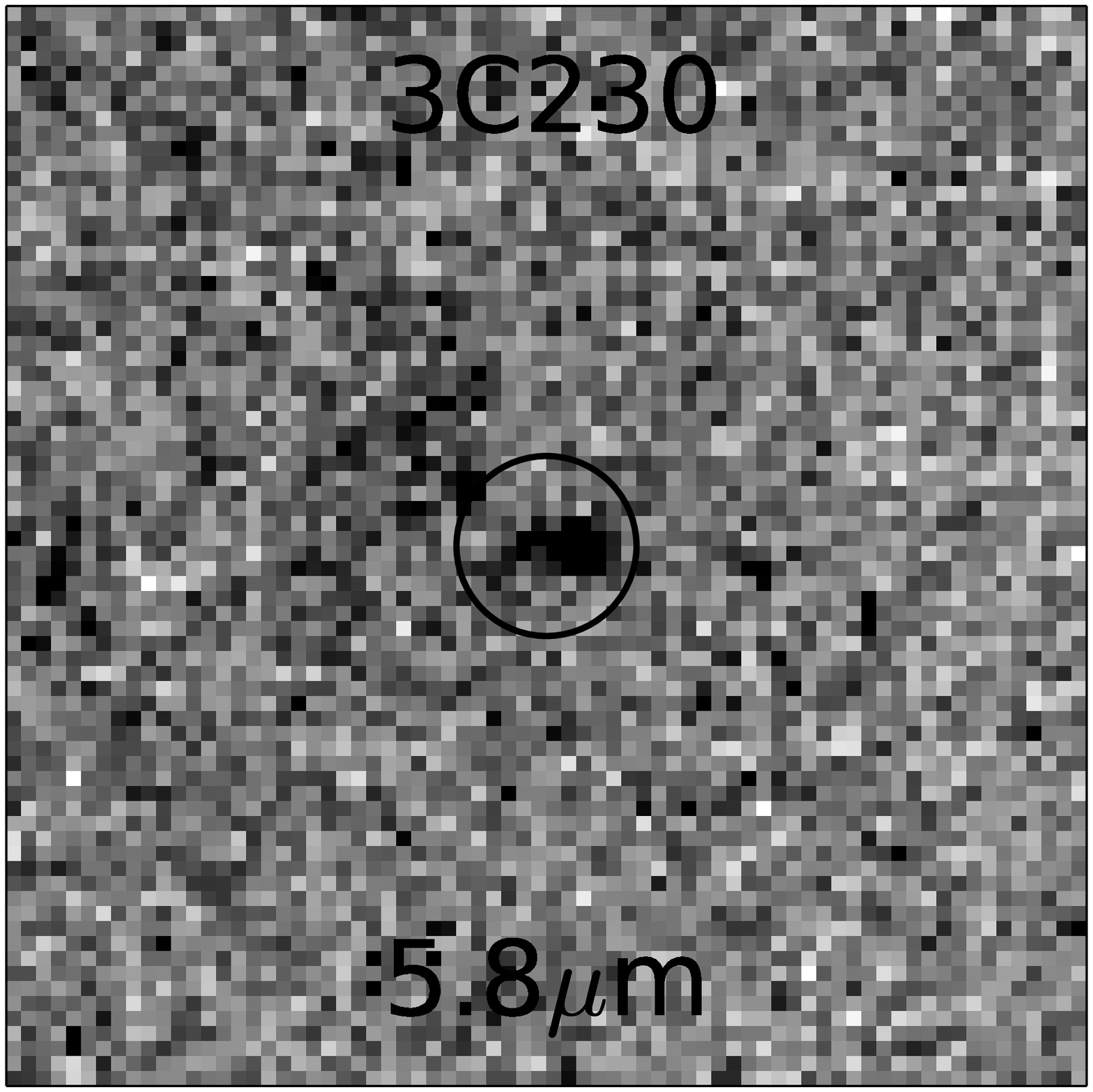}
      \includegraphics[width=1.5cm]{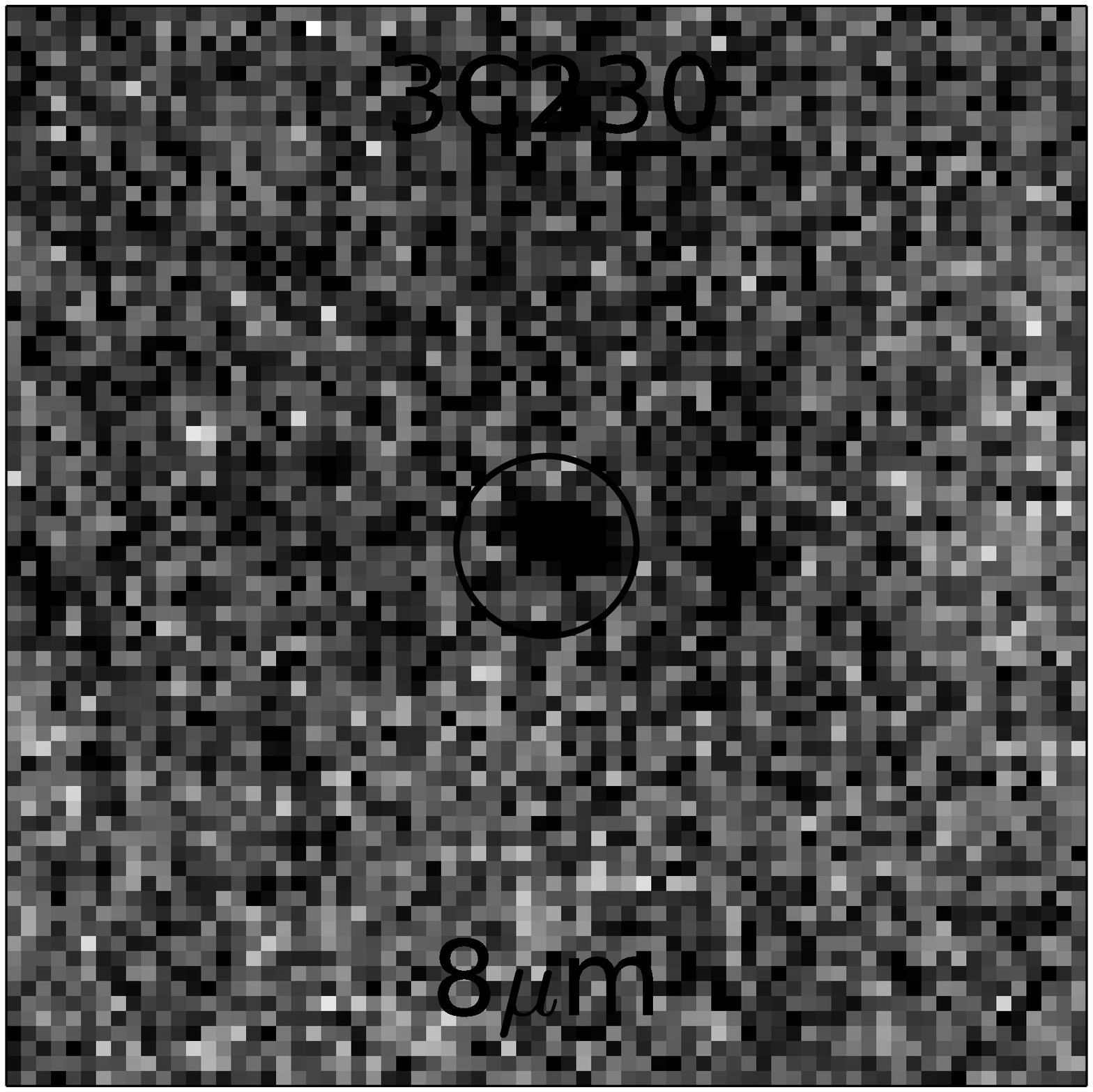}
      \includegraphics[width=1.5cm]{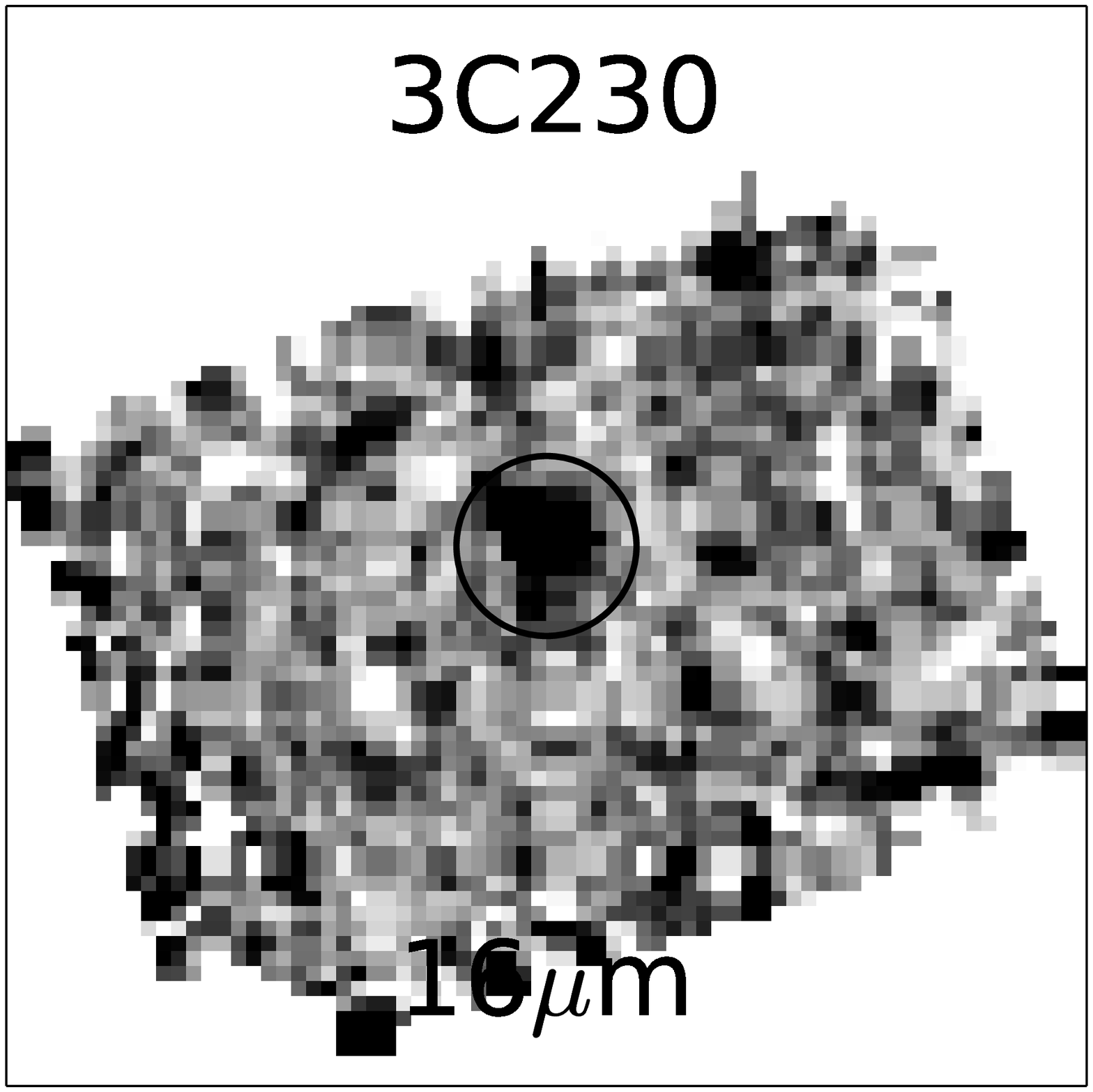}
      \includegraphics[width=1.5cm]{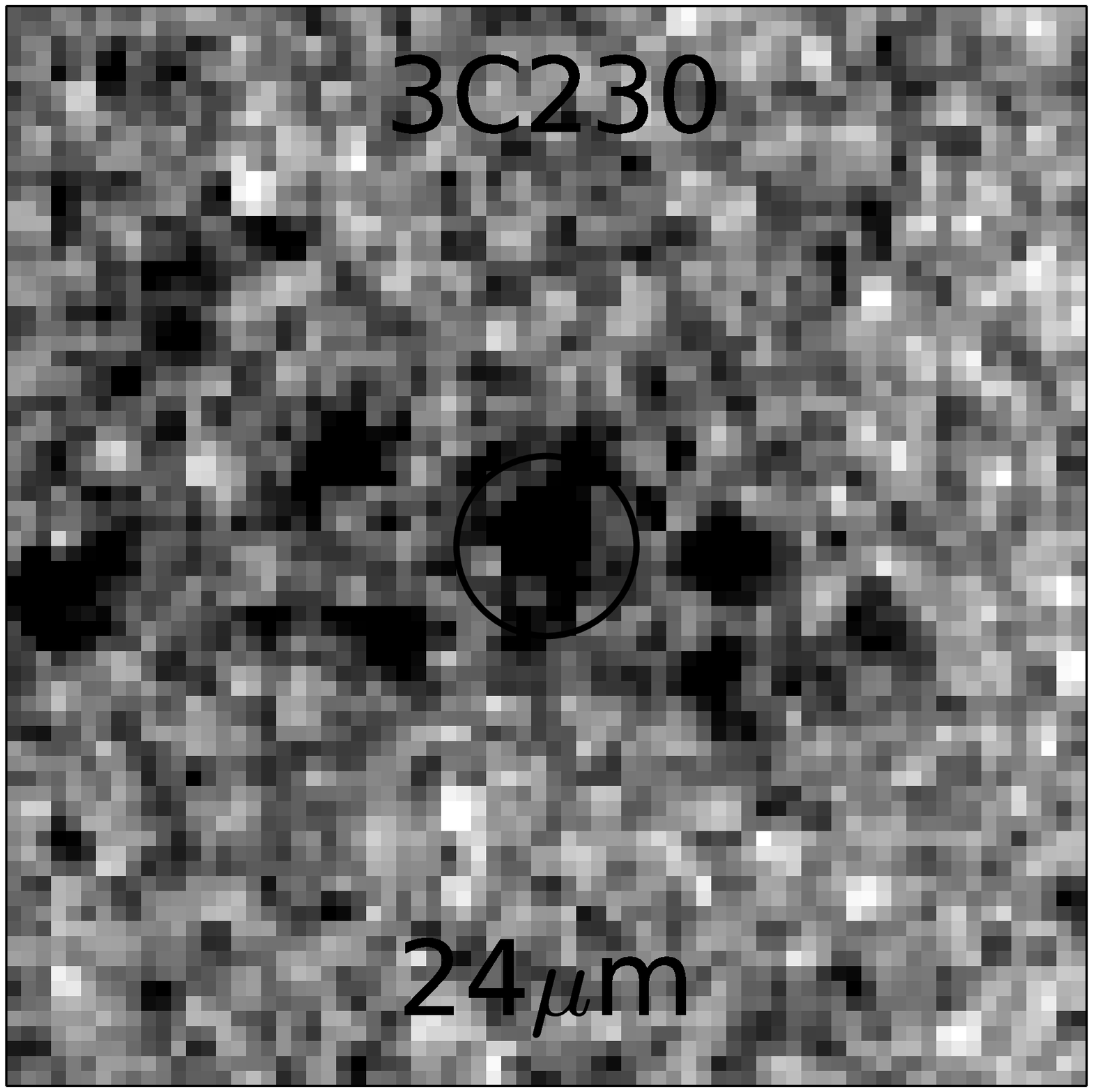}
      \includegraphics[width=1.5cm]{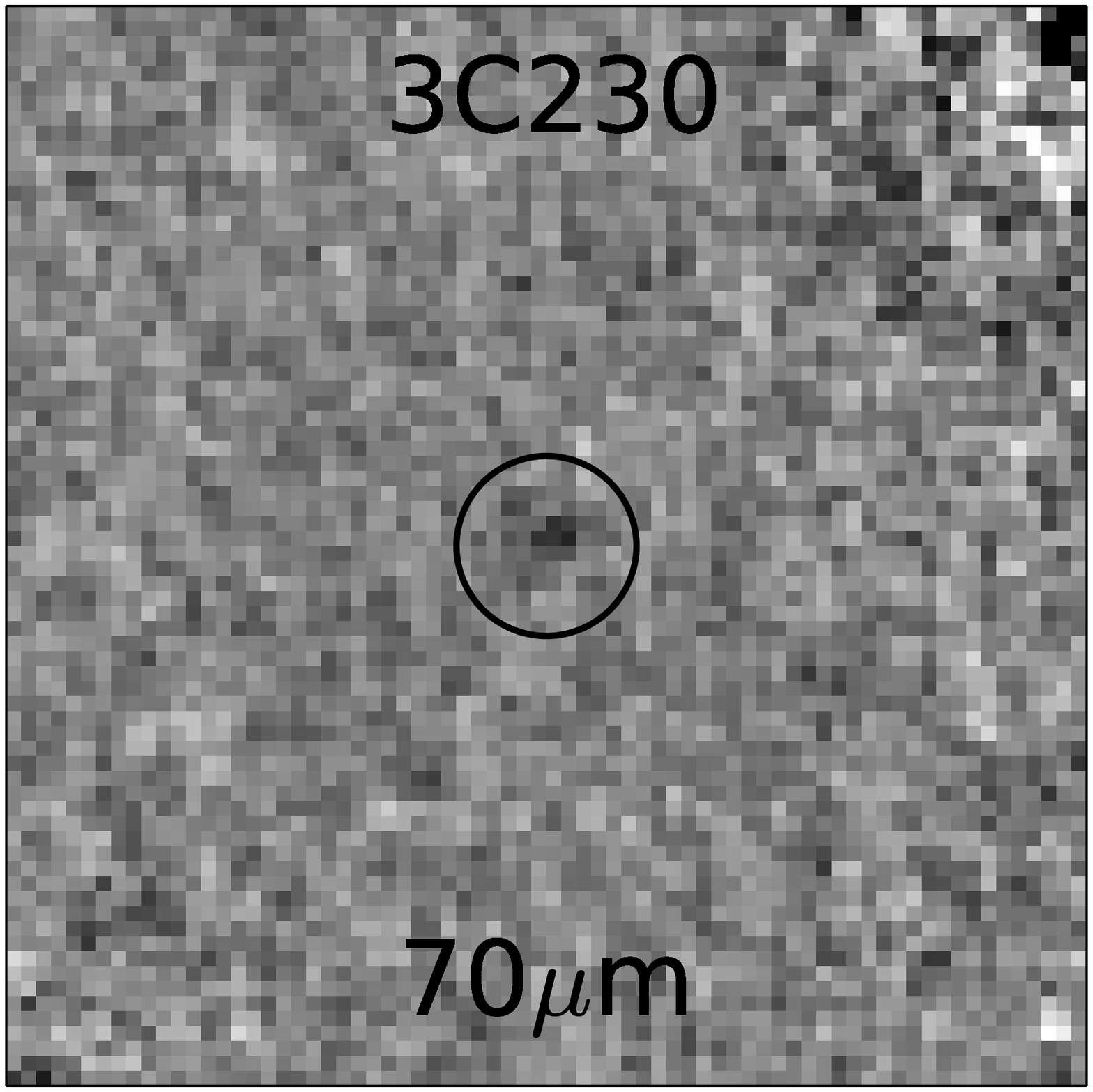}
      \includegraphics[width=1.5cm]{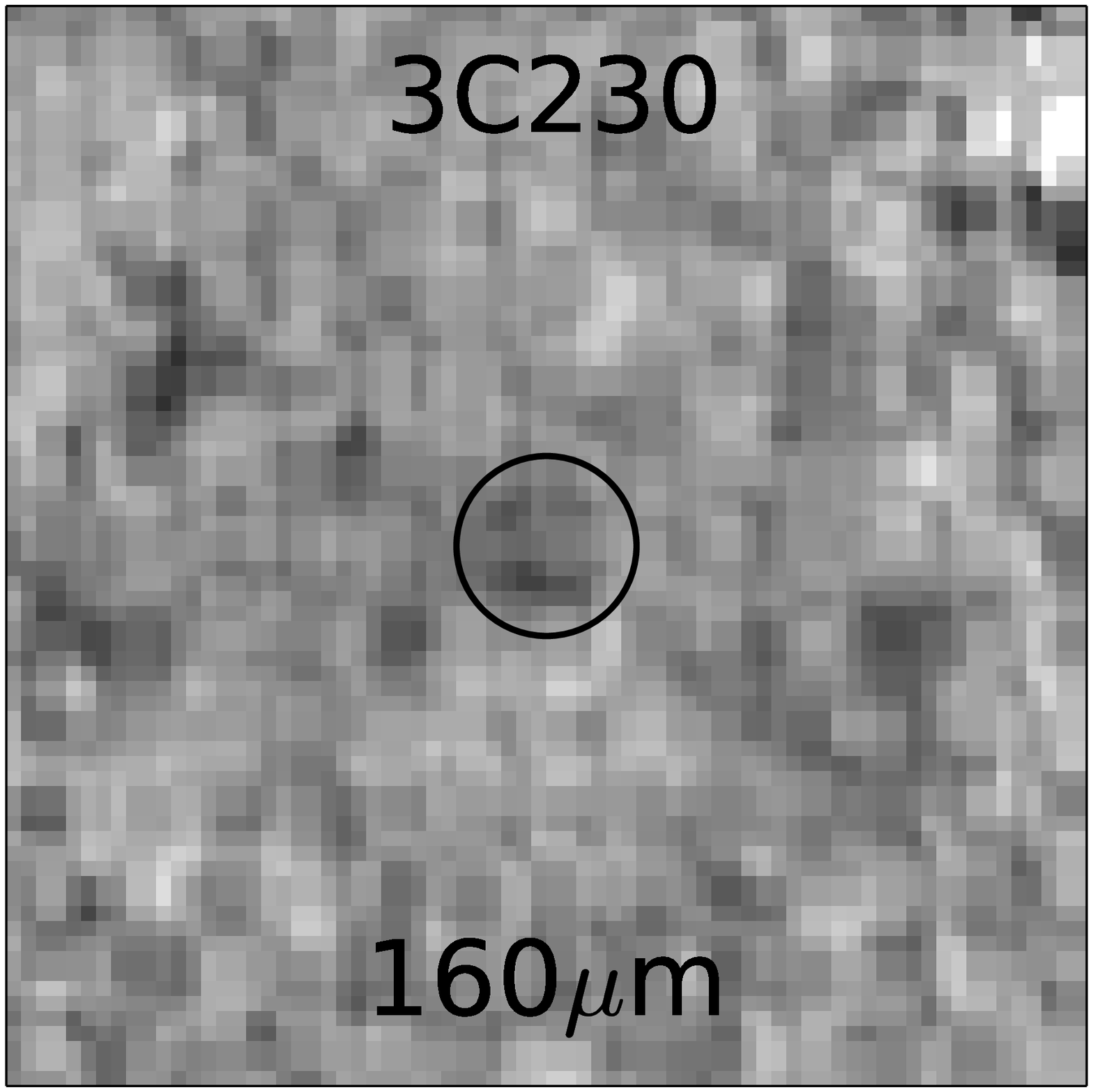}
      \includegraphics[width=1.5cm]{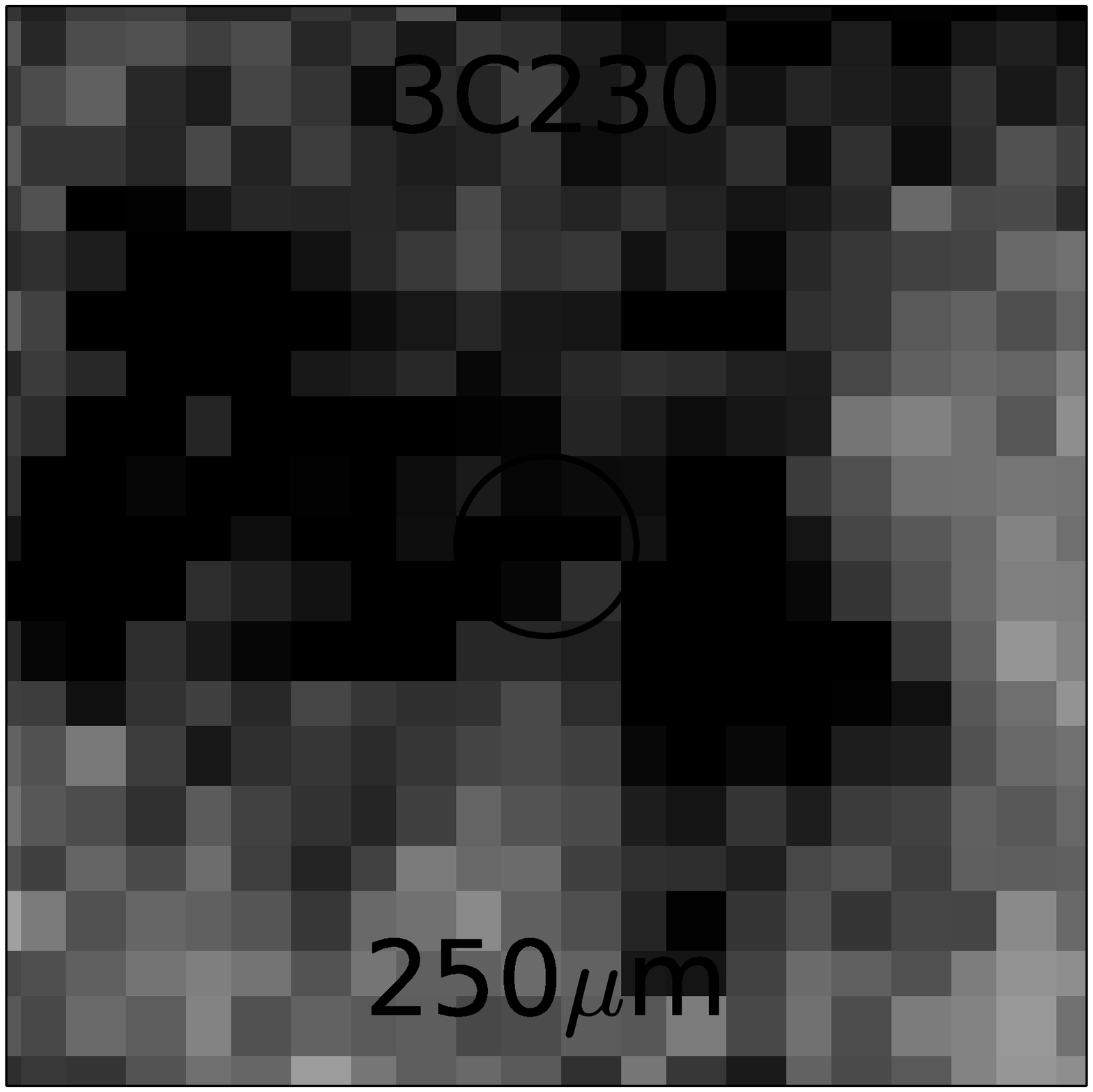}
      \includegraphics[width=1.5cm]{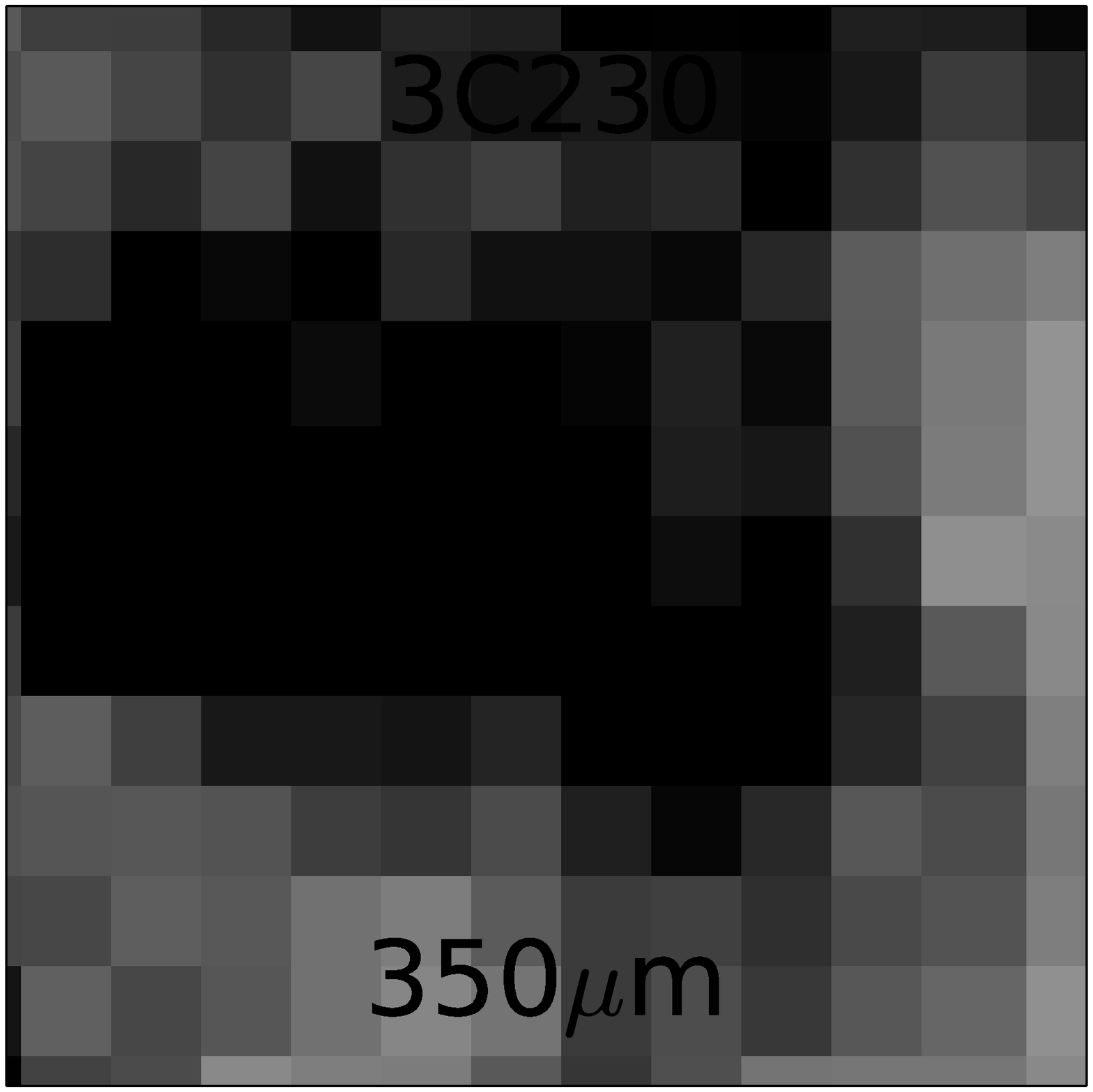}
      \includegraphics[width=1.5cm]{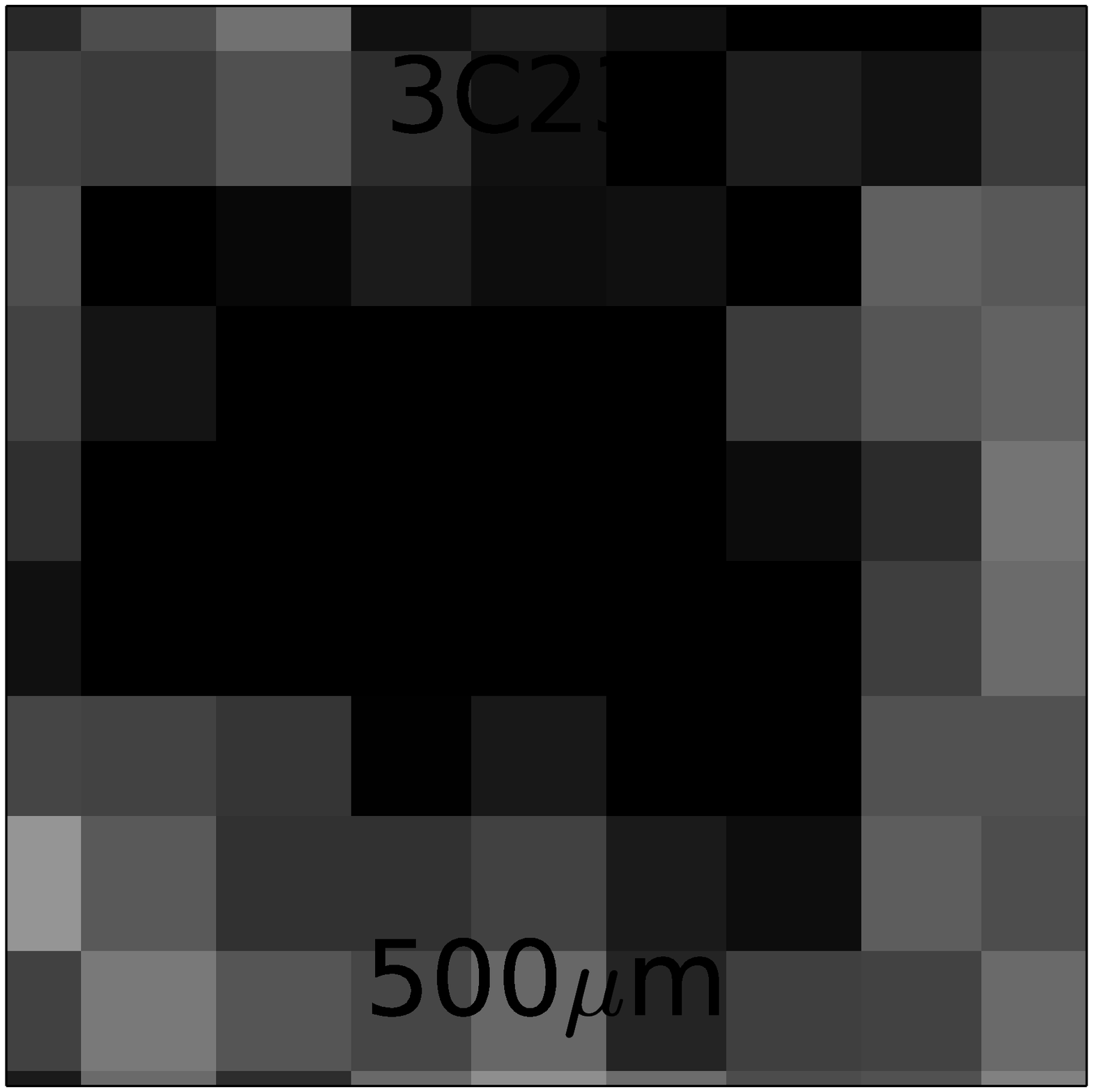}
      \\
      \includegraphics[width=1.5cm]{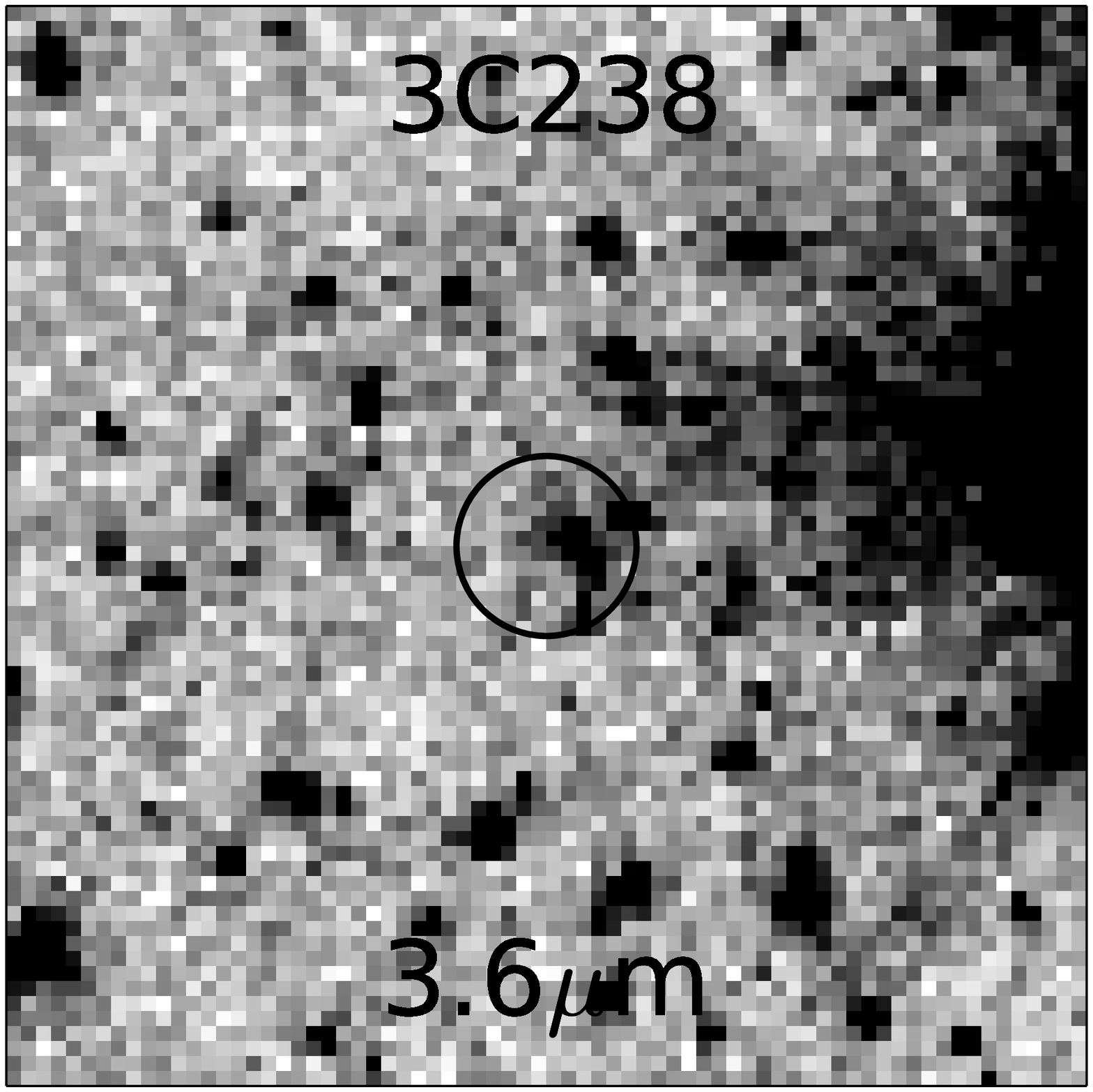}
      \includegraphics[width=1.5cm]{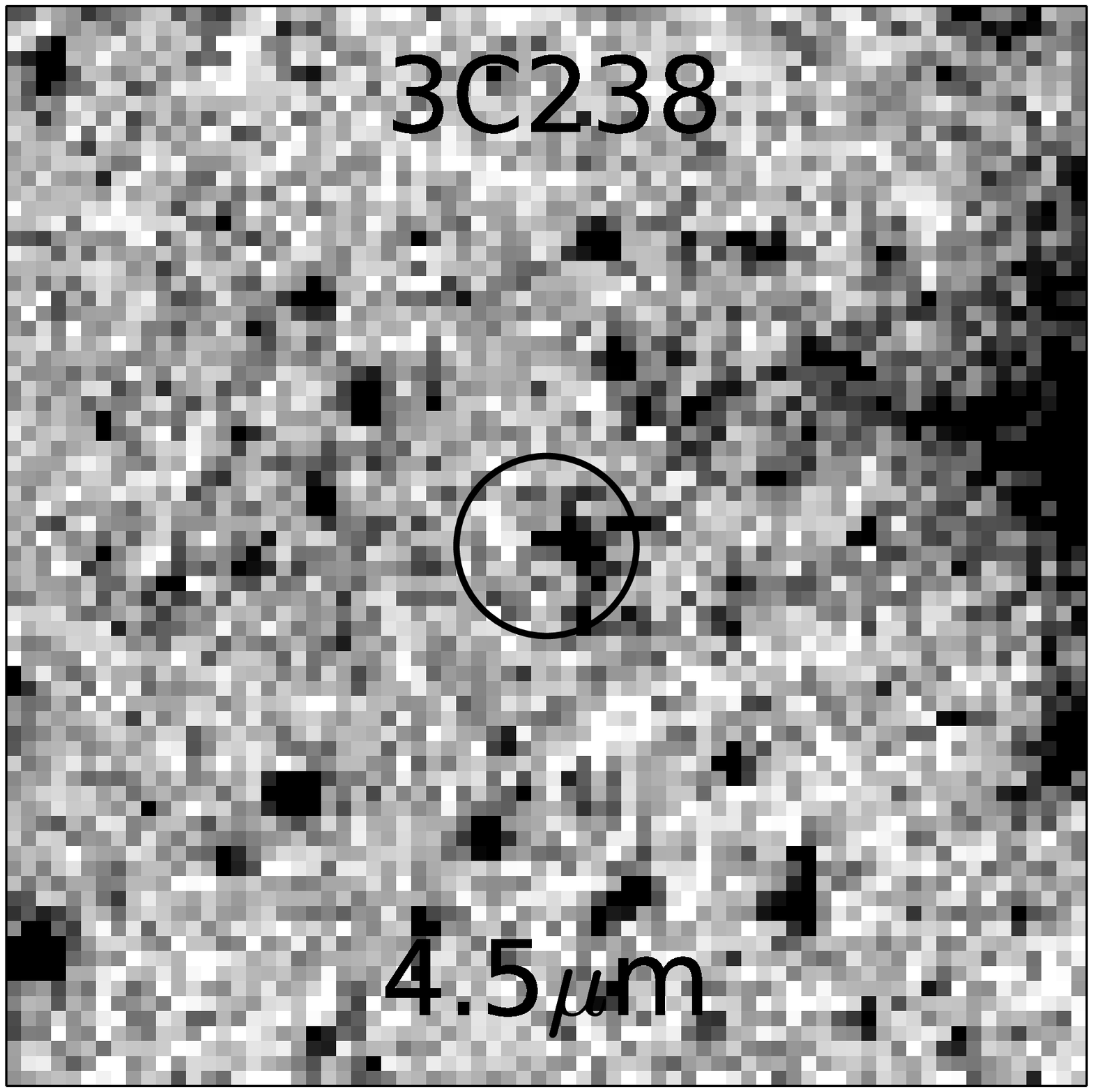}
      \includegraphics[width=1.5cm]{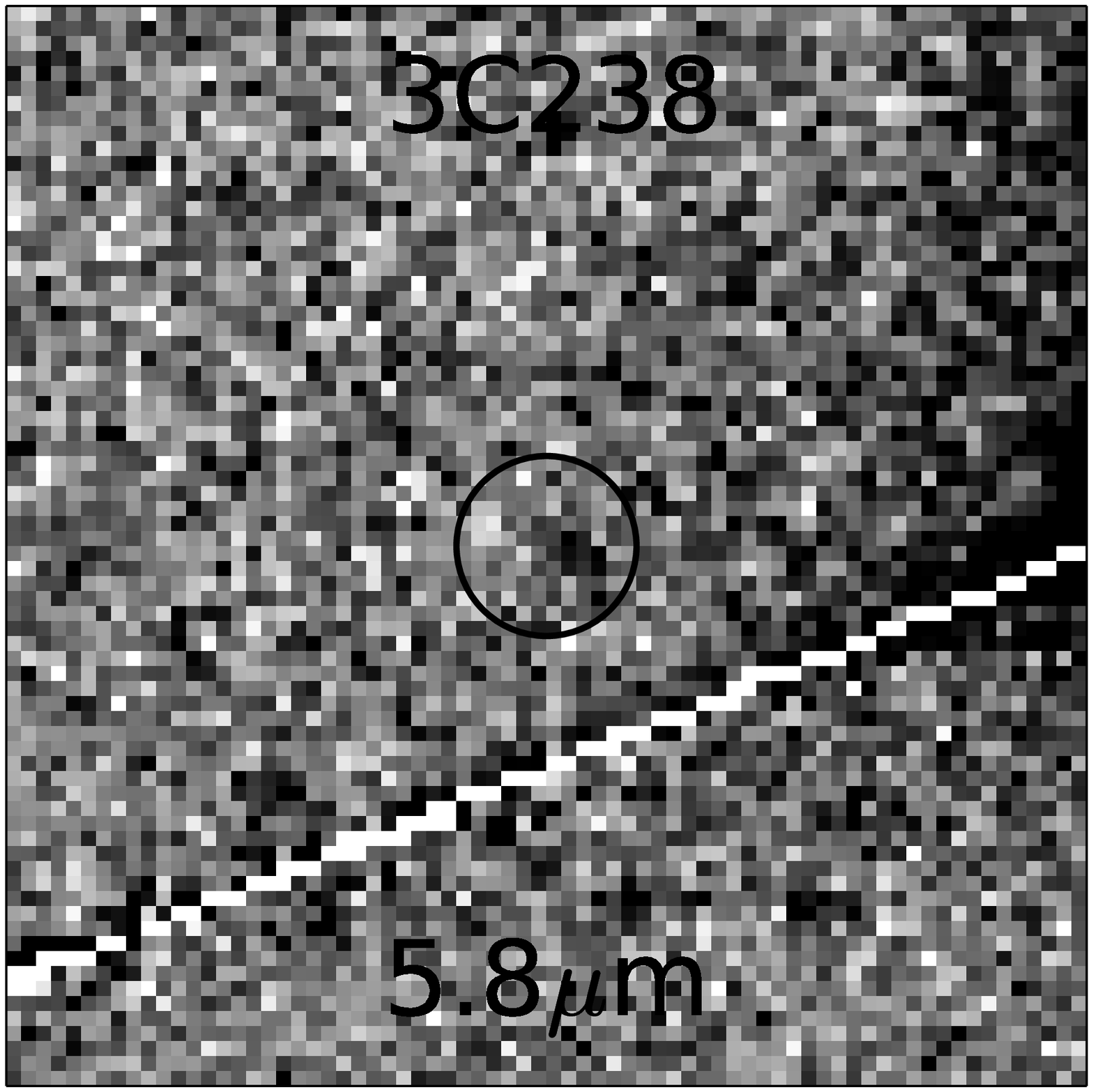}
      \includegraphics[width=1.5cm]{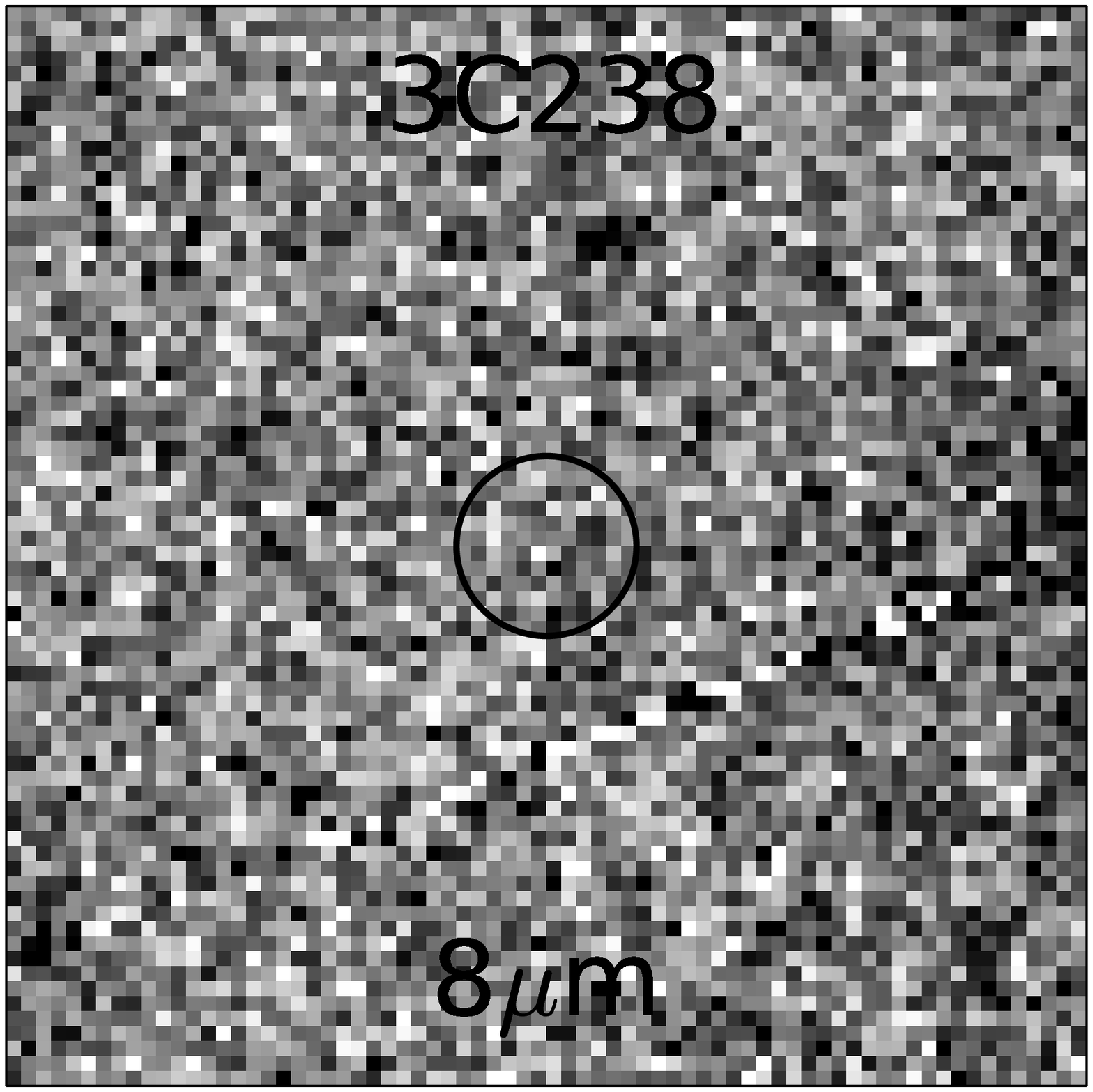}
      \includegraphics[width=1.5cm]{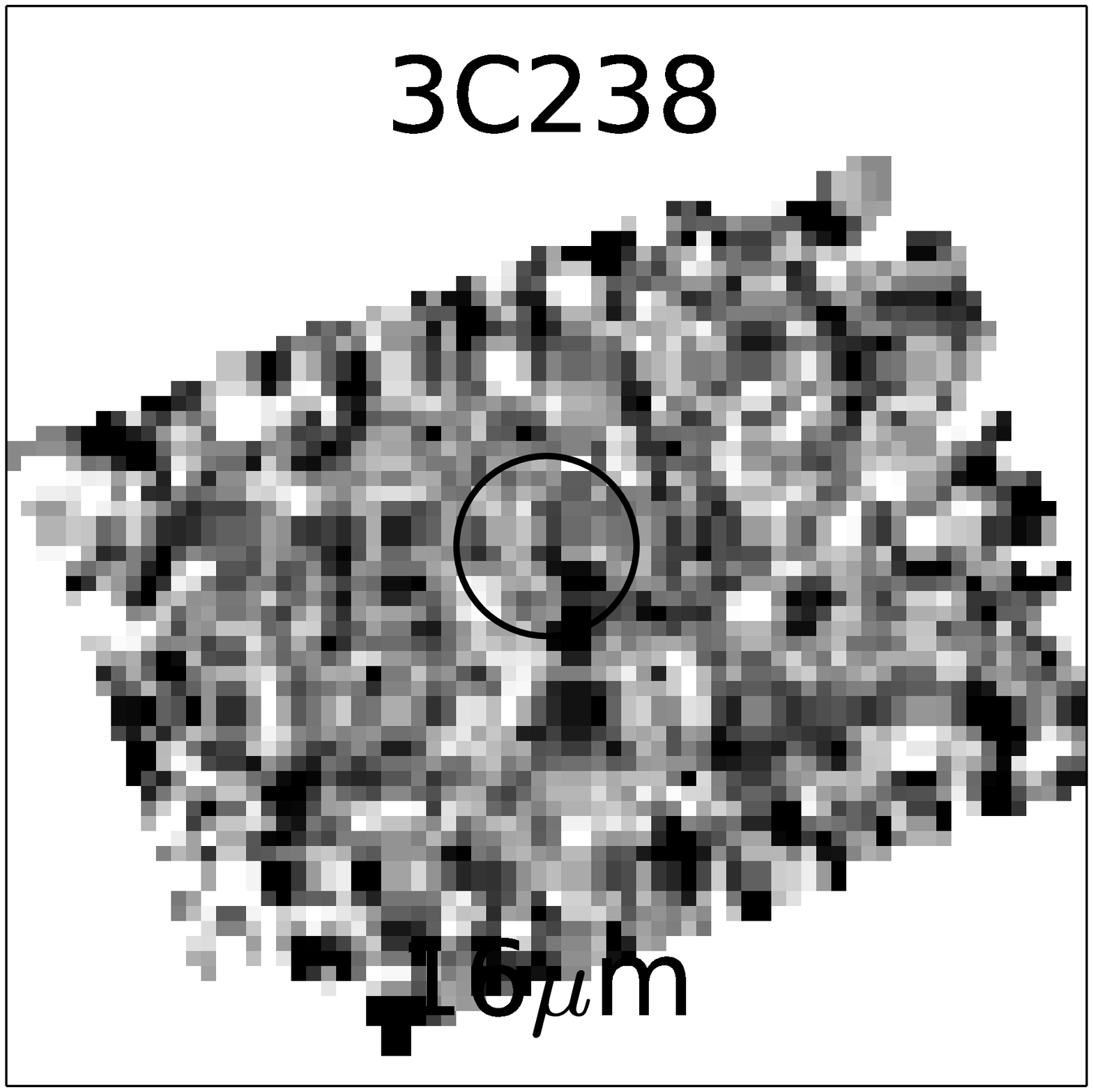}
      \includegraphics[width=1.5cm]{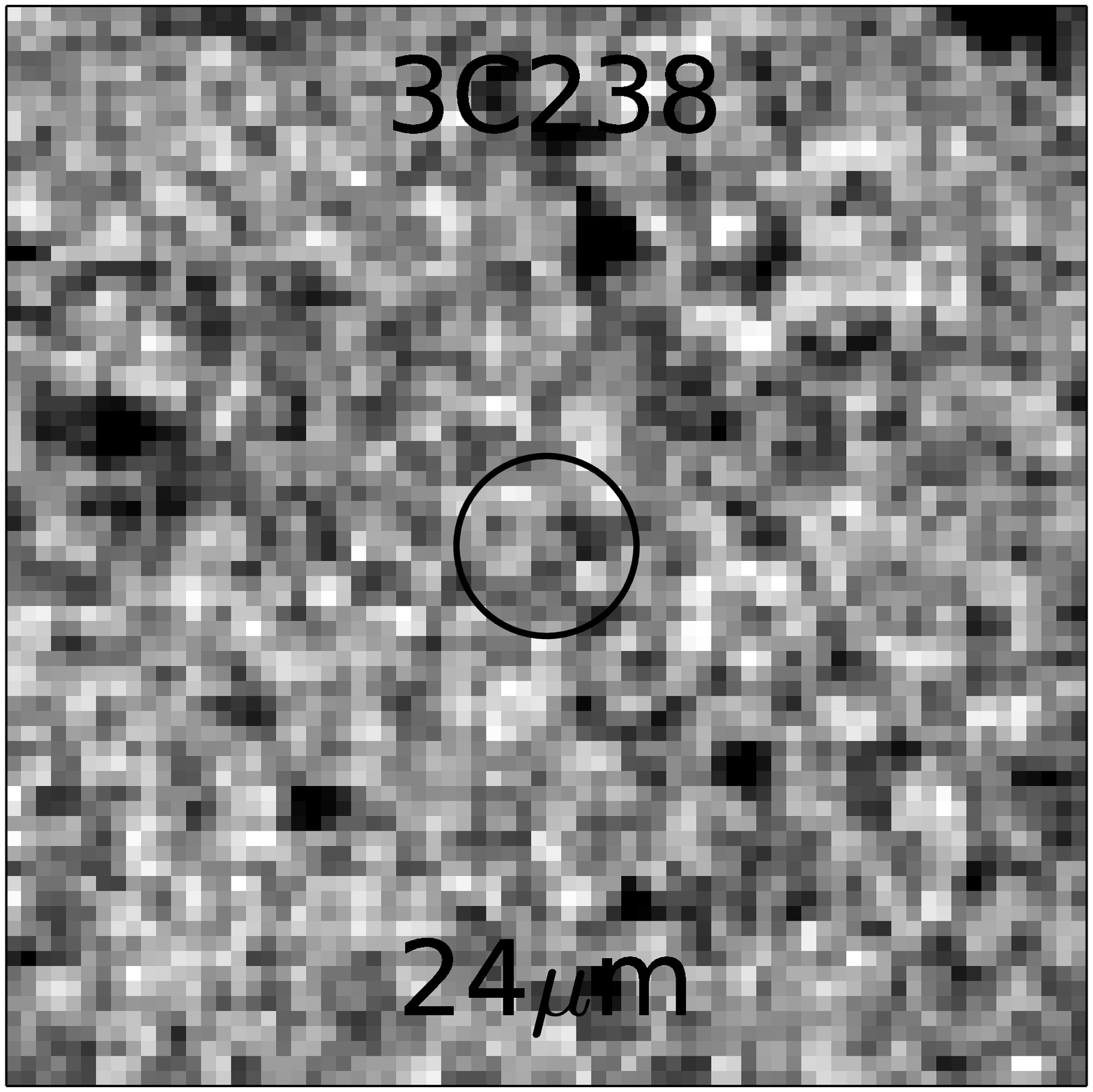}
      \includegraphics[width=1.5cm]{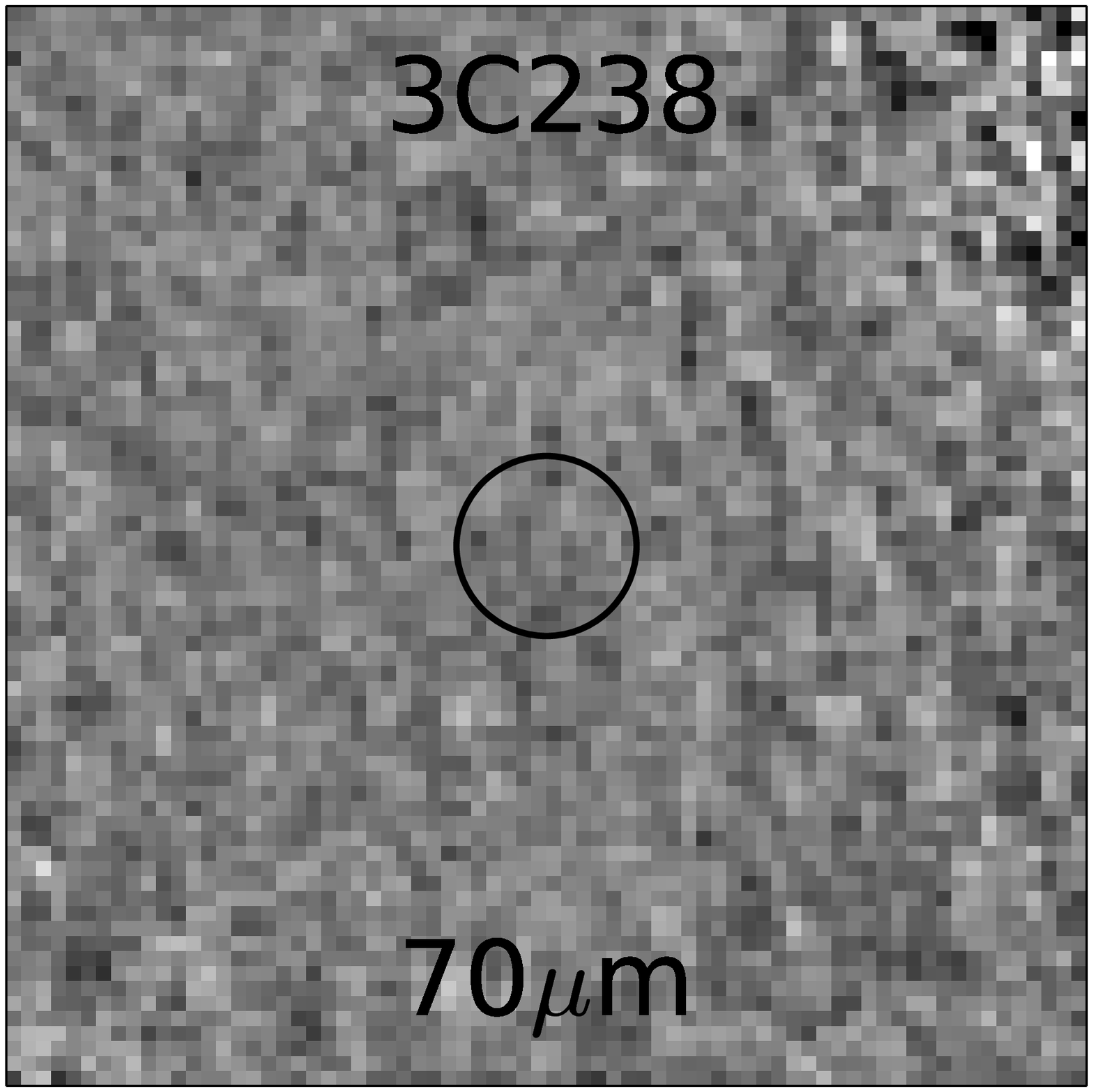}
      \includegraphics[width=1.5cm]{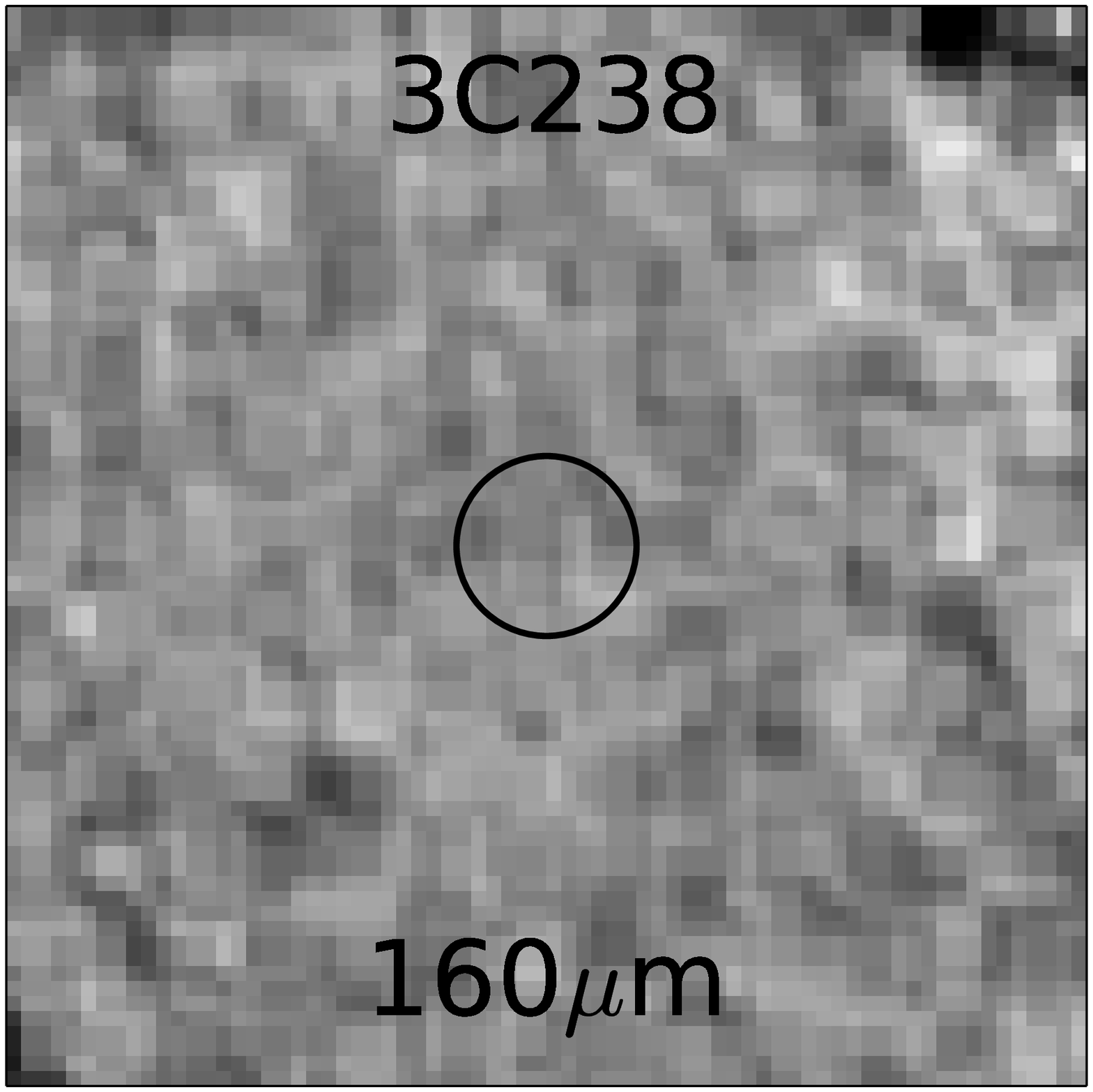}
      \includegraphics[width=1.5cm]{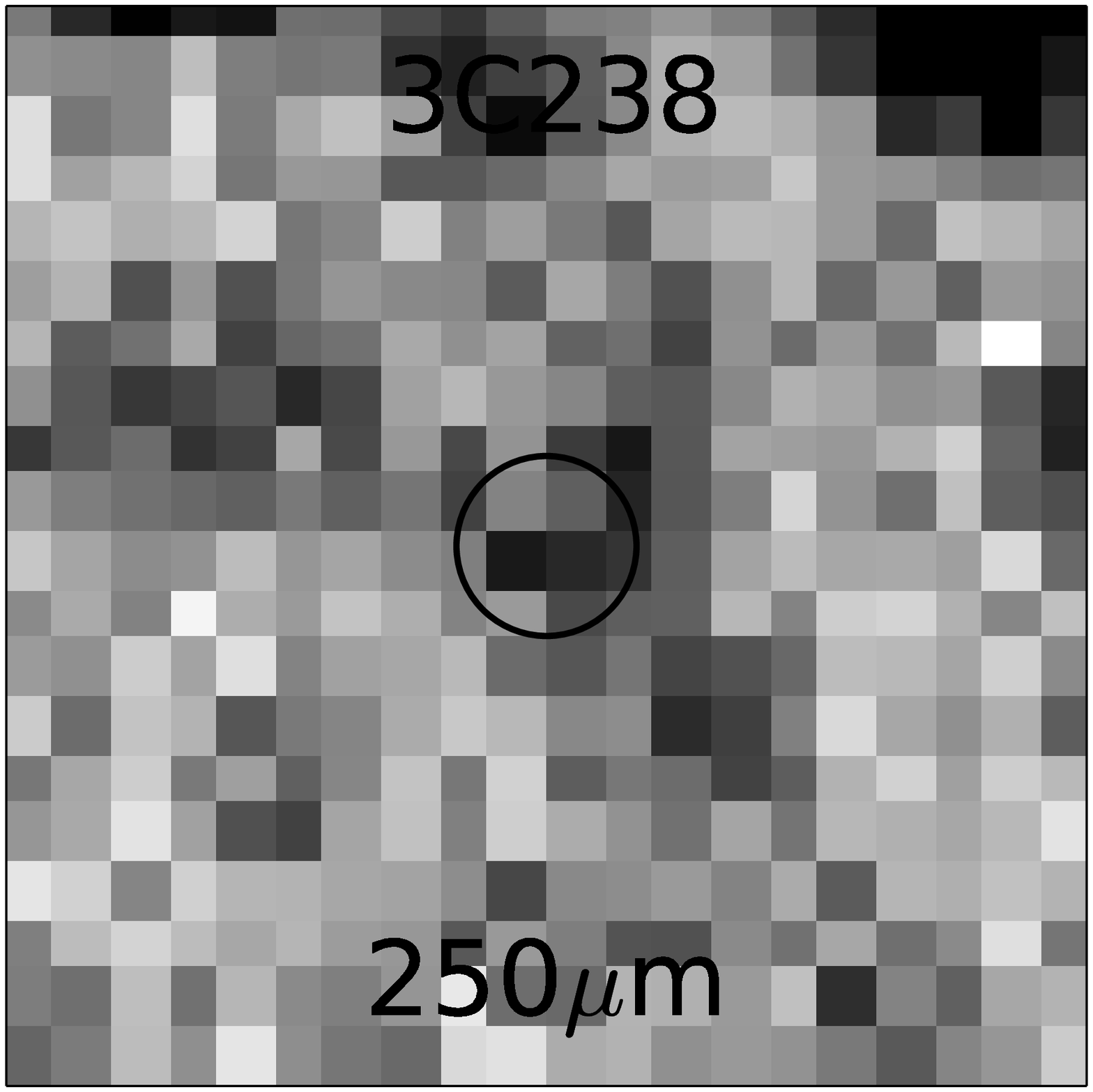}
      \includegraphics[width=1.5cm]{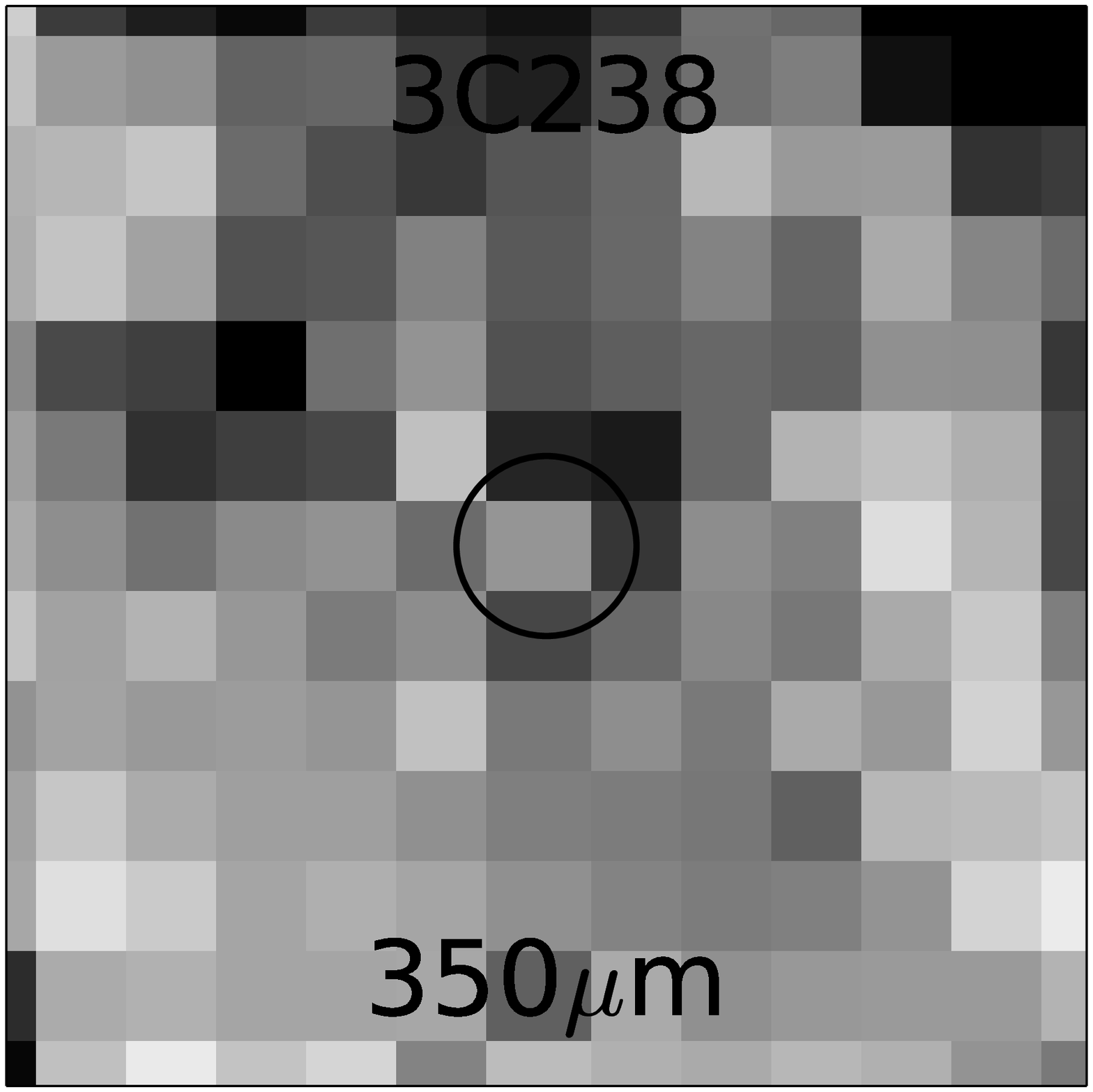}
      \includegraphics[width=1.5cm]{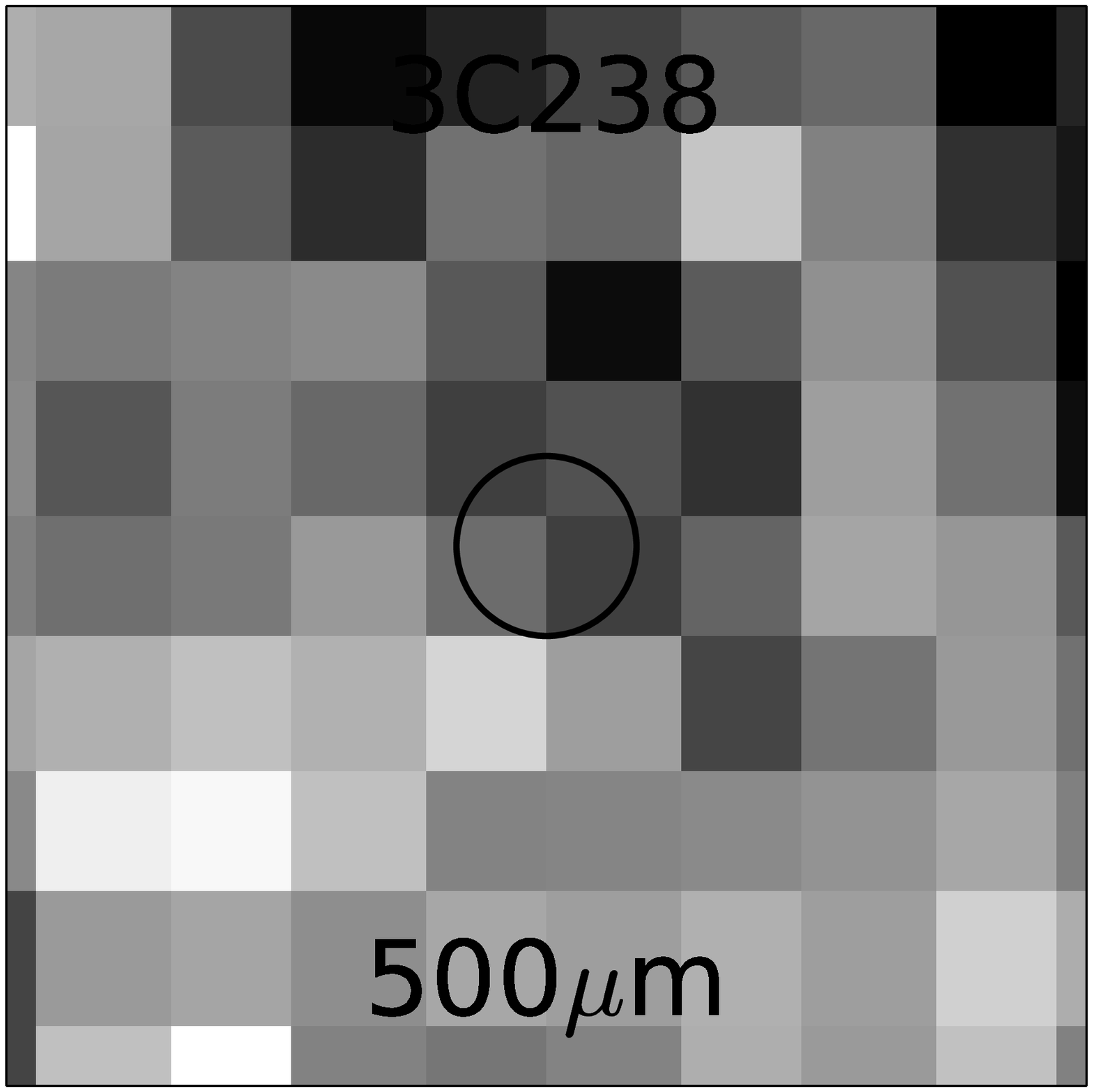}
      \\
      \includegraphics[width=1.5cm]{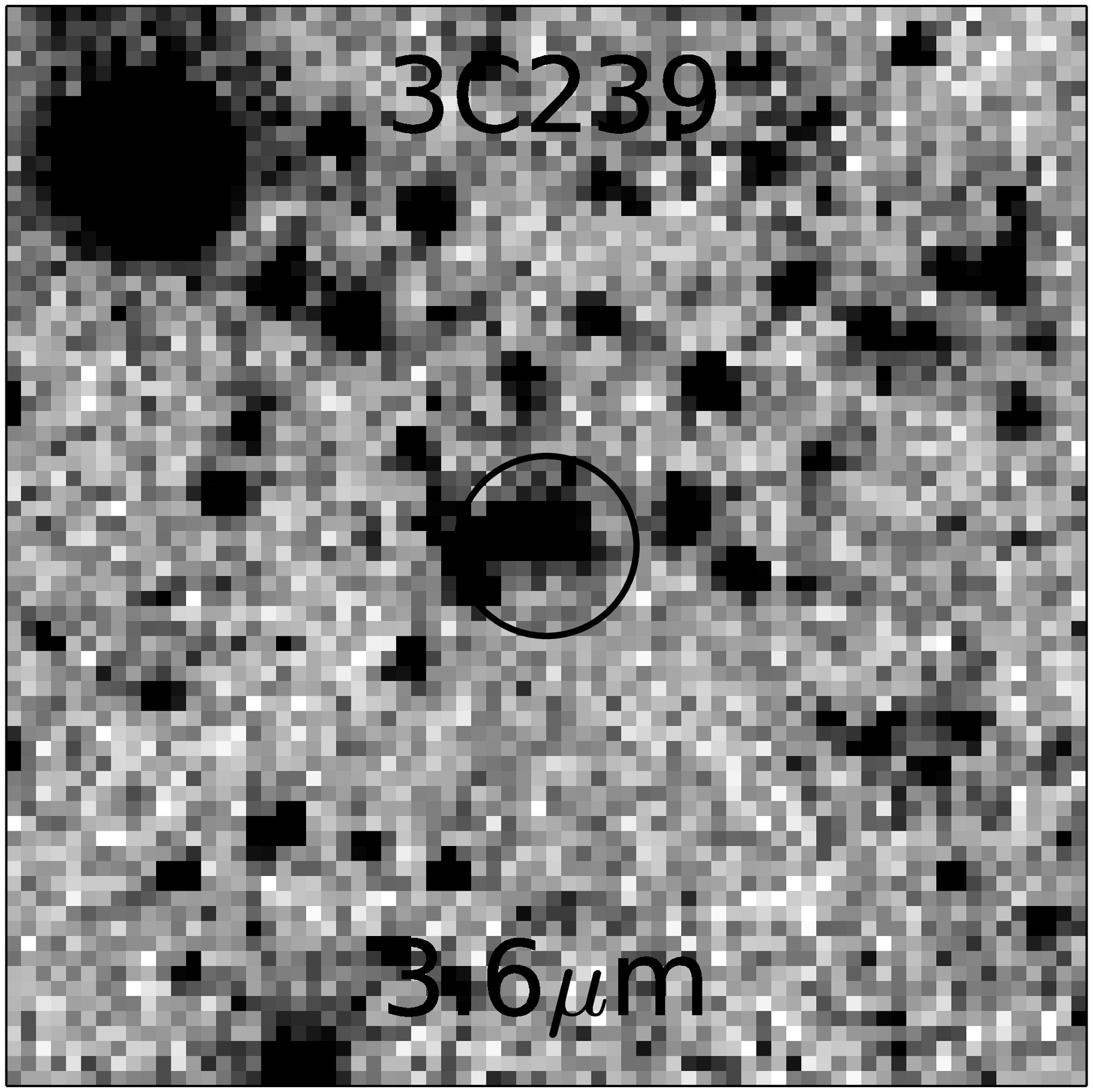}
      \includegraphics[width=1.5cm]{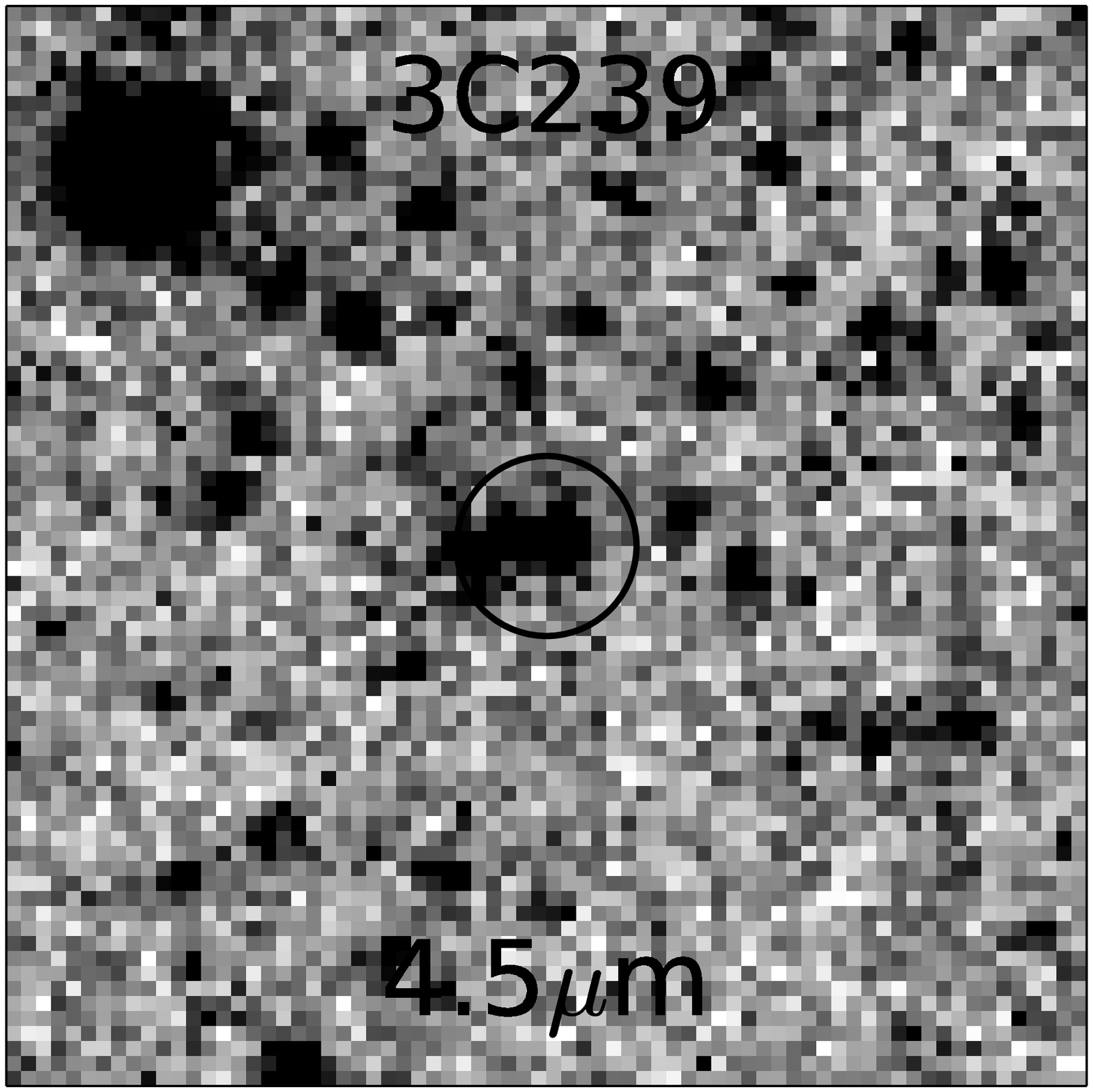}
      \includegraphics[width=1.5cm]{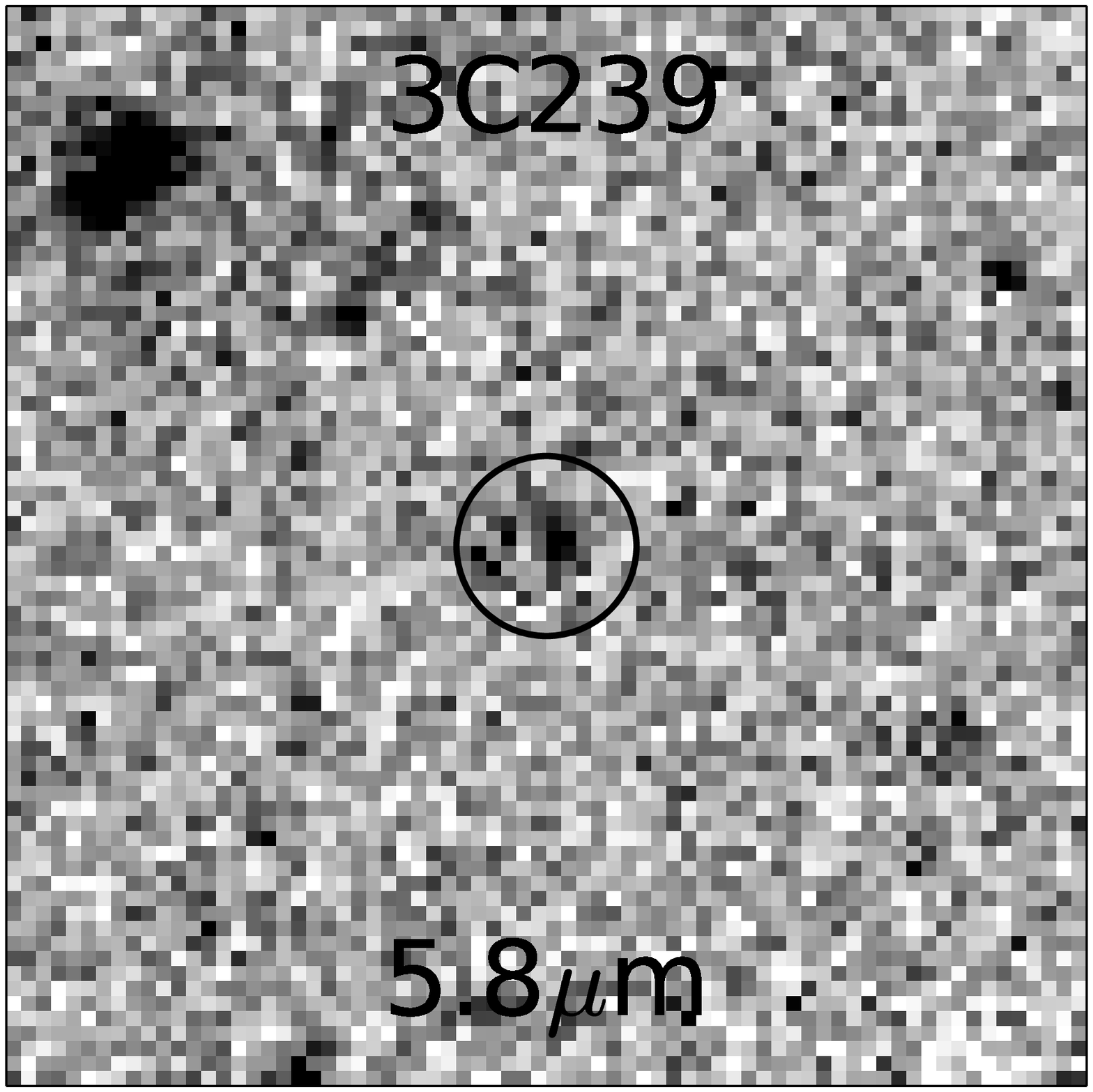}
      \includegraphics[width=1.5cm]{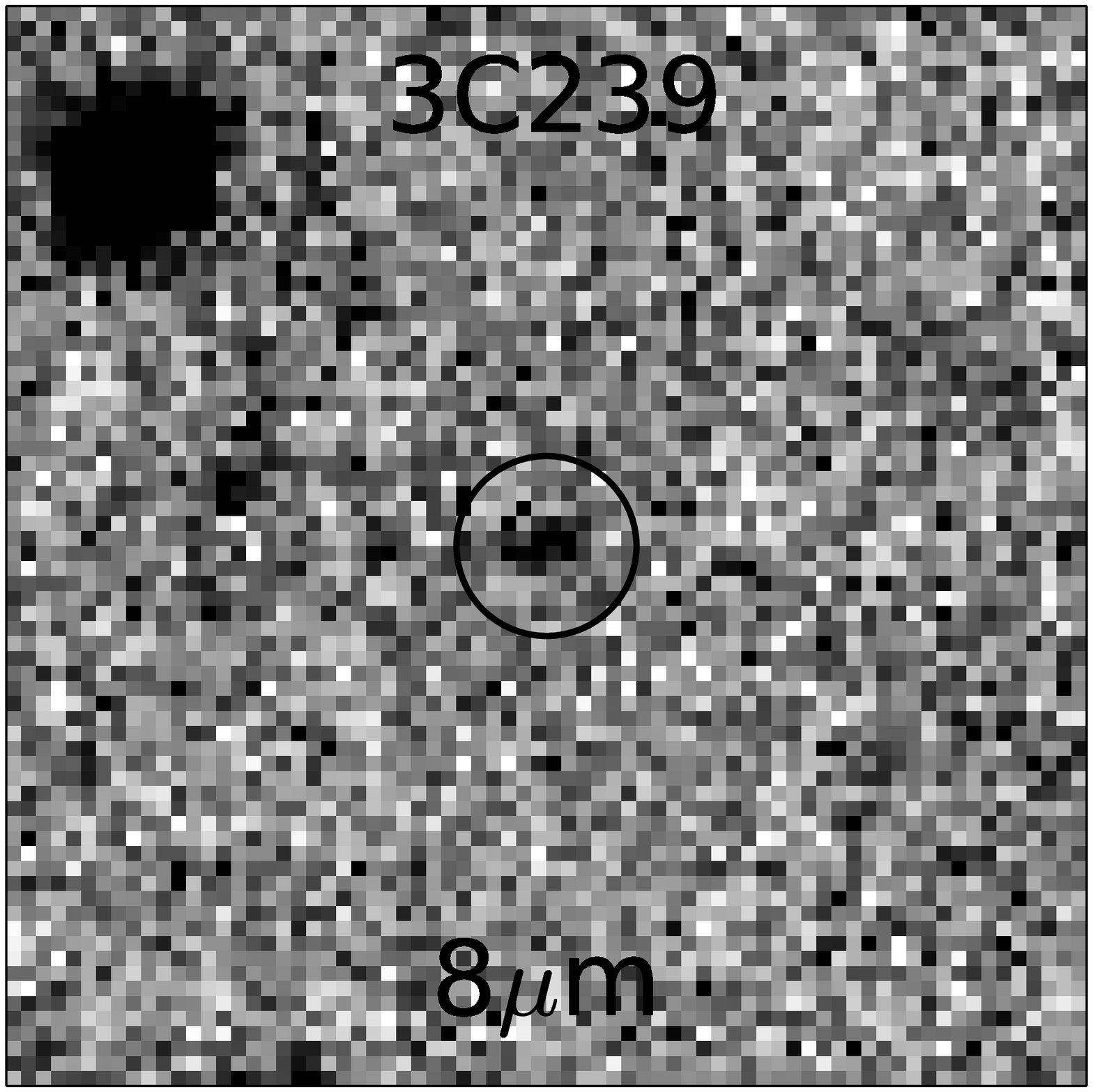}
      \includegraphics[width=1.5cm]{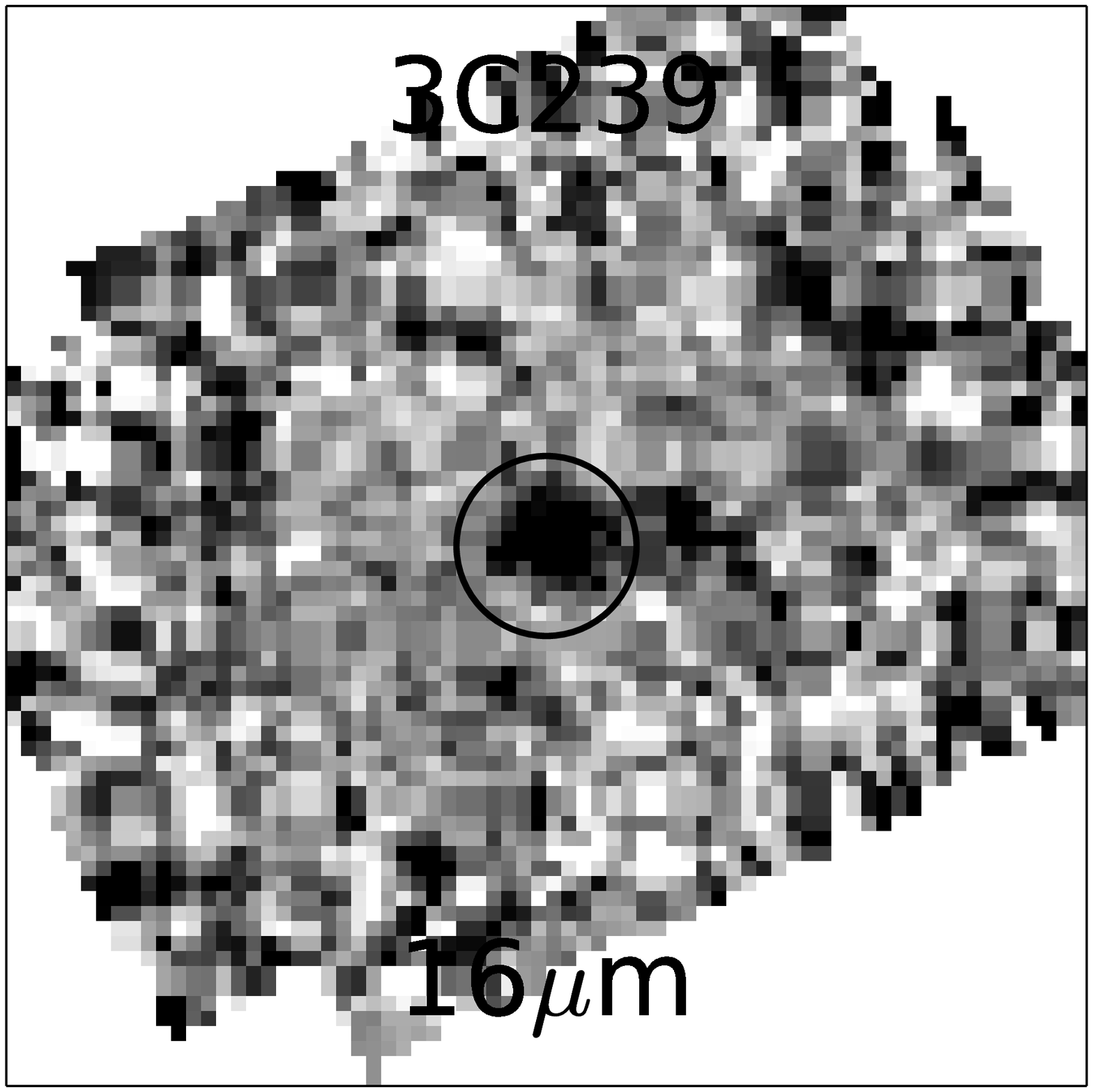}
      \includegraphics[width=1.5cm]{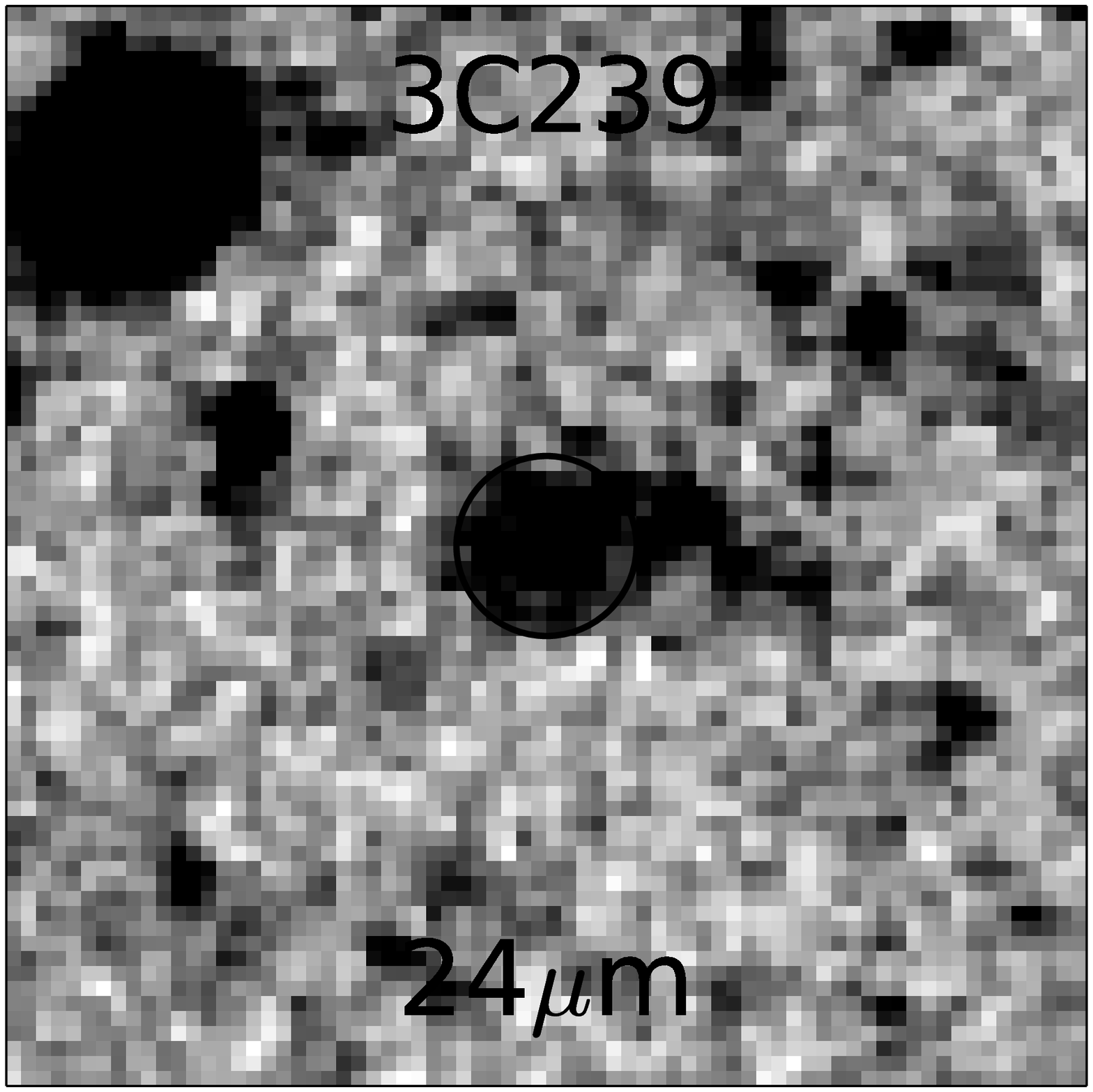}
      \includegraphics[width=1.5cm]{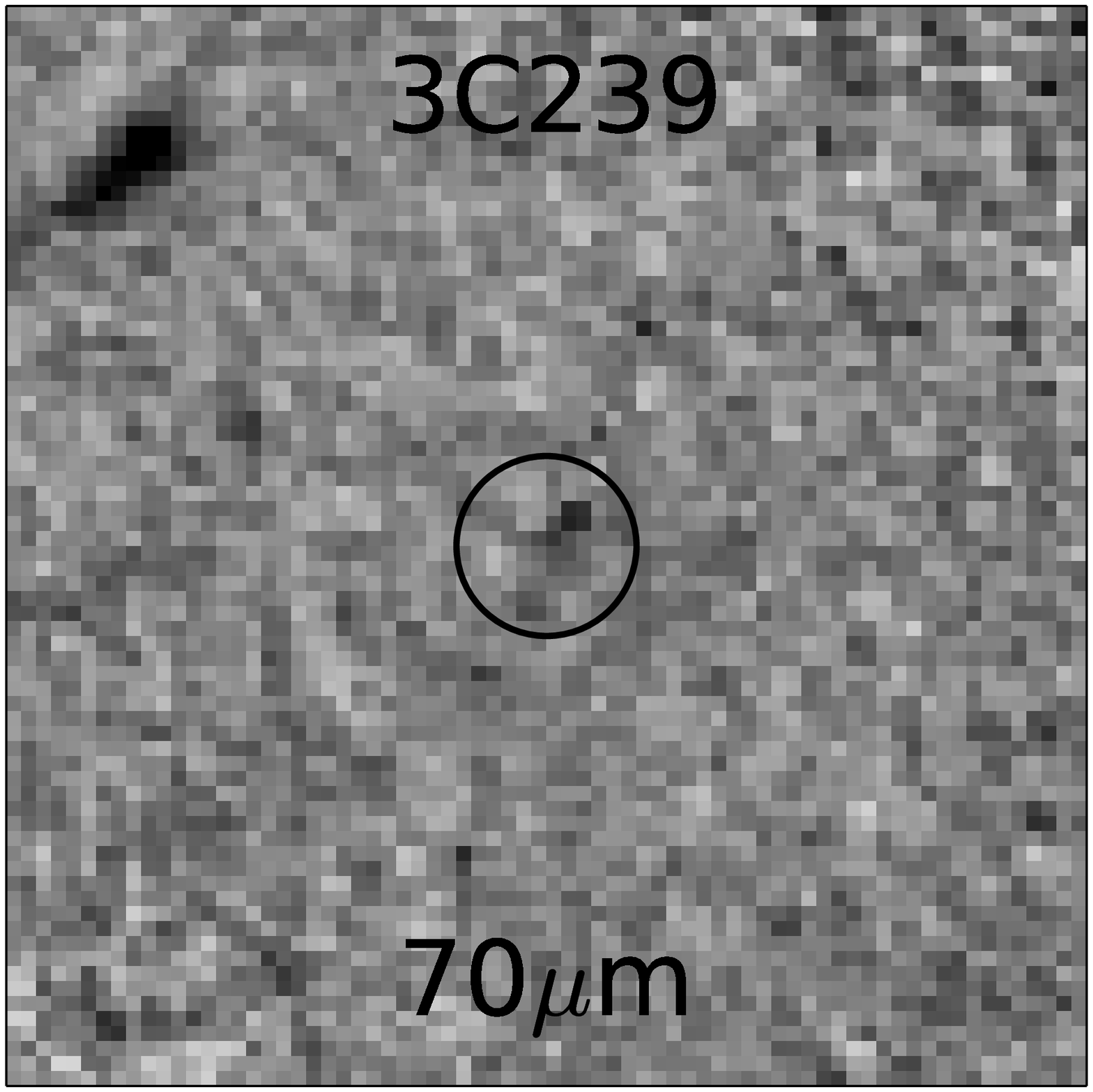}
      \includegraphics[width=1.5cm]{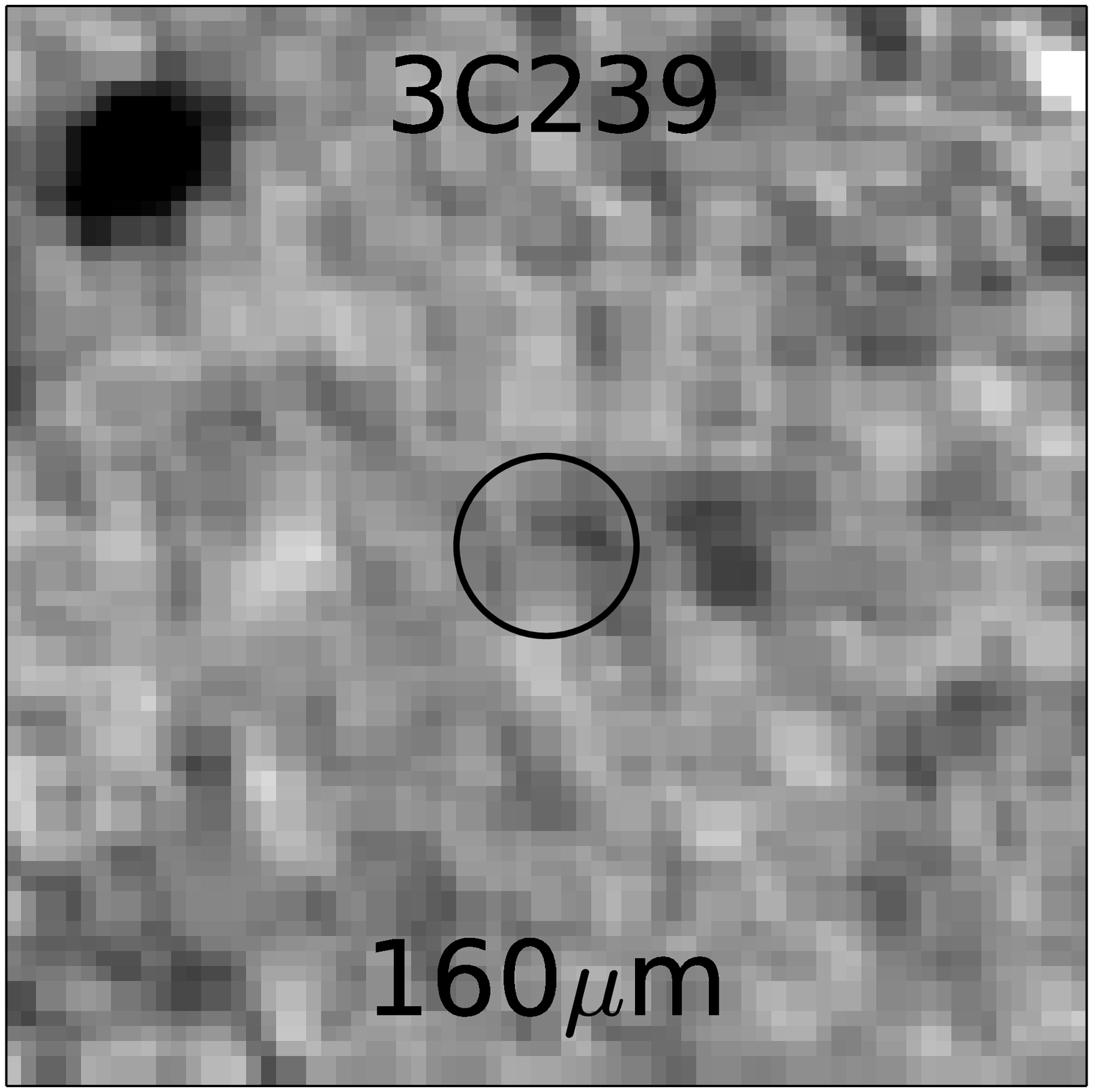}
      \includegraphics[width=1.5cm]{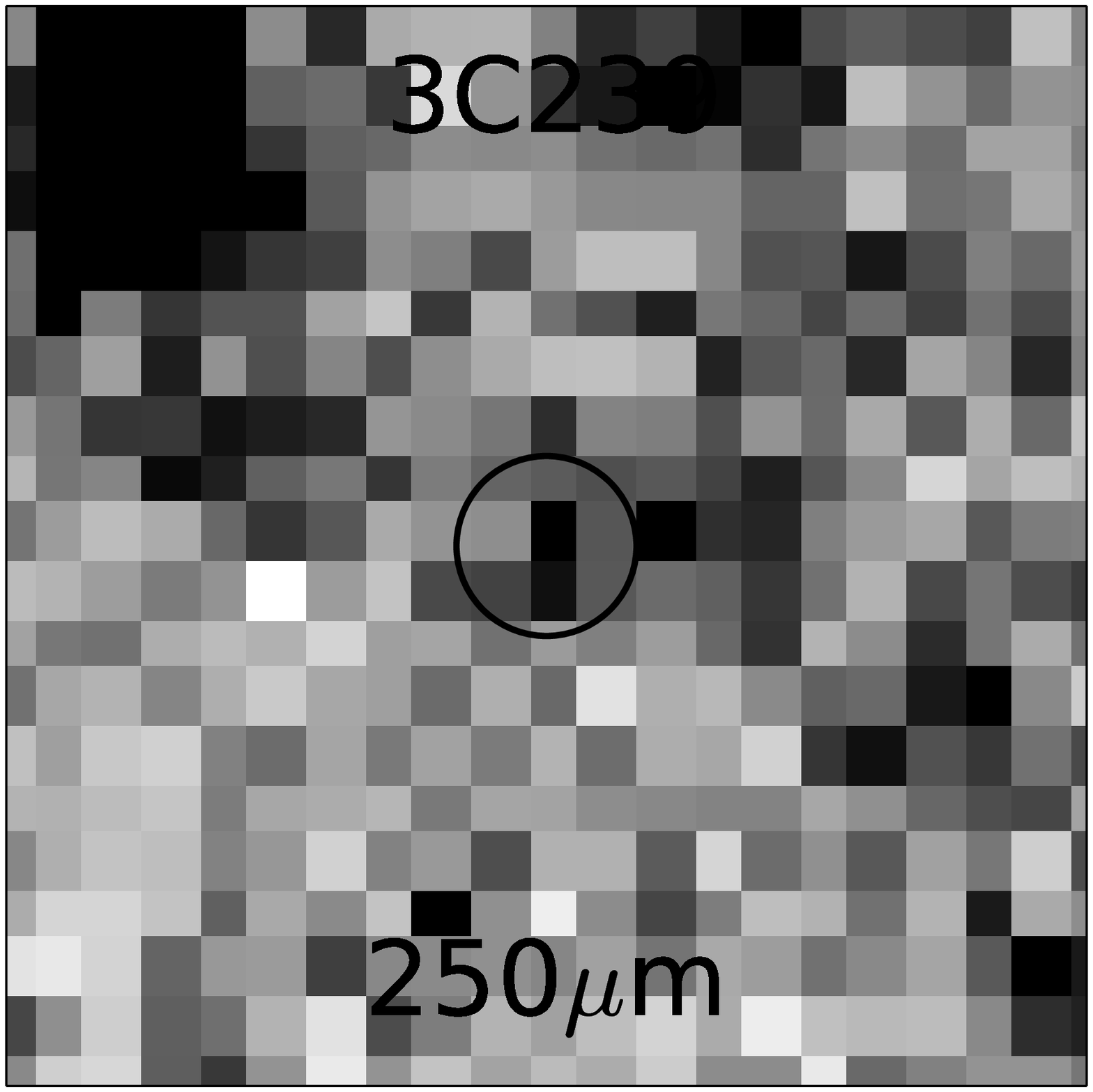}
      \includegraphics[width=1.5cm]{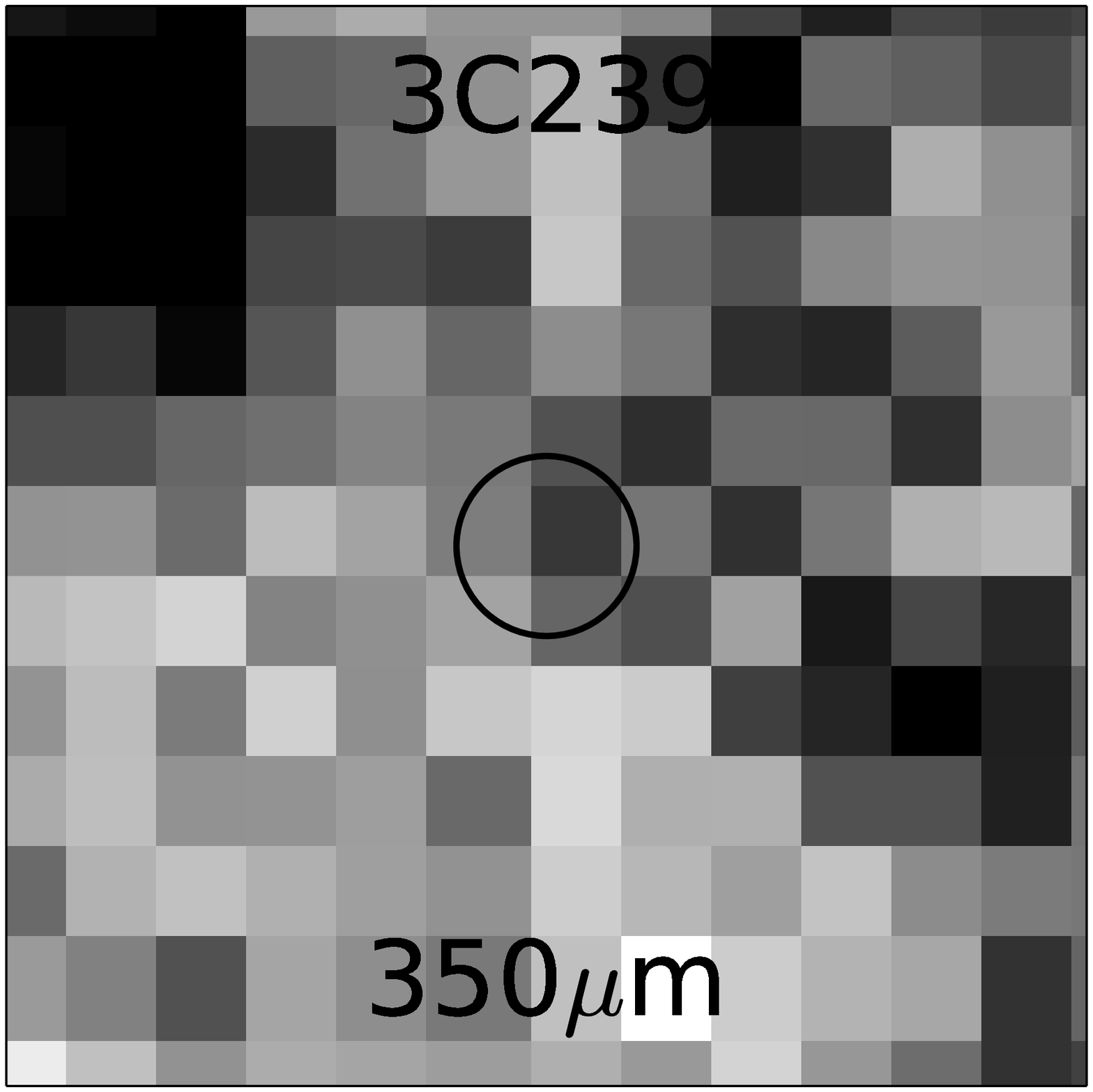}
      \includegraphics[width=1.5cm]{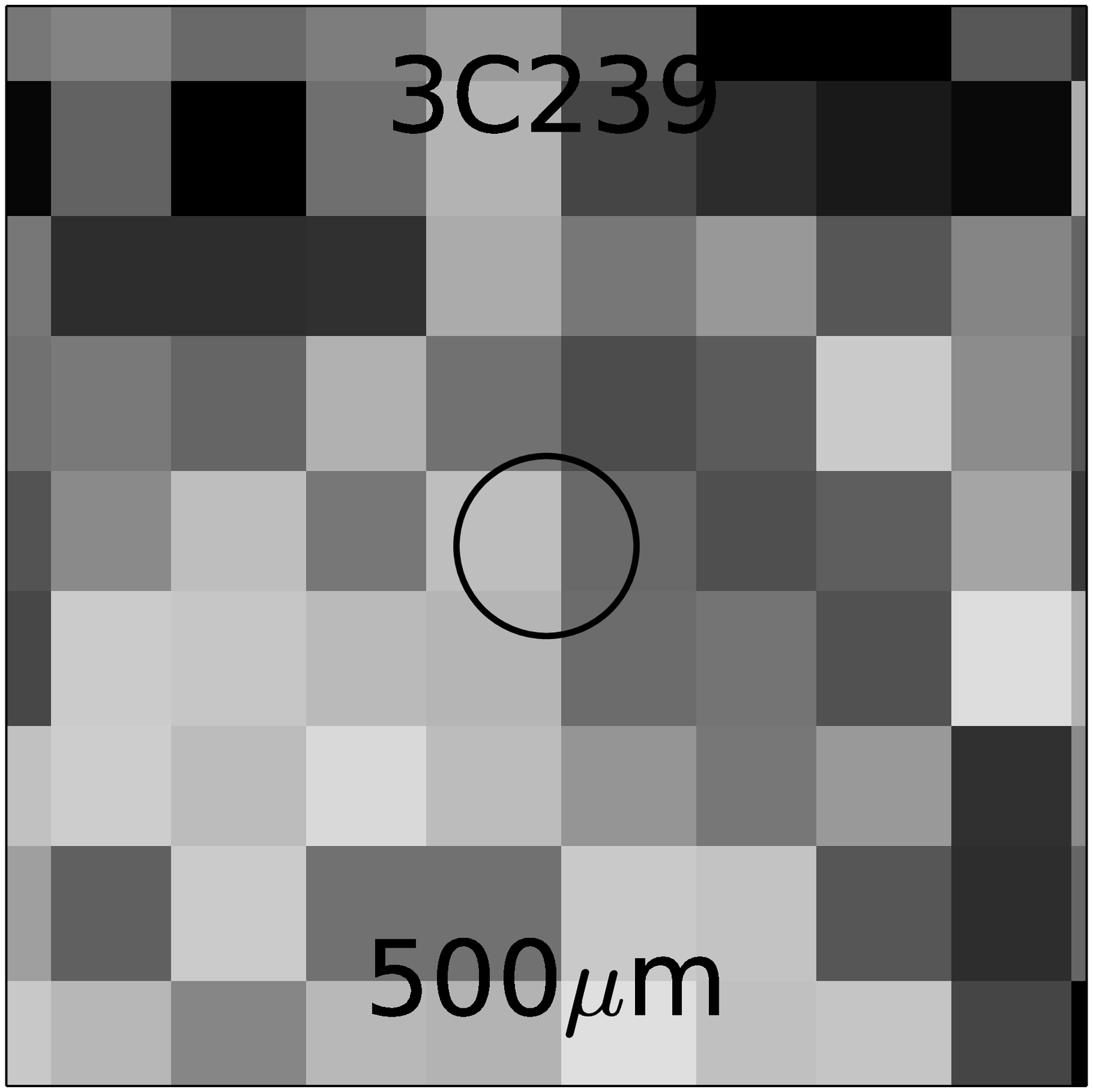}
      \\
      \includegraphics[width=1.5cm]{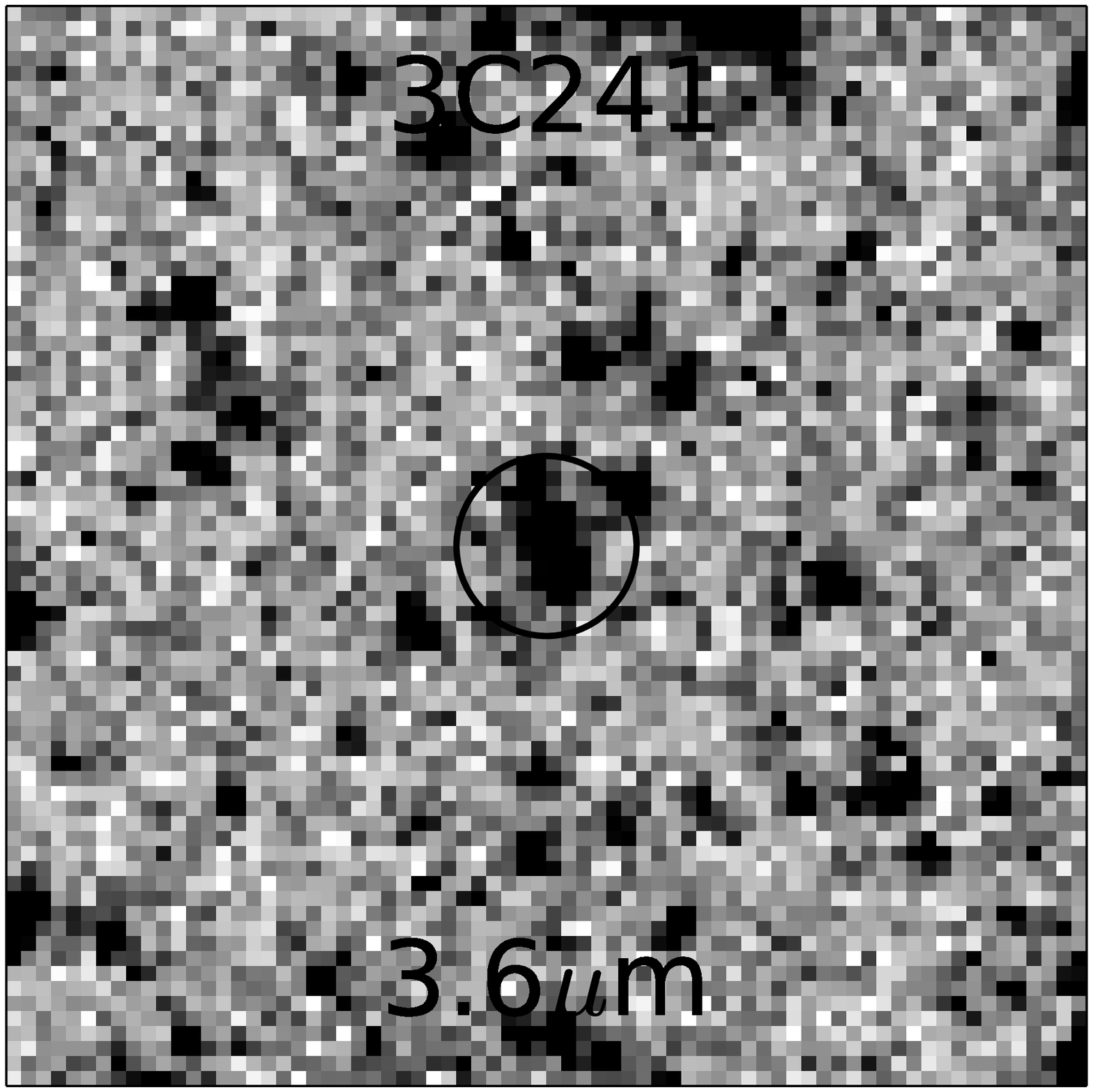}
      \includegraphics[width=1.5cm]{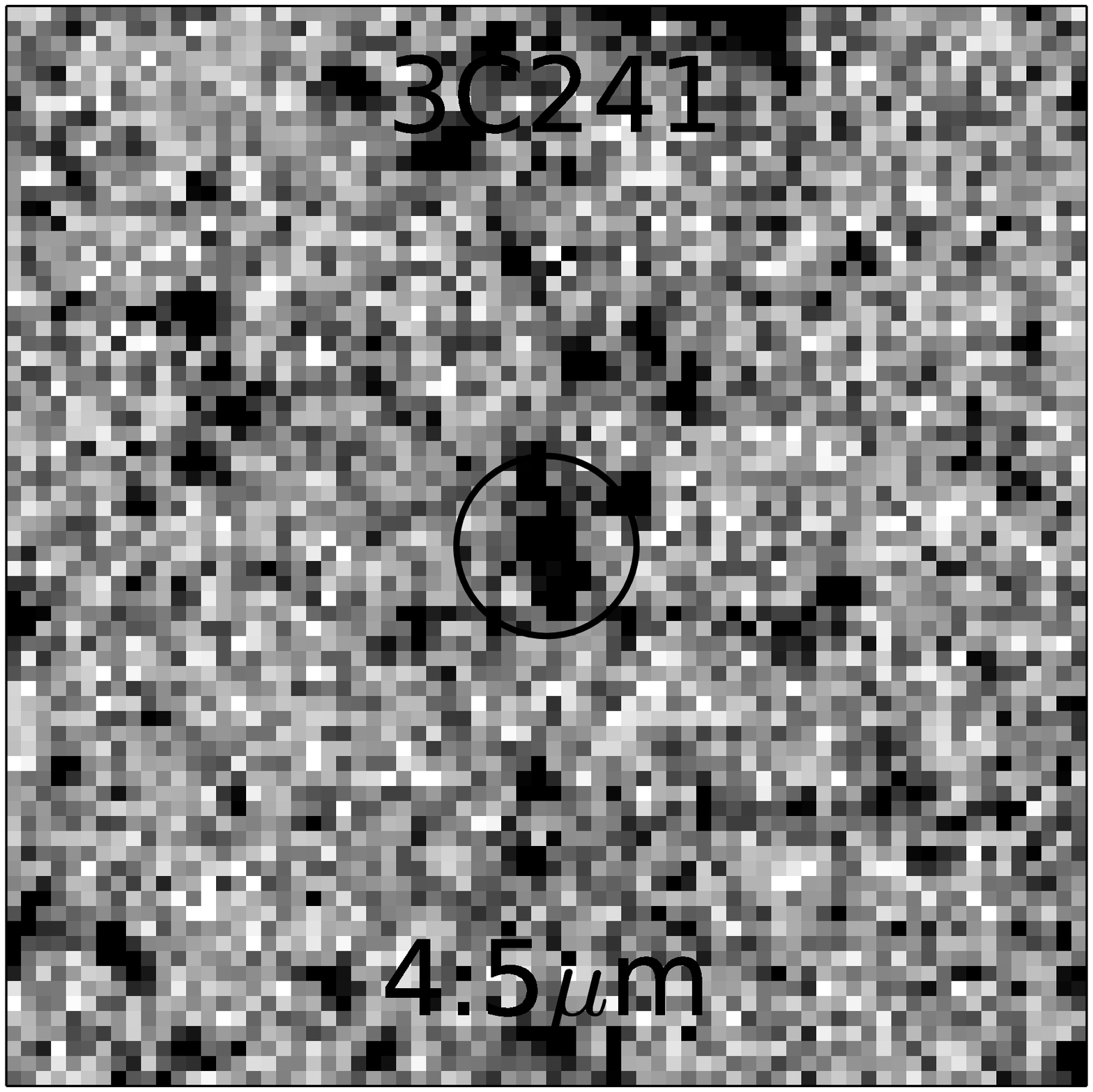}
      \includegraphics[width=1.5cm]{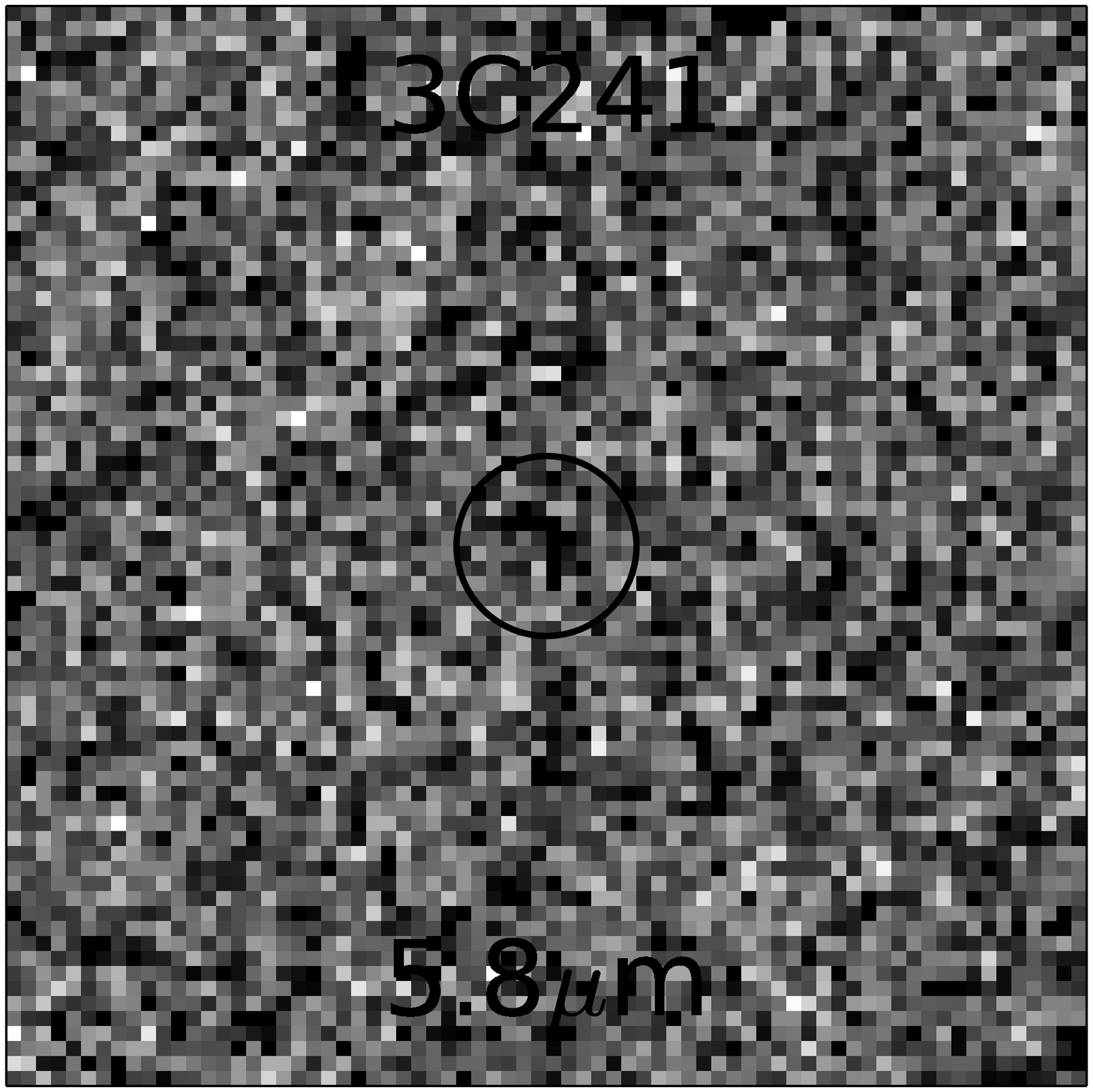}
      \includegraphics[width=1.5cm]{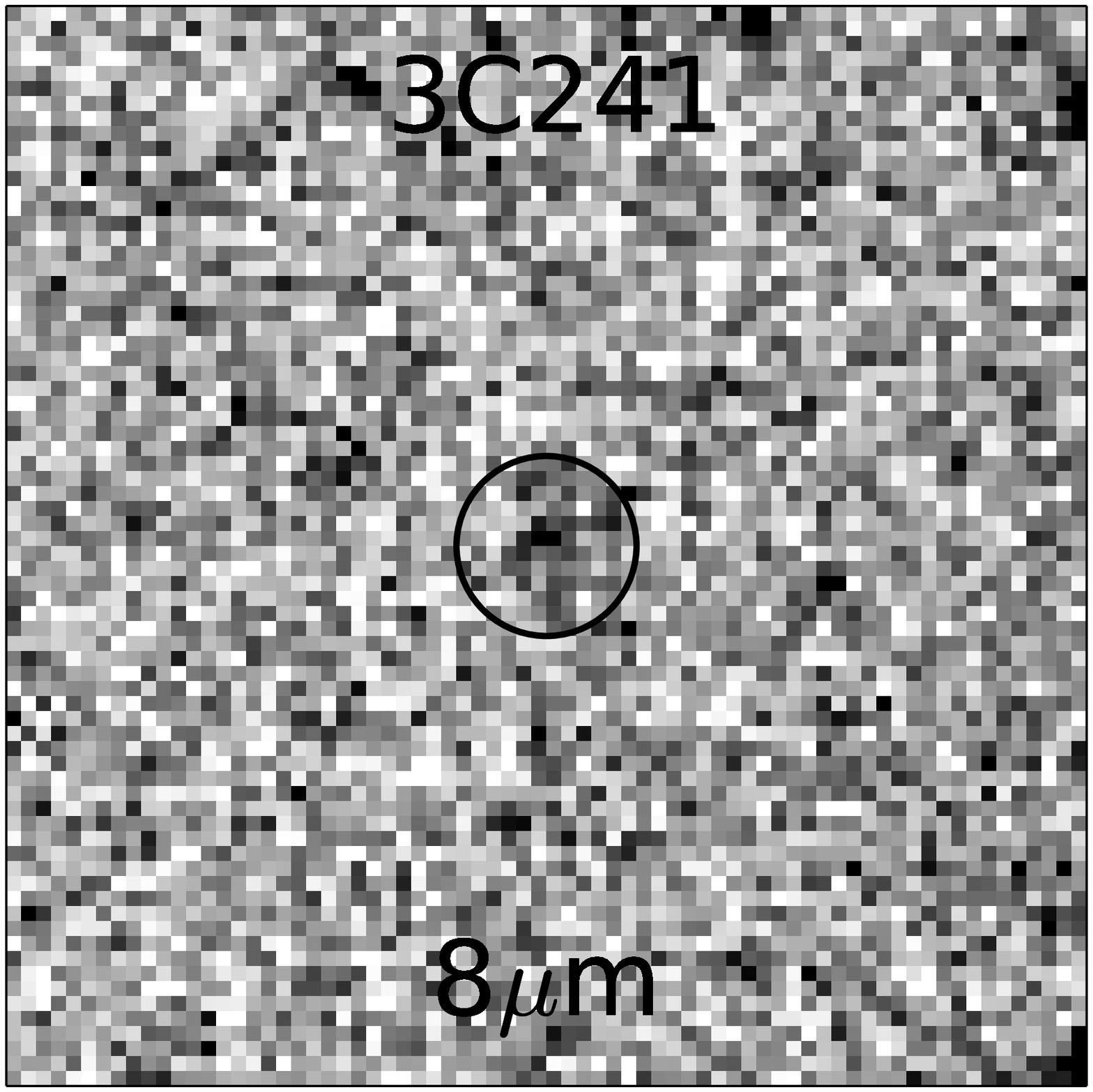}
      \includegraphics[width=1.5cm]{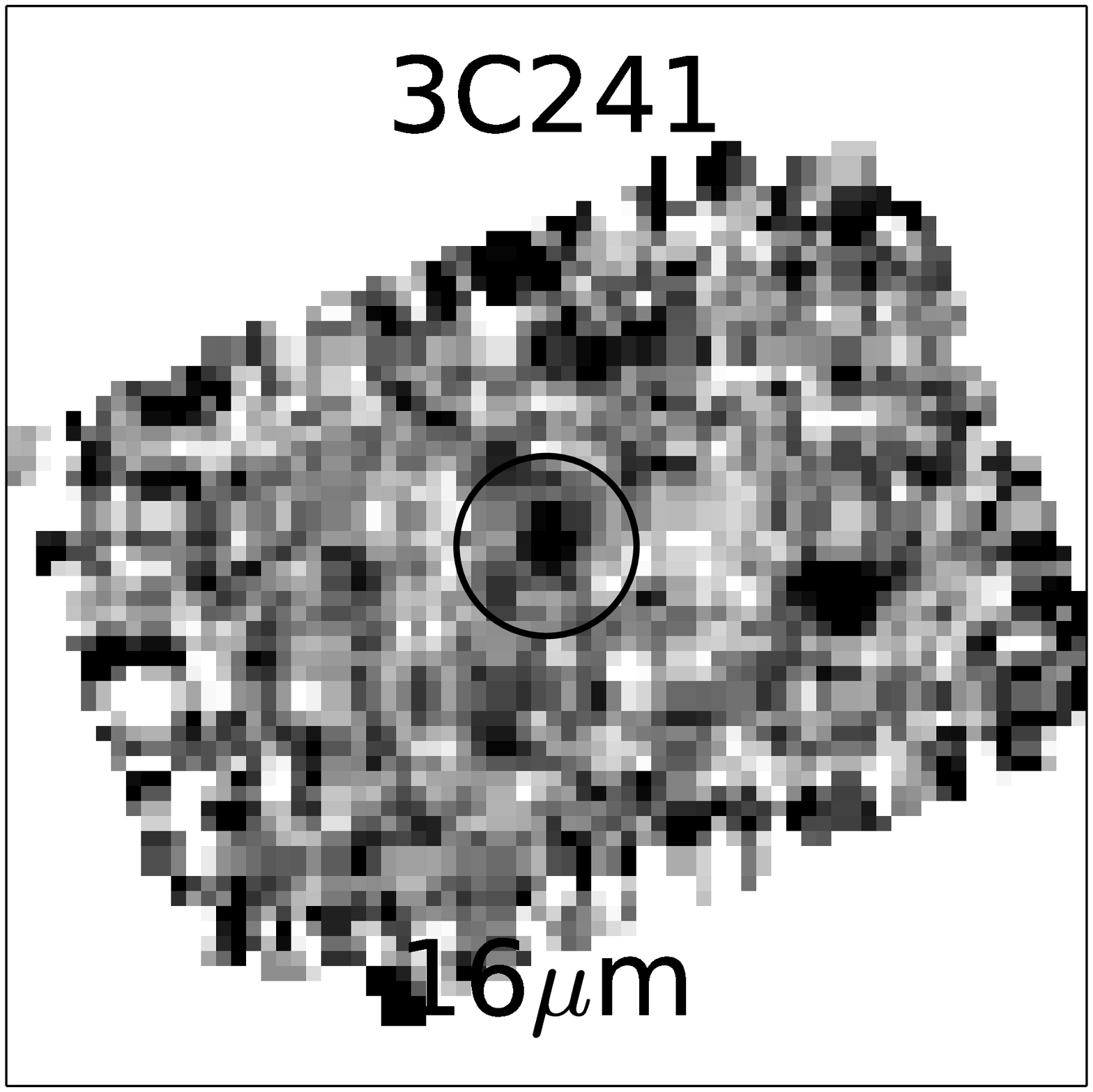}
      \includegraphics[width=1.5cm]{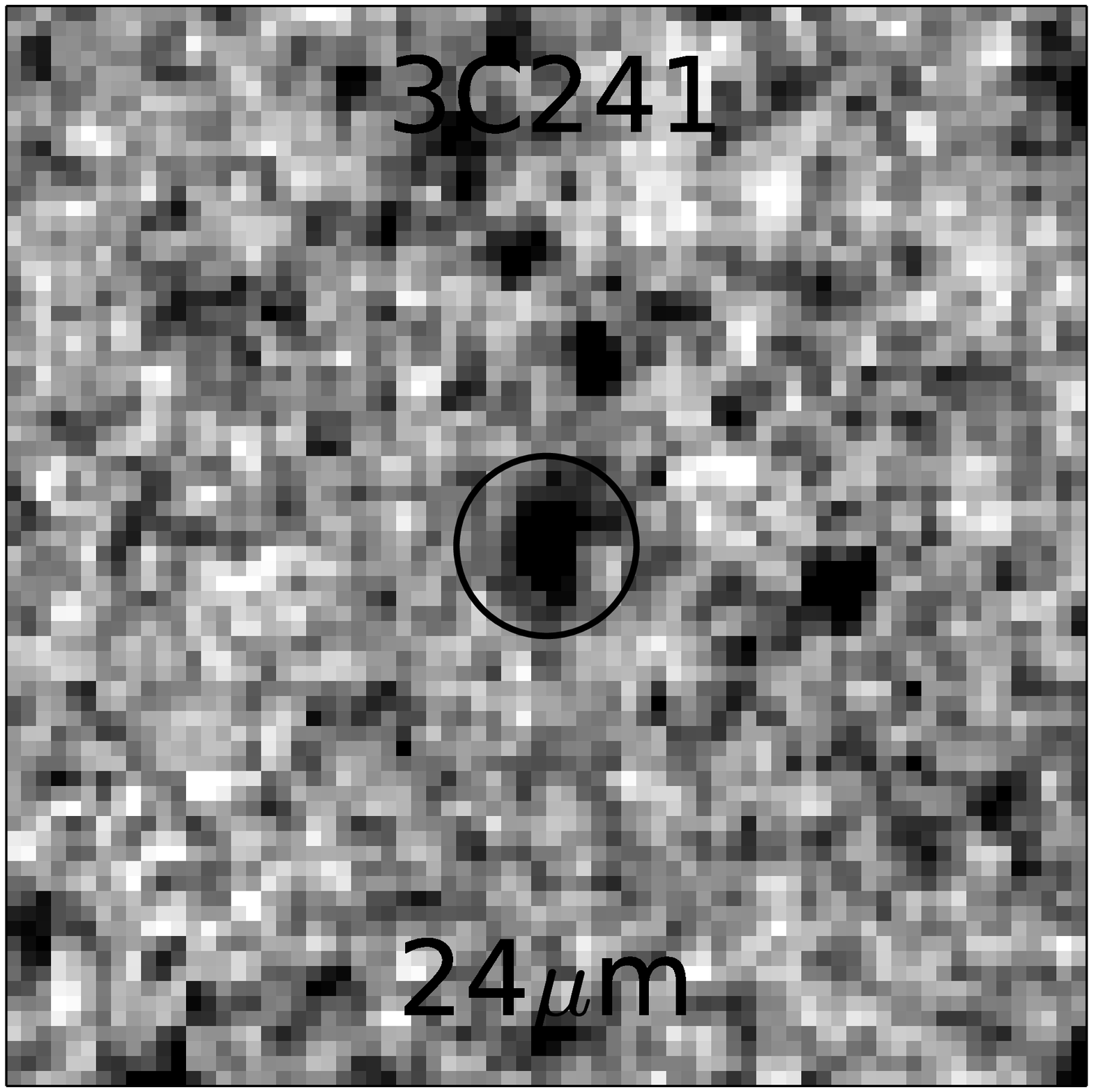}
      \includegraphics[width=1.5cm]{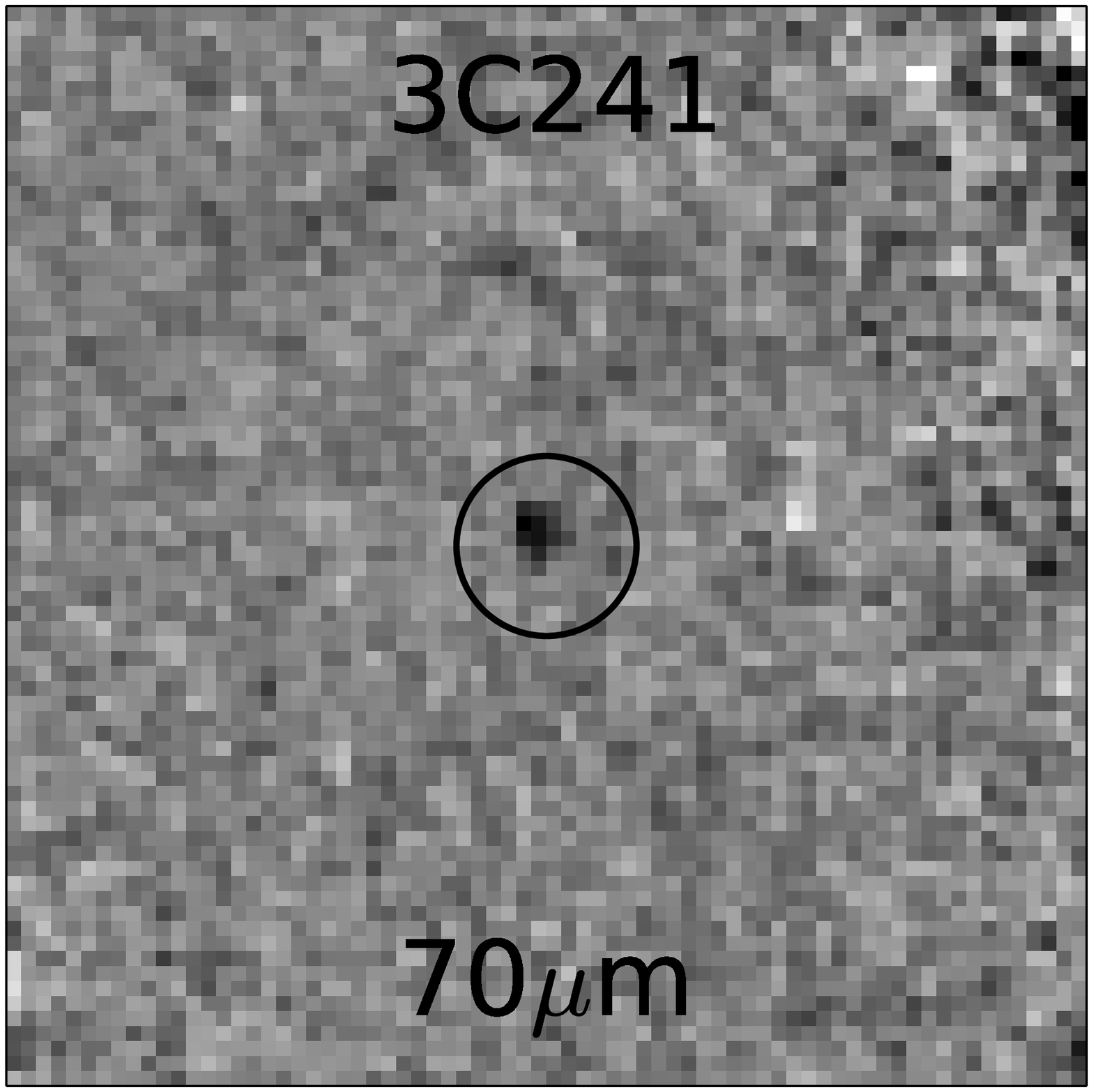}
      \includegraphics[width=1.5cm]{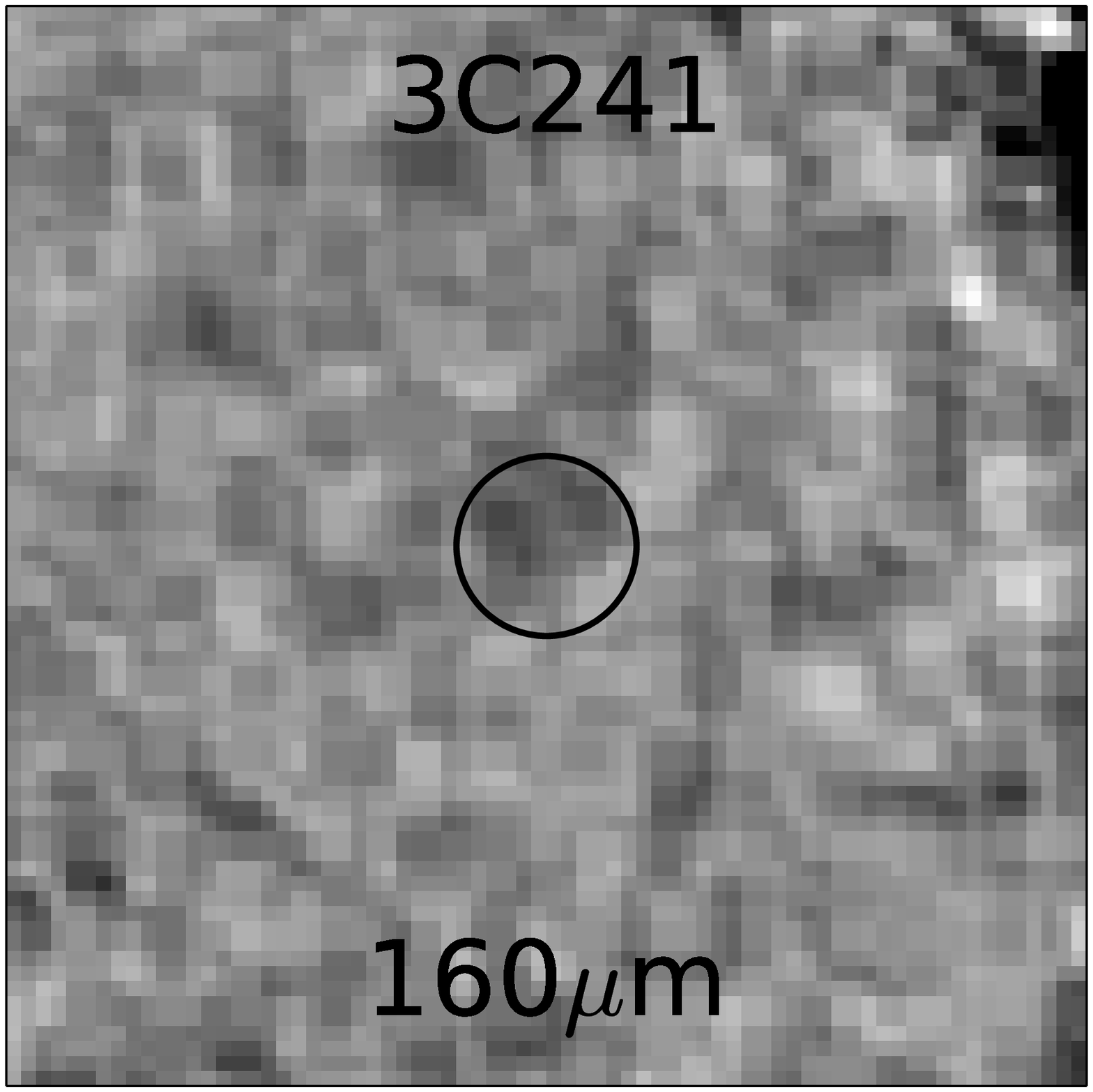}
      \includegraphics[width=1.5cm]{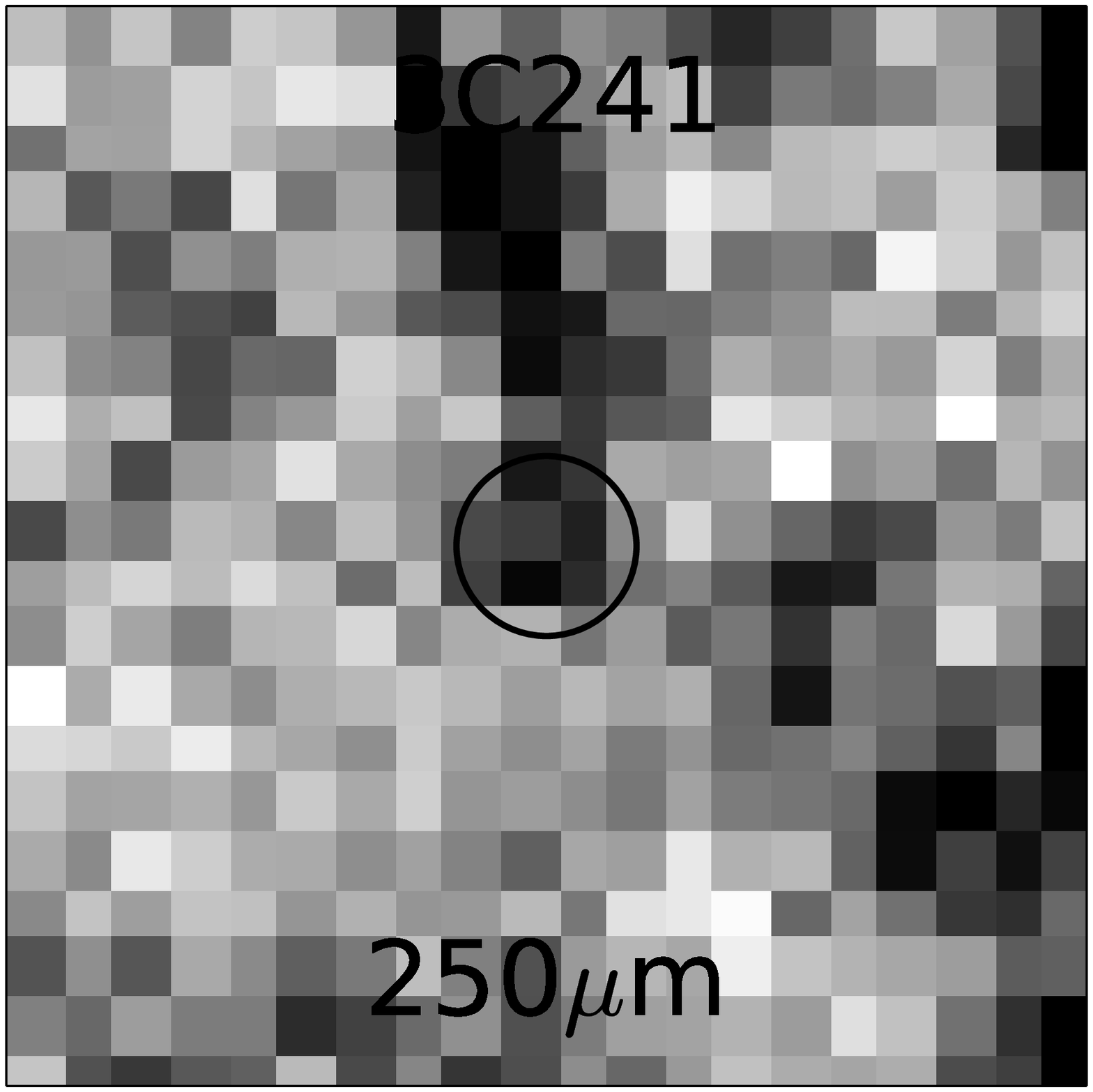}
      \includegraphics[width=1.5cm]{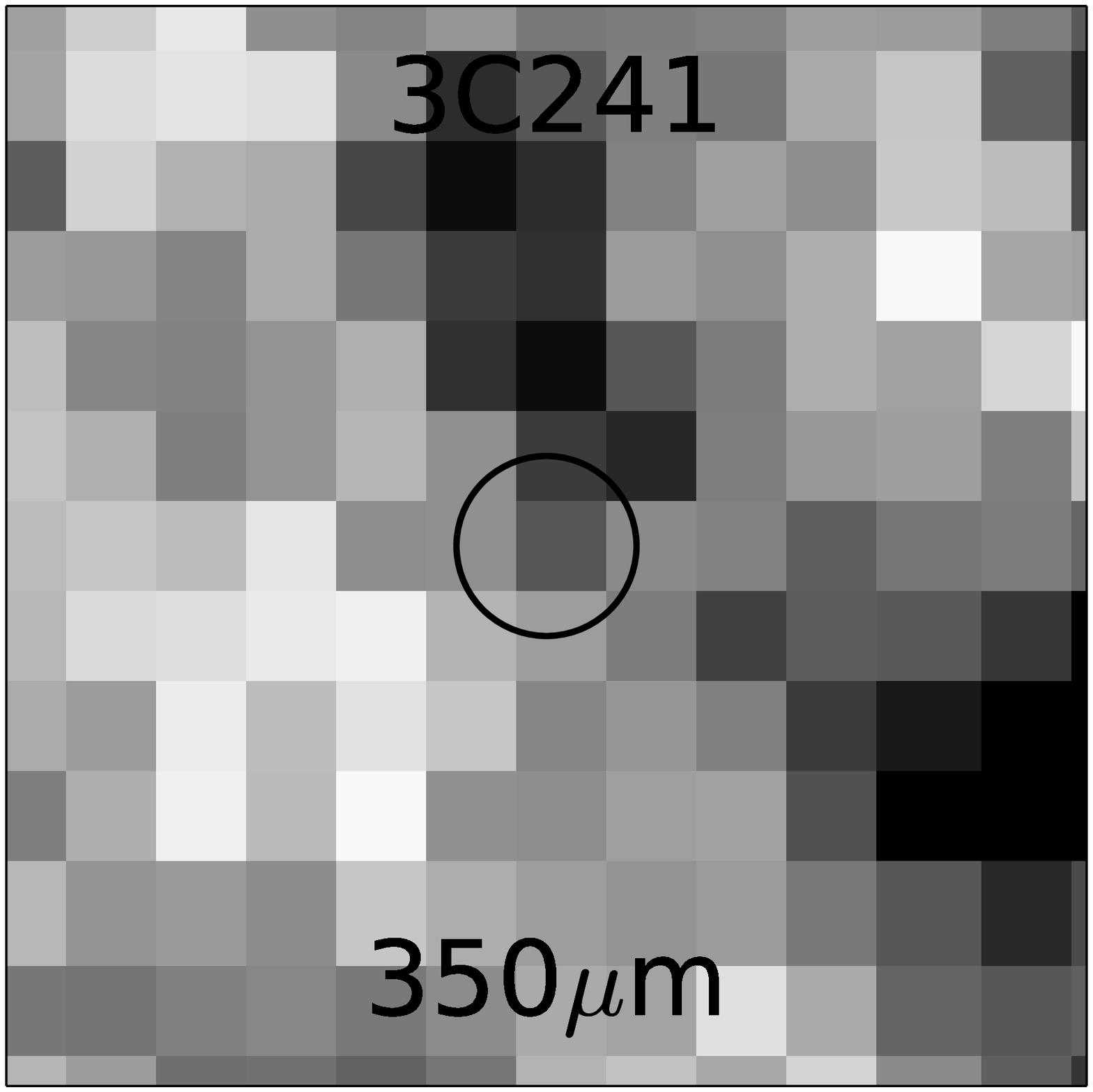}
      \includegraphics[width=1.5cm]{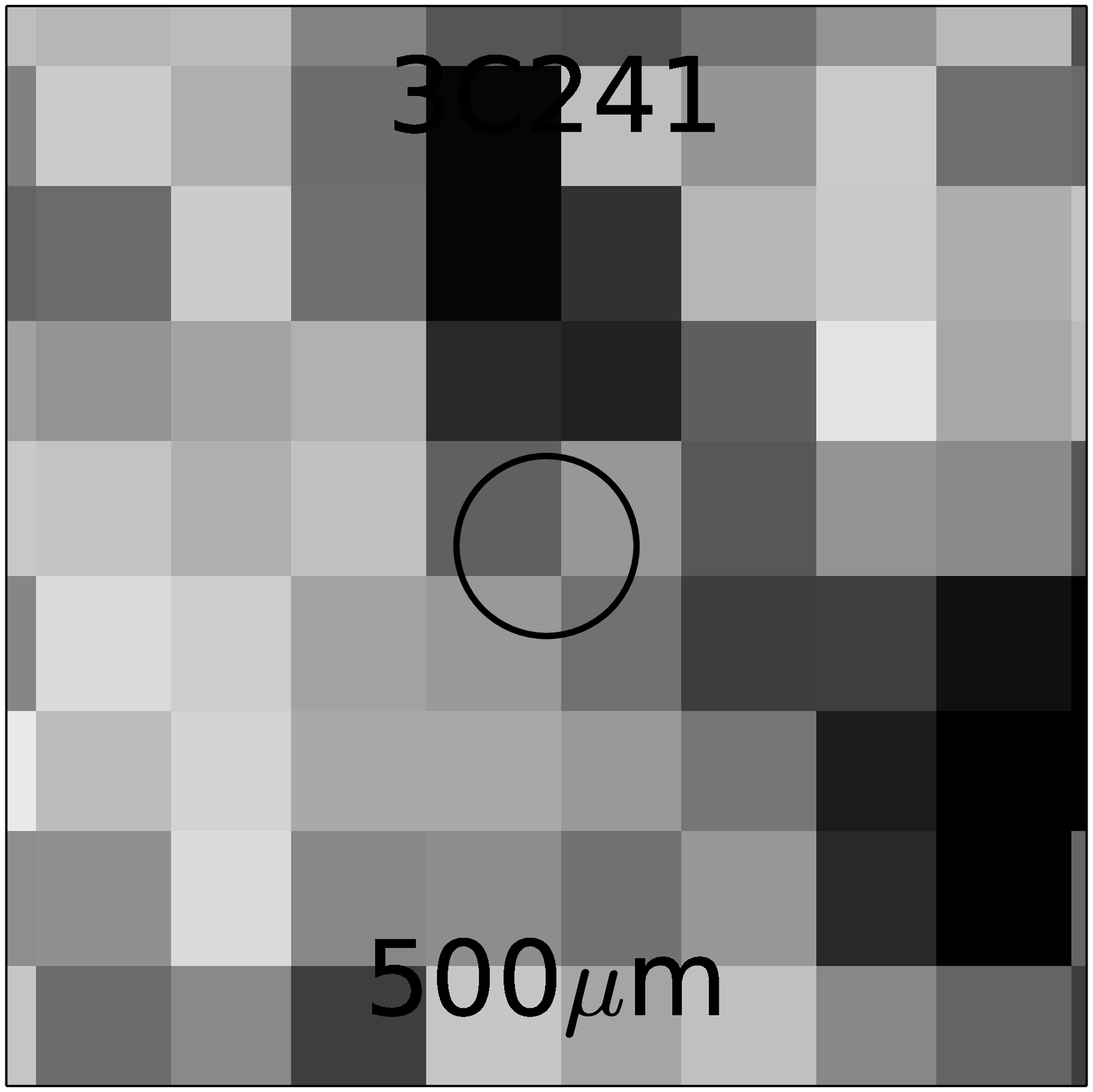}
      \\
      \includegraphics[width=1.5cm]{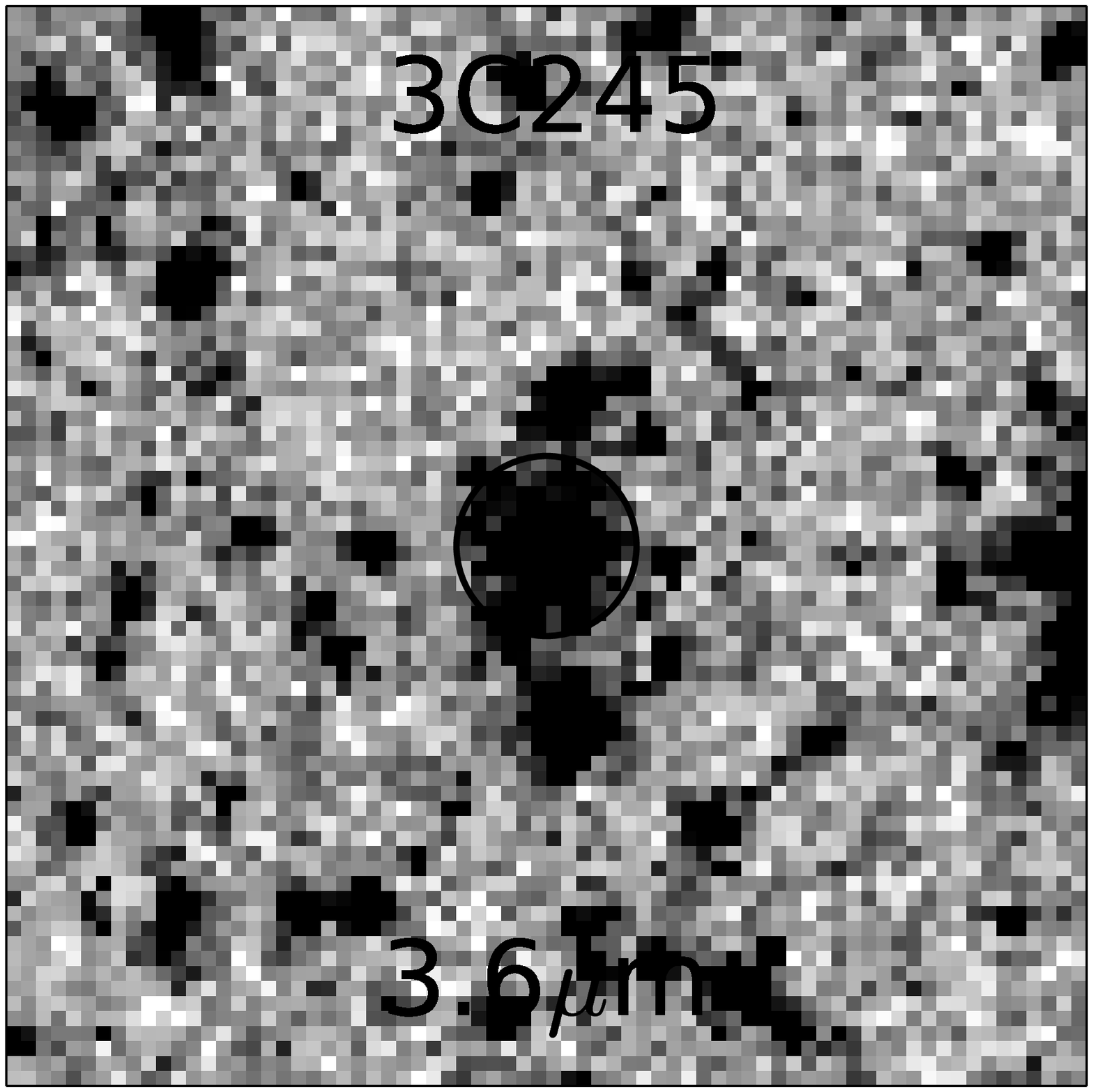}
      \includegraphics[width=1.5cm]{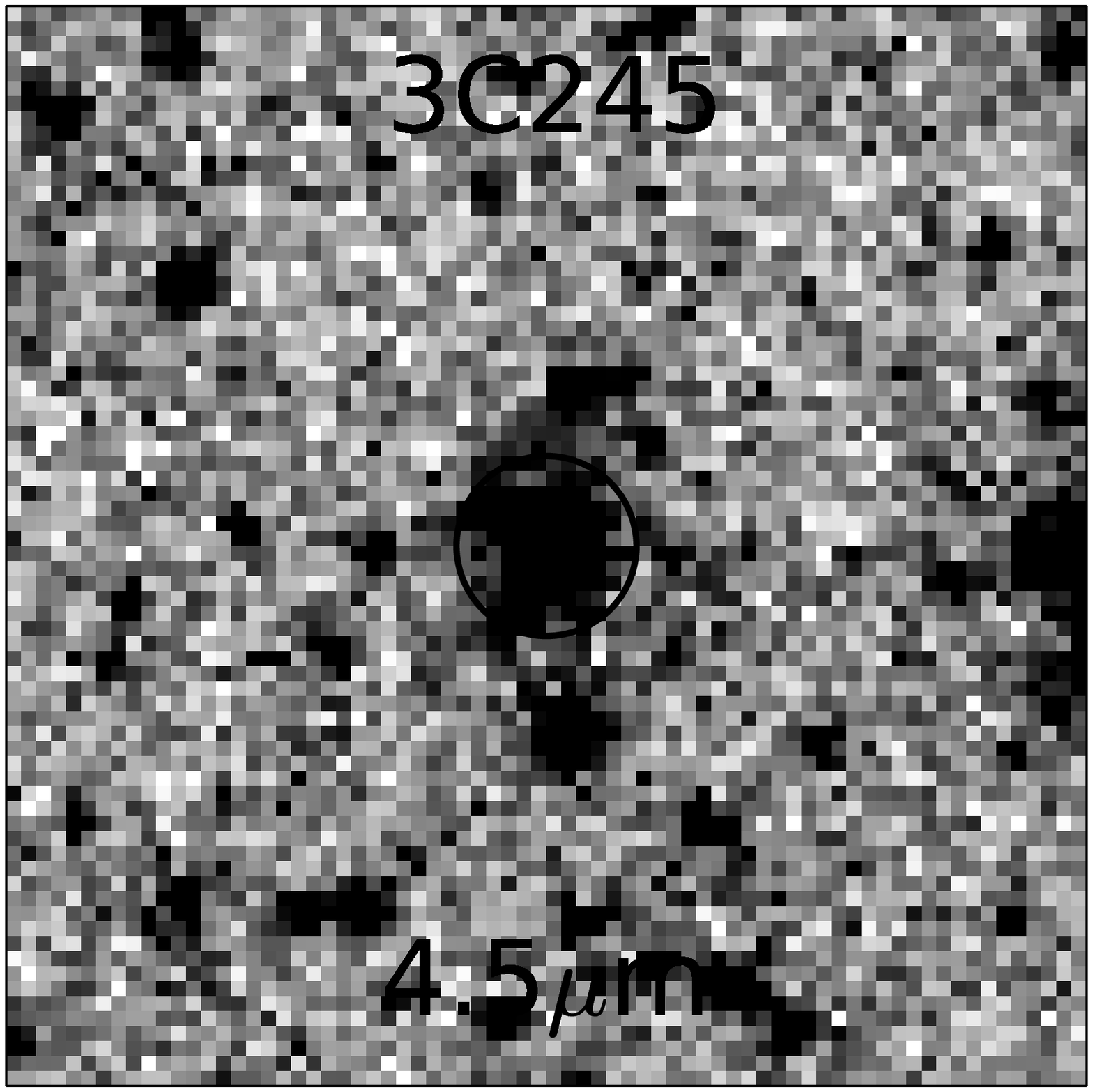}
      \includegraphics[width=1.5cm]{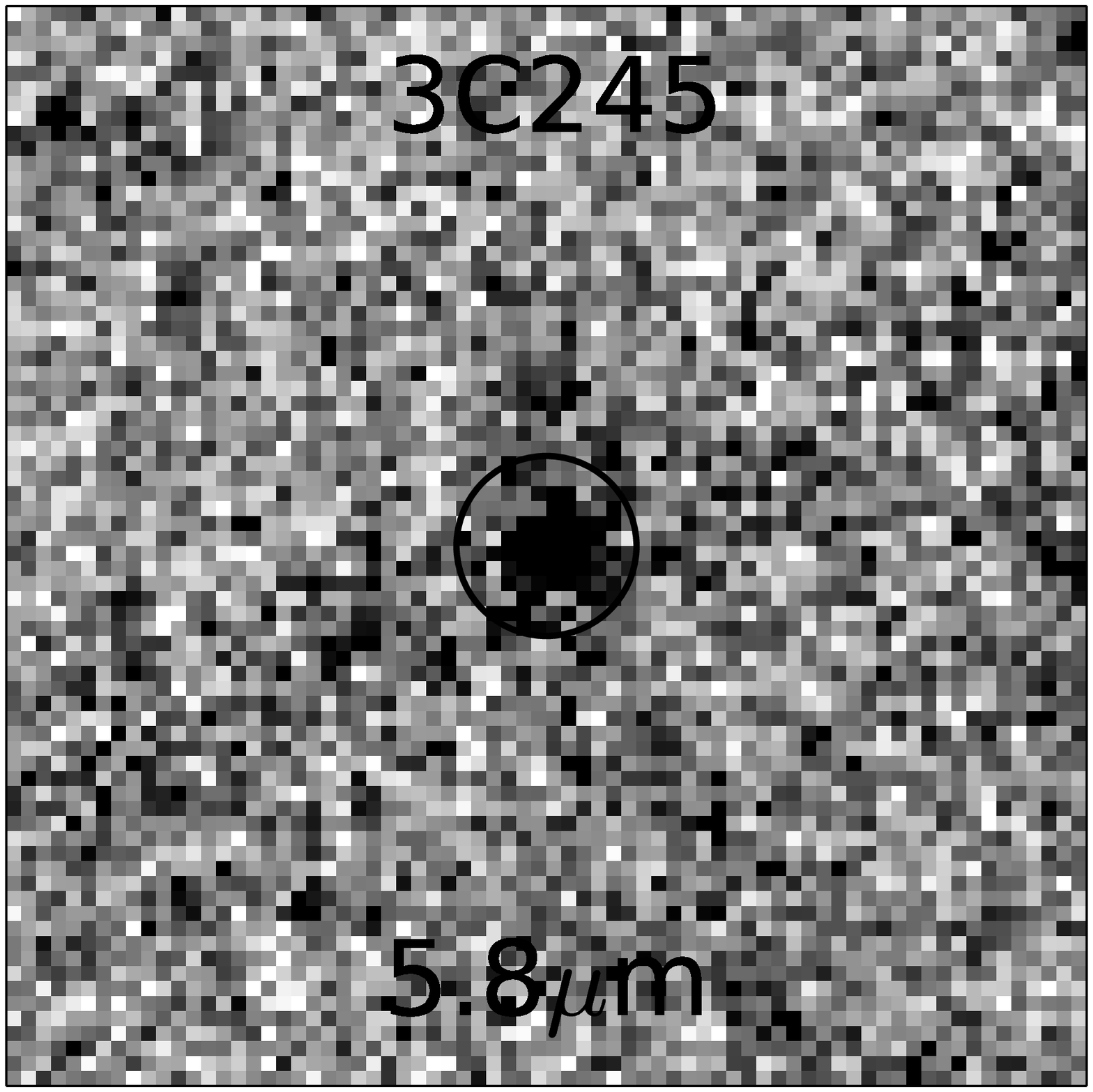}
      \includegraphics[width=1.5cm]{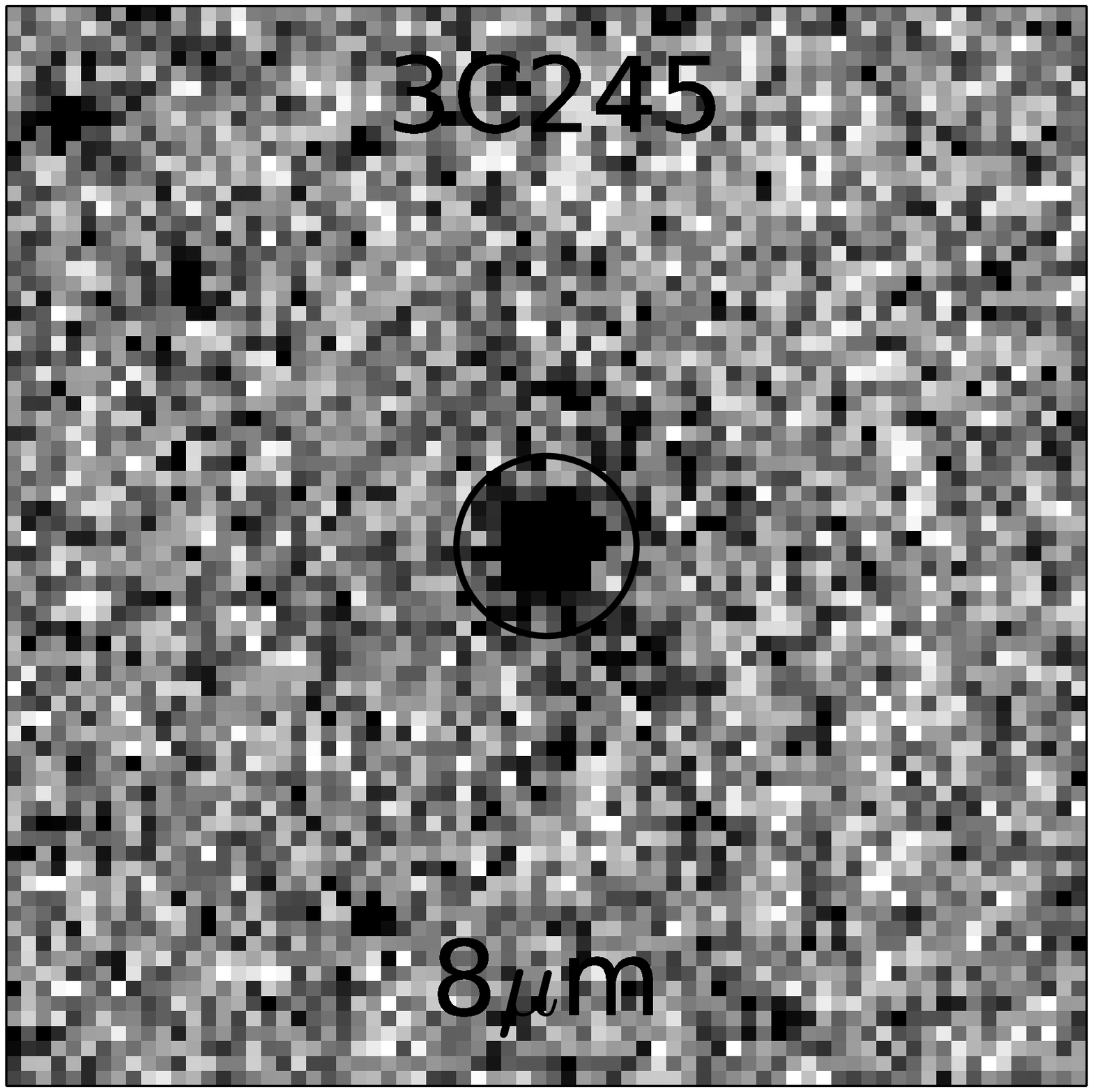}
      \includegraphics[width=1.5cm]{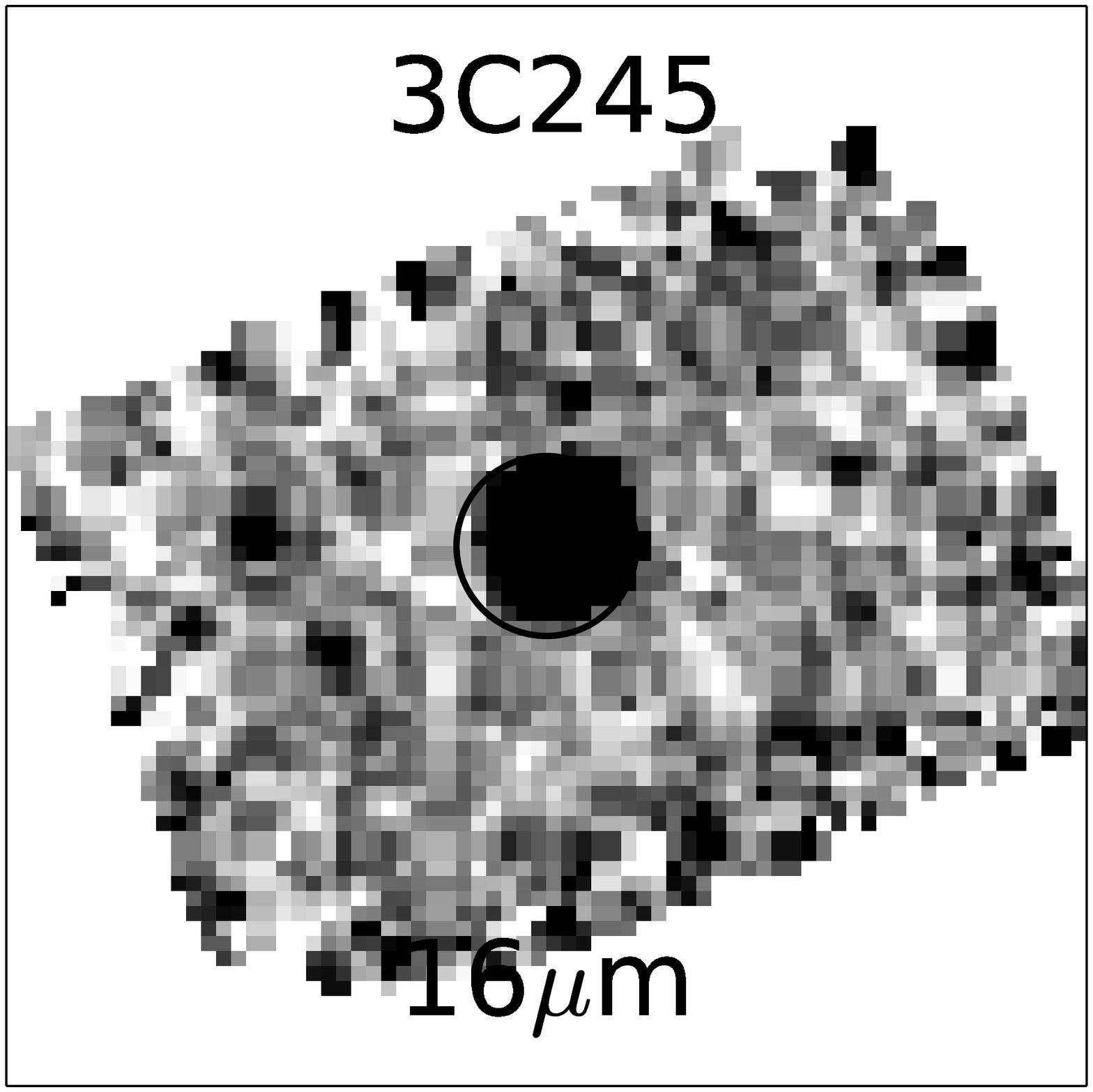}
      \includegraphics[width=1.5cm]{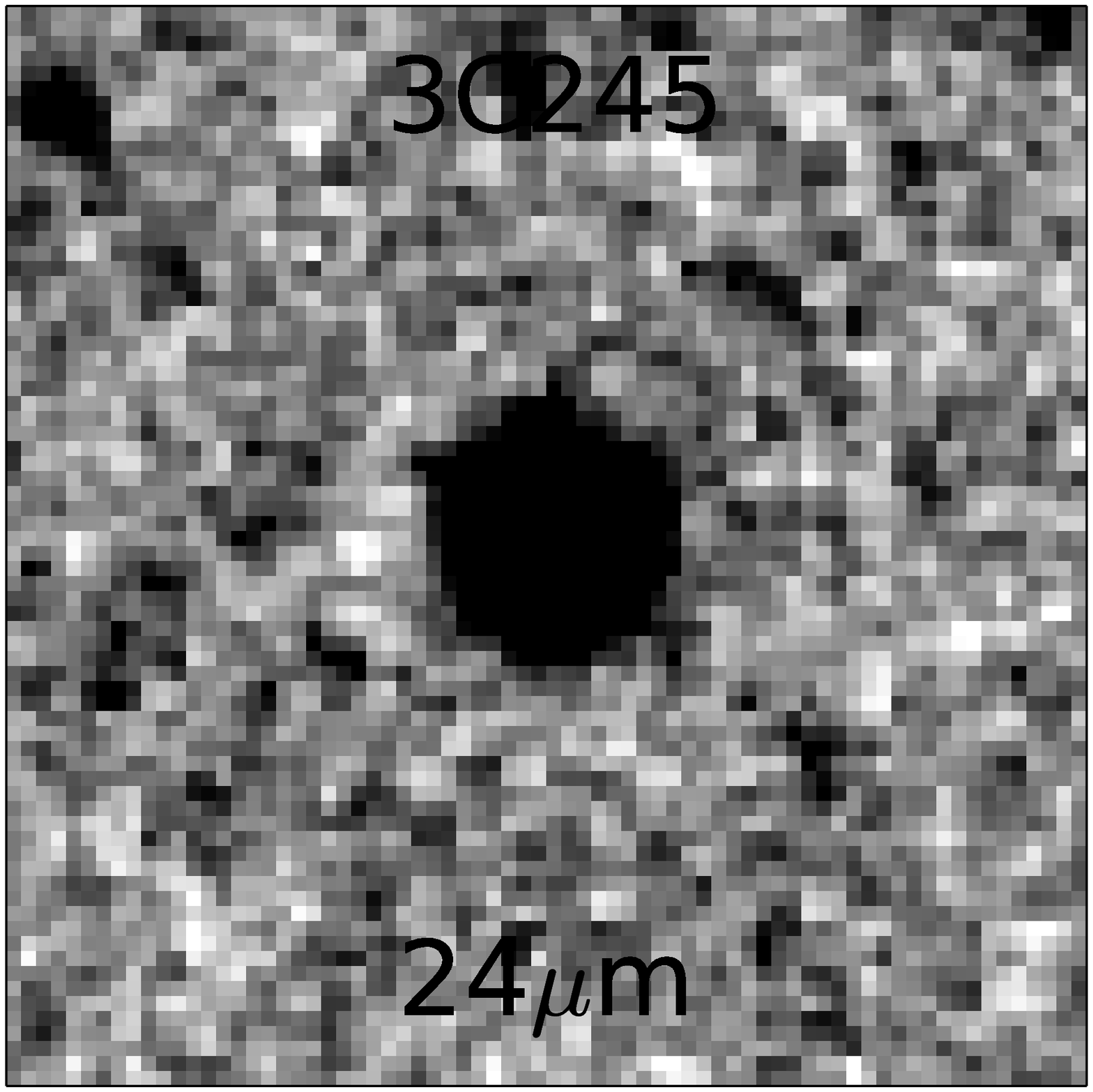}
      \includegraphics[width=1.5cm]{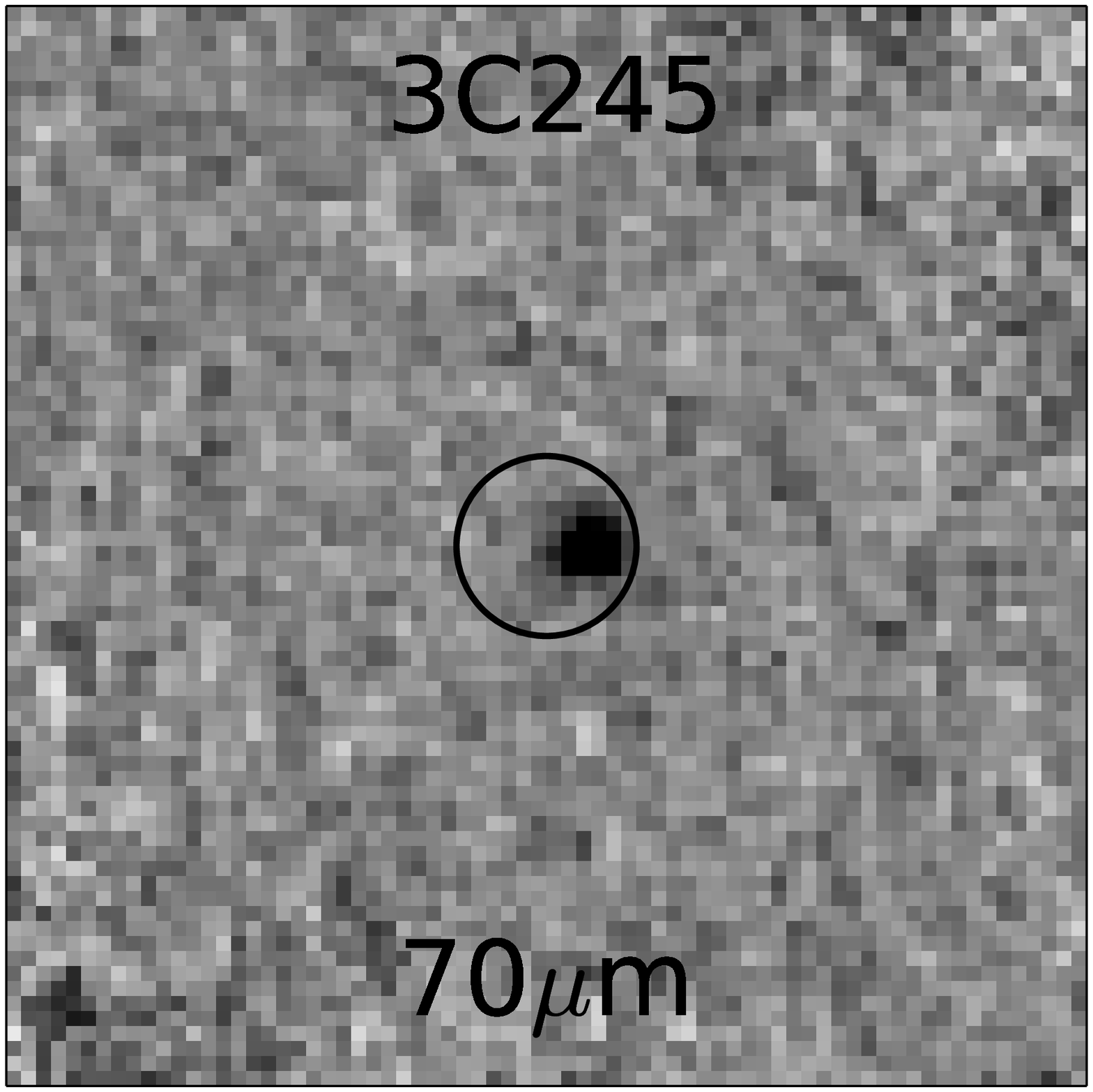}
      \includegraphics[width=1.5cm]{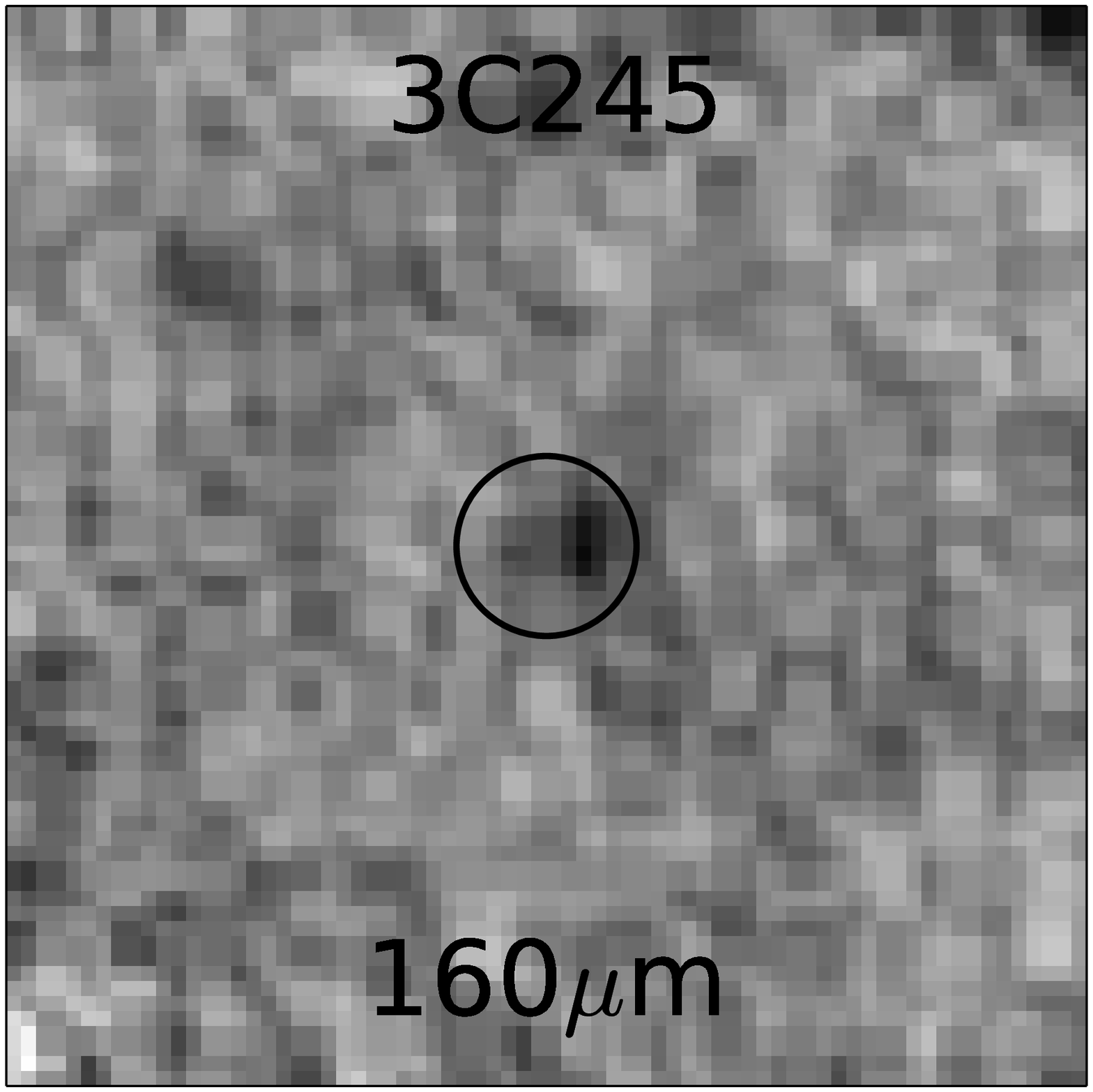}
      \includegraphics[width=1.5cm]{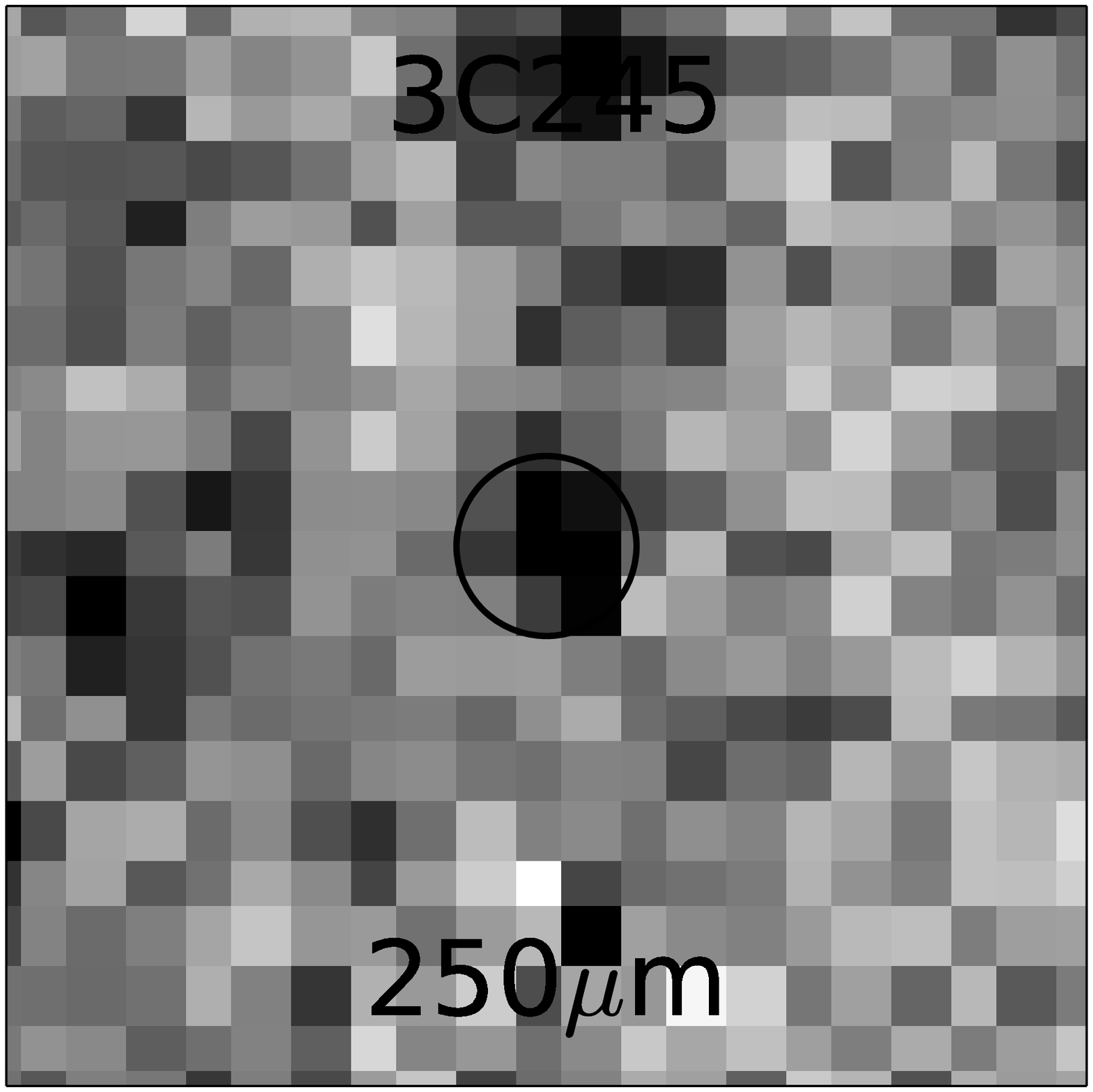}
      \includegraphics[width=1.5cm]{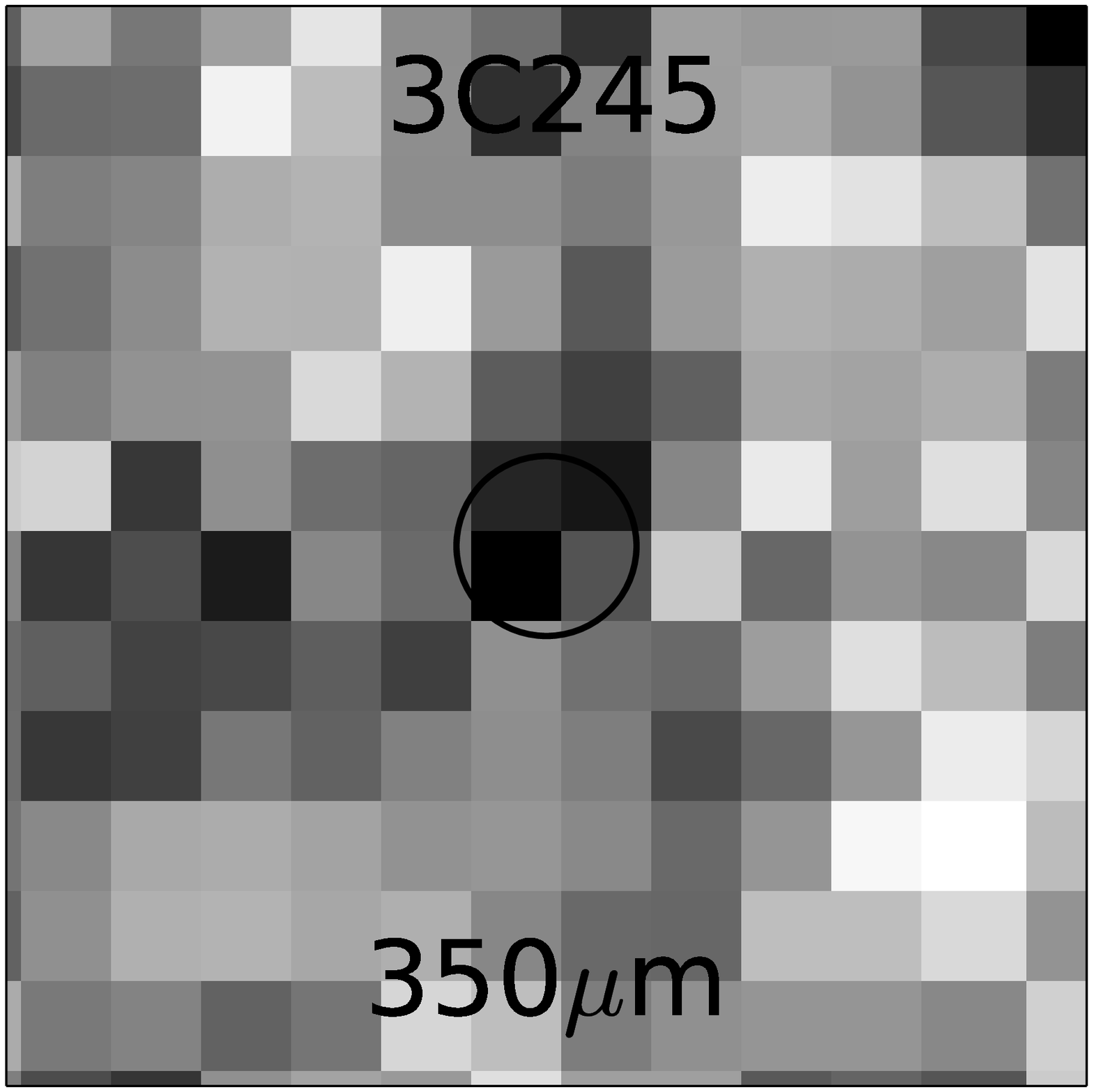}
      \includegraphics[width=1.5cm]{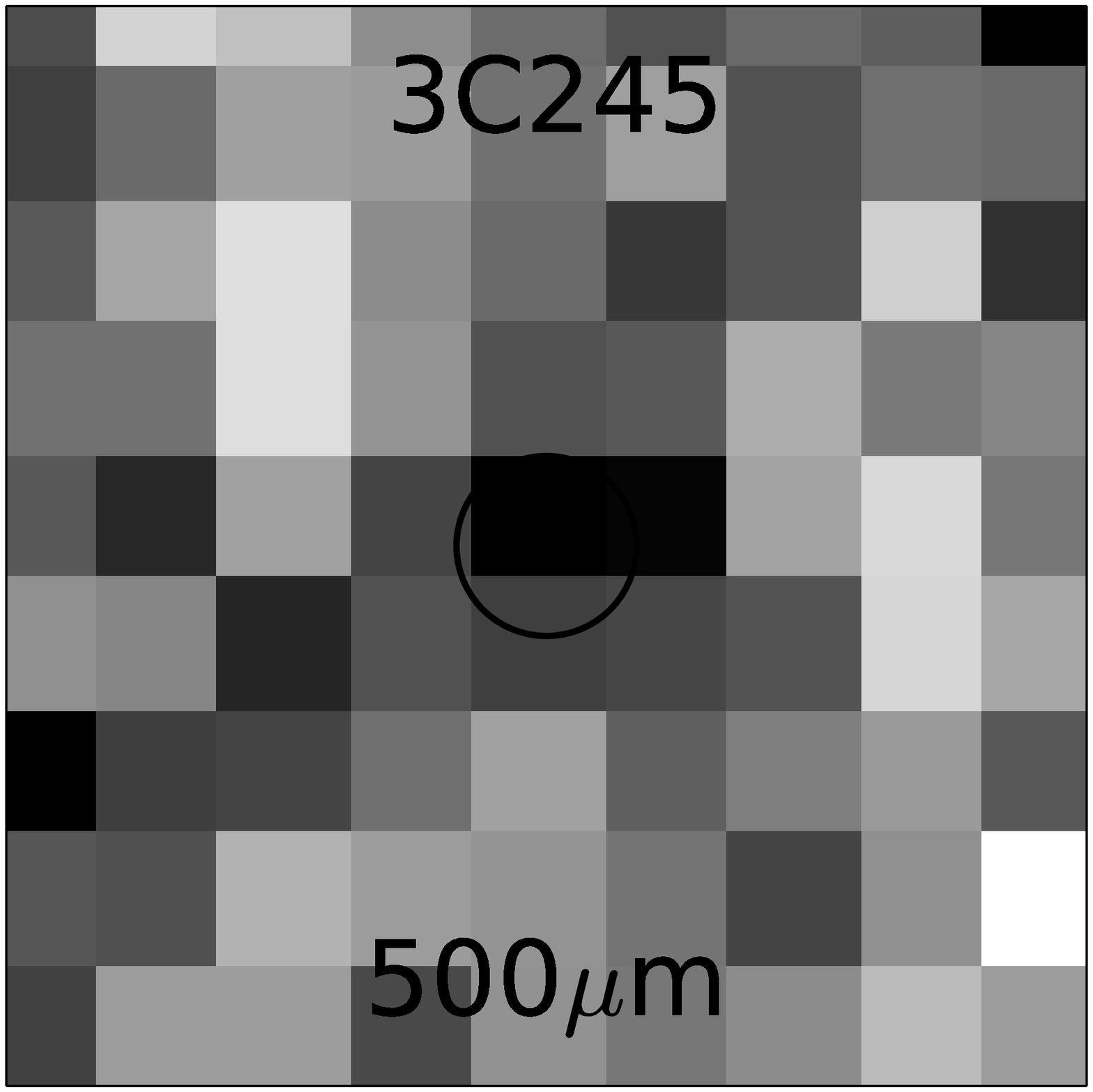}
      \\
      \includegraphics[width=1.5cm]{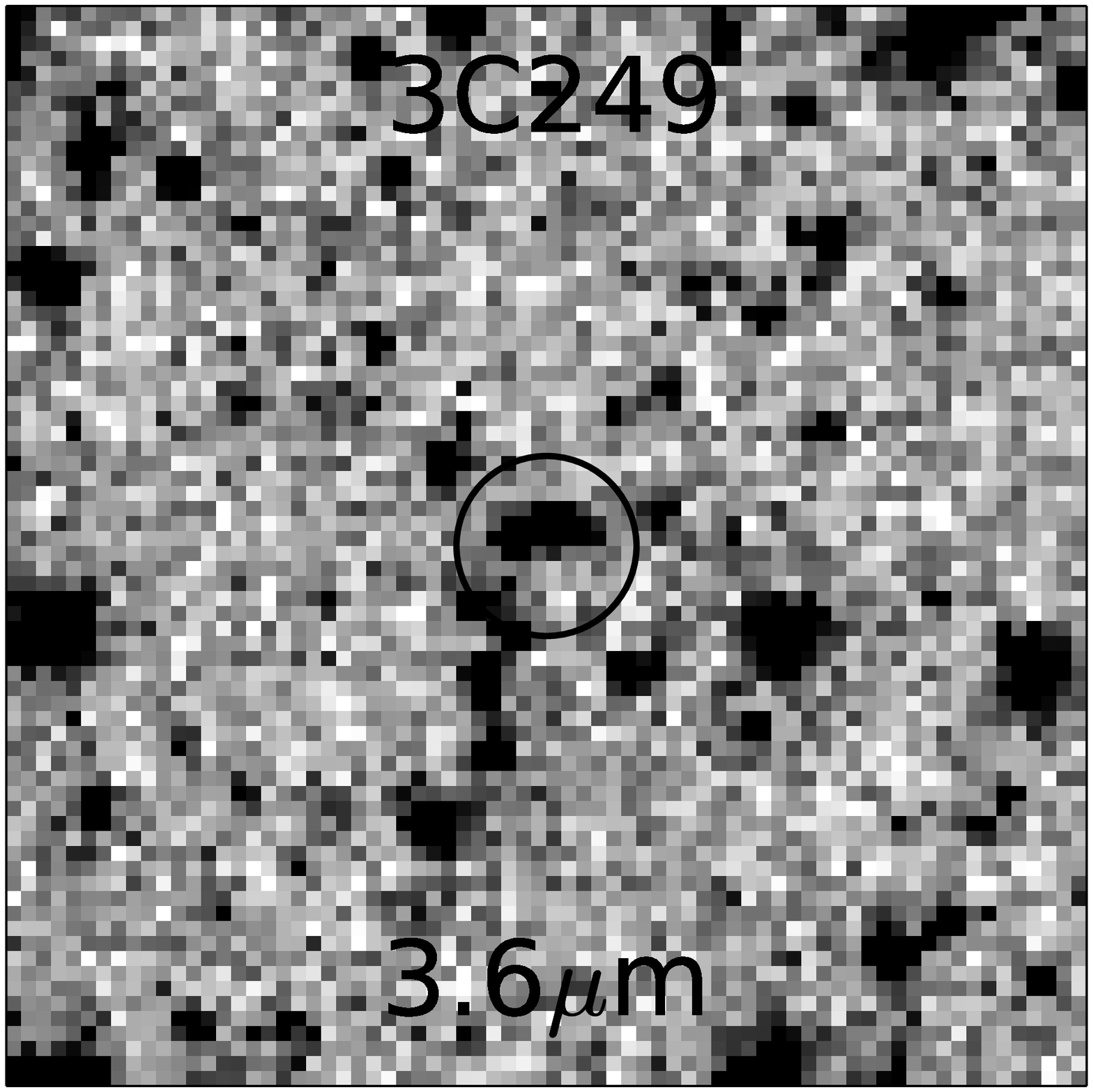}
      \includegraphics[width=1.5cm]{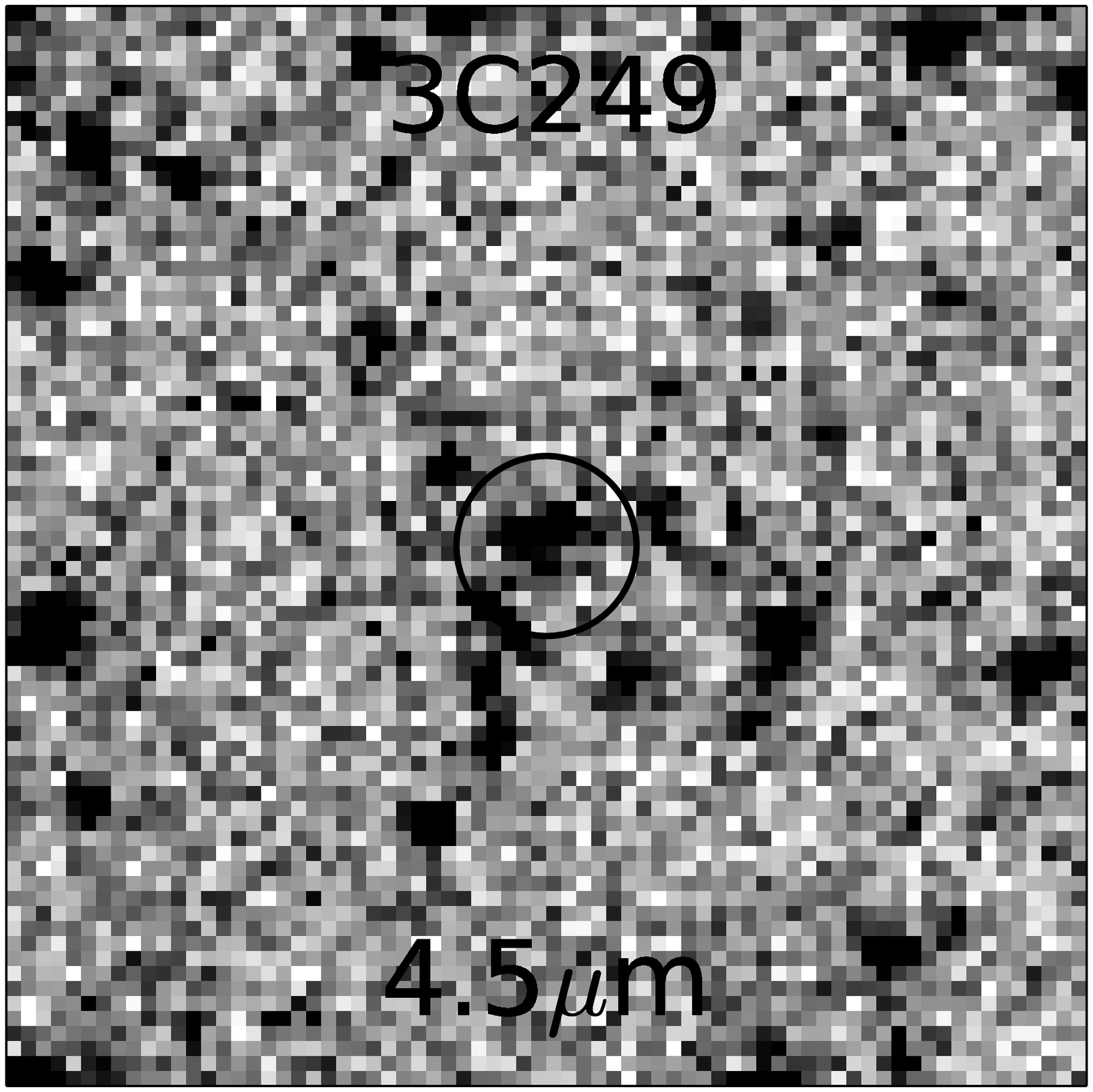}
      \includegraphics[width=1.5cm]{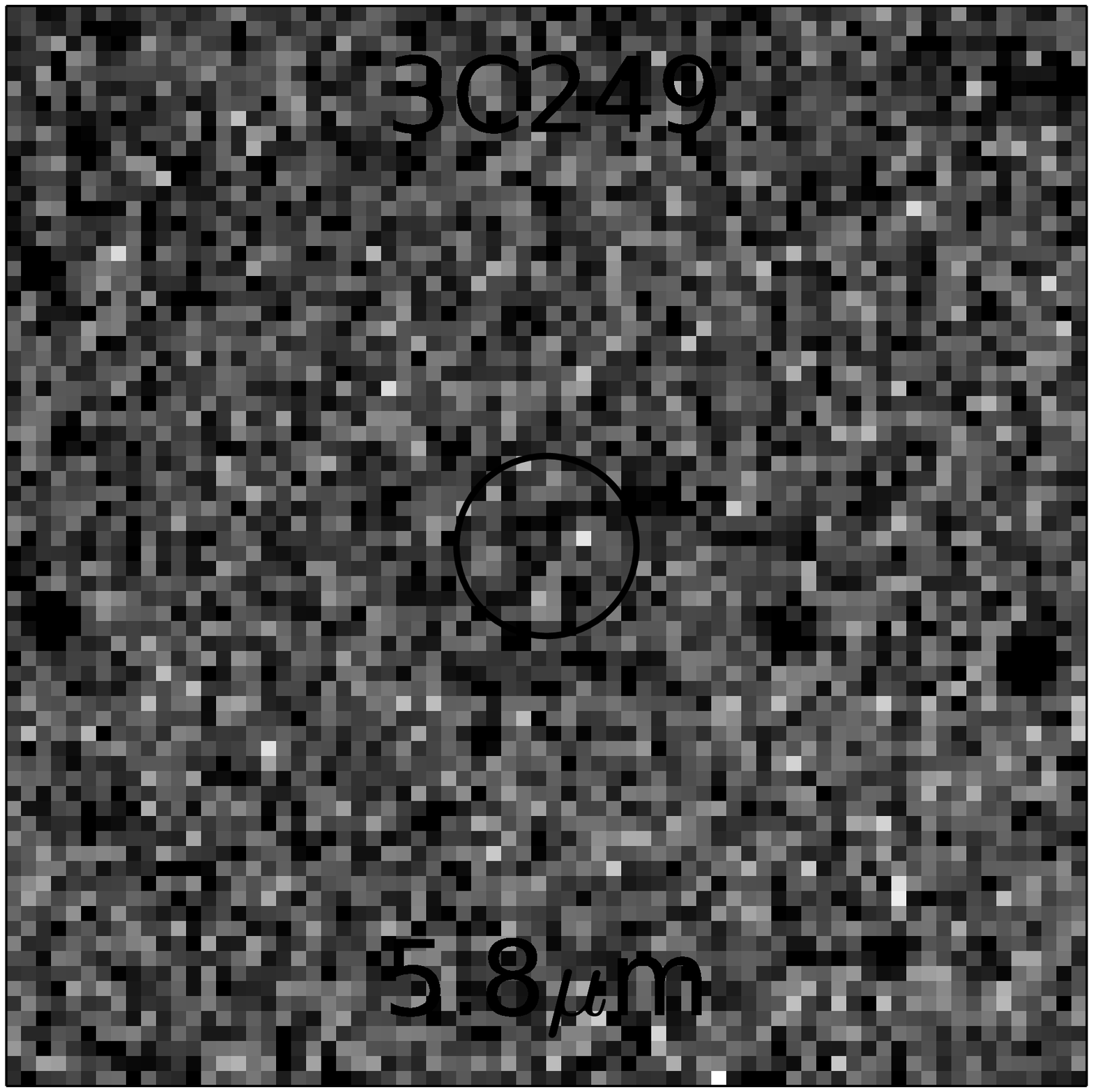}
      \includegraphics[width=1.5cm]{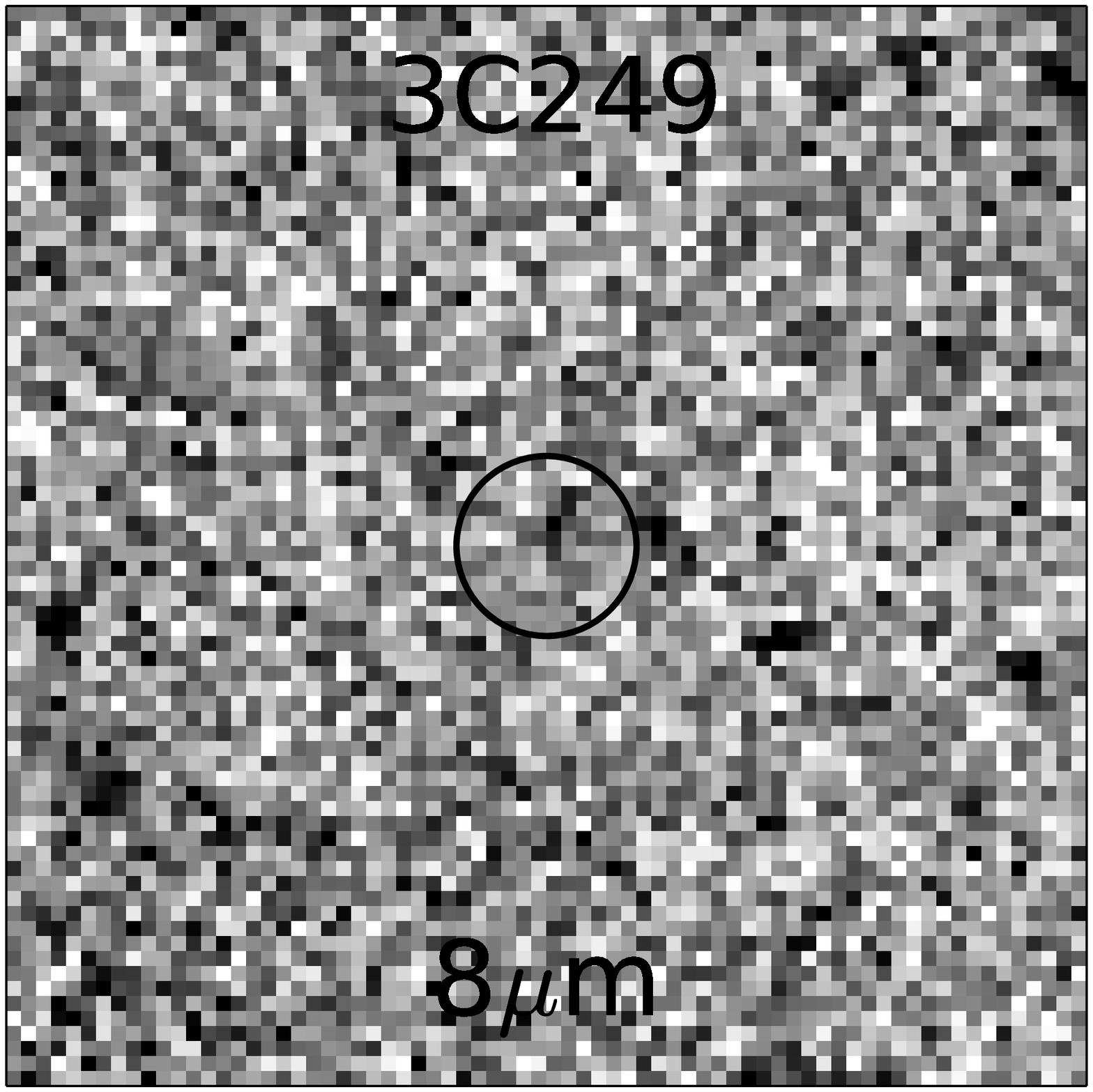}
      \includegraphics[width=1.5cm]{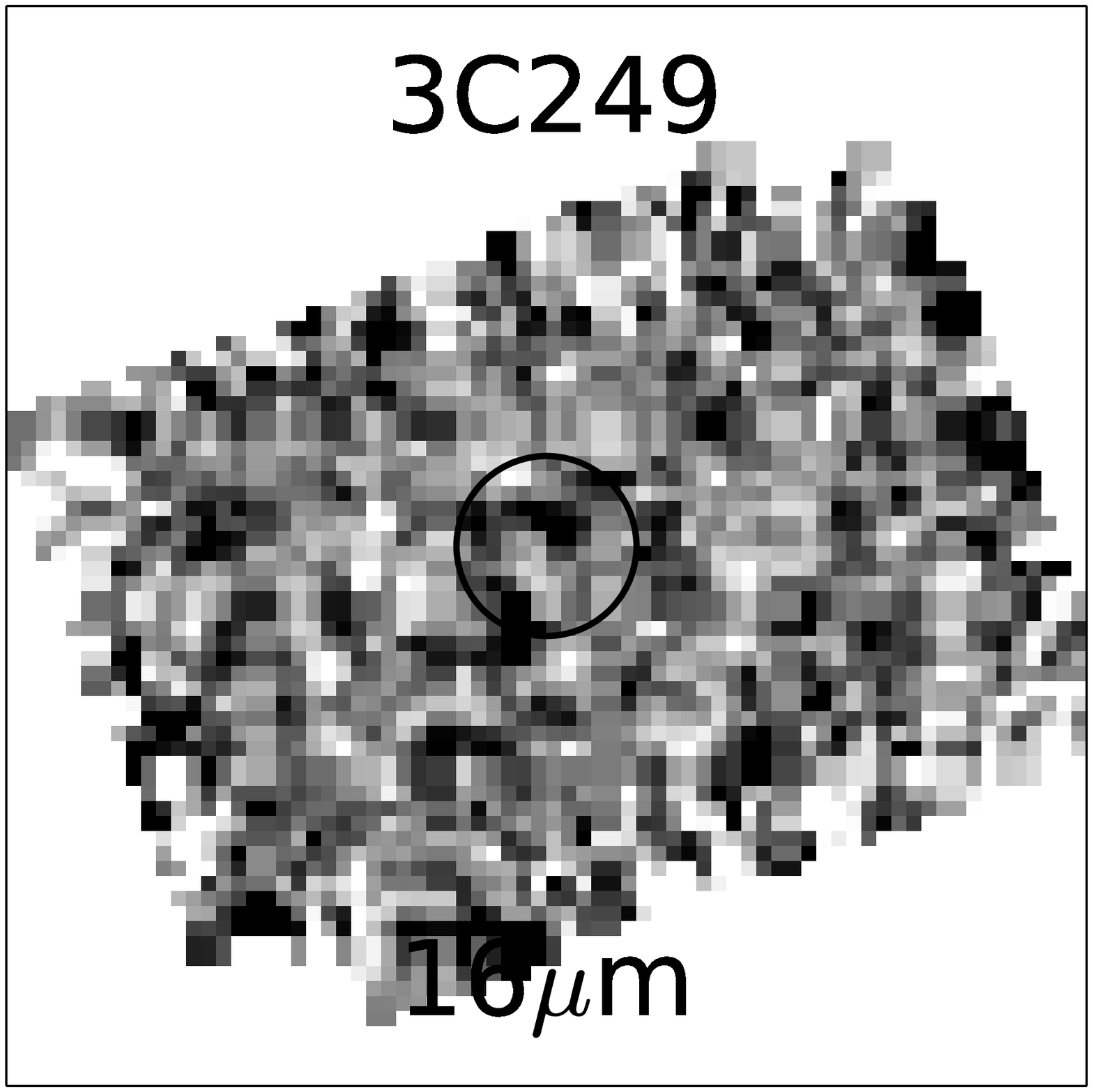}
      \includegraphics[width=1.5cm]{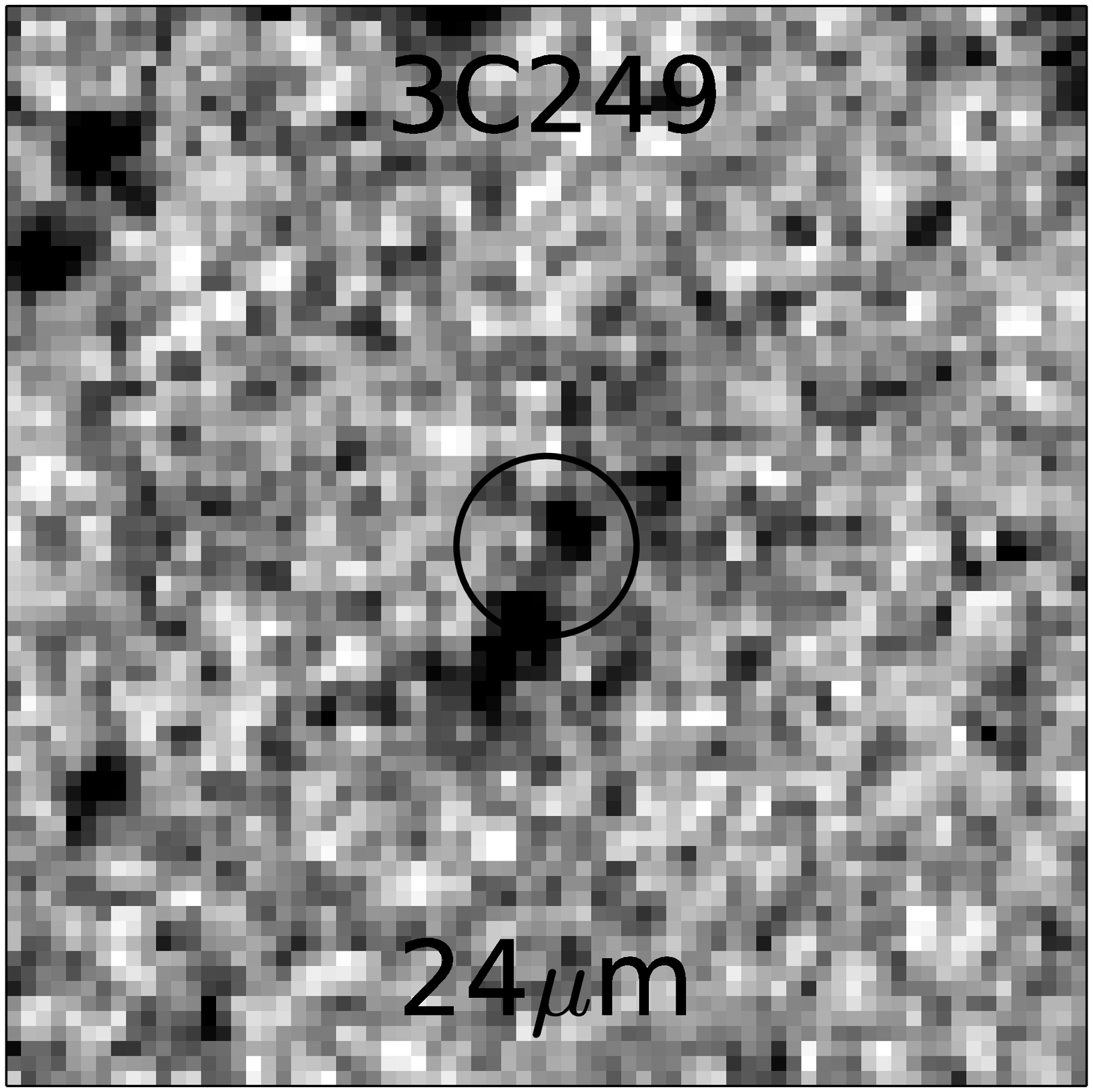}
      \includegraphics[width=1.5cm]{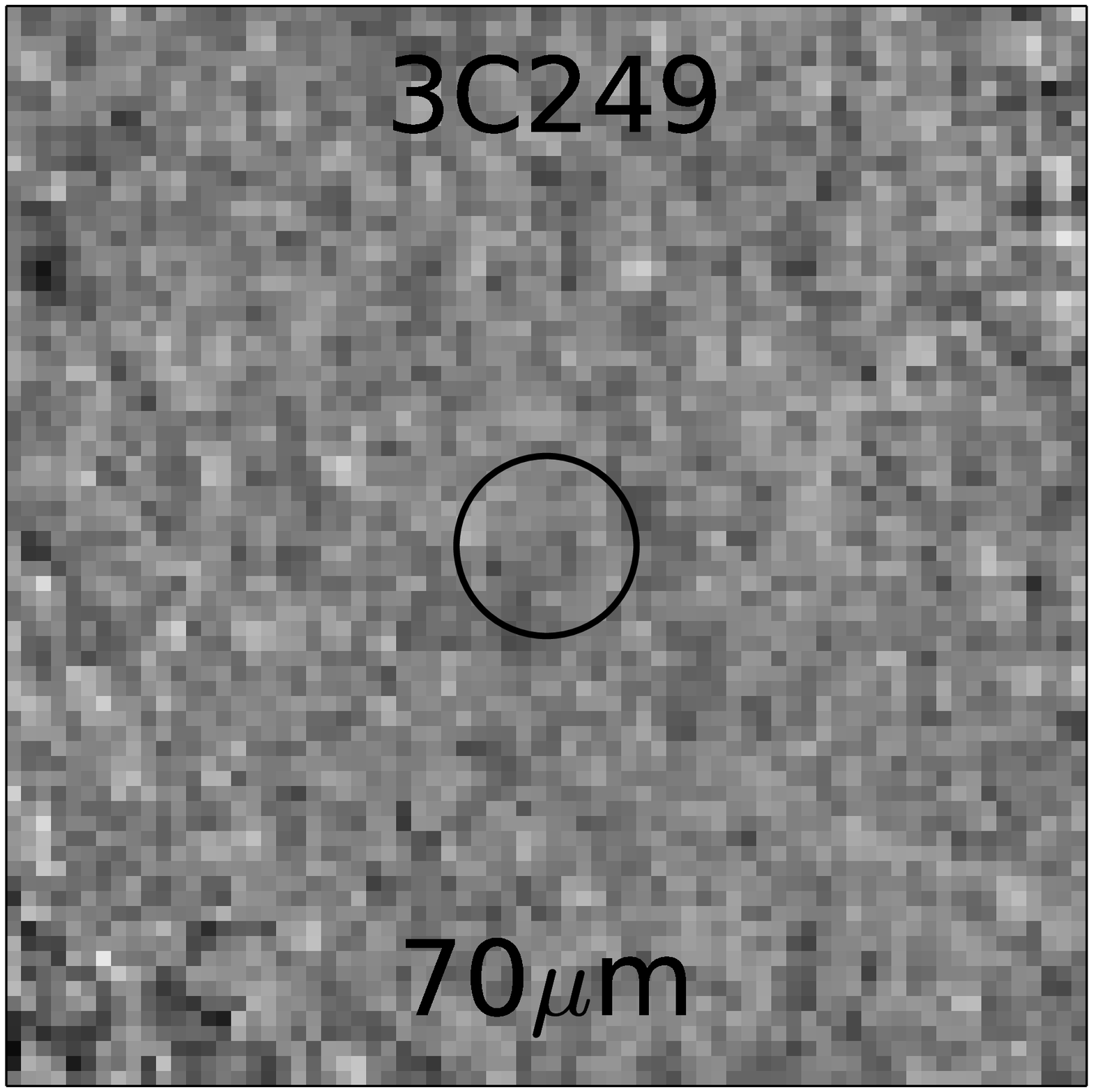}
      \includegraphics[width=1.5cm]{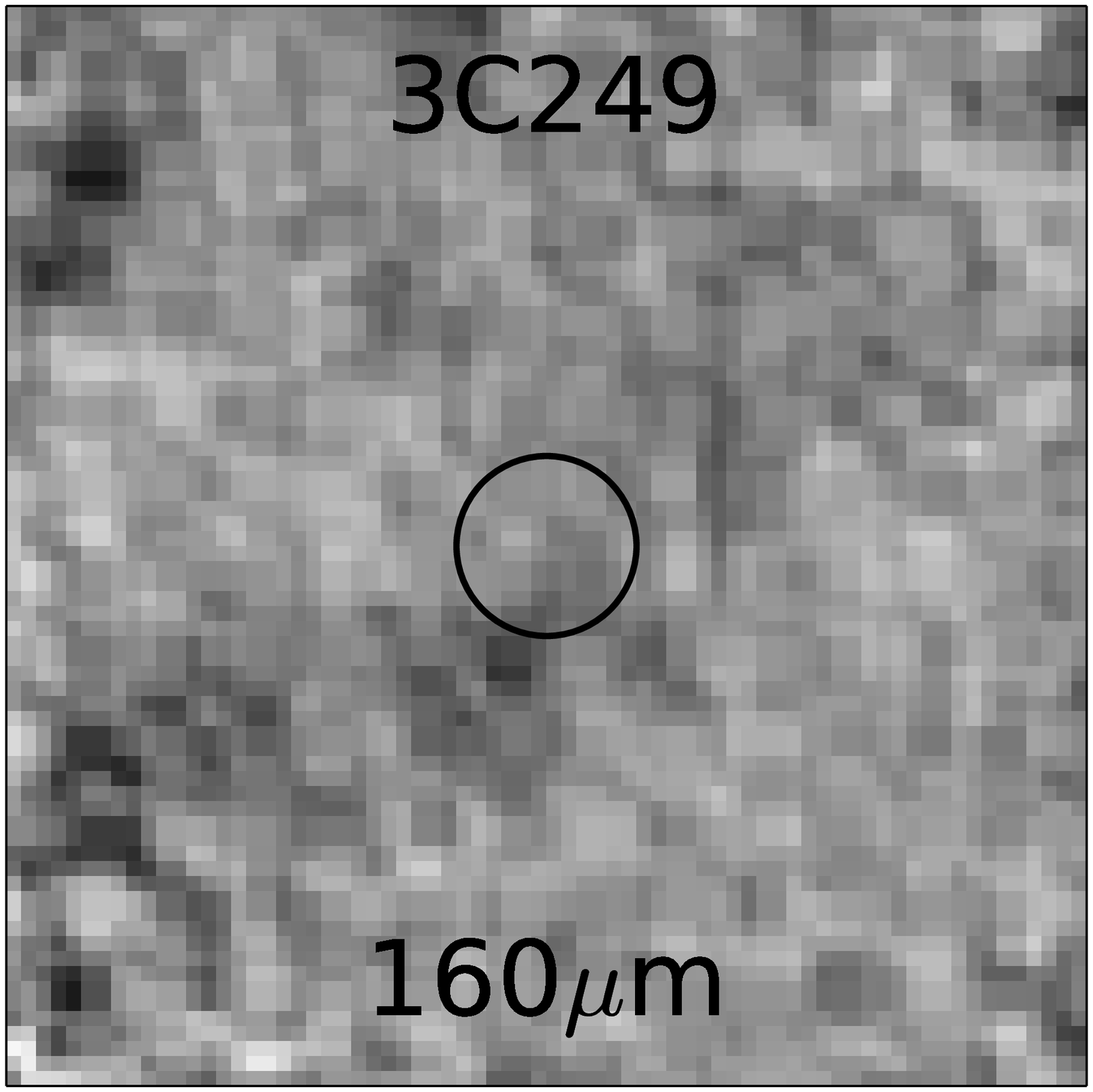}
      \includegraphics[width=1.5cm]{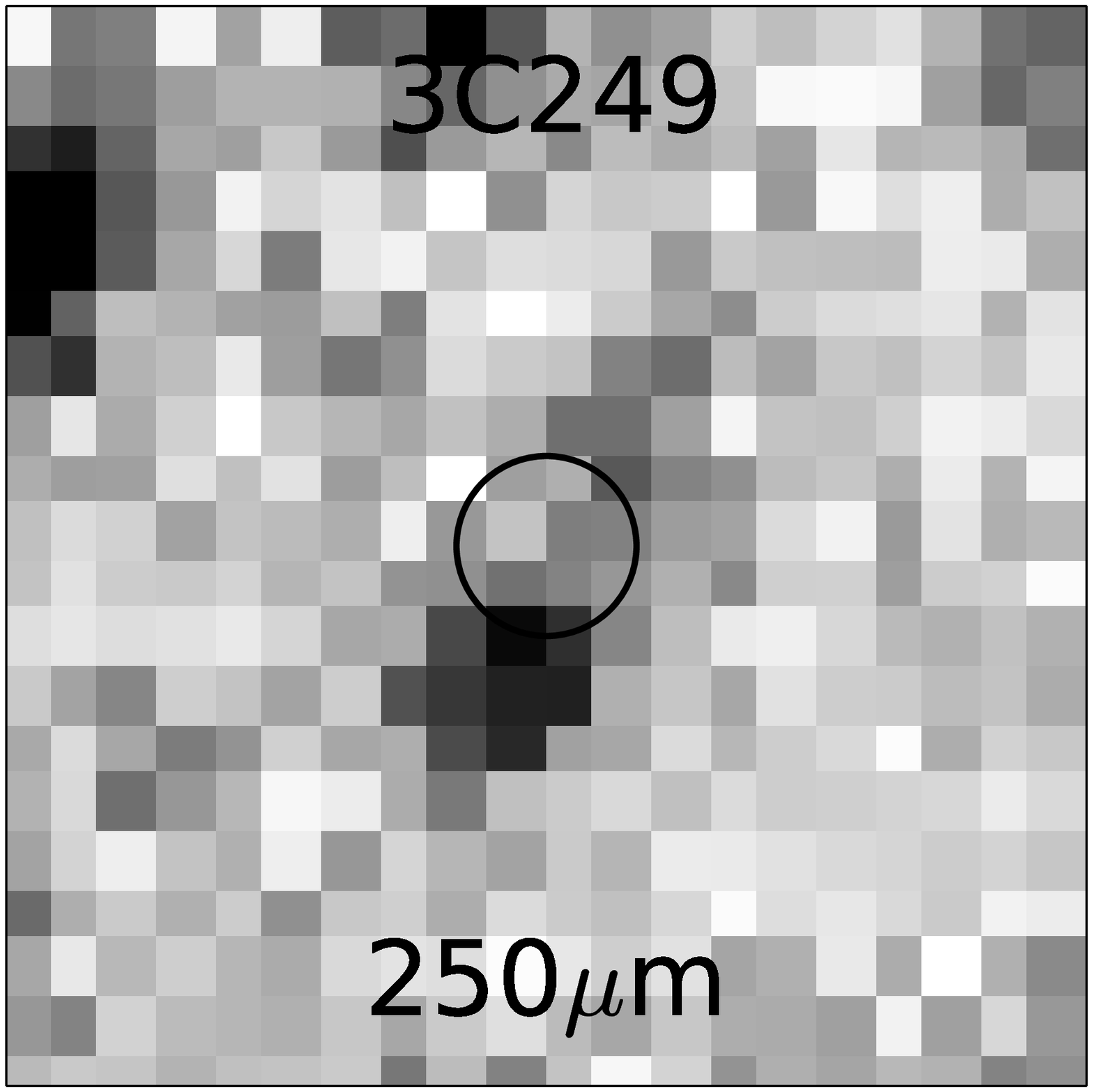}
      \includegraphics[width=1.5cm]{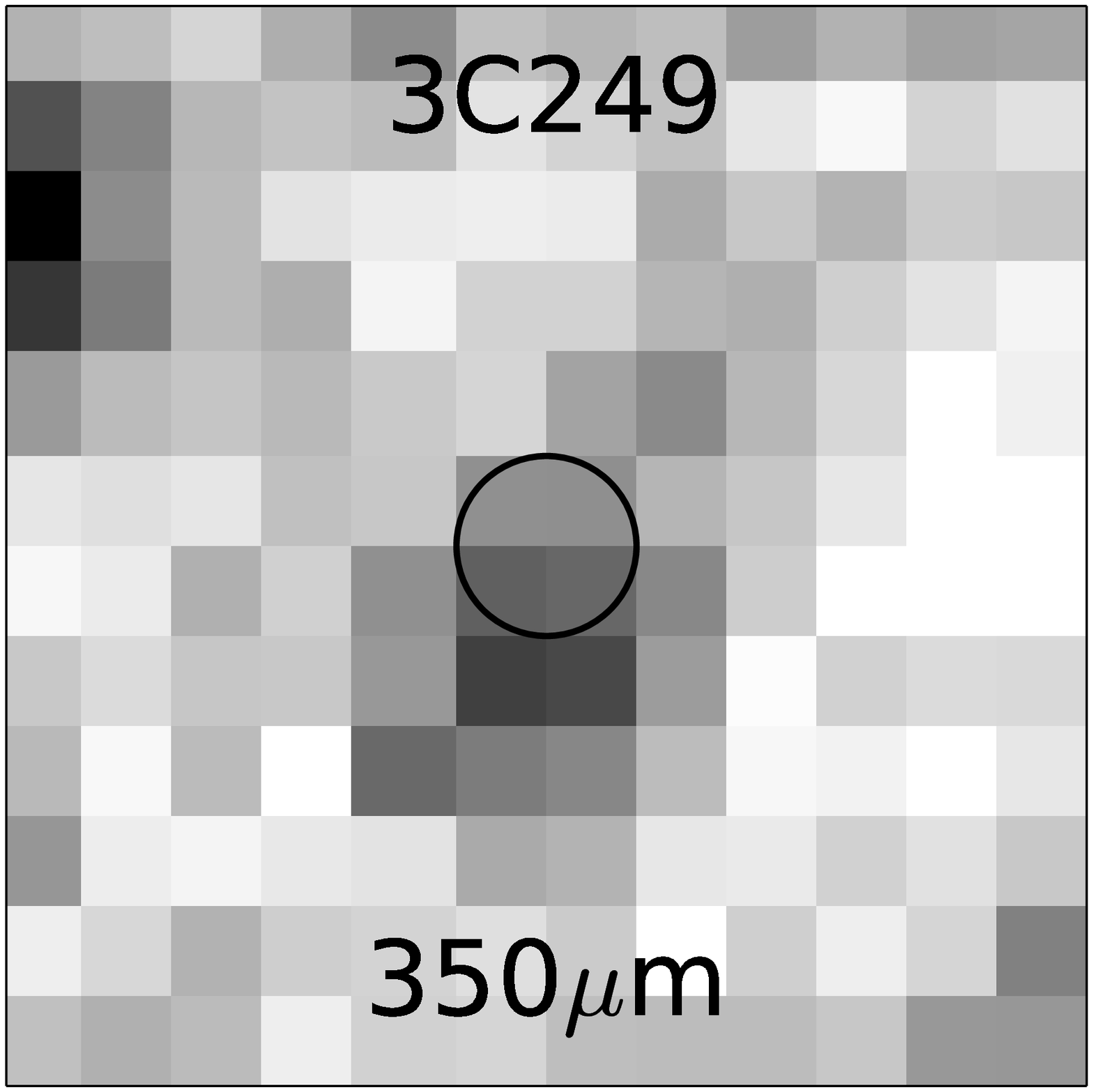}
      \includegraphics[width=1.5cm]{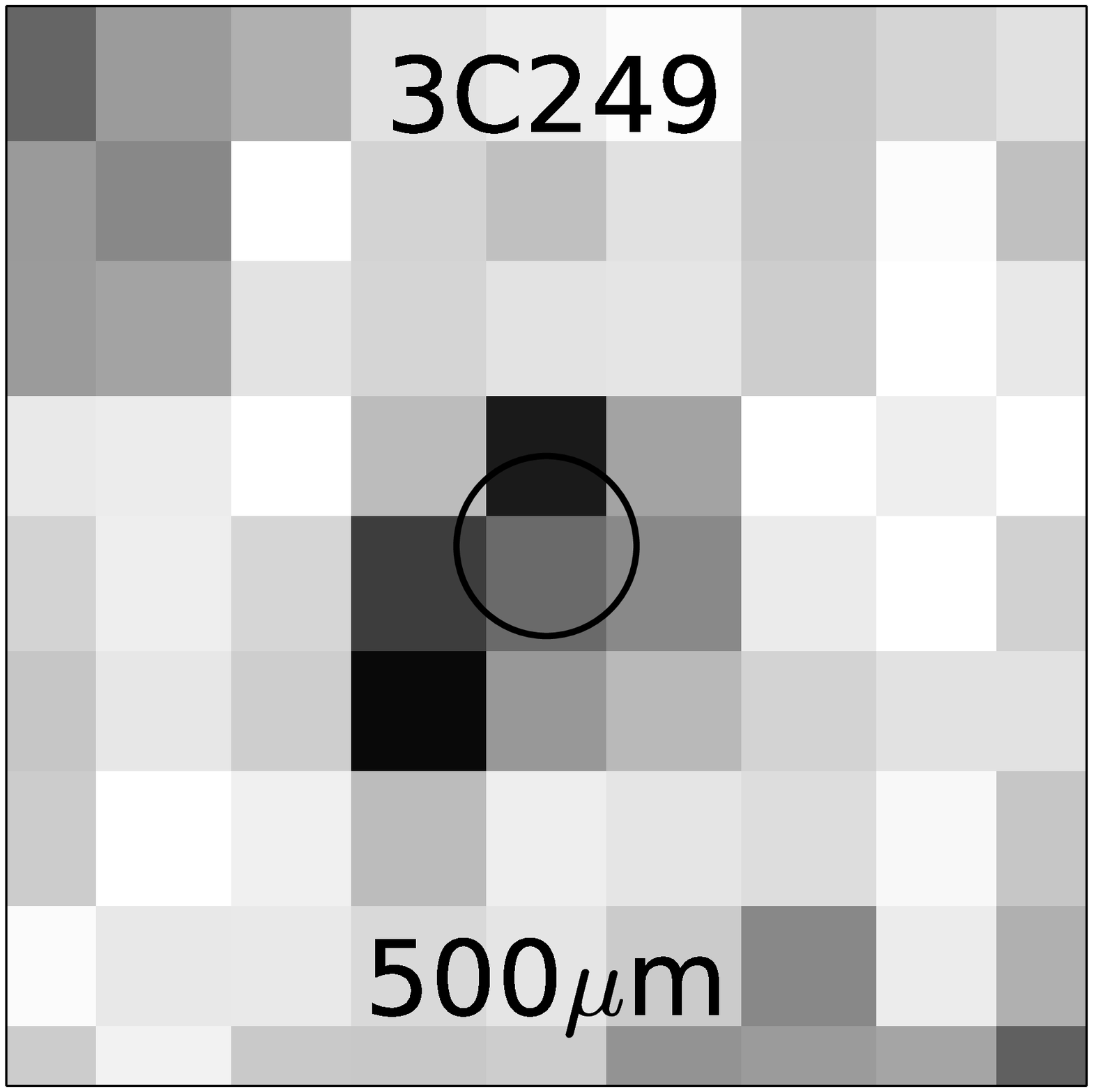}
      \\
      \includegraphics[width=1.5cm]{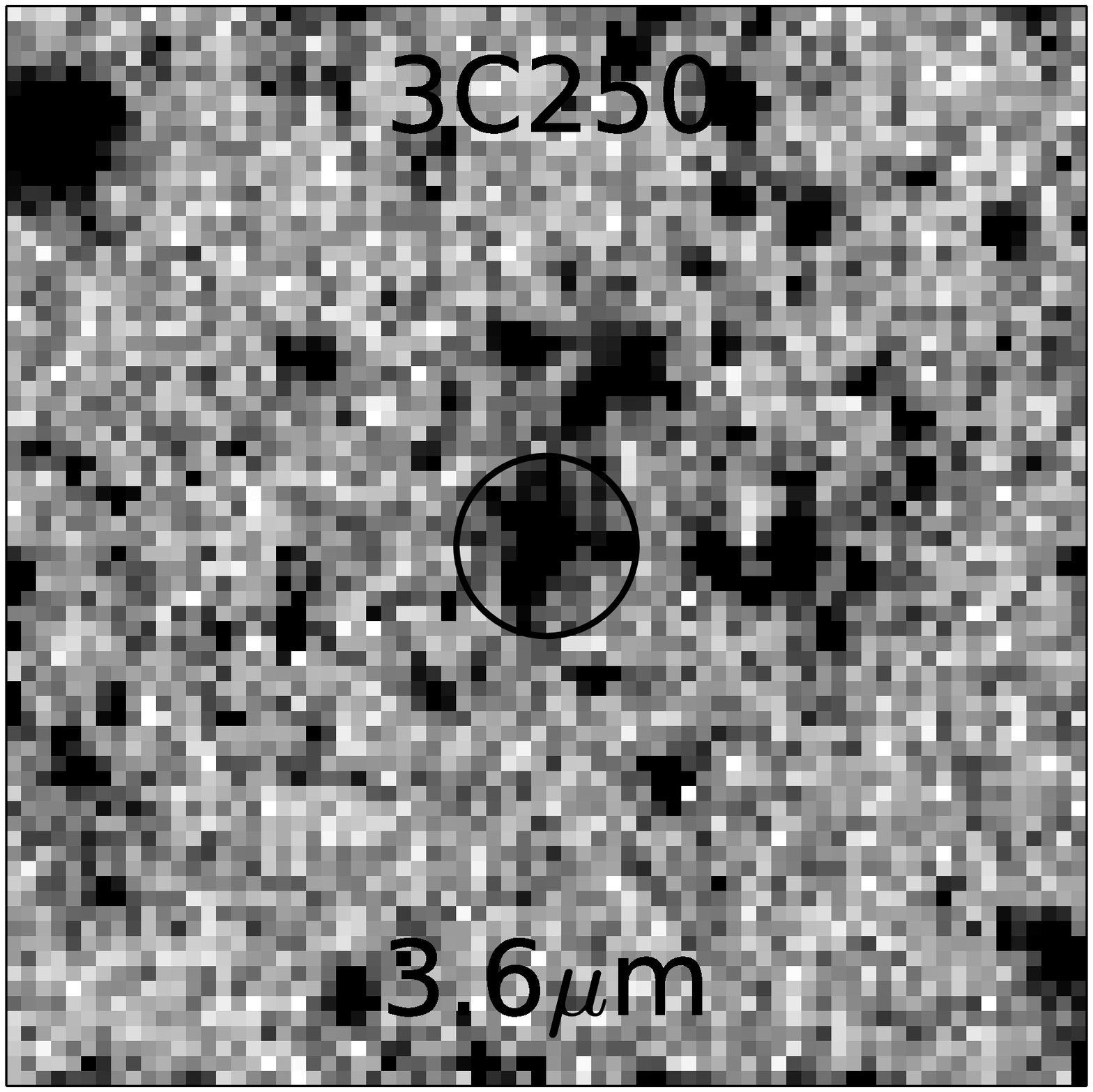}
      \includegraphics[width=1.5cm]{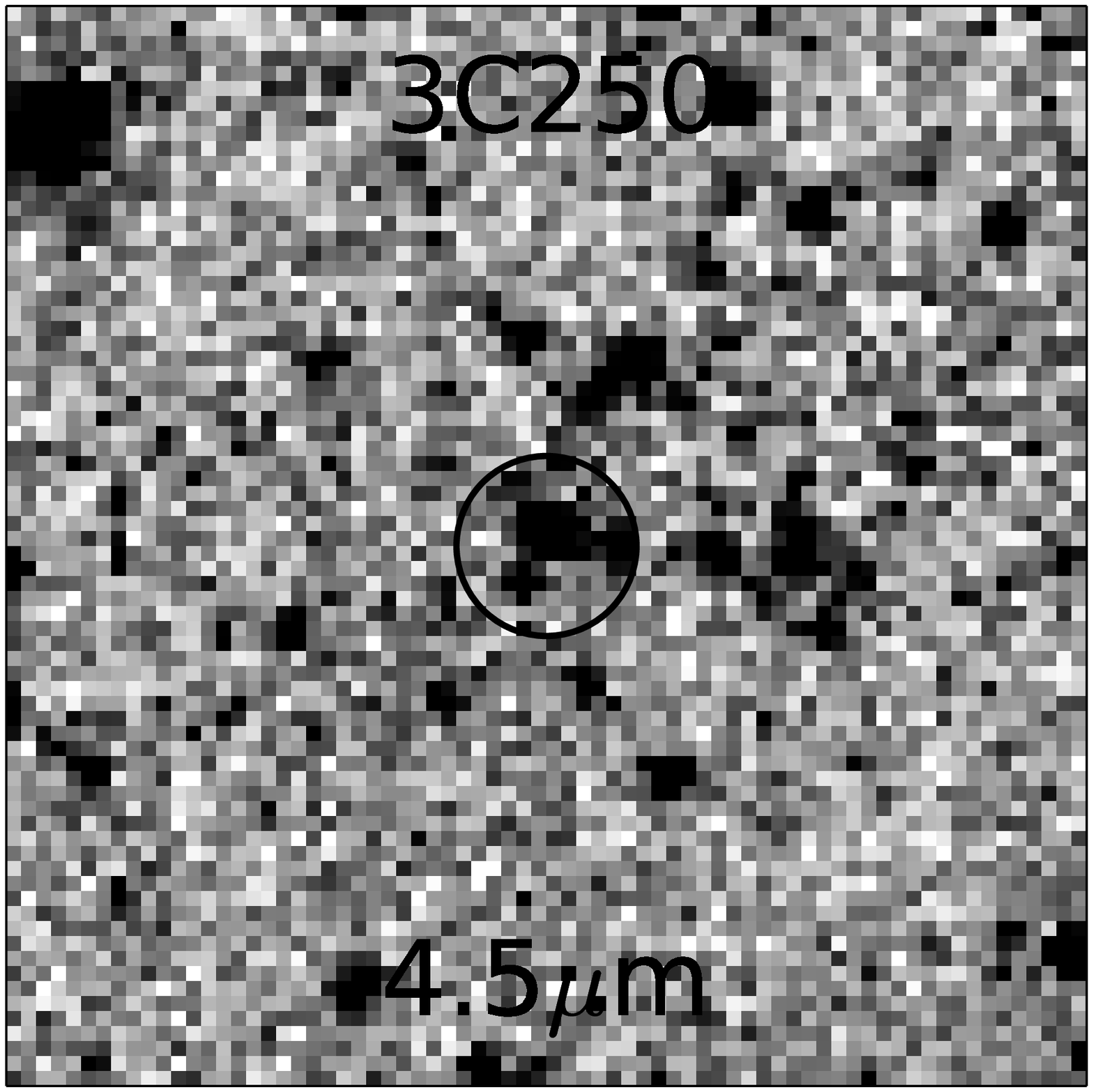}
      \includegraphics[width=1.5cm]{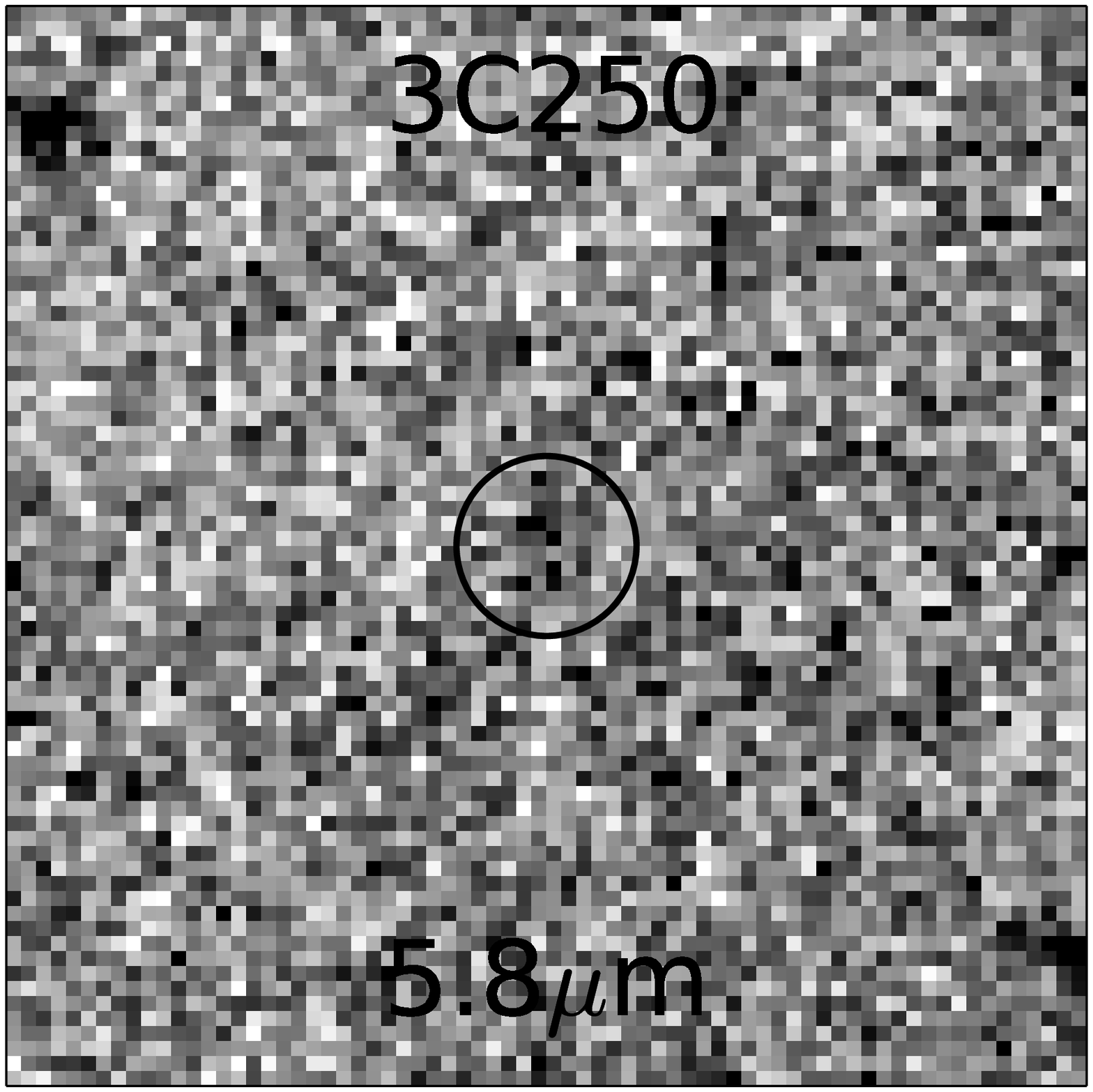}
      \includegraphics[width=1.5cm]{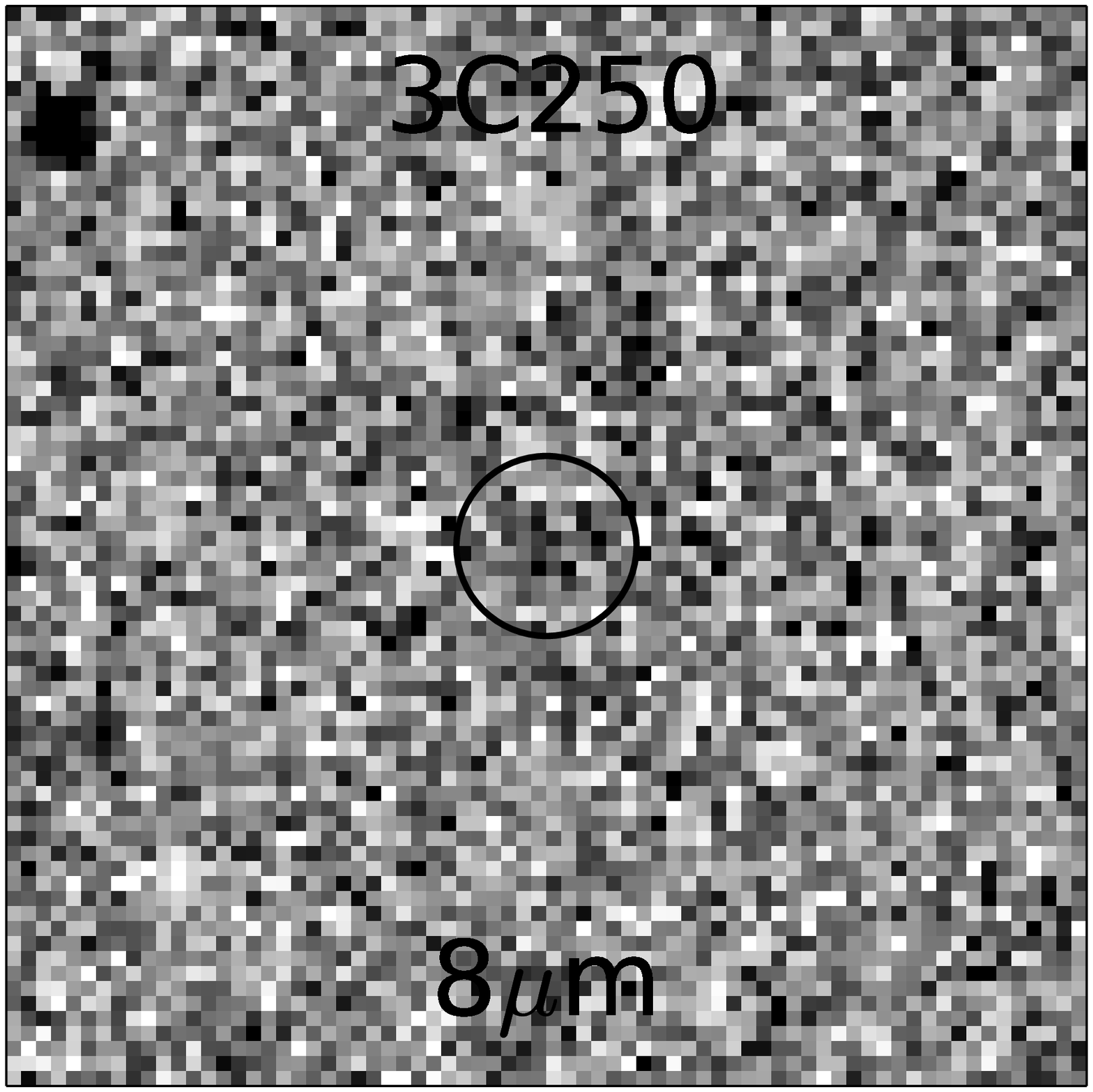}
      \includegraphics[width=1.5cm]{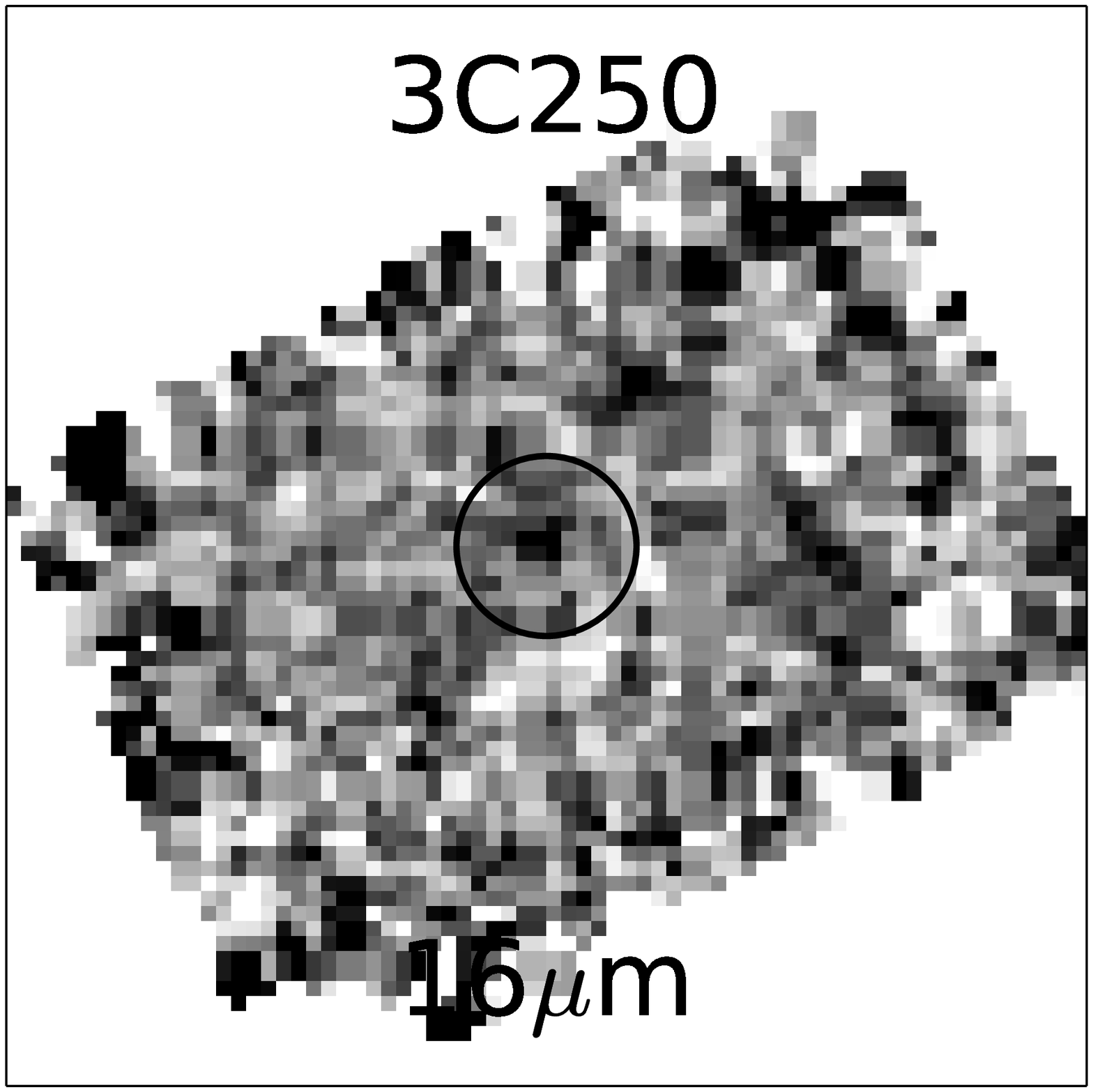}
      \includegraphics[width=1.5cm]{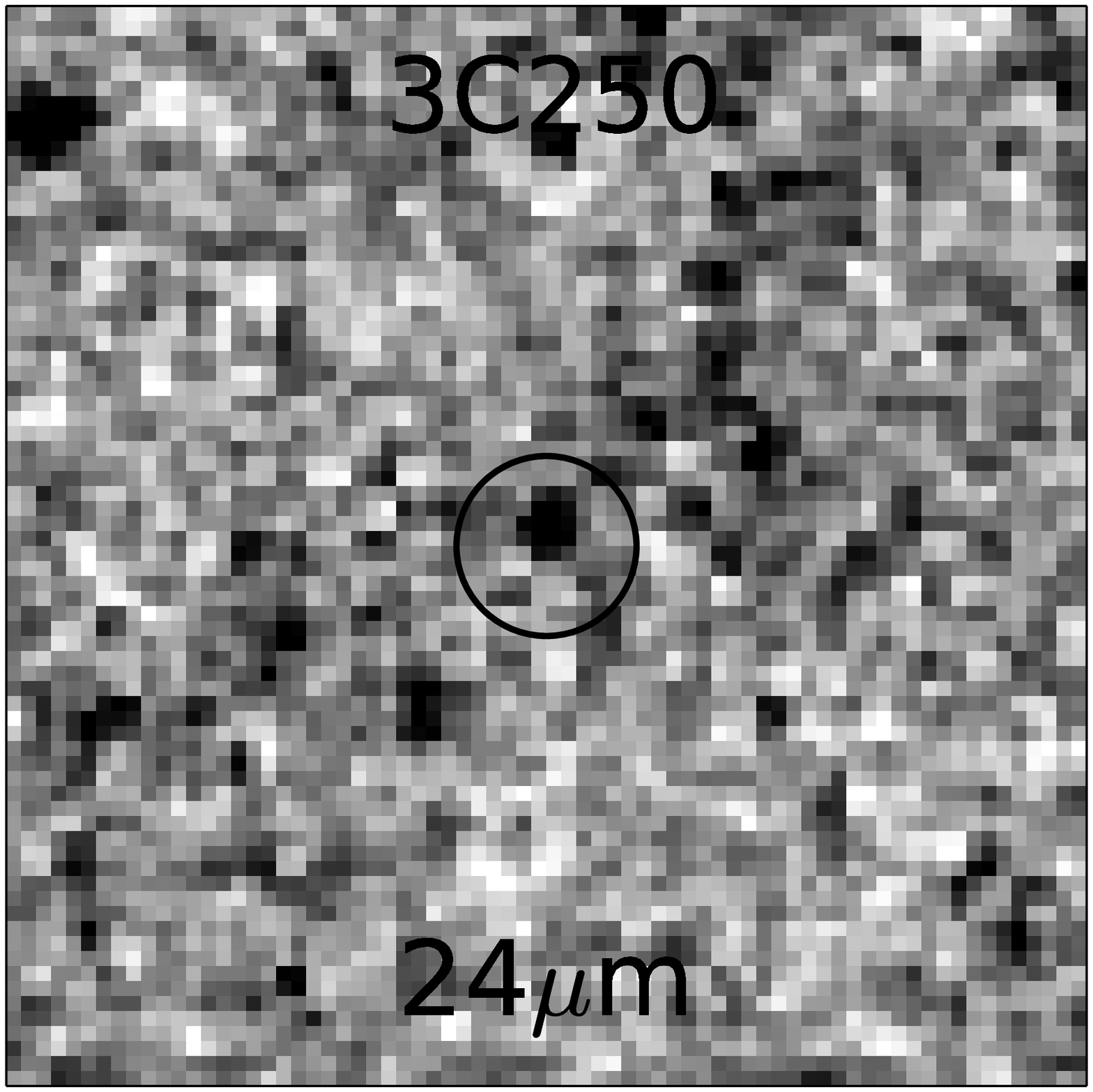}
      \includegraphics[width=1.5cm]{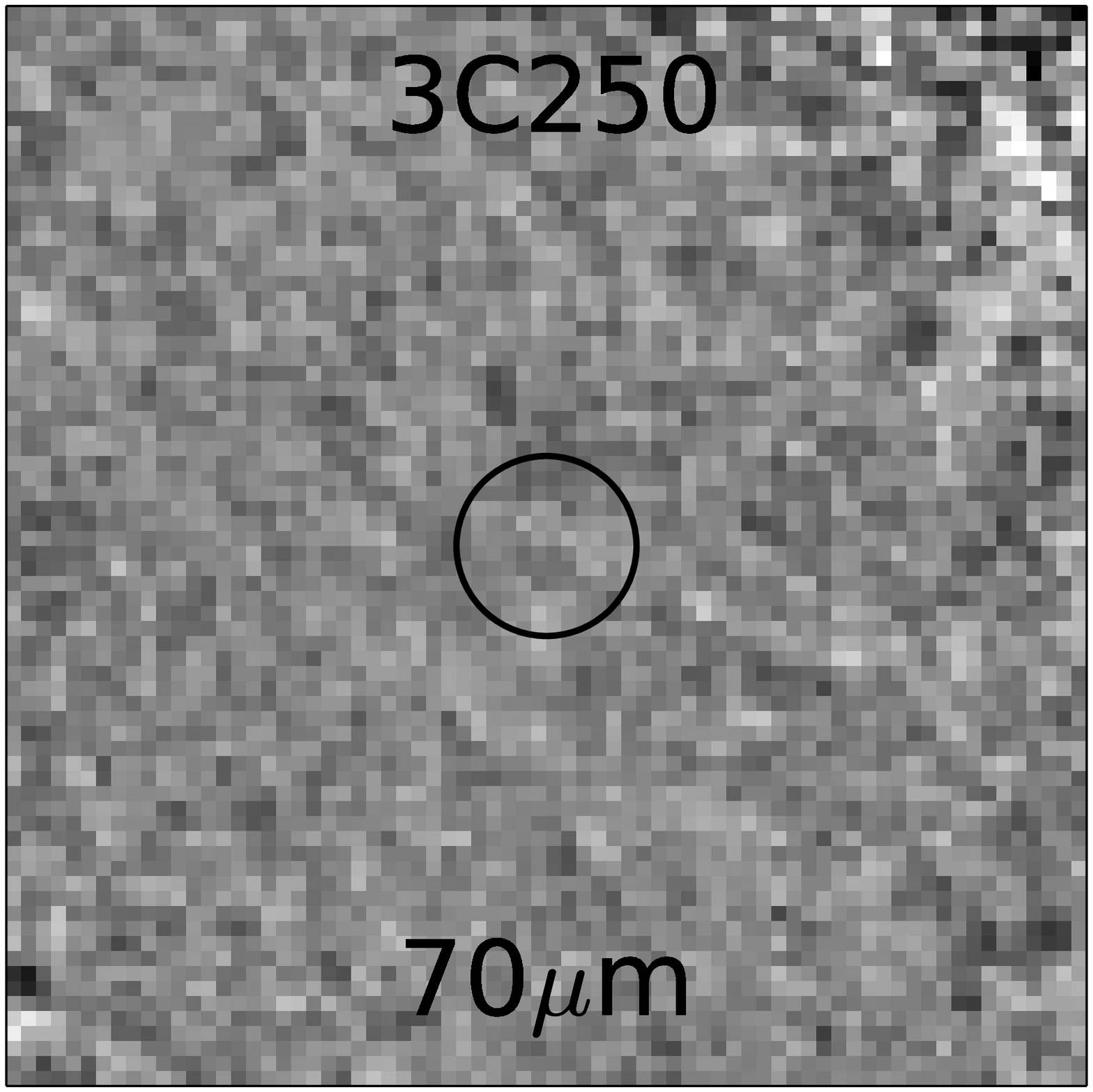}
      \includegraphics[width=1.5cm]{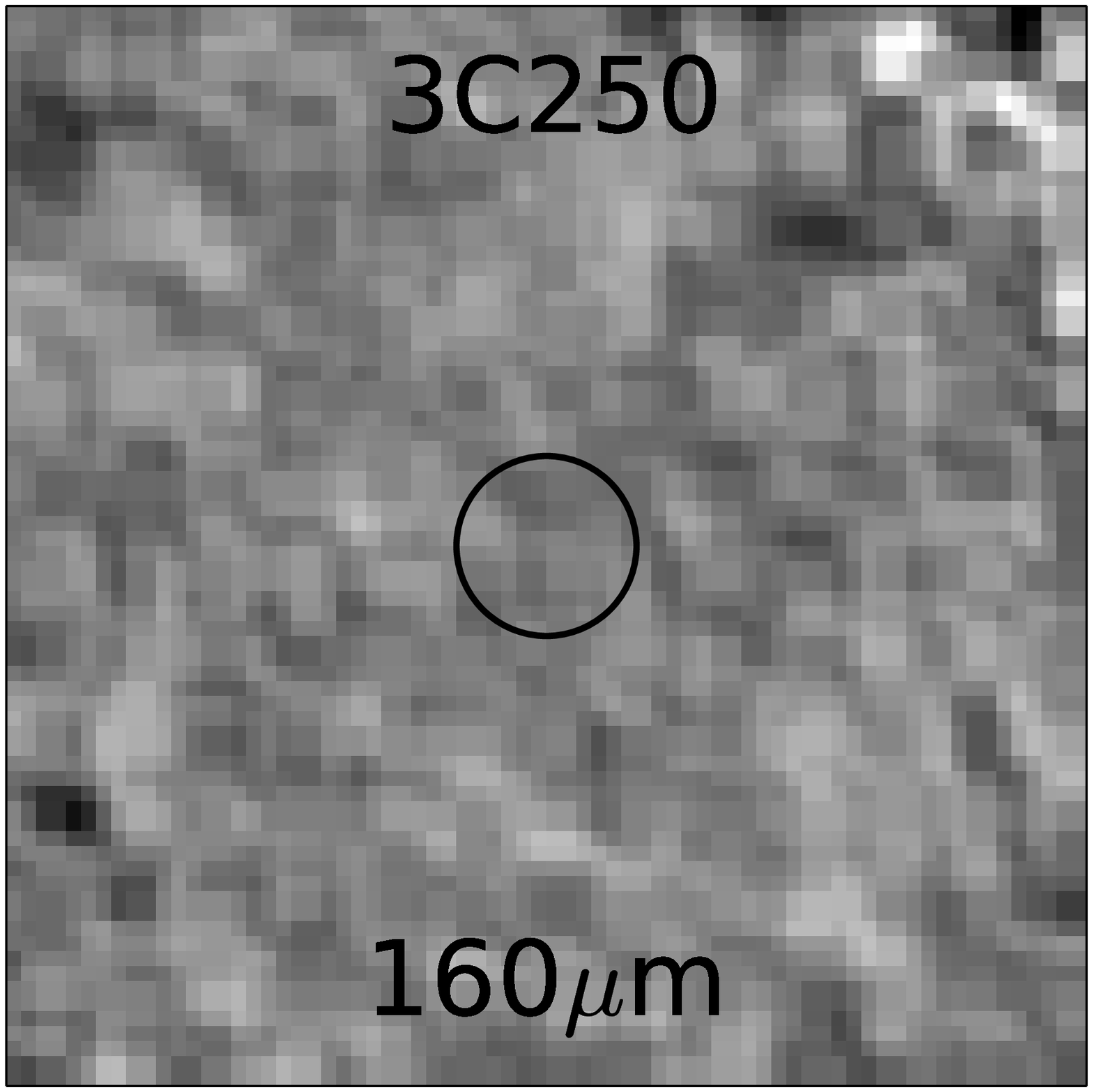}
      \includegraphics[width=1.5cm]{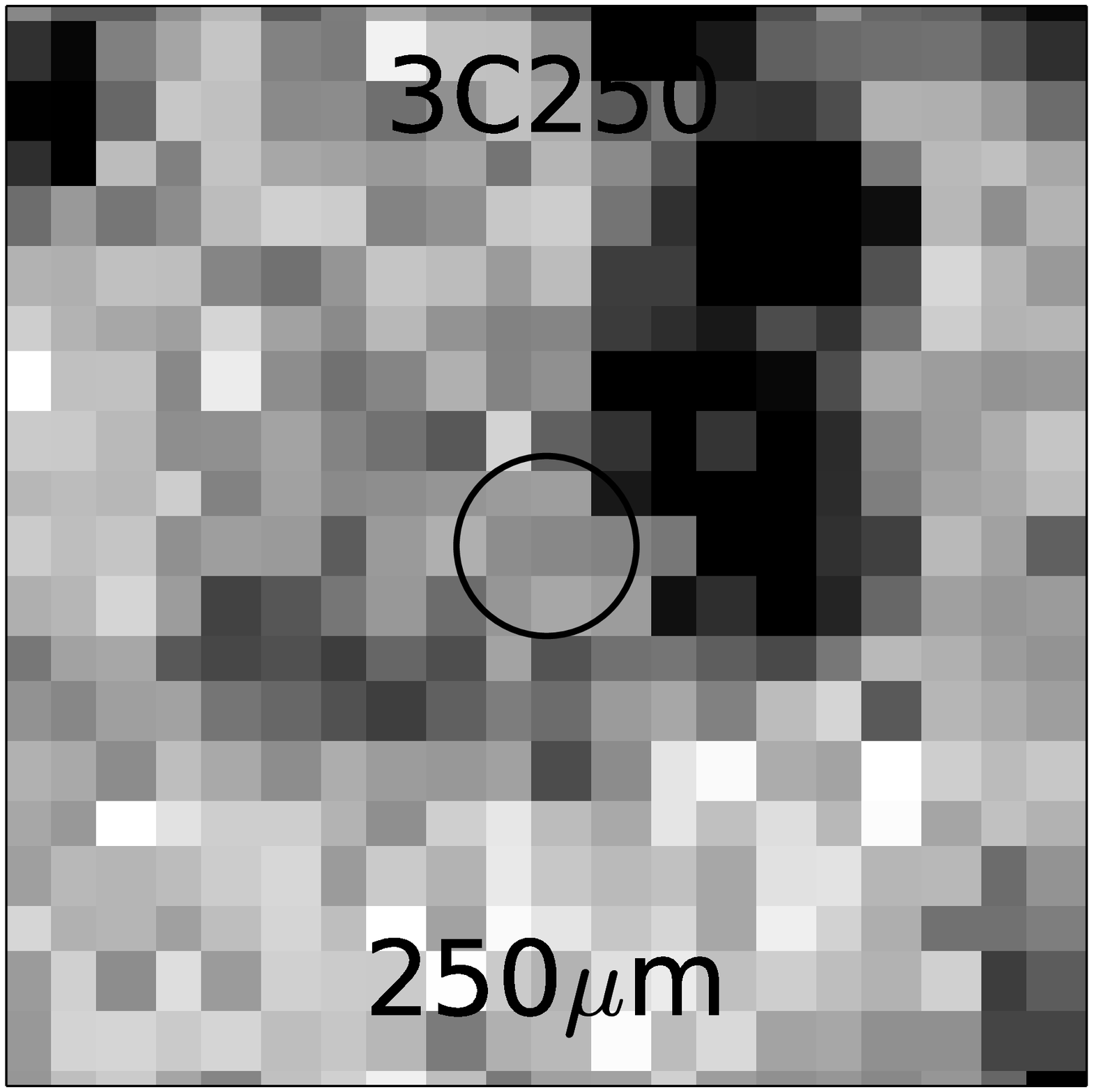}
      \includegraphics[width=1.5cm]{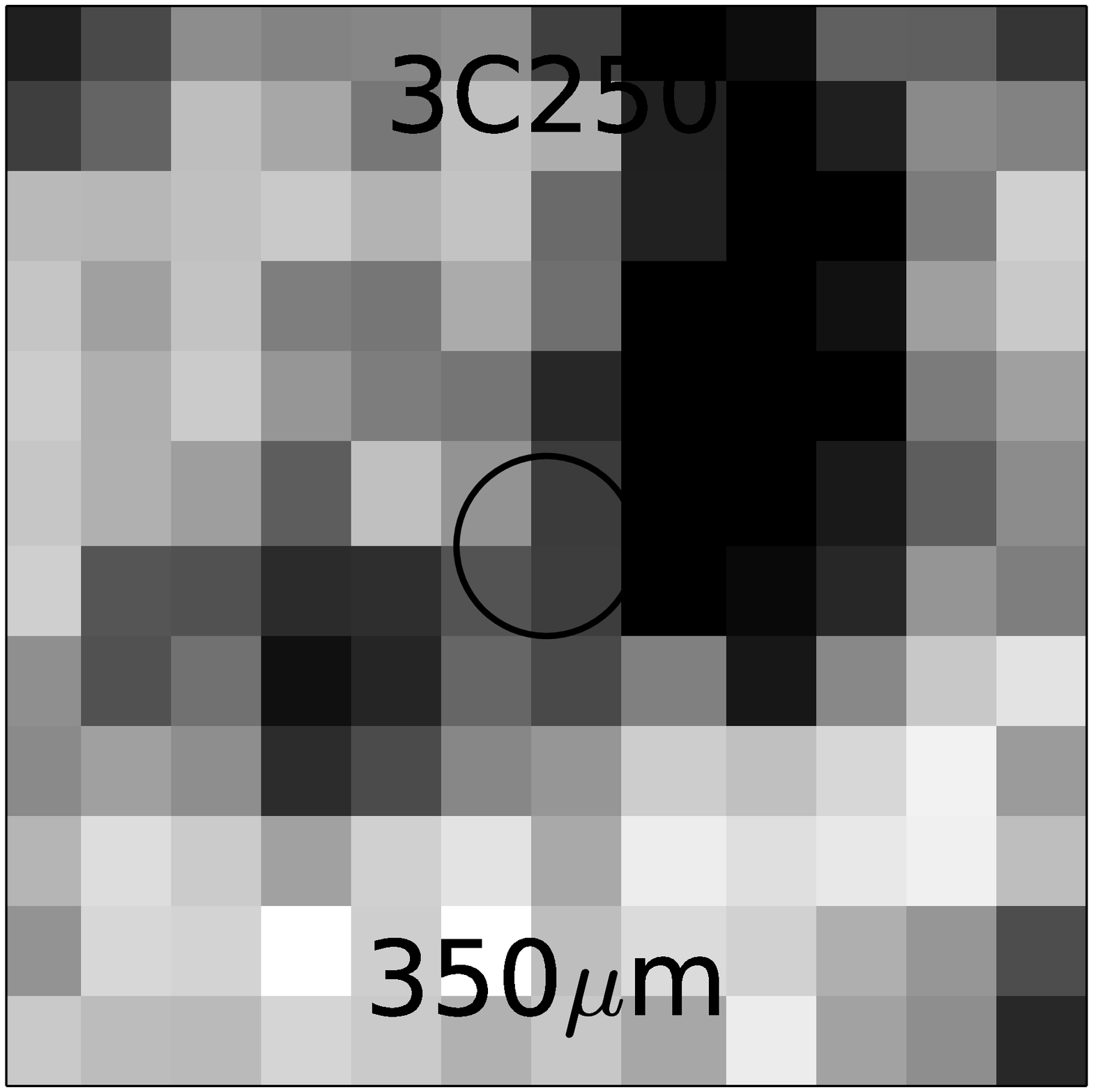}
      \includegraphics[width=1.5cm]{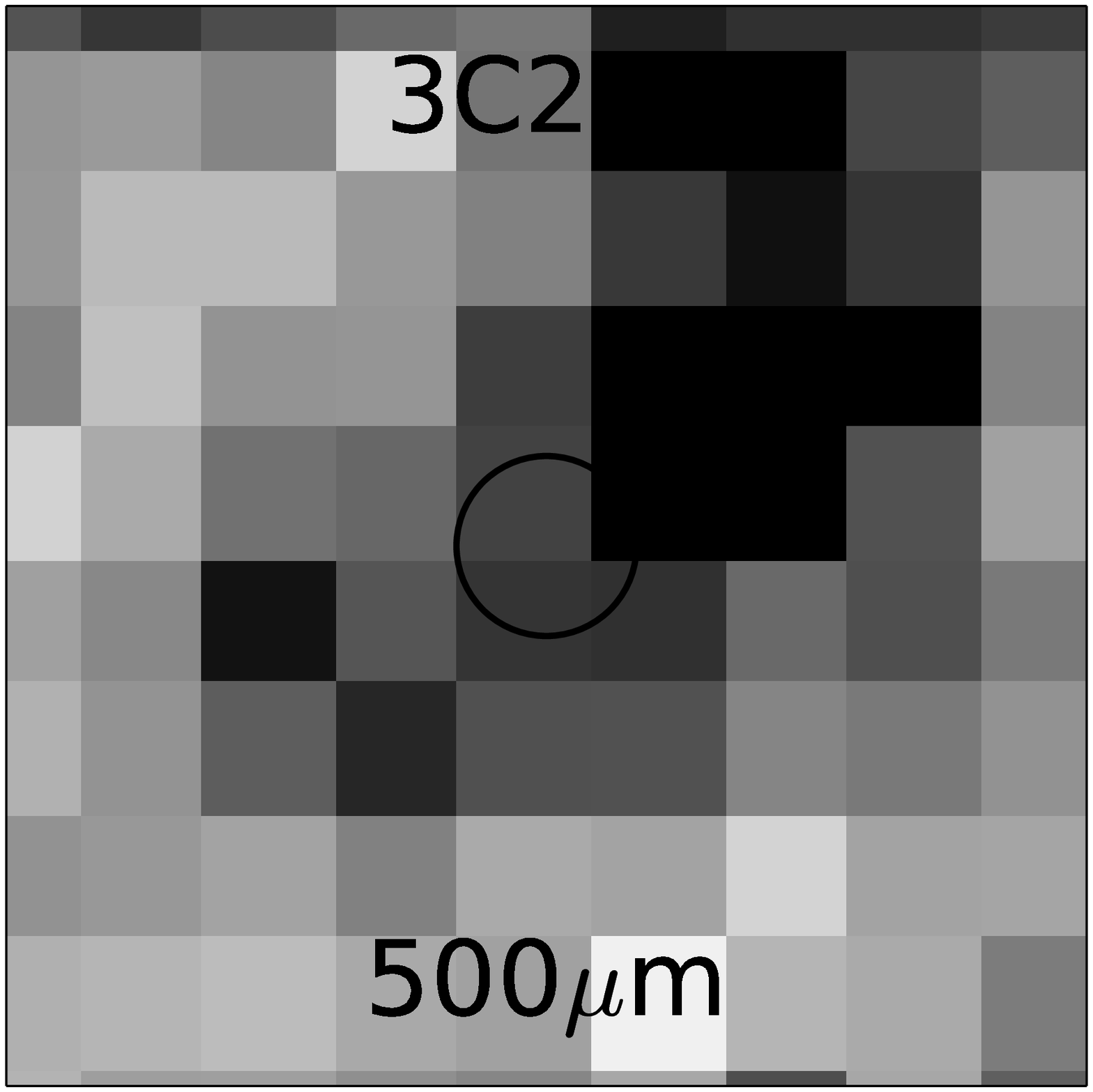}
      \\
      \includegraphics[width=1.5cm]{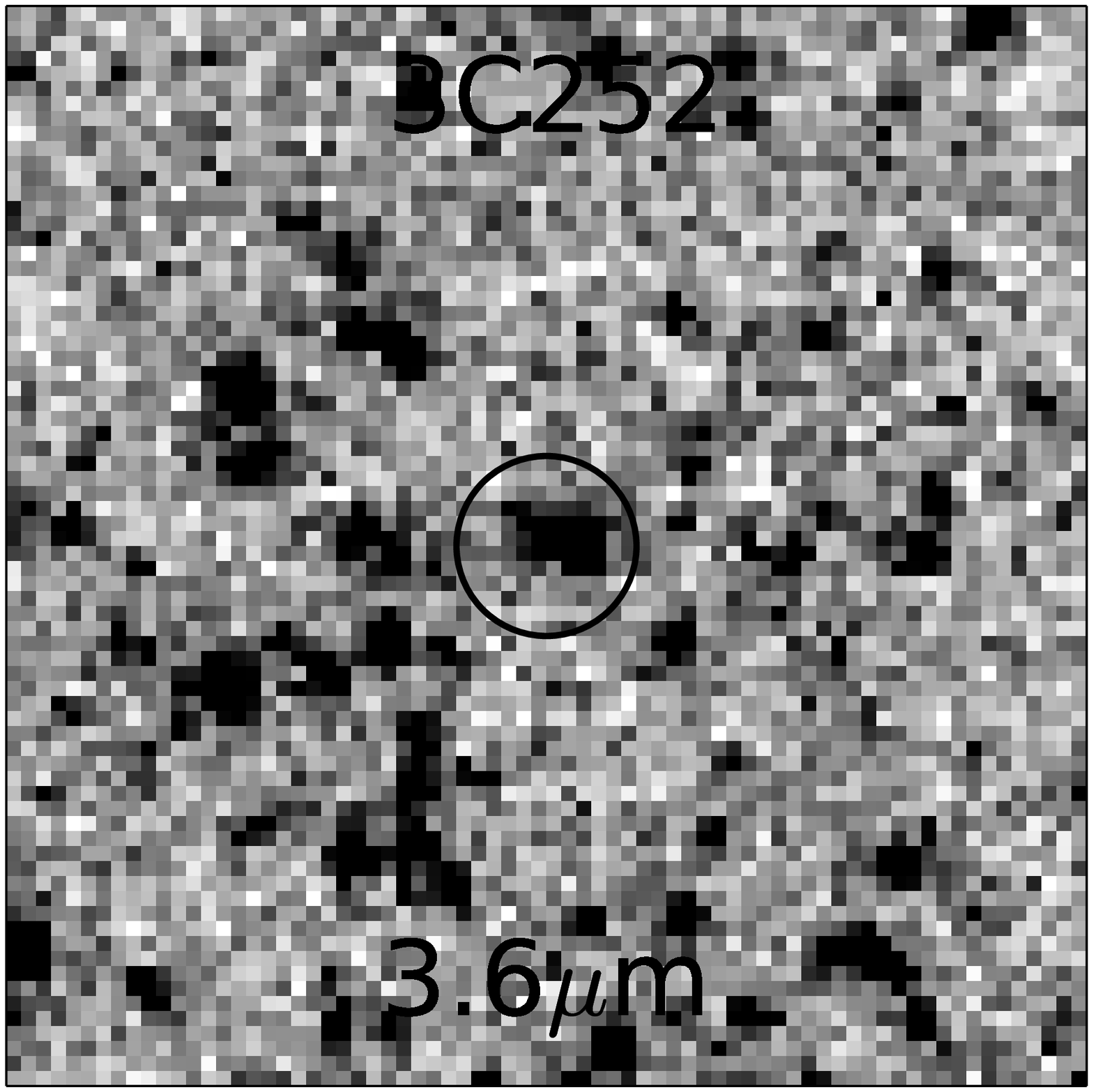}
      \includegraphics[width=1.5cm]{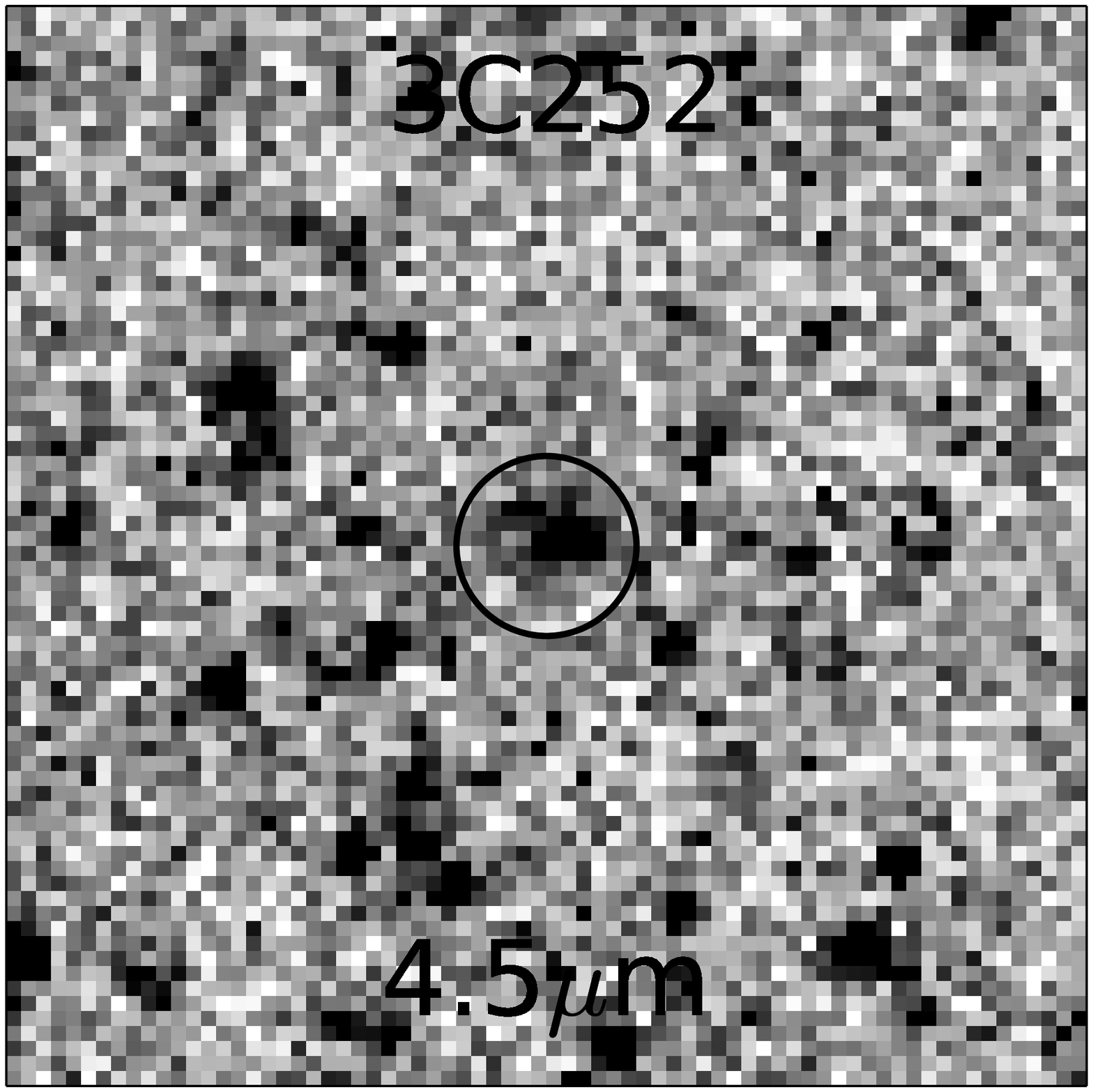}
      \includegraphics[width=1.5cm]{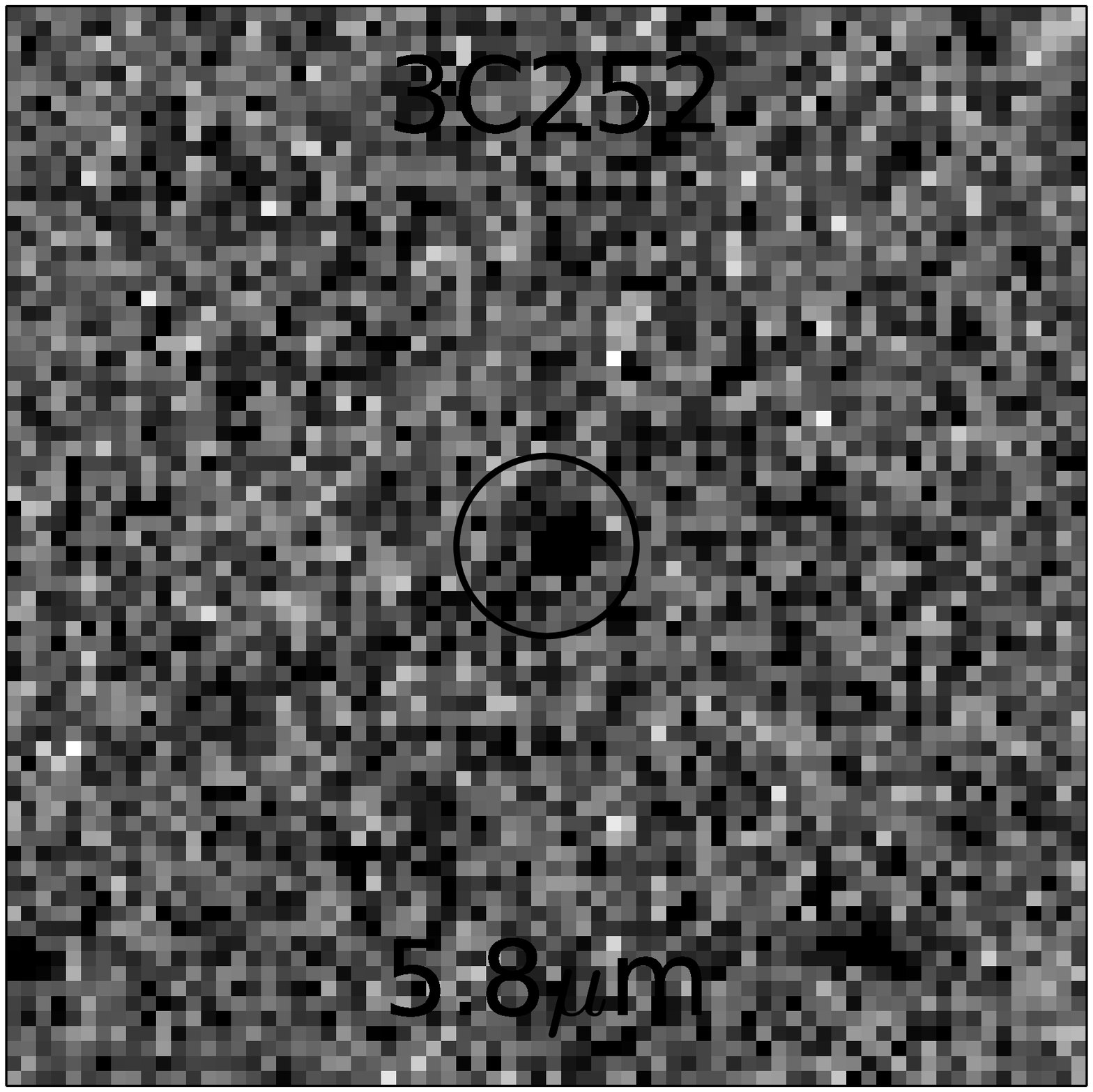}
      \includegraphics[width=1.5cm]{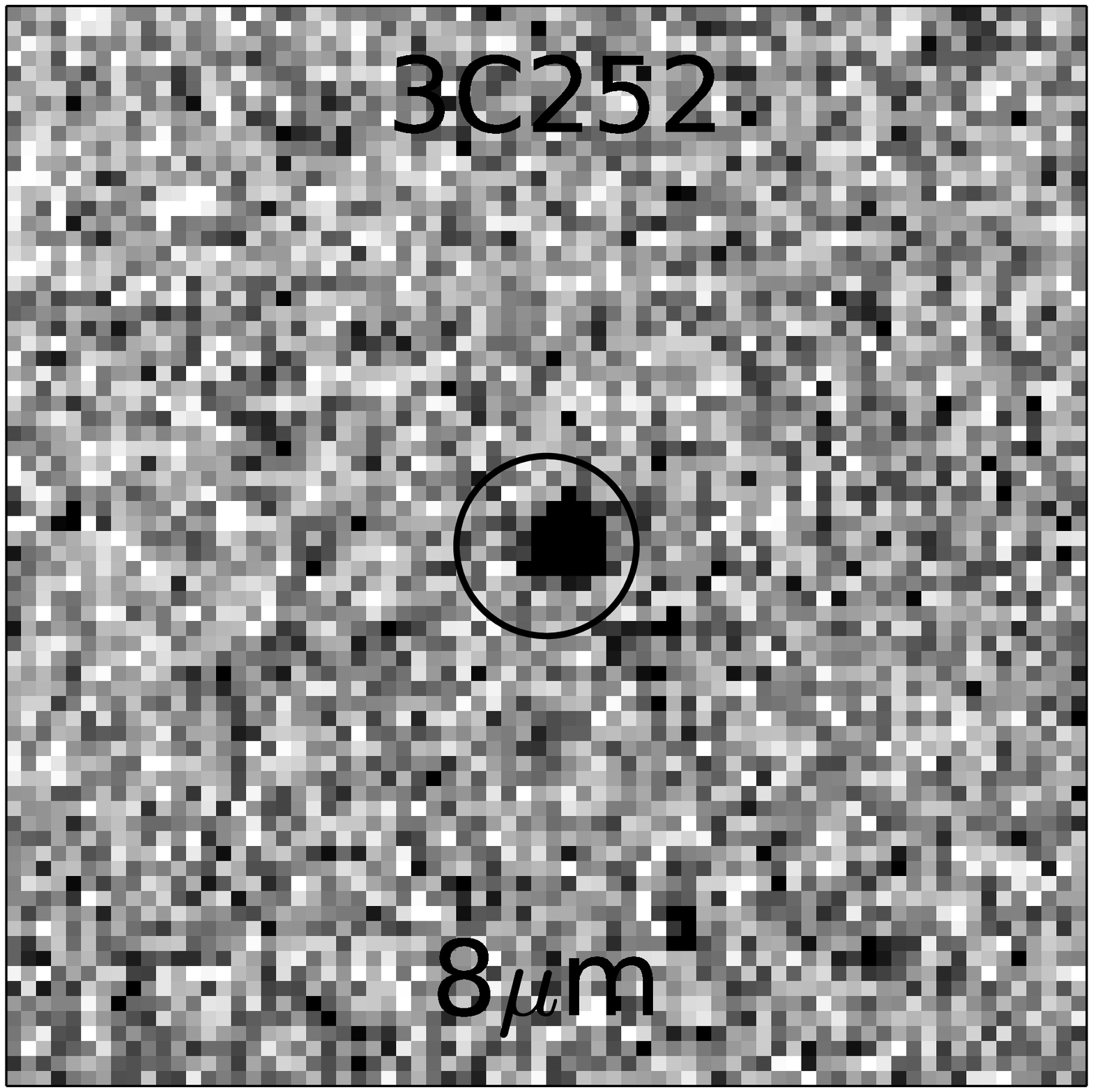}
      \includegraphics[width=1.5cm]{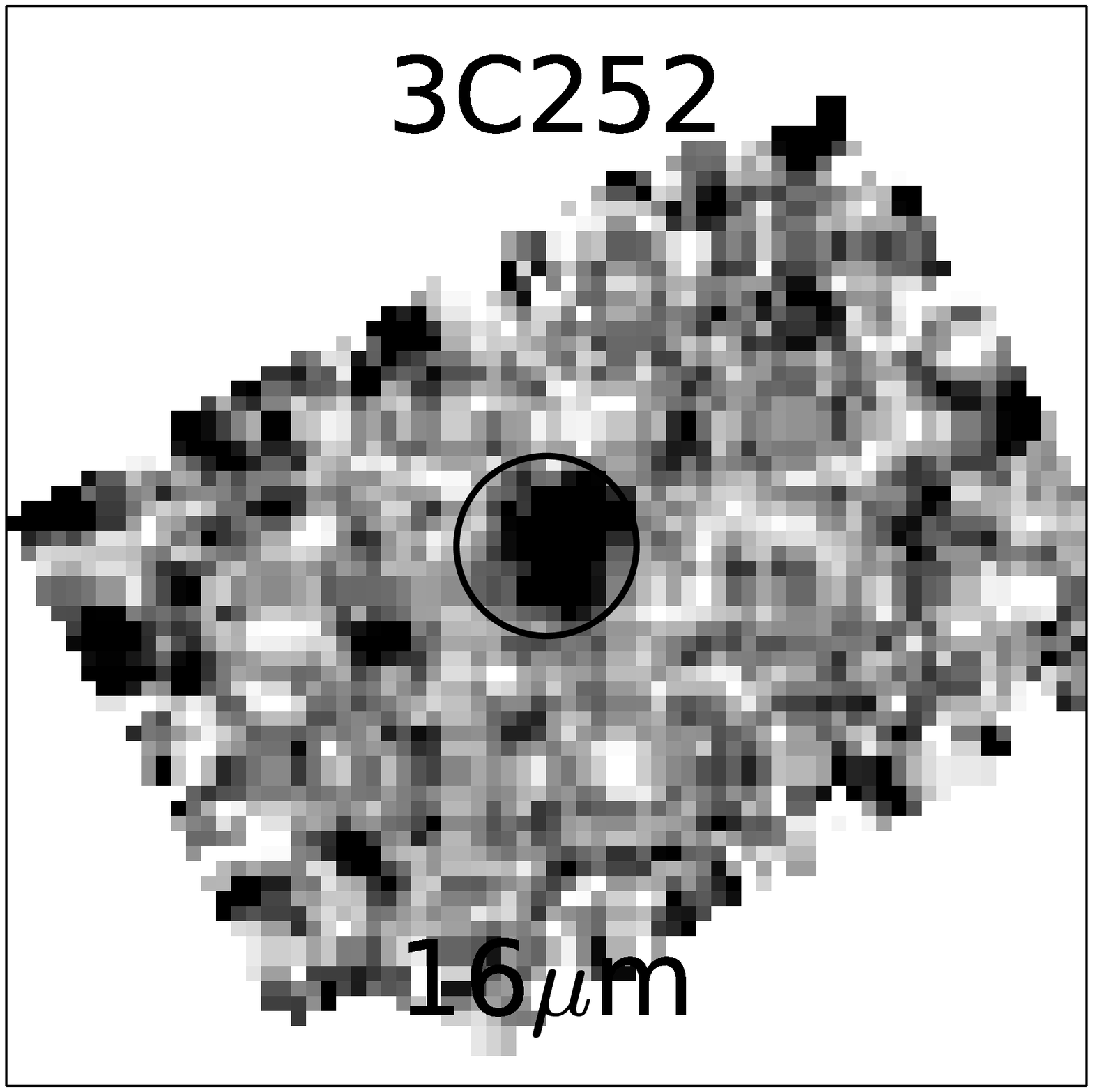}
      \includegraphics[width=1.5cm]{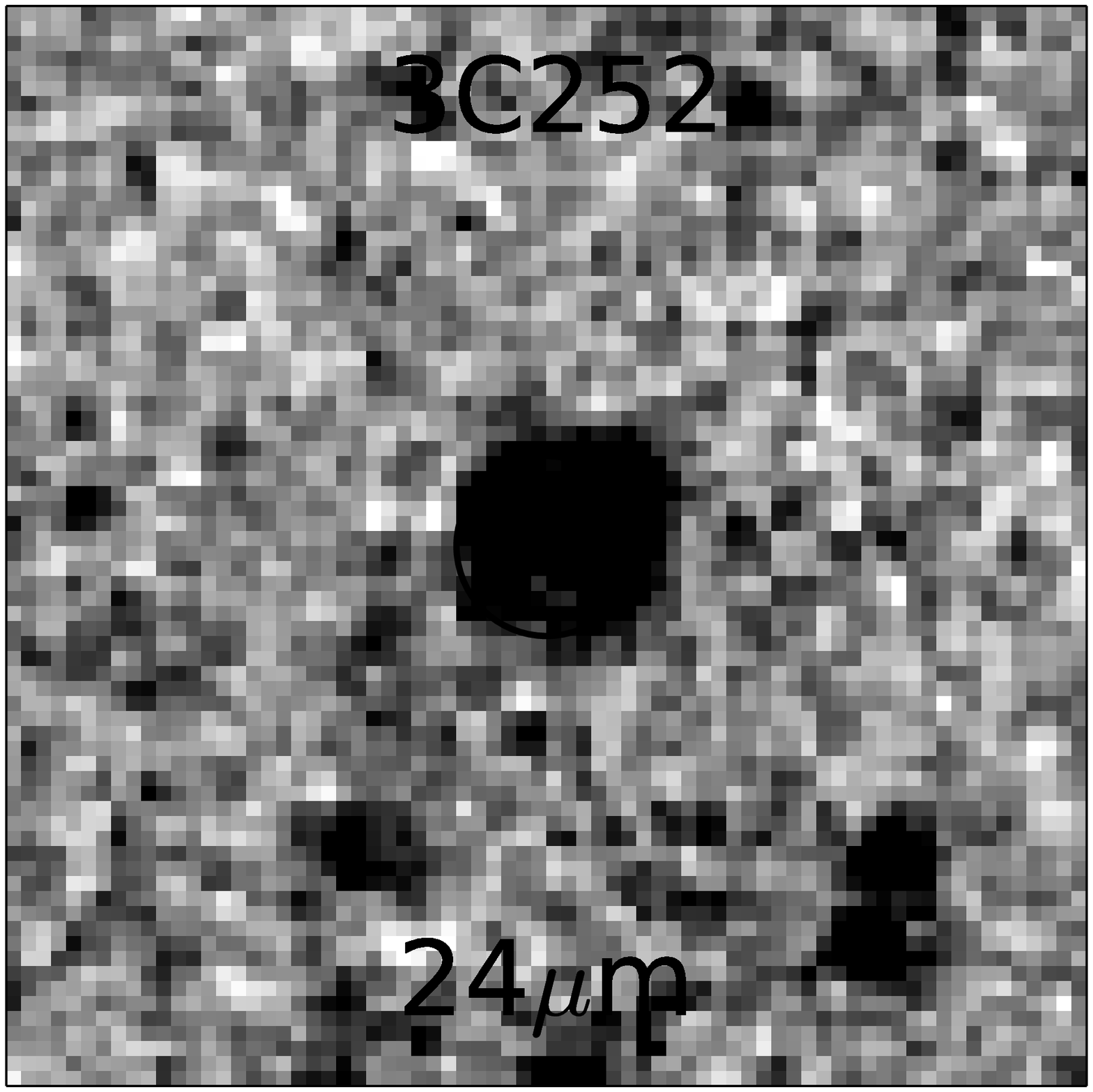}
      \includegraphics[width=1.5cm]{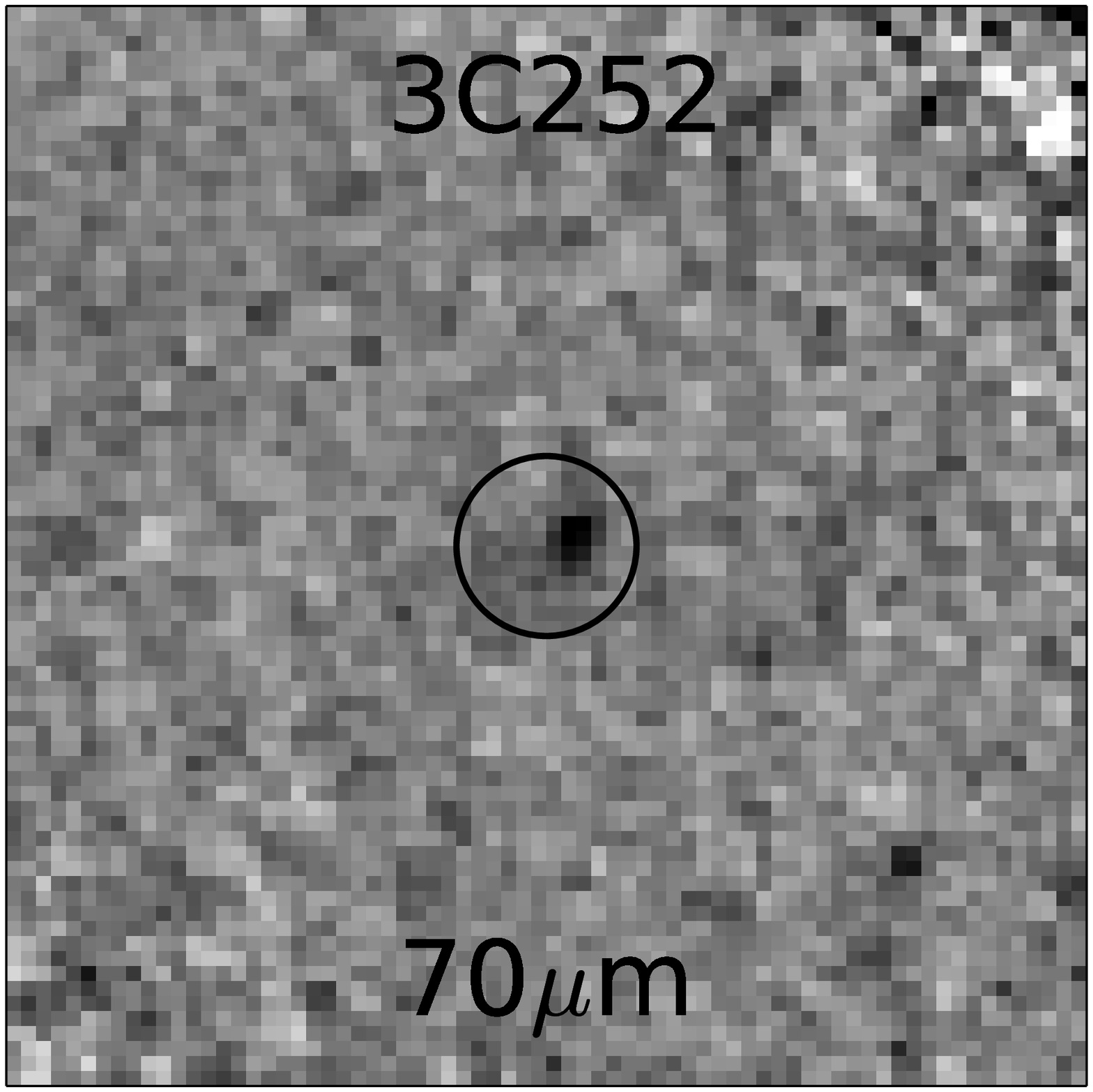}
      \includegraphics[width=1.5cm]{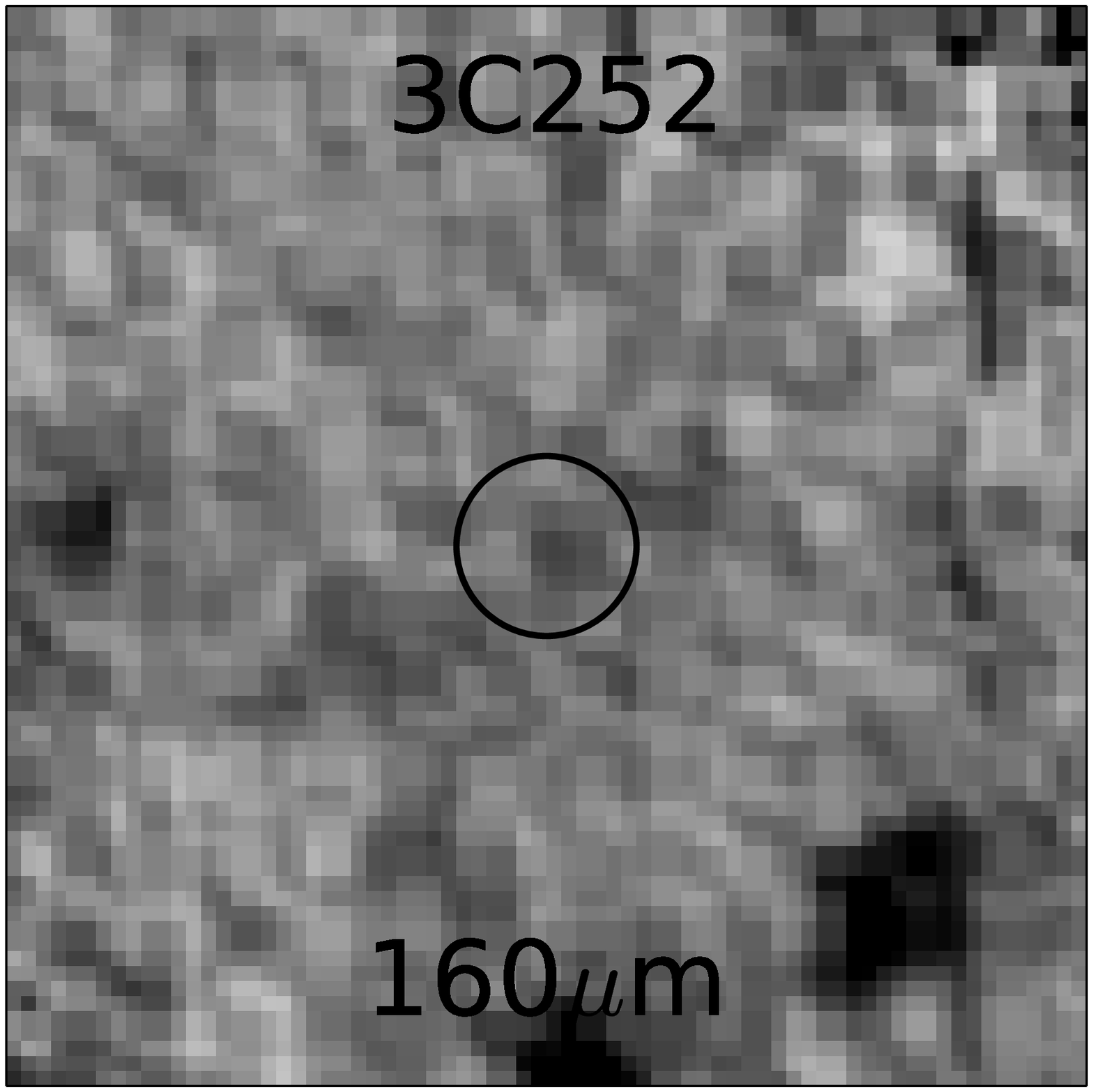}
      \includegraphics[width=1.5cm]{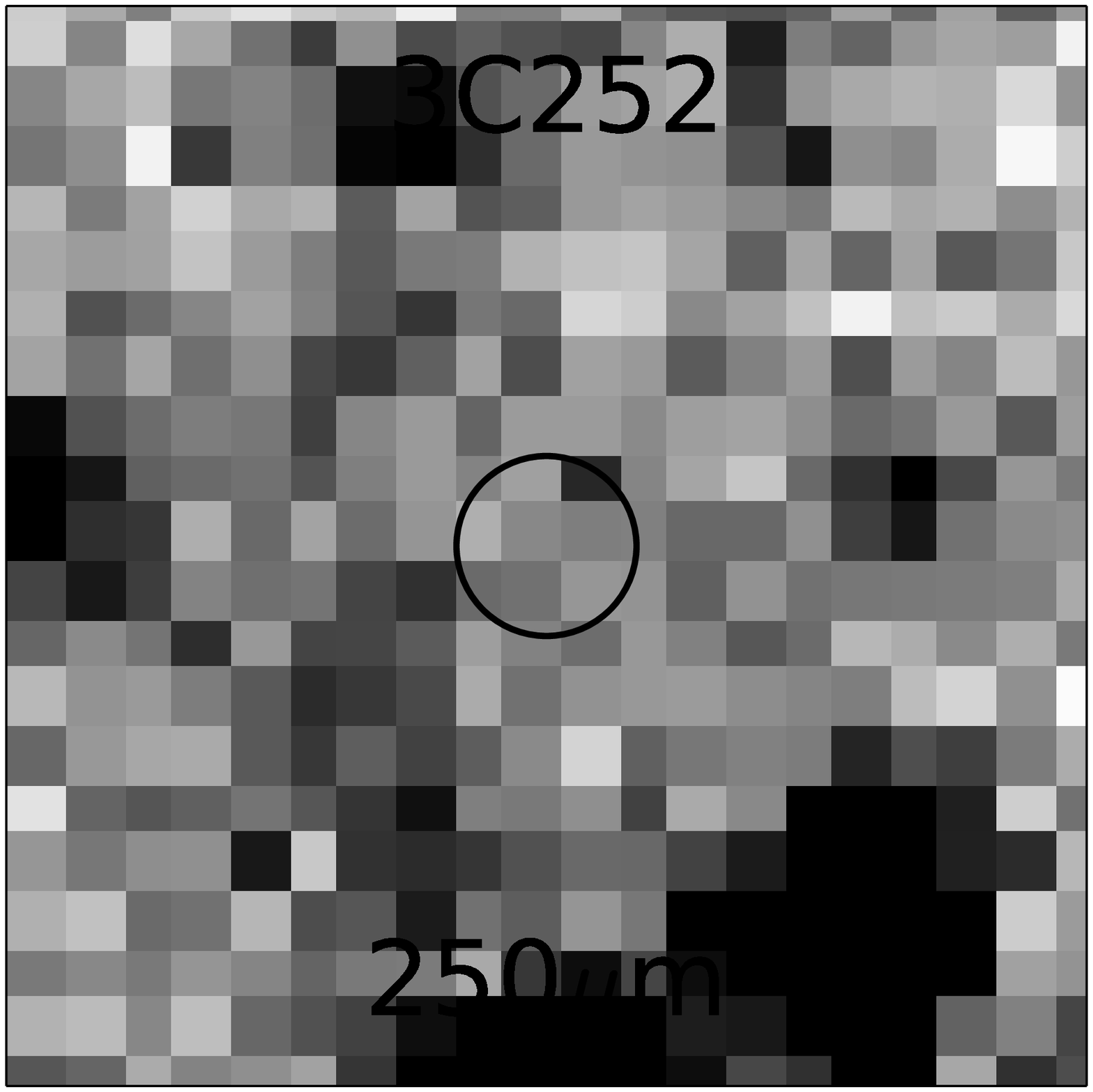}
      \includegraphics[width=1.5cm]{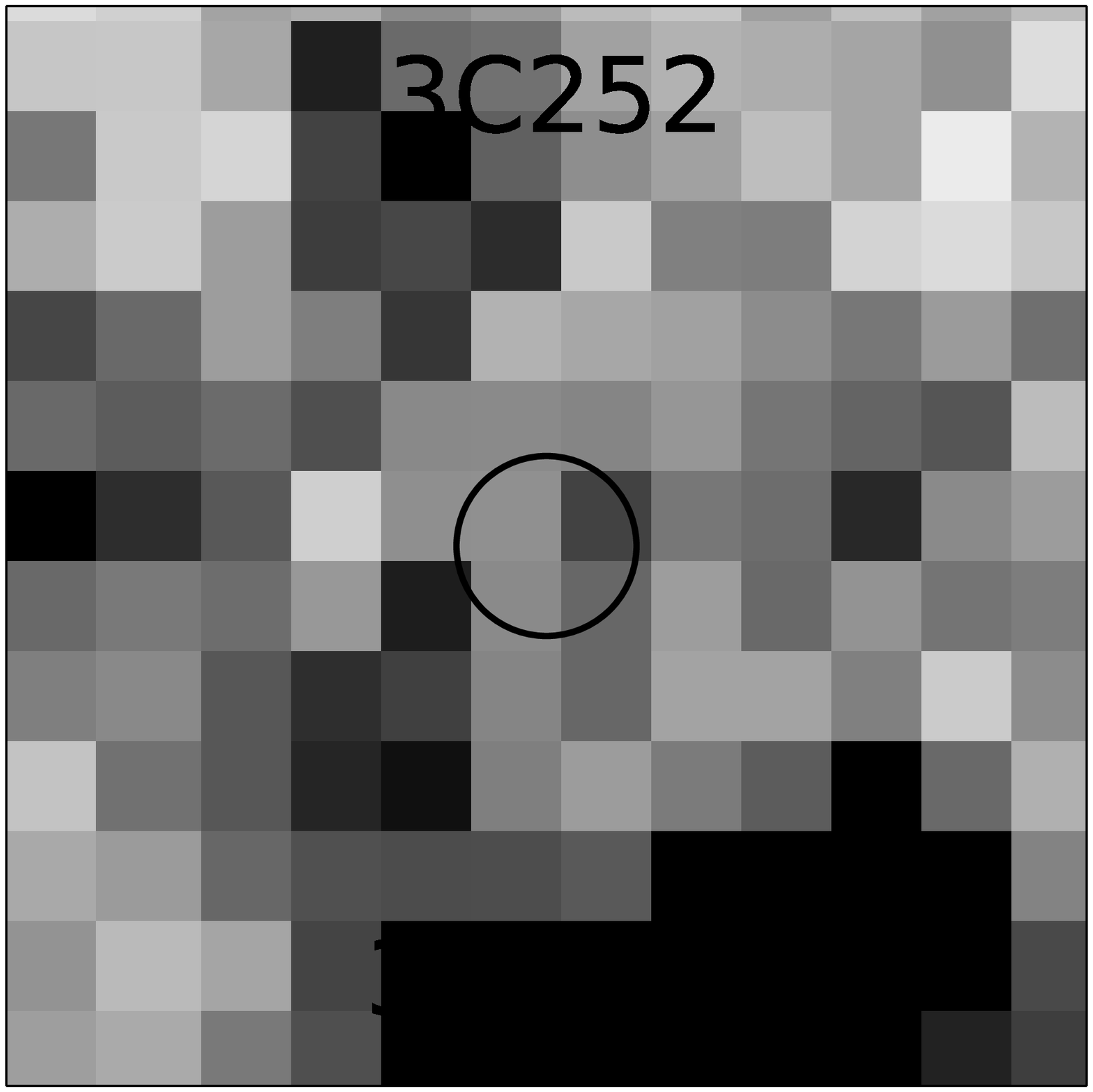}
      \includegraphics[width=1.5cm]{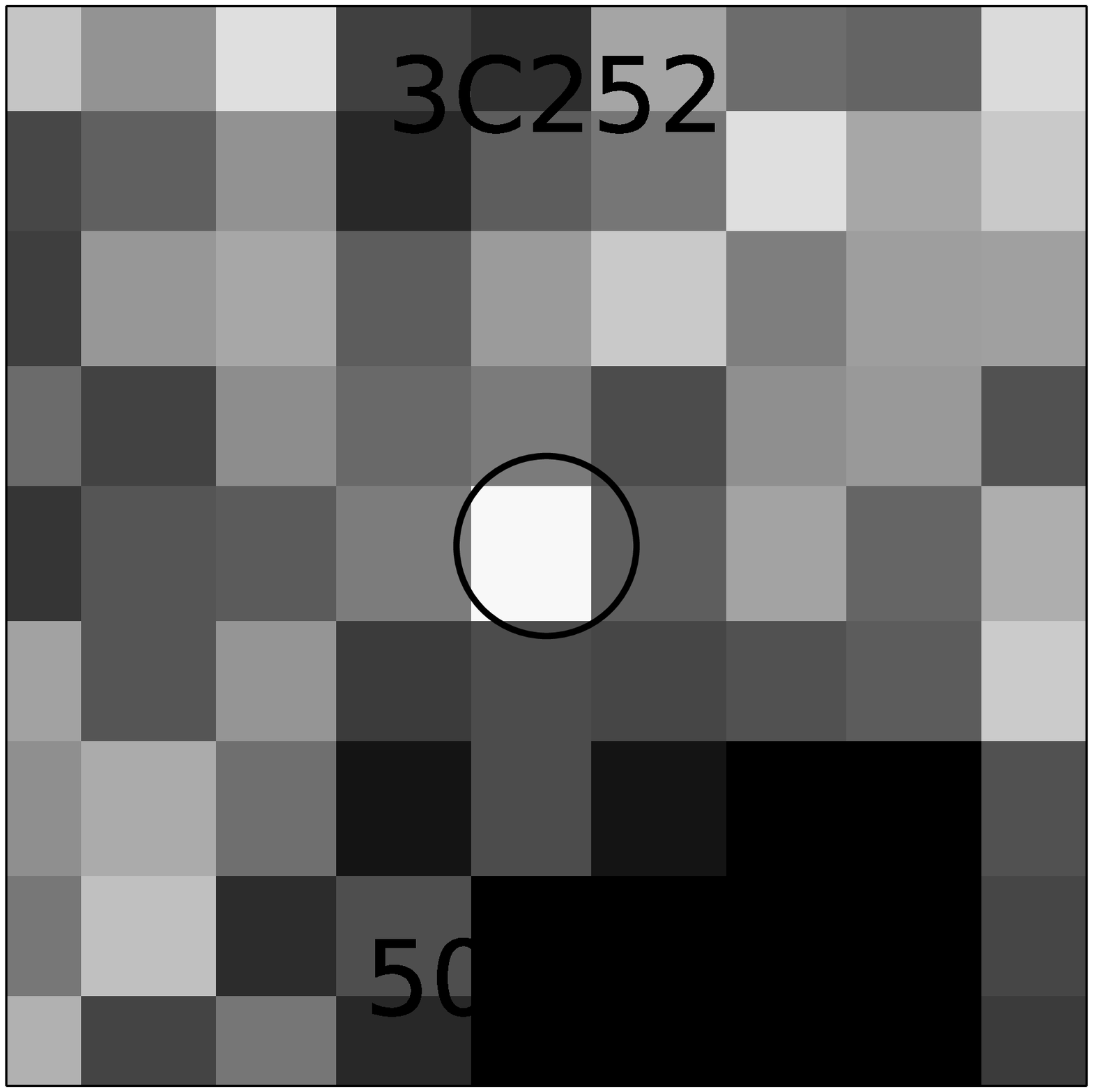}
      \\
      \includegraphics[width=1.5cm]{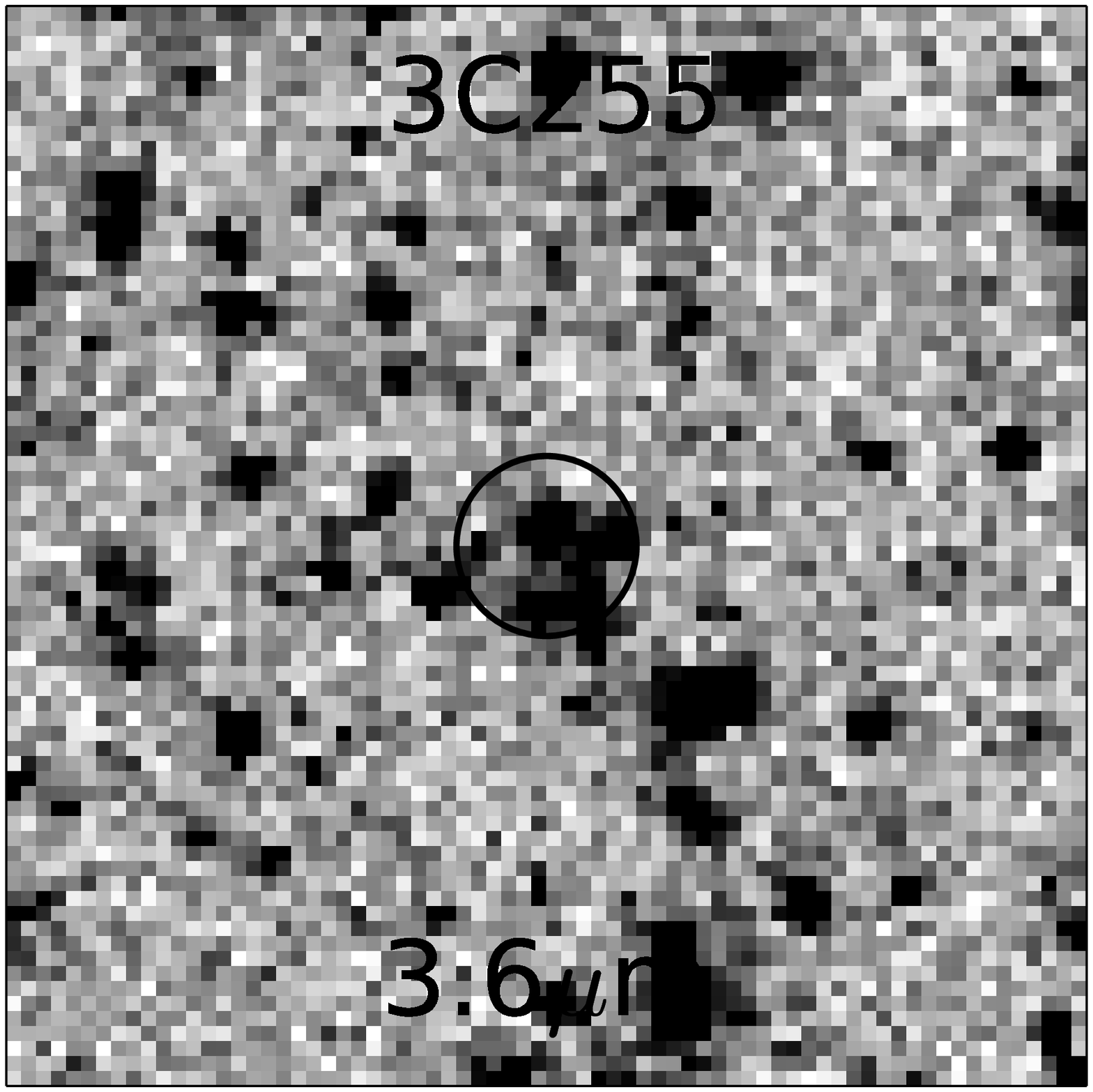}
      \includegraphics[width=1.5cm]{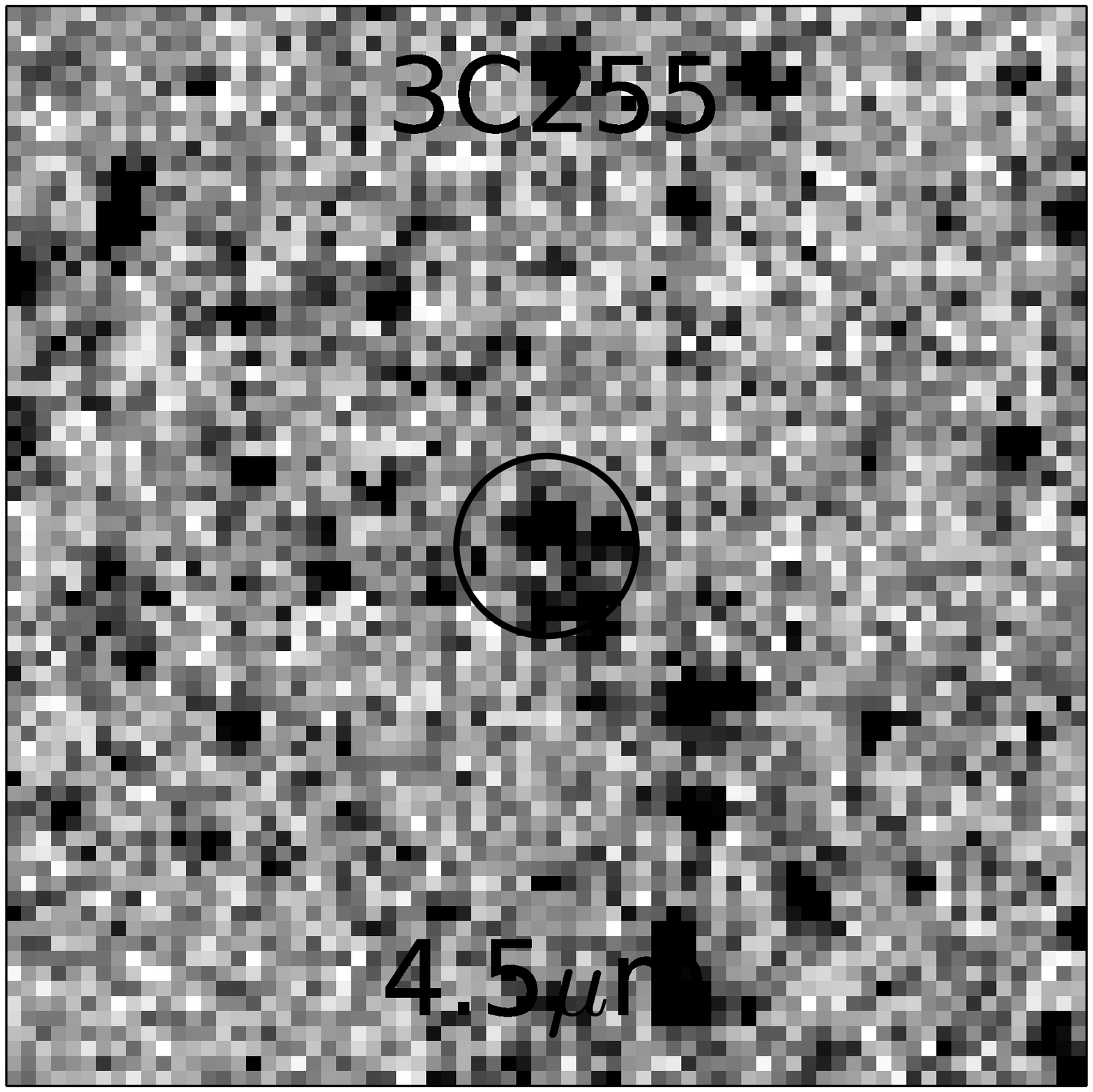}
      \includegraphics[width=1.5cm]{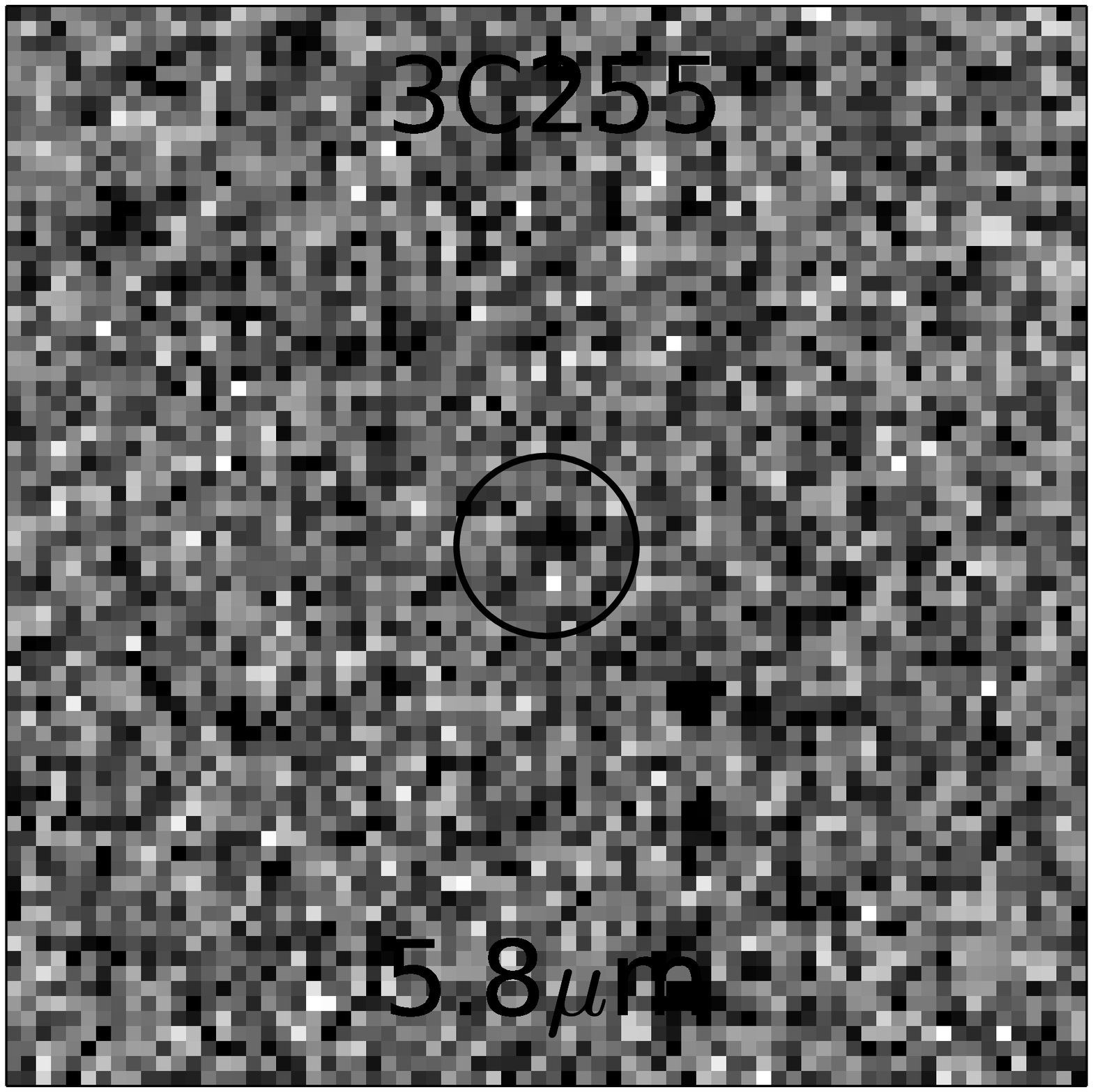}
      \includegraphics[width=1.5cm]{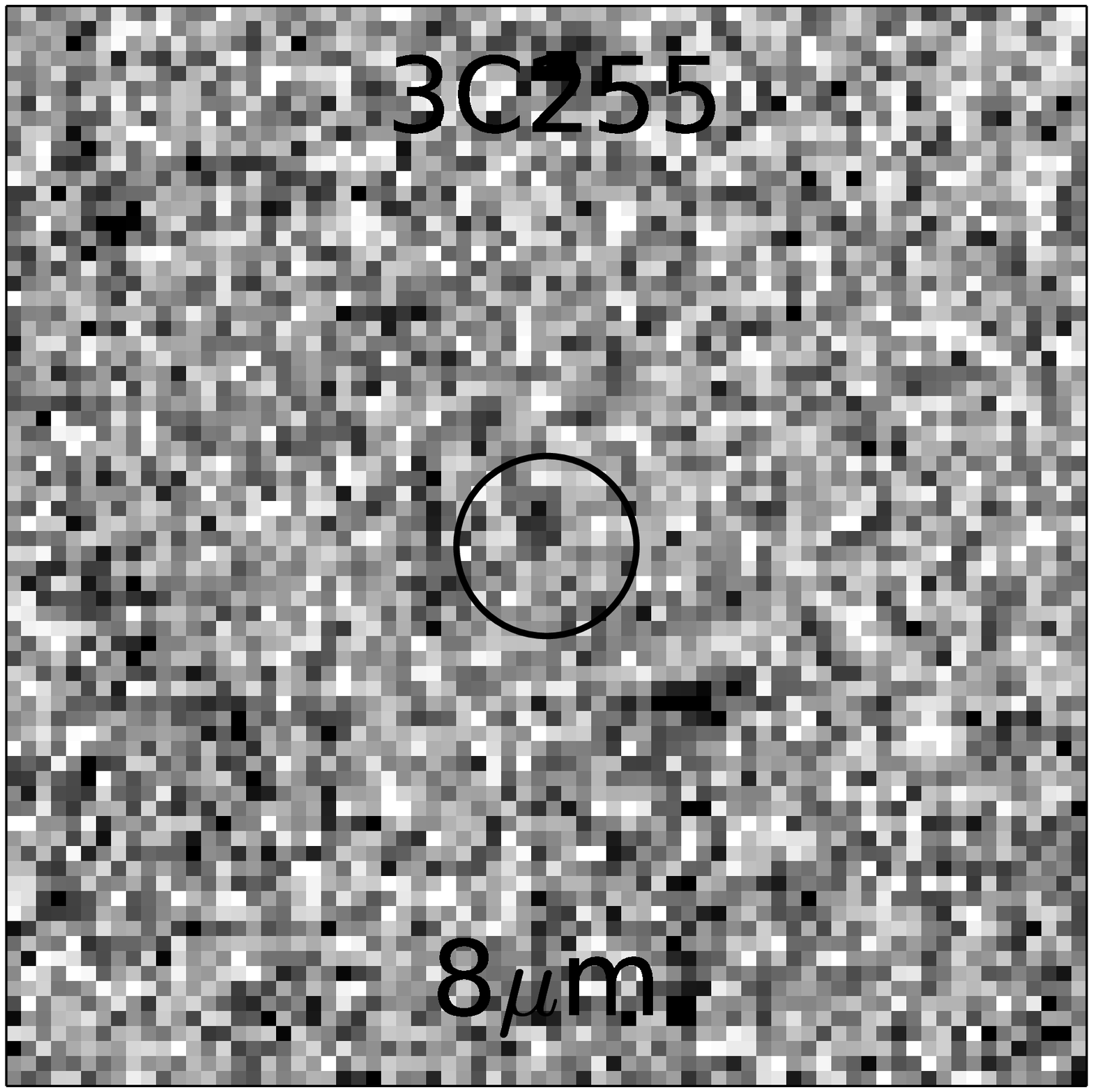}
      \includegraphics[width=1.5cm]{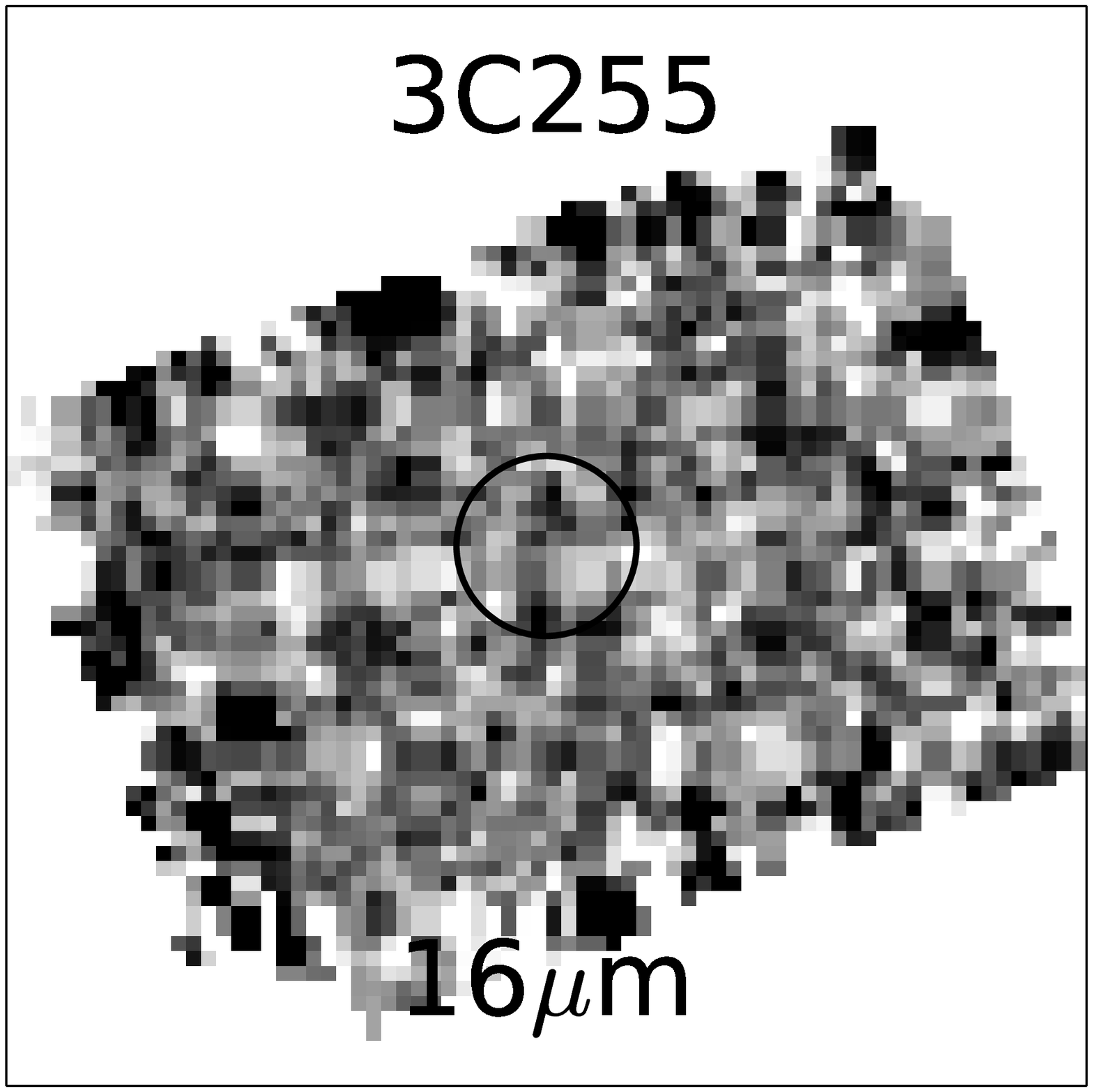}
      \includegraphics[width=1.5cm]{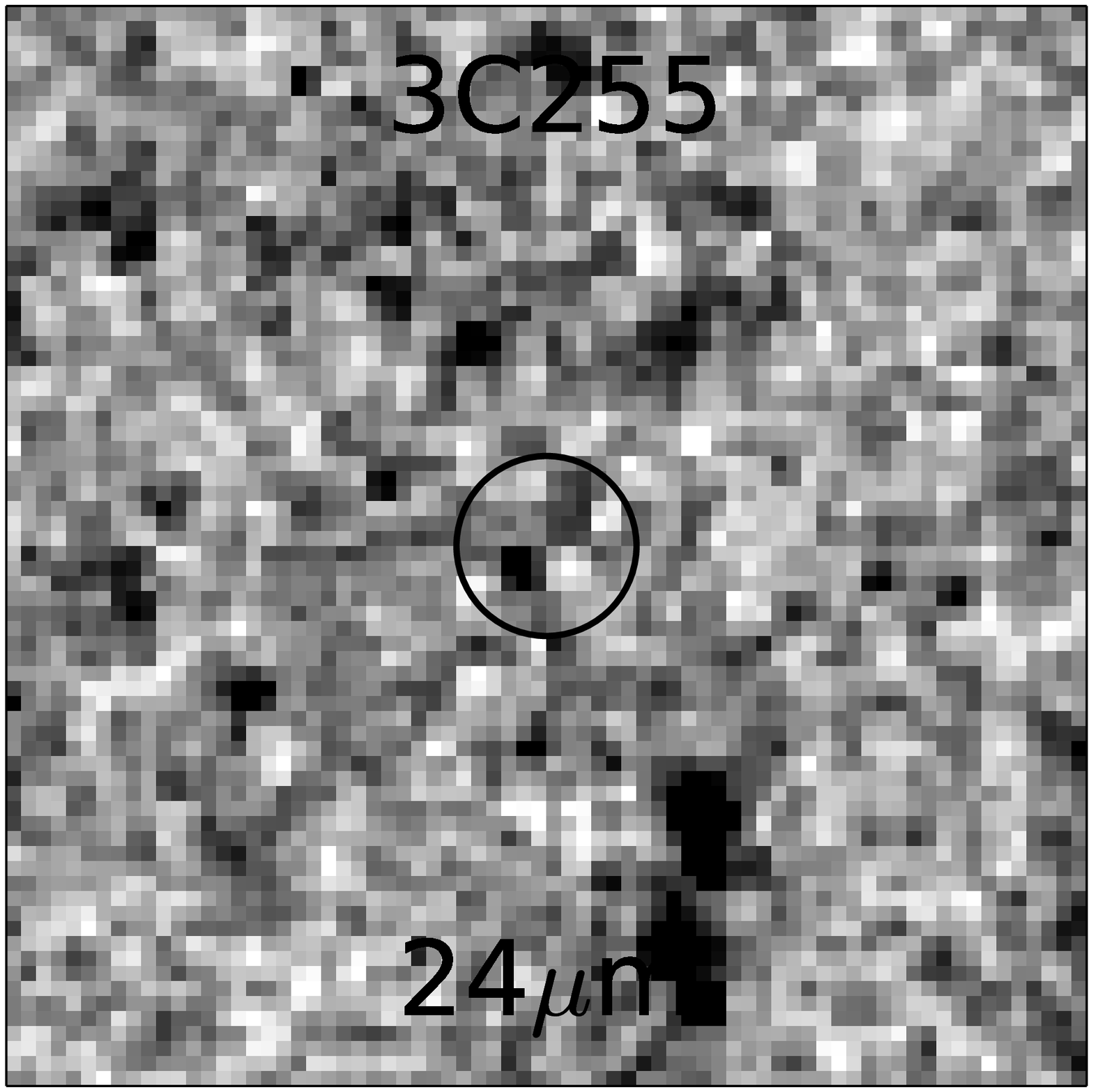}
      \includegraphics[width=1.5cm]{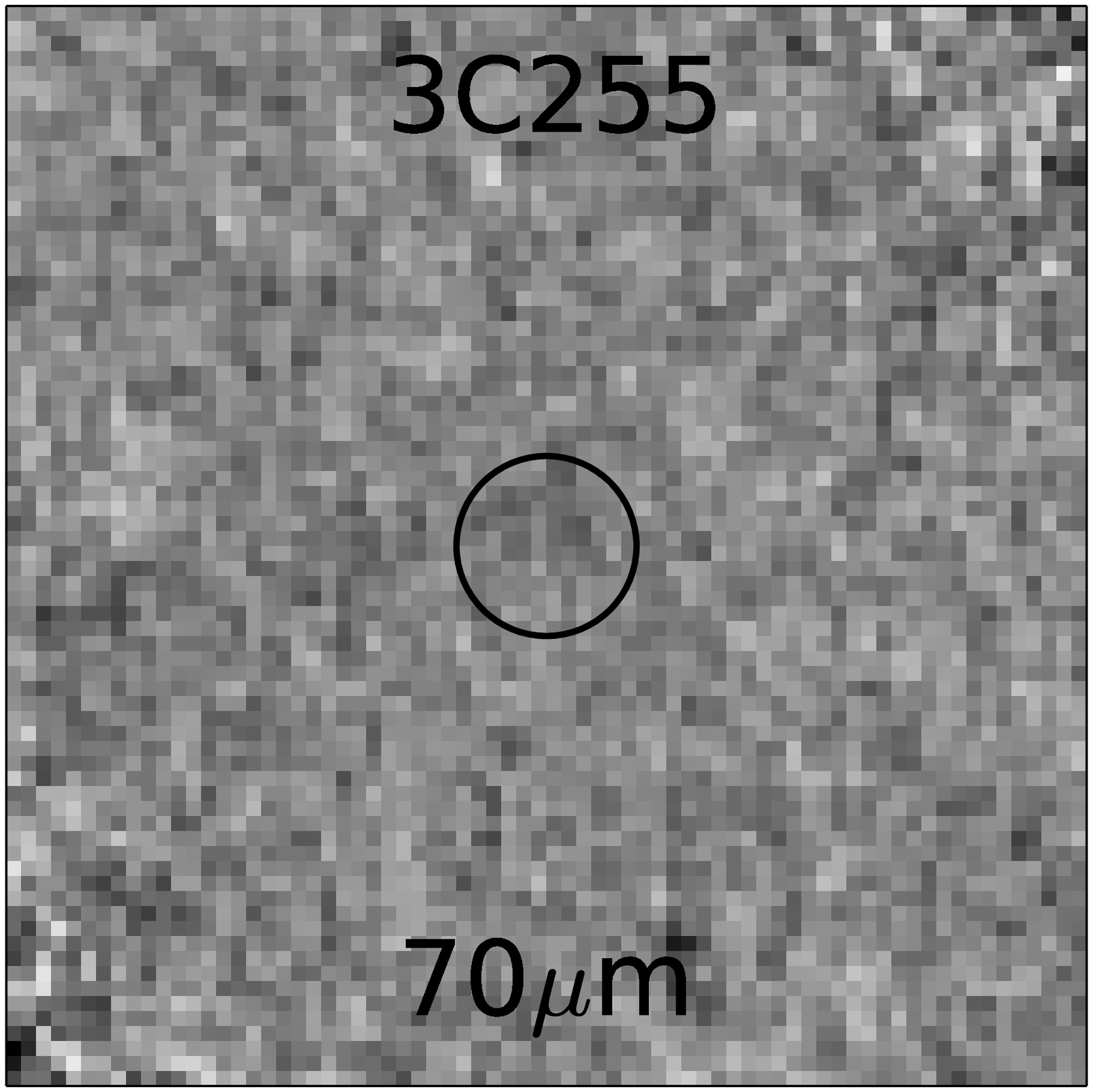}
      \includegraphics[width=1.5cm]{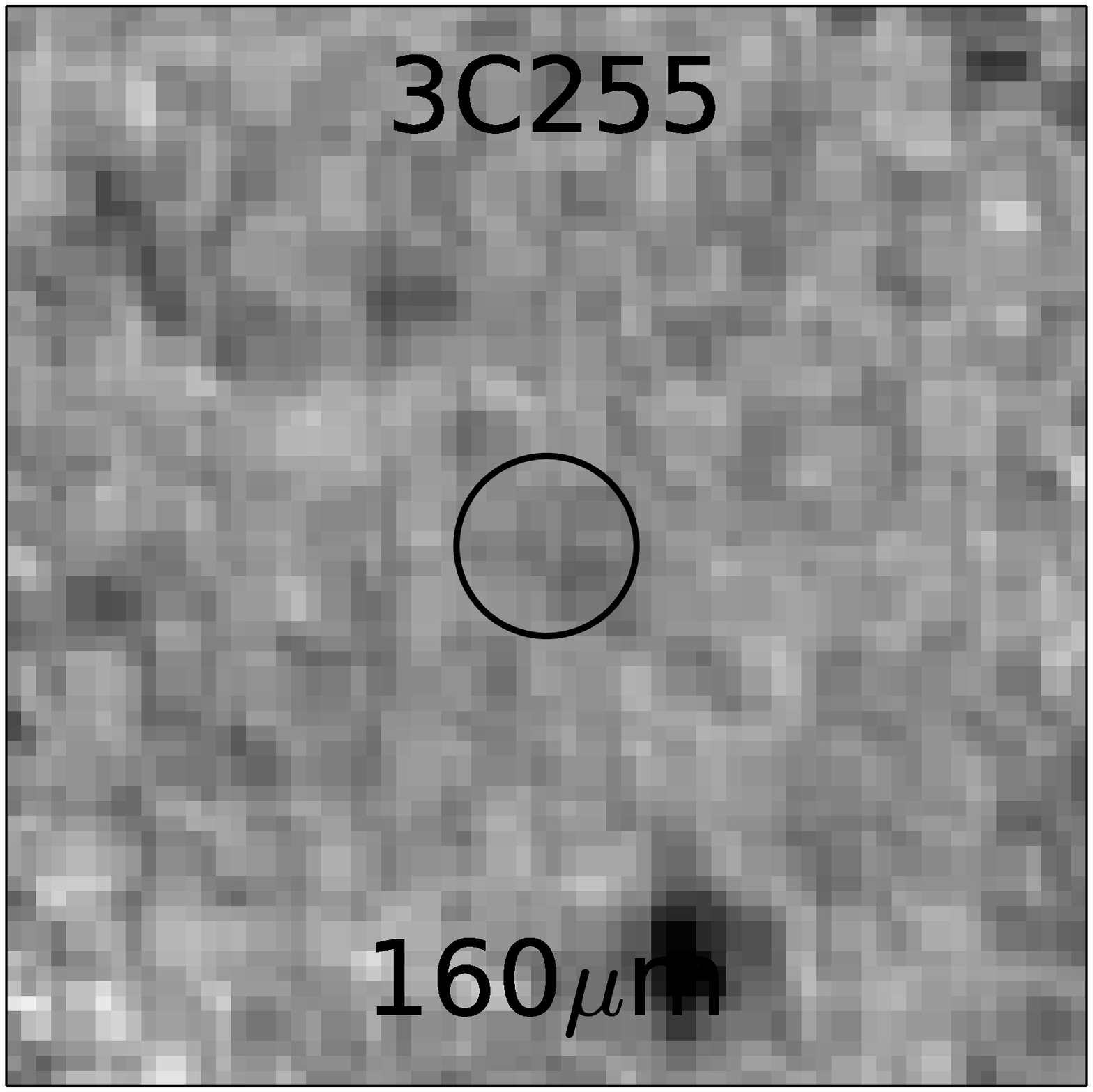}
      \includegraphics[width=1.5cm]{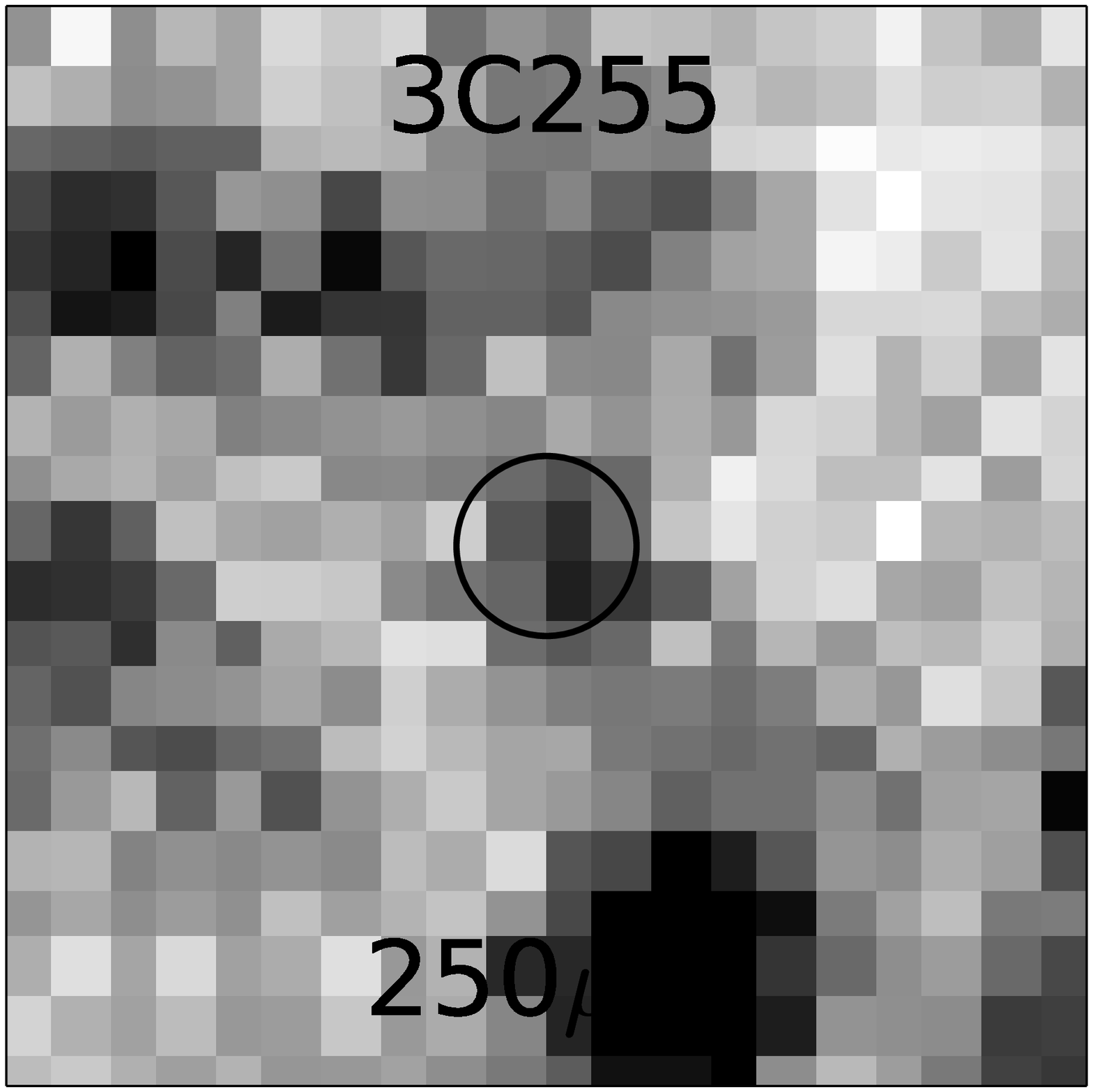}
      \includegraphics[width=1.5cm]{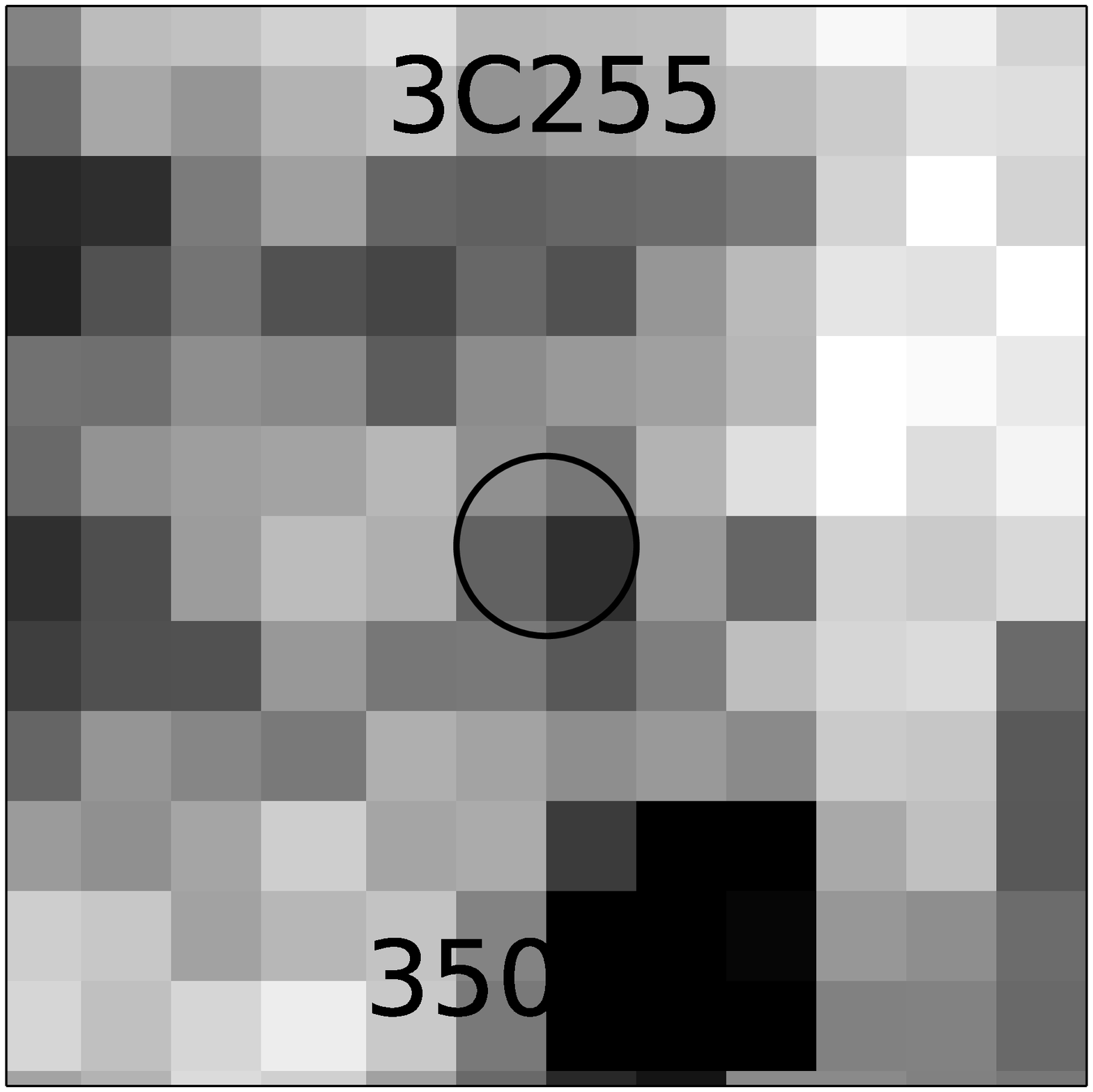}
      \includegraphics[width=1.5cm]{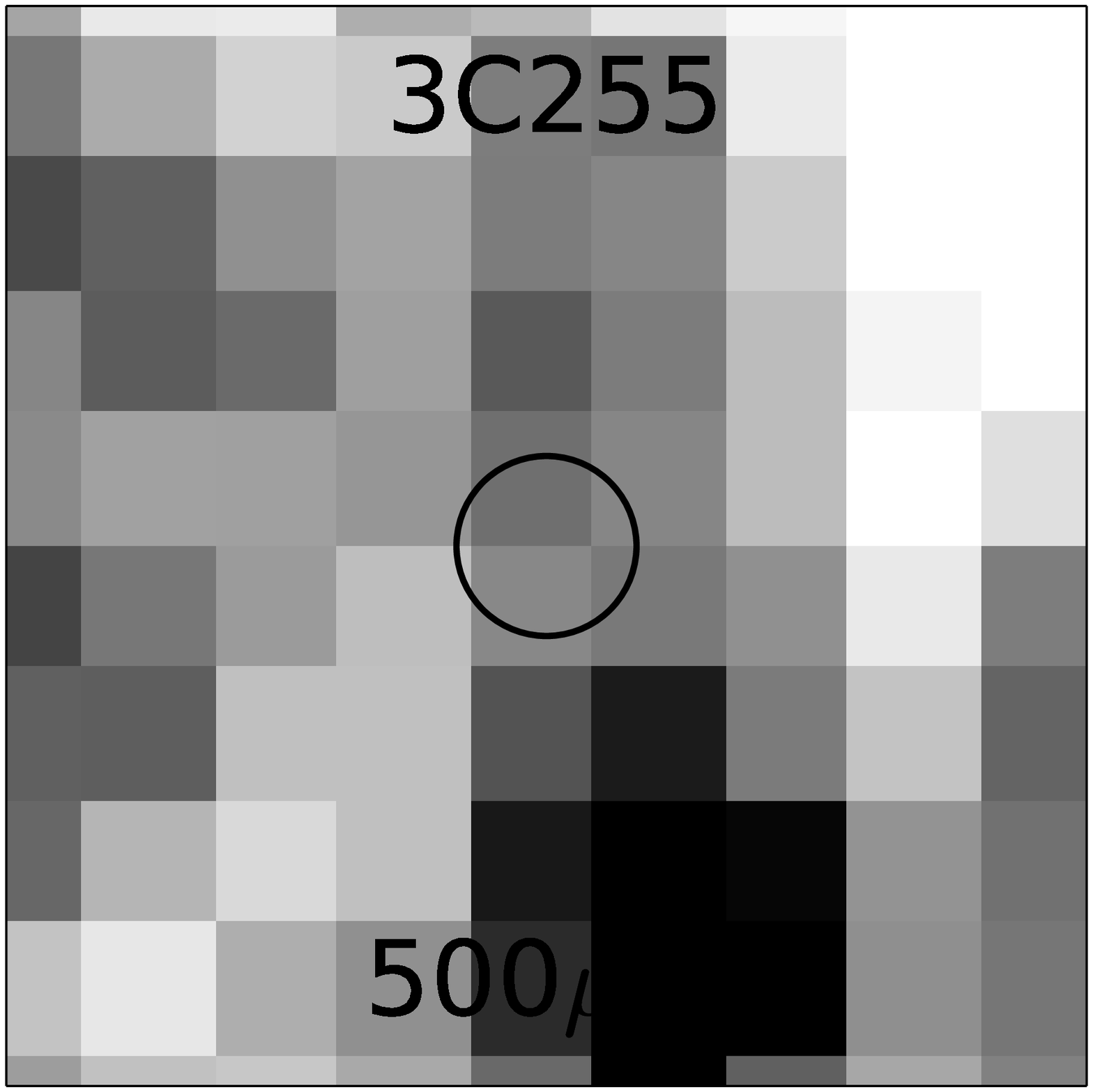}
      \\
      \includegraphics[width=1.5cm]{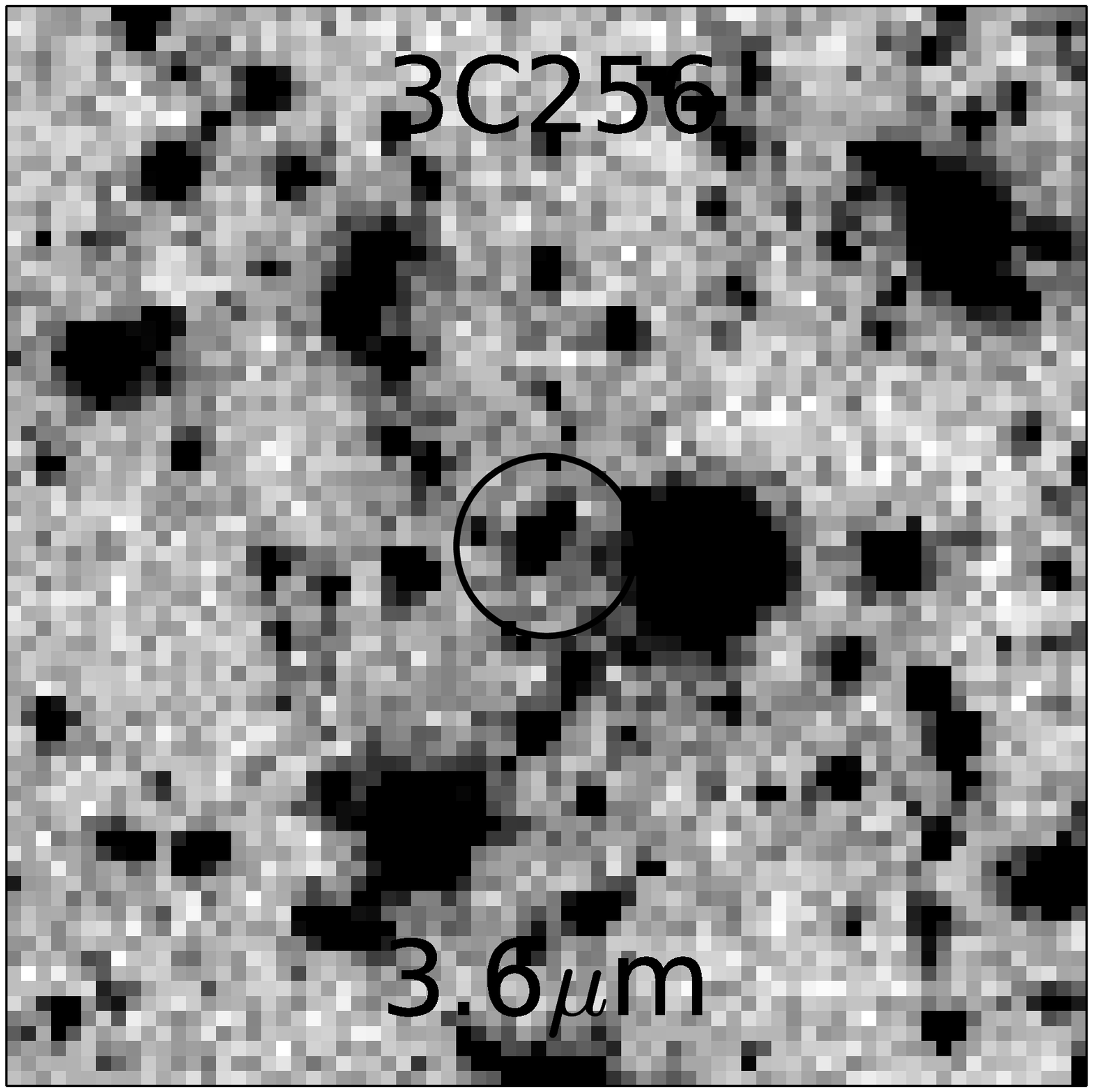}
      \includegraphics[width=1.5cm]{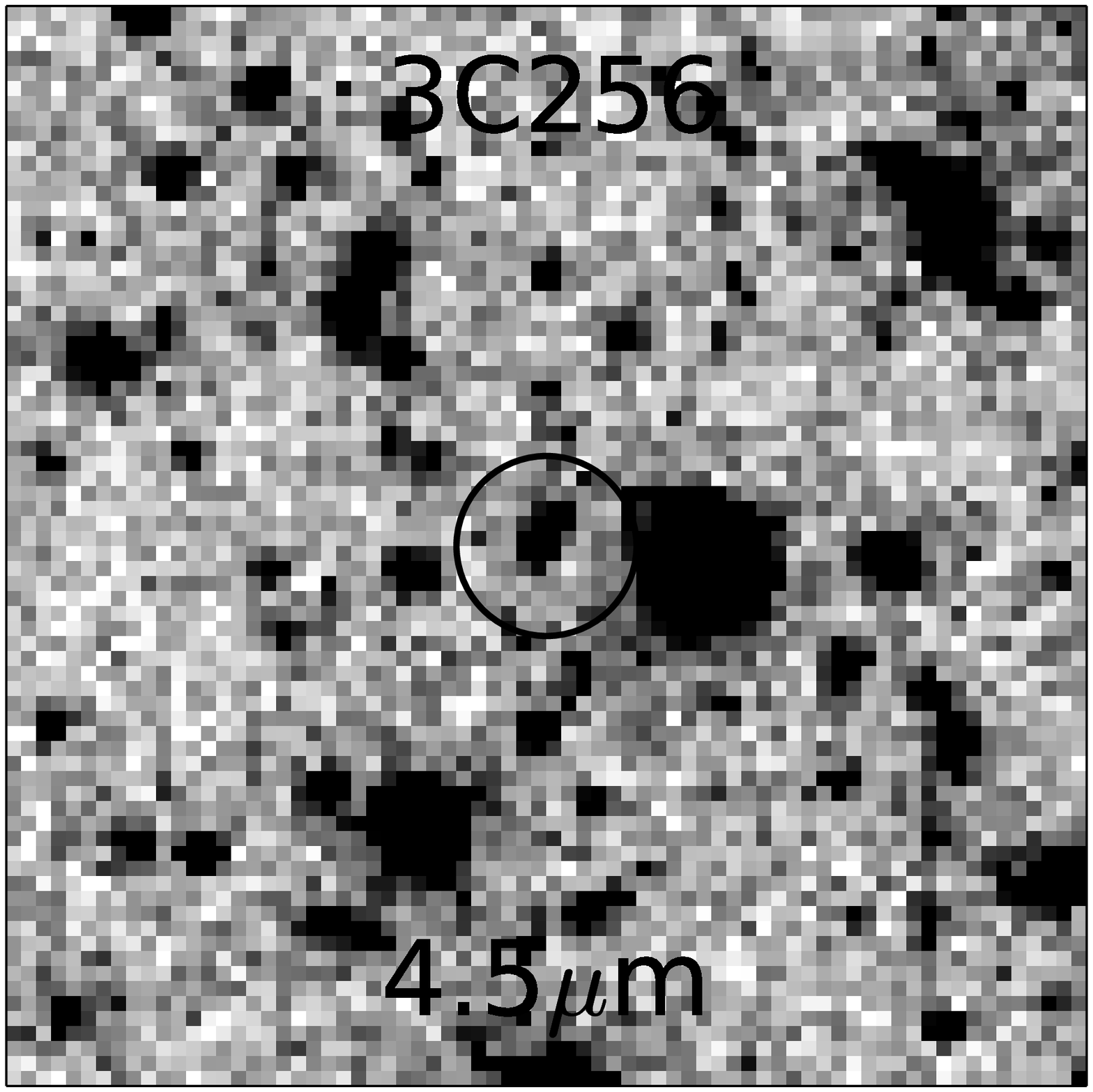}
      \includegraphics[width=1.5cm]{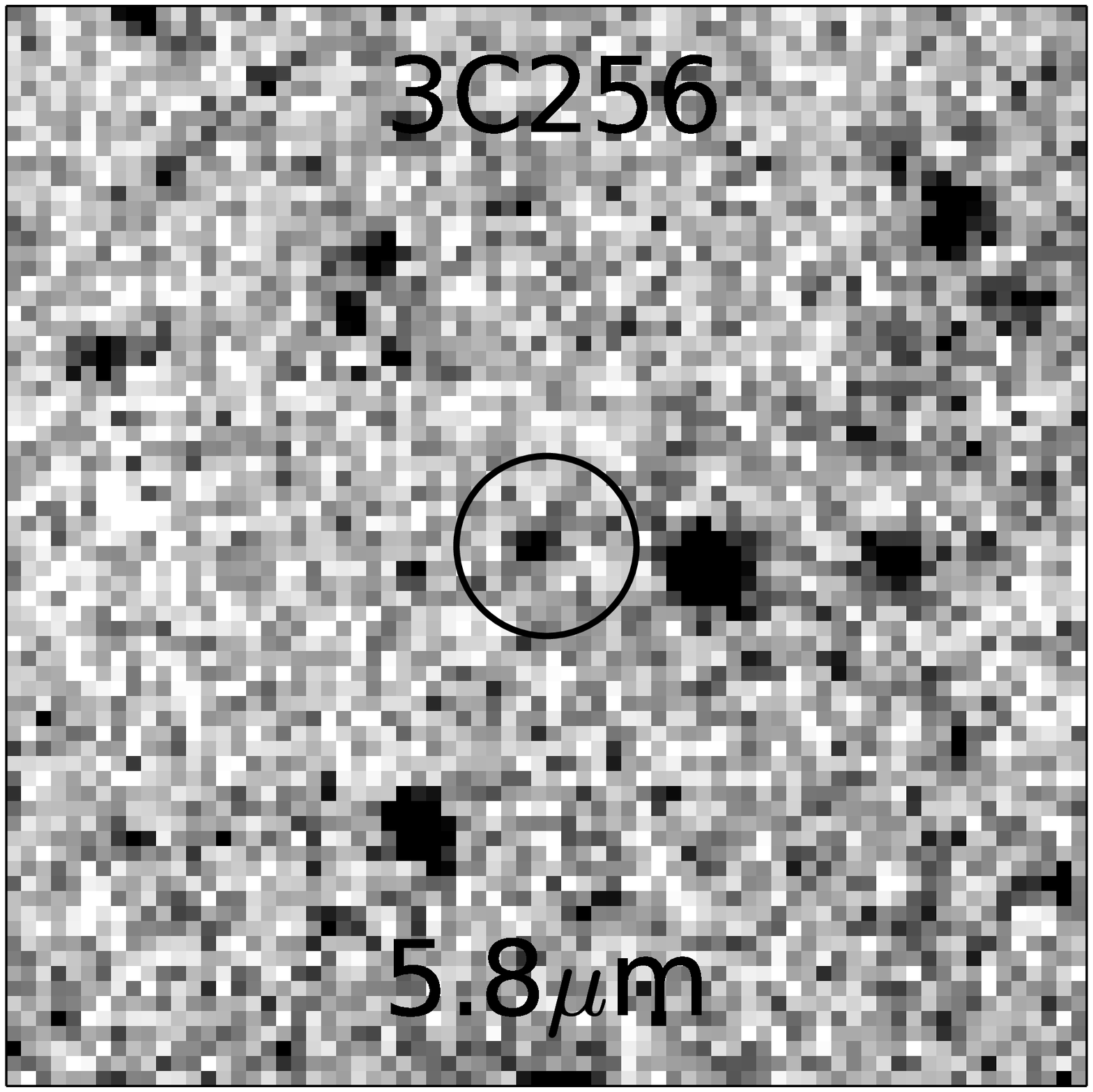}
      \includegraphics[width=1.5cm]{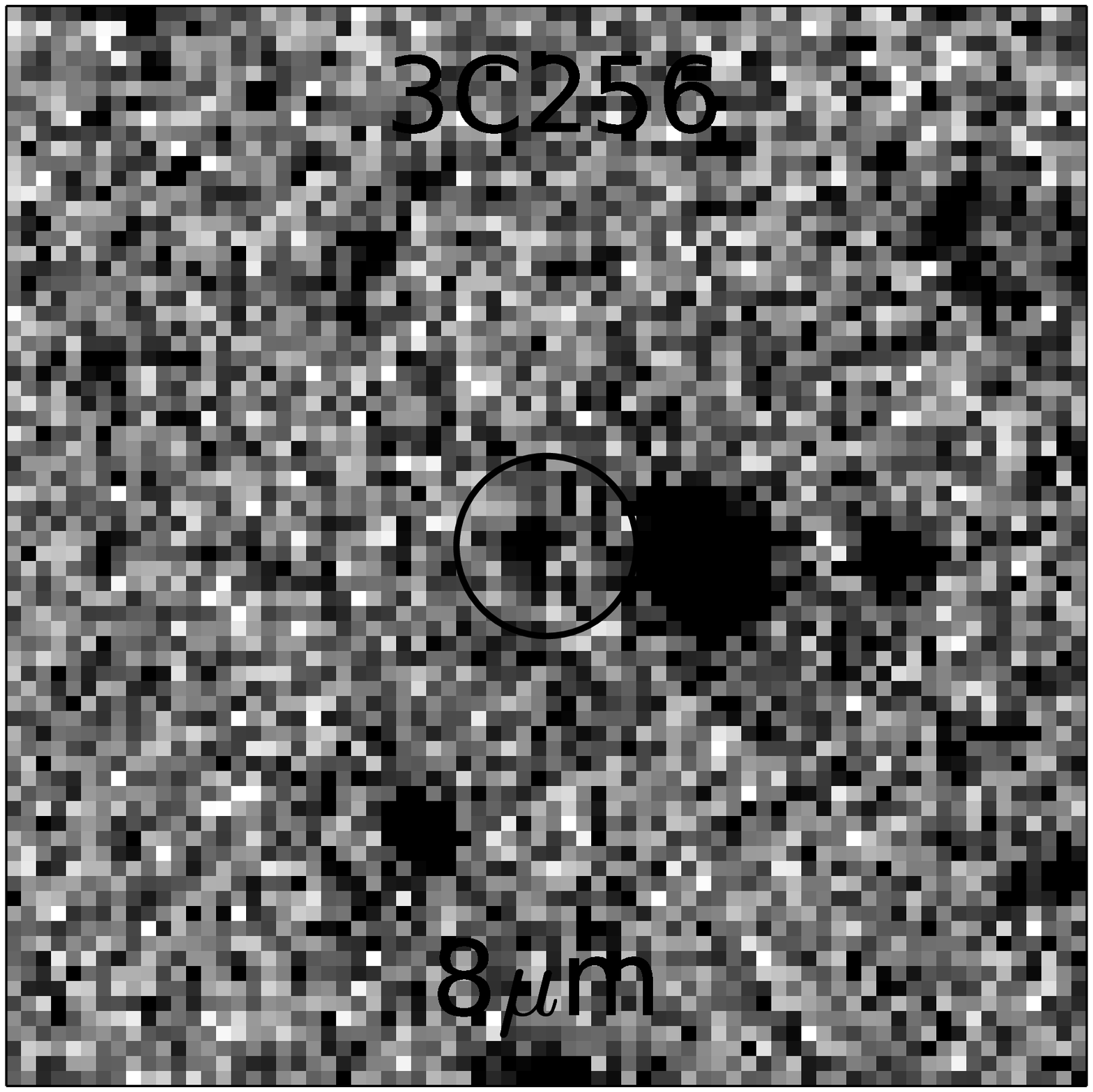}
      \includegraphics[width=1.5cm]{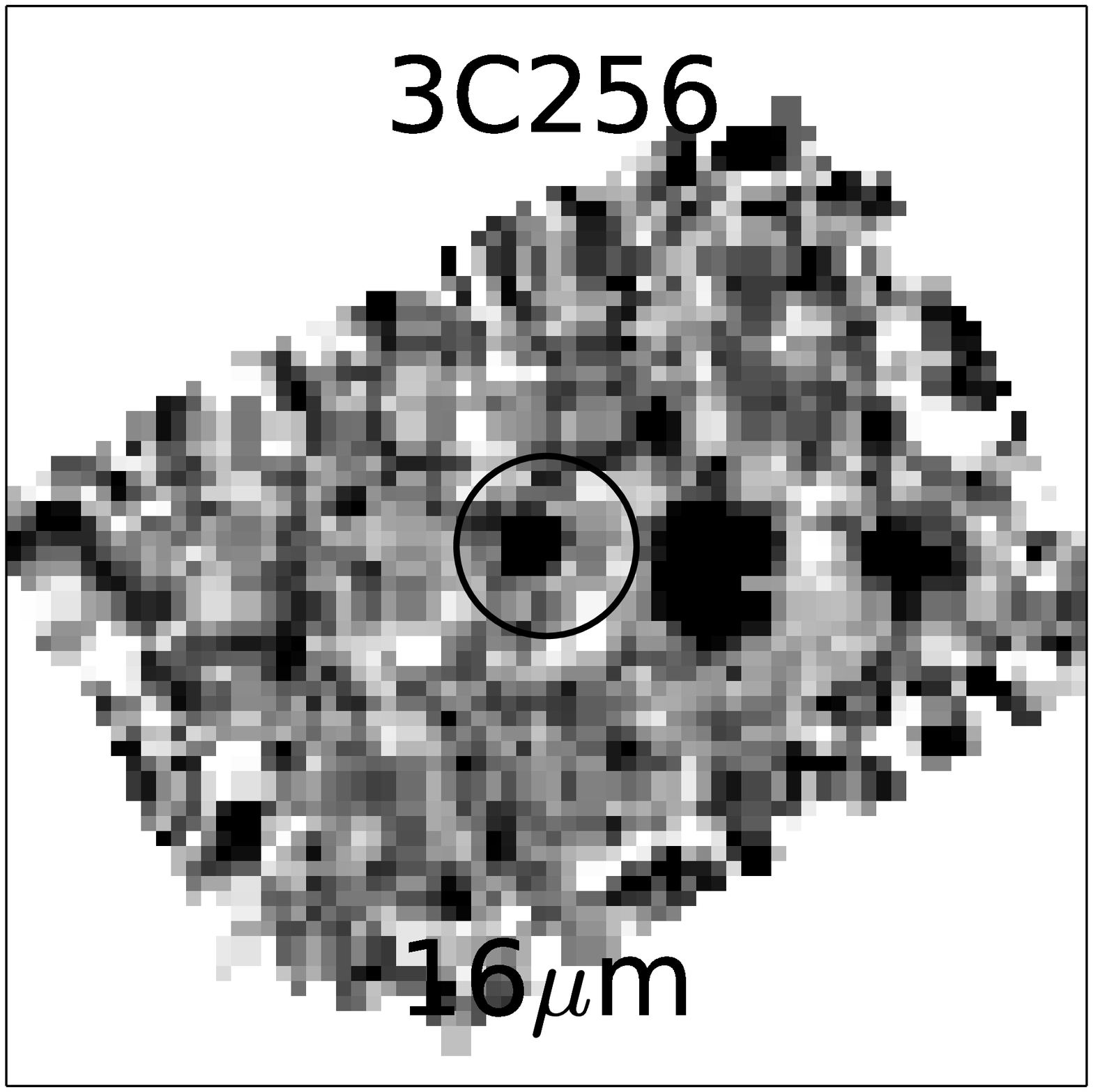}
      \includegraphics[width=1.5cm]{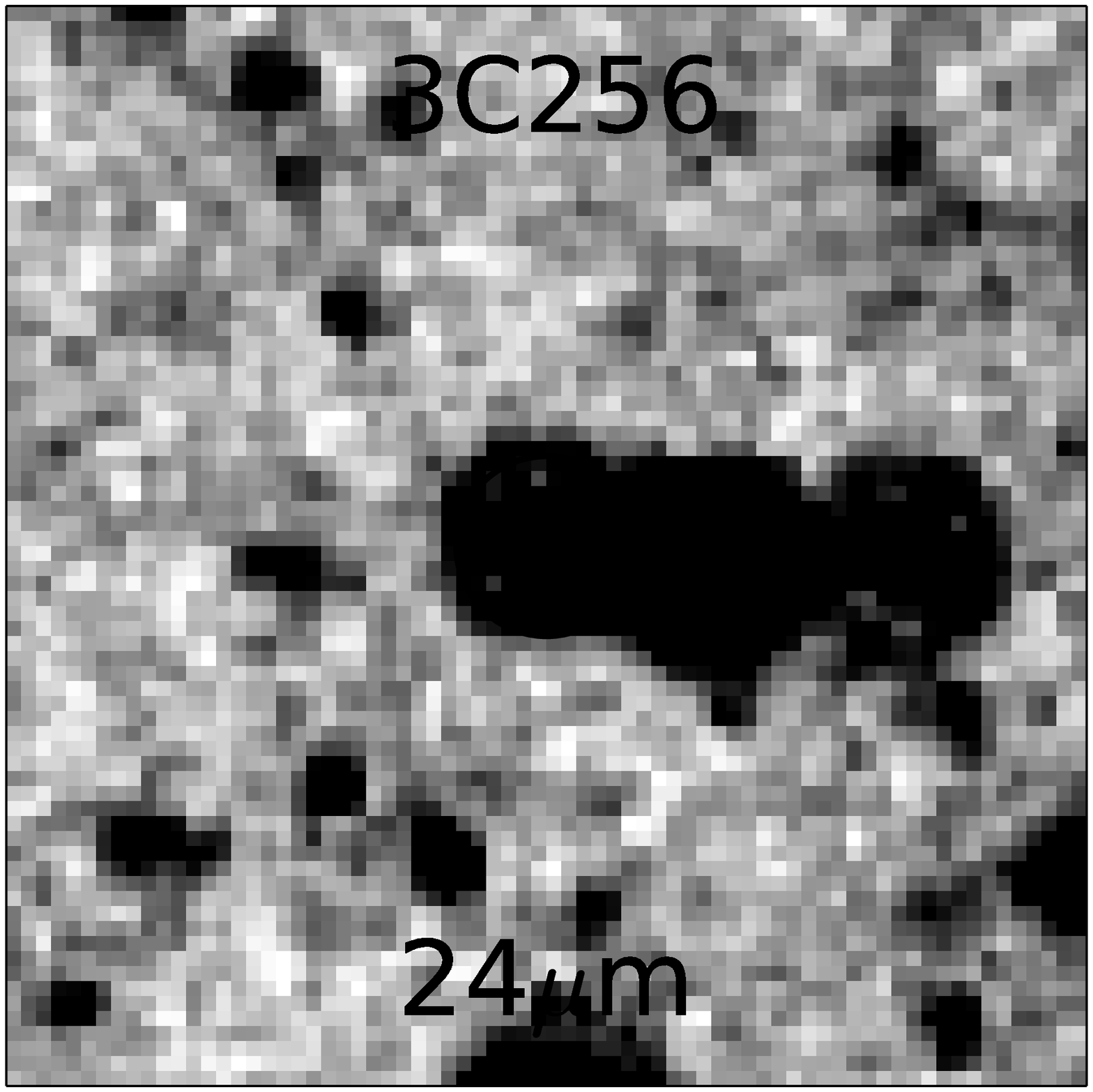}
      \includegraphics[width=1.5cm]{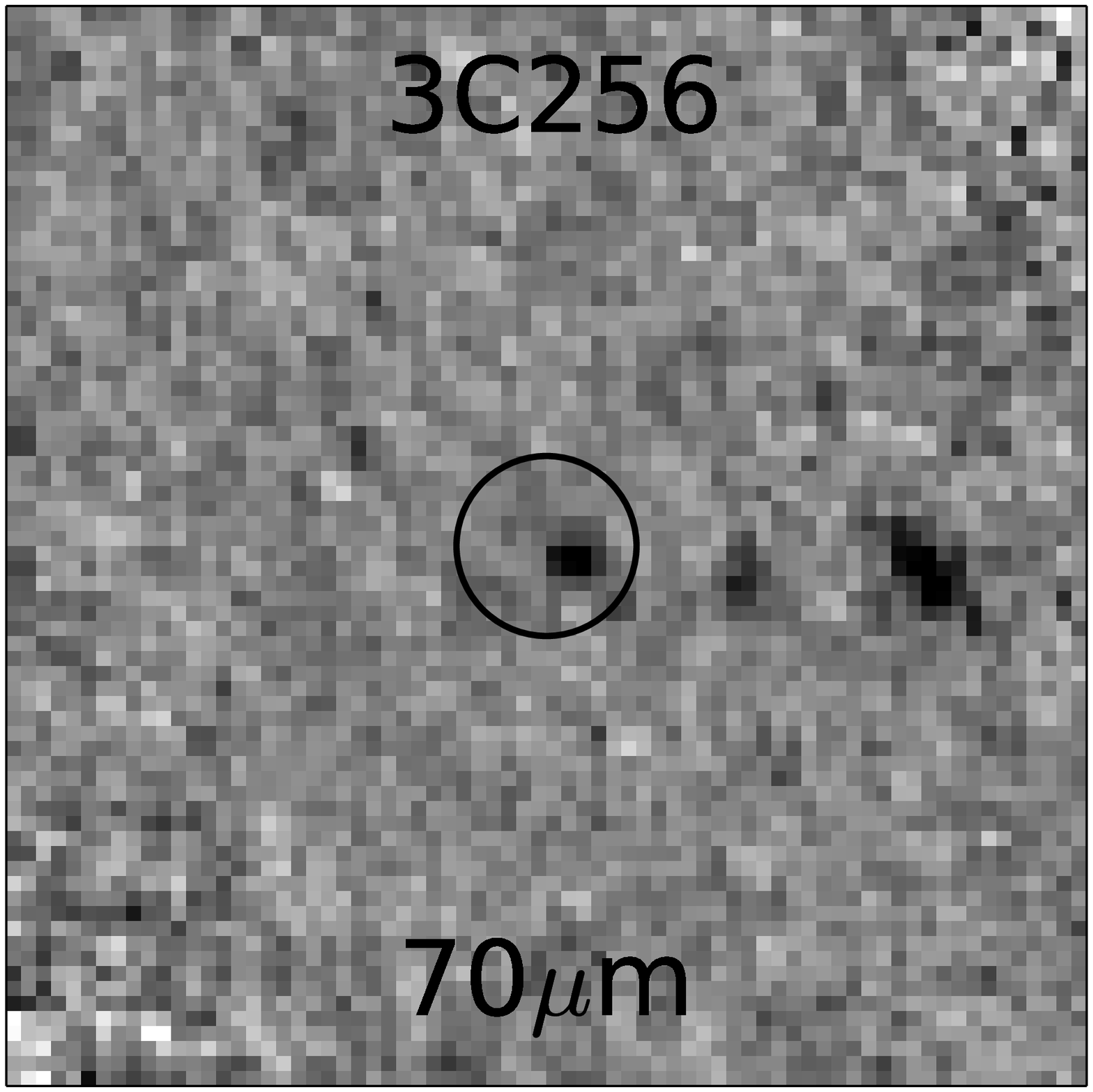}
      \includegraphics[width=1.5cm]{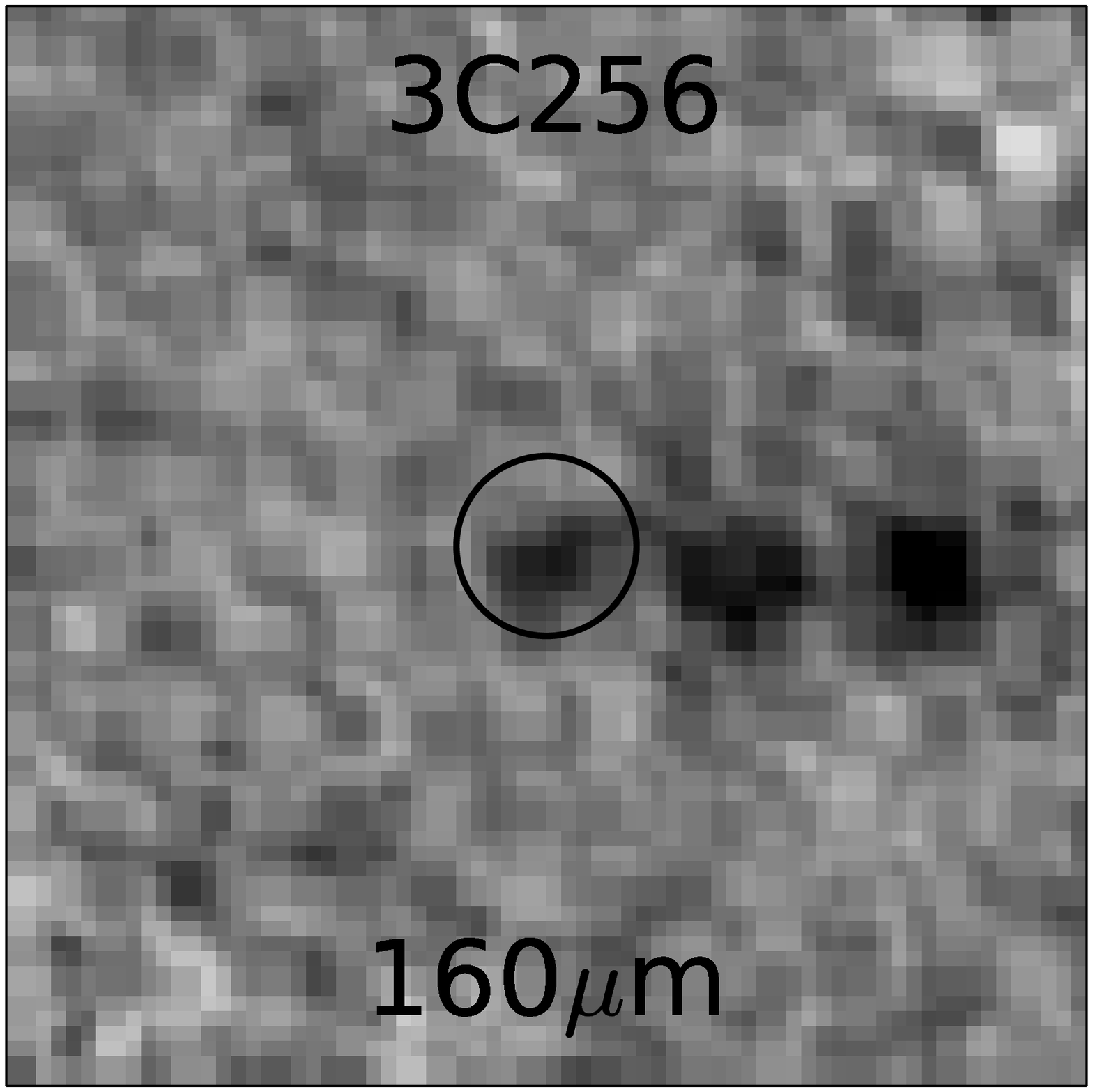}
      \includegraphics[width=1.5cm]{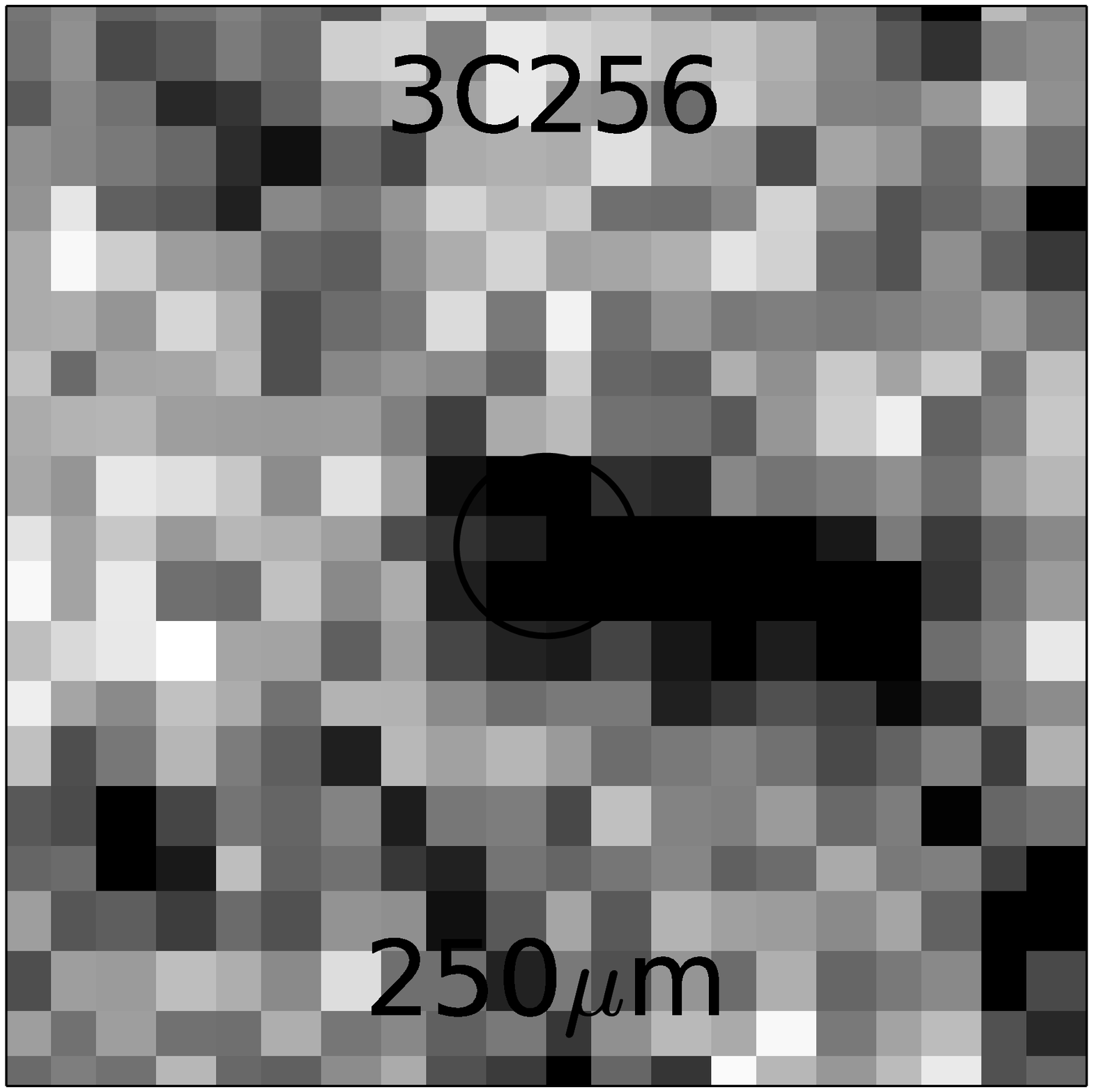}
      \includegraphics[width=1.5cm]{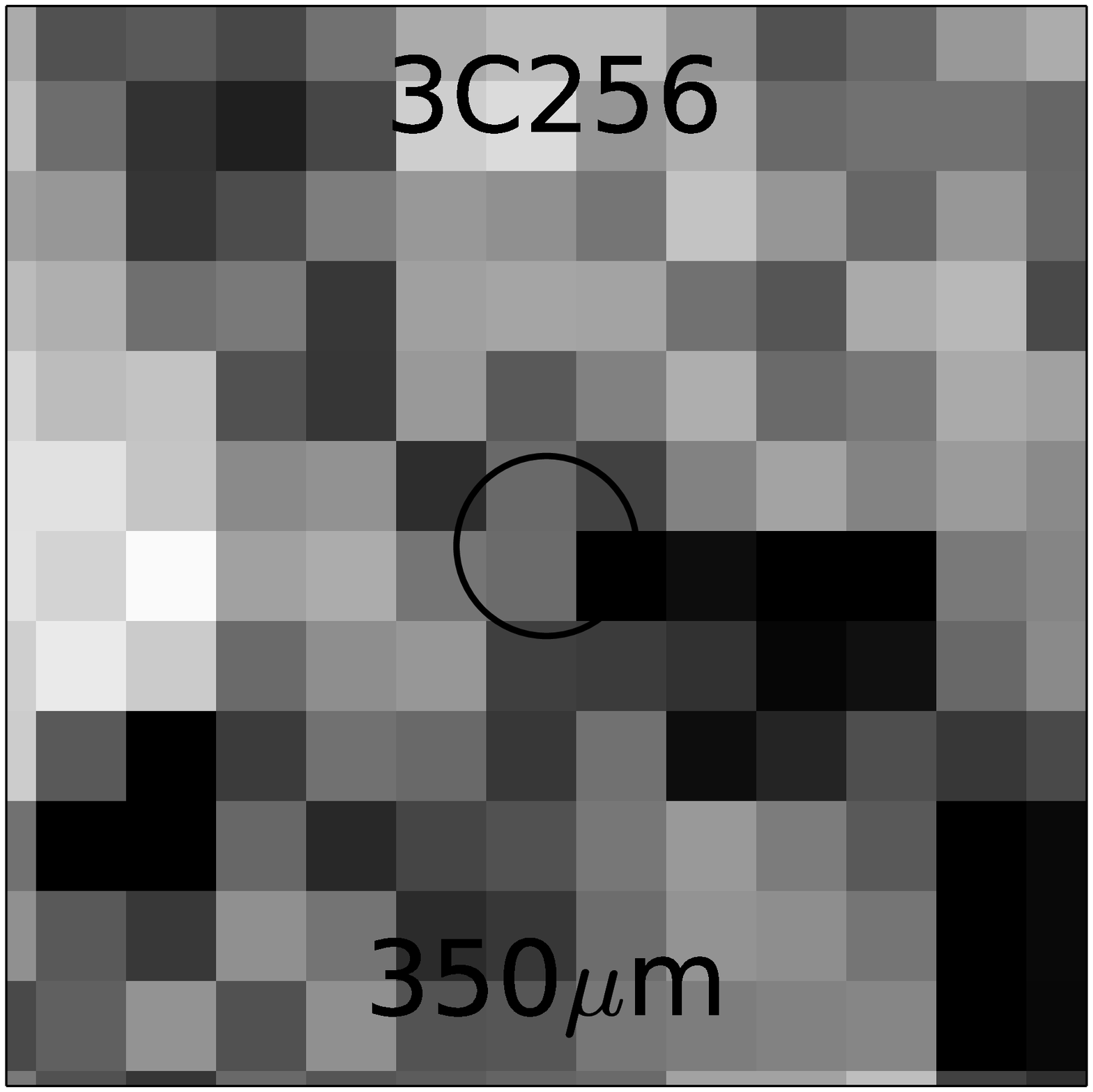}
      \includegraphics[width=1.5cm]{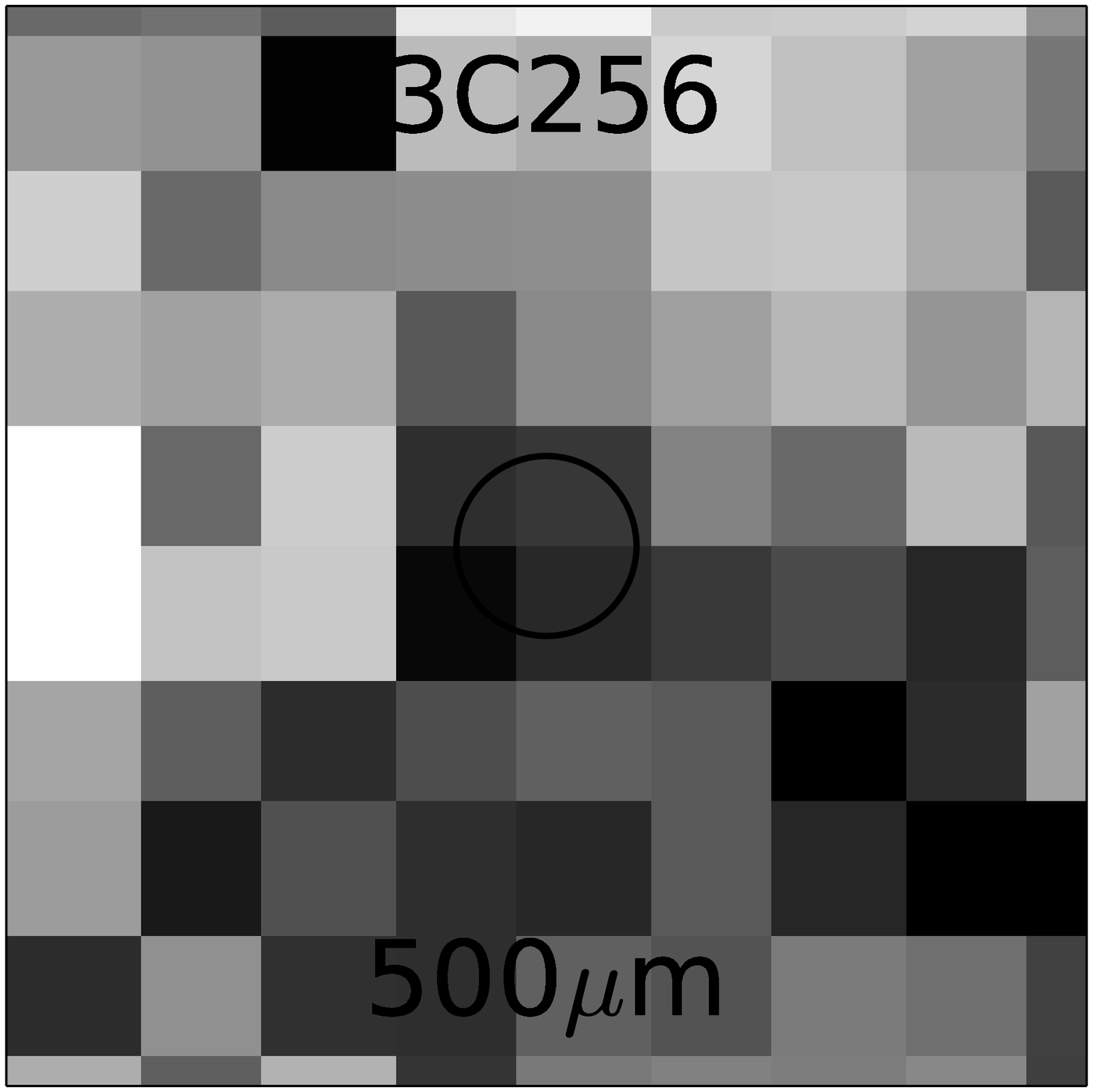}
      \\
      \includegraphics[width=1.5cm]{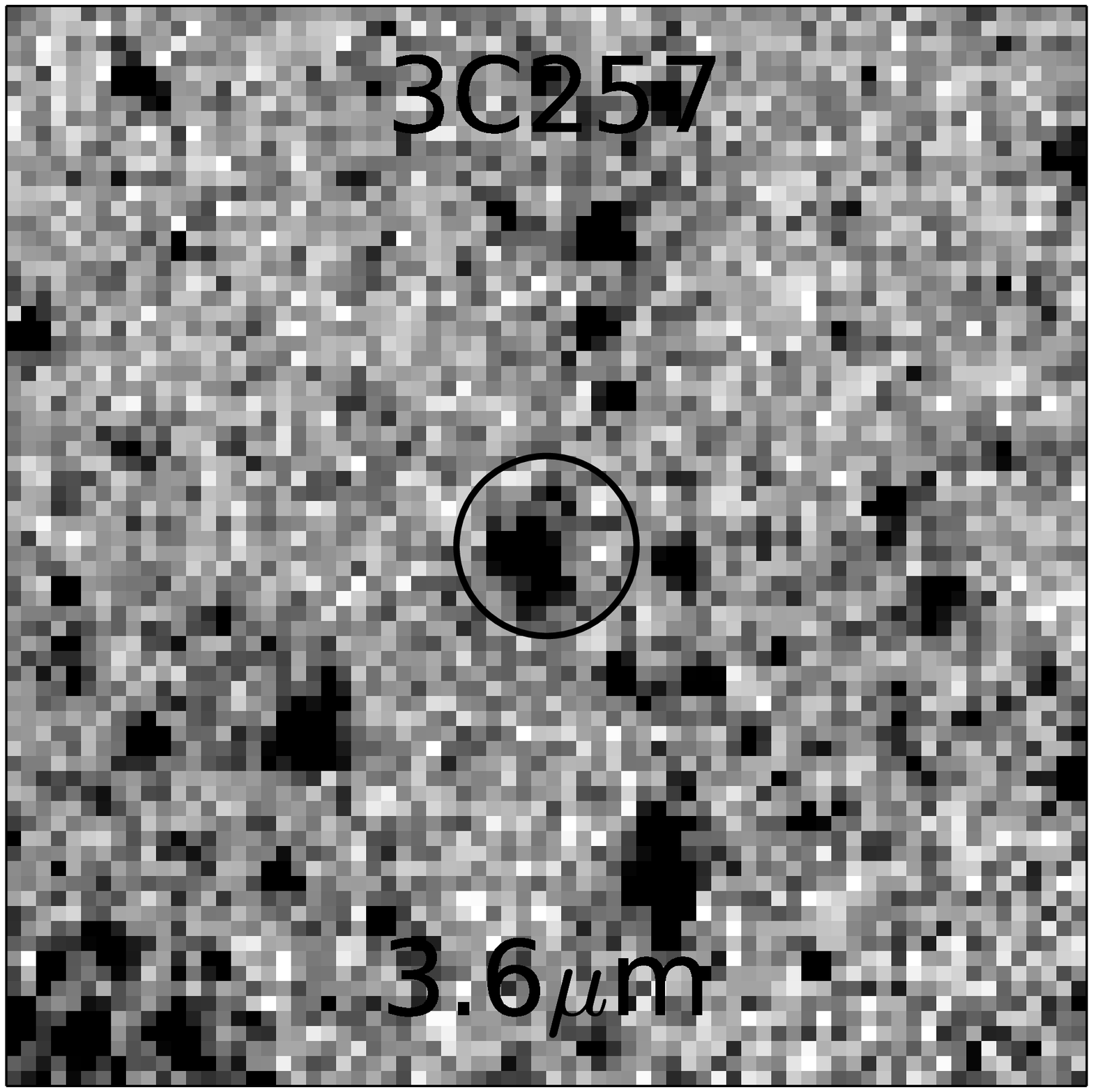}
      \includegraphics[width=1.5cm]{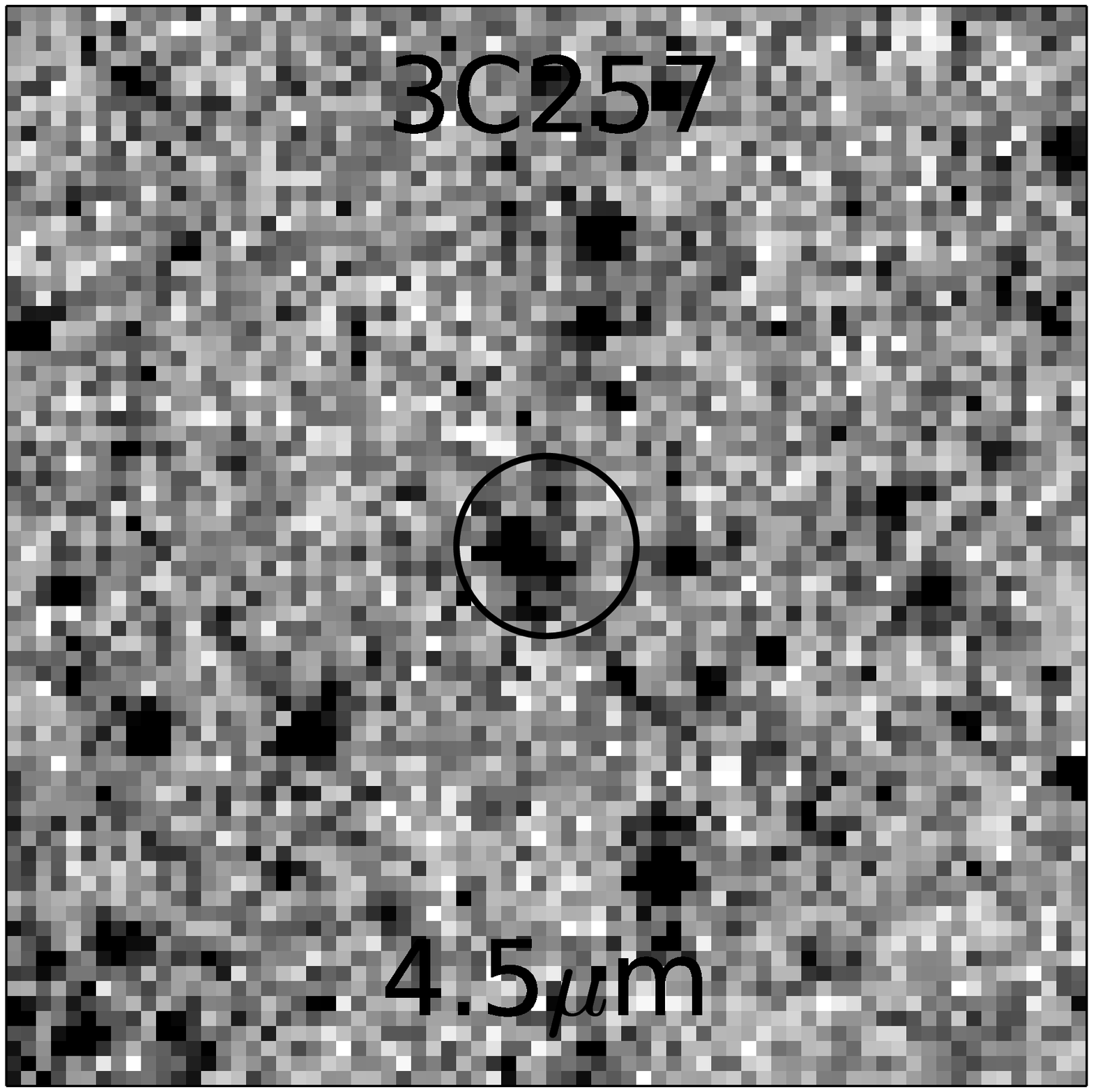}
      \includegraphics[width=1.5cm]{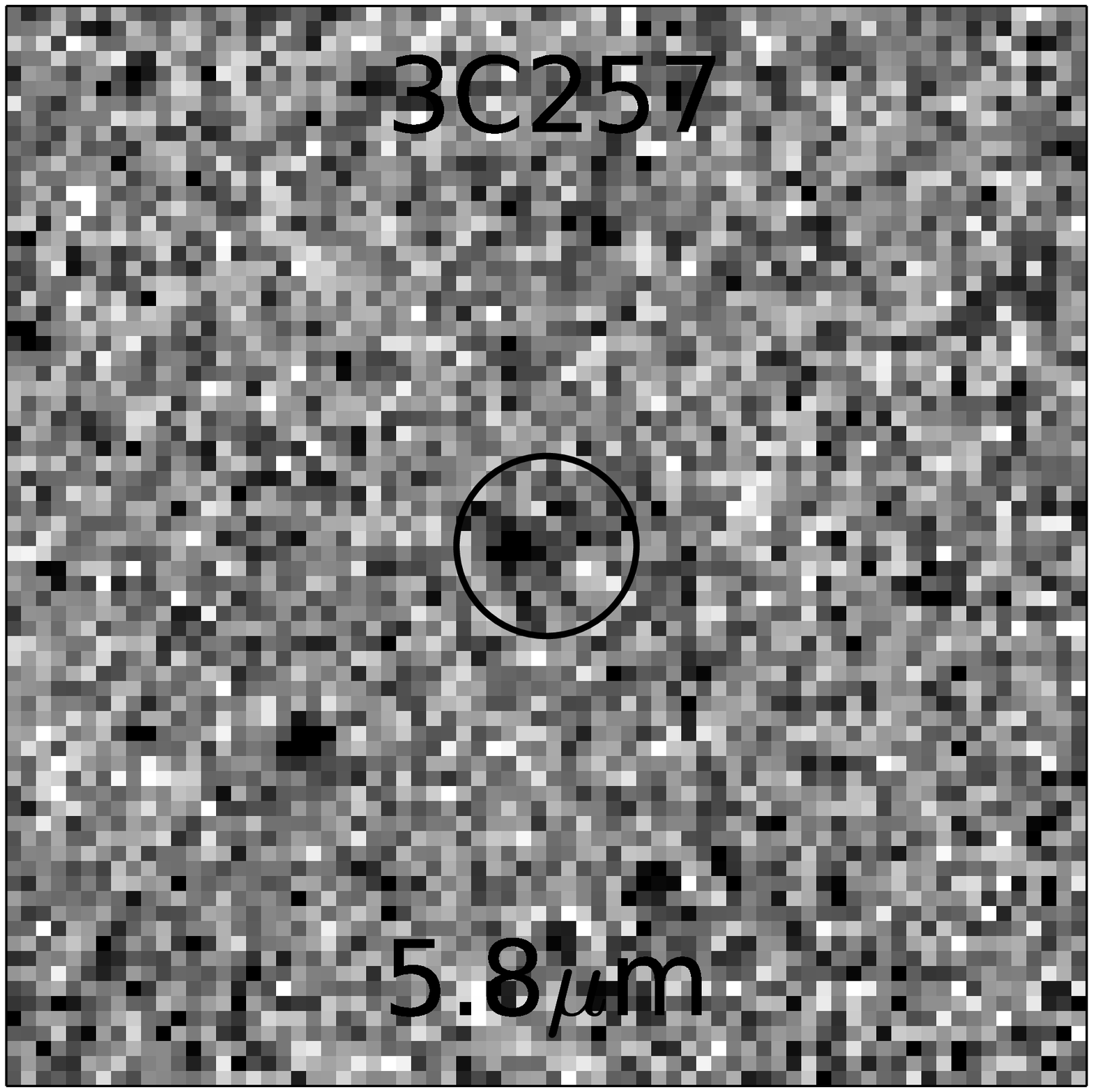}
      \includegraphics[width=1.5cm]{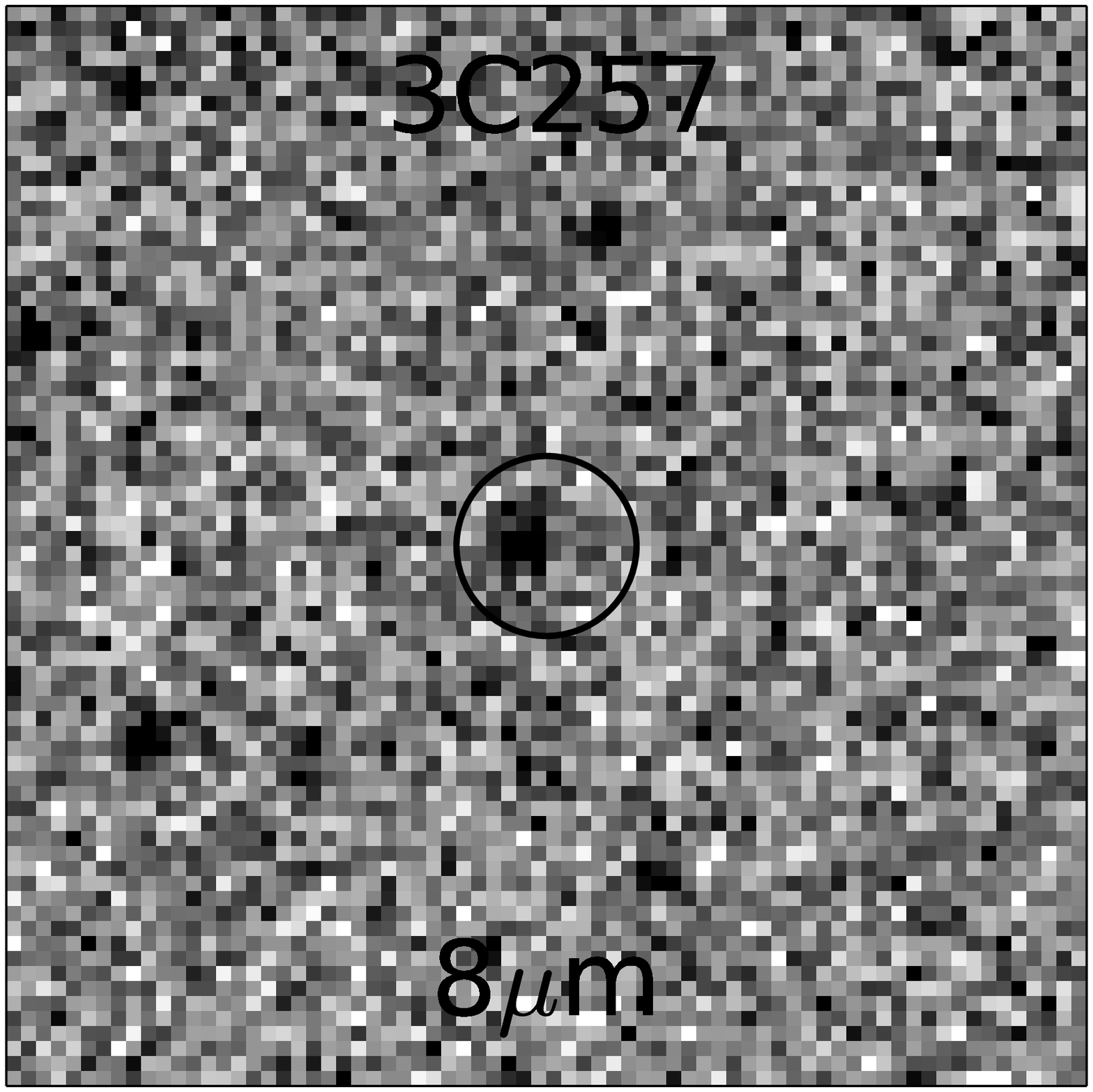}
      \includegraphics[width=1.5cm]{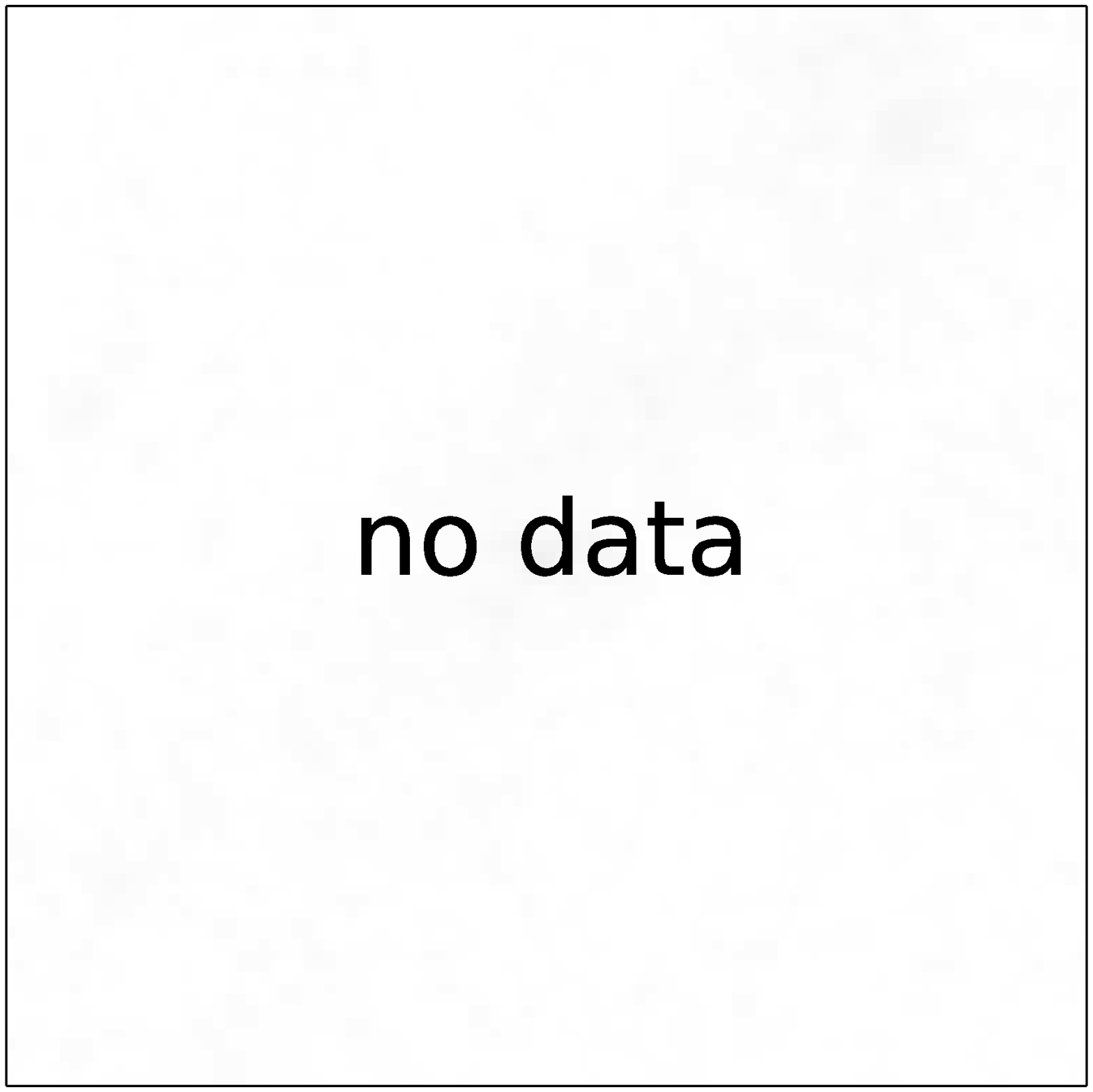}
      \includegraphics[width=1.5cm]{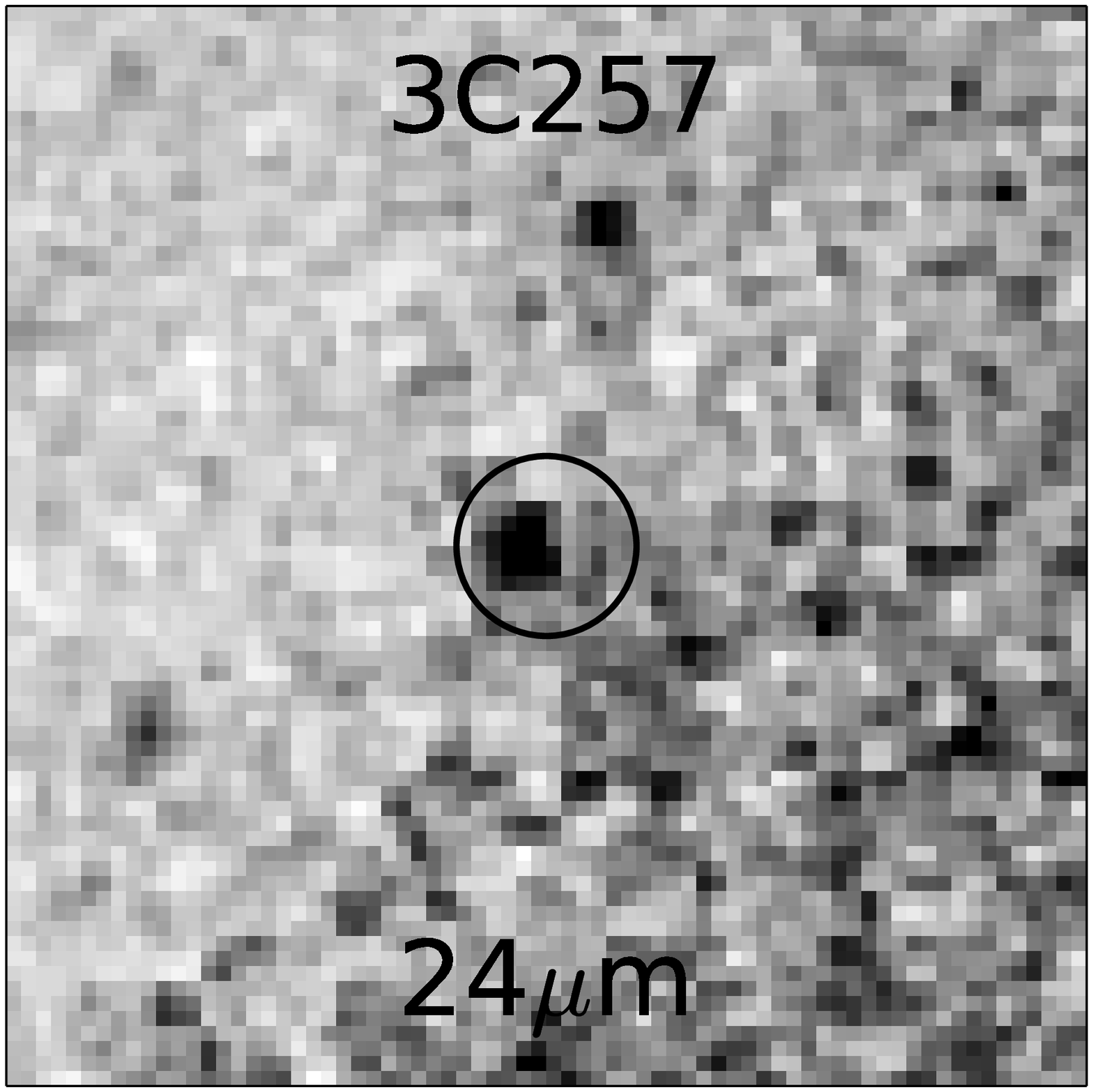}
      \includegraphics[width=1.5cm]{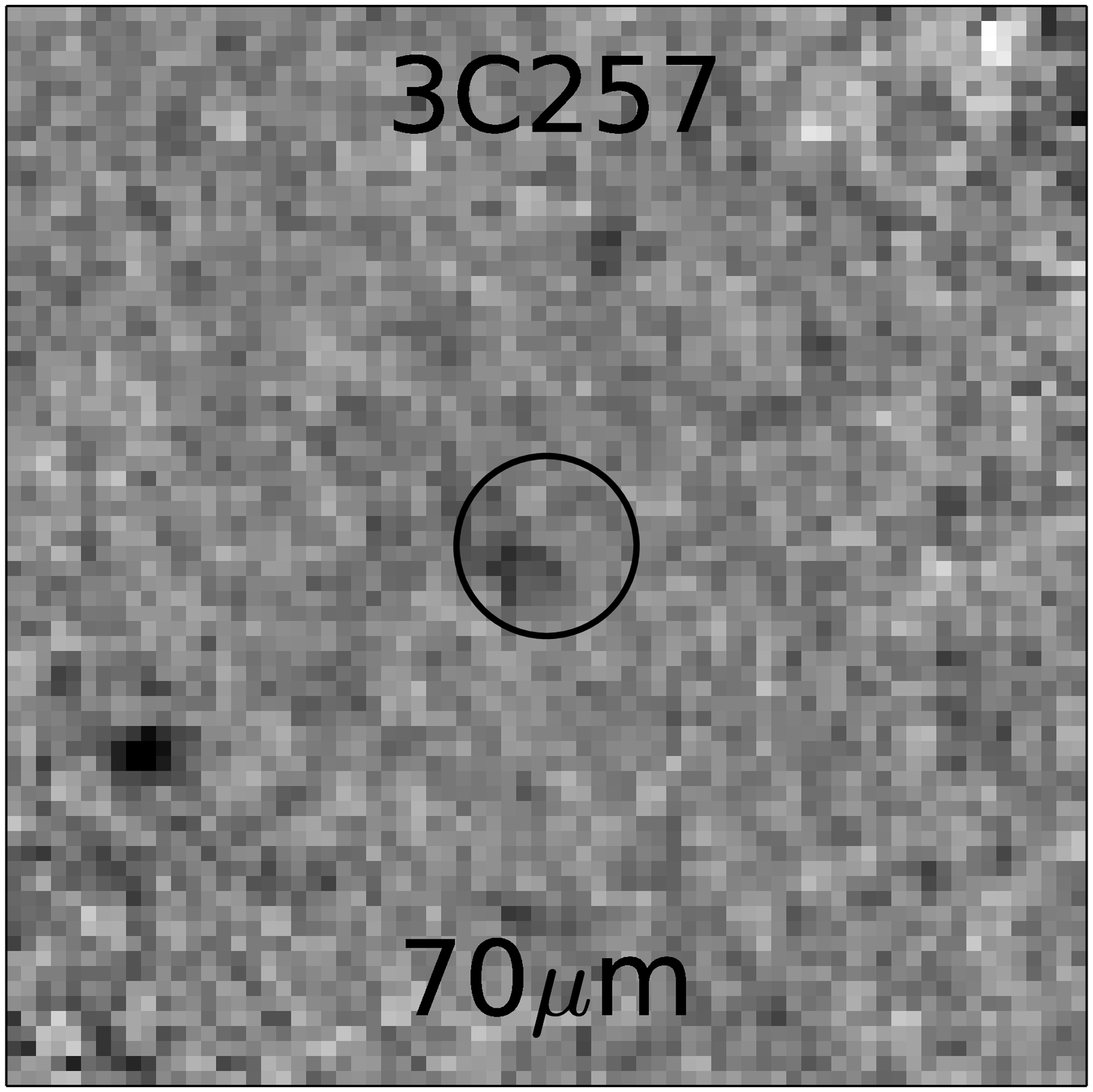}
      \includegraphics[width=1.5cm]{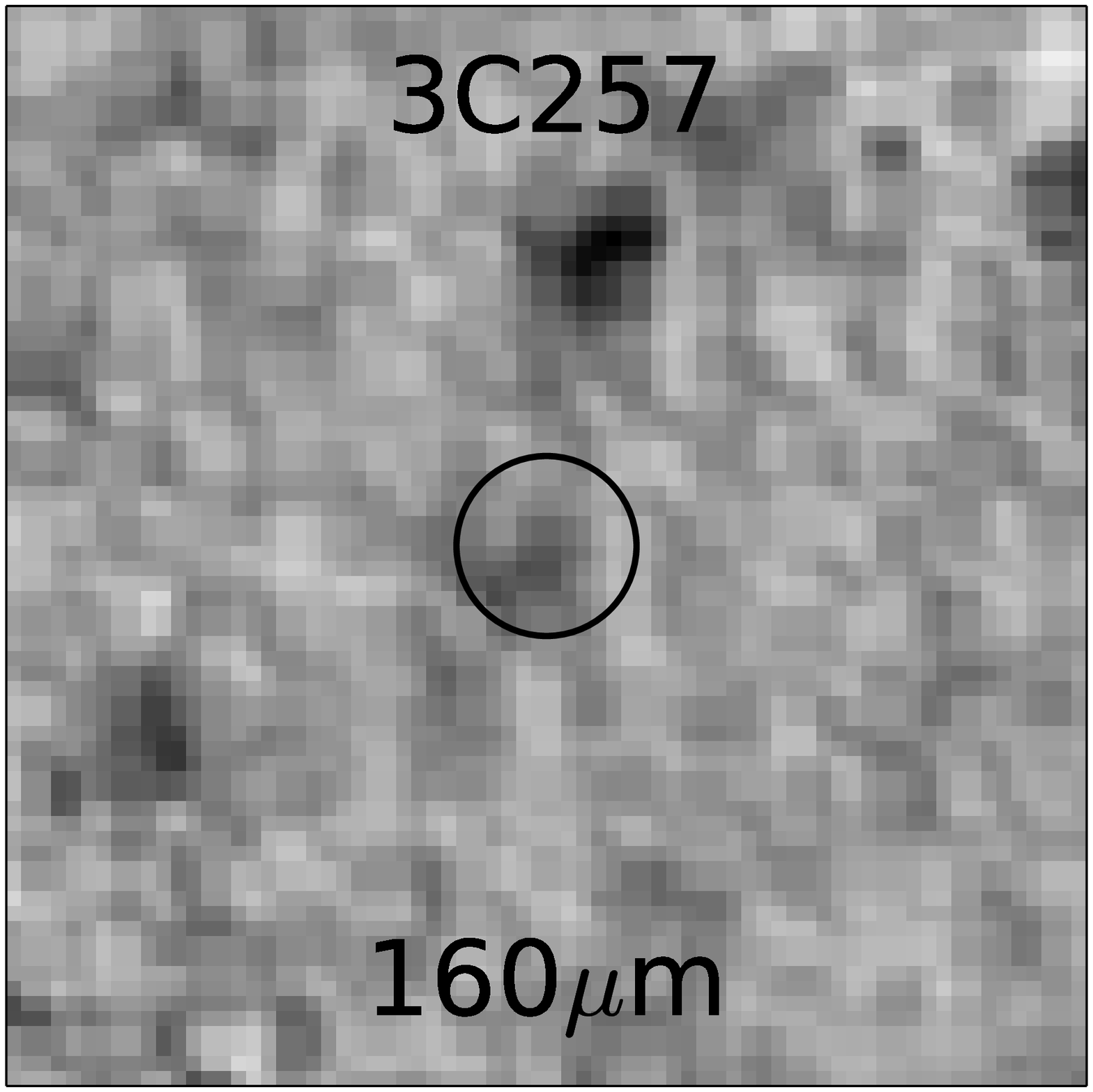}
      \includegraphics[width=1.5cm]{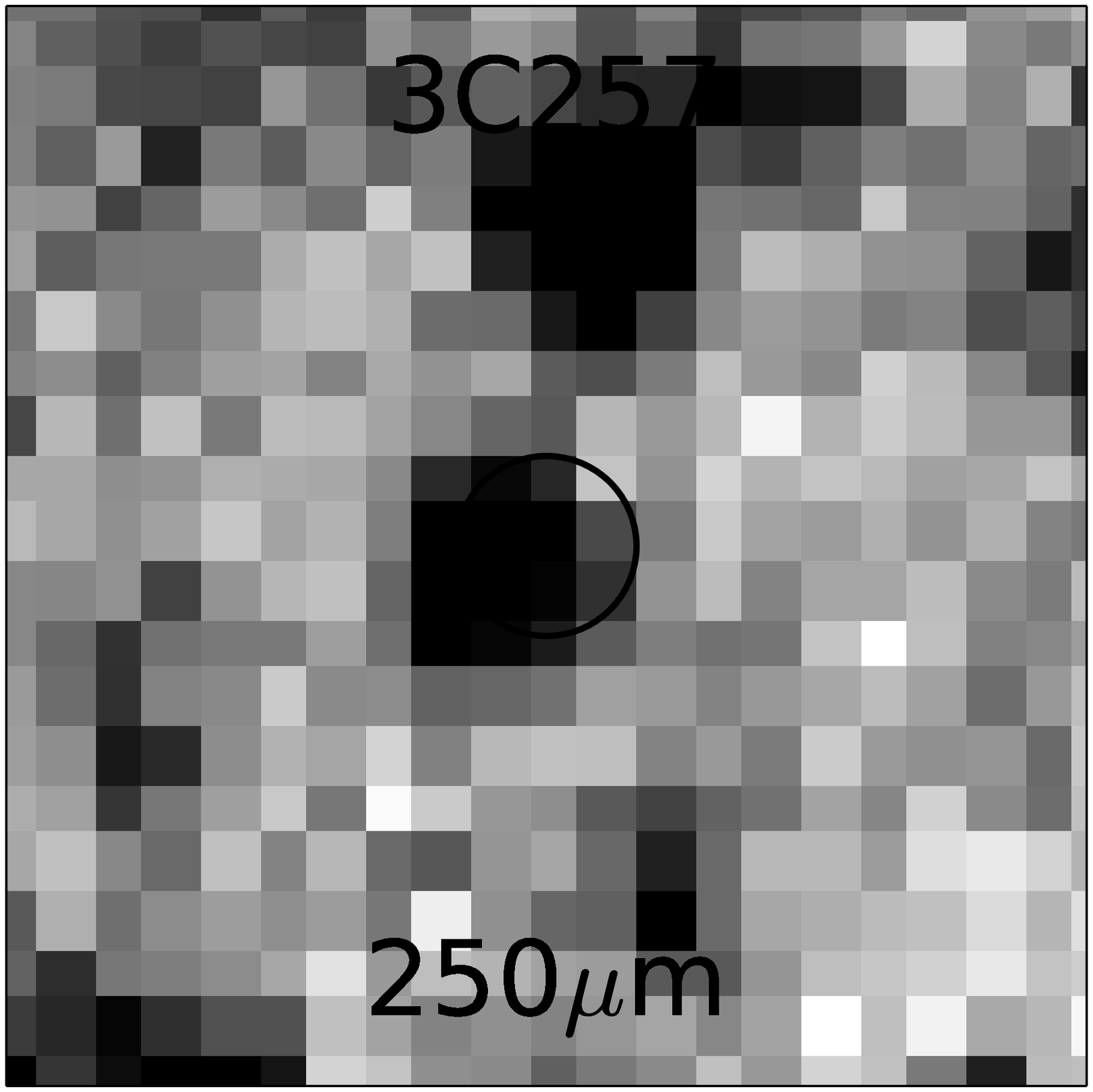}
      \includegraphics[width=1.5cm]{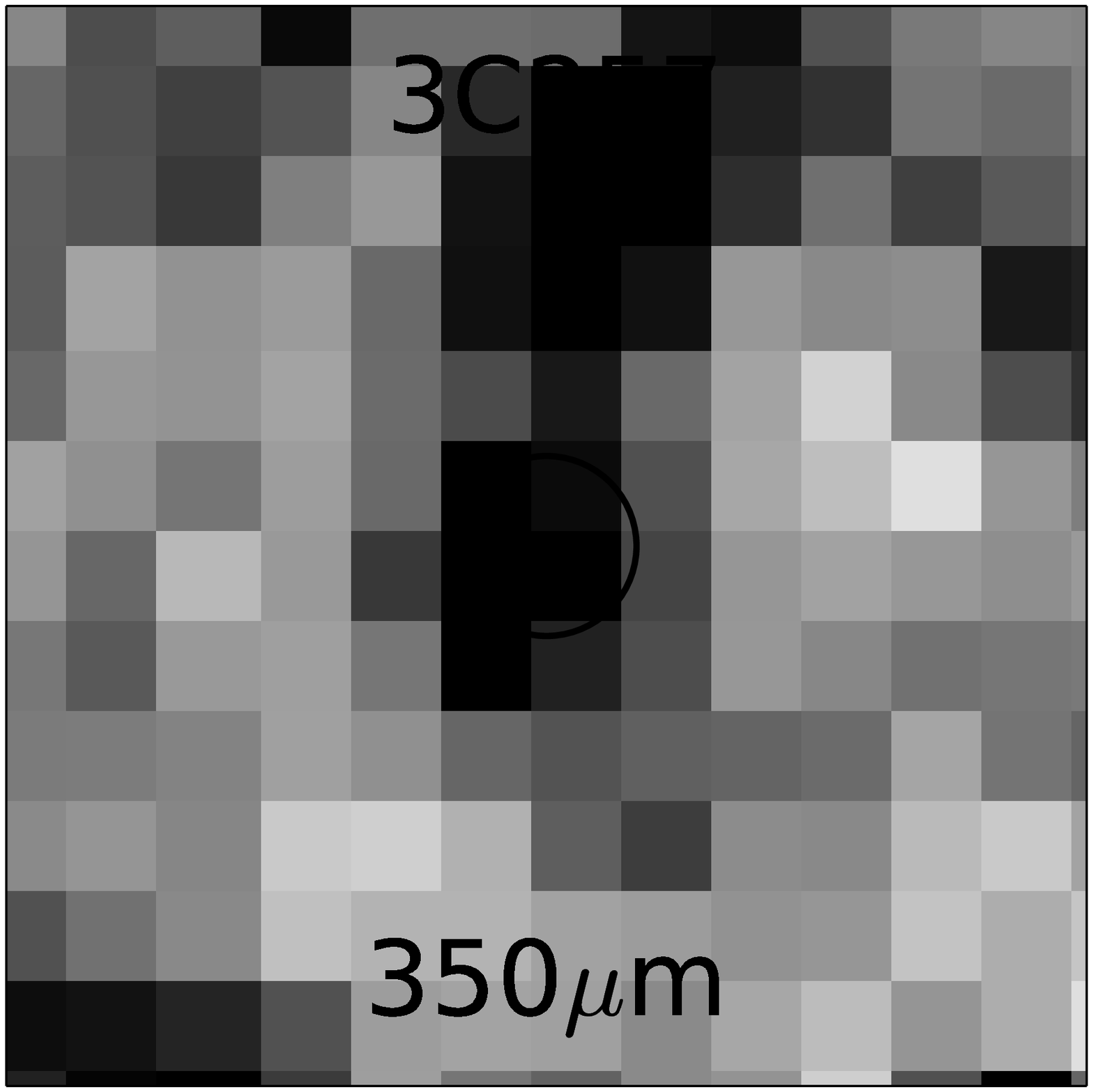}
      \includegraphics[width=1.5cm]{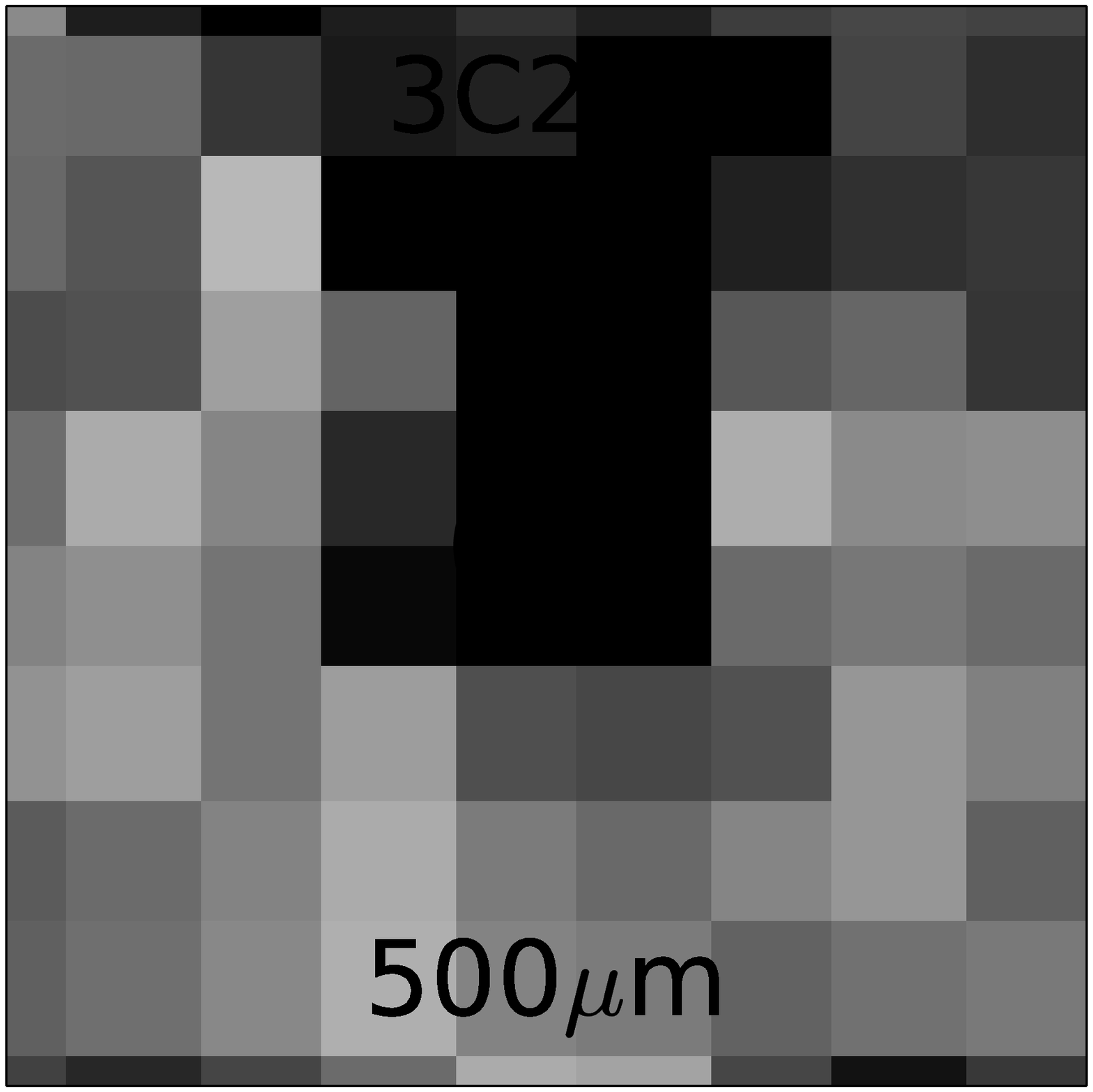}
      \\
      \includegraphics[width=1.5cm]{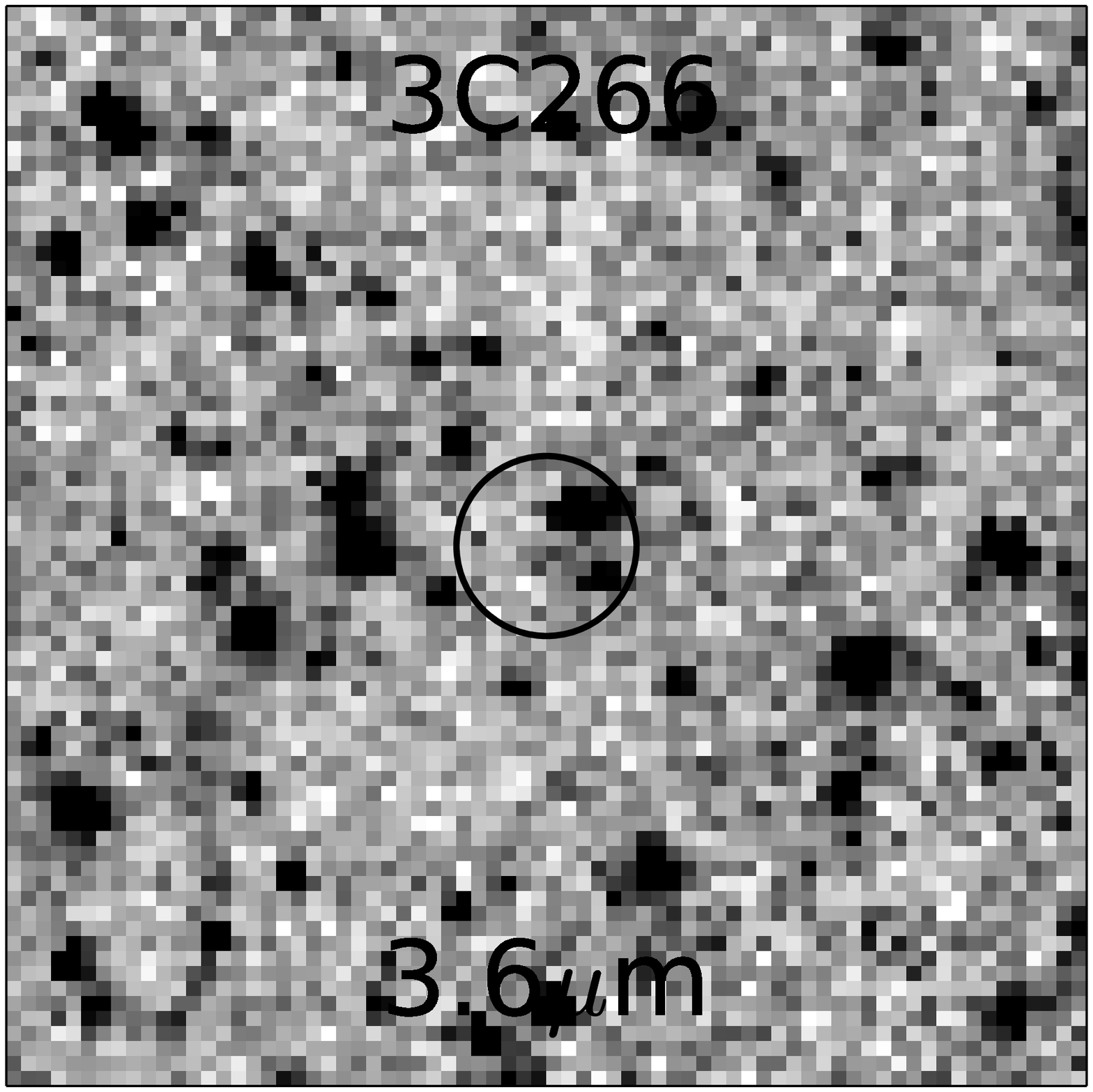}
      \includegraphics[width=1.5cm]{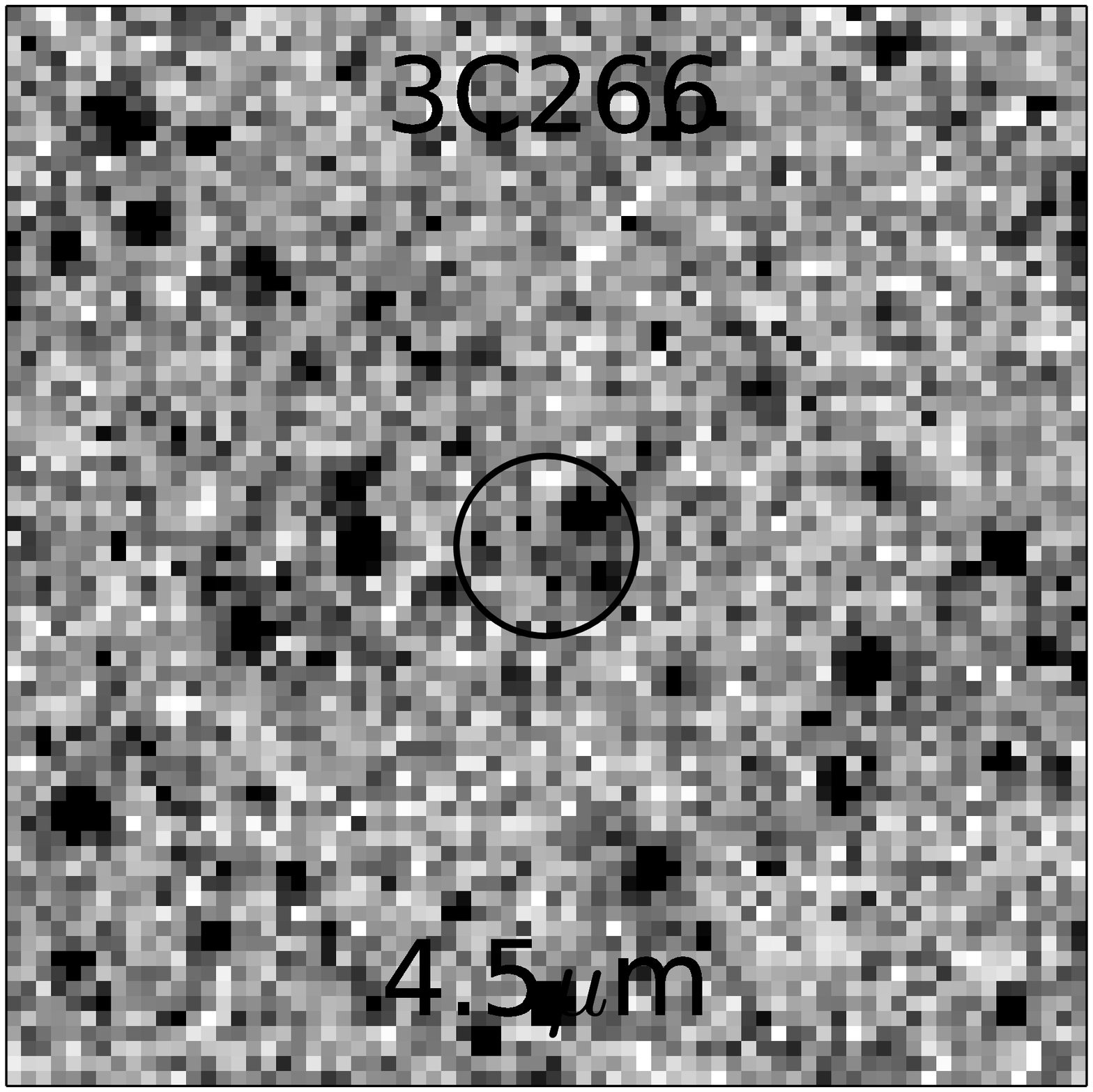}
      \includegraphics[width=1.5cm]{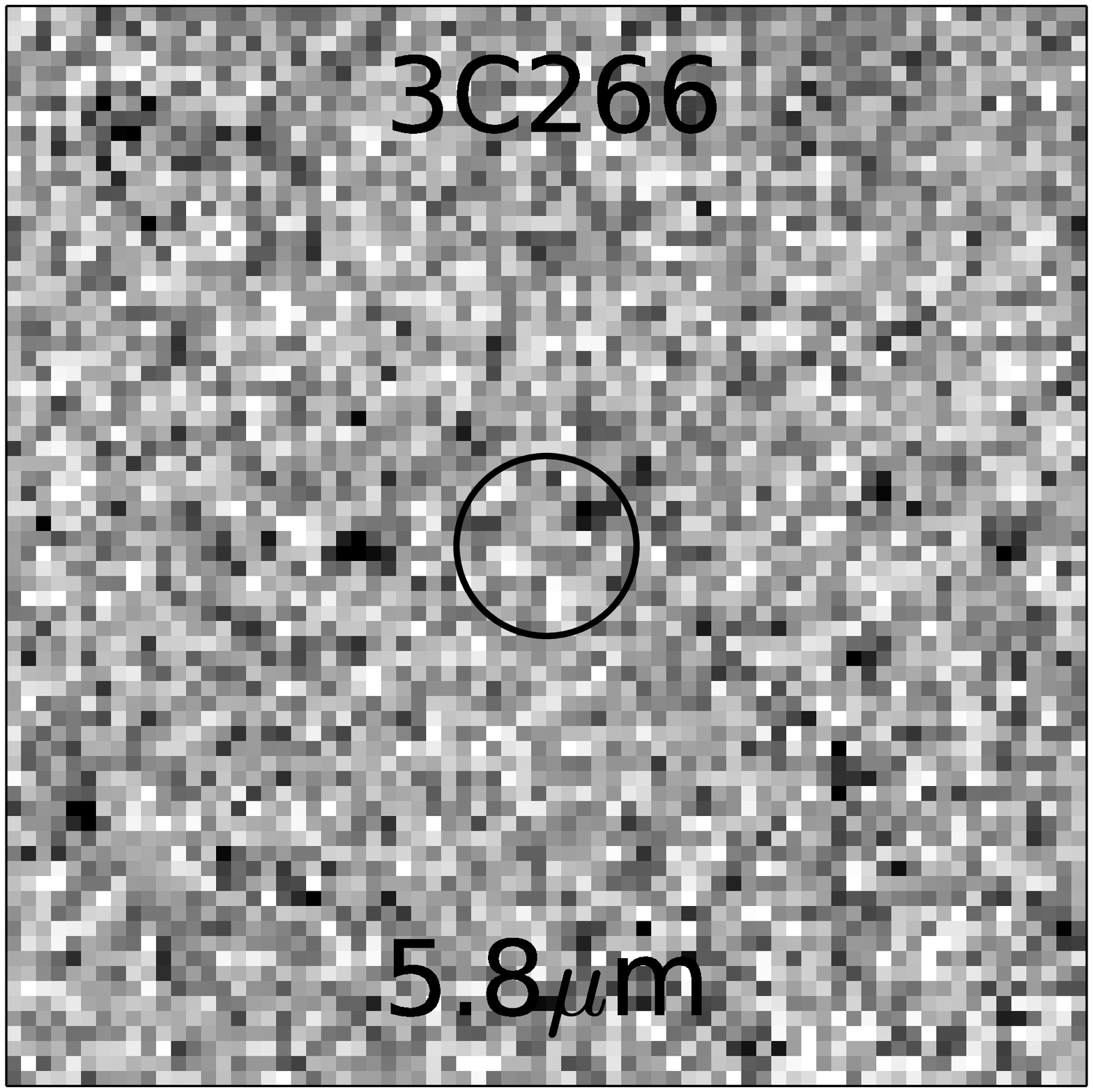}
      \includegraphics[width=1.5cm]{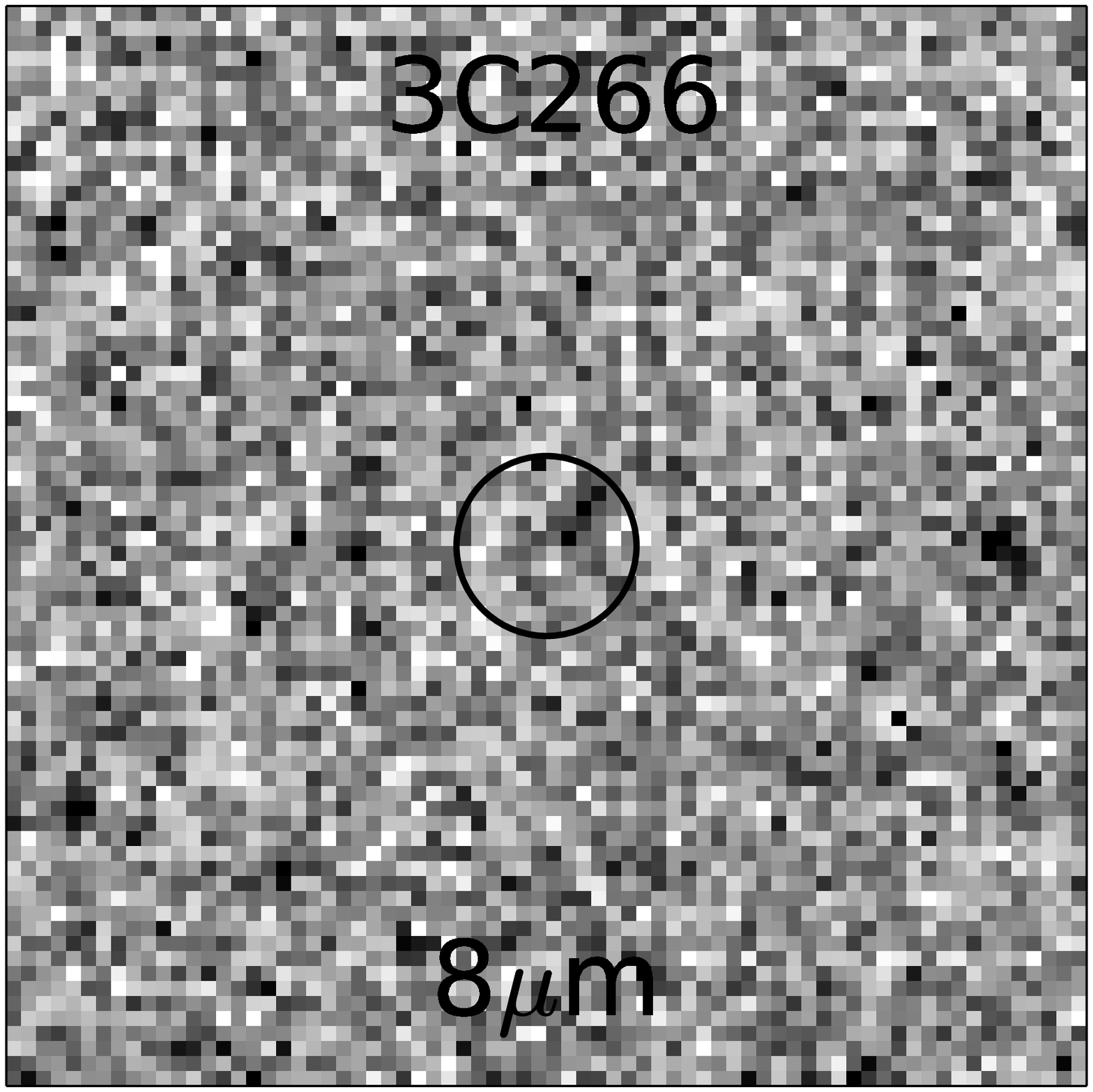}
      \includegraphics[width=1.5cm]{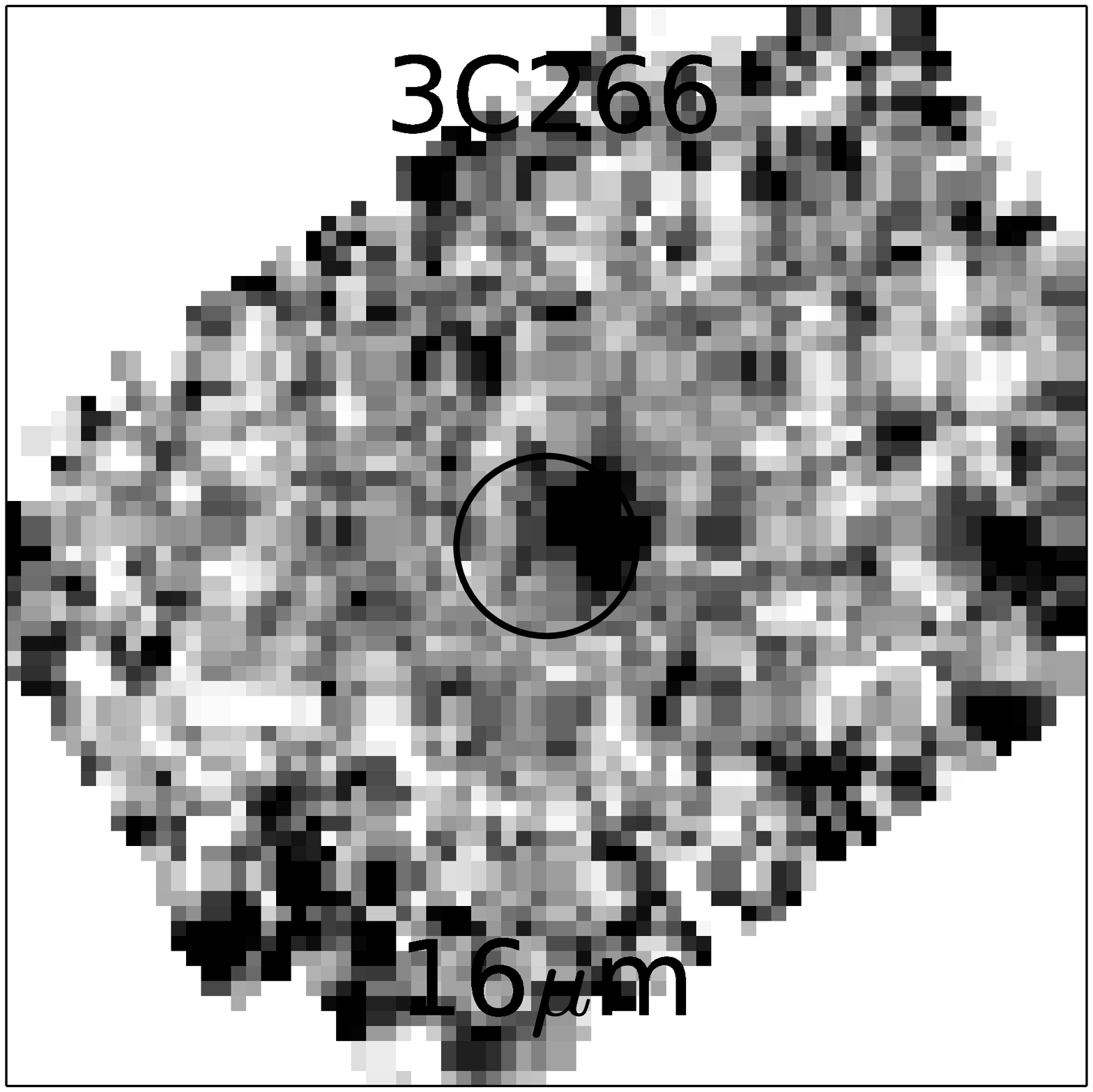}
      \includegraphics[width=1.5cm]{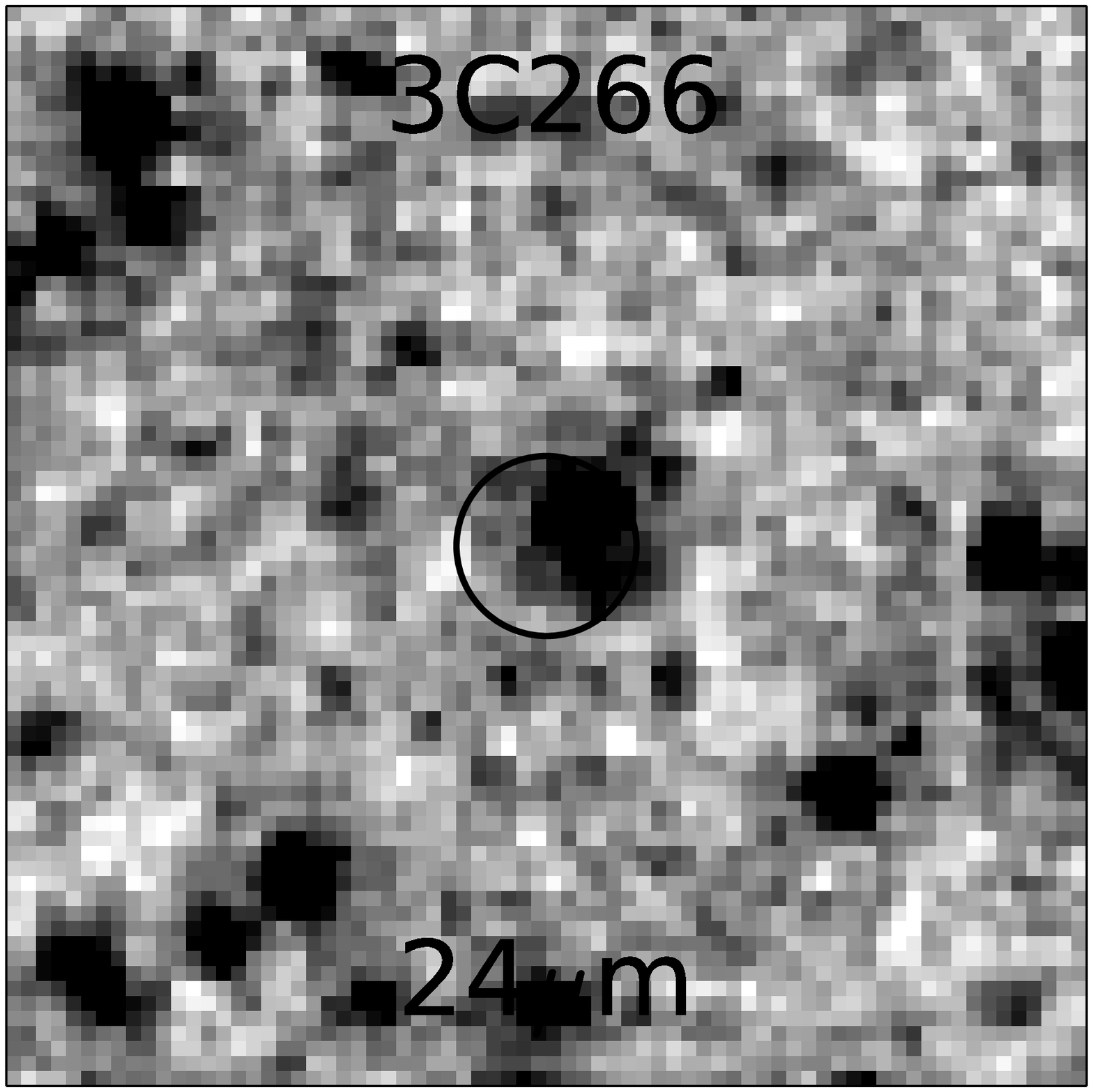}
      \includegraphics[width=1.5cm]{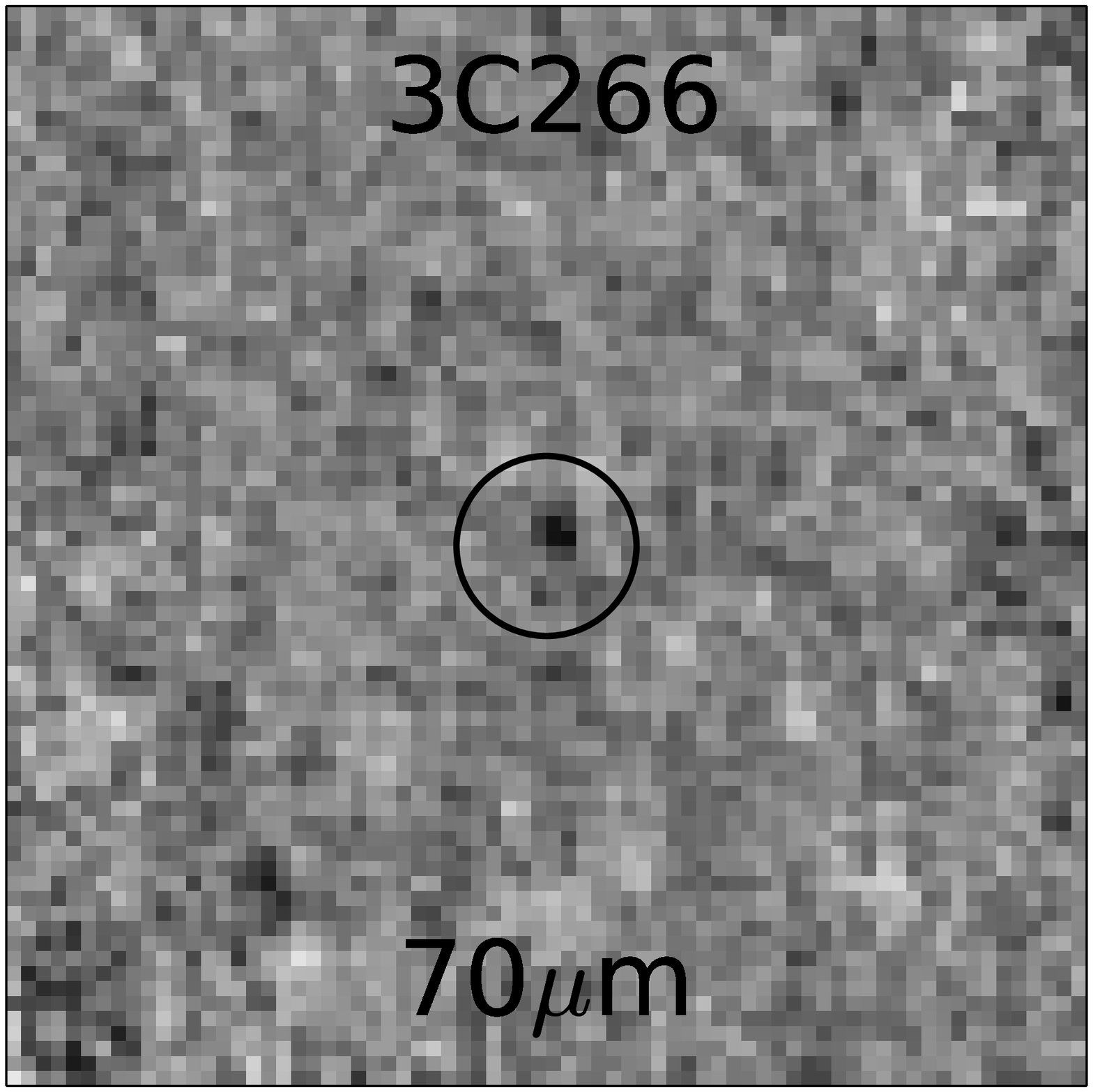}
      \includegraphics[width=1.5cm]{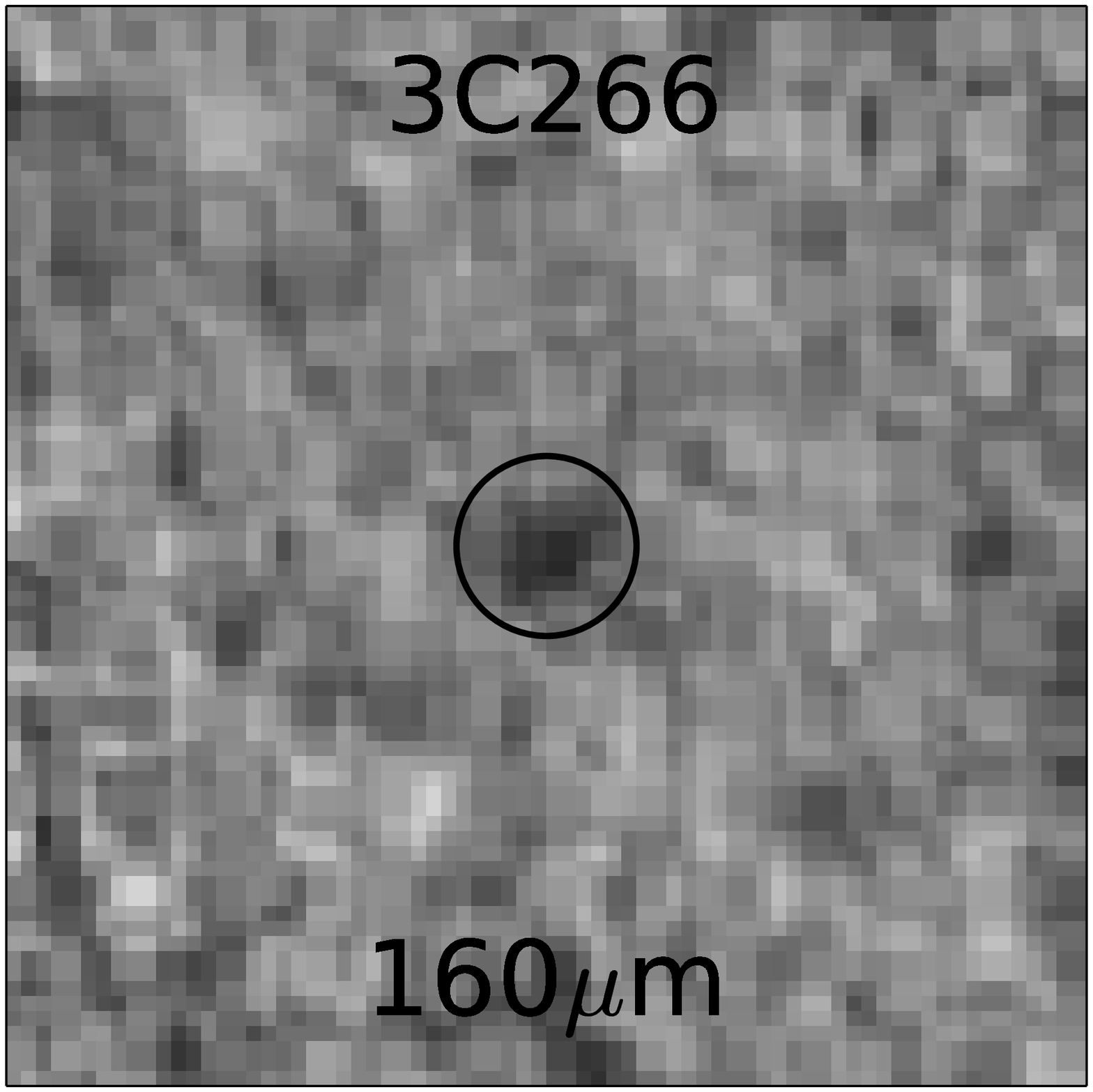}
      \includegraphics[width=1.5cm]{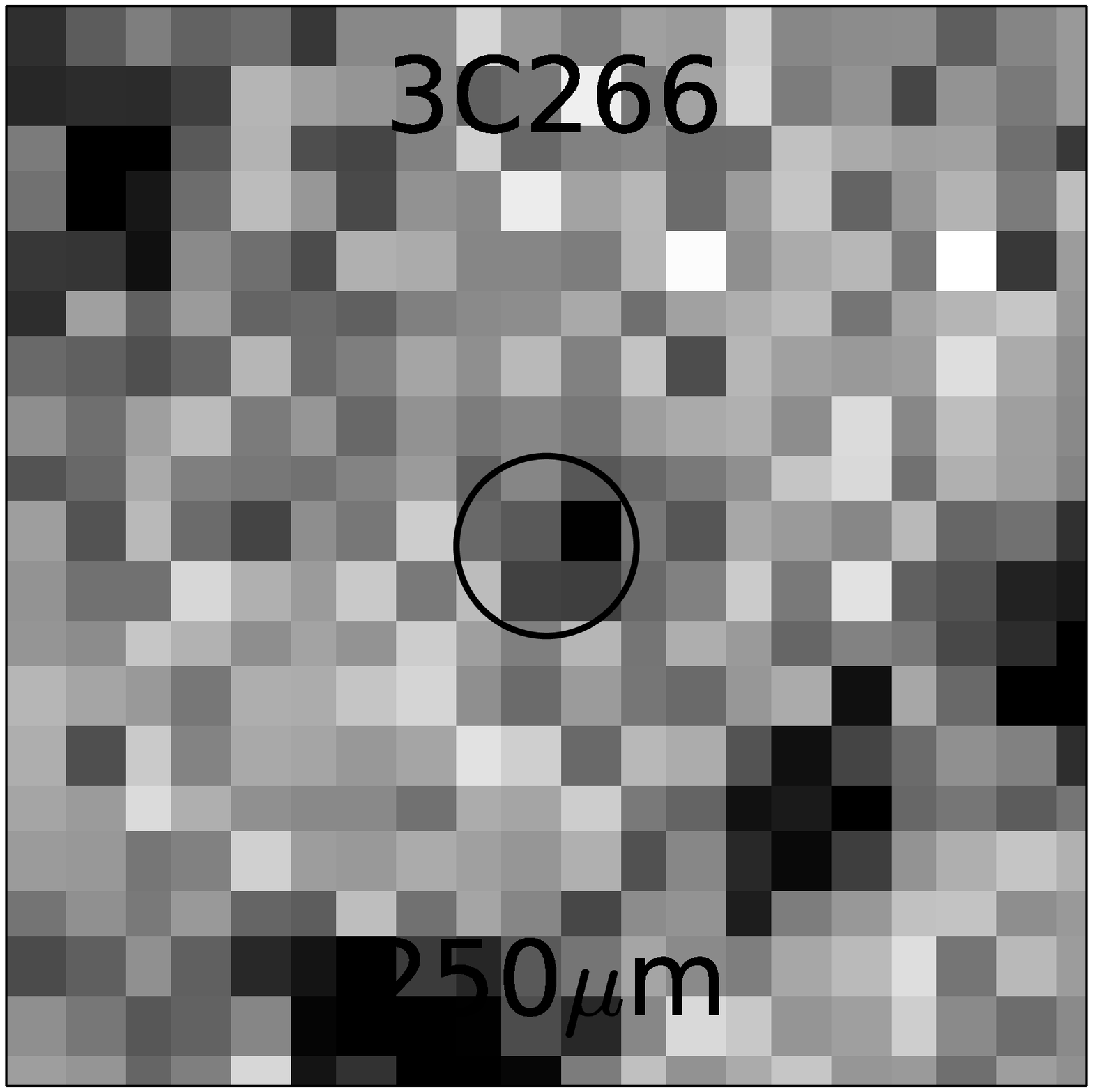}
      \includegraphics[width=1.5cm]{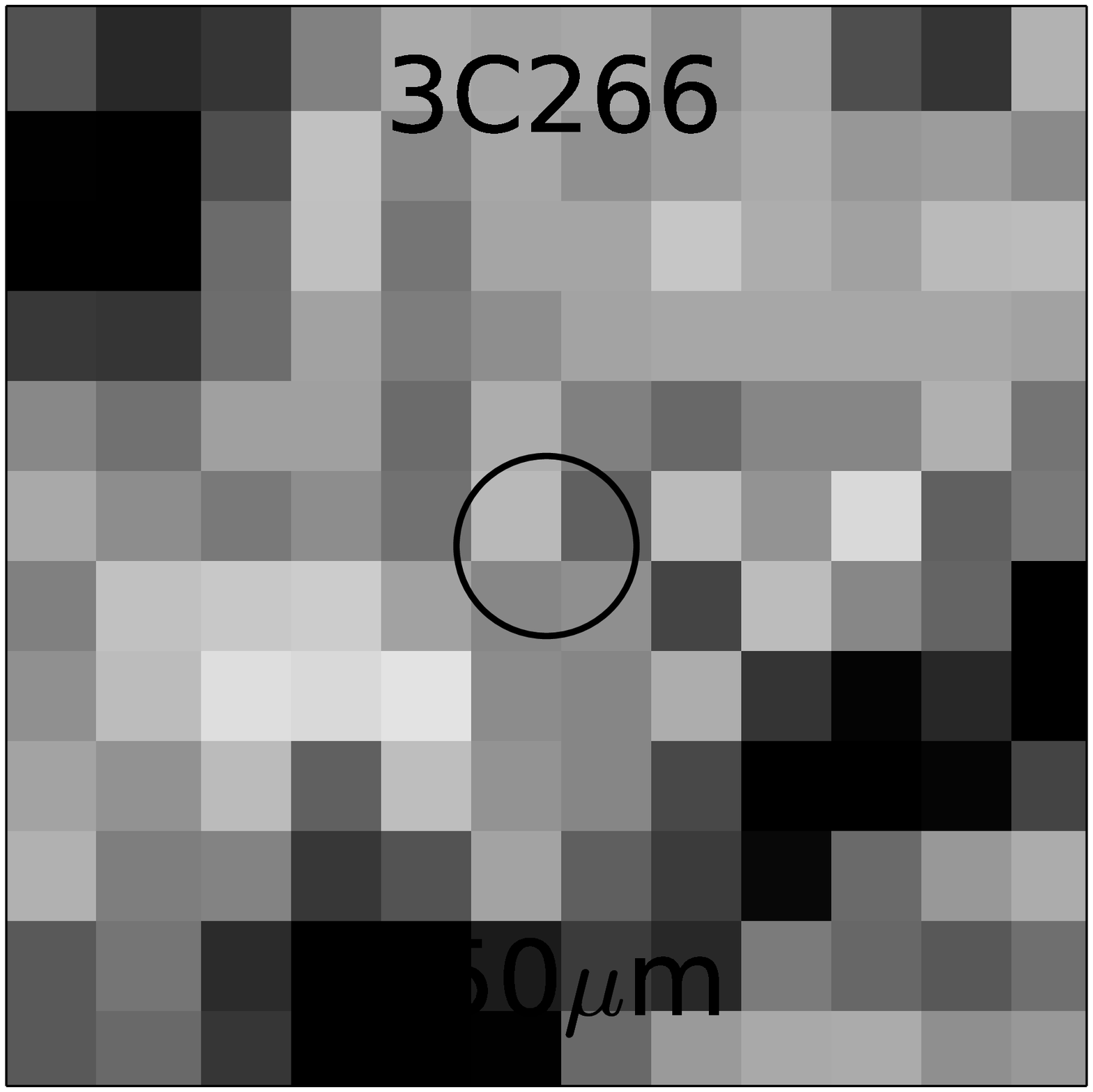}
      \includegraphics[width=1.5cm]{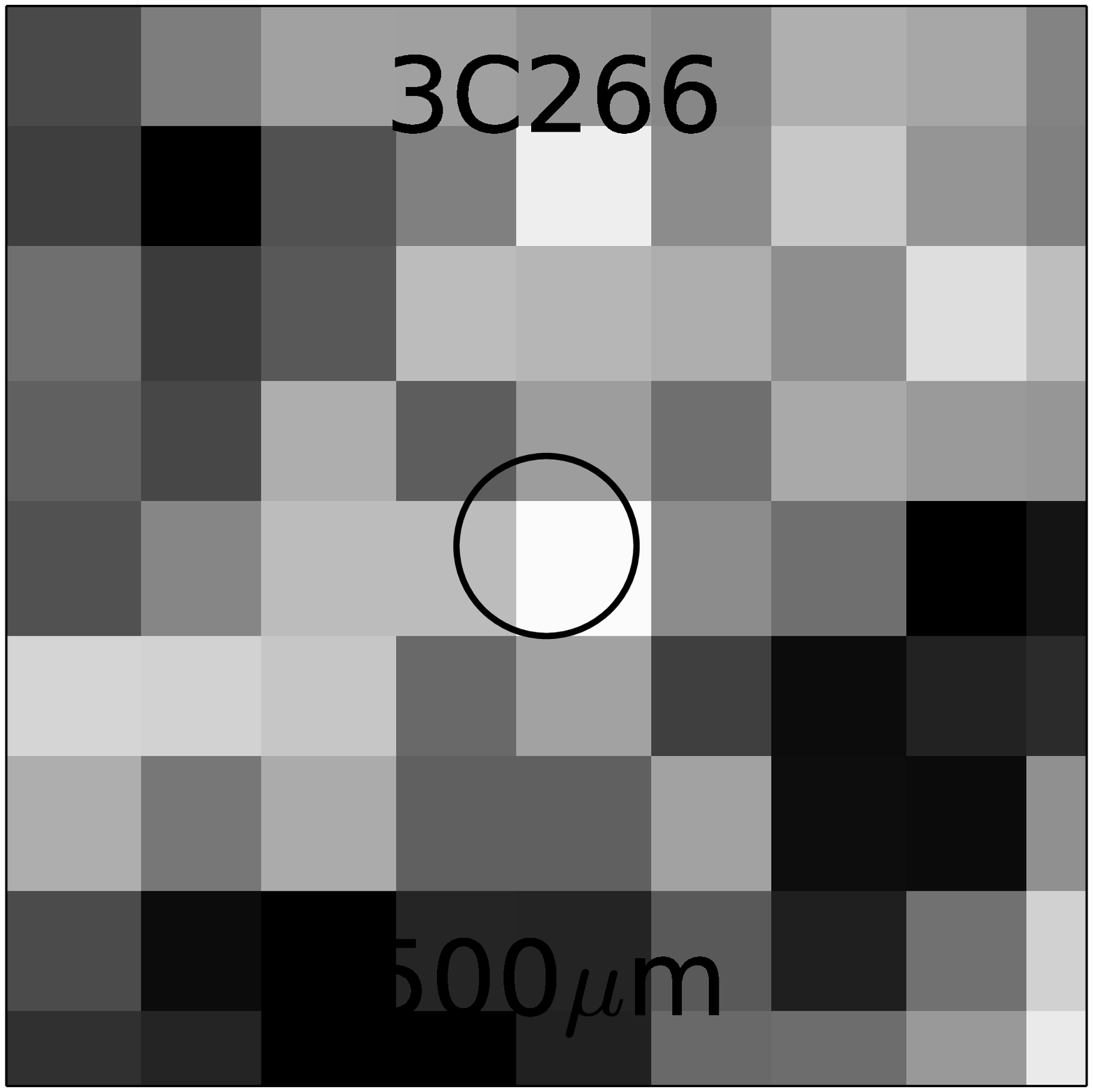}
      \\
      \includegraphics[width=1.5cm]{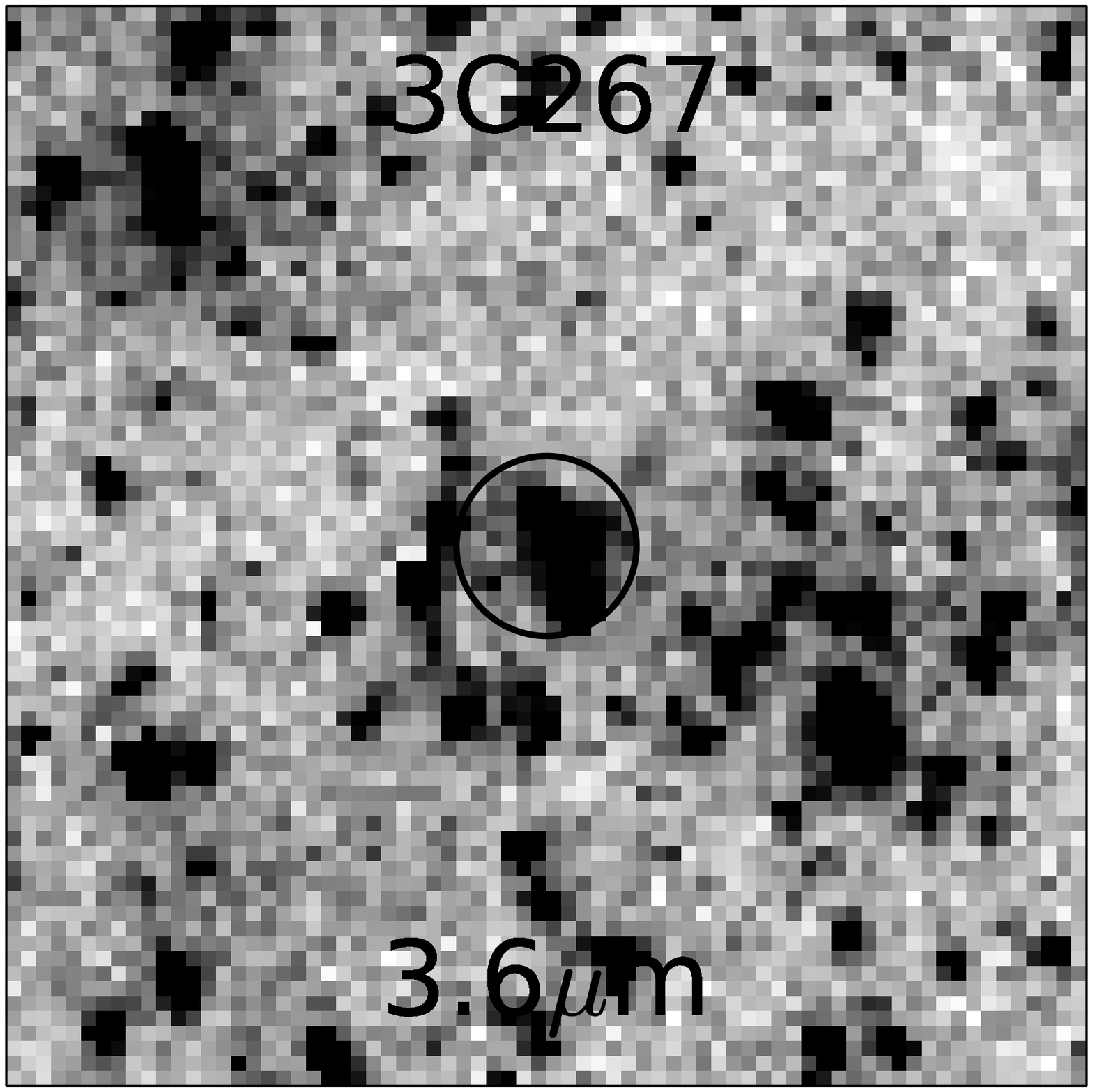}
      \includegraphics[width=1.5cm]{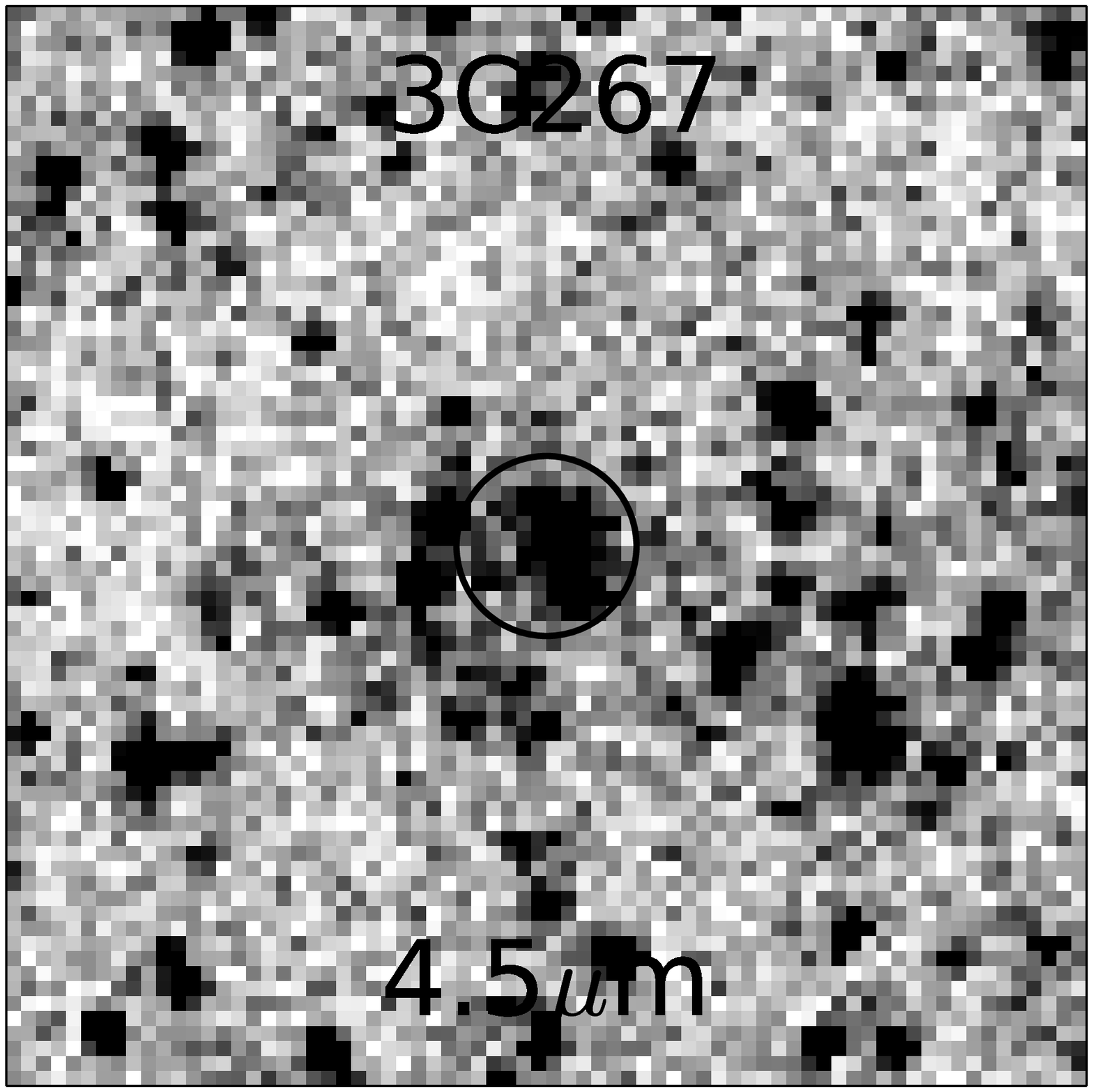}
      \includegraphics[width=1.5cm]{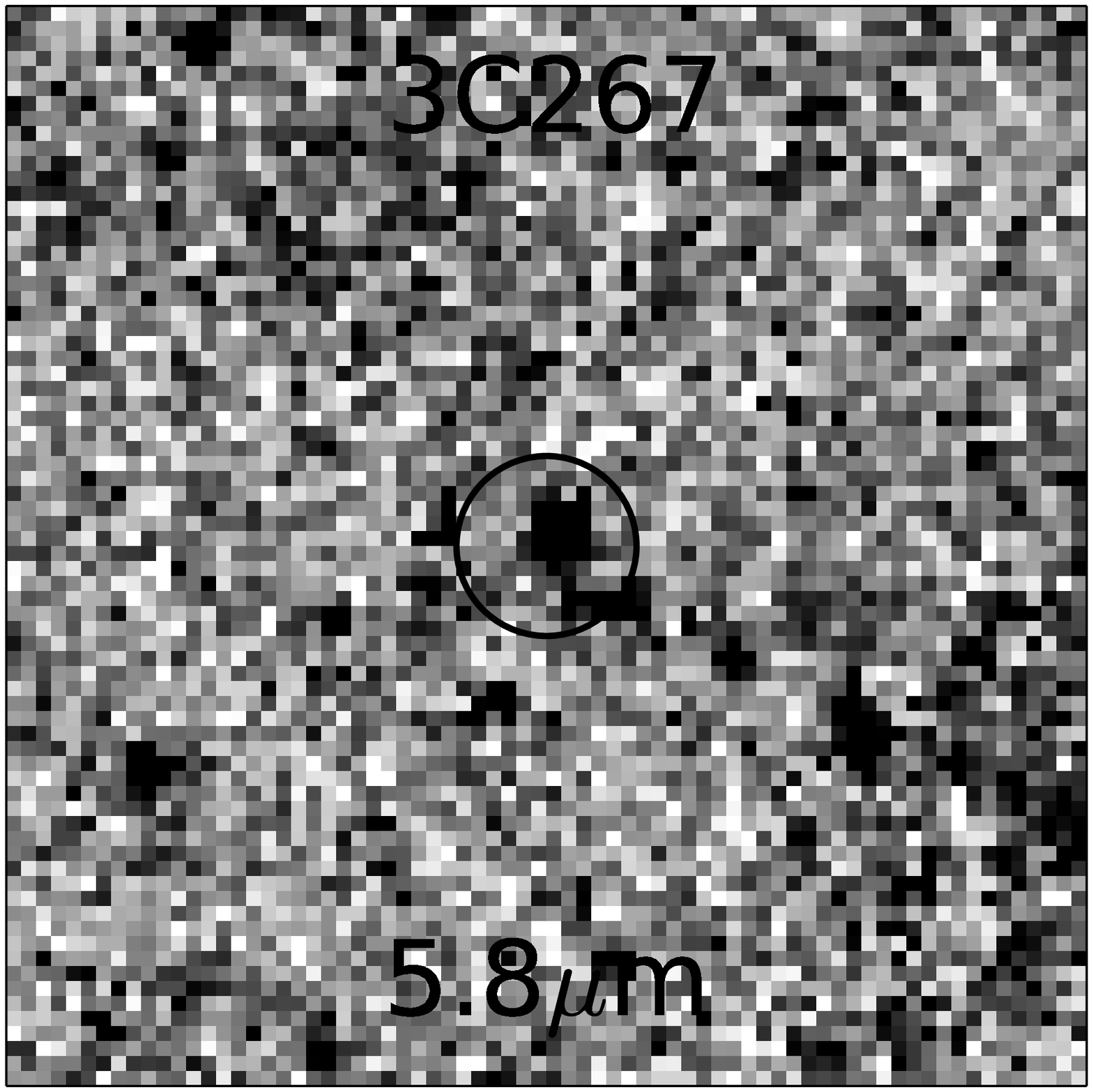}
      \includegraphics[width=1.5cm]{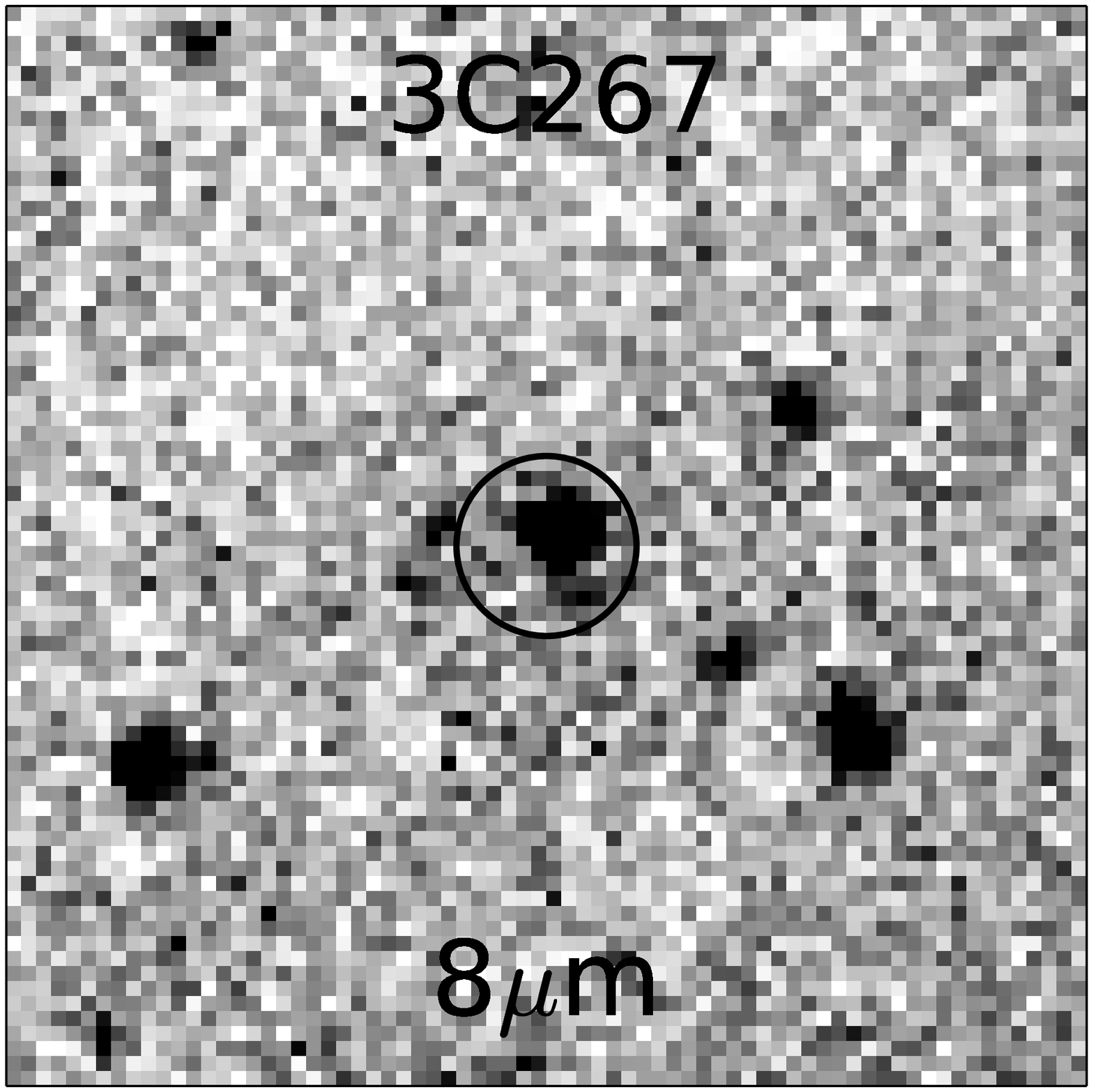}
      \includegraphics[width=1.5cm]{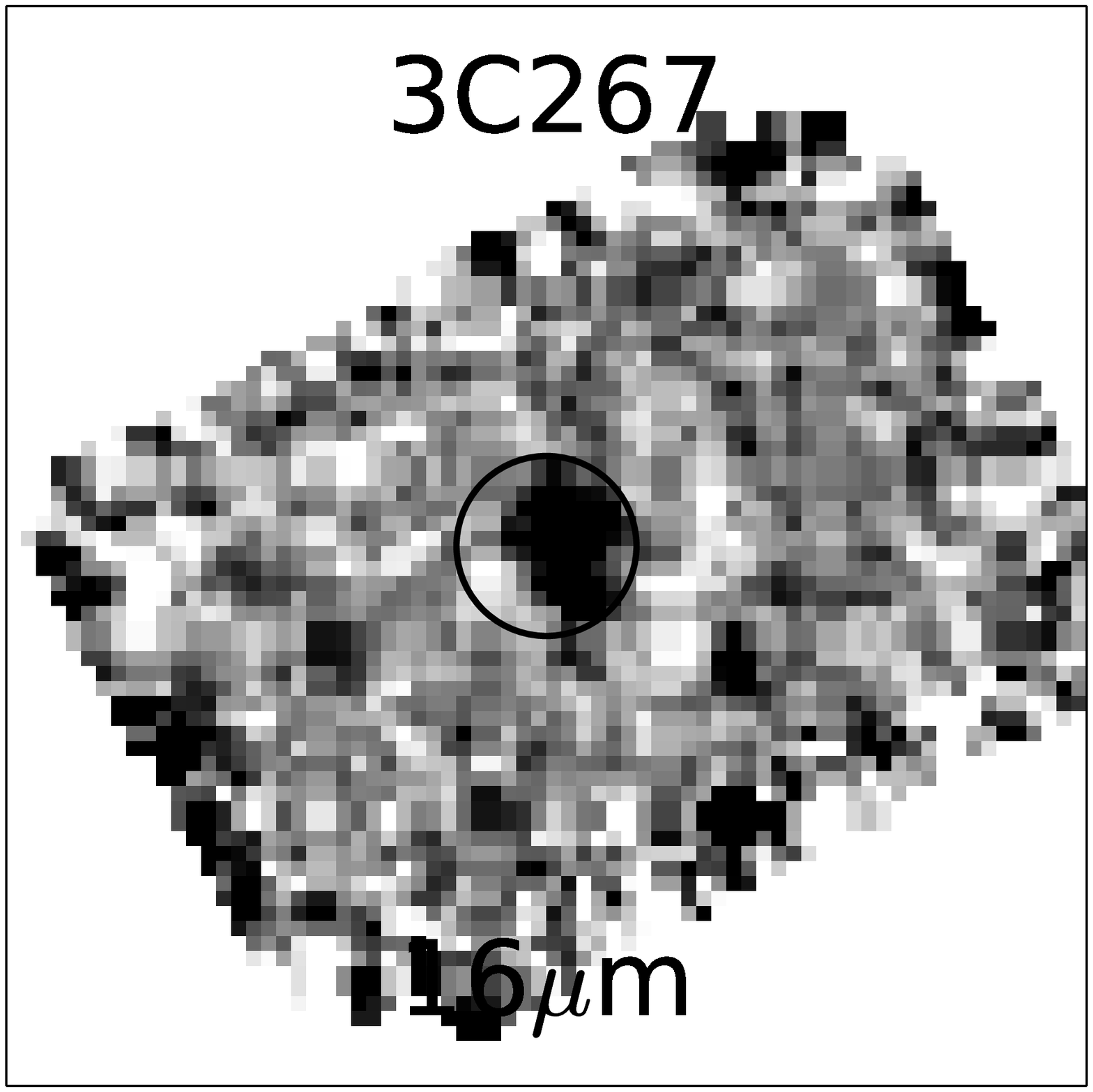}
      \includegraphics[width=1.5cm]{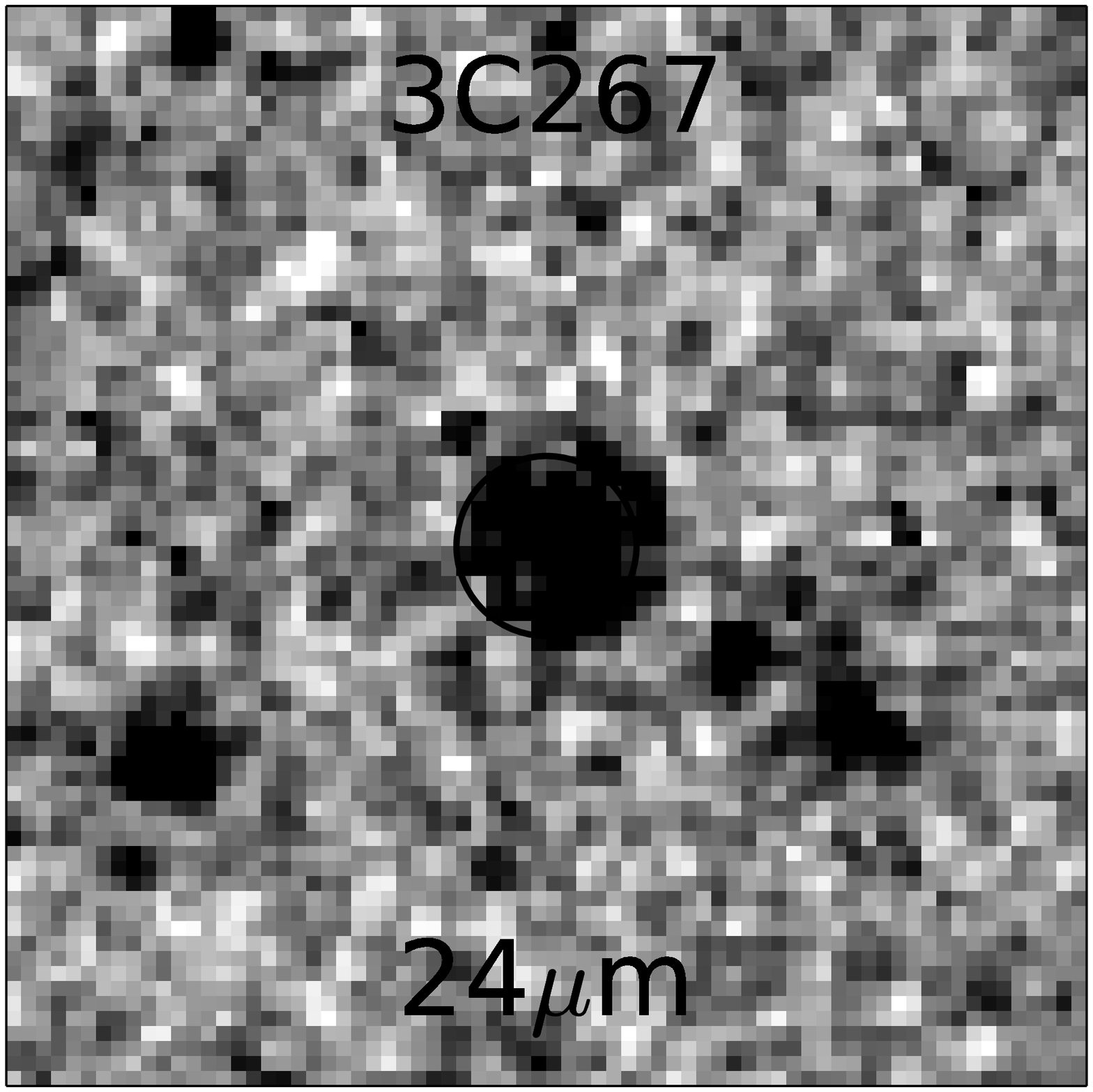}
      \includegraphics[width=1.5cm]{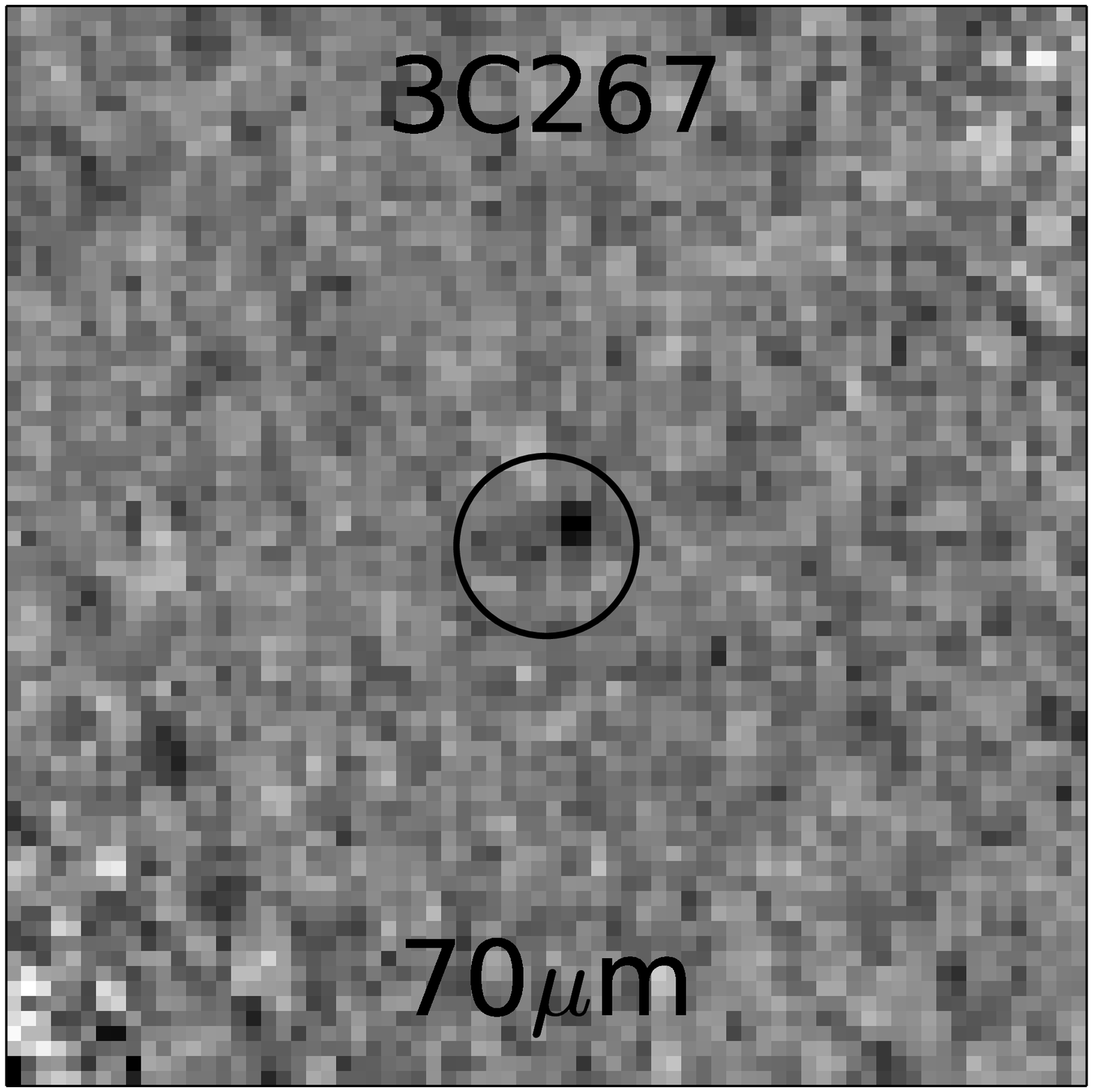}
      \includegraphics[width=1.5cm]{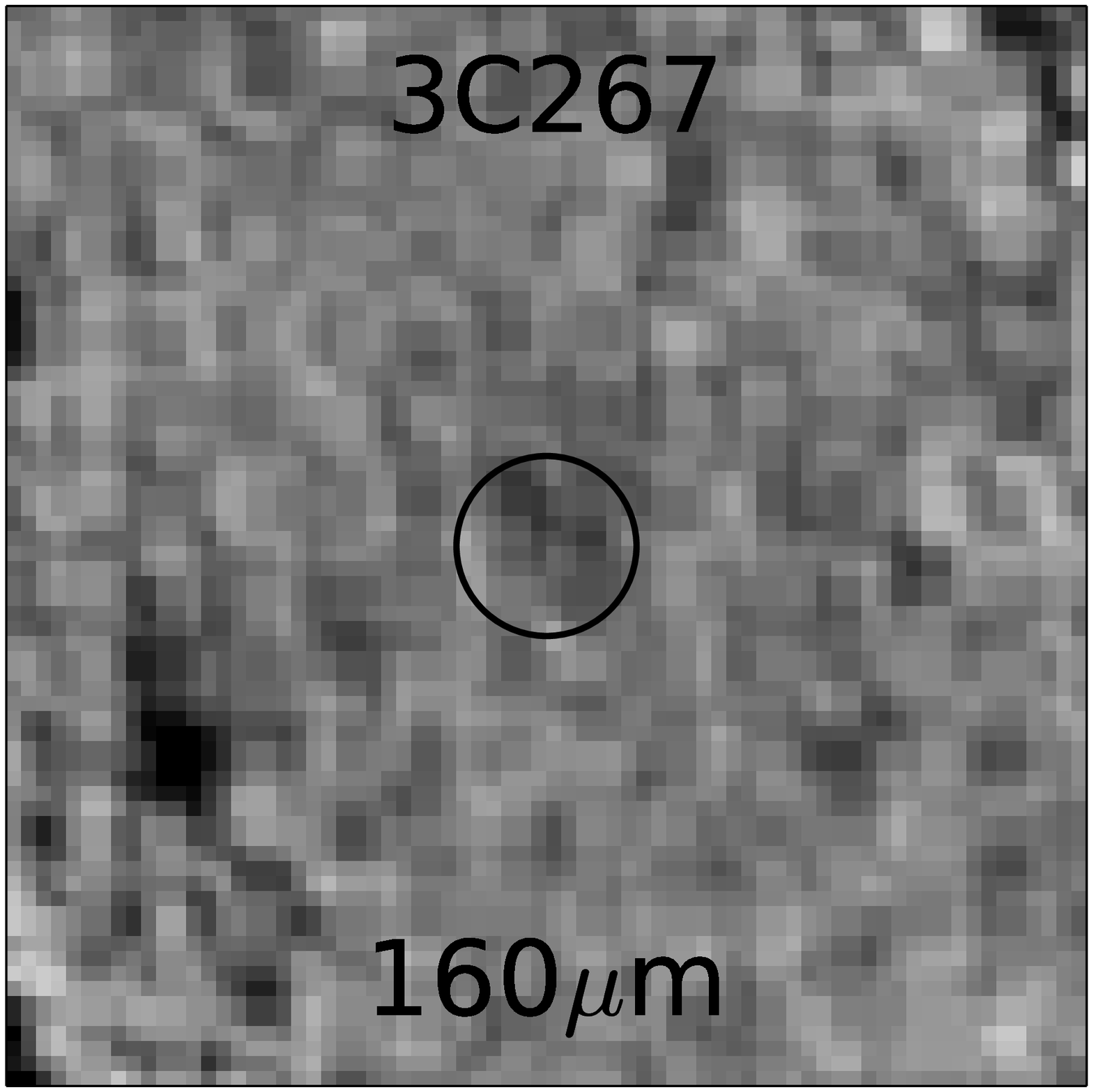}
      \includegraphics[width=1.5cm]{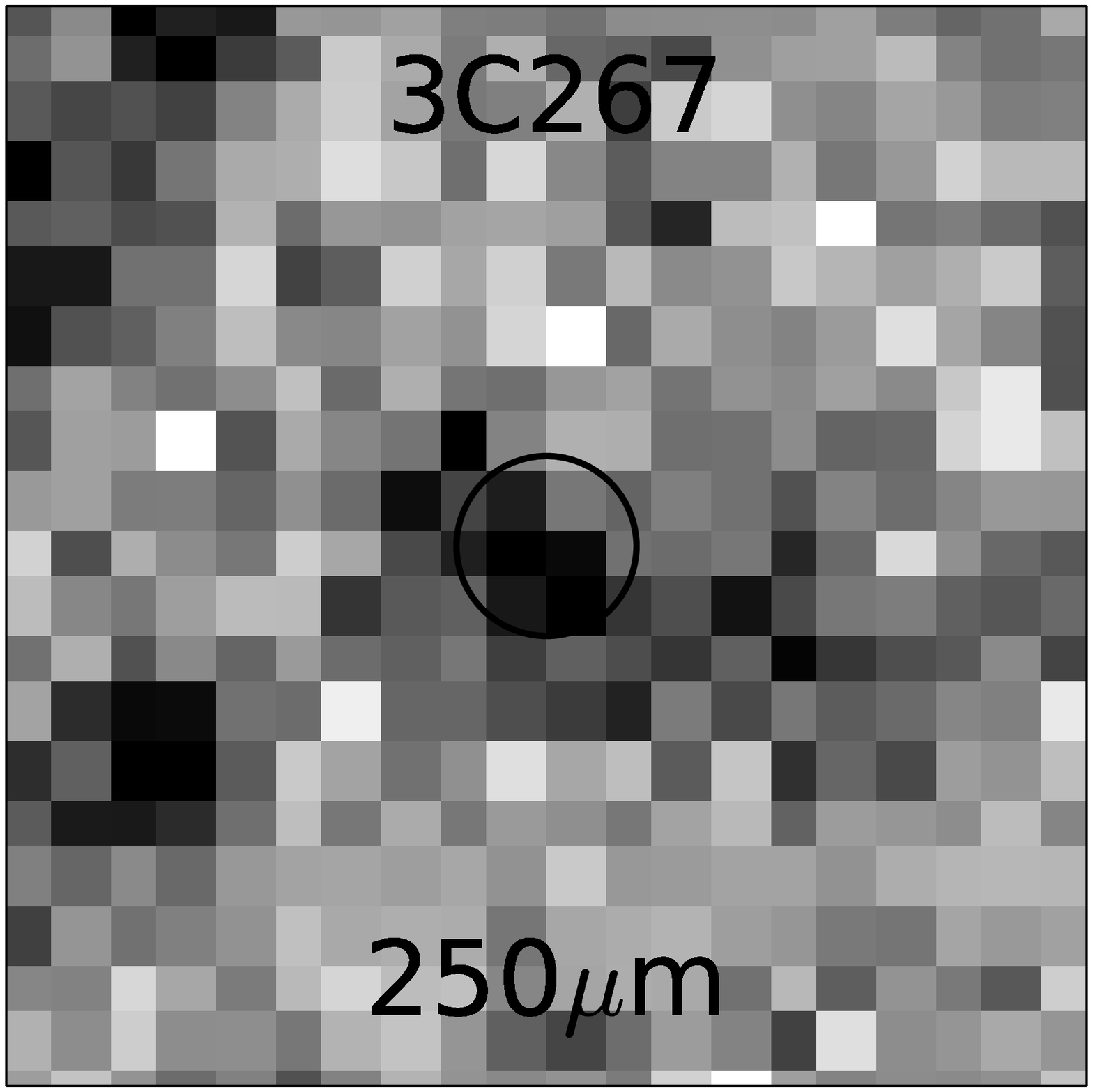}
      \includegraphics[width=1.5cm]{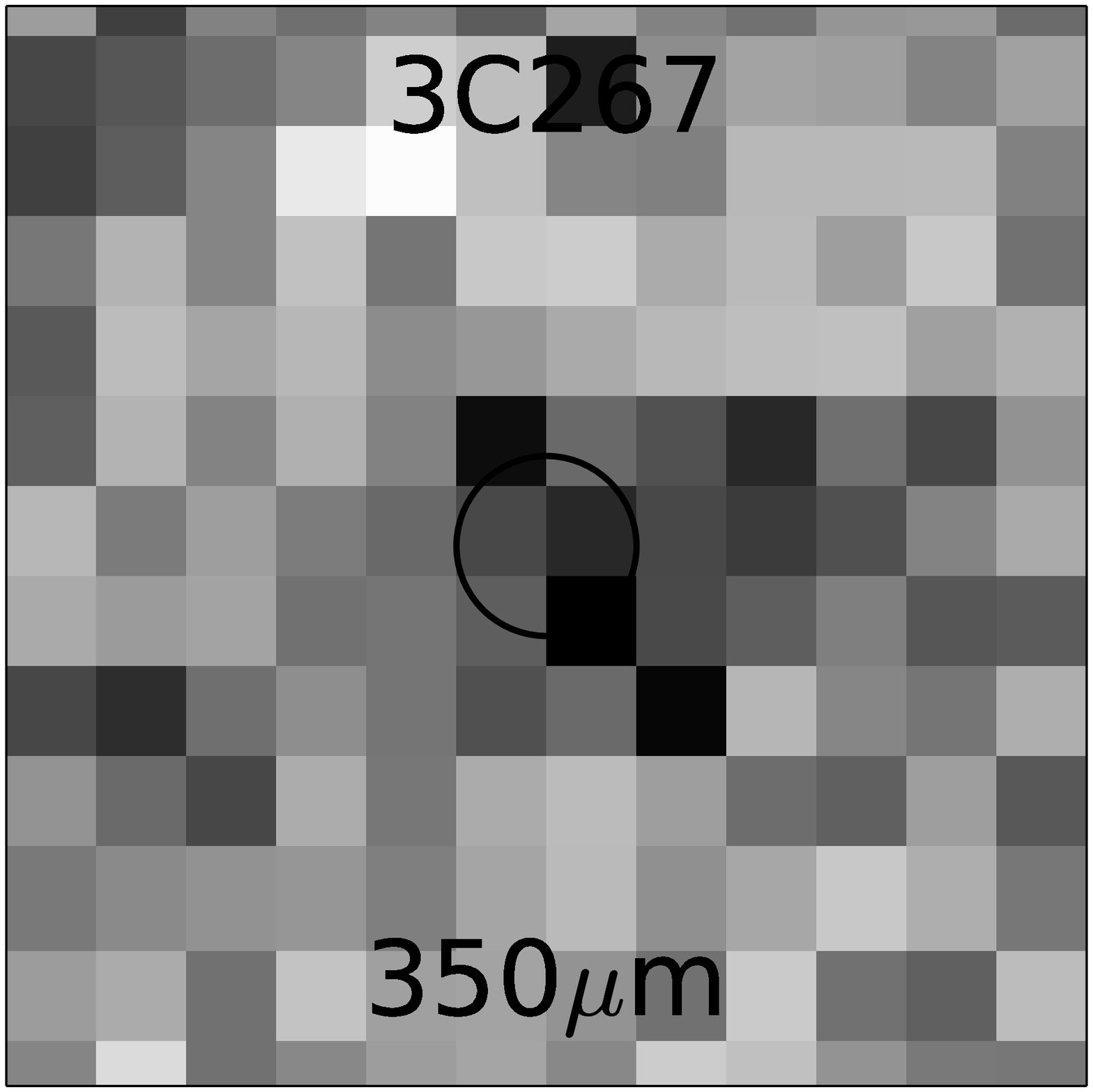}
      \includegraphics[width=1.5cm]{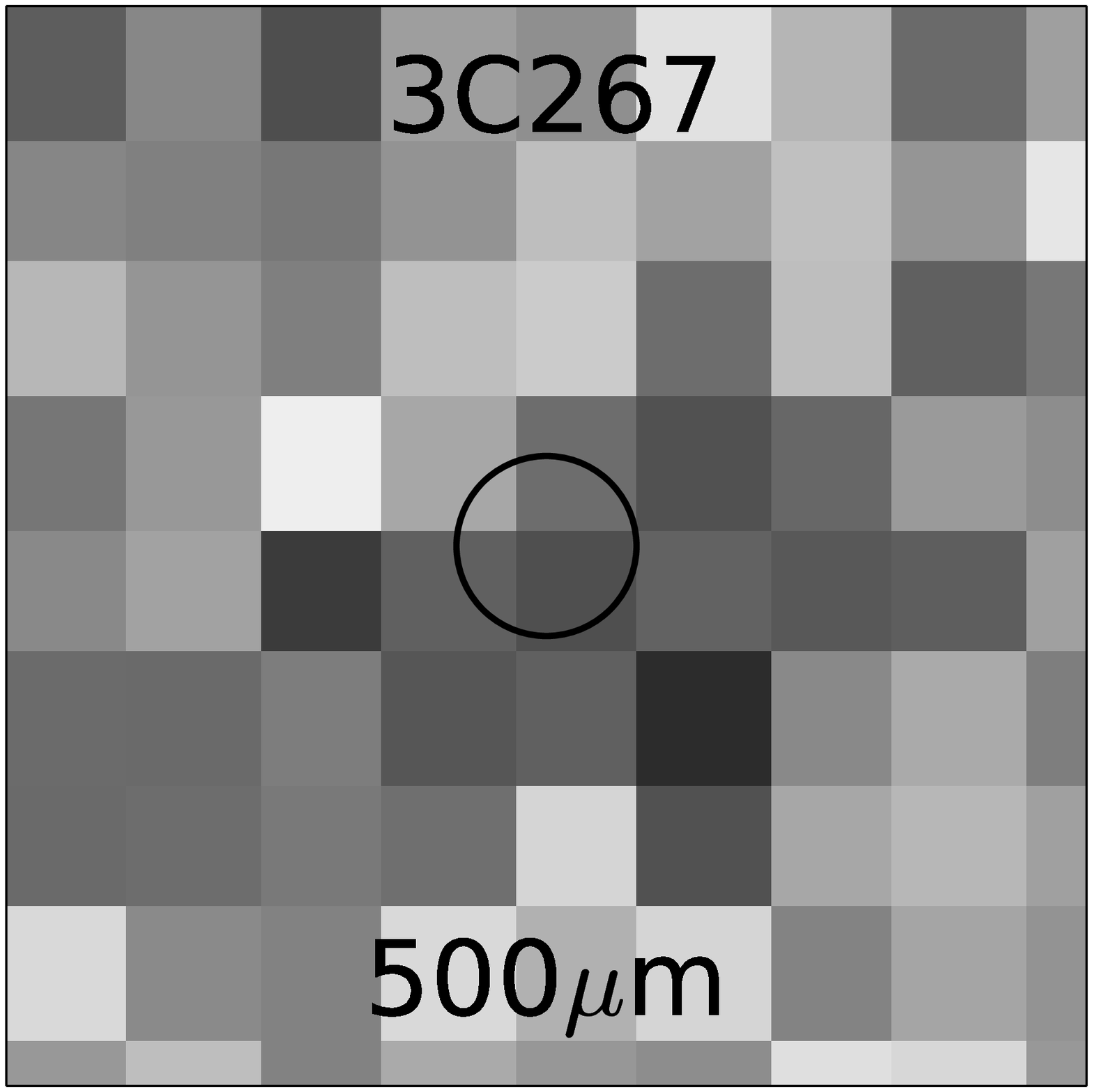}
      \caption{Continued.}
   \end{figure*}    
   \addtocounter{figure}{-1}
   \begin{figure*}
      \includegraphics[width=1.5cm]{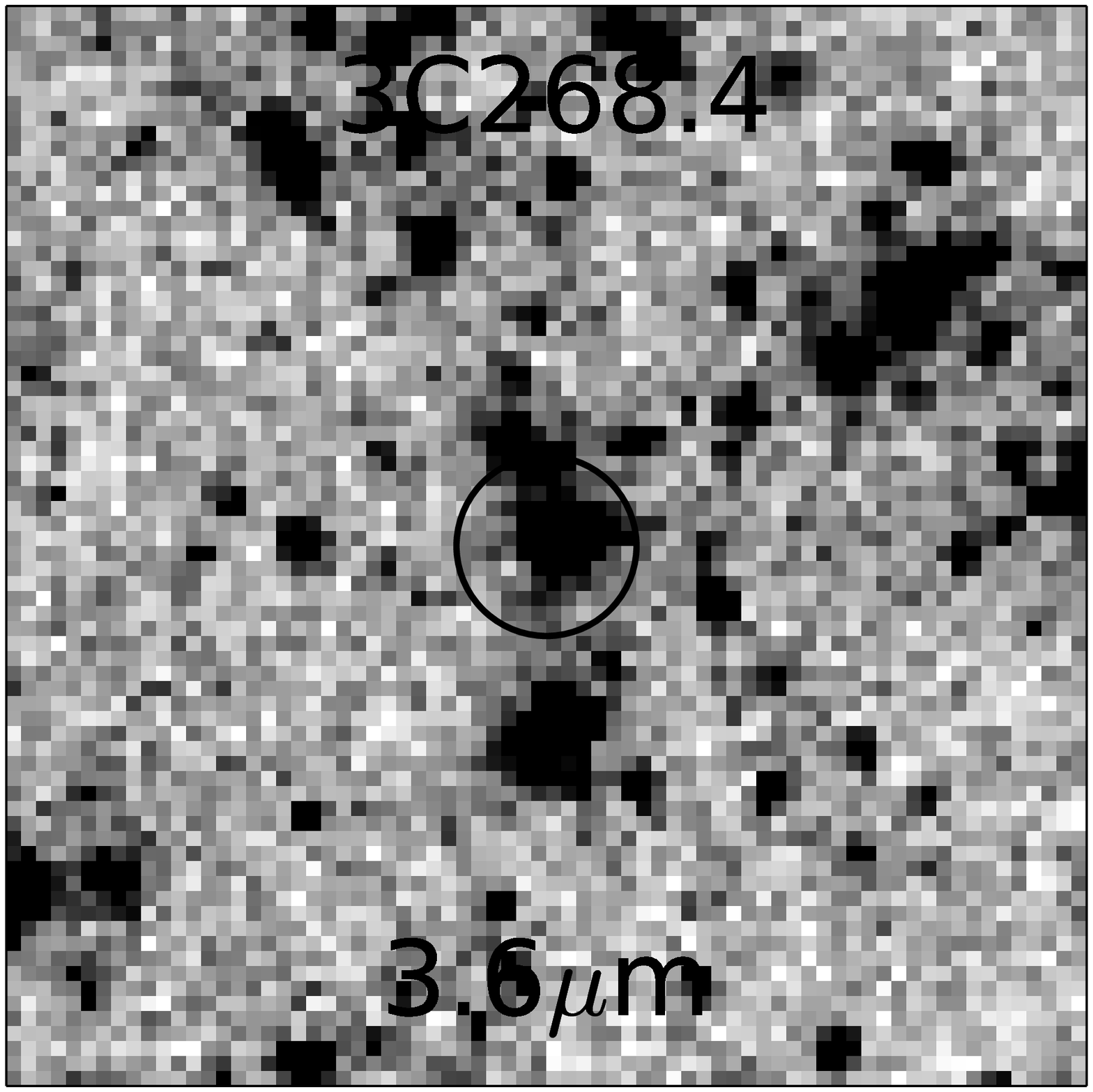}
      \includegraphics[width=1.5cm]{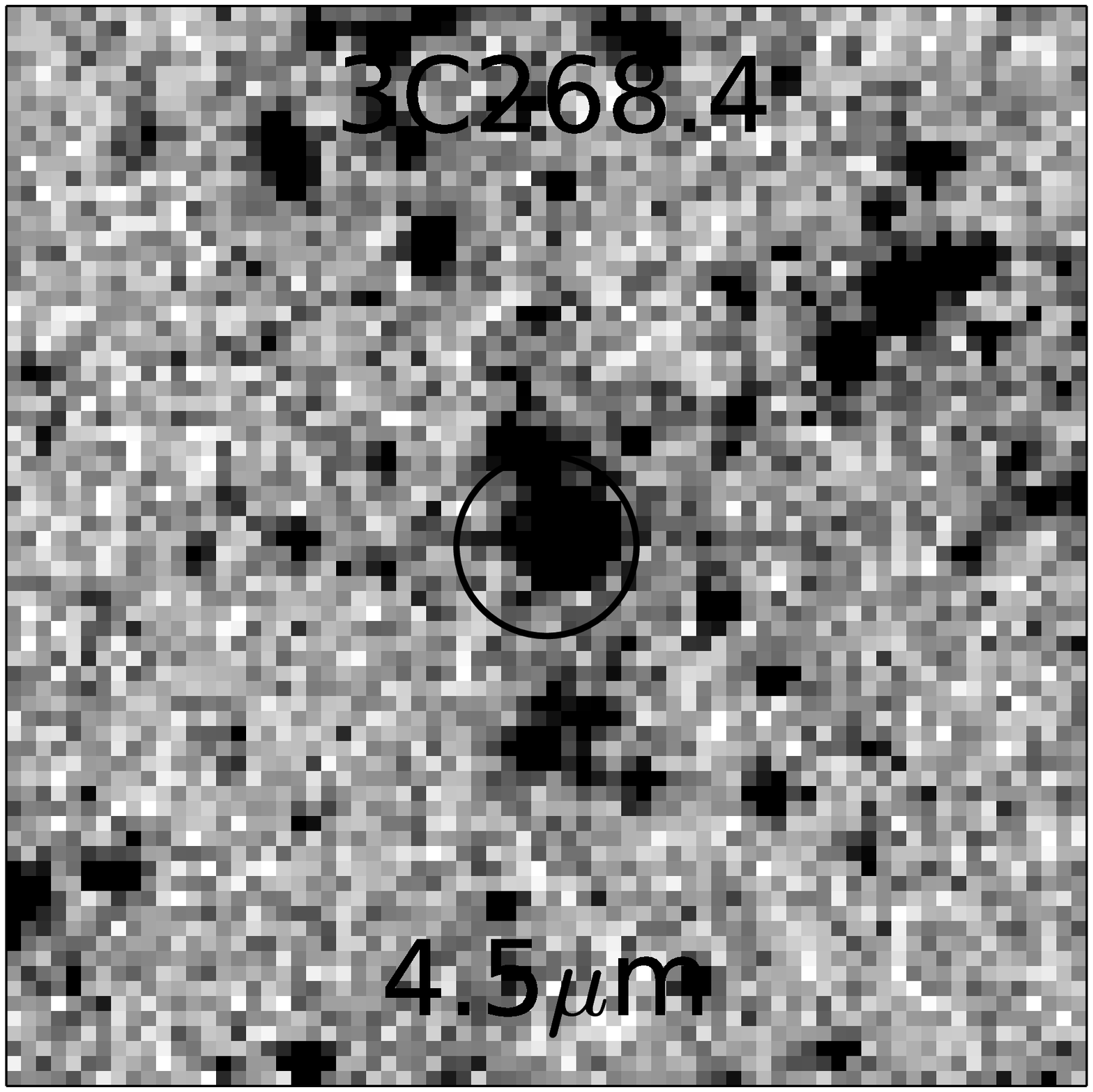}
      \includegraphics[width=1.5cm]{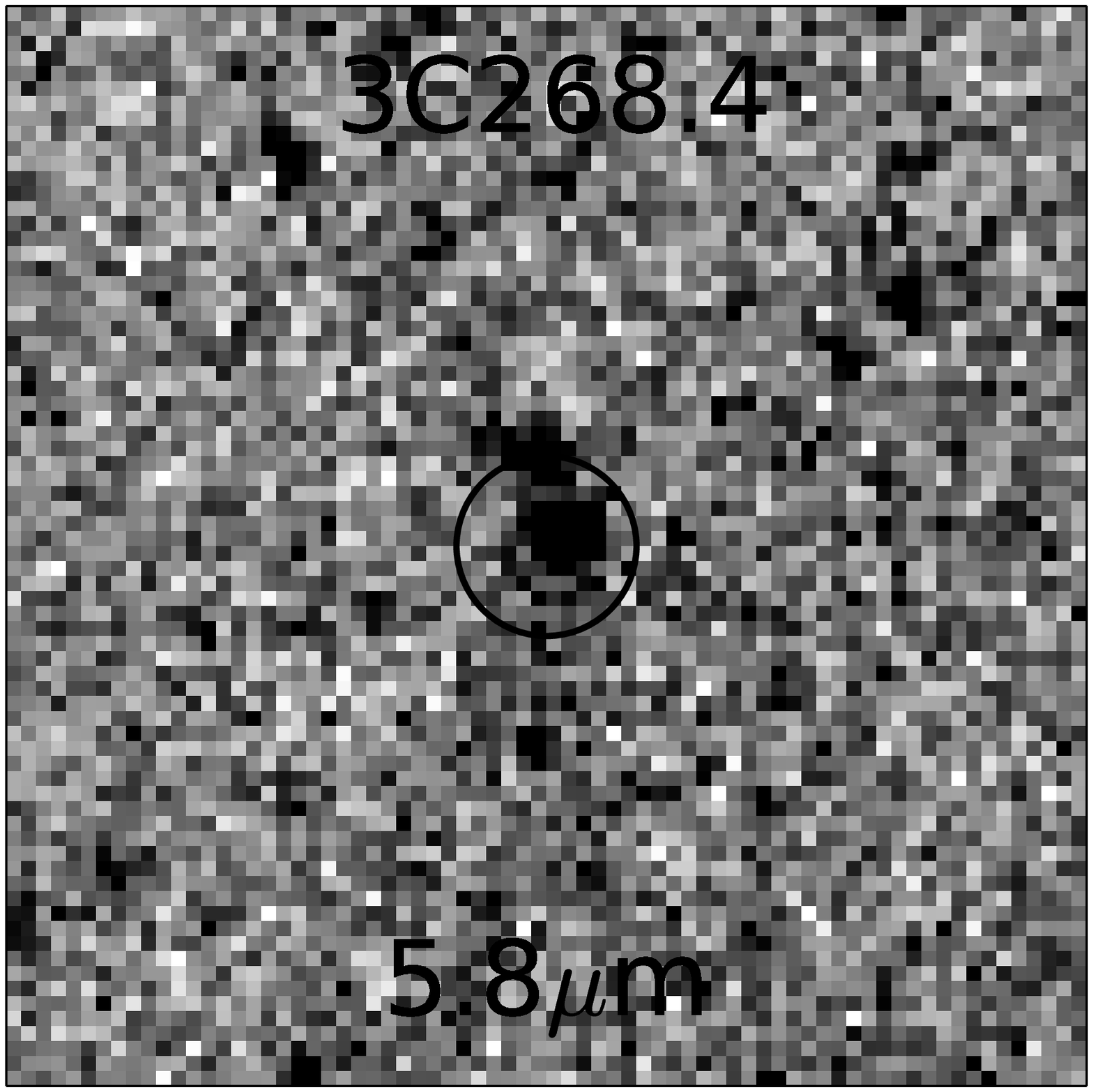}
      \includegraphics[width=1.5cm]{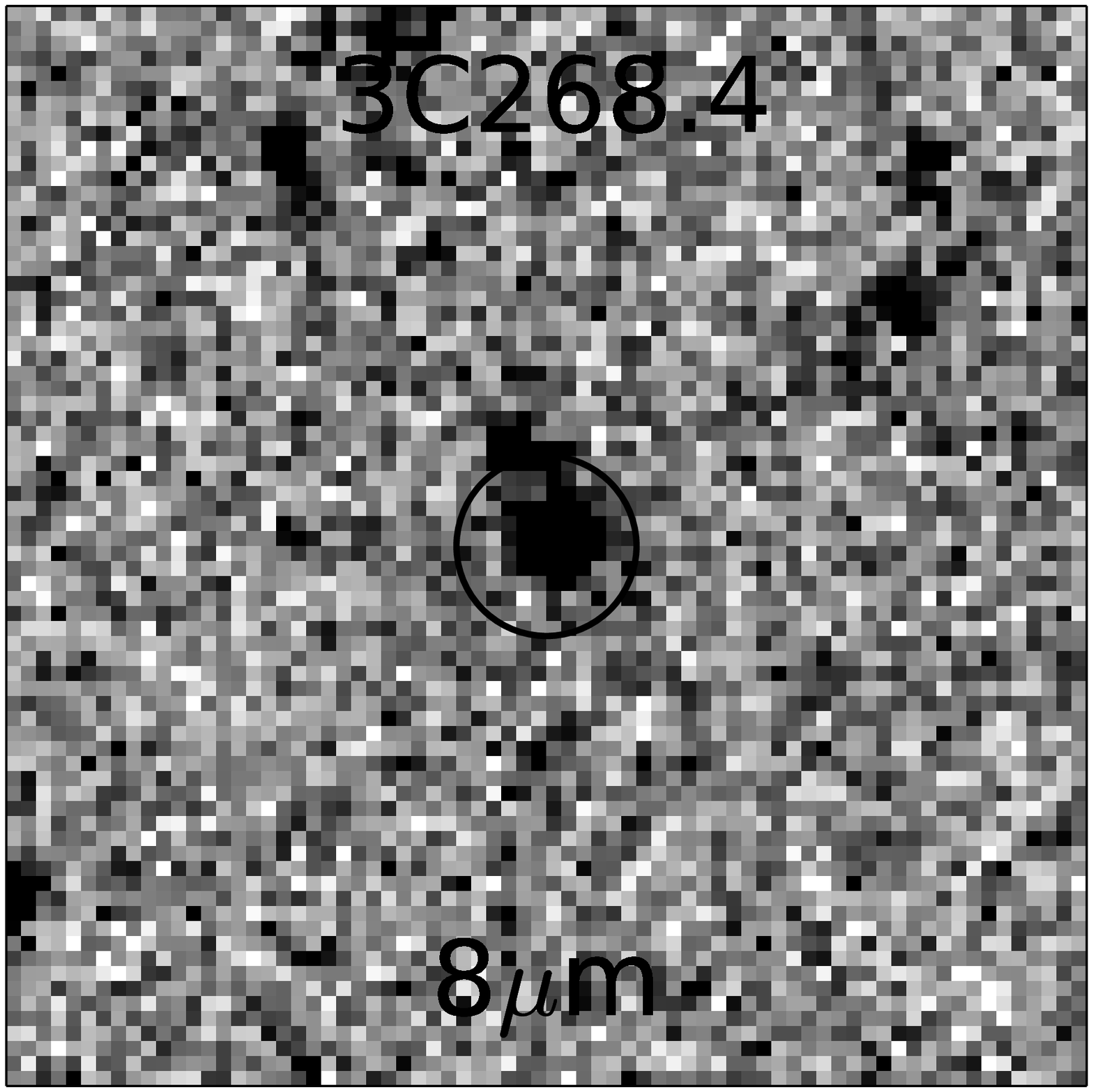}
      \includegraphics[width=1.5cm]{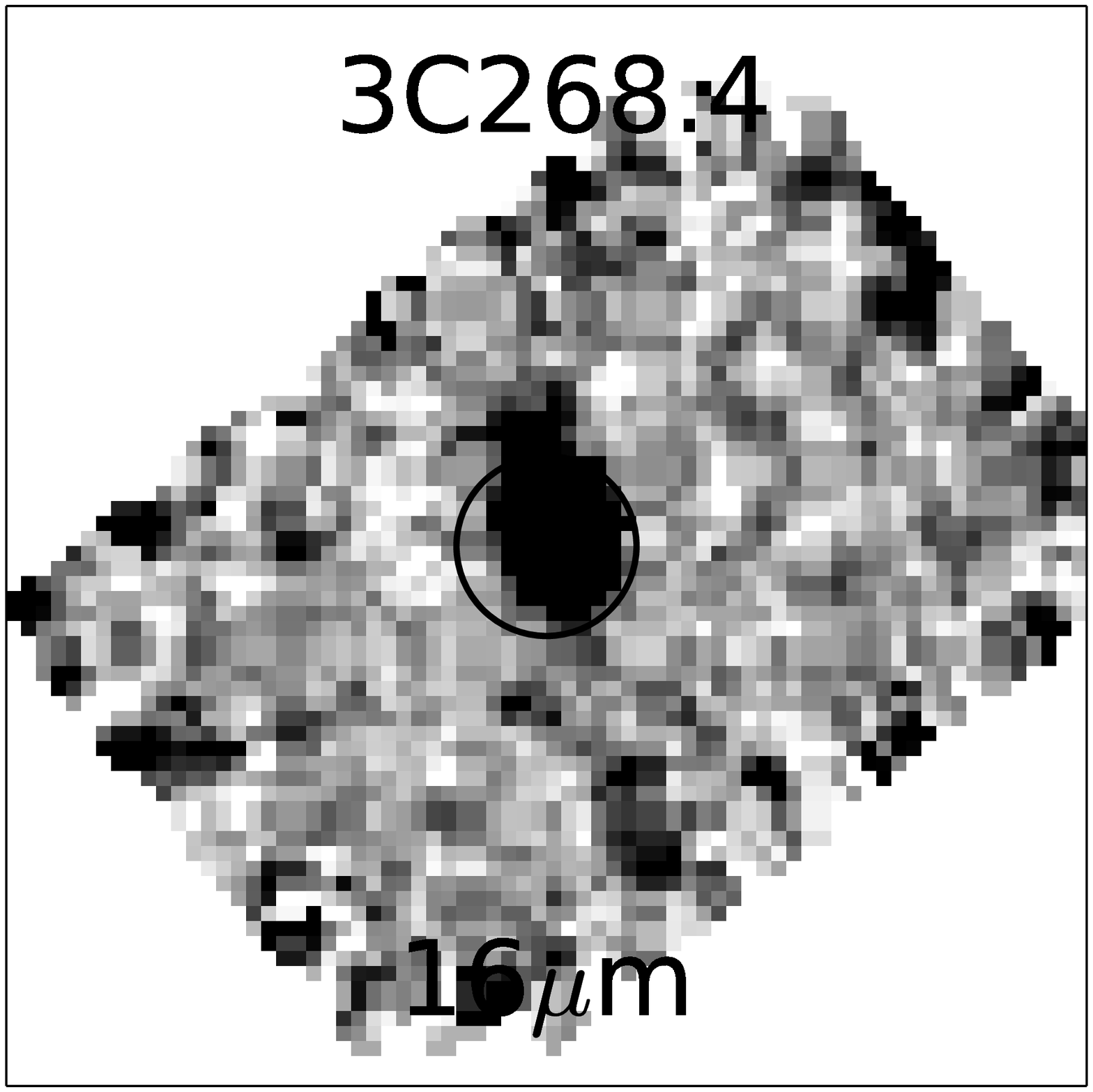}
      \includegraphics[width=1.5cm]{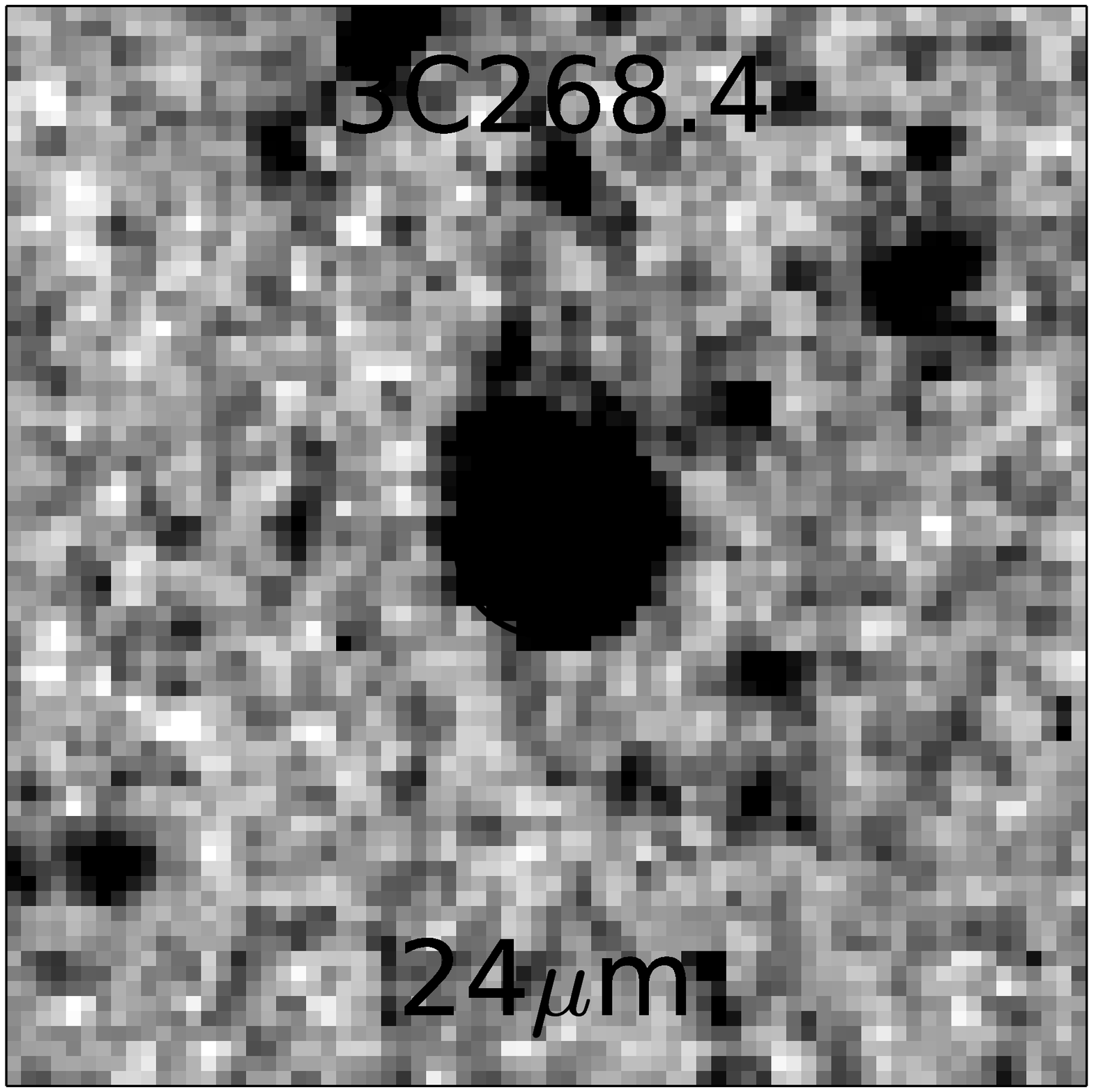}
      \includegraphics[width=1.5cm]{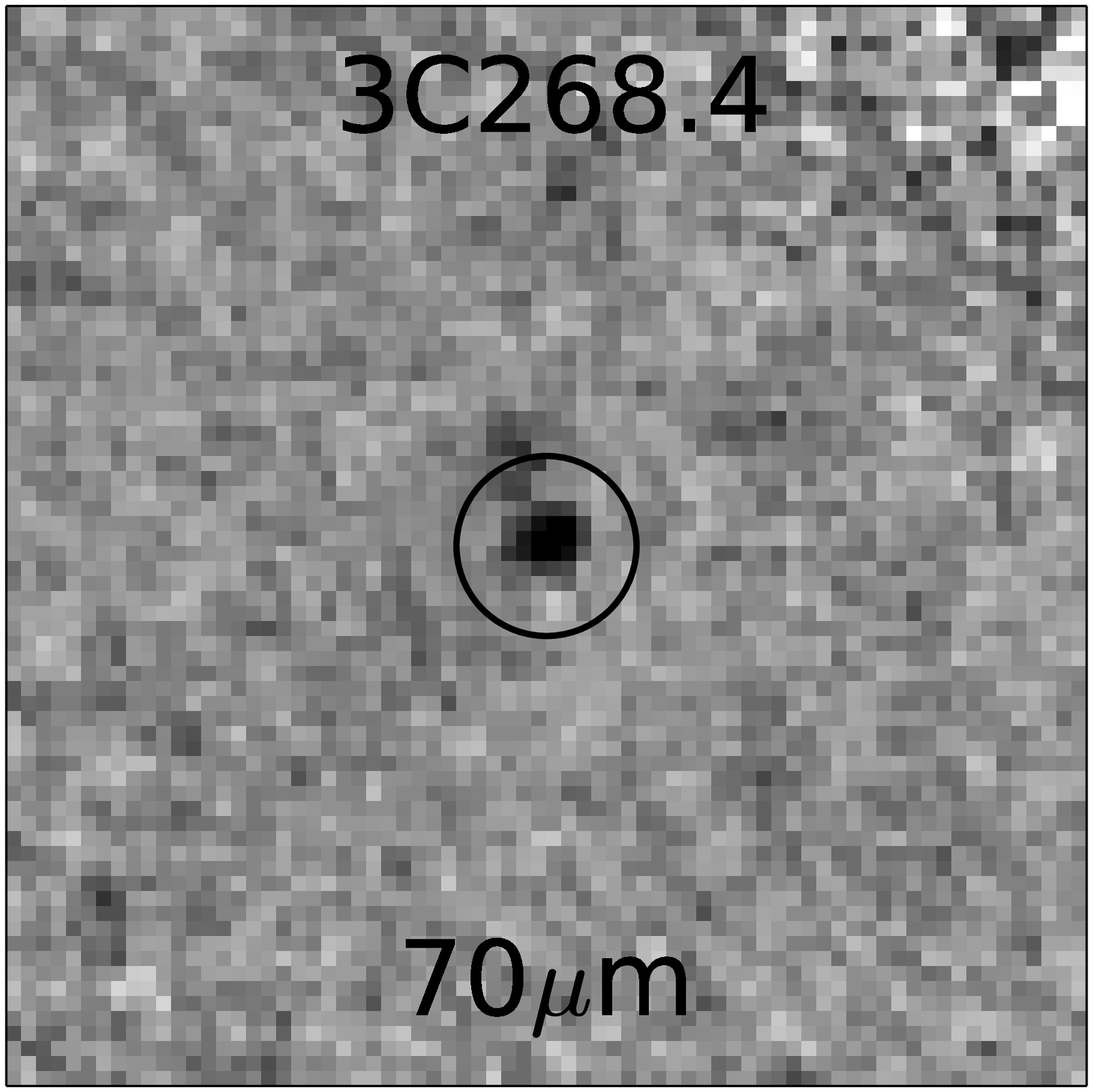}
      \includegraphics[width=1.5cm]{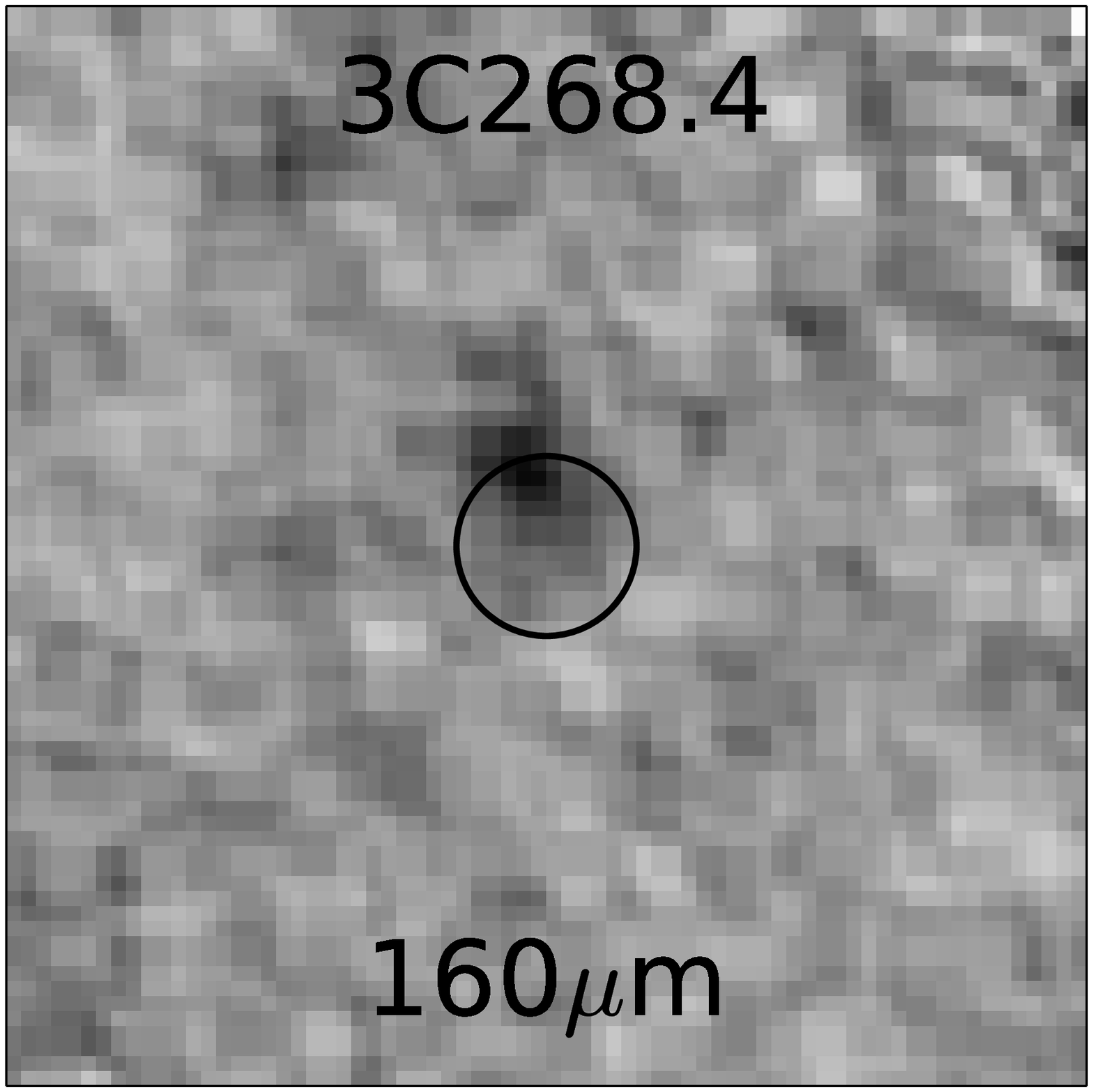}
      \includegraphics[width=1.5cm]{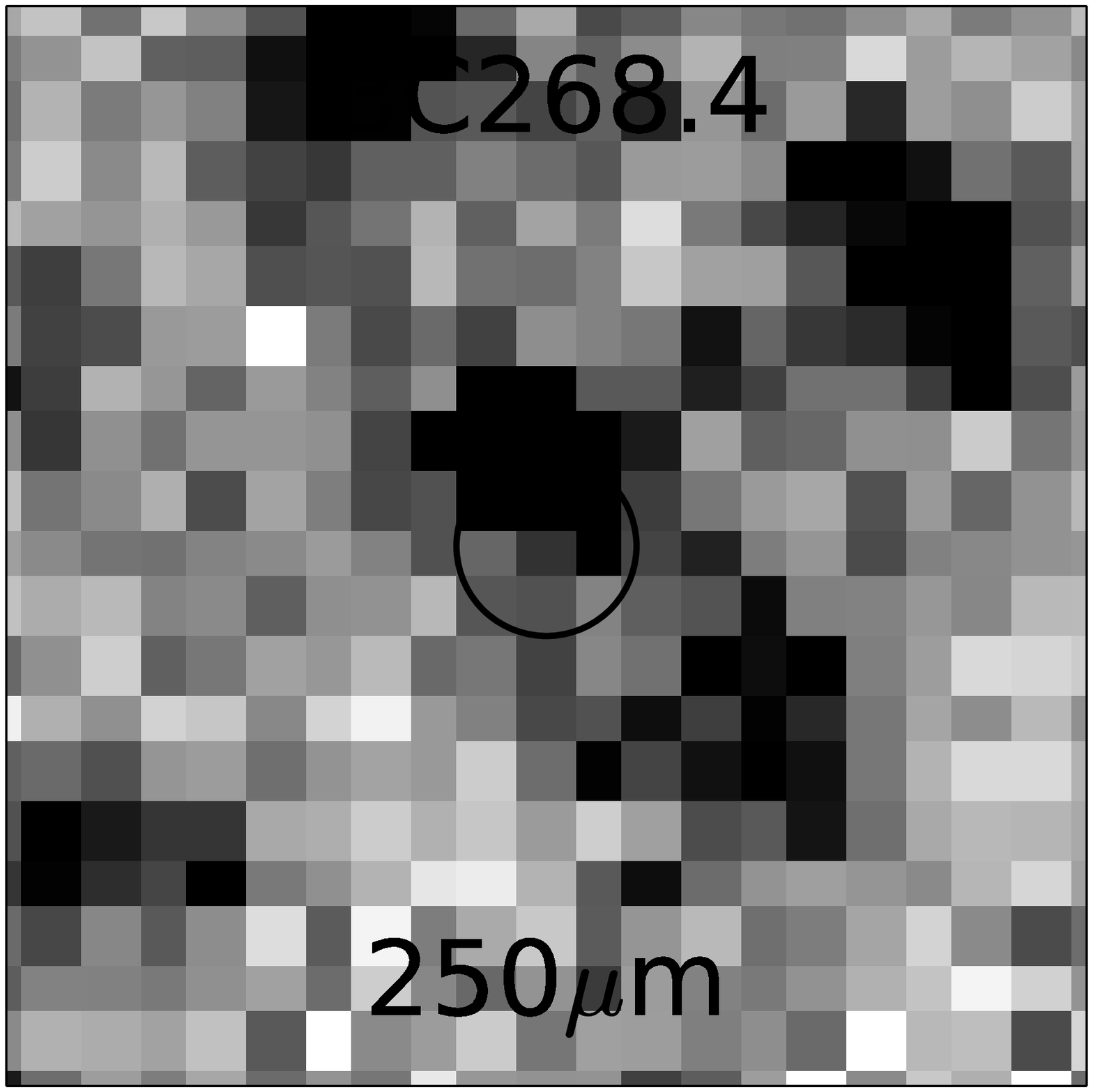}
      \includegraphics[width=1.5cm]{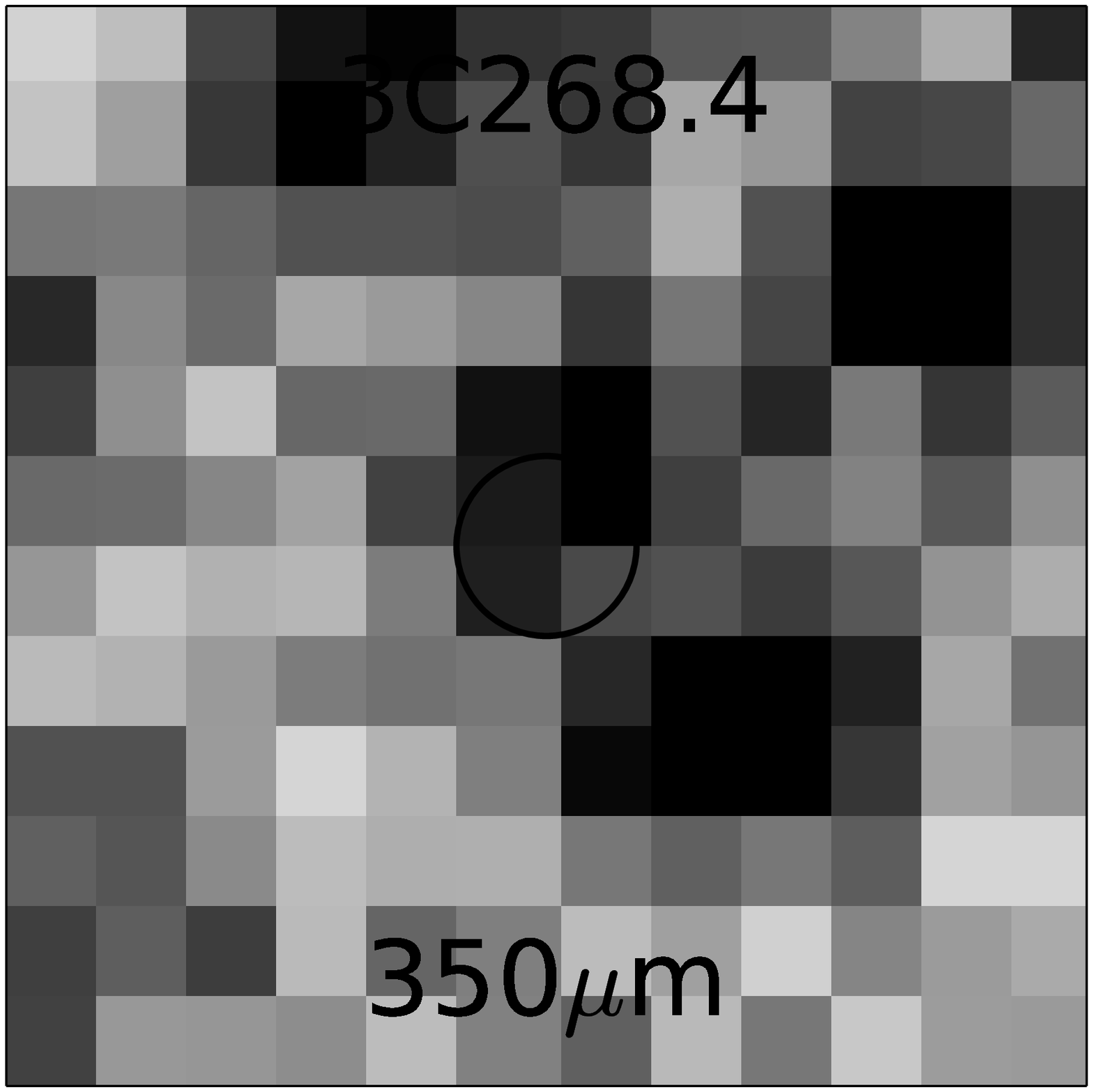}
      \includegraphics[width=1.5cm]{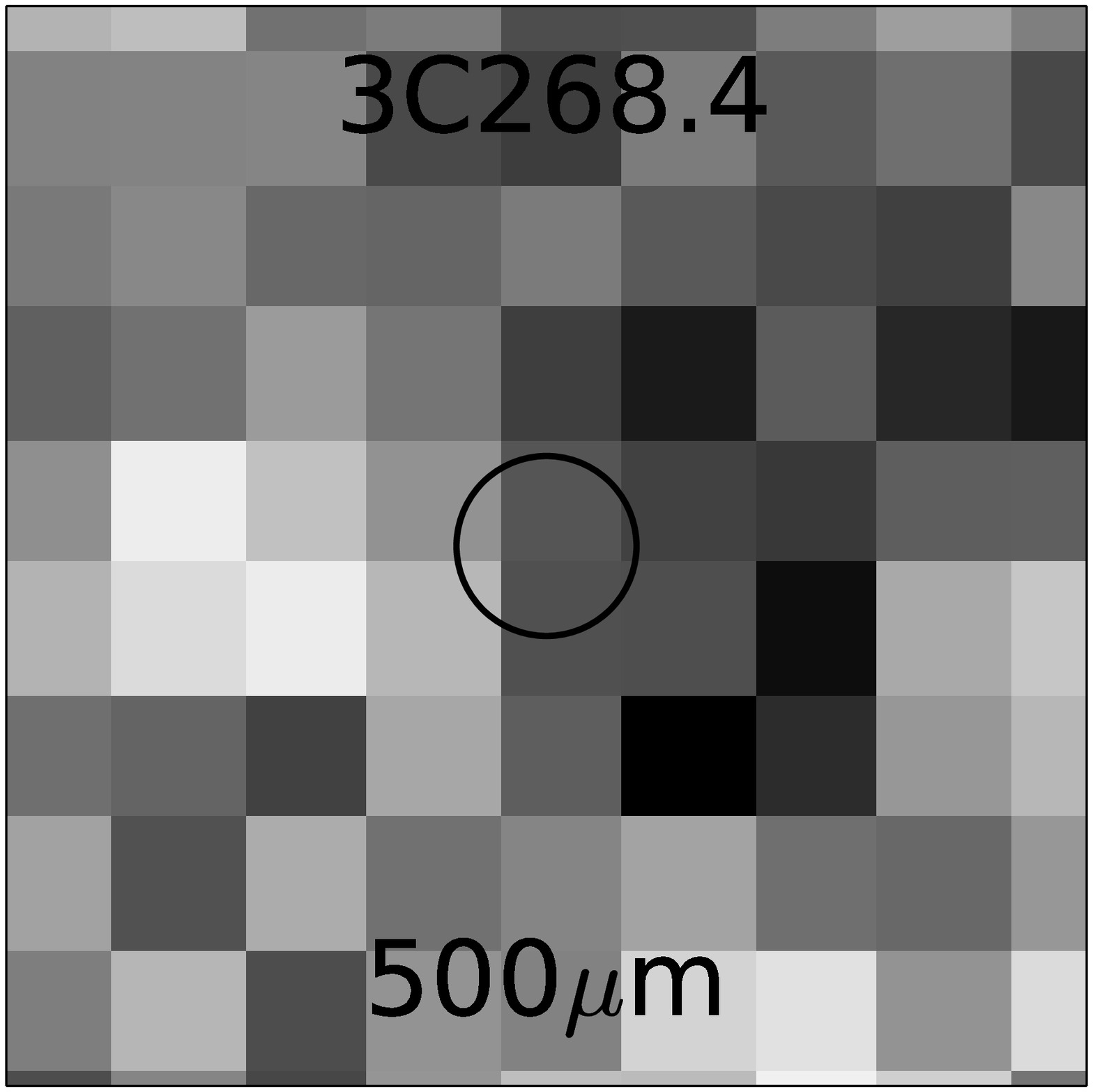}
      \\
      \includegraphics[width=1.5cm]{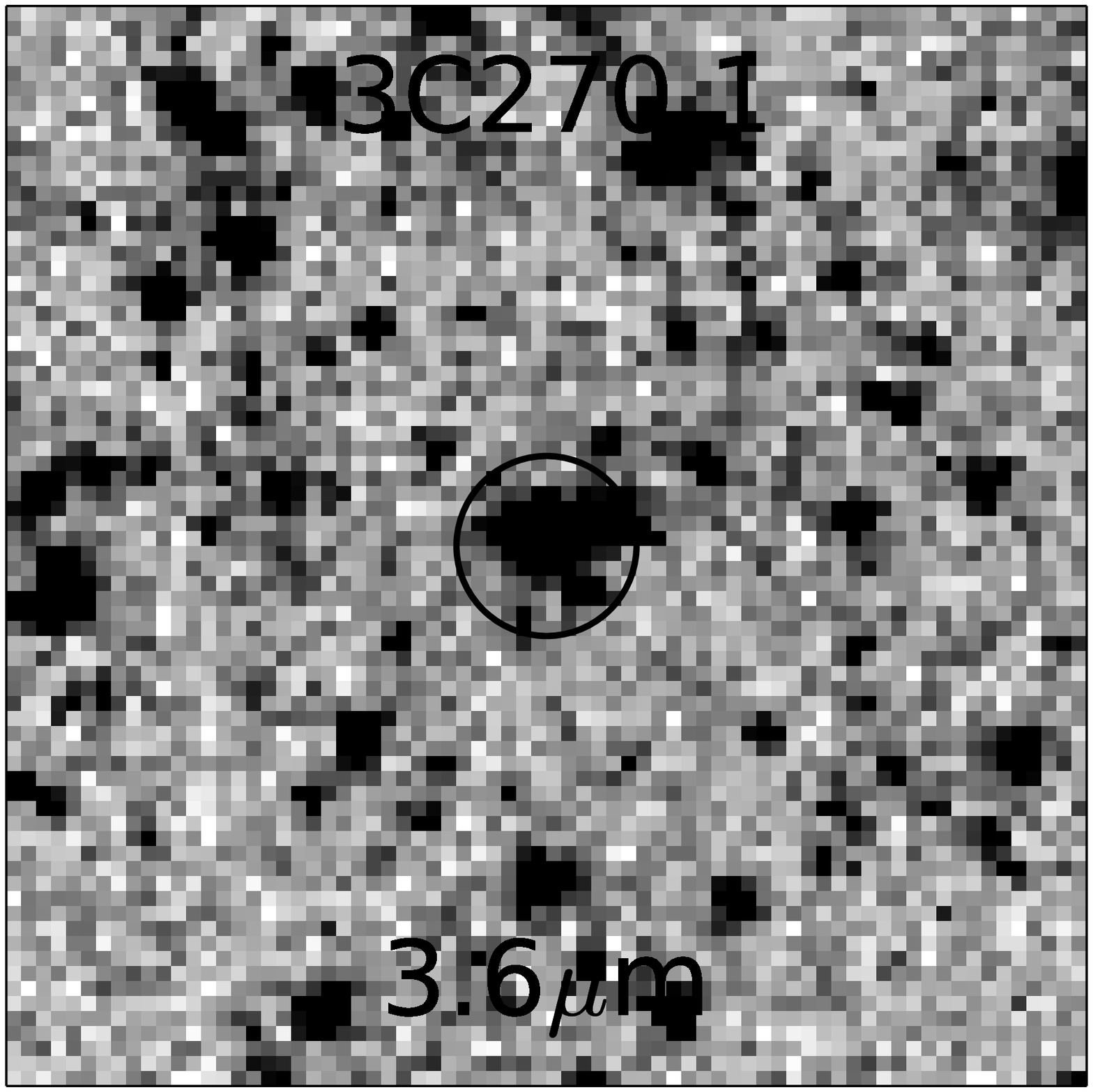}
      \includegraphics[width=1.5cm]{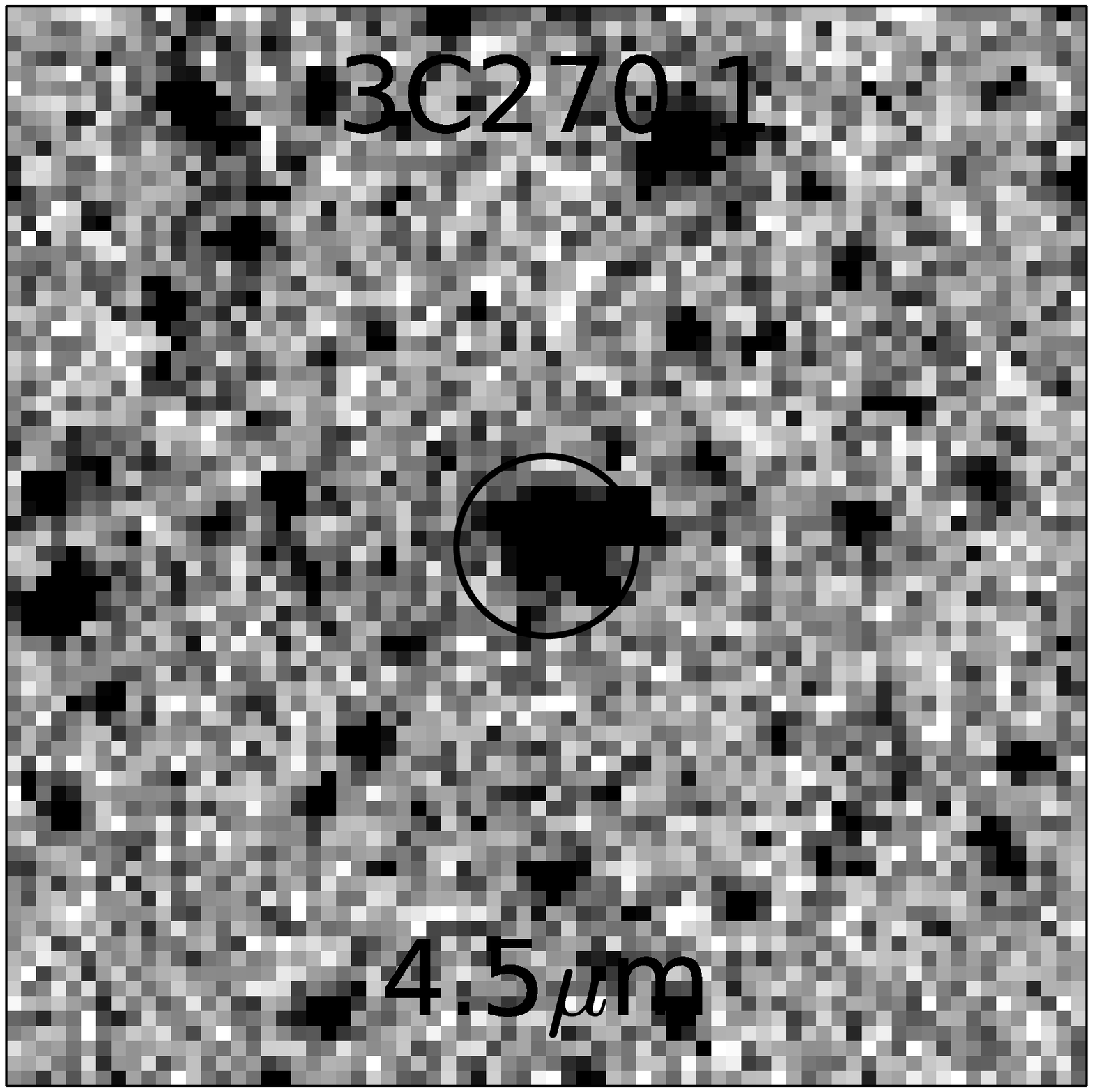}
      \includegraphics[width=1.5cm]{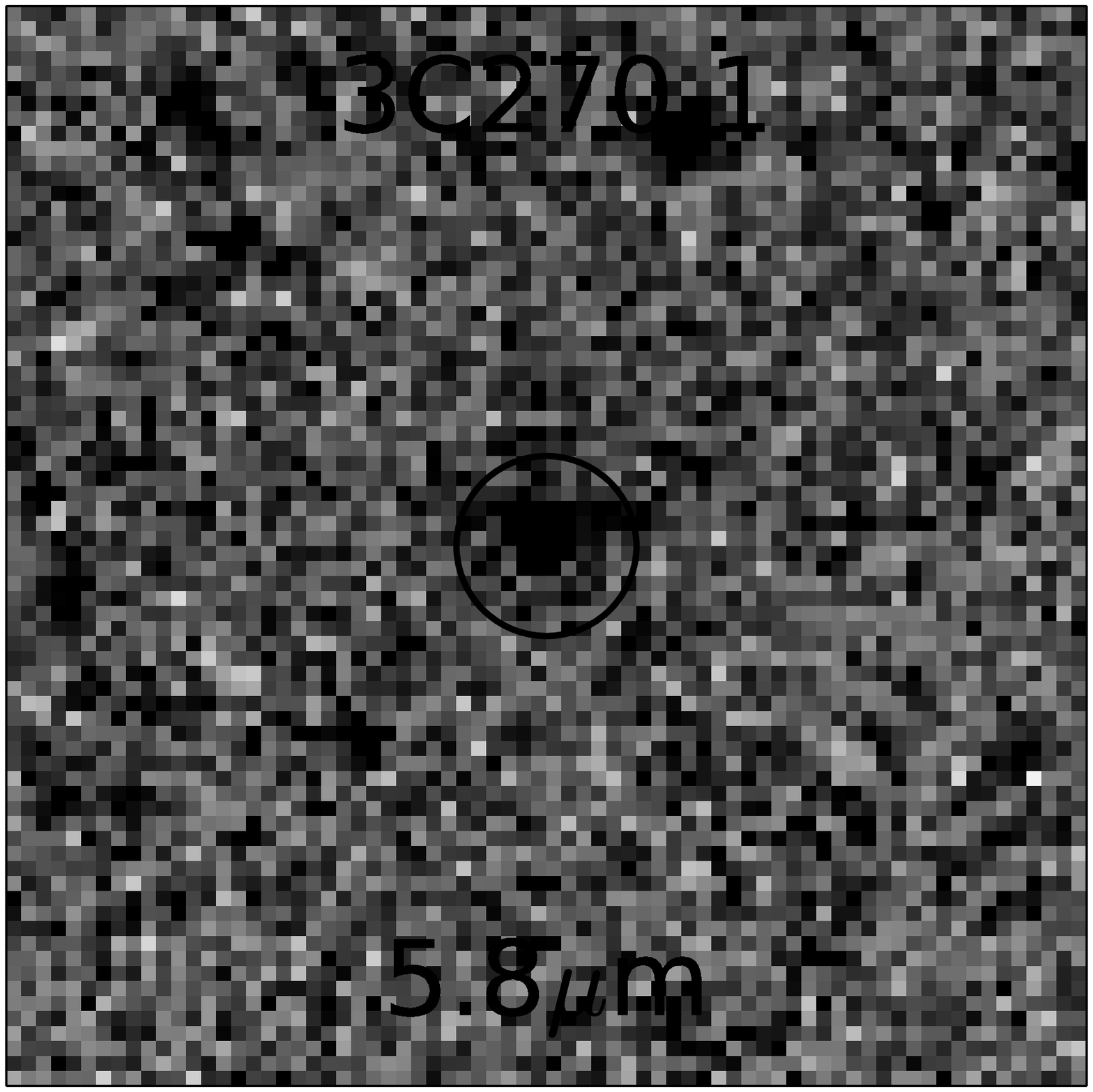}
      \includegraphics[width=1.5cm]{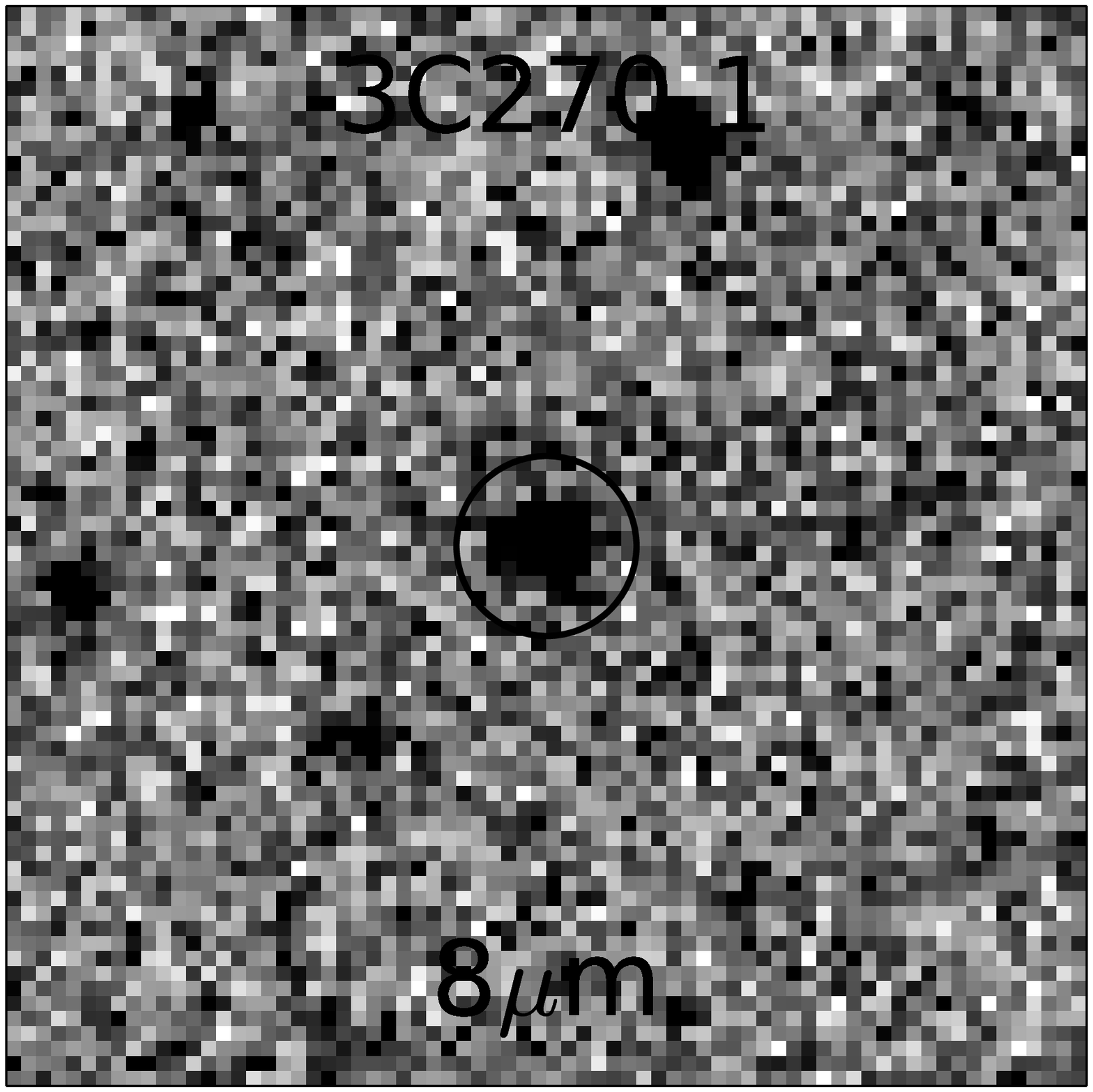}
      \includegraphics[width=1.5cm]{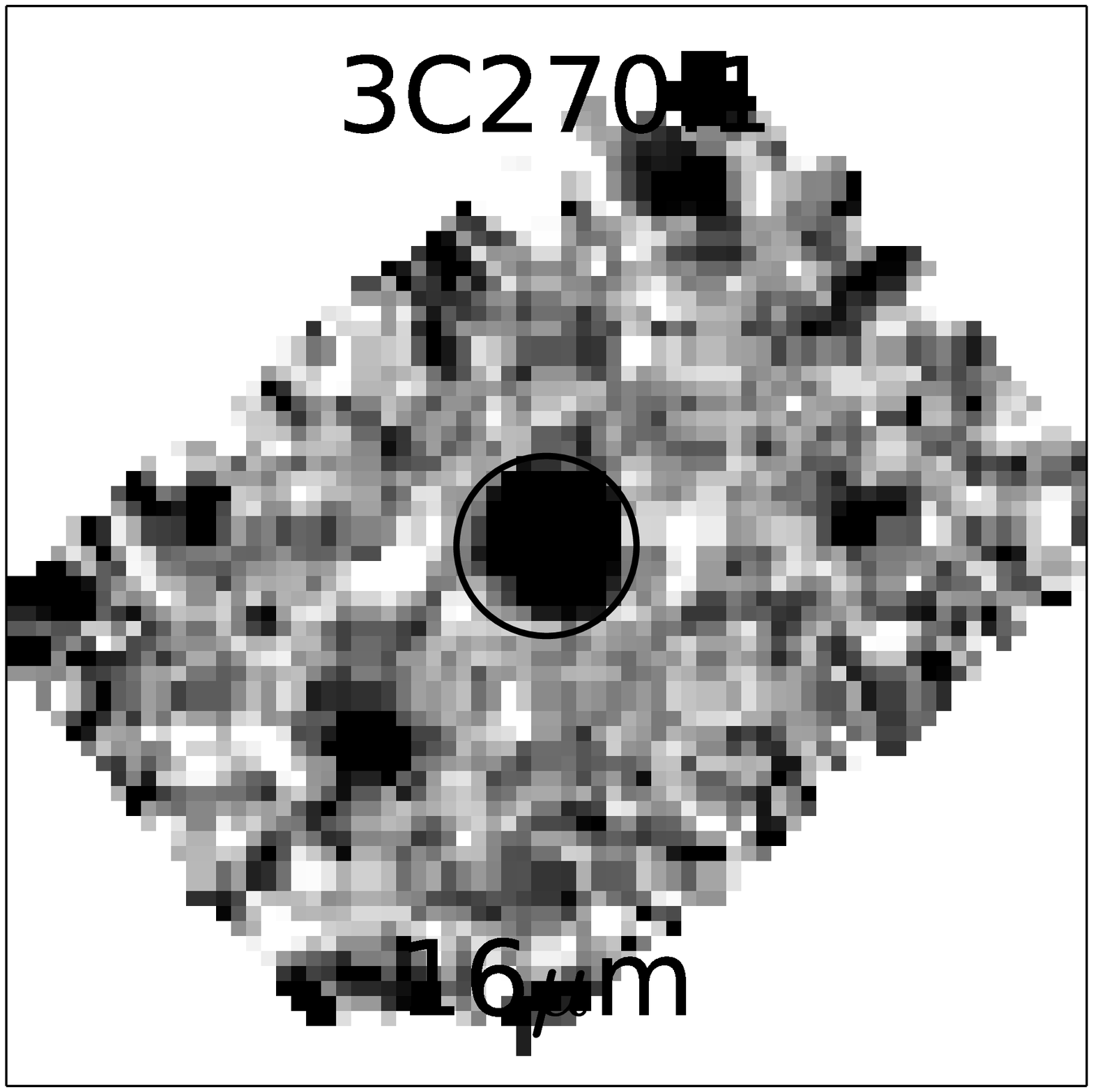}
      \includegraphics[width=1.5cm]{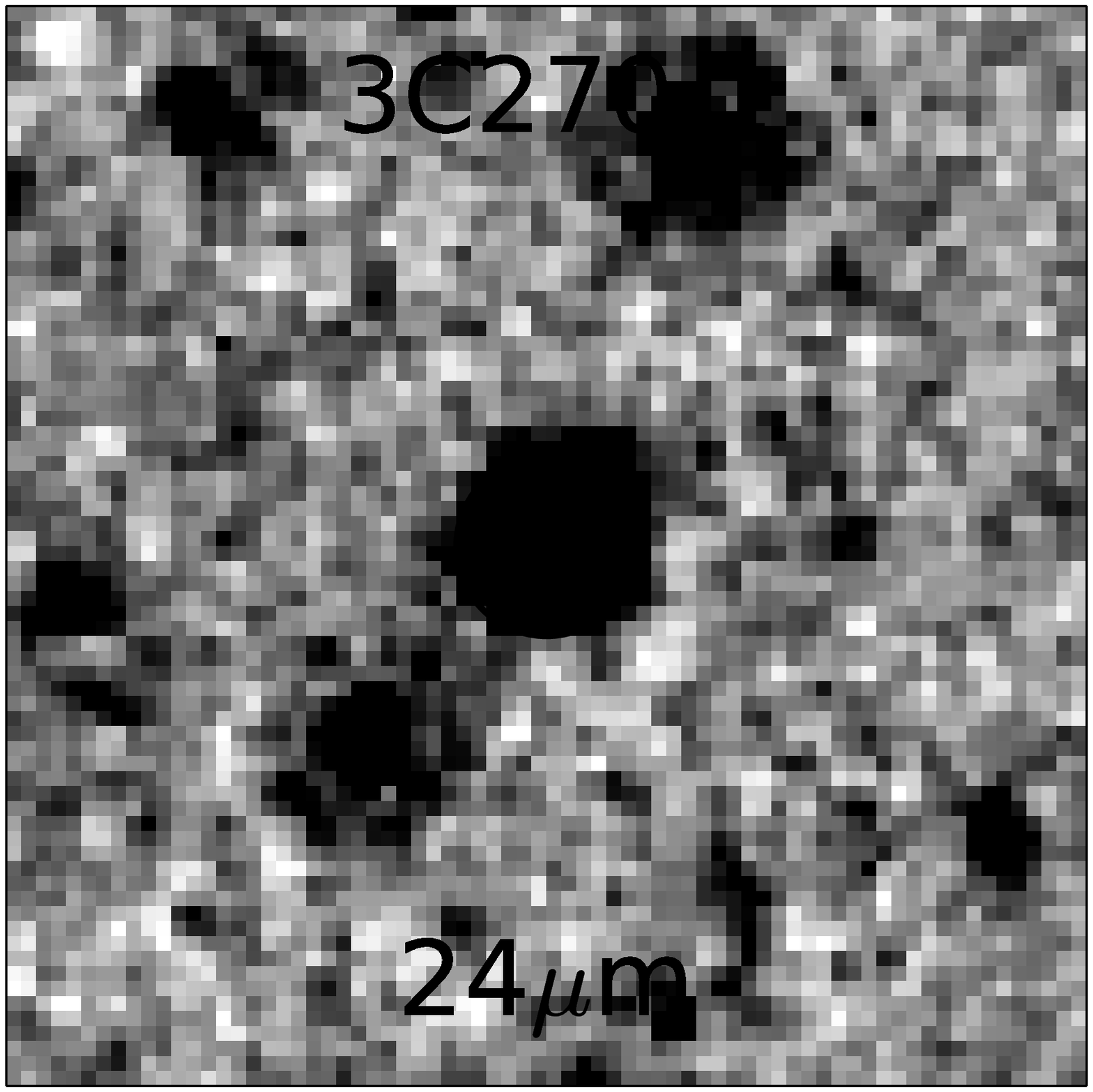} 
      \includegraphics[width=1.5cm]{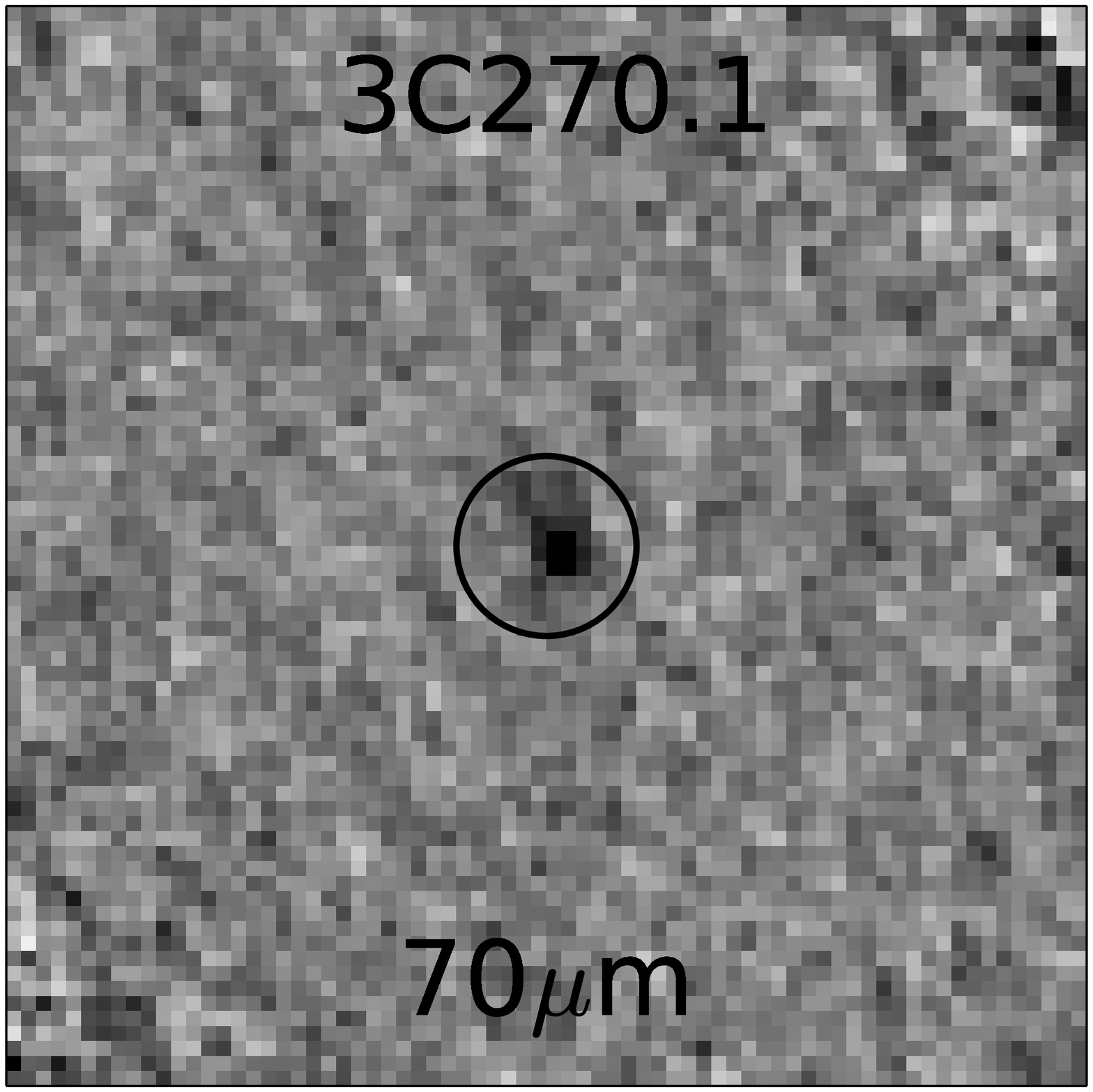}
      \includegraphics[width=1.5cm]{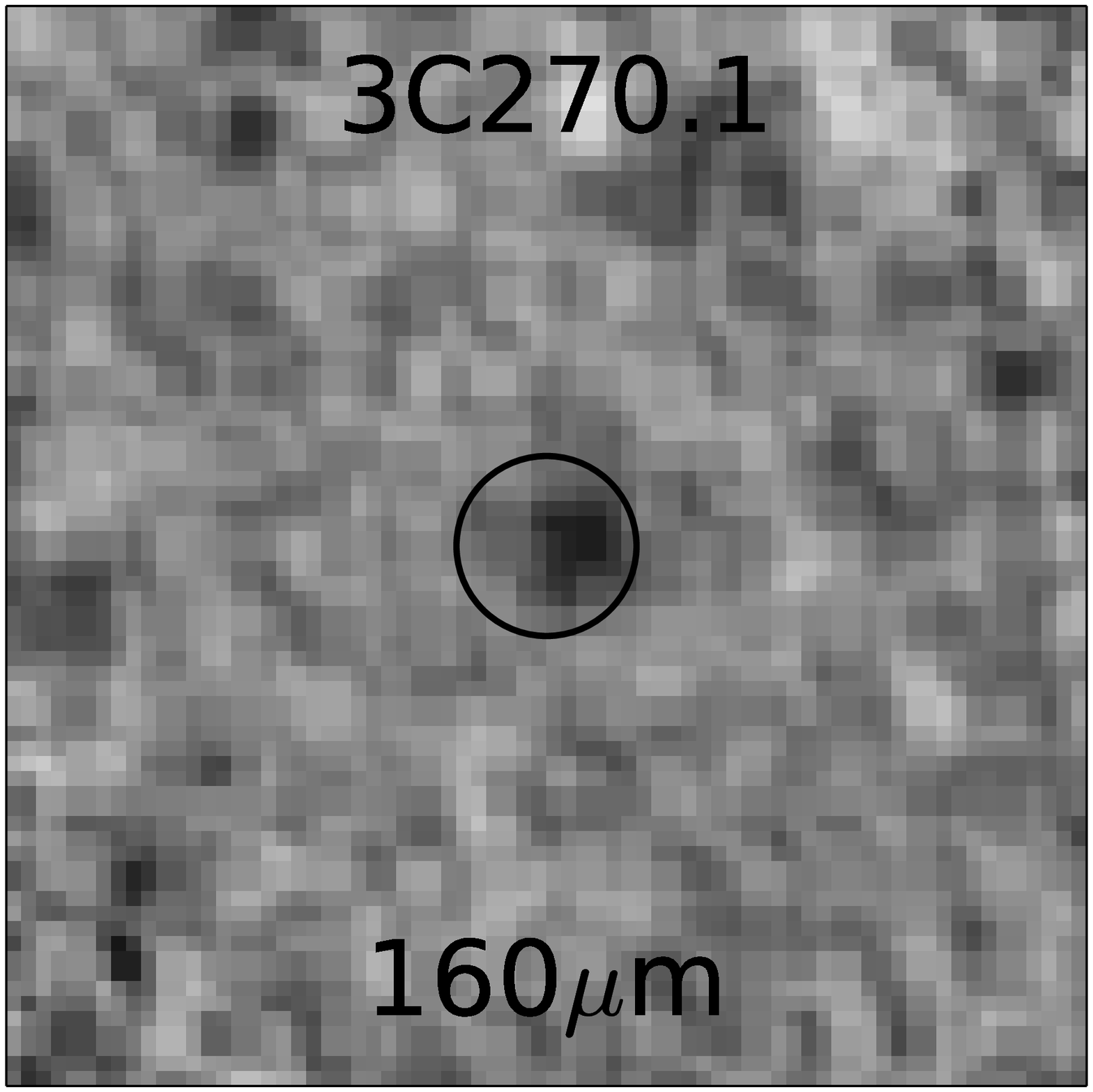}
      \includegraphics[width=1.5cm]{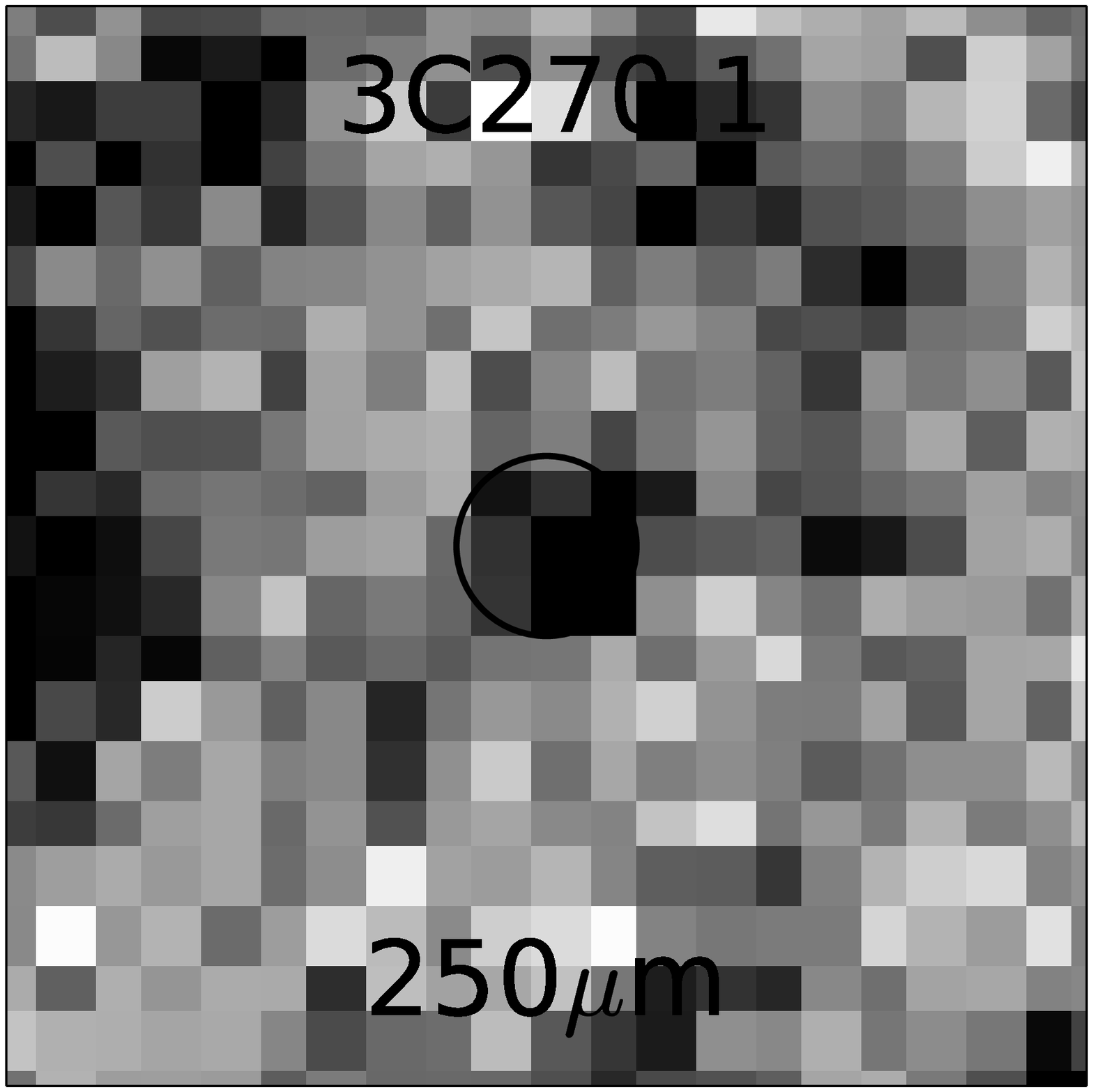}
      \includegraphics[width=1.5cm]{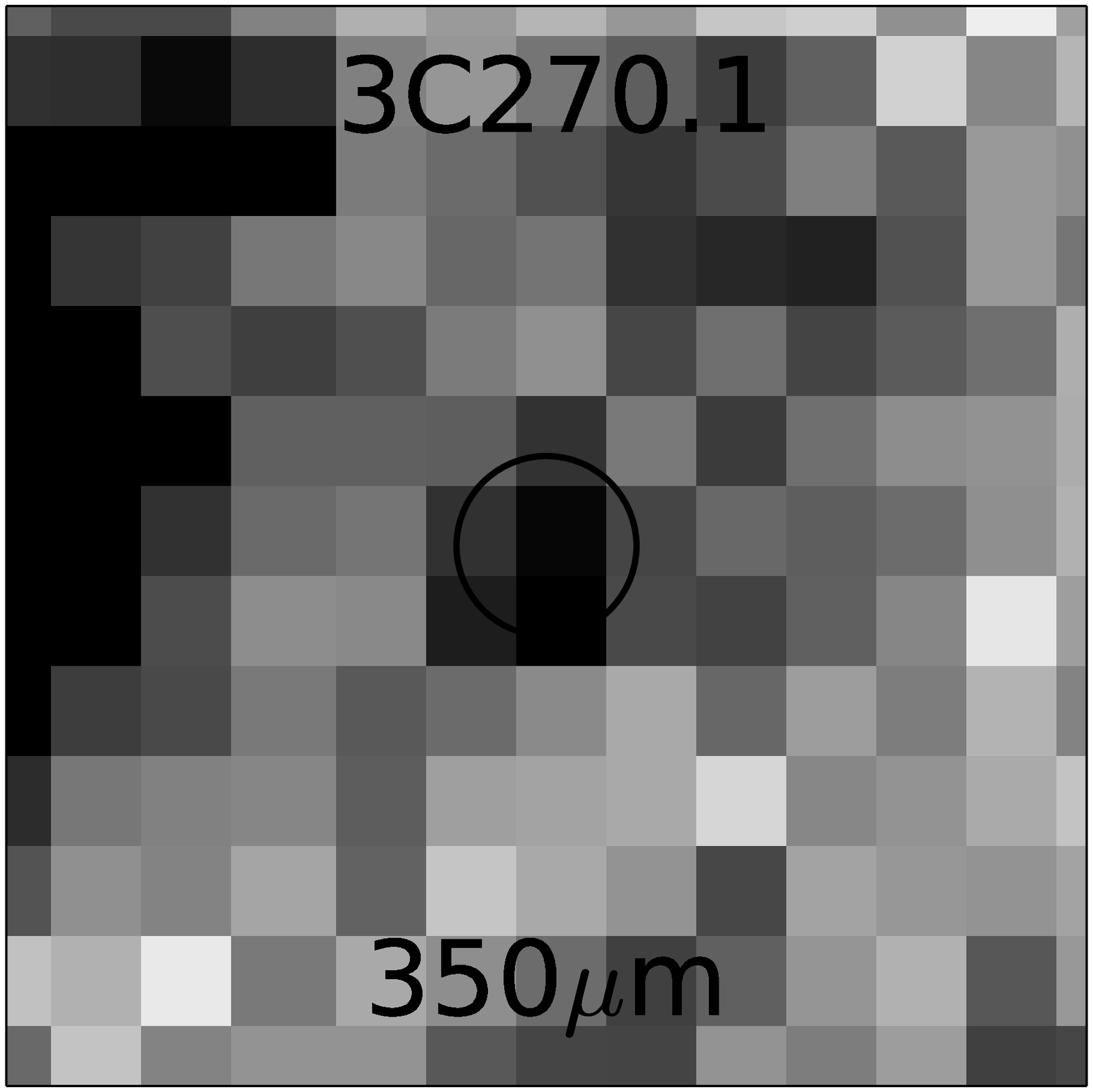}
      \includegraphics[width=1.5cm]{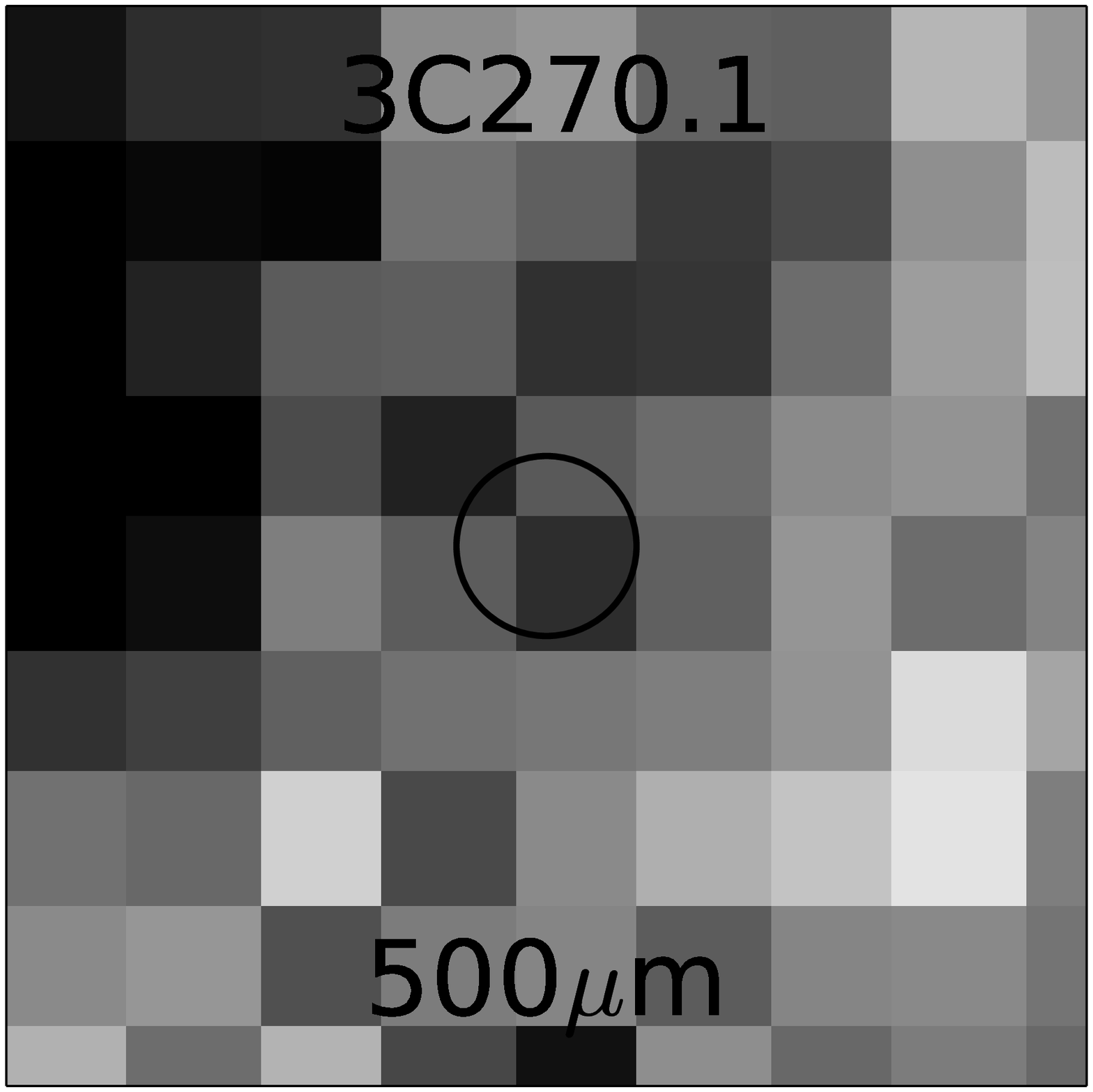}
      \\
      \includegraphics[width=1.5cm]{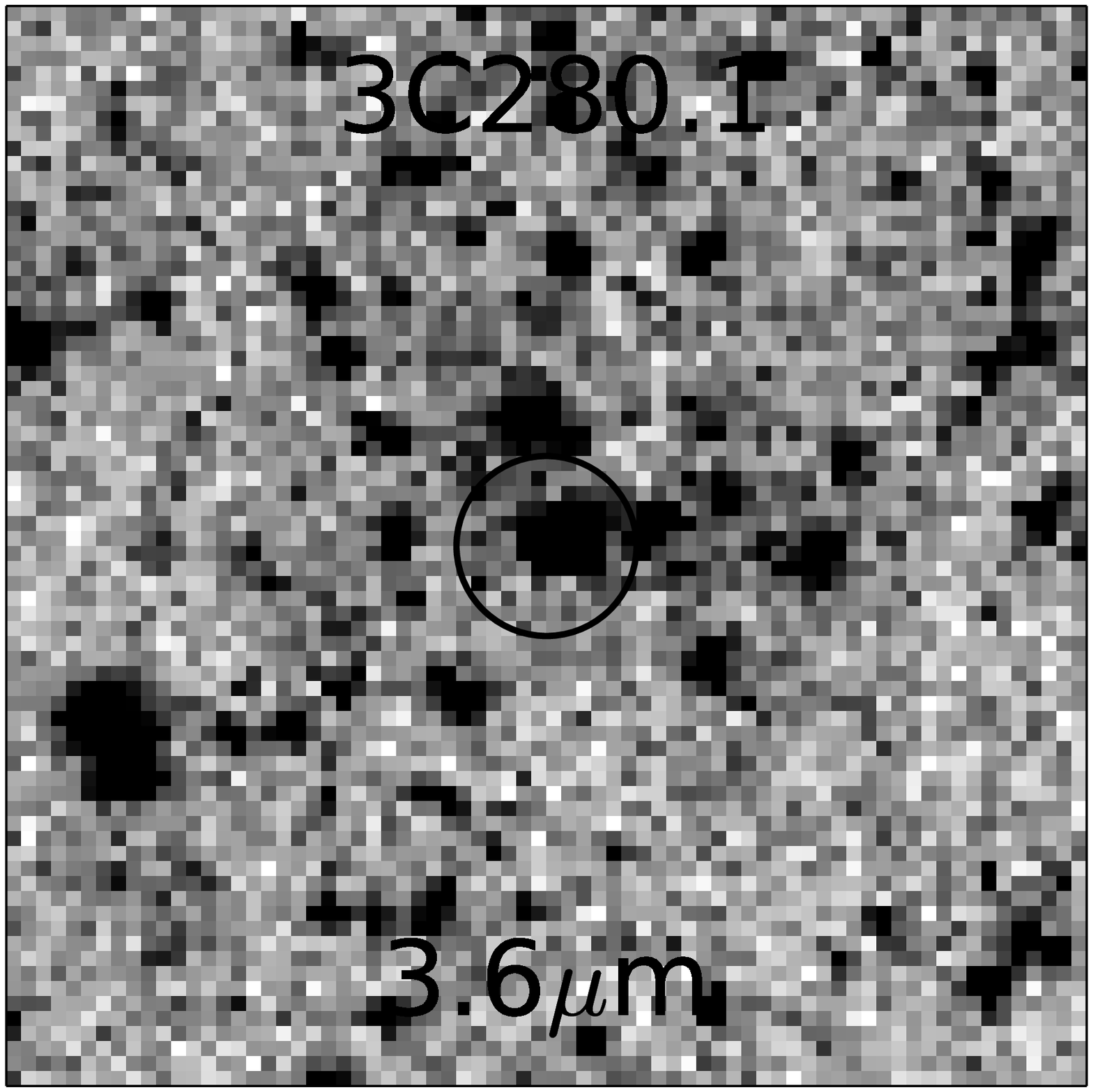}
      \includegraphics[width=1.5cm]{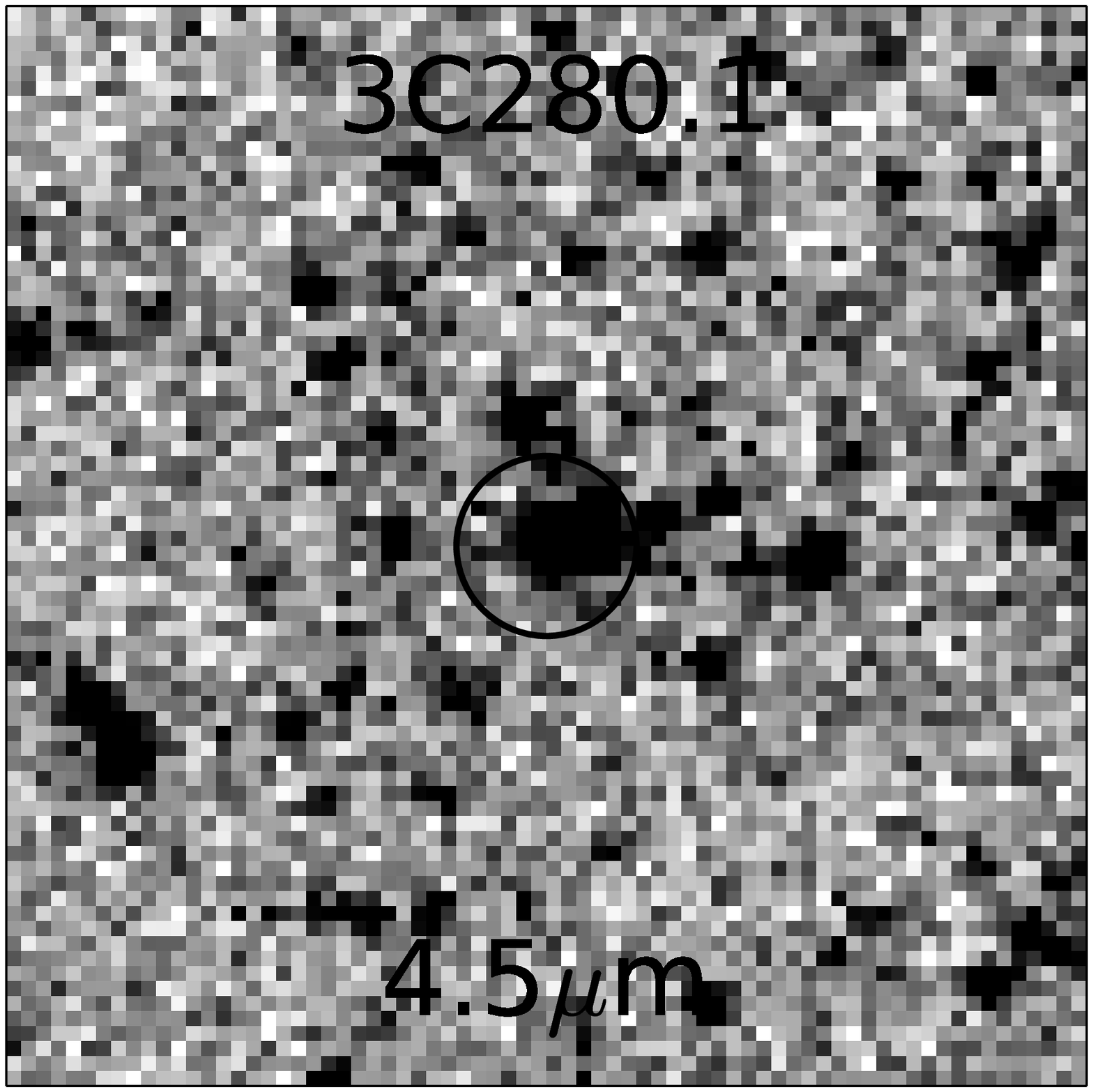}
      \includegraphics[width=1.5cm]{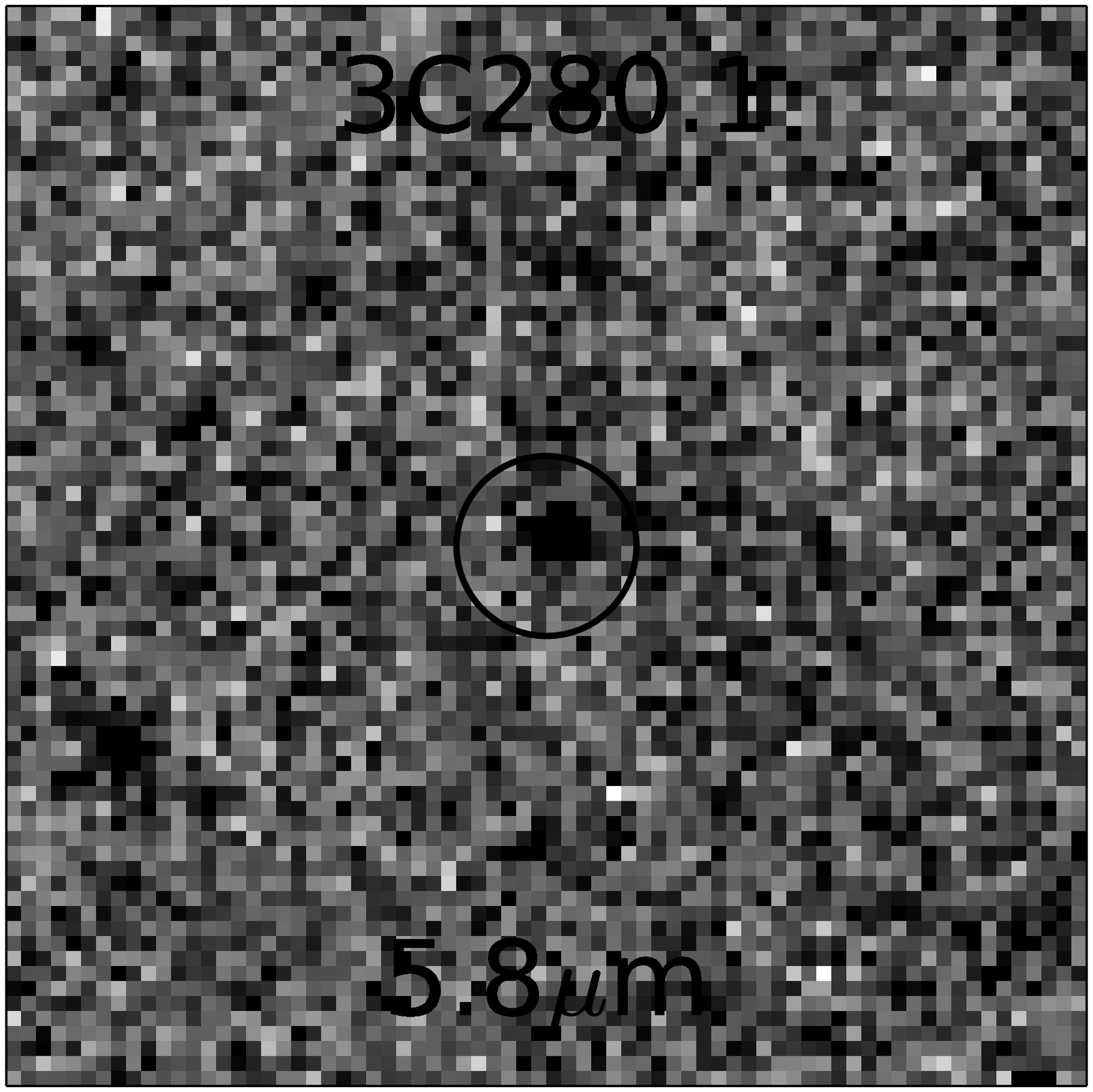}
      \includegraphics[width=1.5cm]{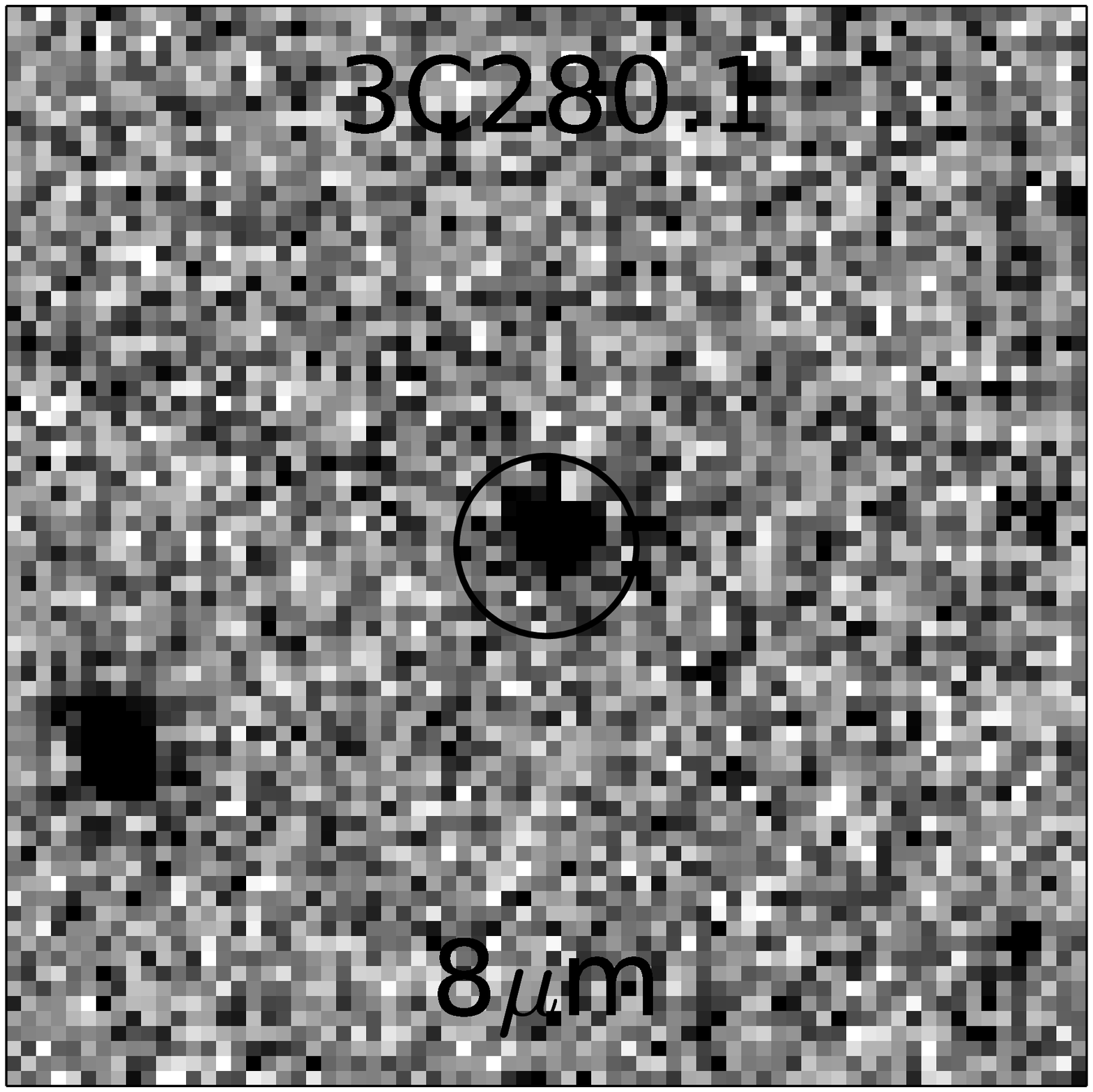}
      \includegraphics[width=1.5cm]{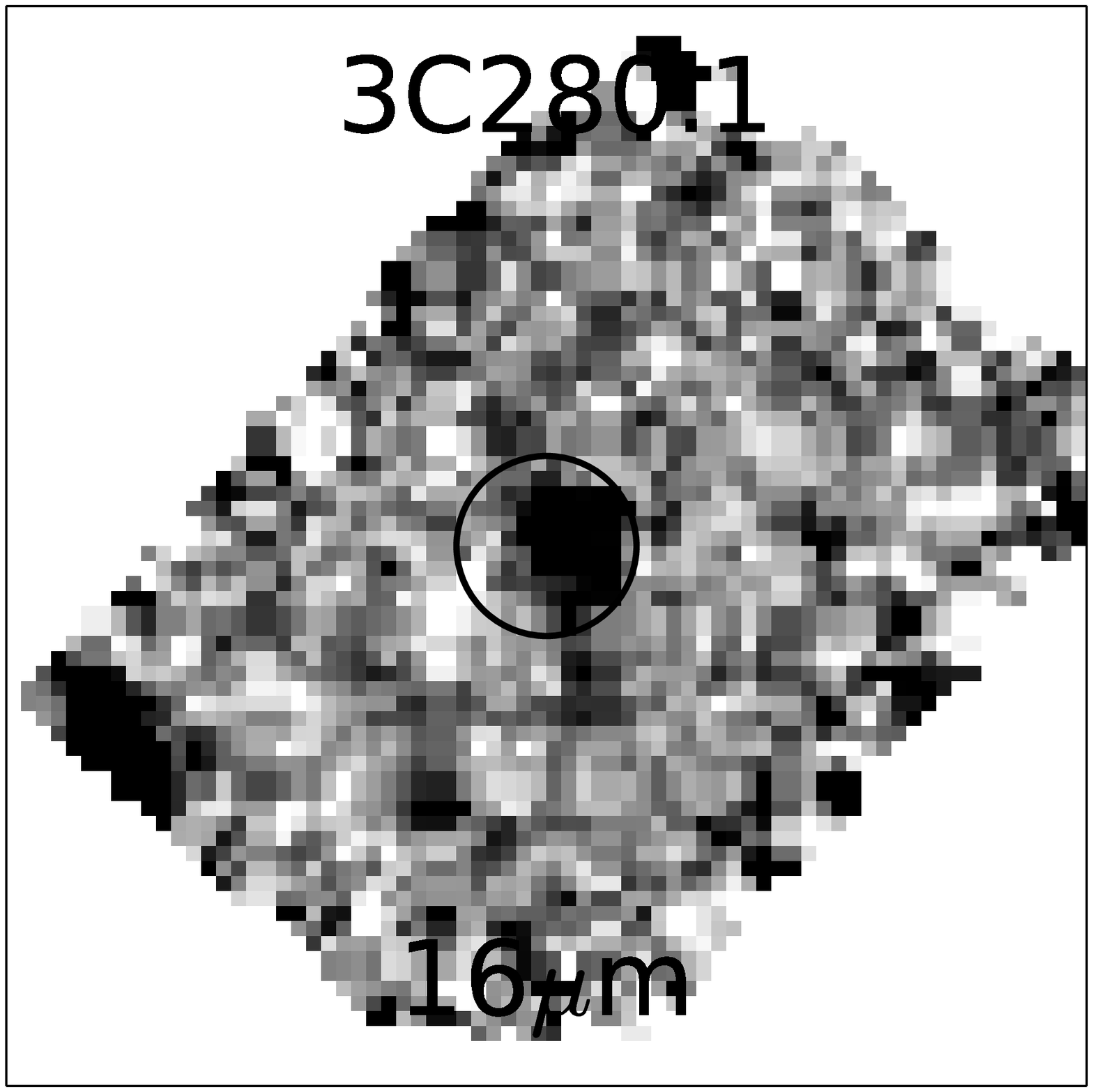}
      \includegraphics[width=1.5cm]{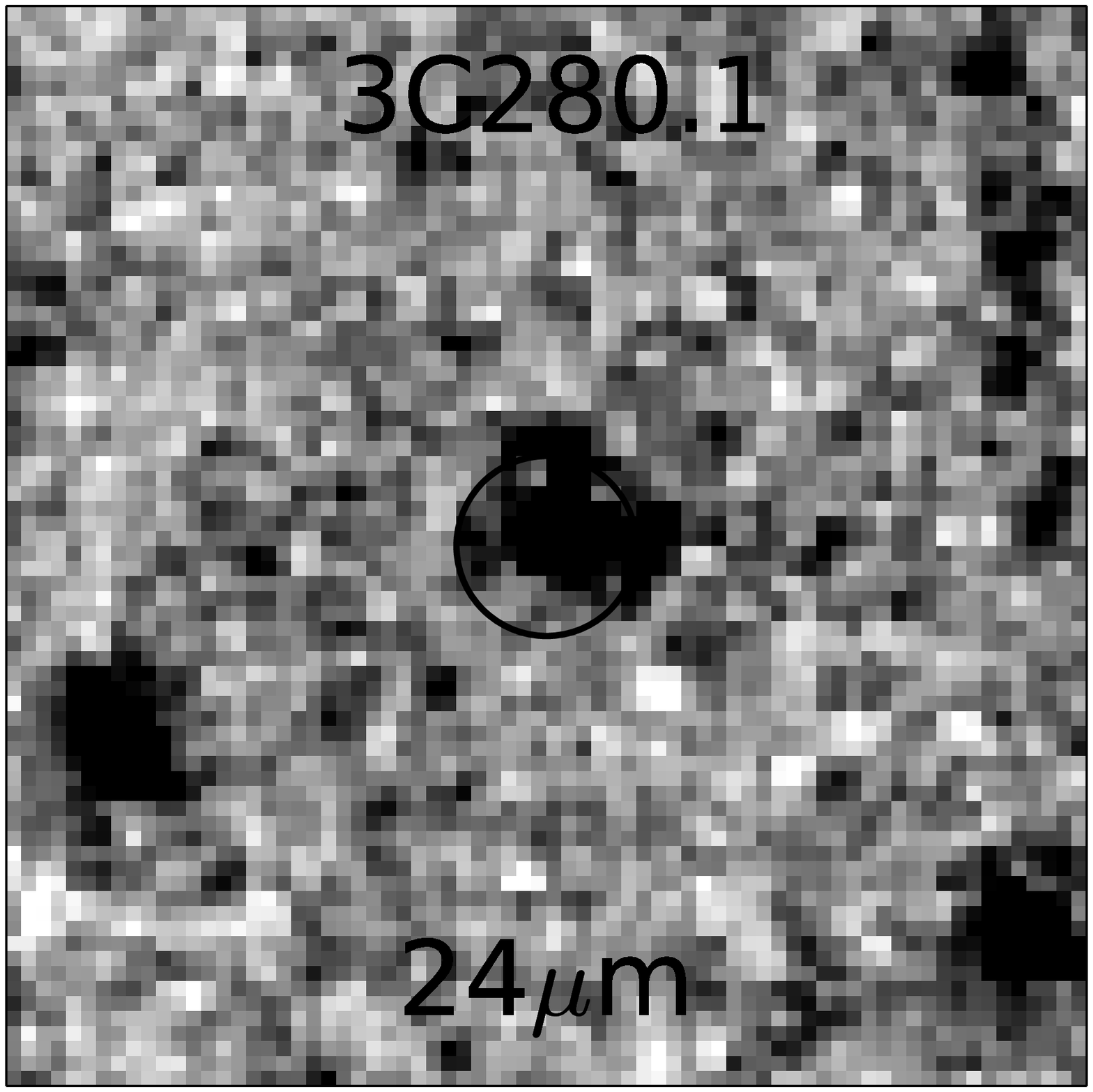}
      \includegraphics[width=1.5cm]{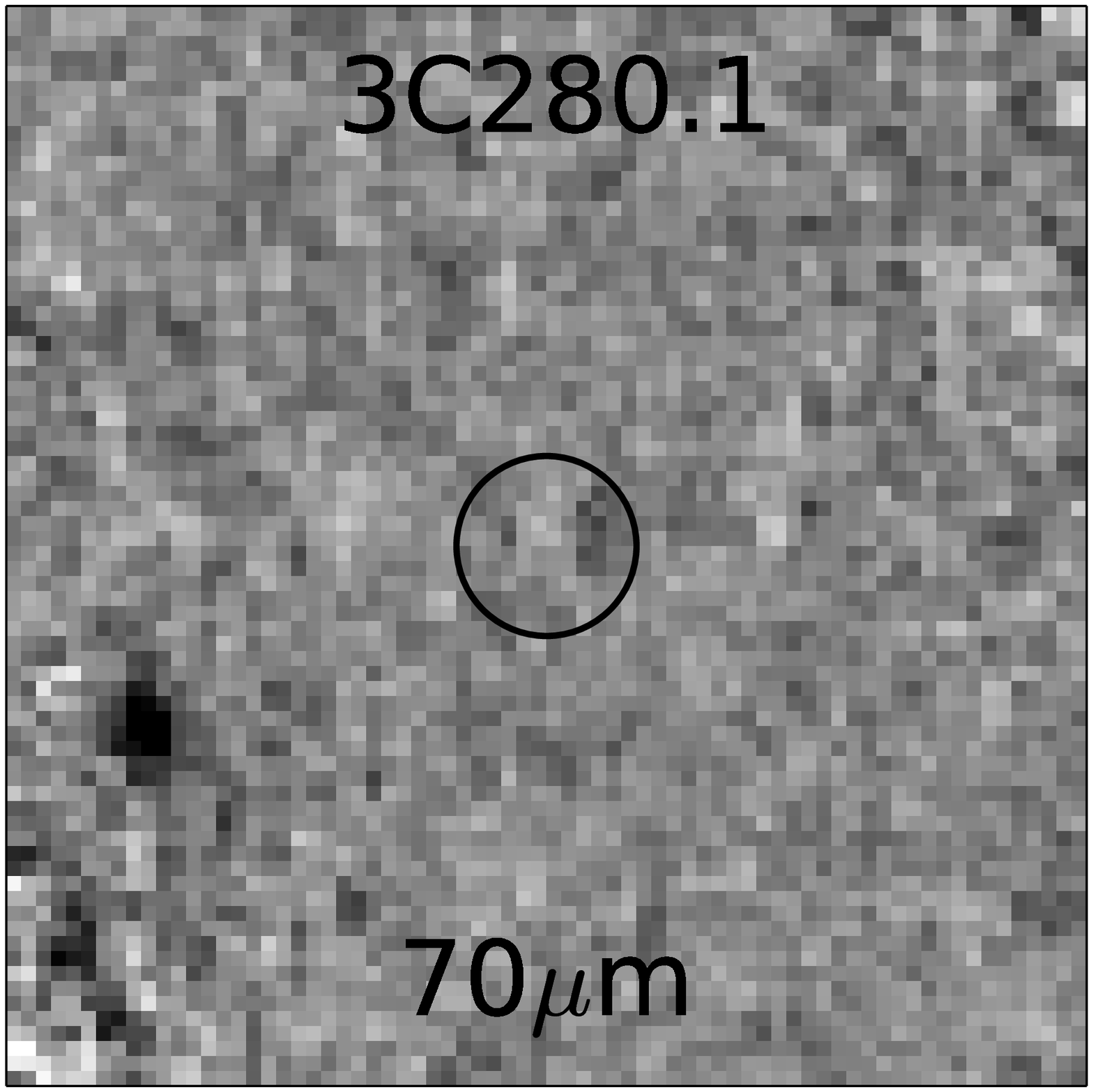}
      \includegraphics[width=1.5cm]{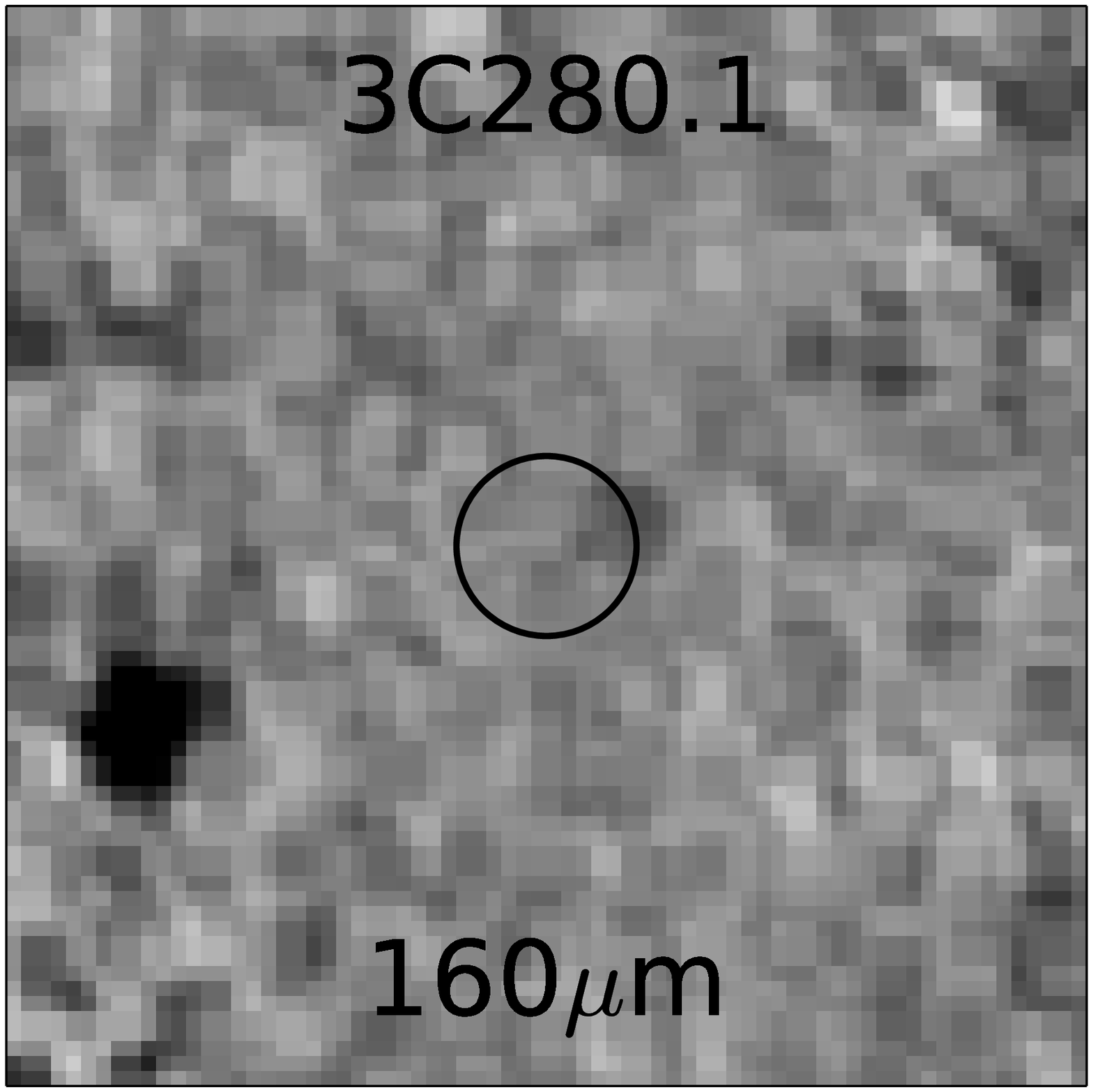}
      \includegraphics[width=1.5cm]{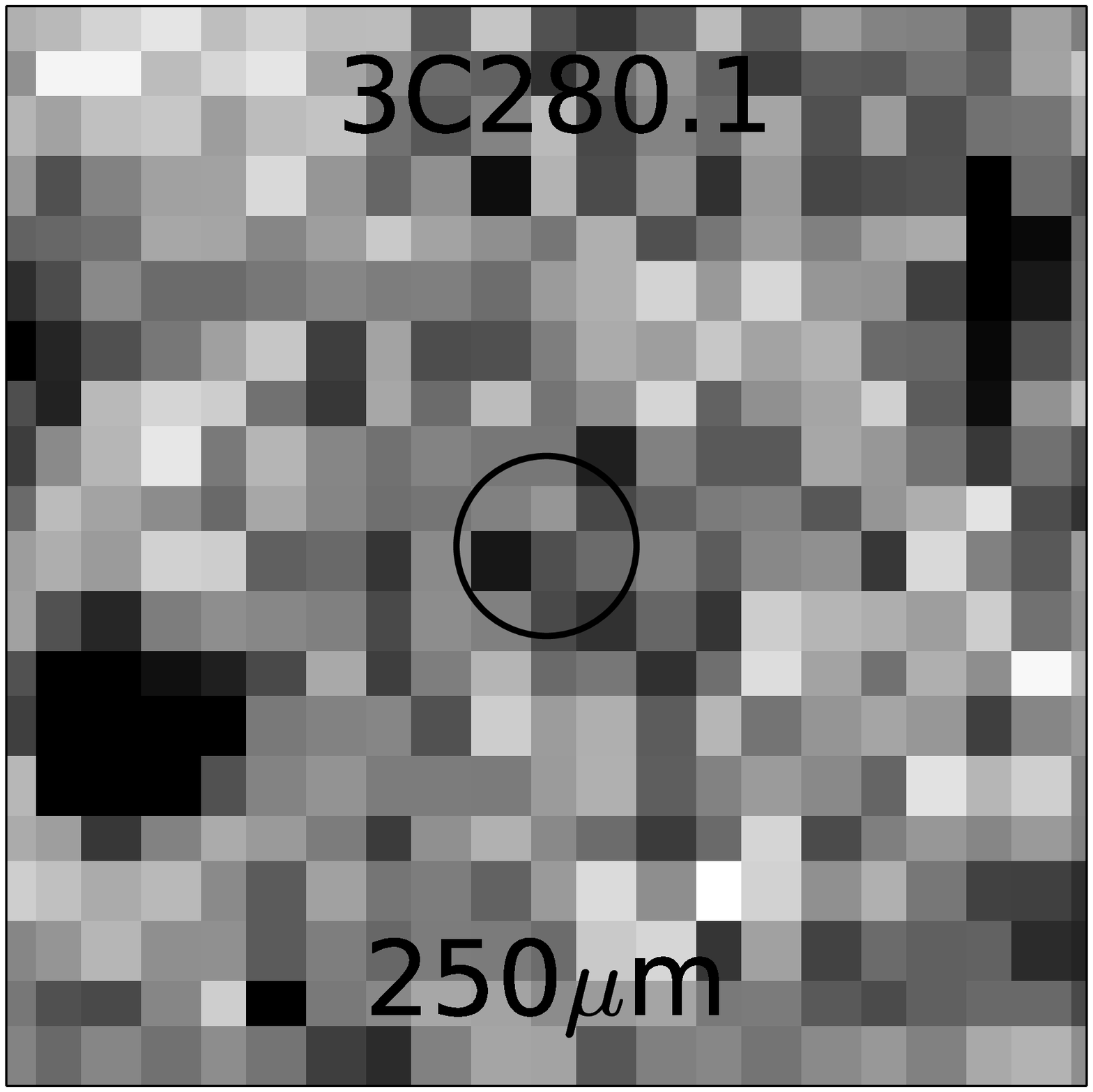}
      \includegraphics[width=1.5cm]{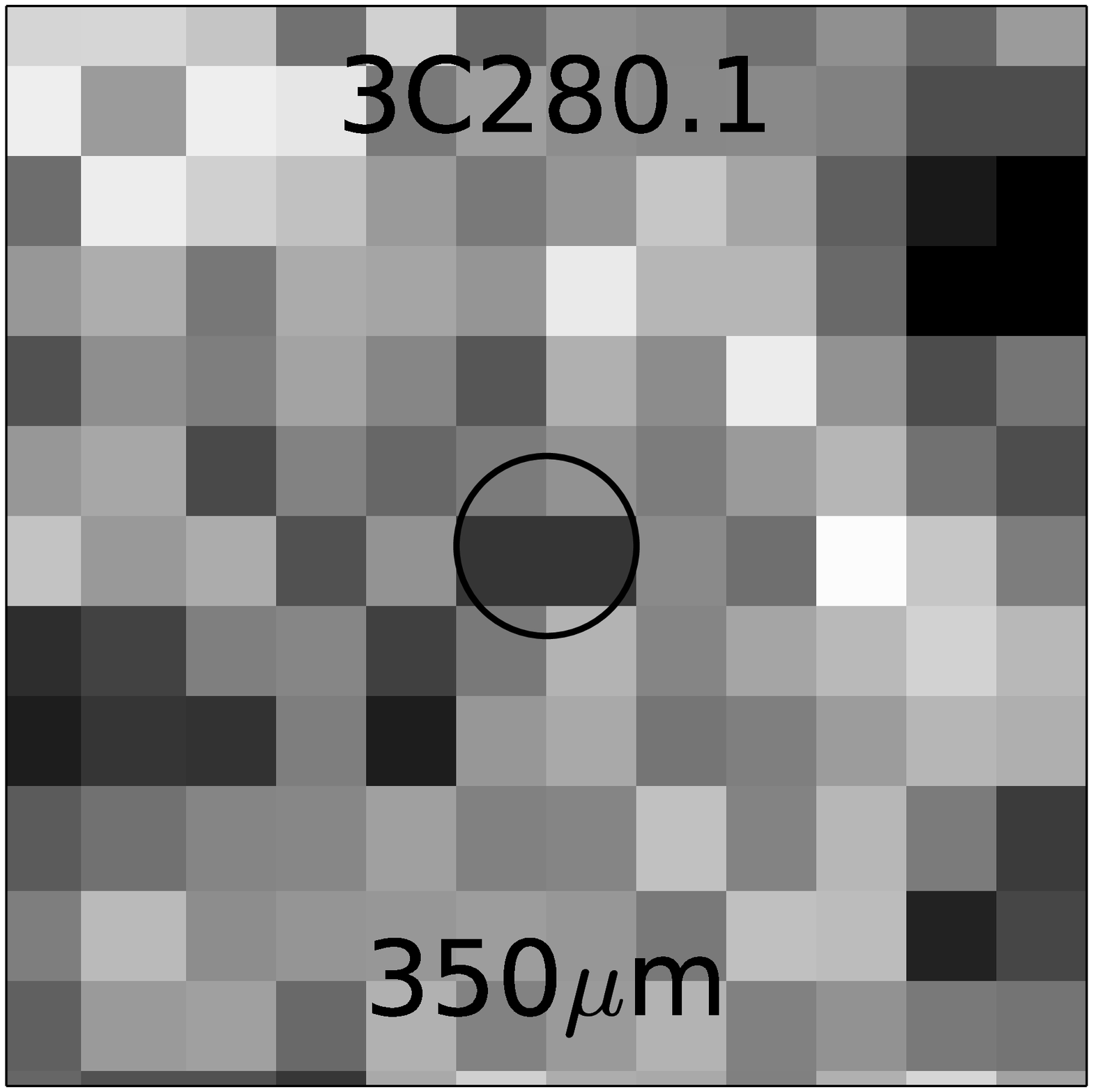}
      \includegraphics[width=1.5cm]{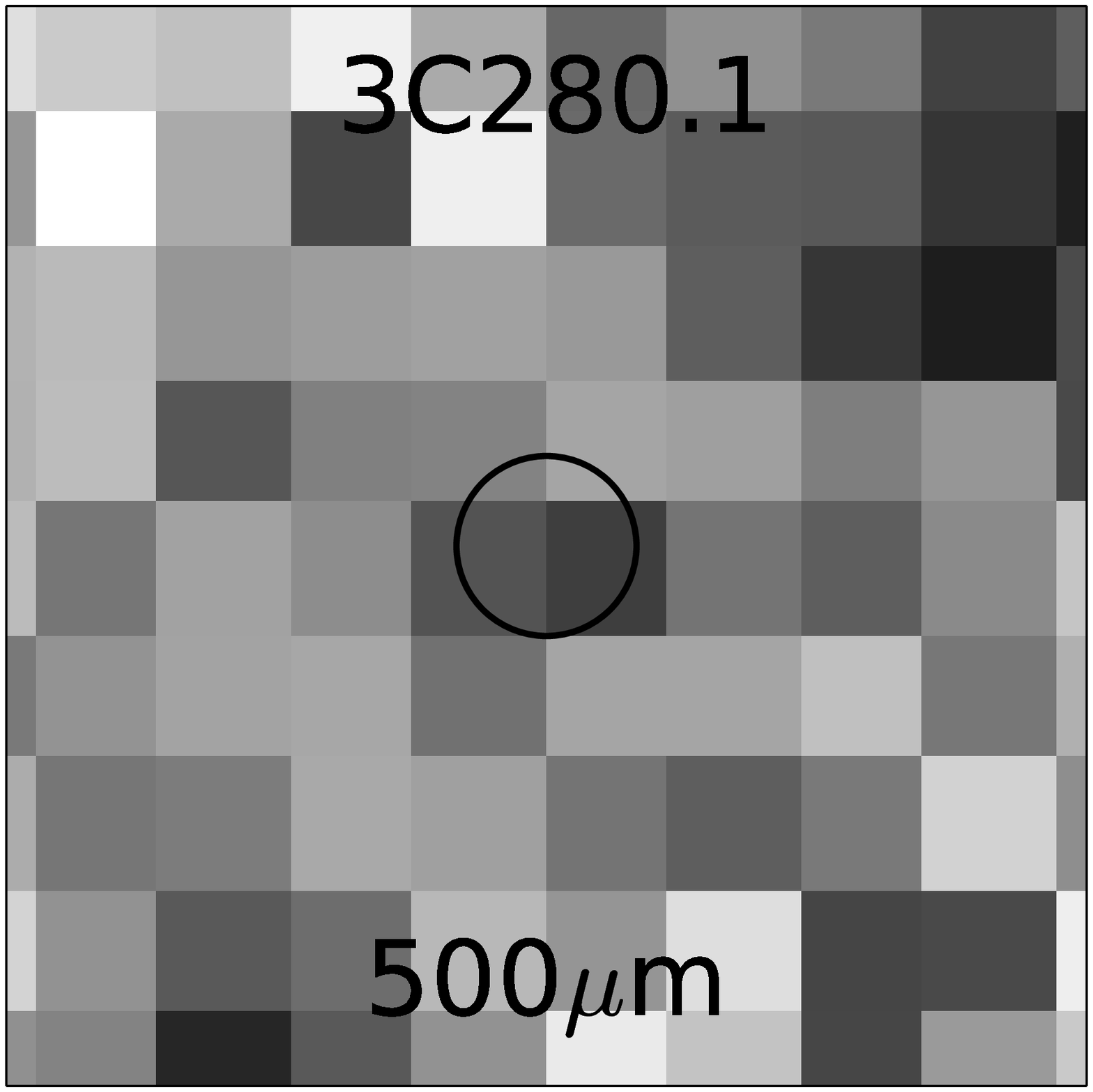}
      \\
      \includegraphics[width=1.5cm]{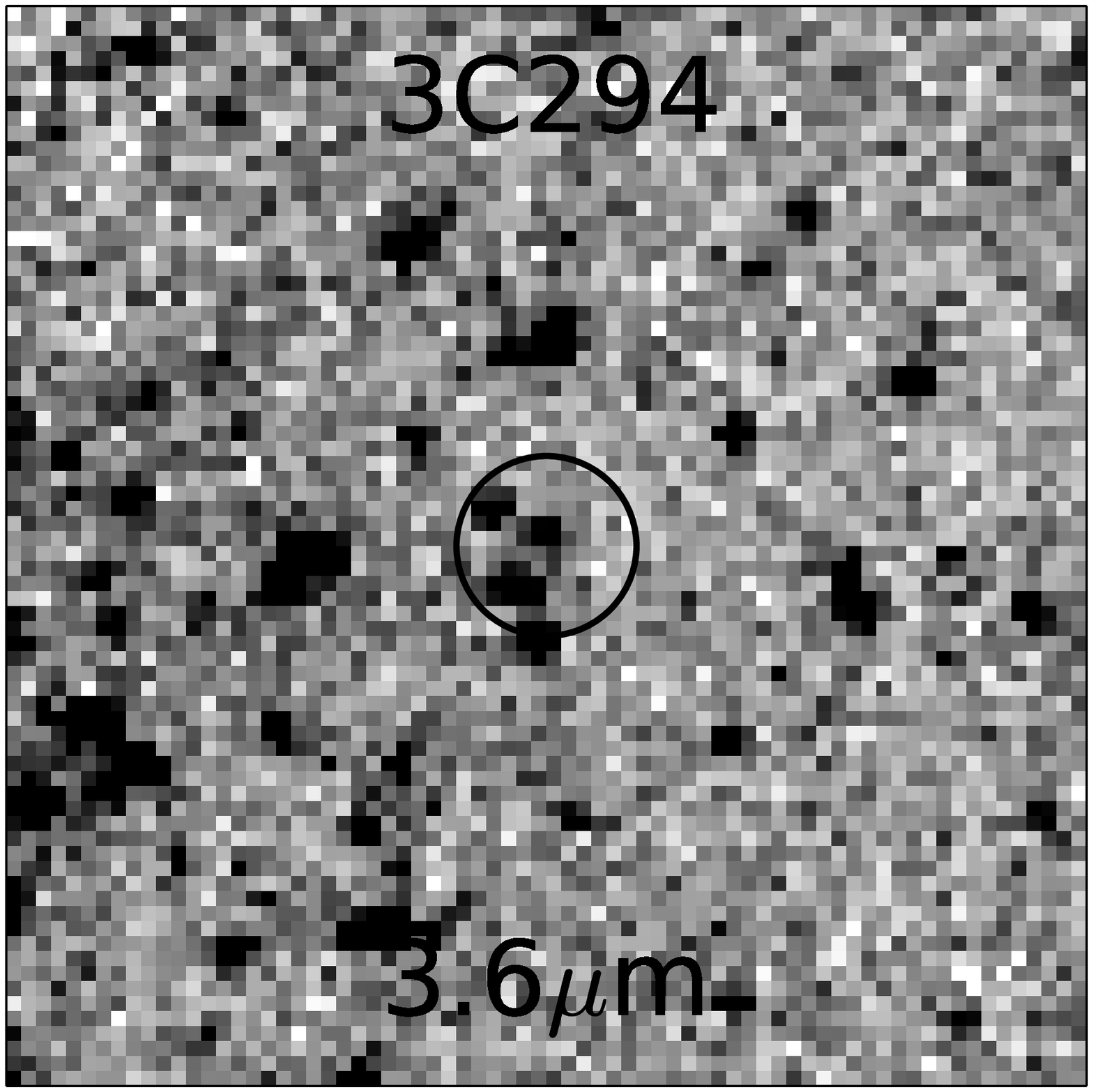}
      \includegraphics[width=1.5cm]{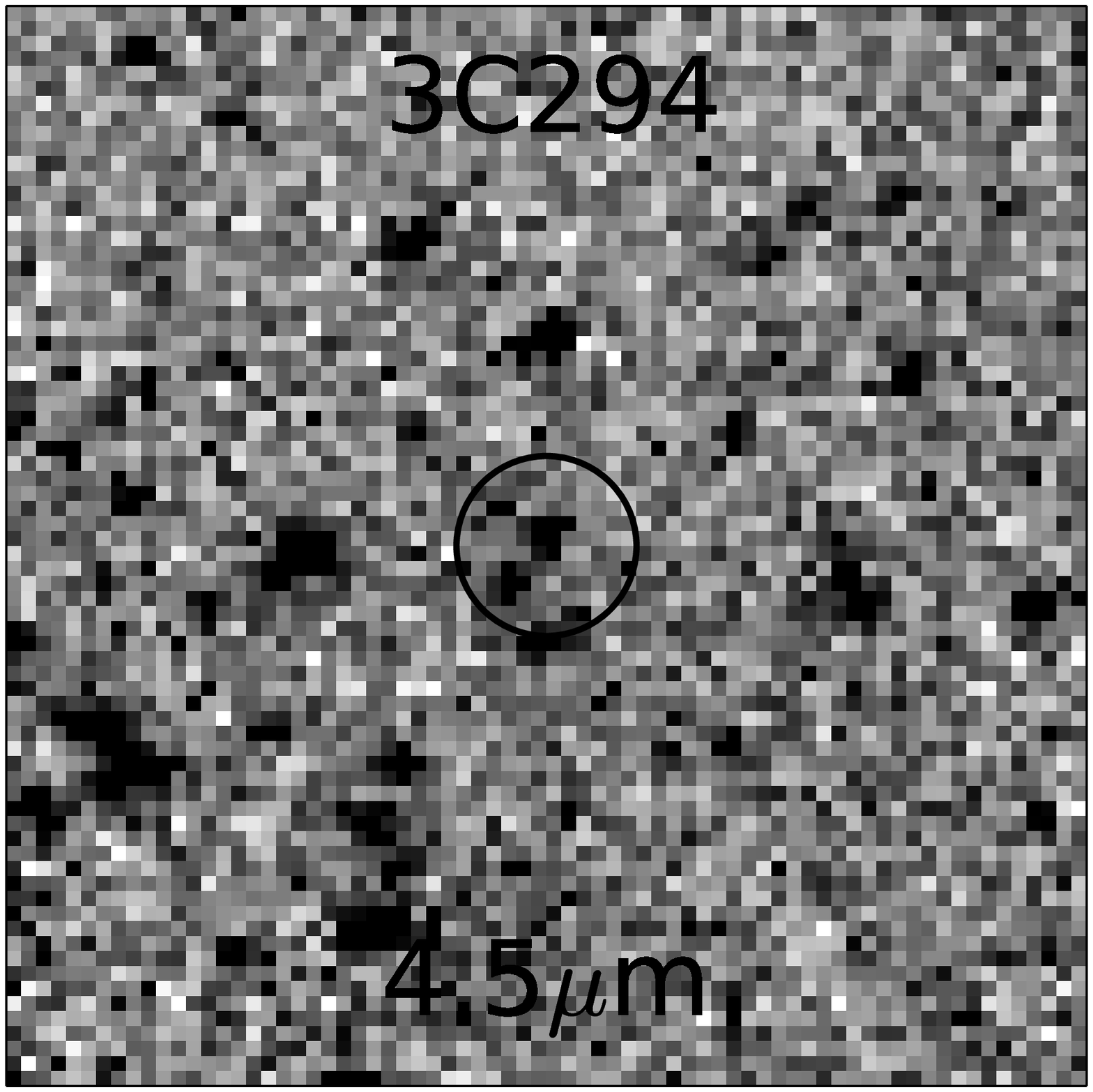}
      \includegraphics[width=1.5cm]{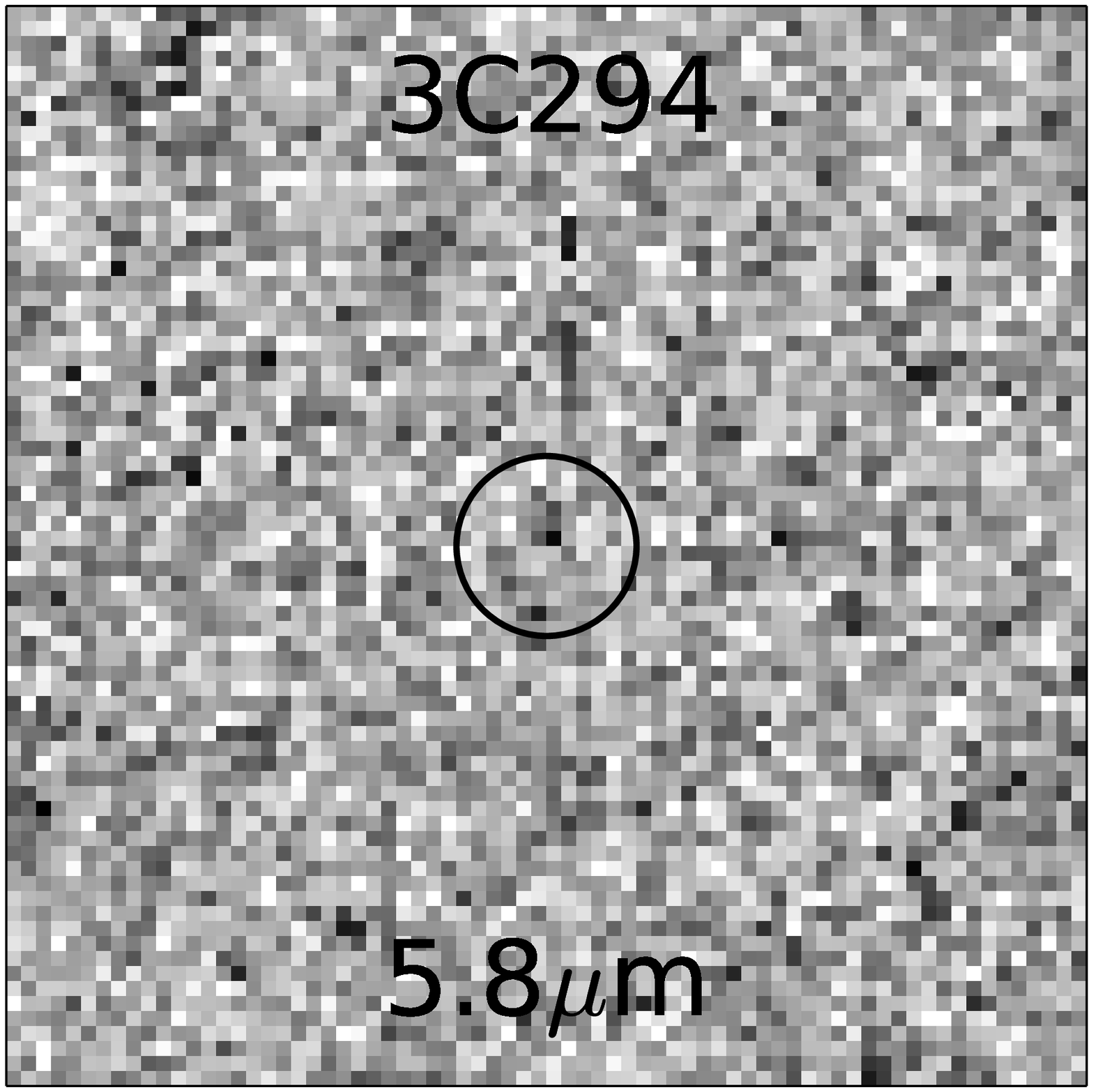}
      \includegraphics[width=1.5cm]{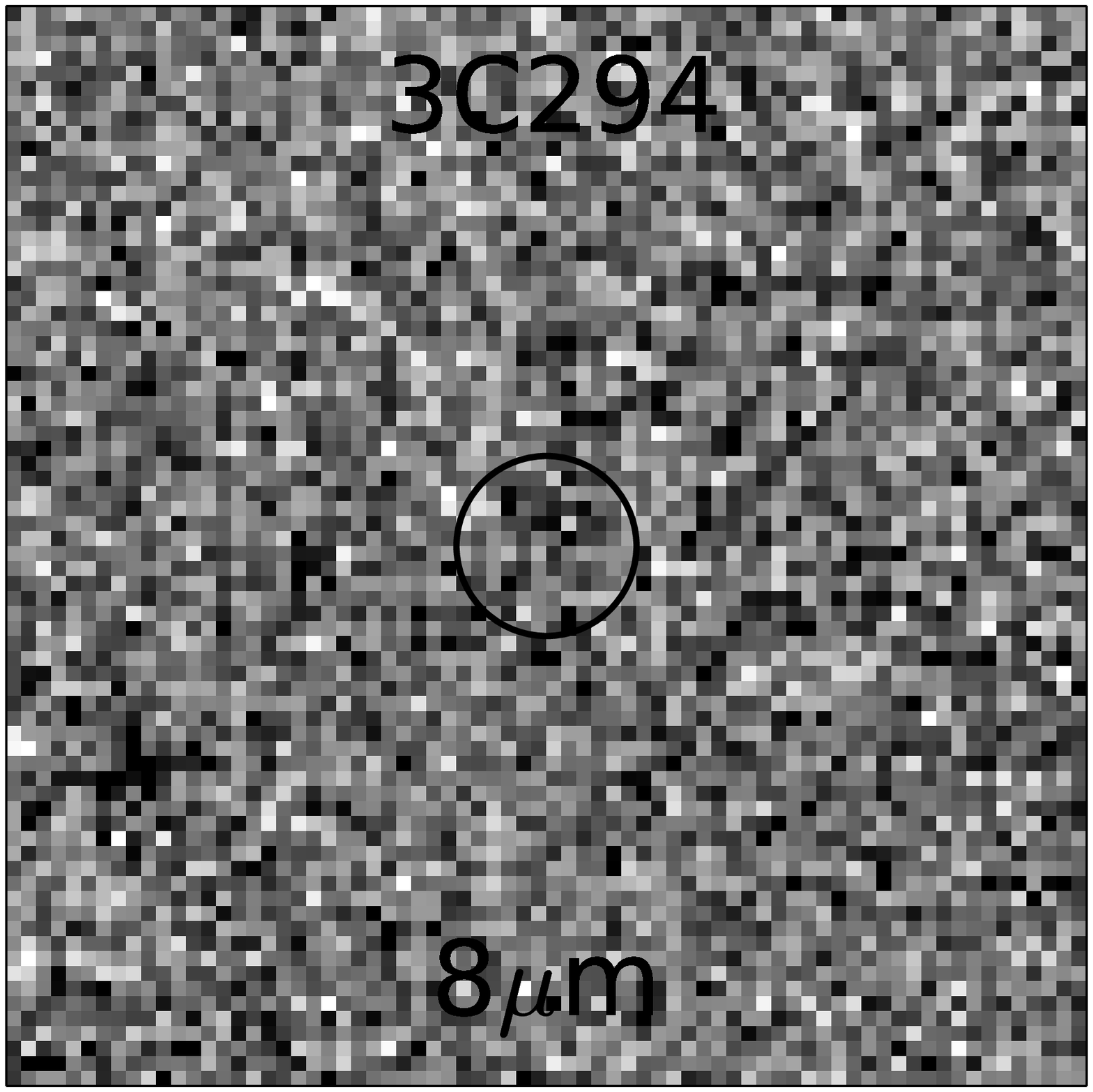}
      \includegraphics[width=1.5cm]{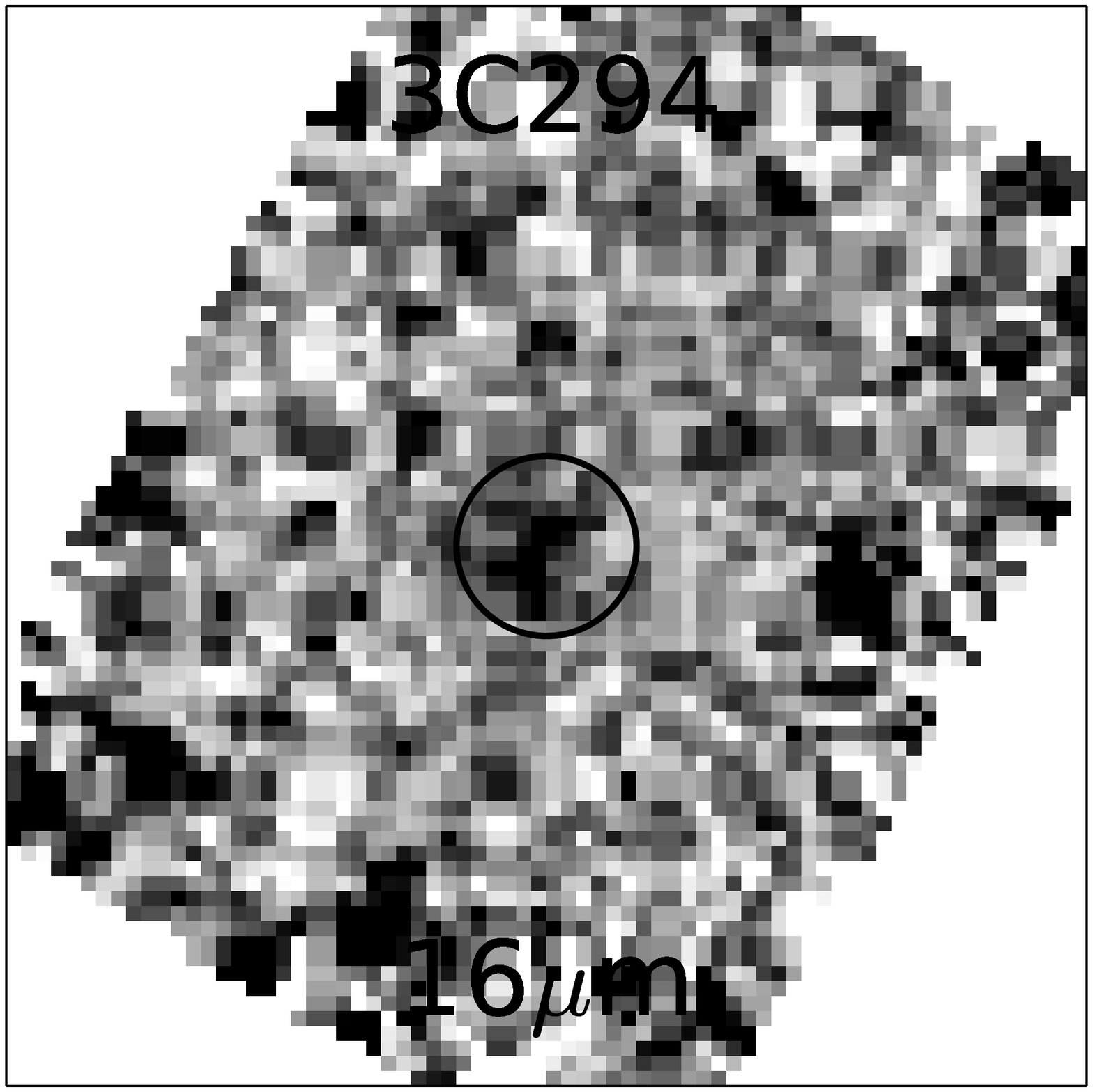}
      \includegraphics[width=1.5cm]{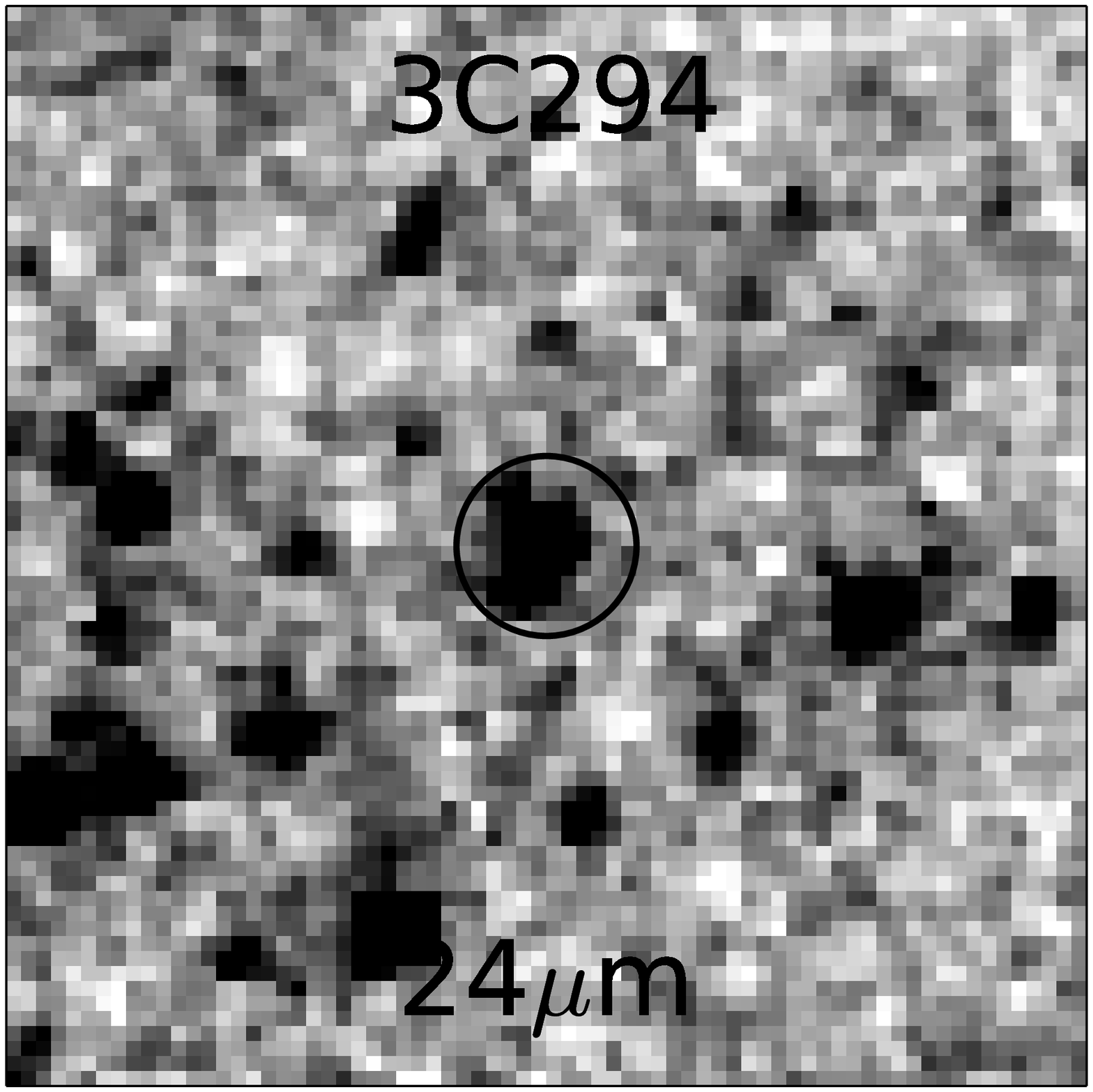}
      \includegraphics[width=1.5cm]{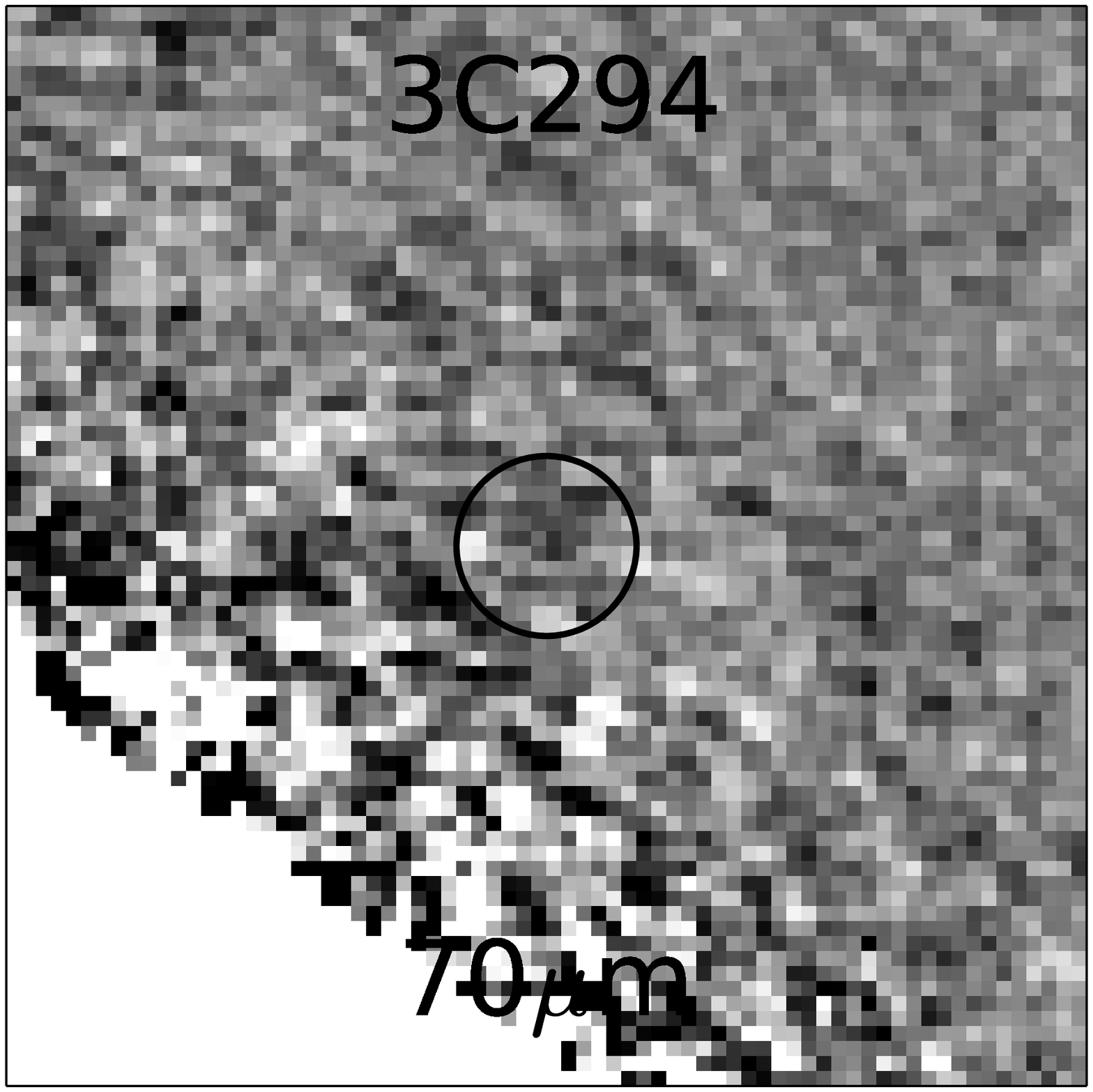}
      \includegraphics[width=1.5cm]{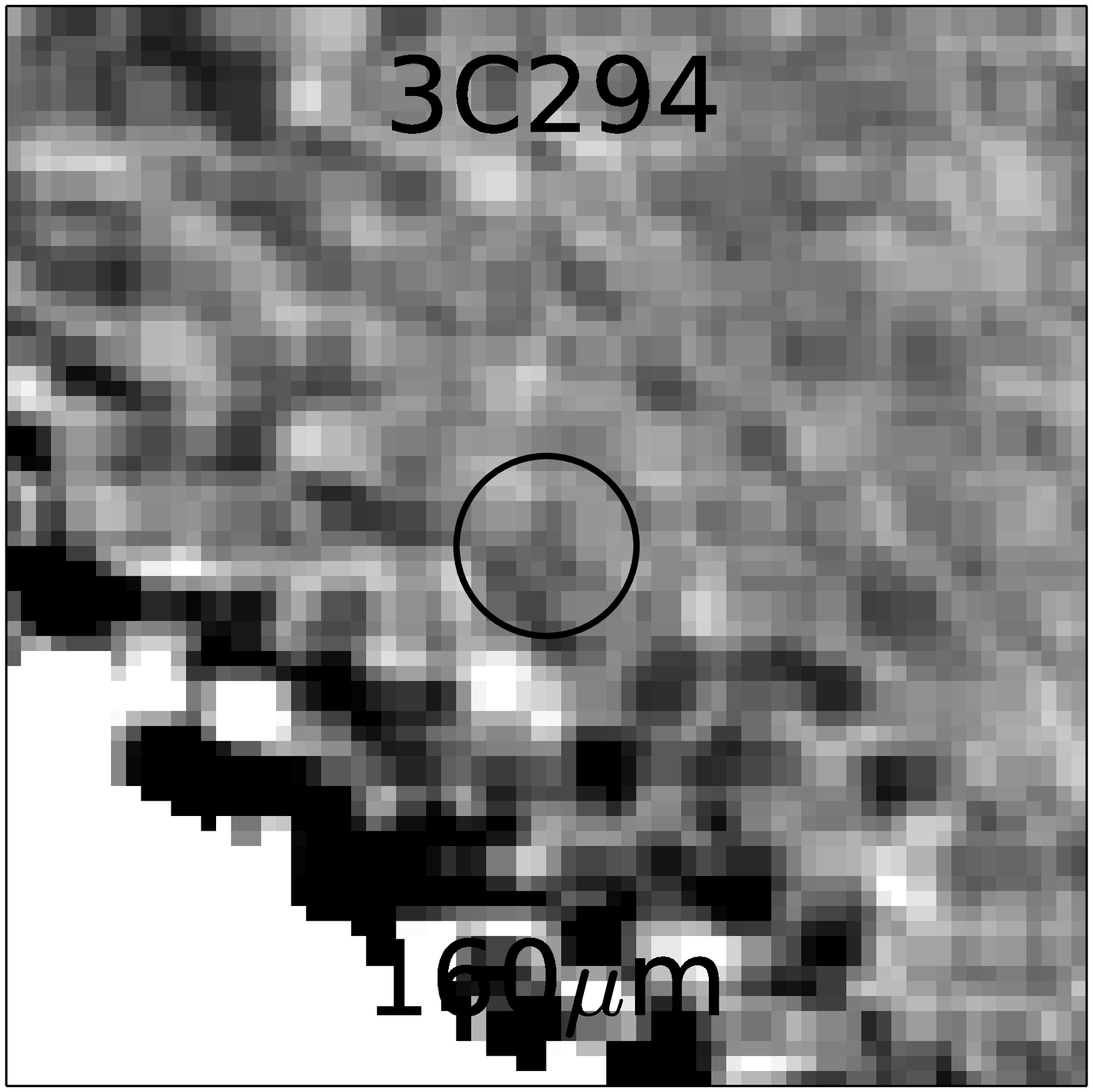}
      \includegraphics[width=1.5cm]{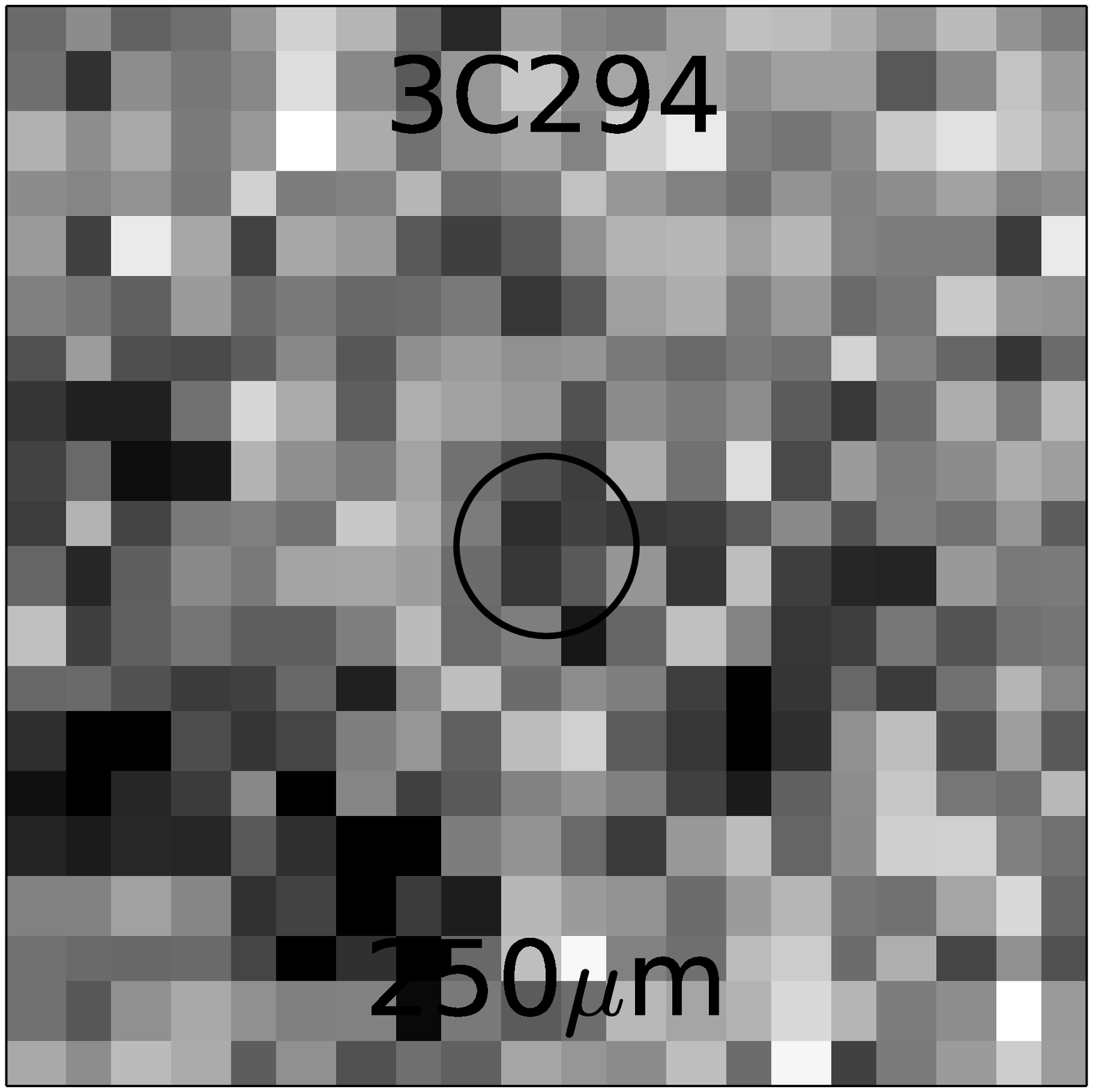}
      \includegraphics[width=1.5cm]{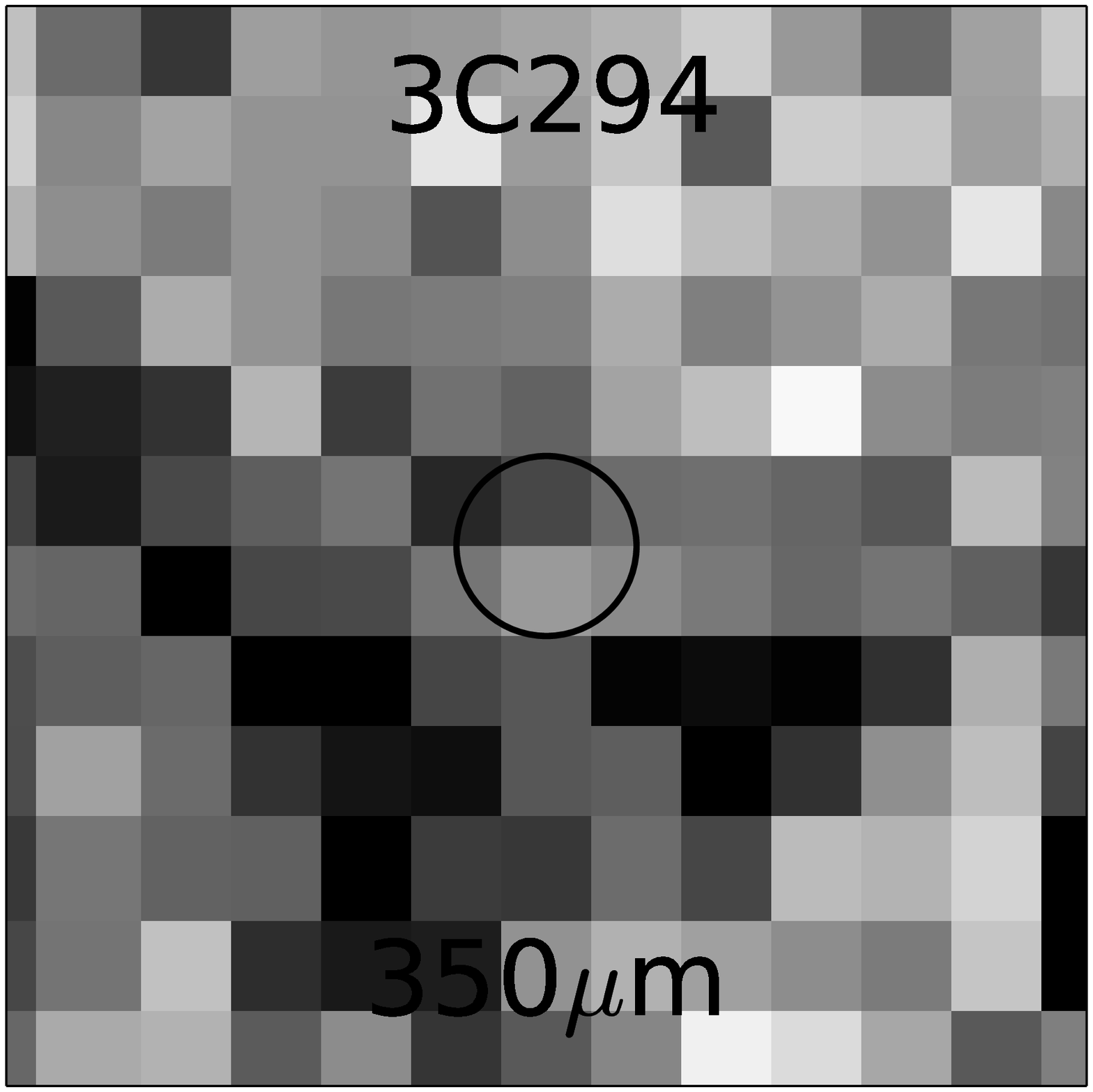}
      \includegraphics[width=1.5cm]{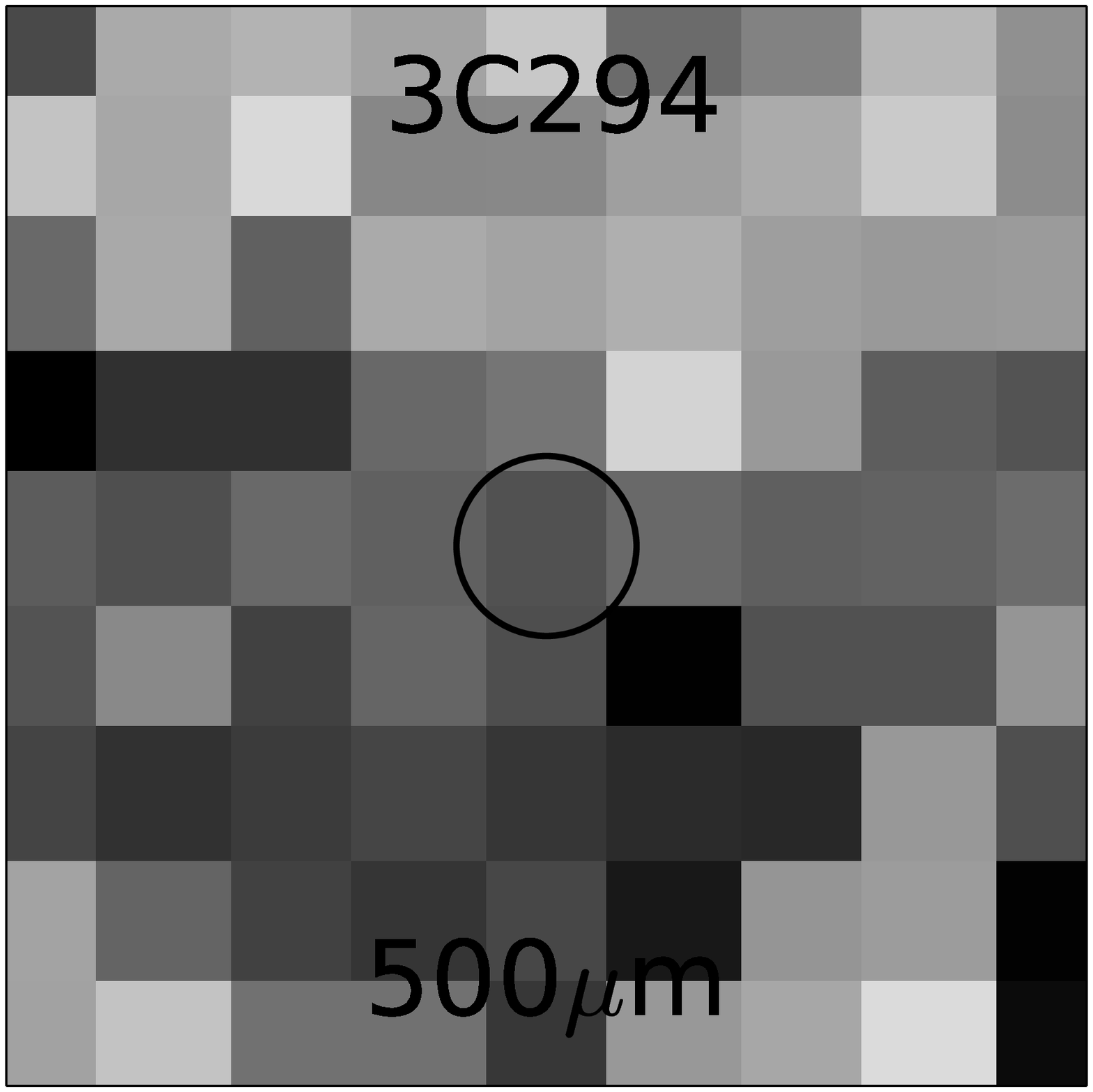}
      \\
      \includegraphics[width=1.5cm]{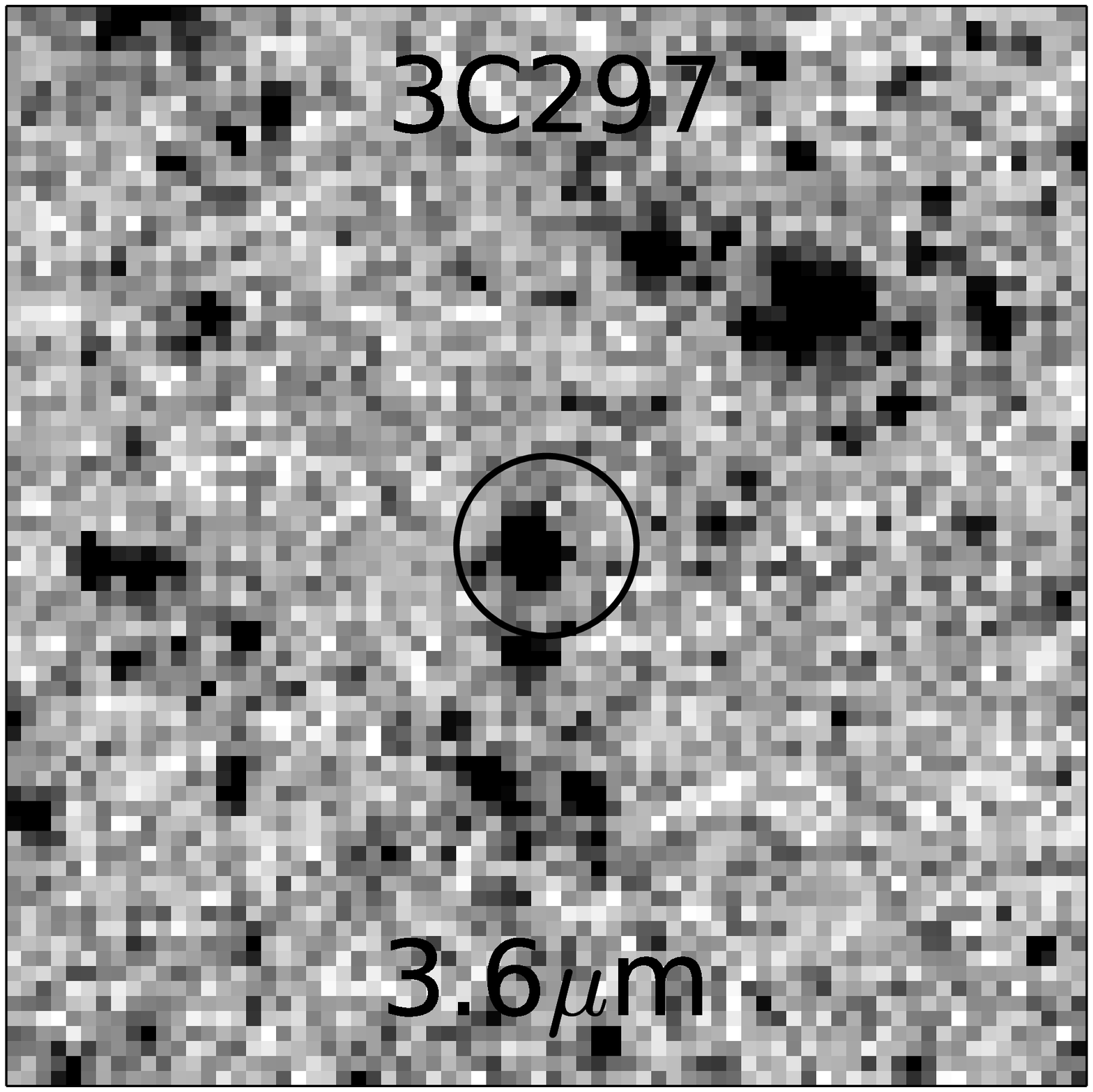}
      \includegraphics[width=1.5cm]{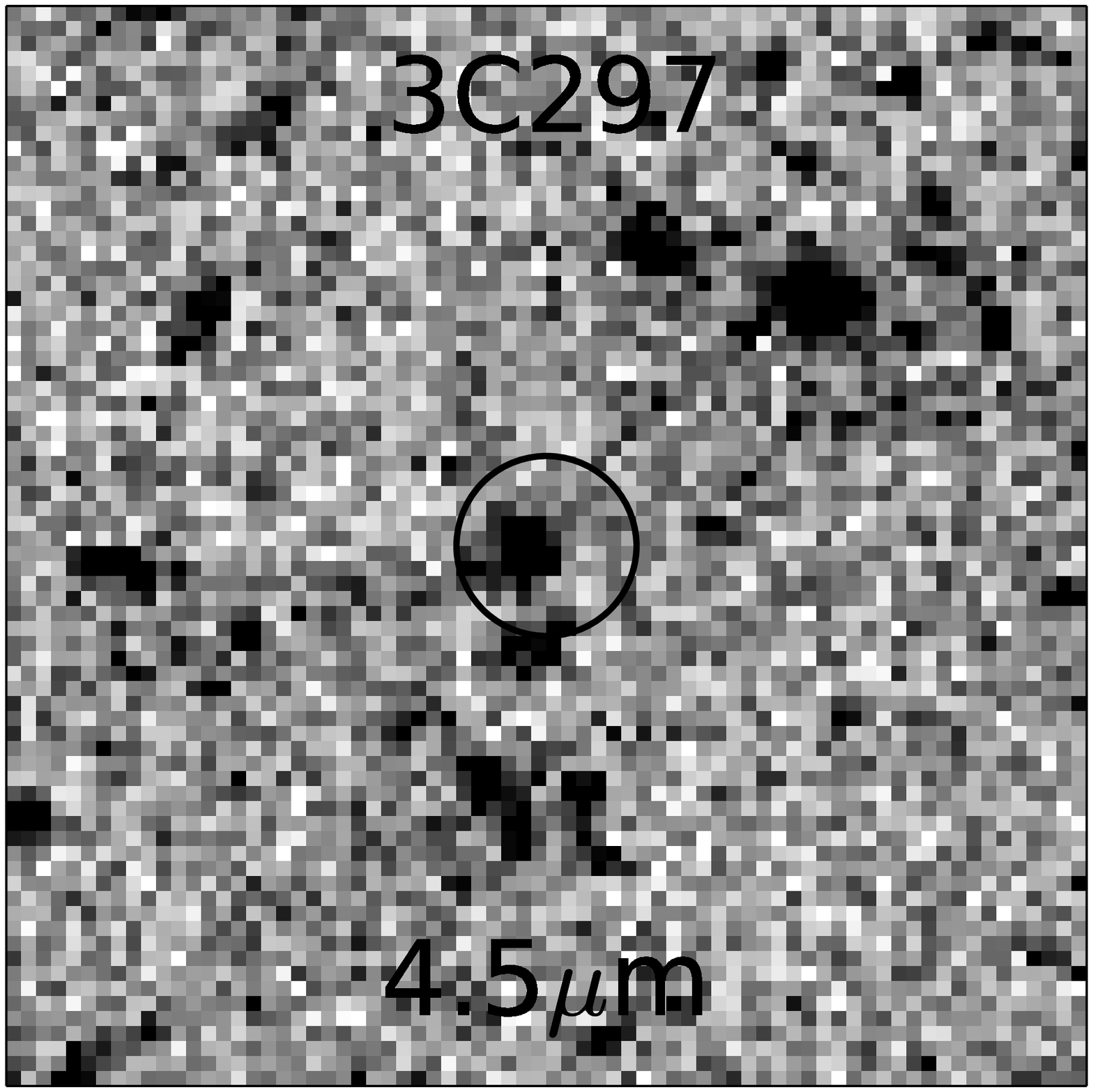}
      \includegraphics[width=1.5cm]{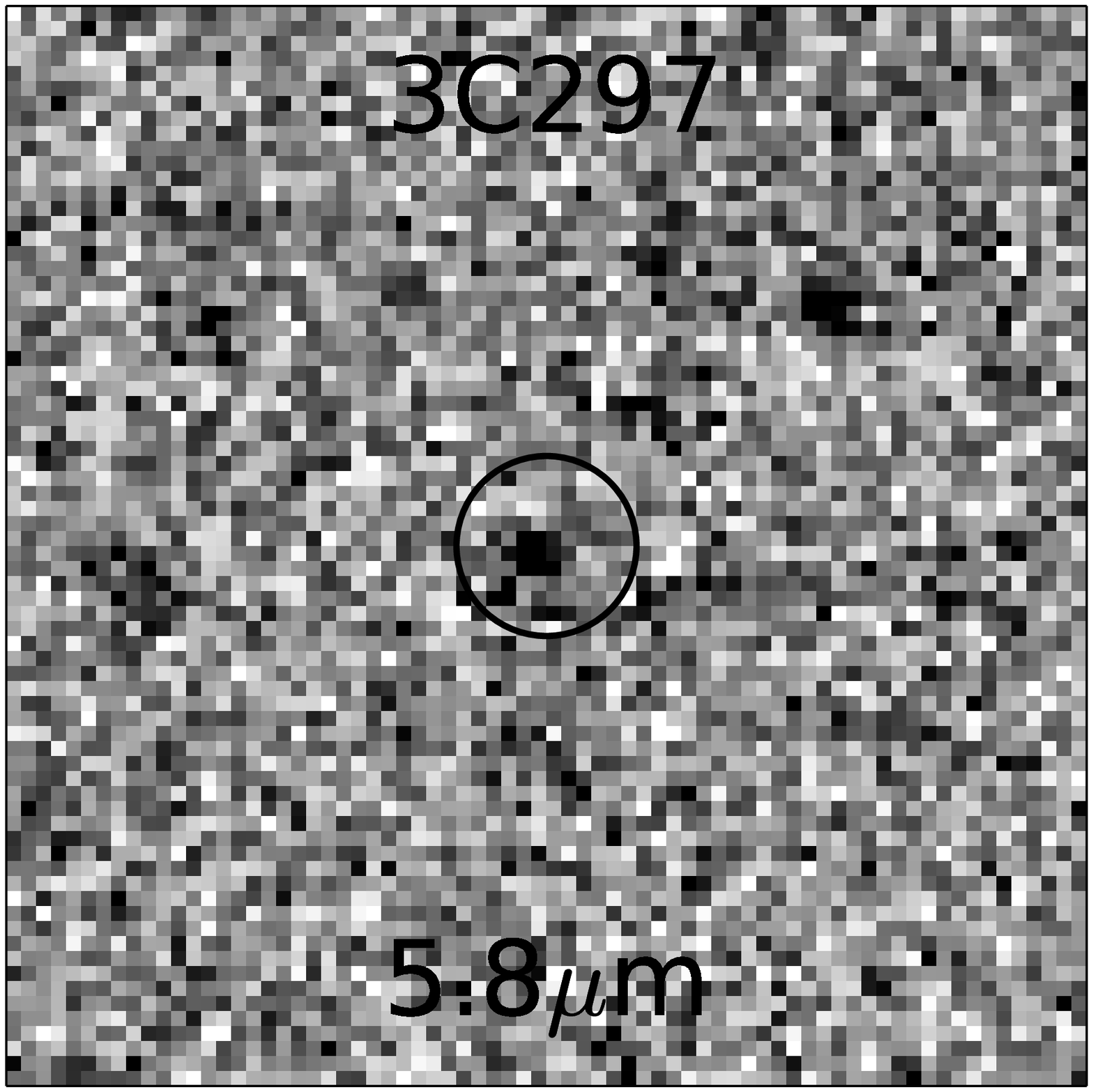}
      \includegraphics[width=1.5cm]{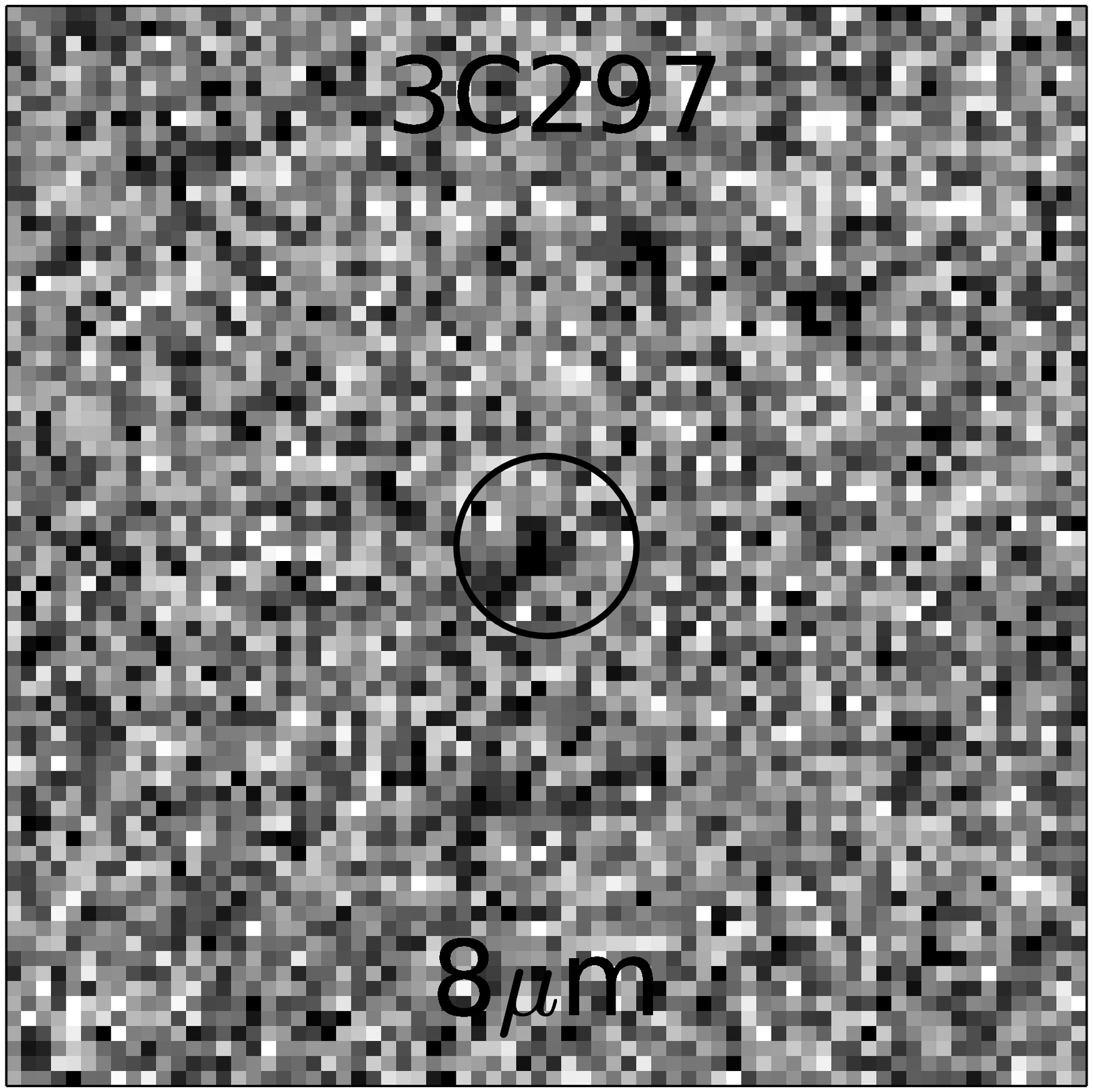}
      \includegraphics[width=1.5cm]{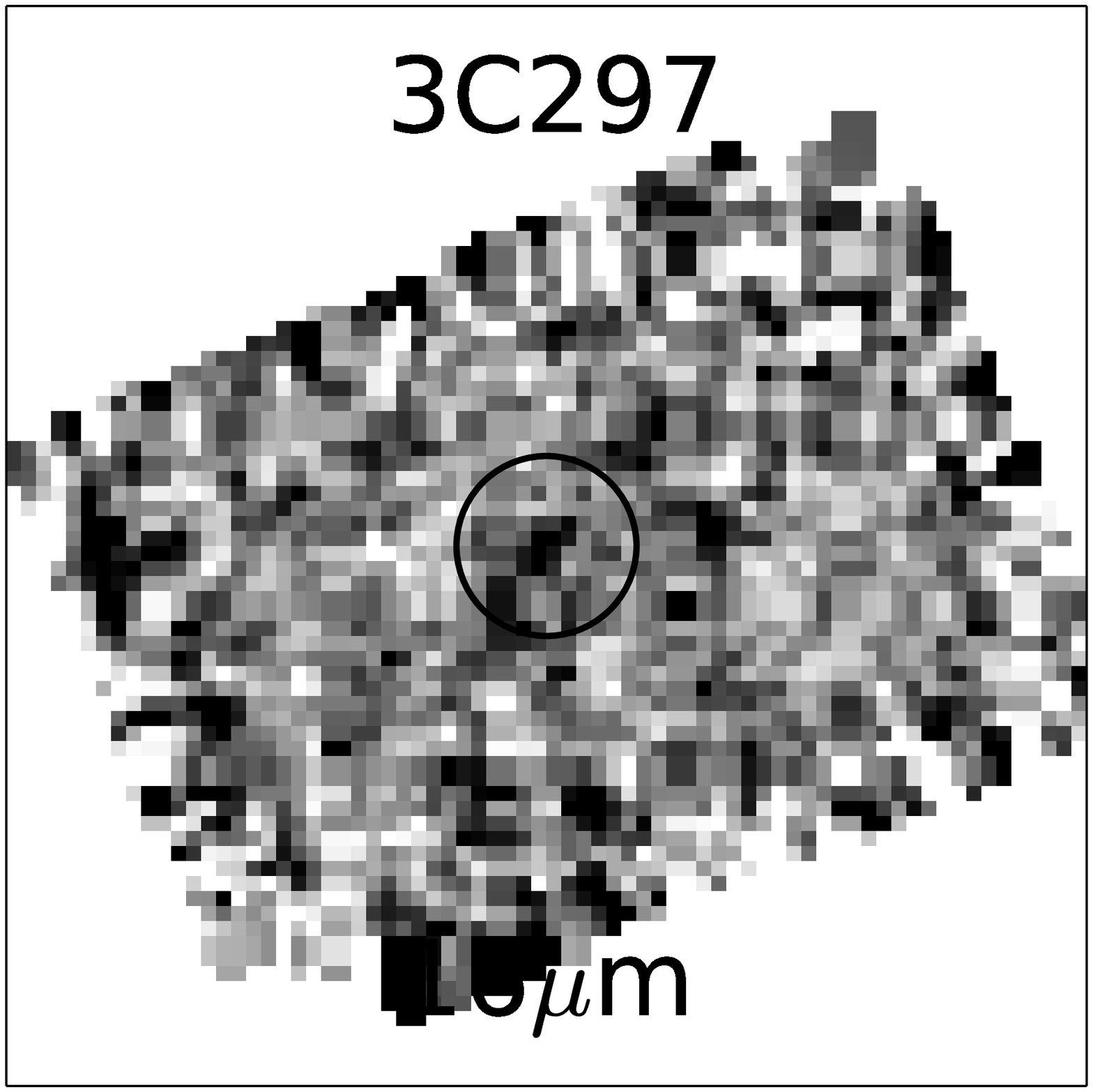}
      \includegraphics[width=1.5cm]{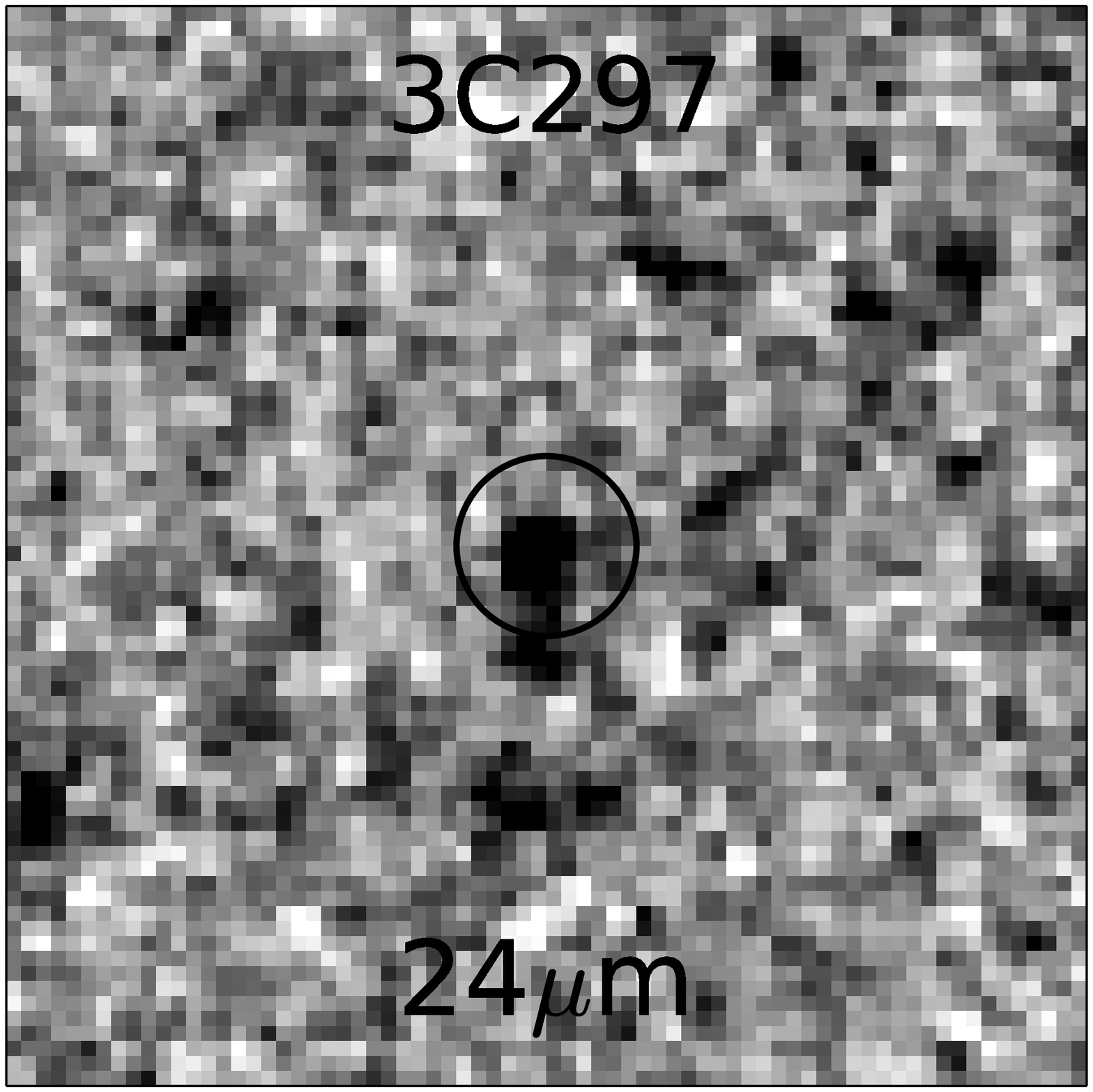}
      \includegraphics[width=1.5cm]{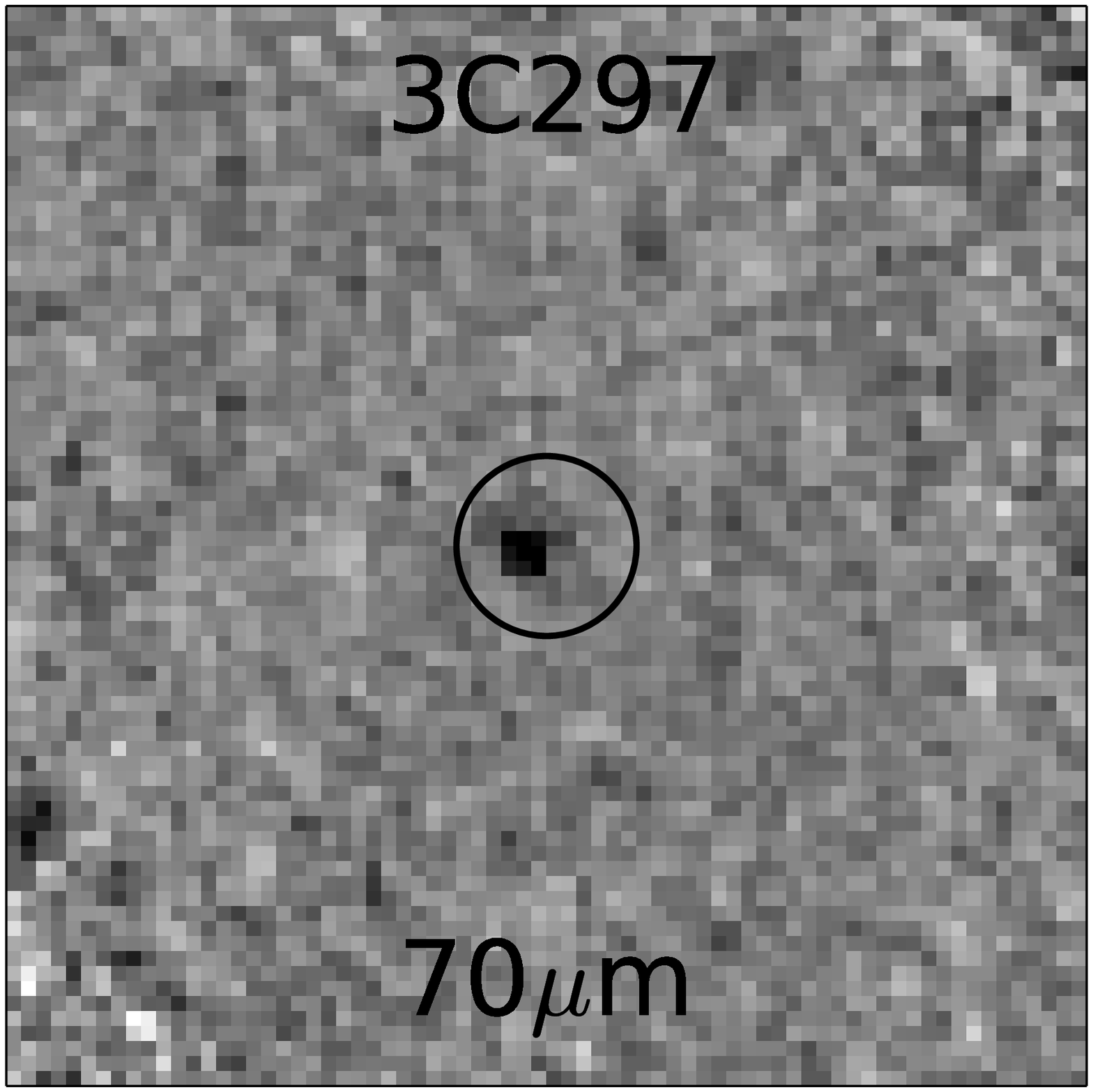}
      \includegraphics[width=1.5cm]{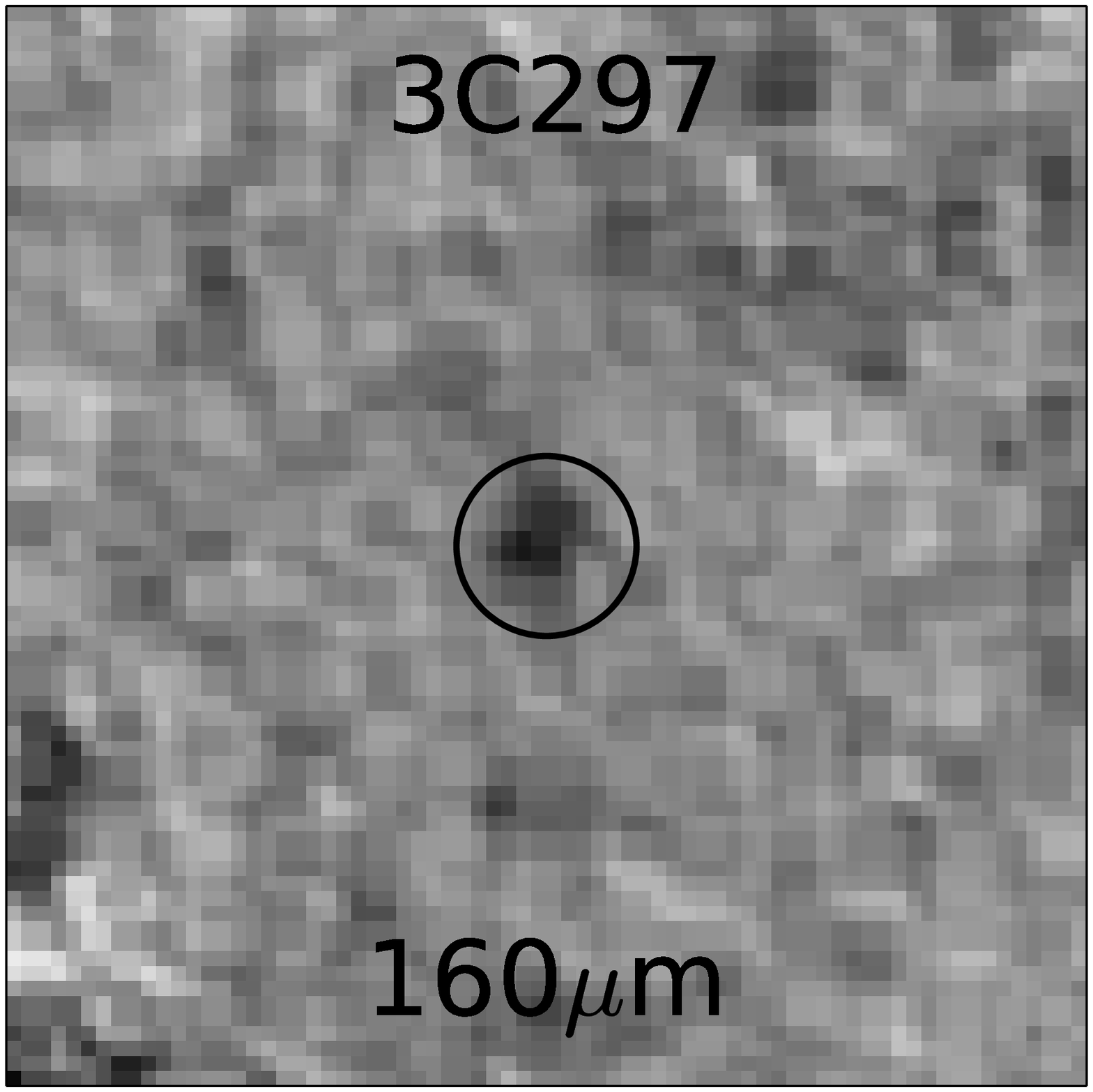}
      \includegraphics[width=1.5cm]{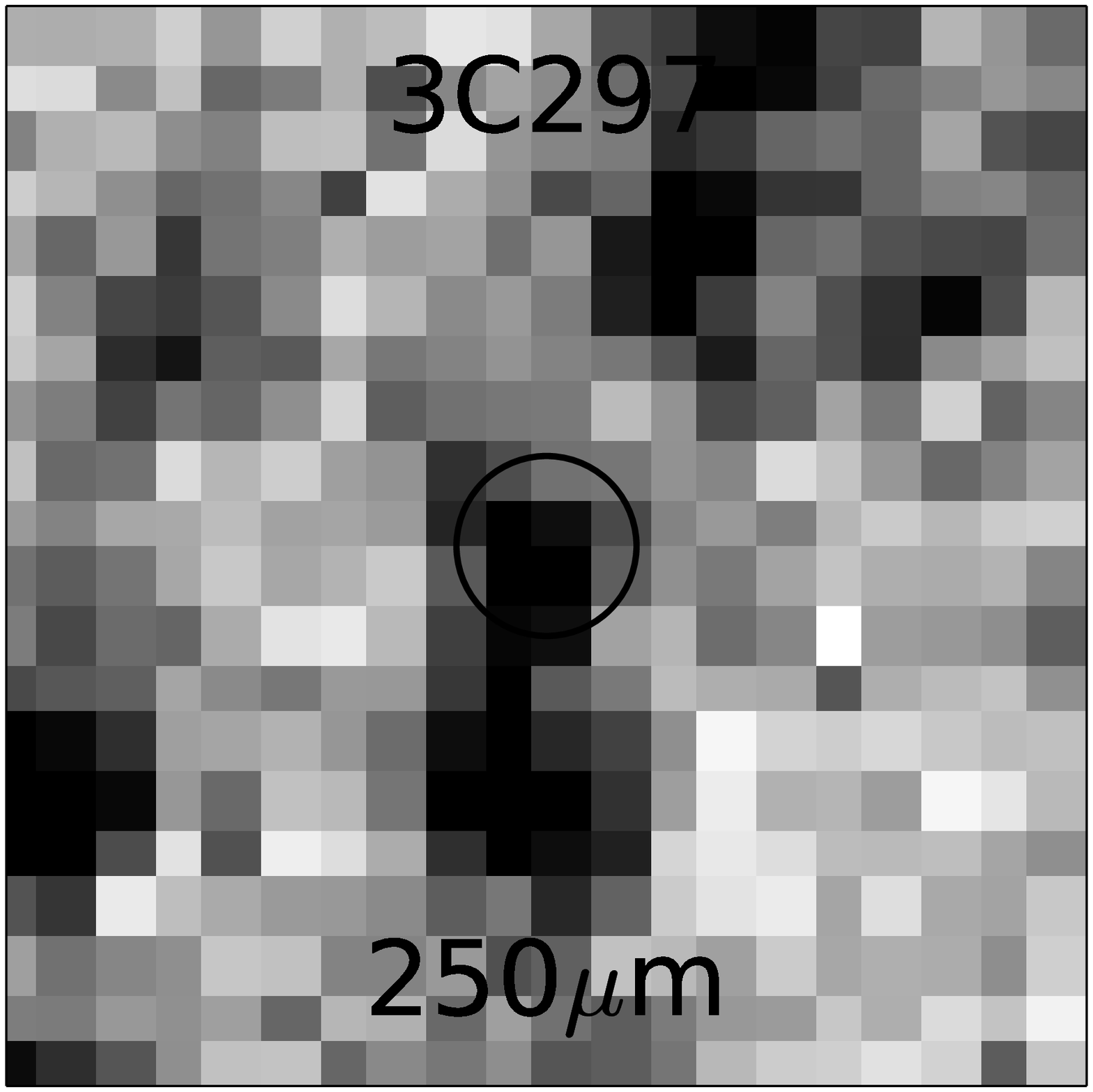}
      \includegraphics[width=1.5cm]{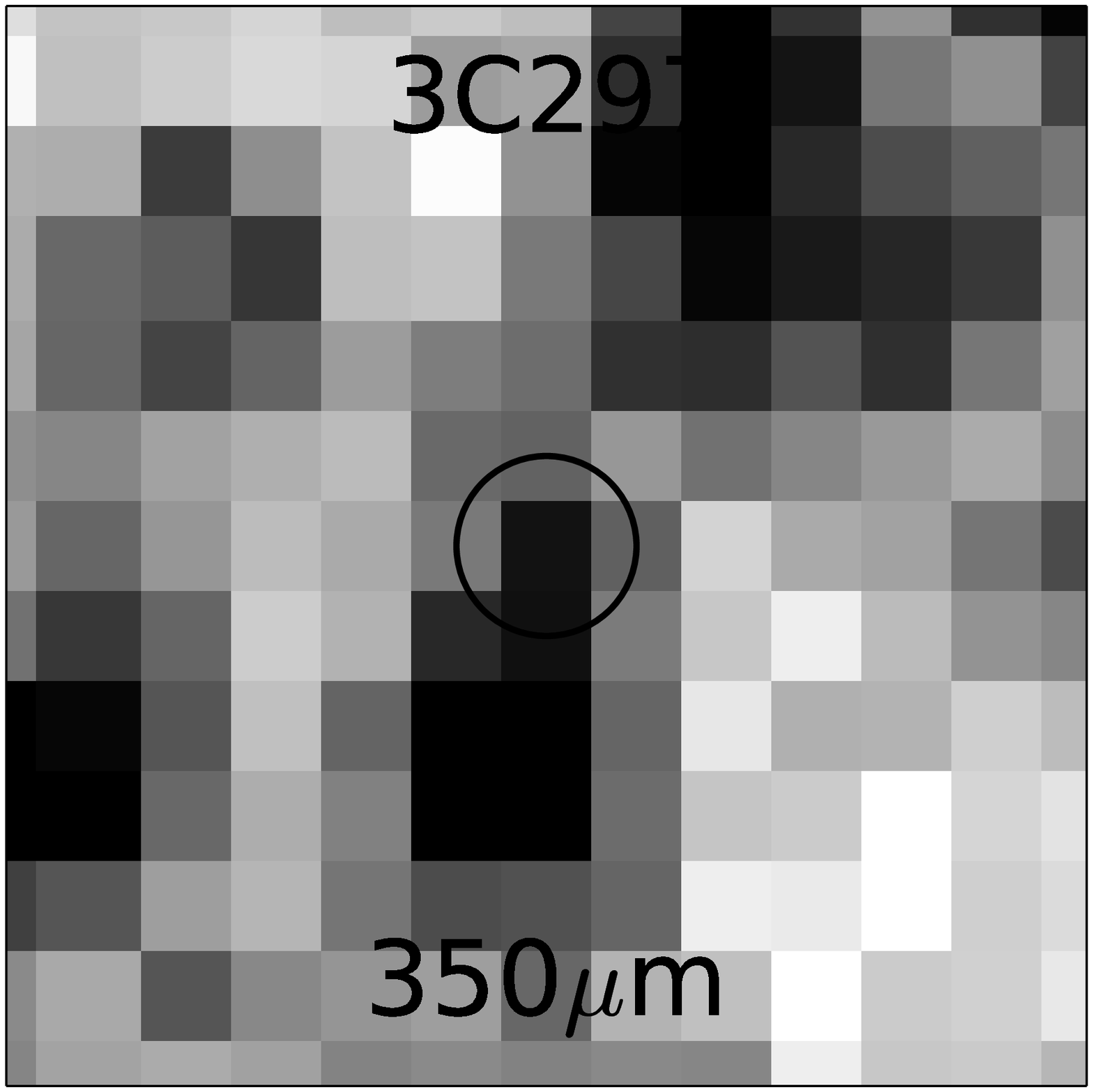}
      \includegraphics[width=1.5cm]{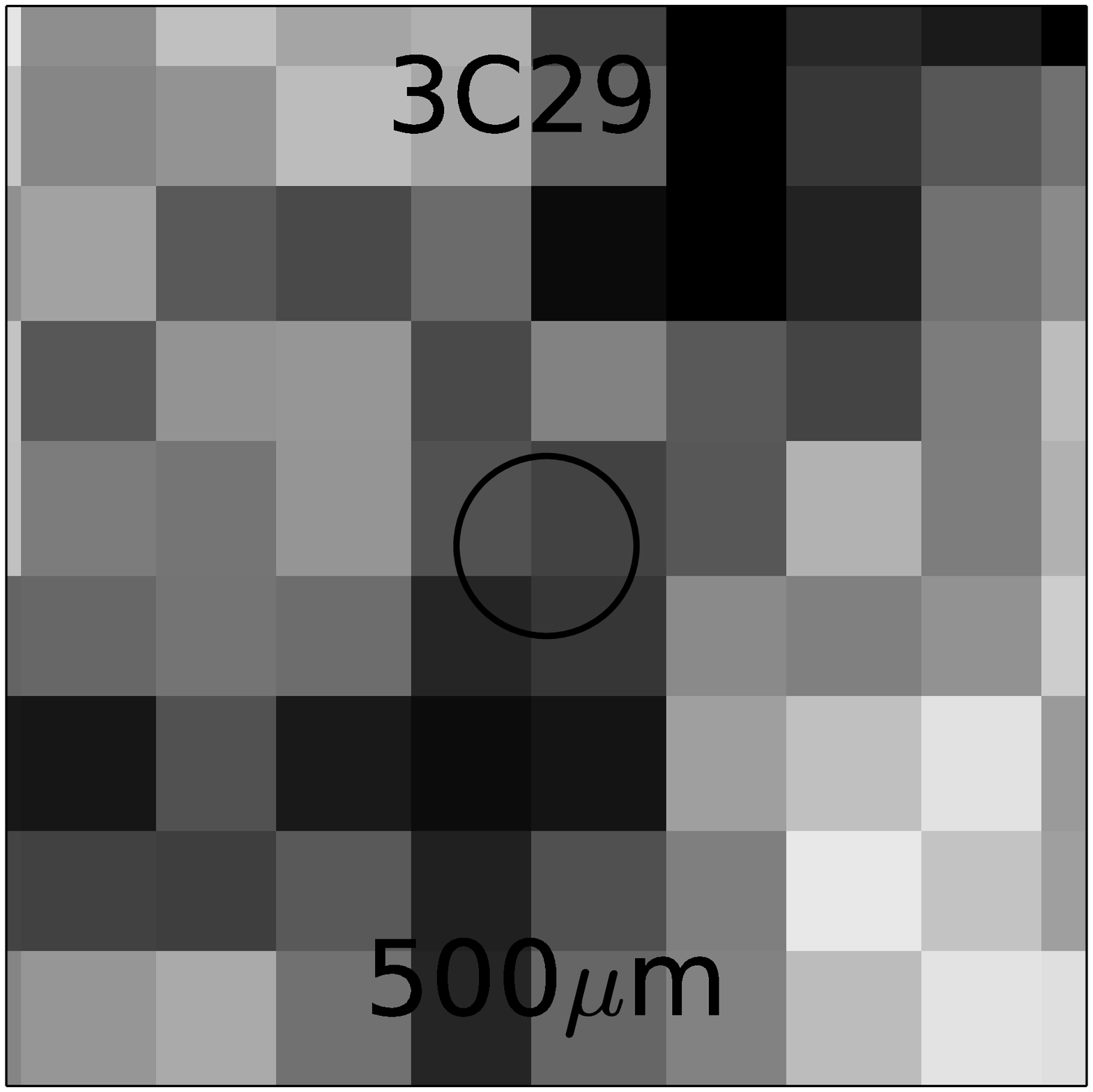}
      \\
      \includegraphics[width=1.5cm]{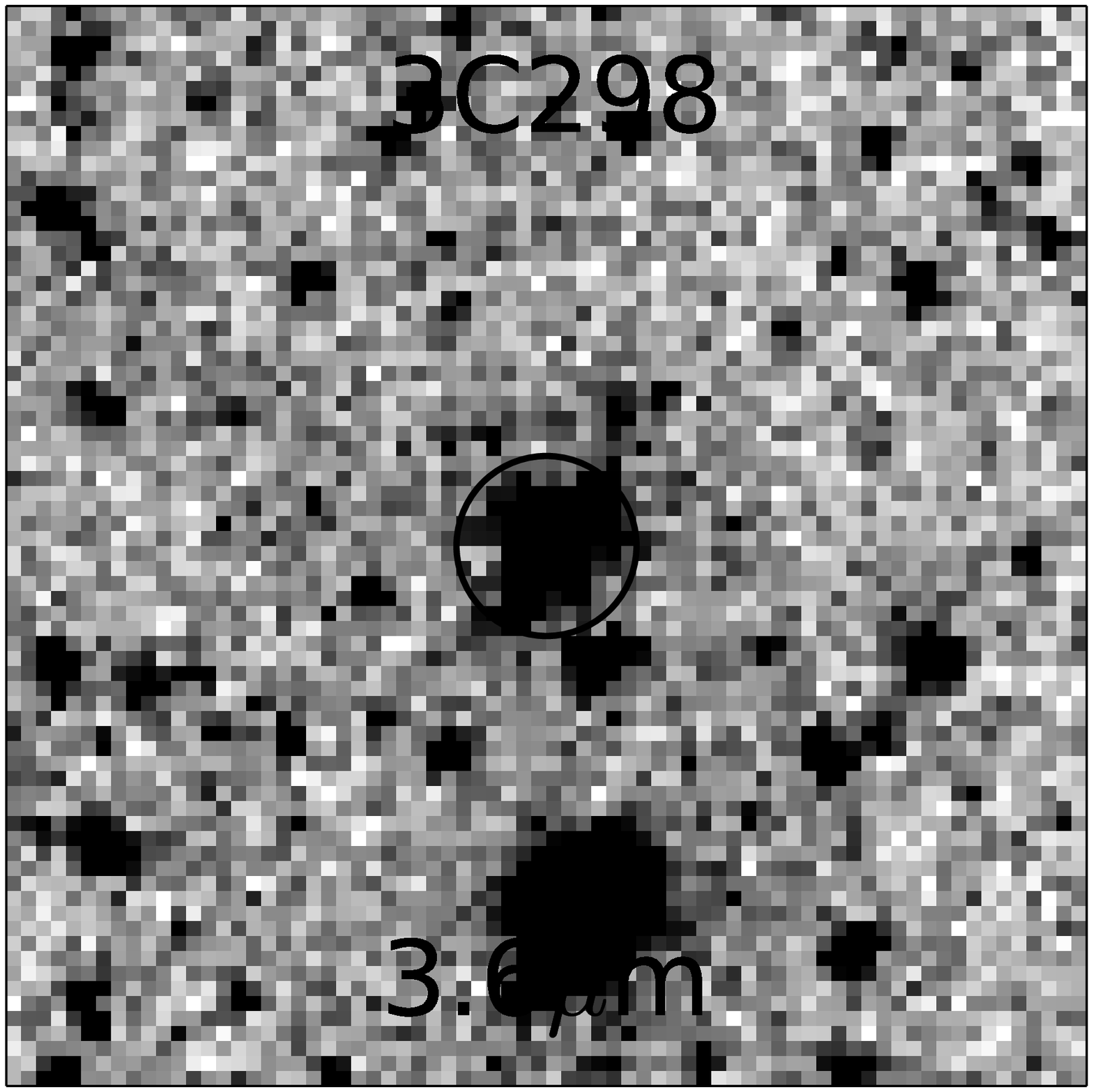}
      \includegraphics[width=1.5cm]{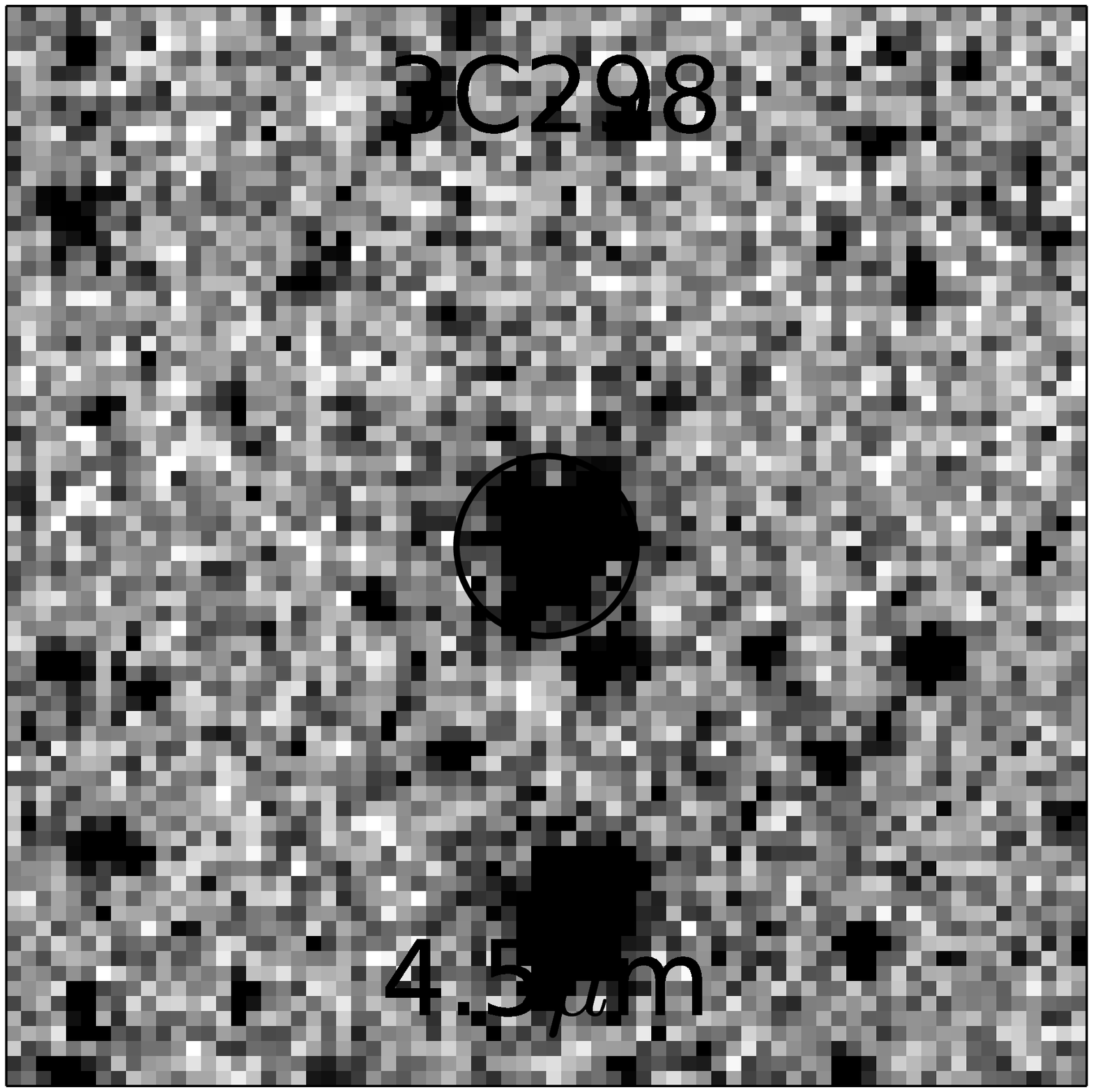}
      \includegraphics[width=1.5cm]{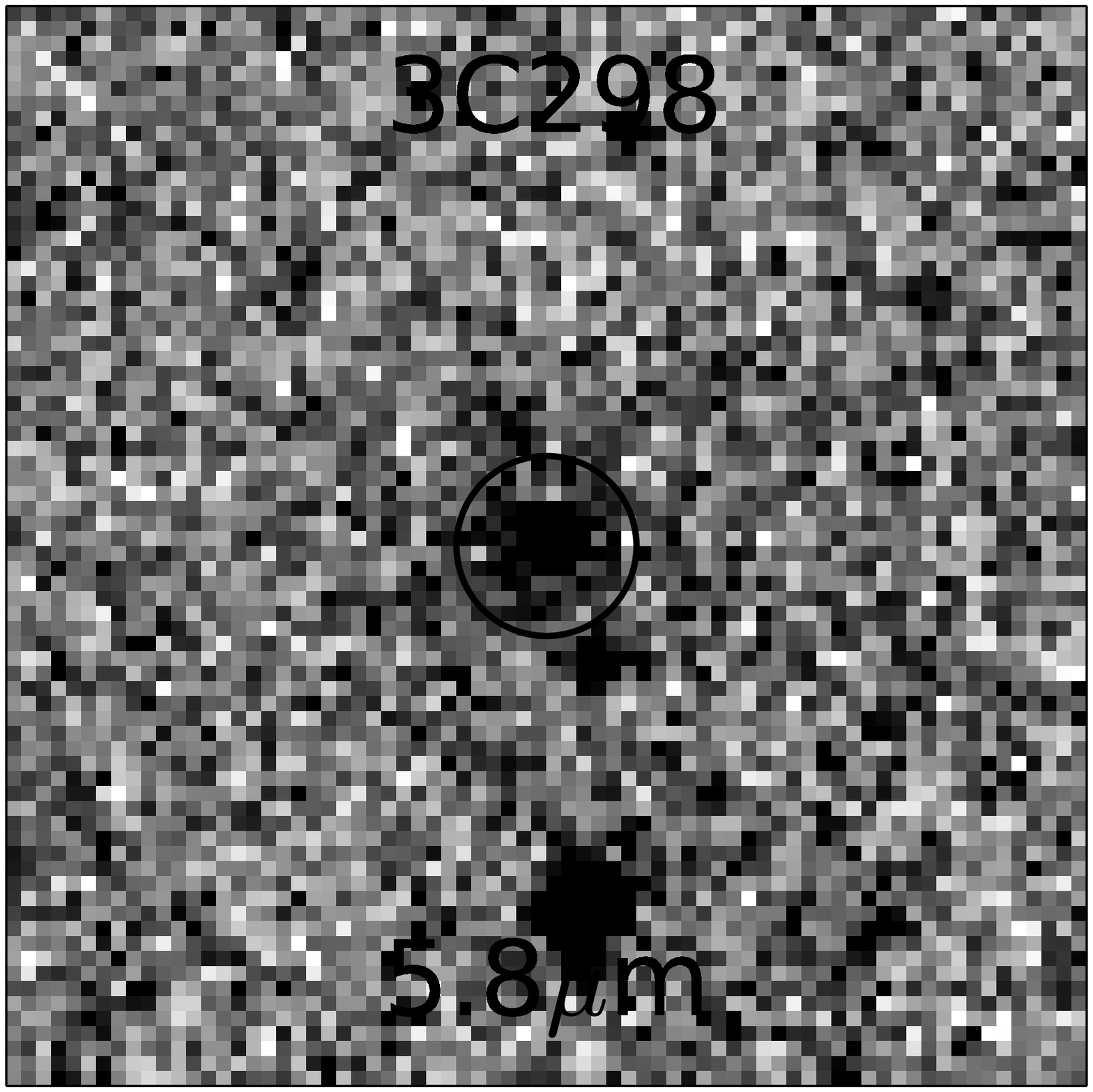}
      \includegraphics[width=1.5cm]{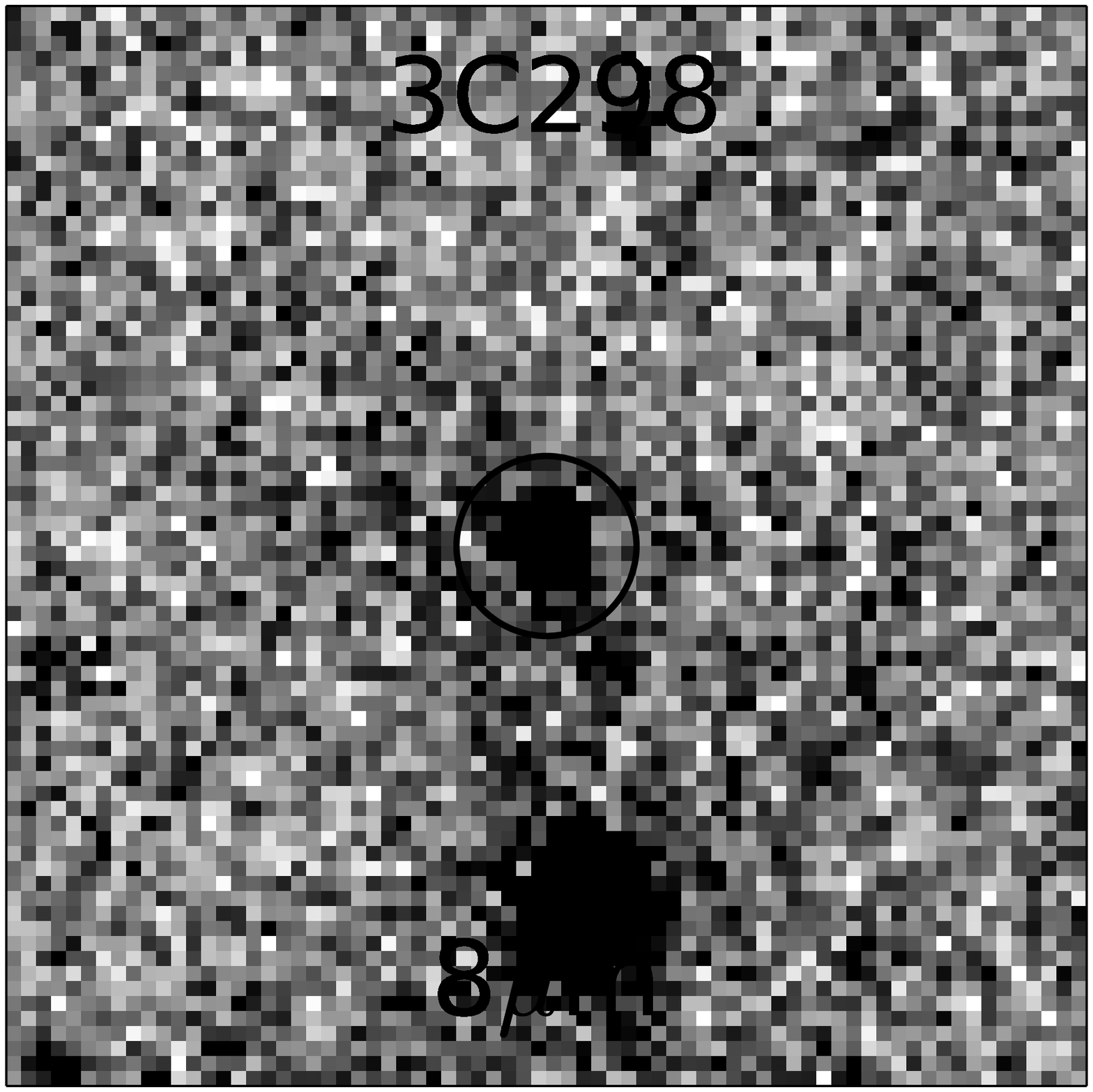}
      \includegraphics[width=1.5cm]{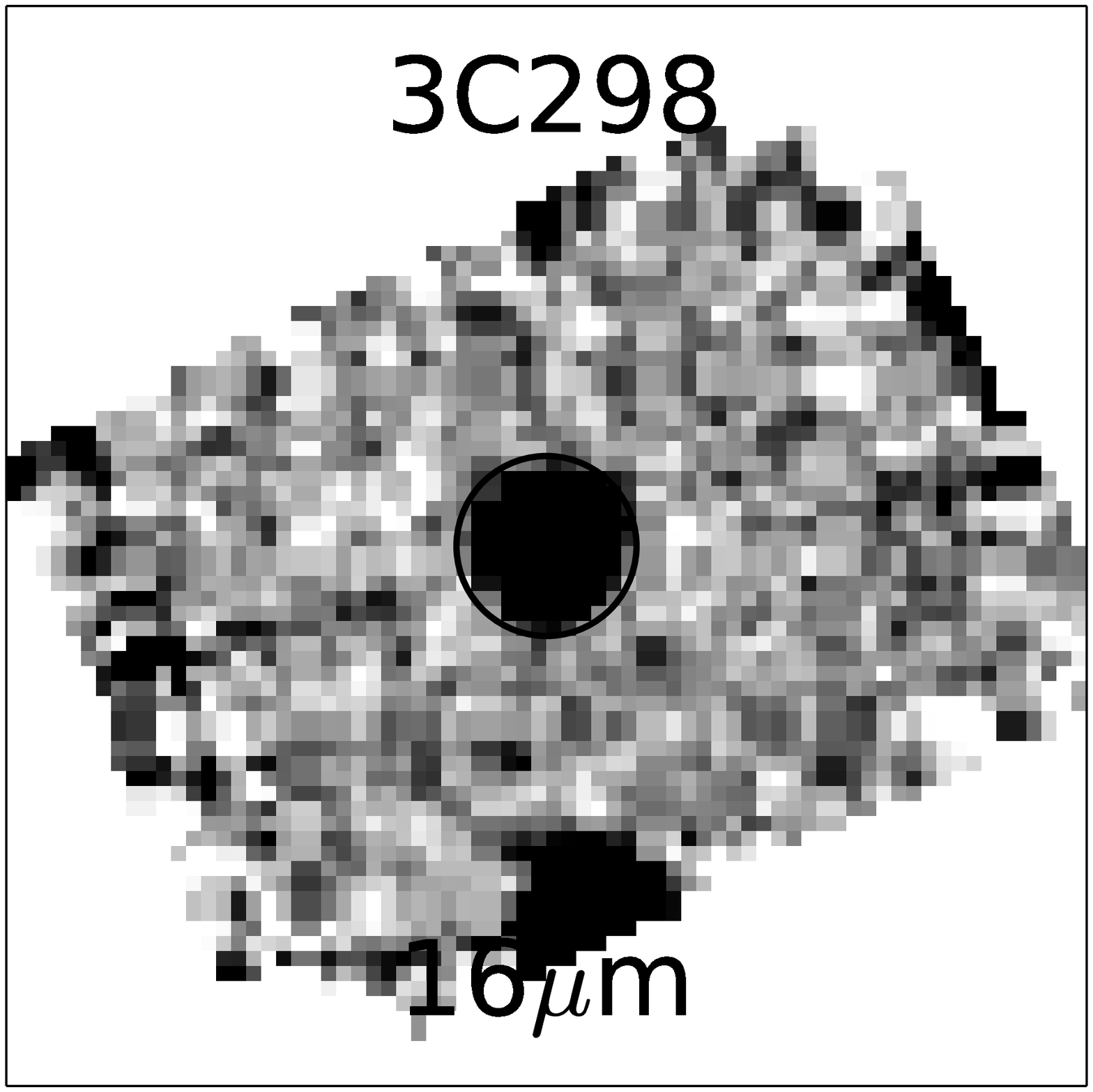}
      \includegraphics[width=1.5cm]{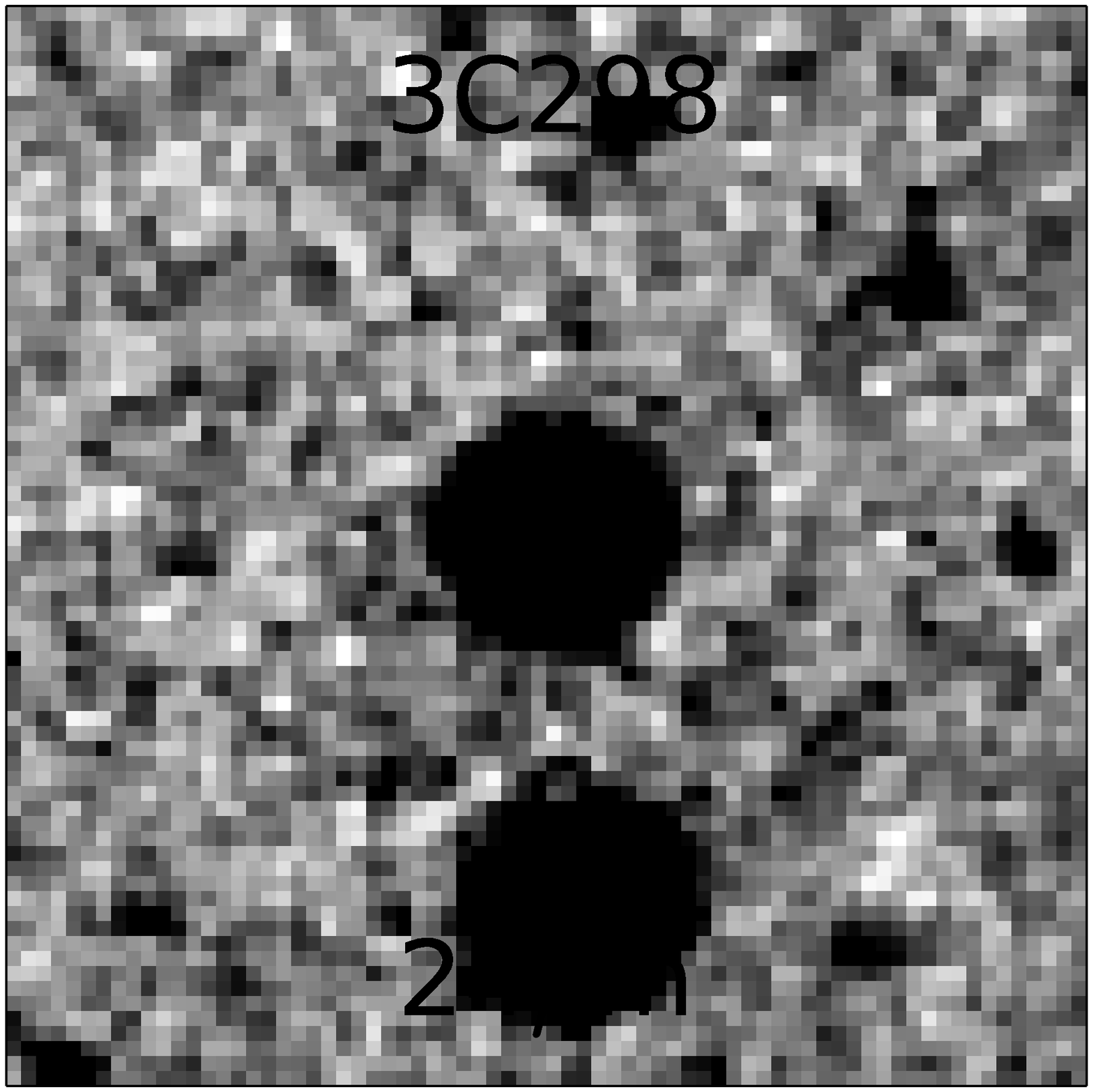}
      \includegraphics[width=1.5cm]{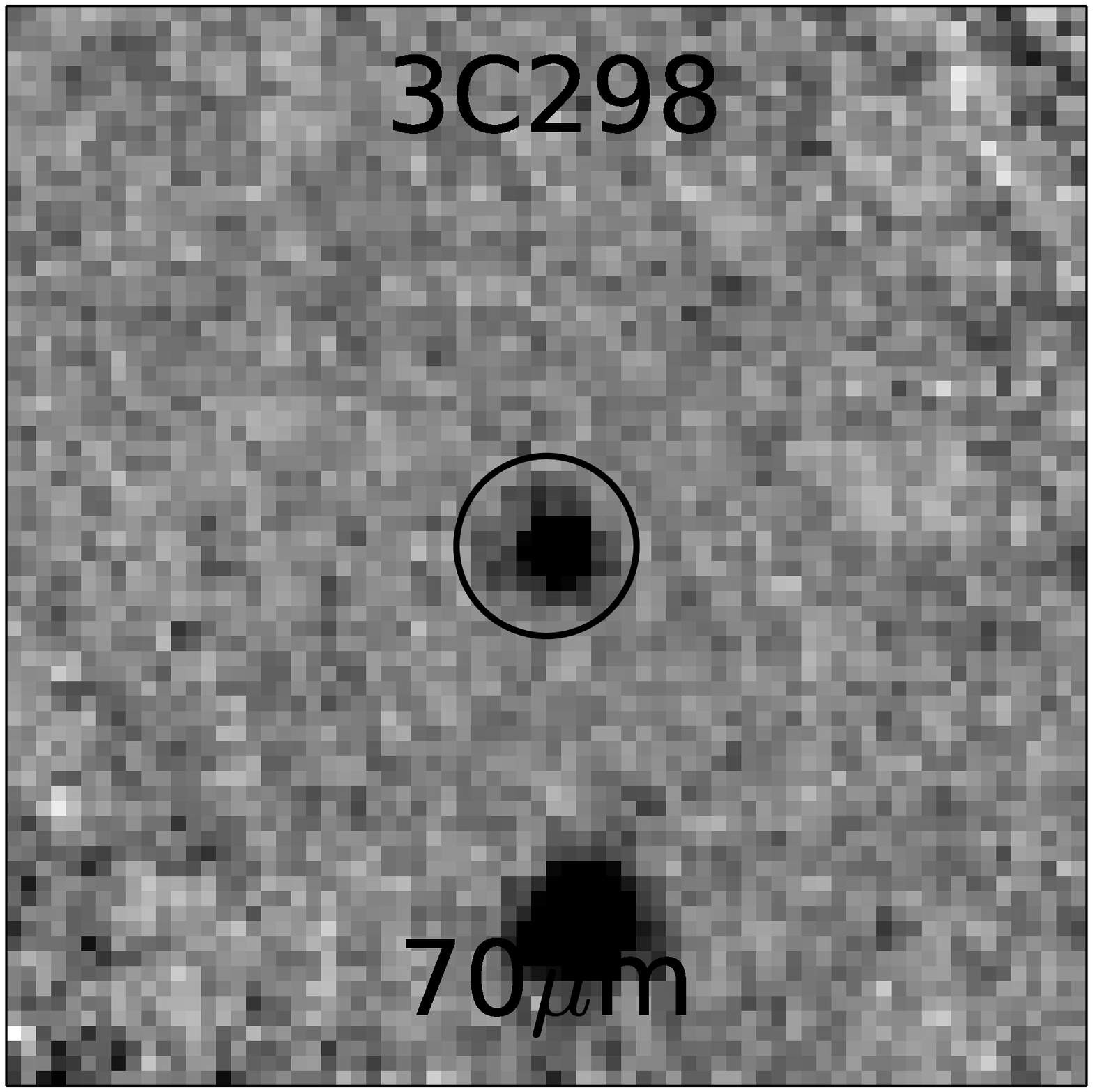}
      \includegraphics[width=1.5cm]{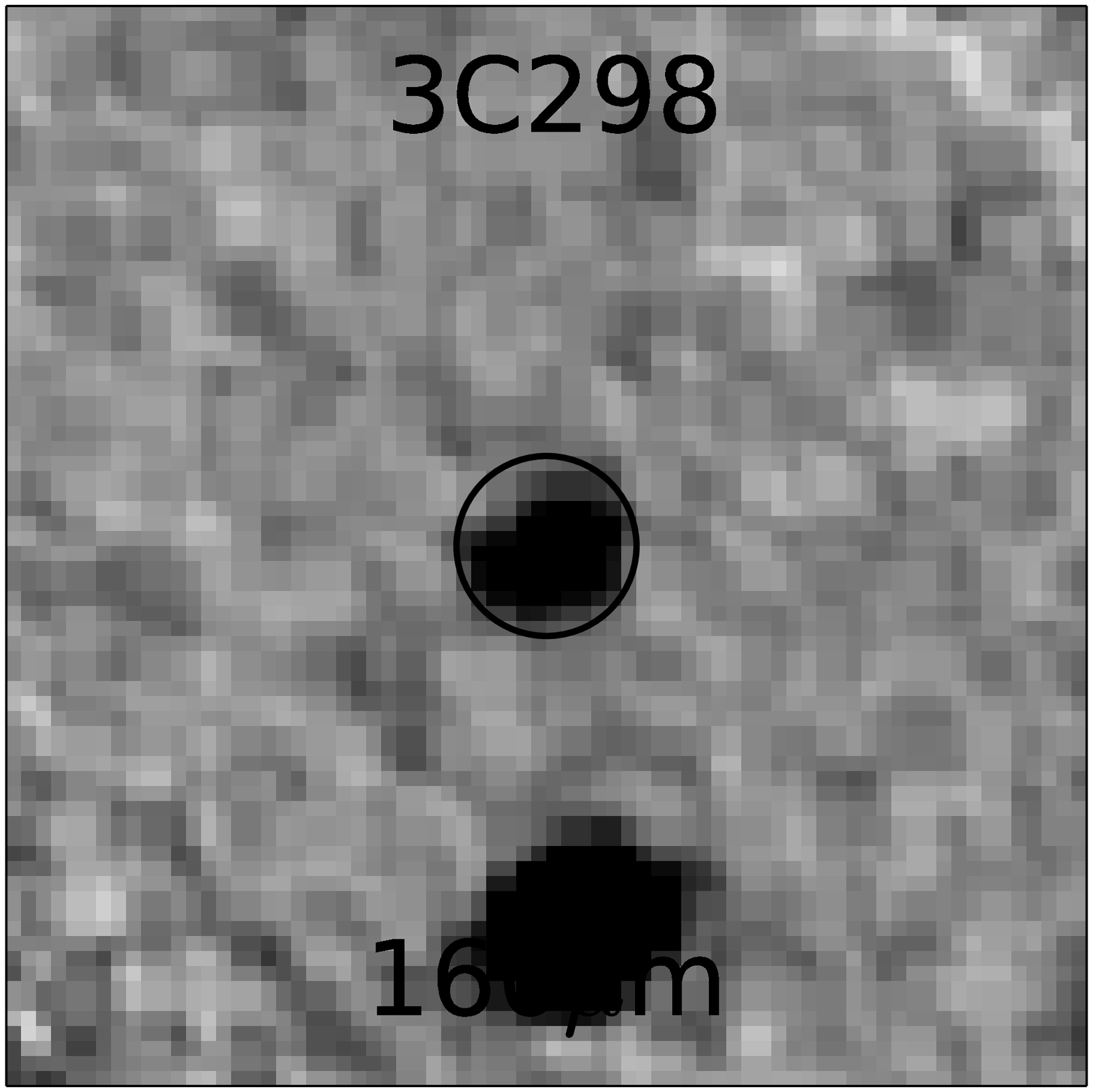}
      \includegraphics[width=1.5cm]{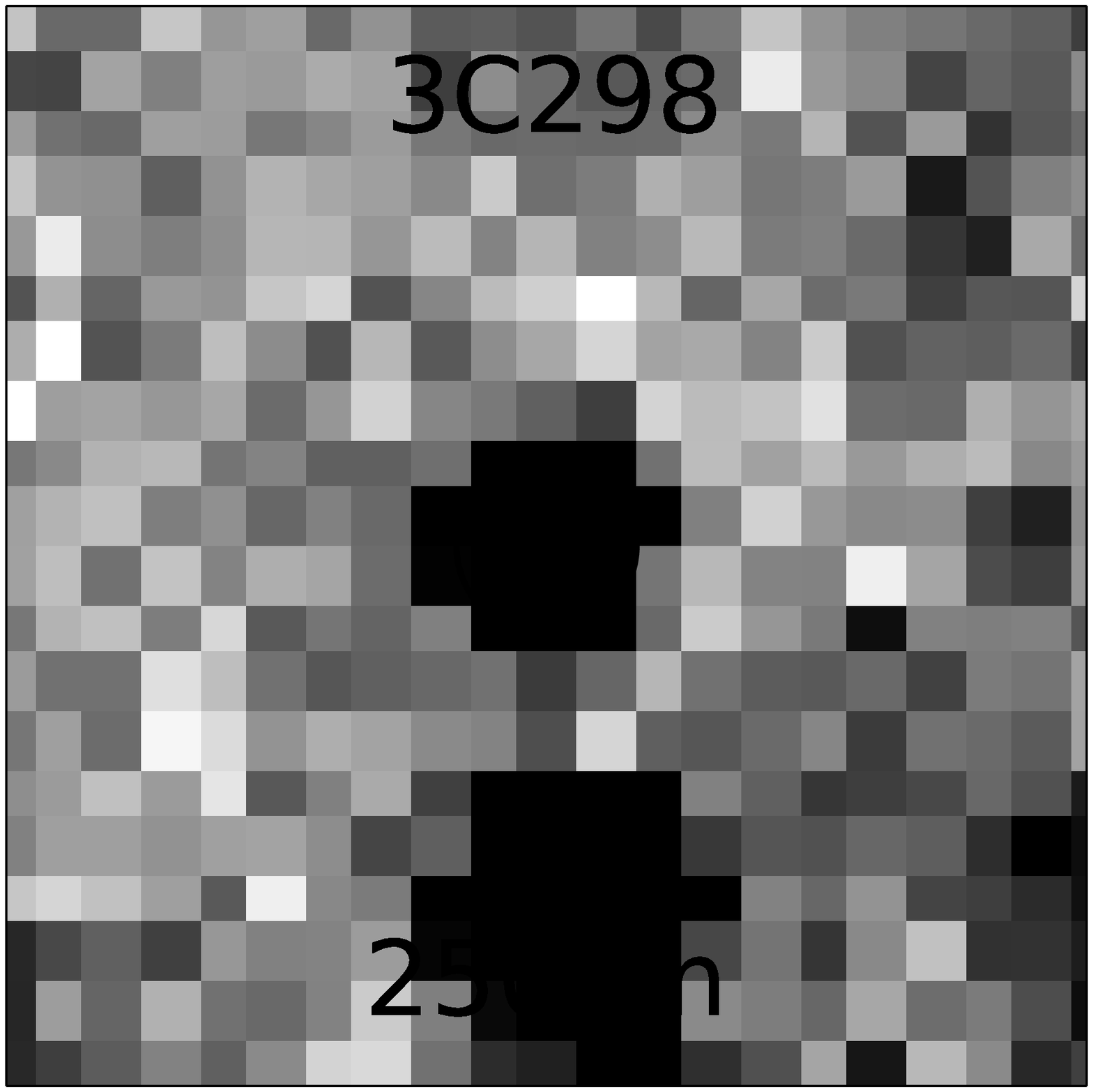}
      \includegraphics[width=1.5cm]{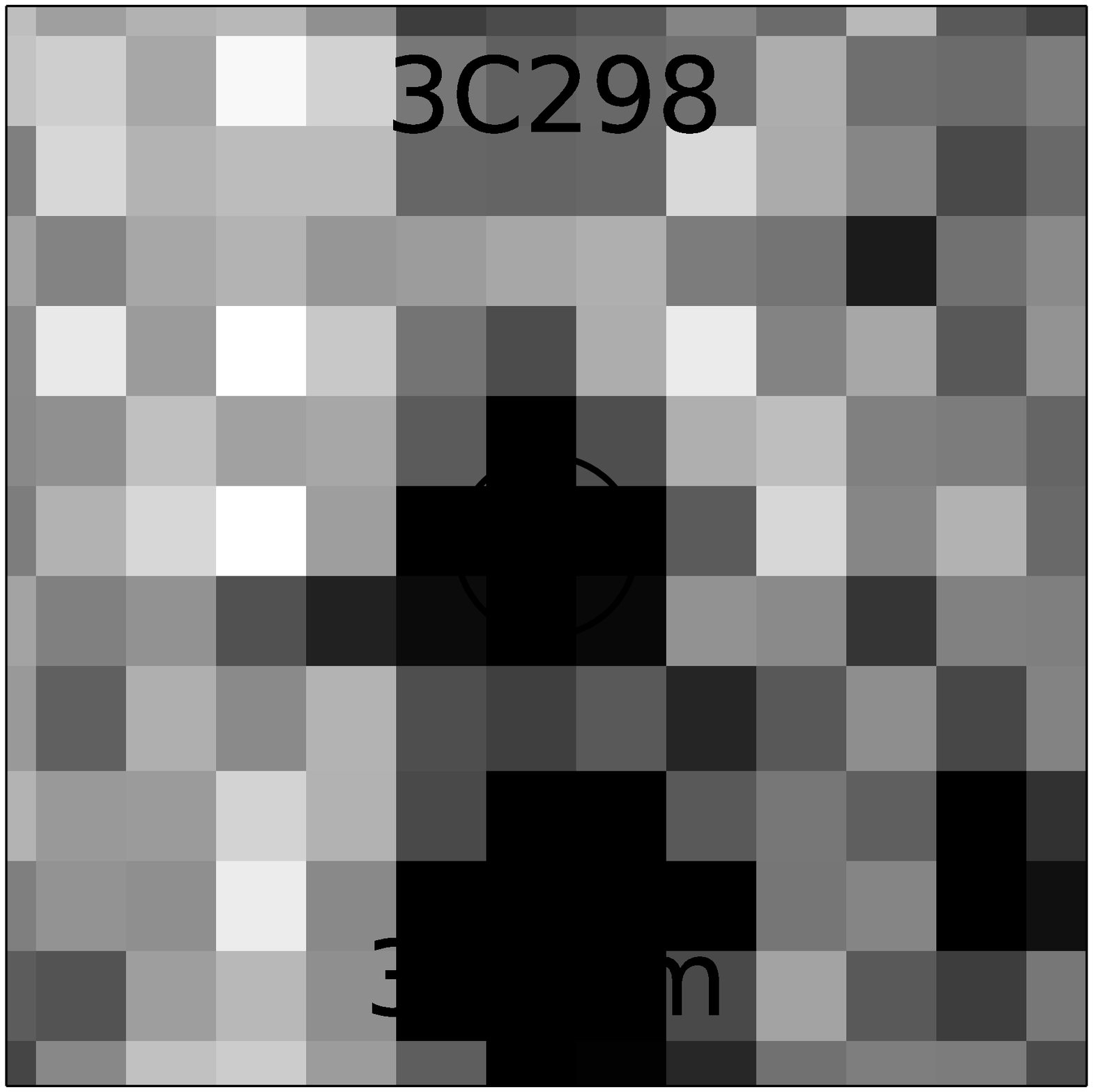}
      \includegraphics[width=1.5cm]{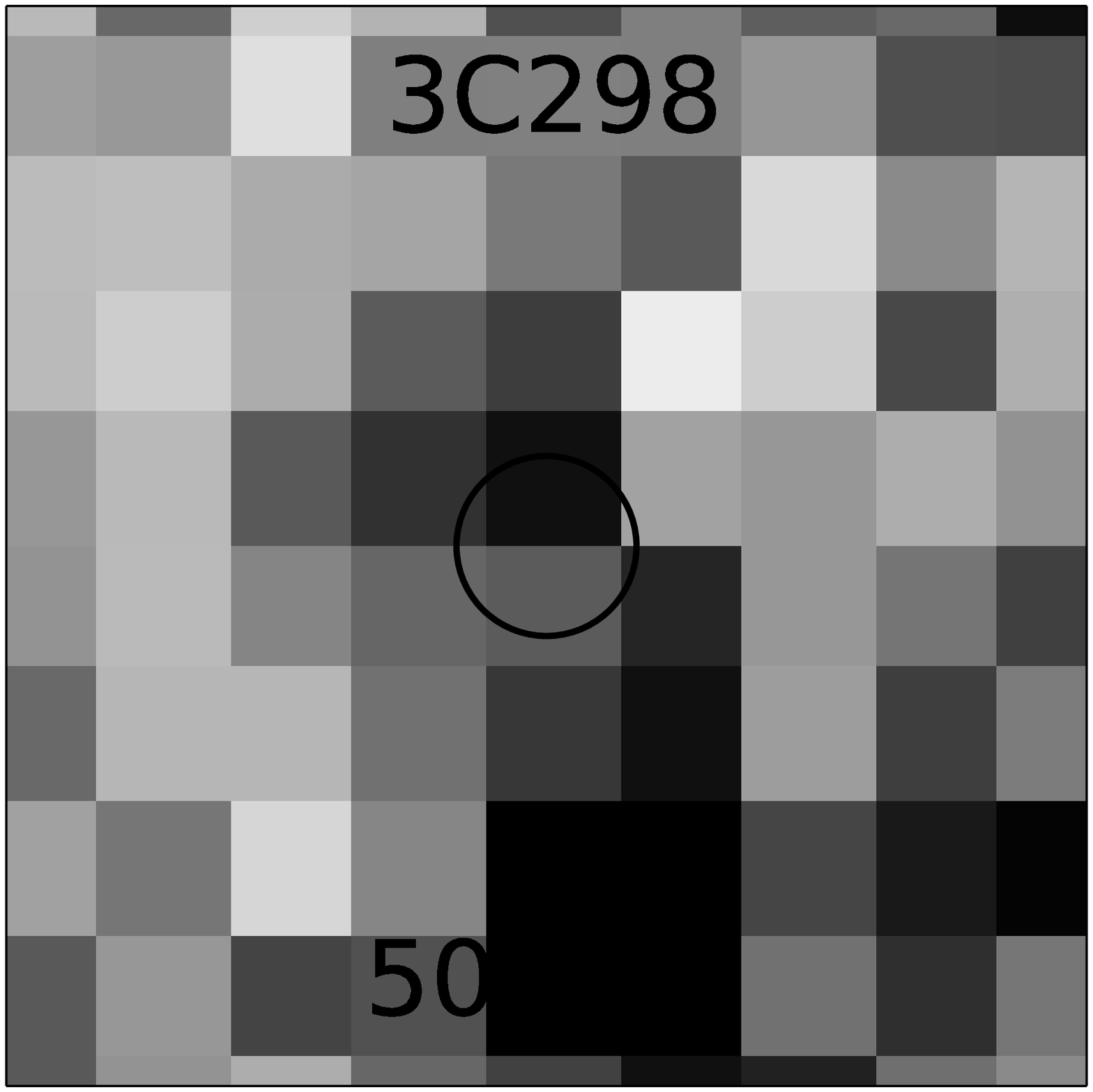}
      \\
      \includegraphics[width=1.5cm]{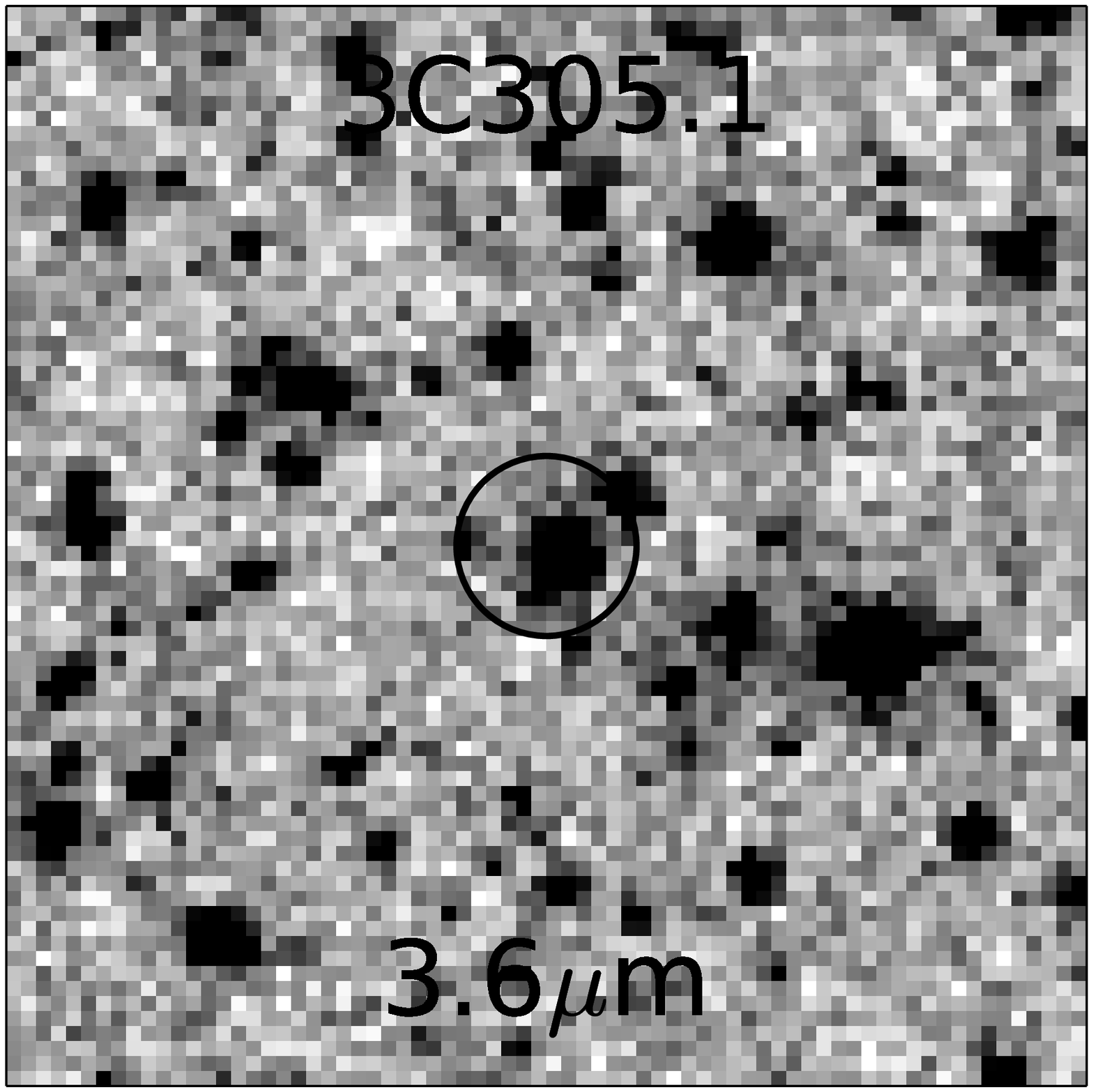}
      \includegraphics[width=1.5cm]{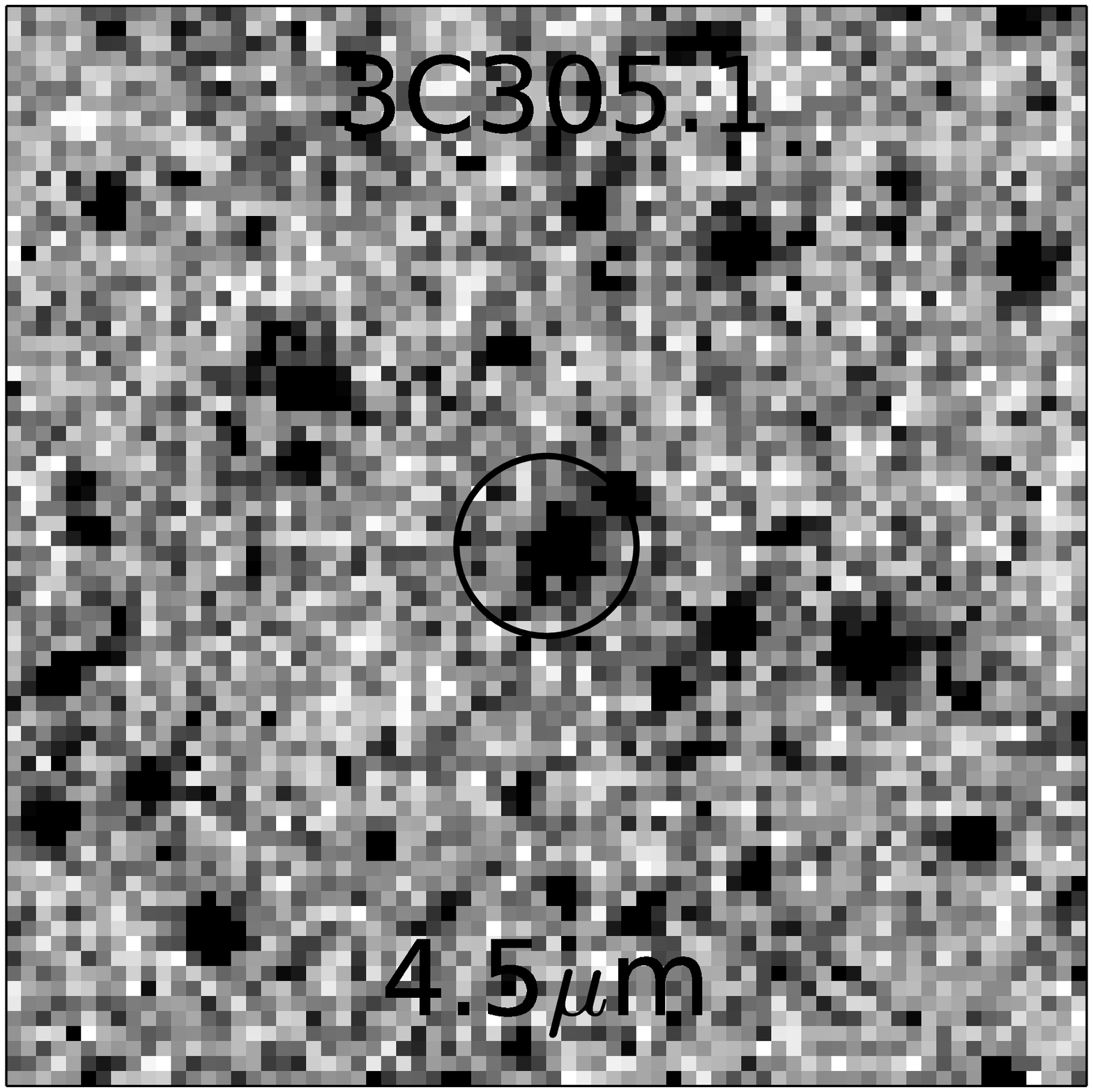}
      \includegraphics[width=1.5cm]{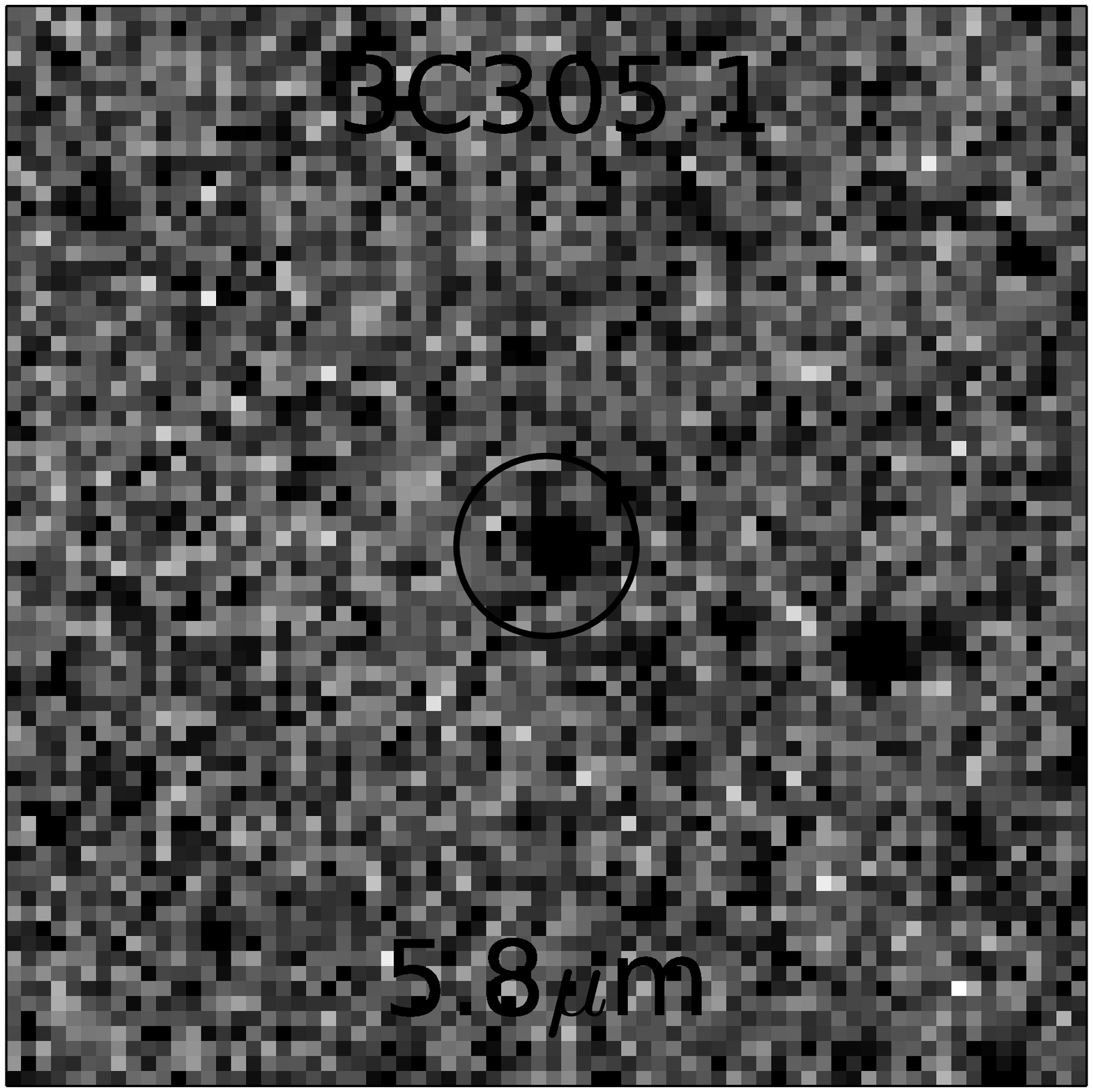}
      \includegraphics[width=1.5cm]{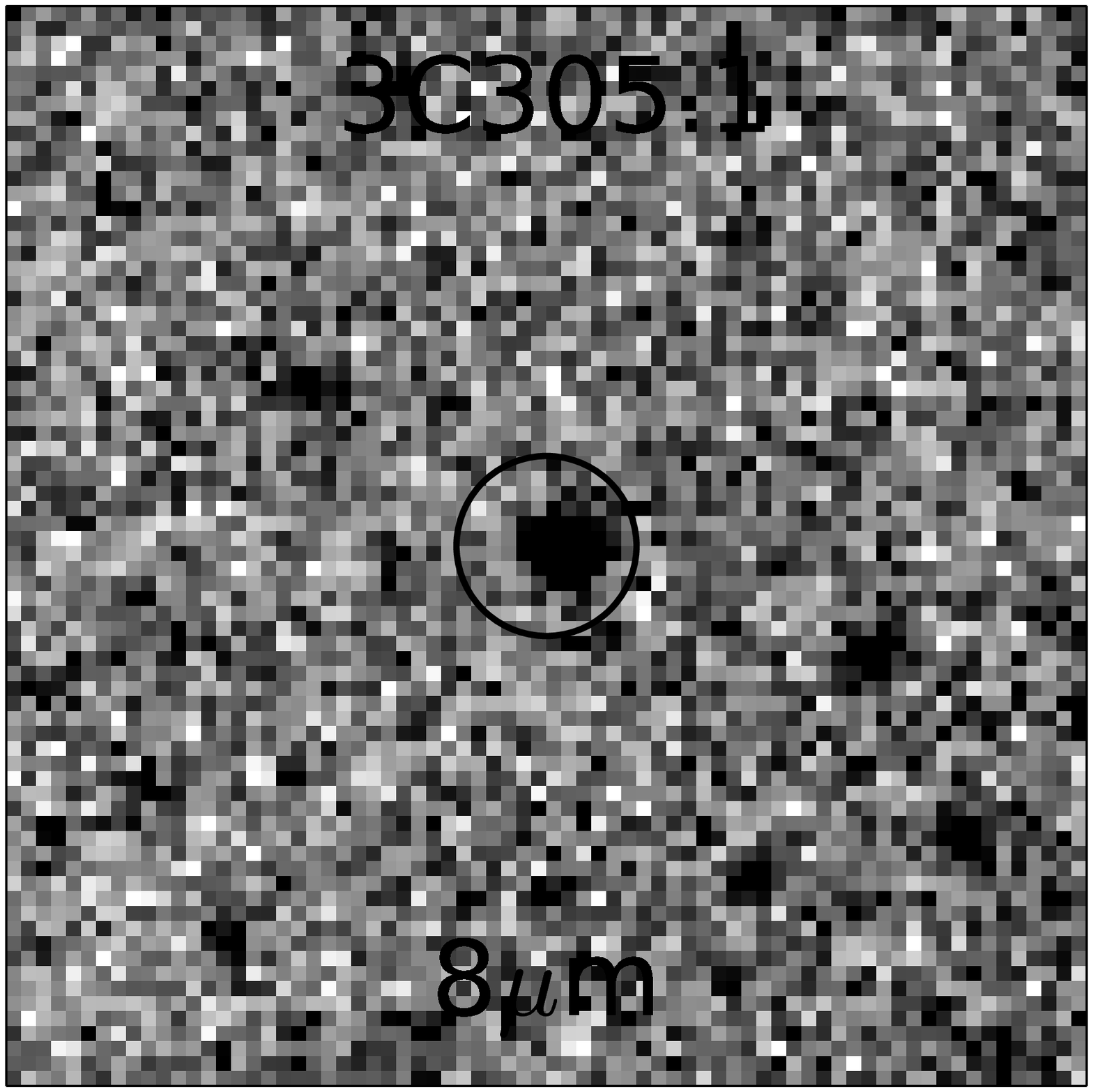}
      \includegraphics[width=1.5cm]{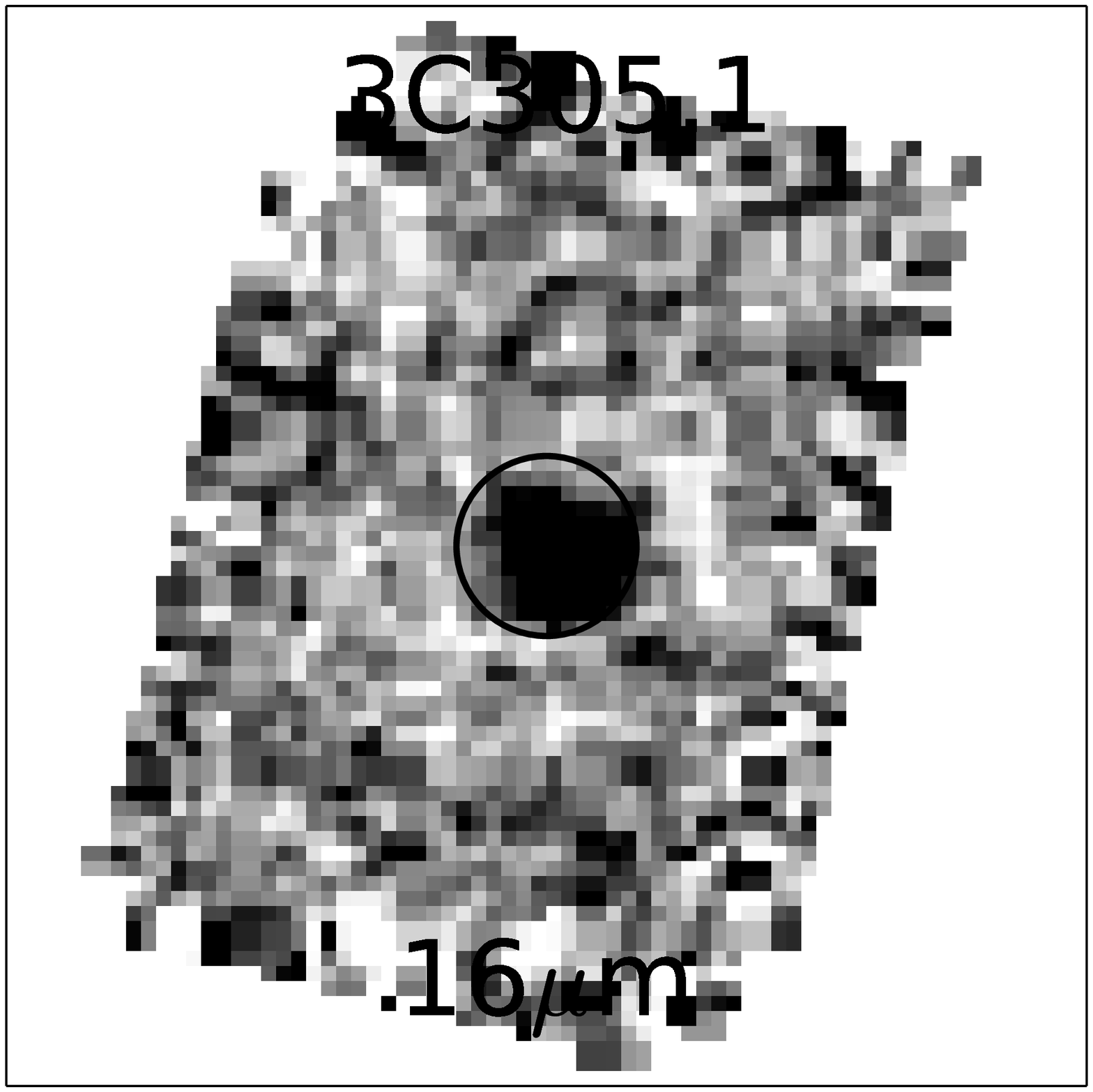}
      \includegraphics[width=1.5cm]{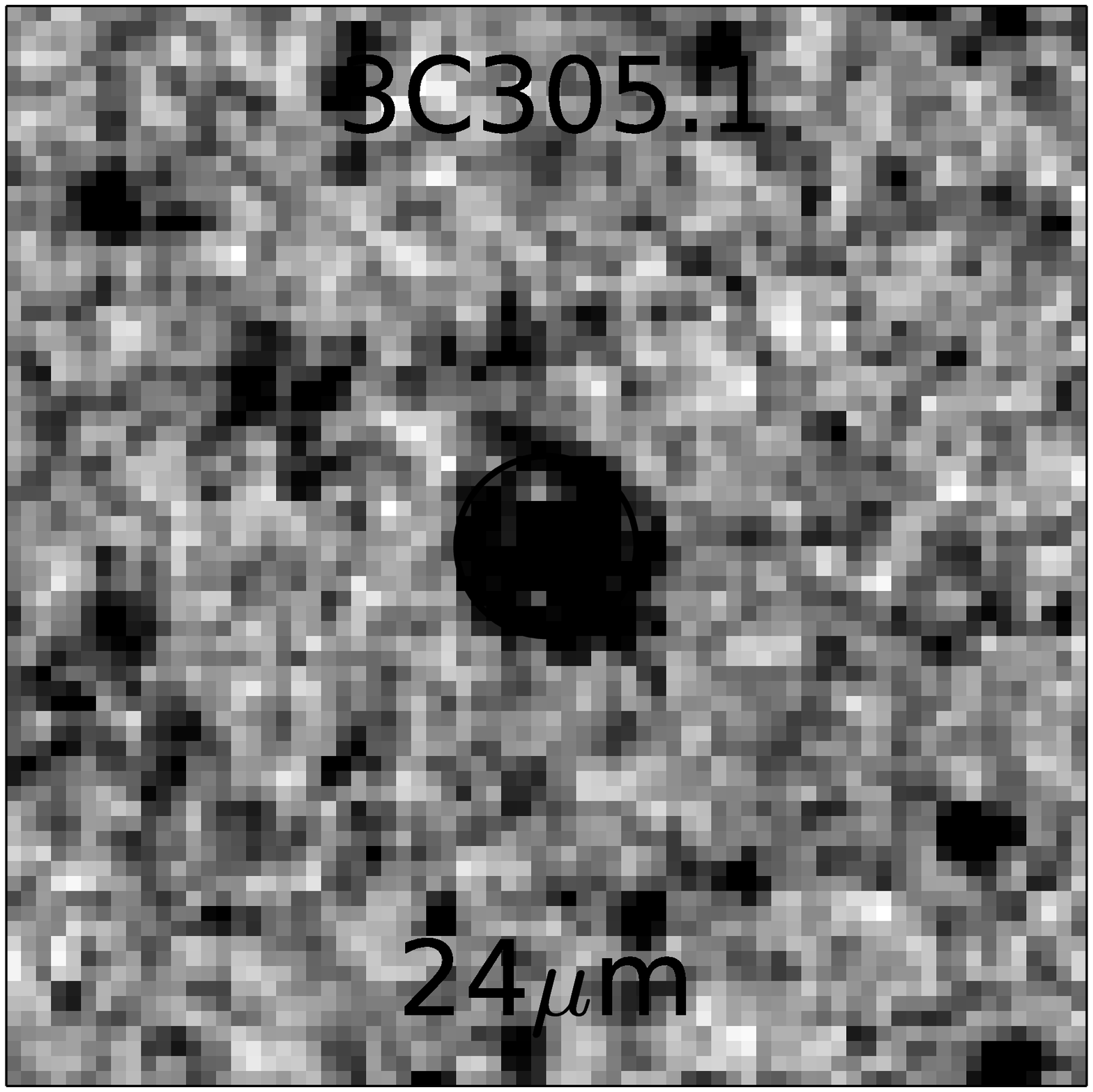}
      \includegraphics[width=1.5cm]{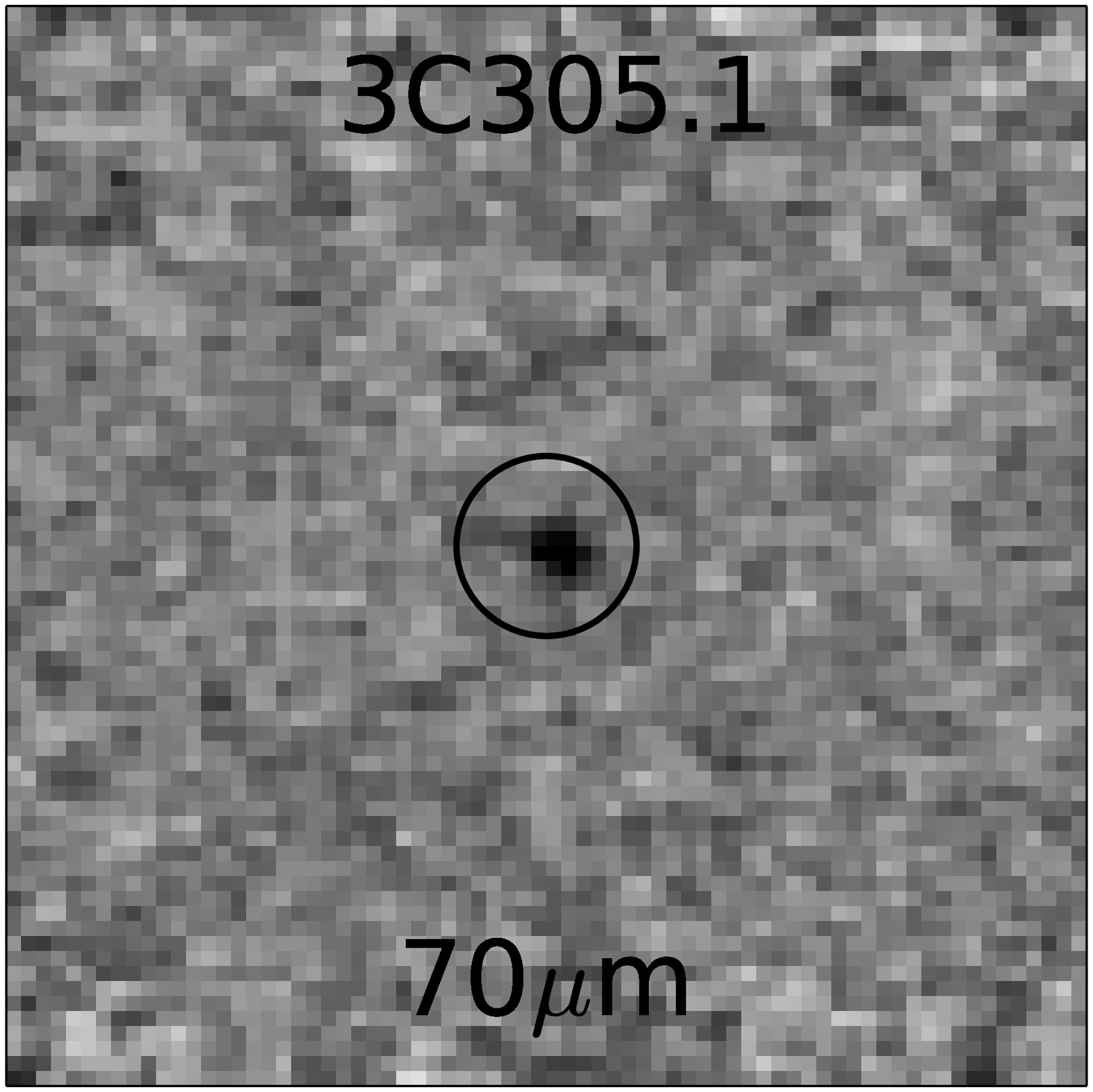}
      \includegraphics[width=1.5cm]{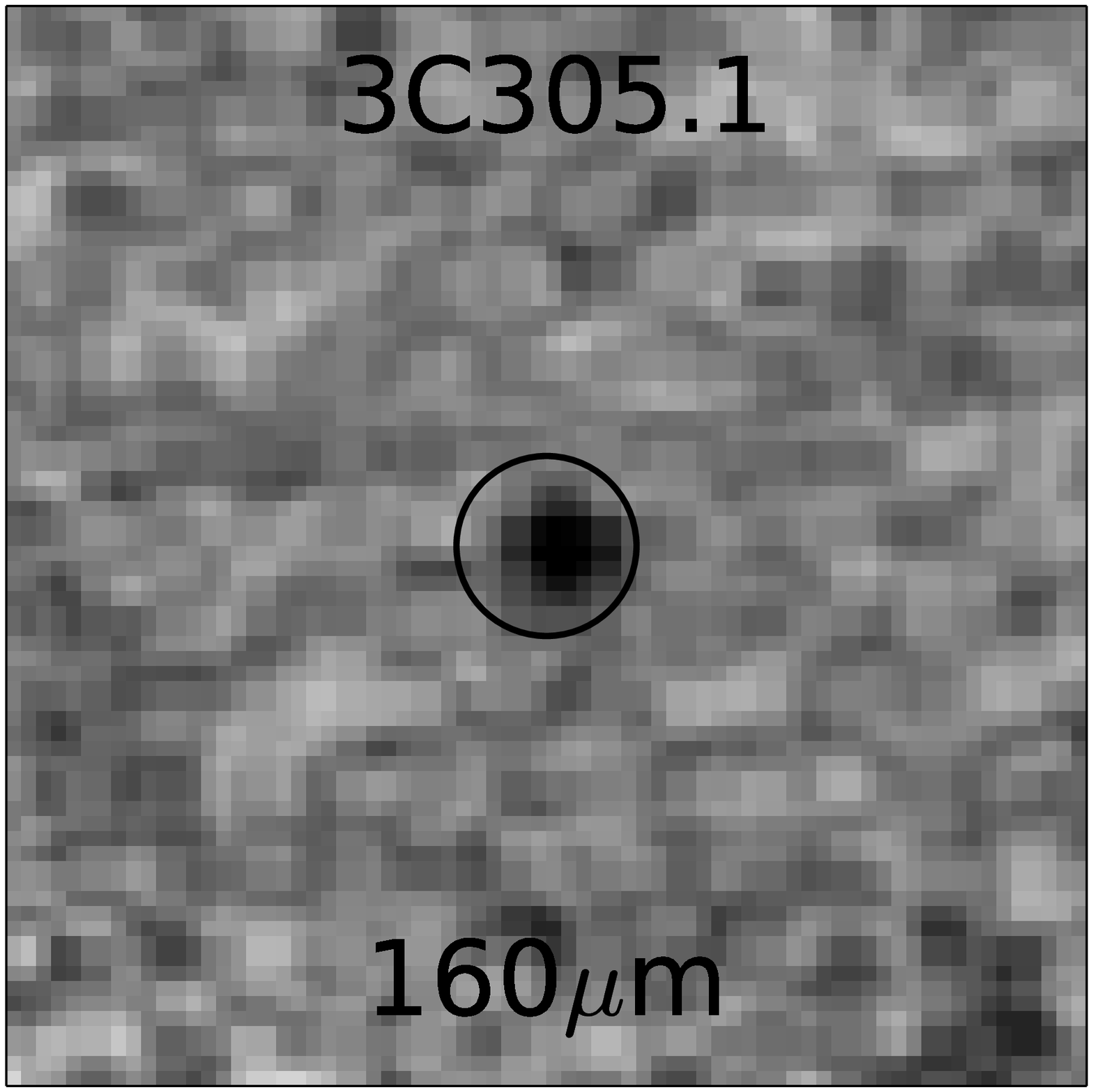}
      \includegraphics[width=1.5cm]{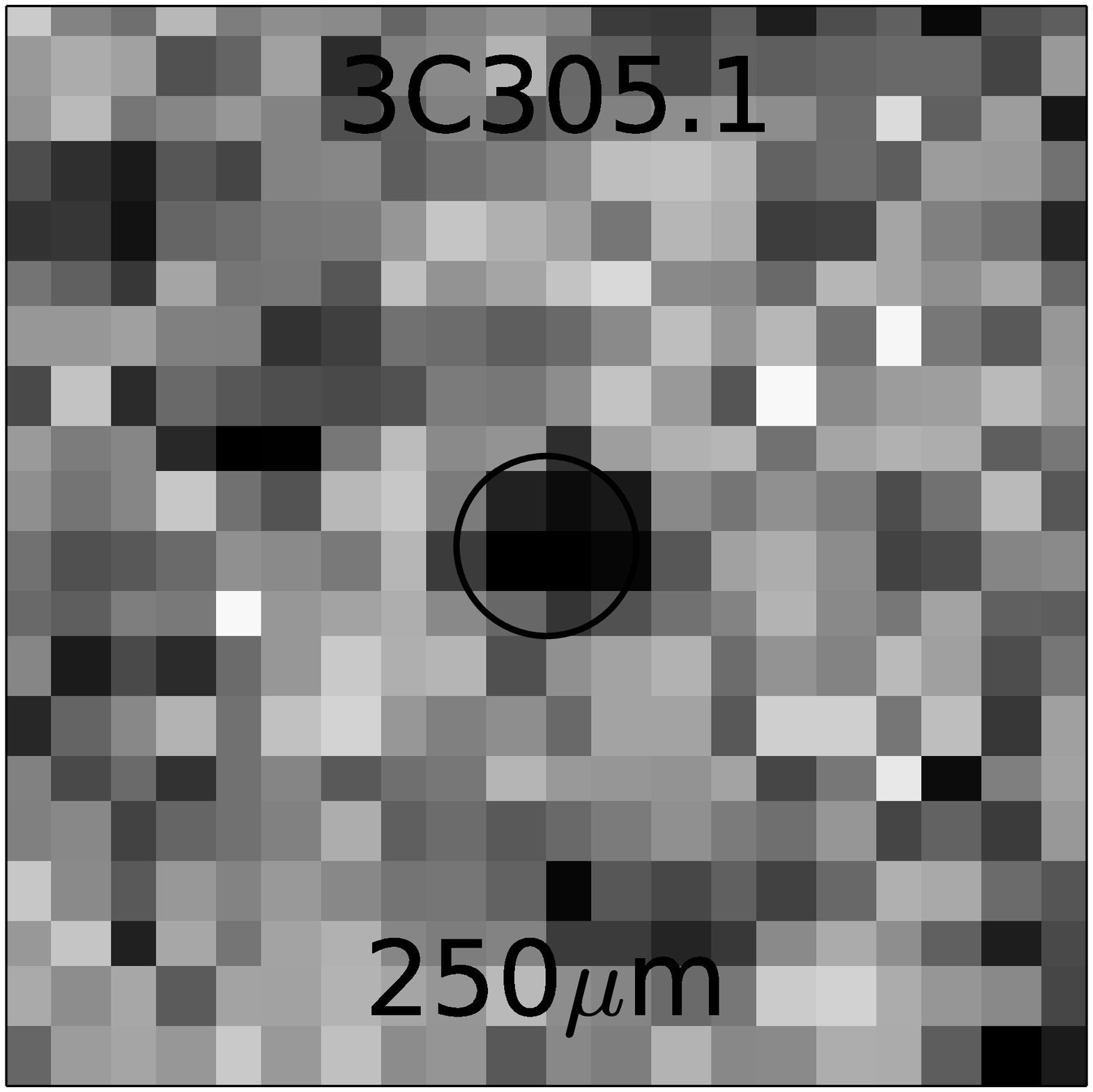}
      \includegraphics[width=1.5cm]{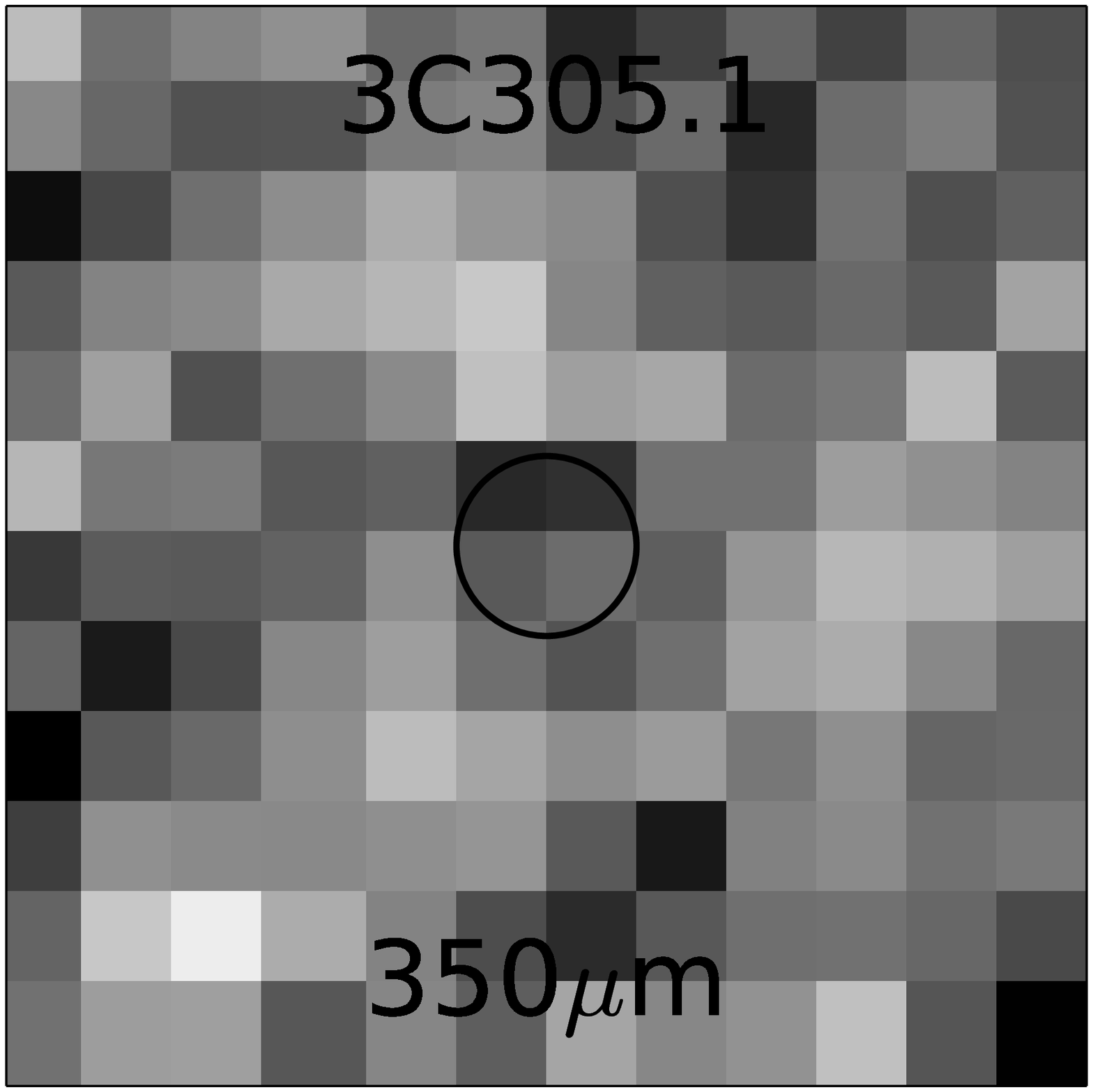}
      \includegraphics[width=1.5cm]{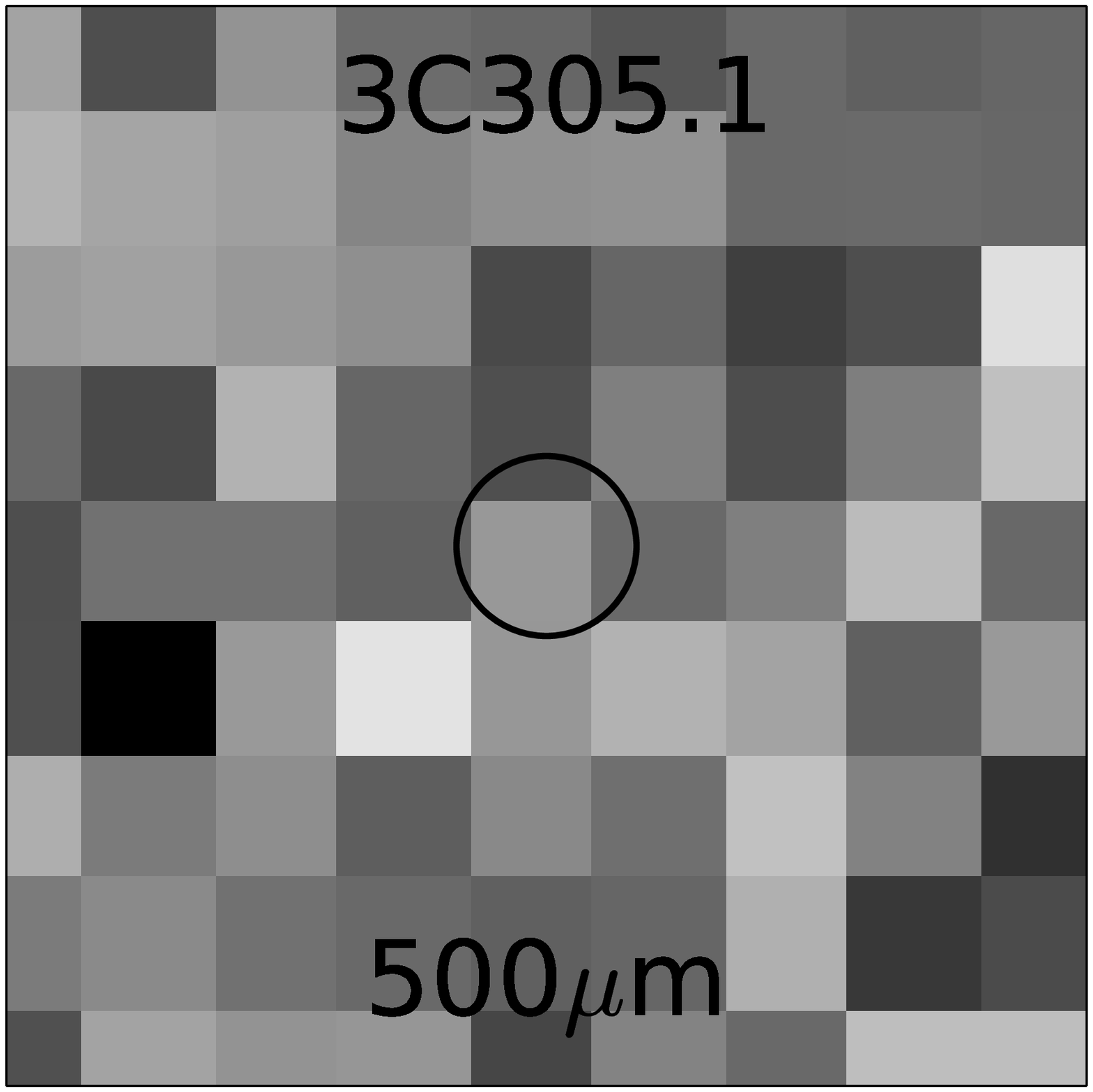}
      \\
      \includegraphics[width=1.5cm]{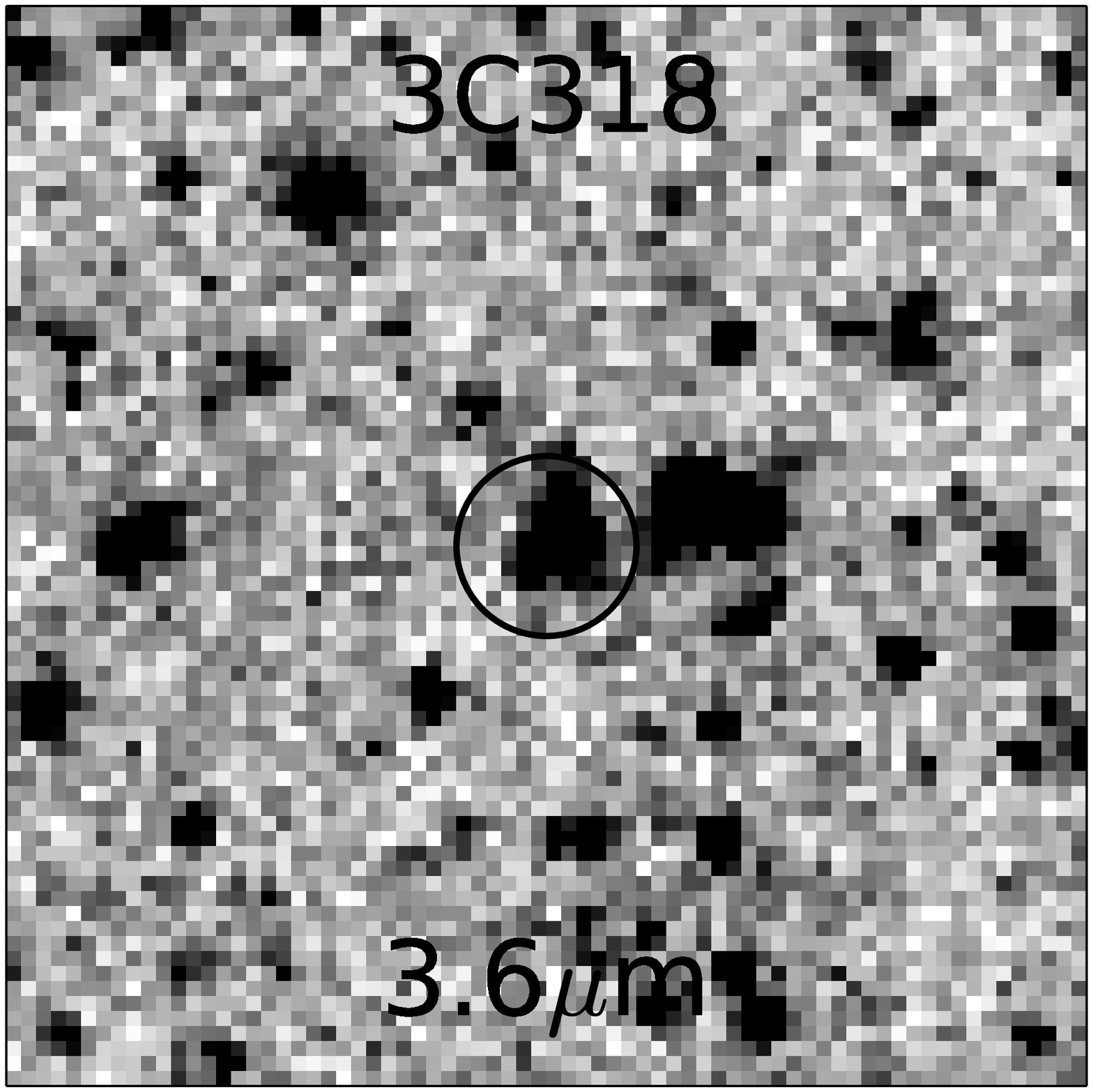}
      \includegraphics[width=1.5cm]{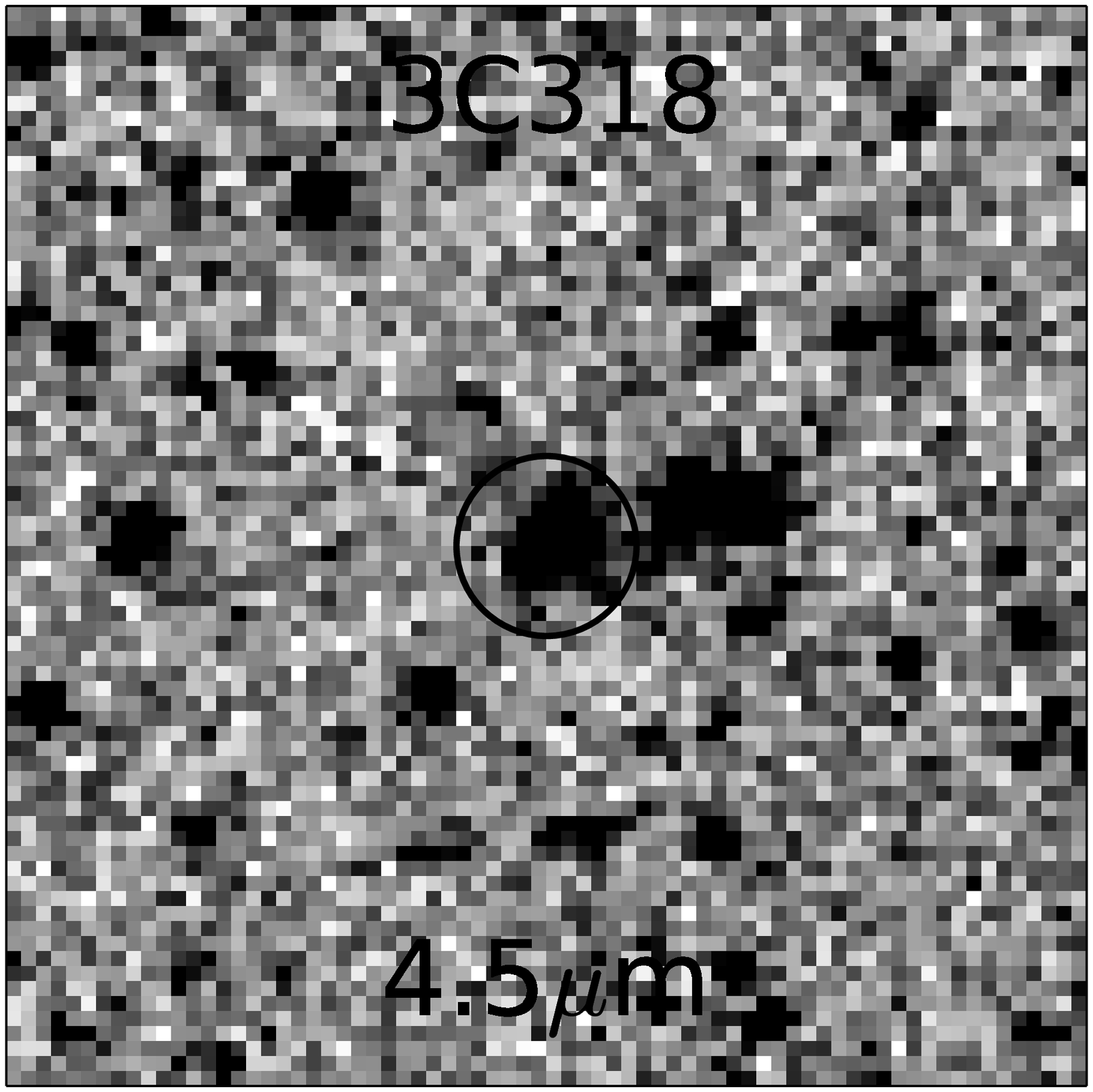}
      \includegraphics[width=1.5cm]{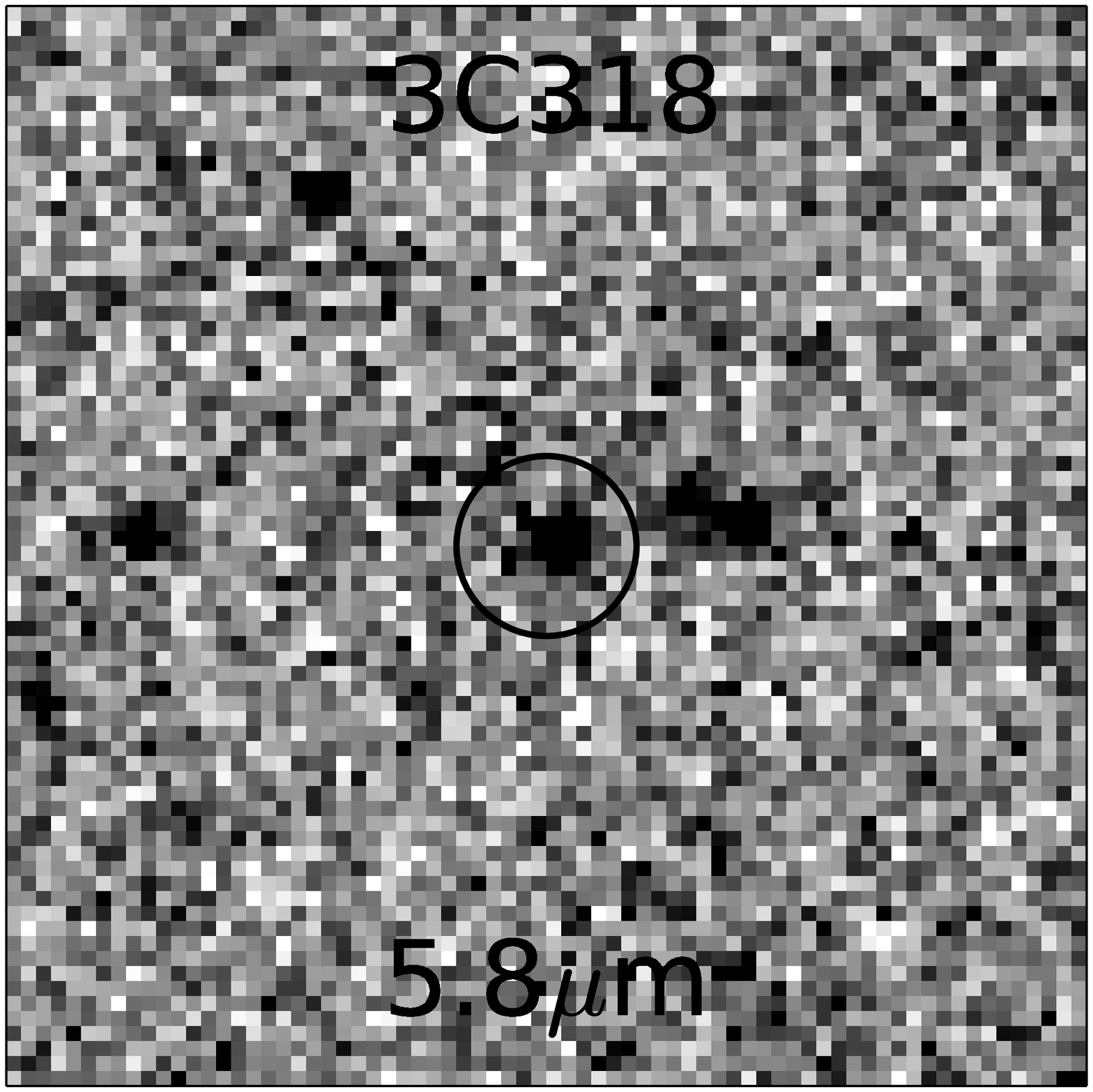}
      \includegraphics[width=1.5cm]{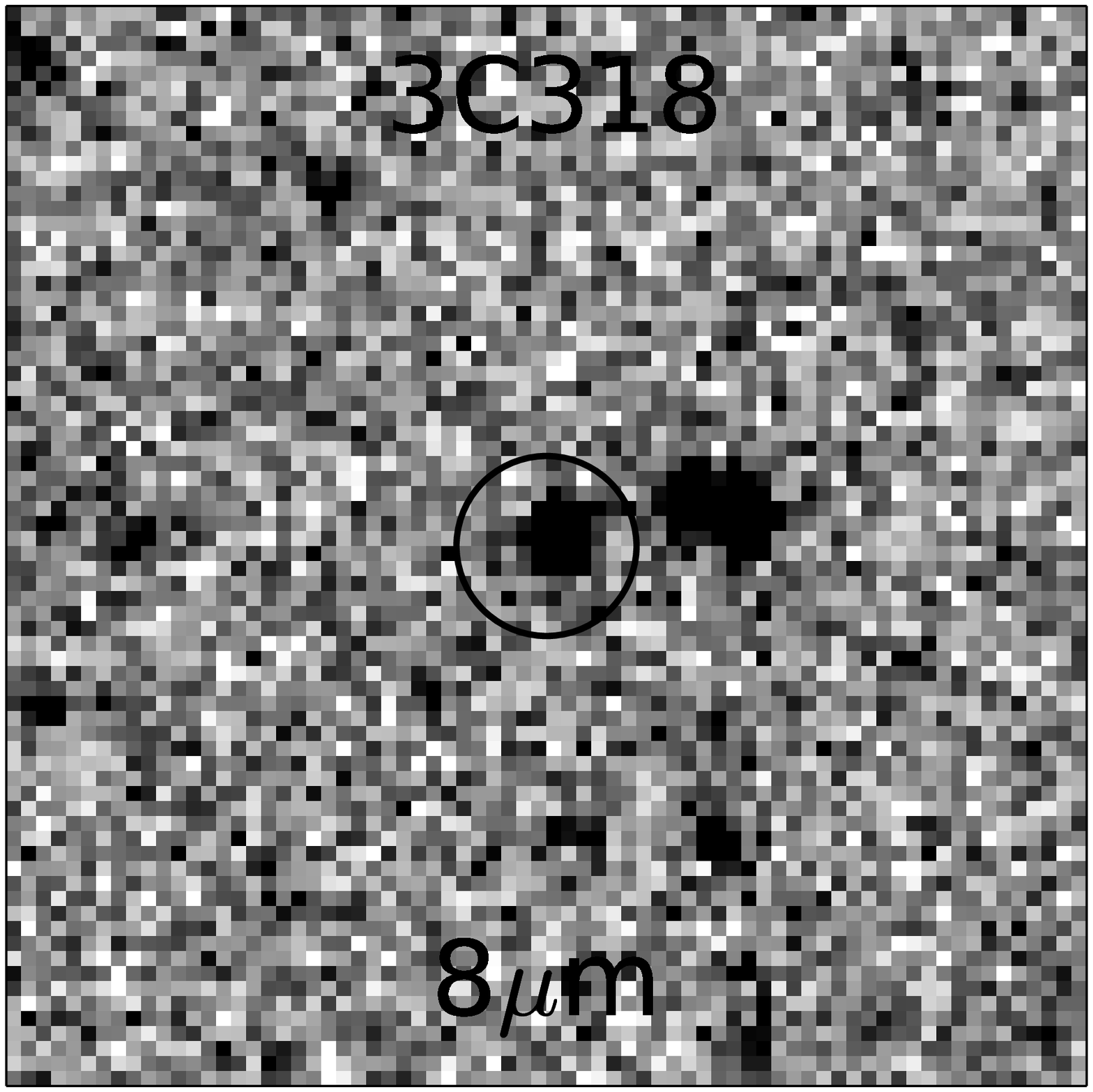}
      \includegraphics[width=1.5cm]{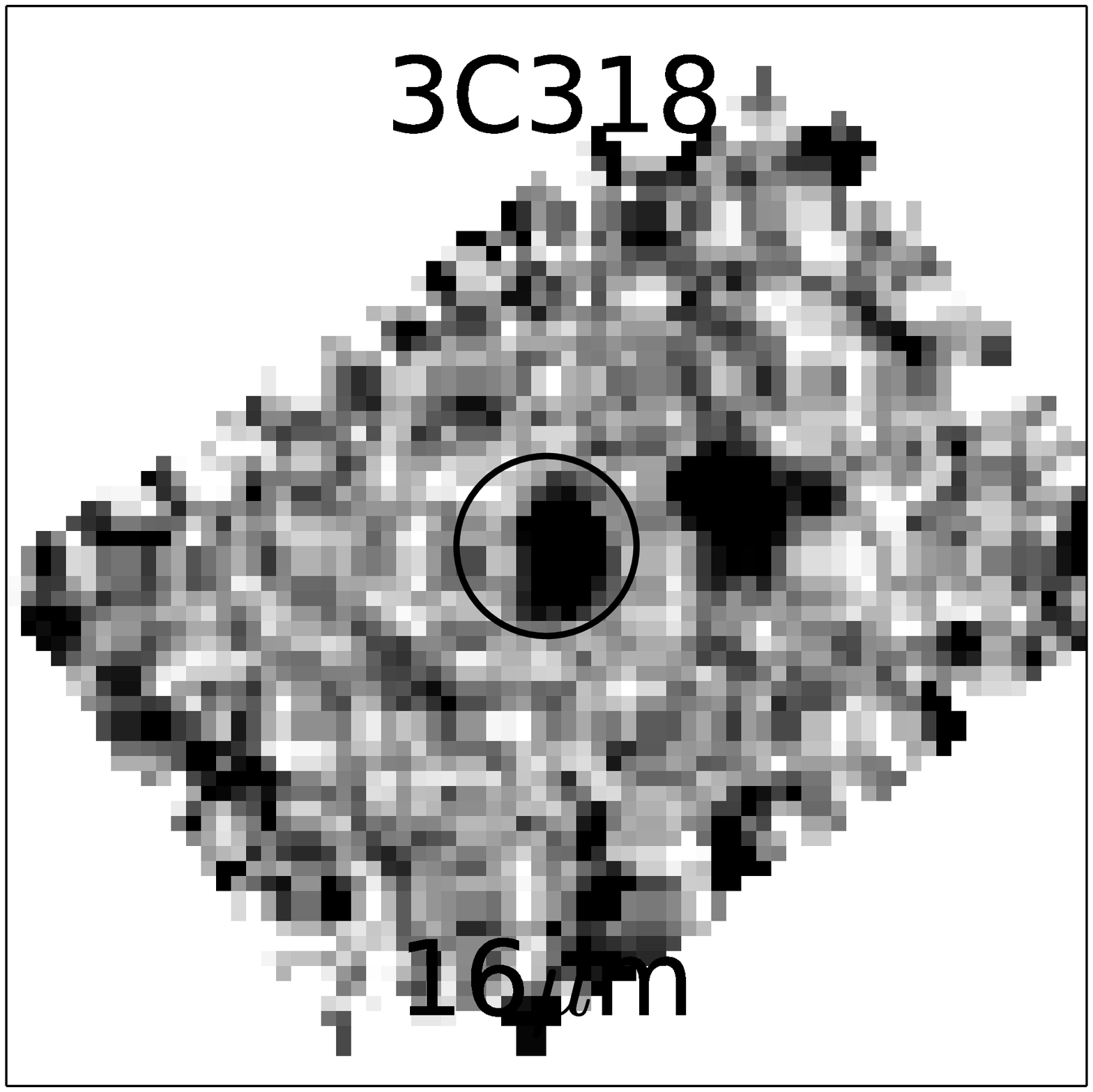}
      \includegraphics[width=1.5cm]{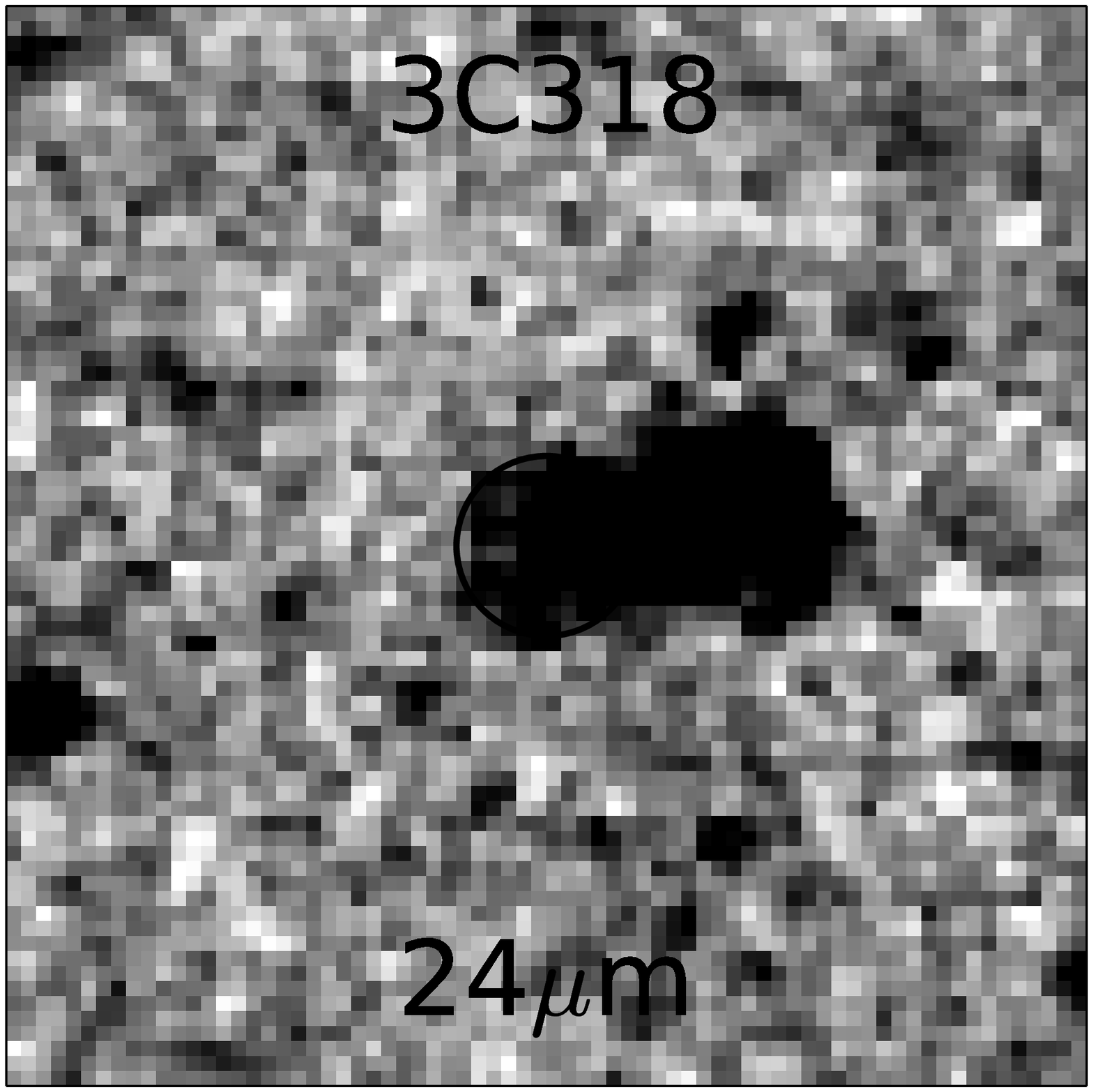}
      \includegraphics[width=1.5cm]{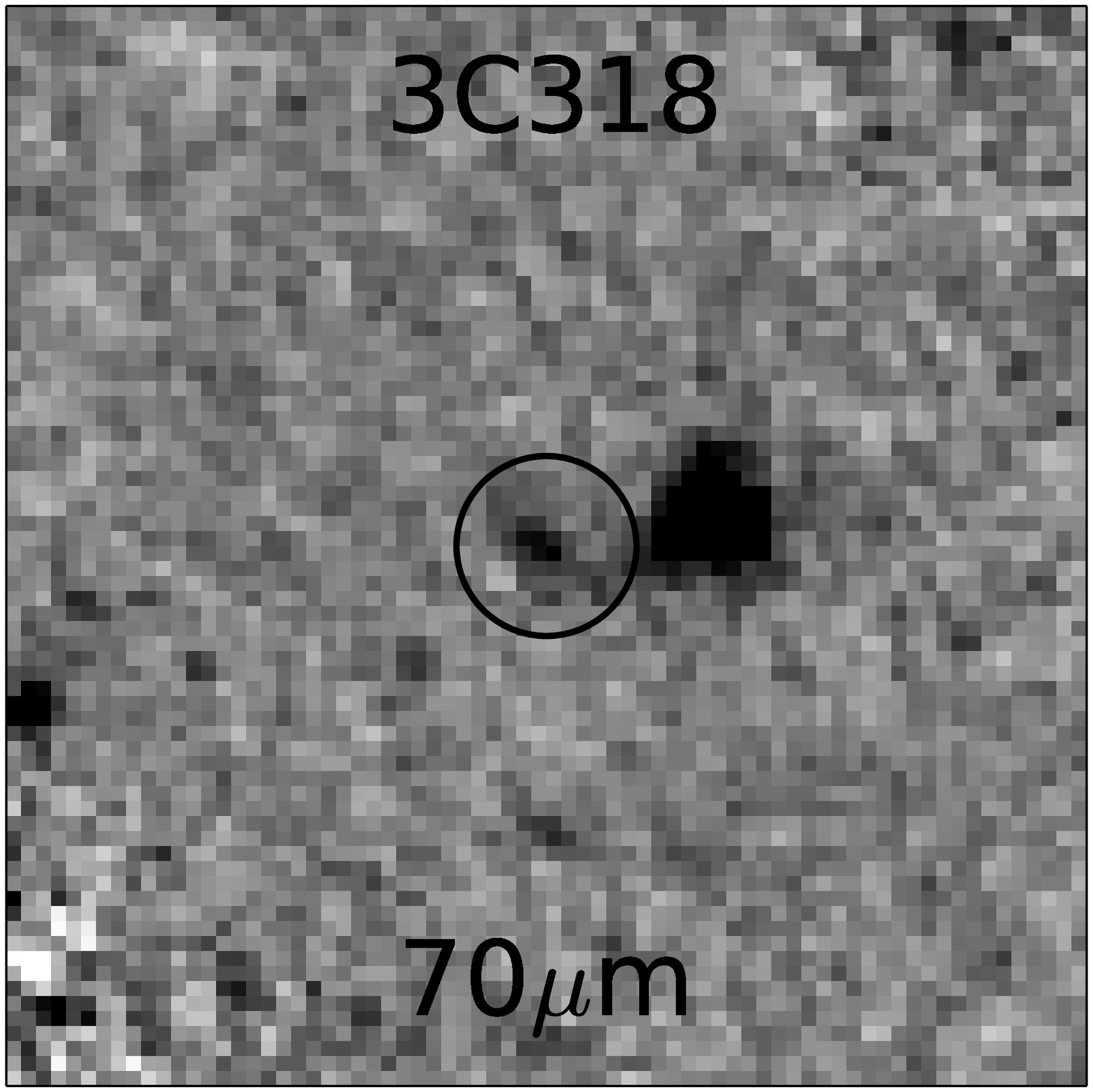}
      \includegraphics[width=1.5cm]{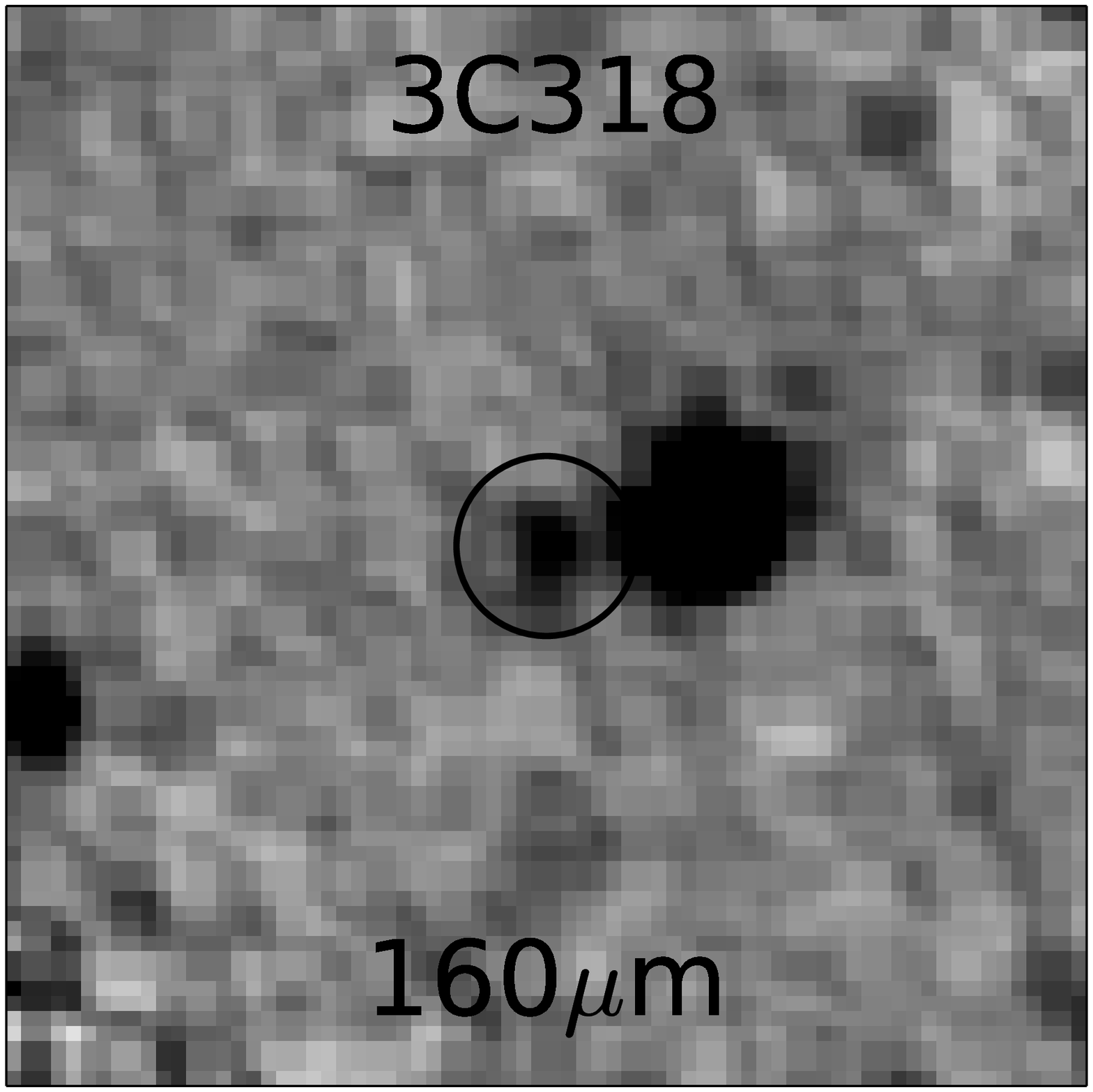}
      \includegraphics[width=1.5cm]{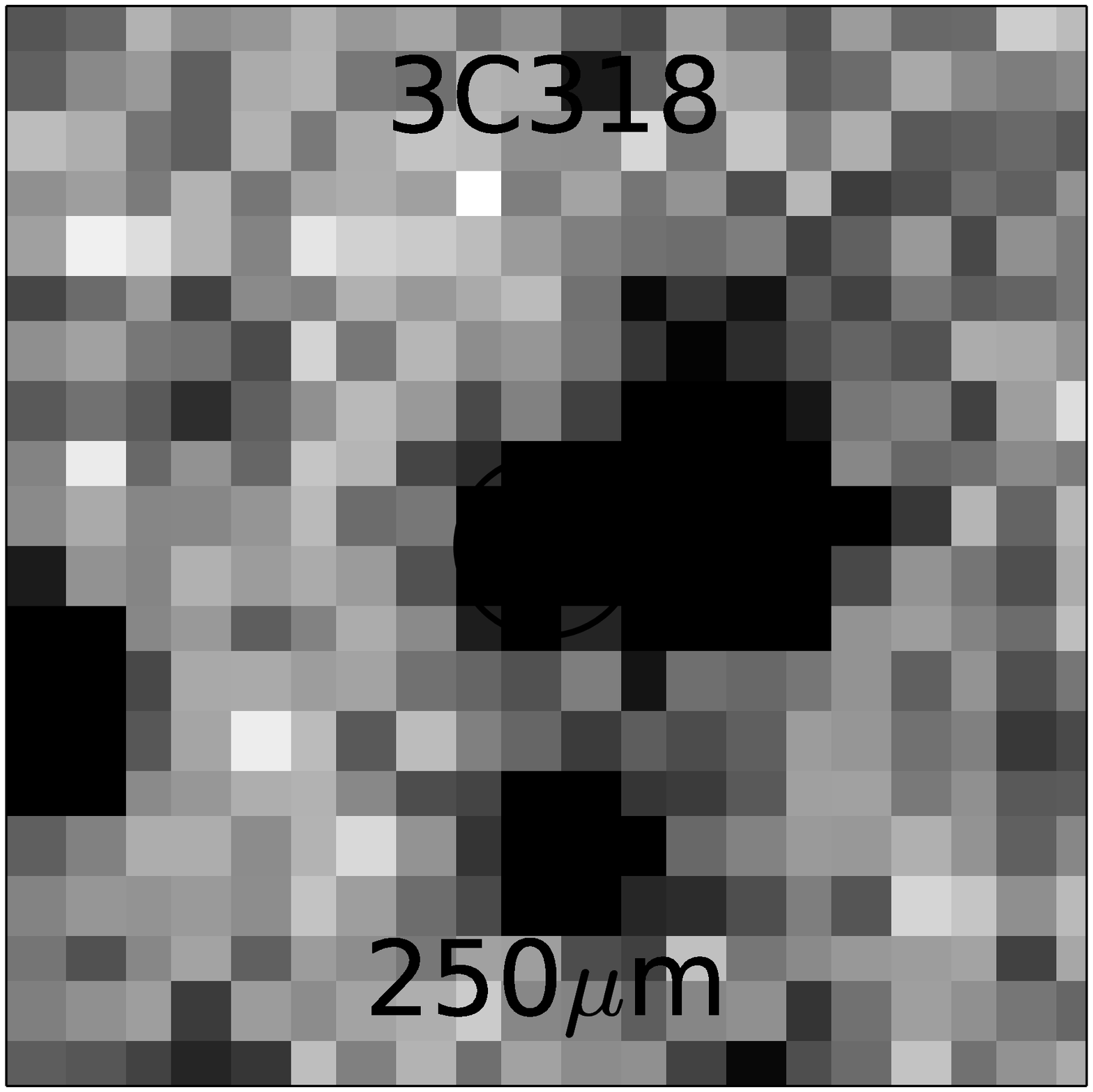}
      \includegraphics[width=1.5cm]{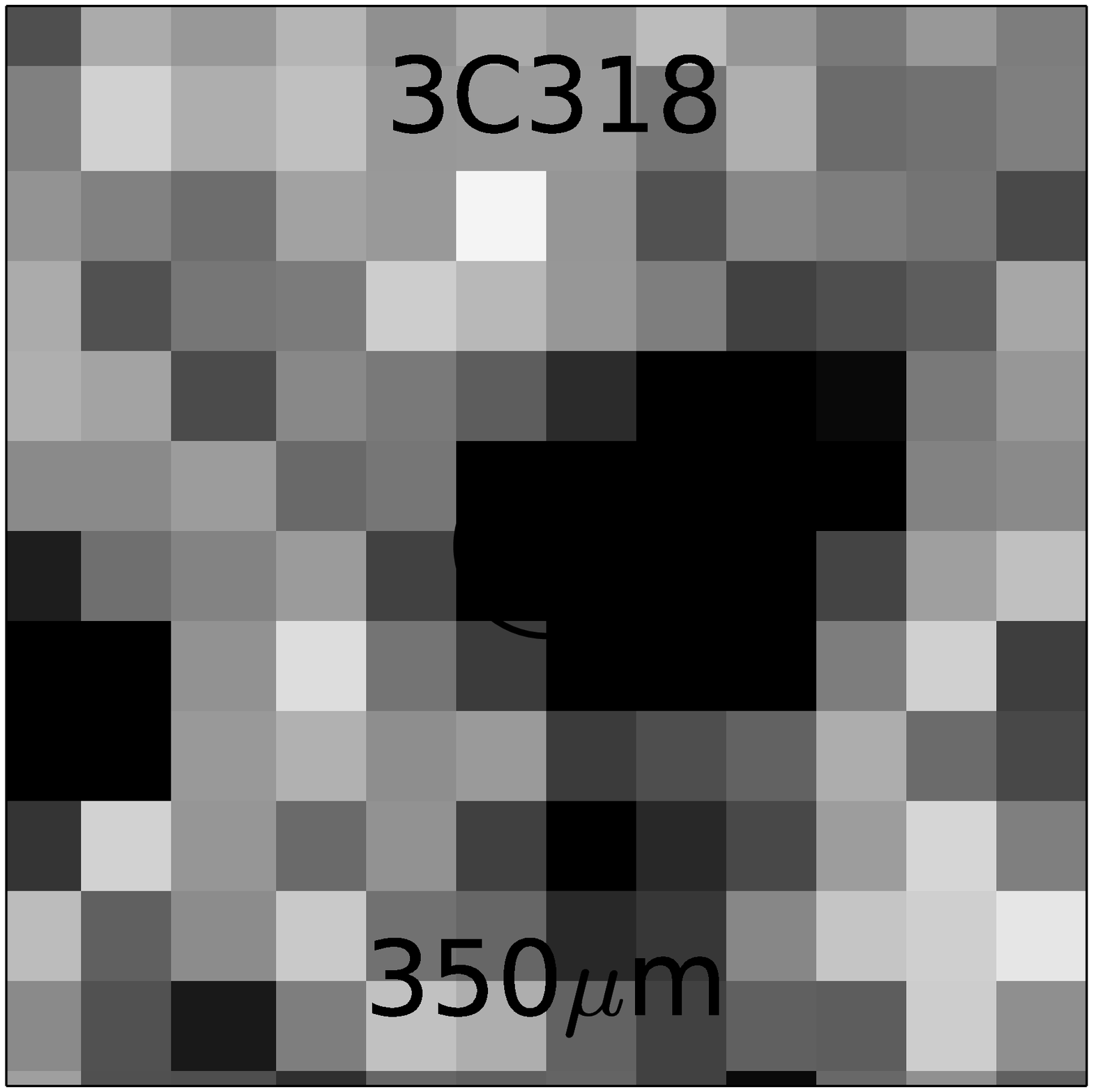}
      \includegraphics[width=1.5cm]{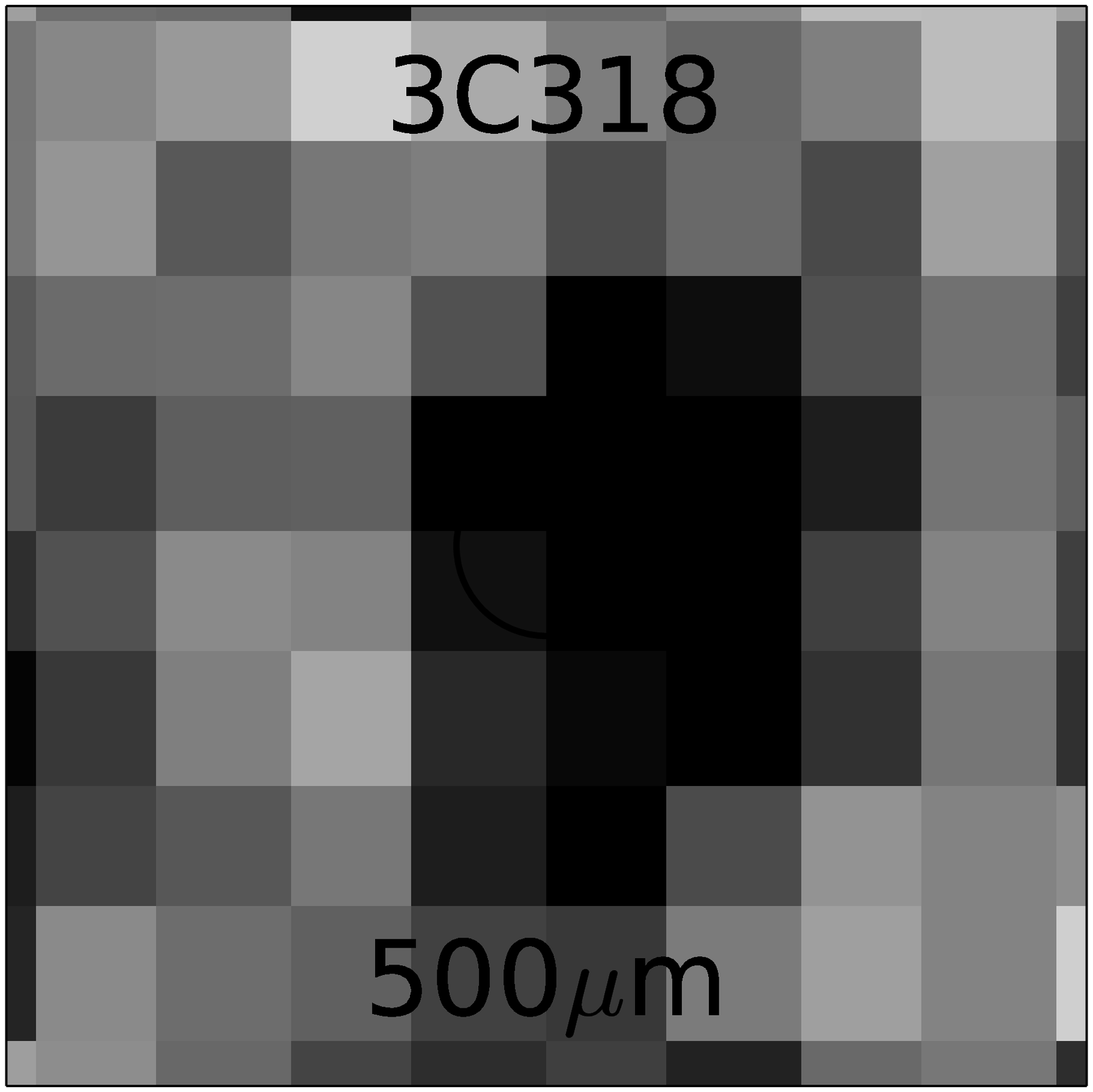}
      \\
      \includegraphics[width=1.5cm]{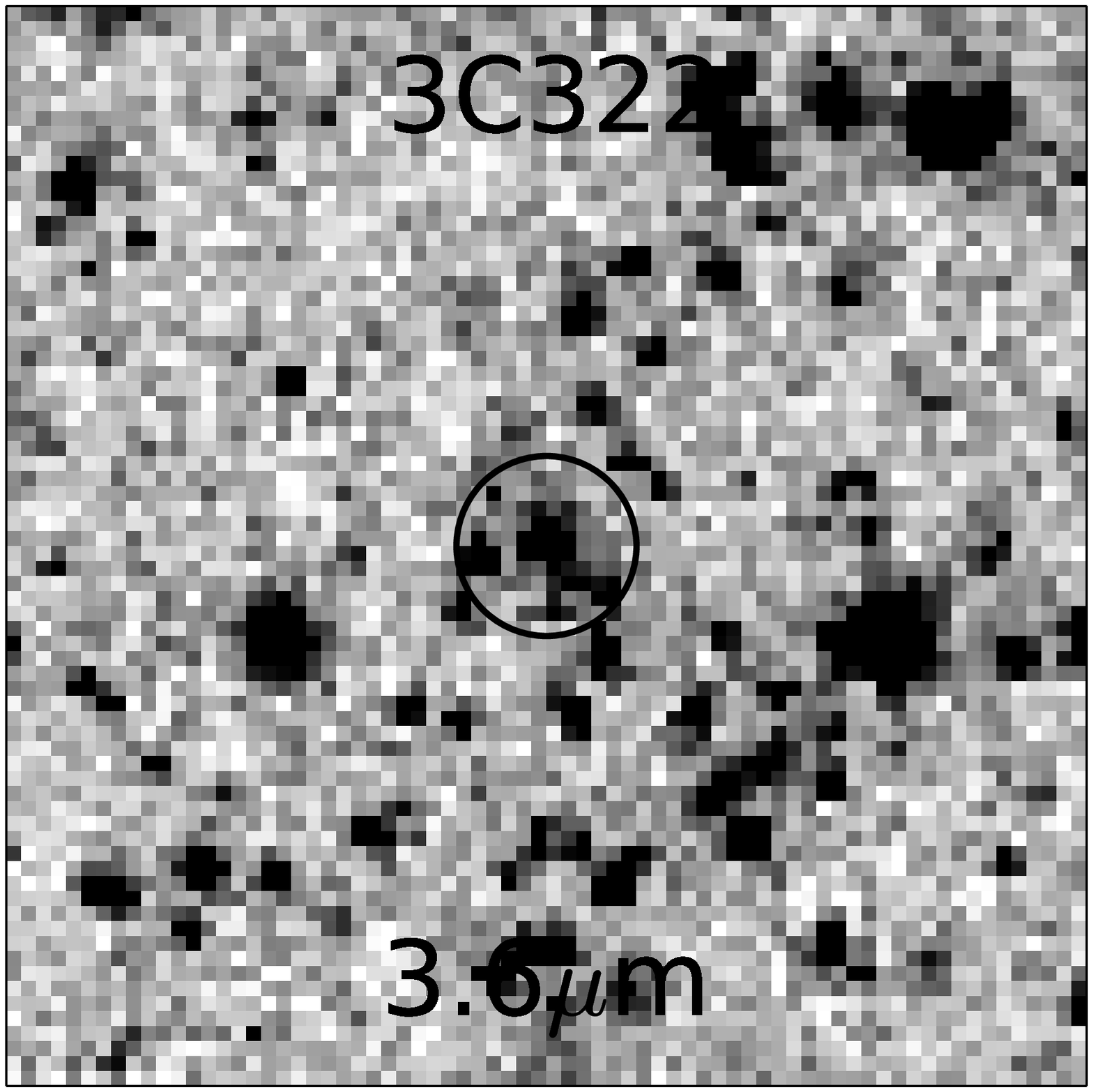}
      \includegraphics[width=1.5cm]{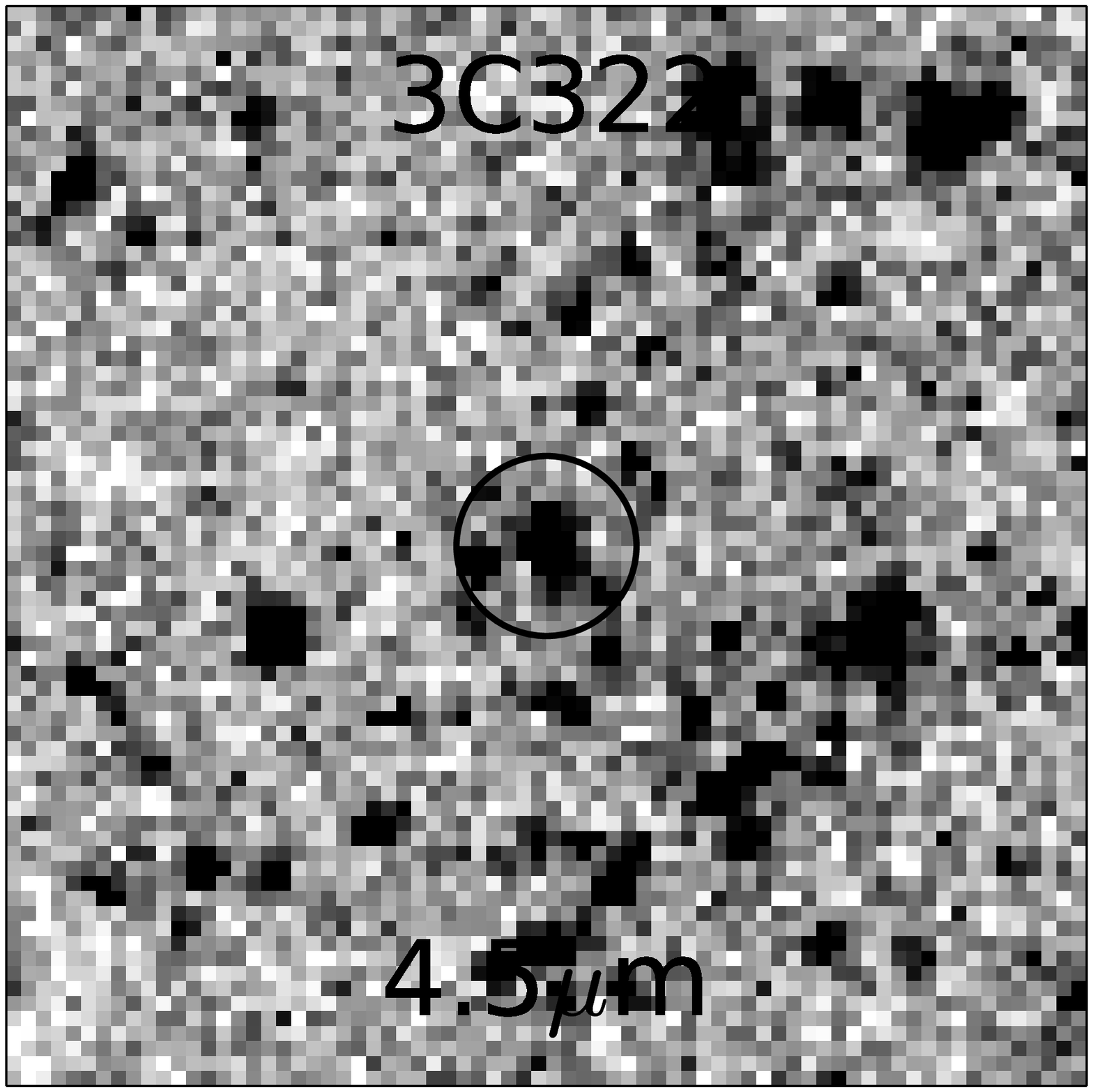}
      \includegraphics[width=1.5cm]{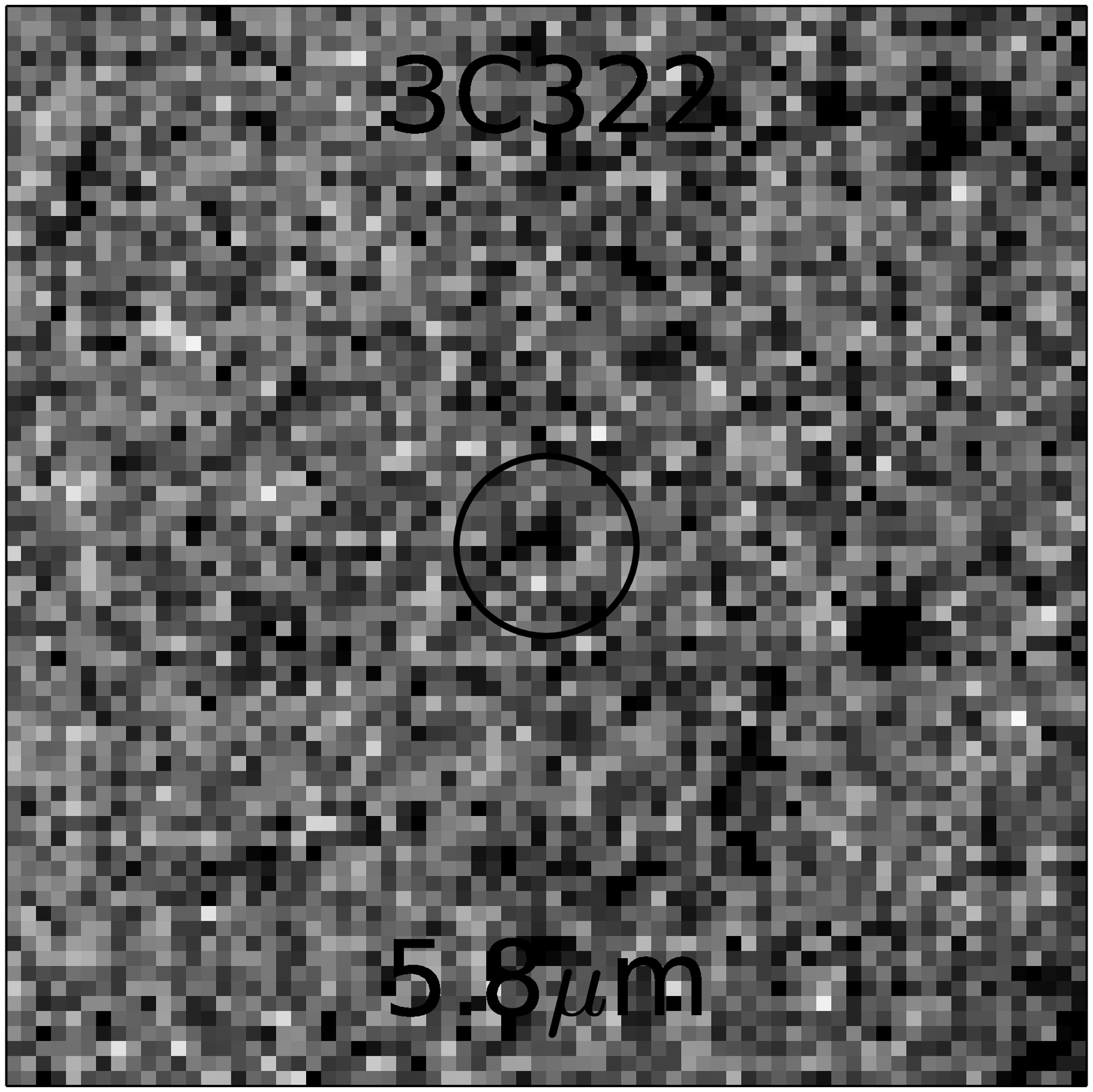}
      \includegraphics[width=1.5cm]{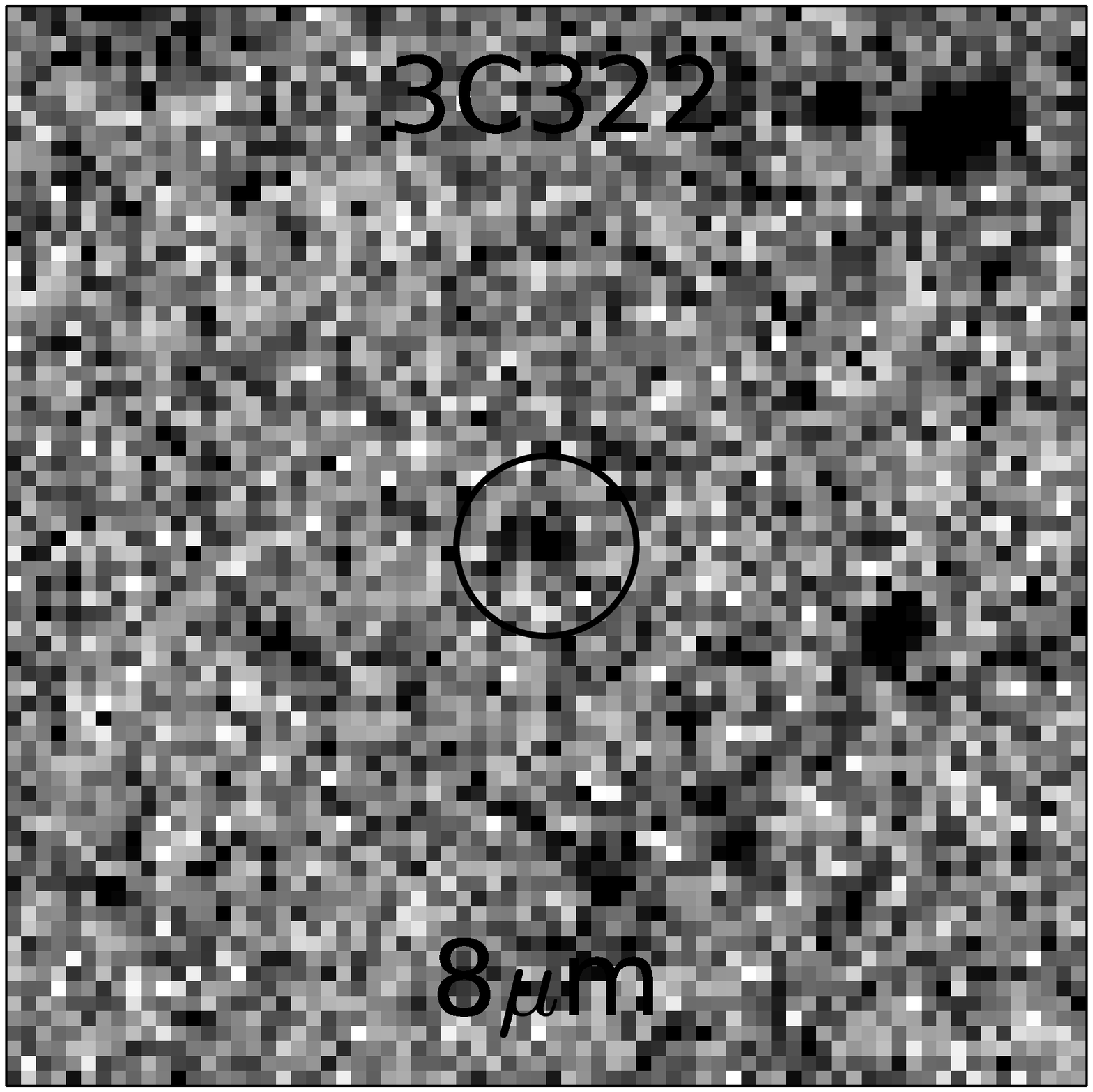}
      \includegraphics[width=1.5cm]{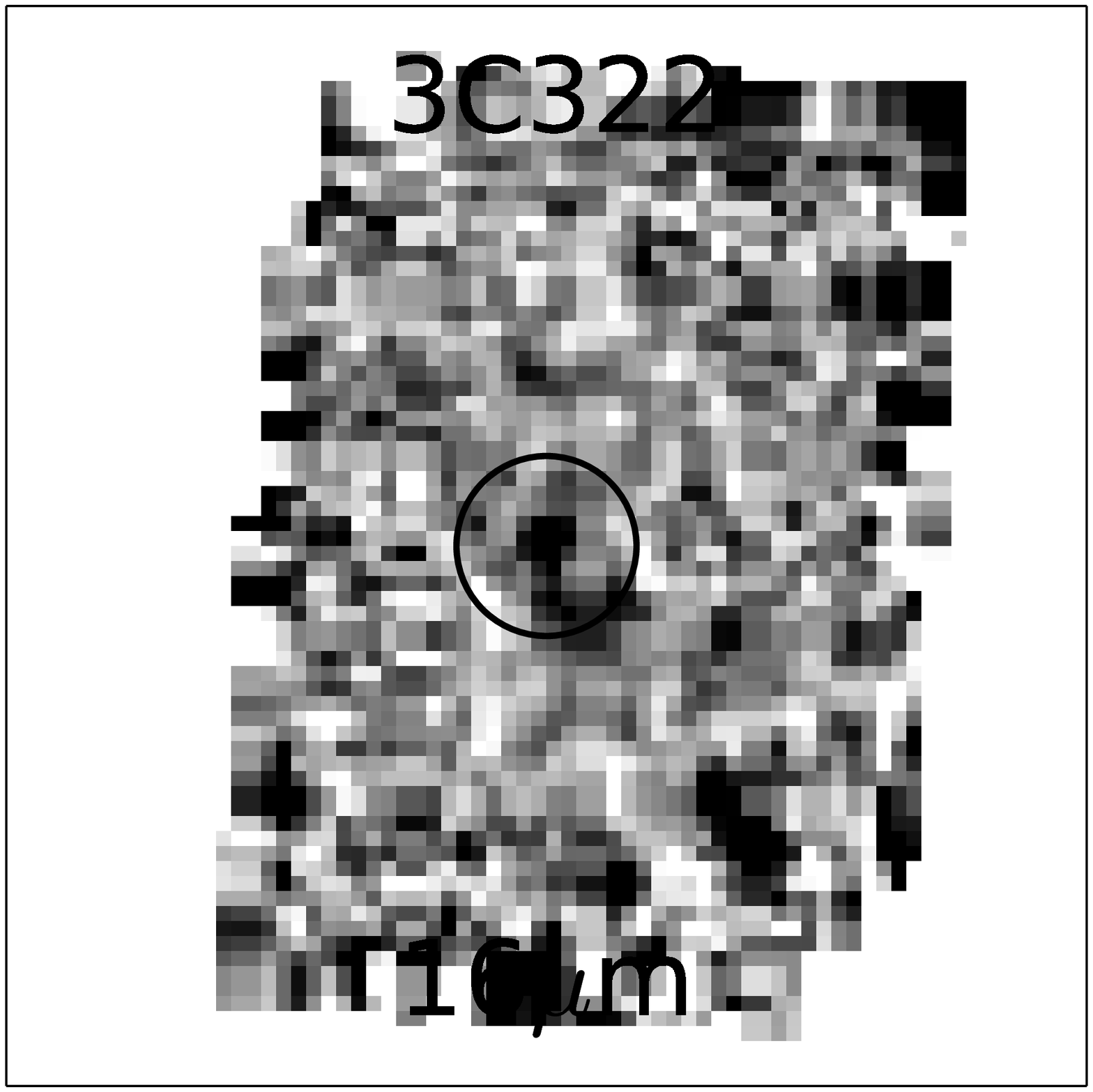}
      \includegraphics[width=1.5cm]{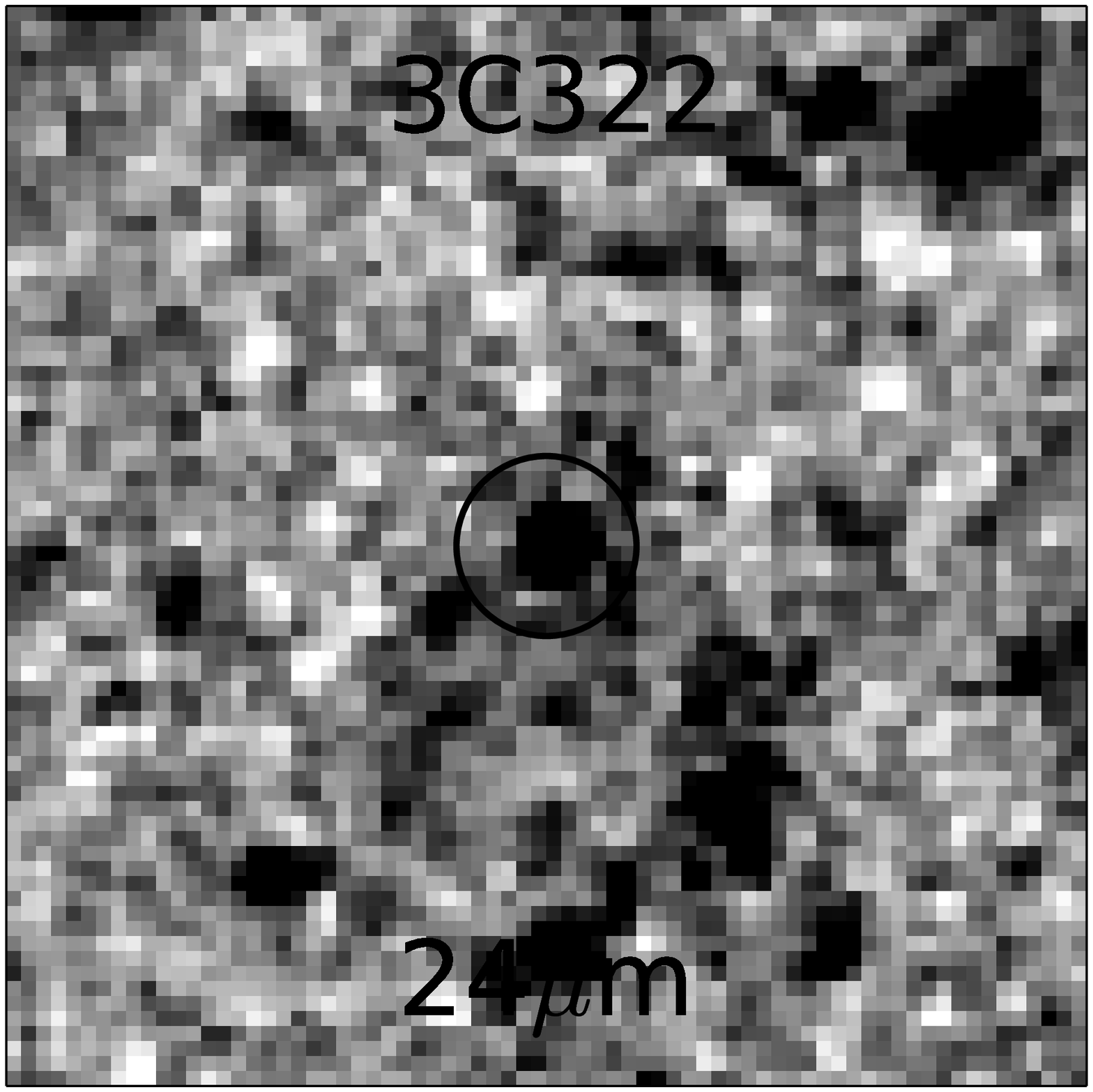}
      \includegraphics[width=1.5cm]{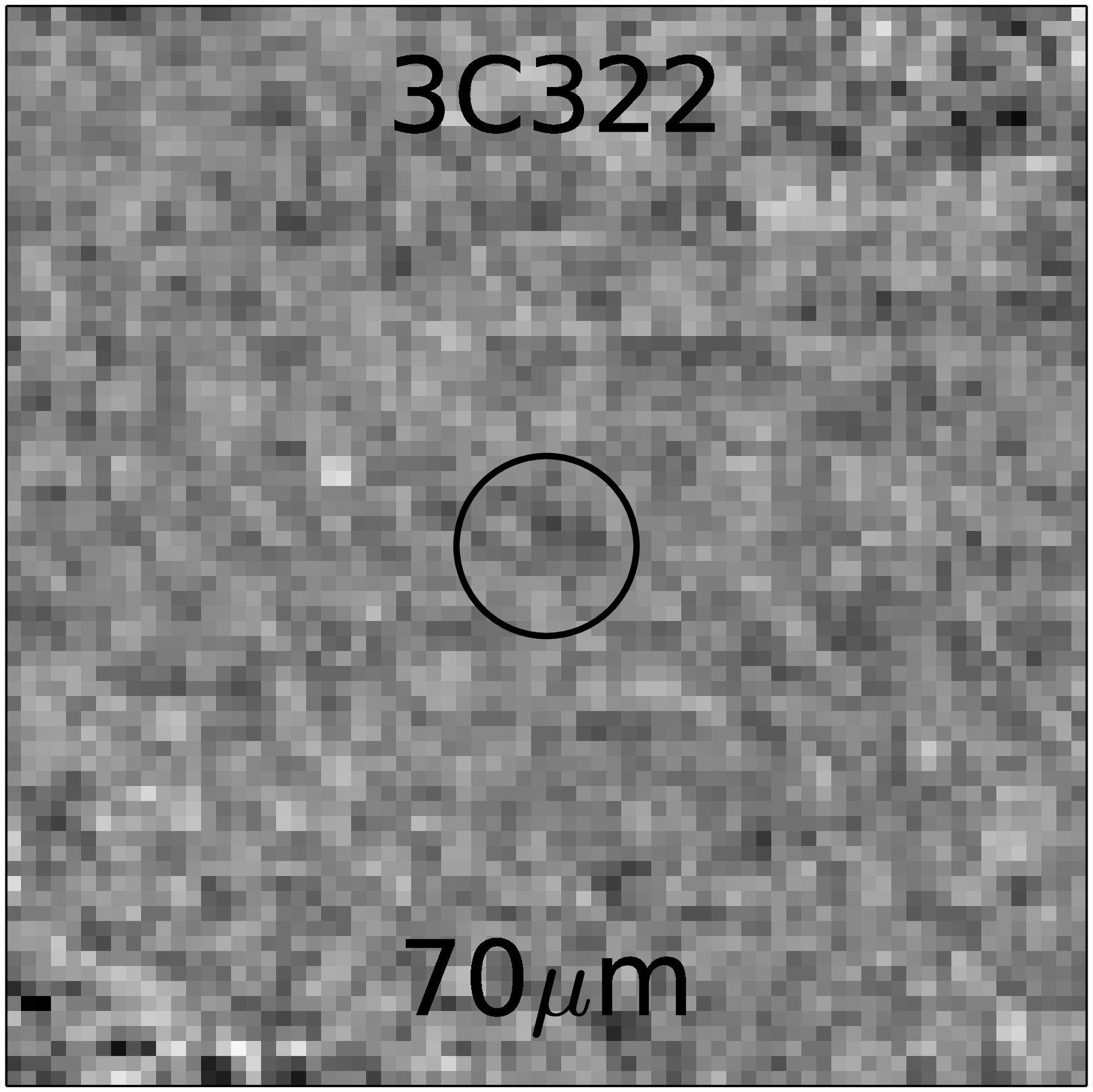}
      \includegraphics[width=1.5cm]{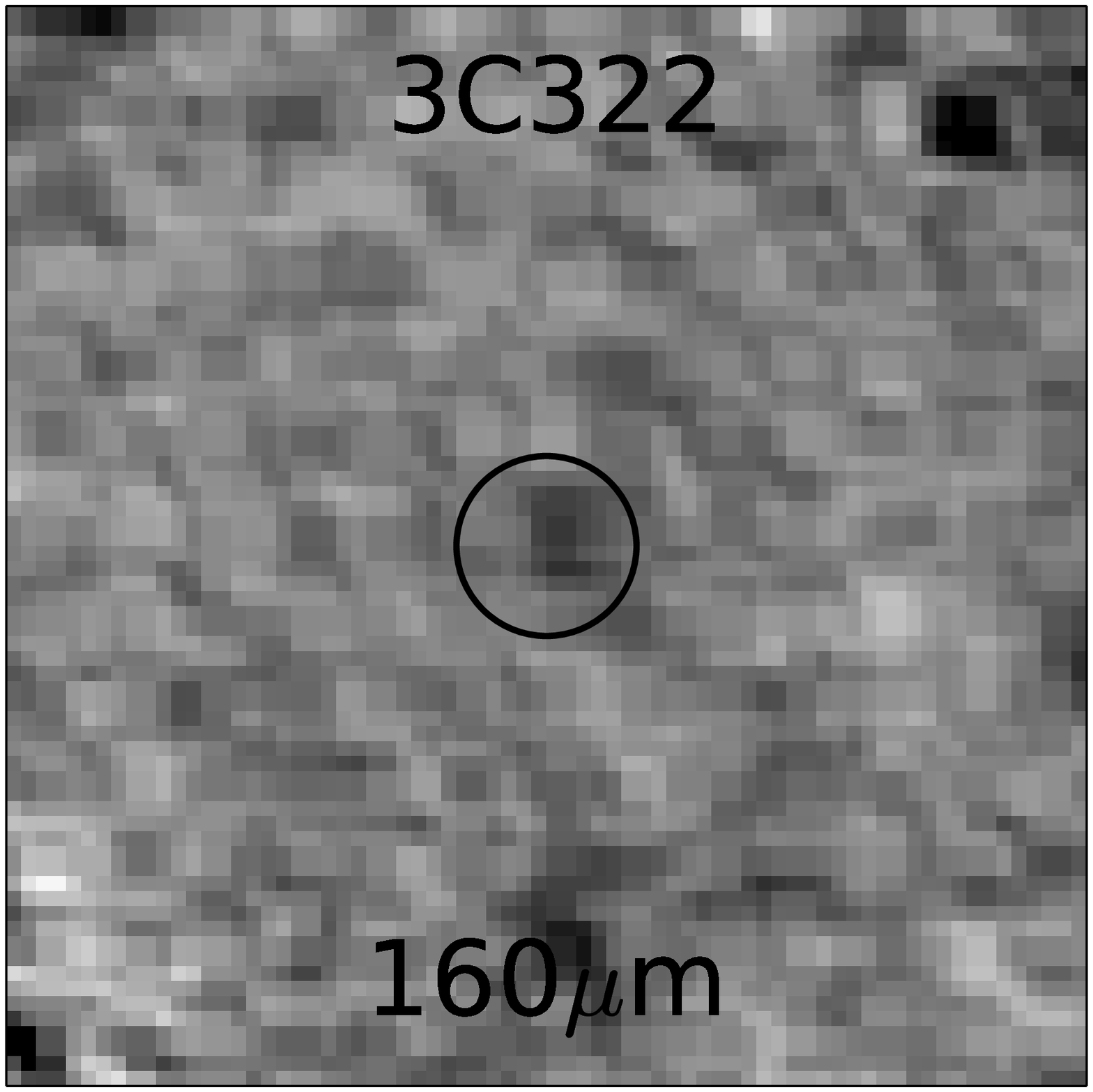}
      \includegraphics[width=1.5cm]{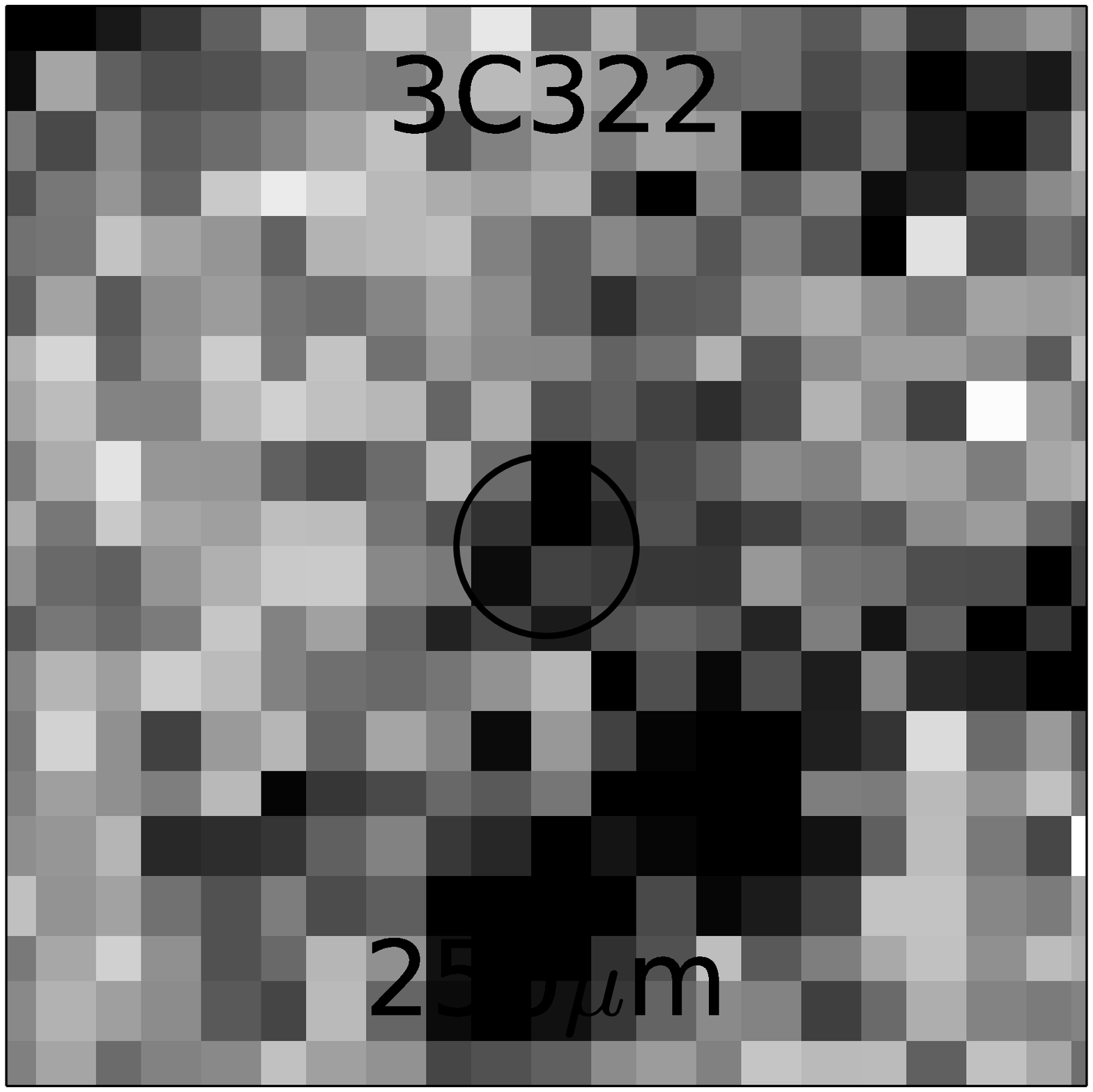}
      \includegraphics[width=1.5cm]{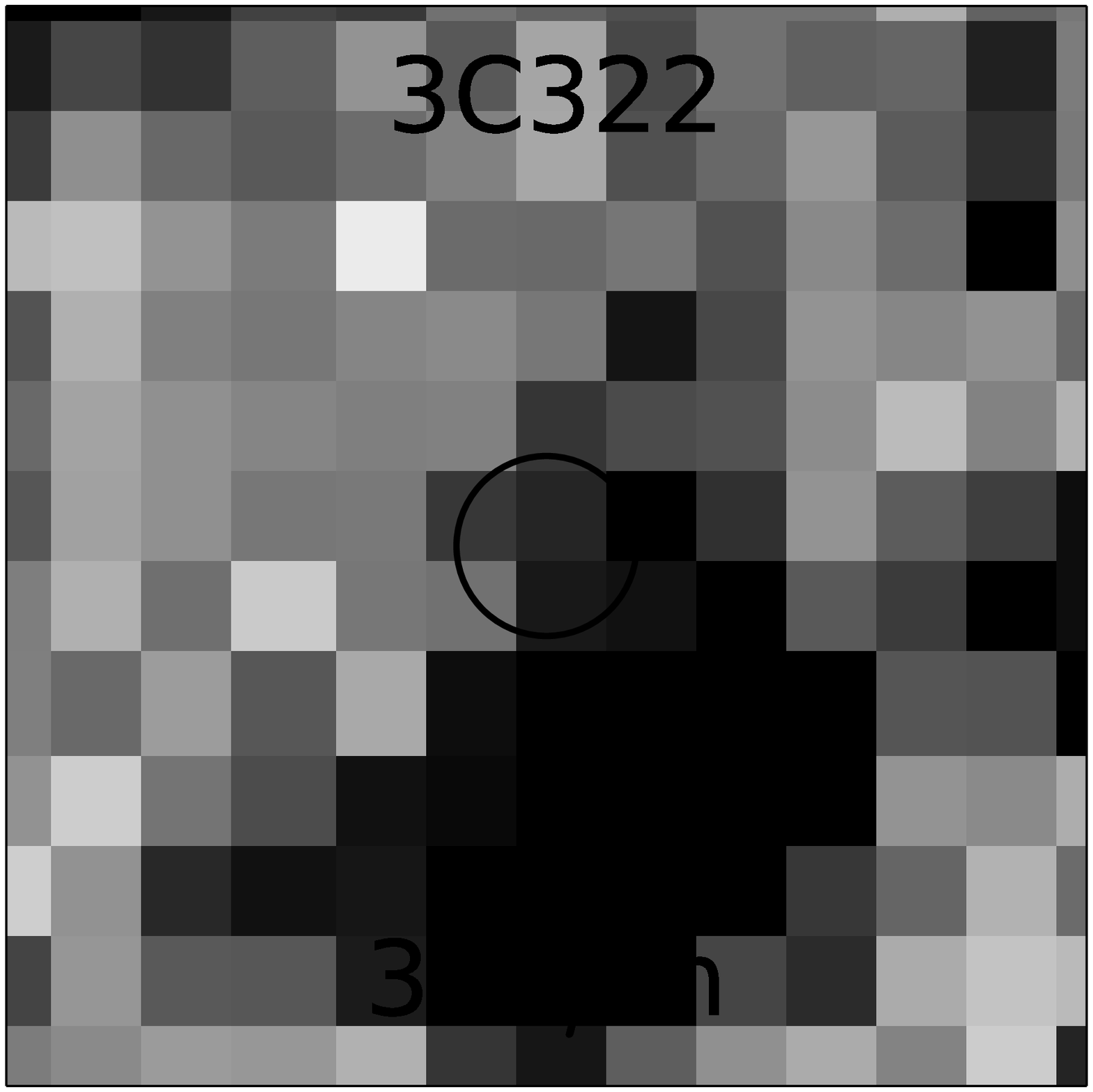}
      \includegraphics[width=1.5cm]{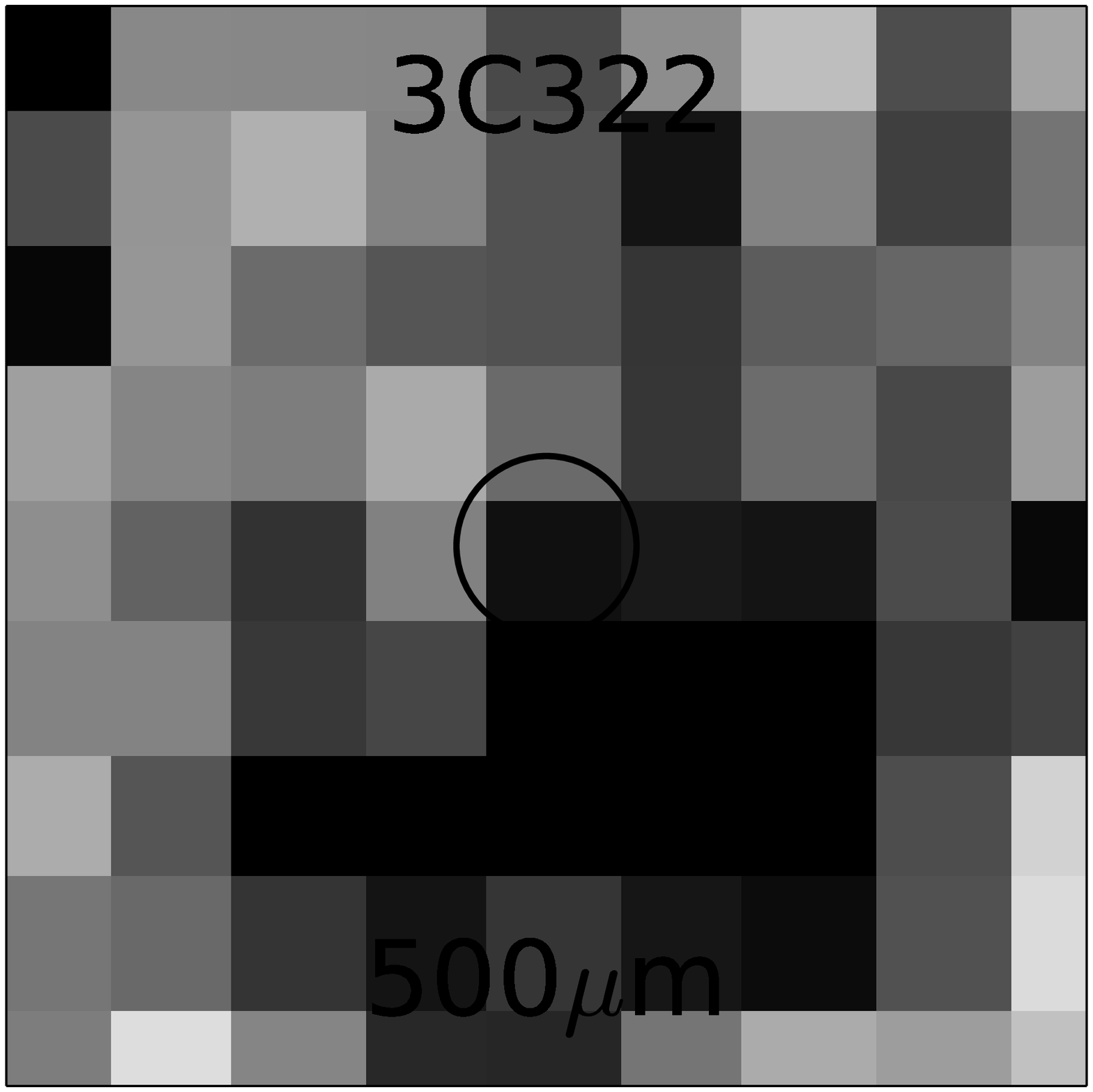}
      \\
      \includegraphics[width=1.5cm]{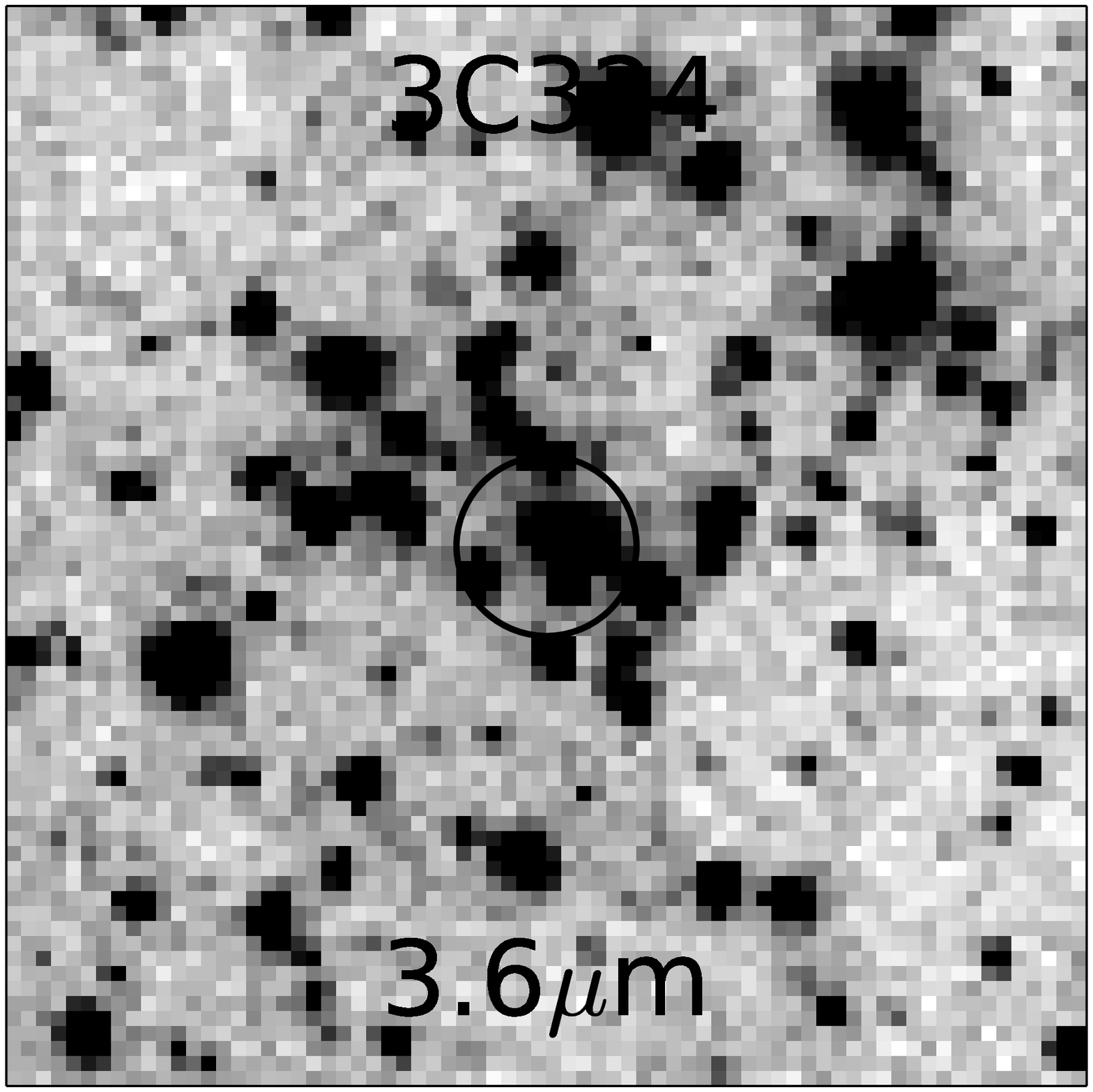}
      \includegraphics[width=1.5cm]{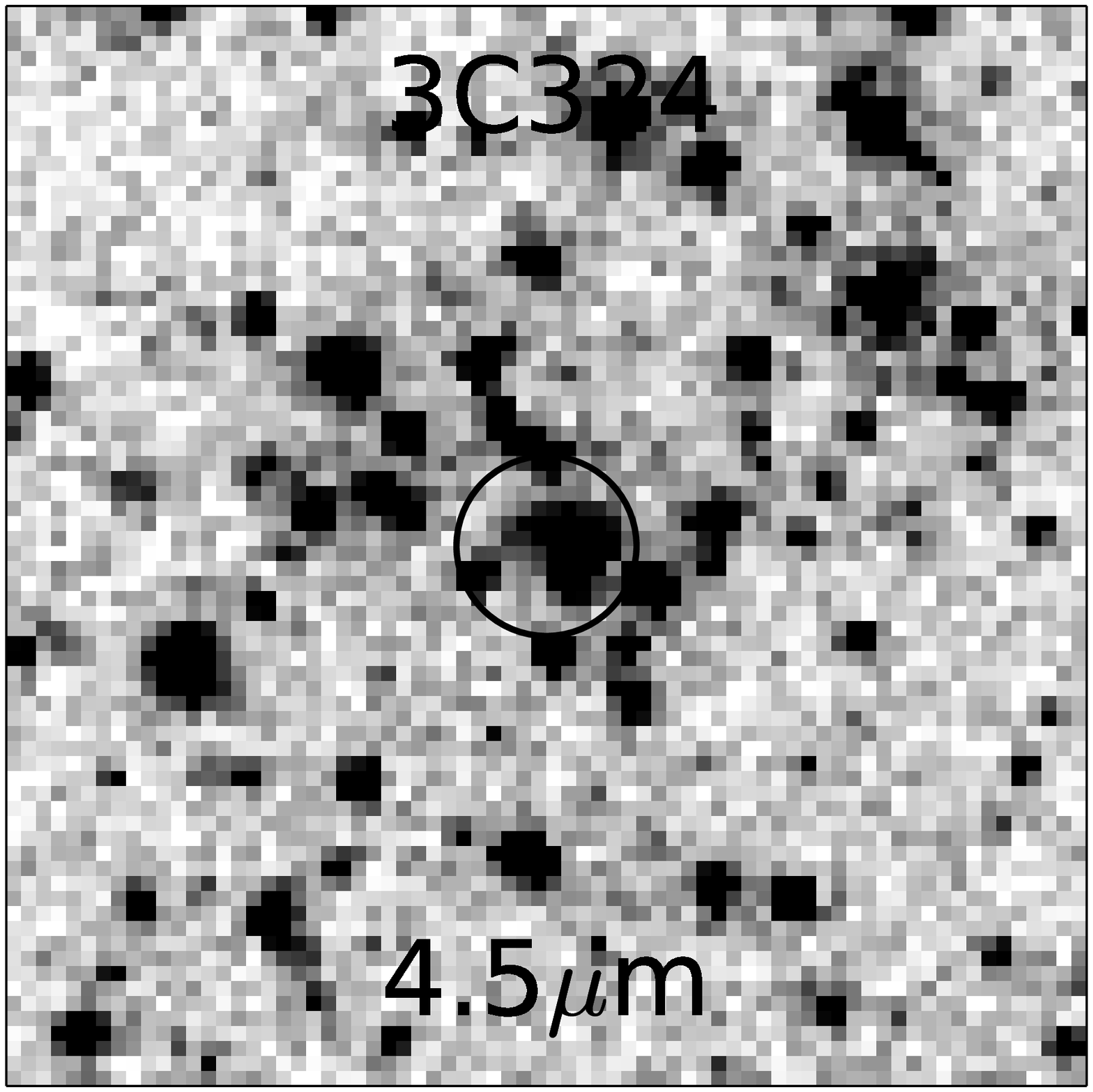}
      \includegraphics[width=1.5cm]{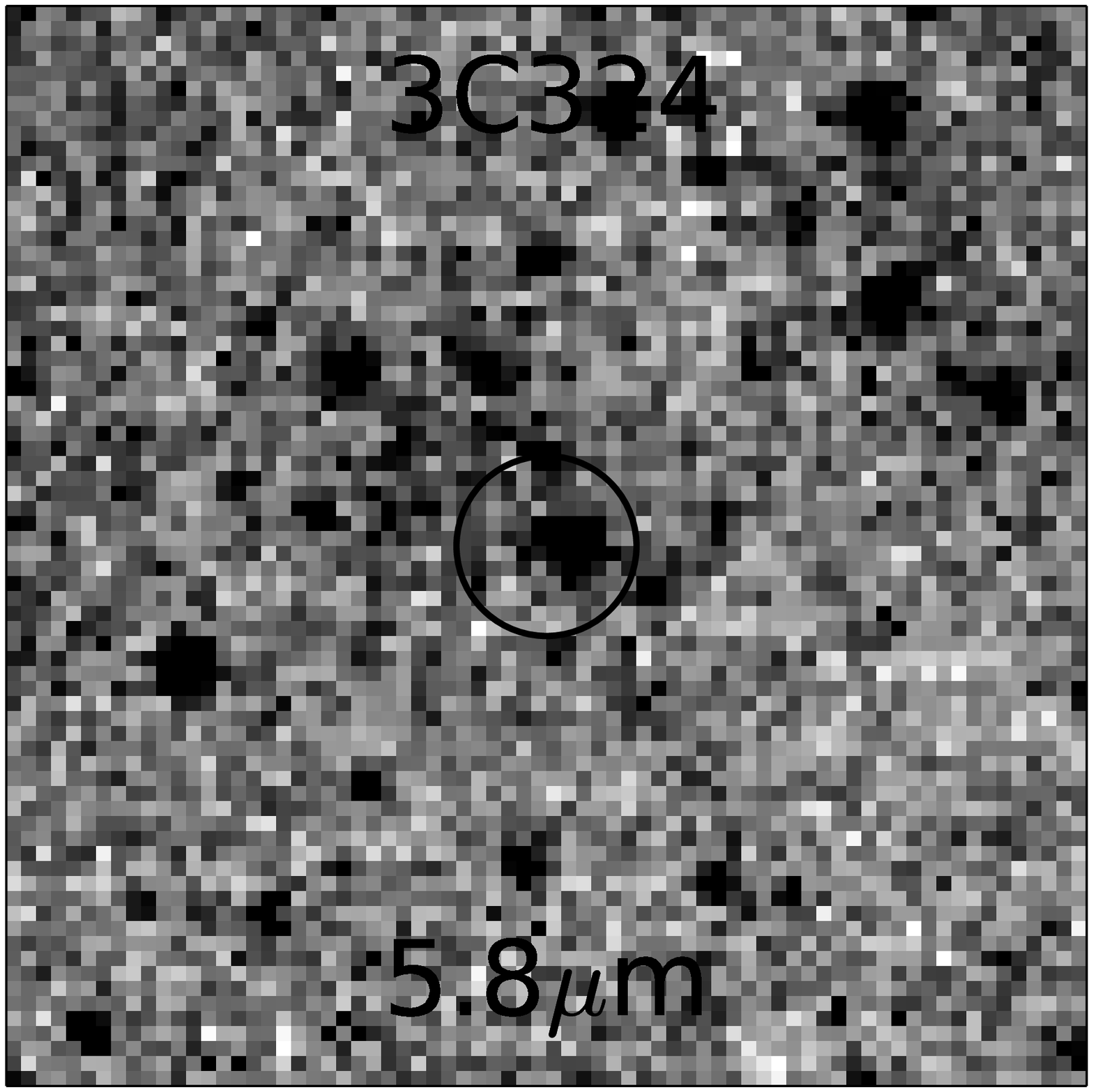}
      \includegraphics[width=1.5cm]{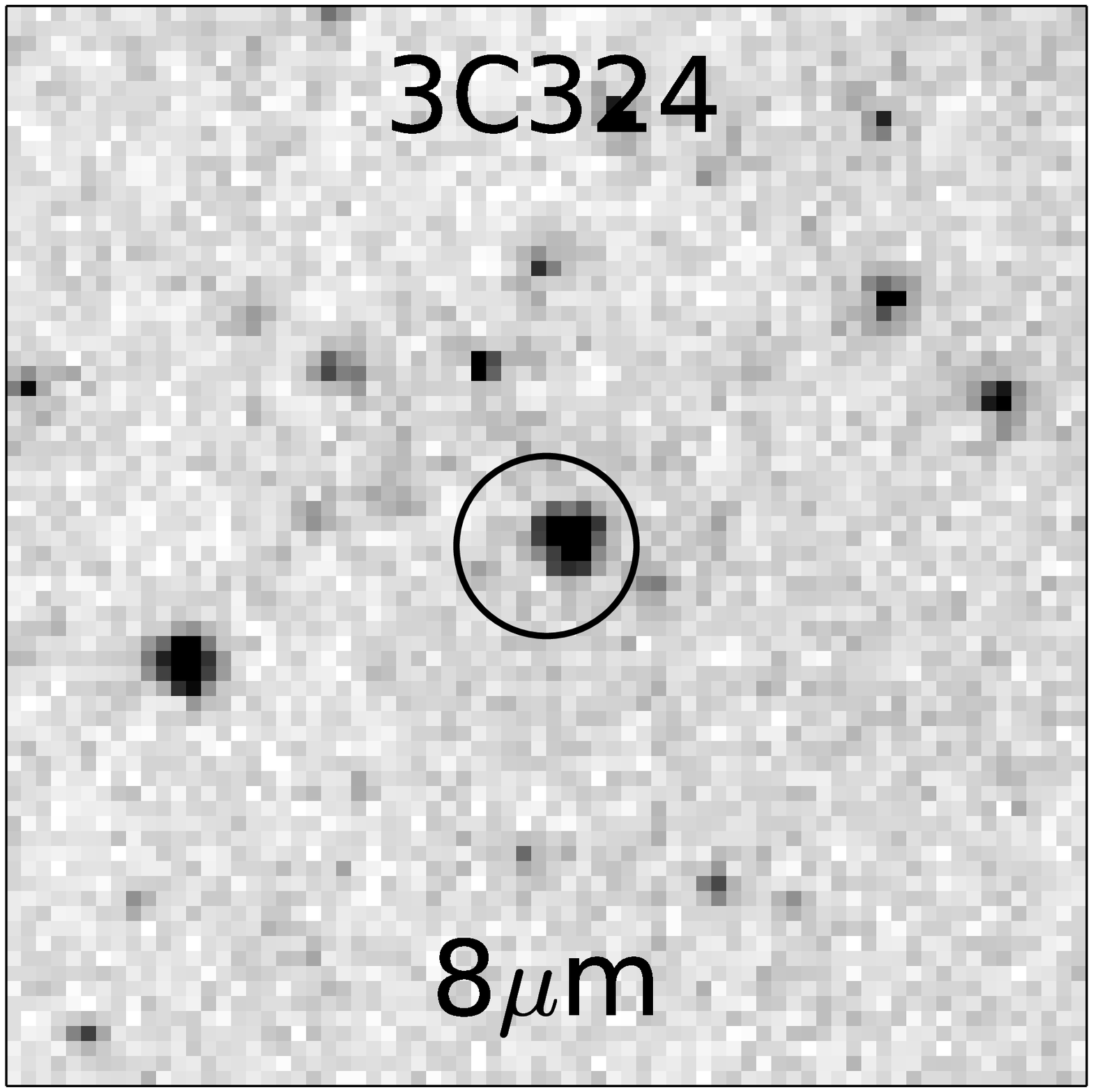}
      \includegraphics[width=1.5cm]{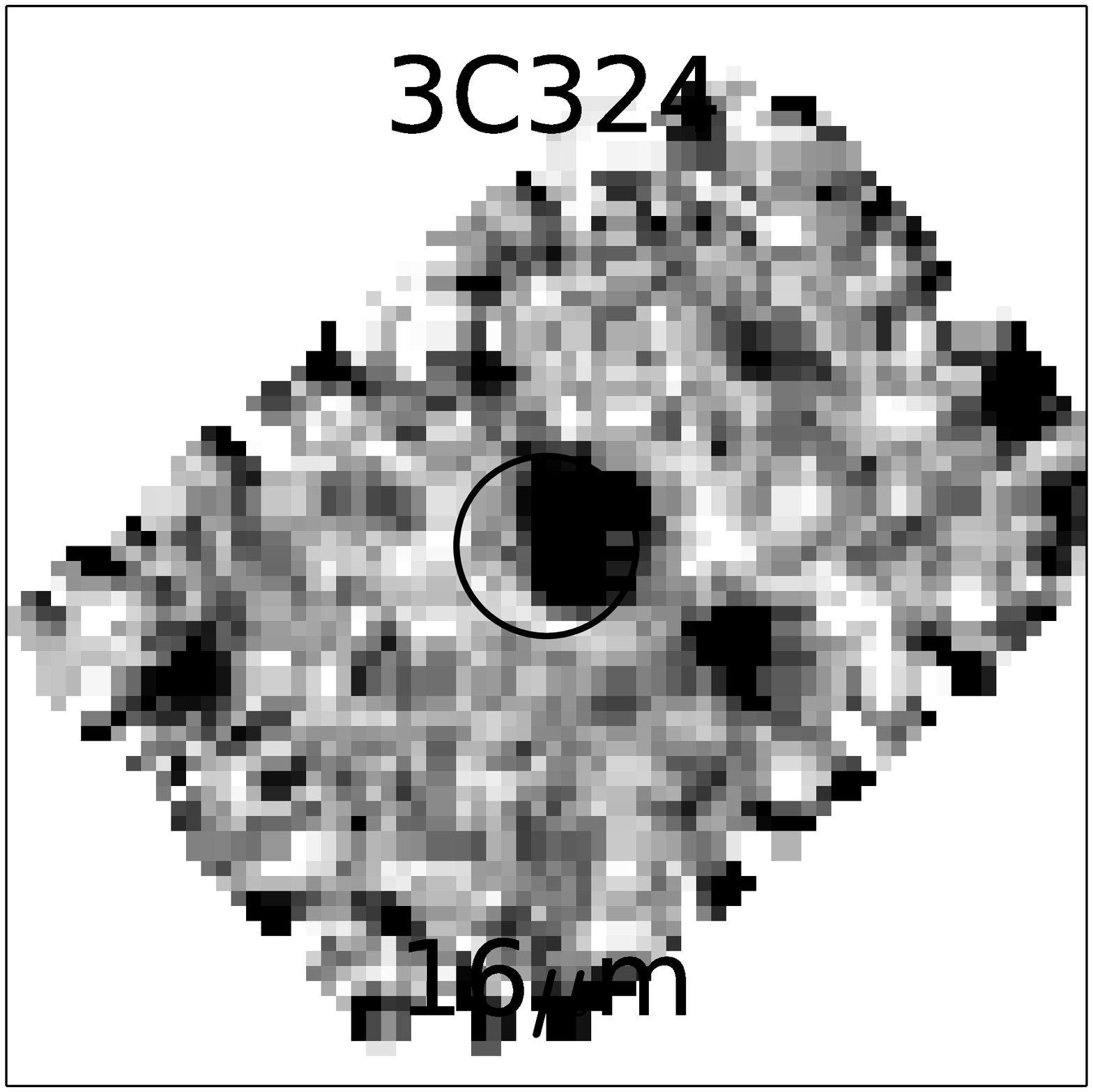}
      \includegraphics[width=1.5cm]{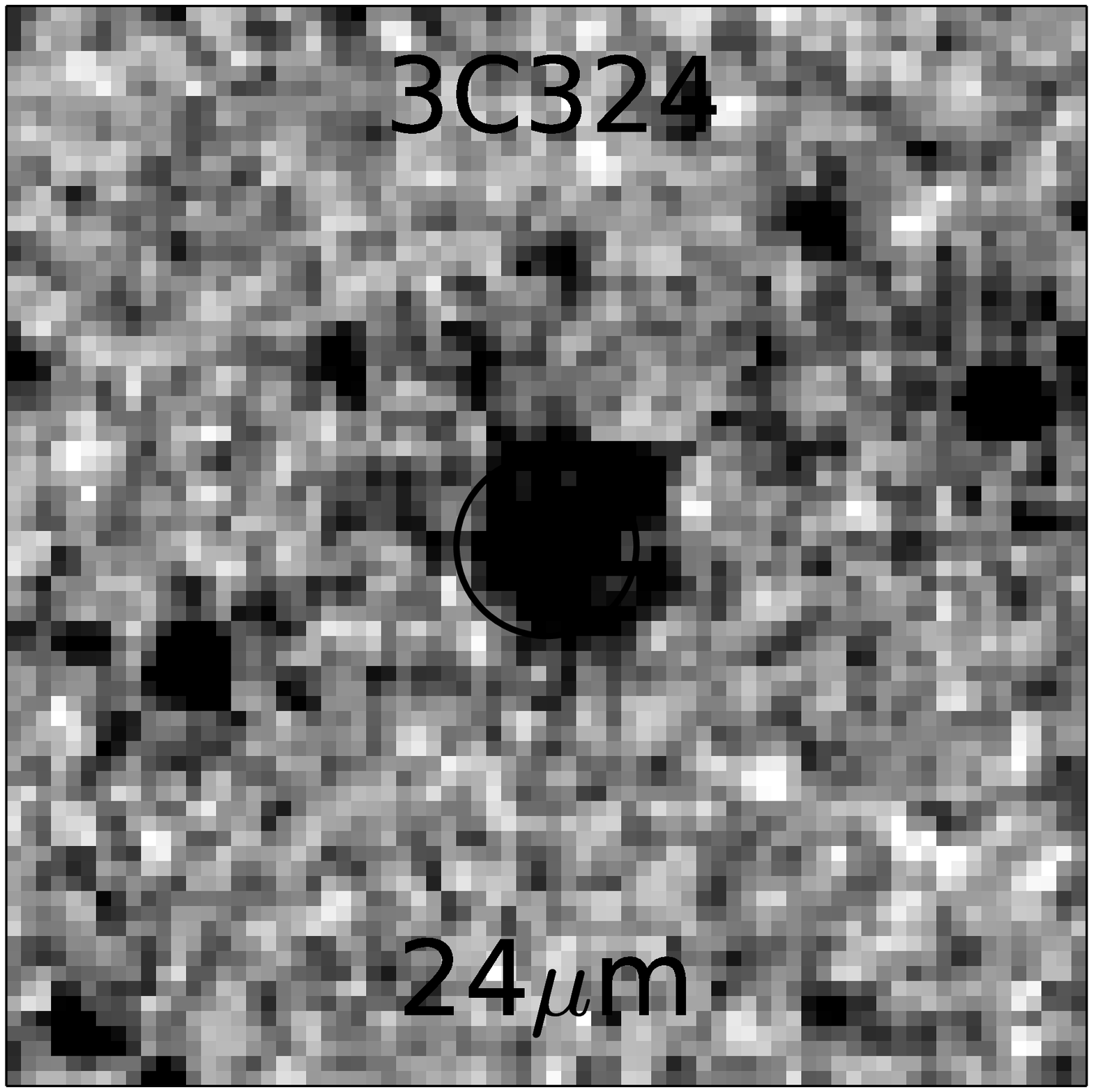}
      \includegraphics[width=1.5cm]{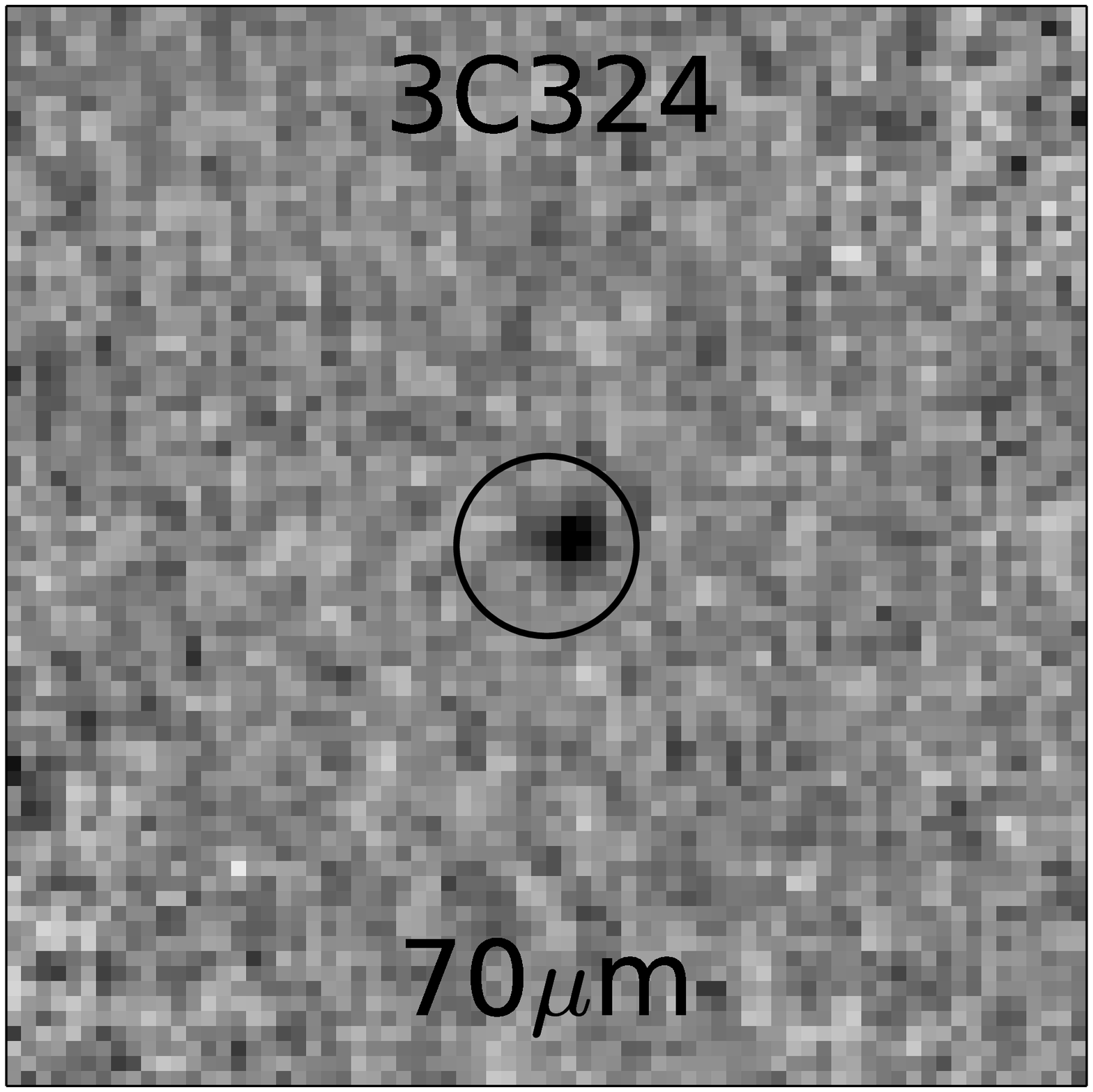}
      \includegraphics[width=1.5cm]{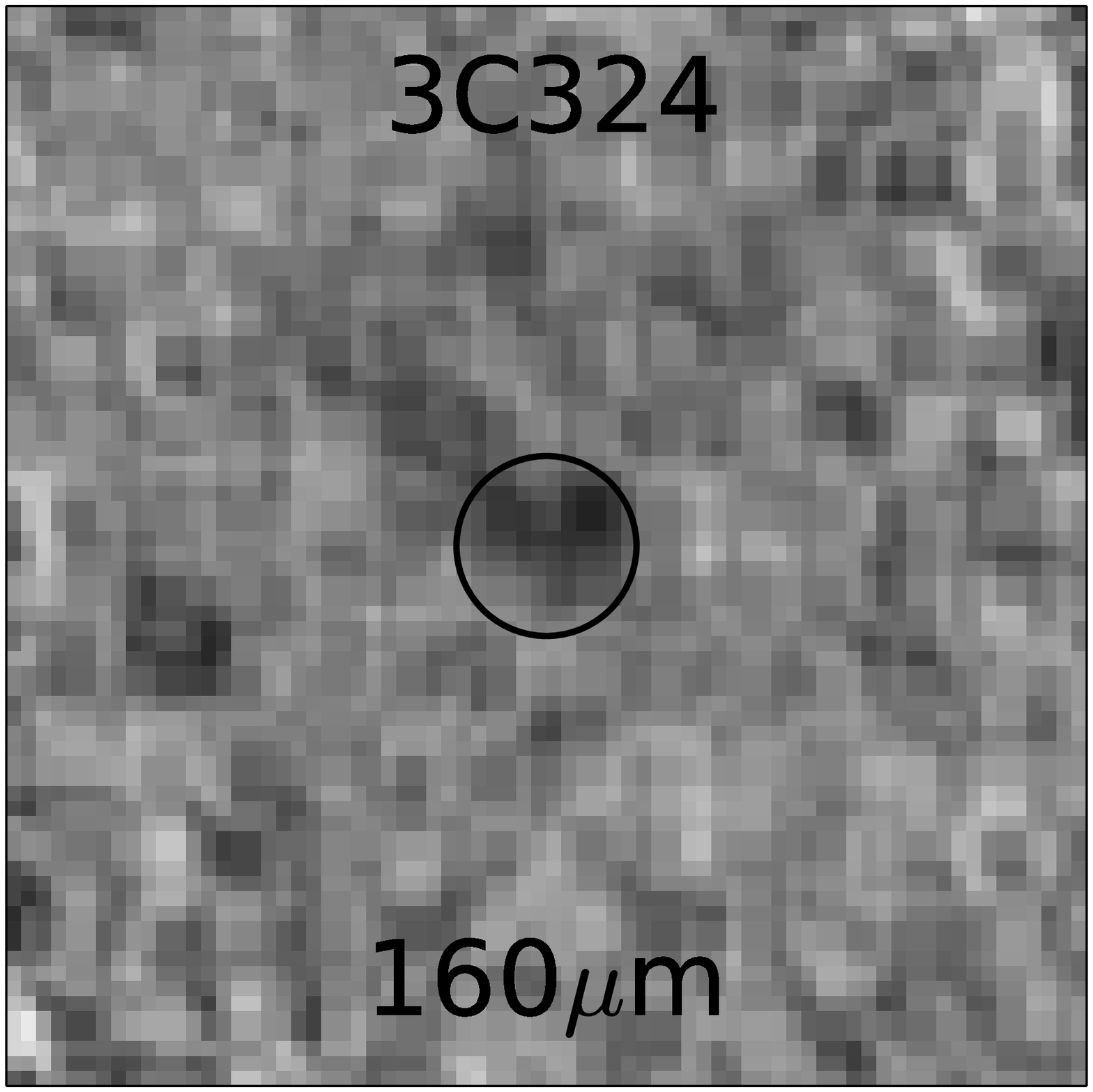}
      \includegraphics[width=1.5cm]{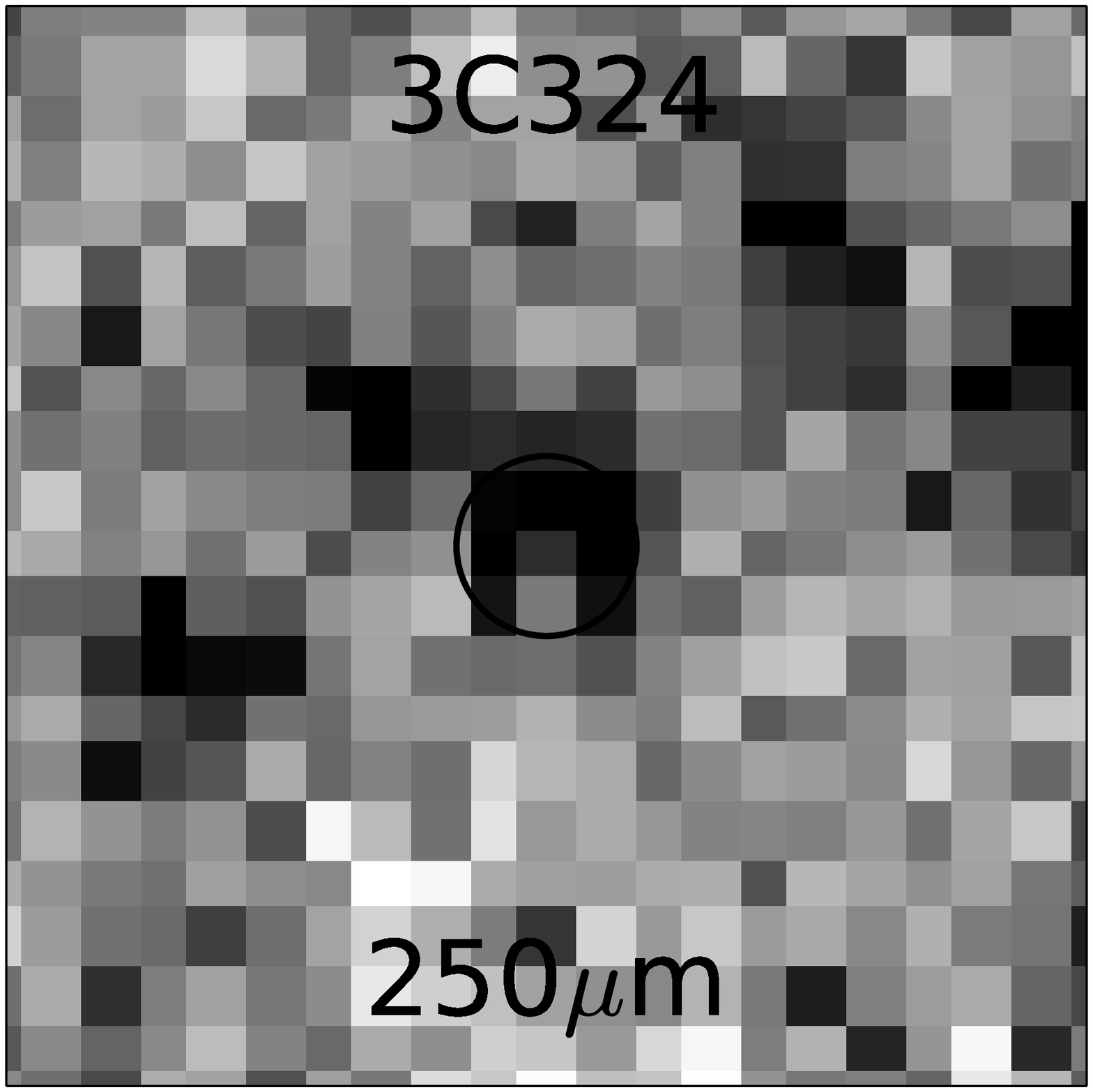}
      \includegraphics[width=1.5cm]{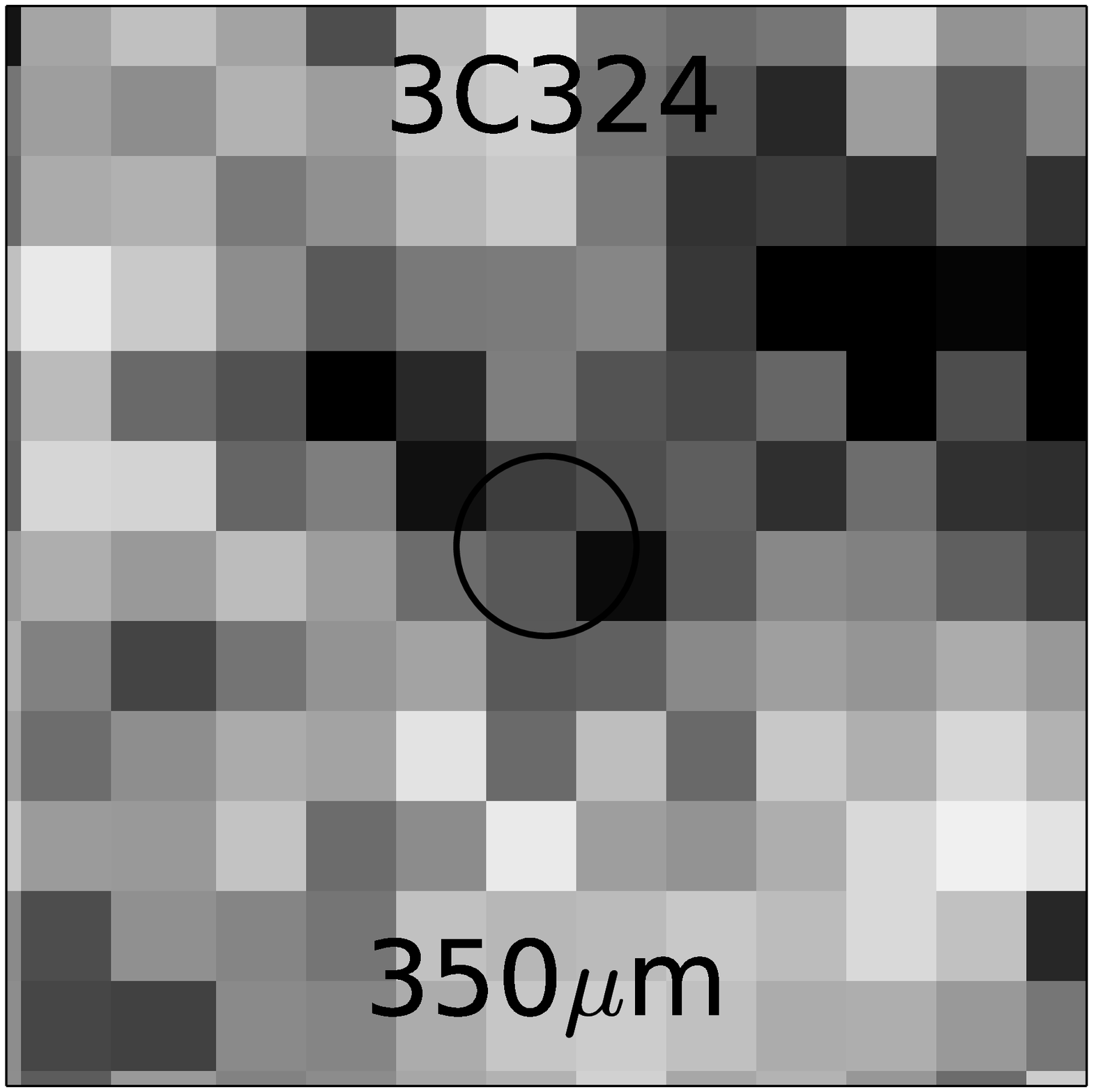}
      \includegraphics[width=1.5cm]{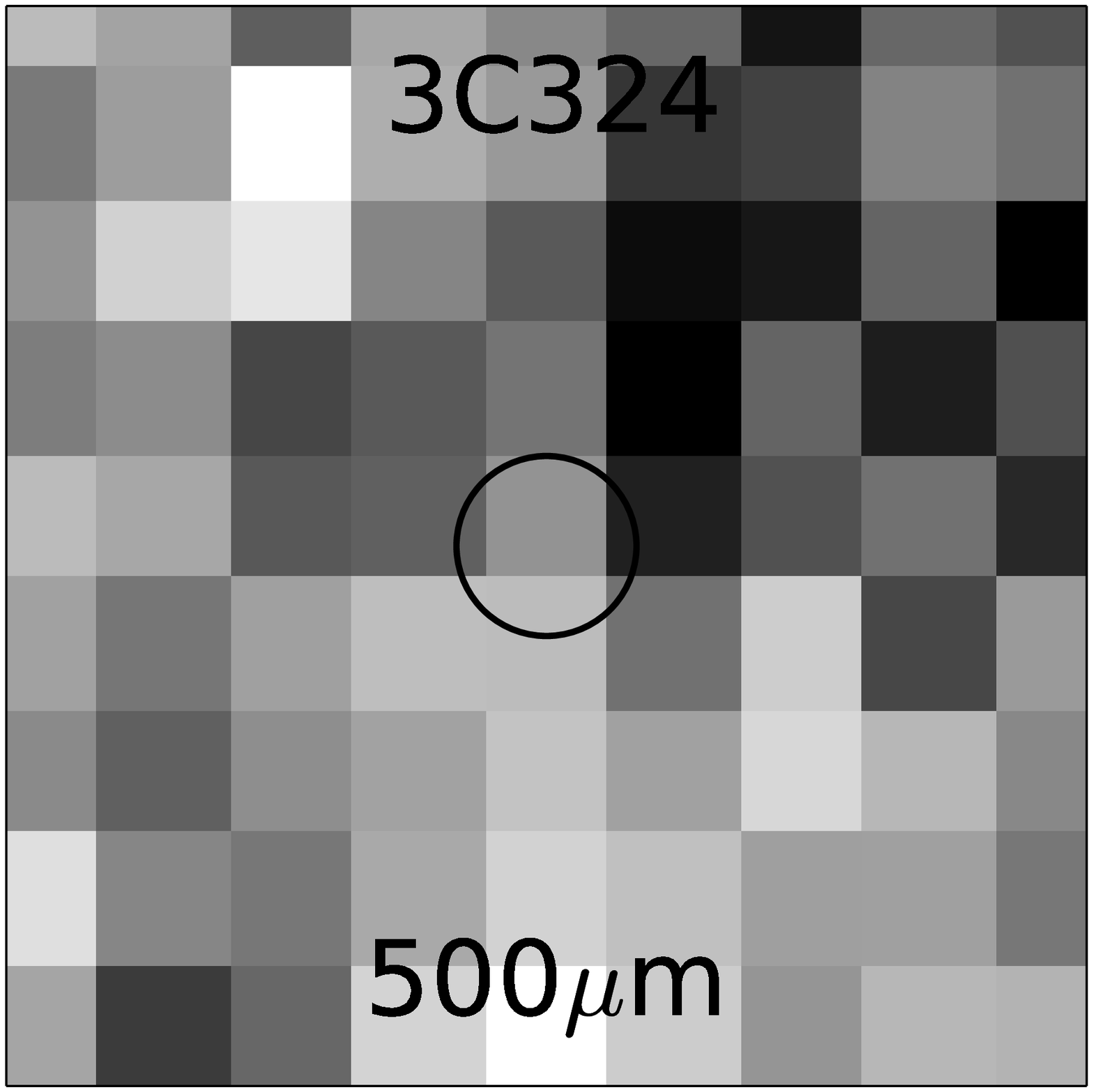}
      \\
      \includegraphics[width=1.5cm]{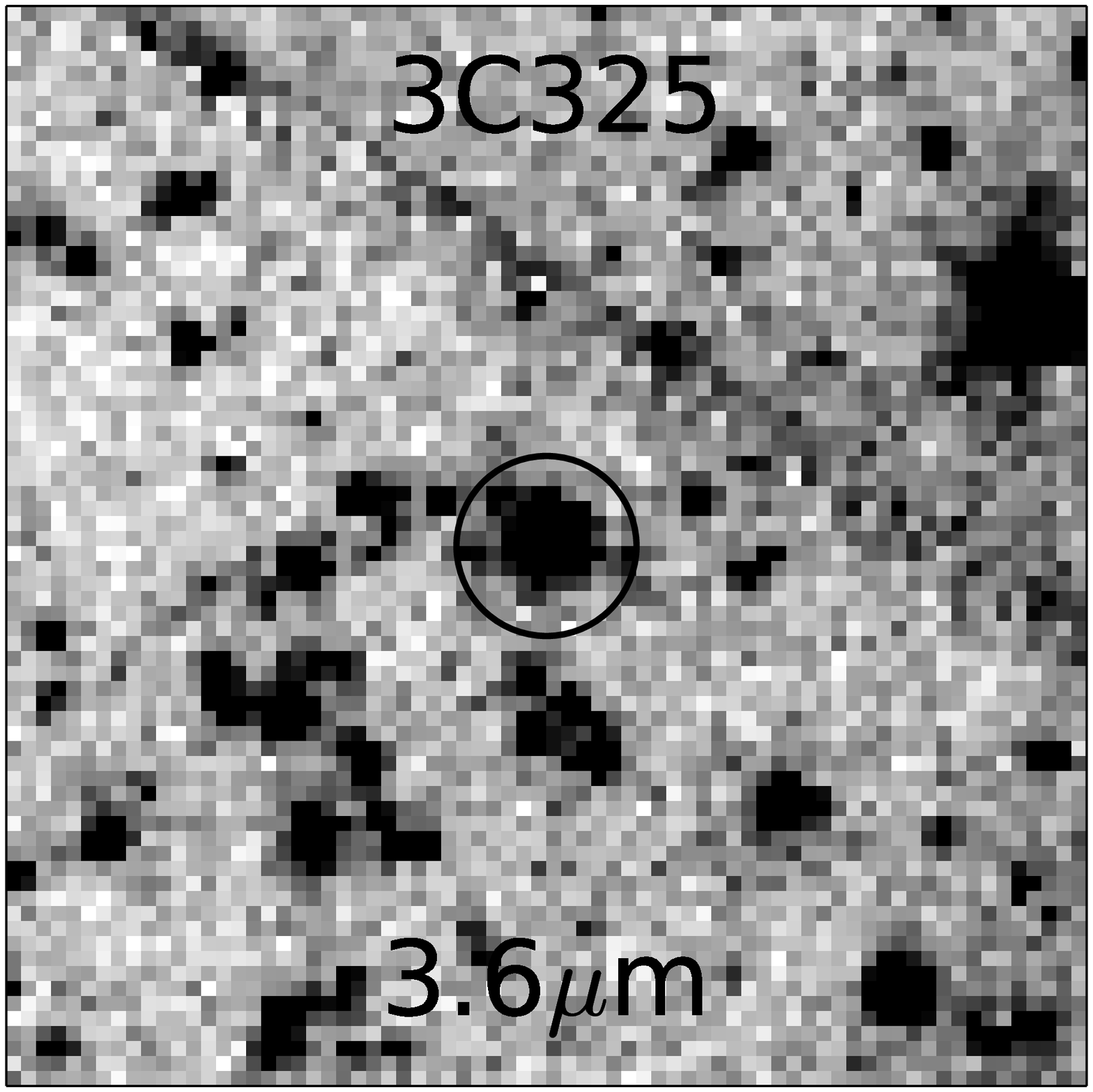}
      \includegraphics[width=1.5cm]{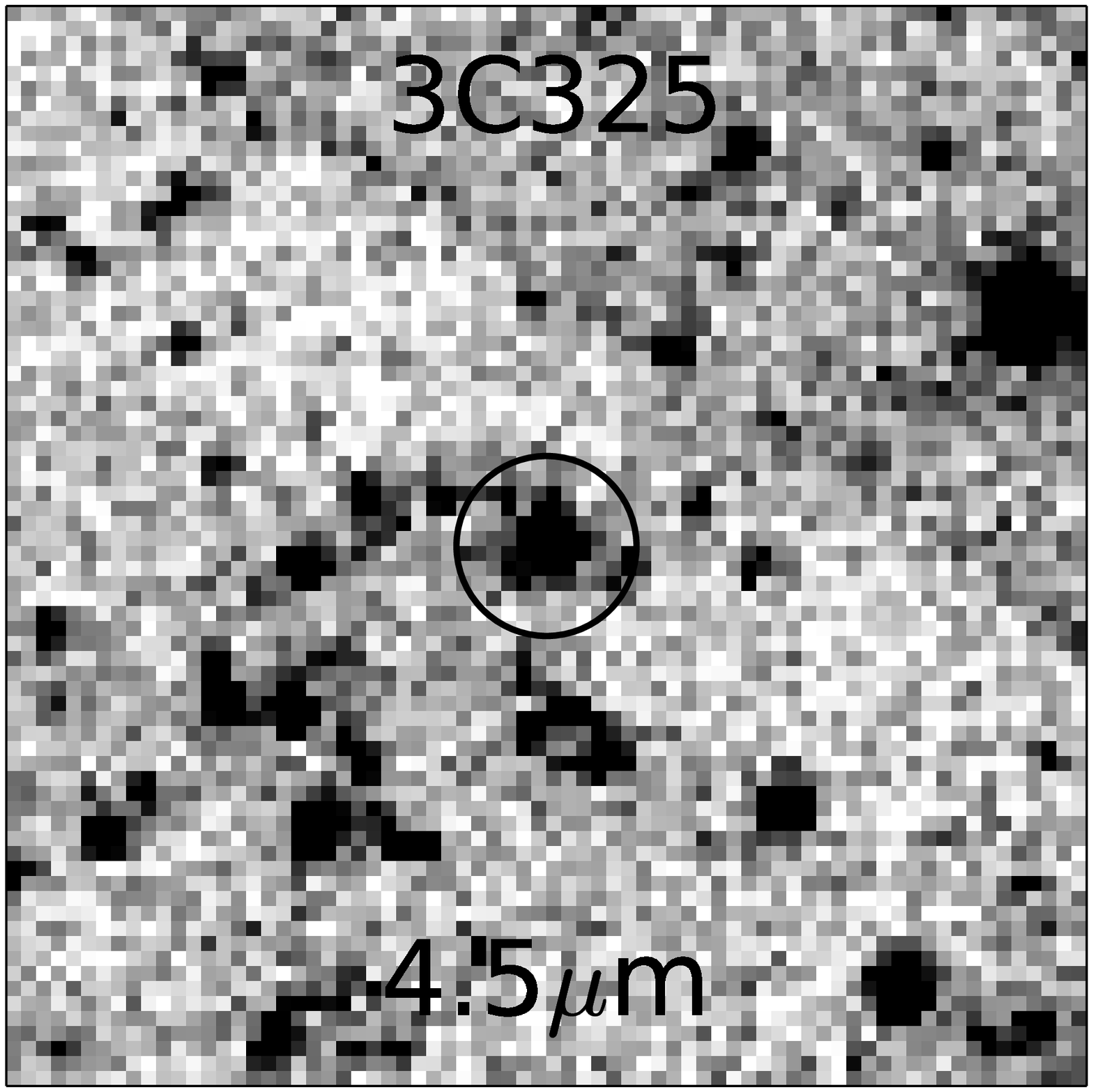}
      \includegraphics[width=1.5cm]{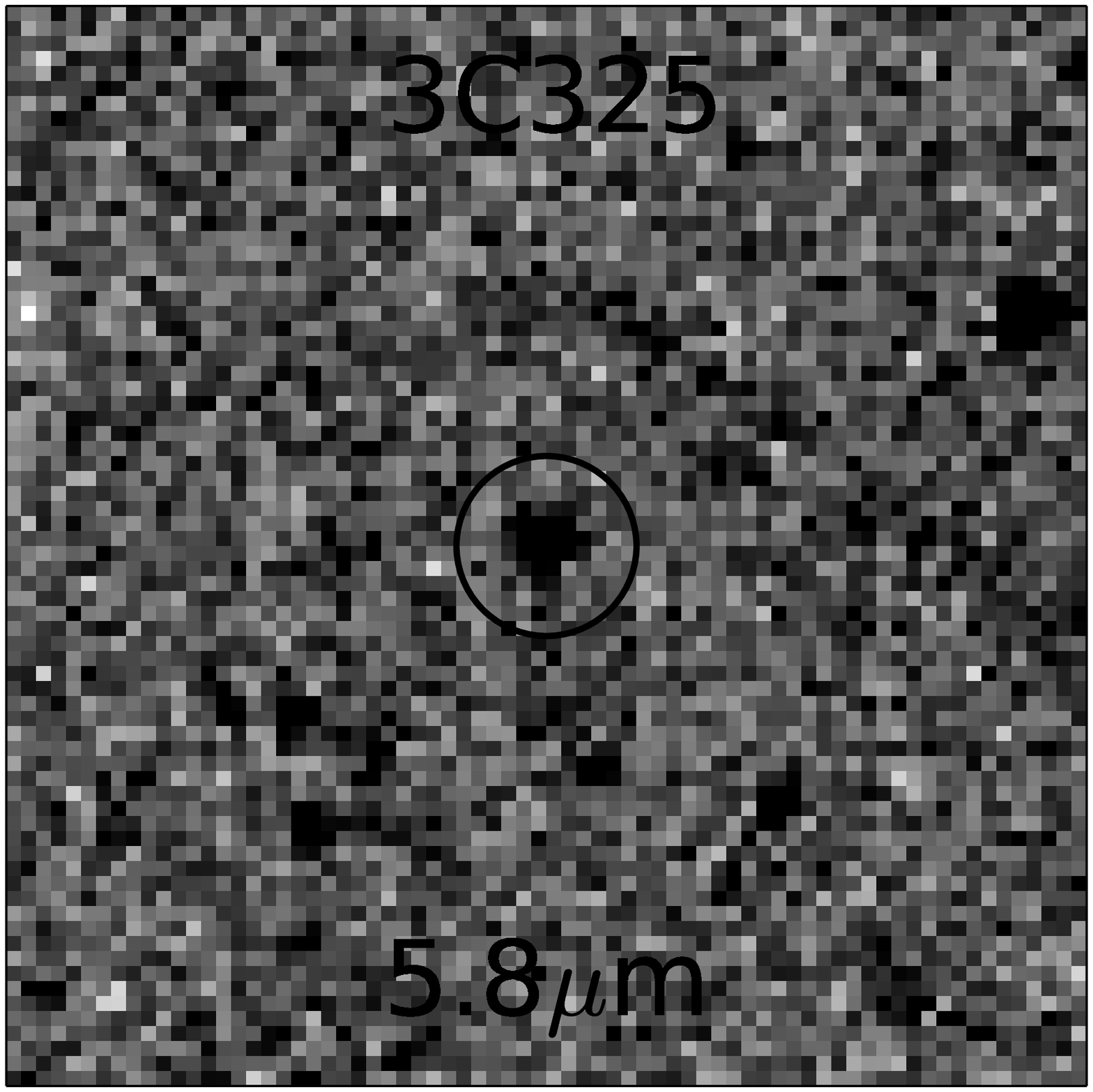}
      \includegraphics[width=1.5cm]{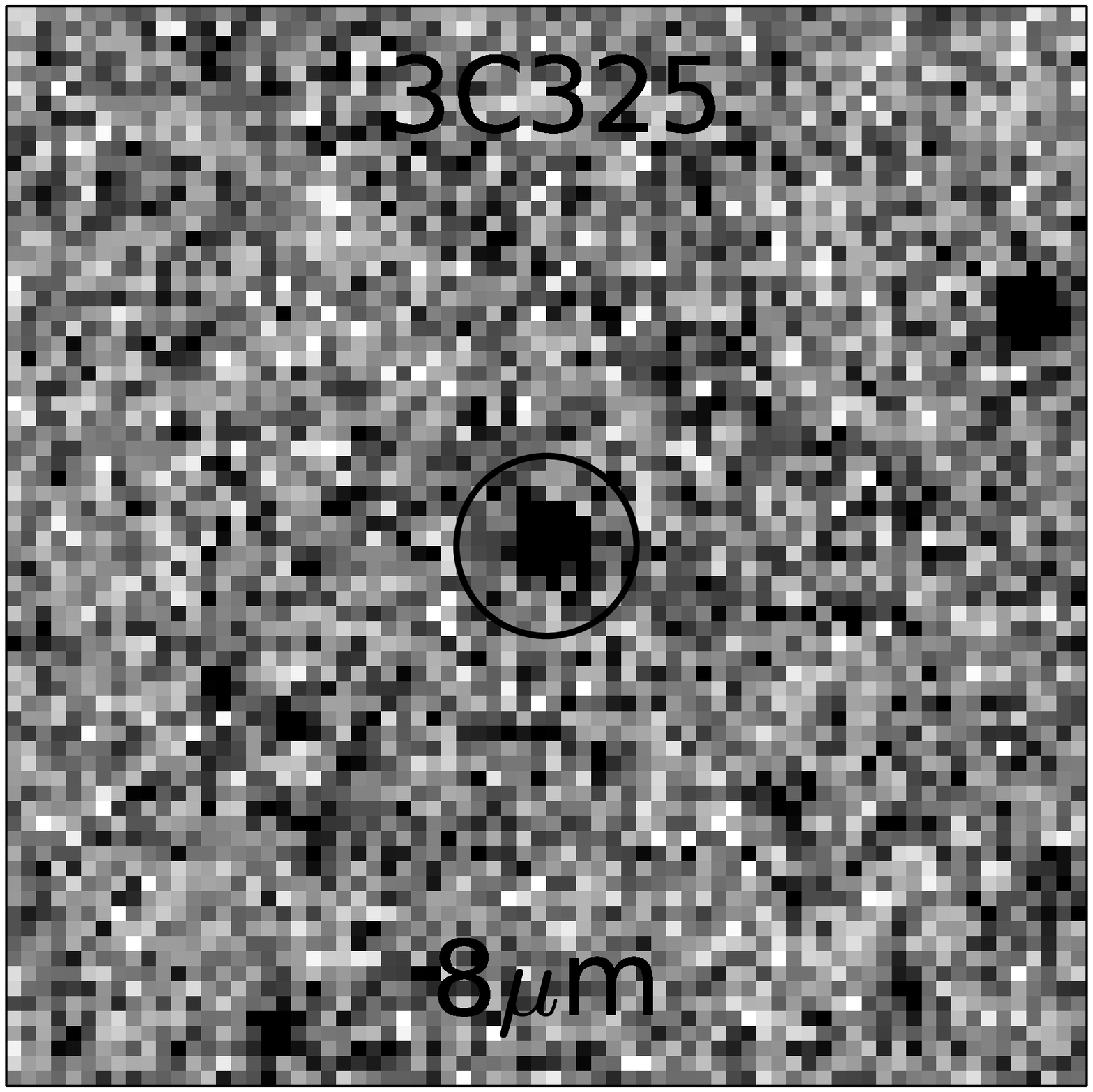}
      \includegraphics[width=1.5cm]{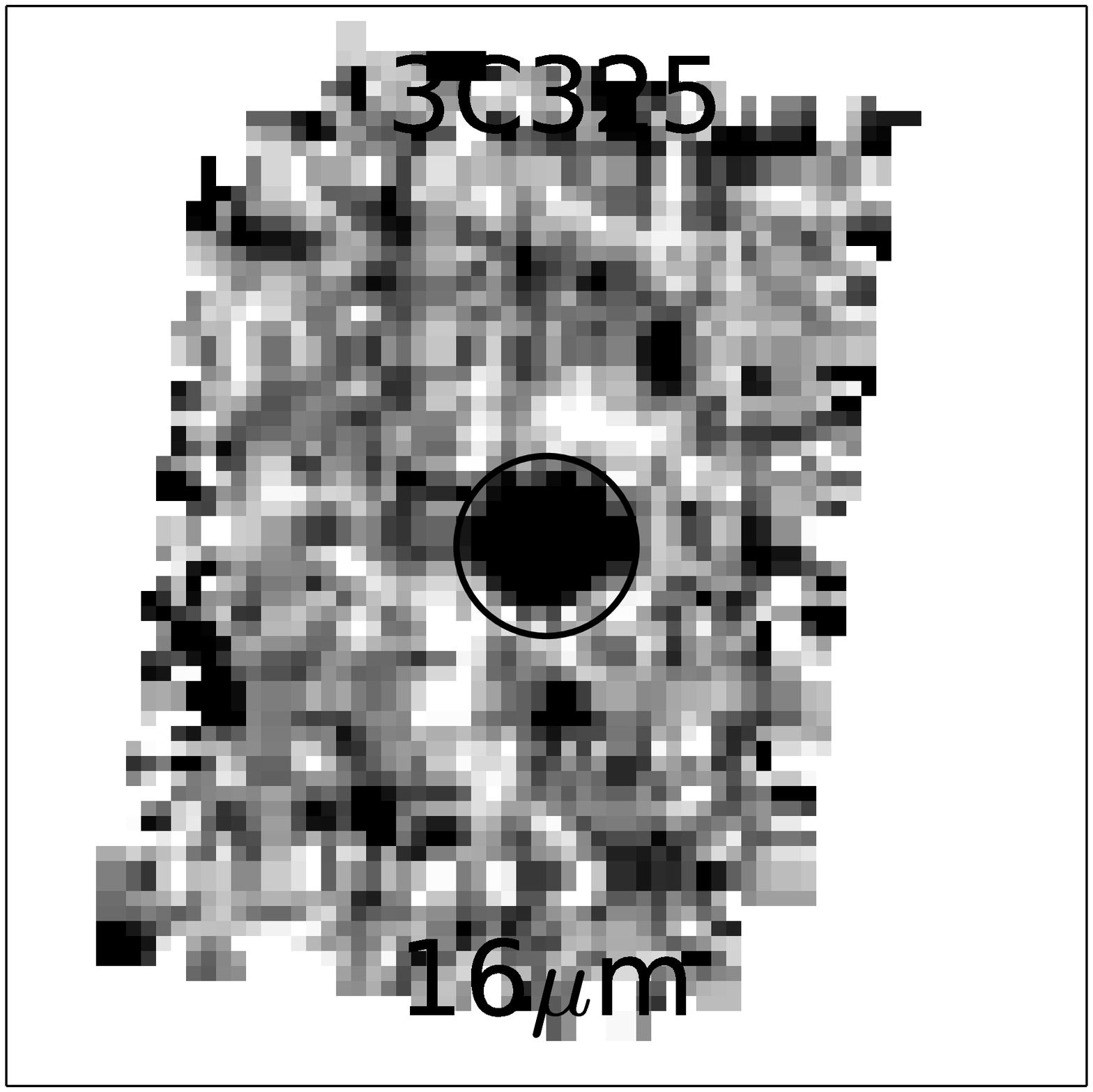}
      \includegraphics[width=1.5cm]{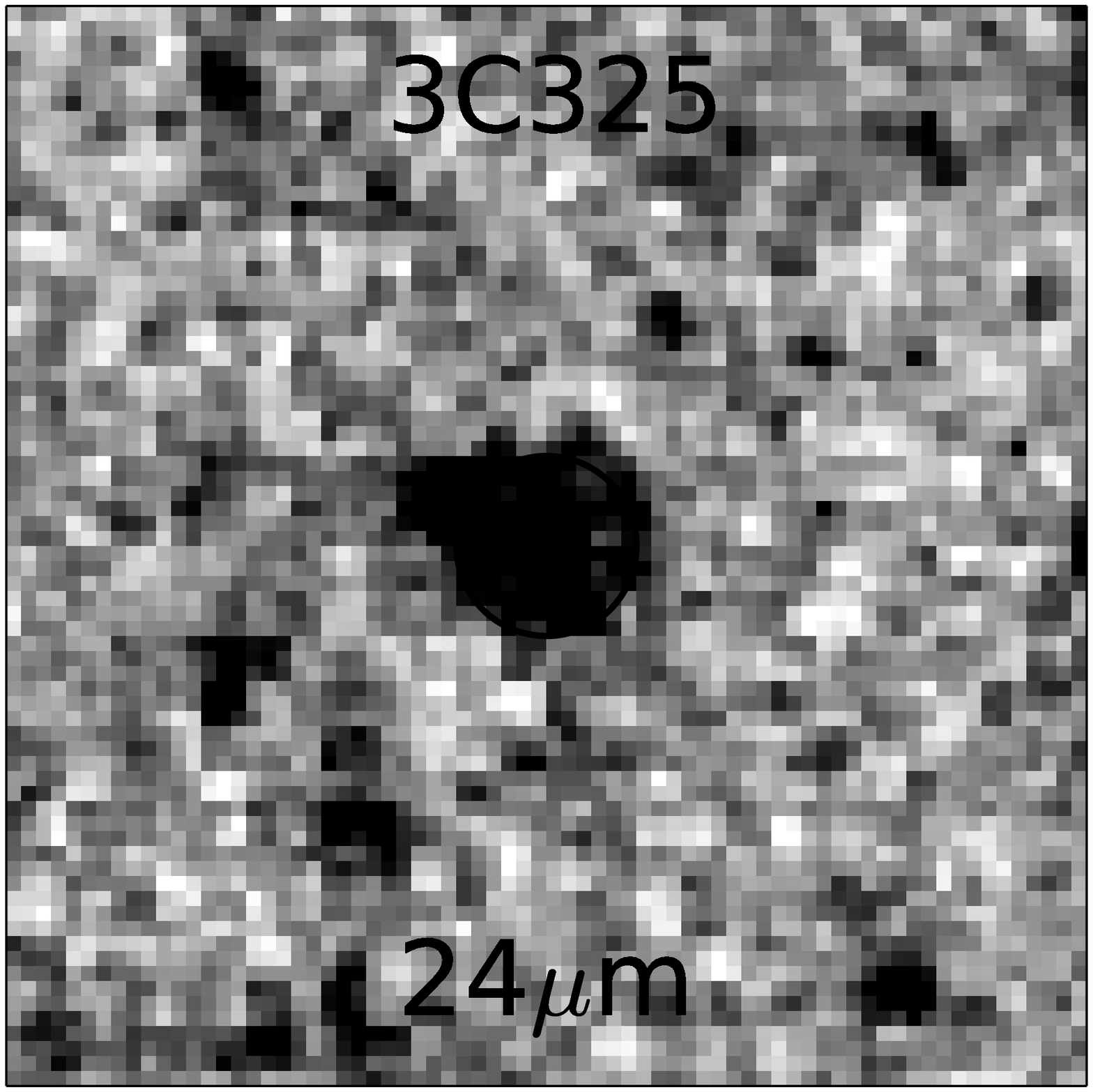}
      \includegraphics[width=1.5cm]{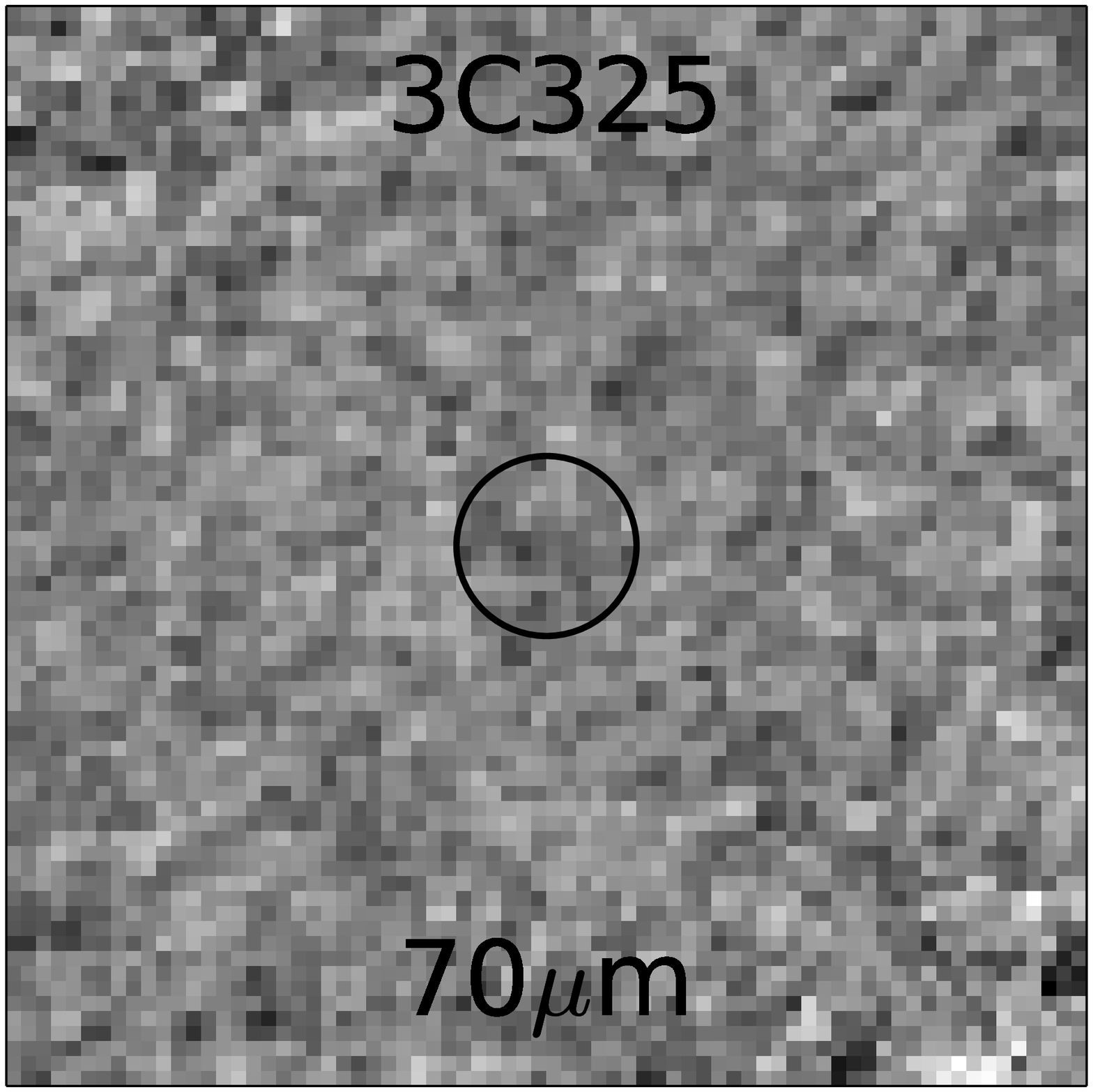}
      \includegraphics[width=1.5cm]{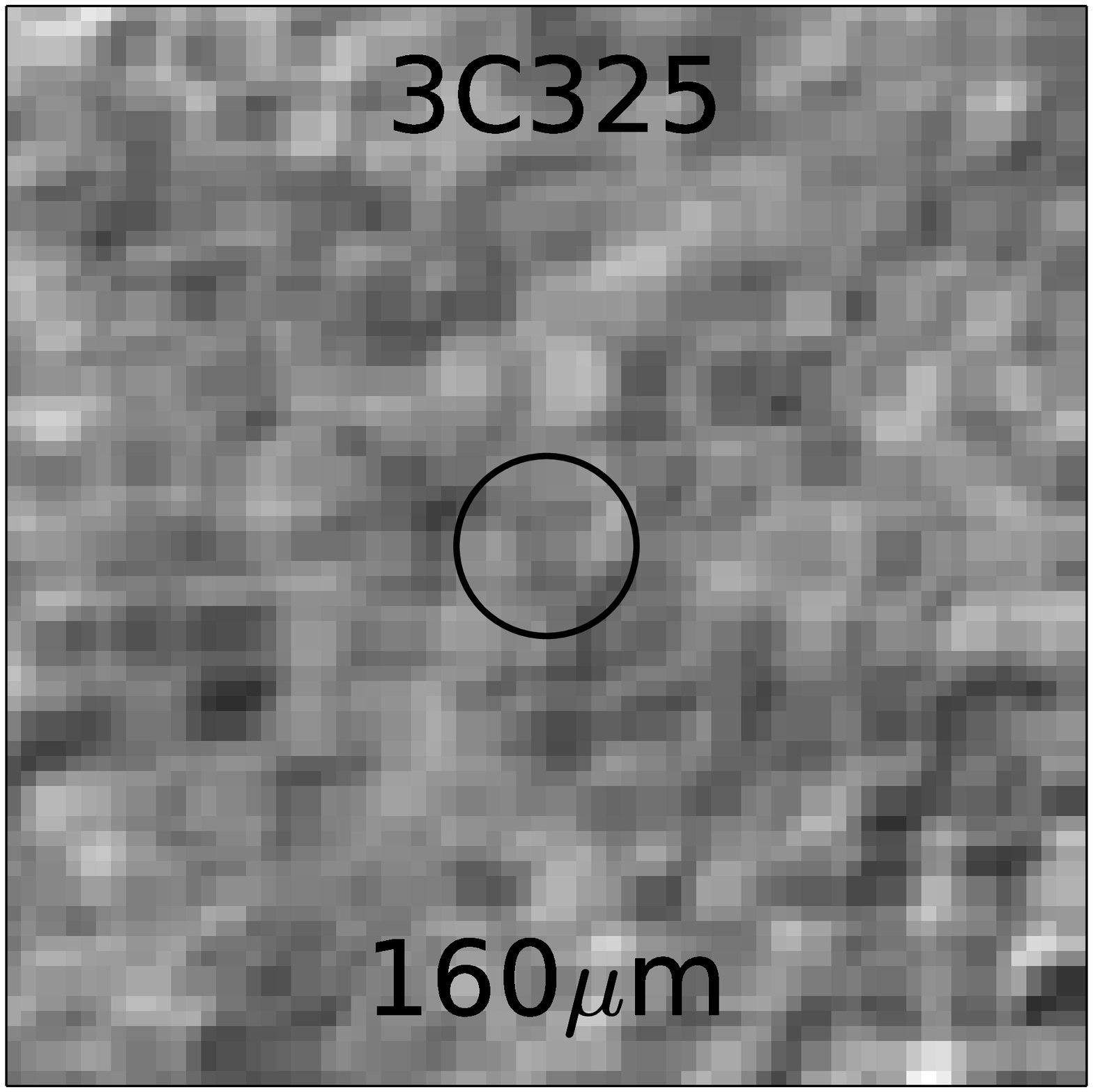}
      \includegraphics[width=1.5cm]{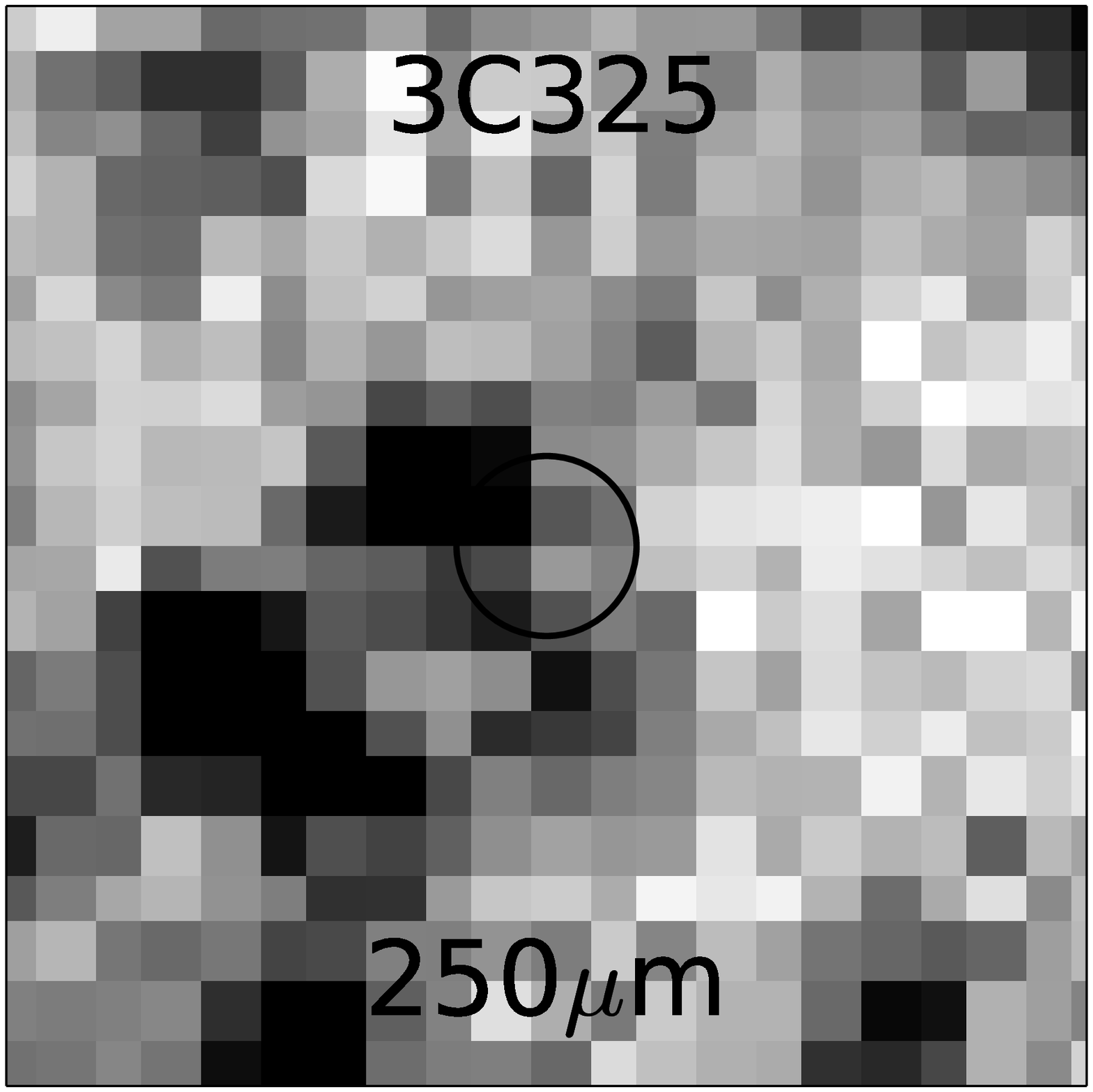}
      \includegraphics[width=1.5cm]{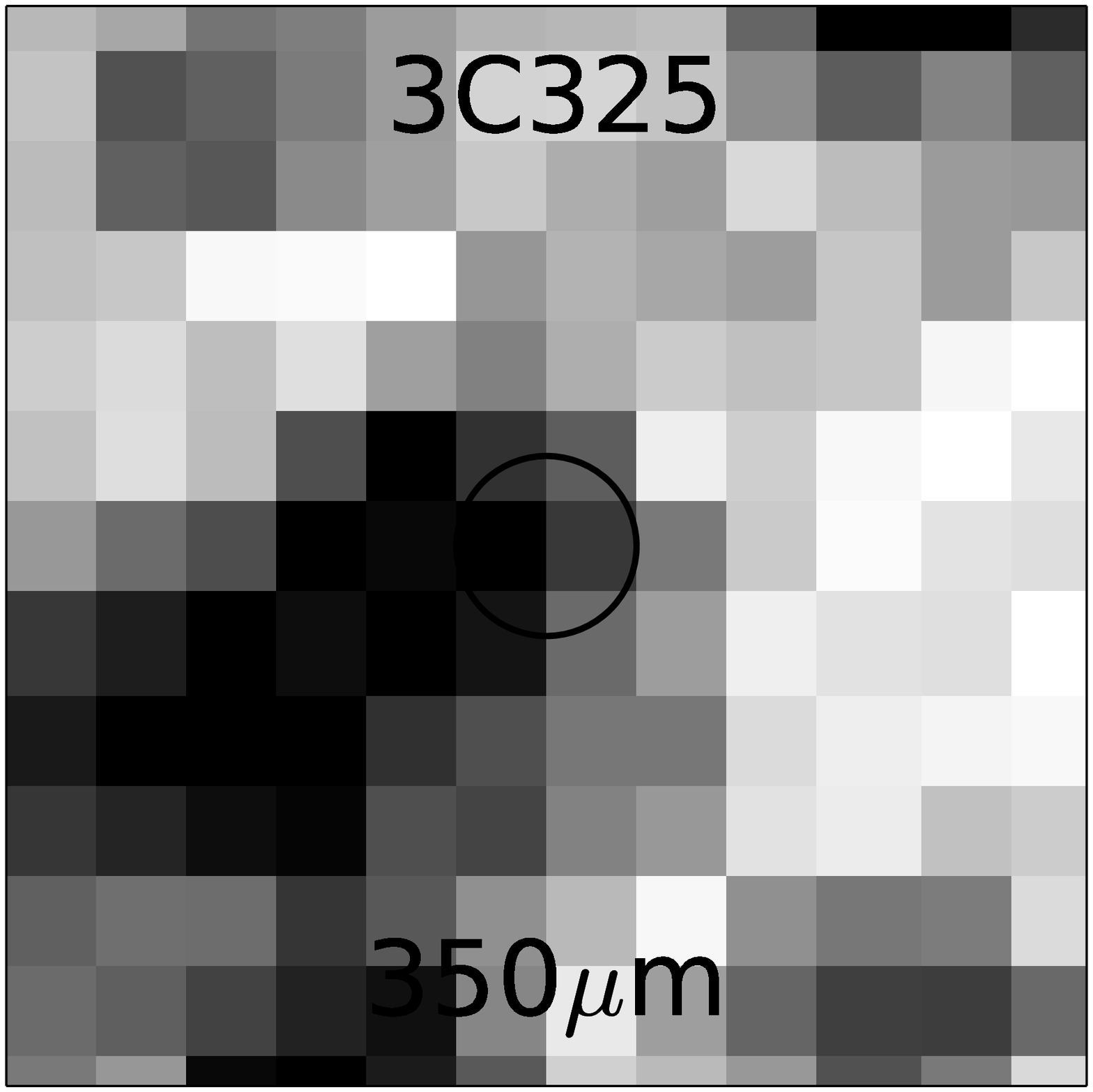}
      \includegraphics[width=1.5cm]{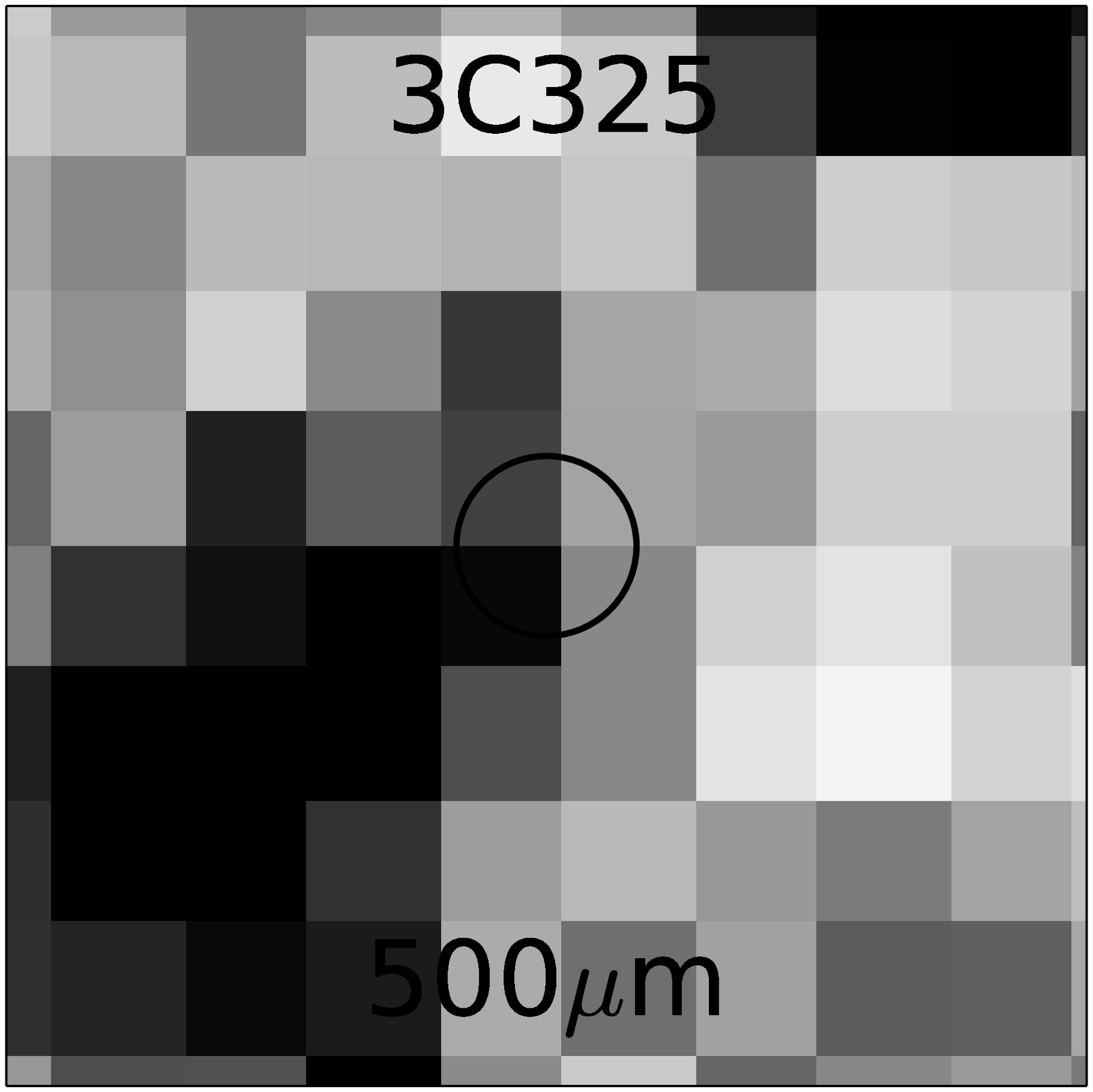}
      \\
      \includegraphics[width=1.5cm]{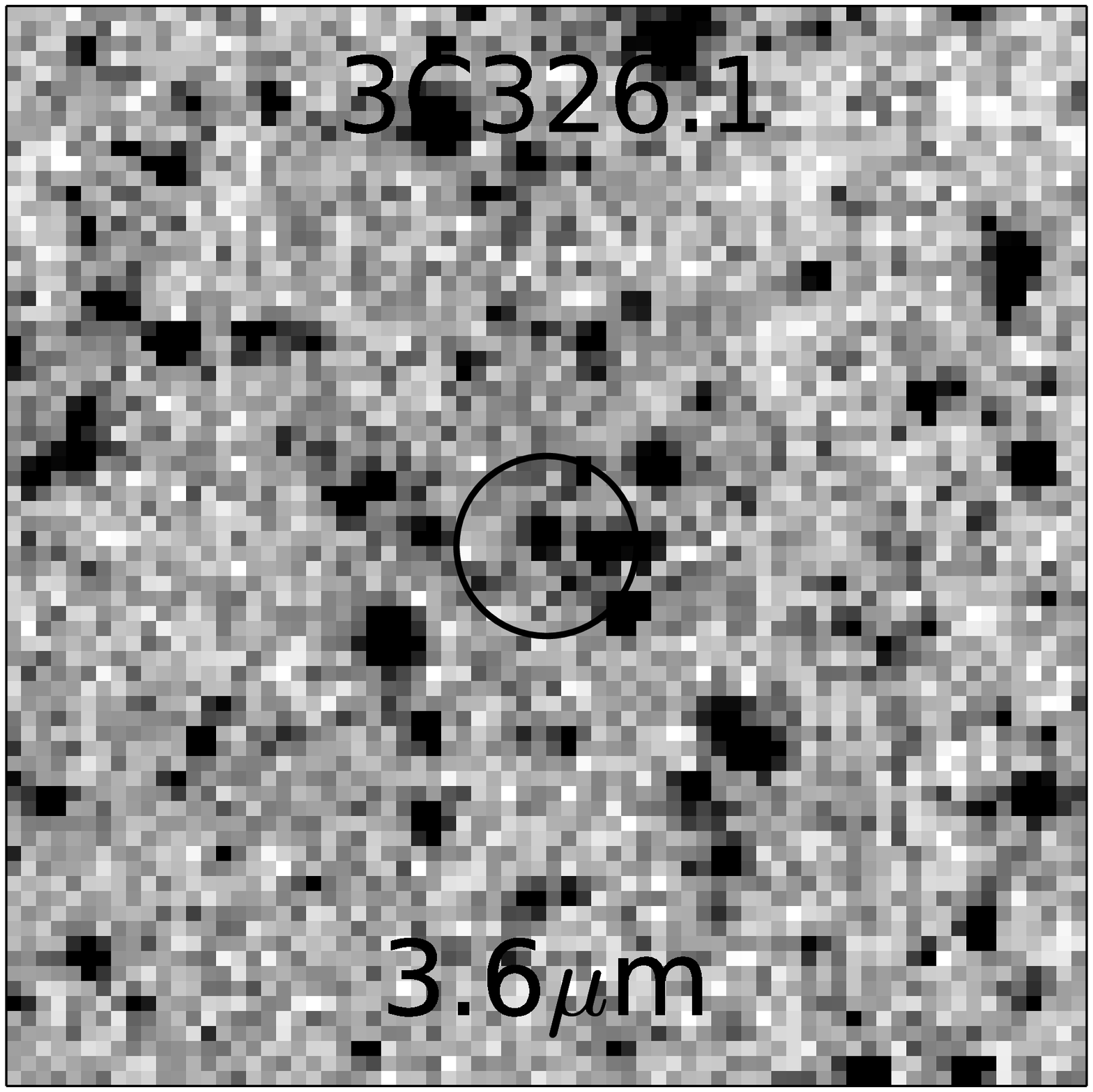}
      \includegraphics[width=1.5cm]{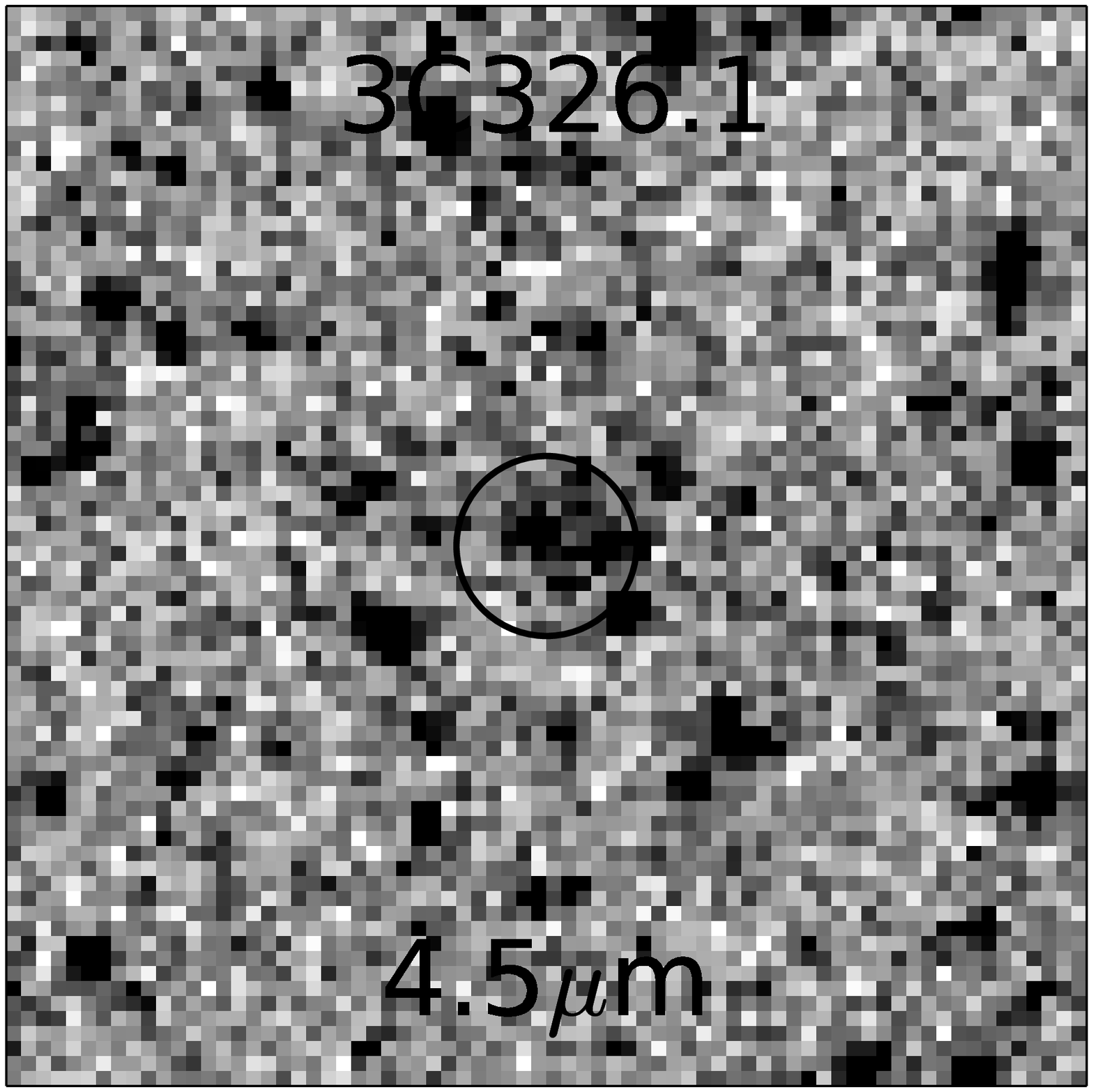}
      \includegraphics[width=1.5cm]{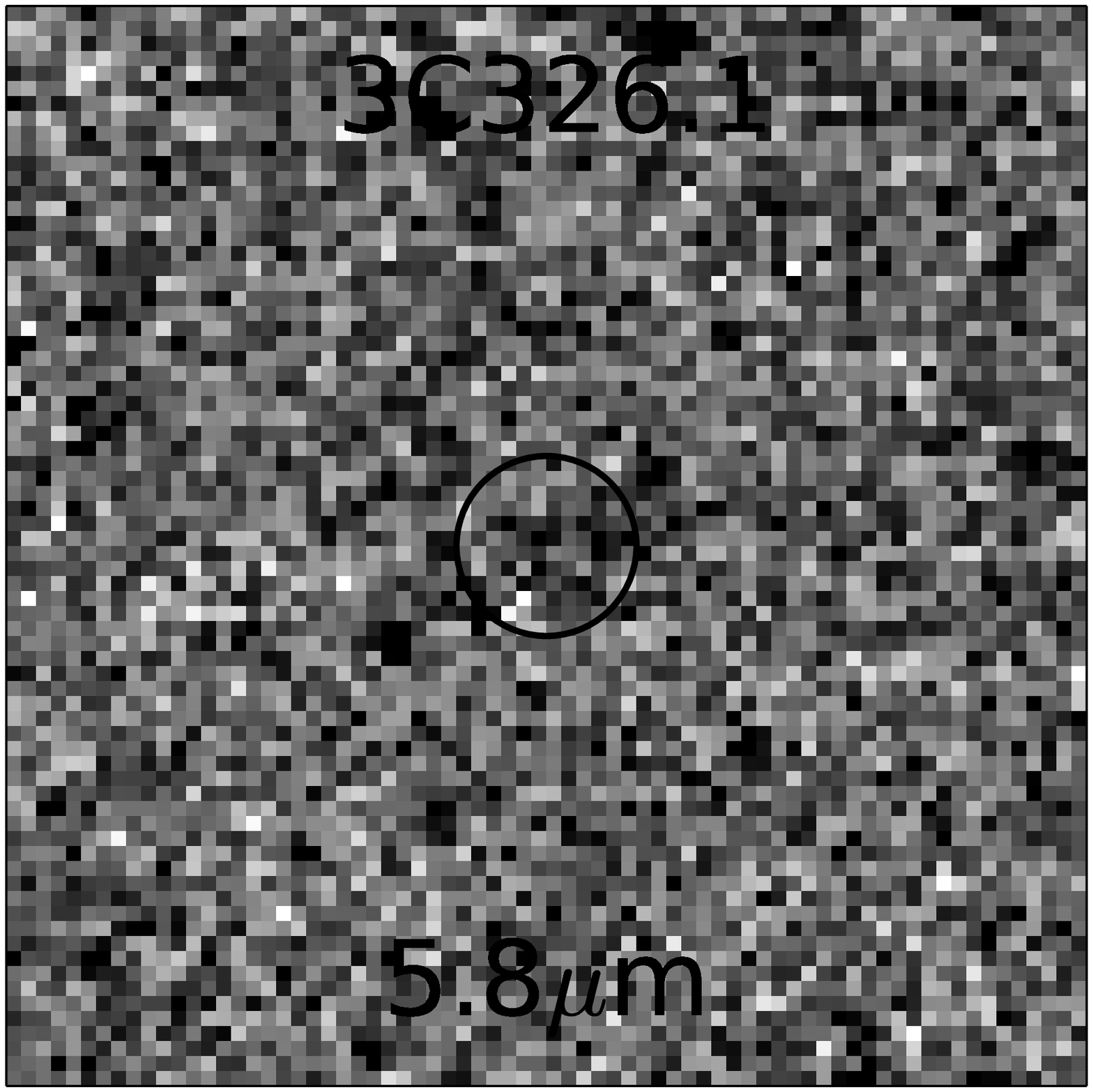}
      \includegraphics[width=1.5cm]{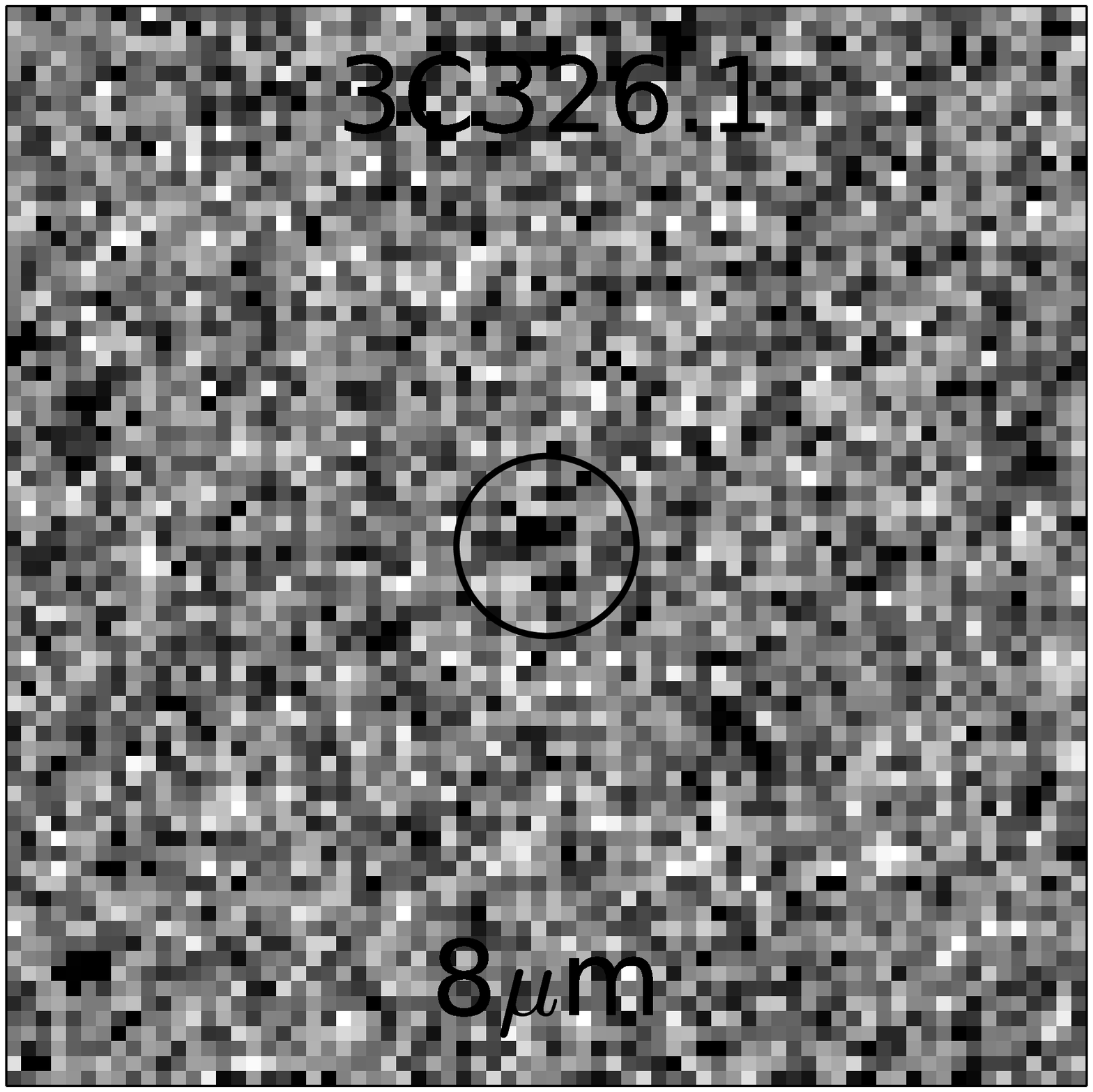}
      \includegraphics[width=1.5cm]{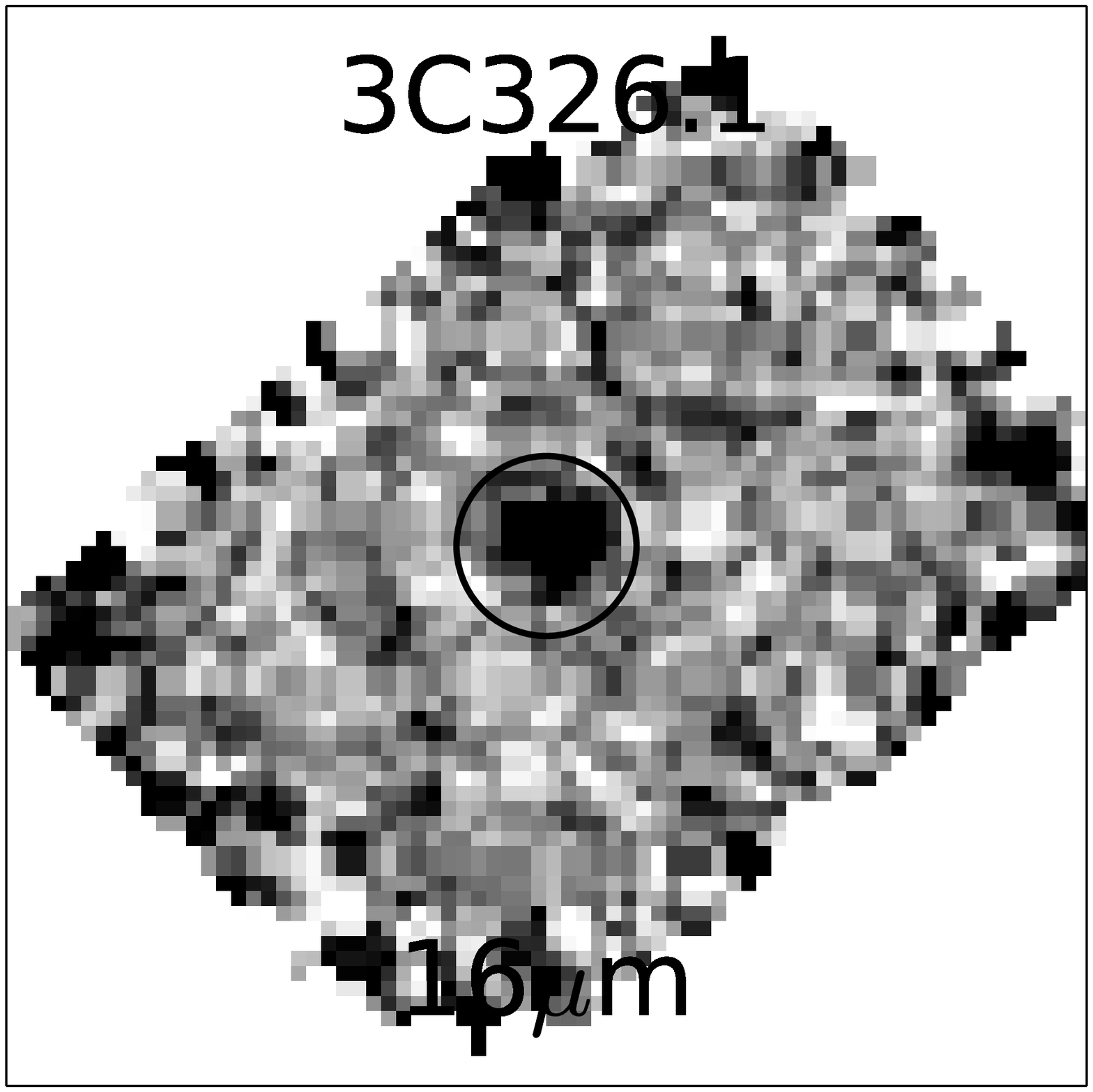}
      \includegraphics[width=1.5cm]{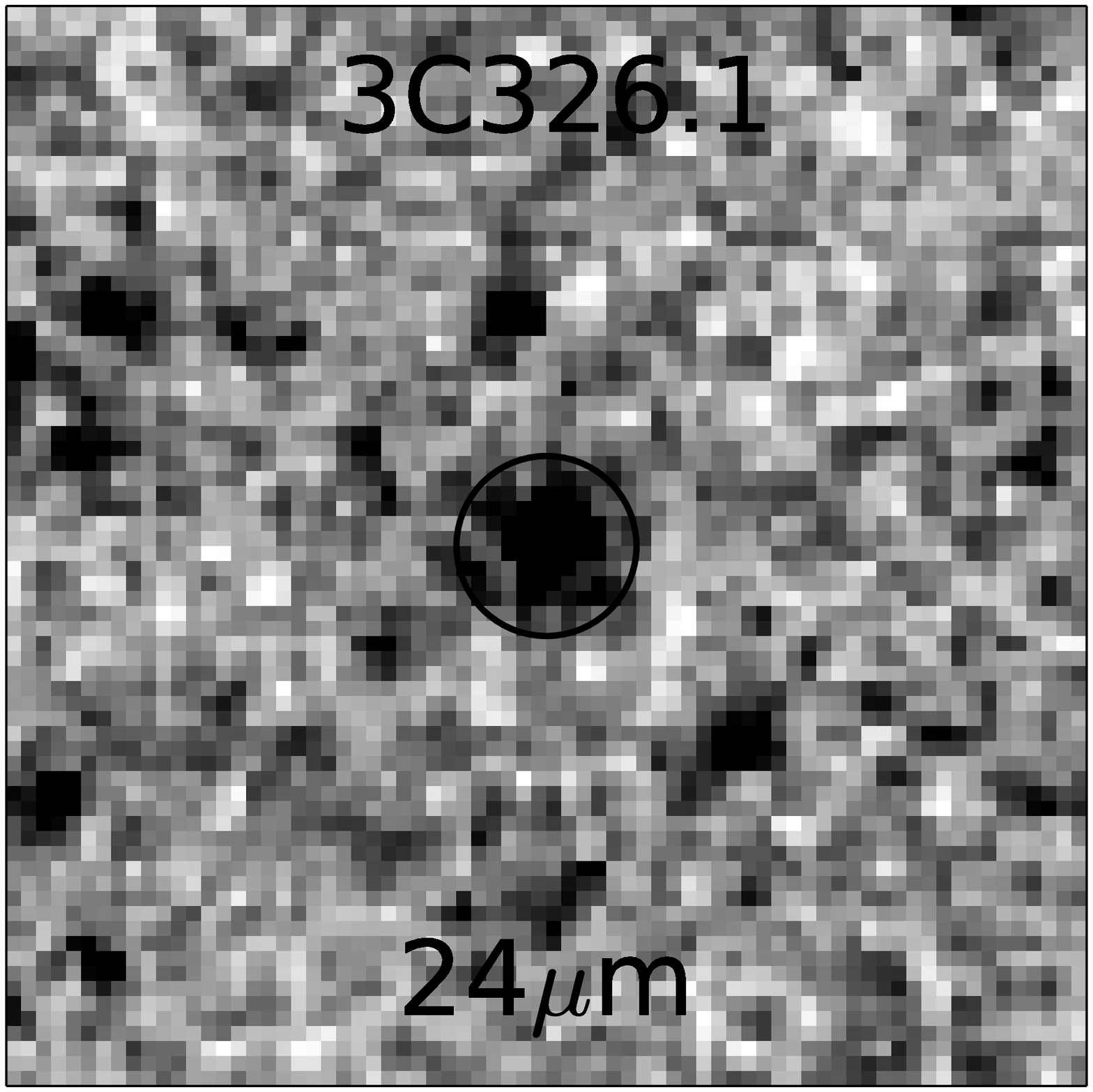}
      \includegraphics[width=1.5cm]{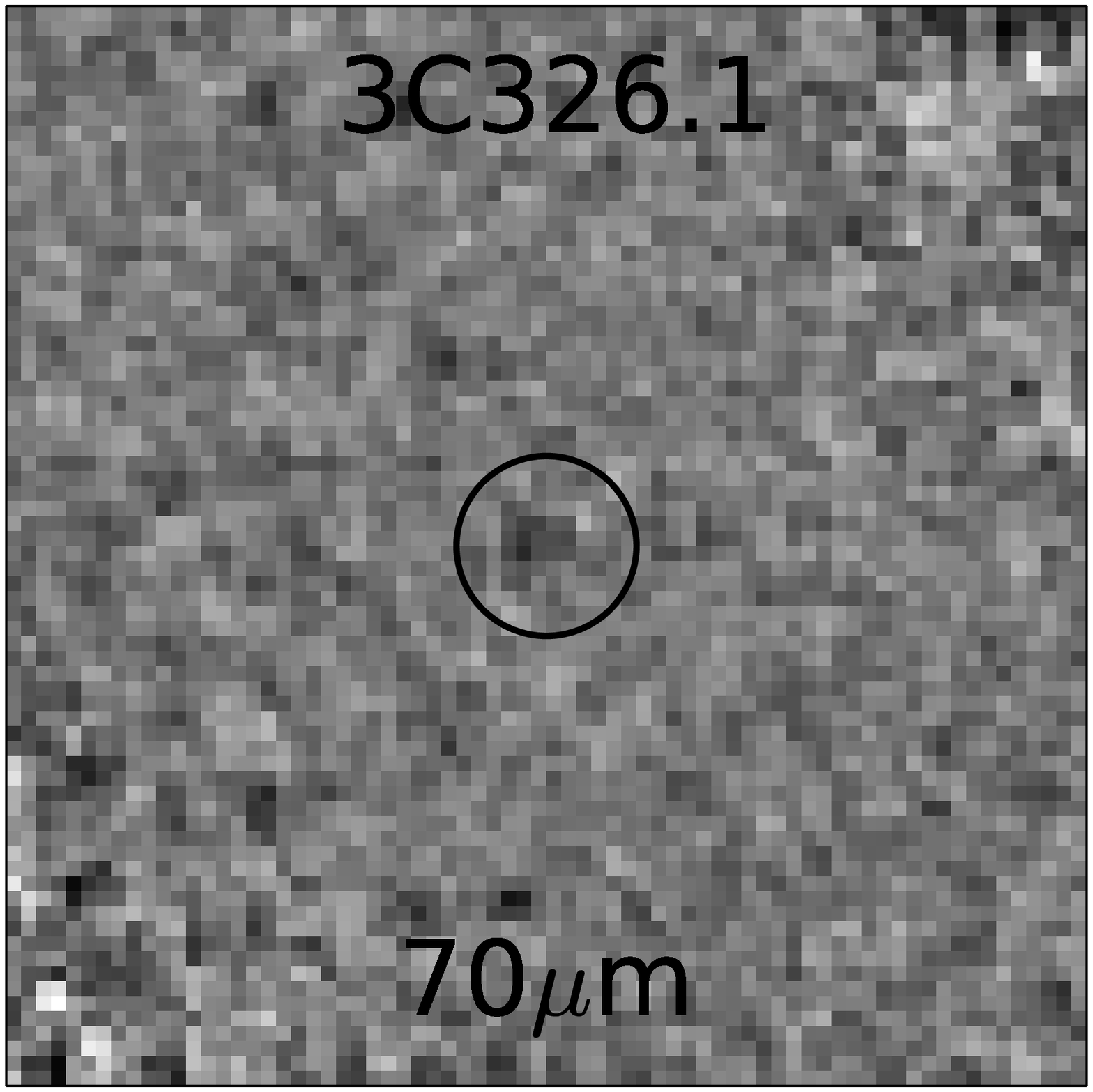}
      \includegraphics[width=1.5cm]{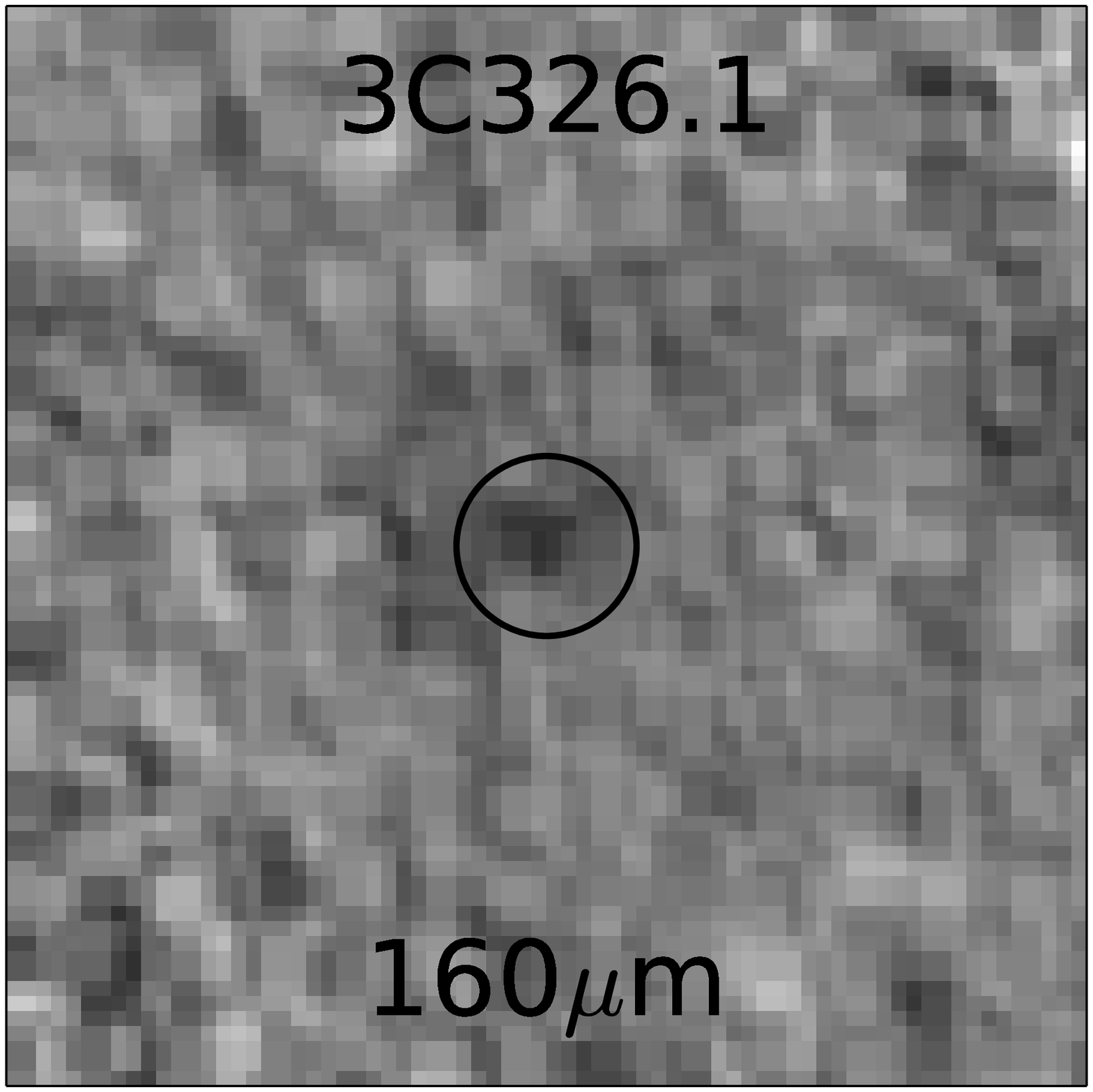}
      \includegraphics[width=1.5cm]{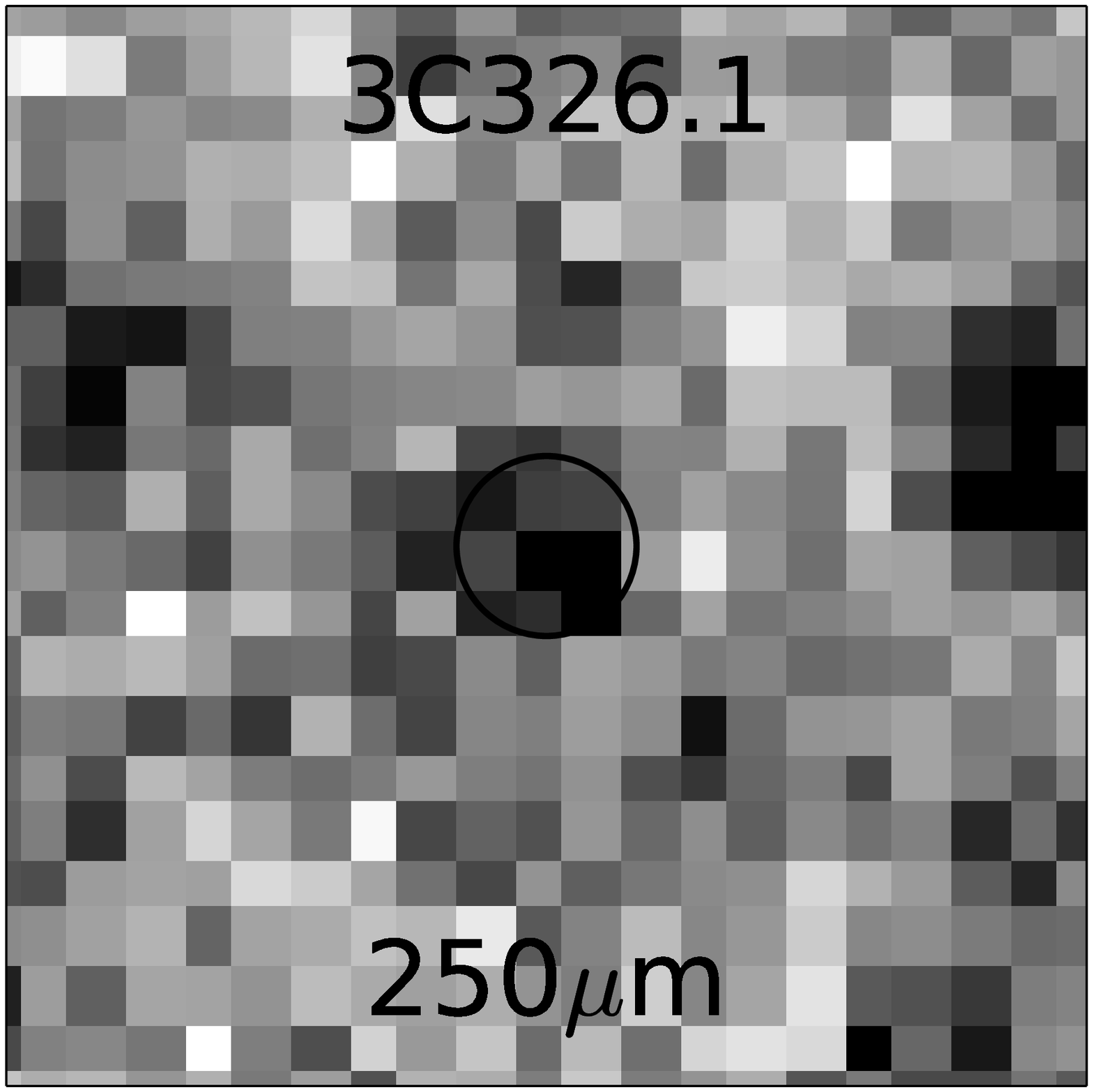}
      \includegraphics[width=1.5cm]{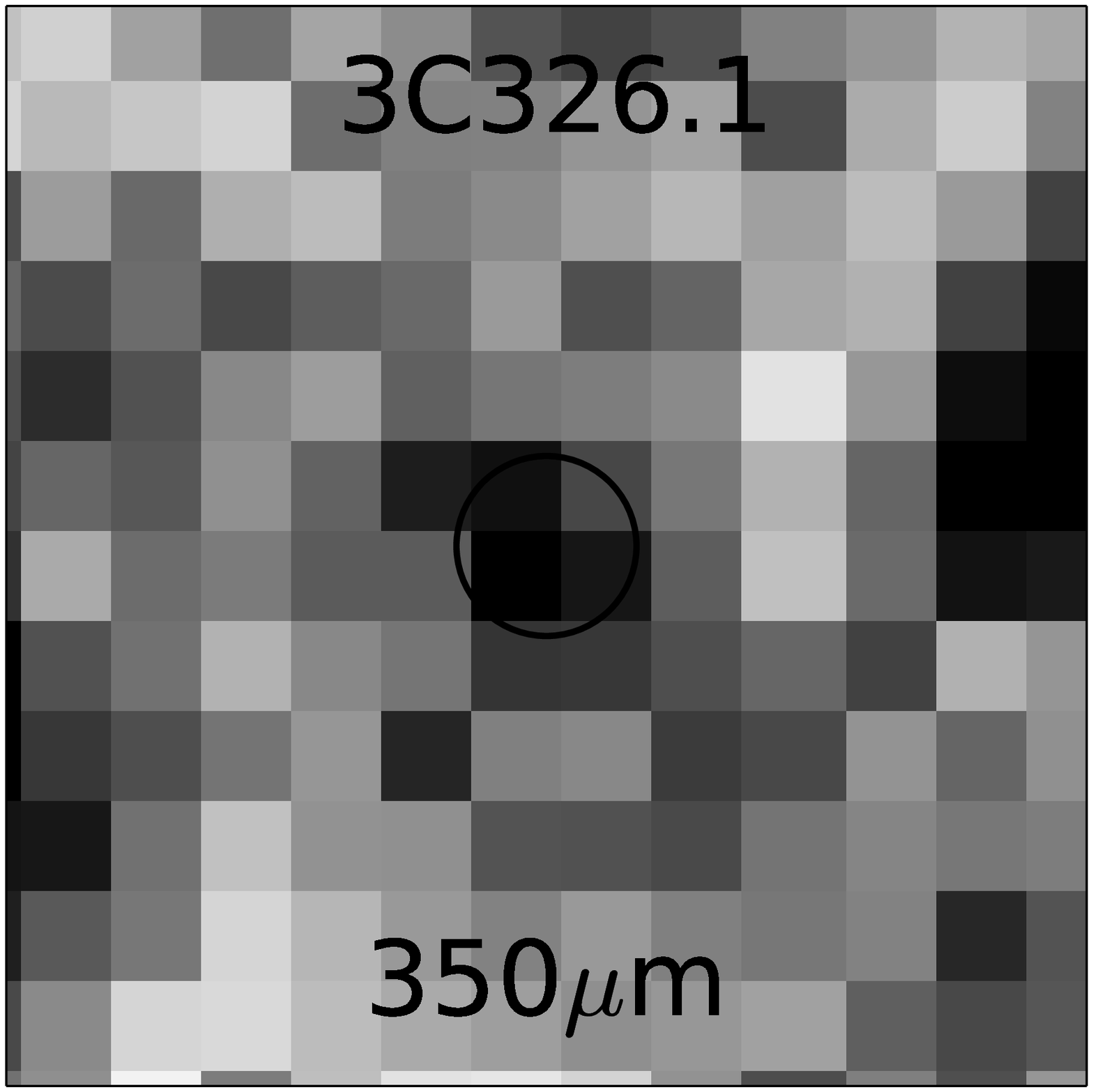}
      \includegraphics[width=1.5cm]{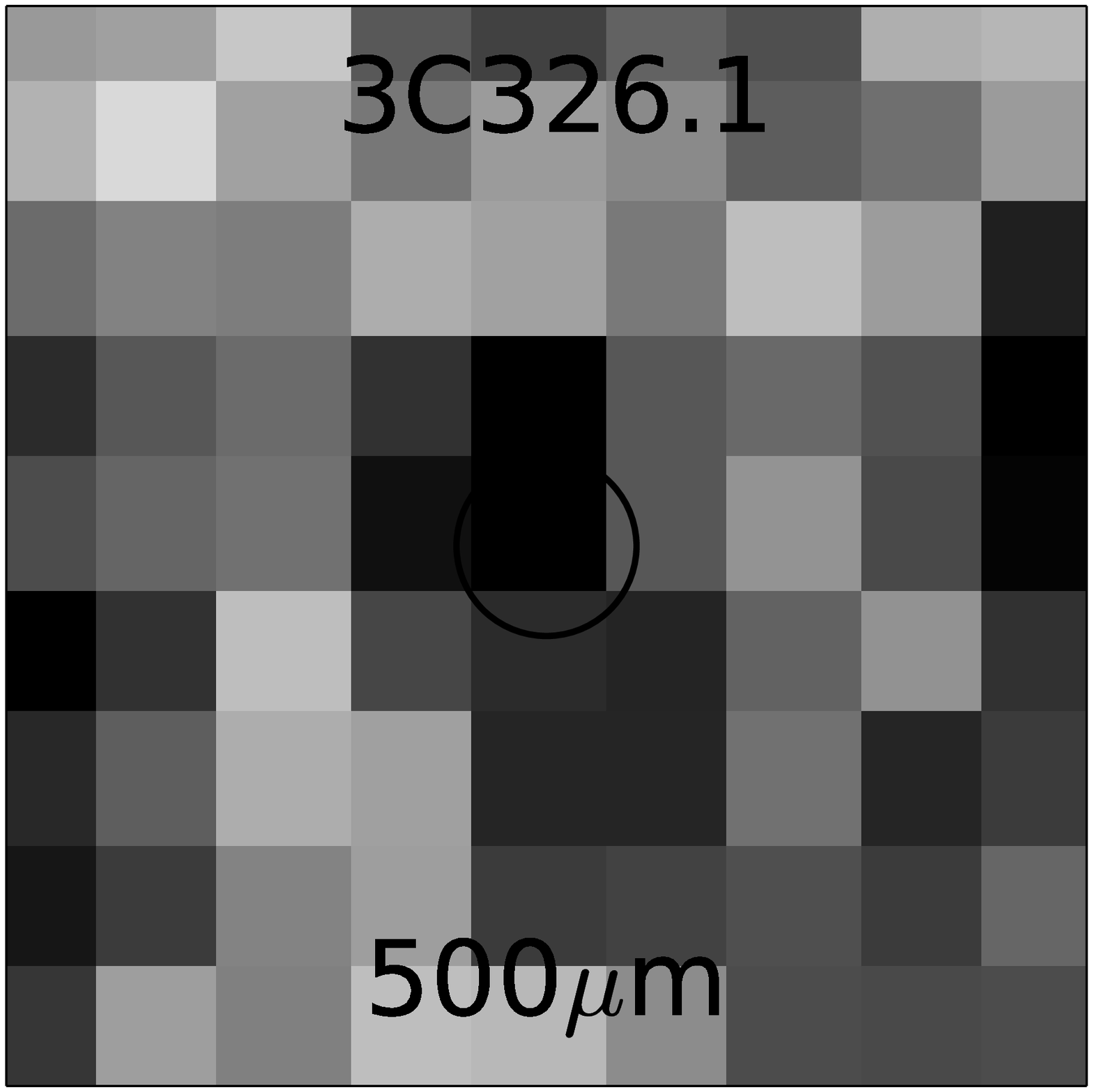}
      \\
      \includegraphics[width=1.5cm]{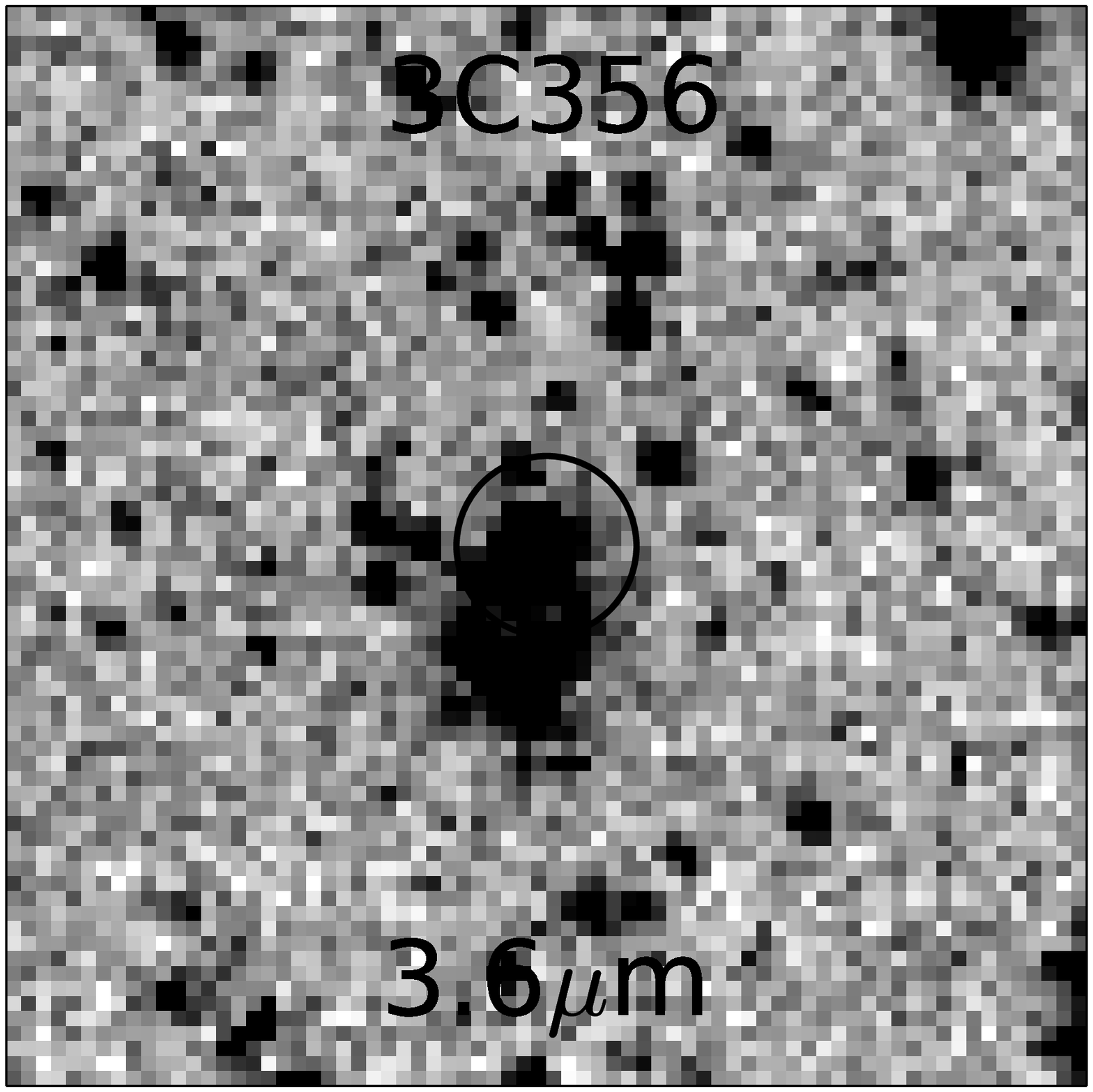}
      \includegraphics[width=1.5cm]{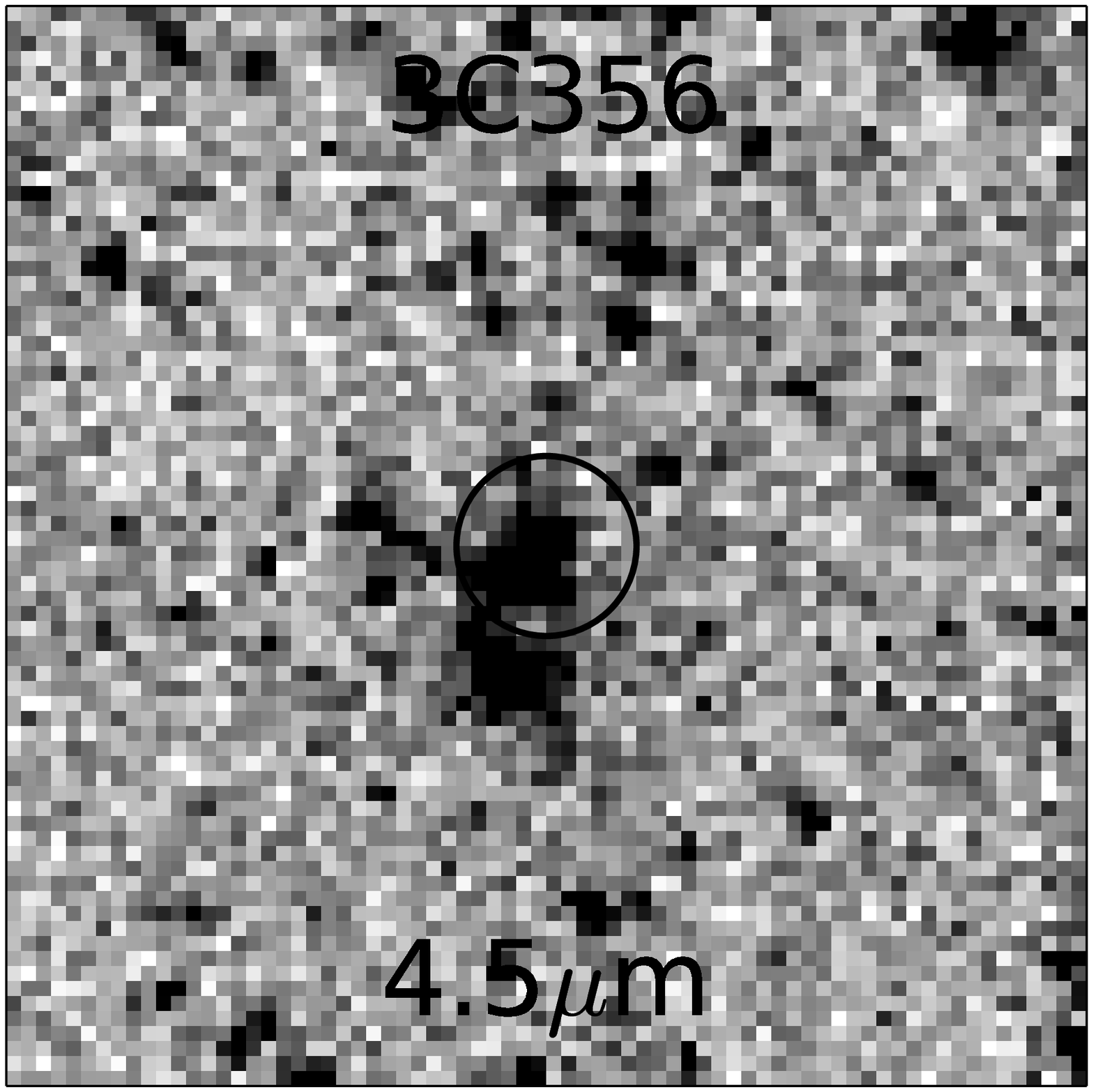}
      \includegraphics[width=1.5cm]{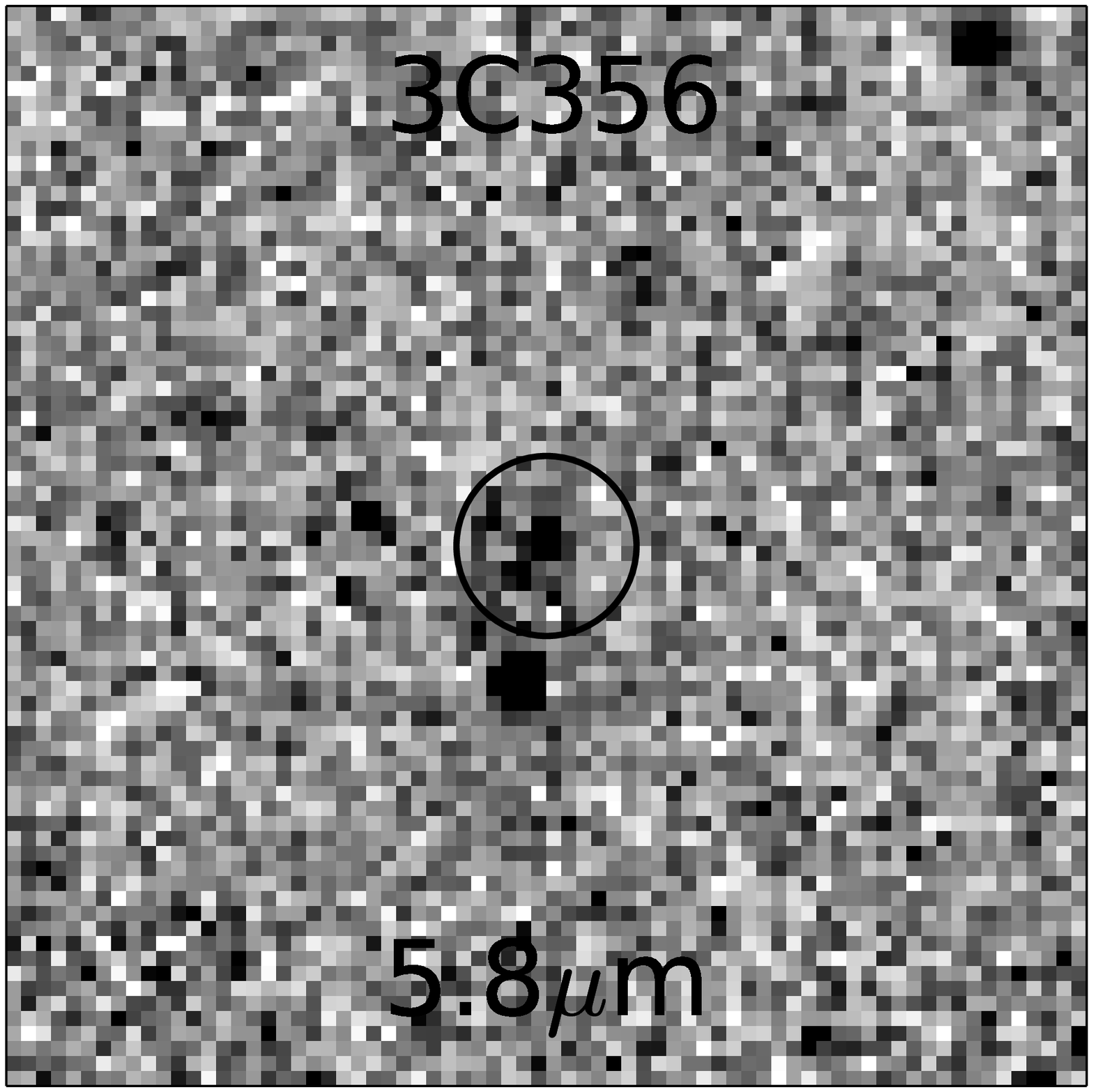}
      \includegraphics[width=1.5cm]{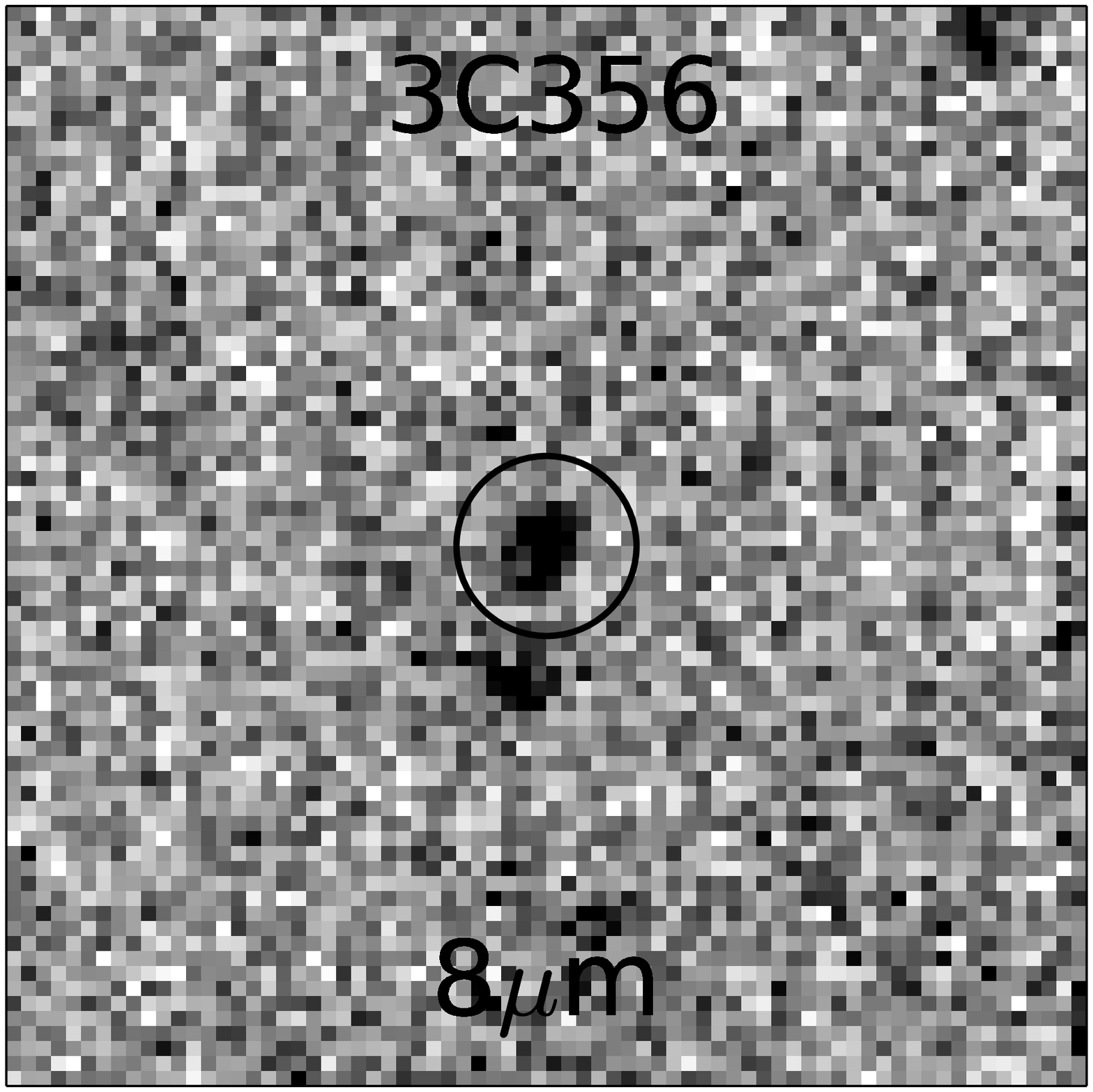}
      \includegraphics[width=1.5cm]{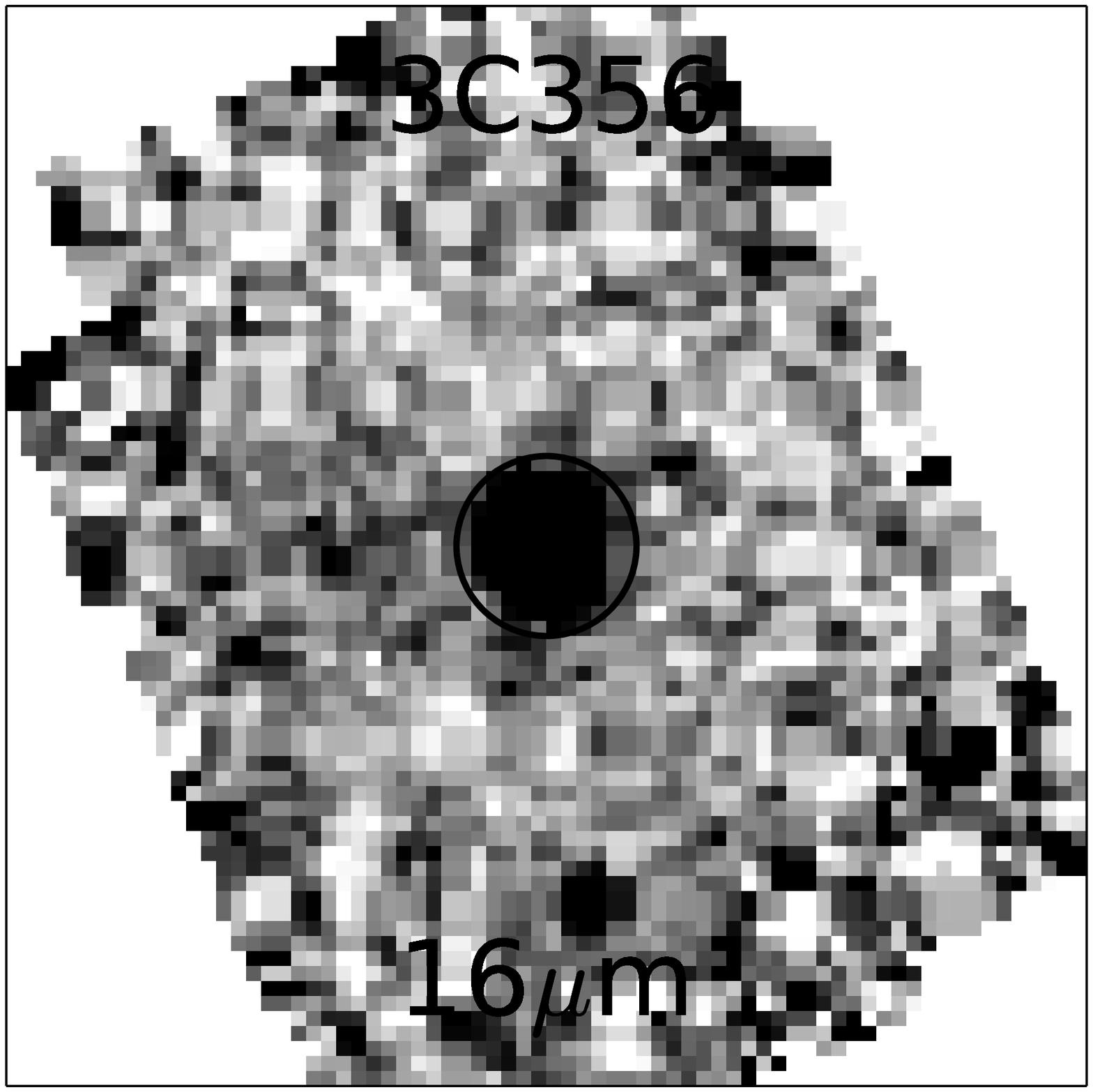}
      \includegraphics[width=1.5cm]{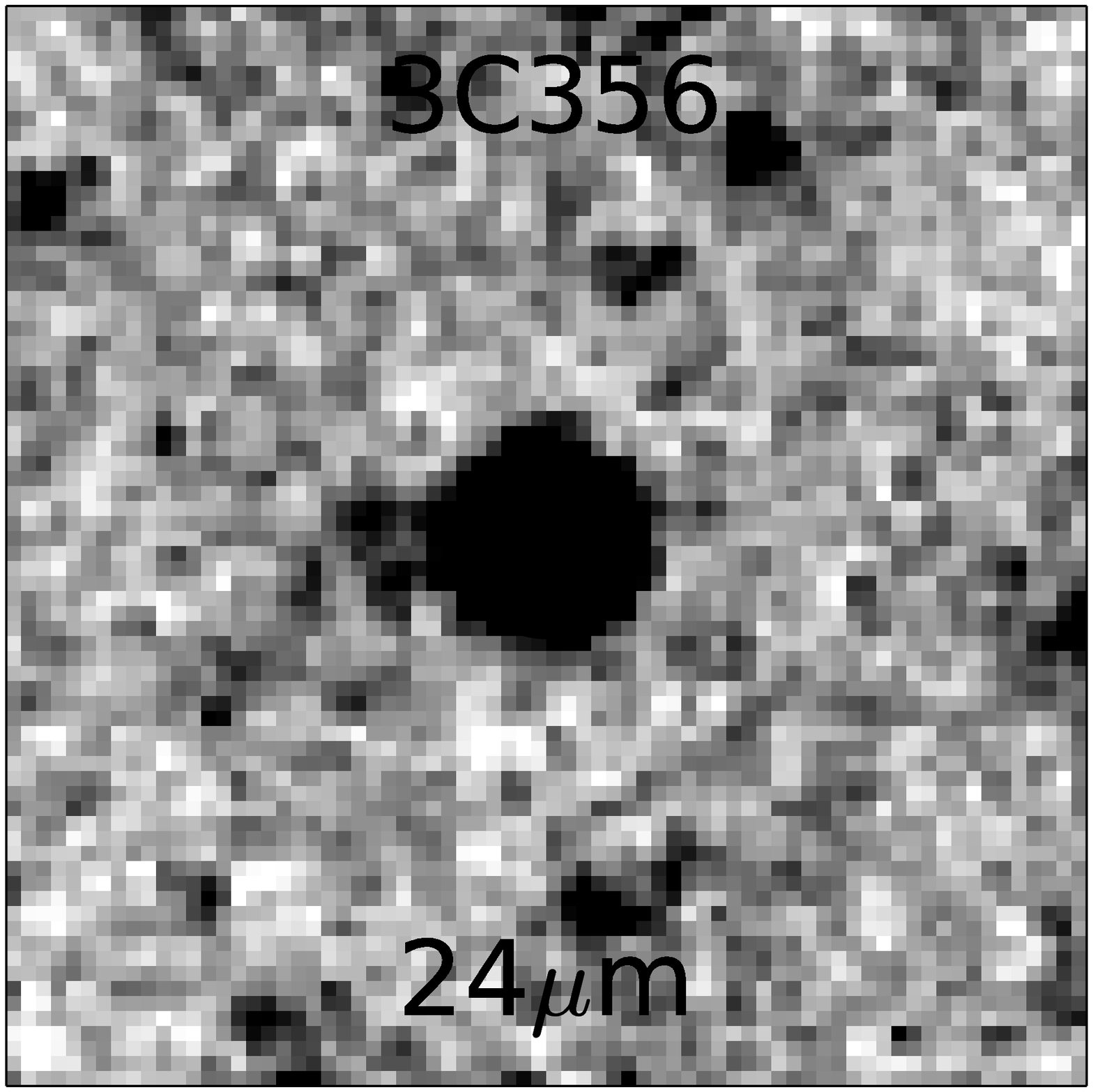}
      \includegraphics[width=1.5cm]{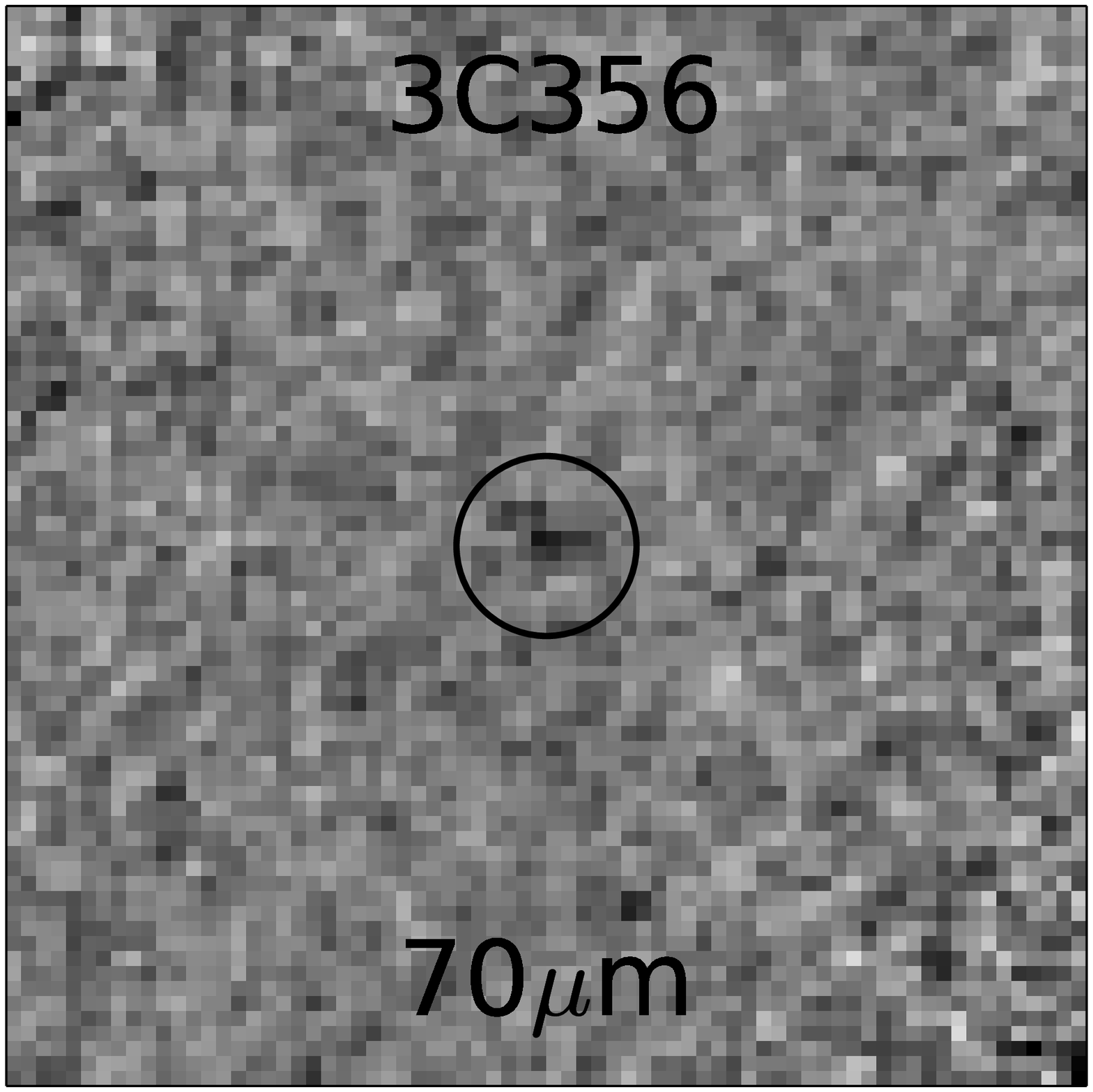}
      \includegraphics[width=1.5cm]{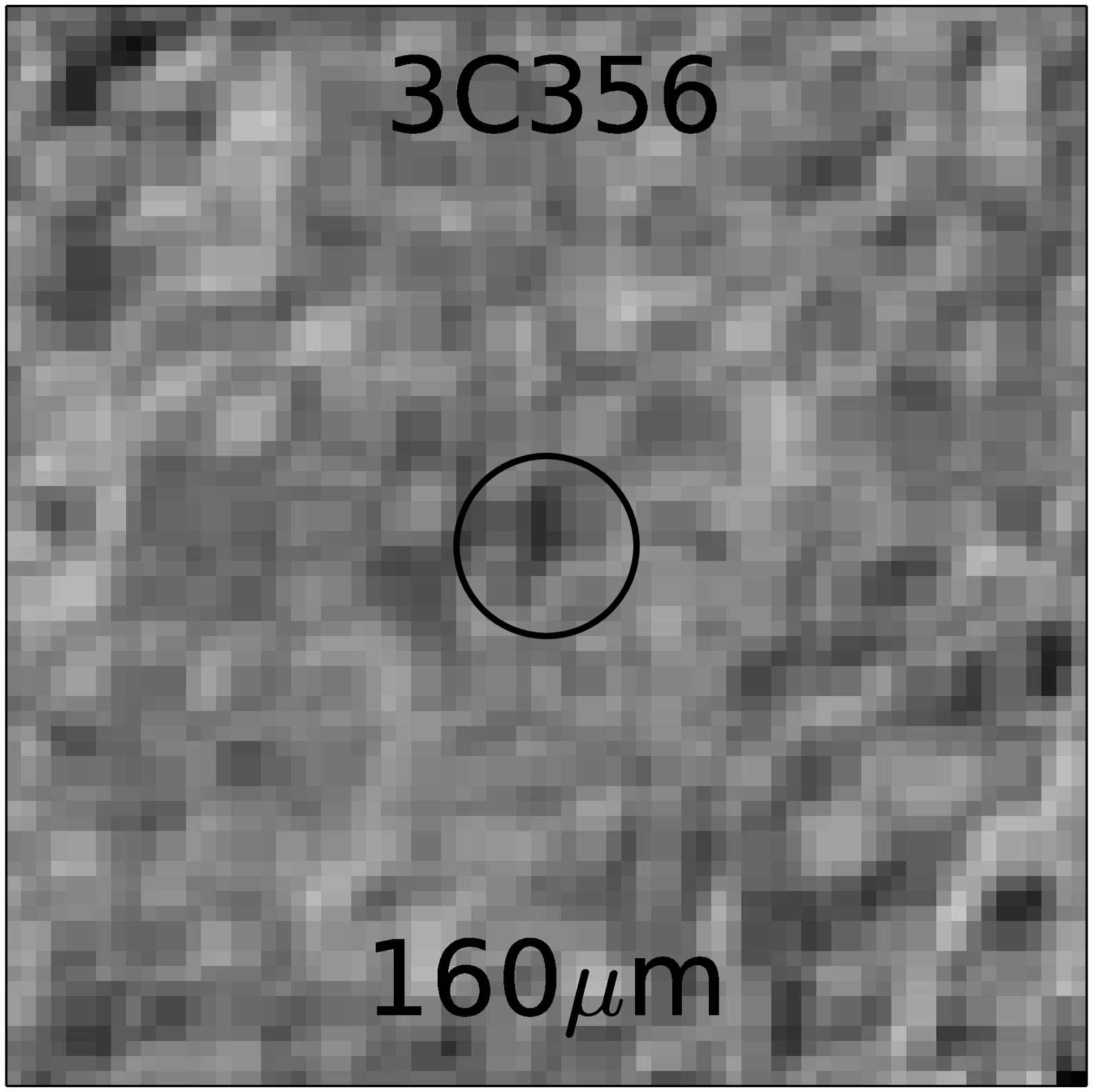}
      \includegraphics[width=1.5cm]{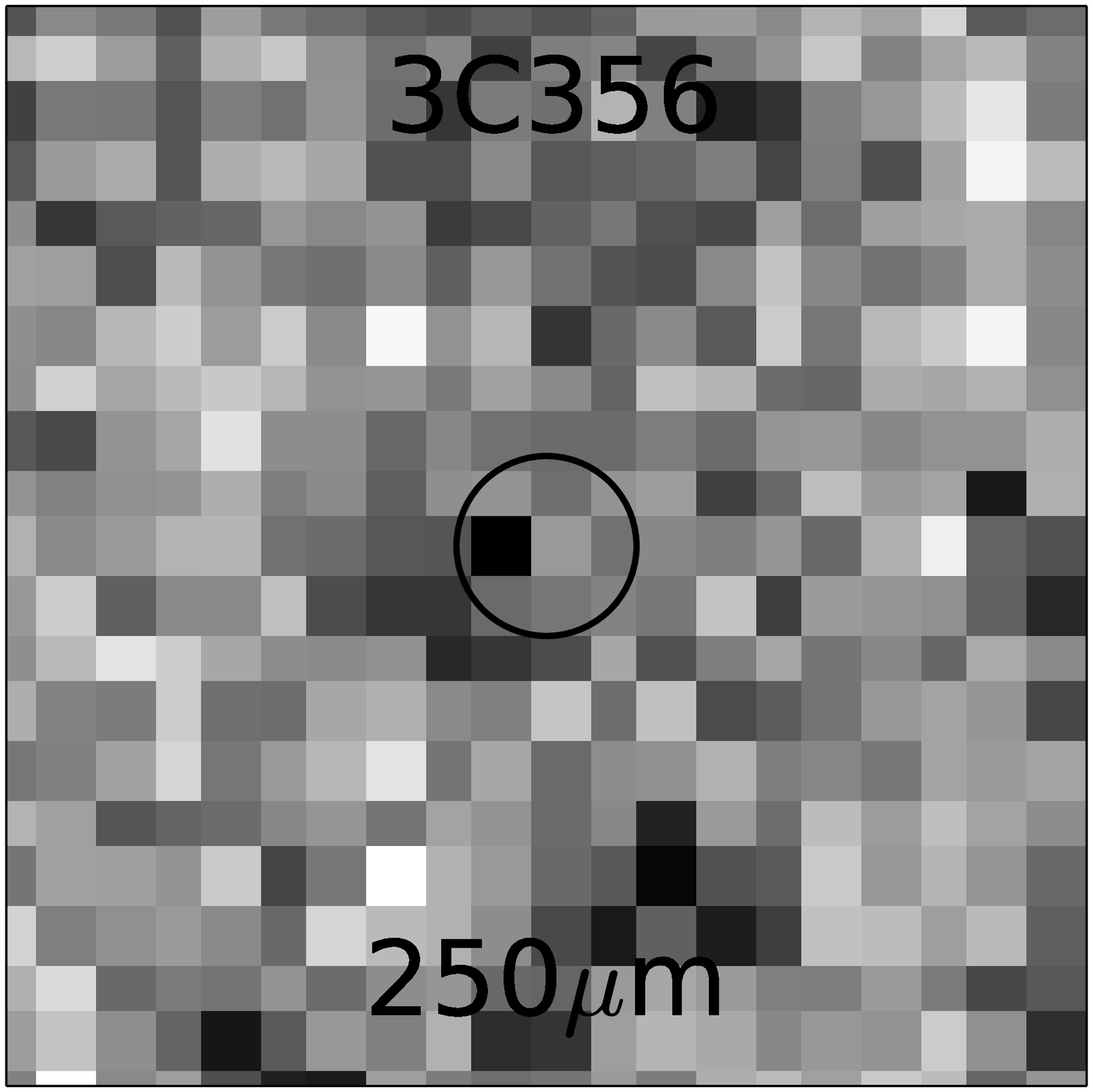}
      \includegraphics[width=1.5cm]{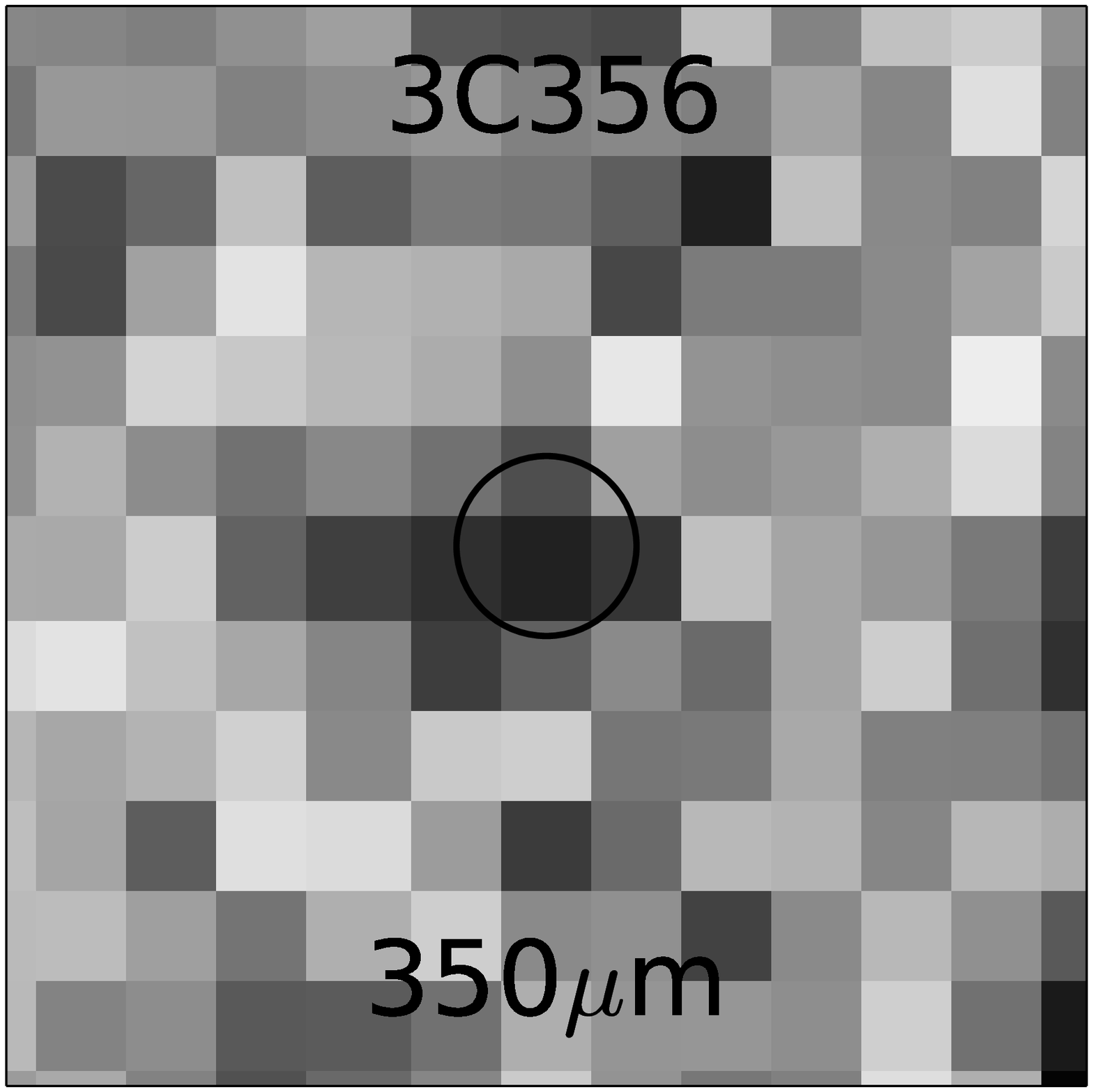}
      \includegraphics[width=1.5cm]{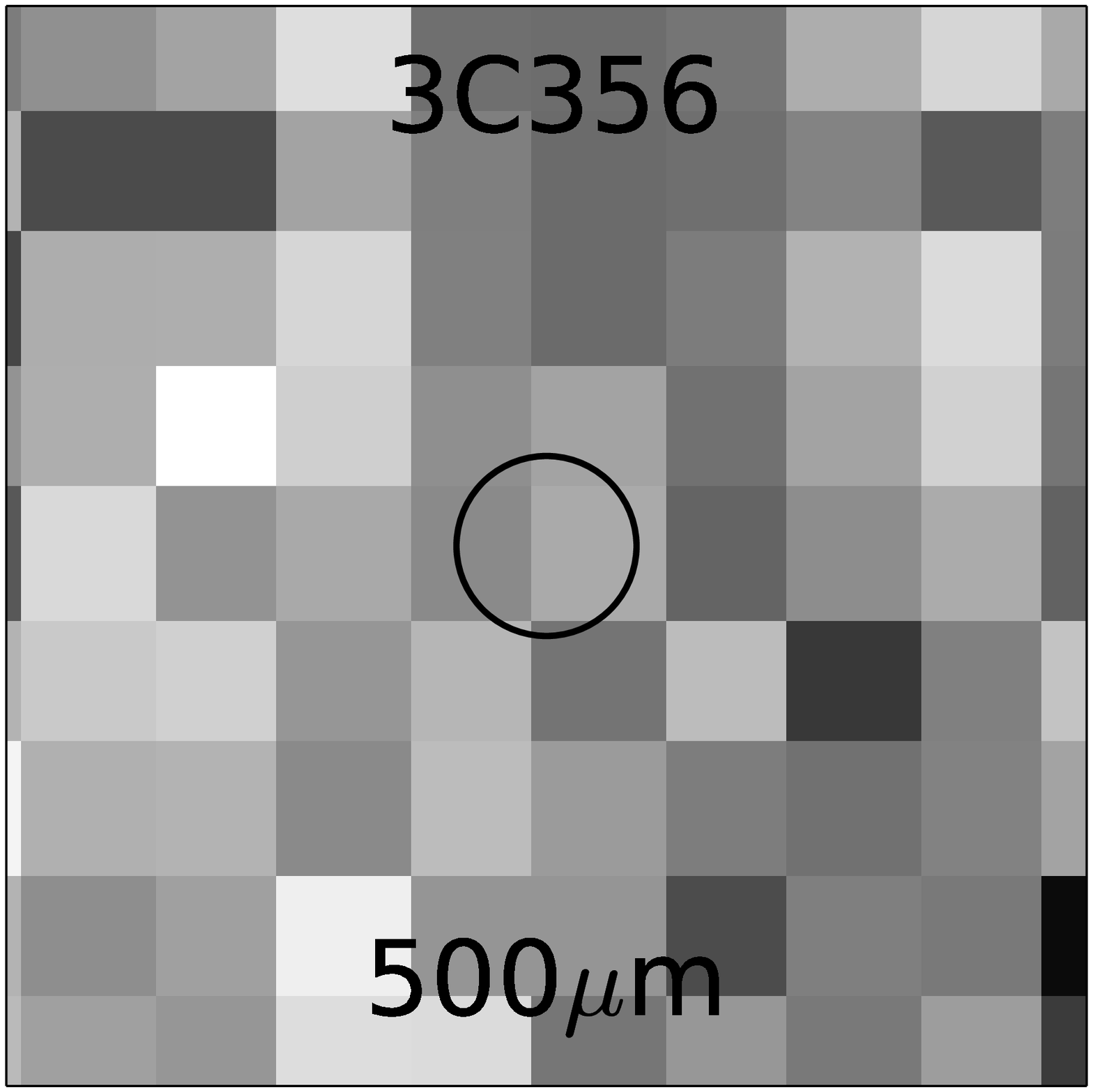}
      \caption{Continued.}
   \end{figure*}    
   \addtocounter{figure}{-1}
   \begin{figure*}
      \includegraphics[width=1.5cm]{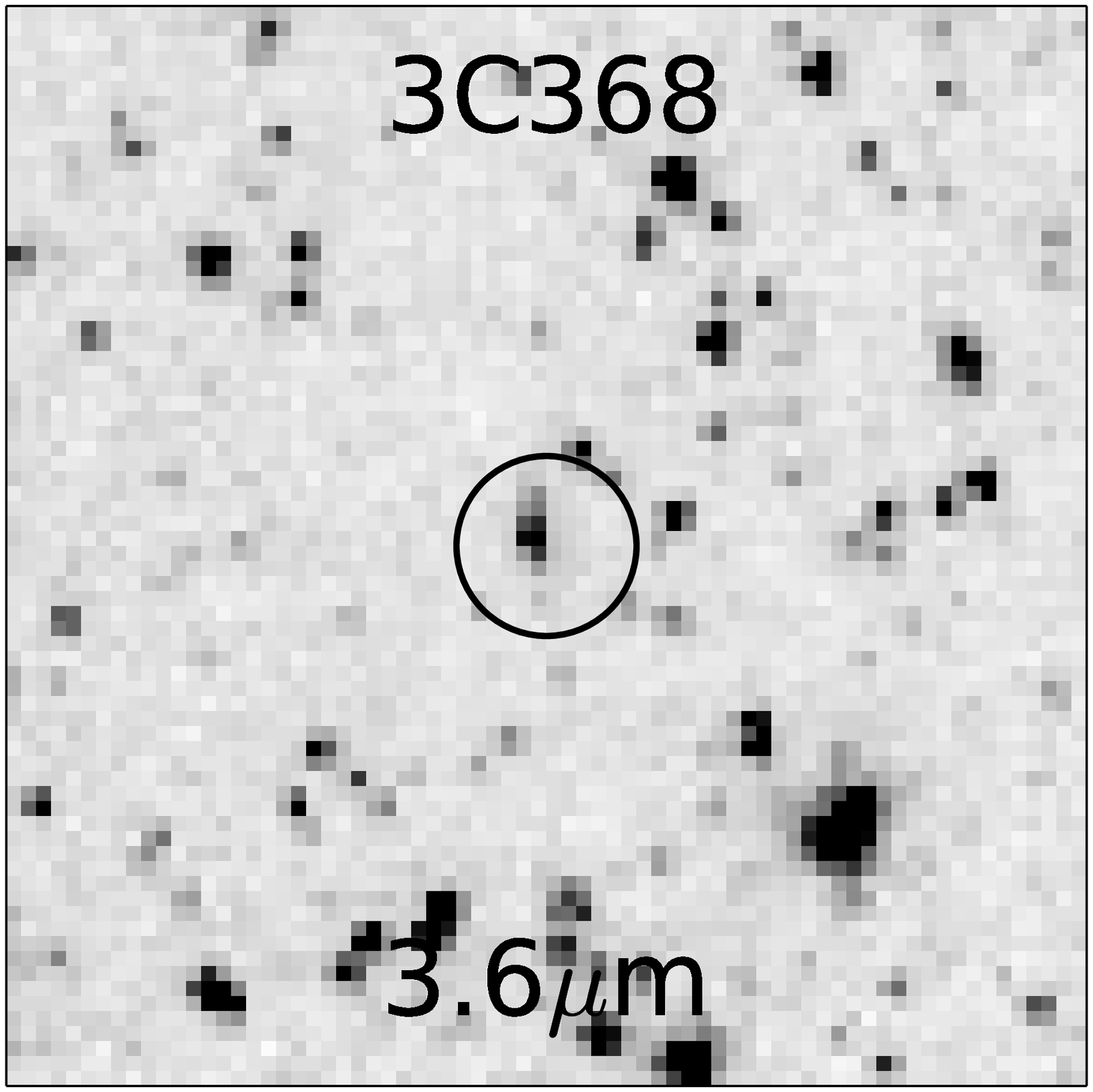}
      \includegraphics[width=1.5cm]{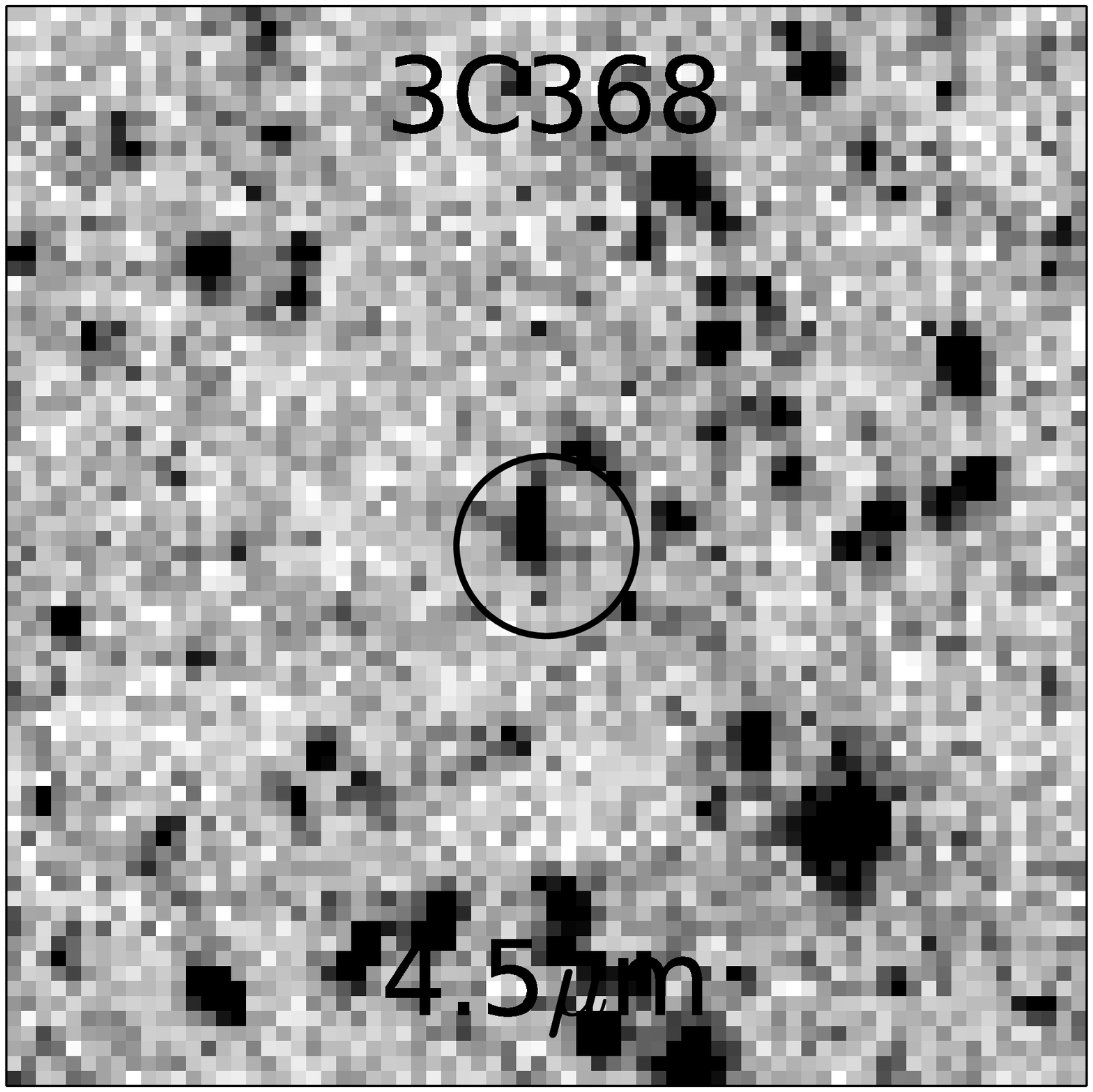}
      \includegraphics[width=1.5cm]{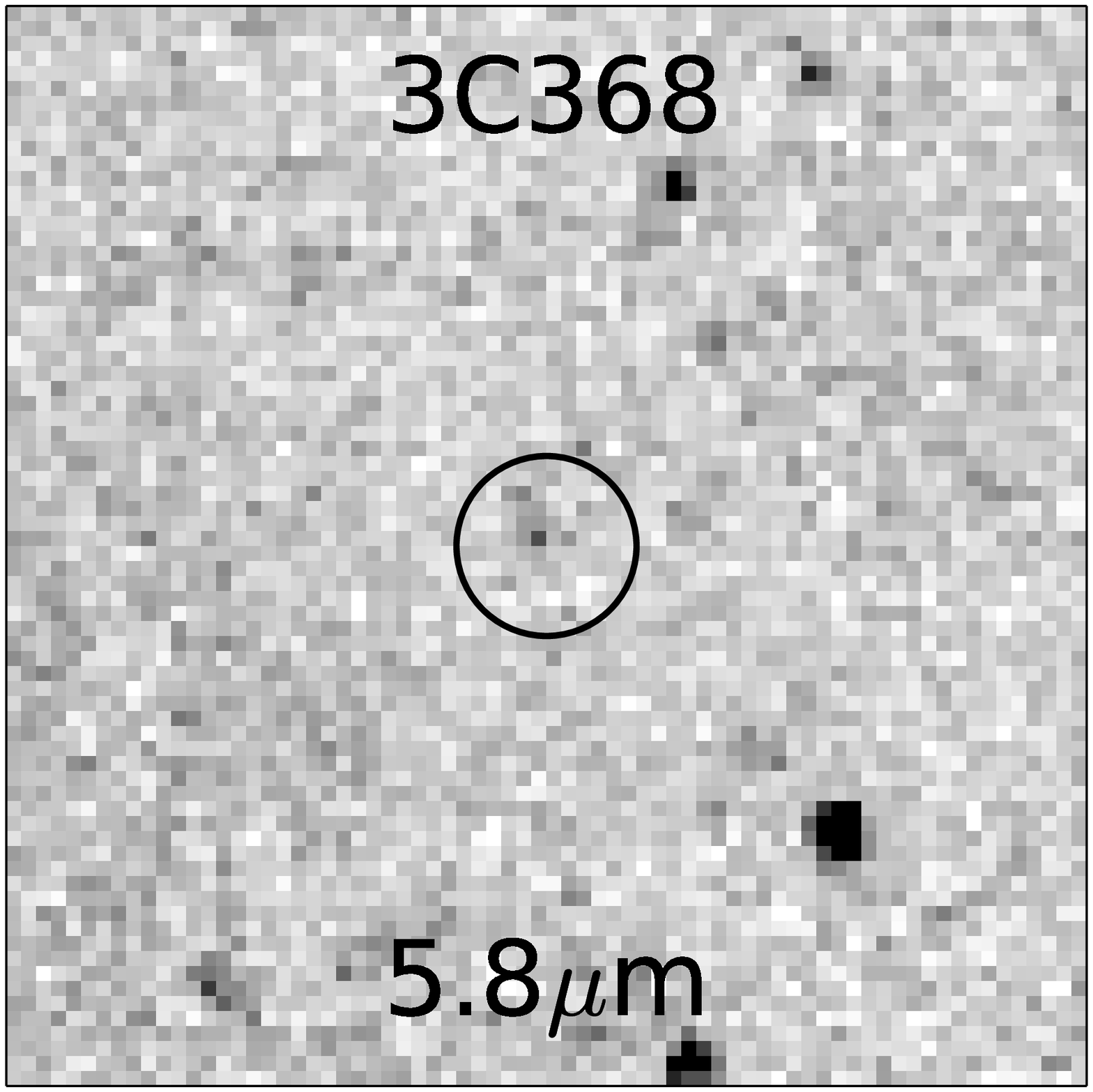}
      \includegraphics[width=1.5cm]{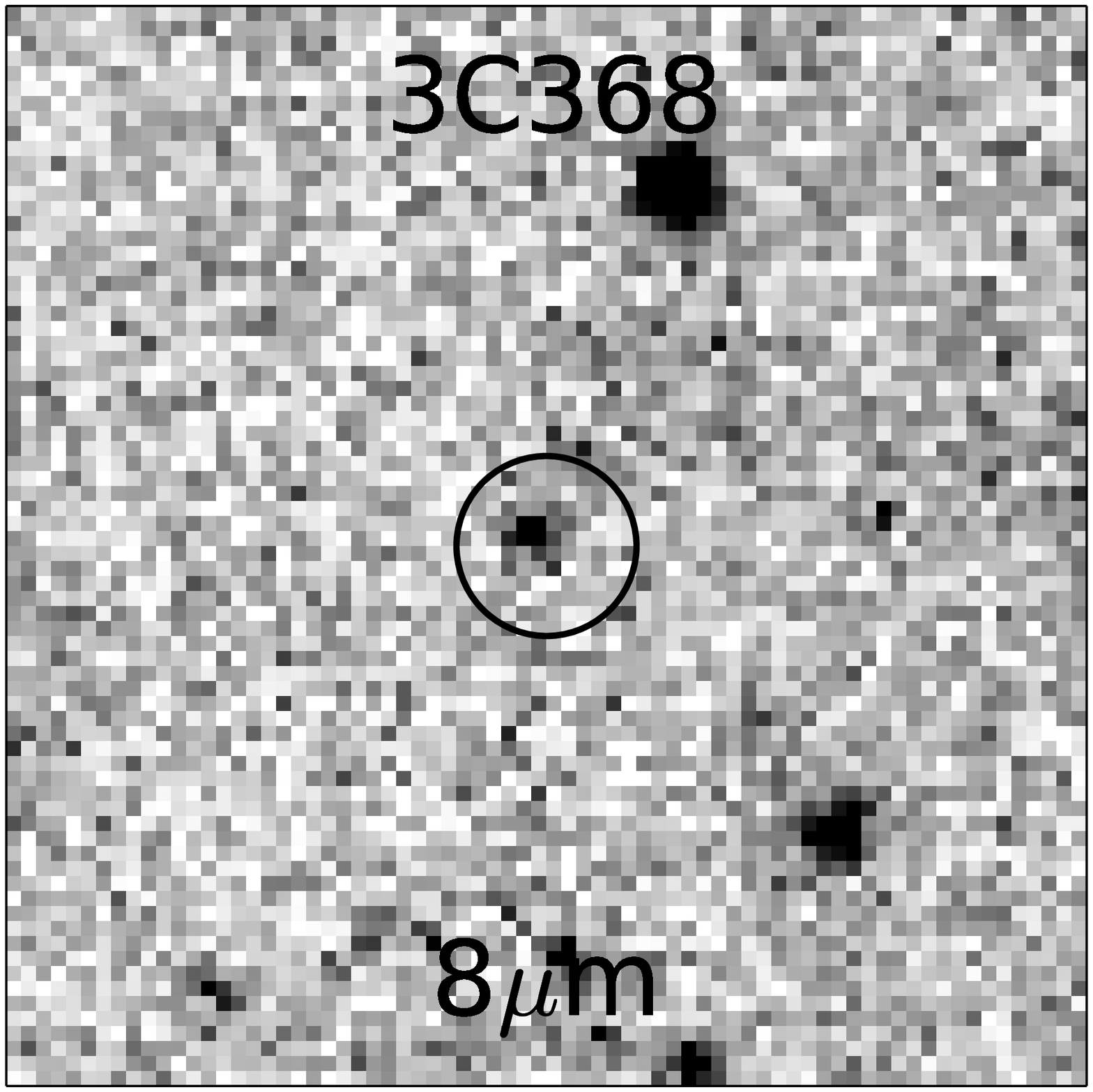}
      \includegraphics[width=1.5cm]{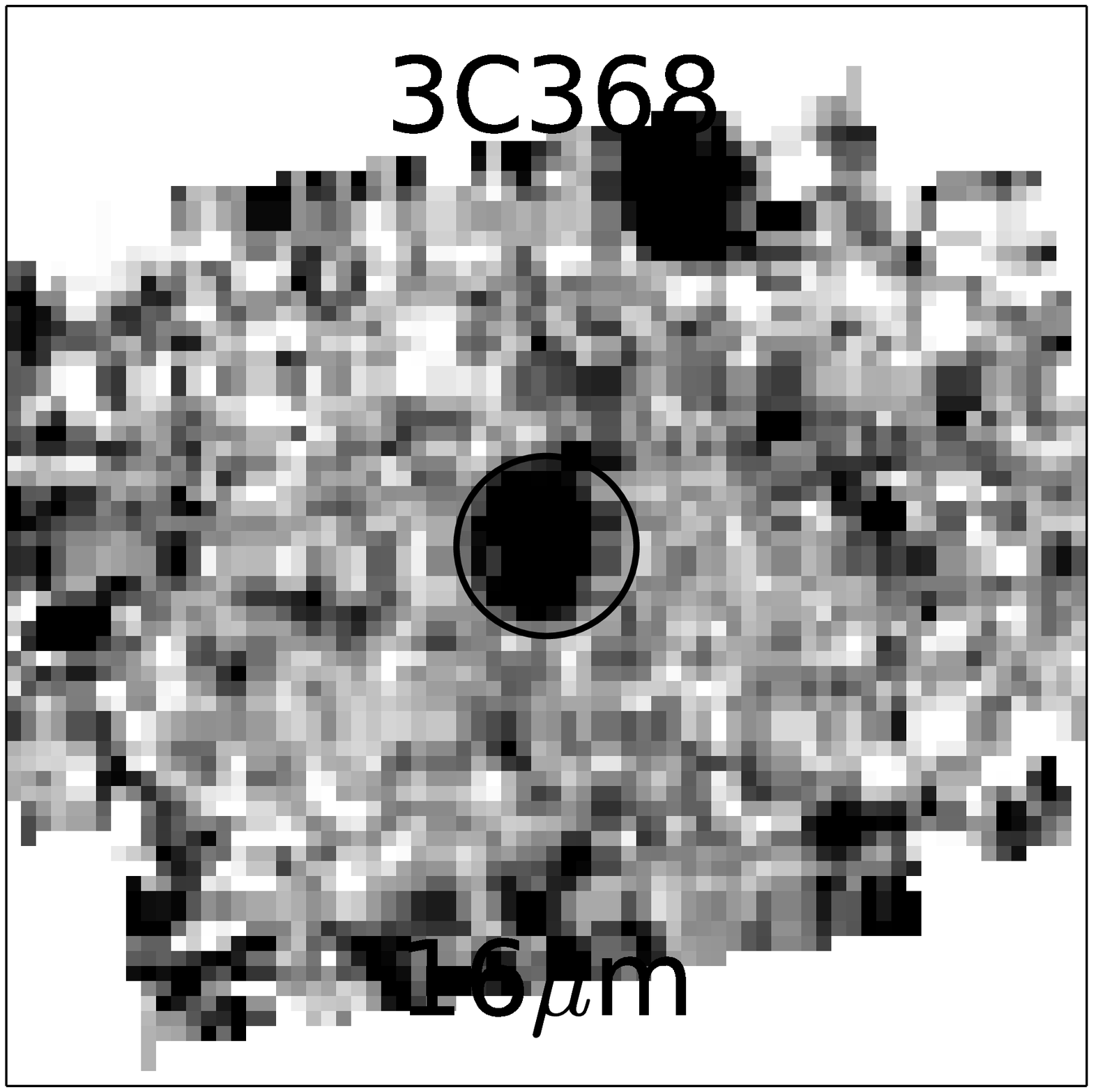}
      \includegraphics[width=1.5cm]{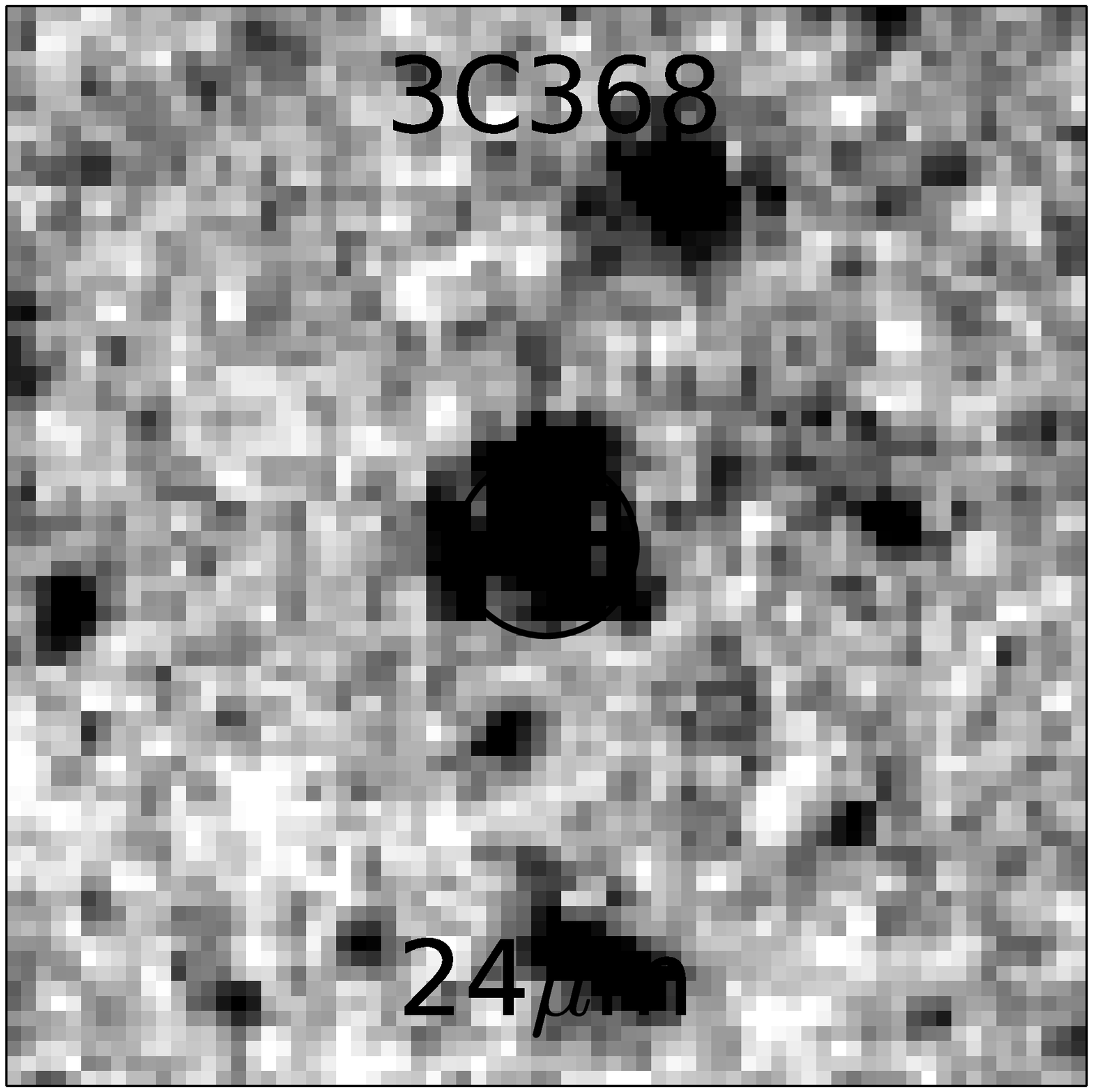}
      \includegraphics[width=1.5cm]{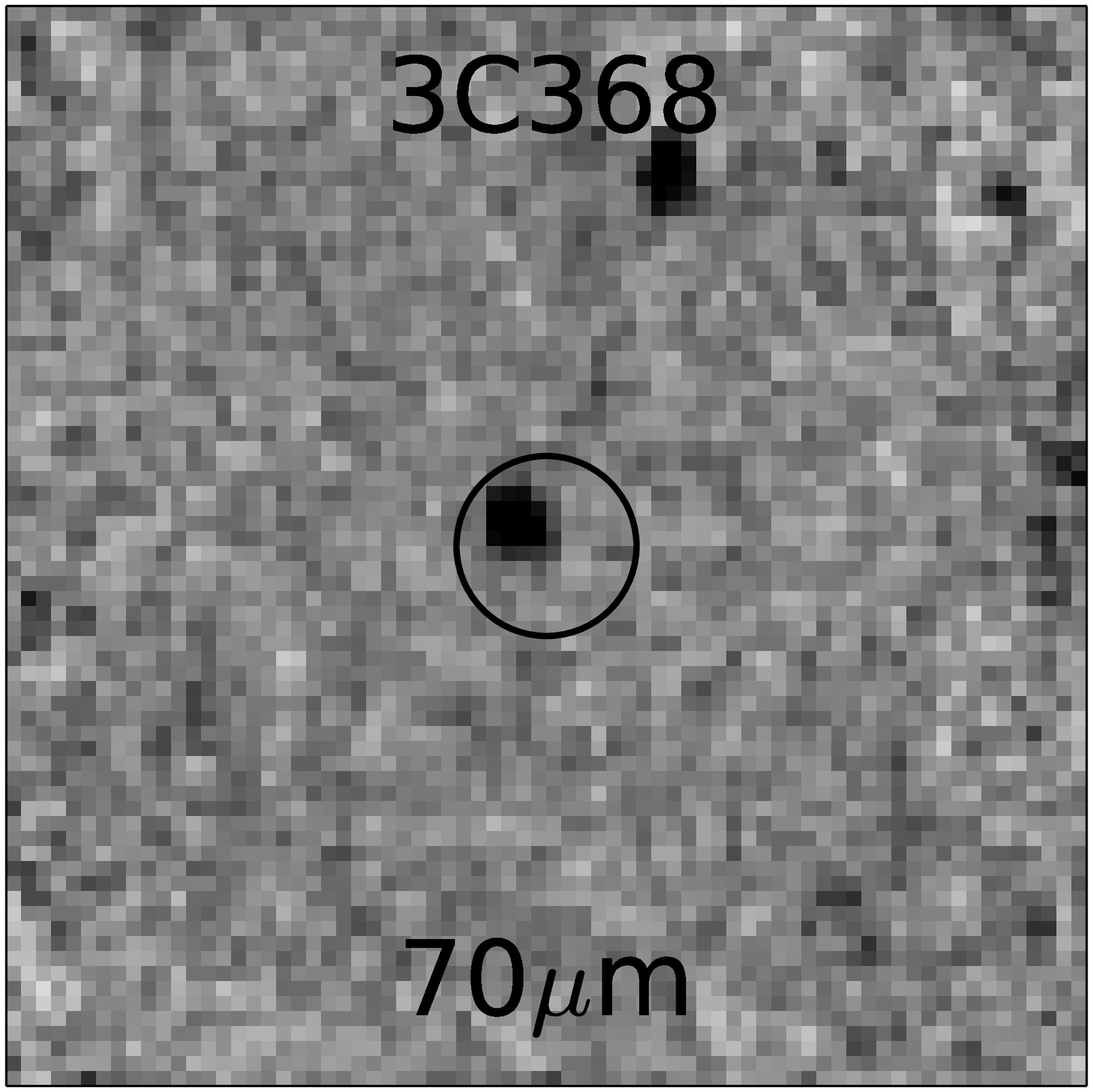}
      \includegraphics[width=1.5cm]{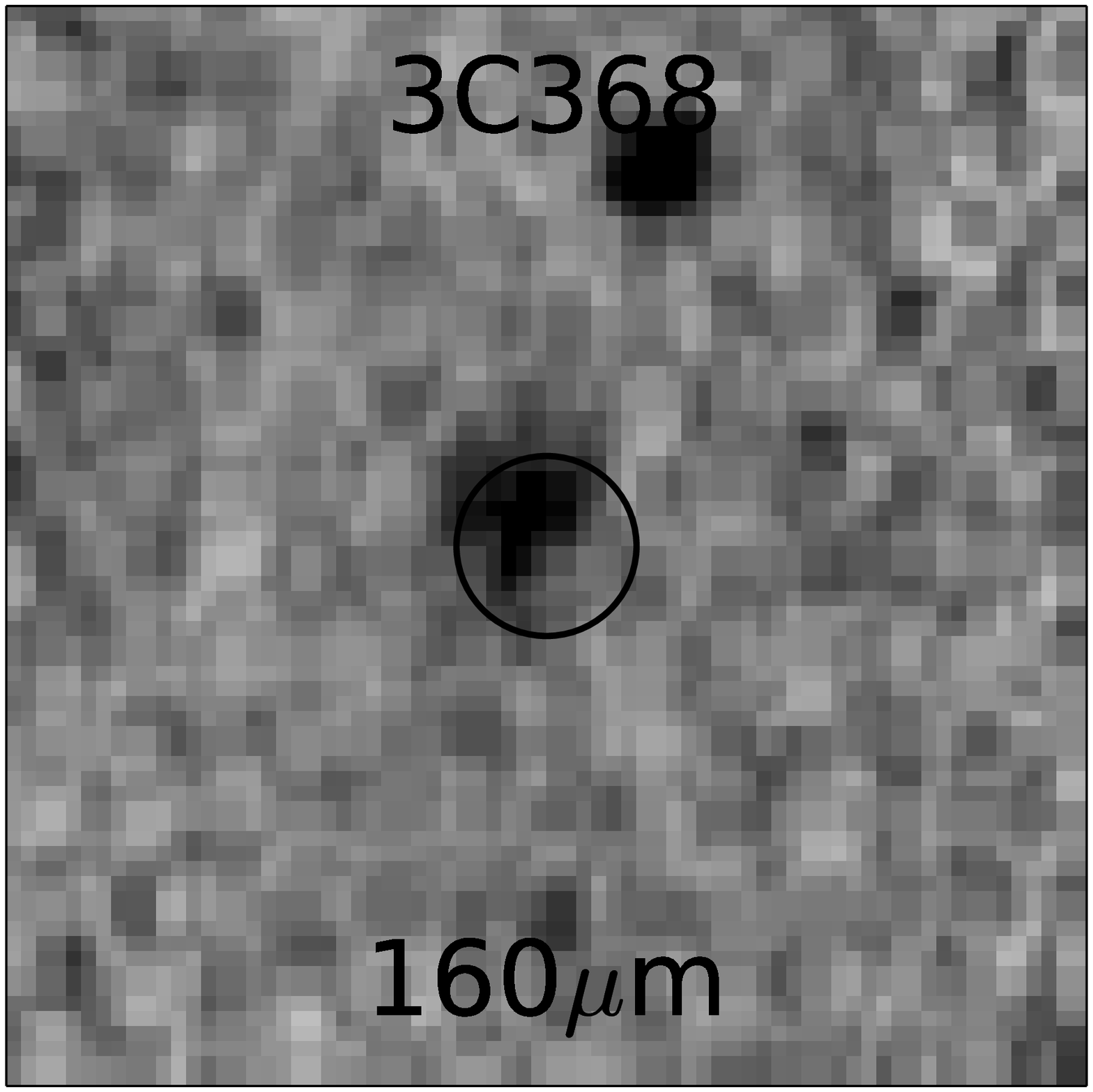}
      \includegraphics[width=1.5cm]{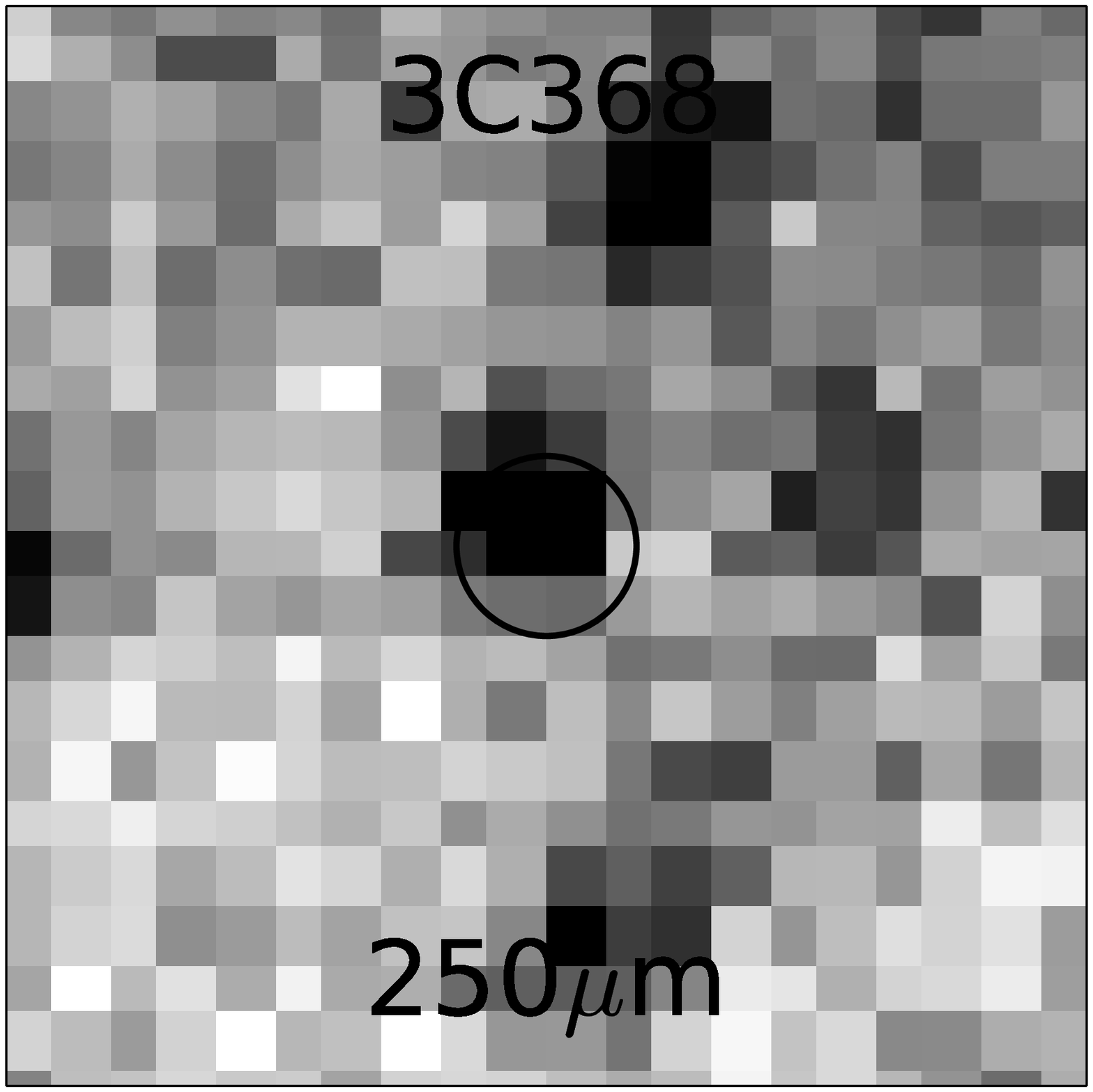}
      \includegraphics[width=1.5cm]{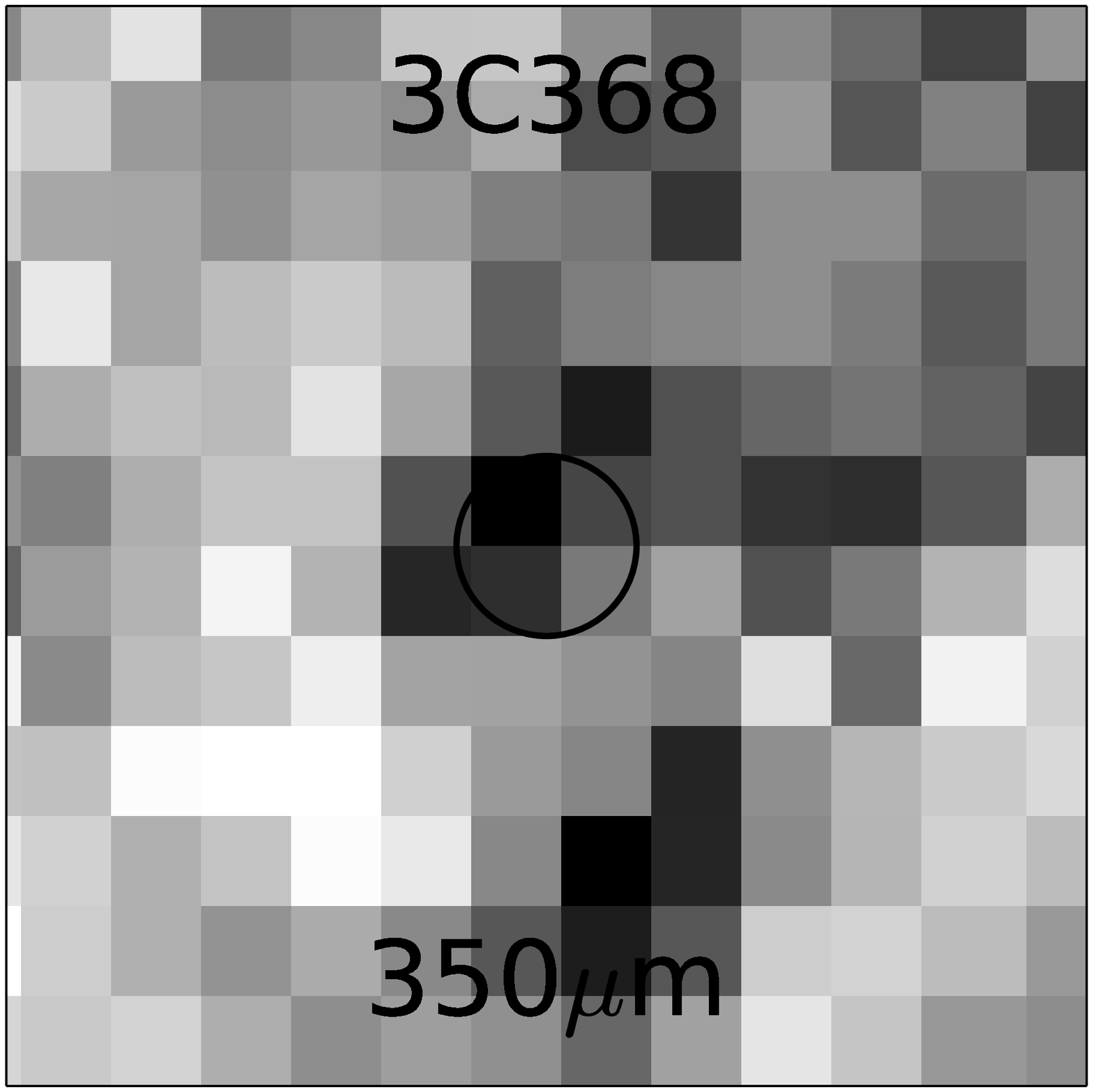}
      \includegraphics[width=1.5cm]{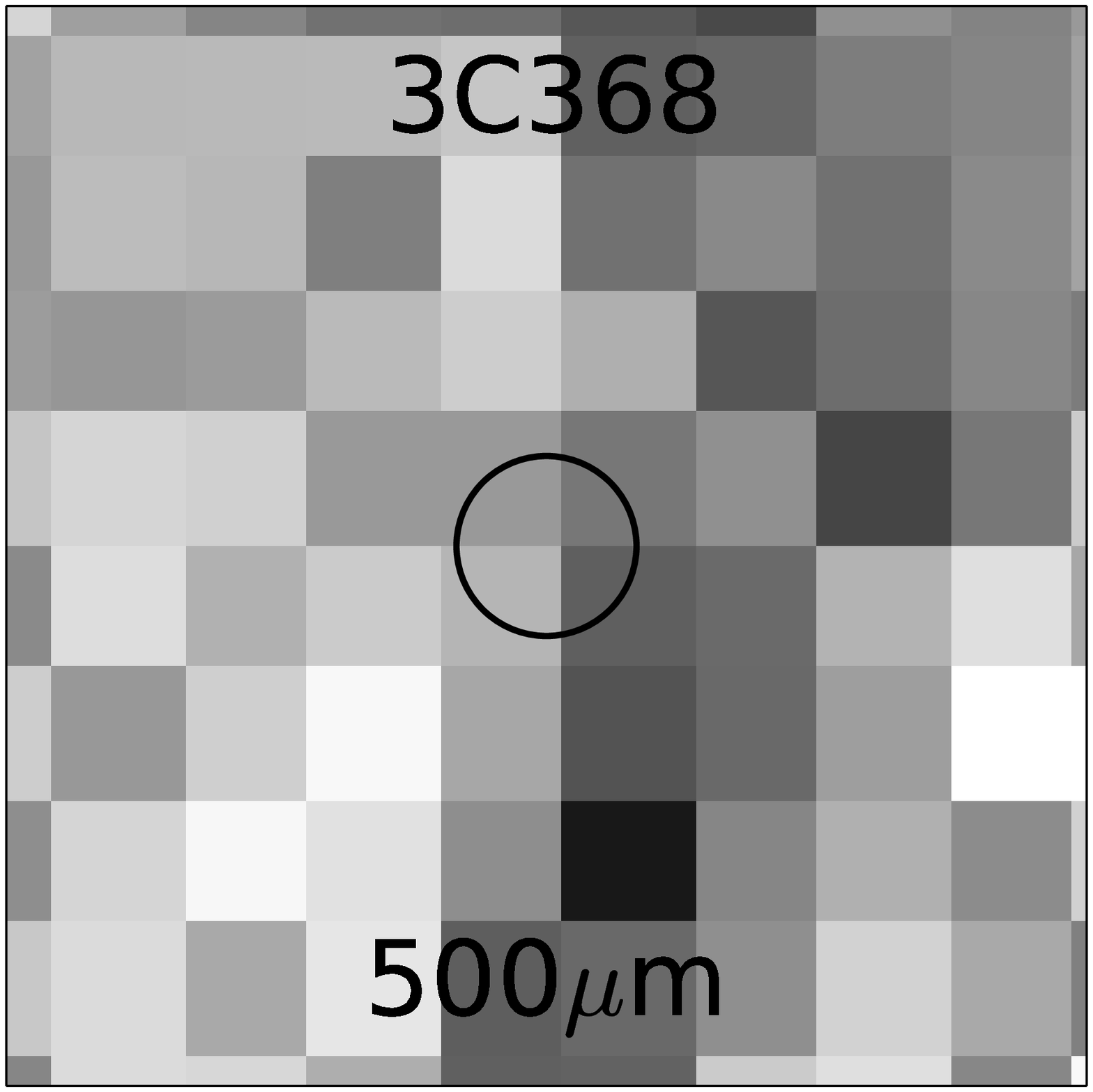}
      \\
      \includegraphics[width=1.5cm]{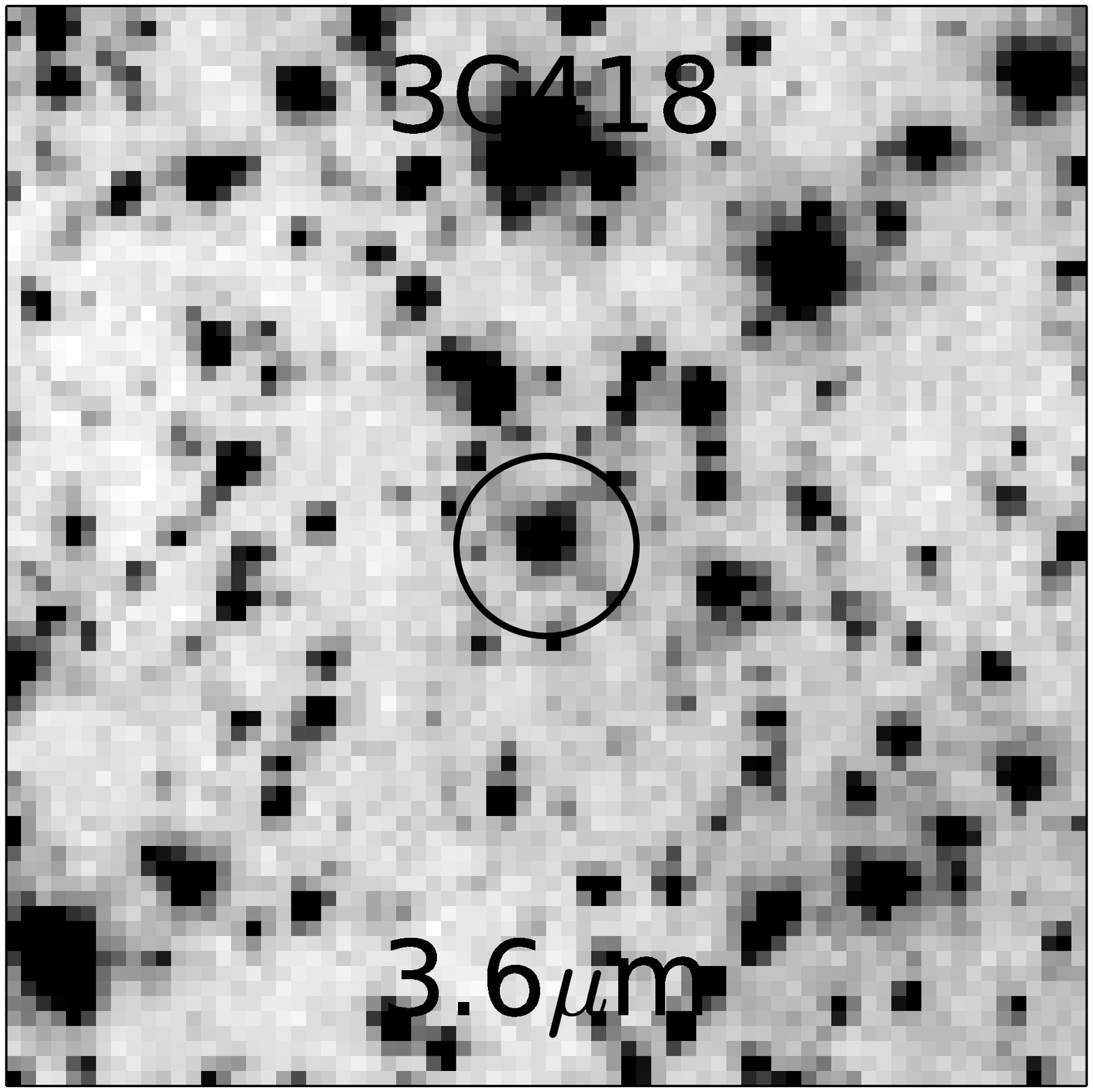}
      \includegraphics[width=1.5cm]{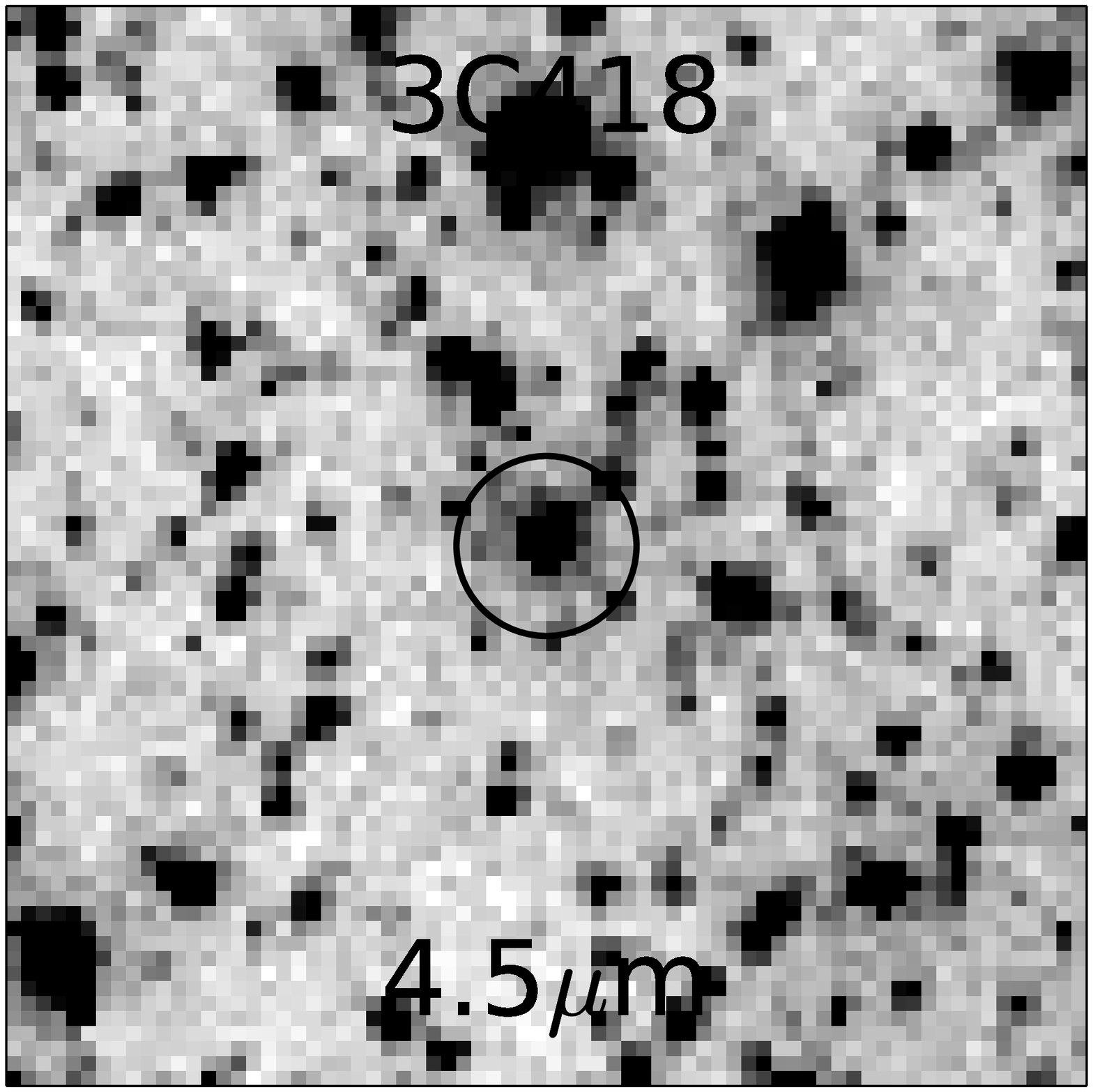}
      \includegraphics[width=1.5cm]{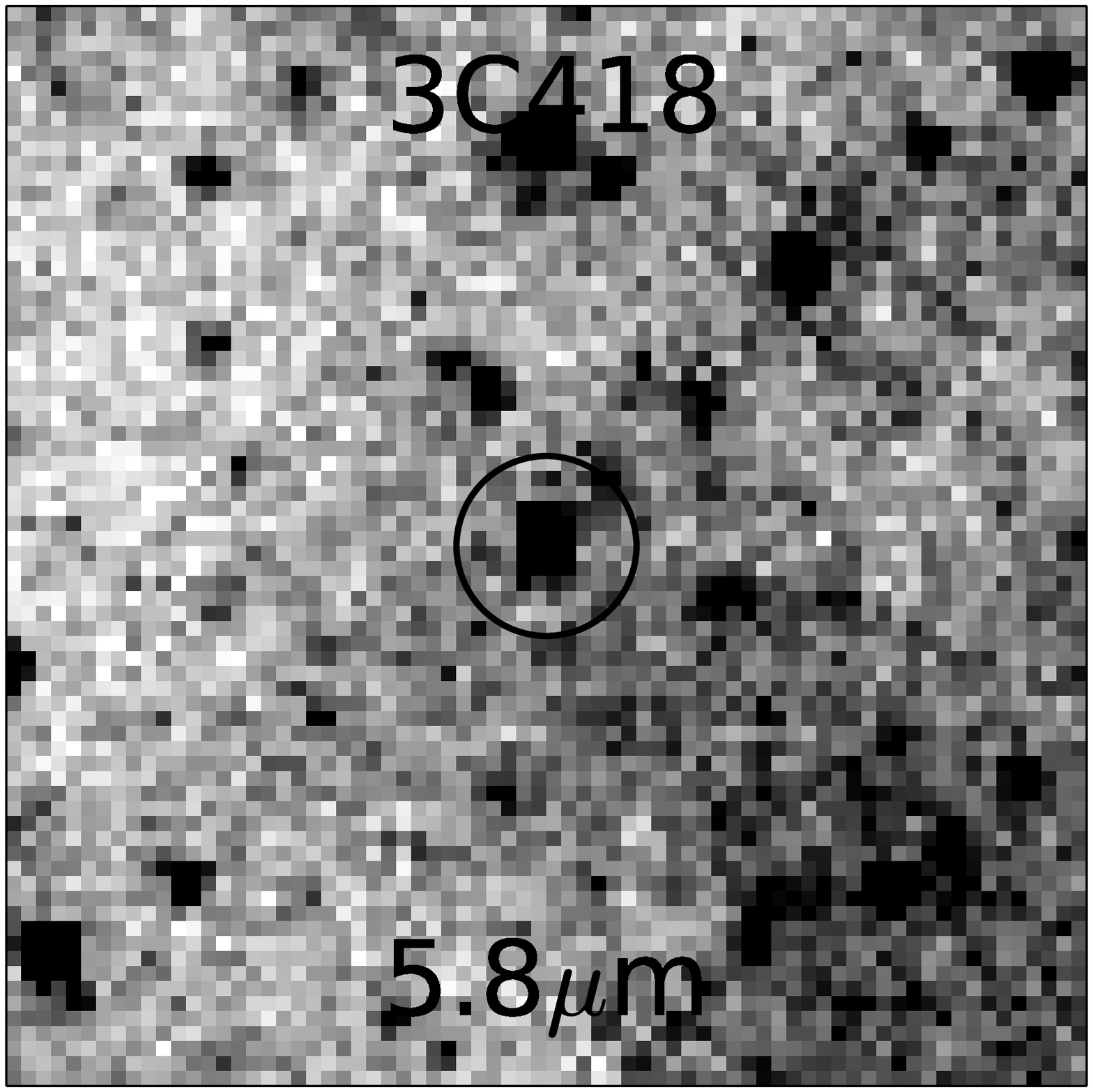}
      \includegraphics[width=1.5cm]{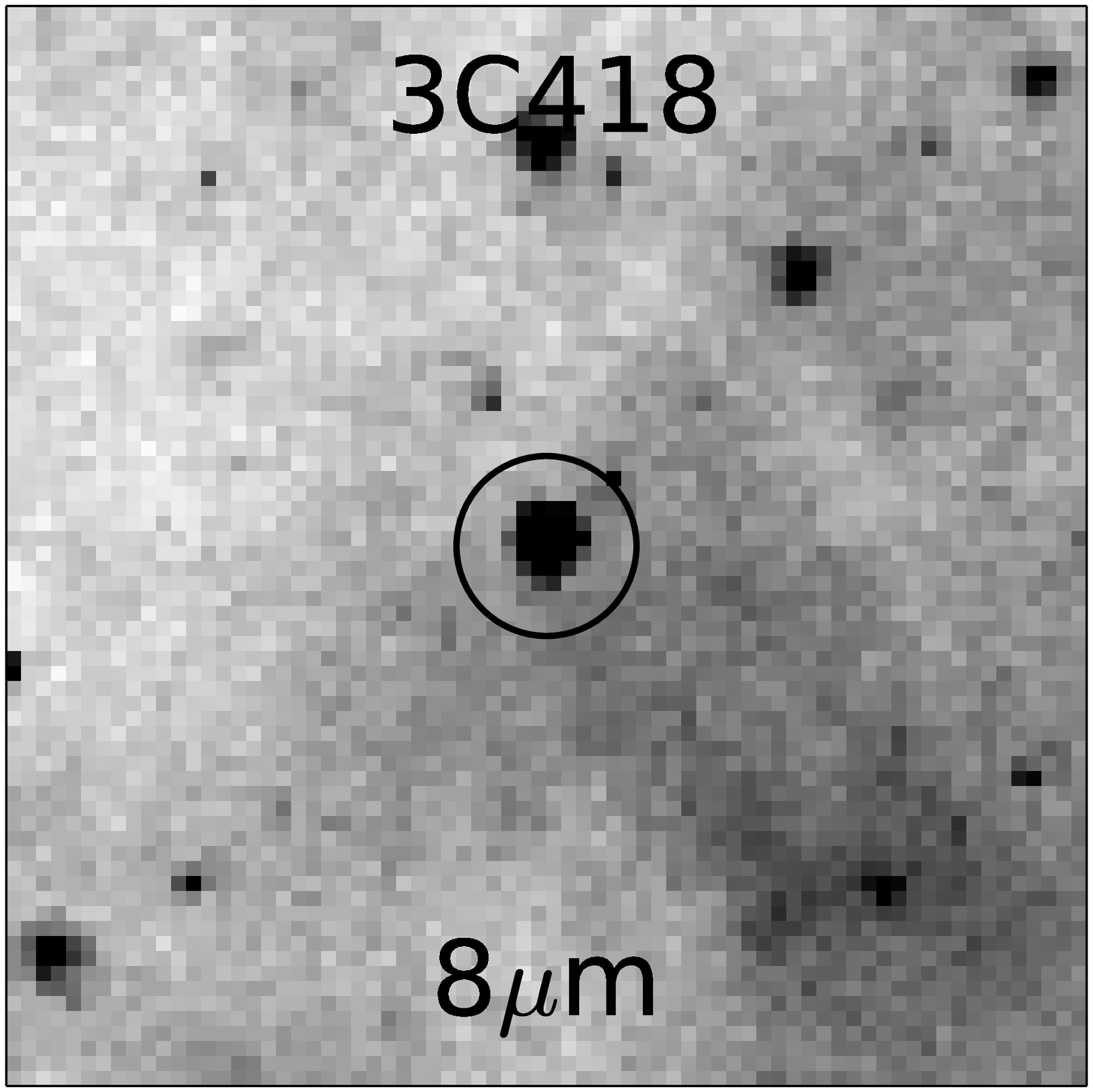}
      \includegraphics[width=1.5cm]{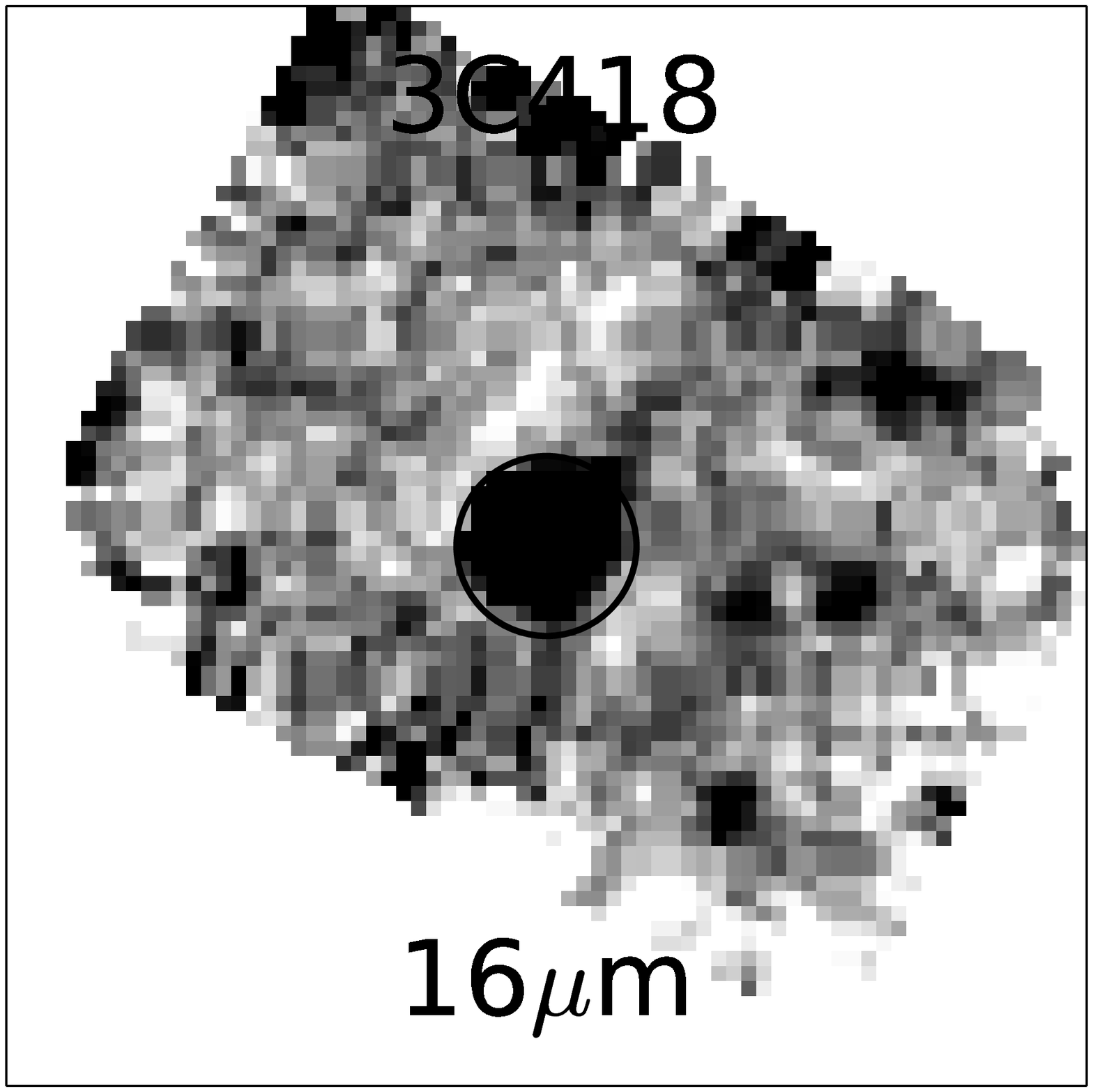}
      \includegraphics[width=1.5cm]{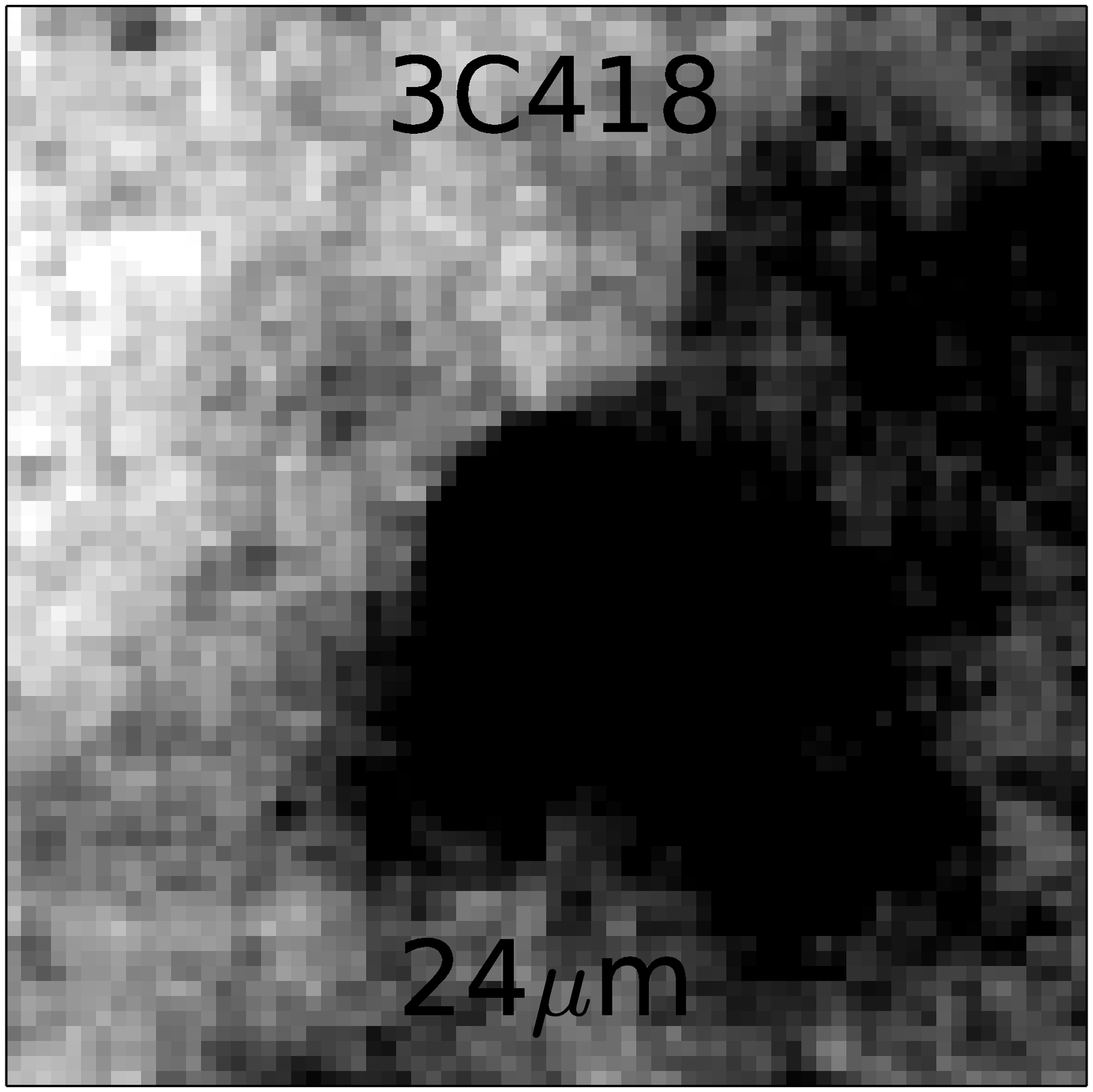}
      \includegraphics[width=1.5cm]{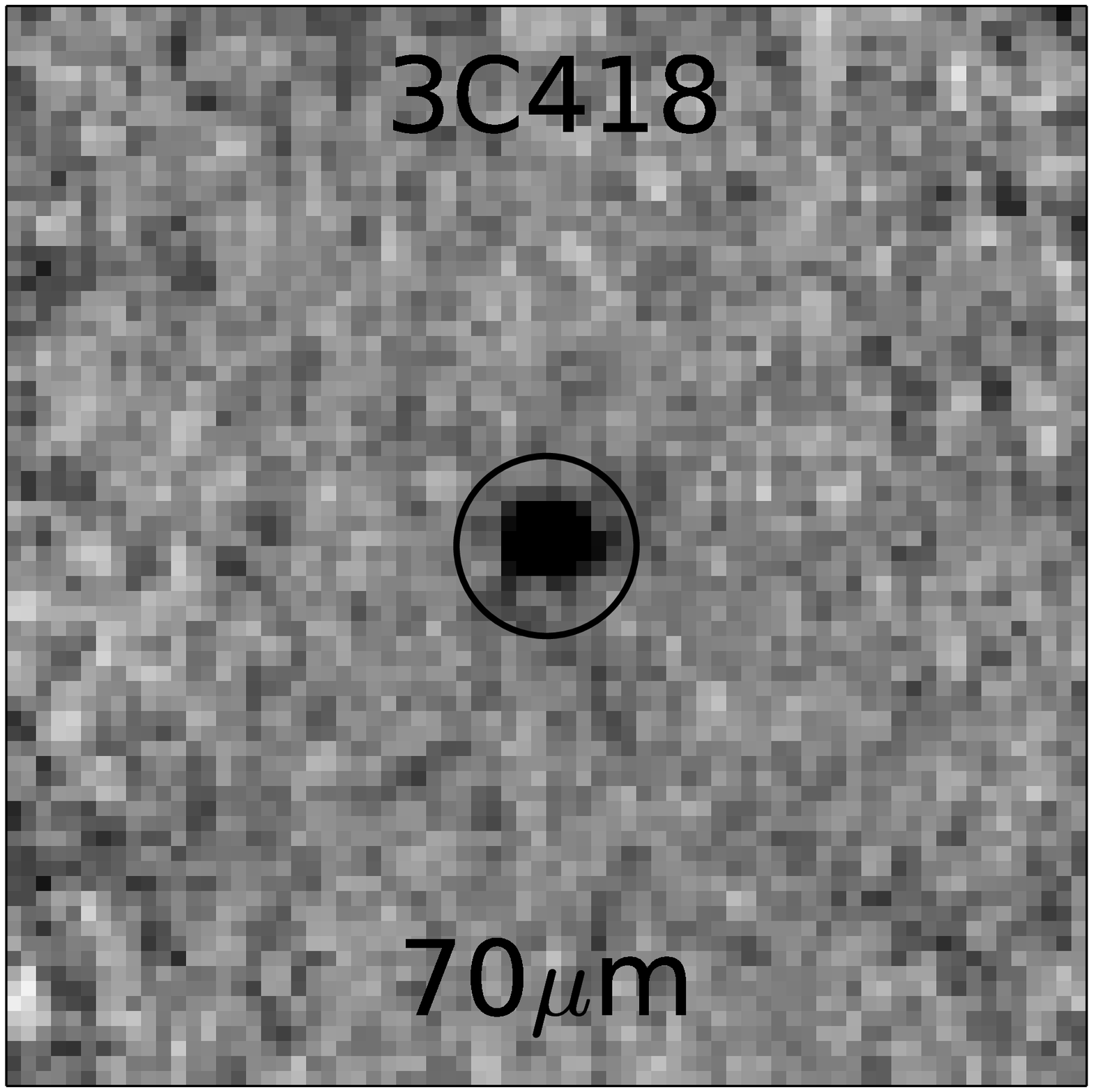}
      \includegraphics[width=1.5cm]{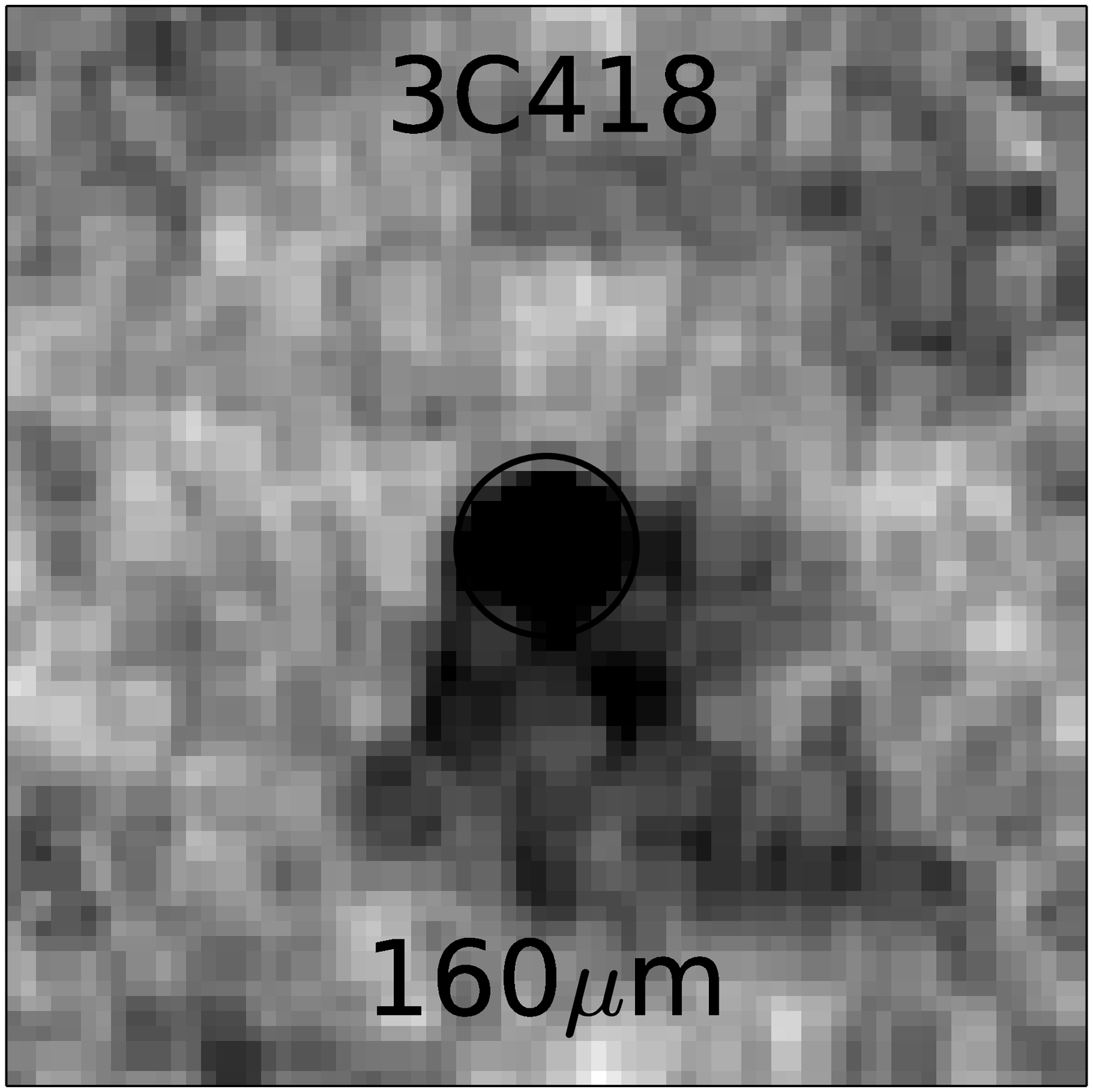}
      \includegraphics[width=1.5cm]{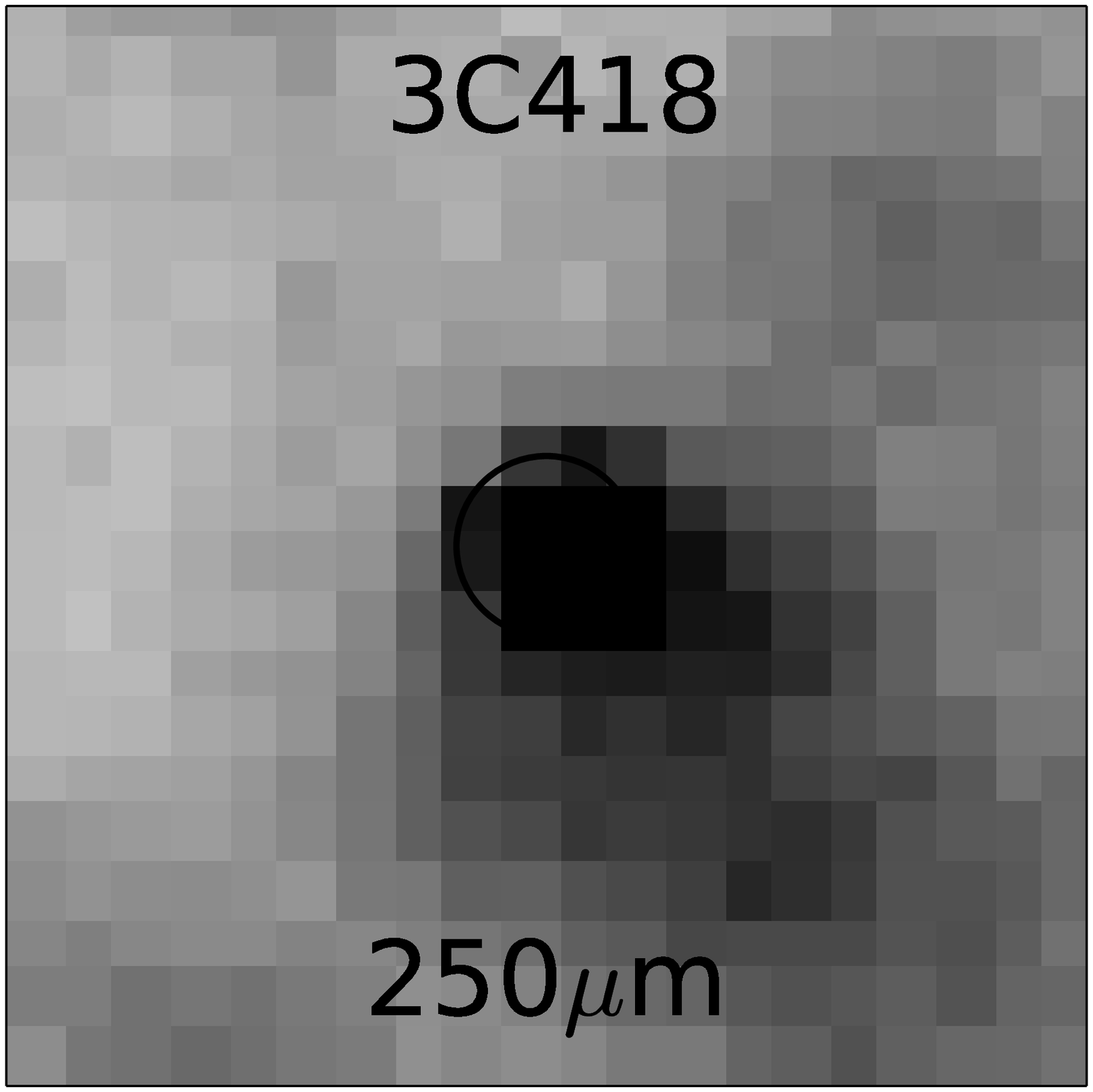}
      \includegraphics[width=1.5cm]{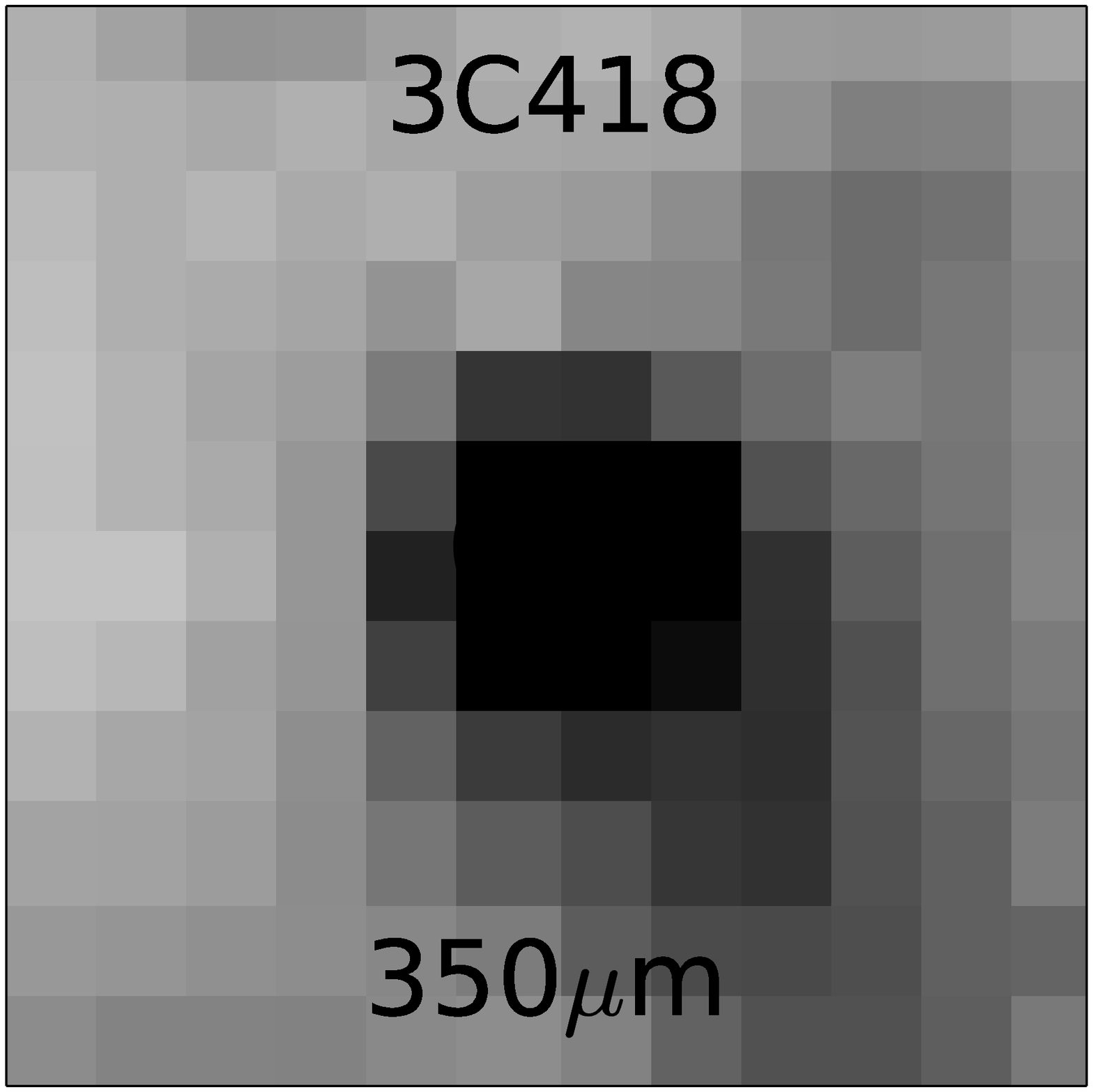}
      \includegraphics[width=1.5cm]{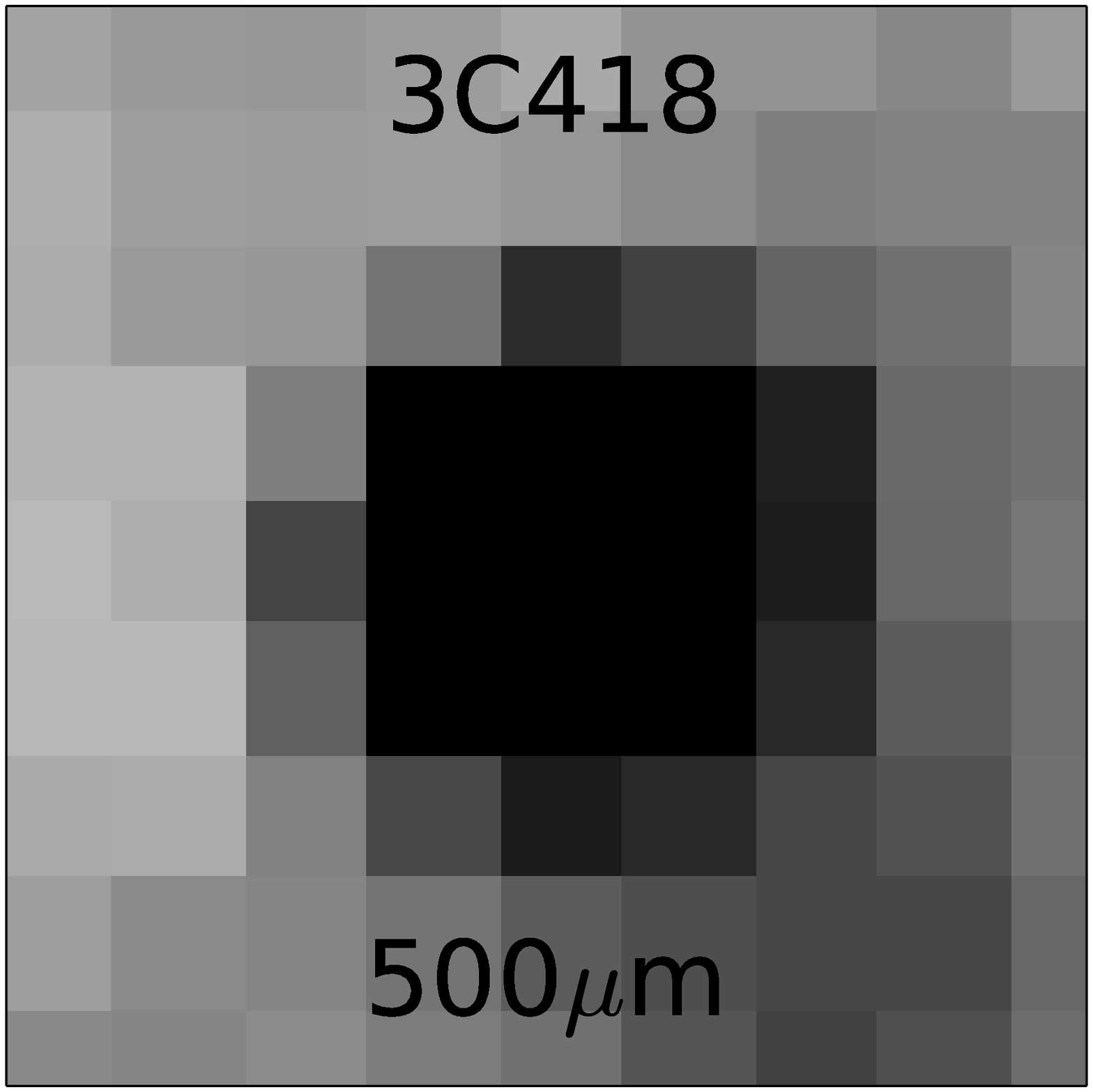}
      \\
      \includegraphics[width=1.5cm]{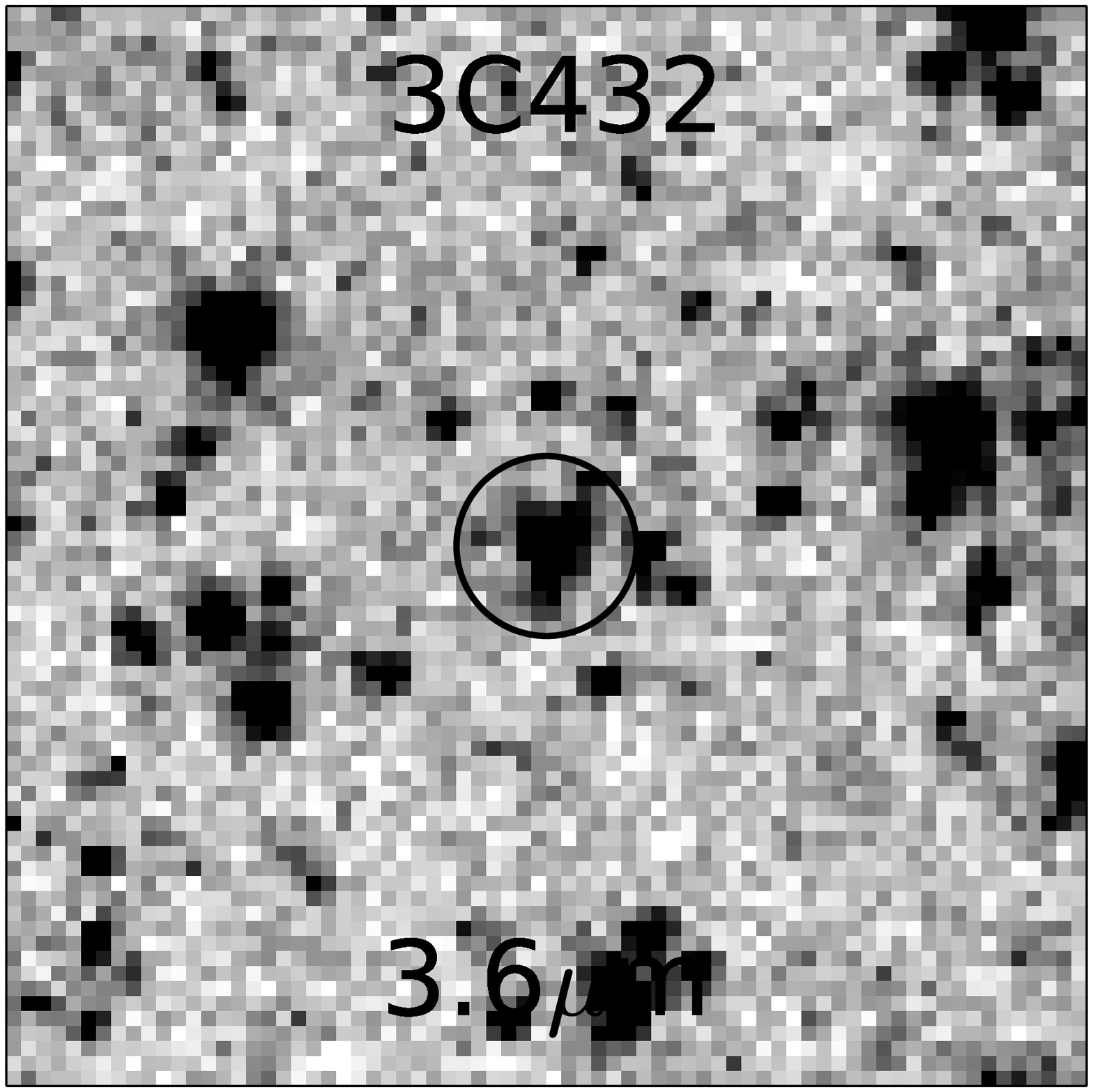}
      \includegraphics[width=1.5cm]{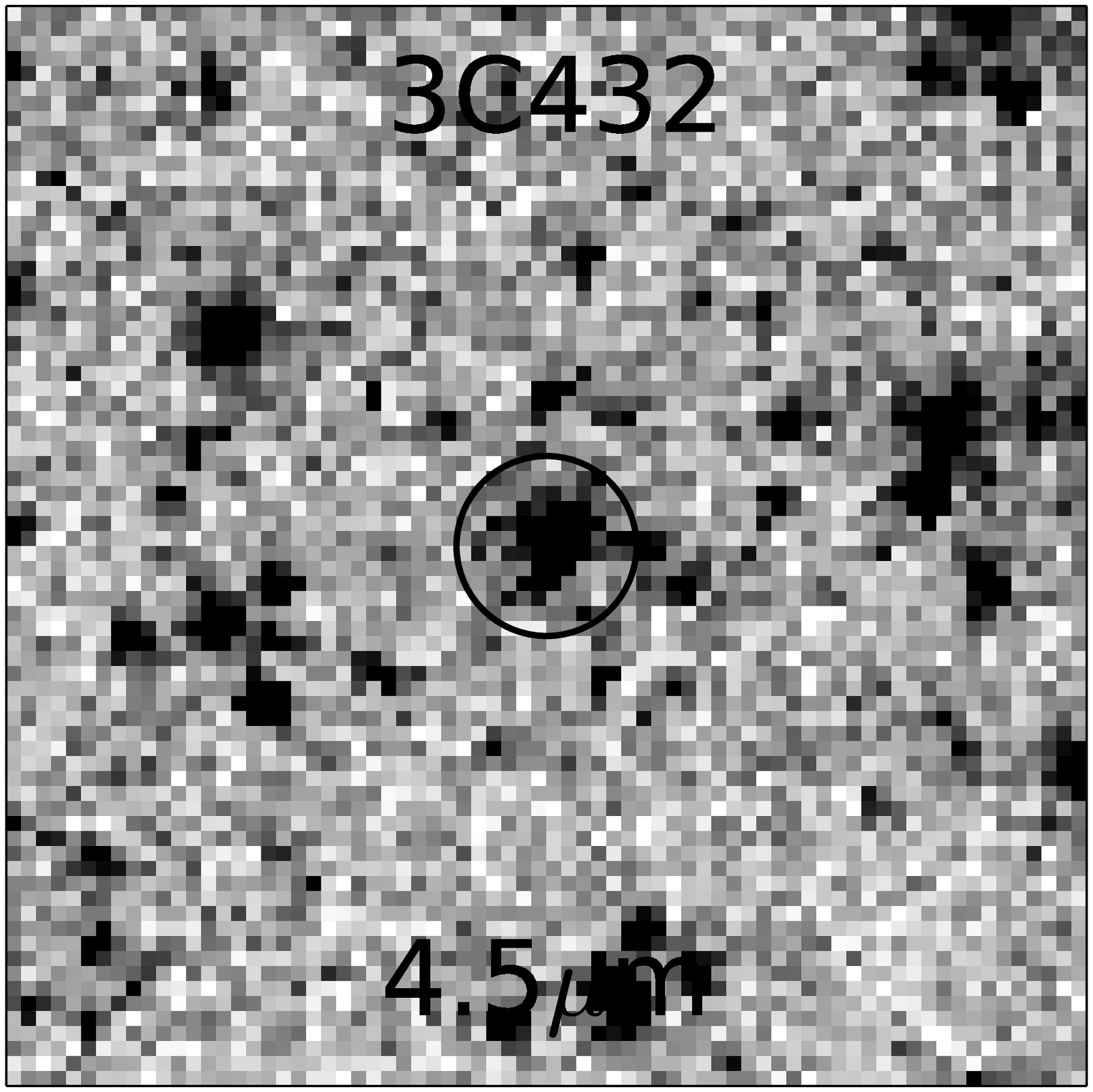}
      \includegraphics[width=1.5cm]{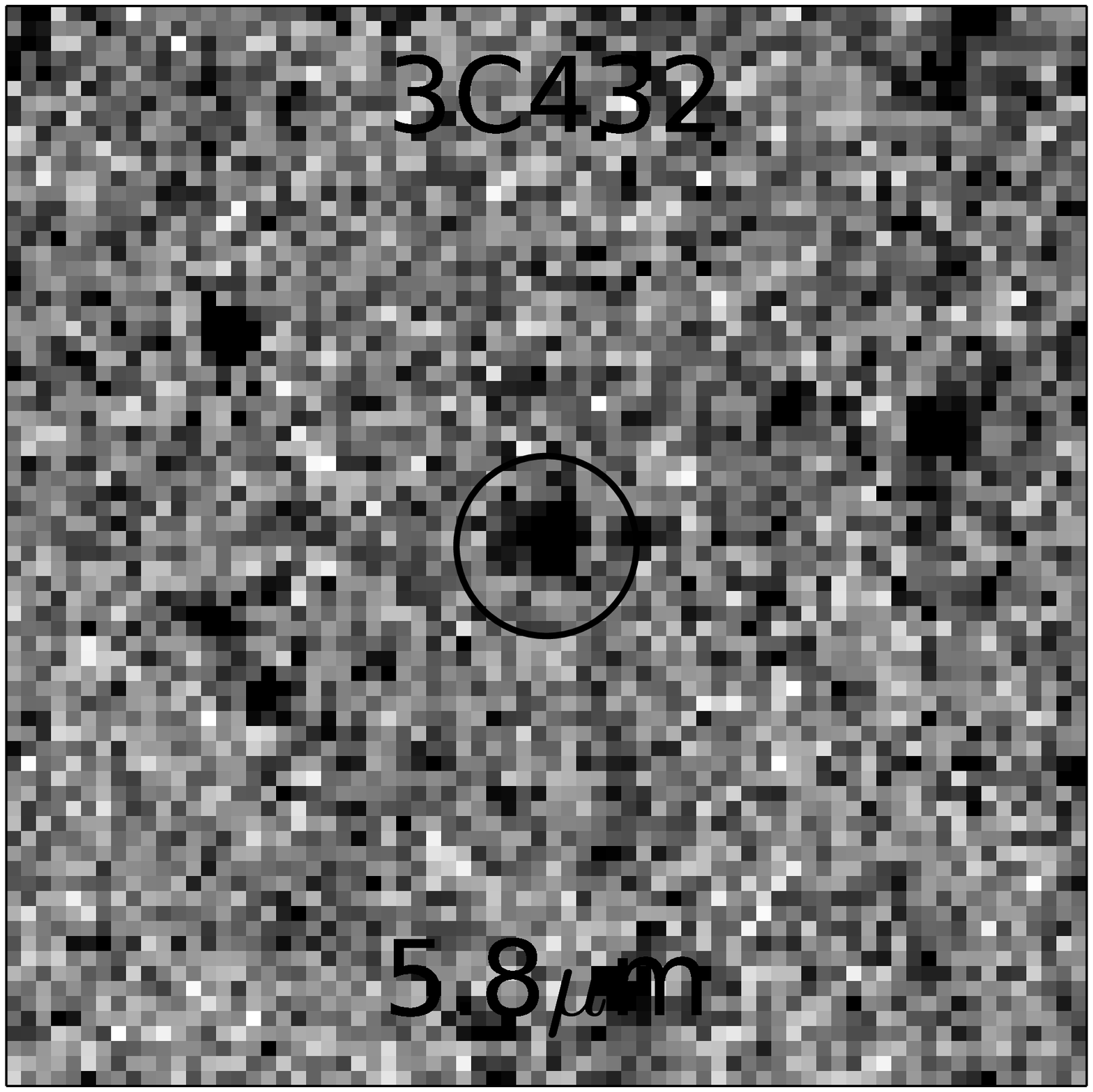}
      \includegraphics[width=1.5cm]{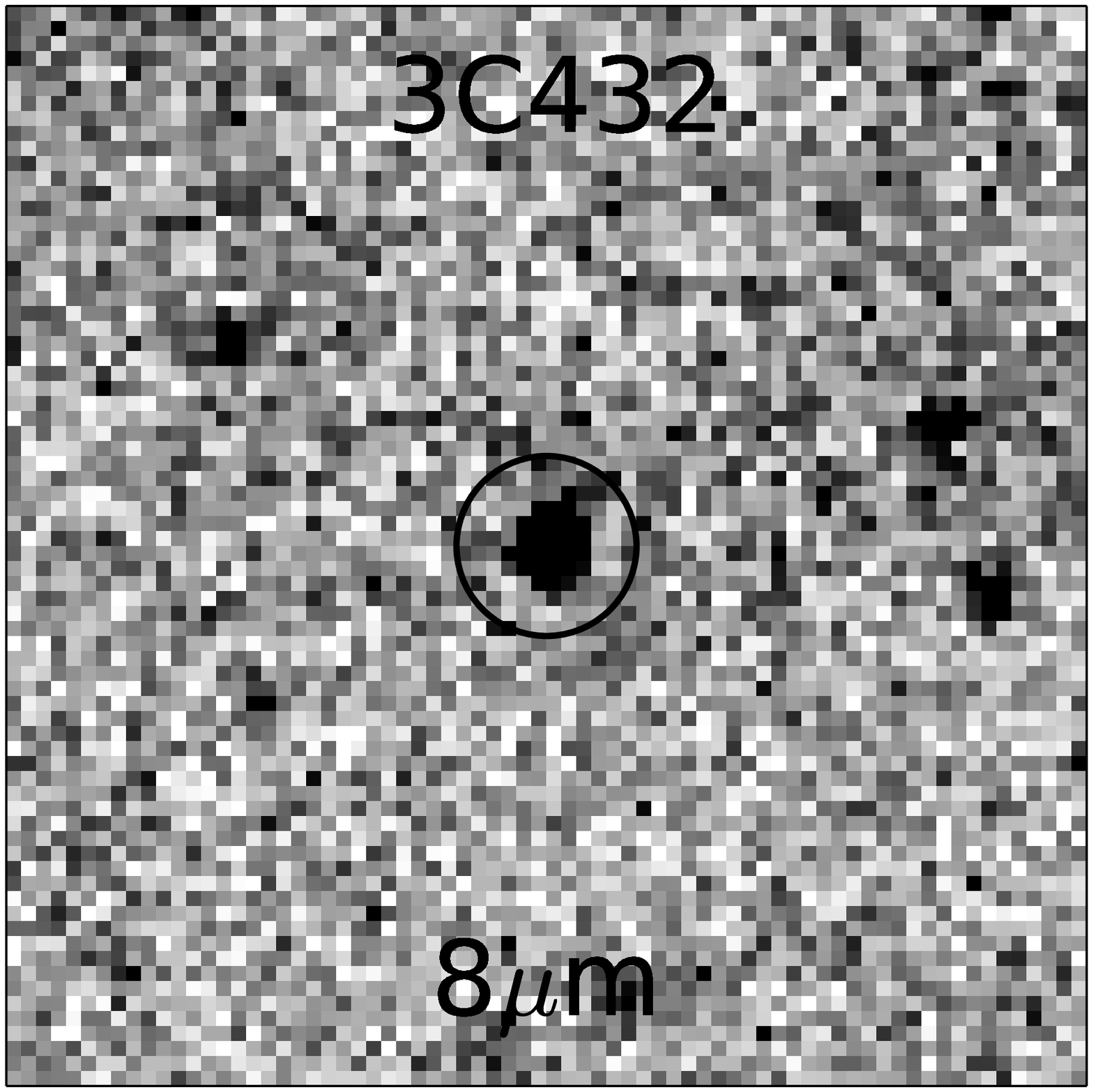}
      \includegraphics[width=1.5cm]{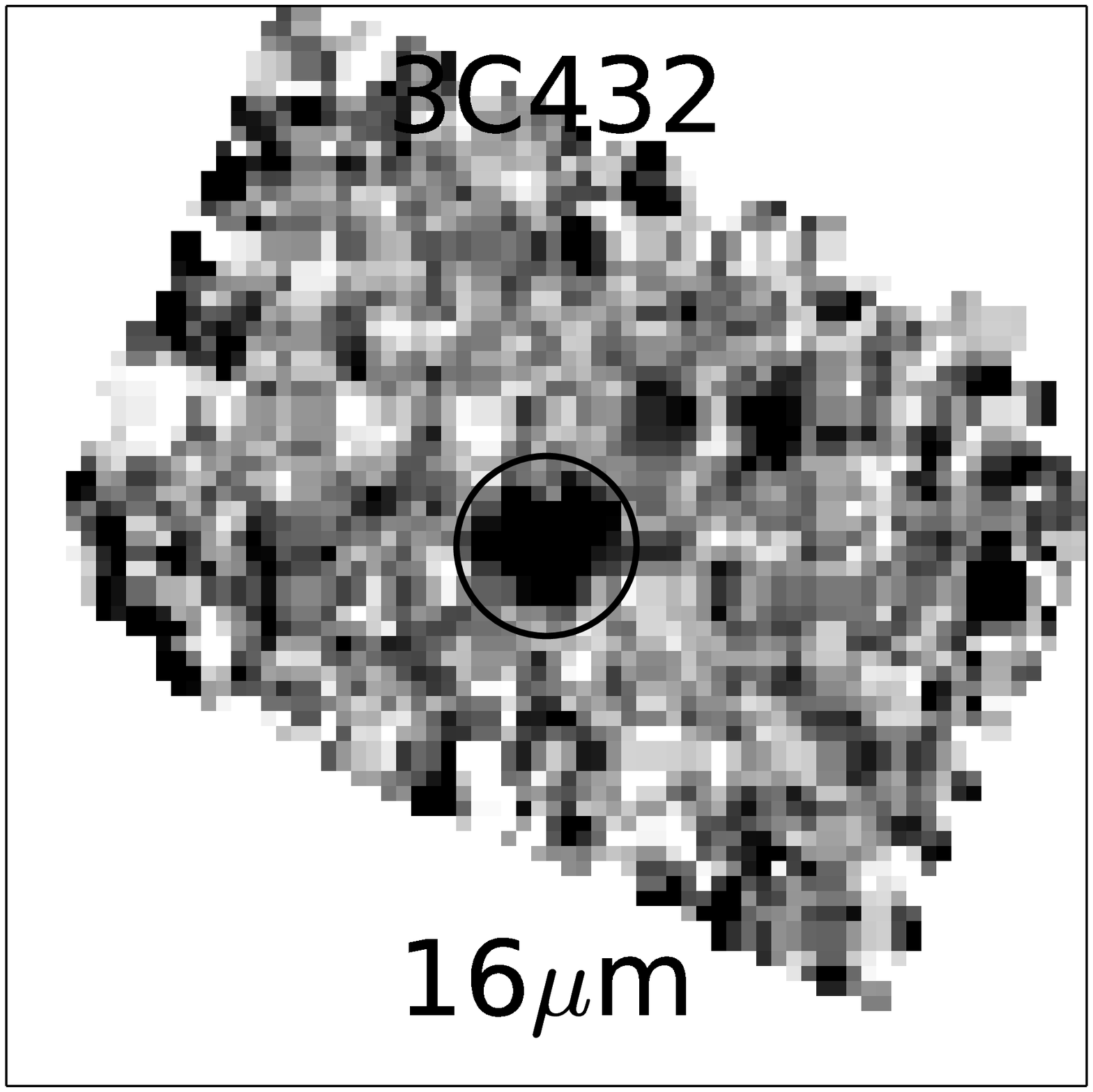}
      \includegraphics[width=1.5cm]{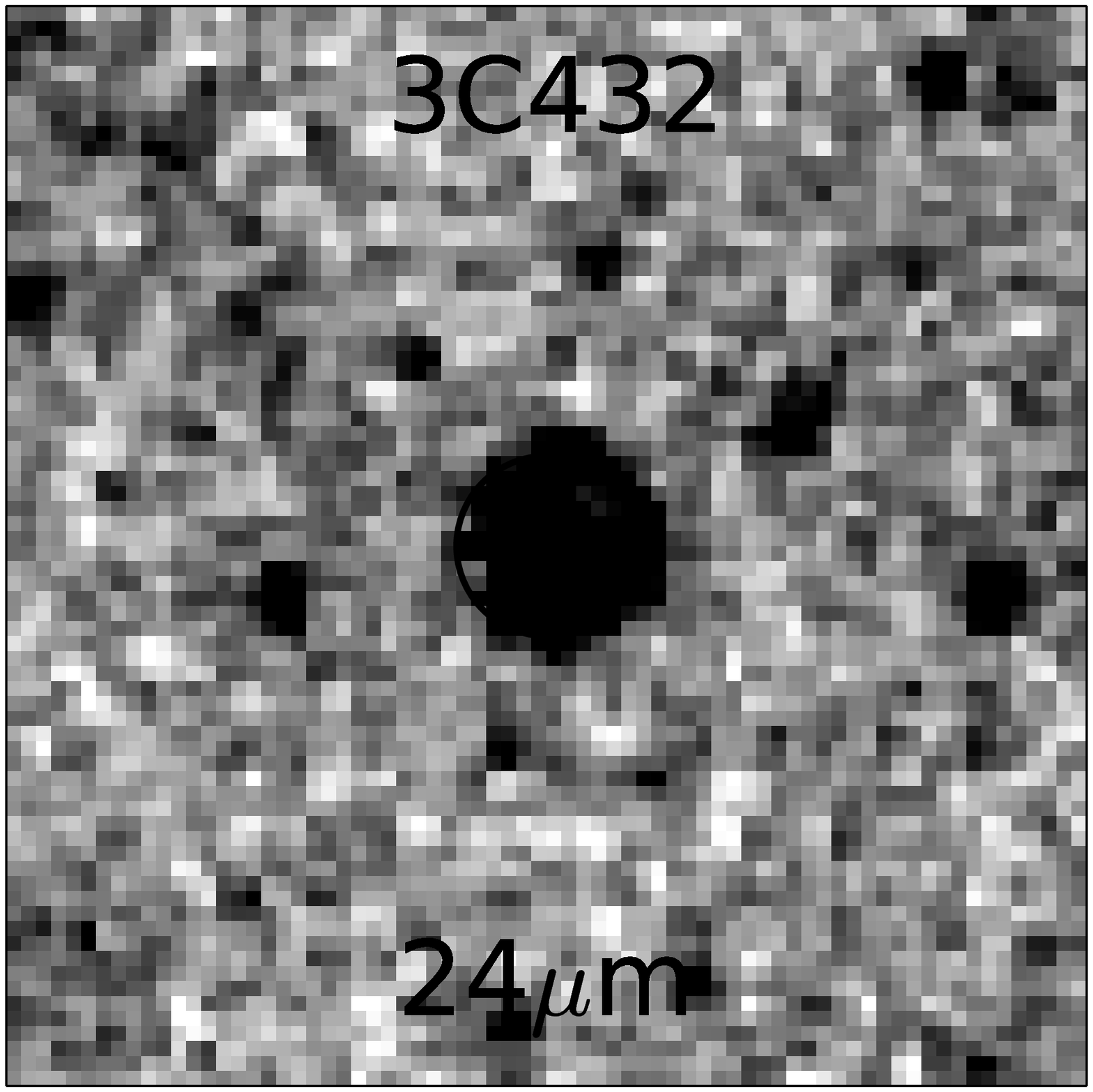}
      \includegraphics[width=1.5cm]{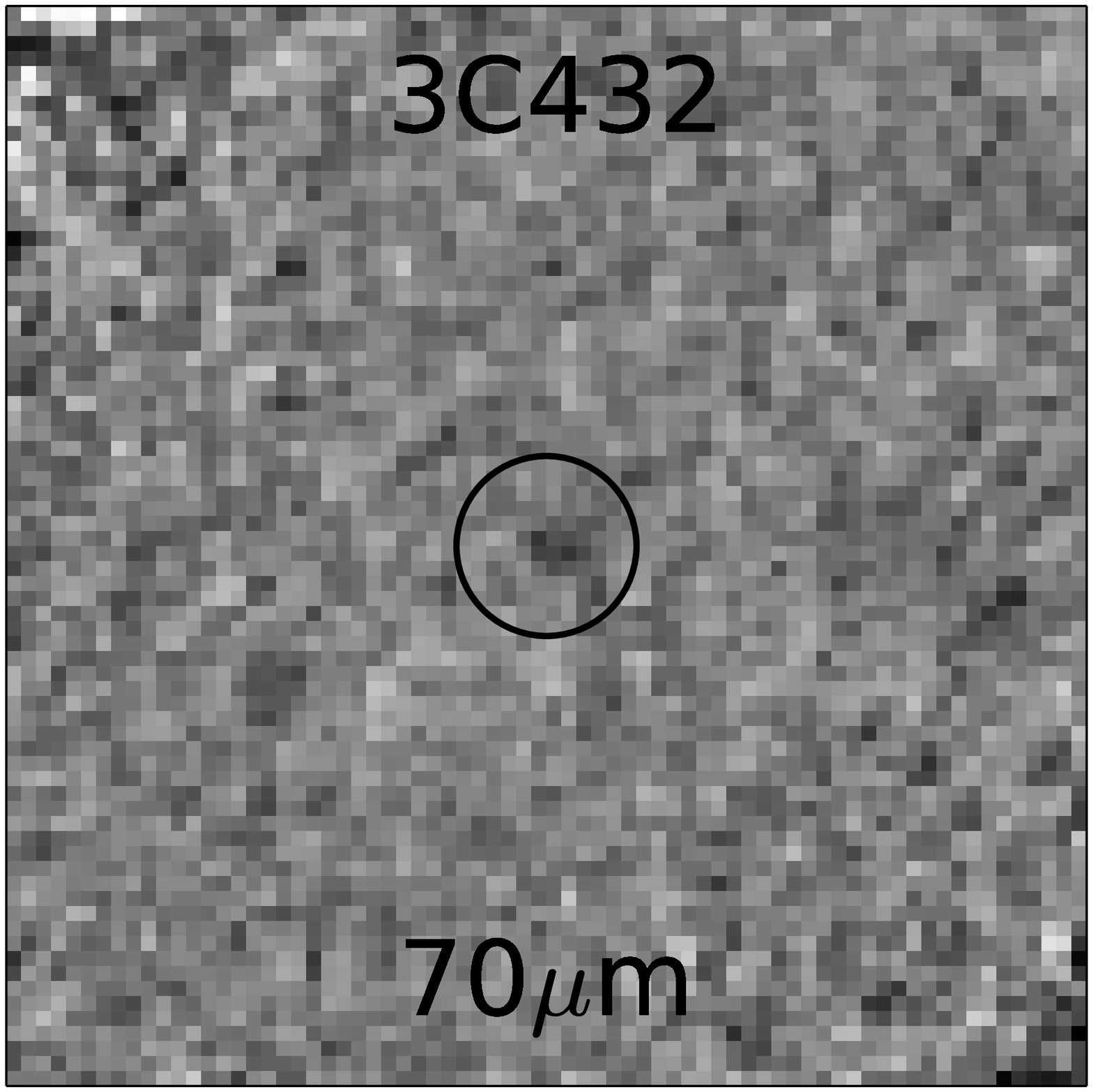}
      \includegraphics[width=1.5cm]{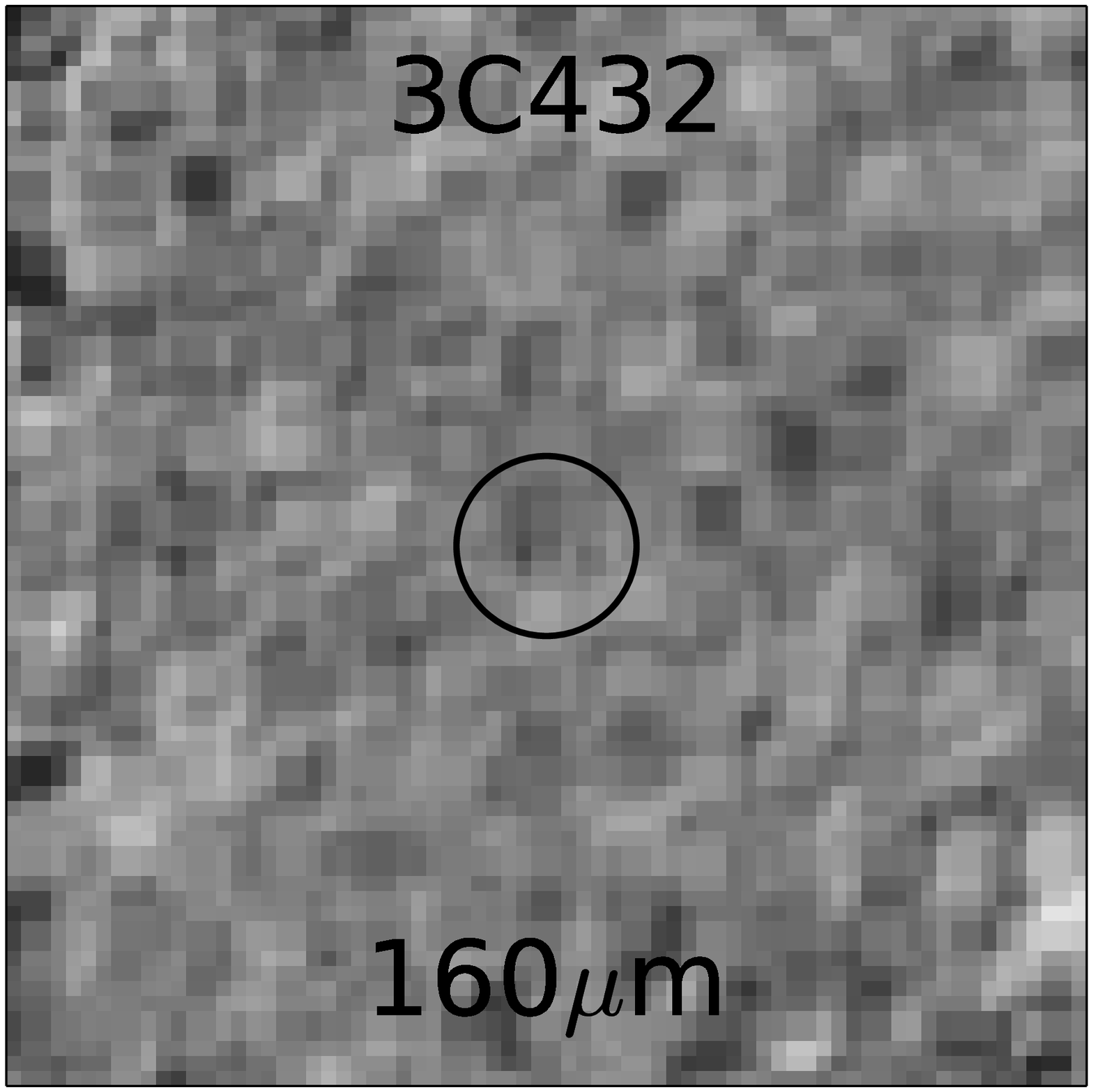}
      \includegraphics[width=1.5cm]{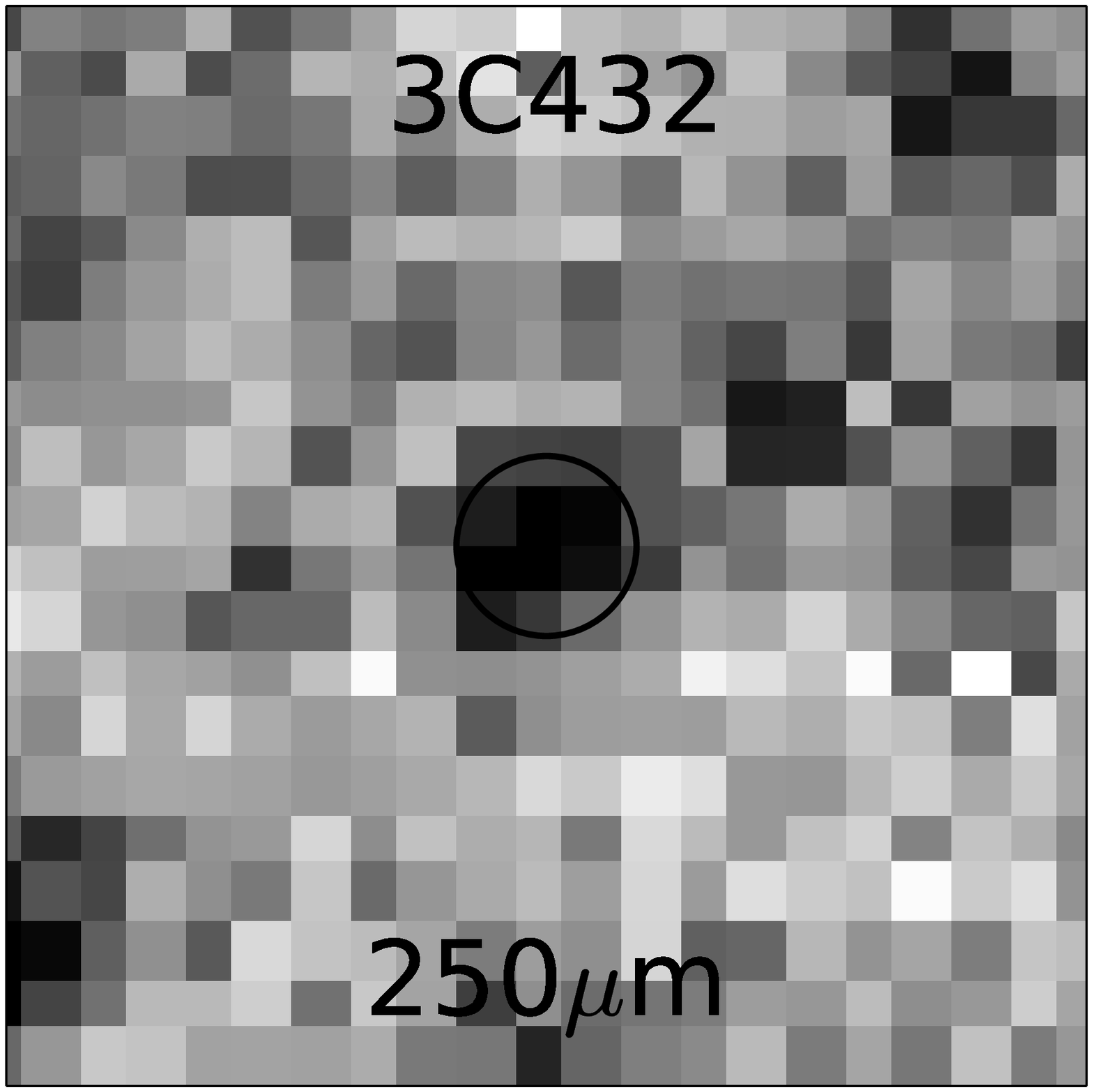}
      \includegraphics[width=1.5cm]{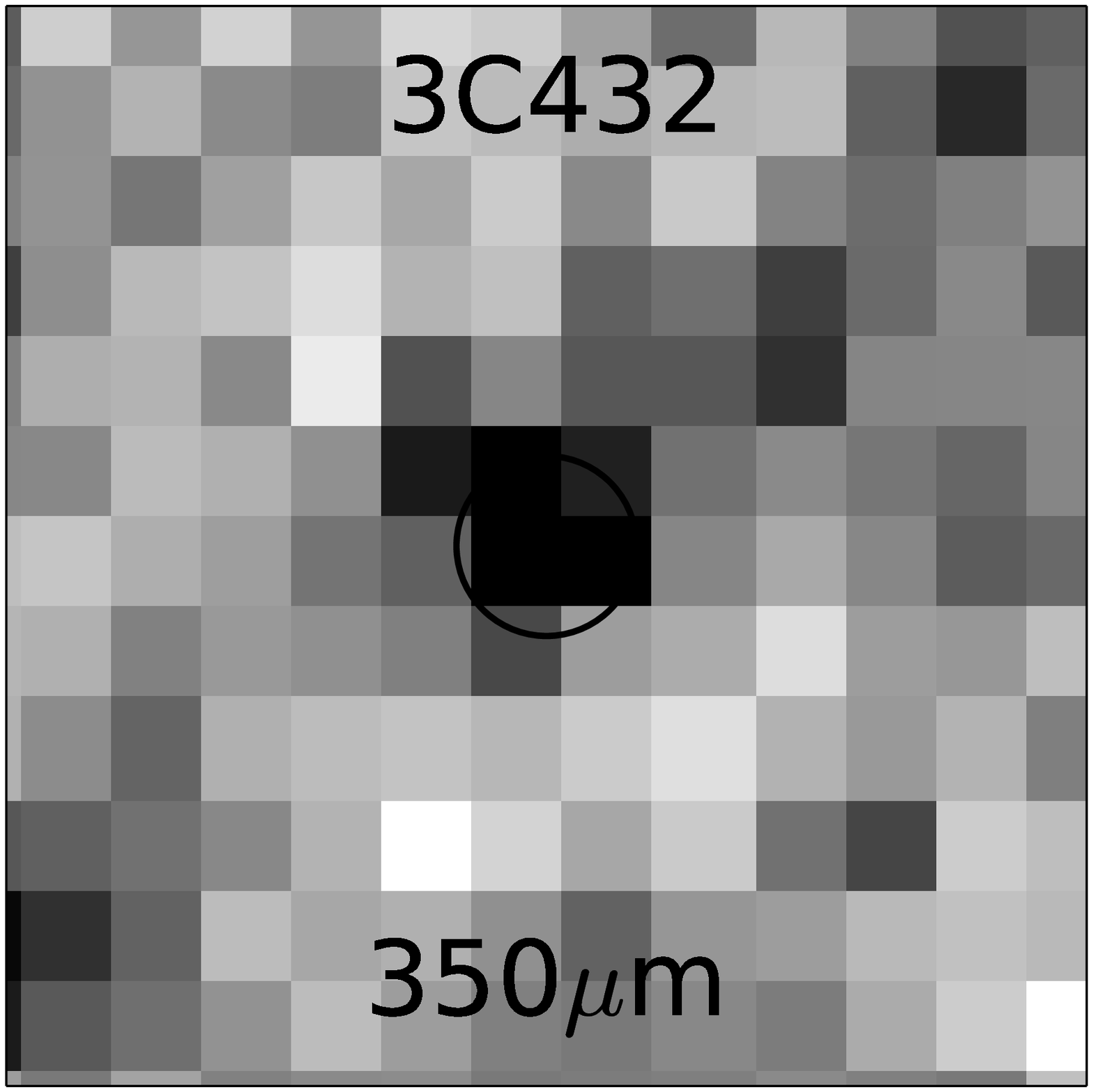}
      \includegraphics[width=1.5cm]{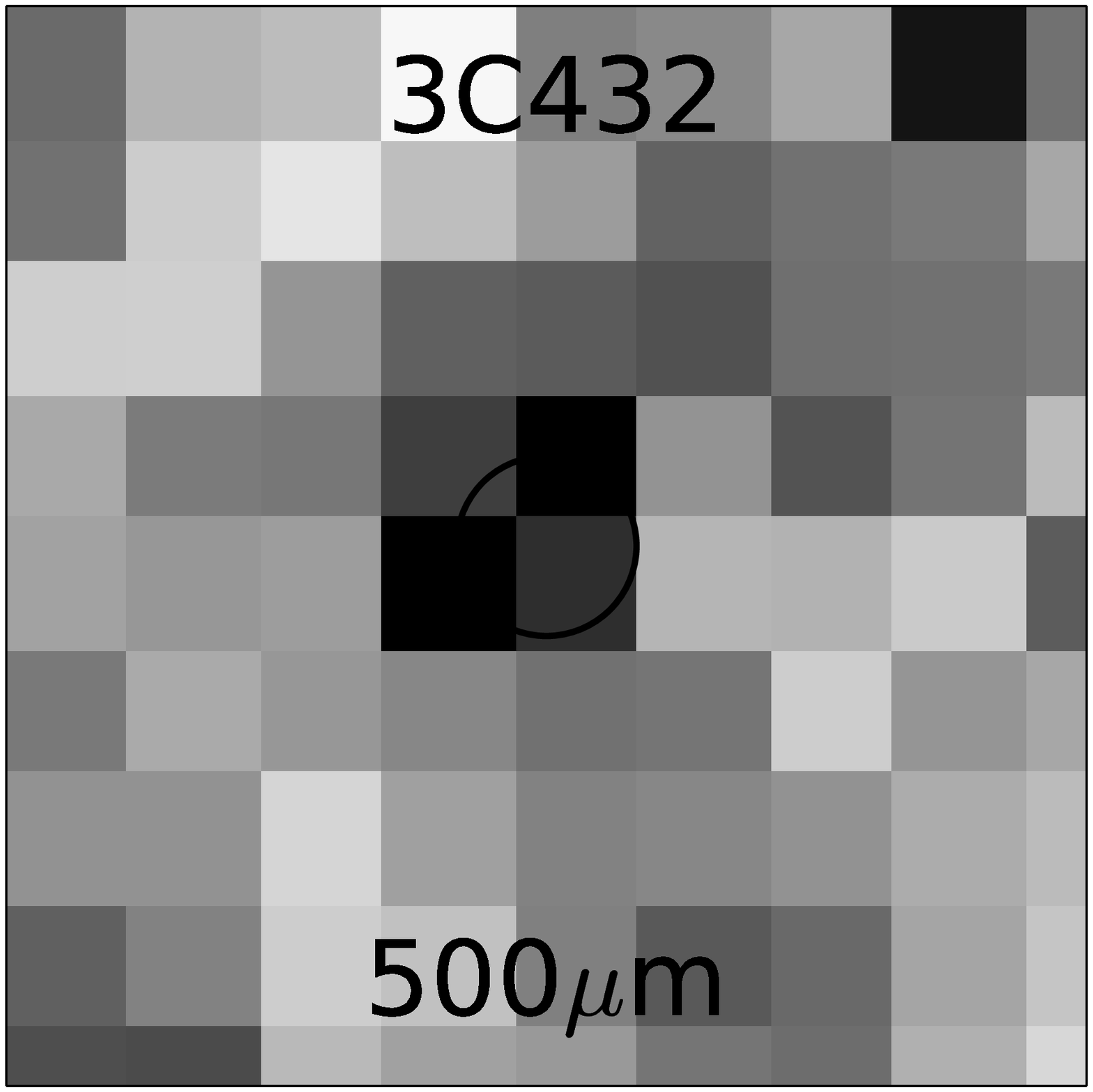}
      \\
      \includegraphics[width=1.5cm]{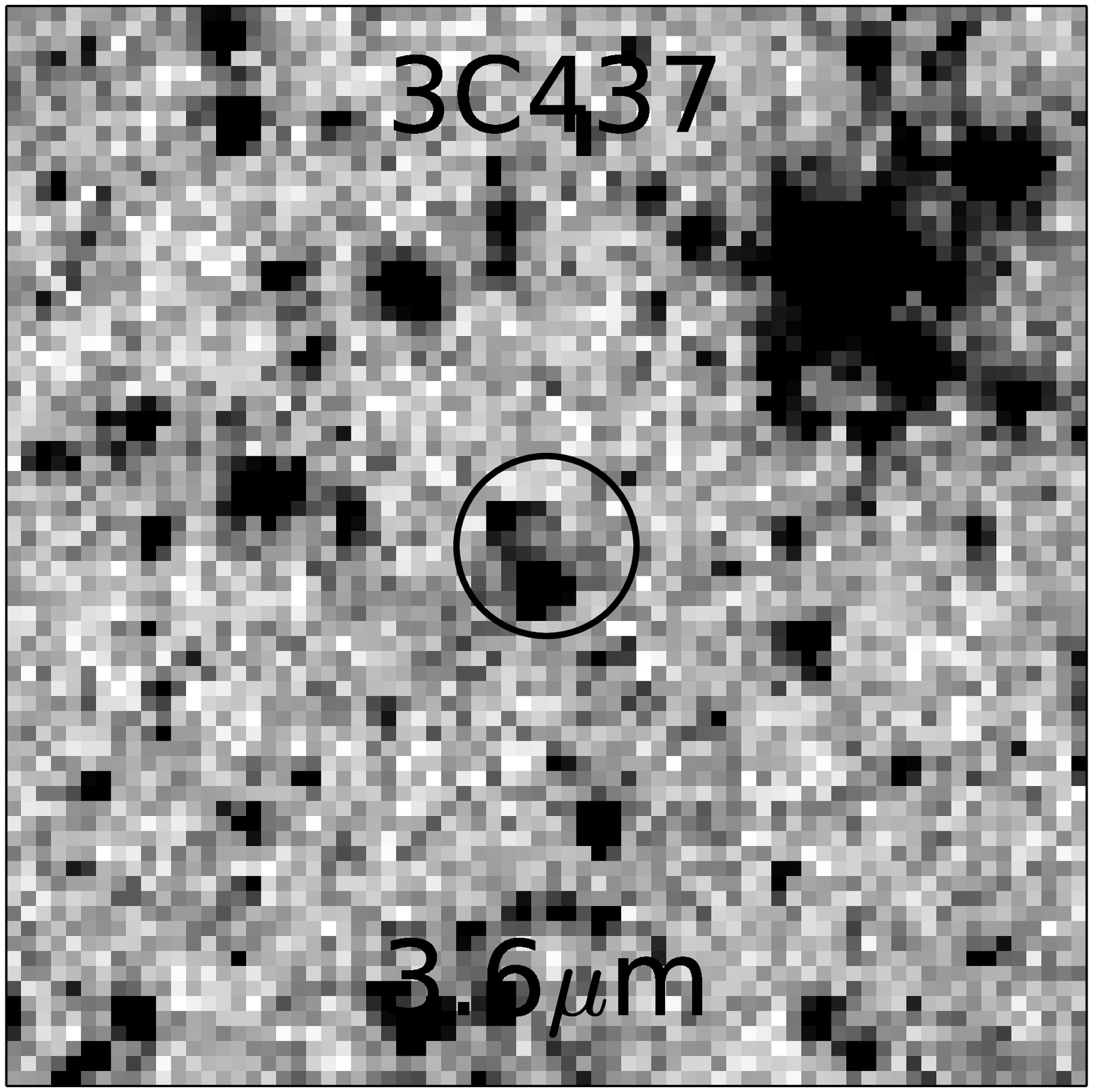}
      \includegraphics[width=1.5cm]{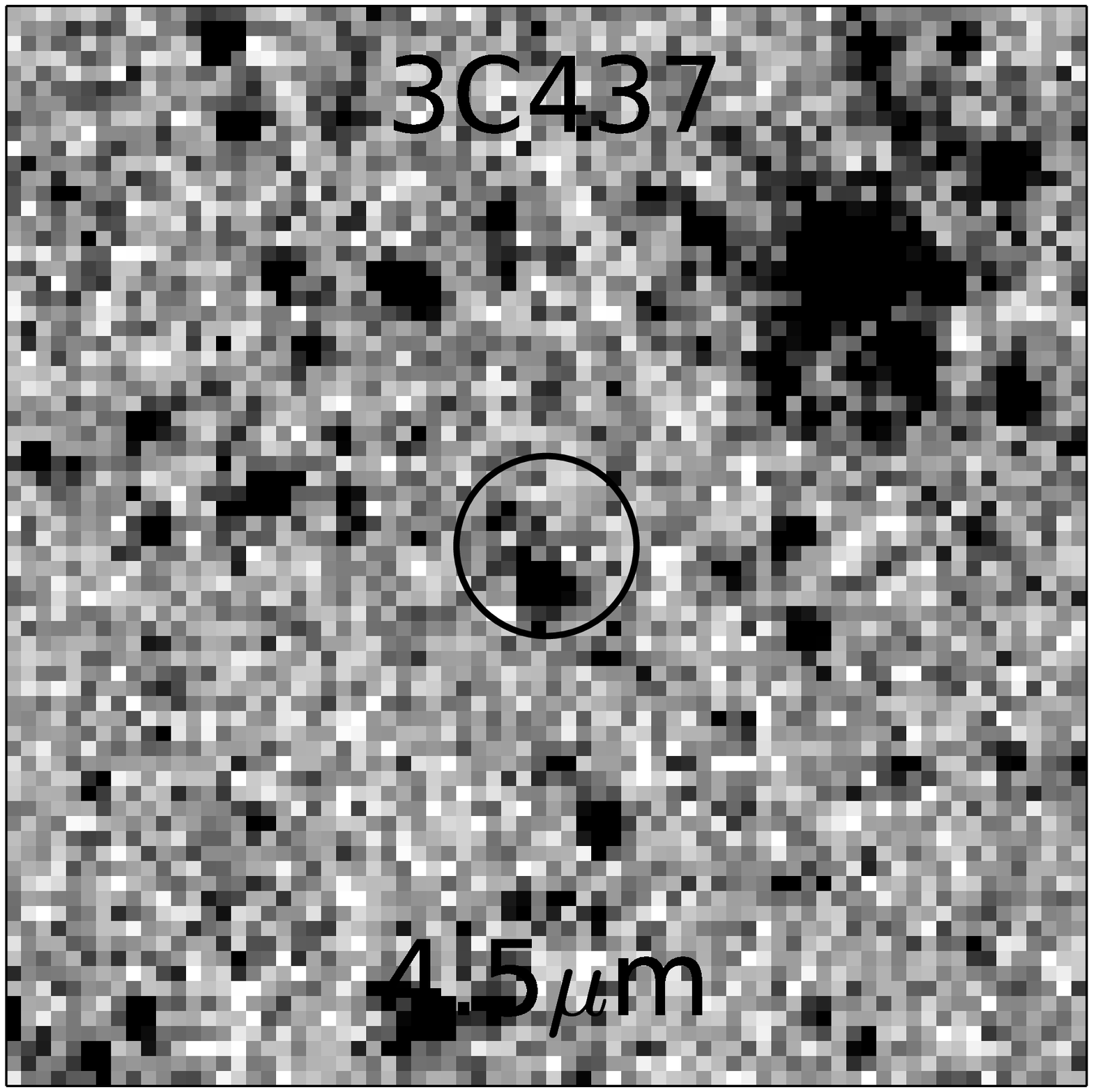}
      \includegraphics[width=1.5cm]{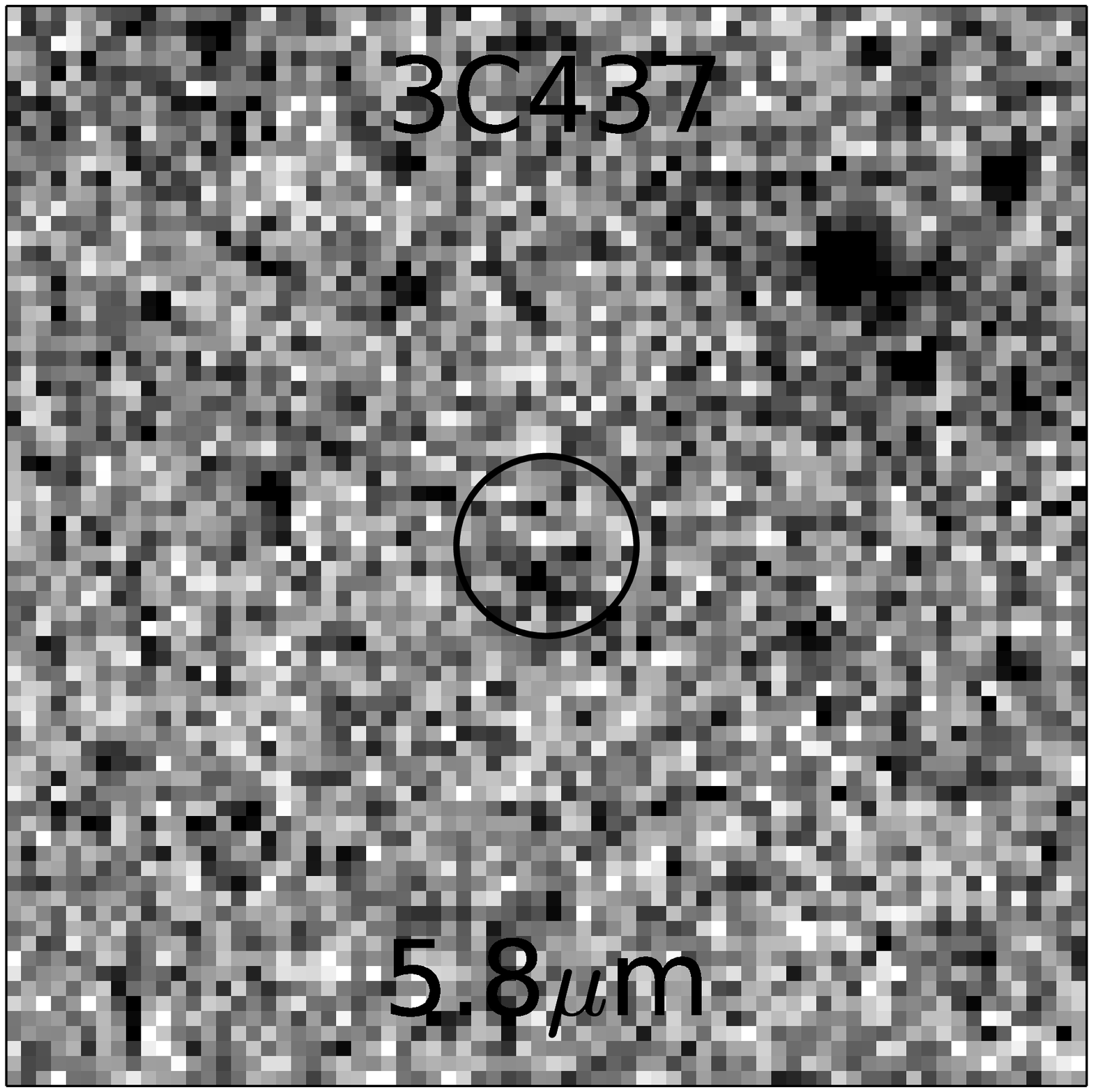}
      \includegraphics[width=1.5cm]{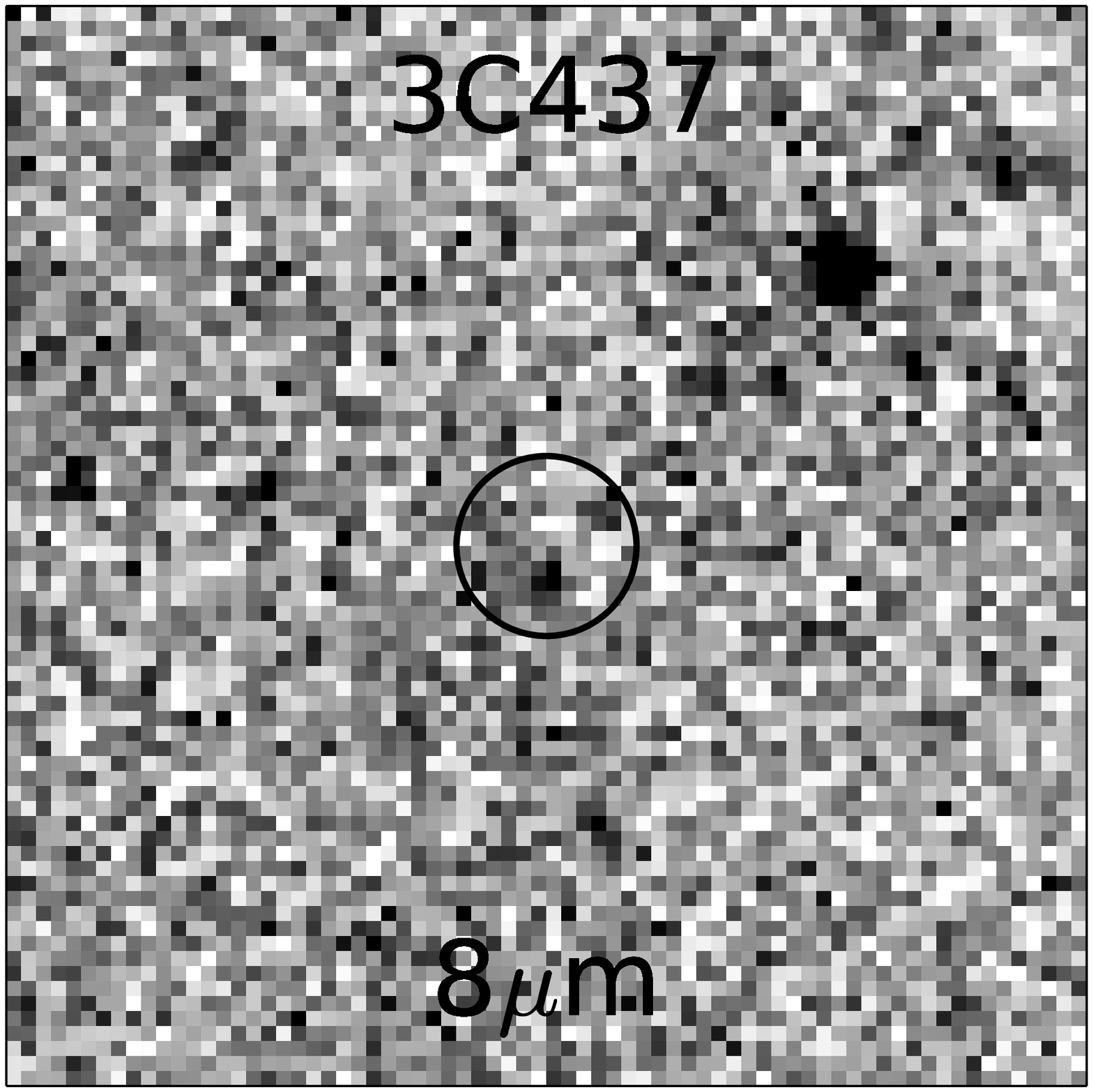}
      \includegraphics[width=1.5cm]{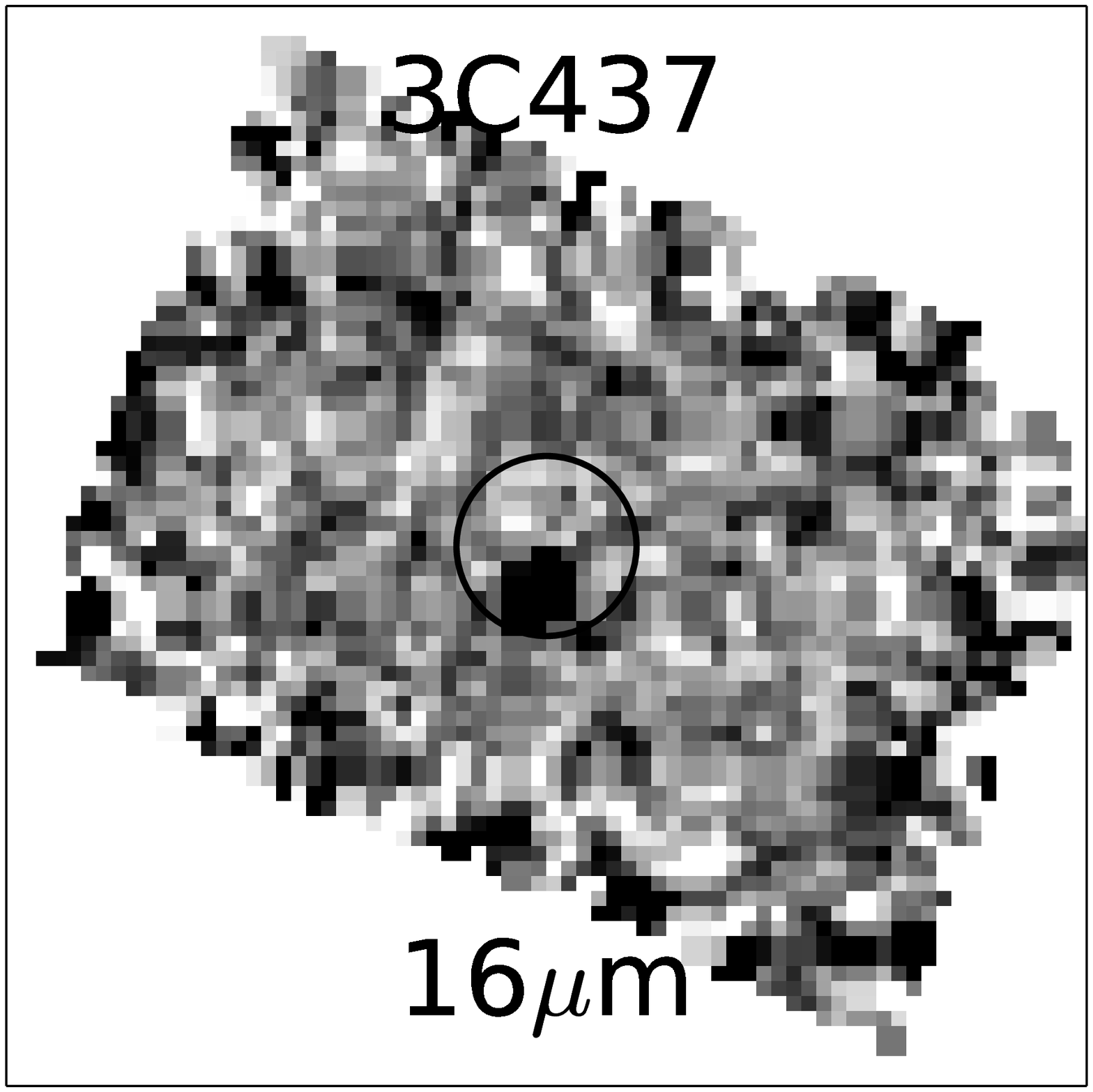}
      \includegraphics[width=1.5cm]{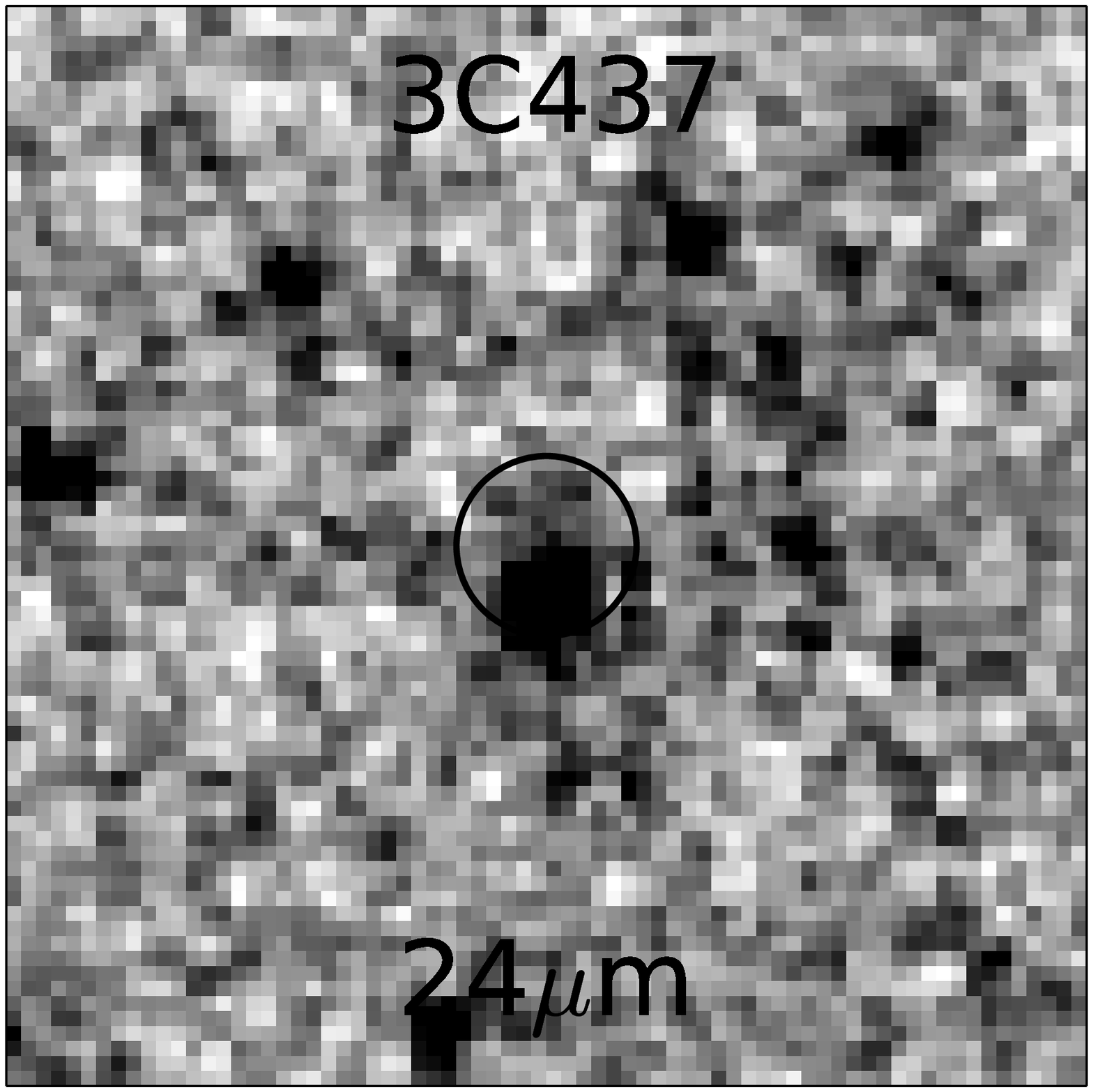}
      \includegraphics[width=1.5cm]{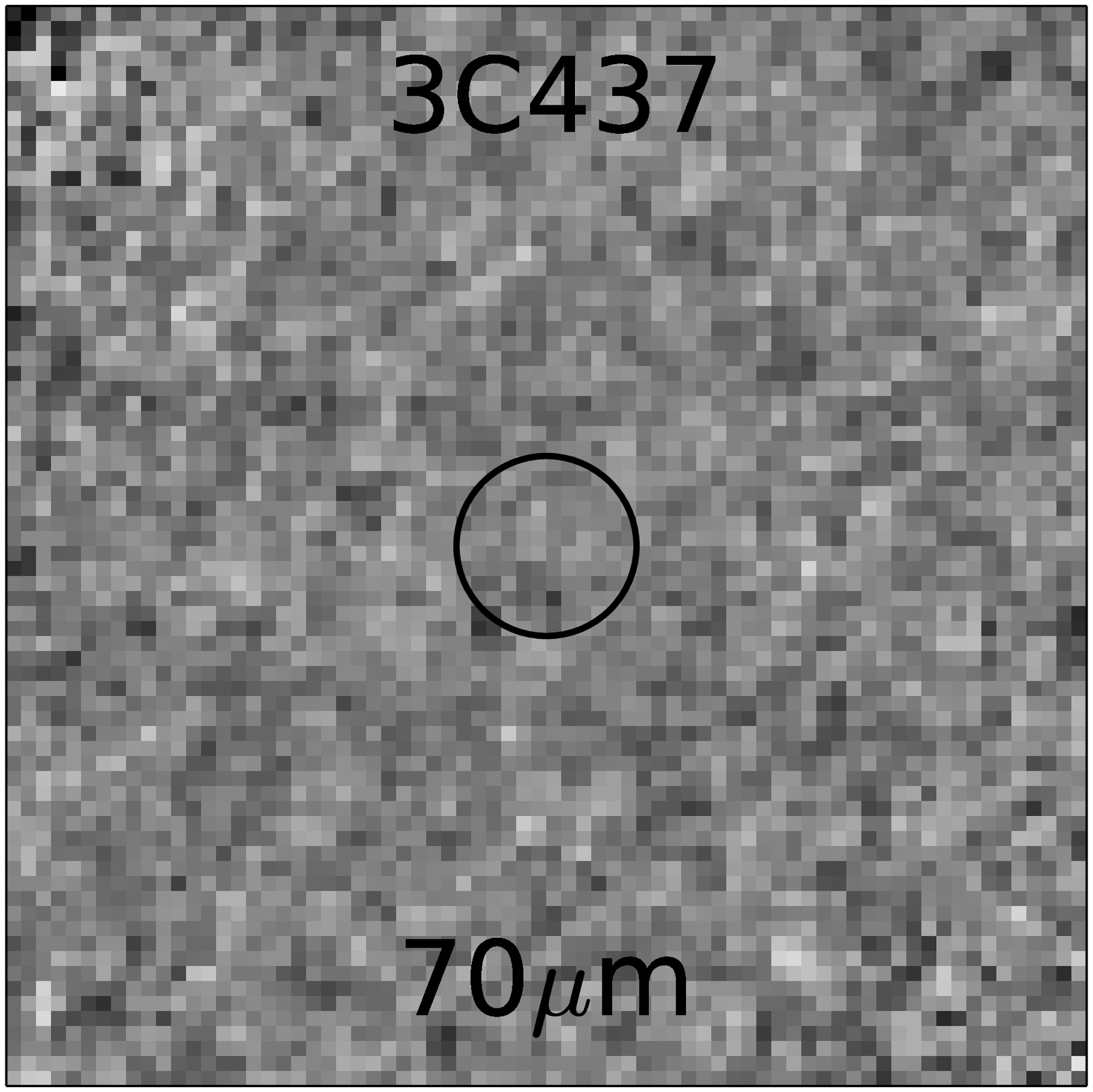}
      \includegraphics[width=1.5cm]{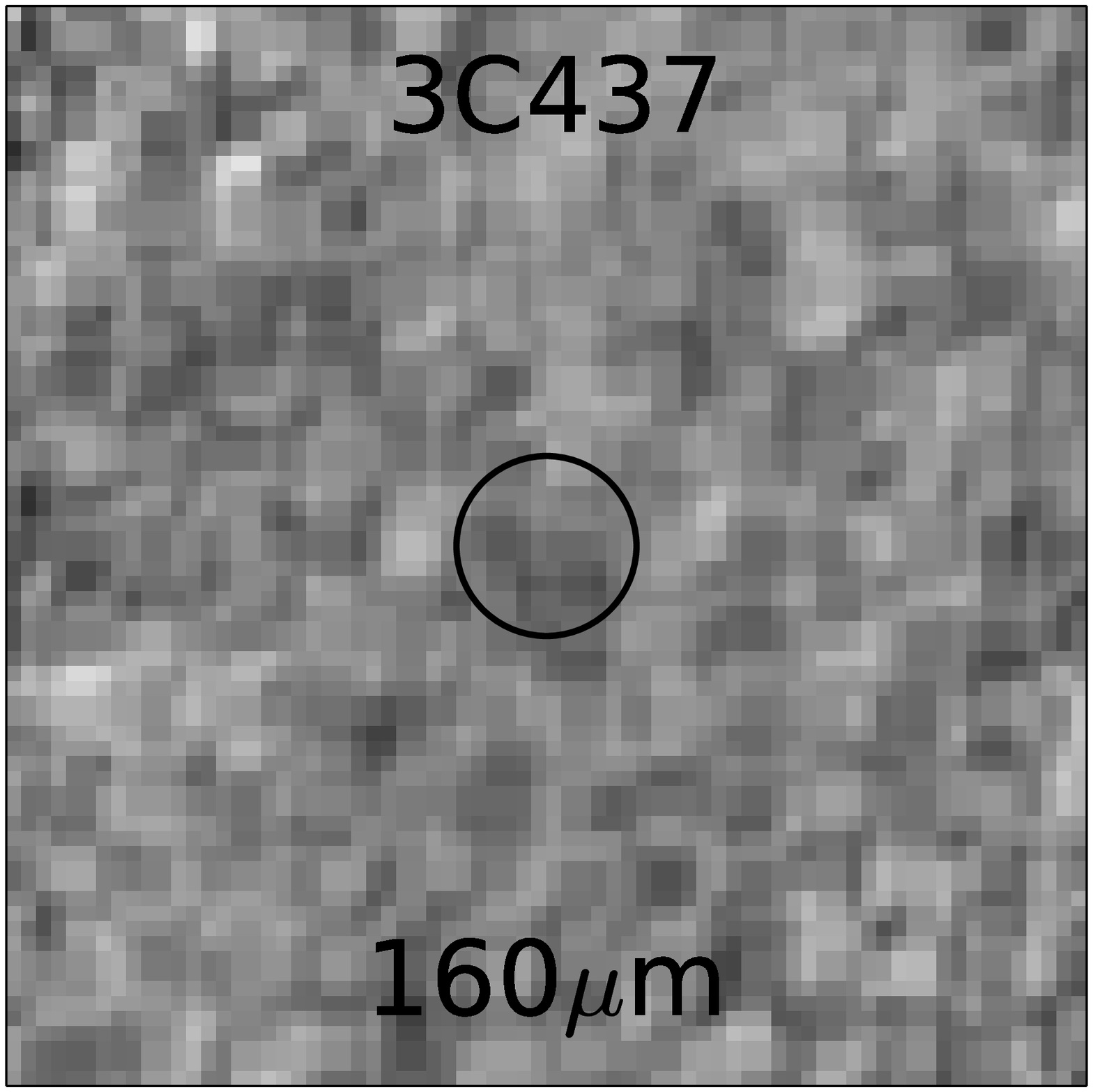}
      \includegraphics[width=1.5cm]{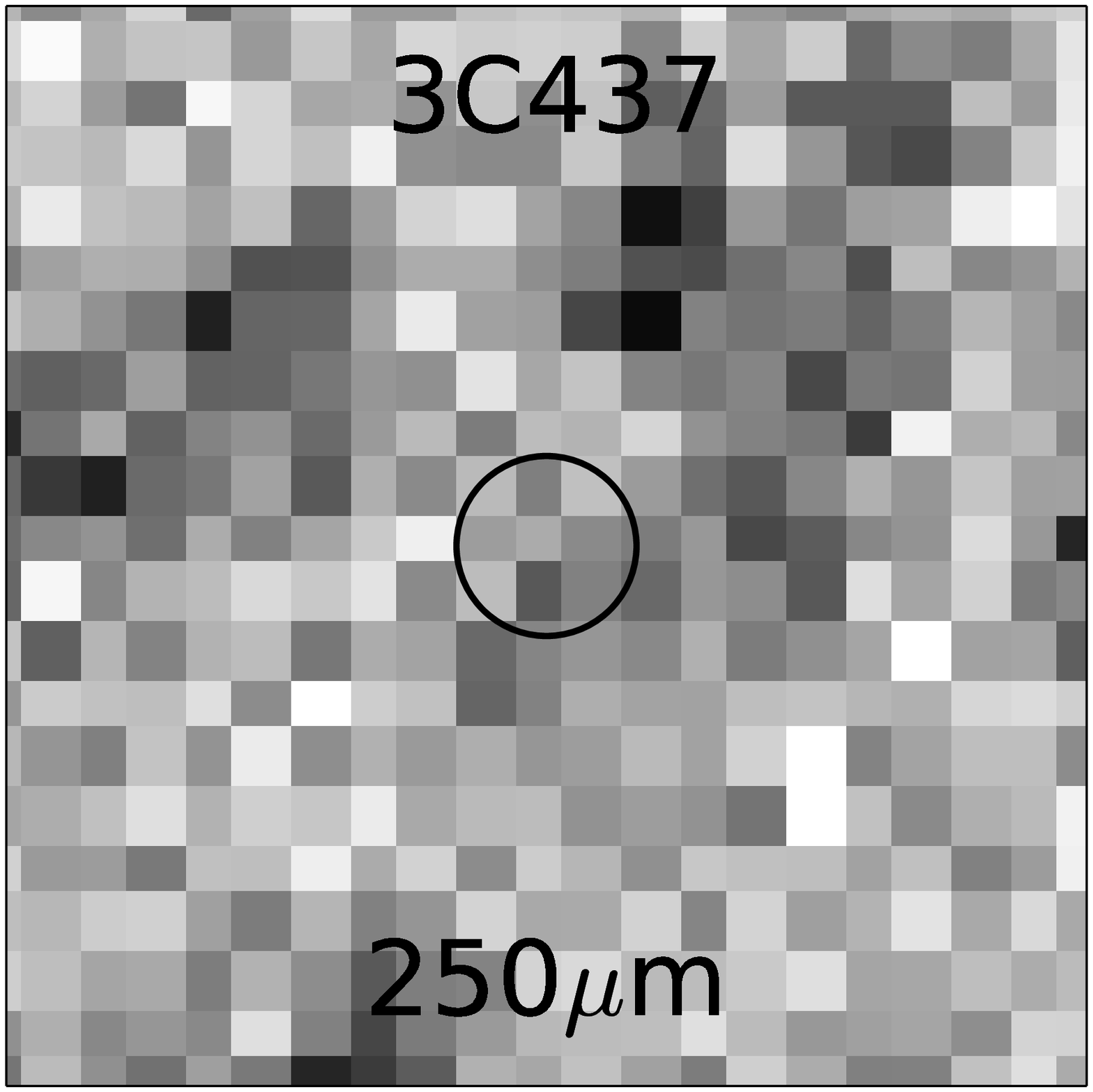}
      \includegraphics[width=1.5cm]{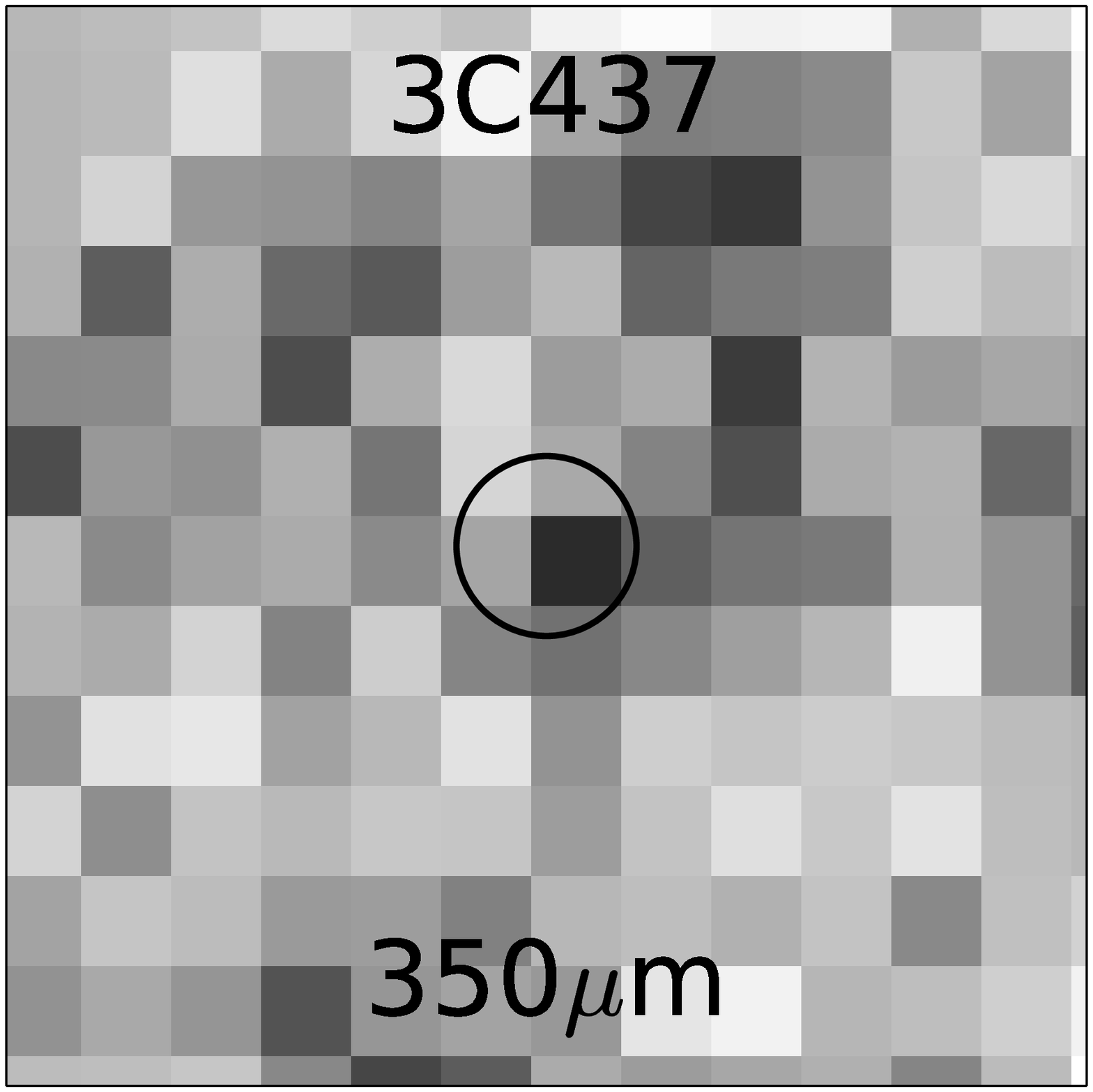}
      \includegraphics[width=1.5cm]{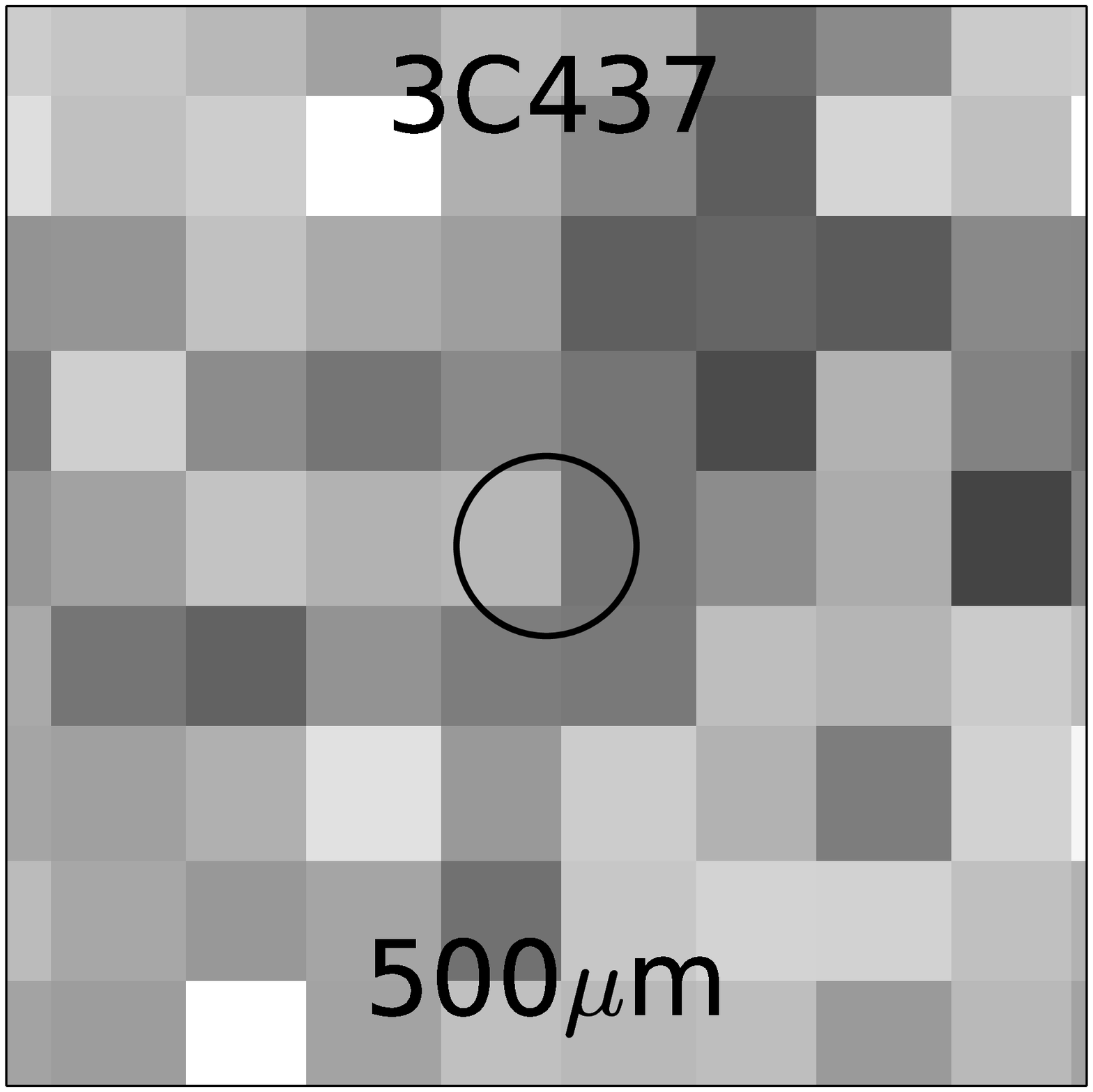}
      \\
      \includegraphics[width=1.5cm]{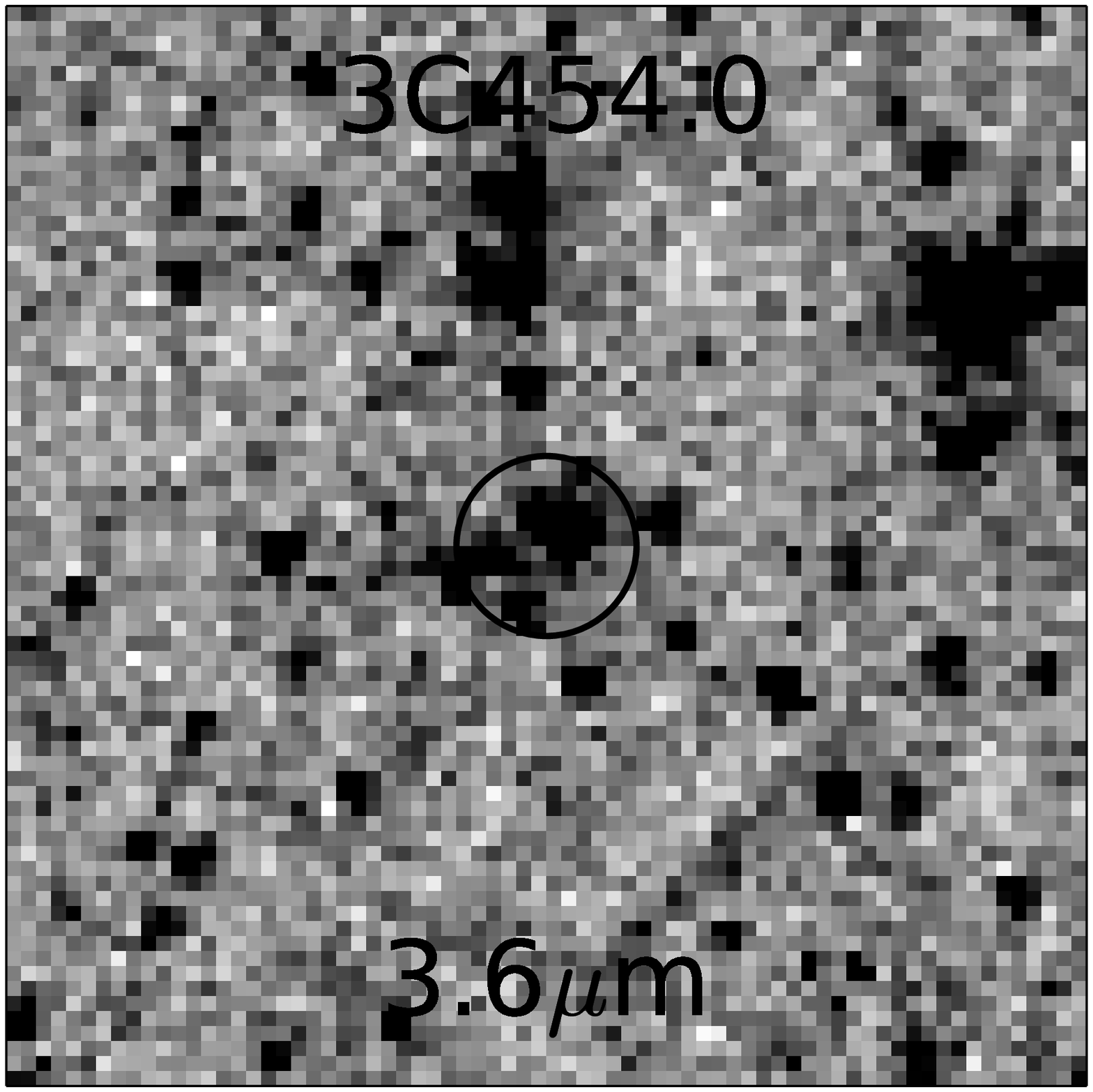}
      \includegraphics[width=1.5cm]{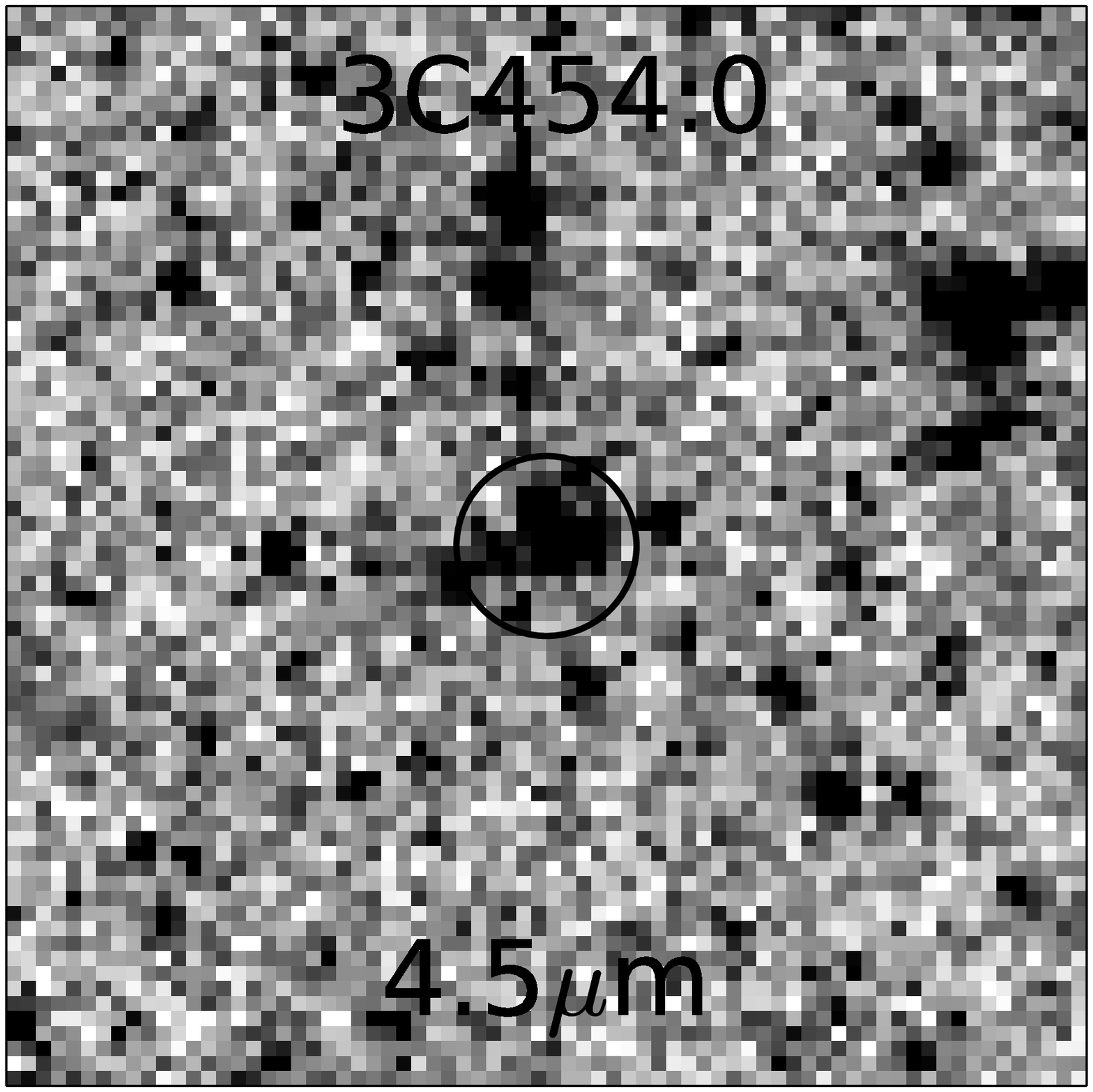}
      \includegraphics[width=1.5cm]{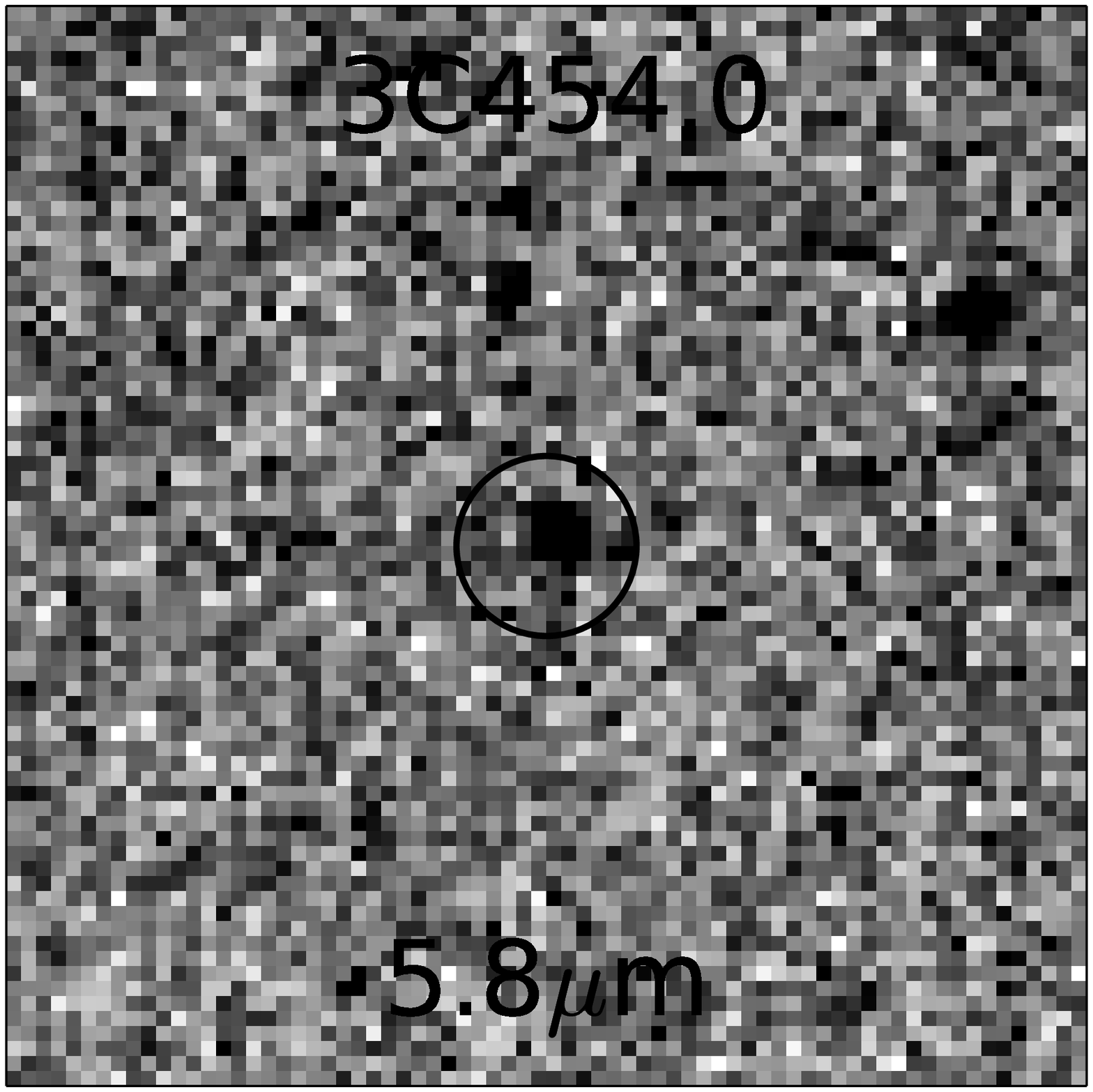}
      \includegraphics[width=1.5cm]{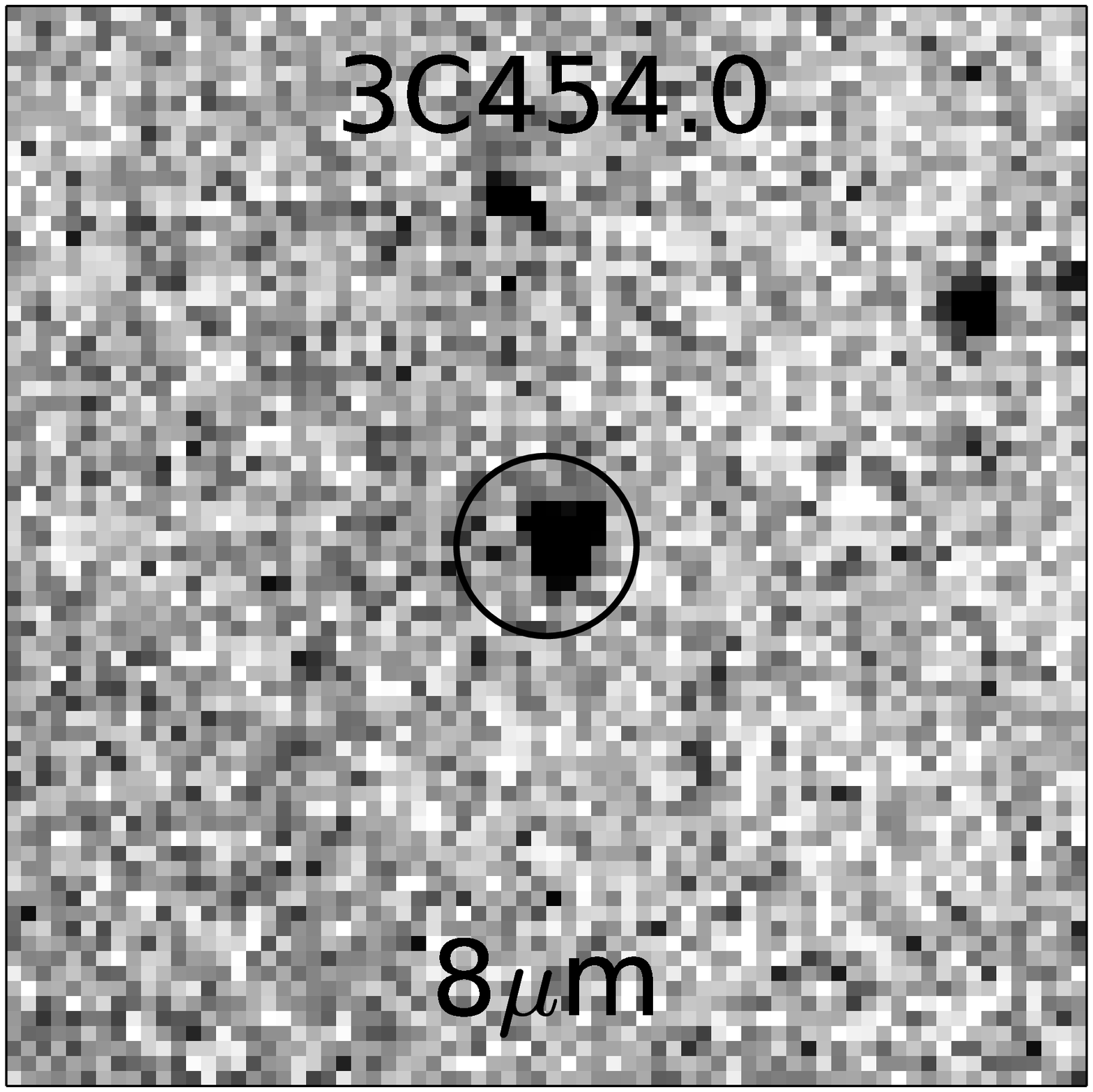}
      \includegraphics[width=1.5cm]{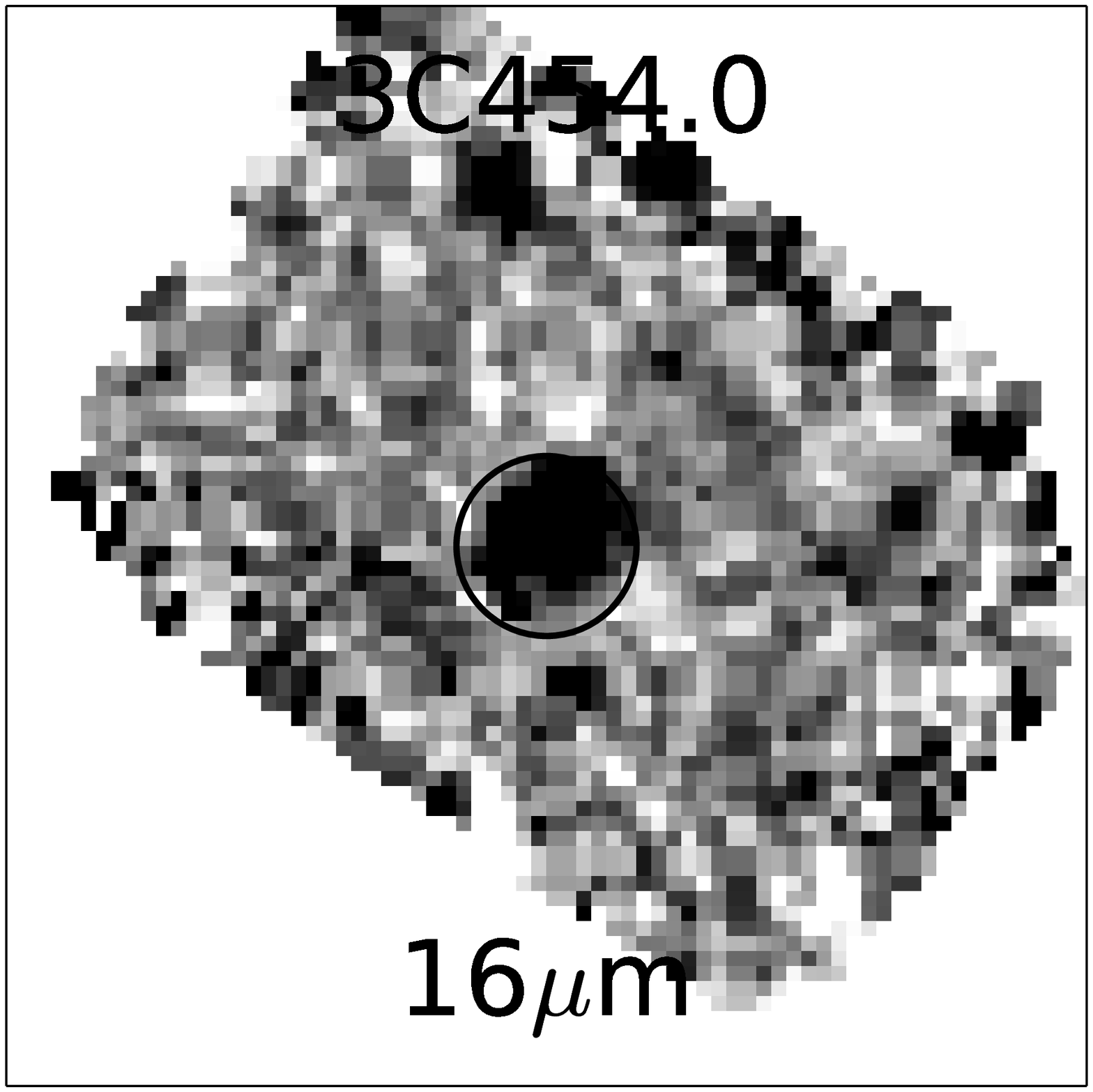}
      \includegraphics[width=1.5cm]{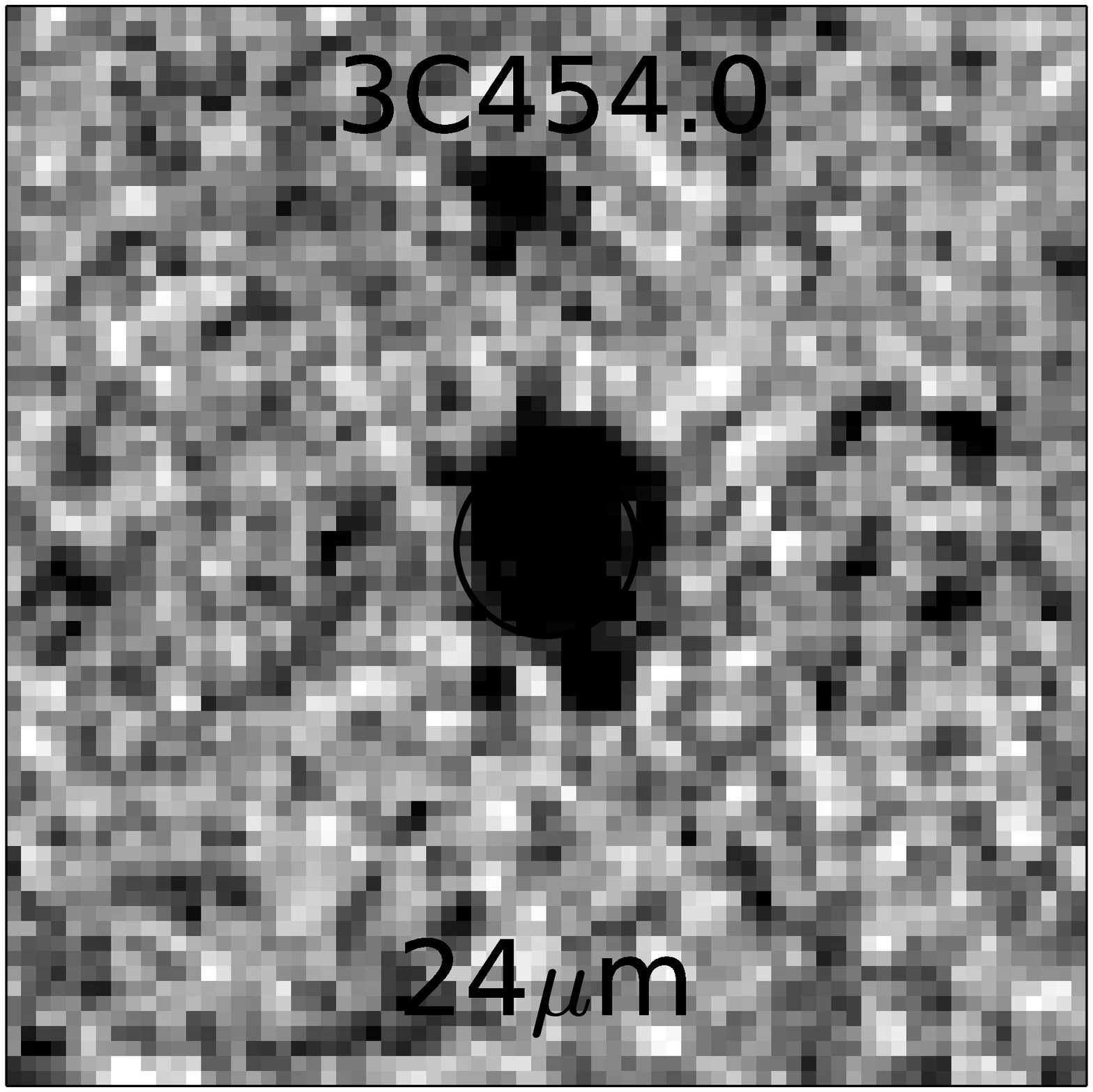}
      \includegraphics[width=1.5cm]{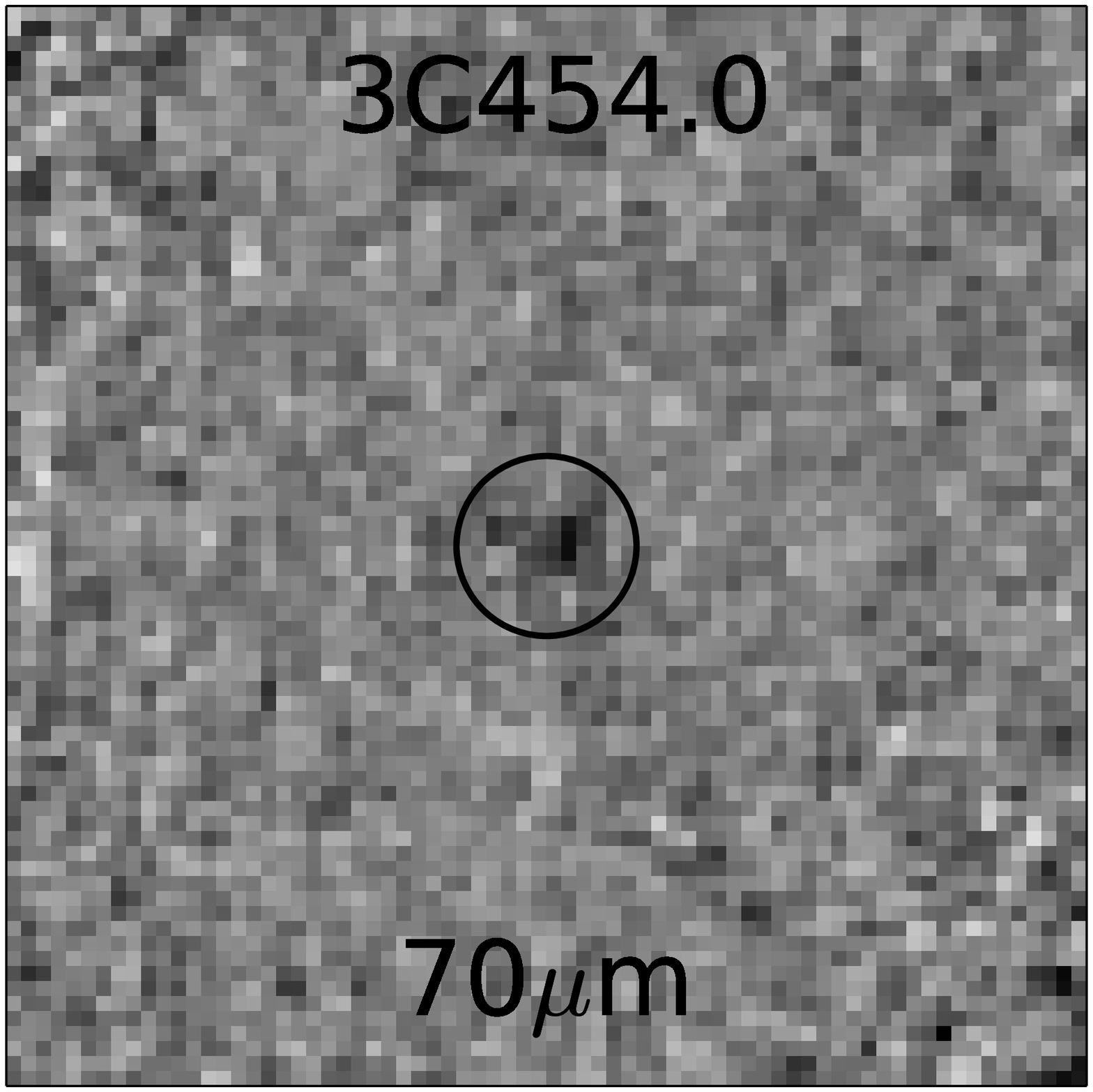}
      \includegraphics[width=1.5cm]{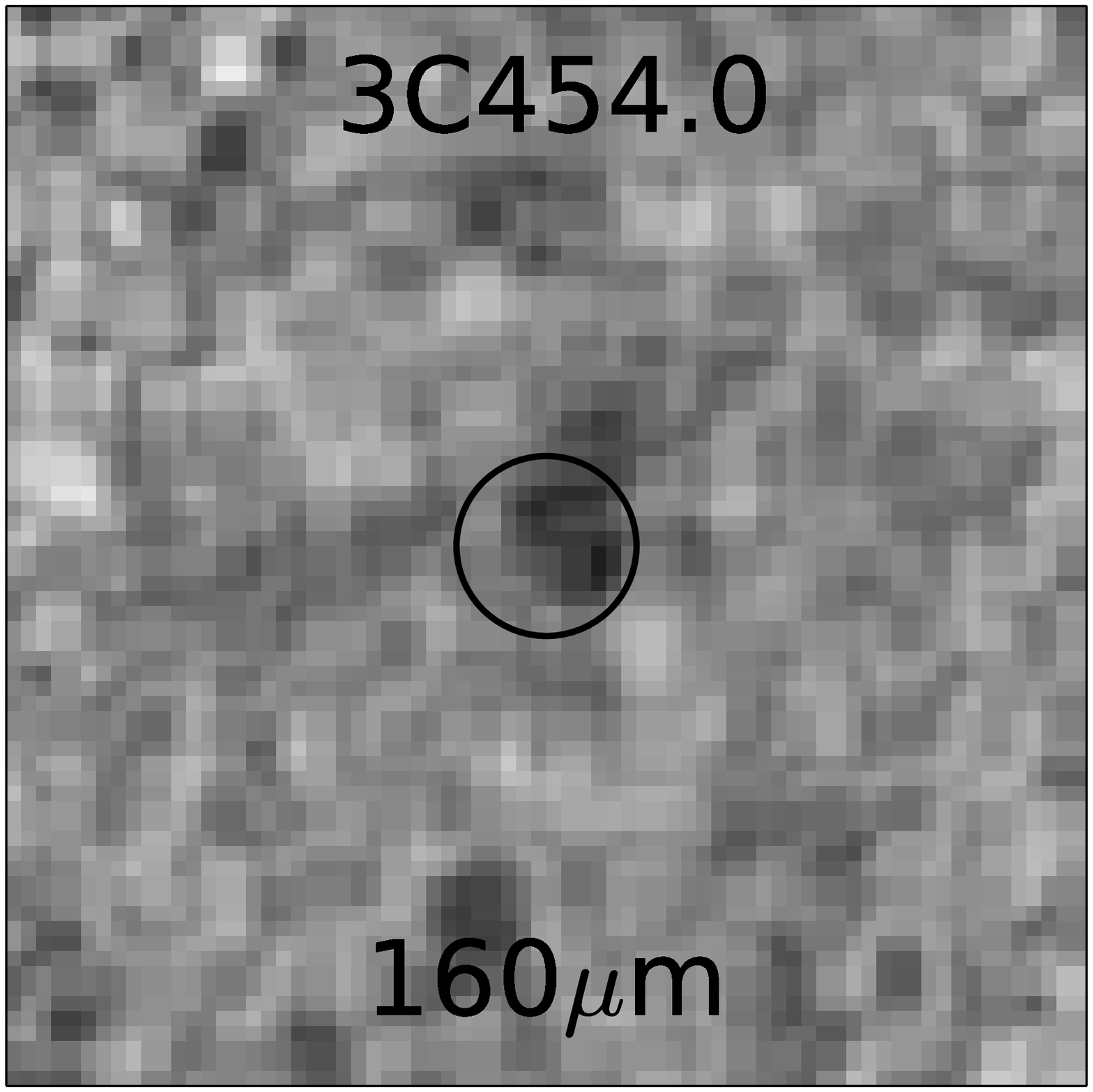}
      \includegraphics[width=1.5cm]{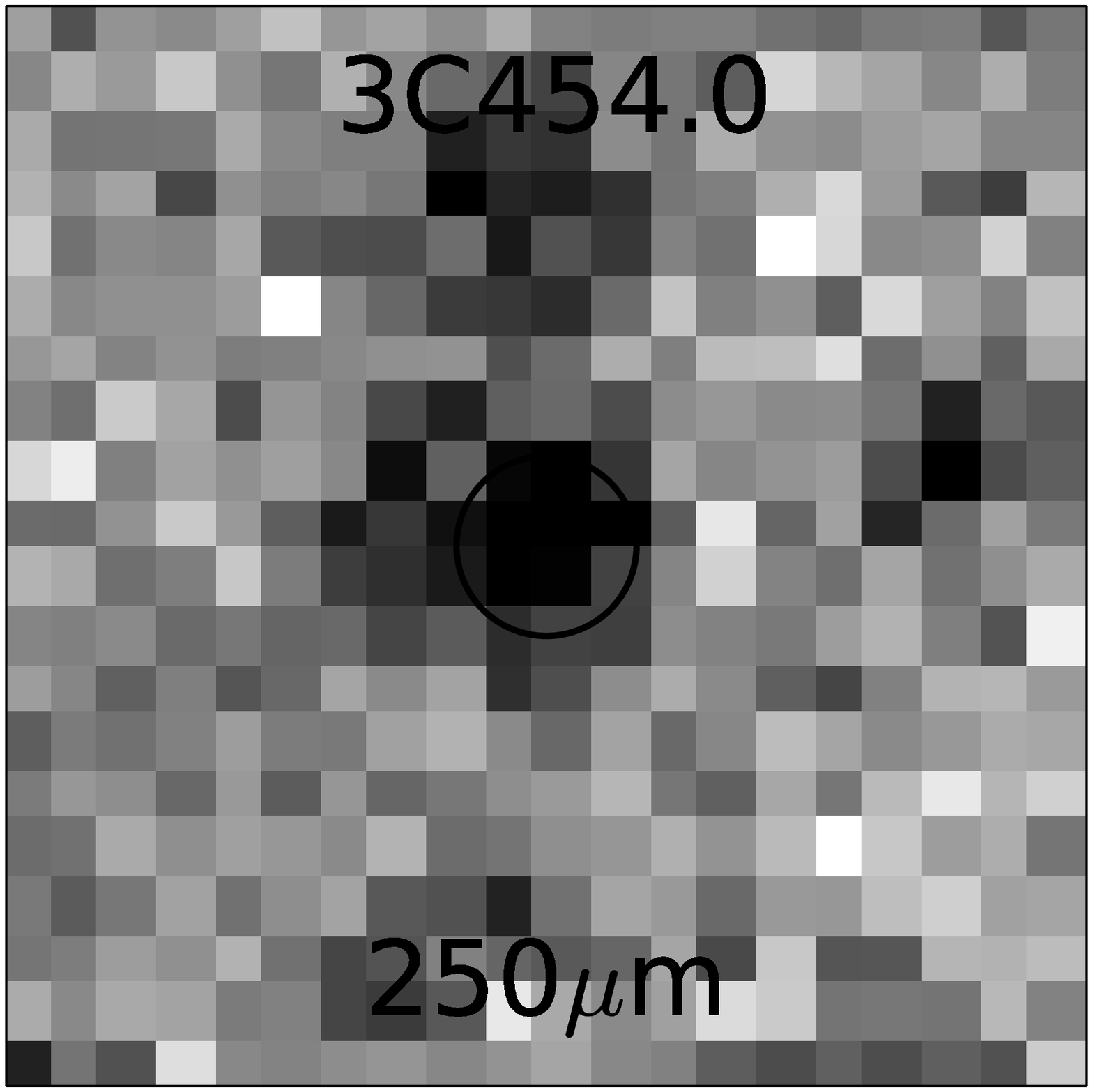}
      \includegraphics[width=1.5cm]{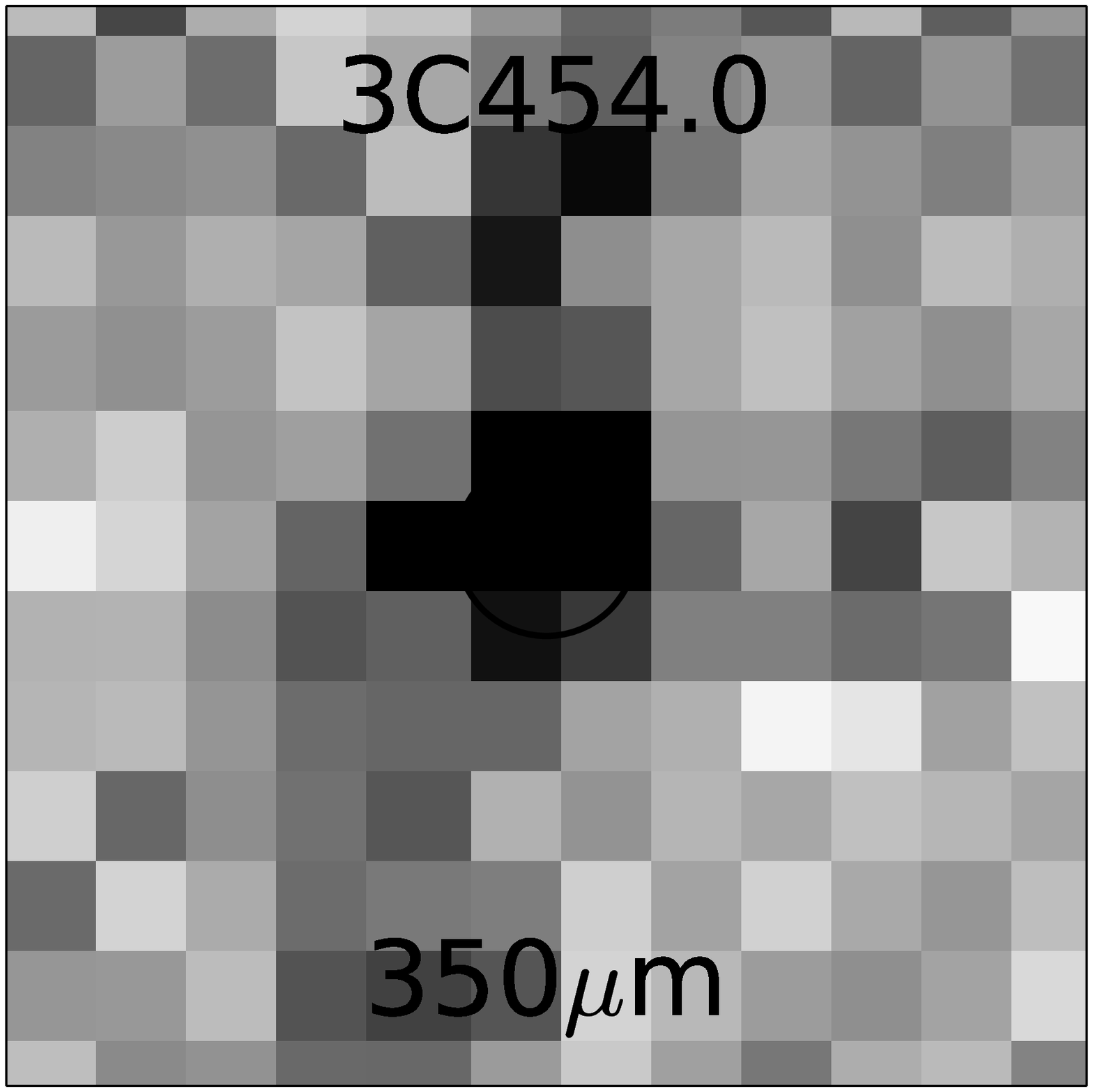}
      \includegraphics[width=1.5cm]{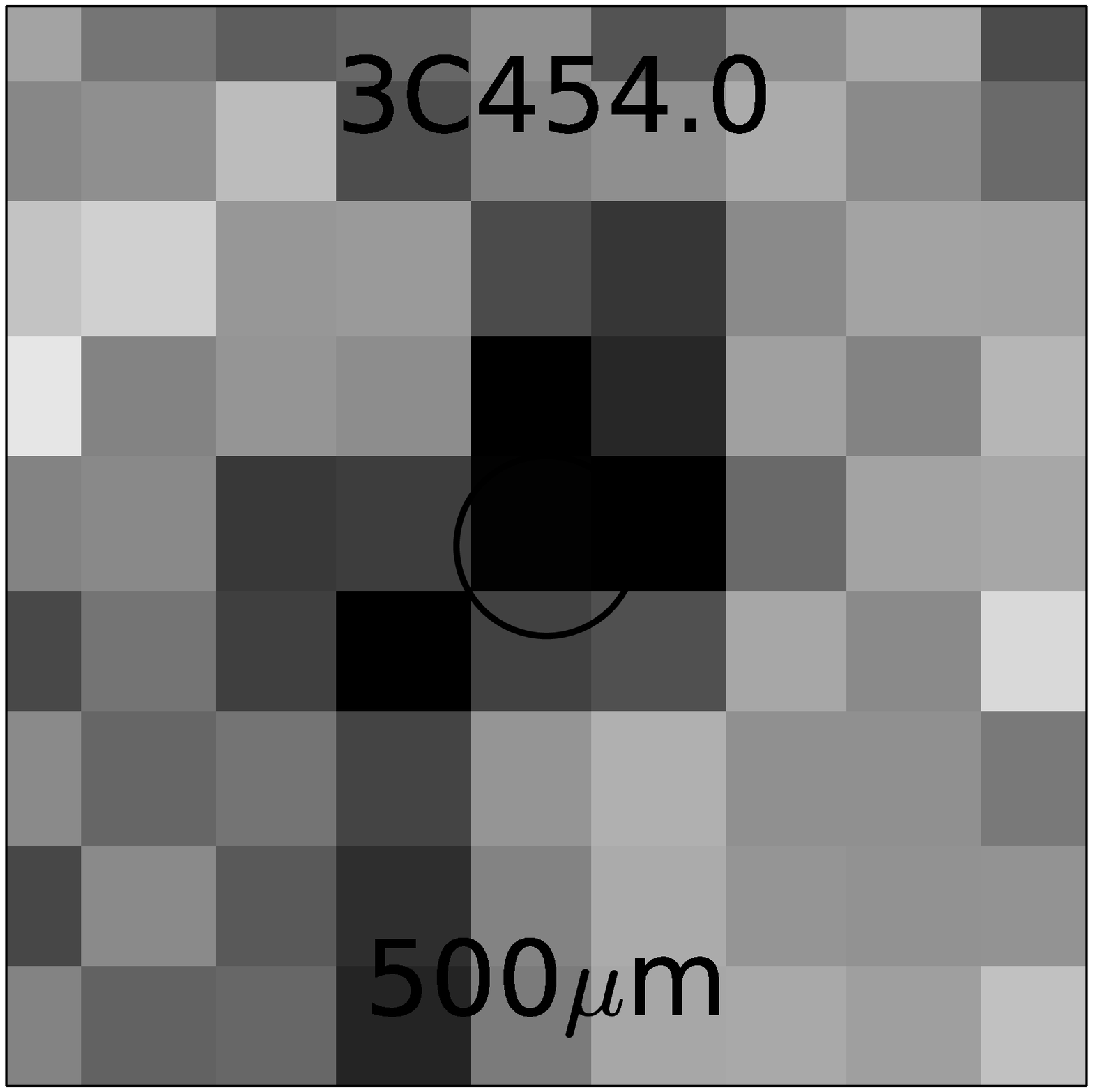}
      \\
      \includegraphics[width=1.5cm]{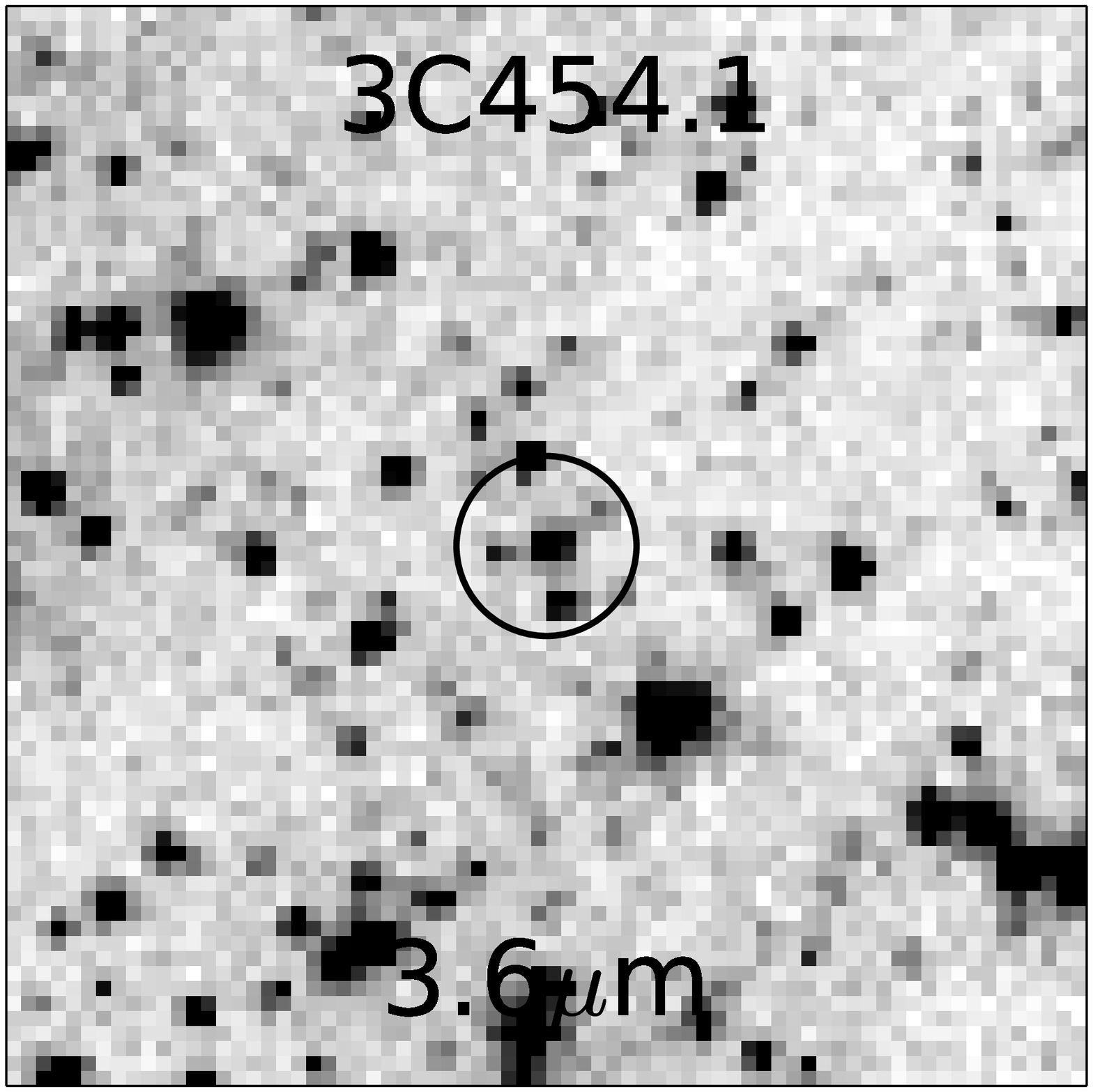}
      \includegraphics[width=1.5cm]{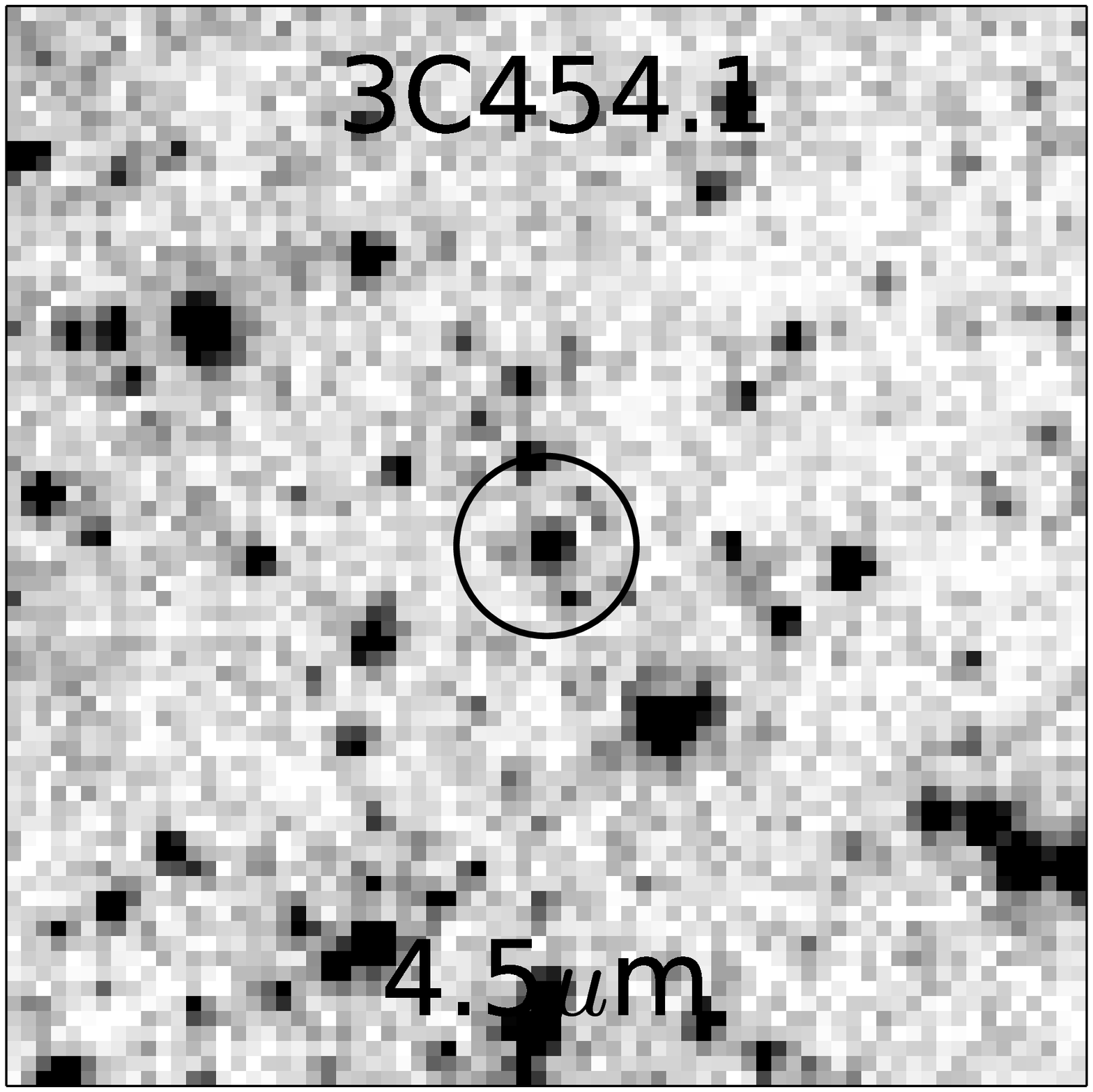}
      \includegraphics[width=1.5cm]{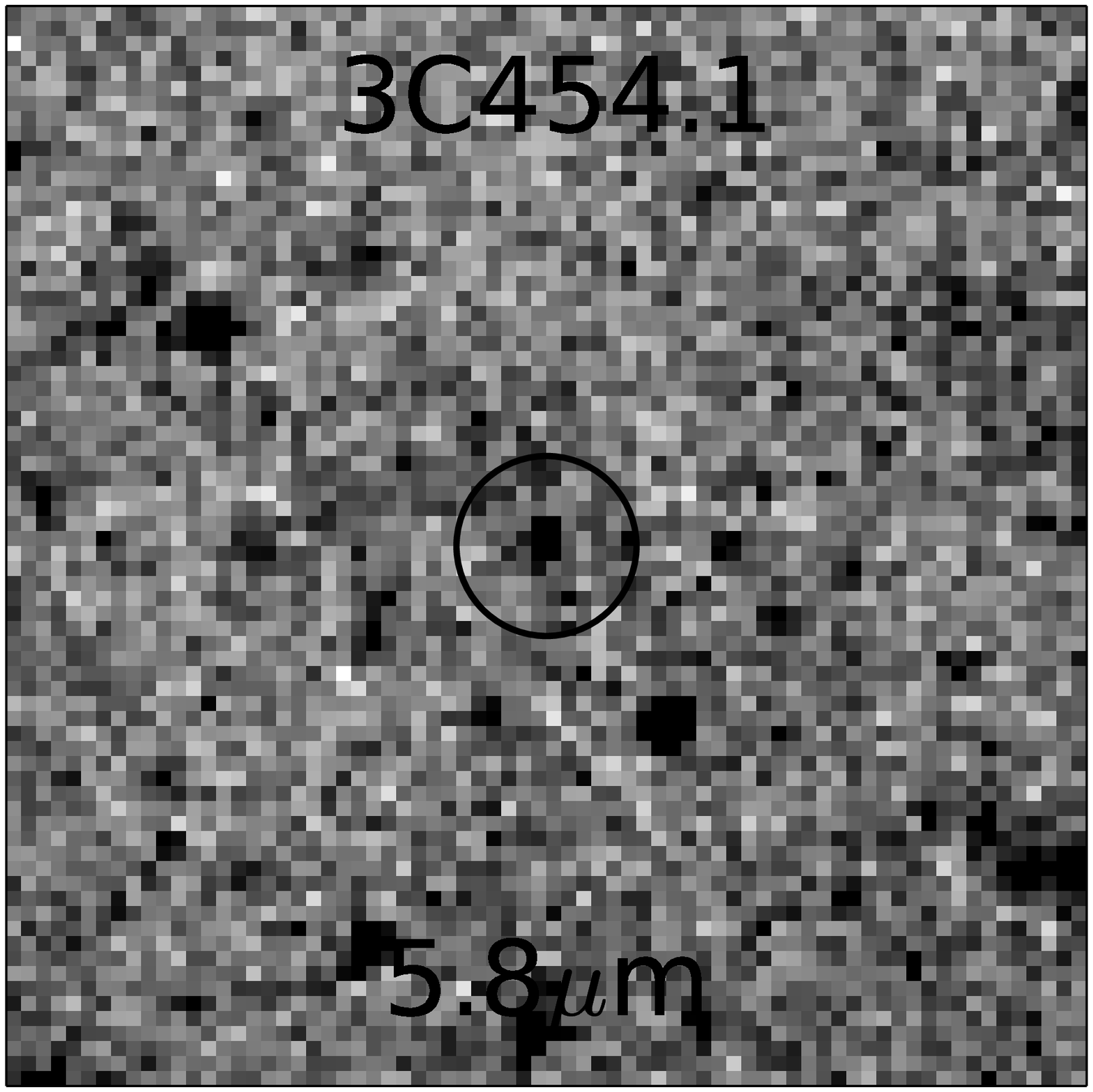}
      \includegraphics[width=1.5cm]{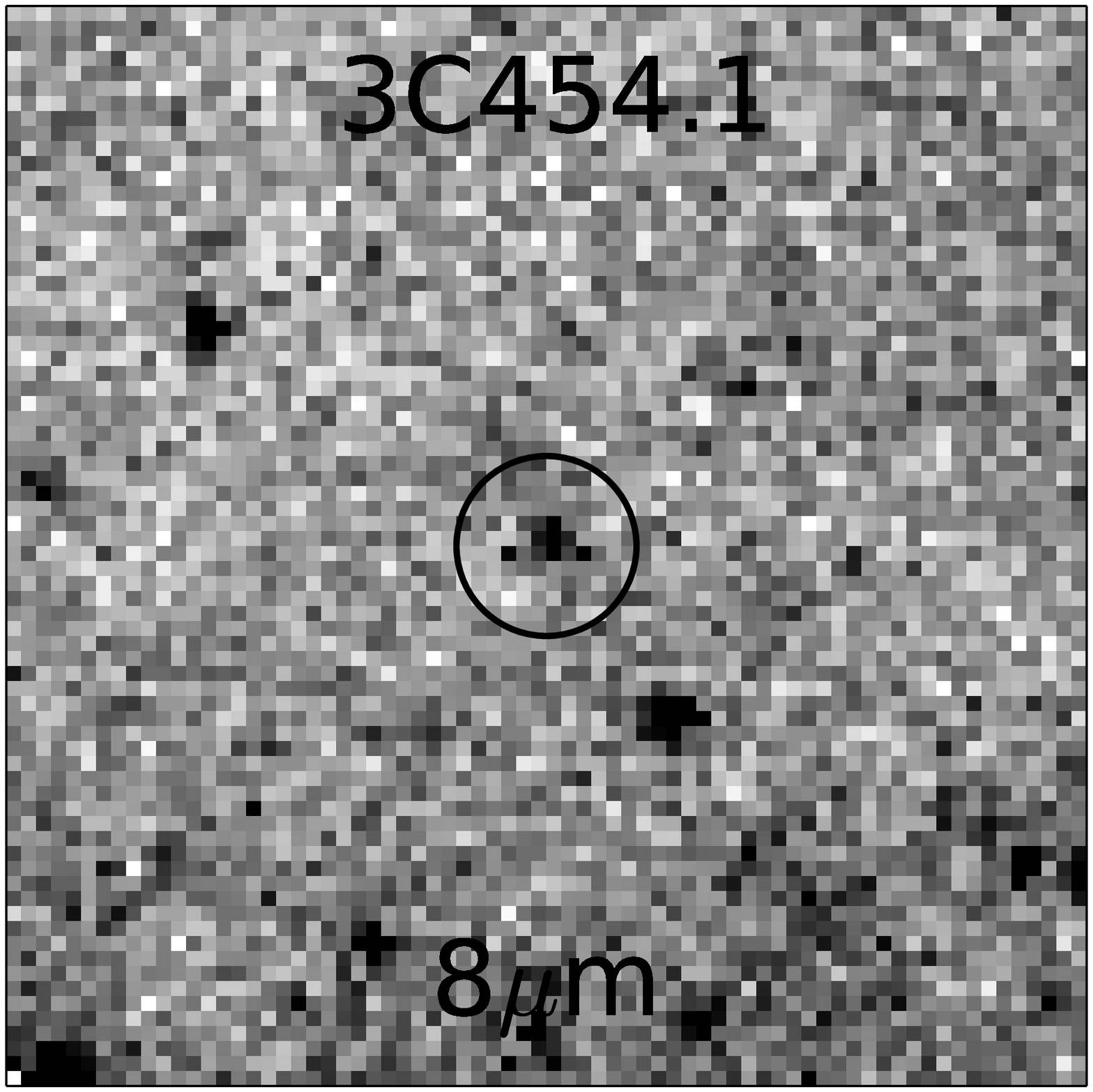}
      \includegraphics[width=1.5cm]{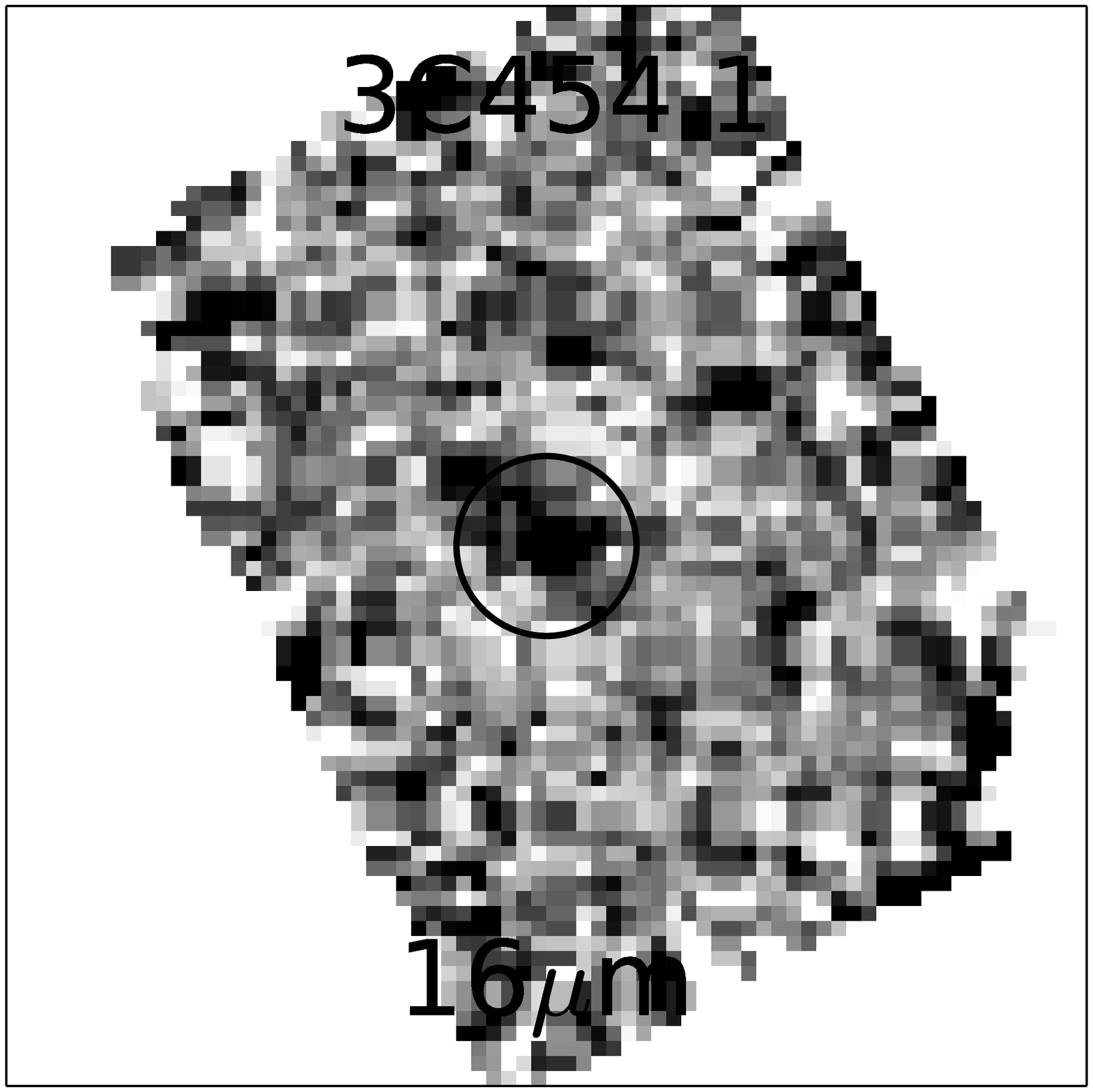}
      \includegraphics[width=1.5cm]{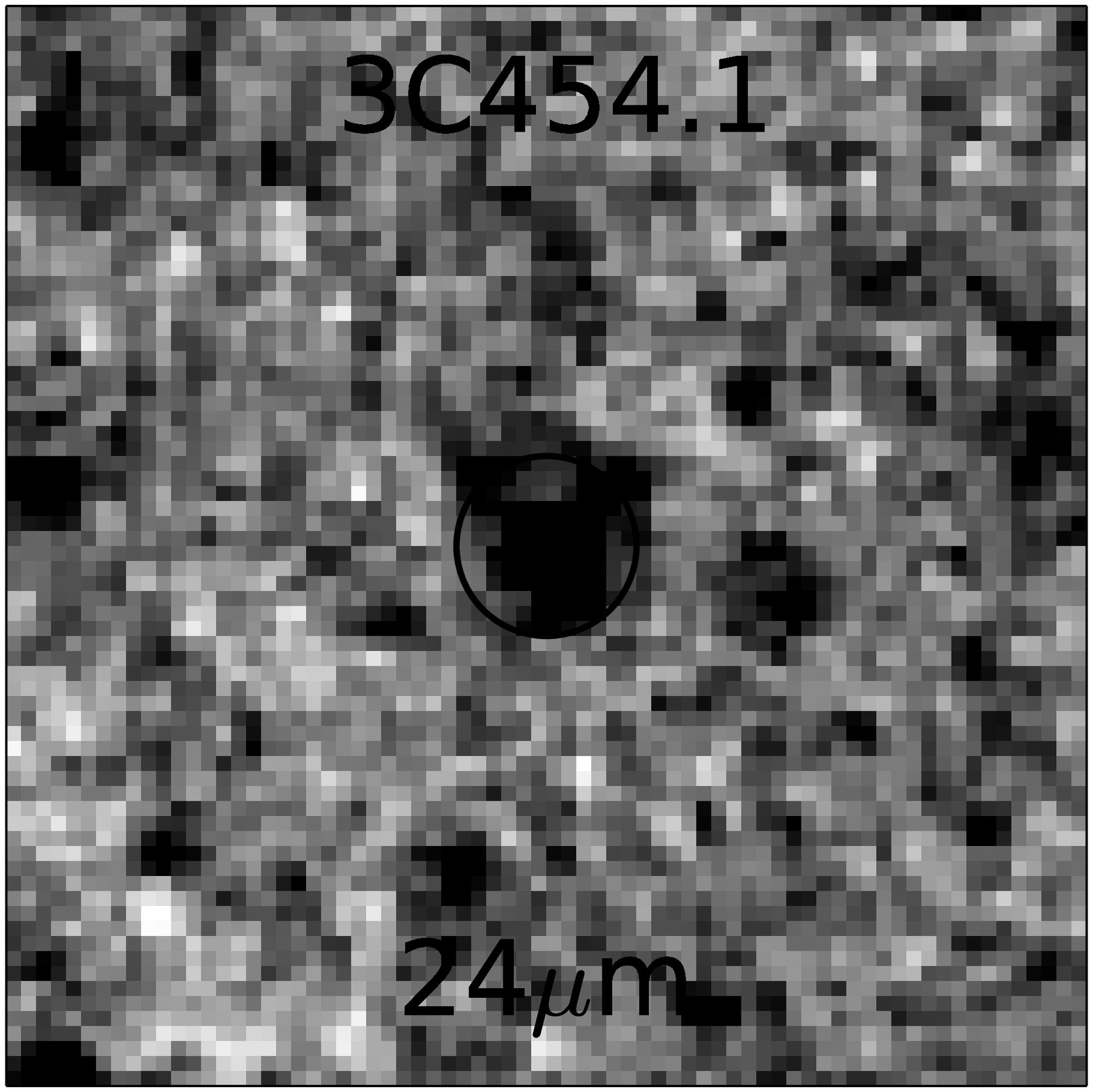}
      \includegraphics[width=1.5cm]{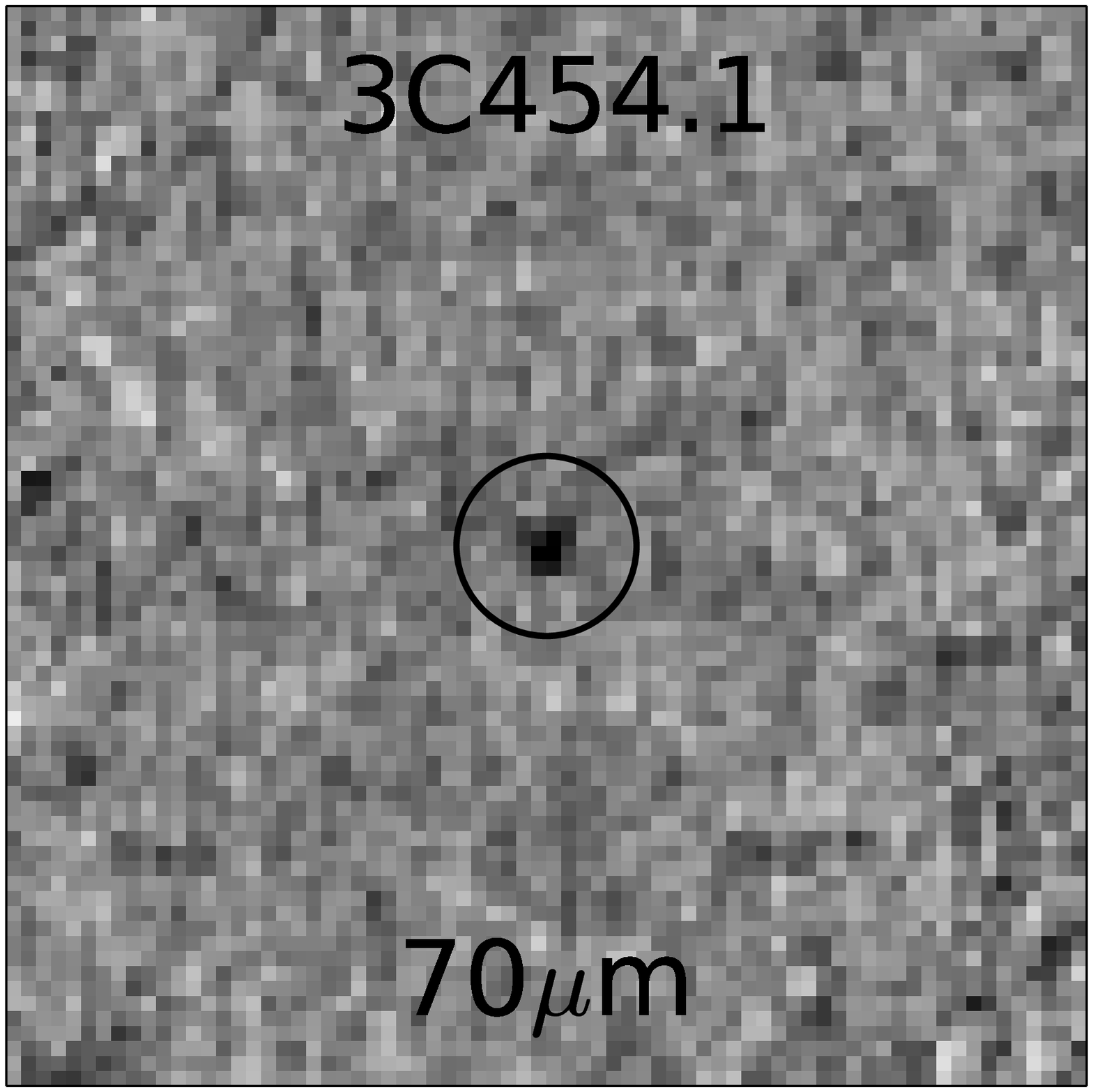}
      \includegraphics[width=1.5cm]{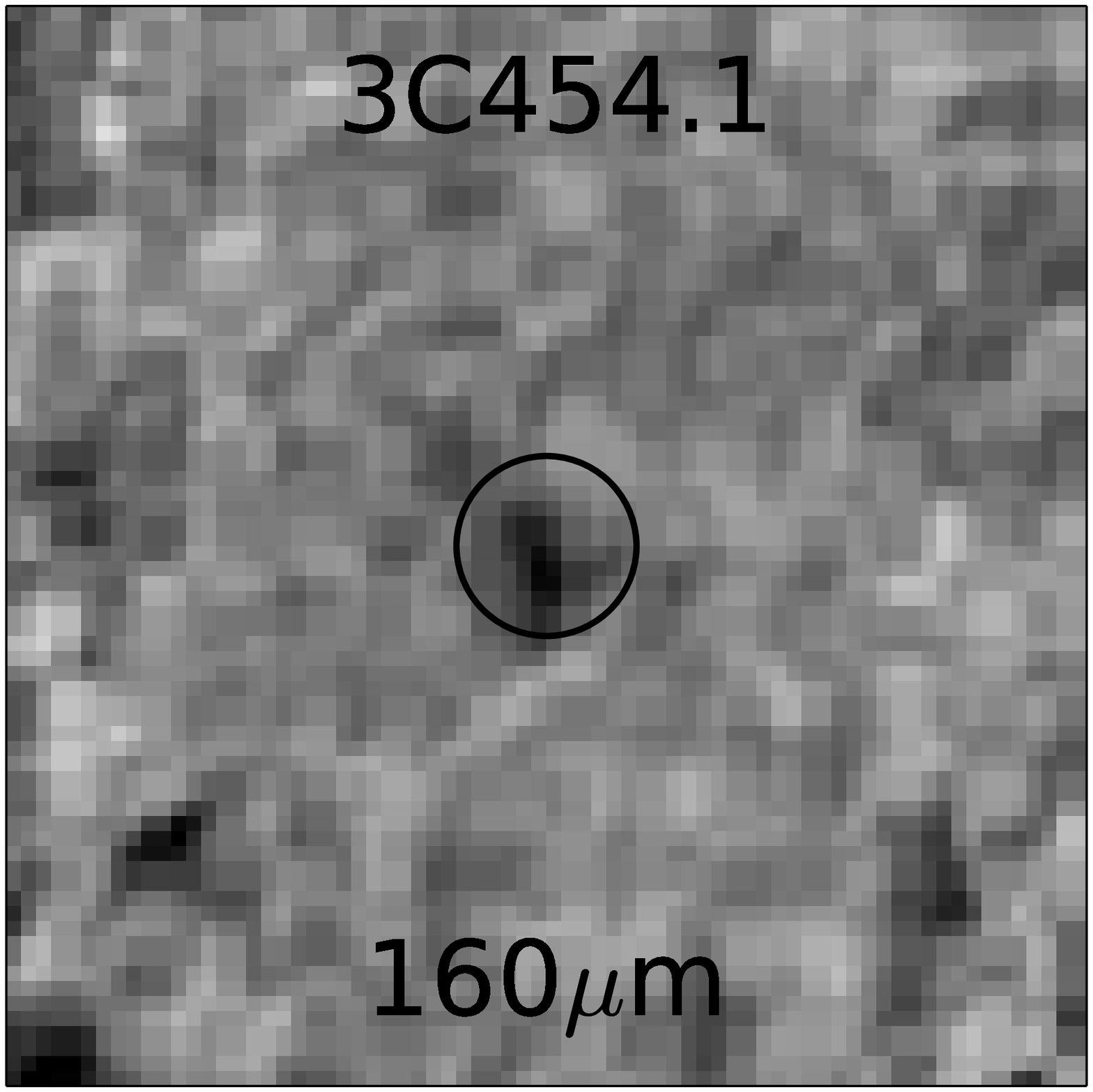}
      \includegraphics[width=1.5cm]{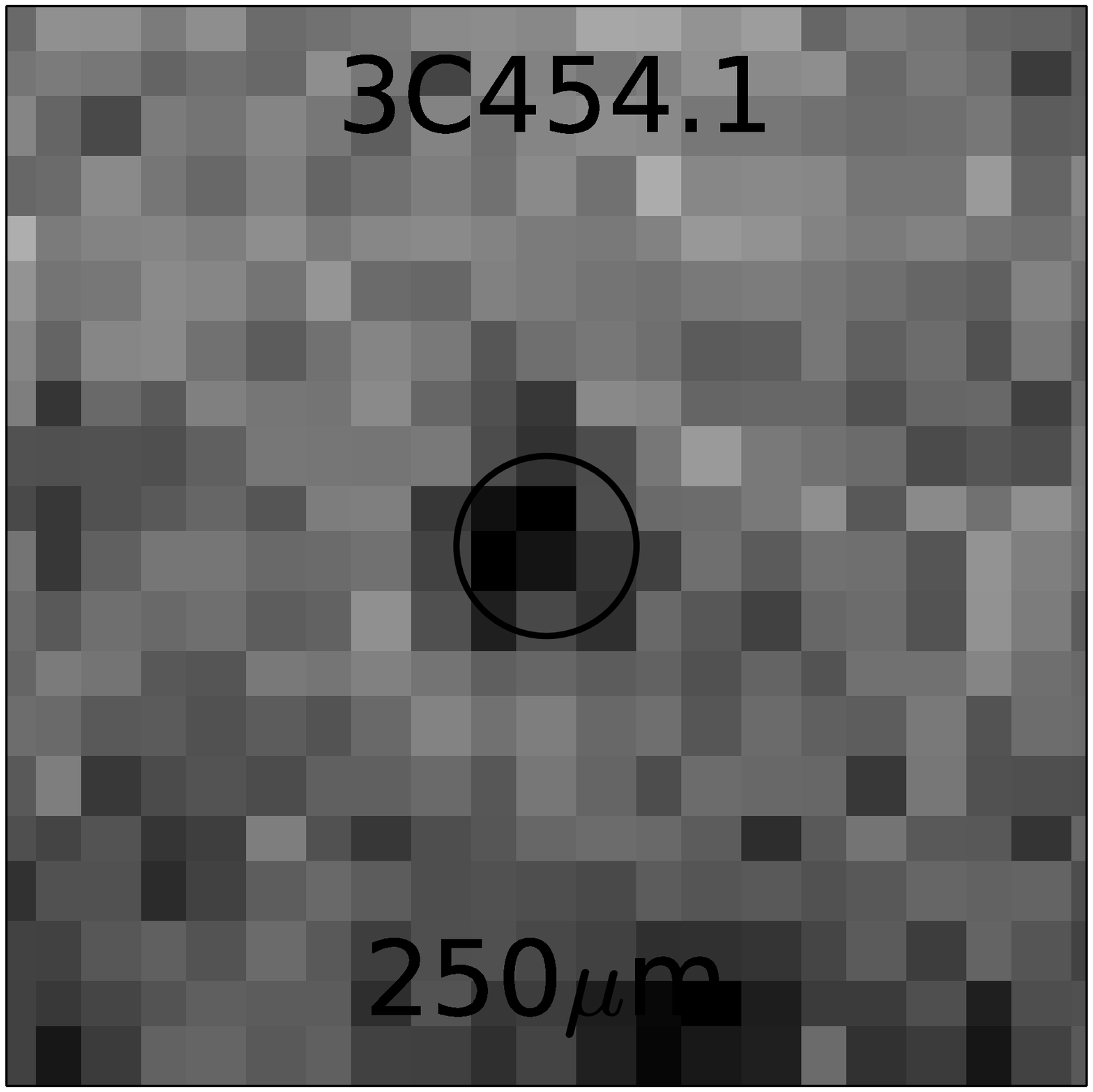}
      \includegraphics[width=1.5cm]{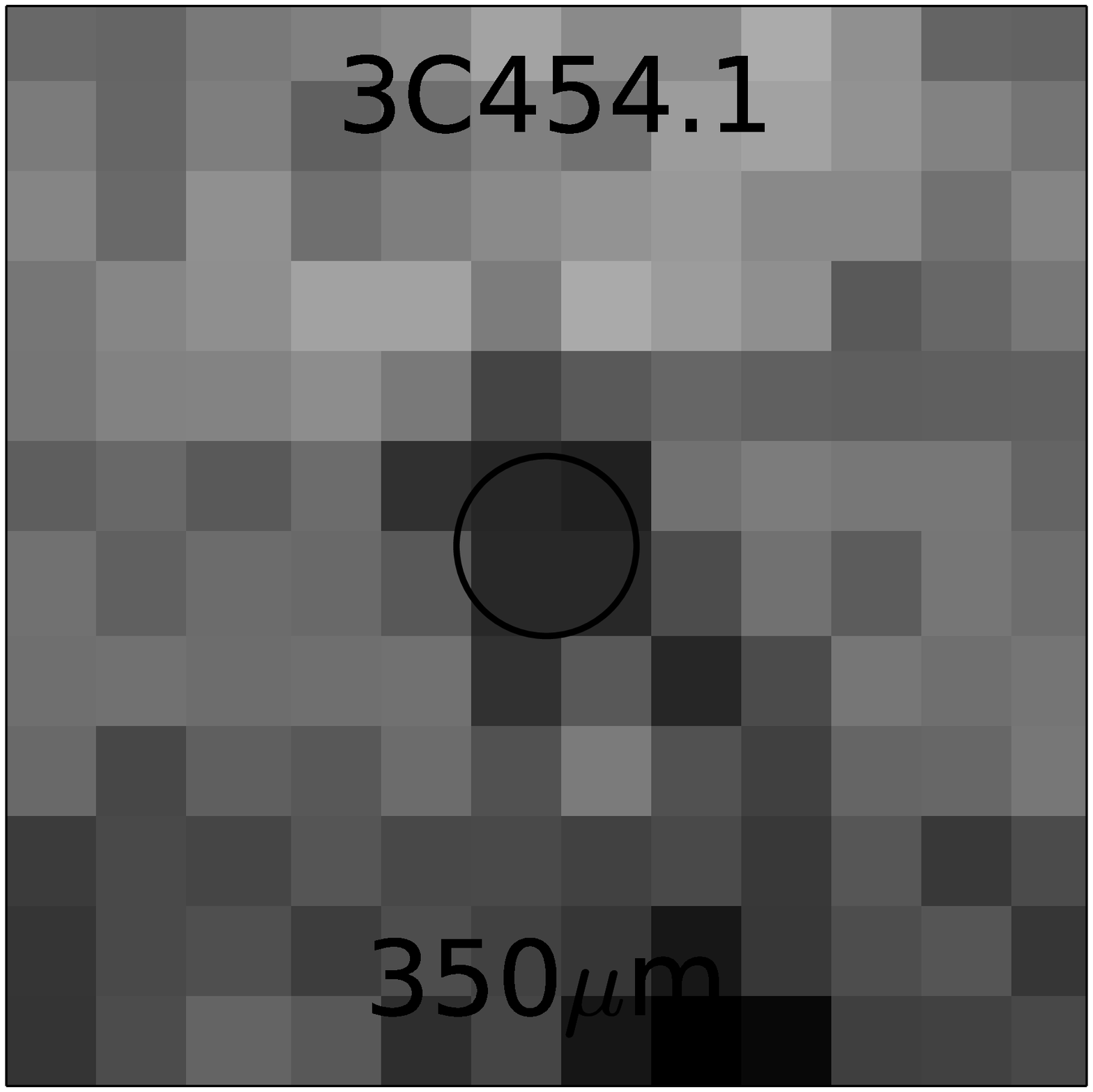}
      \includegraphics[width=1.5cm]{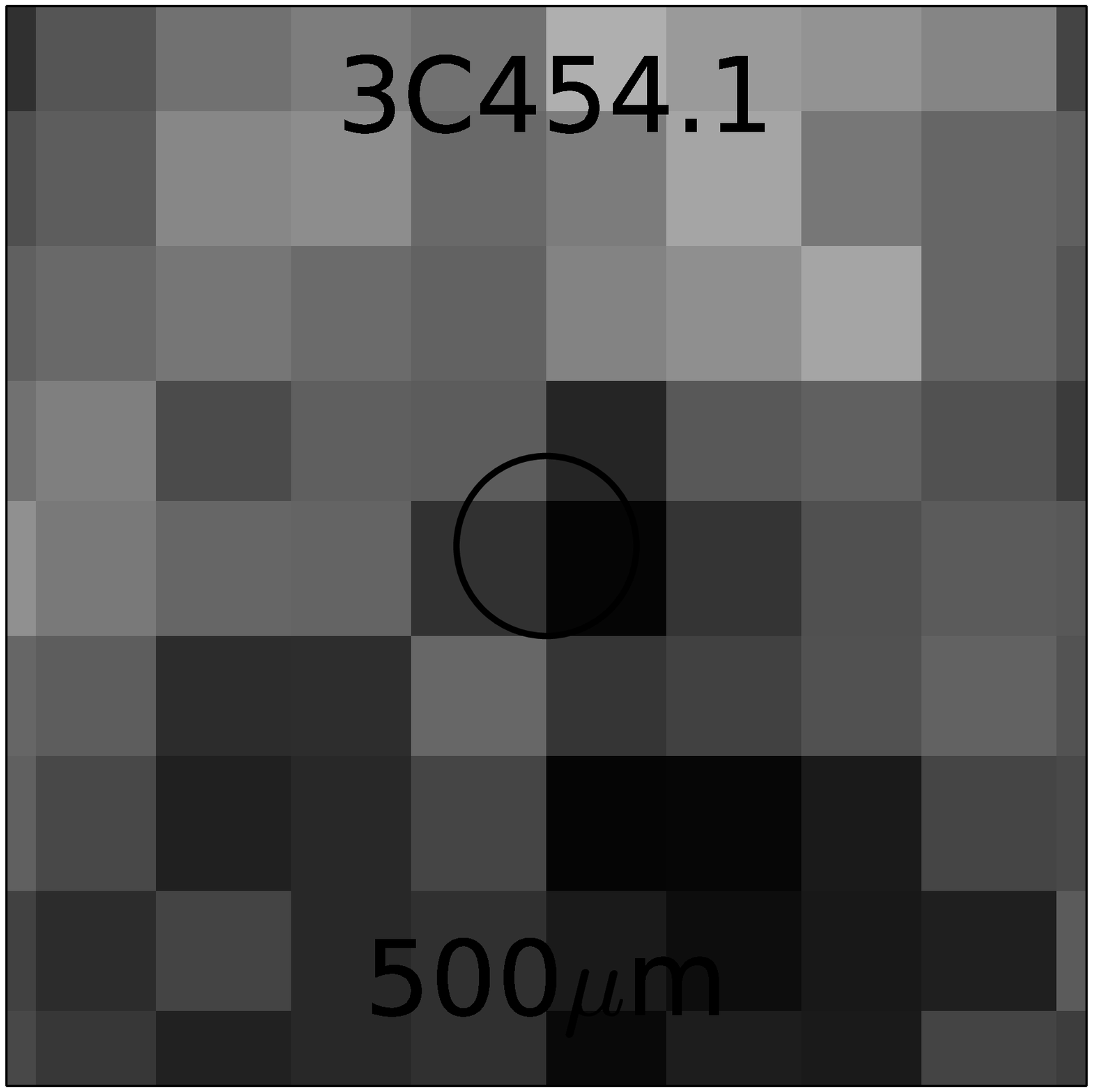}
      \\
      \includegraphics[width=1.5cm]{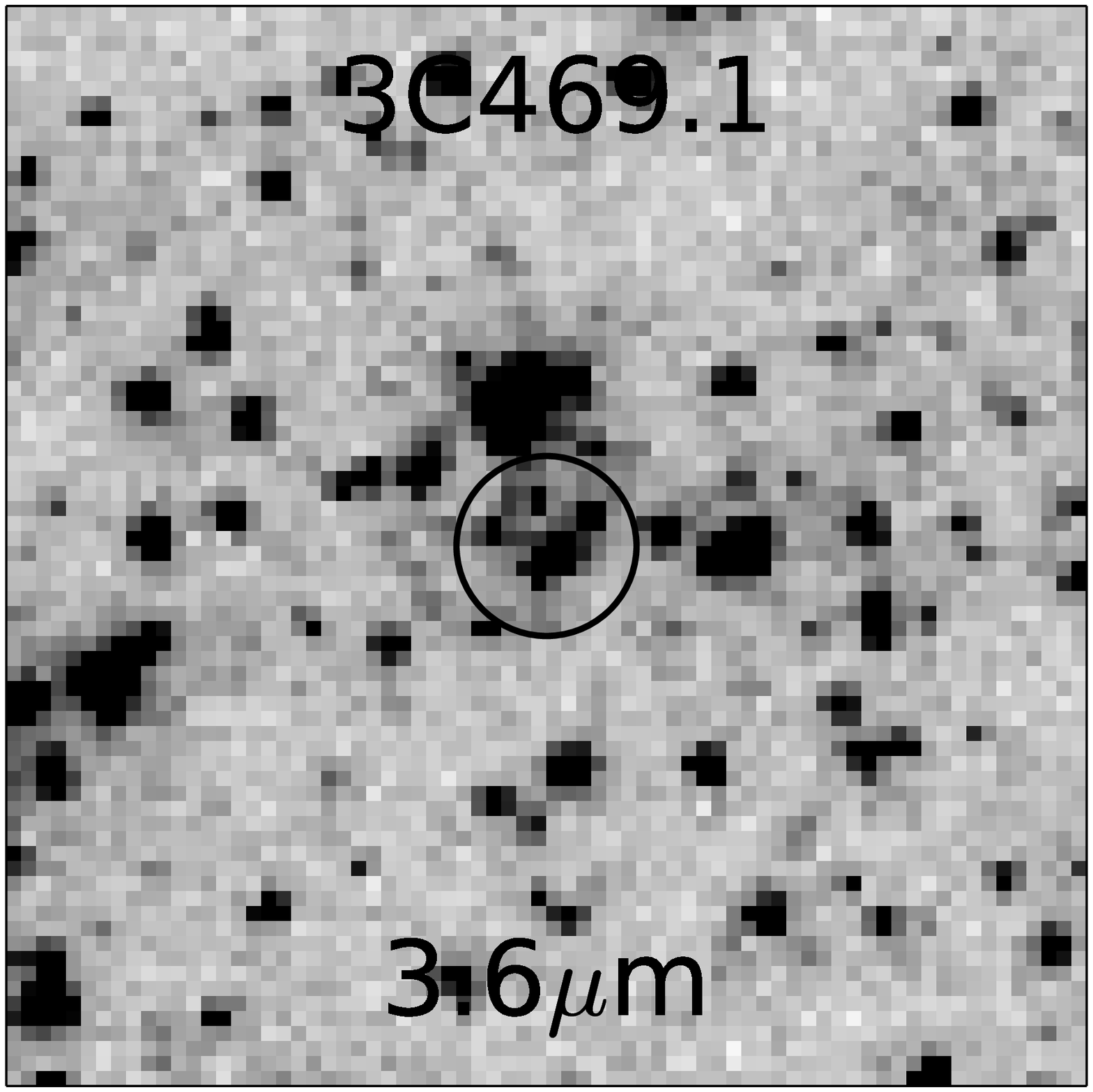}
      \includegraphics[width=1.5cm]{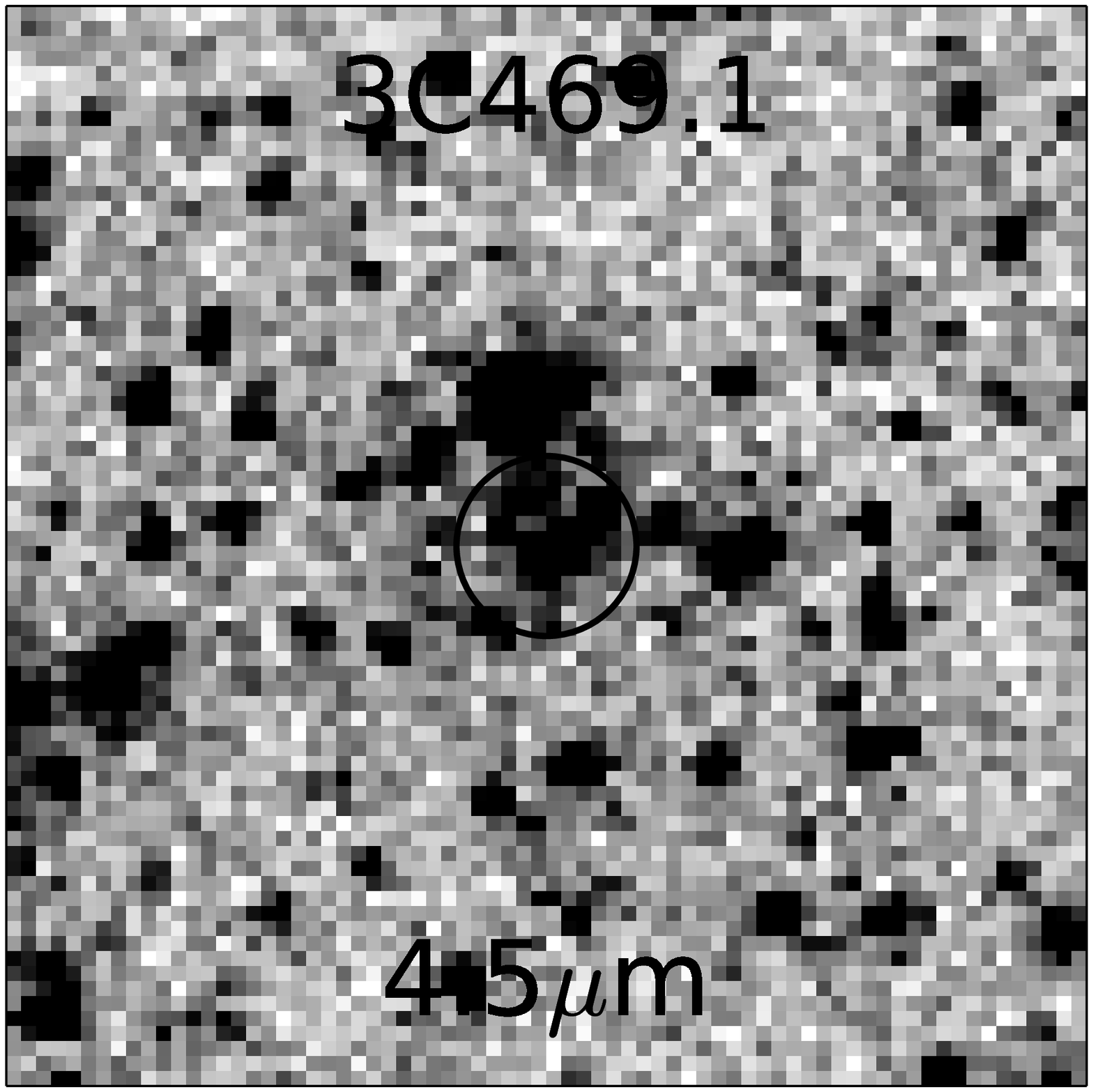}
      \includegraphics[width=1.5cm]{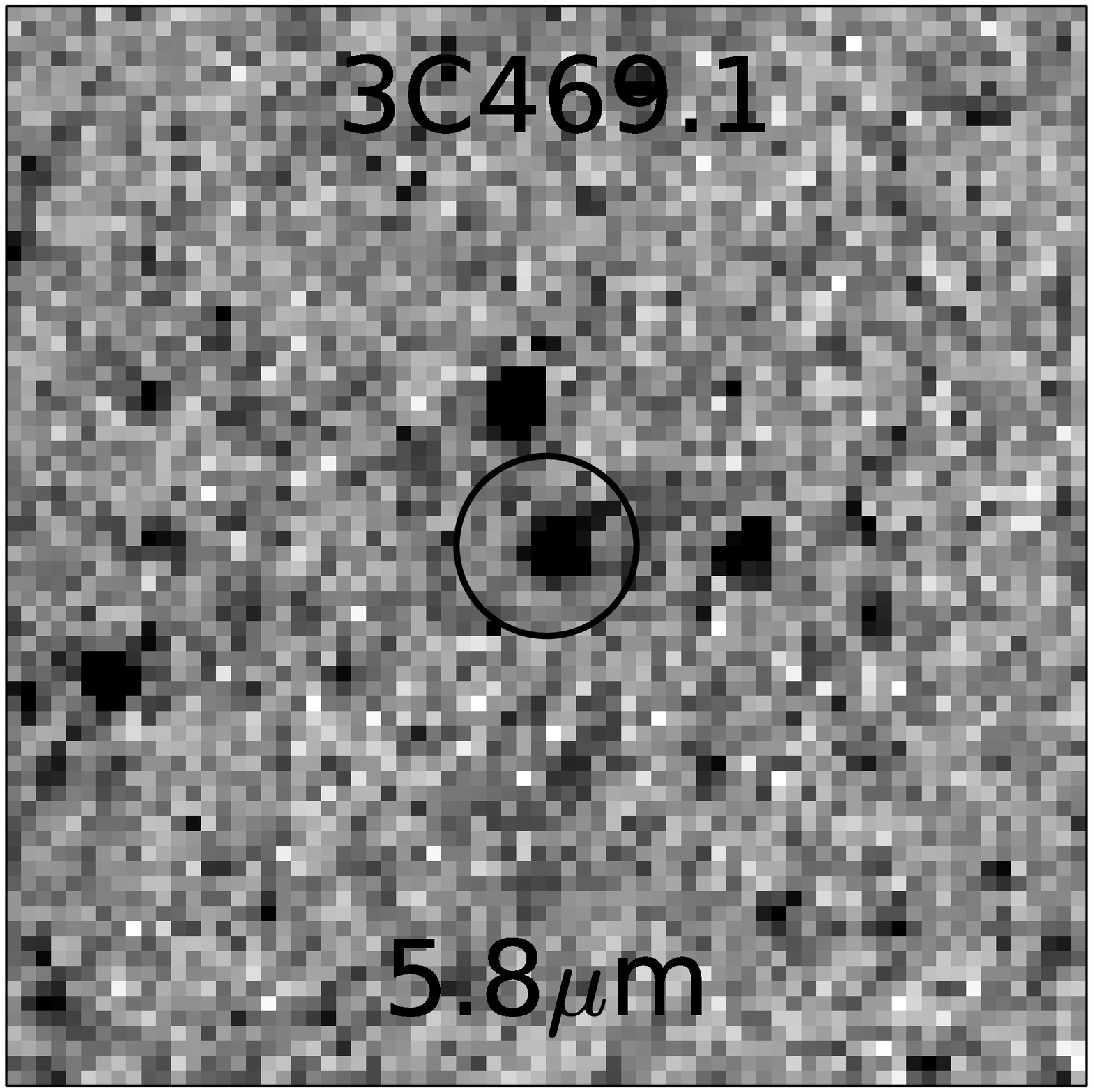}
      \includegraphics[width=1.5cm]{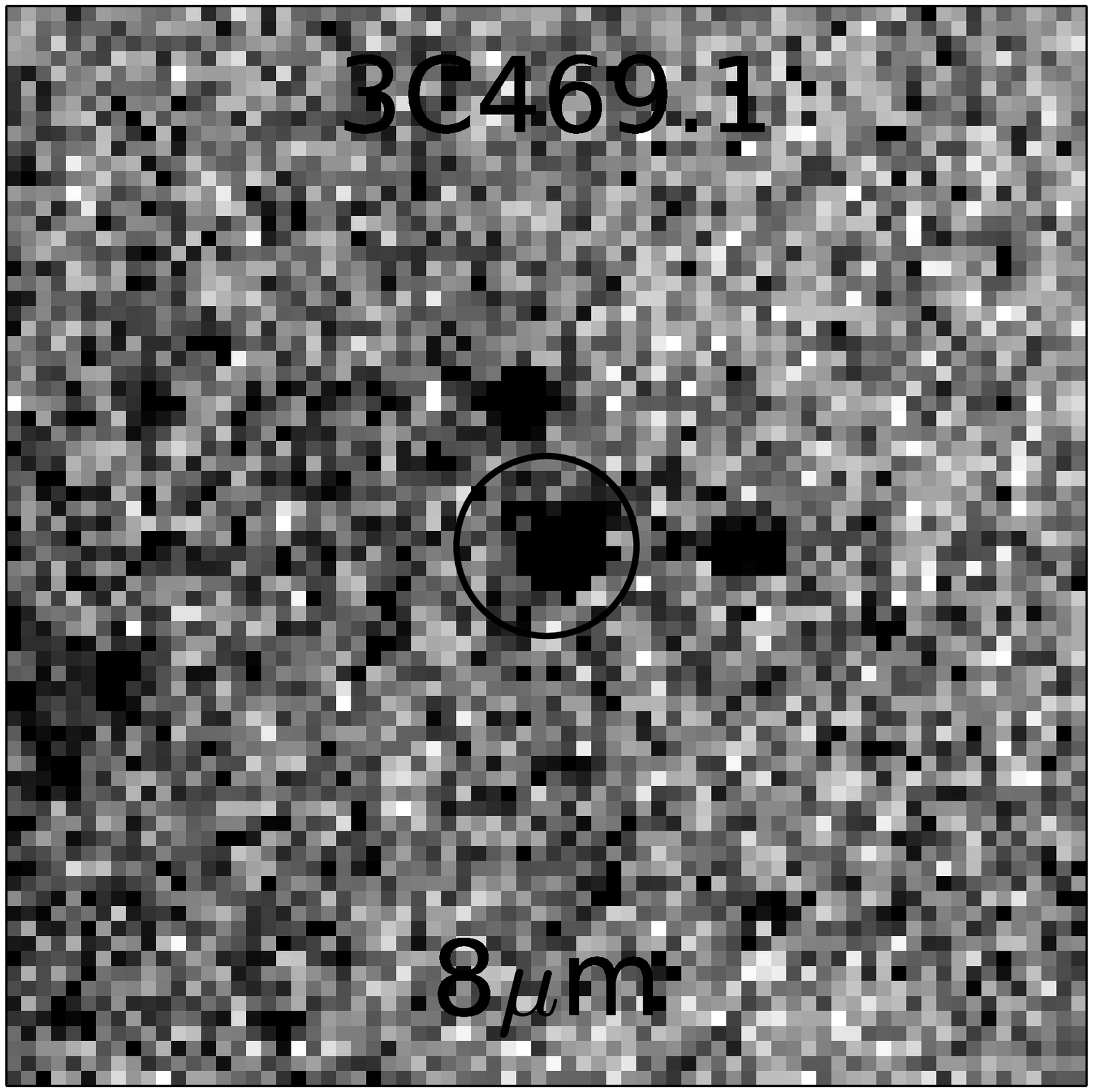}
      \includegraphics[width=1.5cm]{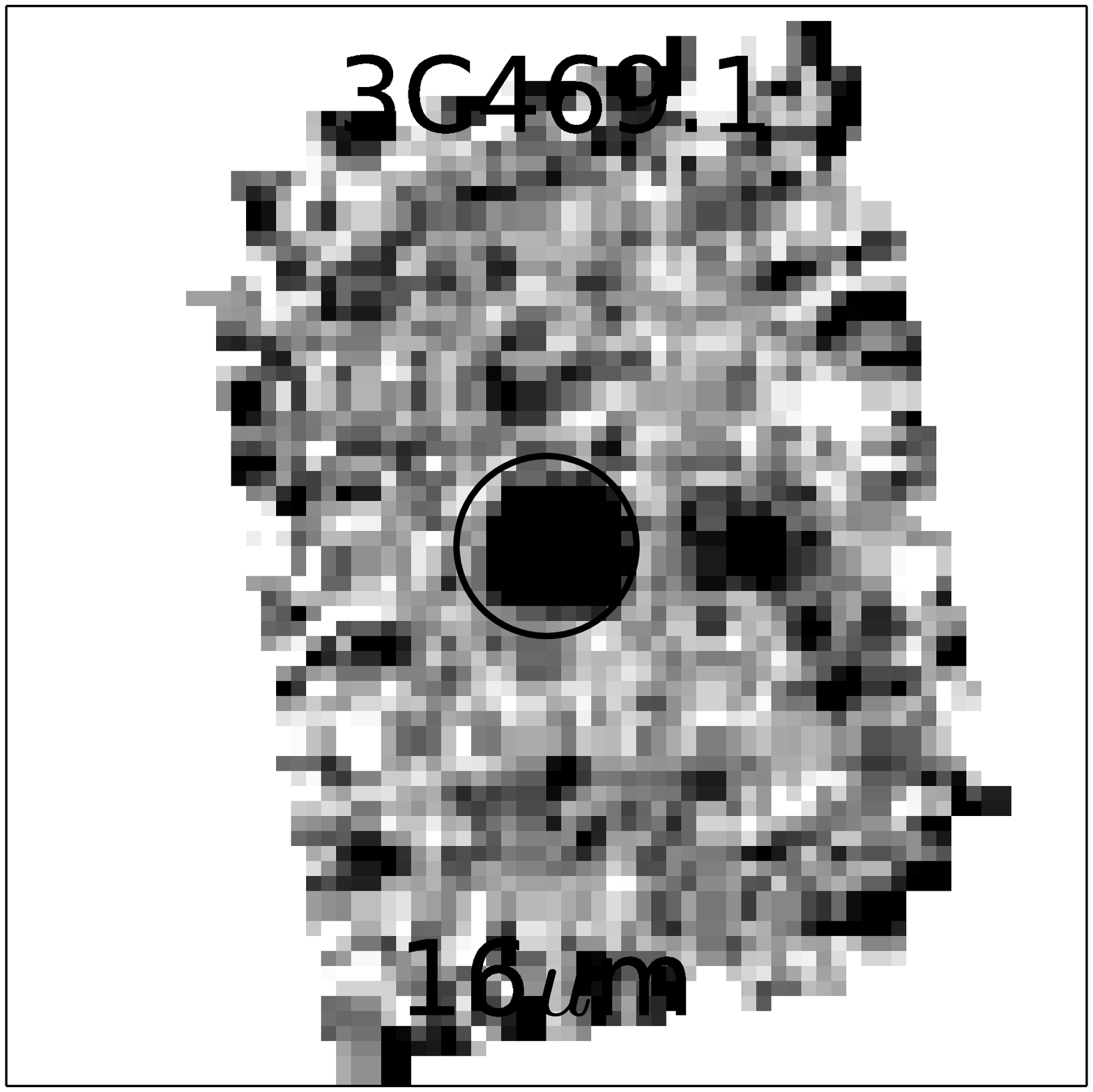}
      \includegraphics[width=1.5cm]{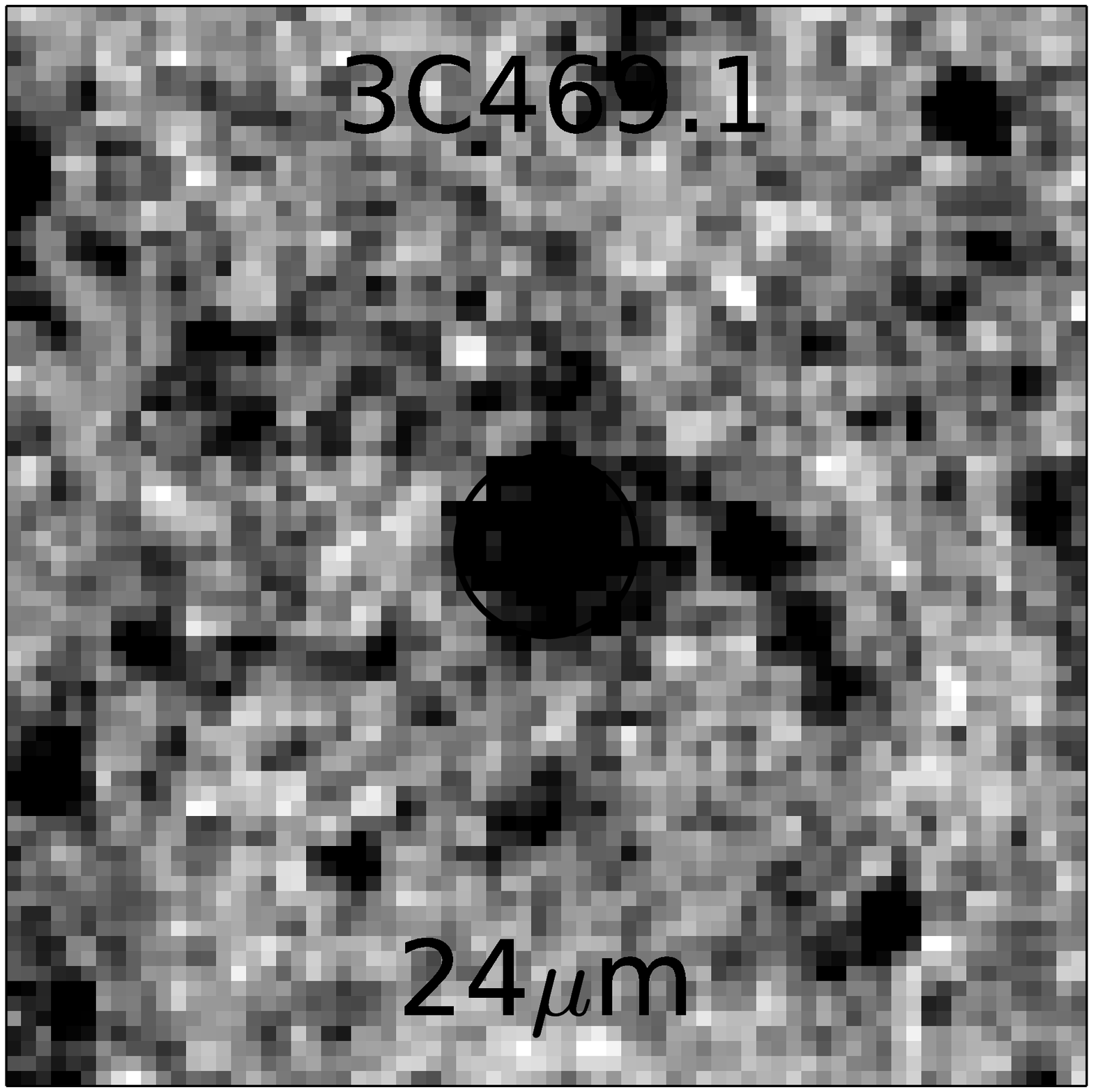}
      \includegraphics[width=1.5cm]{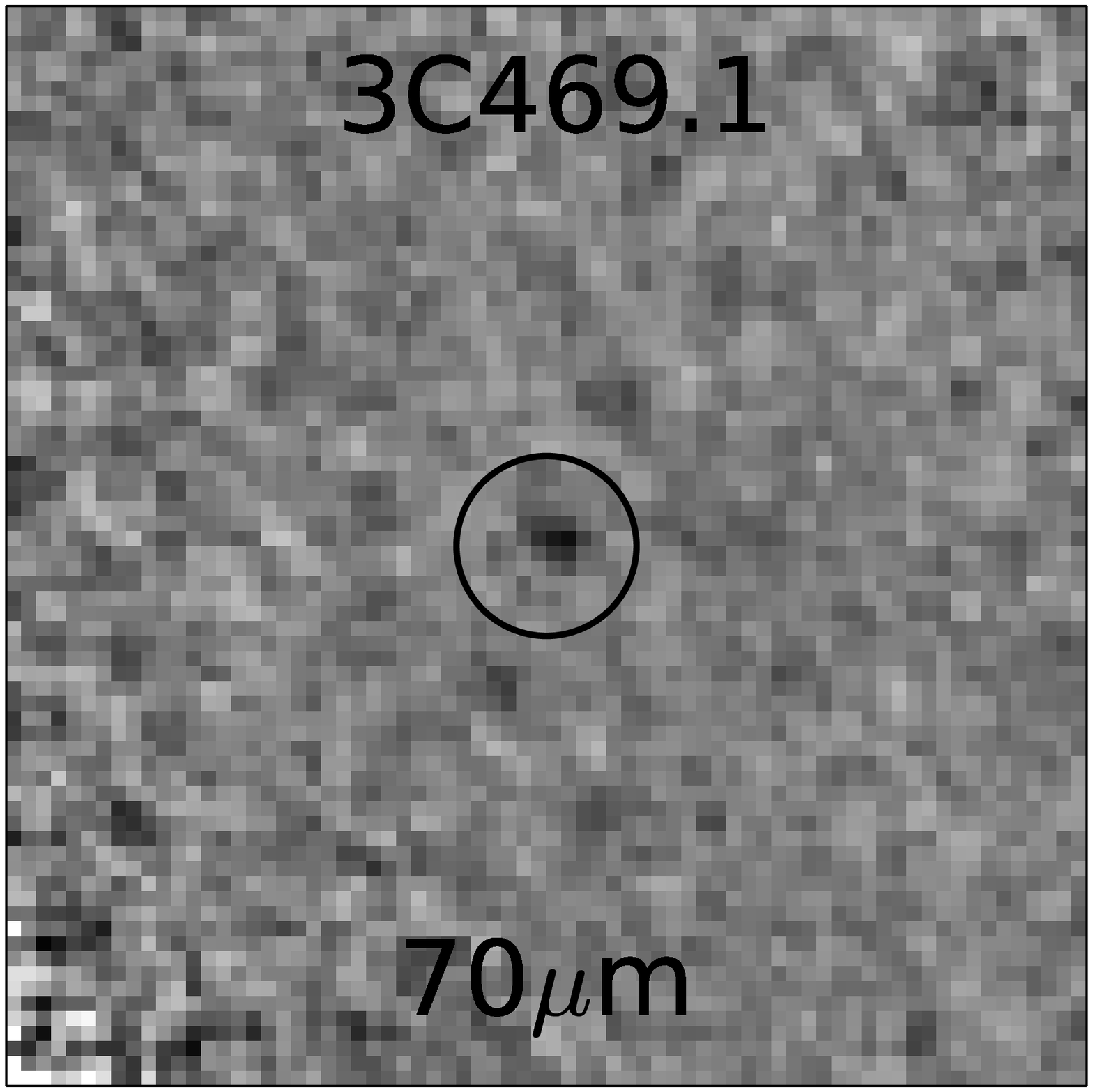}
      \includegraphics[width=1.5cm]{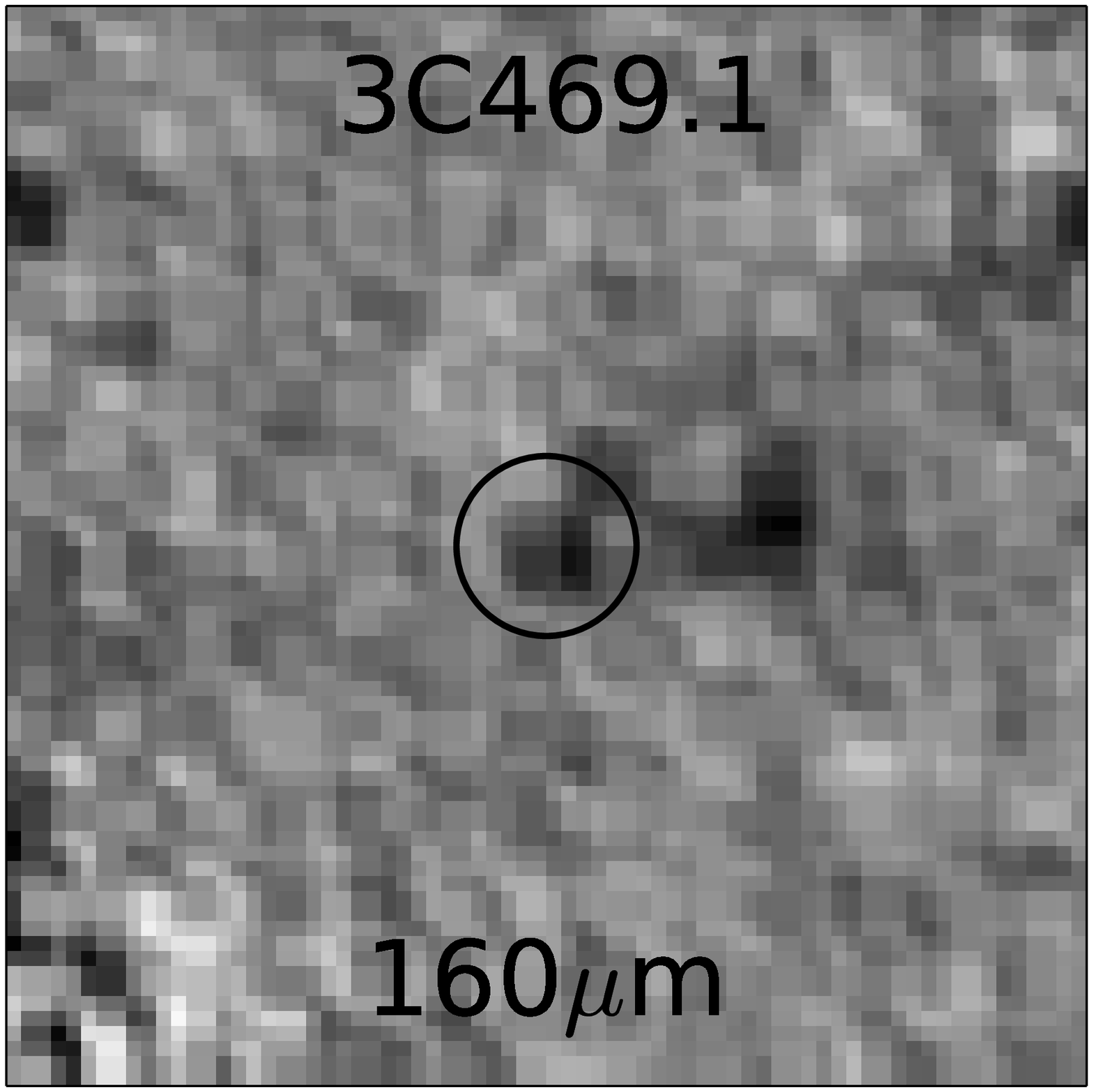}
      \includegraphics[width=1.5cm]{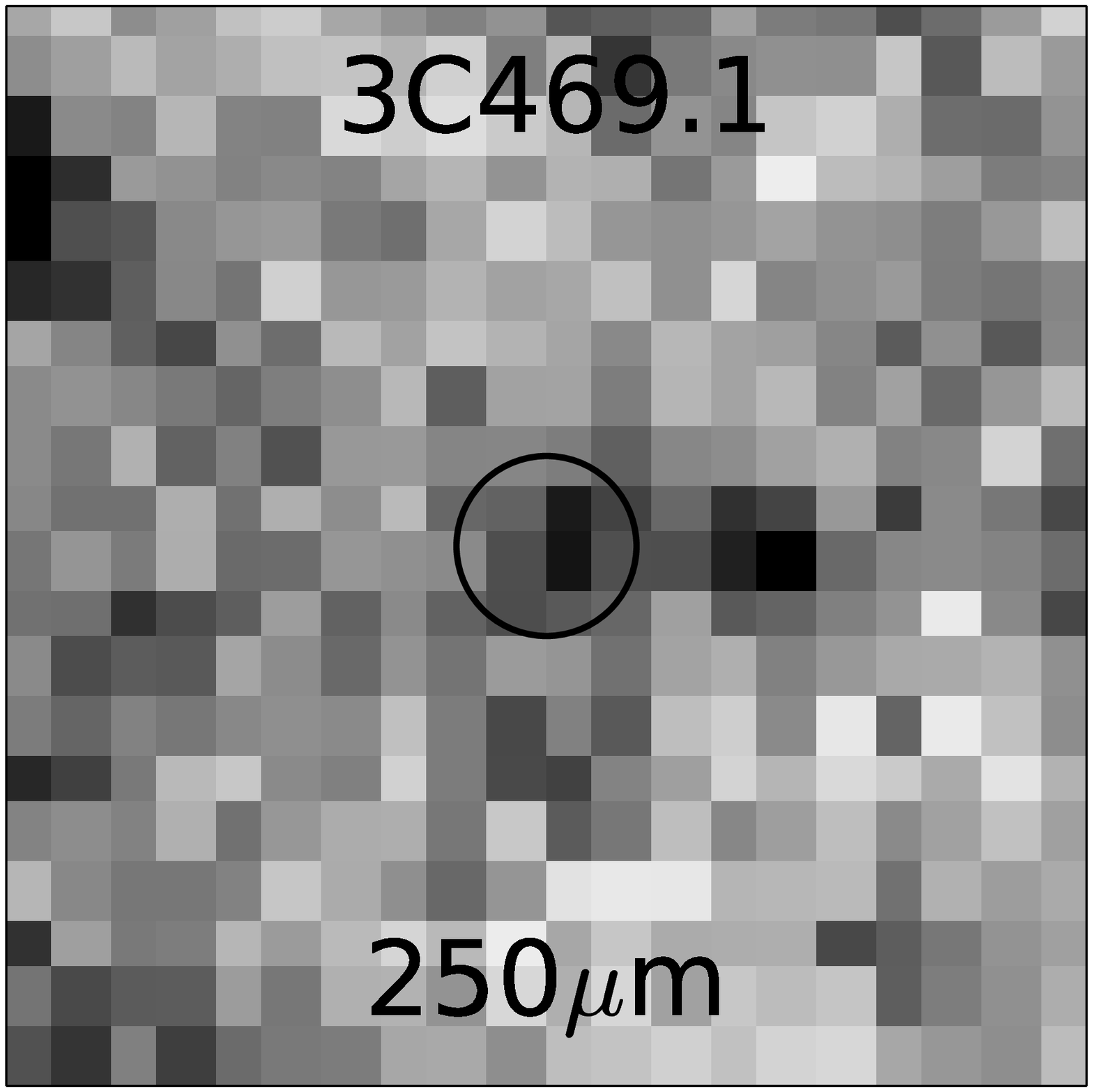}
      \includegraphics[width=1.5cm]{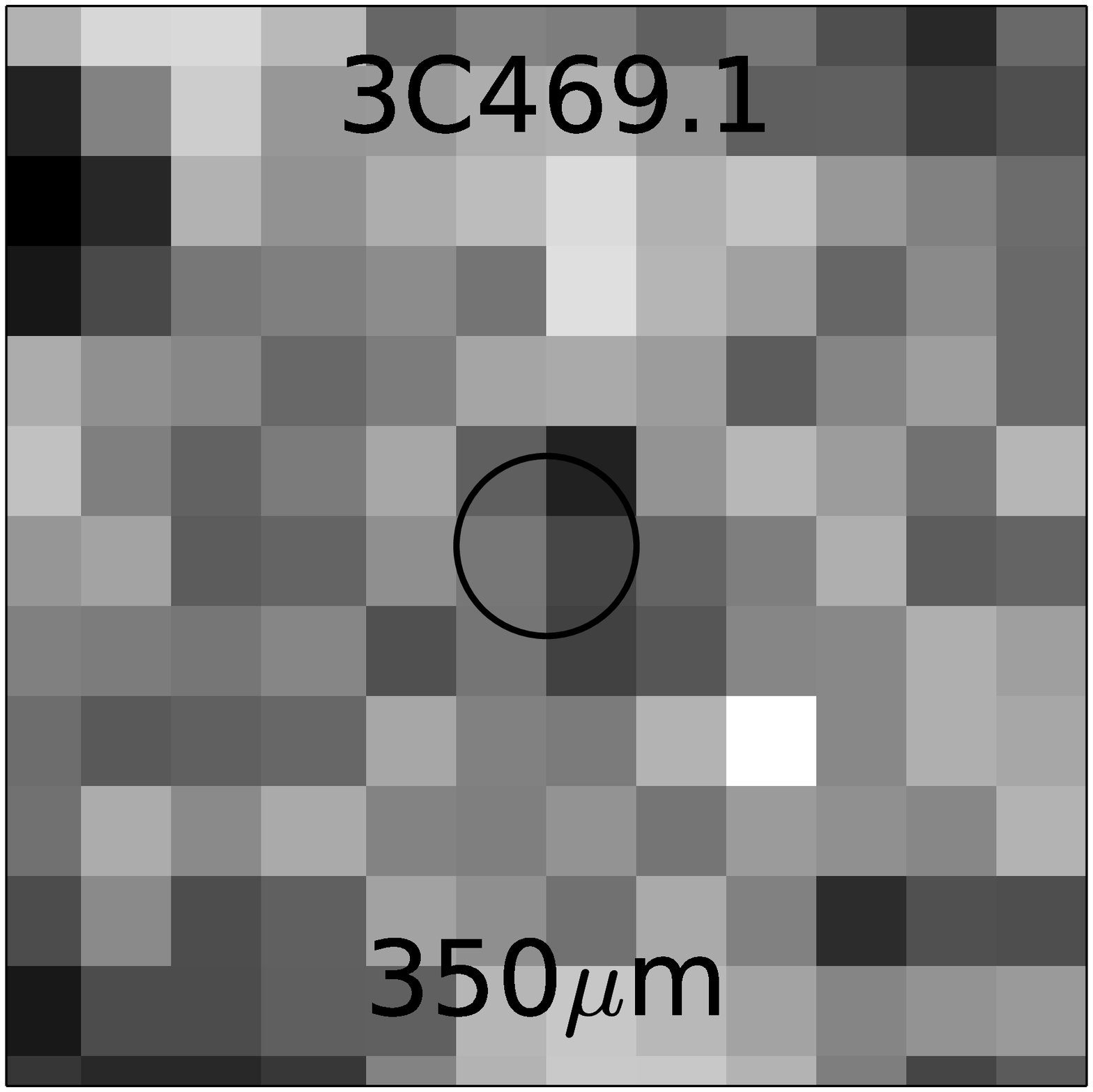}
      \includegraphics[width=1.5cm]{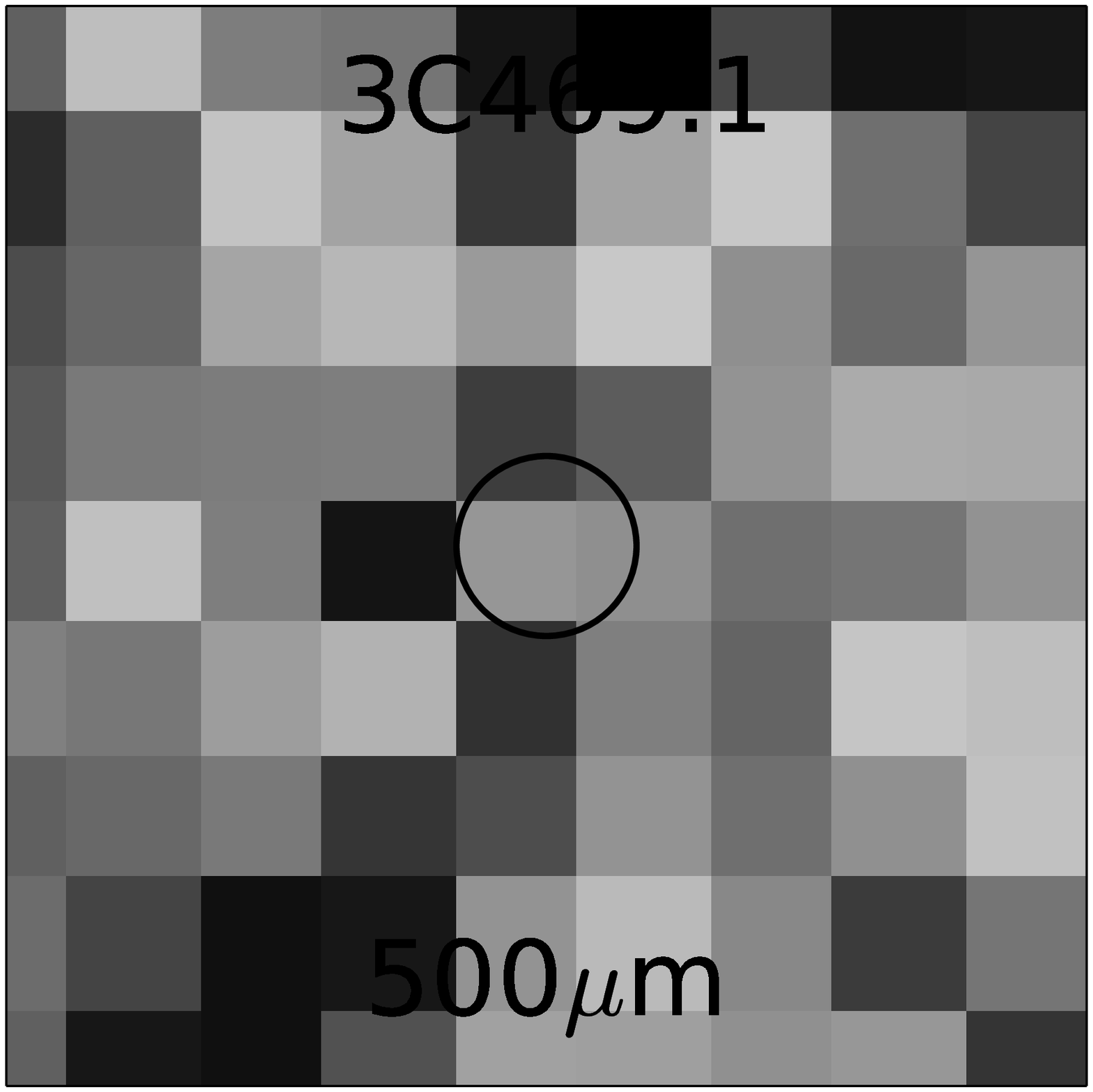}
      \\
      \includegraphics[width=1.5cm]{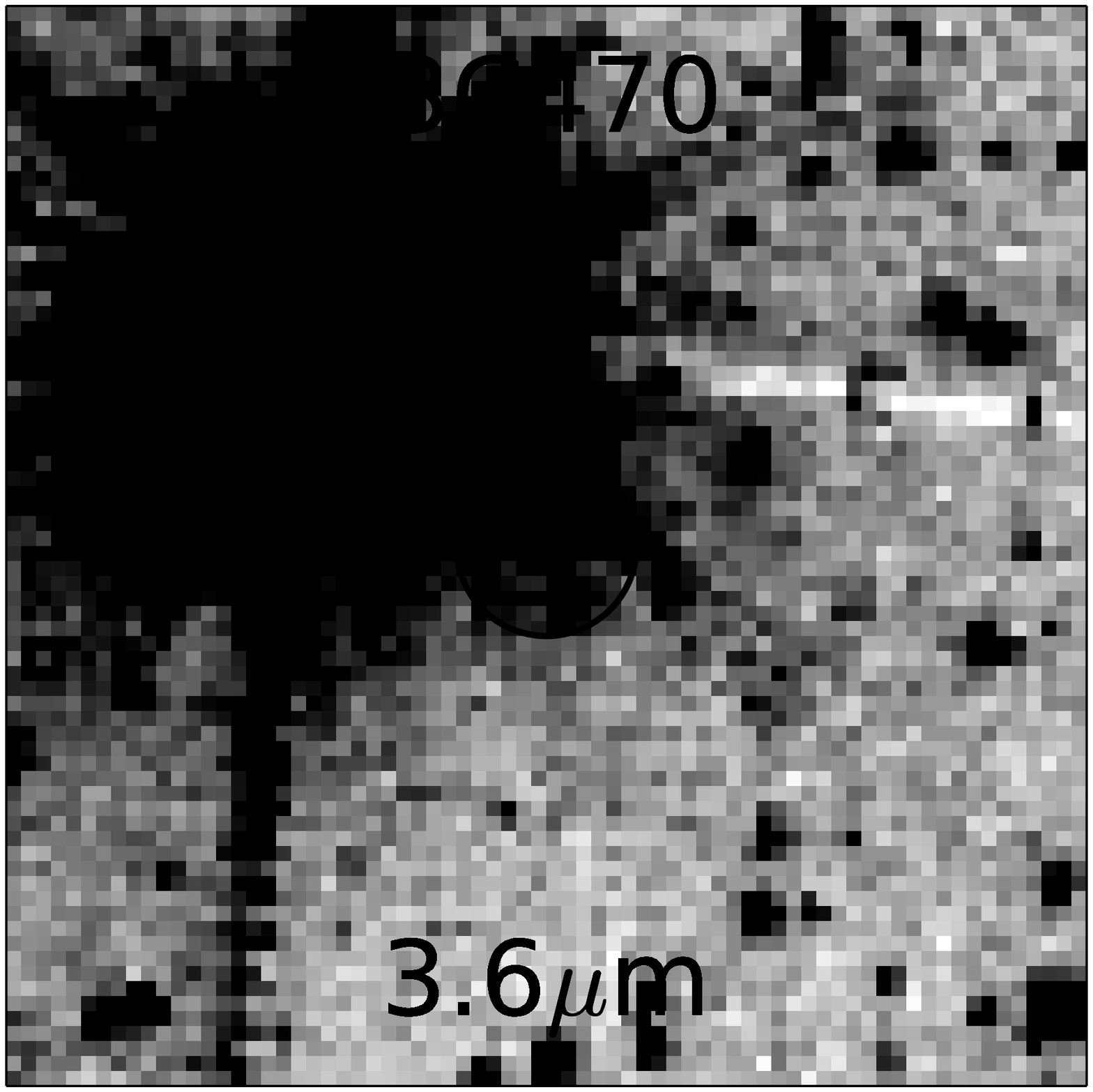}
      \includegraphics[width=1.5cm]{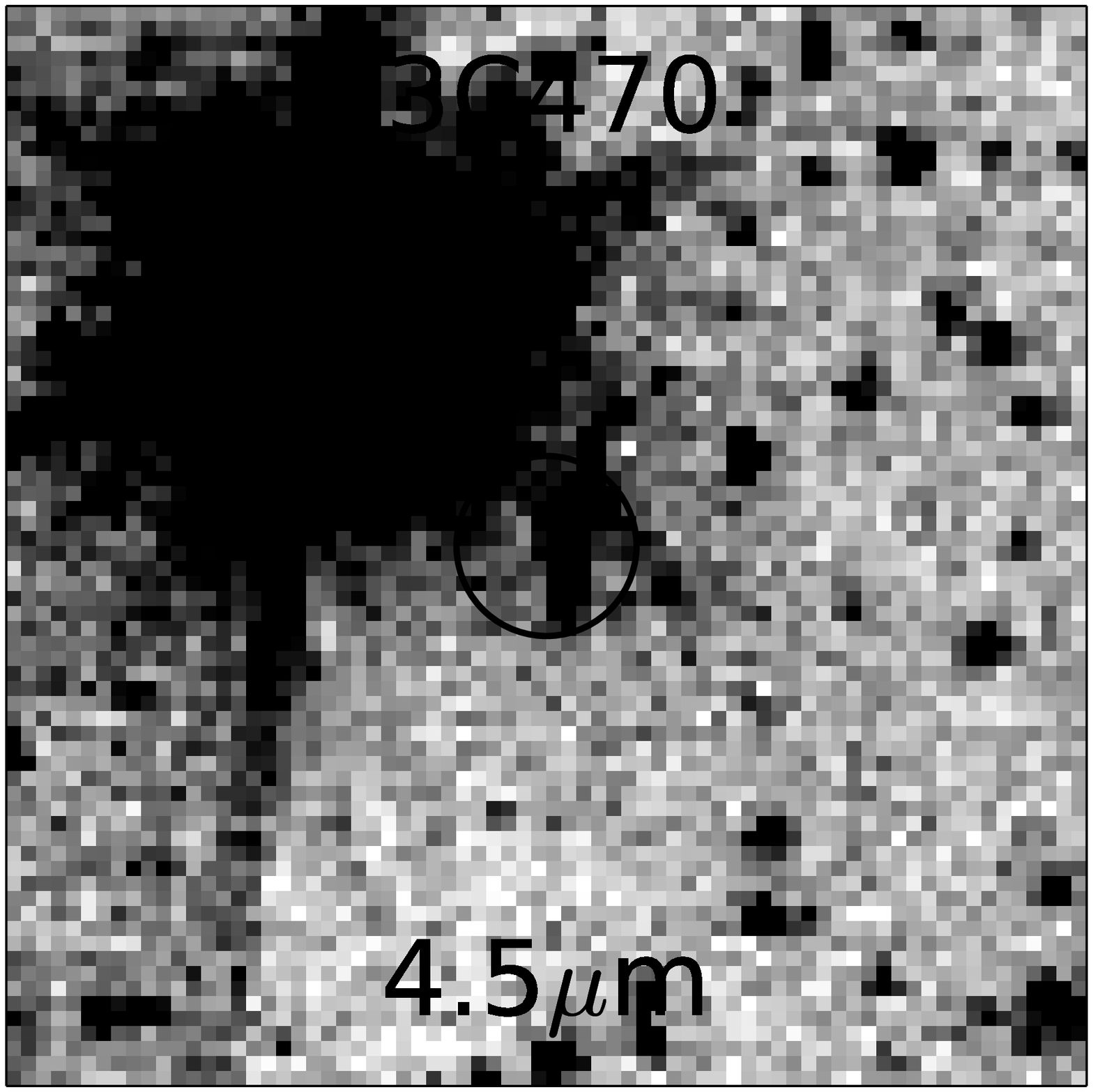}
      \includegraphics[width=1.5cm]{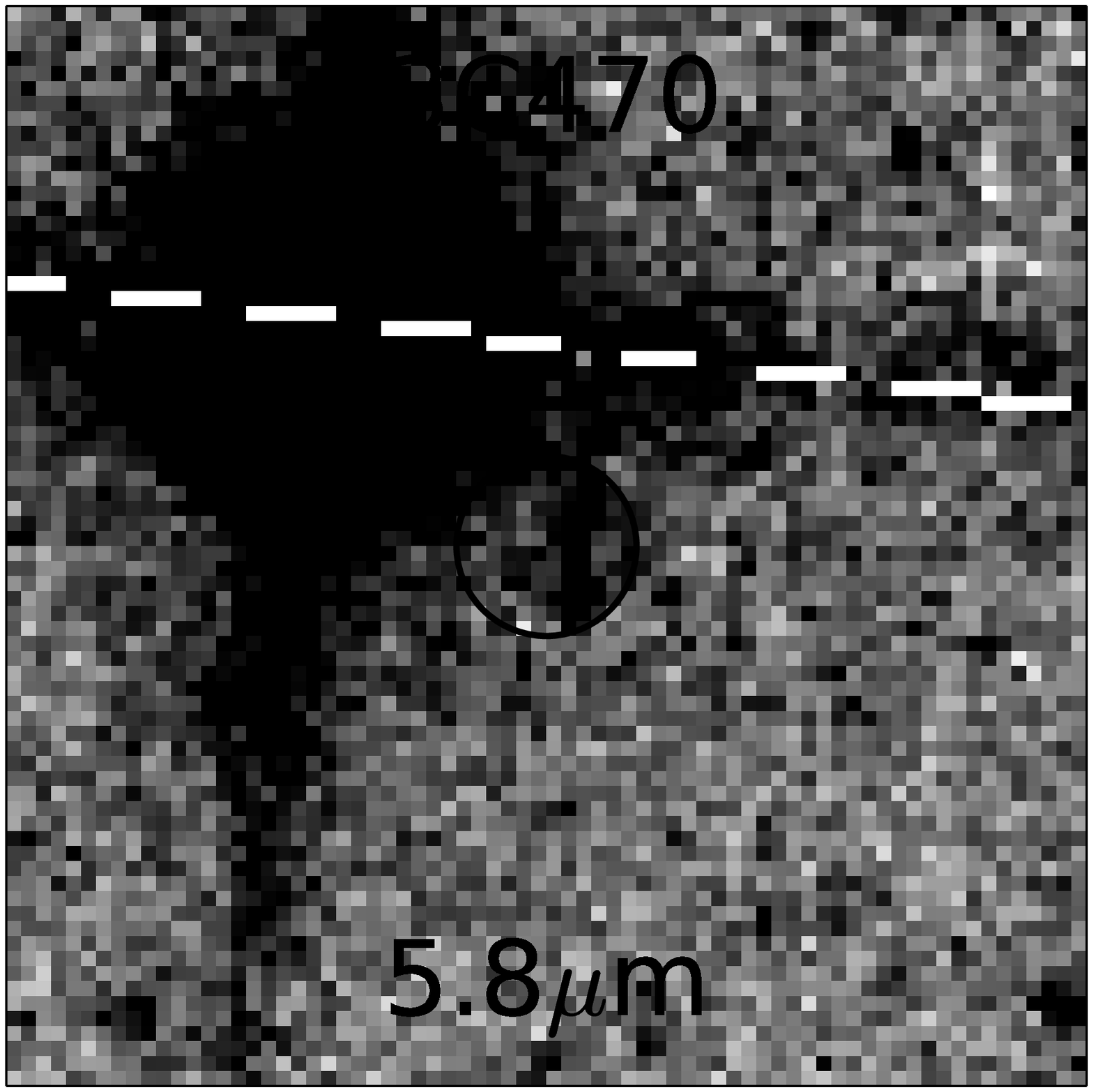}
      \includegraphics[width=1.5cm]{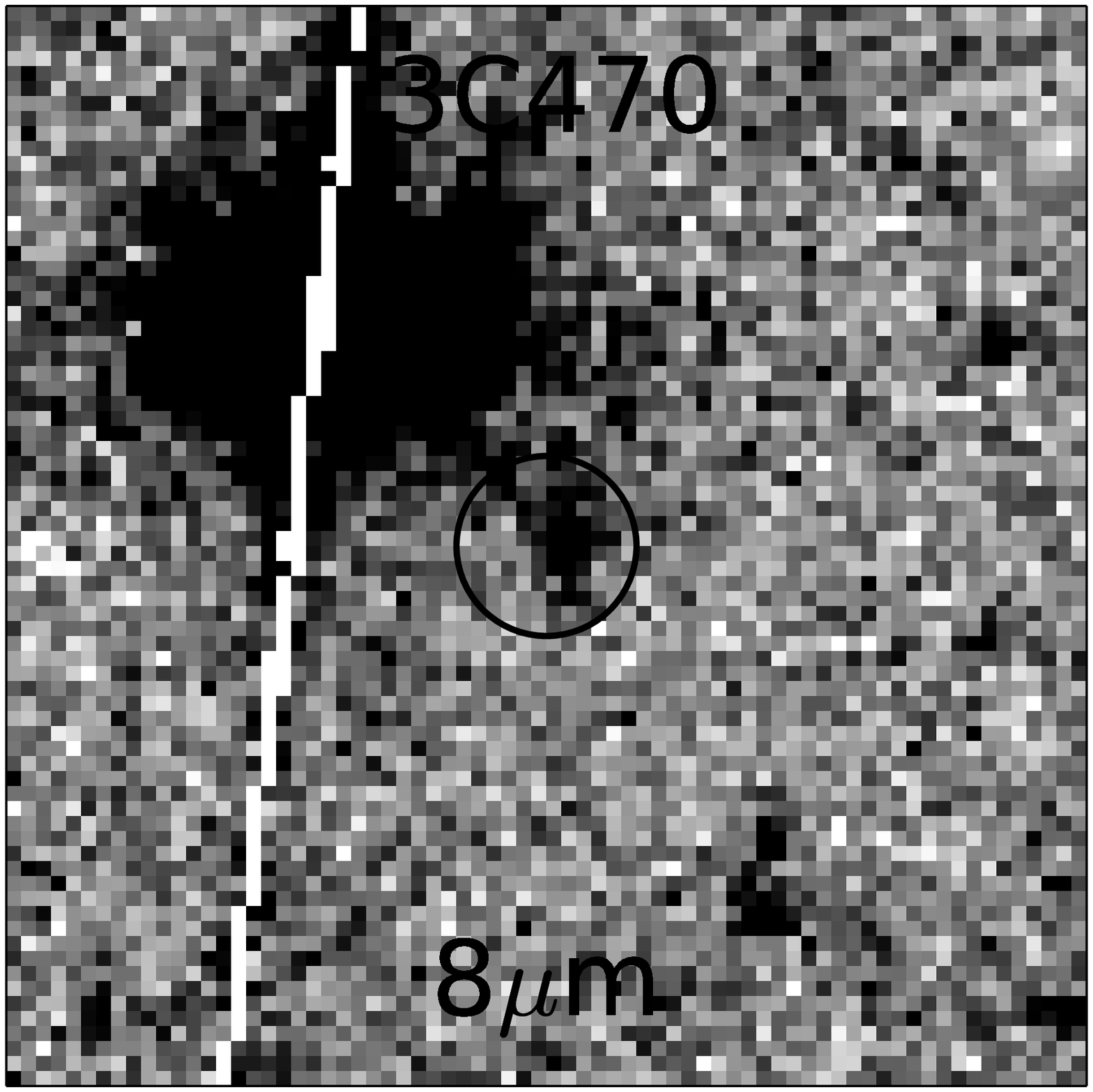}
      \includegraphics[width=1.5cm]{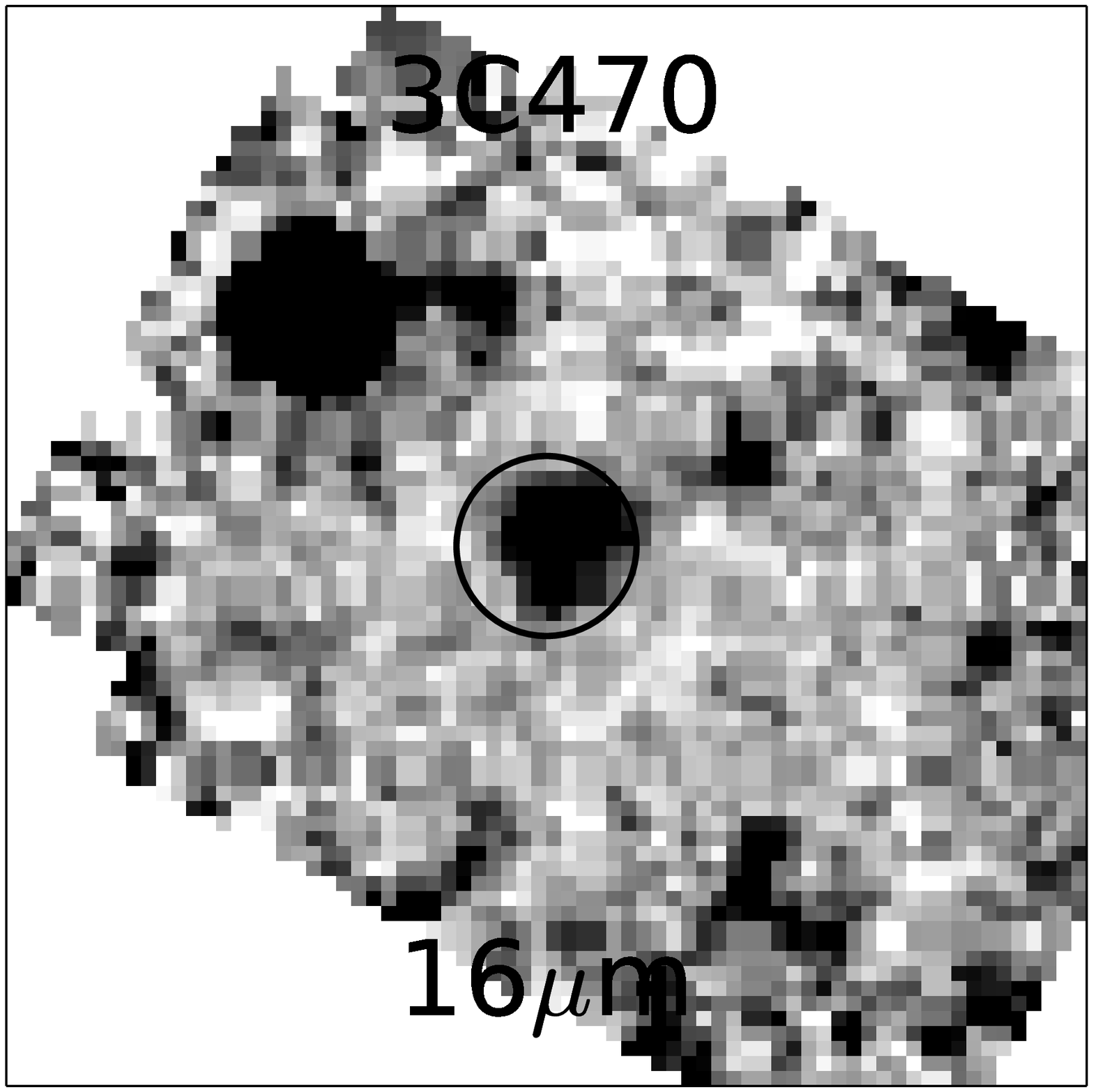}
      \includegraphics[width=1.5cm]{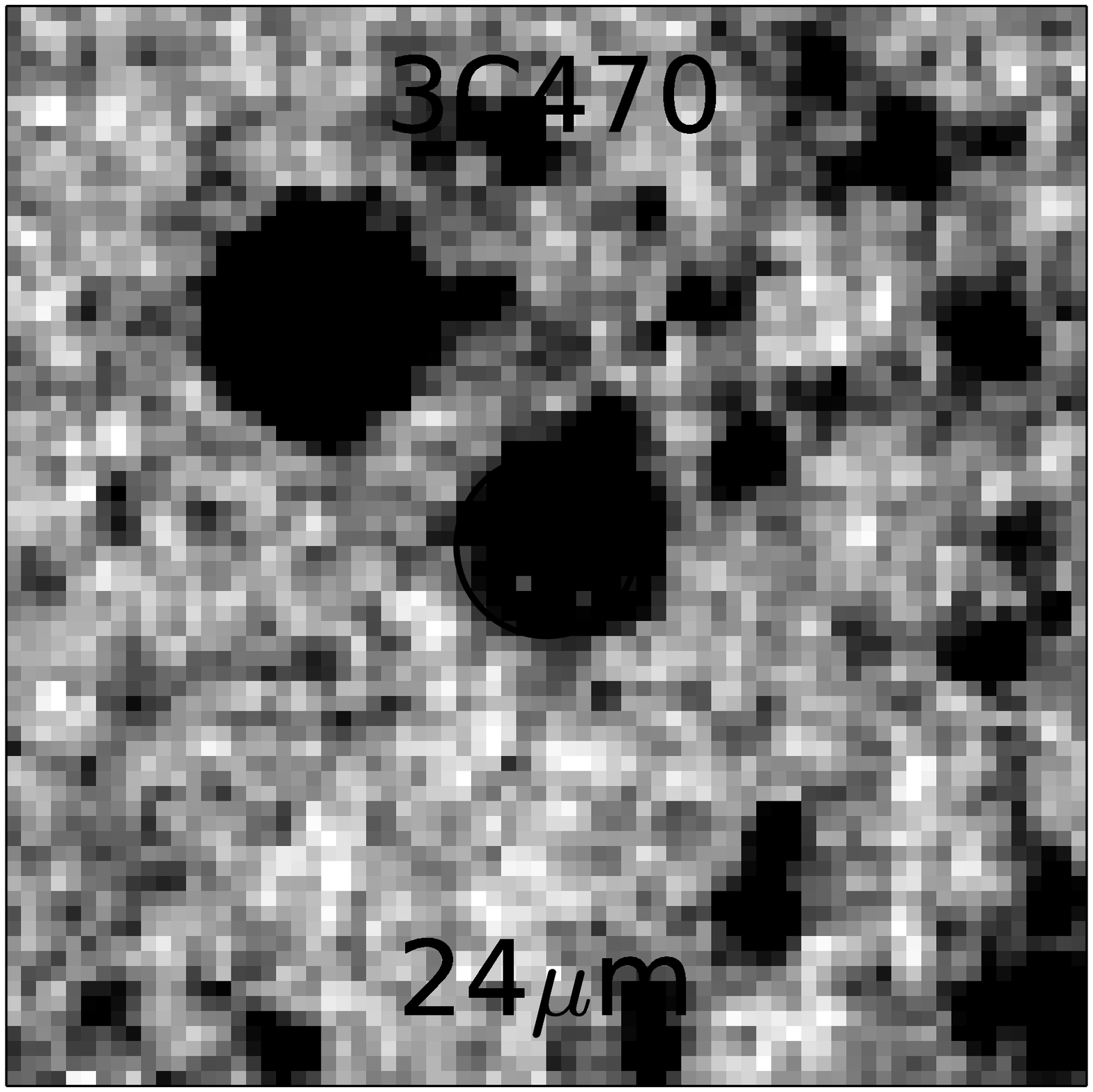}
      \includegraphics[width=1.5cm]{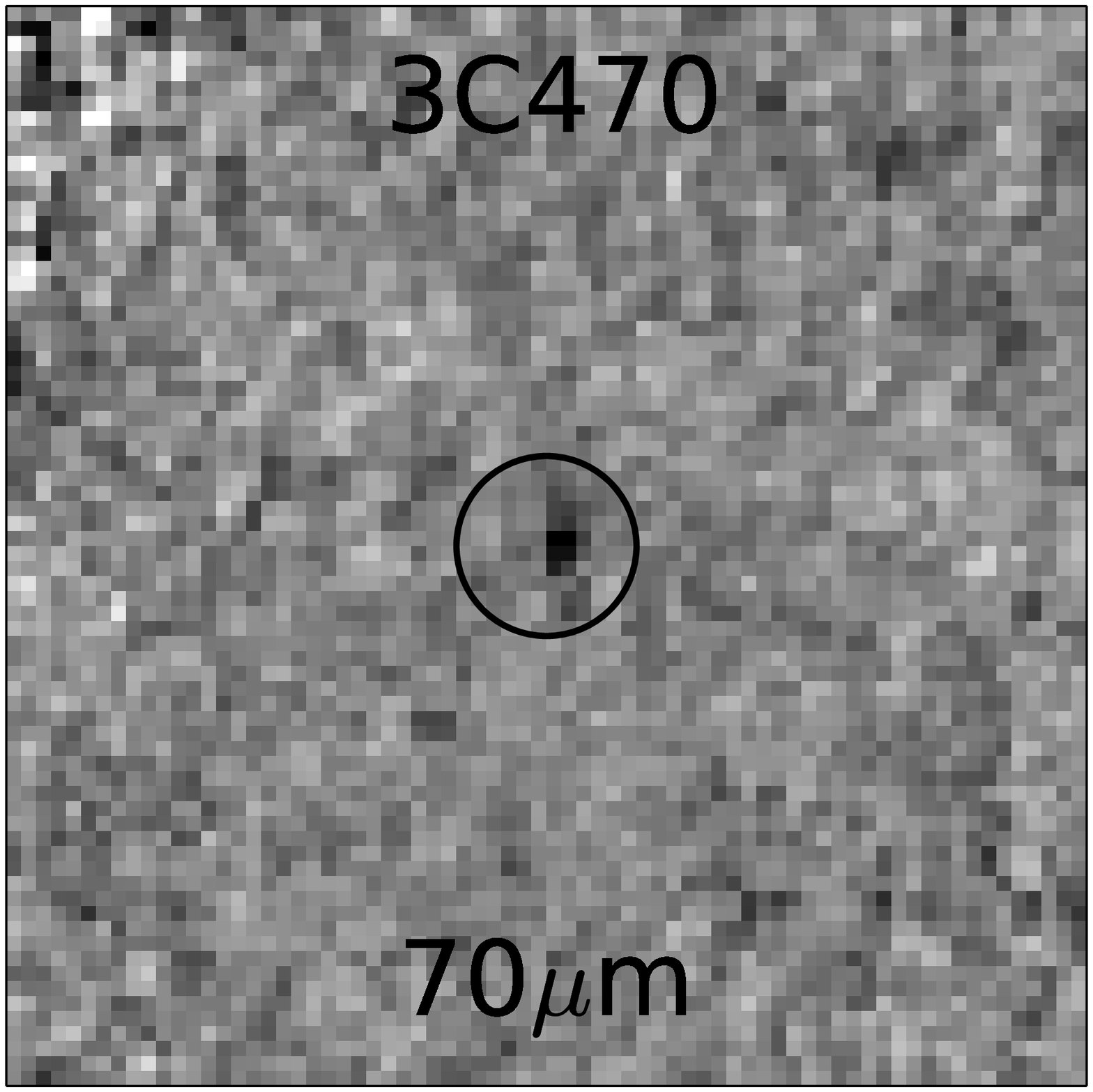}
      \includegraphics[width=1.5cm]{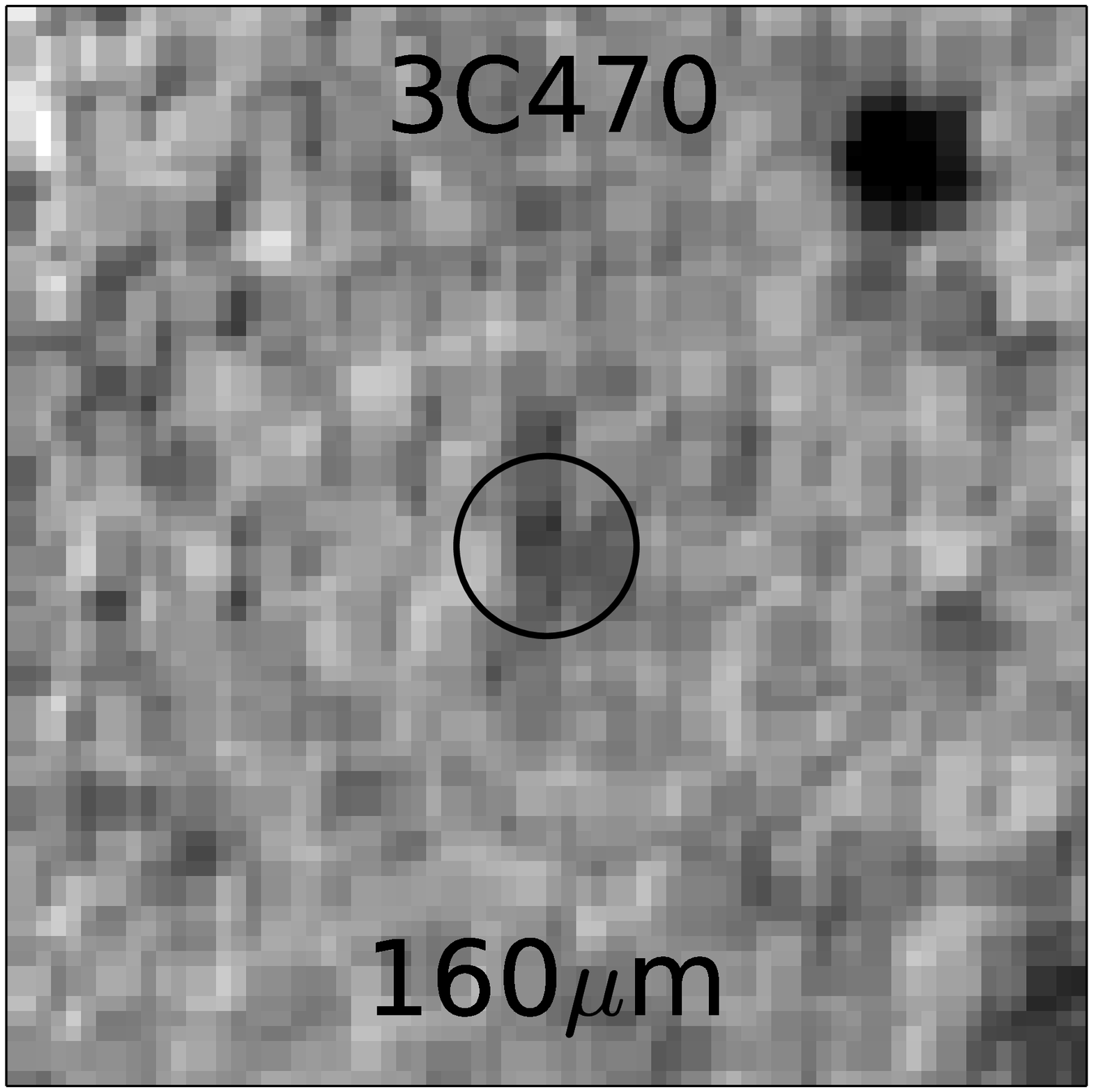}
      \includegraphics[width=1.5cm]{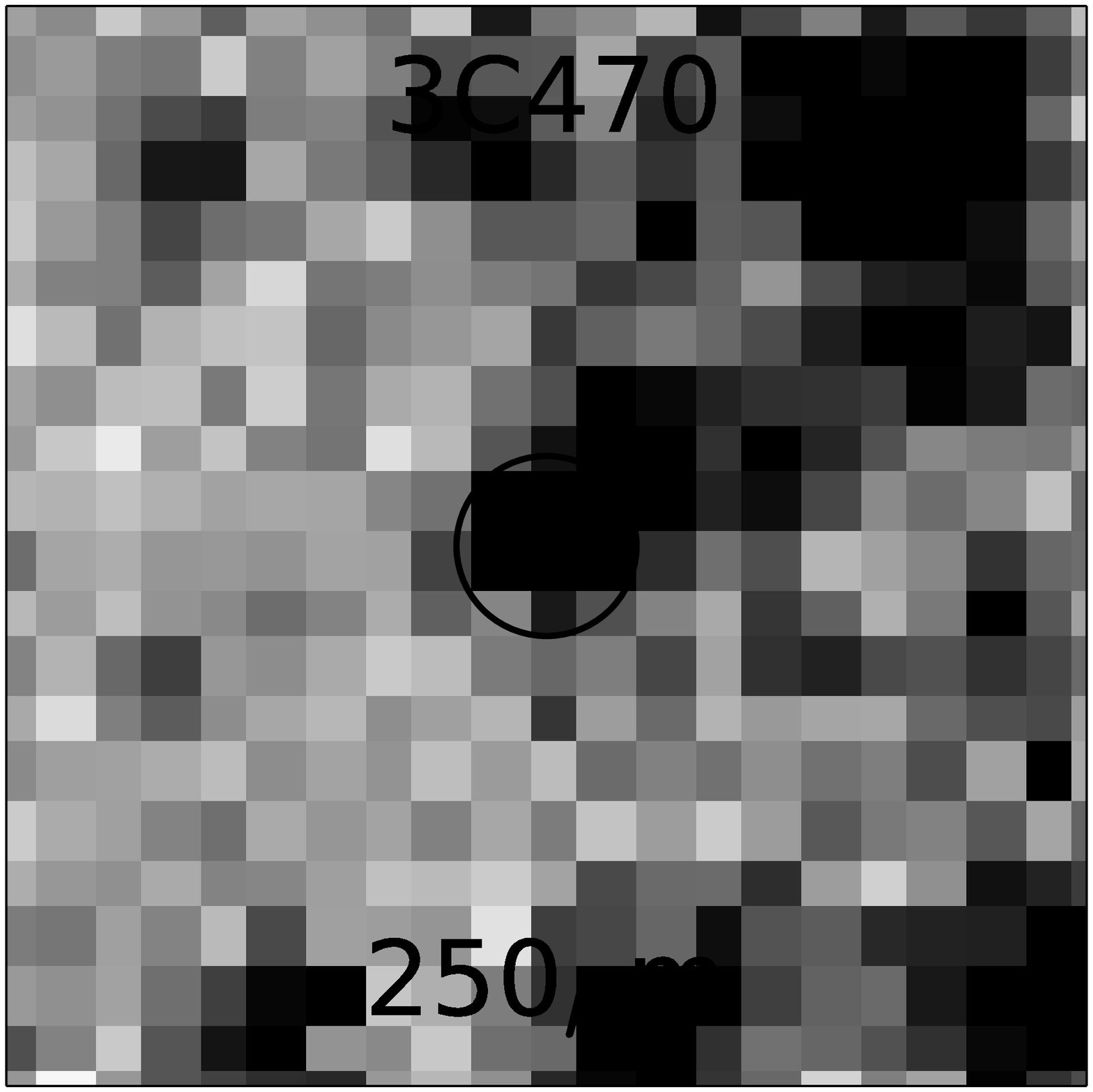}
      \includegraphics[width=1.5cm]{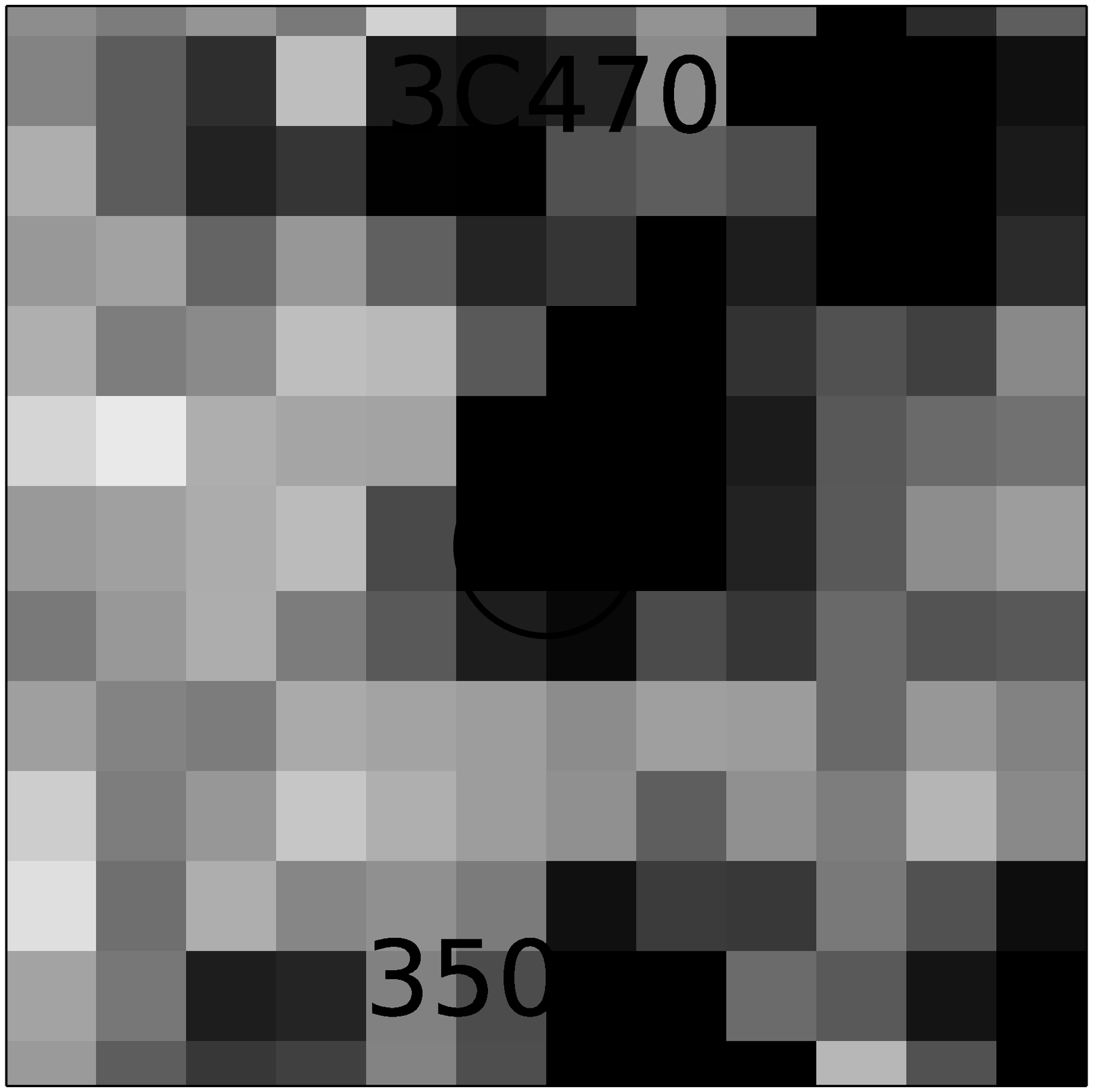}
      \includegraphics[width=1.5cm]{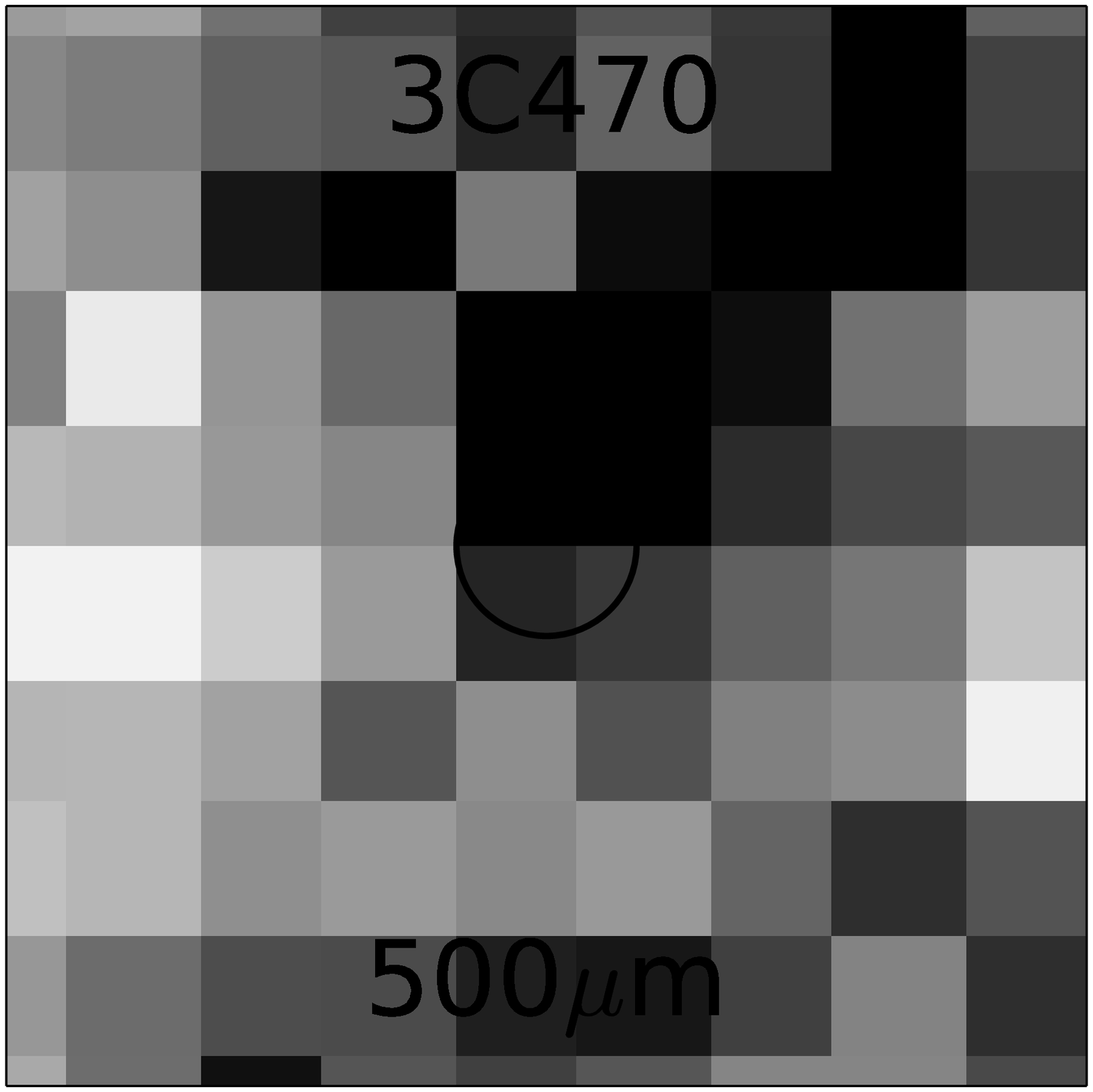}
      \\
      \includegraphics[width=1.5cm]{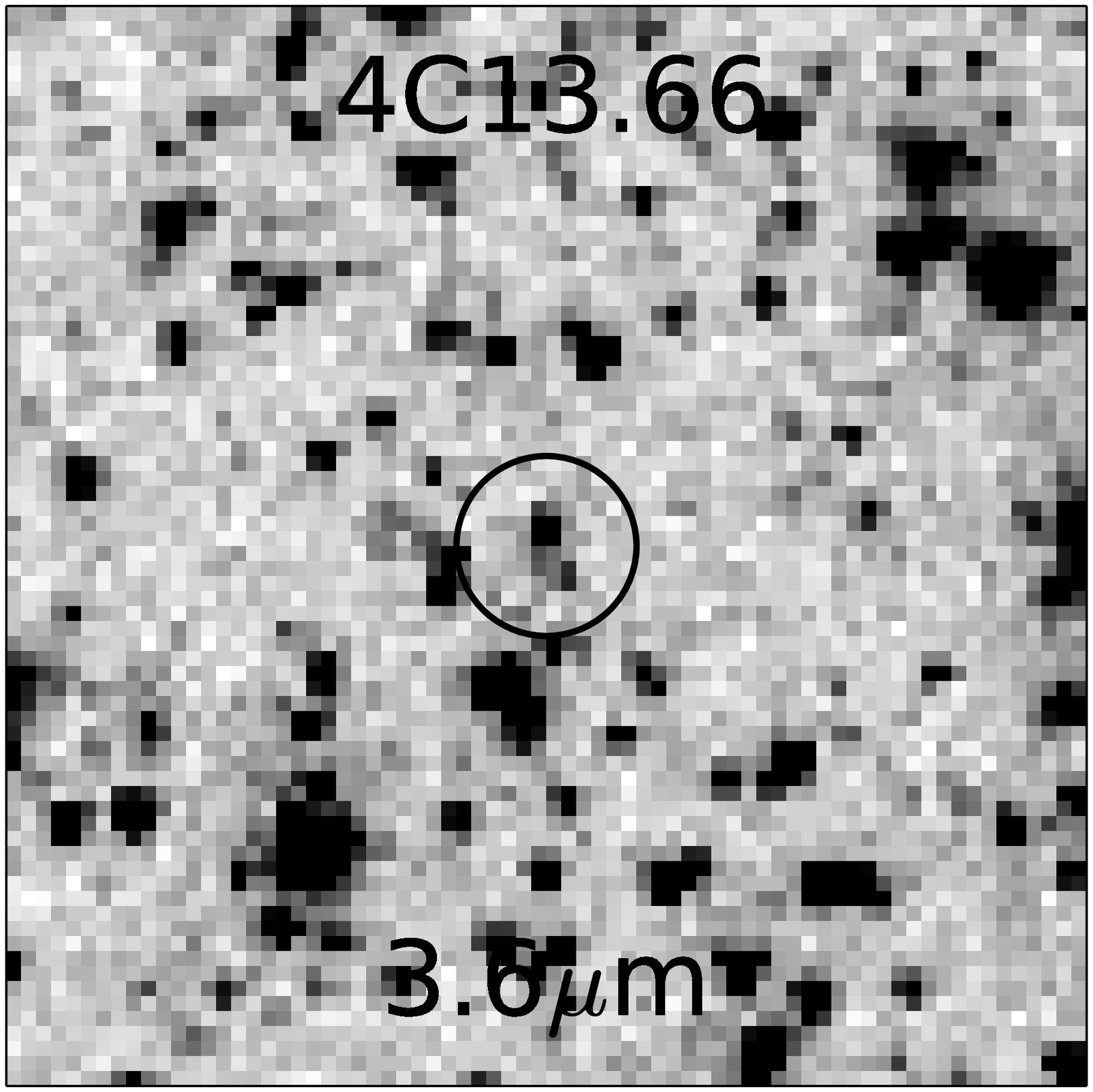}
      \includegraphics[width=1.5cm]{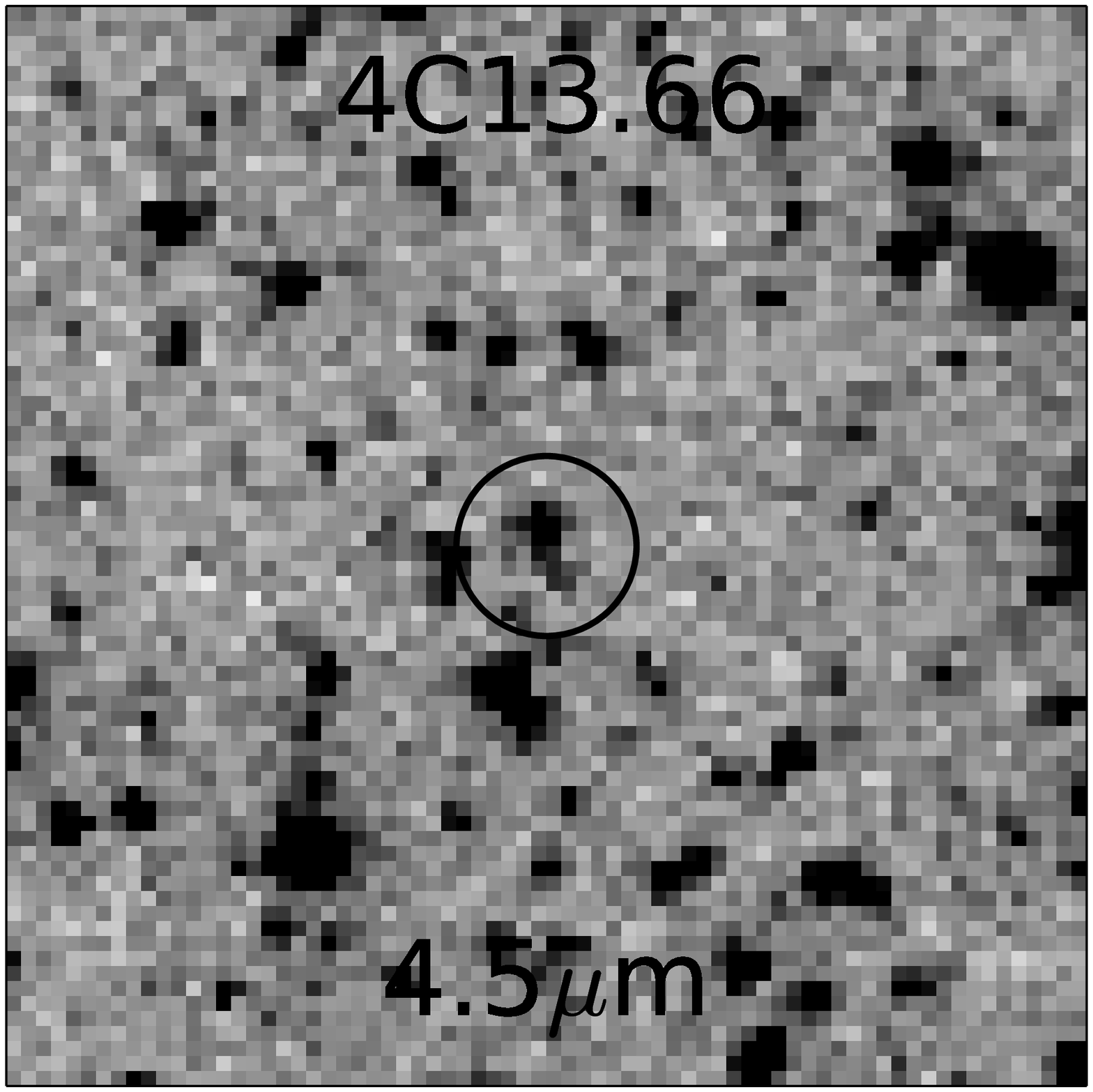}
      \includegraphics[width=1.5cm]{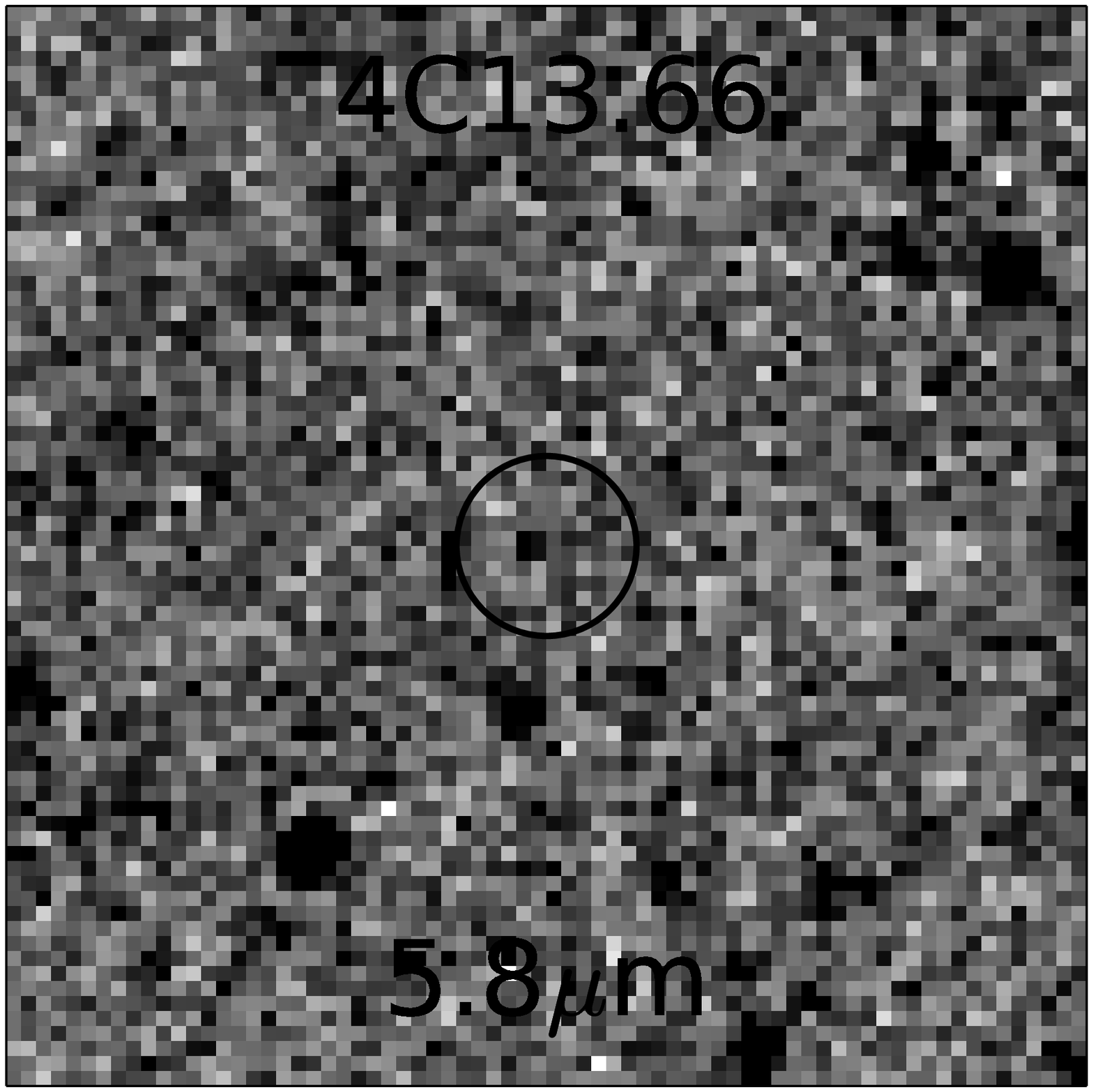}
      \includegraphics[width=1.5cm]{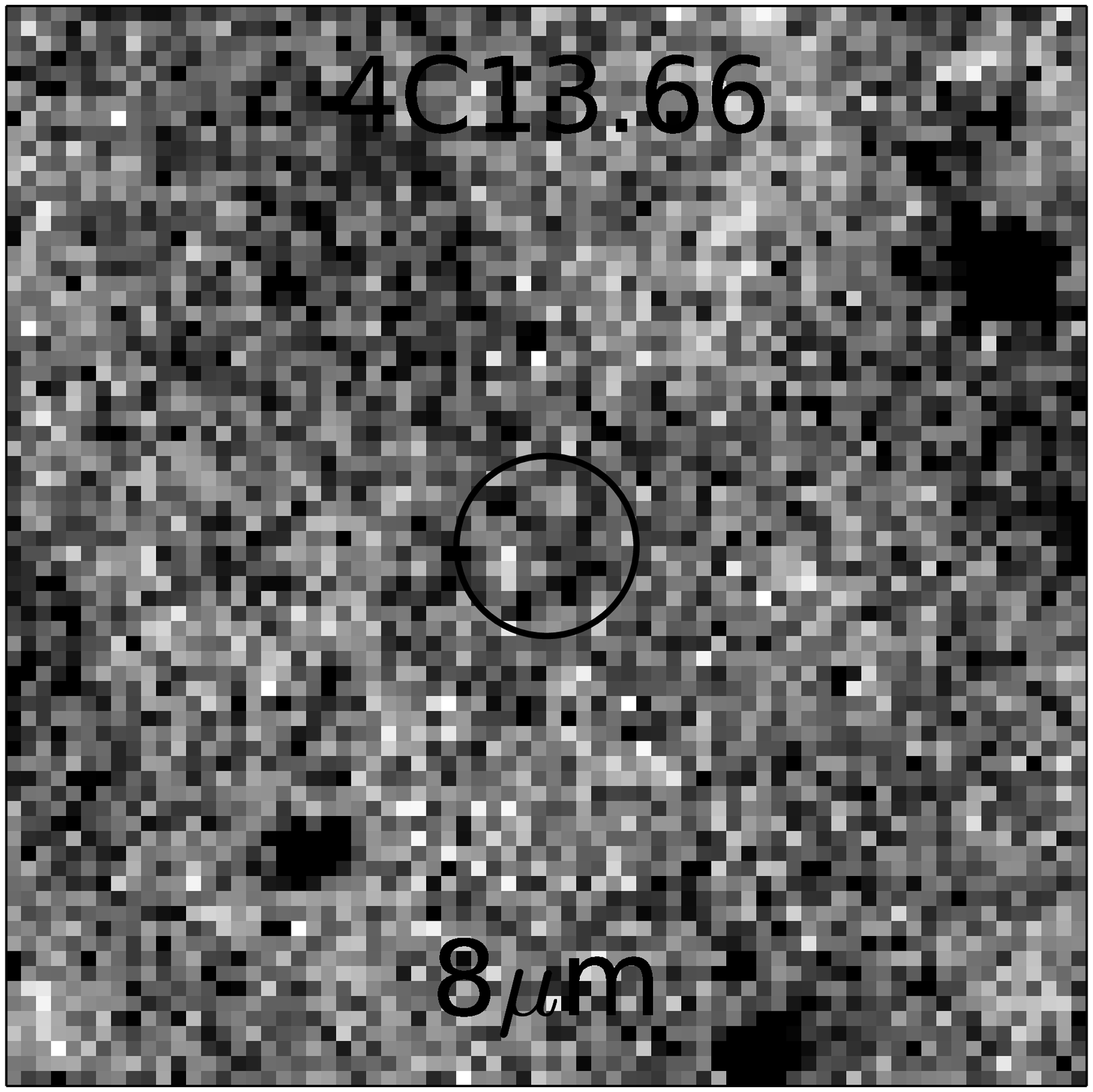}
      \includegraphics[width=1.5cm]{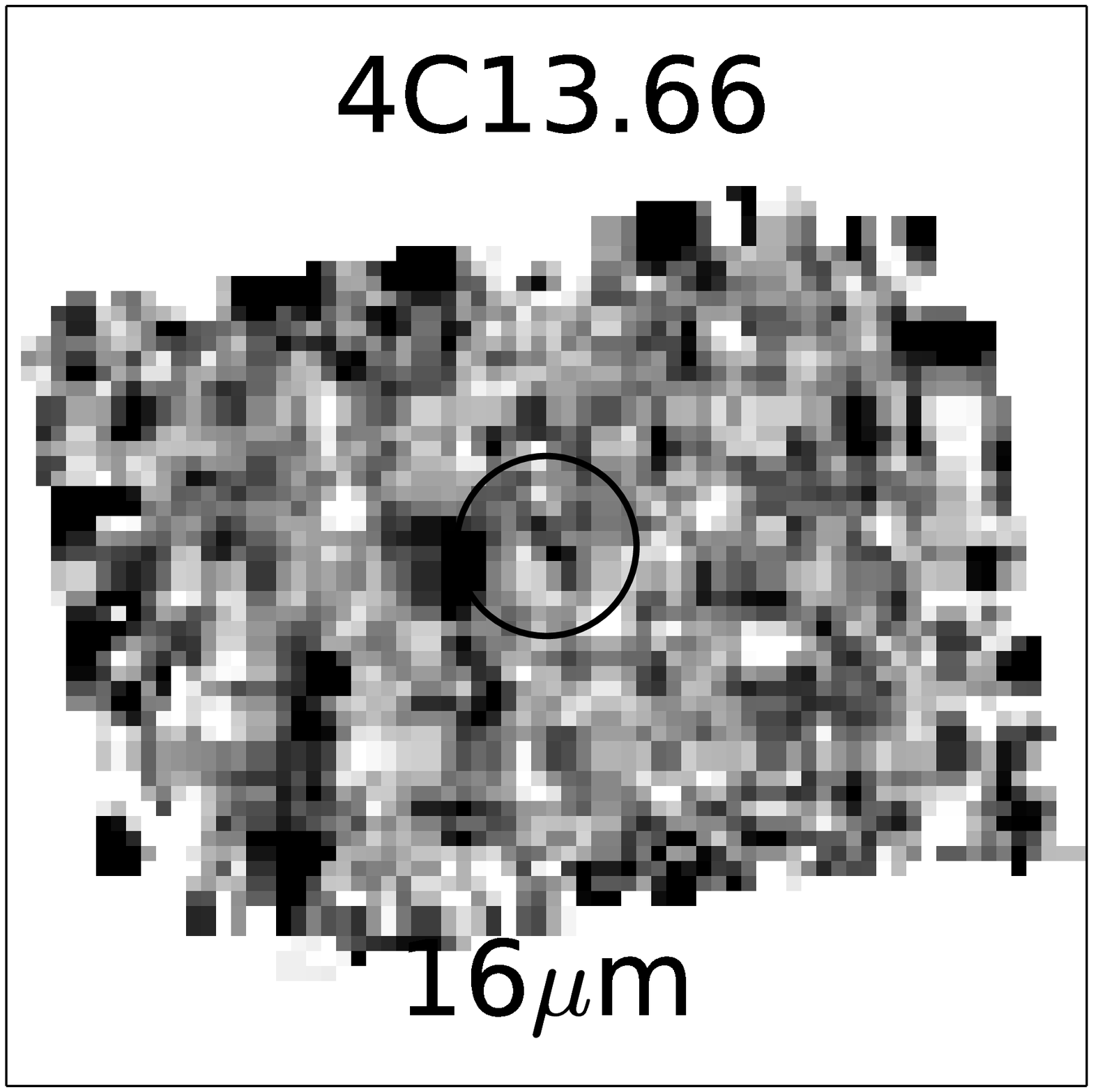}
      \includegraphics[width=1.5cm]{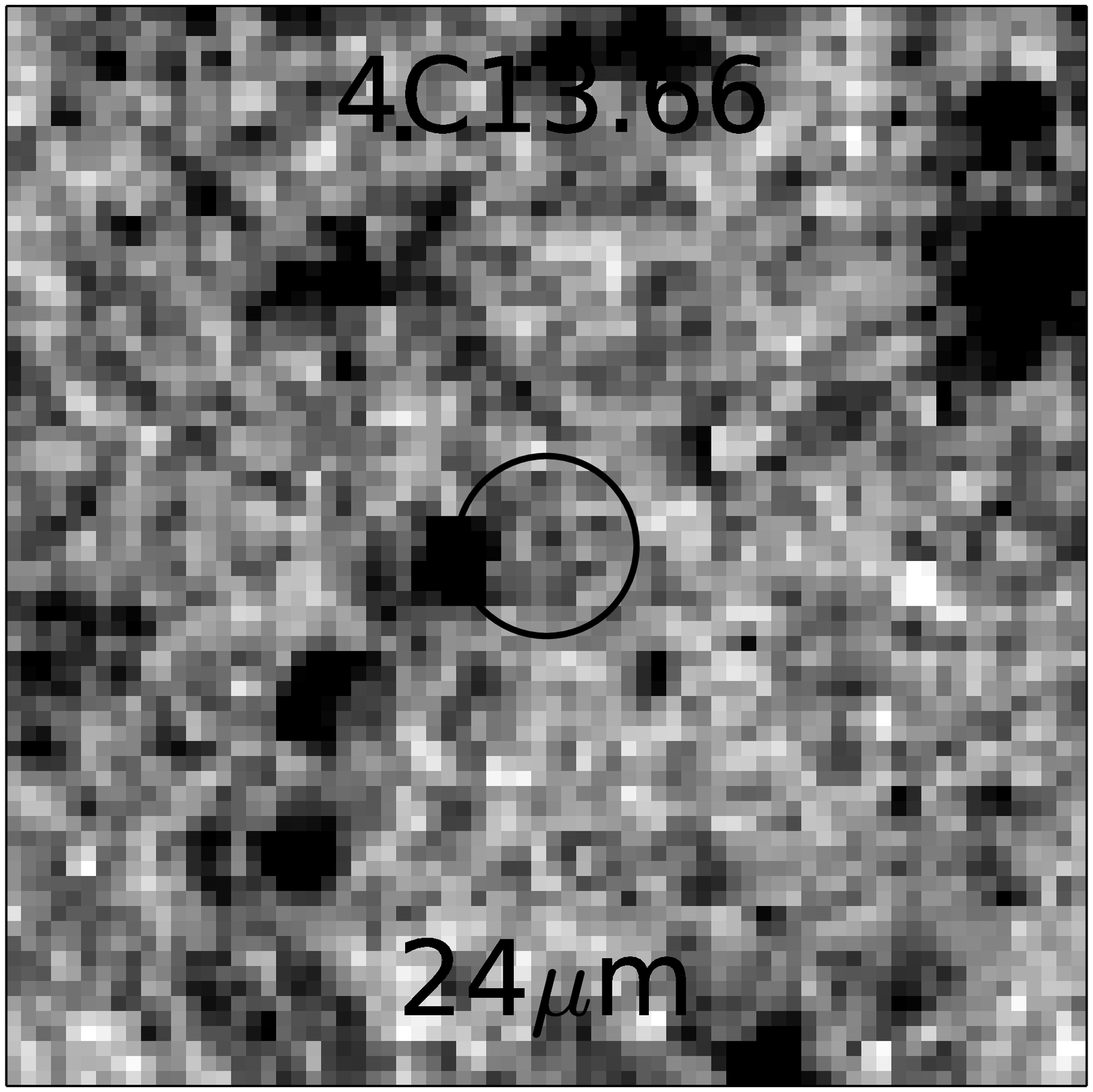}
      \includegraphics[width=1.5cm]{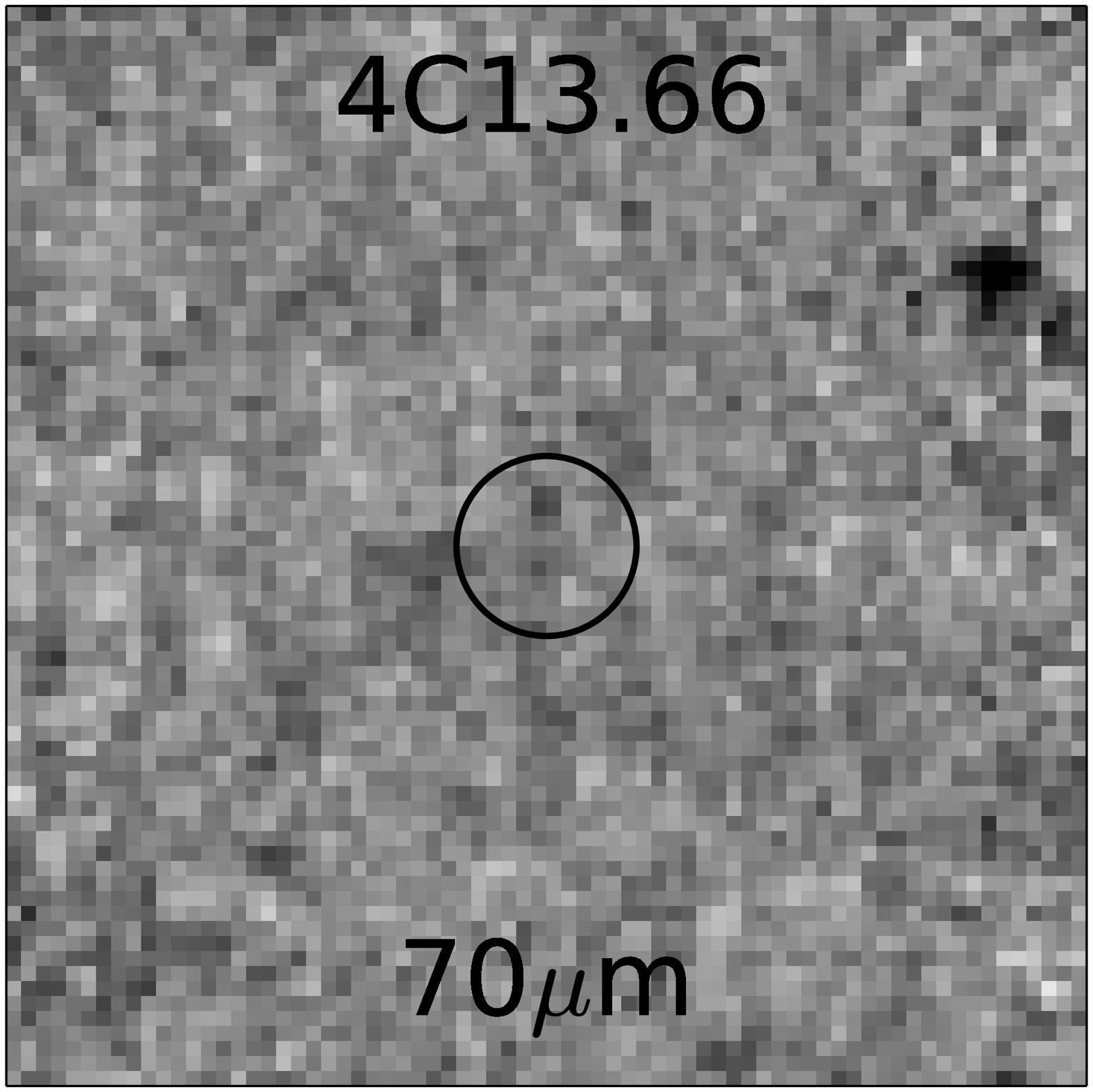}
      \includegraphics[width=1.5cm]{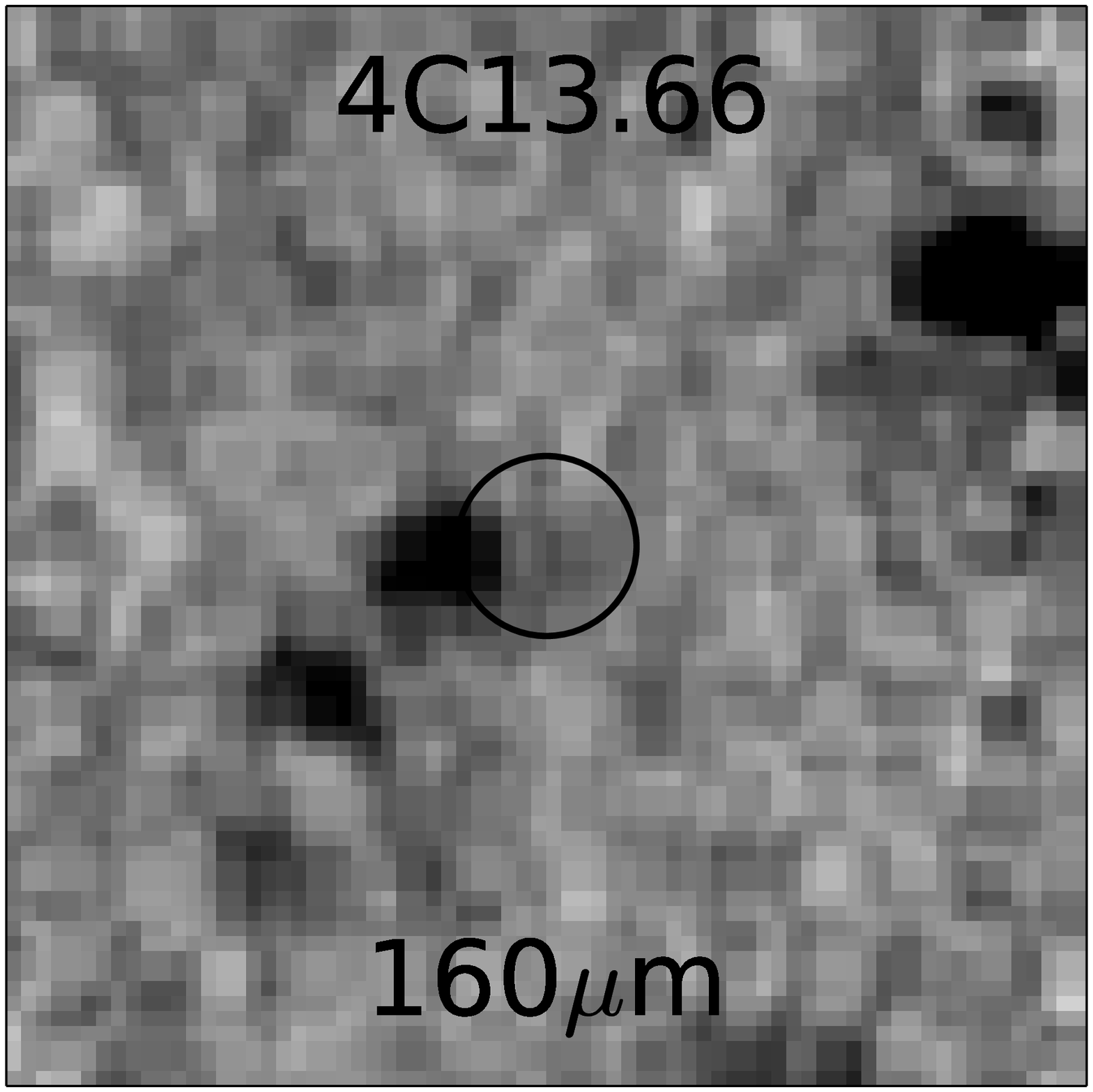}
      \includegraphics[width=1.5cm]{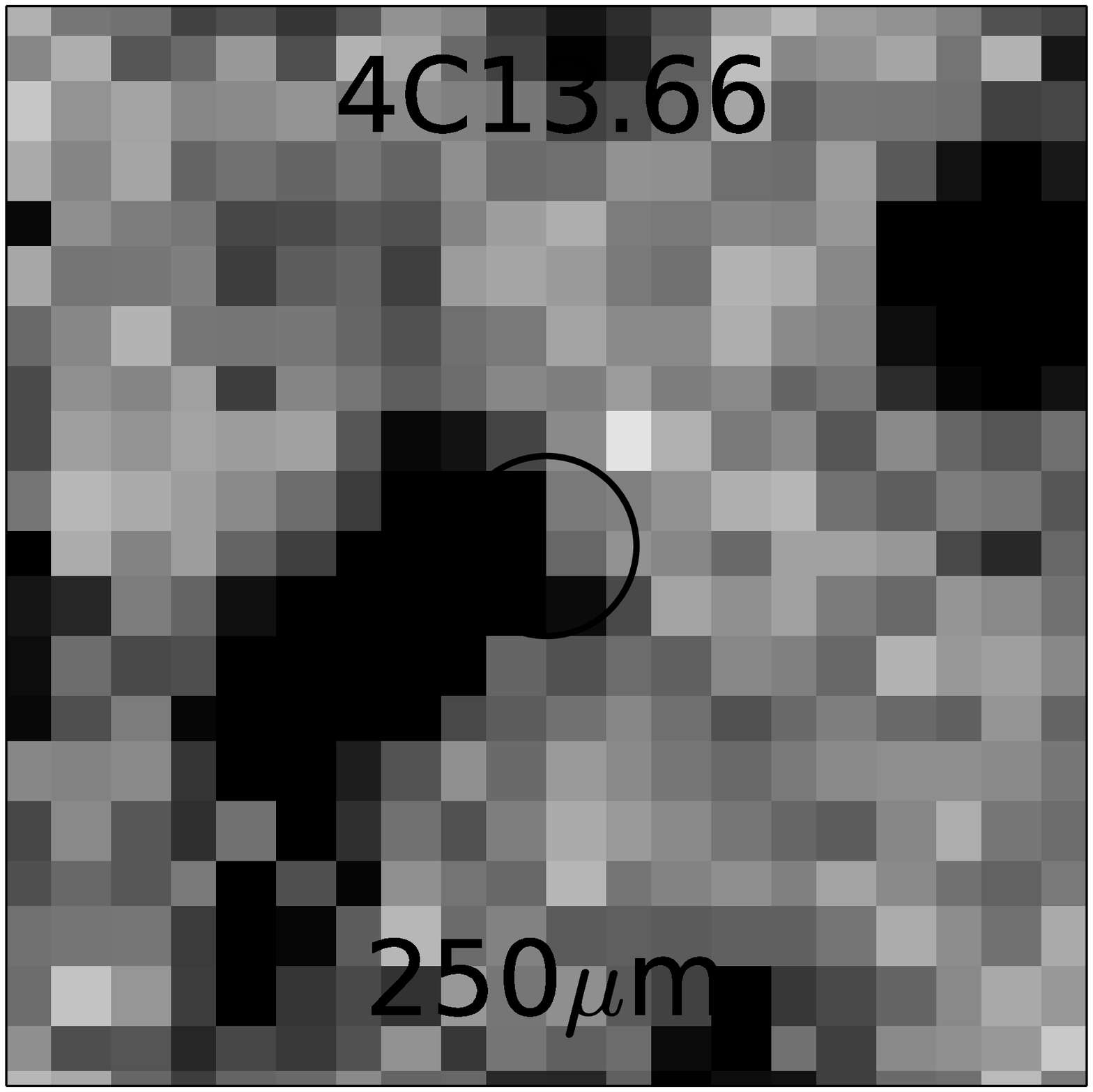}
      \includegraphics[width=1.5cm]{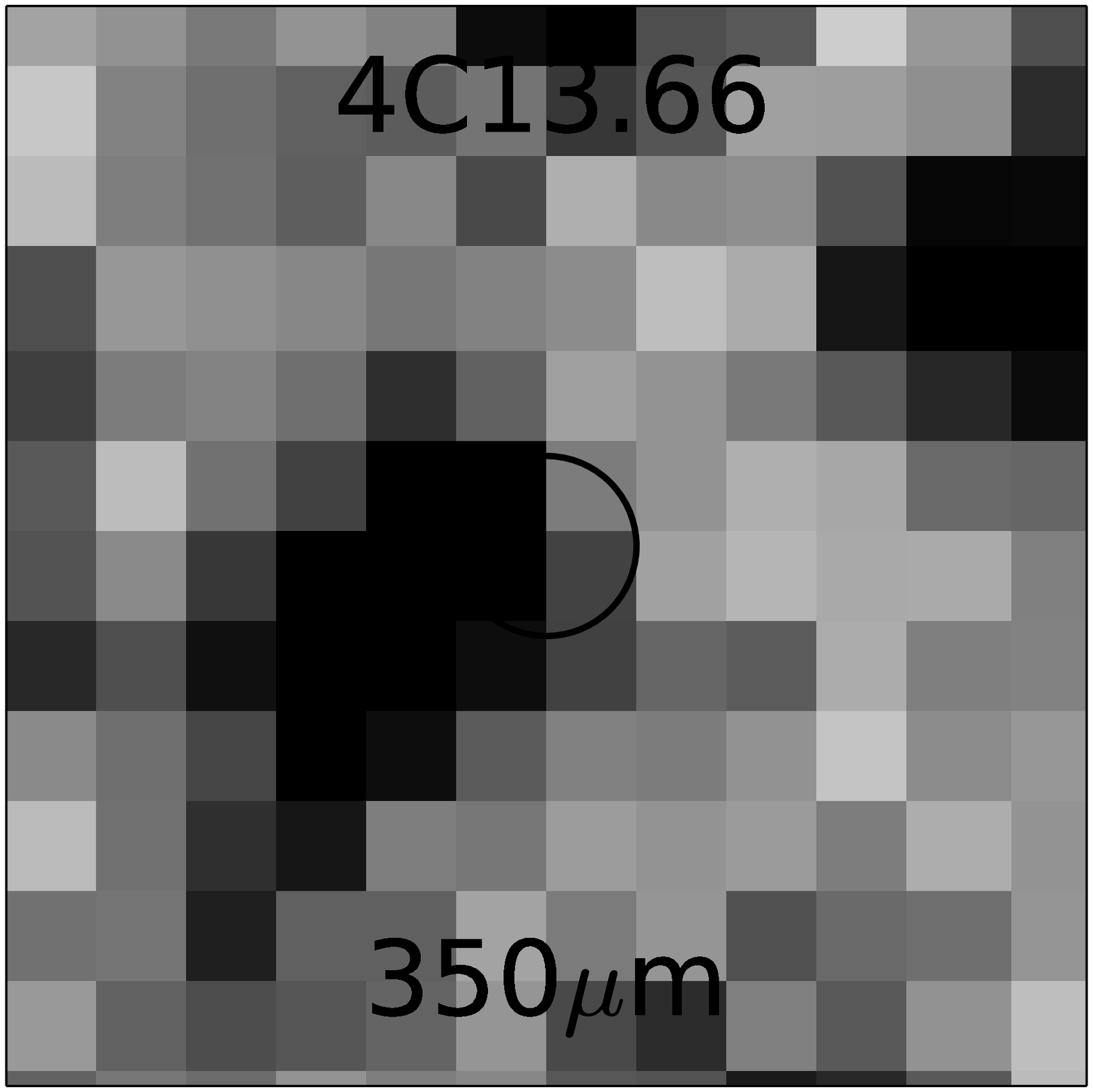}
      \includegraphics[width=1.5cm]{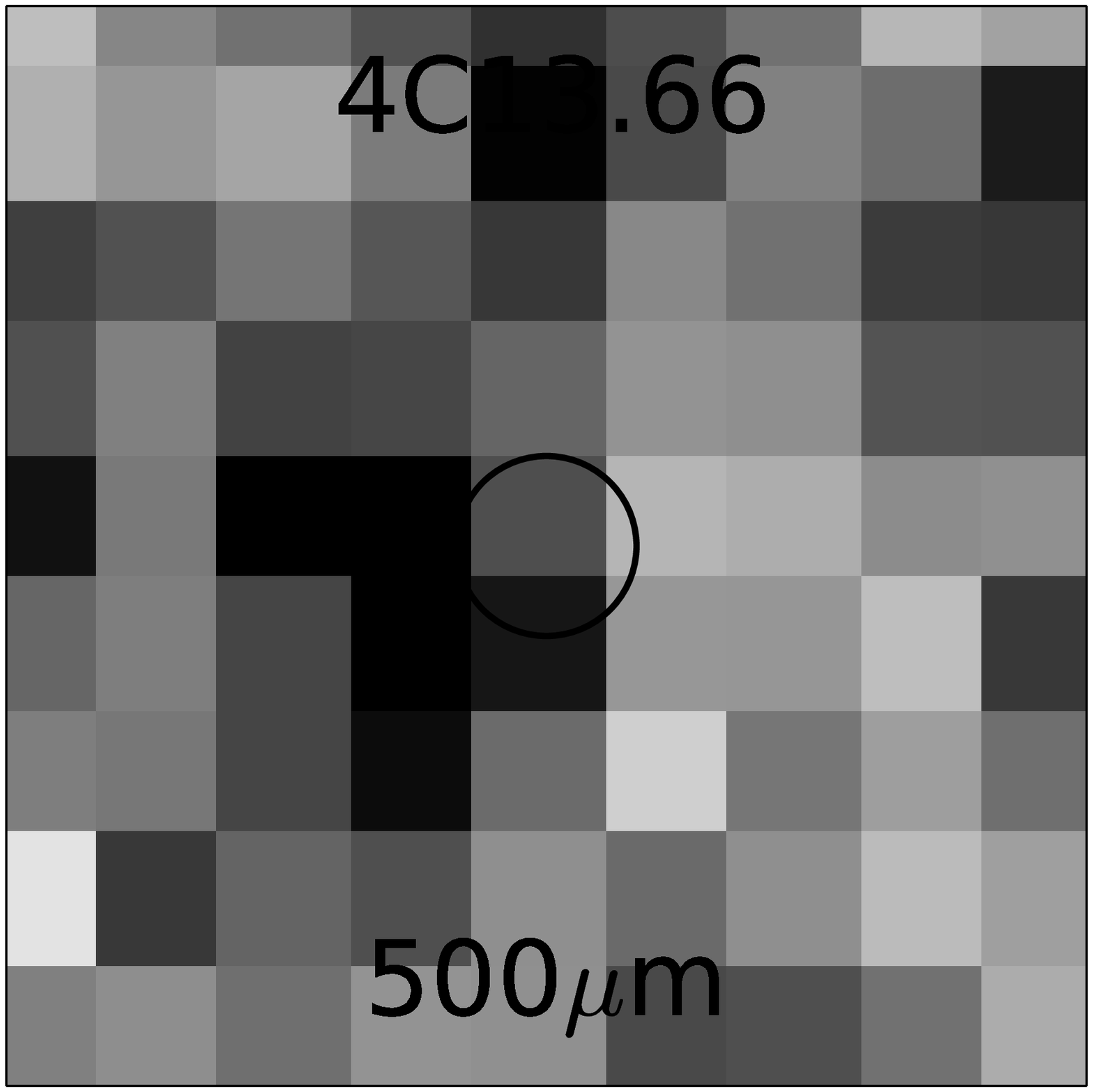}
      \\
      \includegraphics[width=1.5cm]{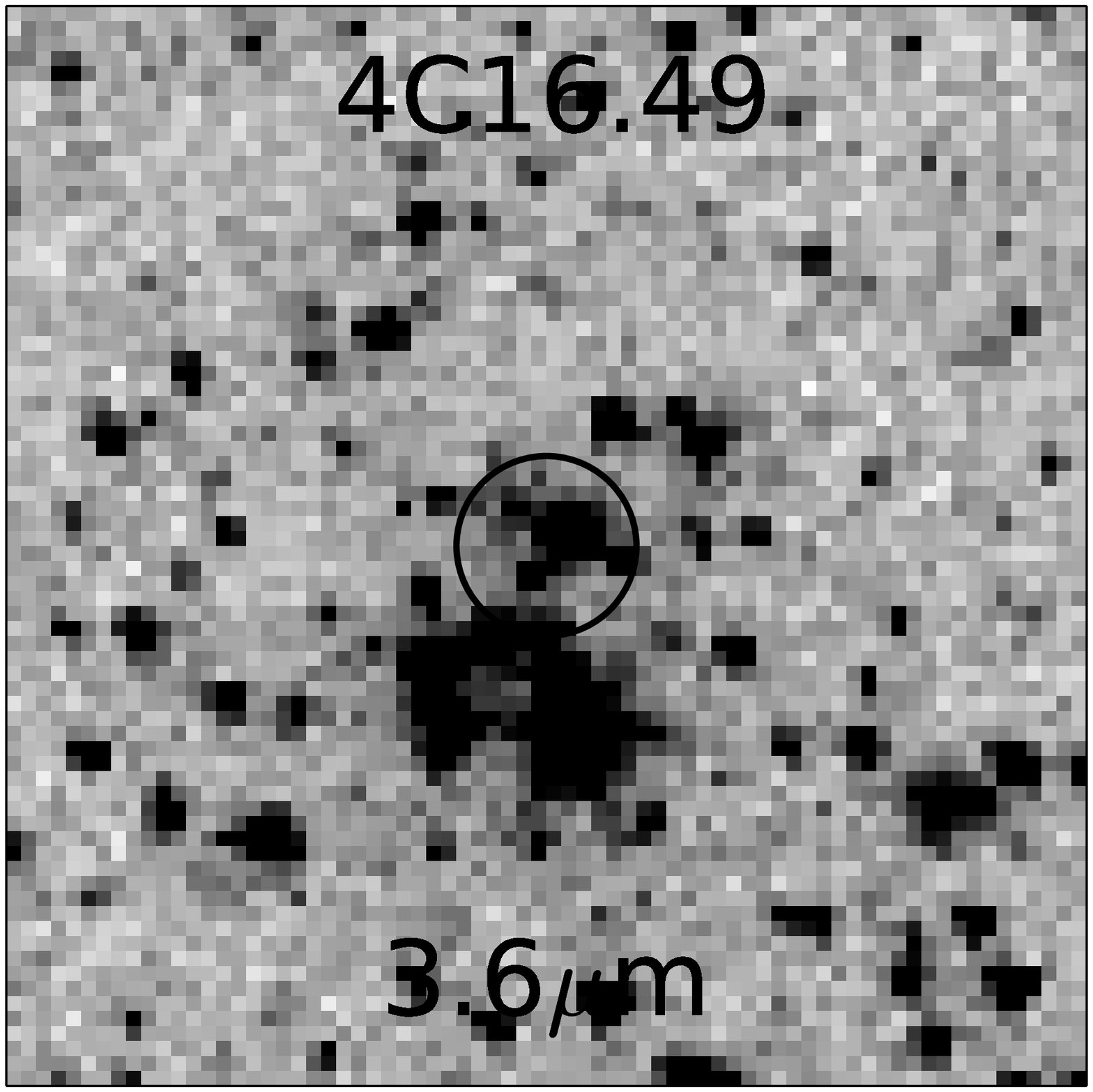}
      \includegraphics[width=1.5cm]{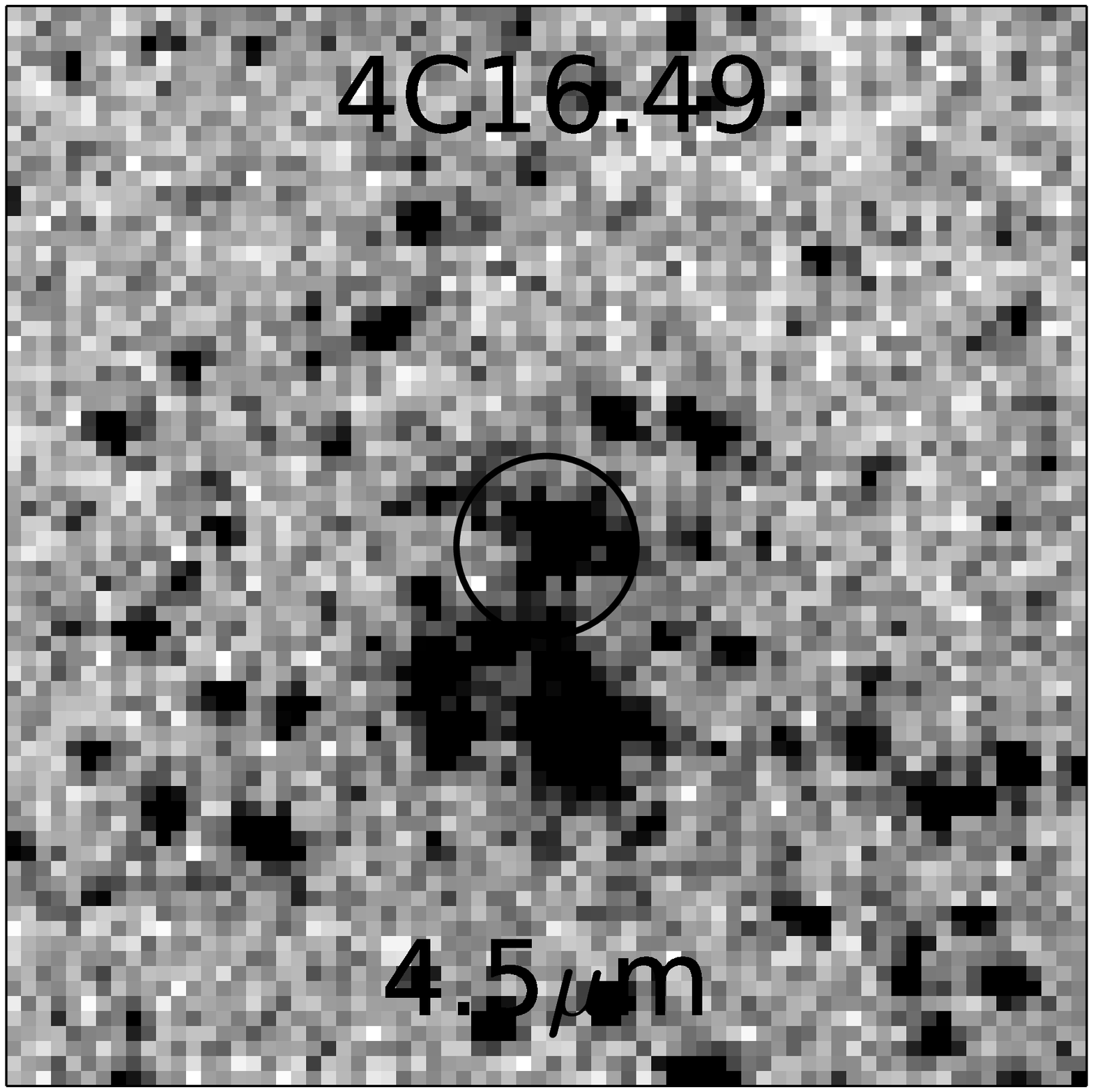}
      \includegraphics[width=1.5cm]{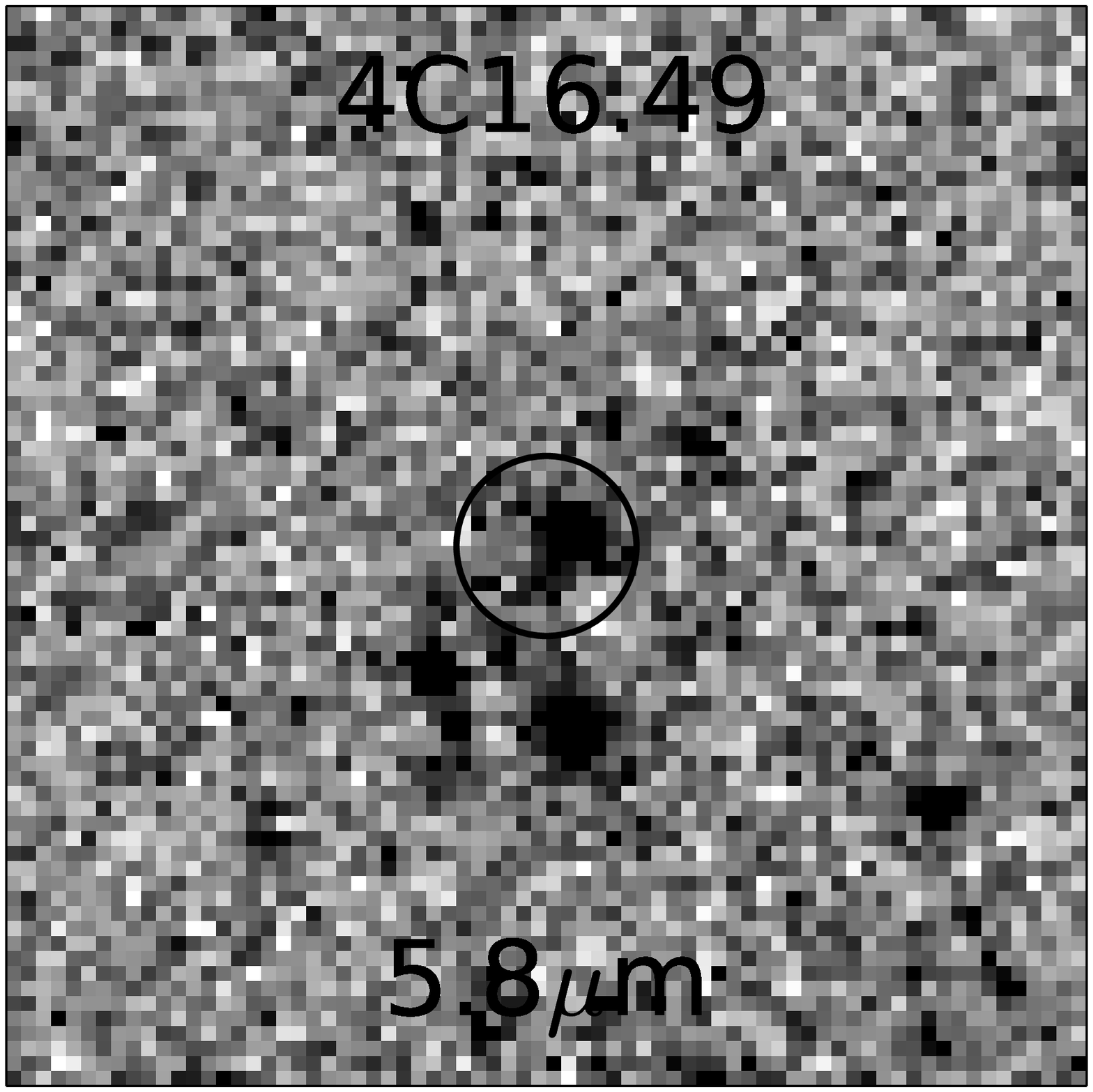}
      \includegraphics[width=1.5cm]{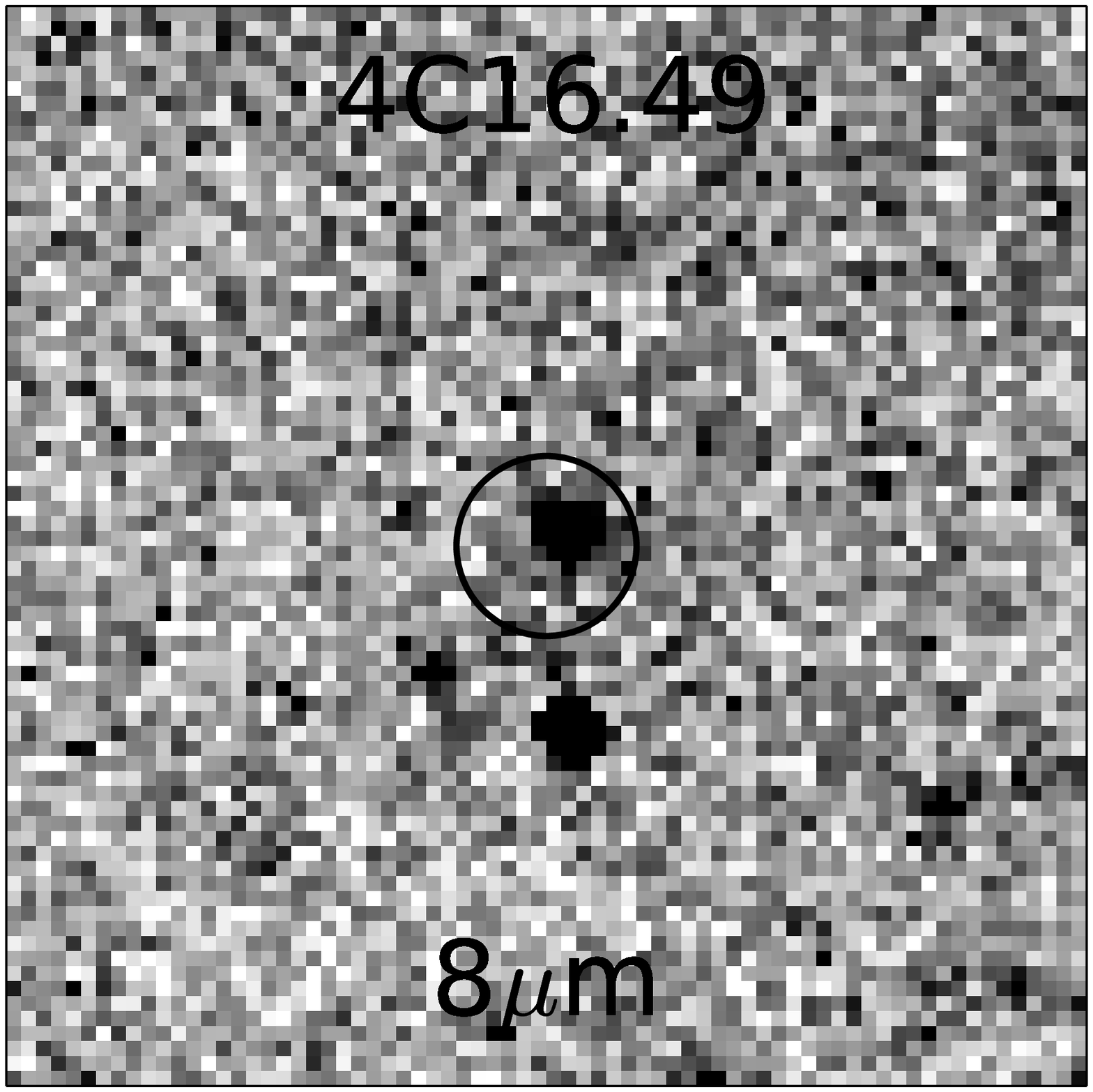}
      \includegraphics[width=1.5cm]{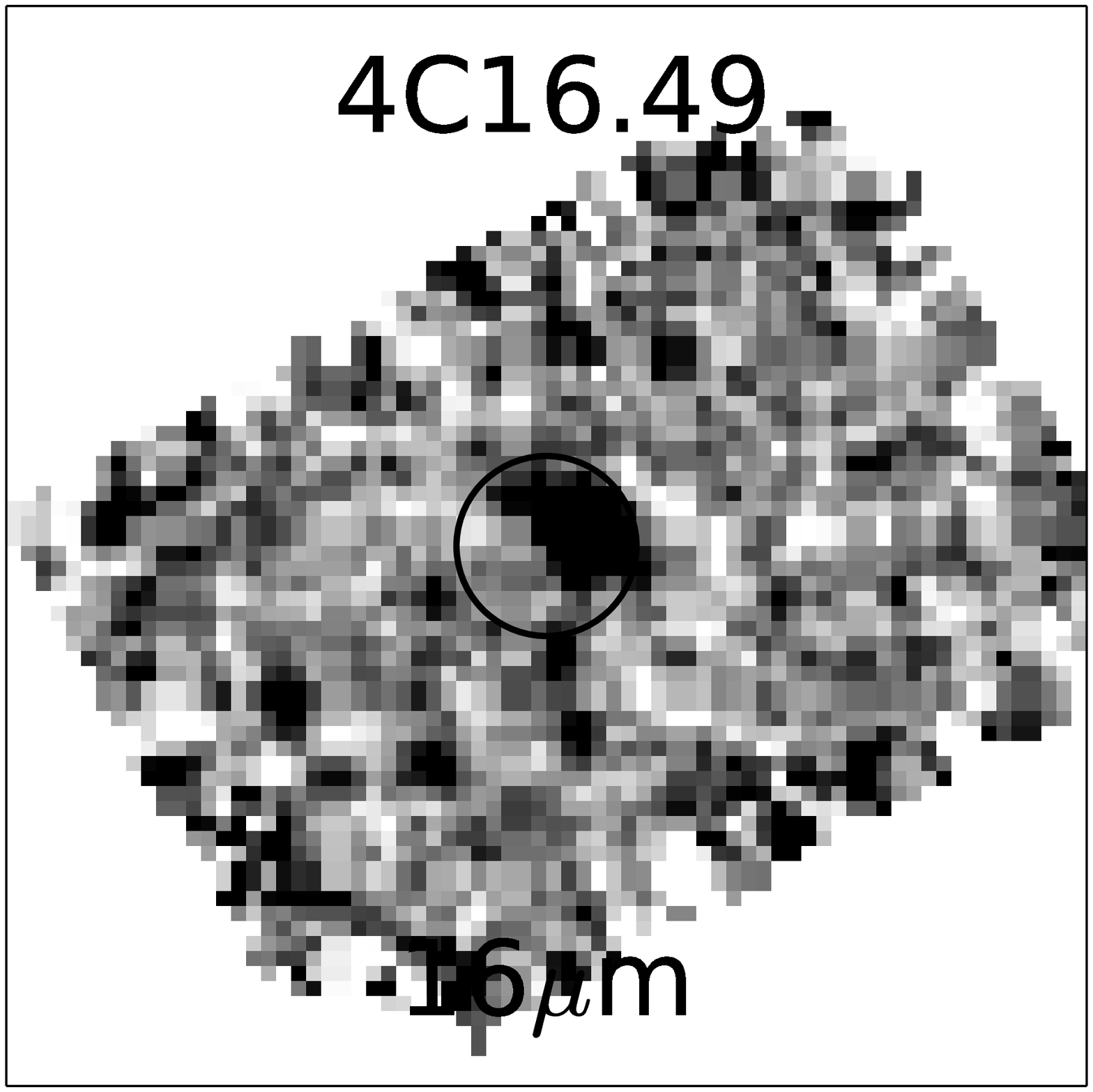}
      \includegraphics[width=1.5cm]{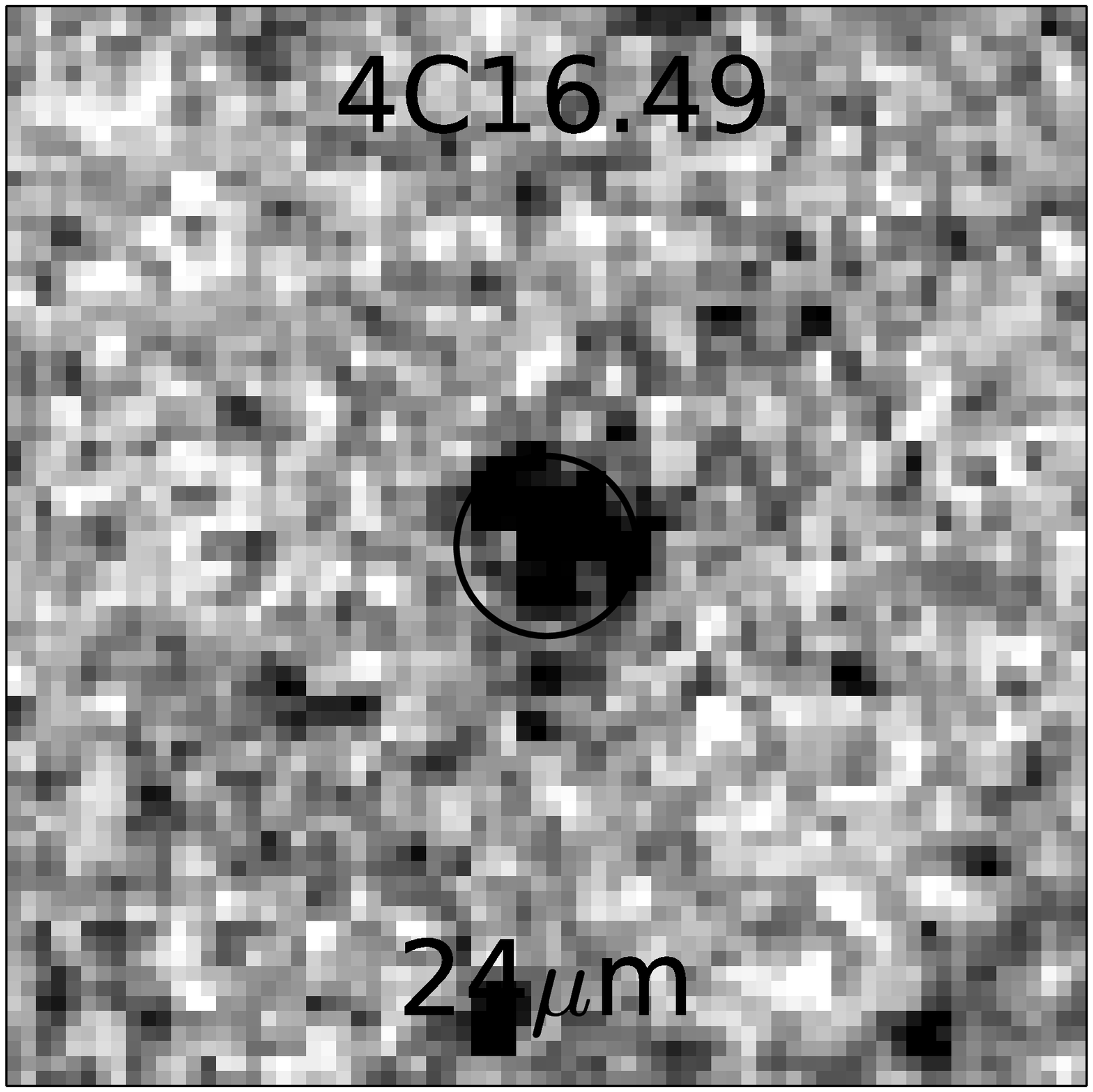}
      \includegraphics[width=1.5cm]{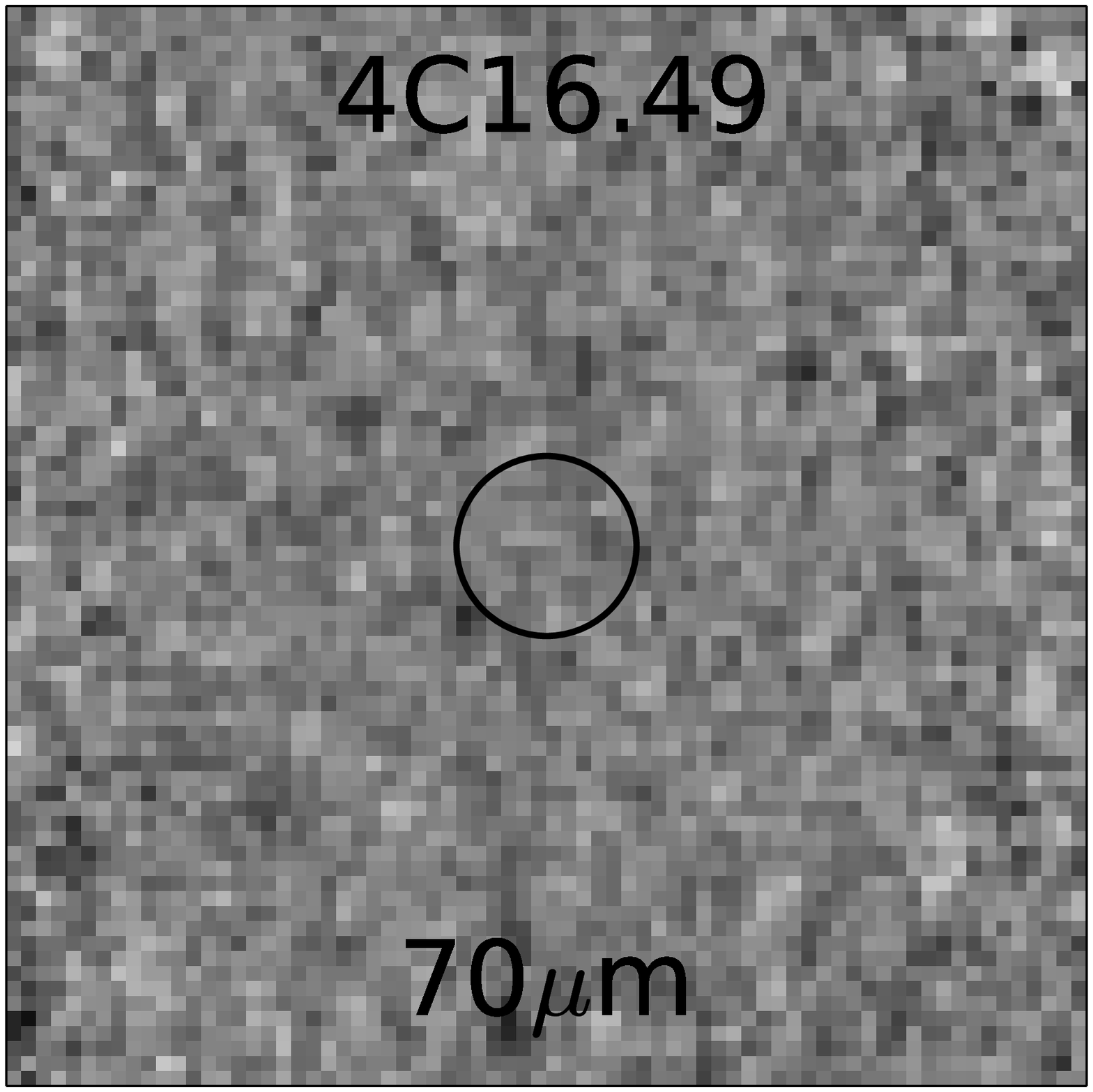}
      \includegraphics[width=1.5cm]{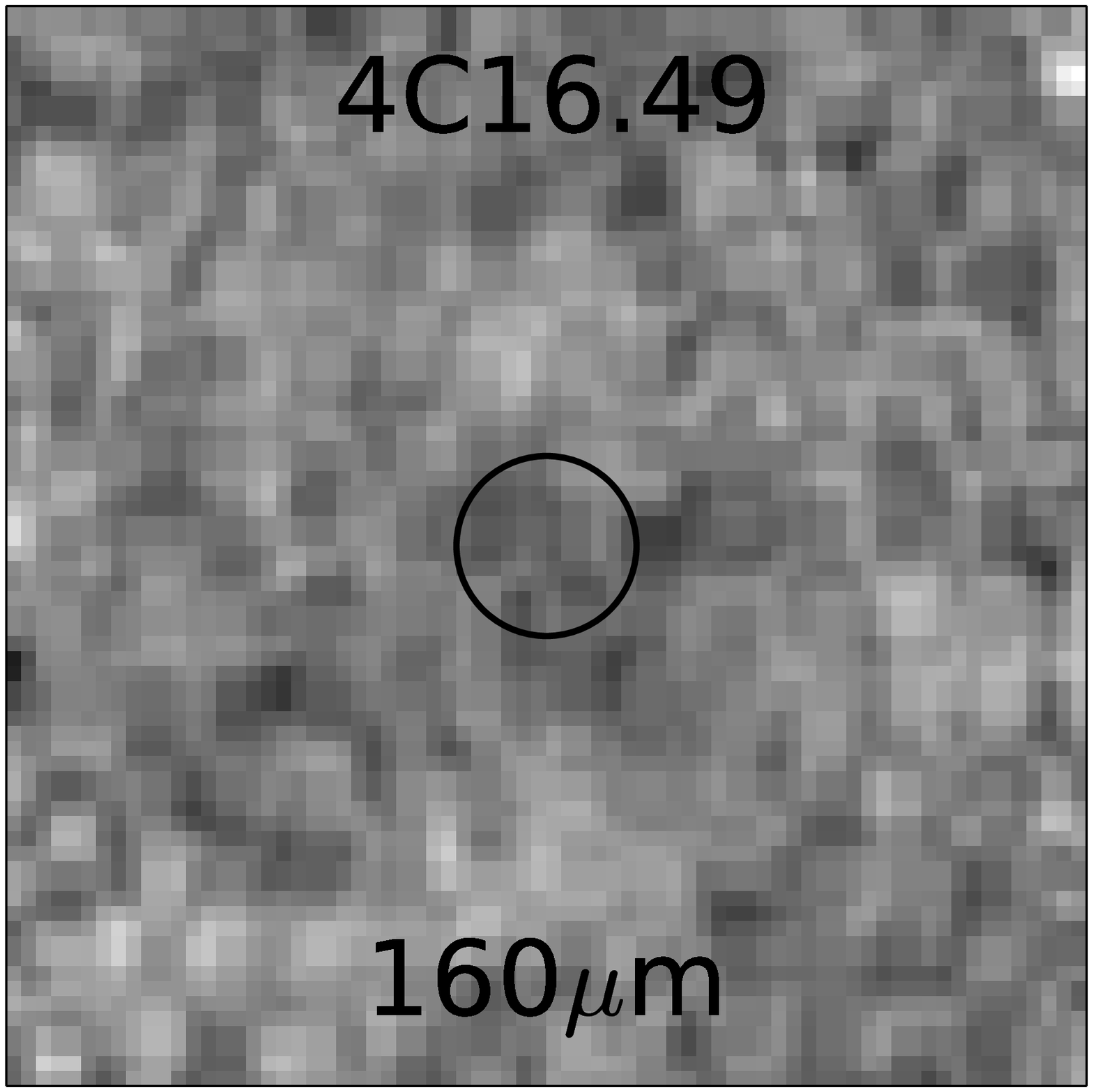}
      \includegraphics[width=1.5cm]{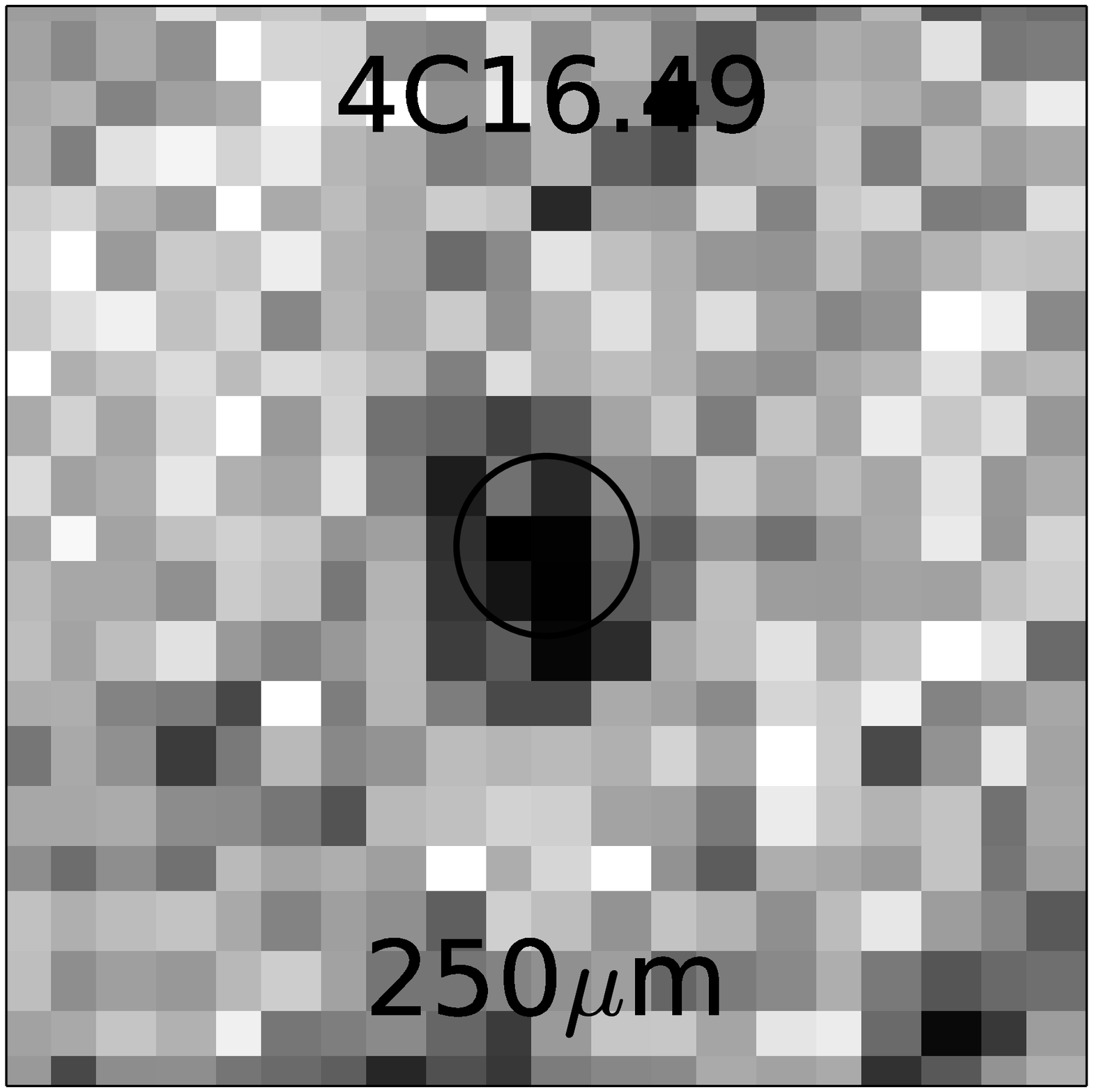}
      \includegraphics[width=1.5cm]{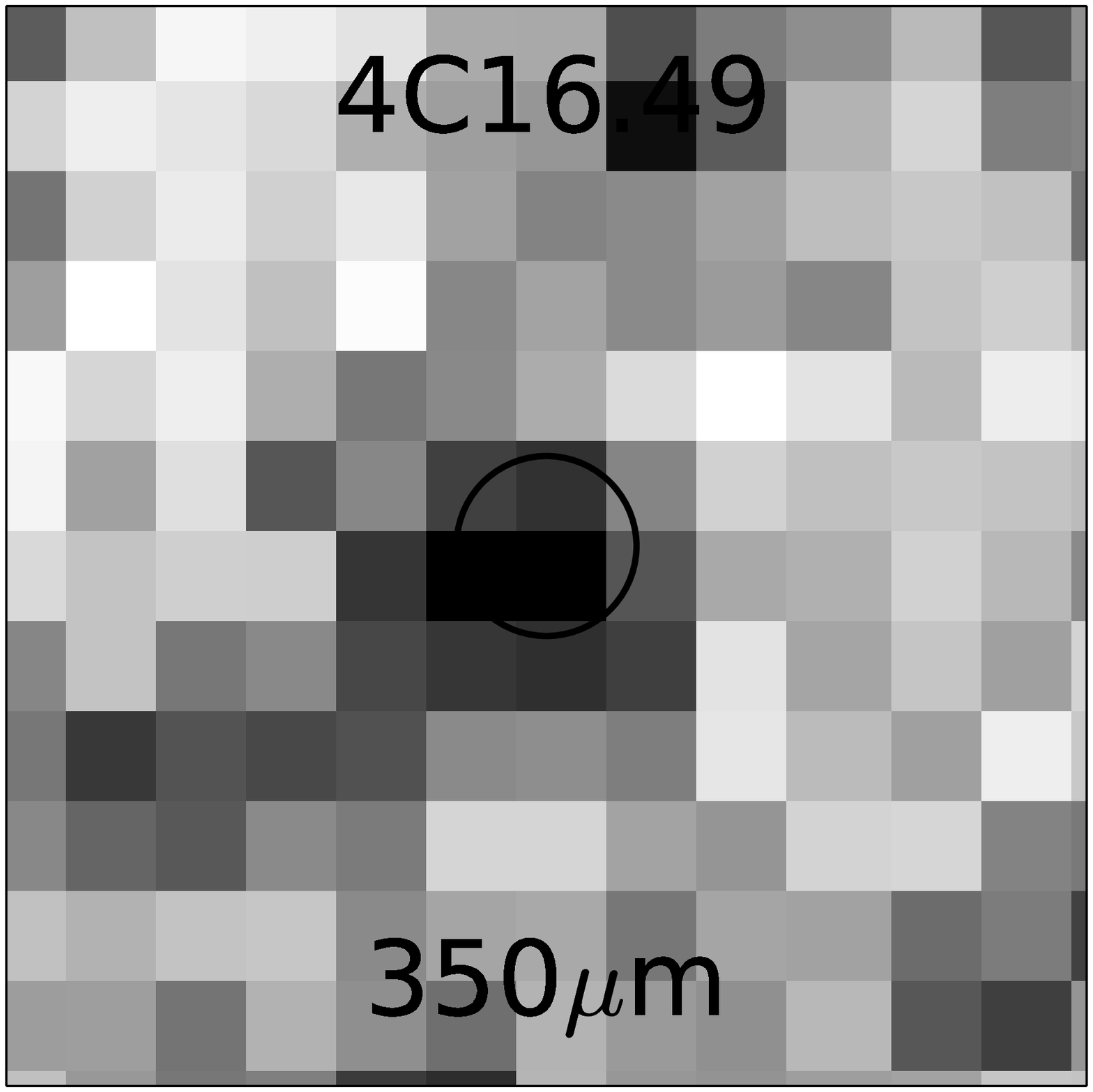}
      \includegraphics[width=1.5cm]{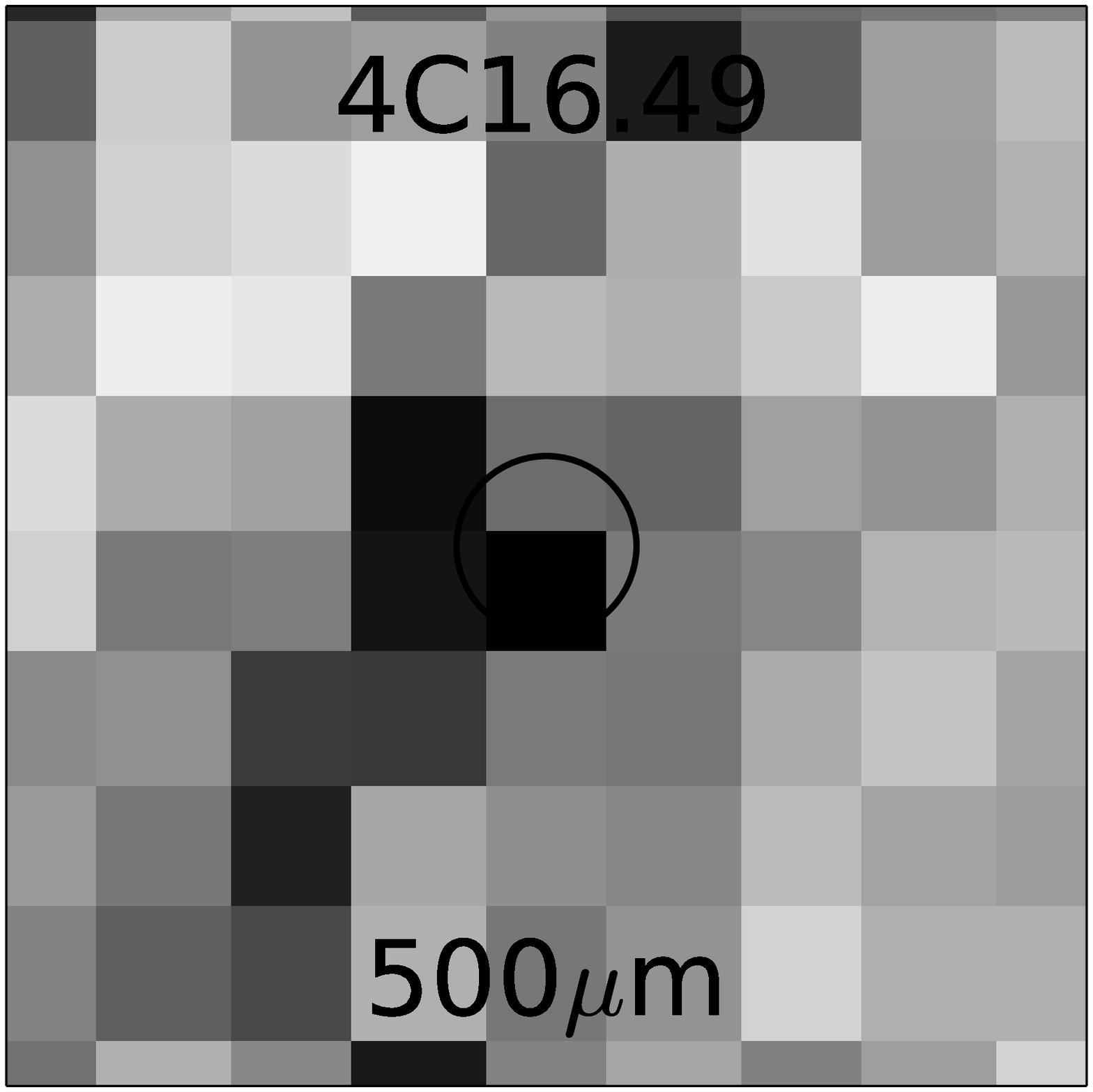}
      \caption{Continued.}
   \end{figure*}    
\end{document}